\pdfoutput=1

\documentclass[11pt,twoside,a4paper,cmspaper,final,collab]{cms-tdr}
\def\svnVersion{166414}\def\cmsCernNo{2012-086}\def\cmsCernDate{\today}\def\cmsMessage{Submitted to the Journal of High Energy Physics}

\begin{document}\cmsNoteHeader{BPH-11-020}

\hyphenation{had-ron-i-za-tion}
\hyphenation{cal-or-i-me-ter}
\hyphenation{de-vices}

\RCS$Revision: 111217 $
\RCS$HeadURL: svn+ssh://svn.cern.ch/reps/tdr2/papers/BPH-11-020/trunk/BPH-11-020.tex $
\RCS$Id: BPH-11-020.tex 111217 2012-03-18 17:26:22Z alverson $
\newlength\cmsFigWidth
\ifthenelse{\boolean{cms@external}}{\setlength\cmsFigWidth{0.85\columnwidth}}{\setlength\cmsFigWidth{0.4\textwidth}}
\ifthenelse{\boolean{cms@external}}{\providecommand{\cmsLeft}{top}}{\providecommand{\cmsLeft}{left}}
\ifthenelse{\boolean{cms@external}}{\providecommand{\cmsRight}{bottom}}{\providecommand{\cmsRight}{right}}
\newlength\fwidth
\setlength\fwidth{0.3\textwidth}
\cmsNoteHeader{BPH-11-020} % This is over-written in the CMS environment: useful as preprint no. for export versions

\def\vdef #1{\expandafter\def\csname #1\endcsname}
\def\vuse #1{\csname #1\endcsname}
\def\vu   #1{\csname #1\endcsname}

\def\hbabar{\mbox{{\huge\bf\sl B}\hspace{-0.1em}{\LARGE\bf\sl A}\hspace{-0.03em}{\huge\bf\sl B}\hspace{-0.1em}{\LARGE\bf\sl A\hspace{-0.03em}R}}}
\def\Lbabar{\mbox{{\LARGE\sl B}\hspace{-0.15em}{\Large\sl A}\hspace{-0.07em}{\LARGE\sl B}\hspace{-0.15em}{\Large\sl A\hspace{-0.02em}R}}}
\def\lbabar{\mbox{{\large\sl B}\hspace{-0.6em} {\normalsize\sl A}\hspace{-0.07em}{\large\sl B}\hspace{-0.6em} {\normalsize\sl A\hspace{-0.05em}R}}}
\def\sbabar{\mbox{{\small\sl B}\hspace{-0.6em} {\tiny\sl A}\hspace{-0.07em}{\small\sl B}\hspace{-0.6em} {\tiny\sl A\hspace{-0.05em}R}}}
\def\mbabar{\mbox{\sl B\hspace{-0.4em} {\small\sl A}\hspace{-0.37em} \sl B\hspace{-0.4em} {\small\sl A\hspace{-0.02em}R}}}
\def\babar{\mbox{\slshape B\kern-0.1em{\smaller A}\kern-0.1em
    B\kern-0.1em{\smaller A\kern-0.2em R}}}

\def\rmm   {\ensuremath{\Delta R(\mu\mu)}}

\def\fls      {\ensuremath{\ell_{3D}/\sigma(\ell_{3D})}}
\def\chidof   {\ensuremath{\chi^2/\mathrm{dof}}}  
\def\closetrk {\ensuremath{N_{\mathrm{trk}}^{\mathrm{close}}}}
\def\docatrk  {\ensuremath{d^{\mathrm{0}}_\mathrm{ca}}}
\def\dca      {\ensuremath{d_\mathrm{ca}}}
\def\ip       {\ensuremath{\delta_{3D}}}
\def\ips      {\ensuremath{\delta_{3D}/\sigma(\delta_{3D})}}

\def\mll   {\ensuremath{m_{\mu\mu}}}

\def\ket#1        {\ensuremath|{#1}\rangle}
\def\bra#1        {\ensuremath\langle{#1}|}
\def\braket#1#2   {\ensuremath\langle{#1}|{#2}\rangle}
\def\tfi#1#2#3    {\ensuremath\langle{#1}|{#2}|{#3}\rangle}
\def\vtwo#1#2     {\ensuremath\left(\begin{array}{c}{#1}\\{#2}\end{array}\right)}
\def\vthree#1#2#3 {\ensuremath\left(\begin{array}{c}{#1}\\{#2}\\{#3}\end{array}\right)}
\def\me           {\ensuremath\mathcal{M}}
\def\ame          {\ensuremath|\mathcal{M}|^2}
\def\asme         {\ensuremath\overline{|\mathcal{M}|^2}}

\def\psib         {\ensuremath\overline{\psi}}

\newcommand{\mat}[2][cccccccccccccccccccccccccc]{\left(
   \begin{array}{#1} 
    #2\\
   \end{array}
  \right)
}
 
\def\vvA        {\ensuremath\left(\begin{matrix} 0 \\ 0 \end{matrix}\right)}
\def\vvB        {\ensuremath\left(\begin{matrix} 1 \\ 0 \end{matrix}\right)}
\def\vvC        {\ensuremath\left(\begin{matrix} 0 \\ 1 \end{matrix}\right)}
\def\vvD        {\ensuremath\left(\begin{matrix} 1 \\ 1 \end{matrix}\right)}

\def\vvvA        {\ensuremath\left(\begin{matrix} 0 \\ 0 \\ 0\end{matrix}\right)}
\def\vvvB        {\ensuremath\left(\begin{matrix} 1 \\ 0 \\ 0\end{matrix}\right)}
\def\vvvC        {\ensuremath\left(\begin{matrix} 0 \\ 1 \\ 0\end{matrix}\right)}
\def\vvvD        {\ensuremath\left(\begin{matrix} 0 \\ 0 \\ 1\end{matrix}\right)}

\def\cpt   {\ensuremath{C\kern-0.2em P\kern-0.1em T}}
\def\cp    {\ensuremath{C\kern-0.2em P}}
\def\cpv   {\ensuremath{C\kern-0.2em P\kern-1.0em / }}
\def\CPV   {\cp-violation}
\def\CPTV  {\cpt-violation}
\def\bfsx  {$B$-physics}
\def\ETm   {\ensuremath{E_T\kern-1.2em/\kern0.6em}}
\def\ET    {\ensuremath{E_T}}
\def\kT    {\ensuremath{k_T}}
\def\ptm   {\ensuremath{p_\perp\kern-1.1em/\kern0.5em}}
\def\pvecm {\ensuremath{\vec{p} \kern-0.4em/\kern0.1em}}
\def\pvec  {\ensuremath{\vec{p}}}

\def\dsj   {\ensuremath{D_{sJ}}}
\def\vxb   {\ensuremath{|V_{xb}|}}
\def\vud   {\ensuremath{|V_{ud}|}}
\def\vus   {\ensuremath{|V_{us}|}}
\def\vub   {\ensuremath{|V_{ub}|}}
\def\vcd   {\ensuremath{|V_{cd}|}}
\def\vcs   {\ensuremath{|V_{cs}|}}
\def\vcb   {\ensuremath{|V_{cb}|}}
\def\vtd   {\ensuremath{|V_{td}|}}
\def\vts   {\ensuremath{|V_{ts}|}}
\def\vtb   {\ensuremath{|V_{tb}|}}

\def\deltam{\ensuremath{\delta m}}
\def\dm    {\ensuremath{\Delta m}}
\def\dt    {\ensuremath{\Delta t}}
\def\dg    {\ensuremath{\Delta \gamma}}
\def\dG    {\ensuremath{\Delta \Gamma}}
\def\dmt   {\ensuremath{\Delta mt}}
\def\dmdt  {\ensuremath{\Delta m \Delta t}}
\def\dms   {\ensuremath{\Delta m_s}}
\def\dmst {\ensuremath{\Delta m_s t}}
\def\dmm   {\ensuremath{\Delta m^2}}
\def\TBY   {\ensuremath{\theta_{\Bz, D^*\ell}}}

\def\de    {\ensuremath{\Delta E}}
\def\mes   {\ensuremath{m_{ES}}}

\def\msd{\ensuremath{\overline{m}_D^2}}
\def\lbar{\ensuremath{\overline{\Lambda}}}
\def\lone{\ensuremath{\lambda_1}}
\def\ltwo{\ensuremath{\lambda_2}}

\def\MUP   {\ensuremath{\mu_\pi^2}}
\def\MUG   {\ensuremath{\mu_G^2}}
\def\RHOD  {\ensuremath{\rho_D^3}}
\def\RHOLS {\ensuremath{\rho_{LS}^3}}

\def\cbf {\ensuremath{{\cal B}}}
\def\clu {\ensuremath{{\cal L}}}
\def\cor {\ensuremath{{\cal O}}}

\def\mmiss{\ensuremath{{m_{miss}^2}}}
\def\rusl{\ensuremath{{R_{u}}}}
\def\mh{\ensuremath{{m_{had}}}}
\def\mmxx {\ensuremath{\langle m_X^2 \rangle~}}
\def\mmx {\ensuremath{\langle m_X \rangle~}}
\def\mxqq{\ensuremath{(m_X, Q^2)}}
\def\mX{\ensuremath{{m_X}}}
\def\mx{\ensuremath{{m_X}}}
\def\mxcut{\ensuremath{{m_X^{cut}}}}
\def\pstar{\ensuremath{{p^*}}}
\def\qtot{\ensuremath{{Q_{tot}}}}
\def\pt{\ensuremath{{p_T}}}
\def\mt{\ensuremath{{m_\perp}}}

\def\pslash{\ensuremath{{p\kern-0.45em /}}}
\def\pvecslash{\ensuremath{{\vec{p} \kern-0.45em /}}}

\def\meanmxx   {\ensuremath{\langle m_X^2 \rangle}}
\def\meanmx    {\ensuremath{\langle m_{X} \rangle}}
\def\mean#1    {\ensuremath{\langle #1 \rangle}}

\def\Bpilnu    {\ensuremath{\Bb\to \pi\ell\nub}}
\def\Bmpilnu   {\ensuremath{\Bm\to \pi^0\ell^-\nub}}
\def\Betalnu   {\ensuremath{\Bb\to \eta\ell\nub}}
\def\Brholnu   {\ensuremath{\Bb\to \rho\ell\nub}}
\def\Bmrholnu  {\ensuremath{\Bb\to \rho^0\ell^-\nub}}
\def\Bomegalnu {\ensuremath{\Bb\to \omega\ell\nub}}
\def\Brhoenu   {\ensuremath{\Bb\to \rho e\nub}}
\def\Bzrhoenu  {\ensuremath{\Bz\to \rho^- e^+\nub}}

\def\Bdlnu     {\ensuremath{\Bb\to D\ell\nub}}
\def\Bdstarlnu {\ensuremath{\Bb\to \Dstar \ell \nub}}
\def\Bzdstarlnu {\ensuremath{\Bzb\to \Dstarp \ell^- \nub}}
\def\Bzdstarenu {\ensuremath{\Bzb\to \Dstarp e^- \nub}}

\def\bll     {\ensuremath{\Bz\to \ell^+\ell^-}}
\def\bee     {\ensuremath{\Bz\to e^+e^-}}
\def\bmm     {\ensuremath{\Bz\to \mu^+\mu^-}}
\def\bem     {\ensuremath{\Bz\to e^\pm\mu^\mp}}
\def\btt     {\ensuremath{\Bs\to \tau^+\tau^-}}
\def\bmt     {\ensuremath{\Bs\to \mu^\pm\tau^\mp}}
\def\bmn     {\ensuremath{B^+\to \mu^+\nu_\mu}}
\def\btn     {\ensuremath{B^+\to \tau^+\nu_\tau}}
\def\bln     {\ensuremath{B^+\to \ell^+\nu_\ell}}
\def\bgen    {\ensuremath{B^-\to \gamma e\nub}}
\def\bgee    {\ensuremath{B^-\to \gamma e^+e^-}}
\def\bgg     {\ensuremath{B^-\to \gamma \gamma}}

\def\jpsitomu {\ensuremath{\jpsi\to \mu^+\mu^-}}
\def\bdmm     {\ensuremath{B^0\to  \mu^+\mu^-}}
\def\bdpipi   {\ensuremath{B^0\to  \pip\pim}}
\def\bdpik    {\ensuremath{B^0\to  \Kp\pim}}
\def\bdkk    {\ensuremath{B^0\to  K^+K^-}}
\def\bdpimunu {\ensuremath{B^0\to  \pim\mup\nu}}
\def\bdmumupz {\ensuremath{B^0\to  \mup\mun\piz}}
\def\lbppi    {\ensuremath{\Lambda_b\to \pim p}}
\def\lbpk     {\ensuremath{\Lambda_b\to K^- p}}
\def\butrmunu {\ensuremath{B^+\to \mup\mun\mup \nu}}
\def\bctrmunu {\ensuremath{B_c\to \mup\mun\mup \nu}}
\def\bcpsimunu{\ensuremath{B_c\to \jpsi(\to\mup\mun)\mup \nu}}

\def\bodmm     {\ensuremath{B^0_d  \to \mu^+\mu^-}}
\def\bsmm     {\ensuremath{B^0_s\to \mu^+\mu^-}}
\def\bskk     {\ensuremath{B^0_s\to K^+K^-}}
\def\bspipi   {\ensuremath{B^0_s\to \pip\pim}}
\def\bspik    {\ensuremath{B^0_s\to K^-\pip}}
\def\bskmunu  {\ensuremath{B^0_s\to K^-\mup\nu}}
\def\bsmumug  {\ensuremath{B^0_s\to \mup\mun\gamma}}
\def\bsmmg    {\ensuremath{B^0_s\to \mu^+\mu^-\gamma}}
\def\bdll     {\ensuremath{B^0_d\to \ell^+\ell^-}}
\def\bsll     {\ensuremath{B^0_s\to \ell^+\ell^-}}
\def\bstt     {\ensuremath{B^0_s\to \tau^+\tau^-}}
\def\bdmt     {\ensuremath{B^0  \to \mu^\pm\tau^\mp}}
\def\bsmt     {\ensuremath{B^0_s\to \mu^\pm\tau^\mp}}
\def\bsdmm    {\ensuremath{B^0_{s (d)}\to \mu^+\mu^-}}
\def\bsdll    {\ensuremath{B^0_{s (d)}\to \ell^+\ell^-}}
\def\bsdtt    {\ensuremath{B^0_{s (d)}\to \tau^+\tau^-}}
\def\tmmm     {\ensuremath{\tau\to \mu\mu\mu}}

\def\hbb      {\ensuremath{H\to \bbbar}}
\def\htt      {\ensuremath{H\to \taup\taum}}
\def\ttH      {\ensuremath{\ttbar H}}

\def\meg      {\ensuremath{\mu\to e\gamma}}
\def\meee     {\ensuremath{\mu\to eee}}
\def\pgg      {\ensuremath{\piz\to \g\g}}

\def\bsg     {\ensuremath{b\to s\gamma}}
\def\bulnu   {\ensuremath{b\to u\ell\nub}}
\def\bclnu   {\ensuremath{b\to c\ell\nub}}
\def\bcenu   {\ensuremath{b\to c e\nub}}
\def\buenu   {\ensuremath{b\to u e\nub}}

\def\bxlnu   {\ensuremath{b\to X\ell^-\nub}}
\def\Bxenu   {\ensuremath{\Bb\to Xe^-\nu}}

\newcommand {\Bxlnu}{\ensuremath{\Bb \rightarrow X \ell \bar{\nu}}}
\newcommand {\Bxclnu}{\ensuremath{\Bb \rightarrow X_c \ell \bar{\nu}}}
\newcommand {\Bxulnu}{\ensuremath{\Bb \rightarrow X_u \ell \bar{\nu}}}
\def\Bpxenu {\ensuremath{\Bp\to Xe^+\nu}}
\def\Bzxenu {\ensuremath{\Bz\to Xe^+\nu}}

\def\Bulnu   {\ensuremath{\Bb\to X_{u}\ell\nub}}
\def\Buenu   {\ensuremath{\Bb\to X_{u} e\nub}}
\def\Bclnu   {\ensuremath{\Bb\to X_{c}\ell\nub}}
\def\Bxulnu  {\ensuremath{\Bb\to X_{u}\ell\nub}}
\def\Bxuenu  {\ensuremath{\Bb\to X_{u} e\nub}}
\def\Bxclnu  {\ensuremath{\Bb\to X_{c}\ell\nub}}

\def\bfactory  {{{\sl B}-factory}}
\def\bfactories{{{\sl B}-factories}}
\def\bFactory  {{{\sl B}-Factory}}
\def\breco     {\ensuremath{B_{reco}}}
\def\btag      {\ensuremath{B_{tag}}}
\def\bdecay    {{$B$-decay}}
\def\bDecay    {{$B$-Decay}}
\def\bdecays   {{$B$-decays}}
\def\bDecays   {{$B$-Decays}}
\def\bhadron   {{$b$-hadron}}
\def\bhadrons  {{$B$-hadrons}}
\def\bmeson    {{$B$-meson}}
\def\bmesons   {{$B$-mesons}}
\def\bquark    {{$b$-quark}}
\def\bquarks   {{$b$-quarks}}
\def\bphysics  {{$b$-physics}}
\def\Bphysics  {{$B$-physics}}

\def\ie   {{\it i.e.}}
\def\cf   {{\it cf.}}
\def\eg   {{\it e.g.}}
\def\etal {{\it et~al.}}
\def\etc  {{\it etc.}}

\def\rtr    {{$\red\triangleright\black$}}
\def\barrow {{$\blue\to\black$}}
\def\bpoint {{$\blue\bullet\black$}}
\def\npoint {{\phantom{$ \blue\bullet\black$}}}

\newcommand\bfac   {$B$-Factories}

\newcommand\bu   {\ensuremath{b\to u}}
\newcommand\bc   {\ensuremath{b\to c}}

\newcommand\islbcd {inclusive semileptonic $B\to c\ell\nu$}
\newcommand\islbud {inclusive semileptonic $B\to u\ell\nu$}

\def\tg     {\ensuremath {\theta^{*}_T}}
\def\ctg     {\ensuremath {\cos{\tg}}}
\def\cth     {\ensuremath {\cos{\theta_{H}}}}
\def\cthe    {\ensuremath {\cos{\theta_{H\,\eta'}}}}
\def\cthr    {\ensuremath {\cos{\theta_{H\,\rho}}}}
\def\ctb     {\ensuremath {\cos{\theta^{*}_{B}}}}
\def\ebeam     {\ensuremath {E^{*}_{b}}}
\def\egcms     {\ensuremath {E^{*}_{\gamma}}}
\def\mkpi      {\ensuremath {M_{\Kp \pim}}}

\let\emi\en
\def\electron   {\ensuremath{e}\xspace}
\def\en         {\ensuremath{e^-}\xspace}   % electron negative (\em is taken)
\def\ep         {\ensuremath{e^+}\xspace}
\def\epm        {\ensuremath{e^\pm}\xspace} 
\def\epem       {\ensuremath{e^+e^-}\xspace}
\def\ee         {\ensuremath{e^-e^-}\xspace}

\def\mmu        {\ensuremath{\mu}\xspace}
\def\mup        {\ensuremath{\mu^+}\xspace}
\def\mun        {\ensuremath{\mu^-}\xspace} % muon negative (\mum is taken)
\def\mumu       {\ensuremath{\mu^+\mu^-}\xspace}
\def\mtau       {\ensuremath{\tau}\xspace}

\def\taup       {\ensuremath{\tau^+}\xspace}
\def\taum       {\ensuremath{\tau^-}\xspace}
\def\tautau     {\ensuremath{\tau^+\tau^-}\xspace}

\def\ellm       {\ensuremath{\ell^-}\xspace}
\def\ellp       {\ensuremath{\ell^+}\xspace}
\def\ellell     {\ensuremath{\ell^+ \ell^-}\xspace}

\def\ellb        {\ensuremath{\bar{\ell}}\xspace}
\def\nub        {\ensuremath{\bar{\nu}}\xspace}
\def\nunub      {\ensuremath{\nu{\bar{\nu}}}\xspace}
\def\nunub      {\ensuremath{\nu{\bar{\nu}}}\xspace}
\def\nue        {\ensuremath{\nu_e}\xspace}
\def\nueb       {\ensuremath{\nub_e}\xspace}
\def\nuenueb    {\ensuremath{\nue\nueb}\xspace}
\def\num        {\ensuremath{\nu_\mu}\xspace}
\def\numb       {\ensuremath{\nub_\mu}\xspace}
\def\numnumb    {\ensuremath{\num\numb}\xspace}
\def\nut        {\ensuremath{\nu_\tau}\xspace}
\def\nutb       {\ensuremath{\nub_\tau}\xspace}
\def\nutnutb    {\ensuremath{\nut\nutb}\xspace}
\def\nul        {\ensuremath{\nu_\ell}\xspace}
\def\nulb       {\ensuremath{\nub_\ell}\xspace}
\def\nulnulb    {\ensuremath{\nul\nulb}\xspace}

\def\g     {\ensuremath{\gamma}\xspace}
\def\gaga  {\ensuremath{\gamma\gamma}\xspace}  %% changed from \gg, which is >>
\def\ggstar{\ensuremath{\gamma\gamma^*}\xspace}

\def\ega    {\ensuremath{e\gamma}\xspace}
\def\game   {\ensuremath{\gamma e^-}\xspace}
\def\epemg  {\ensuremath{e^+e^-\gamma}\xspace}

\def\H      {\ensuremath{H^0}\xspace}
\def\Hp     {\ensuremath{H^+}\xspace}
\def\Hm     {\ensuremath{H^-}\xspace}
\def\Hpm    {\ensuremath{H^\pm}\xspace}
\def\W      {\ensuremath{W}\xspace}
\def\Wp     {\ensuremath{W^+}\xspace}
\def\Wm     {\ensuremath{W^-}\xspace}
\def\Wpm    {\ensuremath{W^\pm}\xspace}
\def\Z      {\ensuremath{Z^0}\xspace}

\def\K     {\ensuremath{K}\xspace}
\def\p     {\ensuremath{p}\xspace}
\def\q     {\ensuremath{q}\xspace}
\def\qbar  {\ensuremath{\overline q}\xspace}
\def\Qbar  {\ensuremath{\overline Q}\xspace}
\def\ffbar {\ensuremath{f\overline f}\xspace}
\def\qqbar {\ensuremath{q\overline q}\xspace}
\def\QQbar {\ensuremath{Q\overline Q}\xspace}
\def\u     {\ensuremath{u}\xspace}
\def\ubar  {\ensuremath{\overline u}\xspace}
\def\uubar {\ensuremath{u\overline u}\xspace}
\def\d     {\ensuremath{d}\xspace}
\def\dbar  {\ensuremath{\overline d}\xspace}
\def\ddbar {\ensuremath{d\overline d}\xspace}
\def\s     {\ensuremath{s}\xspace}
\def\sbar  {\ensuremath{\overline s}\xspace}
\def\ssbar {\ensuremath{s\overline s}\xspace}
\def\c     {\ensuremath{c}\xspace}
\def\cbar  {\ensuremath{\overline c}\xspace}
\def\ccbar {\ensuremath{c\overline c}\xspace}
\def\b     {\ensuremath{b}\xspace}
\def\bbar  {\ensuremath{\overline b}\xspace}
\def\bbbar {\ensuremath{b\overline b}\xspace}
\def\t     {\ensuremath{t}\xspace}
\def\tbar  {\ensuremath{\overline t}\xspace}
\def\tbar  {\ensuremath{\overline t}\xspace}
\def\ttbar {\ensuremath{t\overline t}\xspace}
\def\pbar  {\ensuremath{\overline p}\xspace}
\def\ppbar {\ensuremath{p\overline p}\xspace}

\def\piz   {\ensuremath{\pi^0}\xspace}
\def\pizs  {\ensuremath{\pi^0\mbox\,\rm{s}}\xspace}
\def\ppz   {\ensuremath{\pi^0\pi^0}\xspace}
\def\pip   {\ensuremath{\pi^+}\xspace}
\def\pim   {\ensuremath{\pi^-}\xspace}
\def\pipi  {\ensuremath{\pi^+\pi^-}\xspace}
\def\pipm  {\ensuremath{\pi^\pm}\xspace}
\def\pimp  {\ensuremath{\pi^\mp}\xspace}

\def\kaon  {\ensuremath{K}\xspace}
\def\Kbar  {\kern 0.2em\bar{\kern -0.2em K}{}\xspace}
\def\Kb    {\ensuremath{\Kbar}\xspace}
\def\Kz    {\ensuremath{K^0}\xspace}
\def\Kzb   {\ensuremath{\Kbar^0}\xspace}
\def\KzKzb {\ensuremath{\Kz \kern -0.16em \Kzb}\xspace}
\def\Kp    {\ensuremath{K^+}\xspace}
\def\Km    {\ensuremath{K^-}\xspace}
\def\Kpm   {\ensuremath{K^\pm}\xspace}
\def\Kmp   {\ensuremath{K^\mp}\xspace}
\def\KpKm  {\ensuremath{\Kp \kern -0.16em \Km}\xspace}
\def\KS    {\ensuremath{K^0_{\scriptscriptstyle S}}\xspace} 
\def\KL    {\ensuremath{K^0_{\scriptscriptstyle L}}\xspace} 
\def\Kstarz  {\ensuremath{K^{*0}}\xspace}
\def\Kstarzb {\ensuremath{\Kbar^{*0}}\xspace}
\def\Kstar   {\ensuremath{K^*}\xspace}
\def\Kstarb  {\ensuremath{\Kbar^*}\xspace}
\def\Kstarp  {\ensuremath{K^{*+}}\xspace}
\def\Kstarm  {\ensuremath{K^{*-}}\xspace}
\def\Kstarpm {\ensuremath{K^{*\pm}}\xspace}
\def\Kstarmp {\ensuremath{K^{*\mp}}\xspace}

\newcommand{\etapr}{\ensuremath{\eta^{\prime}}\xspace}

\def\Dbar    {\kern 0.2em\bar{\kern -0.2em D}{}\xspace}
\def\Db      {\ensuremath{\Dbar}\xspace}
\def\Dz      {\ensuremath{D^0}\xspace}
\def\Dzb     {\ensuremath{\Dbar^0}\xspace}
\def\DzDzb   {\ensuremath{\Dz {\kern -0.16em \Dzb}}\xspace}
\def\Dp      {\ensuremath{D^+}\xspace}
\def\Dm      {\ensuremath{D^-}\xspace}
\def\Dpm     {\ensuremath{D^\pm}\xspace}
\def\Dmp     {\ensuremath{D^\mp}\xspace}
\def\DpDm    {\ensuremath{\Dp {\kern -0.16em \Dm}}\xspace}
\def\Dstar   {\ensuremath{D^*}\xspace}
\def\Dstarb  {\ensuremath{\Dbar^*}\xspace}
\def\Dstarz  {\ensuremath{D^{*0}}\xspace}
\def\Dstarzb {\ensuremath{\Dbar^{*0}}\xspace}
\def\Dstarp  {\ensuremath{D^{*+}}\xspace}
\def\Dstarm  {\ensuremath{D^{*-}}\xspace}
\def\Dstarpm {\ensuremath{D^{*\pm}}\xspace}
\def\Dstarmp {\ensuremath{D^{*\mp}}\xspace}
\def\Ds      {\ensuremath{D^+_s}\xspace}
\def\Dsb     {\ensuremath{\Dbar^+_s}\xspace}
\def\Dss     {\ensuremath{D^{*+}_s}\xspace}

\newcommand{\dstr}{\ensuremath{\Dstar}\xspace}
\newcommand{\dstrstr}{\ensuremath{D^{**}}\xspace}
\newcommand{\dsp}{\ensuremath{\Dstarp}\xspace}
\newcommand{\dsm}{\ensuremath{\Dstarm}\xspace}
\newcommand{\dsz}{\ensuremath{\Dstarz}\xspace}

\def\B       {\ensuremath{B}\xspace}
\def\Bbar    {\kern 0.18em\bar{\kern -0.18em B}{}\xspace}
\def\Bb      {\ensuremath{\Bbar}\xspace}
\def\BB      {\ensuremath{B\Bbar}\xspace} 
\def\Bz      {\ensuremath{B^0}\xspace}
\def\Bzb     {\ensuremath{\Bbar^0}\xspace}
\def\BzBzb   {\ensuremath{\Bz {\kern -0.16em \Bzb}}\xspace}
\def\BsBsb   {\ensuremath{\Bs {\kern -0.16em \Bsb}}\xspace}
\def\Bu      {\ensuremath{B^+}\xspace}
\def\Bub     {\ensuremath{B^-}\xspace}
\def\Bp      {\ensuremath{\Bu}\xspace}
\def\Bm      {\ensuremath{\Bub}\xspace}
\def\Bpm     {\ensuremath{B^\pm}\xspace}
\def\Bmp     {\ensuremath{B^\mp}\xspace}
\def\BpBm    {\ensuremath{\Bu {\kern -0.16em \Bub}}\xspace}
\def\Bd      {\ensuremath{B^0_d}\xspace}
\def\Bs      {\ensuremath{B^0_s}\xspace}
\def\Bc      {\ensuremath{B^+_c}\xspace}
\def\Bsb     {\ensuremath{\Bzb_s}\xspace}
\def\Nz      {\ensuremath{M^0}\xspace}
\def\Nbar    {\kern 0.18em\bar{\kern -0.18em M}{}}
\def\Nzb     {\ensuremath{\Nbar^0}}
\def\NzNzb   {\ensuremath{\Nz {\kern -0.16em \Nzb}}}
\def\Nh      {\ensuremath{M_H}\xspace}
\def\Nl      {\ensuremath{M_L}\xspace}
\def\Nphys   {\ensuremath{\Nz_{phys}(t)}\xspace}
\def\Nbphys  {\ensuremath{\Nzb_{phys}(t)}\xspace}
\def\gh      {\ensuremath{\gamma_H}\xspace}
\def\gl      {\ensuremath{\gamma_L}\xspace}

\def\Bzd     {\ensuremath{B_d^0}\xspace}
\def\Bzs     {\ensuremath{B_s^0}\xspace}
\def\Bsd     {\ensuremath{B_{s(d)}}\xspace}

\def\jpsi     {\ensuremath{{J\mskip -3mu/\mskip -2mu\psi\mskip 2mu}}\xspace}
\def\psitwos  {\ensuremath{\psi{(2S)}}\xspace}
\def\psiprpr  {\ensuremath{\psi(3770)}\xspace}
\def\etac     {\ensuremath{\eta_c}\xspace}
\def\chiczero {\ensuremath{\chi_{c0}}\xspace}
\def\chicone  {\ensuremath{\chi_{c1}}\xspace}
\def\chictwo  {\ensuremath{\chi_{c2}}\xspace}
\mathchardef\Upsilon="7107
\def\Y#1S{\ensuremath{\Upsilon{(#1S)}}\xspace}% no space before {...}!
\def\OneS  {\Y1S}
\def\TwoS  {\Y2S}
\def\ThreeS{\Y3S}
\def\FourS {\Y4S}
\def\FiveS {\Y5S}

\def\chic#1{\ensuremath{\chi_{c#1}}\xspace} % dbm

\def\proton      {\ensuremath{p}\xspace}
\def\antiproton  {\ensuremath{\overline p}\xspace}
\def\neutron     {\ensuremath{n}\xspace}
\def\antineutron {\ensuremath{\overline n}\xspace}

\mathchardef\Deltares="7101
\mathchardef\Xi="7104
\mathchardef\Lambda="7103
\mathchardef\Sigma="7106
\mathchardef\Omega="710A

\def\Deltabar{\kern 0.25em\overline{\kern -0.25em \Deltares}{}\xspace}
\def\Lbar{\kern 0.2em\overline{\kern -0.2em\Lambda\kern 0.05em}\kern-0.05em{}\xspace}
\def\Sigbar{\kern 0.2em\overline{\kern -0.2em \Sigma}{}\xspace}
\def\Xibar{\kern 0.2em\overline{\kern -0.2em \Xi}{}\xspace}
\def\Obar{\kern 0.2em\overline{\kern -0.2em \Omega}{}\xspace}
\def\Xb{\kern 0.2em\overline{\kern -0.2em X}{}\xspace}

\def\X {\ensuremath{X}\xspace}

\def\BR         {{\ensuremath{\cal B}\xspace}}
\def\BRtauptoe  {\ensuremath{\BR(\taup \to \ep)}\xspace}
\def\BRtaumtoe  {\ensuremath{\BR(\taum \to \en)}\xspace}
\def\BRtauptomu {\ensuremath{\BR(\taup \to \mup)}\xspace}
\def\BRtaumtomu {\ensuremath{\BR(\taum \to \mun)}\xspace}

\newcommand{\etaprepp}{\ensuremath{\etapr \to \eta \pipi}\xspace}
\newcommand{\etaprrg} {\ensuremath{\etapr \to \rho^0 \g}\xspace}

\def\bdpsikstar {\ensuremath{\Bd \to \jpsi \Kstarz}\xspace}
\def\bspsiphi   {\ensuremath{\Bs \to \jpsi \phi}\xspace}
\def\lblmumu    {\ensuremath{\Lambda_b \to \mup\mun \Lambda}\xspace}
\def\bsphimm    {\ensuremath{\Bs \to \phi\mup\mun}\xspace}
\def\bdkmm      {\ensuremath{\Bz \to K\mup\mun}\xspace}
\def\bpsiks     {\ensuremath{\Bz \to \jpsi \KS}\xspace}
\def\bpsikst    {\ensuremath{\Bz \to \jpsi \Kstar}\xspace}
\def\bpsikl     {\ensuremath{\Bz \to \jpsi \KL}\xspace}
\def\bpsikzeropi{\ensuremath{\Bz \to \jpsi \Kstarz (\to \KL \piz)}\xspace}
\def\bpsikpluspi{\ensuremath{\Bu \to \jpsi \Kstarp (\to \KL \pip)}\xspace}
\def\bpsikpi    {\ensuremath{\Bz/\Bzb \to \jpsi (\to \mumu) \Kpm \pimp}\xspace}
\def\bspsiphi   {\ensuremath{\Bs \to \jpsi \phi}\xspace}
\def\bupsik     {\ensuremath{\Bp \to \jpsi \Kp}\xspace}
\def\bupsimmk   {\ensuremath{\Bpm \to \jpsi (\to \mumu) \Kpm}\xspace}
\def\bpsiX      {\ensuremath{\Bz \to \jpsi \X}\xspace}

\def\Bzbtomu    {\ensuremath{\Bzb \to \mu \X}\xspace}
\def\Bzbtox     {\ensuremath{\Bzb \to \X}\xspace}
\def\Bztopipi   {\ensuremath{\Bz \to \pipi}\xspace}
\def\Bztokpi    {\ensuremath{\Bz \to \Kpm \pimp}\xspace}
\def\Bztorhopi  {\ensuremath{\Bz \to \rho^+ \pim}\xspace}
\def\Bztorhorho {\ensuremath{\Bz \to \rho \rho}\xspace}
\def\Bztokrho   {\ensuremath{\Bz \to K \rho}\xspace}
\def\Bztokstpi  {\ensuremath{\Bz \to \Kstar \pi}\xspace}
\def\Bztoapi    {\ensuremath{\Bz \to a_1 \pi}\xspace}
\def\Bztodd     {\ensuremath{\Bz \to \DpDm}\xspace}
\def\Bztodstd   {\ensuremath{\Bz \to \Dstarp \Dm}\xspace}
\def\Bztodstdst {\ensuremath{\Bz \to \Dstarp \Dstarm}\xspace}

\def\BtoDK      {\ensuremath{B \to DK}\xspace}
\def\Btodstlnu  {\ensuremath{B \to \Dstar \ell \nu}\xspace}
\def\Btodstdlnu {\ensuremath{B \to \Dstar(D) \ell \nu}\xspace}
\def\Btorholnu  {\ensuremath{B \to \rho \ell \nu}\xspace}
\def\Btopilnu   {\ensuremath{B \to \pi \ell \nu}\xspace}

\def\Btoetah    {\ensuremath{B \to \eta h}\xspace}
\def\Btoetaph   {\ensuremath{B \to \etapr h}\xspace}

\newcommand{\Betaprks}{\ensuremath{\Bz \to \etapr \KS}\xspace}
\newcommand{\Betaprkz}{\ensuremath{\Bz \to \etapr \Kz}\xspace}

\def\btosgam    {\ensuremath{b \to s \g}\xspace}
\def\btodgam    {\ensuremath{b \to d \g}\xspace}
\def\btosll     {\ensuremath{b \to s \ellell}\xspace}
\def\btosmm     {\ensuremath{b \to s \mup\mun}\xspace}
\def\btosnunu   {\ensuremath{b \to s \nunub}\xspace}
\def\btosgaga   {\ensuremath{b \to s \gaga}\xspace}
\def\btosglue   {\ensuremath{b \to s g}\xspace}

\def\bbmumuX    {\ensuremath{\bbbar\to \mu^+\mu^-+X}\xspace}
\def\ccmumuX    {\ensuremath{\ccbar\to \mu^+\mu^-+X}\xspace}
\def\bpsimmX    {\ensuremath{b\to \jpsi(\to\mup\mun) X}\xspace}

\def\upsbb   {\ensuremath{\FourS \to \BB}\xspace}
\def\upsbzbz {\ensuremath{\FourS \to \BzBzb}\xspace}
\def\upsbpbm {\ensuremath{\FourS \to \BpBm}\xspace}
\def\upspsikl{\ensuremath{\FourS \to (\bpsikl) (\Bzbtox)}\xspace}

\def\tauptoe    {\ensuremath{\taup \to \ep \nunub}\xspace}
\def\taumtoe    {\ensuremath{\taum \to \en \nunub}\xspace}
\def\tauptomu   {\ensuremath{\taup \to \mup \nunub}\xspace}
\def\taumtomu   {\ensuremath{\taum \to \mun \nunub}\xspace}
\def\tauptopi   {\ensuremath{\taup \to \pip \nub}\xspace}
\def\taumtopi   {\ensuremath{\taum \to \pim \nu}\xspace}

\def\ggtopi     {\ensuremath{\gaga \to \pipi}\xspace}
\def\ggtopiz    {\ensuremath{\gaga \to \ppz}\xspace}
\def\ggstox     {\ensuremath{\ggstar \to \X(1420) \to \kaon \kaon \pi}\xspace}
\def\ggstoeta   {\ensuremath{\ggstar \to \eta(550) \to \pipi \piz}\xspace}

\def\ptot       {\mbox{$p$}\xspace}
\def\pxy        {\mbox{$p_T$}\xspace}
\def\ptrel      {\mbox{$p_\perp^{\mathrm{rel}}$}\xspace}
\def\mes        {\mbox{$m_{\rm ES}$}\xspace}
\def\mec        {\mbox{$m_{\rm EC}$}\xspace}
\def\DeltaE     {\mbox{$\Delta E$}\xspace}

\def\pbcm {\ensuremath{p^*_{\Bz}}\xspace}

\def\mphi       {\mbox{$\phi$}\xspace}
\def\mtheta     {\mbox{$\theta$}\xspace}
\def\ctheta     {\mbox{$\cos\theta$}\xspace}

\newcommand{\ke}{\ensuremath{\mbox{\,ke}}\xspace}

\newcommand{\eev}{\ensuremath{\mbox{\,Ee\kern -0.1em V}}\xspace}
\newcommand{\pev}{\ensuremath{\mbox{\,Pe\kern -0.1em V}}\xspace}
\newcommand{\tev}{\ensuremath{\mbox{\,Te\kern -0.1em V}}\xspace}
\renewcommand{\gev}{\ensuremath{\mbox{\,Ge\kern -0.1em V}}\xspace}
\newcommand{\mev}{\ensuremath{\mbox{\,Me\kern -0.1em V}}\xspace}
\newcommand{\kev}{\ensuremath{\mbox{\,ke\kern -0.1em V}}\xspace}
\newcommand{\ev} {\ensuremath{\mbox{\,e\kern -0.1em V}}\xspace}
\def\microEv         {\ensuremath{\,\mu\mbox{eV}}\xspace}  %% micro eV
\def\milliEv     {\ensuremath{\,\mbox{meV}}\xspace}  %% micro eV
\newcommand{\gevc}{\ensuremath{{\mbox{\,Ge\kern -0.1em V\!/}c}}\xspace}
\newcommand{\mevc}{\ensuremath{{\mbox{\,Me\kern -0.1em V\!/}c}}\xspace}
\newcommand{\gevcc}{\ensuremath{{\mbox{\,Ge\kern -0.1em V\!/}c^2}}\xspace}
\newcommand{\mevcc}{\ensuremath{{\mbox{\,Me\kern -0.1em V\!/}c^2}}\xspace}

\def\N   {\ensuremath{\mbox{\,N}\xspace}}

\def\syin {\ensuremath{^{\prime\prime}}\xspace}
\def\inch {\ensuremath{\rm \,in}\xspace} % \in is taken
\def\ft   {\ensuremath{\rm \,ft}\xspace}
\def\km   {\ensuremath{\mbox{\,km}}\xspace}
\def\m    {\ensuremath{\mbox{\,m}}\xspace}
\def\cm   {\ensuremath{\mbox{\,cm}}\xspace}
\def\sr   {\ensuremath{\mbox{\,sr}}\xspace}
\def\cma  {\ensuremath{\mbox{\,cm}^2}\xspace}
\def\mm   {\ensuremath{\mbox{\,mm}\xspace}}
\def\mma  {\ensuremath{\mbox{\,mm}^2}\xspace}
\def\mum  {\ensuremath{\,\mu\mbox{m}\xspace}}
\def\muma {\ensuremath{\,\mu\mbox{m^2}}\xspace}
\def\fm   {\ensuremath{\mbox{\,fm}}\xspace}
\def\nm   {\ensuremath{\mbox{\,nm}}\xspace}   %% nanometer
\def\nb   {\ensuremath{\mbox{\,nb}}\xspace}
\def\barn      {\ensuremath{\mbox{\,b}}\xspace}
\def\mbarn     {\ensuremath{\mbox{\,mb}}\xspace}
\def\mb        {\ensuremath{\mbox{\,mb}}\xspace}
\def\pb        {\ensuremath{\mbox{\,pb}}\xspace}
\def\invmb     {\ensuremath{\mbox{\,mb}^{-1}}\xspace}
\def\invnb     {\ensuremath{\mbox{\,nb}^{-1}}\xspace}
\def\invpb     {\ensuremath{\mbox{\,pb}^{-1}}\xspace}
\def\ub        {\ensuremath{\,\mu\mbox{b}}\xspace}
\def\invub     {\ensuremath{\mbox{\,\ub}^{-1}}\xspace}
\def\fb        {\ensuremath{\mbox{\,fb}}\xspace}
\def\invfb     {\ensuremath{\mbox{\,fb}^{-1}}\xspace}
\def\ab        {\ensuremath{\mbox{\,ab}}\xspace}
\def\invab     {\ensuremath{\mbox{\,ab}^{-1}}\xspace}
\def\cms       {\ensuremath{\mbox{\,cm}^{-2}\mbox{s}^{-1}}\xspace}
\def\sqrts     {\ensuremath{\sqrt{s}}}

\def\mpc     {\ensuremath{\mbox{\,Mpc}}\xspace}

\def\kW   {\ensuremath{\mbox{\,kW}}\xspace}
\def\MW   {\ensuremath{\mbox{\,MW}}\xspace}
\def\mW   {\ensuremath{\mbox{\,mW}}\xspace}
\def\GW   {\ensuremath{\mbox{\,GW}}\xspace}

\def\Hz  {\ensuremath{\mbox{\, Hz}}\xspace}
\def\kHz {\ensuremath{\mbox{\, kHz}}\xspace}
\def\MHz {\ensuremath{\mbox{\, MHz}}\xspace}

\def\us   {\ensuremath{\,\mu\mbox{s}}\xspace}
\def\ns   {\ensuremath{\mbox{\,ns}}\xspace}
\def\ms   {\ensuremath{\mbox{\,ms}}\xspace}
\def\ps   {\ensuremath{\mbox{\,ps}}\xspace}
\def\fs   {\ensuremath{\mbox{\,fs}}\xspace}
\def\gm   {\ensuremath{\mbox{\,g}}\xspace}

\def\Gy{\ensuremath{\mbox{\,Gy}}\xspace}
\def\sec{\ensuremath{\mbox{\,s}}\xspace}       %% second - this works - jw 4/19
\def\msec{\ensuremath{\mbox{\,ms}}\xspace}       %% second - this works - jw 4/19
\def\usec{\ensuremath{\,\mu \mbox{s}}\xspace}       %% second - this works - jw 4/19
\def\h          {\ensuremath{\mbox{\,h}\xspace}}

\def\kg         {\ensuremath{\mbox{\,kg}}\xspace}  %% kilogram
\def\gram       {\ensuremath{\mbox{\,g}}\xspace}  %% gram

\def\uTesla     {\ensuremath{\,\mu\mbox{T}}\xspace}  %% nanoTesla
\def\mTesla     {\ensuremath{\mbox{\,mT}}\xspace}  %% nanoTesla
\def\nTesla     {\ensuremath{\mbox{\,nT}}\xspace}  %% nanoTesla
\def\Tesla      {\ensuremath{\mbox{\,T}}\xspace}  %% Tesla
\def\Gauss      {\ensuremath{\mbox{\,G}}\xspace}  %% 

\def\mA     {\ensuremath{\mbox{\,mA}}\xspace}  %% mili Ampere
\def\Ampere     {\ensuremath{\mbox{\,A}}\xspace}  %% Ampere
\def\Amp     {\ensuremath{\mbox{\,A}}\xspace}  %% Ampere
\def\Watt     {\ensuremath{\mbox{\,W}}\xspace}  %% Ampere

\def\Xrad {\ensuremath{X_0}\xspace}
\def\NIL{\ensuremath{\lambda_{int}}\xspace}
\let\dgr\degrees

\def\mbar        {\ensuremath{\mbox{\,mbar}}}   %% milibar

\def\MVolts {\ensuremath{\mbox{\, MV}}\xspace}
\def\kVolts {\ensuremath{\mbox{\, kV}}\xspace}
\def\Volts {\ensuremath{\mbox{\, V}}\xspace}
\def\Volt  {\ensuremath{\mbox{\, V}}\xspace}
\def\atm   {\ensuremath{\mbox{\,atm}}\xspace}
\def\Ke    {\ensuremath{\mbox{\, K}}\xspace}
\def\mKe   {\ensuremath{\mbox{\, mK}}\xspace}
\def\mic  {\ensuremath{\,\mu{\rm C}}\xspace}
\def\krad {\ensuremath{\rm \,krad}\xspace}
\def\cmc  {\ensuremath{{\rm \,cm}^3}\xspace}
\def\yr   {\ensuremath{\rm \,yr}\xspace}
\def\hr   {\ensuremath{\rm \,hr}\xspace}
\def\degc {\ensuremath{^\circ}{C}\xspace}
\def\degk {\ensuremath{\mbox{K}}\xspace}
\def\degrees {\ensuremath{^{\circ}}\xspace}
\def\mrad {\ensuremath{\,\mbox{mrad}}\xspace}               %% milliradian
\def\urad {\ensuremath{\,\mu\mbox{rad}}\xspace}               %% milliradian
\def\rad{\ensuremath{\mbox{\,rad}}\xspace}
\def\mradhyph{\ensuremath{\rm -mr}\xspace}
\def\sx    {\ensuremath{\sigma_x}\xspace}    
\def\sy    {\ensuremath{\sigma_y}\xspace}   
\def\sz    {\ensuremath{\sigma_z}\xspace}

\def\order{{\ensuremath{\cal O}}\xspace}
\def\L{{\ensuremath{\cal L}}\xspace}
\def\calL{{\ensuremath{\cal L}}\xspace}
\def\calS{{\ensuremath{\cal S}}\xspace}
\def\calA{{\ensuremath{\cal A}}\xspace}
\def\calD{{\ensuremath{\cal D}}\xspace}
\def\calR{{\ensuremath{\cal R}}\xspace}

\def\ra                 {\ensuremath{\rightarrow}\xspace}
\def\to                 {\ensuremath{\rightarrow}\xspace}

\renewcommand{\stat}{\ensuremath{\mathrm{(stat)}}\xspace}
\renewcommand{\syst}{\ensuremath{\mathrm{(syst)}}\xspace}

\newcommand{\sstat}{\ensuremath{\sigma_{\mathrm{stat}}}\xspace}
\newcommand{\ssyst}{\ensuremath{\sigma_{\mathrm{syst}}}\xspace}

\def\pep2{PEP-II}
\def\BF{$B$ Factory}
\def\abf {asymmetric \BF}
\def\sx    {\ensuremath{\sigma_x}\xspace}     
\def\sy    {\ensuremath{\sigma_y}\xspace}   
\def\sz    {\ensuremath{\sigma_z}\xspace}

\newcommand{\inverse}{\ensuremath{^{-1}}\xspace}
\newcommand{\dedx}{\ensuremath{\mathrm{d}\hspace{-0.1em}E/\mathrm{d}x}\xspace}
\newcommand{\chisq}{\ensuremath{\chi^2}\xspace}
\newcommand{\delm}{\ensuremath{m_{\dstr}-m_{\dz}}\xspace}
\newcommand{\lum} {\ensuremath{\mathcal{L}}\xspace}

\def\gsim{{~\raise.15em\hbox{$>$}\kern-.85em
          \lower.35em\hbox{$\sim$}~}\xspace}
\def\lsim{{~\raise.15em\hbox{$<$}\kern-.85em
          \lower.35em\hbox{$\sim$}~}\xspace}

\def\qsq                {\ensuremath{q^2}\xspace}

\def\kbytes     {\ensuremath{{\rm \,kbytes}}\xspace}
\def\kbsps      {\ensuremath{{\rm \,kbytes/s}}\xspace}
\def\kbits      {\ensuremath{{\rm \,kbits}}\xspace}
\def\kbsps      {\ensuremath{{\rm \,kbits/s}}\xspace}
\def\mbsps      {\ensuremath{{\rm \,Mbits/s}}\xspace}
\def\mbytes     {\ensuremath{{\rm \,Mbytes}}\xspace}
\def\mbps       {\ensuremath{{\rm \,Mbyte/s}}\xspace}
\def\mbsps      {\ensuremath{{\rm \,Mbytes/s}}\xspace}
\def\gbsps      {\ensuremath{{\rm \,Gbits/s}}\xspace}
\def\gbytes     {\ensuremath{{\rm \,Gbytes}}\xspace}
\def\gbsps      {\ensuremath{{\rm \,Gbytes/s}}\xspace}
\def\tbytes     {\ensuremath{{\rm \,Tbytes}}\xspace}
\def\tbpy       {\ensuremath{{\rm \,Tbytes/yr}}\xspace}

\def\tb         {\ensuremath{\tan\beta}\xspace}

\newcommand{\as}{\ensuremath{\alpha_{\scriptscriptstyle S}}\xspace}
\newcommand{\asp}{\ensuremath{{\alpha_{\scriptscriptstyle S}\over\pi}}\xspace}
\newcommand{\MSb}{\ensuremath{\overline{\mathrm{MS}}}\xspace}
\newcommand{\LMSb}{%
  \ensuremath{\Lambda_{\overline{\scriptscriptstyle\mathrm{MS}}}}\xspace
}

\newcommand{\tw}{\ensuremath{\theta_{\scriptscriptstyle W}}\xspace}
\newcommand{\twb}{%
  \ensuremath{\overline{\theta}_{\scriptscriptstyle W}}\xspace
}
\newcommand{\Afb}[1]{{\ensuremath{A_{\scriptscriptstyle FB}^{#1}}}\xspace}
\newcommand{\gv}[1]{{\ensuremath{g_{\scriptscriptstyle V}^{#1}}}\xspace}
\renewcommand{\ga}[1]{{\ensuremath{g_{\scriptscriptstyle A}^{#1}}}\xspace}
\newcommand{\gvb}[1]{{\ensuremath{\overline{g}_{\scriptscriptstyle V}^{#1}}}\xspace}
\newcommand{\gab}[1]{{\ensuremath{\overline{g}_{\scriptscriptstyle A}^{#1}}}\xspace}

\def\eps{\varepsilon\xspace}
\def\epsK{\varepsilon_K\xspace}
\def\epsB{\varepsilon_B\xspace}
\def\epsp{\varepsilon^\prime_K\xspace}

\def\CP                {\ensuremath{C\!P}\xspace}
\def\CPT               {\ensuremath{C\!PT}\xspace} % Looks better without \!

\def\epstag  {\ensuremath{\varepsilon_{\rm tag}}\xspace}
\def\tagfac  {\ensuremath{\epstag(1-2w)^2}\xspace}

\def\rhobar {\ensuremath{\overline \rho}\xspace}
\def\etabar {\ensuremath{\overline \eta}\xspace}
\def\meas {\ensuremath{|V_{cb}|, |\frac{V_{ub}}{V_{cb}}|, 
|\varepsilon_K|, \Delta m_{B_d}}\xspace}

\def\Vud  {\ensuremath{|V_{ud}|}\xspace}
\def\Vcd  {\ensuremath{|V_{cd}|}\xspace}
\def\Vtd  {\ensuremath{|V_{td}|}\xspace}
\def\Vus  {\ensuremath{|V_{us}|}\xspace}
\def\Vcs  {\ensuremath{|V_{cs}|}\xspace}
\def\Vts  {\ensuremath{|V_{ts}|}\xspace}
\def\Vtd  {\ensuremath{|V_{td}|}\xspace}
\def\Vub  {\ensuremath{|V_{ub}|}\xspace}
\def\Vcb  {\ensuremath{|V_{cb}|}\xspace}
\def\Vtb  {\ensuremath{|V_{tb}|}\xspace}

\def\stwoa{\ensuremath{\sin\! 2 \alpha  }\xspace}
\def\stwob{\ensuremath{\sin\! 2 \beta   }\xspace}
\def\stwog{\ensuremath{\sin\! 2 \gamma  }\xspace}
\def\mistag{\ensuremath{w}\xspace}
\def\dilution{\ensuremath{\cal D}\xspace}
\def\deltaz{\ensuremath{{\rm \Delta}z}\xspace}
\def\deltat{\ensuremath{{\rm \Delta}t}\xspace}
\def\deltamd{\ensuremath{{\rm \Delta}m_d}\xspace}

\newcommand{\fsubd}{\ensuremath{f_D}}\xspace
\newcommand{\fds}{\ensuremath{f_{D_s}}\xspace}
\newcommand{\fsubb}{\ensuremath{f_B}\xspace}
\newcommand{\fbd}{\ensuremath{f_{B_d}}\xspace}
\newcommand{\fbs}{\ensuremath{f_{B_s}}\xspace}
\newcommand{\bsubb}{\ensuremath{B_B}\xspace}
\newcommand{\bbd}{\ensuremath{B_{B_d}}\xspace}
\newcommand{\bbs}{\ensuremath{B_{B_s}}\xspace}
\newcommand{\rgbb}{\ensuremath{\hat{B}_B}\xspace}
\newcommand{\rgbbd}{\ensuremath{\hat{B}_{B_d}}\xspace}
\newcommand{\rgbbs}{\ensuremath{\hat{B}_{B_s}}\xspace}
\newcommand{\rgbk}{\ensuremath{\hat{B}_K}\xspace}
\newcommand{\lqcd}{\ensuremath{\Lambda_{\mathrm{QCD}}}\xspace}

\newcommand{\secref}[1]{Section~\ref{sec:#1}}
\newcommand{\subsecref}[1]{Section~\ref{subsec:#1}}
\newcommand{\figref}[1]{Figure~\ref{fig:#1}}
\newcommand{\tabref}[1]{Table~\ref{tab:#1}}

\newcommand{\epjBase}        {Eur.\ Phys.\ Jour.\xspace}
\newcommand{\jprlBase}       {Phys.\ Rev.\ Lett.\xspace}
\newcommand{\jprBase}        {Phys.\ Rev.\xspace}
\newcommand{\jplBase}        {Phys.\ Lett.\xspace}
\newcommand{\nimBaseA}       {Nucl.\ Instr.\ Meth.\xspace}
\newcommand{\nimBaseB}       {Nucl.\ Instr.\ and Meth.\xspace}
\newcommand{\nimBaseC}       {Nucl.\ Instr.\ and Methods\xspace}
\newcommand{\nimBaseD}       {Nucl.\ Instrum.\ Methods\xspace}
\newcommand{\npBase}         {Nucl.\ Phys.\xspace}
\newcommand{\zpBase}         {Z.\ Phys.\xspace}

\newcommand{\apas}      [1]  {{Acta Phys.\ Austr.\ Suppl.\ {\bf #1}}}
\newcommand{\app}       [1]  {{Acta Phys.\ Polon.\ {\bf #1}}}
\newcommand{\ace}       [1]  {{Adv.\ Cry.\ Eng.\ {\bf #1}}}
\newcommand{\anp}       [1]  {{Adv.\ Nucl.\ Phys.\ {\bf #1}}}
\newcommand{\annp}      [1]  {{Ann.\ Phys.\ {\bf #1}}}
\newcommand{\araa}      [1]  {{Ann.\ Rev.\ Astr.\ Ap.\ {\bf #1}}}
\newcommand{\arnps}     [1]  {{Ann.\ Rev.\ Nucl.\ Part.\ Sci.\ {\bf #1}}}
\newcommand{\arns}      [1]  {{Ann.\ Rev.\ Nucl.\ Sci.\ {\bf #1}}}
\newcommand{\appopt}    [1]  {{Appl.\ Opt.\ {\bf #1}}}
\newcommand{\japj}      [1]  {{Astro.\ Phys.\ J.\ {\bf #1}}}
\newcommand{\baps}      [1]  {{Bull.\ Am.\ Phys.\ Soc.\ {\bf #1}}}
\newcommand{\seis}      [1]  {{Bull.\ Seismological Soc.\ of Am.\ {\bf #1}}}
\newcommand{\cmp}       [1]  {{Commun.\ Math.\ Phys.\ {\bf #1}}}
\newcommand{\cnpp}      [1]  {{Comm.\ Nucl.\ Part.\ Phys.\ {\bf #1}}}
\newcommand{\cpc}       [1]  {{Comput.\ Phys.\ Commun.\ {\bf #1}}}
\newcommand{\epj}       [1]  {\epjBase\ {\bf #1}}
\newcommand{\epjc}      [1]  {\epjBase\ C~{\bf #1}}
\newcommand{\fizika}    [1]  {{Fizika~{\bf #1}}}
\newcommand{\fp}        [1]  {{Fortschr.\ Phys.\ {\bf #1}}}
\newcommand{\ited}      [1]  {{IEEE Trans.\ Electron.\ Devices~{\bf #1}}}
\newcommand{\itns}      [1]  {{IEEE Trans.\ Nucl.\ Sci.\ {\bf #1}}}
\newcommand{\ijqe}      [1]  {{IEEE J.\ Quantum Electron.\ {\bf #1}}}
\newcommand{\ijmp}      [1]  {{Int.\ Jour.\ Mod.\ Phys.\ {\bf #1}}}
\newcommand{\ijmpa}     [1]  {{Int.\ J.\ Mod.\ Phys.\ {\bf A{\bf #1}}}}
\newcommand{\jl}        [1]  {{JETP Lett.\ {\bf #1}}}
\newcommand{\jetp}      [1]  {{JETP~{\bf #1}}}
\newcommand{\jpg}       [1]  {{J.\ Phys.\ {\bf G{\bf #1}}}}
\newcommand{\jap}       [1]  {{J.\ Appl.\ Phys.\ {\bf #1}}}
\newcommand{\jmp}       [1]  {{J.\ Math.\ Phys.\ {\bf #1}}}
\newcommand{\jmes}      [1]  {{J.\ Micro.\ Elec.\ Sys.\ {\bf #1}}}
\newcommand{\mpl}       [1]  {{Mod.\ Phys.\ Lett.\ {\bf #1}}}

\newcommand{\nim}       [1]  {\nimBaseC~{\bf #1}}
\newcommand{\nima}      [1]  {\nimBaseC~A~{\bf #1}}

\newcommand{\np}        [1]  {\npBase\ {\bf #1}}
\newcommand{\npb}       [1]  {\npBase\ B~{\bf #1}}
\newcommand{\npps}      [1]  {{Nucl.\ Phys.\ Proc.\ Suppl.\ {\bf #1}}}
\newcommand{\npaps}     [1]  {{Nucl.\ Phys.\ A~Proc.\ Suppl.\ {\bf #1}}}
\newcommand{\npbps}     [1]  {{Nucl.\ Phys.\ B~Proc.\ Suppl.\ {\bf #1}}}

\newcommand{\ncim}      [1]  {{Nuo.\ Cim.\ {\bf #1}}}
\newcommand{\optl}      [1]  {{Opt.\ Lett.\ {\bf #1}}}
\newcommand{\optcom}    [1]  {{Opt.\ Commun.\ {\bf #1}}}
\newcommand{\partacc}   [1]  {{Particle Acclerators~{\bf #1}}}
\newcommand{\pan}       [1]  {{Phys.\ Atom.\ Nuclei~{\bf #1}}}
\newcommand{\pflu}      [1]  {{Physics of Fluids~{\bf #1}}}
\newcommand{\ptoday}    [1]  {{Physics Today~{\bf #1}}}

\newcommand{\jpl}       [1]  {\jplBase\ {\bf #1}}
\newcommand{\plb}       [1]  {\jplBase\ B~{\bf #1}}
\newcommand{\prep}      [1]  {{Phys.\ Rep.\ {\bf #1}}}
\newcommand{\jprl}      [1]  {\jprlBase\ {\bf #1}}
\newcommand{\pr}        [1]  {\jprBase\ {\bf #1}}
\newcommand{\jpra}      [1]  {\jprBase\ A~{\bf #1}}
\newcommand{\jprd}      [1]  {\jprBase\ D~{\bf #1}}
\newcommand{\jpre}      [1]  {\jprBase\ E~{\bf #1}}

\newcommand{\prsl}      [1]  {{Proc.\ Roy.\ Soc.\ Lond.\ {\bf #1}}}
\newcommand{\ppnp}      [1]  {{Prog.\ Part.\ Nucl.\ Phys.\ {\bf #1}}}
\newcommand{\progtp}    [1]  {{Prog.\ Th.\ Phys.\ {\bf #1}}}
\newcommand{\rpp}       [1]  {{Rep.\ Prog.\ Phys.\ {\bf #1}}}
\newcommand{\jrmp}      [1]  {{Rev.\ Mod.\ Phys.\ {\bf #1}}}  % dbm
\newcommand{\rsi}       [1]  {{Rev.\ Sci.\ Instr.\ {\bf #1}}}
\newcommand{\sci}       [1]  {{Science~{\bf #1}}}
\newcommand{\sjnp}      [1]  {{Sov.\ J.\ Nucl.\ Phys.\ {\bf #1}}}
\newcommand{\spd}       [1]  {{Sov.\ Phys.\ Dokl.\ {\bf #1}}}
\newcommand{\spu}       [1]  {{Sov.\ Phys.\ Usp.\ {\bf #1}}}
\newcommand{\tmf}       [1]  {{Teor.\ Mat.\ Fiz.\ {\bf #1}}}
\newcommand{\yf}        [1]  {{Yad.\ Fiz.\ {\bf #1}}}
\renewcommand{\zp}        [1]  {\zpBase\ {\bf #1}}
\newcommand{\zpc}       [1]  {\zpBase\ C~{\bf #1}}
\newcommand{\zpr}       [1]  {{ZhETF Pis.\ Red.\ {\bf #1}}}

\newcommand{\hepex}     [1]  {hep-ex/{#1}}
\newcommand{\hepph}     [1]  {hep-ph/{#1}}
\newcommand{\hepth}     [1]  {hep-th/{#1}}

\def\aslund     {\mbox{\tt Aslund}\xspace}
\def\bbsim      {\mbox{\tt BBsim}\xspace}
\def\beast      {\mbox{\tt Beast}\xspace}
\def\beget      {\mbox{\tt Beget}\xspace}
\def\Bta        {\mbox{\tt Beta}\xspace}
\def\betakfit   {\mbox{\tt BetaKfit}\xspace}
\def\cornelius  {\mbox{\tt Cornelius}\xspace}
\def\evtgen     {\mbox{\tt EvtGen}\xspace}
\def\euclid     {\mbox{\tt Euclid}\xspace}
\def\fitver     {\mbox{\tt FitVer}\xspace}
\def\fluka      {\mbox{\tt Fluka}\xspace}
\def\fortran    {\mbox{\tt Fortran}\xspace}
\def\gcalor     {\mbox{\tt GCalor}\xspace}
\def\geant      {\mbox{\tt GEANT}\xspace}
\def\gheisha    {\mbox{\tt Gheisha}\xspace}
\def\hemicosm   {\mbox{\tt HemiCosm}\xspace}
\def\hepevt     {\mbox{\tt{/HepEvt/}}\xspace}
\def\jetset74   {\mbox{\tt Jetset \hspace{-0.5em}7.\hspace{-0.2em}4}\xspace}
\def\koralb     {\mbox{\tt KoralB}\xspace}
\def\minuit     {\mbox{\tt Minuit}\xspace}
\def\objegs     {\mbox{\tt Objegs}\xspace}
\def\paw        {\mbox{\tt Paw}\xspace}
\def\root       {\mbox{\tt Root}\xspace}
\def\squaw      {\mbox{\tt Squaw}\xspace}
\def\stdhep     {\mbox{\tt StdHep}\xspace}
\def\trackerr   {\mbox{\tt TrackErr}\xspace}
\def\turtle     {\mbox{\tt Decay~Turtle}\xspace}

\vdef{lumiTot} {\ensuremath{ 5 } }

\vdef{default-11:sampleName:BdMc}   {\ensuremath{{B^{0} \rightarrow \mu^{+}\mu^{-} (MC) } } }
\vdef{default-11:lumi:BdMc}   {\ensuremath{{8200.0 } } }
\vdef{default-11:ngen:BdMc}   {\ensuremath{{40000000} } }
\vdef{default-11:nfilt:BdMc}   {\ensuremath{{241197} } }
\vdef{default-11:sampleName:BdMcAcc}   {\ensuremath{{B^{0} \rightarrow \mu^{+}\mu^{-} (acc) } } }
\vdef{default-11:lumi:BdMcAcc}   {\ensuremath{{8200.0 } } }
\vdef{default-11:ngen:BdMcAcc}   {\ensuremath{{n/a } } }
\vdef{default-11:nfilt:BdMcAcc}   {\ensuremath{{241197} } }
\vdef{default-11:sampleName:BmtHT}   {\ensuremath{{HT } } }
\vdef{default-11:lumi:BmtHT}   {\ensuremath{{ 4.9 } } }
\vdef{default-11:ngen:BmtHT}   {\ensuremath{{n/a } } }
\vdef{default-11:nfilt:BmtHT}   {\ensuremath{{13810557} } }
\vdef{default-11:sampleName:BmtJet}   {\ensuremath{{Jet } } }
\vdef{default-11:lumi:BmtJet}   {\ensuremath{{ 4.9 } } }
\vdef{default-11:ngen:BmtJet}   {\ensuremath{{n/a } } }
\vdef{default-11:nfilt:BmtJet}   {\ensuremath{{5759664} } }
\vdef{default-11:sampleName:BmtPhoton}   {\ensuremath{{Photon } } }
\vdef{default-11:lumi:BmtPhoton}   {\ensuremath{{ 4.9 } } }
\vdef{default-11:ngen:BmtPhoton}   {\ensuremath{{n/a } } }
\vdef{default-11:nfilt:BmtPhoton}   {\ensuremath{{2448101} } }
\vdef{default-11:sampleName:CsData}   {\ensuremath{{Data } } }
\vdef{default-11:lumi:CsData}   {\ensuremath{{ 4.9 } } }
\vdef{default-11:ngen:CsData}   {\ensuremath{{n/a } } }
\vdef{default-11:nfilt:CsData}   {\ensuremath{{30614681} } }
\vdef{default-11:sampleName:CsMc}   {\ensuremath{{MC simulation } } }
\vdef{default-11:lumi:CsMc}   {\ensuremath{{ 5.4 } } }
\vdef{default-11:ngen:CsMc}   {\ensuremath{{9970000000} } }
\vdef{default-11:nfilt:CsMc}   {\ensuremath{{328303} } }
\vdef{default-11:sampleName:CsMc1e33}   {\ensuremath{{B_{s}^{0} \rightarrow J/\psi \phi (1e33) } } }
\vdef{default-11:lumi:CsMc1e33}   {\ensuremath{{ 5.4 } } }
\vdef{default-11:ngen:CsMc1e33}   {\ensuremath{{n/a } } }
\vdef{default-11:nfilt:CsMc1e33}   {\ensuremath{{348061} } }
\vdef{default-11:sampleName:CsMc2e33}   {\ensuremath{{B_{s}^{0} \rightarrow J/\psi \phi (2e33) } } }
\vdef{default-11:lumi:CsMc2e33}   {\ensuremath{{ 5.4 } } }
\vdef{default-11:ngen:CsMc2e33}   {\ensuremath{{n/a } } }
\vdef{default-11:nfilt:CsMc2e33}   {\ensuremath{{348136} } }
\vdef{default-11:sampleName:CsMc3e33}   {\ensuremath{{B_{s}^{0} \rightarrow J/\psi \phi (3e33) } } }
\vdef{default-11:lumi:CsMc3e33}   {\ensuremath{{ 5.4 } } }
\vdef{default-11:ngen:CsMc3e33}   {\ensuremath{{n/a } } }
\vdef{default-11:nfilt:CsMc3e33}   {\ensuremath{{348809} } }
\vdef{default-11:sampleName:CsMcAcc}   {\ensuremath{{B_{s}^{0} \rightarrow J/\psi \phi (acc) } } }
\vdef{default-11:lumi:CsMcAcc}   {\ensuremath{{ 5.4 } } }
\vdef{default-11:ngen:CsMcAcc}   {\ensuremath{{   1} } }
\vdef{default-11:nfilt:CsMcAcc}   {\ensuremath{{2463146} } }
\vdef{default-11:sampleName:CsMcPU}   {\ensuremath{{B_{s}^{0} \rightarrow J/\psi \phi (PU) } } }
\vdef{default-11:lumi:CsMcPU}   {\ensuremath{{ 5.0 } } }
\vdef{default-11:ngen:CsMcPU}   {\ensuremath{{n/a } } }
\vdef{default-11:nfilt:CsMcPU}   {\ensuremath{{325788} } }
\vdef{default-11:sampleName:NoData}   {\ensuremath{{Data } } }
\vdef{default-11:lumi:NoData}   {\ensuremath{{ 4.9 } } }
\vdef{default-11:ngen:NoData}   {\ensuremath{{n/a } } }
\vdef{default-11:nfilt:NoData}   {\ensuremath{{30614681} } }
\vdef{default-11:sampleName:NoMc}   {\ensuremath{{MC simulation } } }
\vdef{default-11:lumi:NoMc}   {\ensuremath{{ 5.2 } } }
\vdef{default-11:ngen:NoMc}   {\ensuremath{{15166000000} } }
\vdef{default-11:nfilt:NoMc}   {\ensuremath{{1578601} } }
\vdef{default-11:sampleName:NoMc1e33}   {\ensuremath{{B^{+} \rightarrow J/\psi K^{+} (1e33) } } }
\vdef{default-11:lumi:NoMc1e33}   {\ensuremath{{ 5.2 } } }
\vdef{default-11:ngen:NoMc1e33}   {\ensuremath{{n/a } } }
\vdef{default-11:nfilt:NoMc1e33}   {\ensuremath{{1663223} } }
\vdef{default-11:sampleName:NoMc2e33}   {\ensuremath{{B^{+} \rightarrow J/\psi K^{+} (2e33) } } }
\vdef{default-11:lumi:NoMc2e33}   {\ensuremath{{ 5.2 } } }
\vdef{default-11:ngen:NoMc2e33}   {\ensuremath{{n/a } } }
\vdef{default-11:nfilt:NoMc2e33}   {\ensuremath{{1663449} } }
\vdef{default-11:sampleName:NoMc3e33}   {\ensuremath{{B^{+} \rightarrow J/\psi K^{+} (3e33) } } }
\vdef{default-11:lumi:NoMc3e33}   {\ensuremath{{ 5.2 } } }
\vdef{default-11:ngen:NoMc3e33}   {\ensuremath{{n/a } } }
\vdef{default-11:nfilt:NoMc3e33}   {\ensuremath{{1666741} } }
\vdef{default-11:sampleName:NoMcAcc}   {\ensuremath{{B^{+} \rightarrow J/\psi K^{+} (acc) } } }
\vdef{default-11:lumi:NoMcAcc}   {\ensuremath{{ 5.2 } } }
\vdef{default-11:ngen:NoMcAcc}   {\ensuremath{{   1} } }
\vdef{default-11:nfilt:NoMcAcc}   {\ensuremath{{695478} } }
\vdef{default-11:sampleName:NoMcCMS}   {\ensuremath{{B^{+} \rightarrow J/\psi K^{+} (CMS) } } }
\vdef{default-11:lumi:NoMcCMS}   {\ensuremath{{ 5.0 } } }
\vdef{default-11:ngen:NoMcCMS}   {\ensuremath{{n/a } } }
\vdef{default-11:nfilt:NoMcCMS}   {\ensuremath{{460883} } }
\vdef{default-11:sampleName:NoMcPU}   {\ensuremath{{B^{+} \rightarrow J/\psi K^{+} (PU) } } }
\vdef{default-11:lumi:NoMcPU}   {\ensuremath{{ 5.0 } } }
\vdef{default-11:ngen:NoMcPU}   {\ensuremath{{n/a } } }
\vdef{default-11:nfilt:NoMcPU}   {\ensuremath{{606722} } }
\vdef{default-11:sampleName:SgData}   {\ensuremath{{Data } } }
\vdef{default-11:lumi:SgData}   {\ensuremath{{ 4.9 } } }
\vdef{default-11:ngen:SgData}   {\ensuremath{{n/a } } }
\vdef{default-11:nfilt:SgData}   {\ensuremath{{30614681} } }
\vdef{default-11:sampleName:SgMc}   {\ensuremath{{B_{s}^{0} \rightarrow \mu^{+}\mu^{-} (MC) } } }
\vdef{default-11:lumi:SgMc}   {\ensuremath{{1600.0 } } }
\vdef{default-11:ngen:SgMc}   {\ensuremath{{249000000} } }
\vdef{default-11:nfilt:SgMc}   {\ensuremath{{538721} } }
\vdef{default-11:sampleName:SgMc1e33}   {\ensuremath{{B_{s}^{0} \rightarrow \mu^{+}\mu^{-} (1e33) } } }
\vdef{default-11:lumi:SgMc1e33}   {\ensuremath{{1600.0 } } }
\vdef{default-11:ngen:SgMc1e33}   {\ensuremath{{n/a } } }
\vdef{default-11:nfilt:SgMc1e33}   {\ensuremath{{538721} } }
\vdef{default-11:sampleName:SgMc2e33}   {\ensuremath{{B_{s}^{0} \rightarrow \mu^{+}\mu^{-} (2e33) } } }
\vdef{default-11:lumi:SgMc2e33}   {\ensuremath{{1600.0 } } }
\vdef{default-11:ngen:SgMc2e33}   {\ensuremath{{n/a } } }
\vdef{default-11:nfilt:SgMc2e33}   {\ensuremath{{538721} } }
\vdef{default-11:sampleName:SgMc3e33}   {\ensuremath{{B_{s}^{0} \rightarrow \mu^{+}\mu^{-} (3e33) } } }
\vdef{default-11:lumi:SgMc3e33}   {\ensuremath{{1600.0 } } }
\vdef{default-11:ngen:SgMc3e33}   {\ensuremath{{n/a } } }
\vdef{default-11:nfilt:SgMc3e33}   {\ensuremath{{538721} } }
\vdef{default-11:sampleName:SgMcAcc}   {\ensuremath{{B_{s}^{0} \rightarrow \mu^{+}\mu^{-} (acc) } } }
\vdef{default-11:lumi:SgMcAcc}   {\ensuremath{{1600.0 } } }
\vdef{default-11:ngen:SgMcAcc}   {\ensuremath{{n/a } } }
\vdef{default-11:nfilt:SgMcAcc}   {\ensuremath{{538721} } }
\vdef{default-11:sampleName:SgMcPU}   {\ensuremath{{B_{s}^{0} \rightarrow \mu^{+}\mu^{-} (PU) } } }
\vdef{default-11:lumi:SgMcPU}   {\ensuremath{{1600.0 } } }
\vdef{default-11:ngen:SgMcPU}   {\ensuremath{{n/a } } }
\vdef{default-11:nfilt:SgMcPU}   {\ensuremath{{110321} } }
\vdef{default-11:sampleName:bgBd2KK}   {\ensuremath{{B^{0} \rightarrow K^{+}K^{-} } } }
\vdef{default-11:lumi:bgBd2KK}   {\ensuremath{{27.6 } } }
\vdef{default-11:ngen:bgBd2KK}   {\ensuremath{{200000000} } }
\vdef{default-11:nfilt:bgBd2KK}   {\ensuremath{{72795} } }
\vdef{default-11:sampleName:bgBd2KPi}   {\ensuremath{{B^{0} \rightarrow K^{+}\pi^{-} } } }
\vdef{default-11:lumi:bgBd2KPi}   {\ensuremath{{ 1.1 } } }
\vdef{default-11:ngen:bgBd2KPi}   {\ensuremath{{996000000} } }
\vdef{default-11:nfilt:bgBd2KPi}   {\ensuremath{{362296} } }
\vdef{default-11:sampleName:bgBd2PiMuNu}   {\ensuremath{{B^{0} \rightarrow \pi^{-}\mu^{+}\nu } } }
\vdef{default-11:lumi:bgBd2PiMuNu}   {\ensuremath{{ 1.0 } } }
\vdef{default-11:ngen:bgBd2PiMuNu}   {\ensuremath{{6742000000} } }
\vdef{default-11:nfilt:bgBd2PiMuNu}   {\ensuremath{{992010} } }
\vdef{default-11:sampleName:bgBd2PiPi}   {\ensuremath{{B^{0} \rightarrow \pi^{+}\pi^{-} } } }
\vdef{default-11:lumi:bgBd2PiPi}   {\ensuremath{{ 1.6 } } }
\vdef{default-11:ngen:bgBd2PiPi}   {\ensuremath{{394000000} } }
\vdef{default-11:nfilt:bgBd2PiPi}   {\ensuremath{{142115} } }
\vdef{default-11:sampleName:bgBs2KK}   {\ensuremath{{B_{s}^{0} \rightarrow K^{+}K^{-} } } }
\vdef{default-11:lumi:bgBs2KK}   {\ensuremath{{ 1.0 } } }
\vdef{default-11:ngen:bgBs2KK}   {\ensuremath{{1392000000} } }
\vdef{default-11:nfilt:bgBs2KK}   {\ensuremath{{185683} } }
\vdef{default-11:sampleName:bgBs2KMuNu}   {\ensuremath{{B_{s}^{0} \rightarrow K^{-}\mu^{+}\nu } } }
\vdef{default-11:lumi:bgBs2KMuNu}   {\ensuremath{{ 1.1 } } }
\vdef{default-11:ngen:bgBs2KMuNu}   {\ensuremath{{6702000000} } }
\vdef{default-11:nfilt:bgBs2KMuNu}   {\ensuremath{{373955} } }
\vdef{default-11:sampleName:bgBs2KPi}   {\ensuremath{{B_{s}^{0} \rightarrow \pi^{+}K^{-} } } }
\vdef{default-11:lumi:bgBs2KPi}   {\ensuremath{{ 2.5 } } }
\vdef{default-11:ngen:bgBs2KPi}   {\ensuremath{{598000000} } }
\vdef{default-11:nfilt:bgBs2KPi}   {\ensuremath{{79951} } }
\vdef{default-11:sampleName:bgBs2PiPi}   {\ensuremath{{B_{s}^{0} \rightarrow \pi^{+}\pi^{-} } } }
\vdef{default-11:lumi:bgBs2PiPi}   {\ensuremath{{ 2.5 } } }
\vdef{default-11:ngen:bgBs2PiPi}   {\ensuremath{{148000000} } }
\vdef{default-11:nfilt:bgBs2PiPi}   {\ensuremath{{19480} } }
\vdef{default-11:sampleName:bgLb2KP}   {\ensuremath{{\Lambda_{b}^{0} \rightarrow p K^{-} } } }
\vdef{default-11:lumi:bgLb2KP}   {\ensuremath{{ 1.3 } } }
\vdef{default-11:ngen:bgLb2KP}   {\ensuremath{{356000000} } }
\vdef{default-11:nfilt:bgLb2KP}   {\ensuremath{{30104} } }
\vdef{default-11:sampleName:bgLb2PMuNu}   {\ensuremath{{\Lambda^{0}_{b} \rightarrow p\mu^{-}\bar{\nu} } } }
\vdef{default-11:lumi:bgLb2PMuNu}   {\ensuremath{{ 1.1 } } }
\vdef{default-11:ngen:bgLb2PMuNu}   {\ensuremath{{6910000000} } }
\vdef{default-11:nfilt:bgLb2PMuNu}   {\ensuremath{{249206} } }
\vdef{default-11:sampleName:bgLb2PiP}   {\ensuremath{{\Lambda_{b}^{0} \rightarrow p \pi^{-} } } }
\vdef{default-11:lumi:bgLb2PiP}   {\ensuremath{{ 2.2 } } }
\vdef{default-11:ngen:bgLb2PiP}   {\ensuremath{{374000000} } }
\vdef{default-11:nfilt:bgLb2PiP}   {\ensuremath{{31375} } }
\vdef{default-11:TYPE:cutLine}   {\ensuremath{{Candidate type } } }
\vdef{default-11:TYPE:cutValue}  {\ensuremath{{ Candidate type } } }
\vdef{default-11:TRIGRANGE:cutLine}   {\ensuremath{{HLT_DoubleMu4_Dimuon6_Bs_v1 } } }
\vdef{default-11:TRIGRANGE:cutValue}  {\ensuremath{{99999999) } } }
\vdef{default-11:TRUTHCAND:cutLine}   {\ensuremath{{Candidate type } } }
\vdef{default-11:TRUTHCAND:cutValue}  {\ensuremath{{ Candidate type } } }
\vdef{default-11:CANDPTLO:cutLine}   {\ensuremath{{p_{T}^{min}(B cand) } } }
\vdef{default-11:CANDPTLO:cutValue}  {\ensuremath{{ 7.5 } } }
\vdef{default-11:CANDETALO:cutLine}   {\ensuremath{{\eta^{min}(B cand) } } }
\vdef{default-11:CANDETALO:cutValue}  {\ensuremath{{ -24.0 } } }
\vdef{default-11:CANDETAHI:cutLine}   {\ensuremath{{\eta^{max}(B cand) } } }
\vdef{default-11:CANDETAHI:cutValue}  {\ensuremath{{ 24.0 } } }
\vdef{default-11:CANDALPHA:cutLine}   {\ensuremath{{\alpha } } }
\vdef{default-11:CANDALPHA:cutValue}  {\ensuremath{{ 0.0500 } } }
\vdef{default-11:CANDFLS3D:cutLine}   {\ensuremath{{l_{3d}/\sigma(l_{3d}) } } }
\vdef{default-11:CANDFLS3D:cutValue}  {\ensuremath{{ 15.0 } } }
\vdef{default-11:CANDFLSXY:cutLine}   {\ensuremath{{l_{xy}/\sigma(l_{xy}) } } }
\vdef{default-11:CANDFLSXY:cutValue}  {\ensuremath{{ -99.0 } } }
\vdef{default-11:CANDVTXCHI2:cutLine}   {\ensuremath{{\chi^{2} } } }
\vdef{default-11:CANDVTXCHI2:cutValue}  {\ensuremath{{ 2.0 } } }
\vdef{default-11:CANDISOLATION:cutLine}   {\ensuremath{{I_{trk} } } }
\vdef{default-11:CANDISOLATION:cutValue}  {\ensuremath{{ 0.80 } } }
\vdef{default-11:CANDDOCATRK:cutLine}   {\ensuremath{{doca_{trk} } } }
\vdef{default-11:CANDDOCATRK:cutValue}  {\ensuremath{{ 0.015 } } }
\vdef{default-11:CANDCLOSETRK:cutLine}   {\ensuremath{{N_{close tracks} } } }
\vdef{default-11:CANDCLOSETRK:cutValue}  {\ensuremath{{ 2.00 } } }
\vdef{default-11:PVAVEW8:cutLine}   {\ensuremath{{<w^{PV}_{trk}> } } }
\vdef{default-11:PVAVEW8:cutValue}  {\ensuremath{{ 0.600 } } }
\vdef{default-11:CANDLIP:cutLine}   {\ensuremath{{l_{z} } } }
\vdef{default-11:CANDLIP:cutValue}  {\ensuremath{{ 1.000 } } }
\vdef{default-11:CANDLIPS:cutLine}   {\ensuremath{{l_{z}/\sigma(l_{z}) } } }
\vdef{default-11:CANDLIPS:cutValue}  {\ensuremath{{ 5.000 } } }
\vdef{default-11:CAND2LIP:cutLine}   {\ensuremath{{l_{z,2} } } }
\vdef{default-11:CAND2LIP:cutValue}  {\ensuremath{{ -0.200 } } }
\vdef{default-11:CAND2LIPS:cutLine}   {\ensuremath{{l_{z,2}/\sigma(l_{z,2}) } } }
\vdef{default-11:CAND2LIPS:cutValue}  {\ensuremath{{ -3.000 } } }
\vdef{default-11:CANDIP:cutLine}   {\ensuremath{{l_{3d} } } }
\vdef{default-11:CANDIP:cutValue}  {\ensuremath{{ 0.008 } } }
\vdef{default-11:CANDIPS:cutLine}   {\ensuremath{{l_{3d}/\sigma(l_{3d}) } } }
\vdef{default-11:CANDIPS:cutValue}  {\ensuremath{{ 2.000 } } }
\vdef{default-11:MAXDOCA:cutLine}   {\ensuremath{{d } } }
\vdef{default-11:MAXDOCA:cutValue}  {\ensuremath{{ 0.050 } } }
\vdef{default-11:SIGBOXMIN:cutLine}   {\ensuremath{{SIGBOXMIN } } }
\vdef{default-11:SIGBOXMIN:cutValue}  {\ensuremath{{  5.200 } } }
\vdef{default-11:SIGBOXMAX:cutLine}   {\ensuremath{{SIGBOXMAX } } }
\vdef{default-11:SIGBOXMAX:cutValue}  {\ensuremath{{  5.450 } } }
\vdef{default-11:BGLBOXMIN:cutLine}   {\ensuremath{{BGLBOXMIN } } }
\vdef{default-11:BGLBOXMIN:cutValue}  {\ensuremath{{  4.800 } } }
\vdef{default-11:BGLBOXMAX:cutLine}   {\ensuremath{{BGLBOXMAX } } }
\vdef{default-11:BGLBOXMAX:cutValue}  {\ensuremath{{  5.200 } } }
\vdef{default-11:BGHBOXMIN:cutLine}   {\ensuremath{{BGHBOXMIN } } }
\vdef{default-11:BGHBOXMIN:cutValue}  {\ensuremath{{  5.450 } } }
\vdef{default-11:BGHBOXMAX:cutLine}   {\ensuremath{{BGHBOXMAX } } }
\vdef{default-11:BGHBOXMAX:cutValue}  {\ensuremath{{  6.000 } } }
\vdef{default-11:TRACKQUALITY:cutLine}   {\ensuremath{{track quality } } }
\vdef{default-11:TRACKQUALITY:cutValue}  {\ensuremath{{ 4 } } }
\vdef{default-11:TRACKPTLO:cutLine}   {\ensuremath{{p_{T}^{min}(track) } } }
\vdef{default-11:TRACKPTLO:cutValue}  {\ensuremath{{ 0.5 } } }
\vdef{default-11:TRACKPTHI:cutLine}   {\ensuremath{{p_{T}^{max}(track) } } }
\vdef{default-11:TRACKPTHI:cutValue}  {\ensuremath{{ 9999.0 } } }
\vdef{default-11:TRACKTIP:cutLine}   {\ensuremath{{doca_{xy}(track) } } }
\vdef{default-11:TRACKTIP:cutValue}  {\ensuremath{{ 1.0 } } }
\vdef{default-11:TRACKLIP:cutLine}   {\ensuremath{{doca_{z}(track) } } }
\vdef{default-11:TRACKLIP:cutValue}  {\ensuremath{{ 25.0 } } }
\vdef{default-11:TRACKETALO:cutLine}   {\ensuremath{{\eta_{min}(track) } } }
\vdef{default-11:TRACKETALO:cutValue}  {\ensuremath{{ -2.4 } } }
\vdef{default-11:TRACKETAHI:cutLine}   {\ensuremath{{\eta_{max}(track) } } }
\vdef{default-11:TRACKETAHI:cutValue}  {\ensuremath{{ 2.4 } } }
\vdef{default-11:MUIDMASK:cutLine}   {\ensuremath{{MuIDMask } } }
\vdef{default-11:MUIDMASK:cutValue}  {\ensuremath{{ 80 } } }
\vdef{default-11:MUIDRESULT:cutLine}   {\ensuremath{{MuIDResult } } }
\vdef{default-11:MUIDRESULT:cutValue}  {\ensuremath{{ 80 } } }
\vdef{default-11:MUPTLO:cutLine}   {\ensuremath{{p_{T}^{min}(\mu) } } }
\vdef{default-11:MUPTLO:cutValue}  {\ensuremath{{ 4.0 } } }
\vdef{default-11:MUPTHI:cutLine}   {\ensuremath{{p_{T}^{max}(\mu) } } }
\vdef{default-11:MUPTHI:cutValue}  {\ensuremath{{ 9999.0 } } }
\vdef{default-11:MUETALO:cutLine}   {\ensuremath{{\eta^{min}(\mu) } } }
\vdef{default-11:MUETALO:cutValue}  {\ensuremath{{ -2.4 } } }
\vdef{default-11:MUETAHI:cutLine}   {\ensuremath{{\eta^{max}(\mu) } } }
\vdef{default-11:MUETAHI:cutValue}  {\ensuremath{{ 2.4 } } }
\vdef{default-11:MUIP:cutLine}   {\ensuremath{{IP(\mu) } } }
\vdef{default-11:MUIP:cutValue}  {\ensuremath{{ 99.0 } } }

\vdef{default-11:SgData-A:osiso:loEff}   {\ensuremath{{1.029 } } }
\vdef{default-11:SgData-A:osiso:loEffE}   {\ensuremath{{\mathrm{NaN} } } }
\vdef{default-11:SgData-A:osiso:hiEff}   {\ensuremath{{1.000 } } }
\vdef{default-11:SgData-A:osiso:hiEffE}   {\ensuremath{{0.014 } } }
\vdef{default-11:SgMc-A:osiso:loEff}   {\ensuremath{{1.001 } } }
\vdef{default-11:SgMc-A:osiso:loEffE}   {\ensuremath{{0.014 } } }
\vdef{default-11:SgMc-A:osiso:hiEff}   {\ensuremath{{1.000 } } }
\vdef{default-11:SgMc-A:osiso:hiEffE}   {\ensuremath{{0.014 } } }
\vdef{default-11:SgMc-A:osiso:loDelta}   {\ensuremath{{+0.028 } } }
\vdef{default-11:SgMc-A:osiso:loDeltaE}   {\ensuremath{{\mathrm{NaN} } } }
\vdef{default-11:SgMc-A:osiso:hiDelta}   {\ensuremath{{+0.000 } } }
\vdef{default-11:SgMc-A:osiso:hiDeltaE}   {\ensuremath{{0.020 } } }
\vdef{default-11:SgData-A:osreliso:loEff}   {\ensuremath{{0.179 } } }
\vdef{default-11:SgData-A:osreliso:loEffE}   {\ensuremath{{0.047 } } }
\vdef{default-11:SgData-A:osreliso:hiEff}   {\ensuremath{{0.821 } } }
\vdef{default-11:SgData-A:osreliso:hiEffE}   {\ensuremath{{0.047 } } }
\vdef{default-11:SgMc-A:osreliso:loEff}   {\ensuremath{{0.264 } } }
\vdef{default-11:SgMc-A:osreliso:loEffE}   {\ensuremath{{0.053 } } }
\vdef{default-11:SgMc-A:osreliso:hiEff}   {\ensuremath{{0.736 } } }
\vdef{default-11:SgMc-A:osreliso:hiEffE}   {\ensuremath{{0.053 } } }
\vdef{default-11:SgMc-A:osreliso:loDelta}   {\ensuremath{{-0.383 } } }
\vdef{default-11:SgMc-A:osreliso:loDeltaE}   {\ensuremath{{0.317 } } }
\vdef{default-11:SgMc-A:osreliso:hiDelta}   {\ensuremath{{+0.109 } } }
\vdef{default-11:SgMc-A:osreliso:hiDeltaE}   {\ensuremath{{0.092 } } }
\vdef{default-11:SgData-A:osmuonpt:loEff}   {\ensuremath{{0.000 } } }
\vdef{default-11:SgData-A:osmuonpt:loEffE}   {\ensuremath{{0.236 } } }
\vdef{default-11:SgData-A:osmuonpt:hiEff}   {\ensuremath{{1.000 } } }
\vdef{default-11:SgData-A:osmuonpt:hiEffE}   {\ensuremath{{0.236 } } }
\vdef{default-11:SgMc-A:osmuonpt:loEff}   {\ensuremath{{0.000 } } }
\vdef{default-11:SgMc-A:osmuonpt:loEffE}   {\ensuremath{{0.236 } } }
\vdef{default-11:SgMc-A:osmuonpt:hiEff}   {\ensuremath{{1.000 } } }
\vdef{default-11:SgMc-A:osmuonpt:hiEffE}   {\ensuremath{{0.236 } } }
\vdef{default-11:SgMc-A:osmuonpt:loDelta}   {\ensuremath{{\mathrm{NaN} } } }
\vdef{default-11:SgMc-A:osmuonpt:loDeltaE}   {\ensuremath{{\mathrm{NaN} } } }
\vdef{default-11:SgMc-A:osmuonpt:hiDelta}   {\ensuremath{{+0.000 } } }
\vdef{default-11:SgMc-A:osmuonpt:hiDeltaE}   {\ensuremath{{0.333 } } }
\vdef{default-11:SgData-A:osmuondr:loEff}   {\ensuremath{{0.000 } } }
\vdef{default-11:SgData-A:osmuondr:loEffE}   {\ensuremath{{0.236 } } }
\vdef{default-11:SgData-A:osmuondr:hiEff}   {\ensuremath{{1.000 } } }
\vdef{default-11:SgData-A:osmuondr:hiEffE}   {\ensuremath{{0.236 } } }
\vdef{default-11:SgMc-A:osmuondr:loEff}   {\ensuremath{{0.013 } } }
\vdef{default-11:SgMc-A:osmuondr:loEffE}   {\ensuremath{{0.236 } } }
\vdef{default-11:SgMc-A:osmuondr:hiEff}   {\ensuremath{{0.987 } } }
\vdef{default-11:SgMc-A:osmuondr:hiEffE}   {\ensuremath{{0.236 } } }
\vdef{default-11:SgMc-A:osmuondr:loDelta}   {\ensuremath{{-2.000 } } }
\vdef{default-11:SgMc-A:osmuondr:loDeltaE}   {\ensuremath{{71.653 } } }
\vdef{default-11:SgMc-A:osmuondr:hiDelta}   {\ensuremath{{+0.013 } } }
\vdef{default-11:SgMc-A:osmuondr:hiDeltaE}   {\ensuremath{{0.336 } } }
\vdef{default-11:SgData-A:hlt:loEff}   {\ensuremath{{0.132 } } }
\vdef{default-11:SgData-A:hlt:loEffE}   {\ensuremath{{0.056 } } }
\vdef{default-11:SgData-A:hlt:hiEff}   {\ensuremath{{0.868 } } }
\vdef{default-11:SgData-A:hlt:hiEffE}   {\ensuremath{{0.056 } } }
\vdef{default-11:SgMc-A:hlt:loEff}   {\ensuremath{{0.210 } } }
\vdef{default-11:SgMc-A:hlt:loEffE}   {\ensuremath{{0.062 } } }
\vdef{default-11:SgMc-A:hlt:hiEff}   {\ensuremath{{0.790 } } }
\vdef{default-11:SgMc-A:hlt:hiEffE}   {\ensuremath{{0.065 } } }
\vdef{default-11:SgMc-A:hlt:loDelta}   {\ensuremath{{-0.461 } } }
\vdef{default-11:SgMc-A:hlt:loDeltaE}   {\ensuremath{{0.490 } } }
\vdef{default-11:SgMc-A:hlt:hiDelta}   {\ensuremath{{+0.095 } } }
\vdef{default-11:SgMc-A:hlt:hiDeltaE}   {\ensuremath{{0.104 } } }
\vdef{default-11:SgData-A:muonsid:loEff}   {\ensuremath{{0.233 } } }
\vdef{default-11:SgData-A:muonsid:loEffE}   {\ensuremath{{0.063 } } }
\vdef{default-11:SgData-A:muonsid:hiEff}   {\ensuremath{{0.767 } } }
\vdef{default-11:SgData-A:muonsid:hiEffE}   {\ensuremath{{0.063 } } }
\vdef{default-11:SgMc-A:muonsid:loEff}   {\ensuremath{{0.142 } } }
\vdef{default-11:SgMc-A:muonsid:loEffE}   {\ensuremath{{0.053 } } }
\vdef{default-11:SgMc-A:muonsid:hiEff}   {\ensuremath{{0.858 } } }
\vdef{default-11:SgMc-A:muonsid:hiEffE}   {\ensuremath{{0.056 } } }
\vdef{default-11:SgMc-A:muonsid:loDelta}   {\ensuremath{{+0.482 } } }
\vdef{default-11:SgMc-A:muonsid:loDeltaE}   {\ensuremath{{0.437 } } }
\vdef{default-11:SgMc-A:muonsid:hiDelta}   {\ensuremath{{-0.111 } } }
\vdef{default-11:SgMc-A:muonsid:hiDeltaE}   {\ensuremath{{0.105 } } }
\vdef{default-11:SgData-A:tracksqual:loEff}   {\ensuremath{{0.000 } } }
\vdef{default-11:SgData-A:tracksqual:loEffE}   {\ensuremath{{0.028 } } }
\vdef{default-11:SgData-A:tracksqual:hiEff}   {\ensuremath{{1.000 } } }
\vdef{default-11:SgData-A:tracksqual:hiEffE}   {\ensuremath{{0.028 } } }
\vdef{default-11:SgMc-A:tracksqual:loEff}   {\ensuremath{{0.000 } } }
\vdef{default-11:SgMc-A:tracksqual:loEffE}   {\ensuremath{{0.028 } } }
\vdef{default-11:SgMc-A:tracksqual:hiEff}   {\ensuremath{{1.000 } } }
\vdef{default-11:SgMc-A:tracksqual:hiEffE}   {\ensuremath{{0.039 } } }
\vdef{default-11:SgMc-A:tracksqual:loDelta}   {\ensuremath{{-2.000 } } }
\vdef{default-11:SgMc-A:tracksqual:loDeltaE}   {\ensuremath{{259.894 } } }
\vdef{default-11:SgMc-A:tracksqual:hiDelta}   {\ensuremath{{+0.000 } } }
\vdef{default-11:SgMc-A:tracksqual:hiDeltaE}   {\ensuremath{{0.048 } } }
\vdef{default-11:SgData-A:pvz:loEff}   {\ensuremath{{0.526 } } }
\vdef{default-11:SgData-A:pvz:loEffE}   {\ensuremath{{0.078 } } }
\vdef{default-11:SgData-A:pvz:hiEff}   {\ensuremath{{0.474 } } }
\vdef{default-11:SgData-A:pvz:hiEffE}   {\ensuremath{{0.078 } } }
\vdef{default-11:SgMc-A:pvz:loEff}   {\ensuremath{{0.464 } } }
\vdef{default-11:SgMc-A:pvz:loEffE}   {\ensuremath{{0.078 } } }
\vdef{default-11:SgMc-A:pvz:hiEff}   {\ensuremath{{0.536 } } }
\vdef{default-11:SgMc-A:pvz:hiEffE}   {\ensuremath{{0.078 } } }
\vdef{default-11:SgMc-A:pvz:loDelta}   {\ensuremath{{+0.127 } } }
\vdef{default-11:SgMc-A:pvz:loDeltaE}   {\ensuremath{{0.223 } } }
\vdef{default-11:SgMc-A:pvz:hiDelta}   {\ensuremath{{-0.124 } } }
\vdef{default-11:SgMc-A:pvz:hiDeltaE}   {\ensuremath{{0.219 } } }
\vdef{default-11:SgData-A:pvn:loEff}   {\ensuremath{{1.027 } } }
\vdef{default-11:SgData-A:pvn:loEffE}   {\ensuremath{{0.000 } } }
\vdef{default-11:SgData-A:pvn:hiEff}   {\ensuremath{{1.000 } } }
\vdef{default-11:SgData-A:pvn:hiEffE}   {\ensuremath{{0.025 } } }
\vdef{default-11:SgMc-A:pvn:loEff}   {\ensuremath{{1.000 } } }
\vdef{default-11:SgMc-A:pvn:loEffE}   {\ensuremath{{0.025 } } }
\vdef{default-11:SgMc-A:pvn:hiEff}   {\ensuremath{{1.000 } } }
\vdef{default-11:SgMc-A:pvn:hiEffE}   {\ensuremath{{0.025 } } }
\vdef{default-11:SgMc-A:pvn:loDelta}   {\ensuremath{{+0.027 } } }
\vdef{default-11:SgMc-A:pvn:loDeltaE}   {\ensuremath{{0.025 } } }
\vdef{default-11:SgMc-A:pvn:hiDelta}   {\ensuremath{{+0.000 } } }
\vdef{default-11:SgMc-A:pvn:hiDeltaE}   {\ensuremath{{0.035 } } }
\vdef{default-11:SgData-A:pvavew8:loEff}   {\ensuremath{{0.061 } } }
\vdef{default-11:SgData-A:pvavew8:loEffE}   {\ensuremath{{0.047 } } }
\vdef{default-11:SgData-A:pvavew8:hiEff}   {\ensuremath{{0.939 } } }
\vdef{default-11:SgData-A:pvavew8:hiEffE}   {\ensuremath{{0.047 } } }
\vdef{default-11:SgMc-A:pvavew8:loEff}   {\ensuremath{{0.004 } } }
\vdef{default-11:SgMc-A:pvavew8:loEffE}   {\ensuremath{{0.028 } } }
\vdef{default-11:SgMc-A:pvavew8:hiEff}   {\ensuremath{{0.996 } } }
\vdef{default-11:SgMc-A:pvavew8:hiEffE}   {\ensuremath{{0.039 } } }
\vdef{default-11:SgMc-A:pvavew8:loDelta}   {\ensuremath{{+1.761 } } }
\vdef{default-11:SgMc-A:pvavew8:loDeltaE}   {\ensuremath{{1.630 } } }
\vdef{default-11:SgMc-A:pvavew8:hiDelta}   {\ensuremath{{-0.059 } } }
\vdef{default-11:SgMc-A:pvavew8:hiDeltaE}   {\ensuremath{{0.063 } } }
\vdef{default-11:SgData-A:pvntrk:loEff}   {\ensuremath{{1.000 } } }
\vdef{default-11:SgData-A:pvntrk:loEffE}   {\ensuremath{{0.024 } } }
\vdef{default-11:SgData-A:pvntrk:hiEff}   {\ensuremath{{1.000 } } }
\vdef{default-11:SgData-A:pvntrk:hiEffE}   {\ensuremath{{0.024 } } }
\vdef{default-11:SgMc-A:pvntrk:loEff}   {\ensuremath{{1.000 } } }
\vdef{default-11:SgMc-A:pvntrk:loEffE}   {\ensuremath{{0.025 } } }
\vdef{default-11:SgMc-A:pvntrk:hiEff}   {\ensuremath{{1.000 } } }
\vdef{default-11:SgMc-A:pvntrk:hiEffE}   {\ensuremath{{0.025 } } }
\vdef{default-11:SgMc-A:pvntrk:loDelta}   {\ensuremath{{+0.000 } } }
\vdef{default-11:SgMc-A:pvntrk:loDeltaE}   {\ensuremath{{0.035 } } }
\vdef{default-11:SgMc-A:pvntrk:hiDelta}   {\ensuremath{{+0.000 } } }
\vdef{default-11:SgMc-A:pvntrk:hiDeltaE}   {\ensuremath{{0.035 } } }
\vdef{default-11:SgData-A:muon1pt:loEff}   {\ensuremath{{1.000 } } }
\vdef{default-11:SgData-A:muon1pt:loEffE}   {\ensuremath{{0.026 } } }
\vdef{default-11:SgData-A:muon1pt:hiEff}   {\ensuremath{{1.000 } } }
\vdef{default-11:SgData-A:muon1pt:hiEffE}   {\ensuremath{{0.026 } } }
\vdef{default-11:SgMc-A:muon1pt:loEff}   {\ensuremath{{1.013 } } }
\vdef{default-11:SgMc-A:muon1pt:loEffE}   {\ensuremath{{0.026 } } }
\vdef{default-11:SgMc-A:muon1pt:hiEff}   {\ensuremath{{1.000 } } }
\vdef{default-11:SgMc-A:muon1pt:hiEffE}   {\ensuremath{{0.026 } } }
\vdef{default-11:SgMc-A:muon1pt:loDelta}   {\ensuremath{{-0.012 } } }
\vdef{default-11:SgMc-A:muon1pt:loDeltaE}   {\ensuremath{{0.036 } } }
\vdef{default-11:SgMc-A:muon1pt:hiDelta}   {\ensuremath{{+0.000 } } }
\vdef{default-11:SgMc-A:muon1pt:hiDeltaE}   {\ensuremath{{0.036 } } }
\vdef{default-11:SgData-A:muon2pt:loEff}   {\ensuremath{{0.083 } } }
\vdef{default-11:SgData-A:muon2pt:loEffE}   {\ensuremath{{0.049 } } }
\vdef{default-11:SgData-A:muon2pt:hiEff}   {\ensuremath{{0.917 } } }
\vdef{default-11:SgData-A:muon2pt:hiEffE}   {\ensuremath{{0.049 } } }
\vdef{default-11:SgMc-A:muon2pt:loEff}   {\ensuremath{{0.095 } } }
\vdef{default-11:SgMc-A:muon2pt:loEffE}   {\ensuremath{{0.049 } } }
\vdef{default-11:SgMc-A:muon2pt:hiEff}   {\ensuremath{{0.905 } } }
\vdef{default-11:SgMc-A:muon2pt:hiEffE}   {\ensuremath{{0.054 } } }
\vdef{default-11:SgMc-A:muon2pt:loDelta}   {\ensuremath{{-0.135 } } }
\vdef{default-11:SgMc-A:muon2pt:loDeltaE}   {\ensuremath{{0.779 } } }
\vdef{default-11:SgMc-A:muon2pt:hiDelta}   {\ensuremath{{+0.013 } } }
\vdef{default-11:SgMc-A:muon2pt:hiDeltaE}   {\ensuremath{{0.080 } } }
\vdef{default-11:SgData-A:muonseta:loEff}   {\ensuremath{{0.606 } } }
\vdef{default-11:SgData-A:muonseta:loEffE}   {\ensuremath{{0.059 } } }
\vdef{default-11:SgData-A:muonseta:hiEff}   {\ensuremath{{0.394 } } }
\vdef{default-11:SgData-A:muonseta:hiEffE}   {\ensuremath{{0.059 } } }
\vdef{default-11:SgMc-A:muonseta:loEff}   {\ensuremath{{0.685 } } }
\vdef{default-11:SgMc-A:muonseta:loEffE}   {\ensuremath{{0.056 } } }
\vdef{default-11:SgMc-A:muonseta:hiEff}   {\ensuremath{{0.315 } } }
\vdef{default-11:SgMc-A:muonseta:hiEffE}   {\ensuremath{{0.056 } } }
\vdef{default-11:SgMc-A:muonseta:loDelta}   {\ensuremath{{-0.122 } } }
\vdef{default-11:SgMc-A:muonseta:loDeltaE}   {\ensuremath{{0.127 } } }
\vdef{default-11:SgMc-A:muonseta:hiDelta}   {\ensuremath{{+0.223 } } }
\vdef{default-11:SgMc-A:muonseta:hiDeltaE}   {\ensuremath{{0.229 } } }
\vdef{default-11:SgData-A:pt:loEff}   {\ensuremath{{0.000 } } }
\vdef{default-11:SgData-A:pt:loEffE}   {\ensuremath{{0.014 } } }
\vdef{default-11:SgData-A:pt:hiEff}   {\ensuremath{{1.000 } } }
\vdef{default-11:SgData-A:pt:hiEffE}   {\ensuremath{{0.014 } } }
\vdef{default-11:SgMc-A:pt:loEff}   {\ensuremath{{0.000 } } }
\vdef{default-11:SgMc-A:pt:loEffE}   {\ensuremath{{0.014 } } }
\vdef{default-11:SgMc-A:pt:hiEff}   {\ensuremath{{1.000 } } }
\vdef{default-11:SgMc-A:pt:hiEffE}   {\ensuremath{{0.014 } } }
\vdef{default-11:SgMc-A:pt:loDelta}   {\ensuremath{{\mathrm{NaN} } } }
\vdef{default-11:SgMc-A:pt:loDeltaE}   {\ensuremath{{\mathrm{NaN} } } }
\vdef{default-11:SgMc-A:pt:hiDelta}   {\ensuremath{{+0.000 } } }
\vdef{default-11:SgMc-A:pt:hiDeltaE}   {\ensuremath{{0.020 } } }
\vdef{default-11:SgData-A:p:loEff}   {\ensuremath{{1.030 } } }
\vdef{default-11:SgData-A:p:loEffE}   {\ensuremath{{0.000 } } }
\vdef{default-11:SgData-A:p:hiEff}   {\ensuremath{{1.000 } } }
\vdef{default-11:SgData-A:p:hiEffE}   {\ensuremath{{0.028 } } }
\vdef{default-11:SgMc-A:p:loEff}   {\ensuremath{{1.009 } } }
\vdef{default-11:SgMc-A:p:loEffE}   {\ensuremath{{0.028 } } }
\vdef{default-11:SgMc-A:p:hiEff}   {\ensuremath{{1.000 } } }
\vdef{default-11:SgMc-A:p:hiEffE}   {\ensuremath{{0.028 } } }
\vdef{default-11:SgMc-A:p:loDelta}   {\ensuremath{{+0.021 } } }
\vdef{default-11:SgMc-A:p:loDeltaE}   {\ensuremath{{0.028 } } }
\vdef{default-11:SgMc-A:p:hiDelta}   {\ensuremath{{+0.000 } } }
\vdef{default-11:SgMc-A:p:hiDeltaE}   {\ensuremath{{0.039 } } }
\vdef{default-11:SgData-A:eta:loEff}   {\ensuremath{{0.576 } } }
\vdef{default-11:SgData-A:eta:loEffE}   {\ensuremath{{0.082 } } }
\vdef{default-11:SgData-A:eta:hiEff}   {\ensuremath{{0.424 } } }
\vdef{default-11:SgData-A:eta:hiEffE}   {\ensuremath{{0.082 } } }
\vdef{default-11:SgMc-A:eta:loEff}   {\ensuremath{{0.673 } } }
\vdef{default-11:SgMc-A:eta:loEffE}   {\ensuremath{{0.079 } } }
\vdef{default-11:SgMc-A:eta:hiEff}   {\ensuremath{{0.327 } } }
\vdef{default-11:SgMc-A:eta:hiEffE}   {\ensuremath{{0.077 } } }
\vdef{default-11:SgMc-A:eta:loDelta}   {\ensuremath{{-0.155 } } }
\vdef{default-11:SgMc-A:eta:loDeltaE}   {\ensuremath{{0.184 } } }
\vdef{default-11:SgMc-A:eta:hiDelta}   {\ensuremath{{+0.257 } } }
\vdef{default-11:SgMc-A:eta:hiDeltaE}   {\ensuremath{{0.301 } } }
\vdef{default-11:SgData-A:bdt:loEff}   {\ensuremath{{0.921 } } }
\vdef{default-11:SgData-A:bdt:loEffE}   {\ensuremath{{0.047 } } }
\vdef{default-11:SgData-A:bdt:hiEff}   {\ensuremath{{0.079 } } }
\vdef{default-11:SgData-A:bdt:hiEffE}   {\ensuremath{{0.047 } } }
\vdef{default-11:SgMc-A:bdt:loEff}   {\ensuremath{{0.939 } } }
\vdef{default-11:SgMc-A:bdt:loEffE}   {\ensuremath{{0.047 } } }
\vdef{default-11:SgMc-A:bdt:hiEff}   {\ensuremath{{0.061 } } }
\vdef{default-11:SgMc-A:bdt:hiEffE}   {\ensuremath{{0.041 } } }
\vdef{default-11:SgMc-A:bdt:loDelta}   {\ensuremath{{-0.020 } } }
\vdef{default-11:SgMc-A:bdt:loDeltaE}   {\ensuremath{{0.071 } } }
\vdef{default-11:SgMc-A:bdt:hiDelta}   {\ensuremath{{+0.260 } } }
\vdef{default-11:SgMc-A:bdt:hiDeltaE}   {\ensuremath{{0.885 } } }
\vdef{default-11:SgData-A:fl3d:loEff}   {\ensuremath{{0.990 } } }
\vdef{default-11:SgData-A:fl3d:loEffE}   {\ensuremath{{0.006 } } }
\vdef{default-11:SgData-A:fl3d:hiEff}   {\ensuremath{{0.010 } } }
\vdef{default-11:SgData-A:fl3d:hiEffE}   {\ensuremath{{0.006 } } }
\vdef{default-11:SgMc-A:fl3d:loEff}   {\ensuremath{{0.874 } } }
\vdef{default-11:SgMc-A:fl3d:loEffE}   {\ensuremath{{0.017 } } }
\vdef{default-11:SgMc-A:fl3d:hiEff}   {\ensuremath{{0.126 } } }
\vdef{default-11:SgMc-A:fl3d:hiEffE}   {\ensuremath{{0.017 } } }
\vdef{default-11:SgMc-A:fl3d:loDelta}   {\ensuremath{{+0.124 } } }
\vdef{default-11:SgMc-A:fl3d:loDeltaE}   {\ensuremath{{0.020 } } }
\vdef{default-11:SgMc-A:fl3d:hiDelta}   {\ensuremath{{-1.704 } } }
\vdef{default-11:SgMc-A:fl3d:hiDeltaE}   {\ensuremath{{0.156 } } }
\vdef{default-11:SgData-A:fl3de:loEff}   {\ensuremath{{1.000 } } }
\vdef{default-11:SgData-A:fl3de:loEffE}   {\ensuremath{{0.003 } } }
\vdef{default-11:SgData-A:fl3de:hiEff}   {\ensuremath{{0.036 } } }
\vdef{default-11:SgData-A:fl3de:hiEffE}   {\ensuremath{{0.010 } } }
\vdef{default-11:SgMc-A:fl3de:loEff}   {\ensuremath{{1.000 } } }
\vdef{default-11:SgMc-A:fl3de:loEffE}   {\ensuremath{{0.003 } } }
\vdef{default-11:SgMc-A:fl3de:hiEff}   {\ensuremath{{0.001 } } }
\vdef{default-11:SgMc-A:fl3de:hiEffE}   {\ensuremath{{0.003 } } }
\vdef{default-11:SgMc-A:fl3de:loDelta}   {\ensuremath{{+0.000 } } }
\vdef{default-11:SgMc-A:fl3de:loDeltaE}   {\ensuremath{{0.004 } } }
\vdef{default-11:SgMc-A:fl3de:hiDelta}   {\ensuremath{{+1.924 } } }
\vdef{default-11:SgMc-A:fl3de:hiDeltaE}   {\ensuremath{{0.274 } } }
\vdef{default-11:SgData-A:fls3d:loEff}   {\ensuremath{{0.867 } } }
\vdef{default-11:SgData-A:fls3d:loEffE}   {\ensuremath{{0.017 } } }
\vdef{default-11:SgData-A:fls3d:hiEff}   {\ensuremath{{0.133 } } }
\vdef{default-11:SgData-A:fls3d:hiEffE}   {\ensuremath{{0.017 } } }
\vdef{default-11:SgMc-A:fls3d:loEff}   {\ensuremath{{0.092 } } }
\vdef{default-11:SgMc-A:fls3d:loEffE}   {\ensuremath{{0.015 } } }
\vdef{default-11:SgMc-A:fls3d:hiEff}   {\ensuremath{{0.908 } } }
\vdef{default-11:SgMc-A:fls3d:hiEffE}   {\ensuremath{{0.015 } } }
\vdef{default-11:SgMc-A:fls3d:loDelta}   {\ensuremath{{+1.617 } } }
\vdef{default-11:SgMc-A:fls3d:loDeltaE}   {\ensuremath{{0.055 } } }
\vdef{default-11:SgMc-A:fls3d:hiDelta}   {\ensuremath{{-1.489 } } }
\vdef{default-11:SgMc-A:fls3d:hiDeltaE}   {\ensuremath{{0.058 } } }
\vdef{default-11:SgData-A:flsxy:loEff}   {\ensuremath{{1.000 } } }
\vdef{default-11:SgData-A:flsxy:loEffE}   {\ensuremath{{0.002 } } }
\vdef{default-11:SgData-A:flsxy:hiEff}   {\ensuremath{{1.000 } } }
\vdef{default-11:SgData-A:flsxy:hiEffE}   {\ensuremath{{0.002 } } }
\vdef{default-11:SgMc-A:flsxy:loEff}   {\ensuremath{{1.003 } } }
\vdef{default-11:SgMc-A:flsxy:loEffE}   {\ensuremath{{\mathrm{NaN} } } }
\vdef{default-11:SgMc-A:flsxy:hiEff}   {\ensuremath{{1.000 } } }
\vdef{default-11:SgMc-A:flsxy:hiEffE}   {\ensuremath{{0.002 } } }
\vdef{default-11:SgMc-A:flsxy:loDelta}   {\ensuremath{{-0.003 } } }
\vdef{default-11:SgMc-A:flsxy:loDeltaE}   {\ensuremath{{\mathrm{NaN} } } }
\vdef{default-11:SgMc-A:flsxy:hiDelta}   {\ensuremath{{+0.000 } } }
\vdef{default-11:SgMc-A:flsxy:hiDeltaE}   {\ensuremath{{0.004 } } }
\vdef{default-11:SgData-A:chi2dof:loEff}   {\ensuremath{{0.767 } } }
\vdef{default-11:SgData-A:chi2dof:loEffE}   {\ensuremath{{0.063 } } }
\vdef{default-11:SgData-A:chi2dof:hiEff}   {\ensuremath{{0.233 } } }
\vdef{default-11:SgData-A:chi2dof:hiEffE}   {\ensuremath{{0.063 } } }
\vdef{default-11:SgMc-A:chi2dof:loEff}   {\ensuremath{{0.909 } } }
\vdef{default-11:SgMc-A:chi2dof:loEffE}   {\ensuremath{{0.046 } } }
\vdef{default-11:SgMc-A:chi2dof:hiEff}   {\ensuremath{{0.091 } } }
\vdef{default-11:SgMc-A:chi2dof:hiEffE}   {\ensuremath{{0.042 } } }
\vdef{default-11:SgMc-A:chi2dof:loDelta}   {\ensuremath{{-0.169 } } }
\vdef{default-11:SgMc-A:chi2dof:loDeltaE}   {\ensuremath{{0.096 } } }
\vdef{default-11:SgMc-A:chi2dof:hiDelta}   {\ensuremath{{+0.879 } } }
\vdef{default-11:SgMc-A:chi2dof:hiDeltaE}   {\ensuremath{{0.434 } } }
\vdef{default-11:SgData-A:pchi2dof:loEff}   {\ensuremath{{0.867 } } }
\vdef{default-11:SgData-A:pchi2dof:loEffE}   {\ensuremath{{0.036 } } }
\vdef{default-11:SgData-A:pchi2dof:hiEff}   {\ensuremath{{0.133 } } }
\vdef{default-11:SgData-A:pchi2dof:hiEffE}   {\ensuremath{{0.036 } } }
\vdef{default-11:SgMc-A:pchi2dof:loEff}   {\ensuremath{{0.704 } } }
\vdef{default-11:SgMc-A:pchi2dof:loEffE}   {\ensuremath{{0.048 } } }
\vdef{default-11:SgMc-A:pchi2dof:hiEff}   {\ensuremath{{0.296 } } }
\vdef{default-11:SgMc-A:pchi2dof:hiEffE}   {\ensuremath{{0.048 } } }
\vdef{default-11:SgMc-A:pchi2dof:loDelta}   {\ensuremath{{+0.208 } } }
\vdef{default-11:SgMc-A:pchi2dof:loDeltaE}   {\ensuremath{{0.079 } } }
\vdef{default-11:SgMc-A:pchi2dof:hiDelta}   {\ensuremath{{-0.759 } } }
\vdef{default-11:SgMc-A:pchi2dof:hiDeltaE}   {\ensuremath{{0.270 } } }
\vdef{default-11:SgData-A:alpha:loEff}   {\ensuremath{{0.750 } } }
\vdef{default-11:SgData-A:alpha:loEffE}   {\ensuremath{{0.064 } } }
\vdef{default-11:SgData-A:alpha:hiEff}   {\ensuremath{{0.250 } } }
\vdef{default-11:SgData-A:alpha:hiEffE}   {\ensuremath{{0.064 } } }
\vdef{default-11:SgMc-A:alpha:loEff}   {\ensuremath{{0.984 } } }
\vdef{default-11:SgMc-A:alpha:loEffE}   {\ensuremath{{0.030 } } }
\vdef{default-11:SgMc-A:alpha:hiEff}   {\ensuremath{{0.016 } } }
\vdef{default-11:SgMc-A:alpha:hiEffE}   {\ensuremath{{0.021 } } }
\vdef{default-11:SgMc-A:alpha:loDelta}   {\ensuremath{{-0.270 } } }
\vdef{default-11:SgMc-A:alpha:loDeltaE}   {\ensuremath{{0.089 } } }
\vdef{default-11:SgMc-A:alpha:hiDelta}   {\ensuremath{{+1.760 } } }
\vdef{default-11:SgMc-A:alpha:hiDeltaE}   {\ensuremath{{0.306 } } }
\vdef{default-11:SgData-A:iso:loEff}   {\ensuremath{{0.529 } } }
\vdef{default-11:SgData-A:iso:loEffE}   {\ensuremath{{0.058 } } }
\vdef{default-11:SgData-A:iso:hiEff}   {\ensuremath{{0.471 } } }
\vdef{default-11:SgData-A:iso:hiEffE}   {\ensuremath{{0.058 } } }
\vdef{default-11:SgMc-A:iso:loEff}   {\ensuremath{{0.106 } } }
\vdef{default-11:SgMc-A:iso:loEffE}   {\ensuremath{{0.037 } } }
\vdef{default-11:SgMc-A:iso:hiEff}   {\ensuremath{{0.894 } } }
\vdef{default-11:SgMc-A:iso:hiEffE}   {\ensuremath{{0.039 } } }
\vdef{default-11:SgMc-A:iso:loDelta}   {\ensuremath{{+1.334 } } }
\vdef{default-11:SgMc-A:iso:loDeltaE}   {\ensuremath{{0.203 } } }
\vdef{default-11:SgMc-A:iso:hiDelta}   {\ensuremath{{-0.619 } } }
\vdef{default-11:SgMc-A:iso:hiDeltaE}   {\ensuremath{{0.119 } } }
\vdef{default-11:SgData-A:docatrk:loEff}   {\ensuremath{{0.195 } } }
\vdef{default-11:SgData-A:docatrk:loEffE}   {\ensuremath{{0.061 } } }
\vdef{default-11:SgData-A:docatrk:hiEff}   {\ensuremath{{0.805 } } }
\vdef{default-11:SgData-A:docatrk:hiEffE}   {\ensuremath{{0.061 } } }
\vdef{default-11:SgMc-A:docatrk:loEff}   {\ensuremath{{0.098 } } }
\vdef{default-11:SgMc-A:docatrk:loEffE}   {\ensuremath{{0.049 } } }
\vdef{default-11:SgMc-A:docatrk:hiEff}   {\ensuremath{{0.902 } } }
\vdef{default-11:SgMc-A:docatrk:hiEffE}   {\ensuremath{{0.049 } } }
\vdef{default-11:SgMc-A:docatrk:loDelta}   {\ensuremath{{+0.661 } } }
\vdef{default-11:SgMc-A:docatrk:loDeltaE}   {\ensuremath{{0.528 } } }
\vdef{default-11:SgMc-A:docatrk:hiDelta}   {\ensuremath{{-0.114 } } }
\vdef{default-11:SgMc-A:docatrk:hiDeltaE}   {\ensuremath{{0.094 } } }
\vdef{default-11:SgData-A:isotrk:loEff}   {\ensuremath{{1.000 } } }
\vdef{default-11:SgData-A:isotrk:loEffE}   {\ensuremath{{0.014 } } }
\vdef{default-11:SgData-A:isotrk:hiEff}   {\ensuremath{{1.000 } } }
\vdef{default-11:SgData-A:isotrk:hiEffE}   {\ensuremath{{0.014 } } }
\vdef{default-11:SgMc-A:isotrk:loEff}   {\ensuremath{{1.000 } } }
\vdef{default-11:SgMc-A:isotrk:loEffE}   {\ensuremath{{0.014 } } }
\vdef{default-11:SgMc-A:isotrk:hiEff}   {\ensuremath{{1.000 } } }
\vdef{default-11:SgMc-A:isotrk:hiEffE}   {\ensuremath{{0.014 } } }
\vdef{default-11:SgMc-A:isotrk:loDelta}   {\ensuremath{{+0.000 } } }
\vdef{default-11:SgMc-A:isotrk:loDeltaE}   {\ensuremath{{0.019 } } }
\vdef{default-11:SgMc-A:isotrk:hiDelta}   {\ensuremath{{+0.000 } } }
\vdef{default-11:SgMc-A:isotrk:hiDeltaE}   {\ensuremath{{0.019 } } }
\vdef{default-11:SgData-A:closetrk:loEff}   {\ensuremath{{0.825 } } }
\vdef{default-11:SgData-A:closetrk:loEffE}   {\ensuremath{{0.060 } } }
\vdef{default-11:SgData-A:closetrk:hiEff}   {\ensuremath{{0.175 } } }
\vdef{default-11:SgData-A:closetrk:hiEffE}   {\ensuremath{{0.060 } } }
\vdef{default-11:SgMc-A:closetrk:loEff}   {\ensuremath{{0.969 } } }
\vdef{default-11:SgMc-A:closetrk:loEffE}   {\ensuremath{{0.033 } } }
\vdef{default-11:SgMc-A:closetrk:hiEff}   {\ensuremath{{0.031 } } }
\vdef{default-11:SgMc-A:closetrk:hiEffE}   {\ensuremath{{0.033 } } }
\vdef{default-11:SgMc-A:closetrk:loDelta}   {\ensuremath{{-0.161 } } }
\vdef{default-11:SgMc-A:closetrk:loDeltaE}   {\ensuremath{{0.080 } } }
\vdef{default-11:SgMc-A:closetrk:hiDelta}   {\ensuremath{{+1.404 } } }
\vdef{default-11:SgMc-A:closetrk:hiDeltaE}   {\ensuremath{{0.577 } } }
\vdef{default-11:SgData-A:lip:loEff}   {\ensuremath{{1.000 } } }
\vdef{default-11:SgData-A:lip:loEffE}   {\ensuremath{{0.028 } } }
\vdef{default-11:SgData-A:lip:hiEff}   {\ensuremath{{0.000 } } }
\vdef{default-11:SgData-A:lip:hiEffE}   {\ensuremath{{0.028 } } }
\vdef{default-11:SgMc-A:lip:loEff}   {\ensuremath{{1.000 } } }
\vdef{default-11:SgMc-A:lip:loEffE}   {\ensuremath{{0.028 } } }
\vdef{default-11:SgMc-A:lip:hiEff}   {\ensuremath{{0.000 } } }
\vdef{default-11:SgMc-A:lip:hiEffE}   {\ensuremath{{0.028 } } }
\vdef{default-11:SgMc-A:lip:loDelta}   {\ensuremath{{+0.000 } } }
\vdef{default-11:SgMc-A:lip:loDeltaE}   {\ensuremath{{0.039 } } }
\vdef{default-11:SgMc-A:lip:hiDelta}   {\ensuremath{{\mathrm{NaN} } } }
\vdef{default-11:SgMc-A:lip:hiDeltaE}   {\ensuremath{{\mathrm{NaN} } } }
\vdef{default-11:SgData-A:lip:inEff}   {\ensuremath{{1.000 } } }
\vdef{default-11:SgData-A:lip:inEffE}   {\ensuremath{{0.028 } } }
\vdef{default-11:SgMc-A:lip:inEff}   {\ensuremath{{1.000 } } }
\vdef{default-11:SgMc-A:lip:inEffE}   {\ensuremath{{0.028 } } }
\vdef{default-11:SgMc-A:lip:inDelta}   {\ensuremath{{+0.000 } } }
\vdef{default-11:SgMc-A:lip:inDeltaE}   {\ensuremath{{0.039 } } }
\vdef{default-11:SgData-A:lips:loEff}   {\ensuremath{{1.000 } } }
\vdef{default-11:SgData-A:lips:loEffE}   {\ensuremath{{0.028 } } }
\vdef{default-11:SgData-A:lips:hiEff}   {\ensuremath{{0.000 } } }
\vdef{default-11:SgData-A:lips:hiEffE}   {\ensuremath{{0.028 } } }
\vdef{default-11:SgMc-A:lips:loEff}   {\ensuremath{{1.000 } } }
\vdef{default-11:SgMc-A:lips:loEffE}   {\ensuremath{{0.028 } } }
\vdef{default-11:SgMc-A:lips:hiEff}   {\ensuremath{{0.000 } } }
\vdef{default-11:SgMc-A:lips:hiEffE}   {\ensuremath{{0.028 } } }
\vdef{default-11:SgMc-A:lips:loDelta}   {\ensuremath{{+0.000 } } }
\vdef{default-11:SgMc-A:lips:loDeltaE}   {\ensuremath{{0.039 } } }
\vdef{default-11:SgMc-A:lips:hiDelta}   {\ensuremath{{\mathrm{NaN} } } }
\vdef{default-11:SgMc-A:lips:hiDeltaE}   {\ensuremath{{\mathrm{NaN} } } }
\vdef{default-11:SgData-A:lips:inEff}   {\ensuremath{{1.000 } } }
\vdef{default-11:SgData-A:lips:inEffE}   {\ensuremath{{0.028 } } }
\vdef{default-11:SgMc-A:lips:inEff}   {\ensuremath{{1.000 } } }
\vdef{default-11:SgMc-A:lips:inEffE}   {\ensuremath{{0.028 } } }
\vdef{default-11:SgMc-A:lips:inDelta}   {\ensuremath{{+0.000 } } }
\vdef{default-11:SgMc-A:lips:inDeltaE}   {\ensuremath{{0.039 } } }
\vdef{default-11:SgData-A:ip:loEff}   {\ensuremath{{0.750 } } }
\vdef{default-11:SgData-A:ip:loEffE}   {\ensuremath{{0.064 } } }
\vdef{default-11:SgData-A:ip:hiEff}   {\ensuremath{{0.250 } } }
\vdef{default-11:SgData-A:ip:hiEffE}   {\ensuremath{{0.064 } } }
\vdef{default-11:SgMc-A:ip:loEff}   {\ensuremath{{0.959 } } }
\vdef{default-11:SgMc-A:ip:loEffE}   {\ensuremath{{0.036 } } }
\vdef{default-11:SgMc-A:ip:hiEff}   {\ensuremath{{0.041 } } }
\vdef{default-11:SgMc-A:ip:hiEffE}   {\ensuremath{{0.030 } } }
\vdef{default-11:SgMc-A:ip:loDelta}   {\ensuremath{{-0.245 } } }
\vdef{default-11:SgMc-A:ip:loDeltaE}   {\ensuremath{{0.092 } } }
\vdef{default-11:SgMc-A:ip:hiDelta}   {\ensuremath{{+1.441 } } }
\vdef{default-11:SgMc-A:ip:hiDeltaE}   {\ensuremath{{0.373 } } }
\vdef{default-11:SgData-A:ips:loEff}   {\ensuremath{{0.717 } } }
\vdef{default-11:SgData-A:ips:loEffE}   {\ensuremath{{0.065 } } }
\vdef{default-11:SgData-A:ips:hiEff}   {\ensuremath{{0.283 } } }
\vdef{default-11:SgData-A:ips:hiEffE}   {\ensuremath{{0.065 } } }
\vdef{default-11:SgMc-A:ips:loEff}   {\ensuremath{{0.978 } } }
\vdef{default-11:SgMc-A:ips:loEffE}   {\ensuremath{{0.035 } } }
\vdef{default-11:SgMc-A:ips:hiEff}   {\ensuremath{{0.022 } } }
\vdef{default-11:SgMc-A:ips:hiEffE}   {\ensuremath{{0.029 } } }
\vdef{default-11:SgMc-A:ips:loDelta}   {\ensuremath{{-0.307 } } }
\vdef{default-11:SgMc-A:ips:loDeltaE}   {\ensuremath{{0.095 } } }
\vdef{default-11:SgMc-A:ips:hiDelta}   {\ensuremath{{+1.709 } } }
\vdef{default-11:SgMc-A:ips:hiDeltaE}   {\ensuremath{{0.353 } } }
\vdef{default-11:SgData-A:maxdoca:loEff}   {\ensuremath{{1.000 } } }
\vdef{default-11:SgData-A:maxdoca:loEffE}   {\ensuremath{{0.028 } } }
\vdef{default-11:SgData-A:maxdoca:hiEff}   {\ensuremath{{0.000 } } }
\vdef{default-11:SgData-A:maxdoca:hiEffE}   {\ensuremath{{0.028 } } }
\vdef{default-11:SgMc-A:maxdoca:loEff}   {\ensuremath{{1.000 } } }
\vdef{default-11:SgMc-A:maxdoca:loEffE}   {\ensuremath{{0.028 } } }
\vdef{default-11:SgMc-A:maxdoca:hiEff}   {\ensuremath{{0.000 } } }
\vdef{default-11:SgMc-A:maxdoca:hiEffE}   {\ensuremath{{0.028 } } }
\vdef{default-11:SgMc-A:maxdoca:loDelta}   {\ensuremath{{+0.000 } } }
\vdef{default-11:SgMc-A:maxdoca:loDeltaE}   {\ensuremath{{0.039 } } }
\vdef{default-11:SgMc-A:maxdoca:hiDelta}   {\ensuremath{{\mathrm{NaN} } } }
\vdef{default-11:SgMc-A:maxdoca:hiDeltaE}   {\ensuremath{{\mathrm{NaN} } } }
\vdef{default-11:SgData-A:osiso:loEff}   {\ensuremath{{1.029 } } }
\vdef{default-11:SgData-A:osiso:loEffE}   {\ensuremath{{\mathrm{NaN} } } }
\vdef{default-11:SgData-A:osiso:hiEff}   {\ensuremath{{1.000 } } }
\vdef{default-11:SgData-A:osiso:hiEffE}   {\ensuremath{{0.014 } } }
\vdef{default-11:SgMcPU-A:osiso:loEff}   {\ensuremath{{1.000 } } }
\vdef{default-11:SgMcPU-A:osiso:loEffE}   {\ensuremath{{0.014 } } }
\vdef{default-11:SgMcPU-A:osiso:hiEff}   {\ensuremath{{1.000 } } }
\vdef{default-11:SgMcPU-A:osiso:hiEffE}   {\ensuremath{{0.014 } } }
\vdef{default-11:SgMcPU-A:osiso:loDelta}   {\ensuremath{{+0.029 } } }
\vdef{default-11:SgMcPU-A:osiso:loDeltaE}   {\ensuremath{{\mathrm{NaN} } } }
\vdef{default-11:SgMcPU-A:osiso:hiDelta}   {\ensuremath{{+0.000 } } }
\vdef{default-11:SgMcPU-A:osiso:hiDeltaE}   {\ensuremath{{0.020 } } }
\vdef{default-11:SgData-A:osreliso:loEff}   {\ensuremath{{0.179 } } }
\vdef{default-11:SgData-A:osreliso:loEffE}   {\ensuremath{{0.047 } } }
\vdef{default-11:SgData-A:osreliso:hiEff}   {\ensuremath{{0.821 } } }
\vdef{default-11:SgData-A:osreliso:hiEffE}   {\ensuremath{{0.047 } } }
\vdef{default-11:SgMcPU-A:osreliso:loEff}   {\ensuremath{{0.248 } } }
\vdef{default-11:SgMcPU-A:osreliso:loEffE}   {\ensuremath{{0.052 } } }
\vdef{default-11:SgMcPU-A:osreliso:hiEff}   {\ensuremath{{0.752 } } }
\vdef{default-11:SgMcPU-A:osreliso:hiEffE}   {\ensuremath{{0.052 } } }
\vdef{default-11:SgMcPU-A:osreliso:loDelta}   {\ensuremath{{-0.324 } } }
\vdef{default-11:SgMcPU-A:osreliso:loDeltaE}   {\ensuremath{{0.326 } } }
\vdef{default-11:SgMcPU-A:osreliso:hiDelta}   {\ensuremath{{+0.088 } } }
\vdef{default-11:SgMcPU-A:osreliso:hiDeltaE}   {\ensuremath{{0.090 } } }
\vdef{default-11:SgData-A:osmuonpt:loEff}   {\ensuremath{{0.000 } } }
\vdef{default-11:SgData-A:osmuonpt:loEffE}   {\ensuremath{{0.236 } } }
\vdef{default-11:SgData-A:osmuonpt:hiEff}   {\ensuremath{{1.000 } } }
\vdef{default-11:SgData-A:osmuonpt:hiEffE}   {\ensuremath{{0.236 } } }
\vdef{default-11:SgMcPU-A:osmuonpt:loEff}   {\ensuremath{{0.000 } } }
\vdef{default-11:SgMcPU-A:osmuonpt:loEffE}   {\ensuremath{{0.289 } } }
\vdef{default-11:SgMcPU-A:osmuonpt:hiEff}   {\ensuremath{{1.000 } } }
\vdef{default-11:SgMcPU-A:osmuonpt:hiEffE}   {\ensuremath{{0.289 } } }
\vdef{default-11:SgMcPU-A:osmuonpt:loDelta}   {\ensuremath{{\mathrm{NaN} } } }
\vdef{default-11:SgMcPU-A:osmuonpt:loDeltaE}   {\ensuremath{{\mathrm{NaN} } } }
\vdef{default-11:SgMcPU-A:osmuonpt:hiDelta}   {\ensuremath{{+0.000 } } }
\vdef{default-11:SgMcPU-A:osmuonpt:hiDeltaE}   {\ensuremath{{0.373 } } }
\vdef{default-11:SgData-A:osmuondr:loEff}   {\ensuremath{{0.000 } } }
\vdef{default-11:SgData-A:osmuondr:loEffE}   {\ensuremath{{0.236 } } }
\vdef{default-11:SgData-A:osmuondr:hiEff}   {\ensuremath{{1.000 } } }
\vdef{default-11:SgData-A:osmuondr:hiEffE}   {\ensuremath{{0.236 } } }
\vdef{default-11:SgMcPU-A:osmuondr:loEff}   {\ensuremath{{0.000 } } }
\vdef{default-11:SgMcPU-A:osmuondr:loEffE}   {\ensuremath{{0.236 } } }
\vdef{default-11:SgMcPU-A:osmuondr:hiEff}   {\ensuremath{{1.000 } } }
\vdef{default-11:SgMcPU-A:osmuondr:hiEffE}   {\ensuremath{{0.236 } } }
\vdef{default-11:SgMcPU-A:osmuondr:loDelta}   {\ensuremath{{\mathrm{NaN} } } }
\vdef{default-11:SgMcPU-A:osmuondr:loDeltaE}   {\ensuremath{{\mathrm{NaN} } } }
\vdef{default-11:SgMcPU-A:osmuondr:hiDelta}   {\ensuremath{{+0.000 } } }
\vdef{default-11:SgMcPU-A:osmuondr:hiDeltaE}   {\ensuremath{{0.333 } } }
\vdef{default-11:SgData-A:hlt:loEff}   {\ensuremath{{0.132 } } }
\vdef{default-11:SgData-A:hlt:loEffE}   {\ensuremath{{0.056 } } }
\vdef{default-11:SgData-A:hlt:hiEff}   {\ensuremath{{0.868 } } }
\vdef{default-11:SgData-A:hlt:hiEffE}   {\ensuremath{{0.056 } } }
\vdef{default-11:SgMcPU-A:hlt:loEff}   {\ensuremath{{0.287 } } }
\vdef{default-11:SgMcPU-A:hlt:loEffE}   {\ensuremath{{0.070 } } }
\vdef{default-11:SgMcPU-A:hlt:hiEff}   {\ensuremath{{0.713 } } }
\vdef{default-11:SgMcPU-A:hlt:hiEffE}   {\ensuremath{{0.072 } } }
\vdef{default-11:SgMcPU-A:hlt:loDelta}   {\ensuremath{{-0.744 } } }
\vdef{default-11:SgMcPU-A:hlt:loDeltaE}   {\ensuremath{{0.421 } } }
\vdef{default-11:SgMcPU-A:hlt:hiDelta}   {\ensuremath{{+0.197 } } }
\vdef{default-11:SgMcPU-A:hlt:hiDeltaE}   {\ensuremath{{0.118 } } }
\vdef{default-11:SgData-A:muonsid:loEff}   {\ensuremath{{0.233 } } }
\vdef{default-11:SgData-A:muonsid:loEffE}   {\ensuremath{{0.063 } } }
\vdef{default-11:SgData-A:muonsid:hiEff}   {\ensuremath{{0.767 } } }
\vdef{default-11:SgData-A:muonsid:hiEffE}   {\ensuremath{{0.063 } } }
\vdef{default-11:SgMcPU-A:muonsid:loEff}   {\ensuremath{{0.144 } } }
\vdef{default-11:SgMcPU-A:muonsid:loEffE}   {\ensuremath{{0.053 } } }
\vdef{default-11:SgMcPU-A:muonsid:hiEff}   {\ensuremath{{0.856 } } }
\vdef{default-11:SgMcPU-A:muonsid:hiEffE}   {\ensuremath{{0.056 } } }
\vdef{default-11:SgMcPU-A:muonsid:loDelta}   {\ensuremath{{+0.473 } } }
\vdef{default-11:SgMcPU-A:muonsid:loDeltaE}   {\ensuremath{{0.435 } } }
\vdef{default-11:SgMcPU-A:muonsid:hiDelta}   {\ensuremath{{-0.110 } } }
\vdef{default-11:SgMcPU-A:muonsid:hiDeltaE}   {\ensuremath{{0.105 } } }
\vdef{default-11:SgData-A:tracksqual:loEff}   {\ensuremath{{0.000 } } }
\vdef{default-11:SgData-A:tracksqual:loEffE}   {\ensuremath{{0.028 } } }
\vdef{default-11:SgData-A:tracksqual:hiEff}   {\ensuremath{{1.000 } } }
\vdef{default-11:SgData-A:tracksqual:hiEffE}   {\ensuremath{{0.028 } } }
\vdef{default-11:SgMcPU-A:tracksqual:loEff}   {\ensuremath{{0.002 } } }
\vdef{default-11:SgMcPU-A:tracksqual:loEffE}   {\ensuremath{{0.028 } } }
\vdef{default-11:SgMcPU-A:tracksqual:hiEff}   {\ensuremath{{0.998 } } }
\vdef{default-11:SgMcPU-A:tracksqual:hiEffE}   {\ensuremath{{0.039 } } }
\vdef{default-11:SgMcPU-A:tracksqual:loDelta}   {\ensuremath{{-2.000 } } }
\vdef{default-11:SgMcPU-A:tracksqual:loDeltaE}   {\ensuremath{{73.637 } } }
\vdef{default-11:SgMcPU-A:tracksqual:hiDelta}   {\ensuremath{{+0.002 } } }
\vdef{default-11:SgMcPU-A:tracksqual:hiDeltaE}   {\ensuremath{{0.048 } } }
\vdef{default-11:SgData-A:pvz:loEff}   {\ensuremath{{0.526 } } }
\vdef{default-11:SgData-A:pvz:loEffE}   {\ensuremath{{0.078 } } }
\vdef{default-11:SgData-A:pvz:hiEff}   {\ensuremath{{0.474 } } }
\vdef{default-11:SgData-A:pvz:hiEffE}   {\ensuremath{{0.078 } } }
\vdef{default-11:SgMcPU-A:pvz:loEff}   {\ensuremath{{0.463 } } }
\vdef{default-11:SgMcPU-A:pvz:loEffE}   {\ensuremath{{0.078 } } }
\vdef{default-11:SgMcPU-A:pvz:hiEff}   {\ensuremath{{0.537 } } }
\vdef{default-11:SgMcPU-A:pvz:hiEffE}   {\ensuremath{{0.078 } } }
\vdef{default-11:SgMcPU-A:pvz:loDelta}   {\ensuremath{{+0.128 } } }
\vdef{default-11:SgMcPU-A:pvz:loDeltaE}   {\ensuremath{{0.223 } } }
\vdef{default-11:SgMcPU-A:pvz:hiDelta}   {\ensuremath{{-0.126 } } }
\vdef{default-11:SgMcPU-A:pvz:hiDeltaE}   {\ensuremath{{0.219 } } }
\vdef{default-11:SgData-A:pvn:loEff}   {\ensuremath{{1.027 } } }
\vdef{default-11:SgData-A:pvn:loEffE}   {\ensuremath{{0.000 } } }
\vdef{default-11:SgData-A:pvn:hiEff}   {\ensuremath{{1.000 } } }
\vdef{default-11:SgData-A:pvn:hiEffE}   {\ensuremath{{0.025 } } }
\vdef{default-11:SgMcPU-A:pvn:loEff}   {\ensuremath{{1.075 } } }
\vdef{default-11:SgMcPU-A:pvn:loEffE}   {\ensuremath{{\mathrm{NaN} } } }
\vdef{default-11:SgMcPU-A:pvn:hiEff}   {\ensuremath{{1.000 } } }
\vdef{default-11:SgMcPU-A:pvn:hiEffE}   {\ensuremath{{0.025 } } }
\vdef{default-11:SgMcPU-A:pvn:loDelta}   {\ensuremath{{-0.046 } } }
\vdef{default-11:SgMcPU-A:pvn:loDeltaE}   {\ensuremath{{\mathrm{NaN} } } }
\vdef{default-11:SgMcPU-A:pvn:hiDelta}   {\ensuremath{{+0.000 } } }
\vdef{default-11:SgMcPU-A:pvn:hiDeltaE}   {\ensuremath{{0.035 } } }
\vdef{default-11:SgData-A:pvavew8:loEff}   {\ensuremath{{0.061 } } }
\vdef{default-11:SgData-A:pvavew8:loEffE}   {\ensuremath{{0.047 } } }
\vdef{default-11:SgData-A:pvavew8:hiEff}   {\ensuremath{{0.939 } } }
\vdef{default-11:SgData-A:pvavew8:hiEffE}   {\ensuremath{{0.047 } } }
\vdef{default-11:SgMcPU-A:pvavew8:loEff}   {\ensuremath{{0.003 } } }
\vdef{default-11:SgMcPU-A:pvavew8:loEffE}   {\ensuremath{{0.028 } } }
\vdef{default-11:SgMcPU-A:pvavew8:hiEff}   {\ensuremath{{0.997 } } }
\vdef{default-11:SgMcPU-A:pvavew8:hiEffE}   {\ensuremath{{0.039 } } }
\vdef{default-11:SgMcPU-A:pvavew8:loDelta}   {\ensuremath{{+1.810 } } }
\vdef{default-11:SgMcPU-A:pvavew8:loDeltaE}   {\ensuremath{{1.669 } } }
\vdef{default-11:SgMcPU-A:pvavew8:hiDelta}   {\ensuremath{{-0.059 } } }
\vdef{default-11:SgMcPU-A:pvavew8:hiDeltaE}   {\ensuremath{{0.063 } } }
\vdef{default-11:SgData-A:pvntrk:loEff}   {\ensuremath{{1.000 } } }
\vdef{default-11:SgData-A:pvntrk:loEffE}   {\ensuremath{{0.024 } } }
\vdef{default-11:SgData-A:pvntrk:hiEff}   {\ensuremath{{1.000 } } }
\vdef{default-11:SgData-A:pvntrk:hiEffE}   {\ensuremath{{0.024 } } }
\vdef{default-11:SgMcPU-A:pvntrk:loEff}   {\ensuremath{{1.000 } } }
\vdef{default-11:SgMcPU-A:pvntrk:loEffE}   {\ensuremath{{0.024 } } }
\vdef{default-11:SgMcPU-A:pvntrk:hiEff}   {\ensuremath{{1.000 } } }
\vdef{default-11:SgMcPU-A:pvntrk:hiEffE}   {\ensuremath{{0.024 } } }
\vdef{default-11:SgMcPU-A:pvntrk:loDelta}   {\ensuremath{{+0.000 } } }
\vdef{default-11:SgMcPU-A:pvntrk:loDeltaE}   {\ensuremath{{0.034 } } }
\vdef{default-11:SgMcPU-A:pvntrk:hiDelta}   {\ensuremath{{+0.000 } } }
\vdef{default-11:SgMcPU-A:pvntrk:hiDeltaE}   {\ensuremath{{0.034 } } }
\vdef{default-11:SgData-A:muon1pt:loEff}   {\ensuremath{{1.000 } } }
\vdef{default-11:SgData-A:muon1pt:loEffE}   {\ensuremath{{0.026 } } }
\vdef{default-11:SgData-A:muon1pt:hiEff}   {\ensuremath{{1.000 } } }
\vdef{default-11:SgData-A:muon1pt:hiEffE}   {\ensuremath{{0.026 } } }
\vdef{default-11:SgMcPU-A:muon1pt:loEff}   {\ensuremath{{1.012 } } }
\vdef{default-11:SgMcPU-A:muon1pt:loEffE}   {\ensuremath{{0.026 } } }
\vdef{default-11:SgMcPU-A:muon1pt:hiEff}   {\ensuremath{{1.000 } } }
\vdef{default-11:SgMcPU-A:muon1pt:hiEffE}   {\ensuremath{{0.026 } } }
\vdef{default-11:SgMcPU-A:muon1pt:loDelta}   {\ensuremath{{-0.012 } } }
\vdef{default-11:SgMcPU-A:muon1pt:loDeltaE}   {\ensuremath{{0.036 } } }
\vdef{default-11:SgMcPU-A:muon1pt:hiDelta}   {\ensuremath{{+0.000 } } }
\vdef{default-11:SgMcPU-A:muon1pt:hiDeltaE}   {\ensuremath{{0.036 } } }
\vdef{default-11:SgData-A:muon2pt:loEff}   {\ensuremath{{0.083 } } }
\vdef{default-11:SgData-A:muon2pt:loEffE}   {\ensuremath{{0.049 } } }
\vdef{default-11:SgData-A:muon2pt:hiEff}   {\ensuremath{{0.917 } } }
\vdef{default-11:SgData-A:muon2pt:hiEffE}   {\ensuremath{{0.049 } } }
\vdef{default-11:SgMcPU-A:muon2pt:loEff}   {\ensuremath{{0.006 } } }
\vdef{default-11:SgMcPU-A:muon2pt:loEffE}   {\ensuremath{{0.026 } } }
\vdef{default-11:SgMcPU-A:muon2pt:hiEff}   {\ensuremath{{0.994 } } }
\vdef{default-11:SgMcPU-A:muon2pt:hiEffE}   {\ensuremath{{0.026 } } }
\vdef{default-11:SgMcPU-A:muon2pt:loDelta}   {\ensuremath{{+1.730 } } }
\vdef{default-11:SgMcPU-A:muon2pt:loDeltaE}   {\ensuremath{{1.108 } } }
\vdef{default-11:SgMcPU-A:muon2pt:hiDelta}   {\ensuremath{{-0.081 } } }
\vdef{default-11:SgMcPU-A:muon2pt:hiDeltaE}   {\ensuremath{{0.060 } } }
\vdef{default-11:SgData-A:muonseta:loEff}   {\ensuremath{{0.606 } } }
\vdef{default-11:SgData-A:muonseta:loEffE}   {\ensuremath{{0.059 } } }
\vdef{default-11:SgData-A:muonseta:hiEff}   {\ensuremath{{0.394 } } }
\vdef{default-11:SgData-A:muonseta:hiEffE}   {\ensuremath{{0.059 } } }
\vdef{default-11:SgMcPU-A:muonseta:loEff}   {\ensuremath{{0.721 } } }
\vdef{default-11:SgMcPU-A:muonseta:loEffE}   {\ensuremath{{0.055 } } }
\vdef{default-11:SgMcPU-A:muonseta:hiEff}   {\ensuremath{{0.279 } } }
\vdef{default-11:SgMcPU-A:muonseta:hiEffE}   {\ensuremath{{0.055 } } }
\vdef{default-11:SgMcPU-A:muonseta:loDelta}   {\ensuremath{{-0.174 } } }
\vdef{default-11:SgMcPU-A:muonseta:loDeltaE}   {\ensuremath{{0.122 } } }
\vdef{default-11:SgMcPU-A:muonseta:hiDelta}   {\ensuremath{{+0.343 } } }
\vdef{default-11:SgMcPU-A:muonseta:hiDeltaE}   {\ensuremath{{0.239 } } }
\vdef{default-11:SgData-A:pt:loEff}   {\ensuremath{{0.000 } } }
\vdef{default-11:SgData-A:pt:loEffE}   {\ensuremath{{0.014 } } }
\vdef{default-11:SgData-A:pt:hiEff}   {\ensuremath{{1.000 } } }
\vdef{default-11:SgData-A:pt:hiEffE}   {\ensuremath{{0.014 } } }
\vdef{default-11:SgMcPU-A:pt:loEff}   {\ensuremath{{0.000 } } }
\vdef{default-11:SgMcPU-A:pt:loEffE}   {\ensuremath{{0.014 } } }
\vdef{default-11:SgMcPU-A:pt:hiEff}   {\ensuremath{{1.000 } } }
\vdef{default-11:SgMcPU-A:pt:hiEffE}   {\ensuremath{{0.014 } } }
\vdef{default-11:SgMcPU-A:pt:loDelta}   {\ensuremath{{\mathrm{NaN} } } }
\vdef{default-11:SgMcPU-A:pt:loDeltaE}   {\ensuremath{{\mathrm{NaN} } } }
\vdef{default-11:SgMcPU-A:pt:hiDelta}   {\ensuremath{{+0.000 } } }
\vdef{default-11:SgMcPU-A:pt:hiDeltaE}   {\ensuremath{{0.020 } } }
\vdef{default-11:SgData-A:p:loEff}   {\ensuremath{{1.030 } } }
\vdef{default-11:SgData-A:p:loEffE}   {\ensuremath{{0.000 } } }
\vdef{default-11:SgData-A:p:hiEff}   {\ensuremath{{1.000 } } }
\vdef{default-11:SgData-A:p:hiEffE}   {\ensuremath{{0.028 } } }
\vdef{default-11:SgMcPU-A:p:loEff}   {\ensuremath{{1.010 } } }
\vdef{default-11:SgMcPU-A:p:loEffE}   {\ensuremath{{0.028 } } }
\vdef{default-11:SgMcPU-A:p:hiEff}   {\ensuremath{{1.000 } } }
\vdef{default-11:SgMcPU-A:p:hiEffE}   {\ensuremath{{0.028 } } }
\vdef{default-11:SgMcPU-A:p:loDelta}   {\ensuremath{{+0.019 } } }
\vdef{default-11:SgMcPU-A:p:loDeltaE}   {\ensuremath{{0.027 } } }
\vdef{default-11:SgMcPU-A:p:hiDelta}   {\ensuremath{{+0.000 } } }
\vdef{default-11:SgMcPU-A:p:hiDeltaE}   {\ensuremath{{0.039 } } }
\vdef{default-11:SgData-A:eta:loEff}   {\ensuremath{{0.576 } } }
\vdef{default-11:SgData-A:eta:loEffE}   {\ensuremath{{0.082 } } }
\vdef{default-11:SgData-A:eta:hiEff}   {\ensuremath{{0.424 } } }
\vdef{default-11:SgData-A:eta:hiEffE}   {\ensuremath{{0.082 } } }
\vdef{default-11:SgMcPU-A:eta:loEff}   {\ensuremath{{0.713 } } }
\vdef{default-11:SgMcPU-A:eta:loEffE}   {\ensuremath{{0.077 } } }
\vdef{default-11:SgMcPU-A:eta:hiEff}   {\ensuremath{{0.287 } } }
\vdef{default-11:SgMcPU-A:eta:hiEffE}   {\ensuremath{{0.075 } } }
\vdef{default-11:SgMcPU-A:eta:loDelta}   {\ensuremath{{-0.213 } } }
\vdef{default-11:SgMcPU-A:eta:loDeltaE}   {\ensuremath{{0.178 } } }
\vdef{default-11:SgMcPU-A:eta:hiDelta}   {\ensuremath{{+0.386 } } }
\vdef{default-11:SgMcPU-A:eta:hiDeltaE}   {\ensuremath{{0.314 } } }
\vdef{default-11:SgData-A:bdt:loEff}   {\ensuremath{{0.921 } } }
\vdef{default-11:SgData-A:bdt:loEffE}   {\ensuremath{{0.047 } } }
\vdef{default-11:SgData-A:bdt:hiEff}   {\ensuremath{{0.079 } } }
\vdef{default-11:SgData-A:bdt:hiEffE}   {\ensuremath{{0.047 } } }
\vdef{default-11:SgMcPU-A:bdt:loEff}   {\ensuremath{{0.920 } } }
\vdef{default-11:SgMcPU-A:bdt:loEffE}   {\ensuremath{{0.052 } } }
\vdef{default-11:SgMcPU-A:bdt:hiEff}   {\ensuremath{{0.080 } } }
\vdef{default-11:SgMcPU-A:bdt:hiEffE}   {\ensuremath{{0.047 } } }
\vdef{default-11:SgMcPU-A:bdt:loDelta}   {\ensuremath{{+0.001 } } }
\vdef{default-11:SgMcPU-A:bdt:loDeltaE}   {\ensuremath{{0.076 } } }
\vdef{default-11:SgMcPU-A:bdt:hiDelta}   {\ensuremath{{-0.009 } } }
\vdef{default-11:SgMcPU-A:bdt:hiDeltaE}   {\ensuremath{{0.836 } } }
\vdef{default-11:SgData-A:fl3d:loEff}   {\ensuremath{{0.990 } } }
\vdef{default-11:SgData-A:fl3d:loEffE}   {\ensuremath{{0.006 } } }
\vdef{default-11:SgData-A:fl3d:hiEff}   {\ensuremath{{0.010 } } }
\vdef{default-11:SgData-A:fl3d:hiEffE}   {\ensuremath{{0.006 } } }
\vdef{default-11:SgMcPU-A:fl3d:loEff}   {\ensuremath{{0.878 } } }
\vdef{default-11:SgMcPU-A:fl3d:loEffE}   {\ensuremath{{0.017 } } }
\vdef{default-11:SgMcPU-A:fl3d:hiEff}   {\ensuremath{{0.122 } } }
\vdef{default-11:SgMcPU-A:fl3d:hiEffE}   {\ensuremath{{0.016 } } }
\vdef{default-11:SgMcPU-A:fl3d:loDelta}   {\ensuremath{{+0.120 } } }
\vdef{default-11:SgMcPU-A:fl3d:loDeltaE}   {\ensuremath{{0.020 } } }
\vdef{default-11:SgMcPU-A:fl3d:hiDelta}   {\ensuremath{{-1.696 } } }
\vdef{default-11:SgMcPU-A:fl3d:hiDeltaE}   {\ensuremath{{0.160 } } }
\vdef{default-11:SgData-A:fl3de:loEff}   {\ensuremath{{1.000 } } }
\vdef{default-11:SgData-A:fl3de:loEffE}   {\ensuremath{{0.003 } } }
\vdef{default-11:SgData-A:fl3de:hiEff}   {\ensuremath{{0.036 } } }
\vdef{default-11:SgData-A:fl3de:hiEffE}   {\ensuremath{{0.010 } } }
\vdef{default-11:SgMcPU-A:fl3de:loEff}   {\ensuremath{{1.000 } } }
\vdef{default-11:SgMcPU-A:fl3de:loEffE}   {\ensuremath{{0.003 } } }
\vdef{default-11:SgMcPU-A:fl3de:hiEff}   {\ensuremath{{0.000 } } }
\vdef{default-11:SgMcPU-A:fl3de:hiEffE}   {\ensuremath{{0.003 } } }
\vdef{default-11:SgMcPU-A:fl3de:loDelta}   {\ensuremath{{+0.000 } } }
\vdef{default-11:SgMcPU-A:fl3de:loDeltaE}   {\ensuremath{{0.004 } } }
\vdef{default-11:SgMcPU-A:fl3de:hiDelta}   {\ensuremath{{+2.000 } } }
\vdef{default-11:SgMcPU-A:fl3de:hiDeltaE}   {\ensuremath{{0.283 } } }
\vdef{default-11:SgData-A:fls3d:loEff}   {\ensuremath{{0.867 } } }
\vdef{default-11:SgData-A:fls3d:loEffE}   {\ensuremath{{0.017 } } }
\vdef{default-11:SgData-A:fls3d:hiEff}   {\ensuremath{{0.133 } } }
\vdef{default-11:SgData-A:fls3d:hiEffE}   {\ensuremath{{0.017 } } }
\vdef{default-11:SgMcPU-A:fls3d:loEff}   {\ensuremath{{0.086 } } }
\vdef{default-11:SgMcPU-A:fls3d:loEffE}   {\ensuremath{{0.014 } } }
\vdef{default-11:SgMcPU-A:fls3d:hiEff}   {\ensuremath{{0.914 } } }
\vdef{default-11:SgMcPU-A:fls3d:hiEffE}   {\ensuremath{{0.014 } } }
\vdef{default-11:SgMcPU-A:fls3d:loDelta}   {\ensuremath{{+1.640 } } }
\vdef{default-11:SgMcPU-A:fls3d:loDeltaE}   {\ensuremath{{0.054 } } }
\vdef{default-11:SgMcPU-A:fls3d:hiDelta}   {\ensuremath{{-1.491 } } }
\vdef{default-11:SgMcPU-A:fls3d:hiDeltaE}   {\ensuremath{{0.057 } } }
\vdef{default-11:SgData-A:flsxy:loEff}   {\ensuremath{{1.000 } } }
\vdef{default-11:SgData-A:flsxy:loEffE}   {\ensuremath{{0.002 } } }
\vdef{default-11:SgData-A:flsxy:hiEff}   {\ensuremath{{1.000 } } }
\vdef{default-11:SgData-A:flsxy:hiEffE}   {\ensuremath{{0.002 } } }
\vdef{default-11:SgMcPU-A:flsxy:loEff}   {\ensuremath{{1.005 } } }
\vdef{default-11:SgMcPU-A:flsxy:loEffE}   {\ensuremath{{\mathrm{NaN} } } }
\vdef{default-11:SgMcPU-A:flsxy:hiEff}   {\ensuremath{{1.000 } } }
\vdef{default-11:SgMcPU-A:flsxy:hiEffE}   {\ensuremath{{0.002 } } }
\vdef{default-11:SgMcPU-A:flsxy:loDelta}   {\ensuremath{{-0.005 } } }
\vdef{default-11:SgMcPU-A:flsxy:loDeltaE}   {\ensuremath{{\mathrm{NaN} } } }
\vdef{default-11:SgMcPU-A:flsxy:hiDelta}   {\ensuremath{{+0.000 } } }
\vdef{default-11:SgMcPU-A:flsxy:hiDeltaE}   {\ensuremath{{0.004 } } }
\vdef{default-11:SgData-A:chi2dof:loEff}   {\ensuremath{{0.767 } } }
\vdef{default-11:SgData-A:chi2dof:loEffE}   {\ensuremath{{0.063 } } }
\vdef{default-11:SgData-A:chi2dof:hiEff}   {\ensuremath{{0.233 } } }
\vdef{default-11:SgData-A:chi2dof:hiEffE}   {\ensuremath{{0.063 } } }
\vdef{default-11:SgMcPU-A:chi2dof:loEff}   {\ensuremath{{0.892 } } }
\vdef{default-11:SgMcPU-A:chi2dof:loEffE}   {\ensuremath{{0.050 } } }
\vdef{default-11:SgMcPU-A:chi2dof:hiEff}   {\ensuremath{{0.108 } } }
\vdef{default-11:SgMcPU-A:chi2dof:hiEffE}   {\ensuremath{{0.046 } } }
\vdef{default-11:SgMcPU-A:chi2dof:loDelta}   {\ensuremath{{-0.150 } } }
\vdef{default-11:SgMcPU-A:chi2dof:loDeltaE}   {\ensuremath{{0.099 } } }
\vdef{default-11:SgMcPU-A:chi2dof:hiDelta}   {\ensuremath{{+0.733 } } }
\vdef{default-11:SgMcPU-A:chi2dof:hiDeltaE}   {\ensuremath{{0.441 } } }
\vdef{default-11:SgData-A:pchi2dof:loEff}   {\ensuremath{{0.867 } } }
\vdef{default-11:SgData-A:pchi2dof:loEffE}   {\ensuremath{{0.036 } } }
\vdef{default-11:SgData-A:pchi2dof:hiEff}   {\ensuremath{{0.133 } } }
\vdef{default-11:SgData-A:pchi2dof:hiEffE}   {\ensuremath{{0.036 } } }
\vdef{default-11:SgMcPU-A:pchi2dof:loEff}   {\ensuremath{{0.695 } } }
\vdef{default-11:SgMcPU-A:pchi2dof:loEffE}   {\ensuremath{{0.048 } } }
\vdef{default-11:SgMcPU-A:pchi2dof:hiEff}   {\ensuremath{{0.305 } } }
\vdef{default-11:SgMcPU-A:pchi2dof:hiEffE}   {\ensuremath{{0.048 } } }
\vdef{default-11:SgMcPU-A:pchi2dof:loDelta}   {\ensuremath{{+0.220 } } }
\vdef{default-11:SgMcPU-A:pchi2dof:loDeltaE}   {\ensuremath{{0.080 } } }
\vdef{default-11:SgMcPU-A:pchi2dof:hiDelta}   {\ensuremath{{-0.783 } } }
\vdef{default-11:SgMcPU-A:pchi2dof:hiDeltaE}   {\ensuremath{{0.265 } } }
\vdef{default-11:SgData-A:alpha:loEff}   {\ensuremath{{0.750 } } }
\vdef{default-11:SgData-A:alpha:loEffE}   {\ensuremath{{0.064 } } }
\vdef{default-11:SgData-A:alpha:hiEff}   {\ensuremath{{0.250 } } }
\vdef{default-11:SgData-A:alpha:hiEffE}   {\ensuremath{{0.064 } } }
\vdef{default-11:SgMcPU-A:alpha:loEff}   {\ensuremath{{0.991 } } }
\vdef{default-11:SgMcPU-A:alpha:loEffE}   {\ensuremath{{0.030 } } }
\vdef{default-11:SgMcPU-A:alpha:hiEff}   {\ensuremath{{0.009 } } }
\vdef{default-11:SgMcPU-A:alpha:hiEffE}   {\ensuremath{{0.021 } } }
\vdef{default-11:SgMcPU-A:alpha:loDelta}   {\ensuremath{{-0.277 } } }
\vdef{default-11:SgMcPU-A:alpha:loDeltaE}   {\ensuremath{{0.089 } } }
\vdef{default-11:SgMcPU-A:alpha:hiDelta}   {\ensuremath{{+1.861 } } }
\vdef{default-11:SgMcPU-A:alpha:hiDeltaE}   {\ensuremath{{0.319 } } }
\vdef{default-11:SgData-A:iso:loEff}   {\ensuremath{{0.529 } } }
\vdef{default-11:SgData-A:iso:loEffE}   {\ensuremath{{0.058 } } }
\vdef{default-11:SgData-A:iso:hiEff}   {\ensuremath{{0.471 } } }
\vdef{default-11:SgData-A:iso:hiEffE}   {\ensuremath{{0.058 } } }
\vdef{default-11:SgMcPU-A:iso:loEff}   {\ensuremath{{0.114 } } }
\vdef{default-11:SgMcPU-A:iso:loEffE}   {\ensuremath{{0.037 } } }
\vdef{default-11:SgMcPU-A:iso:hiEff}   {\ensuremath{{0.886 } } }
\vdef{default-11:SgMcPU-A:iso:hiEffE}   {\ensuremath{{0.039 } } }
\vdef{default-11:SgMcPU-A:iso:loDelta}   {\ensuremath{{+1.291 } } }
\vdef{default-11:SgMcPU-A:iso:loDeltaE}   {\ensuremath{{0.199 } } }
\vdef{default-11:SgMcPU-A:iso:hiDelta}   {\ensuremath{{-0.611 } } }
\vdef{default-11:SgMcPU-A:iso:hiDeltaE}   {\ensuremath{{0.119 } } }
\vdef{default-11:SgData-A:docatrk:loEff}   {\ensuremath{{0.195 } } }
\vdef{default-11:SgData-A:docatrk:loEffE}   {\ensuremath{{0.061 } } }
\vdef{default-11:SgData-A:docatrk:hiEff}   {\ensuremath{{0.805 } } }
\vdef{default-11:SgData-A:docatrk:hiEffE}   {\ensuremath{{0.061 } } }
\vdef{default-11:SgMcPU-A:docatrk:loEff}   {\ensuremath{{0.108 } } }
\vdef{default-11:SgMcPU-A:docatrk:loEffE}   {\ensuremath{{0.049 } } }
\vdef{default-11:SgMcPU-A:docatrk:hiEff}   {\ensuremath{{0.892 } } }
\vdef{default-11:SgMcPU-A:docatrk:hiEffE}   {\ensuremath{{0.049 } } }
\vdef{default-11:SgMcPU-A:docatrk:loDelta}   {\ensuremath{{+0.574 } } }
\vdef{default-11:SgMcPU-A:docatrk:loDeltaE}   {\ensuremath{{0.509 } } }
\vdef{default-11:SgMcPU-A:docatrk:hiDelta}   {\ensuremath{{-0.103 } } }
\vdef{default-11:SgMcPU-A:docatrk:hiDeltaE}   {\ensuremath{{0.094 } } }
\vdef{default-11:SgData-A:isotrk:loEff}   {\ensuremath{{1.000 } } }
\vdef{default-11:SgData-A:isotrk:loEffE}   {\ensuremath{{0.014 } } }
\vdef{default-11:SgData-A:isotrk:hiEff}   {\ensuremath{{1.000 } } }
\vdef{default-11:SgData-A:isotrk:hiEffE}   {\ensuremath{{0.014 } } }
\vdef{default-11:SgMcPU-A:isotrk:loEff}   {\ensuremath{{1.000 } } }
\vdef{default-11:SgMcPU-A:isotrk:loEffE}   {\ensuremath{{0.014 } } }
\vdef{default-11:SgMcPU-A:isotrk:hiEff}   {\ensuremath{{1.000 } } }
\vdef{default-11:SgMcPU-A:isotrk:hiEffE}   {\ensuremath{{0.014 } } }
\vdef{default-11:SgMcPU-A:isotrk:loDelta}   {\ensuremath{{+0.000 } } }
\vdef{default-11:SgMcPU-A:isotrk:loDeltaE}   {\ensuremath{{0.019 } } }
\vdef{default-11:SgMcPU-A:isotrk:hiDelta}   {\ensuremath{{+0.000 } } }
\vdef{default-11:SgMcPU-A:isotrk:hiDeltaE}   {\ensuremath{{0.019 } } }
\vdef{default-11:SgData-A:closetrk:loEff}   {\ensuremath{{0.825 } } }
\vdef{default-11:SgData-A:closetrk:loEffE}   {\ensuremath{{0.060 } } }
\vdef{default-11:SgData-A:closetrk:hiEff}   {\ensuremath{{0.175 } } }
\vdef{default-11:SgData-A:closetrk:hiEffE}   {\ensuremath{{0.060 } } }
\vdef{default-11:SgMcPU-A:closetrk:loEff}   {\ensuremath{{0.969 } } }
\vdef{default-11:SgMcPU-A:closetrk:loEffE}   {\ensuremath{{0.039 } } }
\vdef{default-11:SgMcPU-A:closetrk:hiEff}   {\ensuremath{{0.031 } } }
\vdef{default-11:SgMcPU-A:closetrk:hiEffE}   {\ensuremath{{0.032 } } }
\vdef{default-11:SgMcPU-A:closetrk:loDelta}   {\ensuremath{{-0.161 } } }
\vdef{default-11:SgMcPU-A:closetrk:loDeltaE}   {\ensuremath{{0.083 } } }
\vdef{default-11:SgMcPU-A:closetrk:hiDelta}   {\ensuremath{{+1.402 } } }
\vdef{default-11:SgMcPU-A:closetrk:hiDeltaE}   {\ensuremath{{0.565 } } }
\vdef{default-11:SgData-A:lip:loEff}   {\ensuremath{{1.000 } } }
\vdef{default-11:SgData-A:lip:loEffE}   {\ensuremath{{0.028 } } }
\vdef{default-11:SgData-A:lip:hiEff}   {\ensuremath{{0.000 } } }
\vdef{default-11:SgData-A:lip:hiEffE}   {\ensuremath{{0.028 } } }
\vdef{default-11:SgMcPU-A:lip:loEff}   {\ensuremath{{1.000 } } }
\vdef{default-11:SgMcPU-A:lip:loEffE}   {\ensuremath{{0.028 } } }
\vdef{default-11:SgMcPU-A:lip:hiEff}   {\ensuremath{{0.000 } } }
\vdef{default-11:SgMcPU-A:lip:hiEffE}   {\ensuremath{{0.028 } } }
\vdef{default-11:SgMcPU-A:lip:loDelta}   {\ensuremath{{+0.000 } } }
\vdef{default-11:SgMcPU-A:lip:loDeltaE}   {\ensuremath{{0.039 } } }
\vdef{default-11:SgMcPU-A:lip:hiDelta}   {\ensuremath{{\mathrm{NaN} } } }
\vdef{default-11:SgMcPU-A:lip:hiDeltaE}   {\ensuremath{{\mathrm{NaN} } } }
\vdef{default-11:SgData-A:lip:inEff}   {\ensuremath{{1.000 } } }
\vdef{default-11:SgData-A:lip:inEffE}   {\ensuremath{{0.028 } } }
\vdef{default-11:SgMcPU-A:lip:inEff}   {\ensuremath{{1.000 } } }
\vdef{default-11:SgMcPU-A:lip:inEffE}   {\ensuremath{{0.028 } } }
\vdef{default-11:SgMcPU-A:lip:inDelta}   {\ensuremath{{+0.000 } } }
\vdef{default-11:SgMcPU-A:lip:inDeltaE}   {\ensuremath{{0.039 } } }
\vdef{default-11:SgData-A:lips:loEff}   {\ensuremath{{1.000 } } }
\vdef{default-11:SgData-A:lips:loEffE}   {\ensuremath{{0.028 } } }
\vdef{default-11:SgData-A:lips:hiEff}   {\ensuremath{{0.000 } } }
\vdef{default-11:SgData-A:lips:hiEffE}   {\ensuremath{{0.028 } } }
\vdef{default-11:SgMcPU-A:lips:loEff}   {\ensuremath{{1.000 } } }
\vdef{default-11:SgMcPU-A:lips:loEffE}   {\ensuremath{{0.029 } } }
\vdef{default-11:SgMcPU-A:lips:hiEff}   {\ensuremath{{0.000 } } }
\vdef{default-11:SgMcPU-A:lips:hiEffE}   {\ensuremath{{0.029 } } }
\vdef{default-11:SgMcPU-A:lips:loDelta}   {\ensuremath{{+0.000 } } }
\vdef{default-11:SgMcPU-A:lips:loDeltaE}   {\ensuremath{{0.040 } } }
\vdef{default-11:SgMcPU-A:lips:hiDelta}   {\ensuremath{{\mathrm{NaN} } } }
\vdef{default-11:SgMcPU-A:lips:hiDeltaE}   {\ensuremath{{\mathrm{NaN} } } }
\vdef{default-11:SgData-A:lips:inEff}   {\ensuremath{{1.000 } } }
\vdef{default-11:SgData-A:lips:inEffE}   {\ensuremath{{0.028 } } }
\vdef{default-11:SgMcPU-A:lips:inEff}   {\ensuremath{{1.000 } } }
\vdef{default-11:SgMcPU-A:lips:inEffE}   {\ensuremath{{0.029 } } }
\vdef{default-11:SgMcPU-A:lips:inDelta}   {\ensuremath{{+0.000 } } }
\vdef{default-11:SgMcPU-A:lips:inDeltaE}   {\ensuremath{{0.040 } } }
\vdef{default-11:SgData-A:ip:loEff}   {\ensuremath{{0.750 } } }
\vdef{default-11:SgData-A:ip:loEffE}   {\ensuremath{{0.064 } } }
\vdef{default-11:SgData-A:ip:hiEff}   {\ensuremath{{0.250 } } }
\vdef{default-11:SgData-A:ip:hiEffE}   {\ensuremath{{0.064 } } }
\vdef{default-11:SgMcPU-A:ip:loEff}   {\ensuremath{{0.978 } } }
\vdef{default-11:SgMcPU-A:ip:loEffE}   {\ensuremath{{0.030 } } }
\vdef{default-11:SgMcPU-A:ip:hiEff}   {\ensuremath{{0.022 } } }
\vdef{default-11:SgMcPU-A:ip:hiEffE}   {\ensuremath{{0.021 } } }
\vdef{default-11:SgMcPU-A:ip:loDelta}   {\ensuremath{{-0.264 } } }
\vdef{default-11:SgMcPU-A:ip:loDeltaE}   {\ensuremath{{0.089 } } }
\vdef{default-11:SgMcPU-A:ip:hiDelta}   {\ensuremath{{+1.674 } } }
\vdef{default-11:SgMcPU-A:ip:hiDeltaE}   {\ensuremath{{0.297 } } }
\vdef{default-11:SgData-A:ips:loEff}   {\ensuremath{{0.717 } } }
\vdef{default-11:SgData-A:ips:loEffE}   {\ensuremath{{0.065 } } }
\vdef{default-11:SgData-A:ips:hiEff}   {\ensuremath{{0.283 } } }
\vdef{default-11:SgData-A:ips:hiEffE}   {\ensuremath{{0.065 } } }
\vdef{default-11:SgMcPU-A:ips:loEff}   {\ensuremath{{0.964 } } }
\vdef{default-11:SgMcPU-A:ips:loEffE}   {\ensuremath{{0.029 } } }
\vdef{default-11:SgMcPU-A:ips:hiEff}   {\ensuremath{{0.036 } } }
\vdef{default-11:SgMcPU-A:ips:hiEffE}   {\ensuremath{{0.029 } } }
\vdef{default-11:SgMcPU-A:ips:loDelta}   {\ensuremath{{-0.293 } } }
\vdef{default-11:SgMcPU-A:ips:loDeltaE}   {\ensuremath{{0.093 } } }
\vdef{default-11:SgMcPU-A:ips:hiDelta}   {\ensuremath{{+1.544 } } }
\vdef{default-11:SgMcPU-A:ips:hiDeltaE}   {\ensuremath{{0.337 } } }
\vdef{default-11:SgData-A:maxdoca:loEff}   {\ensuremath{{1.000 } } }
\vdef{default-11:SgData-A:maxdoca:loEffE}   {\ensuremath{{0.028 } } }
\vdef{default-11:SgData-A:maxdoca:hiEff}   {\ensuremath{{0.000 } } }
\vdef{default-11:SgData-A:maxdoca:hiEffE}   {\ensuremath{{0.028 } } }
\vdef{default-11:SgMcPU-A:maxdoca:loEff}   {\ensuremath{{1.000 } } }
\vdef{default-11:SgMcPU-A:maxdoca:loEffE}   {\ensuremath{{0.028 } } }
\vdef{default-11:SgMcPU-A:maxdoca:hiEff}   {\ensuremath{{0.000 } } }
\vdef{default-11:SgMcPU-A:maxdoca:hiEffE}   {\ensuremath{{0.028 } } }
\vdef{default-11:SgMcPU-A:maxdoca:loDelta}   {\ensuremath{{+0.000 } } }
\vdef{default-11:SgMcPU-A:maxdoca:loDeltaE}   {\ensuremath{{0.039 } } }
\vdef{default-11:SgMcPU-A:maxdoca:hiDelta}   {\ensuremath{{\mathrm{NaN} } } }
\vdef{default-11:SgMcPU-A:maxdoca:hiDeltaE}   {\ensuremath{{\mathrm{NaN} } } }
\vdef{default-11:SgData-APV0:osiso:loEff}   {\ensuremath{{1.048 } } }
\vdef{default-11:SgData-APV0:osiso:loEffE}   {\ensuremath{{0.000 } } }
\vdef{default-11:SgData-APV0:osiso:hiEff}   {\ensuremath{{1.000 } } }
\vdef{default-11:SgData-APV0:osiso:hiEffE}   {\ensuremath{{0.042 } } }
\vdef{default-11:SgMcPU-APV0:osiso:loEff}   {\ensuremath{{1.000 } } }
\vdef{default-11:SgMcPU-APV0:osiso:loEffE}   {\ensuremath{{0.042 } } }
\vdef{default-11:SgMcPU-APV0:osiso:hiEff}   {\ensuremath{{1.000 } } }
\vdef{default-11:SgMcPU-APV0:osiso:hiEffE}   {\ensuremath{{0.042 } } }
\vdef{default-11:SgMcPU-APV0:osiso:loDelta}   {\ensuremath{{+0.047 } } }
\vdef{default-11:SgMcPU-APV0:osiso:loDeltaE}   {\ensuremath{{0.042 } } }
\vdef{default-11:SgMcPU-APV0:osiso:hiDelta}   {\ensuremath{{+0.000 } } }
\vdef{default-11:SgMcPU-APV0:osiso:hiDeltaE}   {\ensuremath{{0.059 } } }
\vdef{default-11:SgData-APV0:osreliso:loEff}   {\ensuremath{{0.286 } } }
\vdef{default-11:SgData-APV0:osreliso:loEffE}   {\ensuremath{{0.094 } } }
\vdef{default-11:SgData-APV0:osreliso:hiEff}   {\ensuremath{{0.714 } } }
\vdef{default-11:SgData-APV0:osreliso:hiEffE}   {\ensuremath{{0.094 } } }
\vdef{default-11:SgMcPU-APV0:osreliso:loEff}   {\ensuremath{{0.265 } } }
\vdef{default-11:SgMcPU-APV0:osreliso:loEffE}   {\ensuremath{{0.090 } } }
\vdef{default-11:SgMcPU-APV0:osreliso:hiEff}   {\ensuremath{{0.735 } } }
\vdef{default-11:SgMcPU-APV0:osreliso:hiEffE}   {\ensuremath{{0.094 } } }
\vdef{default-11:SgMcPU-APV0:osreliso:loDelta}   {\ensuremath{{+0.074 } } }
\vdef{default-11:SgMcPU-APV0:osreliso:loDeltaE}   {\ensuremath{{0.471 } } }
\vdef{default-11:SgMcPU-APV0:osreliso:hiDelta}   {\ensuremath{{-0.028 } } }
\vdef{default-11:SgMcPU-APV0:osreliso:hiDeltaE}   {\ensuremath{{0.183 } } }
\vdef{default-11:SgData-APV0:osmuonpt:loEff}   {\ensuremath{{\mathrm{NaN} } } }
\vdef{default-11:SgData-APV0:osmuonpt:loEffE}   {\ensuremath{{0.289 } } }
\vdef{default-11:SgData-APV0:osmuonpt:hiEff}   {\ensuremath{{\mathrm{NaN} } } }
\vdef{default-11:SgData-APV0:osmuonpt:hiEffE}   {\ensuremath{{0.289 } } }
\vdef{default-11:SgMcPU-APV0:osmuonpt:loEff}   {\ensuremath{{\mathrm{NaN} } } }
\vdef{default-11:SgMcPU-APV0:osmuonpt:loEffE}   {\ensuremath{{0.289 } } }
\vdef{default-11:SgMcPU-APV0:osmuonpt:hiEff}   {\ensuremath{{\mathrm{NaN} } } }
\vdef{default-11:SgMcPU-APV0:osmuonpt:hiEffE}   {\ensuremath{{0.289 } } }
\vdef{default-11:SgMcPU-APV0:osmuonpt:loDelta}   {\ensuremath{{\mathrm{NaN} } } }
\vdef{default-11:SgMcPU-APV0:osmuonpt:loDeltaE}   {\ensuremath{{\mathrm{NaN} } } }
\vdef{default-11:SgMcPU-APV0:osmuonpt:hiDelta}   {\ensuremath{{\mathrm{NaN} } } }
\vdef{default-11:SgMcPU-APV0:osmuonpt:hiDeltaE}   {\ensuremath{{\mathrm{NaN} } } }
\vdef{default-11:SgData-APV0:osmuondr:loEff}   {\ensuremath{{\mathrm{NaN} } } }
\vdef{default-11:SgData-APV0:osmuondr:loEffE}   {\ensuremath{{0.289 } } }
\vdef{default-11:SgData-APV0:osmuondr:hiEff}   {\ensuremath{{\mathrm{NaN} } } }
\vdef{default-11:SgData-APV0:osmuondr:hiEffE}   {\ensuremath{{0.289 } } }
\vdef{default-11:SgMcPU-APV0:osmuondr:loEff}   {\ensuremath{{\mathrm{NaN} } } }
\vdef{default-11:SgMcPU-APV0:osmuondr:loEffE}   {\ensuremath{{0.289 } } }
\vdef{default-11:SgMcPU-APV0:osmuondr:hiEff}   {\ensuremath{{\mathrm{NaN} } } }
\vdef{default-11:SgMcPU-APV0:osmuondr:hiEffE}   {\ensuremath{{0.289 } } }
\vdef{default-11:SgMcPU-APV0:osmuondr:loDelta}   {\ensuremath{{\mathrm{NaN} } } }
\vdef{default-11:SgMcPU-APV0:osmuondr:loDeltaE}   {\ensuremath{{\mathrm{NaN} } } }
\vdef{default-11:SgMcPU-APV0:osmuondr:hiDelta}   {\ensuremath{{\mathrm{NaN} } } }
\vdef{default-11:SgMcPU-APV0:osmuondr:hiDeltaE}   {\ensuremath{{\mathrm{NaN} } } }
\vdef{default-11:SgData-APV0:hlt:loEff}   {\ensuremath{{0.000 } } }
\vdef{default-11:SgData-APV0:hlt:loEffE}   {\ensuremath{{0.077 } } }
\vdef{default-11:SgData-APV0:hlt:hiEff}   {\ensuremath{{1.000 } } }
\vdef{default-11:SgData-APV0:hlt:hiEffE}   {\ensuremath{{0.077 } } }
\vdef{default-11:SgMcPU-APV0:hlt:loEff}   {\ensuremath{{0.248 } } }
\vdef{default-11:SgMcPU-APV0:hlt:loEffE}   {\ensuremath{{0.120 } } }
\vdef{default-11:SgMcPU-APV0:hlt:hiEff}   {\ensuremath{{0.752 } } }
\vdef{default-11:SgMcPU-APV0:hlt:hiEffE}   {\ensuremath{{0.131 } } }
\vdef{default-11:SgMcPU-APV0:hlt:loDelta}   {\ensuremath{{-2.000 } } }
\vdef{default-11:SgMcPU-APV0:hlt:loDeltaE}   {\ensuremath{{1.236 } } }
\vdef{default-11:SgMcPU-APV0:hlt:hiDelta}   {\ensuremath{{+0.283 } } }
\vdef{default-11:SgMcPU-APV0:hlt:hiDeltaE}   {\ensuremath{{0.186 } } }
\vdef{default-11:SgData-APV0:muonsid:loEff}   {\ensuremath{{0.000 } } }
\vdef{default-11:SgData-APV0:muonsid:loEffE}   {\ensuremath{{0.077 } } }
\vdef{default-11:SgData-APV0:muonsid:hiEff}   {\ensuremath{{1.000 } } }
\vdef{default-11:SgData-APV0:muonsid:hiEffE}   {\ensuremath{{0.077 } } }
\vdef{default-11:SgMcPU-APV0:muonsid:loEff}   {\ensuremath{{0.132 } } }
\vdef{default-11:SgMcPU-APV0:muonsid:loEffE}   {\ensuremath{{0.103 } } }
\vdef{default-11:SgMcPU-APV0:muonsid:hiEff}   {\ensuremath{{0.868 } } }
\vdef{default-11:SgMcPU-APV0:muonsid:hiEffE}   {\ensuremath{{0.120 } } }
\vdef{default-11:SgMcPU-APV0:muonsid:loDelta}   {\ensuremath{{-2.000 } } }
\vdef{default-11:SgMcPU-APV0:muonsid:loDeltaE}   {\ensuremath{{2.326 } } }
\vdef{default-11:SgMcPU-APV0:muonsid:hiDelta}   {\ensuremath{{+0.141 } } }
\vdef{default-11:SgMcPU-APV0:muonsid:hiDeltaE}   {\ensuremath{{0.157 } } }
\vdef{default-11:SgData-APV0:tracksqual:loEff}   {\ensuremath{{0.000 } } }
\vdef{default-11:SgData-APV0:tracksqual:loEffE}   {\ensuremath{{0.077 } } }
\vdef{default-11:SgData-APV0:tracksqual:hiEff}   {\ensuremath{{1.000 } } }
\vdef{default-11:SgData-APV0:tracksqual:hiEffE}   {\ensuremath{{0.077 } } }
\vdef{default-11:SgMcPU-APV0:tracksqual:loEff}   {\ensuremath{{0.000 } } }
\vdef{default-11:SgMcPU-APV0:tracksqual:loEffE}   {\ensuremath{{0.077 } } }
\vdef{default-11:SgMcPU-APV0:tracksqual:hiEff}   {\ensuremath{{1.000 } } }
\vdef{default-11:SgMcPU-APV0:tracksqual:hiEffE}   {\ensuremath{{0.077 } } }
\vdef{default-11:SgMcPU-APV0:tracksqual:loDelta}   {\ensuremath{{\mathrm{NaN} } } }
\vdef{default-11:SgMcPU-APV0:tracksqual:loDeltaE}   {\ensuremath{{\mathrm{NaN} } } }
\vdef{default-11:SgMcPU-APV0:tracksqual:hiDelta}   {\ensuremath{{+0.000 } } }
\vdef{default-11:SgMcPU-APV0:tracksqual:hiDeltaE}   {\ensuremath{{0.108 } } }
\vdef{default-11:SgData-APV0:pvz:loEff}   {\ensuremath{{0.500 } } }
\vdef{default-11:SgData-APV0:pvz:loEffE}   {\ensuremath{{0.139 } } }
\vdef{default-11:SgData-APV0:pvz:hiEff}   {\ensuremath{{0.500 } } }
\vdef{default-11:SgData-APV0:pvz:hiEffE}   {\ensuremath{{0.139 } } }
\vdef{default-11:SgMcPU-APV0:pvz:loEff}   {\ensuremath{{0.465 } } }
\vdef{default-11:SgMcPU-APV0:pvz:loEffE}   {\ensuremath{{0.144 } } }
\vdef{default-11:SgMcPU-APV0:pvz:hiEff}   {\ensuremath{{0.535 } } }
\vdef{default-11:SgMcPU-APV0:pvz:hiEffE}   {\ensuremath{{0.144 } } }
\vdef{default-11:SgMcPU-APV0:pvz:loDelta}   {\ensuremath{{+0.073 } } }
\vdef{default-11:SgMcPU-APV0:pvz:loDeltaE}   {\ensuremath{{0.415 } } }
\vdef{default-11:SgMcPU-APV0:pvz:hiDelta}   {\ensuremath{{-0.068 } } }
\vdef{default-11:SgMcPU-APV0:pvz:hiDeltaE}   {\ensuremath{{0.386 } } }
\vdef{default-11:SgData-APV0:pvn:loEff}   {\ensuremath{{1.000 } } }
\vdef{default-11:SgData-APV0:pvn:loEffE}   {\ensuremath{{0.077 } } }
\vdef{default-11:SgData-APV0:pvn:hiEff}   {\ensuremath{{1.000 } } }
\vdef{default-11:SgData-APV0:pvn:hiEffE}   {\ensuremath{{0.077 } } }
\vdef{default-11:SgMcPU-APV0:pvn:loEff}   {\ensuremath{{1.000 } } }
\vdef{default-11:SgMcPU-APV0:pvn:loEffE}   {\ensuremath{{0.077 } } }
\vdef{default-11:SgMcPU-APV0:pvn:hiEff}   {\ensuremath{{1.000 } } }
\vdef{default-11:SgMcPU-APV0:pvn:hiEffE}   {\ensuremath{{0.077 } } }
\vdef{default-11:SgMcPU-APV0:pvn:loDelta}   {\ensuremath{{+0.000 } } }
\vdef{default-11:SgMcPU-APV0:pvn:loDeltaE}   {\ensuremath{{0.108 } } }
\vdef{default-11:SgMcPU-APV0:pvn:hiDelta}   {\ensuremath{{+0.000 } } }
\vdef{default-11:SgMcPU-APV0:pvn:hiDeltaE}   {\ensuremath{{0.108 } } }
\vdef{default-11:SgData-APV0:pvavew8:loEff}   {\ensuremath{{0.100 } } }
\vdef{default-11:SgData-APV0:pvavew8:loEffE}   {\ensuremath{{0.103 } } }
\vdef{default-11:SgData-APV0:pvavew8:hiEff}   {\ensuremath{{0.900 } } }
\vdef{default-11:SgData-APV0:pvavew8:hiEffE}   {\ensuremath{{0.103 } } }
\vdef{default-11:SgMcPU-APV0:pvavew8:loEff}   {\ensuremath{{0.005 } } }
\vdef{default-11:SgMcPU-APV0:pvavew8:loEffE}   {\ensuremath{{0.077 } } }
\vdef{default-11:SgMcPU-APV0:pvavew8:hiEff}   {\ensuremath{{0.995 } } }
\vdef{default-11:SgMcPU-APV0:pvavew8:hiEffE}   {\ensuremath{{0.103 } } }
\vdef{default-11:SgMcPU-APV0:pvavew8:loDelta}   {\ensuremath{{+1.801 } } }
\vdef{default-11:SgMcPU-APV0:pvavew8:loDeltaE}   {\ensuremath{{2.776 } } }
\vdef{default-11:SgMcPU-APV0:pvavew8:hiDelta}   {\ensuremath{{-0.100 } } }
\vdef{default-11:SgMcPU-APV0:pvavew8:hiDeltaE}   {\ensuremath{{0.154 } } }
\vdef{default-11:SgData-APV0:pvntrk:loEff}   {\ensuremath{{1.000 } } }
\vdef{default-11:SgData-APV0:pvntrk:loEffE}   {\ensuremath{{0.077 } } }
\vdef{default-11:SgData-APV0:pvntrk:hiEff}   {\ensuremath{{1.000 } } }
\vdef{default-11:SgData-APV0:pvntrk:hiEffE}   {\ensuremath{{0.077 } } }
\vdef{default-11:SgMcPU-APV0:pvntrk:loEff}   {\ensuremath{{1.000 } } }
\vdef{default-11:SgMcPU-APV0:pvntrk:loEffE}   {\ensuremath{{0.077 } } }
\vdef{default-11:SgMcPU-APV0:pvntrk:hiEff}   {\ensuremath{{1.000 } } }
\vdef{default-11:SgMcPU-APV0:pvntrk:hiEffE}   {\ensuremath{{0.077 } } }
\vdef{default-11:SgMcPU-APV0:pvntrk:loDelta}   {\ensuremath{{+0.000 } } }
\vdef{default-11:SgMcPU-APV0:pvntrk:loDeltaE}   {\ensuremath{{0.108 } } }
\vdef{default-11:SgMcPU-APV0:pvntrk:hiDelta}   {\ensuremath{{+0.000 } } }
\vdef{default-11:SgMcPU-APV0:pvntrk:hiDeltaE}   {\ensuremath{{0.108 } } }
\vdef{default-11:SgData-APV0:muon1pt:loEff}   {\ensuremath{{1.000 } } }
\vdef{default-11:SgData-APV0:muon1pt:loEffE}   {\ensuremath{{0.077 } } }
\vdef{default-11:SgData-APV0:muon1pt:hiEff}   {\ensuremath{{1.000 } } }
\vdef{default-11:SgData-APV0:muon1pt:hiEffE}   {\ensuremath{{0.077 } } }
\vdef{default-11:SgMcPU-APV0:muon1pt:loEff}   {\ensuremath{{1.005 } } }
\vdef{default-11:SgMcPU-APV0:muon1pt:loEffE}   {\ensuremath{{0.000 } } }
\vdef{default-11:SgMcPU-APV0:muon1pt:hiEff}   {\ensuremath{{1.000 } } }
\vdef{default-11:SgMcPU-APV0:muon1pt:hiEffE}   {\ensuremath{{0.083 } } }
\vdef{default-11:SgMcPU-APV0:muon1pt:loDelta}   {\ensuremath{{-0.005 } } }
\vdef{default-11:SgMcPU-APV0:muon1pt:loDeltaE}   {\ensuremath{{0.077 } } }
\vdef{default-11:SgMcPU-APV0:muon1pt:hiDelta}   {\ensuremath{{+0.000 } } }
\vdef{default-11:SgMcPU-APV0:muon1pt:hiDeltaE}   {\ensuremath{{0.113 } } }
\vdef{default-11:SgData-APV0:muon2pt:loEff}   {\ensuremath{{0.000 } } }
\vdef{default-11:SgData-APV0:muon2pt:loEffE}   {\ensuremath{{0.077 } } }
\vdef{default-11:SgData-APV0:muon2pt:hiEff}   {\ensuremath{{1.000 } } }
\vdef{default-11:SgData-APV0:muon2pt:hiEffE}   {\ensuremath{{0.077 } } }
\vdef{default-11:SgMcPU-APV0:muon2pt:loEff}   {\ensuremath{{0.000 } } }
\vdef{default-11:SgMcPU-APV0:muon2pt:loEffE}   {\ensuremath{{0.083 } } }
\vdef{default-11:SgMcPU-APV0:muon2pt:hiEff}   {\ensuremath{{1.000 } } }
\vdef{default-11:SgMcPU-APV0:muon2pt:hiEffE}   {\ensuremath{{0.083 } } }
\vdef{default-11:SgMcPU-APV0:muon2pt:loDelta}   {\ensuremath{{\mathrm{NaN} } } }
\vdef{default-11:SgMcPU-APV0:muon2pt:loDeltaE}   {\ensuremath{{\mathrm{NaN} } } }
\vdef{default-11:SgMcPU-APV0:muon2pt:hiDelta}   {\ensuremath{{+0.000 } } }
\vdef{default-11:SgMcPU-APV0:muon2pt:hiDeltaE}   {\ensuremath{{0.113 } } }
\vdef{default-11:SgData-APV0:muonseta:loEff}   {\ensuremath{{0.550 } } }
\vdef{default-11:SgData-APV0:muonseta:loEffE}   {\ensuremath{{0.104 } } }
\vdef{default-11:SgData-APV0:muonseta:hiEff}   {\ensuremath{{0.450 } } }
\vdef{default-11:SgData-APV0:muonseta:hiEffE}   {\ensuremath{{0.104 } } }
\vdef{default-11:SgMcPU-APV0:muonseta:loEff}   {\ensuremath{{0.751 } } }
\vdef{default-11:SgMcPU-APV0:muonseta:loEffE}   {\ensuremath{{0.091 } } }
\vdef{default-11:SgMcPU-APV0:muonseta:hiEff}   {\ensuremath{{0.249 } } }
\vdef{default-11:SgMcPU-APV0:muonseta:hiEffE}   {\ensuremath{{0.091 } } }
\vdef{default-11:SgMcPU-APV0:muonseta:loDelta}   {\ensuremath{{-0.309 } } }
\vdef{default-11:SgMcPU-APV0:muonseta:loDeltaE}   {\ensuremath{{0.219 } } }
\vdef{default-11:SgMcPU-APV0:muonseta:hiDelta}   {\ensuremath{{+0.576 } } }
\vdef{default-11:SgMcPU-APV0:muonseta:hiDeltaE}   {\ensuremath{{0.396 } } }
\vdef{default-11:SgData-APV0:pt:loEff}   {\ensuremath{{0.000 } } }
\vdef{default-11:SgData-APV0:pt:loEffE}   {\ensuremath{{0.043 } } }
\vdef{default-11:SgData-APV0:pt:hiEff}   {\ensuremath{{1.000 } } }
\vdef{default-11:SgData-APV0:pt:hiEffE}   {\ensuremath{{0.043 } } }
\vdef{default-11:SgMcPU-APV0:pt:loEff}   {\ensuremath{{0.000 } } }
\vdef{default-11:SgMcPU-APV0:pt:loEffE}   {\ensuremath{{0.045 } } }
\vdef{default-11:SgMcPU-APV0:pt:hiEff}   {\ensuremath{{1.000 } } }
\vdef{default-11:SgMcPU-APV0:pt:hiEffE}   {\ensuremath{{0.045 } } }
\vdef{default-11:SgMcPU-APV0:pt:loDelta}   {\ensuremath{{\mathrm{NaN} } } }
\vdef{default-11:SgMcPU-APV0:pt:loDeltaE}   {\ensuremath{{\mathrm{NaN} } } }
\vdef{default-11:SgMcPU-APV0:pt:hiDelta}   {\ensuremath{{+0.000 } } }
\vdef{default-11:SgMcPU-APV0:pt:hiDeltaE}   {\ensuremath{{0.063 } } }
\vdef{default-11:SgData-APV0:p:loEff}   {\ensuremath{{1.000 } } }
\vdef{default-11:SgData-APV0:p:loEffE}   {\ensuremath{{0.077 } } }
\vdef{default-11:SgData-APV0:p:hiEff}   {\ensuremath{{1.000 } } }
\vdef{default-11:SgData-APV0:p:hiEffE}   {\ensuremath{{0.077 } } }
\vdef{default-11:SgMcPU-APV0:p:loEff}   {\ensuremath{{1.010 } } }
\vdef{default-11:SgMcPU-APV0:p:loEffE}   {\ensuremath{{0.077 } } }
\vdef{default-11:SgMcPU-APV0:p:hiEff}   {\ensuremath{{1.000 } } }
\vdef{default-11:SgMcPU-APV0:p:hiEffE}   {\ensuremath{{0.077 } } }
\vdef{default-11:SgMcPU-APV0:p:loDelta}   {\ensuremath{{-0.010 } } }
\vdef{default-11:SgMcPU-APV0:p:loDeltaE}   {\ensuremath{{0.108 } } }
\vdef{default-11:SgMcPU-APV0:p:hiDelta}   {\ensuremath{{+0.000 } } }
\vdef{default-11:SgMcPU-APV0:p:hiDeltaE}   {\ensuremath{{0.108 } } }
\vdef{default-11:SgData-APV0:eta:loEff}   {\ensuremath{{0.500 } } }
\vdef{default-11:SgData-APV0:eta:loEffE}   {\ensuremath{{0.139 } } }
\vdef{default-11:SgData-APV0:eta:hiEff}   {\ensuremath{{0.500 } } }
\vdef{default-11:SgData-APV0:eta:hiEffE}   {\ensuremath{{0.139 } } }
\vdef{default-11:SgMcPU-APV0:eta:loEff}   {\ensuremath{{0.764 } } }
\vdef{default-11:SgMcPU-APV0:eta:loEffE}   {\ensuremath{{0.131 } } }
\vdef{default-11:SgMcPU-APV0:eta:hiEff}   {\ensuremath{{0.236 } } }
\vdef{default-11:SgMcPU-APV0:eta:hiEffE}   {\ensuremath{{0.120 } } }
\vdef{default-11:SgMcPU-APV0:eta:loDelta}   {\ensuremath{{-0.418 } } }
\vdef{default-11:SgMcPU-APV0:eta:loDeltaE}   {\ensuremath{{0.312 } } }
\vdef{default-11:SgMcPU-APV0:eta:hiDelta}   {\ensuremath{{+0.719 } } }
\vdef{default-11:SgMcPU-APV0:eta:hiDeltaE}   {\ensuremath{{0.505 } } }
\vdef{default-11:SgData-APV0:bdt:loEff}   {\ensuremath{{0.900 } } }
\vdef{default-11:SgData-APV0:bdt:loEffE}   {\ensuremath{{0.103 } } }
\vdef{default-11:SgData-APV0:bdt:hiEff}   {\ensuremath{{0.100 } } }
\vdef{default-11:SgData-APV0:bdt:hiEffE}   {\ensuremath{{0.103 } } }
\vdef{default-11:SgMcPU-APV0:bdt:loEff}   {\ensuremath{{0.925 } } }
\vdef{default-11:SgMcPU-APV0:bdt:loEffE}   {\ensuremath{{0.083 } } }
\vdef{default-11:SgMcPU-APV0:bdt:hiEff}   {\ensuremath{{0.075 } } }
\vdef{default-11:SgMcPU-APV0:bdt:hiEffE}   {\ensuremath{{0.083 } } }
\vdef{default-11:SgMcPU-APV0:bdt:loDelta}   {\ensuremath{{-0.028 } } }
\vdef{default-11:SgMcPU-APV0:bdt:loDeltaE}   {\ensuremath{{0.146 } } }
\vdef{default-11:SgMcPU-APV0:bdt:hiDelta}   {\ensuremath{{+0.288 } } }
\vdef{default-11:SgMcPU-APV0:bdt:hiDeltaE}   {\ensuremath{{1.485 } } }
\vdef{default-11:SgData-APV0:fl3d:loEff}   {\ensuremath{{1.000 } } }
\vdef{default-11:SgData-APV0:fl3d:loEffE}   {\ensuremath{{0.008 } } }
\vdef{default-11:SgData-APV0:fl3d:hiEff}   {\ensuremath{{0.000 } } }
\vdef{default-11:SgData-APV0:fl3d:hiEffE}   {\ensuremath{{0.008 } } }
\vdef{default-11:SgMcPU-APV0:fl3d:loEff}   {\ensuremath{{0.867 } } }
\vdef{default-11:SgMcPU-APV0:fl3d:loEffE}   {\ensuremath{{0.031 } } }
\vdef{default-11:SgMcPU-APV0:fl3d:hiEff}   {\ensuremath{{0.133 } } }
\vdef{default-11:SgMcPU-APV0:fl3d:hiEffE}   {\ensuremath{{0.031 } } }
\vdef{default-11:SgMcPU-APV0:fl3d:loDelta}   {\ensuremath{{+0.143 } } }
\vdef{default-11:SgMcPU-APV0:fl3d:loDeltaE}   {\ensuremath{{0.036 } } }
\vdef{default-11:SgMcPU-APV0:fl3d:hiDelta}   {\ensuremath{{-2.000 } } }
\vdef{default-11:SgMcPU-APV0:fl3d:hiDeltaE}   {\ensuremath{{0.246 } } }
\vdef{default-11:SgData-APV0:fl3de:loEff}   {\ensuremath{{1.000 } } }
\vdef{default-11:SgData-APV0:fl3de:loEffE}   {\ensuremath{{0.009 } } }
\vdef{default-11:SgData-APV0:fl3de:hiEff}   {\ensuremath{{0.044 } } }
\vdef{default-11:SgData-APV0:fl3de:hiEffE}   {\ensuremath{{0.020 } } }
\vdef{default-11:SgMcPU-APV0:fl3de:loEff}   {\ensuremath{{1.000 } } }
\vdef{default-11:SgMcPU-APV0:fl3de:loEffE}   {\ensuremath{{0.009 } } }
\vdef{default-11:SgMcPU-APV0:fl3de:hiEff}   {\ensuremath{{0.000 } } }
\vdef{default-11:SgMcPU-APV0:fl3de:hiEffE}   {\ensuremath{{0.009 } } }
\vdef{default-11:SgMcPU-APV0:fl3de:loDelta}   {\ensuremath{{+0.000 } } }
\vdef{default-11:SgMcPU-APV0:fl3de:loDeltaE}   {\ensuremath{{0.012 } } }
\vdef{default-11:SgMcPU-APV0:fl3de:hiDelta}   {\ensuremath{{+2.000 } } }
\vdef{default-11:SgMcPU-APV0:fl3de:hiDeltaE}   {\ensuremath{{0.786 } } }
\vdef{default-11:SgData-APV0:fls3d:loEff}   {\ensuremath{{0.891 } } }
\vdef{default-11:SgData-APV0:fls3d:loEffE}   {\ensuremath{{0.029 } } }
\vdef{default-11:SgData-APV0:fls3d:hiEff}   {\ensuremath{{0.109 } } }
\vdef{default-11:SgData-APV0:fls3d:hiEffE}   {\ensuremath{{0.029 } } }
\vdef{default-11:SgMcPU-APV0:fls3d:loEff}   {\ensuremath{{0.099 } } }
\vdef{default-11:SgMcPU-APV0:fls3d:loEffE}   {\ensuremath{{0.027 } } }
\vdef{default-11:SgMcPU-APV0:fls3d:hiEff}   {\ensuremath{{0.901 } } }
\vdef{default-11:SgMcPU-APV0:fls3d:hiEffE}   {\ensuremath{{0.028 } } }
\vdef{default-11:SgMcPU-APV0:fls3d:loDelta}   {\ensuremath{{+1.601 } } }
\vdef{default-11:SgMcPU-APV0:fls3d:loDeltaE}   {\ensuremath{{0.099 } } }
\vdef{default-11:SgMcPU-APV0:fls3d:hiDelta}   {\ensuremath{{-1.568 } } }
\vdef{default-11:SgMcPU-APV0:fls3d:hiDeltaE}   {\ensuremath{{0.103 } } }
\vdef{default-11:SgData-APV0:flsxy:loEff}   {\ensuremath{{1.000 } } }
\vdef{default-11:SgData-APV0:flsxy:loEffE}   {\ensuremath{{0.008 } } }
\vdef{default-11:SgData-APV0:flsxy:hiEff}   {\ensuremath{{1.000 } } }
\vdef{default-11:SgData-APV0:flsxy:hiEffE}   {\ensuremath{{0.008 } } }
\vdef{default-11:SgMcPU-APV0:flsxy:loEff}   {\ensuremath{{1.000 } } }
\vdef{default-11:SgMcPU-APV0:flsxy:loEffE}   {\ensuremath{{0.008 } } }
\vdef{default-11:SgMcPU-APV0:flsxy:hiEff}   {\ensuremath{{1.000 } } }
\vdef{default-11:SgMcPU-APV0:flsxy:hiEffE}   {\ensuremath{{0.008 } } }
\vdef{default-11:SgMcPU-APV0:flsxy:loDelta}   {\ensuremath{{+0.000 } } }
\vdef{default-11:SgMcPU-APV0:flsxy:loDeltaE}   {\ensuremath{{0.012 } } }
\vdef{default-11:SgMcPU-APV0:flsxy:hiDelta}   {\ensuremath{{+0.000 } } }
\vdef{default-11:SgMcPU-APV0:flsxy:hiDeltaE}   {\ensuremath{{0.012 } } }
\vdef{default-11:SgData-APV0:chi2dof:loEff}   {\ensuremath{{0.833 } } }
\vdef{default-11:SgData-APV0:chi2dof:loEffE}   {\ensuremath{{0.106 } } }
\vdef{default-11:SgData-APV0:chi2dof:hiEff}   {\ensuremath{{0.167 } } }
\vdef{default-11:SgData-APV0:chi2dof:hiEffE}   {\ensuremath{{0.106 } } }
\vdef{default-11:SgMcPU-APV0:chi2dof:loEff}   {\ensuremath{{0.905 } } }
\vdef{default-11:SgMcPU-APV0:chi2dof:loEffE}   {\ensuremath{{0.106 } } }
\vdef{default-11:SgMcPU-APV0:chi2dof:hiEff}   {\ensuremath{{0.095 } } }
\vdef{default-11:SgMcPU-APV0:chi2dof:hiEffE}   {\ensuremath{{0.090 } } }
\vdef{default-11:SgMcPU-APV0:chi2dof:loDelta}   {\ensuremath{{-0.083 } } }
\vdef{default-11:SgMcPU-APV0:chi2dof:loDeltaE}   {\ensuremath{{0.173 } } }
\vdef{default-11:SgMcPU-APV0:chi2dof:hiDelta}   {\ensuremath{{+0.550 } } }
\vdef{default-11:SgMcPU-APV0:chi2dof:hiDeltaE}   {\ensuremath{{1.059 } } }
\vdef{default-11:SgData-APV0:pchi2dof:loEff}   {\ensuremath{{0.900 } } }
\vdef{default-11:SgData-APV0:pchi2dof:loEffE}   {\ensuremath{{0.058 } } }
\vdef{default-11:SgData-APV0:pchi2dof:hiEff}   {\ensuremath{{0.100 } } }
\vdef{default-11:SgData-APV0:pchi2dof:hiEffE}   {\ensuremath{{0.058 } } }
\vdef{default-11:SgMcPU-APV0:pchi2dof:loEff}   {\ensuremath{{0.659 } } }
\vdef{default-11:SgMcPU-APV0:pchi2dof:loEffE}   {\ensuremath{{0.085 } } }
\vdef{default-11:SgMcPU-APV0:pchi2dof:hiEff}   {\ensuremath{{0.341 } } }
\vdef{default-11:SgMcPU-APV0:pchi2dof:hiEffE}   {\ensuremath{{0.085 } } }
\vdef{default-11:SgMcPU-APV0:pchi2dof:loDelta}   {\ensuremath{{+0.309 } } }
\vdef{default-11:SgMcPU-APV0:pchi2dof:loDeltaE}   {\ensuremath{{0.140 } } }
\vdef{default-11:SgMcPU-APV0:pchi2dof:hiDelta}   {\ensuremath{{-1.093 } } }
\vdef{default-11:SgMcPU-APV0:pchi2dof:hiDeltaE}   {\ensuremath{{0.440 } } }
\vdef{default-11:SgData-APV0:alpha:loEff}   {\ensuremath{{0.769 } } }
\vdef{default-11:SgData-APV0:alpha:loEffE}   {\ensuremath{{0.111 } } }
\vdef{default-11:SgData-APV0:alpha:hiEff}   {\ensuremath{{0.231 } } }
\vdef{default-11:SgData-APV0:alpha:hiEffE}   {\ensuremath{{0.111 } } }
\vdef{default-11:SgMcPU-APV0:alpha:loEff}   {\ensuremath{{1.000 } } }
\vdef{default-11:SgMcPU-APV0:alpha:loEffE}   {\ensuremath{{0.066 } } }
\vdef{default-11:SgMcPU-APV0:alpha:hiEff}   {\ensuremath{{0.000 } } }
\vdef{default-11:SgMcPU-APV0:alpha:hiEffE}   {\ensuremath{{0.066 } } }
\vdef{default-11:SgMcPU-APV0:alpha:loDelta}   {\ensuremath{{-0.261 } } }
\vdef{default-11:SgMcPU-APV0:alpha:loDeltaE}   {\ensuremath{{0.156 } } }
\vdef{default-11:SgMcPU-APV0:alpha:hiDelta}   {\ensuremath{{+2.000 } } }
\vdef{default-11:SgMcPU-APV0:alpha:hiDeltaE}   {\ensuremath{{1.153 } } }
\vdef{default-11:SgData-APV0:iso:loEff}   {\ensuremath{{0.545 } } }
\vdef{default-11:SgData-APV0:iso:loEffE}   {\ensuremath{{0.100 } } }
\vdef{default-11:SgData-APV0:iso:hiEff}   {\ensuremath{{0.455 } } }
\vdef{default-11:SgData-APV0:iso:hiEffE}   {\ensuremath{{0.100 } } }
\vdef{default-11:SgMcPU-APV0:iso:loEff}   {\ensuremath{{0.095 } } }
\vdef{default-11:SgMcPU-APV0:iso:loEffE}   {\ensuremath{{0.069 } } }
\vdef{default-11:SgMcPU-APV0:iso:hiEff}   {\ensuremath{{0.905 } } }
\vdef{default-11:SgMcPU-APV0:iso:hiEffE}   {\ensuremath{{0.069 } } }
\vdef{default-11:SgMcPU-APV0:iso:loDelta}   {\ensuremath{{+1.408 } } }
\vdef{default-11:SgMcPU-APV0:iso:loDeltaE}   {\ensuremath{{0.377 } } }
\vdef{default-11:SgMcPU-APV0:iso:hiDelta}   {\ensuremath{{-0.663 } } }
\vdef{default-11:SgMcPU-APV0:iso:hiDeltaE}   {\ensuremath{{0.207 } } }
\vdef{default-11:SgData-APV0:docatrk:loEff}   {\ensuremath{{0.231 } } }
\vdef{default-11:SgData-APV0:docatrk:loEffE}   {\ensuremath{{0.111 } } }
\vdef{default-11:SgData-APV0:docatrk:hiEff}   {\ensuremath{{0.769 } } }
\vdef{default-11:SgData-APV0:docatrk:hiEffE}   {\ensuremath{{0.111 } } }
\vdef{default-11:SgMcPU-APV0:docatrk:loEff}   {\ensuremath{{0.074 } } }
\vdef{default-11:SgMcPU-APV0:docatrk:loEffE}   {\ensuremath{{0.062 } } }
\vdef{default-11:SgMcPU-APV0:docatrk:hiEff}   {\ensuremath{{0.926 } } }
\vdef{default-11:SgMcPU-APV0:docatrk:hiEffE}   {\ensuremath{{0.085 } } }
\vdef{default-11:SgMcPU-APV0:docatrk:loDelta}   {\ensuremath{{+1.033 } } }
\vdef{default-11:SgMcPU-APV0:docatrk:loDeltaE}   {\ensuremath{{0.714 } } }
\vdef{default-11:SgMcPU-APV0:docatrk:hiDelta}   {\ensuremath{{-0.185 } } }
\vdef{default-11:SgMcPU-APV0:docatrk:hiDeltaE}   {\ensuremath{{0.169 } } }
\vdef{default-11:SgData-APV0:isotrk:loEff}   {\ensuremath{{1.000 } } }
\vdef{default-11:SgData-APV0:isotrk:loEffE}   {\ensuremath{{0.040 } } }
\vdef{default-11:SgData-APV0:isotrk:hiEff}   {\ensuremath{{1.000 } } }
\vdef{default-11:SgData-APV0:isotrk:hiEffE}   {\ensuremath{{0.040 } } }
\vdef{default-11:SgMcPU-APV0:isotrk:loEff}   {\ensuremath{{1.000 } } }
\vdef{default-11:SgMcPU-APV0:isotrk:loEffE}   {\ensuremath{{0.042 } } }
\vdef{default-11:SgMcPU-APV0:isotrk:hiEff}   {\ensuremath{{1.000 } } }
\vdef{default-11:SgMcPU-APV0:isotrk:hiEffE}   {\ensuremath{{0.042 } } }
\vdef{default-11:SgMcPU-APV0:isotrk:loDelta}   {\ensuremath{{+0.000 } } }
\vdef{default-11:SgMcPU-APV0:isotrk:loDeltaE}   {\ensuremath{{0.058 } } }
\vdef{default-11:SgMcPU-APV0:isotrk:hiDelta}   {\ensuremath{{+0.000 } } }
\vdef{default-11:SgMcPU-APV0:isotrk:hiDeltaE}   {\ensuremath{{0.058 } } }
\vdef{default-11:SgData-APV0:closetrk:loEff}   {\ensuremath{{0.769 } } }
\vdef{default-11:SgData-APV0:closetrk:loEffE}   {\ensuremath{{0.111 } } }
\vdef{default-11:SgData-APV0:closetrk:hiEff}   {\ensuremath{{0.231 } } }
\vdef{default-11:SgData-APV0:closetrk:hiEffE}   {\ensuremath{{0.111 } } }
\vdef{default-11:SgMcPU-APV0:closetrk:loEff}   {\ensuremath{{0.965 } } }
\vdef{default-11:SgMcPU-APV0:closetrk:loEffE}   {\ensuremath{{0.085 } } }
\vdef{default-11:SgMcPU-APV0:closetrk:hiEff}   {\ensuremath{{0.035 } } }
\vdef{default-11:SgMcPU-APV0:closetrk:hiEffE}   {\ensuremath{{0.062 } } }
\vdef{default-11:SgMcPU-APV0:closetrk:loDelta}   {\ensuremath{{-0.225 } } }
\vdef{default-11:SgMcPU-APV0:closetrk:loDeltaE}   {\ensuremath{{0.166 } } }
\vdef{default-11:SgMcPU-APV0:closetrk:hiDelta}   {\ensuremath{{+1.469 } } }
\vdef{default-11:SgMcPU-APV0:closetrk:hiDeltaE}   {\ensuremath{{0.842 } } }
\vdef{default-11:SgData-APV0:lip:loEff}   {\ensuremath{{1.000 } } }
\vdef{default-11:SgData-APV0:lip:loEffE}   {\ensuremath{{0.077 } } }
\vdef{default-11:SgData-APV0:lip:hiEff}   {\ensuremath{{0.000 } } }
\vdef{default-11:SgData-APV0:lip:hiEffE}   {\ensuremath{{0.077 } } }
\vdef{default-11:SgMcPU-APV0:lip:loEff}   {\ensuremath{{1.000 } } }
\vdef{default-11:SgMcPU-APV0:lip:loEffE}   {\ensuremath{{0.083 } } }
\vdef{default-11:SgMcPU-APV0:lip:hiEff}   {\ensuremath{{0.000 } } }
\vdef{default-11:SgMcPU-APV0:lip:hiEffE}   {\ensuremath{{0.083 } } }
\vdef{default-11:SgMcPU-APV0:lip:loDelta}   {\ensuremath{{+0.000 } } }
\vdef{default-11:SgMcPU-APV0:lip:loDeltaE}   {\ensuremath{{0.113 } } }
\vdef{default-11:SgMcPU-APV0:lip:hiDelta}   {\ensuremath{{\mathrm{NaN} } } }
\vdef{default-11:SgMcPU-APV0:lip:hiDeltaE}   {\ensuremath{{\mathrm{NaN} } } }
\vdef{default-11:SgData-APV0:lip:inEff}   {\ensuremath{{1.000 } } }
\vdef{default-11:SgData-APV0:lip:inEffE}   {\ensuremath{{0.077 } } }
\vdef{default-11:SgMcPU-APV0:lip:inEff}   {\ensuremath{{1.000 } } }
\vdef{default-11:SgMcPU-APV0:lip:inEffE}   {\ensuremath{{0.083 } } }
\vdef{default-11:SgMcPU-APV0:lip:inDelta}   {\ensuremath{{+0.000 } } }
\vdef{default-11:SgMcPU-APV0:lip:inDeltaE}   {\ensuremath{{0.113 } } }
\vdef{default-11:SgData-APV0:lips:loEff}   {\ensuremath{{1.000 } } }
\vdef{default-11:SgData-APV0:lips:loEffE}   {\ensuremath{{0.077 } } }
\vdef{default-11:SgData-APV0:lips:hiEff}   {\ensuremath{{0.000 } } }
\vdef{default-11:SgData-APV0:lips:hiEffE}   {\ensuremath{{0.077 } } }
\vdef{default-11:SgMcPU-APV0:lips:loEff}   {\ensuremath{{1.000 } } }
\vdef{default-11:SgMcPU-APV0:lips:loEffE}   {\ensuremath{{0.083 } } }
\vdef{default-11:SgMcPU-APV0:lips:hiEff}   {\ensuremath{{0.000 } } }
\vdef{default-11:SgMcPU-APV0:lips:hiEffE}   {\ensuremath{{0.083 } } }
\vdef{default-11:SgMcPU-APV0:lips:loDelta}   {\ensuremath{{+0.000 } } }
\vdef{default-11:SgMcPU-APV0:lips:loDeltaE}   {\ensuremath{{0.113 } } }
\vdef{default-11:SgMcPU-APV0:lips:hiDelta}   {\ensuremath{{\mathrm{NaN} } } }
\vdef{default-11:SgMcPU-APV0:lips:hiDeltaE}   {\ensuremath{{\mathrm{NaN} } } }
\vdef{default-11:SgData-APV0:lips:inEff}   {\ensuremath{{1.000 } } }
\vdef{default-11:SgData-APV0:lips:inEffE}   {\ensuremath{{0.077 } } }
\vdef{default-11:SgMcPU-APV0:lips:inEff}   {\ensuremath{{1.000 } } }
\vdef{default-11:SgMcPU-APV0:lips:inEffE}   {\ensuremath{{0.083 } } }
\vdef{default-11:SgMcPU-APV0:lips:inDelta}   {\ensuremath{{+0.000 } } }
\vdef{default-11:SgMcPU-APV0:lips:inDeltaE}   {\ensuremath{{0.113 } } }
\vdef{default-11:SgData-APV0:ip:loEff}   {\ensuremath{{0.769 } } }
\vdef{default-11:SgData-APV0:ip:loEffE}   {\ensuremath{{0.111 } } }
\vdef{default-11:SgData-APV0:ip:hiEff}   {\ensuremath{{0.231 } } }
\vdef{default-11:SgData-APV0:ip:hiEffE}   {\ensuremath{{0.111 } } }
\vdef{default-11:SgMcPU-APV0:ip:loEff}   {\ensuremath{{0.974 } } }
\vdef{default-11:SgMcPU-APV0:ip:loEffE}   {\ensuremath{{0.085 } } }
\vdef{default-11:SgMcPU-APV0:ip:hiEff}   {\ensuremath{{0.026 } } }
\vdef{default-11:SgMcPU-APV0:ip:hiEffE}   {\ensuremath{{0.062 } } }
\vdef{default-11:SgMcPU-APV0:ip:loDelta}   {\ensuremath{{-0.235 } } }
\vdef{default-11:SgMcPU-APV0:ip:loDeltaE}   {\ensuremath{{0.166 } } }
\vdef{default-11:SgMcPU-APV0:ip:hiDelta}   {\ensuremath{{+1.602 } } }
\vdef{default-11:SgMcPU-APV0:ip:hiDeltaE}   {\ensuremath{{0.893 } } }
\vdef{default-11:SgData-APV0:ips:loEff}   {\ensuremath{{0.714 } } }
\vdef{default-11:SgData-APV0:ips:loEffE}   {\ensuremath{{0.112 } } }
\vdef{default-11:SgData-APV0:ips:hiEff}   {\ensuremath{{0.286 } } }
\vdef{default-11:SgData-APV0:ips:hiEffE}   {\ensuremath{{0.112 } } }
\vdef{default-11:SgMcPU-APV0:ips:loEff}   {\ensuremath{{0.960 } } }
\vdef{default-11:SgMcPU-APV0:ips:loEffE}   {\ensuremath{{0.080 } } }
\vdef{default-11:SgMcPU-APV0:ips:hiEff}   {\ensuremath{{0.040 } } }
\vdef{default-11:SgMcPU-APV0:ips:hiEffE}   {\ensuremath{{0.059 } } }
\vdef{default-11:SgMcPU-APV0:ips:loDelta}   {\ensuremath{{-0.293 } } }
\vdef{default-11:SgMcPU-APV0:ips:loDeltaE}   {\ensuremath{{0.174 } } }
\vdef{default-11:SgMcPU-APV0:ips:hiDelta}   {\ensuremath{{+1.507 } } }
\vdef{default-11:SgMcPU-APV0:ips:hiDeltaE}   {\ensuremath{{0.654 } } }
\vdef{default-11:SgData-APV0:maxdoca:loEff}   {\ensuremath{{1.000 } } }
\vdef{default-11:SgData-APV0:maxdoca:loEffE}   {\ensuremath{{0.077 } } }
\vdef{default-11:SgData-APV0:maxdoca:hiEff}   {\ensuremath{{0.000 } } }
\vdef{default-11:SgData-APV0:maxdoca:hiEffE}   {\ensuremath{{0.077 } } }
\vdef{default-11:SgMcPU-APV0:maxdoca:loEff}   {\ensuremath{{1.000 } } }
\vdef{default-11:SgMcPU-APV0:maxdoca:loEffE}   {\ensuremath{{0.083 } } }
\vdef{default-11:SgMcPU-APV0:maxdoca:hiEff}   {\ensuremath{{0.000 } } }
\vdef{default-11:SgMcPU-APV0:maxdoca:hiEffE}   {\ensuremath{{0.083 } } }
\vdef{default-11:SgMcPU-APV0:maxdoca:loDelta}   {\ensuremath{{+0.000 } } }
\vdef{default-11:SgMcPU-APV0:maxdoca:loDeltaE}   {\ensuremath{{0.113 } } }
\vdef{default-11:SgMcPU-APV0:maxdoca:hiDelta}   {\ensuremath{{\mathrm{NaN} } } }
\vdef{default-11:SgMcPU-APV0:maxdoca:hiDeltaE}   {\ensuremath{{\mathrm{NaN} } } }
\vdef{default-11:SgData-APV1:osiso:loEff}   {\ensuremath{{1.000 } } }
\vdef{default-11:SgData-APV1:osiso:loEffE}   {\ensuremath{{0.045 } } }
\vdef{default-11:SgData-APV1:osiso:hiEff}   {\ensuremath{{1.000 } } }
\vdef{default-11:SgData-APV1:osiso:hiEffE}   {\ensuremath{{0.045 } } }
\vdef{default-11:SgMcPU-APV1:osiso:loEff}   {\ensuremath{{1.000 } } }
\vdef{default-11:SgMcPU-APV1:osiso:loEffE}   {\ensuremath{{0.048 } } }
\vdef{default-11:SgMcPU-APV1:osiso:hiEff}   {\ensuremath{{1.000 } } }
\vdef{default-11:SgMcPU-APV1:osiso:hiEffE}   {\ensuremath{{0.048 } } }
\vdef{default-11:SgMcPU-APV1:osiso:loDelta}   {\ensuremath{{+0.000 } } }
\vdef{default-11:SgMcPU-APV1:osiso:loDeltaE}   {\ensuremath{{0.066 } } }
\vdef{default-11:SgMcPU-APV1:osiso:hiDelta}   {\ensuremath{{+0.000 } } }
\vdef{default-11:SgMcPU-APV1:osiso:hiDeltaE}   {\ensuremath{{0.066 } } }
\vdef{default-11:SgData-APV1:osreliso:loEff}   {\ensuremath{{0.105 } } }
\vdef{default-11:SgData-APV1:osreliso:loEffE}   {\ensuremath{{0.075 } } }
\vdef{default-11:SgData-APV1:osreliso:hiEff}   {\ensuremath{{0.895 } } }
\vdef{default-11:SgData-APV1:osreliso:hiEffE}   {\ensuremath{{0.075 } } }
\vdef{default-11:SgMcPU-APV1:osreliso:loEff}   {\ensuremath{{0.238 } } }
\vdef{default-11:SgMcPU-APV1:osreliso:loEffE}   {\ensuremath{{0.094 } } }
\vdef{default-11:SgMcPU-APV1:osreliso:hiEff}   {\ensuremath{{0.762 } } }
\vdef{default-11:SgMcPU-APV1:osreliso:hiEffE}   {\ensuremath{{0.094 } } }
\vdef{default-11:SgMcPU-APV1:osreliso:loDelta}   {\ensuremath{{-0.775 } } }
\vdef{default-11:SgMcPU-APV1:osreliso:loDeltaE}   {\ensuremath{{0.690 } } }
\vdef{default-11:SgMcPU-APV1:osreliso:hiDelta}   {\ensuremath{{+0.161 } } }
\vdef{default-11:SgMcPU-APV1:osreliso:hiDeltaE}   {\ensuremath{{0.149 } } }
\vdef{default-11:SgData-APV1:osmuonpt:loEff}   {\ensuremath{{0.000 } } }
\vdef{default-11:SgData-APV1:osmuonpt:loEffE}   {\ensuremath{{0.236 } } }
\vdef{default-11:SgData-APV1:osmuonpt:hiEff}   {\ensuremath{{1.000 } } }
\vdef{default-11:SgData-APV1:osmuonpt:hiEffE}   {\ensuremath{{0.236 } } }
\vdef{default-11:SgMcPU-APV1:osmuonpt:loEff}   {\ensuremath{{0.000 } } }
\vdef{default-11:SgMcPU-APV1:osmuonpt:loEffE}   {\ensuremath{{0.289 } } }
\vdef{default-11:SgMcPU-APV1:osmuonpt:hiEff}   {\ensuremath{{1.000 } } }
\vdef{default-11:SgMcPU-APV1:osmuonpt:hiEffE}   {\ensuremath{{0.289 } } }
\vdef{default-11:SgMcPU-APV1:osmuonpt:loDelta}   {\ensuremath{{\mathrm{NaN} } } }
\vdef{default-11:SgMcPU-APV1:osmuonpt:loDeltaE}   {\ensuremath{{\mathrm{NaN} } } }
\vdef{default-11:SgMcPU-APV1:osmuonpt:hiDelta}   {\ensuremath{{+0.000 } } }
\vdef{default-11:SgMcPU-APV1:osmuonpt:hiDeltaE}   {\ensuremath{{0.373 } } }
\vdef{default-11:SgData-APV1:osmuondr:loEff}   {\ensuremath{{0.000 } } }
\vdef{default-11:SgData-APV1:osmuondr:loEffE}   {\ensuremath{{0.236 } } }
\vdef{default-11:SgData-APV1:osmuondr:hiEff}   {\ensuremath{{1.000 } } }
\vdef{default-11:SgData-APV1:osmuondr:hiEffE}   {\ensuremath{{0.236 } } }
\vdef{default-11:SgMcPU-APV1:osmuondr:loEff}   {\ensuremath{{0.000 } } }
\vdef{default-11:SgMcPU-APV1:osmuondr:loEffE}   {\ensuremath{{0.289 } } }
\vdef{default-11:SgMcPU-APV1:osmuondr:hiEff}   {\ensuremath{{1.000 } } }
\vdef{default-11:SgMcPU-APV1:osmuondr:hiEffE}   {\ensuremath{{0.289 } } }
\vdef{default-11:SgMcPU-APV1:osmuondr:loDelta}   {\ensuremath{{\mathrm{NaN} } } }
\vdef{default-11:SgMcPU-APV1:osmuondr:loDeltaE}   {\ensuremath{{\mathrm{NaN} } } }
\vdef{default-11:SgMcPU-APV1:osmuondr:hiDelta}   {\ensuremath{{+0.000 } } }
\vdef{default-11:SgMcPU-APV1:osmuondr:hiDeltaE}   {\ensuremath{{0.373 } } }
\vdef{default-11:SgData-APV1:hlt:loEff}   {\ensuremath{{0.231 } } }
\vdef{default-11:SgData-APV1:hlt:loEffE}   {\ensuremath{{0.111 } } }
\vdef{default-11:SgData-APV1:hlt:hiEff}   {\ensuremath{{0.769 } } }
\vdef{default-11:SgData-APV1:hlt:hiEffE}   {\ensuremath{{0.111 } } }
\vdef{default-11:SgMcPU-APV1:hlt:loEff}   {\ensuremath{{0.356 } } }
\vdef{default-11:SgMcPU-APV1:hlt:loEffE}   {\ensuremath{{0.118 } } }
\vdef{default-11:SgMcPU-APV1:hlt:hiEff}   {\ensuremath{{0.644 } } }
\vdef{default-11:SgMcPU-APV1:hlt:hiEffE}   {\ensuremath{{0.122 } } }
\vdef{default-11:SgMcPU-APV1:hlt:loDelta}   {\ensuremath{{-0.426 } } }
\vdef{default-11:SgMcPU-APV1:hlt:loDeltaE}   {\ensuremath{{0.556 } } }
\vdef{default-11:SgMcPU-APV1:hlt:hiDelta}   {\ensuremath{{+0.177 } } }
\vdef{default-11:SgMcPU-APV1:hlt:hiDeltaE}   {\ensuremath{{0.236 } } }
\vdef{default-11:SgData-APV1:muonsid:loEff}   {\ensuremath{{0.286 } } }
\vdef{default-11:SgData-APV1:muonsid:loEffE}   {\ensuremath{{0.112 } } }
\vdef{default-11:SgData-APV1:muonsid:hiEff}   {\ensuremath{{0.714 } } }
\vdef{default-11:SgData-APV1:muonsid:hiEffE}   {\ensuremath{{0.112 } } }
\vdef{default-11:SgMcPU-APV1:muonsid:loEff}   {\ensuremath{{0.120 } } }
\vdef{default-11:SgMcPU-APV1:muonsid:loEffE}   {\ensuremath{{0.080 } } }
\vdef{default-11:SgMcPU-APV1:muonsid:hiEff}   {\ensuremath{{0.880 } } }
\vdef{default-11:SgMcPU-APV1:muonsid:hiEffE}   {\ensuremath{{0.095 } } }
\vdef{default-11:SgMcPU-APV1:muonsid:loDelta}   {\ensuremath{{+0.815 } } }
\vdef{default-11:SgMcPU-APV1:muonsid:loDeltaE}   {\ensuremath{{0.646 } } }
\vdef{default-11:SgMcPU-APV1:muonsid:hiDelta}   {\ensuremath{{-0.208 } } }
\vdef{default-11:SgMcPU-APV1:muonsid:hiDeltaE}   {\ensuremath{{0.189 } } }
\vdef{default-11:SgData-APV1:tracksqual:loEff}   {\ensuremath{{0.000 } } }
\vdef{default-11:SgData-APV1:tracksqual:loEffE}   {\ensuremath{{0.077 } } }
\vdef{default-11:SgData-APV1:tracksqual:hiEff}   {\ensuremath{{1.000 } } }
\vdef{default-11:SgData-APV1:tracksqual:hiEffE}   {\ensuremath{{0.077 } } }
\vdef{default-11:SgMcPU-APV1:tracksqual:loEff}   {\ensuremath{{0.005 } } }
\vdef{default-11:SgMcPU-APV1:tracksqual:loEffE}   {\ensuremath{{0.077 } } }
\vdef{default-11:SgMcPU-APV1:tracksqual:hiEff}   {\ensuremath{{0.995 } } }
\vdef{default-11:SgMcPU-APV1:tracksqual:hiEffE}   {\ensuremath{{0.103 } } }
\vdef{default-11:SgMcPU-APV1:tracksqual:loDelta}   {\ensuremath{{-2.000 } } }
\vdef{default-11:SgMcPU-APV1:tracksqual:loDeltaE}   {\ensuremath{{65.311 } } }
\vdef{default-11:SgMcPU-APV1:tracksqual:hiDelta}   {\ensuremath{{+0.005 } } }
\vdef{default-11:SgMcPU-APV1:tracksqual:hiDeltaE}   {\ensuremath{{0.129 } } }
\vdef{default-11:SgData-APV1:pvz:loEff}   {\ensuremath{{0.615 } } }
\vdef{default-11:SgData-APV1:pvz:loEffE}   {\ensuremath{{0.122 } } }
\vdef{default-11:SgData-APV1:pvz:hiEff}   {\ensuremath{{0.385 } } }
\vdef{default-11:SgData-APV1:pvz:hiEffE}   {\ensuremath{{0.122 } } }
\vdef{default-11:SgMcPU-APV1:pvz:loEff}   {\ensuremath{{0.492 } } }
\vdef{default-11:SgMcPU-APV1:pvz:loEffE}   {\ensuremath{{0.129 } } }
\vdef{default-11:SgMcPU-APV1:pvz:hiEff}   {\ensuremath{{0.508 } } }
\vdef{default-11:SgMcPU-APV1:pvz:hiEffE}   {\ensuremath{{0.129 } } }
\vdef{default-11:SgMcPU-APV1:pvz:loDelta}   {\ensuremath{{+0.222 } } }
\vdef{default-11:SgMcPU-APV1:pvz:loDeltaE}   {\ensuremath{{0.325 } } }
\vdef{default-11:SgMcPU-APV1:pvz:hiDelta}   {\ensuremath{{-0.276 } } }
\vdef{default-11:SgMcPU-APV1:pvz:hiDeltaE}   {\ensuremath{{0.400 } } }
\vdef{default-11:SgData-APV1:pvn:loEff}   {\ensuremath{{1.083 } } }
\vdef{default-11:SgData-APV1:pvn:loEffE}   {\ensuremath{{0.000 } } }
\vdef{default-11:SgData-APV1:pvn:hiEff}   {\ensuremath{{1.000 } } }
\vdef{default-11:SgData-APV1:pvn:hiEffE}   {\ensuremath{{0.066 } } }
\vdef{default-11:SgMcPU-APV1:pvn:loEff}   {\ensuremath{{1.246 } } }
\vdef{default-11:SgMcPU-APV1:pvn:loEffE}   {\ensuremath{{\mathrm{NaN} } } }
\vdef{default-11:SgMcPU-APV1:pvn:hiEff}   {\ensuremath{{1.000 } } }
\vdef{default-11:SgMcPU-APV1:pvn:hiEffE}   {\ensuremath{{0.066 } } }
\vdef{default-11:SgMcPU-APV1:pvn:loDelta}   {\ensuremath{{-0.140 } } }
\vdef{default-11:SgMcPU-APV1:pvn:loDeltaE}   {\ensuremath{{\mathrm{NaN} } } }
\vdef{default-11:SgMcPU-APV1:pvn:hiDelta}   {\ensuremath{{+0.000 } } }
\vdef{default-11:SgMcPU-APV1:pvn:hiDeltaE}   {\ensuremath{{0.094 } } }
\vdef{default-11:SgData-APV1:pvavew8:loEff}   {\ensuremath{{0.000 } } }
\vdef{default-11:SgData-APV1:pvavew8:loEffE}   {\ensuremath{{0.077 } } }
\vdef{default-11:SgData-APV1:pvavew8:hiEff}   {\ensuremath{{1.000 } } }
\vdef{default-11:SgData-APV1:pvavew8:hiEffE}   {\ensuremath{{0.077 } } }
\vdef{default-11:SgMcPU-APV1:pvavew8:loEff}   {\ensuremath{{0.005 } } }
\vdef{default-11:SgMcPU-APV1:pvavew8:loEffE}   {\ensuremath{{0.077 } } }
\vdef{default-11:SgMcPU-APV1:pvavew8:hiEff}   {\ensuremath{{0.995 } } }
\vdef{default-11:SgMcPU-APV1:pvavew8:hiEffE}   {\ensuremath{{0.103 } } }
\vdef{default-11:SgMcPU-APV1:pvavew8:loDelta}   {\ensuremath{{-2.000 } } }
\vdef{default-11:SgMcPU-APV1:pvavew8:loDeltaE}   {\ensuremath{{65.004 } } }
\vdef{default-11:SgMcPU-APV1:pvavew8:hiDelta}   {\ensuremath{{+0.005 } } }
\vdef{default-11:SgMcPU-APV1:pvavew8:hiDeltaE}   {\ensuremath{{0.129 } } }
\vdef{default-11:SgData-APV1:pvntrk:loEff}   {\ensuremath{{1.000 } } }
\vdef{default-11:SgData-APV1:pvntrk:loEffE}   {\ensuremath{{0.062 } } }
\vdef{default-11:SgData-APV1:pvntrk:hiEff}   {\ensuremath{{1.000 } } }
\vdef{default-11:SgData-APV1:pvntrk:hiEffE}   {\ensuremath{{0.062 } } }
\vdef{default-11:SgMcPU-APV1:pvntrk:loEff}   {\ensuremath{{1.000 } } }
\vdef{default-11:SgMcPU-APV1:pvntrk:loEffE}   {\ensuremath{{0.062 } } }
\vdef{default-11:SgMcPU-APV1:pvntrk:hiEff}   {\ensuremath{{1.000 } } }
\vdef{default-11:SgMcPU-APV1:pvntrk:hiEffE}   {\ensuremath{{0.062 } } }
\vdef{default-11:SgMcPU-APV1:pvntrk:loDelta}   {\ensuremath{{+0.000 } } }
\vdef{default-11:SgMcPU-APV1:pvntrk:loDeltaE}   {\ensuremath{{0.088 } } }
\vdef{default-11:SgMcPU-APV1:pvntrk:hiDelta}   {\ensuremath{{+0.000 } } }
\vdef{default-11:SgMcPU-APV1:pvntrk:hiDeltaE}   {\ensuremath{{0.088 } } }
\vdef{default-11:SgData-APV1:muon1pt:loEff}   {\ensuremath{{1.000 } } }
\vdef{default-11:SgData-APV1:muon1pt:loEffE}   {\ensuremath{{0.071 } } }
\vdef{default-11:SgData-APV1:muon1pt:hiEff}   {\ensuremath{{1.000 } } }
\vdef{default-11:SgData-APV1:muon1pt:hiEffE}   {\ensuremath{{0.071 } } }
\vdef{default-11:SgMcPU-APV1:muon1pt:loEff}   {\ensuremath{{1.014 } } }
\vdef{default-11:SgMcPU-APV1:muon1pt:loEffE}   {\ensuremath{{0.000 } } }
\vdef{default-11:SgMcPU-APV1:muon1pt:hiEff}   {\ensuremath{{1.000 } } }
\vdef{default-11:SgMcPU-APV1:muon1pt:hiEffE}   {\ensuremath{{0.077 } } }
\vdef{default-11:SgMcPU-APV1:muon1pt:loDelta}   {\ensuremath{{-0.014 } } }
\vdef{default-11:SgMcPU-APV1:muon1pt:loDeltaE}   {\ensuremath{{0.071 } } }
\vdef{default-11:SgMcPU-APV1:muon1pt:hiDelta}   {\ensuremath{{+0.000 } } }
\vdef{default-11:SgMcPU-APV1:muon1pt:hiDeltaE}   {\ensuremath{{0.105 } } }
\vdef{default-11:SgData-APV1:muon2pt:loEff}   {\ensuremath{{0.091 } } }
\vdef{default-11:SgData-APV1:muon2pt:loEffE}   {\ensuremath{{0.096 } } }
\vdef{default-11:SgData-APV1:muon2pt:hiEff}   {\ensuremath{{0.909 } } }
\vdef{default-11:SgData-APV1:muon2pt:hiEffE}   {\ensuremath{{0.096 } } }
\vdef{default-11:SgMcPU-APV1:muon2pt:loEff}   {\ensuremath{{0.014 } } }
\vdef{default-11:SgMcPU-APV1:muon2pt:loEffE}   {\ensuremath{{0.077 } } }
\vdef{default-11:SgMcPU-APV1:muon2pt:hiEff}   {\ensuremath{{0.986 } } }
\vdef{default-11:SgMcPU-APV1:muon2pt:hiEffE}   {\ensuremath{{0.077 } } }
\vdef{default-11:SgMcPU-APV1:muon2pt:loDelta}   {\ensuremath{{+1.468 } } }
\vdef{default-11:SgMcPU-APV1:muon2pt:loDeltaE}   {\ensuremath{{2.582 } } }
\vdef{default-11:SgMcPU-APV1:muon2pt:hiDelta}   {\ensuremath{{-0.081 } } }
\vdef{default-11:SgMcPU-APV1:muon2pt:hiDeltaE}   {\ensuremath{{0.131 } } }
\vdef{default-11:SgData-APV1:muonseta:loEff}   {\ensuremath{{0.900 } } }
\vdef{default-11:SgData-APV1:muonseta:loEffE}   {\ensuremath{{0.072 } } }
\vdef{default-11:SgData-APV1:muonseta:hiEff}   {\ensuremath{{0.100 } } }
\vdef{default-11:SgData-APV1:muonseta:hiEffE}   {\ensuremath{{0.072 } } }
\vdef{default-11:SgMcPU-APV1:muonseta:loEff}   {\ensuremath{{0.675 } } }
\vdef{default-11:SgMcPU-APV1:muonseta:loEffE}   {\ensuremath{{0.100 } } }
\vdef{default-11:SgMcPU-APV1:muonseta:hiEff}   {\ensuremath{{0.325 } } }
\vdef{default-11:SgMcPU-APV1:muonseta:hiEffE}   {\ensuremath{{0.097 } } }
\vdef{default-11:SgMcPU-APV1:muonseta:loDelta}   {\ensuremath{{+0.286 } } }
\vdef{default-11:SgMcPU-APV1:muonseta:loDeltaE}   {\ensuremath{{0.165 } } }
\vdef{default-11:SgMcPU-APV1:muonseta:hiDelta}   {\ensuremath{{-1.060 } } }
\vdef{default-11:SgMcPU-APV1:muonseta:hiDeltaE}   {\ensuremath{{0.558 } } }
\vdef{default-11:SgData-APV1:pt:loEff}   {\ensuremath{{0.000 } } }
\vdef{default-11:SgData-APV1:pt:loEffE}   {\ensuremath{{0.043 } } }
\vdef{default-11:SgData-APV1:pt:hiEff}   {\ensuremath{{1.000 } } }
\vdef{default-11:SgData-APV1:pt:hiEffE}   {\ensuremath{{0.043 } } }
\vdef{default-11:SgMcPU-APV1:pt:loEff}   {\ensuremath{{0.000 } } }
\vdef{default-11:SgMcPU-APV1:pt:loEffE}   {\ensuremath{{0.045 } } }
\vdef{default-11:SgMcPU-APV1:pt:hiEff}   {\ensuremath{{1.000 } } }
\vdef{default-11:SgMcPU-APV1:pt:hiEffE}   {\ensuremath{{0.045 } } }
\vdef{default-11:SgMcPU-APV1:pt:loDelta}   {\ensuremath{{\mathrm{NaN} } } }
\vdef{default-11:SgMcPU-APV1:pt:loDeltaE}   {\ensuremath{{\mathrm{NaN} } } }
\vdef{default-11:SgMcPU-APV1:pt:hiDelta}   {\ensuremath{{+0.000 } } }
\vdef{default-11:SgMcPU-APV1:pt:hiDeltaE}   {\ensuremath{{0.063 } } }
\vdef{default-11:SgData-APV1:p:loEff}   {\ensuremath{{1.000 } } }
\vdef{default-11:SgData-APV1:p:loEffE}   {\ensuremath{{0.077 } } }
\vdef{default-11:SgData-APV1:p:hiEff}   {\ensuremath{{1.000 } } }
\vdef{default-11:SgData-APV1:p:hiEffE}   {\ensuremath{{0.077 } } }
\vdef{default-11:SgMcPU-APV1:p:loEff}   {\ensuremath{{1.009 } } }
\vdef{default-11:SgMcPU-APV1:p:loEffE}   {\ensuremath{{0.077 } } }
\vdef{default-11:SgMcPU-APV1:p:hiEff}   {\ensuremath{{1.000 } } }
\vdef{default-11:SgMcPU-APV1:p:hiEffE}   {\ensuremath{{0.077 } } }
\vdef{default-11:SgMcPU-APV1:p:loDelta}   {\ensuremath{{-0.009 } } }
\vdef{default-11:SgMcPU-APV1:p:loDeltaE}   {\ensuremath{{0.108 } } }
\vdef{default-11:SgMcPU-APV1:p:hiDelta}   {\ensuremath{{+0.000 } } }
\vdef{default-11:SgMcPU-APV1:p:hiDeltaE}   {\ensuremath{{0.108 } } }
\vdef{default-11:SgData-APV1:eta:loEff}   {\ensuremath{{0.900 } } }
\vdef{default-11:SgData-APV1:eta:loEffE}   {\ensuremath{{0.103 } } }
\vdef{default-11:SgData-APV1:eta:hiEff}   {\ensuremath{{0.100 } } }
\vdef{default-11:SgData-APV1:eta:hiEffE}   {\ensuremath{{0.103 } } }
\vdef{default-11:SgMcPU-APV1:eta:loEff}   {\ensuremath{{0.656 } } }
\vdef{default-11:SgMcPU-APV1:eta:loEffE}   {\ensuremath{{0.137 } } }
\vdef{default-11:SgMcPU-APV1:eta:hiEff}   {\ensuremath{{0.344 } } }
\vdef{default-11:SgMcPU-APV1:eta:hiEffE}   {\ensuremath{{0.131 } } }
\vdef{default-11:SgMcPU-APV1:eta:loDelta}   {\ensuremath{{+0.314 } } }
\vdef{default-11:SgMcPU-APV1:eta:loDeltaE}   {\ensuremath{{0.232 } } }
\vdef{default-11:SgMcPU-APV1:eta:hiDelta}   {\ensuremath{{-1.100 } } }
\vdef{default-11:SgMcPU-APV1:eta:hiDeltaE}   {\ensuremath{{0.768 } } }
\vdef{default-11:SgData-APV1:bdt:loEff}   {\ensuremath{{0.923 } } }
\vdef{default-11:SgData-APV1:bdt:loEffE}   {\ensuremath{{0.085 } } }
\vdef{default-11:SgData-APV1:bdt:hiEff}   {\ensuremath{{0.077 } } }
\vdef{default-11:SgData-APV1:bdt:hiEffE}   {\ensuremath{{0.085 } } }
\vdef{default-11:SgMcPU-APV1:bdt:loEff}   {\ensuremath{{0.888 } } }
\vdef{default-11:SgMcPU-APV1:bdt:loEffE}   {\ensuremath{{0.100 } } }
\vdef{default-11:SgMcPU-APV1:bdt:hiEff}   {\ensuremath{{0.112 } } }
\vdef{default-11:SgMcPU-APV1:bdt:hiEffE}   {\ensuremath{{0.085 } } }
\vdef{default-11:SgMcPU-APV1:bdt:loDelta}   {\ensuremath{{+0.039 } } }
\vdef{default-11:SgMcPU-APV1:bdt:loDeltaE}   {\ensuremath{{0.145 } } }
\vdef{default-11:SgMcPU-APV1:bdt:hiDelta}   {\ensuremath{{-0.375 } } }
\vdef{default-11:SgMcPU-APV1:bdt:hiDeltaE}   {\ensuremath{{1.291 } } }
\vdef{default-11:SgData-APV1:fl3d:loEff}   {\ensuremath{{0.967 } } }
\vdef{default-11:SgData-APV1:fl3d:loEffE}   {\ensuremath{{0.021 } } }
\vdef{default-11:SgData-APV1:fl3d:hiEff}   {\ensuremath{{0.033 } } }
\vdef{default-11:SgData-APV1:fl3d:hiEffE}   {\ensuremath{{0.021 } } }
\vdef{default-11:SgMcPU-APV1:fl3d:loEff}   {\ensuremath{{0.874 } } }
\vdef{default-11:SgMcPU-APV1:fl3d:loEffE}   {\ensuremath{{0.036 } } }
\vdef{default-11:SgMcPU-APV1:fl3d:hiEff}   {\ensuremath{{0.126 } } }
\vdef{default-11:SgMcPU-APV1:fl3d:hiEffE}   {\ensuremath{{0.035 } } }
\vdef{default-11:SgMcPU-APV1:fl3d:loDelta}   {\ensuremath{{+0.101 } } }
\vdef{default-11:SgMcPU-APV1:fl3d:loDeltaE}   {\ensuremath{{0.046 } } }
\vdef{default-11:SgMcPU-APV1:fl3d:hiDelta}   {\ensuremath{{-1.168 } } }
\vdef{default-11:SgMcPU-APV1:fl3d:hiDeltaE}   {\ensuremath{{0.456 } } }
\vdef{default-11:SgData-APV1:fl3de:loEff}   {\ensuremath{{1.000 } } }
\vdef{default-11:SgData-APV1:fl3de:loEffE}   {\ensuremath{{0.011 } } }
\vdef{default-11:SgData-APV1:fl3de:hiEff}   {\ensuremath{{0.011 } } }
\vdef{default-11:SgData-APV1:fl3de:hiEffE}   {\ensuremath{{0.015 } } }
\vdef{default-11:SgMcPU-APV1:fl3de:loEff}   {\ensuremath{{1.000 } } }
\vdef{default-11:SgMcPU-APV1:fl3de:loEffE}   {\ensuremath{{0.011 } } }
\vdef{default-11:SgMcPU-APV1:fl3de:hiEff}   {\ensuremath{{0.000 } } }
\vdef{default-11:SgMcPU-APV1:fl3de:hiEffE}   {\ensuremath{{0.011 } } }
\vdef{default-11:SgMcPU-APV1:fl3de:loDelta}   {\ensuremath{{+0.000 } } }
\vdef{default-11:SgMcPU-APV1:fl3de:loDeltaE}   {\ensuremath{{0.015 } } }
\vdef{default-11:SgMcPU-APV1:fl3de:hiDelta}   {\ensuremath{{+2.000 } } }
\vdef{default-11:SgMcPU-APV1:fl3de:hiDeltaE}   {\ensuremath{{3.913 } } }
\vdef{default-11:SgData-APV1:fls3d:loEff}   {\ensuremath{{0.824 } } }
\vdef{default-11:SgData-APV1:fls3d:loEffE}   {\ensuremath{{0.040 } } }
\vdef{default-11:SgData-APV1:fls3d:hiEff}   {\ensuremath{{0.176 } } }
\vdef{default-11:SgData-APV1:fls3d:hiEffE}   {\ensuremath{{0.040 } } }
\vdef{default-11:SgMcPU-APV1:fls3d:loEff}   {\ensuremath{{0.074 } } }
\vdef{default-11:SgMcPU-APV1:fls3d:loEffE}   {\ensuremath{{0.027 } } }
\vdef{default-11:SgMcPU-APV1:fls3d:hiEff}   {\ensuremath{{0.926 } } }
\vdef{default-11:SgMcPU-APV1:fls3d:hiEffE}   {\ensuremath{{0.029 } } }
\vdef{default-11:SgMcPU-APV1:fls3d:loDelta}   {\ensuremath{{+1.670 } } }
\vdef{default-11:SgMcPU-APV1:fls3d:loDeltaE}   {\ensuremath{{0.112 } } }
\vdef{default-11:SgMcPU-APV1:fls3d:hiDelta}   {\ensuremath{{-1.362 } } }
\vdef{default-11:SgMcPU-APV1:fls3d:hiDeltaE}   {\ensuremath{{0.123 } } }
\vdef{default-11:SgData-APV1:flsxy:loEff}   {\ensuremath{{1.000 } } }
\vdef{default-11:SgData-APV1:flsxy:loEffE}   {\ensuremath{{0.011 } } }
\vdef{default-11:SgData-APV1:flsxy:hiEff}   {\ensuremath{{1.000 } } }
\vdef{default-11:SgData-APV1:flsxy:hiEffE}   {\ensuremath{{0.011 } } }
\vdef{default-11:SgMcPU-APV1:flsxy:loEff}   {\ensuremath{{1.012 } } }
\vdef{default-11:SgMcPU-APV1:flsxy:loEffE}   {\ensuremath{{\mathrm{NaN} } } }
\vdef{default-11:SgMcPU-APV1:flsxy:hiEff}   {\ensuremath{{1.000 } } }
\vdef{default-11:SgMcPU-APV1:flsxy:hiEffE}   {\ensuremath{{0.011 } } }
\vdef{default-11:SgMcPU-APV1:flsxy:loDelta}   {\ensuremath{{-0.012 } } }
\vdef{default-11:SgMcPU-APV1:flsxy:loDeltaE}   {\ensuremath{{\mathrm{NaN} } } }
\vdef{default-11:SgMcPU-APV1:flsxy:hiDelta}   {\ensuremath{{+0.000 } } }
\vdef{default-11:SgMcPU-APV1:flsxy:hiDeltaE}   {\ensuremath{{0.015 } } }
\vdef{default-11:SgData-APV1:chi2dof:loEff}   {\ensuremath{{0.714 } } }
\vdef{default-11:SgData-APV1:chi2dof:loEffE}   {\ensuremath{{0.112 } } }
\vdef{default-11:SgData-APV1:chi2dof:hiEff}   {\ensuremath{{0.286 } } }
\vdef{default-11:SgData-APV1:chi2dof:hiEffE}   {\ensuremath{{0.112 } } }
\vdef{default-11:SgMcPU-APV1:chi2dof:loEff}   {\ensuremath{{0.855 } } }
\vdef{default-11:SgMcPU-APV1:chi2dof:loEffE}   {\ensuremath{{0.105 } } }
\vdef{default-11:SgMcPU-APV1:chi2dof:hiEff}   {\ensuremath{{0.145 } } }
\vdef{default-11:SgMcPU-APV1:chi2dof:hiEffE}   {\ensuremath{{0.095 } } }
\vdef{default-11:SgMcPU-APV1:chi2dof:loDelta}   {\ensuremath{{-0.179 } } }
\vdef{default-11:SgMcPU-APV1:chi2dof:loDeltaE}   {\ensuremath{{0.198 } } }
\vdef{default-11:SgMcPU-APV1:chi2dof:hiDelta}   {\ensuremath{{+0.652 } } }
\vdef{default-11:SgMcPU-APV1:chi2dof:hiDeltaE}   {\ensuremath{{0.681 } } }
\vdef{default-11:SgData-APV1:pchi2dof:loEff}   {\ensuremath{{0.880 } } }
\vdef{default-11:SgData-APV1:pchi2dof:loEffE}   {\ensuremath{{0.067 } } }
\vdef{default-11:SgData-APV1:pchi2dof:hiEff}   {\ensuremath{{0.120 } } }
\vdef{default-11:SgData-APV1:pchi2dof:hiEffE}   {\ensuremath{{0.067 } } }
\vdef{default-11:SgMcPU-APV1:pchi2dof:loEff}   {\ensuremath{{0.706 } } }
\vdef{default-11:SgMcPU-APV1:pchi2dof:loEffE}   {\ensuremath{{0.089 } } }
\vdef{default-11:SgMcPU-APV1:pchi2dof:hiEff}   {\ensuremath{{0.294 } } }
\vdef{default-11:SgMcPU-APV1:pchi2dof:hiEffE}   {\ensuremath{{0.089 } } }
\vdef{default-11:SgMcPU-APV1:pchi2dof:loDelta}   {\ensuremath{{+0.220 } } }
\vdef{default-11:SgMcPU-APV1:pchi2dof:loDeltaE}   {\ensuremath{{0.145 } } }
\vdef{default-11:SgMcPU-APV1:pchi2dof:hiDelta}   {\ensuremath{{-0.841 } } }
\vdef{default-11:SgMcPU-APV1:pchi2dof:hiDeltaE}   {\ensuremath{{0.523 } } }
\vdef{default-11:SgData-APV1:alpha:loEff}   {\ensuremath{{0.833 } } }
\vdef{default-11:SgData-APV1:alpha:loEffE}   {\ensuremath{{0.106 } } }
\vdef{default-11:SgData-APV1:alpha:hiEff}   {\ensuremath{{0.167 } } }
\vdef{default-11:SgData-APV1:alpha:hiEffE}   {\ensuremath{{0.106 } } }
\vdef{default-11:SgMcPU-APV1:alpha:loEff}   {\ensuremath{{0.981 } } }
\vdef{default-11:SgMcPU-APV1:alpha:loEffE}   {\ensuremath{{0.071 } } }
\vdef{default-11:SgMcPU-APV1:alpha:hiEff}   {\ensuremath{{0.019 } } }
\vdef{default-11:SgMcPU-APV1:alpha:hiEffE}   {\ensuremath{{0.071 } } }
\vdef{default-11:SgMcPU-APV1:alpha:loDelta}   {\ensuremath{{-0.163 } } }
\vdef{default-11:SgMcPU-APV1:alpha:loDeltaE}   {\ensuremath{{0.145 } } }
\vdef{default-11:SgMcPU-APV1:alpha:hiDelta}   {\ensuremath{{+1.600 } } }
\vdef{default-11:SgMcPU-APV1:alpha:hiDeltaE}   {\ensuremath{{1.403 } } }
\vdef{default-11:SgData-APV1:iso:loEff}   {\ensuremath{{0.474 } } }
\vdef{default-11:SgData-APV1:iso:loEffE}   {\ensuremath{{0.106 } } }
\vdef{default-11:SgData-APV1:iso:hiEff}   {\ensuremath{{0.526 } } }
\vdef{default-11:SgData-APV1:iso:hiEffE}   {\ensuremath{{0.106 } } }
\vdef{default-11:SgMcPU-APV1:iso:loEff}   {\ensuremath{{0.120 } } }
\vdef{default-11:SgMcPU-APV1:iso:loEffE}   {\ensuremath{{0.075 } } }
\vdef{default-11:SgMcPU-APV1:iso:hiEff}   {\ensuremath{{0.880 } } }
\vdef{default-11:SgMcPU-APV1:iso:hiEffE}   {\ensuremath{{0.084 } } }
\vdef{default-11:SgMcPU-APV1:iso:loDelta}   {\ensuremath{{+1.190 } } }
\vdef{default-11:SgMcPU-APV1:iso:loDeltaE}   {\ensuremath{{0.426 } } }
\vdef{default-11:SgMcPU-APV1:iso:hiDelta}   {\ensuremath{{-0.503 } } }
\vdef{default-11:SgMcPU-APV1:iso:hiDeltaE}   {\ensuremath{{0.209 } } }
\vdef{default-11:SgData-APV1:docatrk:loEff}   {\ensuremath{{0.091 } } }
\vdef{default-11:SgData-APV1:docatrk:loEffE}   {\ensuremath{{0.096 } } }
\vdef{default-11:SgData-APV1:docatrk:hiEff}   {\ensuremath{{0.909 } } }
\vdef{default-11:SgData-APV1:docatrk:hiEffE}   {\ensuremath{{0.096 } } }
\vdef{default-11:SgMcPU-APV1:docatrk:loEff}   {\ensuremath{{0.142 } } }
\vdef{default-11:SgMcPU-APV1:docatrk:loEffE}   {\ensuremath{{0.096 } } }
\vdef{default-11:SgMcPU-APV1:docatrk:hiEff}   {\ensuremath{{0.858 } } }
\vdef{default-11:SgMcPU-APV1:docatrk:hiEffE}   {\ensuremath{{0.113 } } }
\vdef{default-11:SgMcPU-APV1:docatrk:loDelta}   {\ensuremath{{-0.437 } } }
\vdef{default-11:SgMcPU-APV1:docatrk:loDeltaE}   {\ensuremath{{1.200 } } }
\vdef{default-11:SgMcPU-APV1:docatrk:hiDelta}   {\ensuremath{{+0.057 } } }
\vdef{default-11:SgMcPU-APV1:docatrk:hiDeltaE}   {\ensuremath{{0.169 } } }
\vdef{default-11:SgData-APV1:isotrk:loEff}   {\ensuremath{{1.000 } } }
\vdef{default-11:SgData-APV1:isotrk:loEffE}   {\ensuremath{{0.045 } } }
\vdef{default-11:SgData-APV1:isotrk:hiEff}   {\ensuremath{{1.000 } } }
\vdef{default-11:SgData-APV1:isotrk:hiEffE}   {\ensuremath{{0.045 } } }
\vdef{default-11:SgMcPU-APV1:isotrk:loEff}   {\ensuremath{{1.000 } } }
\vdef{default-11:SgMcPU-APV1:isotrk:loEffE}   {\ensuremath{{0.048 } } }
\vdef{default-11:SgMcPU-APV1:isotrk:hiEff}   {\ensuremath{{1.000 } } }
\vdef{default-11:SgMcPU-APV1:isotrk:hiEffE}   {\ensuremath{{0.048 } } }
\vdef{default-11:SgMcPU-APV1:isotrk:loDelta}   {\ensuremath{{+0.000 } } }
\vdef{default-11:SgMcPU-APV1:isotrk:loDeltaE}   {\ensuremath{{0.066 } } }
\vdef{default-11:SgMcPU-APV1:isotrk:hiDelta}   {\ensuremath{{+0.000 } } }
\vdef{default-11:SgMcPU-APV1:isotrk:hiDeltaE}   {\ensuremath{{0.066 } } }
\vdef{default-11:SgData-APV1:closetrk:loEff}   {\ensuremath{{1.000 } } }
\vdef{default-11:SgData-APV1:closetrk:loEffE}   {\ensuremath{{0.077 } } }
\vdef{default-11:SgData-APV1:closetrk:hiEff}   {\ensuremath{{0.000 } } }
\vdef{default-11:SgData-APV1:closetrk:hiEffE}   {\ensuremath{{0.077 } } }
\vdef{default-11:SgMcPU-APV1:closetrk:loEff}   {\ensuremath{{0.972 } } }
\vdef{default-11:SgMcPU-APV1:closetrk:loEffE}   {\ensuremath{{0.103 } } }
\vdef{default-11:SgMcPU-APV1:closetrk:hiEff}   {\ensuremath{{0.028 } } }
\vdef{default-11:SgMcPU-APV1:closetrk:hiEffE}   {\ensuremath{{0.077 } } }
\vdef{default-11:SgMcPU-APV1:closetrk:loDelta}   {\ensuremath{{+0.028 } } }
\vdef{default-11:SgMcPU-APV1:closetrk:loDeltaE}   {\ensuremath{{0.131 } } }
\vdef{default-11:SgMcPU-APV1:closetrk:hiDelta}   {\ensuremath{{-2.000 } } }
\vdef{default-11:SgMcPU-APV1:closetrk:hiDeltaE}   {\ensuremath{{11.141 } } }
\vdef{default-11:SgData-APV1:lip:loEff}   {\ensuremath{{1.000 } } }
\vdef{default-11:SgData-APV1:lip:loEffE}   {\ensuremath{{0.077 } } }
\vdef{default-11:SgData-APV1:lip:hiEff}   {\ensuremath{{0.000 } } }
\vdef{default-11:SgData-APV1:lip:hiEffE}   {\ensuremath{{0.077 } } }
\vdef{default-11:SgMcPU-APV1:lip:loEff}   {\ensuremath{{1.000 } } }
\vdef{default-11:SgMcPU-APV1:lip:loEffE}   {\ensuremath{{0.077 } } }
\vdef{default-11:SgMcPU-APV1:lip:hiEff}   {\ensuremath{{0.000 } } }
\vdef{default-11:SgMcPU-APV1:lip:hiEffE}   {\ensuremath{{0.077 } } }
\vdef{default-11:SgMcPU-APV1:lip:loDelta}   {\ensuremath{{+0.000 } } }
\vdef{default-11:SgMcPU-APV1:lip:loDeltaE}   {\ensuremath{{0.108 } } }
\vdef{default-11:SgMcPU-APV1:lip:hiDelta}   {\ensuremath{{\mathrm{NaN} } } }
\vdef{default-11:SgMcPU-APV1:lip:hiDeltaE}   {\ensuremath{{\mathrm{NaN} } } }
\vdef{default-11:SgData-APV1:lip:inEff}   {\ensuremath{{1.000 } } }
\vdef{default-11:SgData-APV1:lip:inEffE}   {\ensuremath{{0.077 } } }
\vdef{default-11:SgMcPU-APV1:lip:inEff}   {\ensuremath{{1.000 } } }
\vdef{default-11:SgMcPU-APV1:lip:inEffE}   {\ensuremath{{0.077 } } }
\vdef{default-11:SgMcPU-APV1:lip:inDelta}   {\ensuremath{{+0.000 } } }
\vdef{default-11:SgMcPU-APV1:lip:inDeltaE}   {\ensuremath{{0.108 } } }
\vdef{default-11:SgData-APV1:lips:loEff}   {\ensuremath{{1.000 } } }
\vdef{default-11:SgData-APV1:lips:loEffE}   {\ensuremath{{0.077 } } }
\vdef{default-11:SgData-APV1:lips:hiEff}   {\ensuremath{{0.000 } } }
\vdef{default-11:SgData-APV1:lips:hiEffE}   {\ensuremath{{0.077 } } }
\vdef{default-11:SgMcPU-APV1:lips:loEff}   {\ensuremath{{1.000 } } }
\vdef{default-11:SgMcPU-APV1:lips:loEffE}   {\ensuremath{{0.077 } } }
\vdef{default-11:SgMcPU-APV1:lips:hiEff}   {\ensuremath{{0.000 } } }
\vdef{default-11:SgMcPU-APV1:lips:hiEffE}   {\ensuremath{{0.077 } } }
\vdef{default-11:SgMcPU-APV1:lips:loDelta}   {\ensuremath{{+0.000 } } }
\vdef{default-11:SgMcPU-APV1:lips:loDeltaE}   {\ensuremath{{0.108 } } }
\vdef{default-11:SgMcPU-APV1:lips:hiDelta}   {\ensuremath{{\mathrm{NaN} } } }
\vdef{default-11:SgMcPU-APV1:lips:hiDeltaE}   {\ensuremath{{\mathrm{NaN} } } }
\vdef{default-11:SgData-APV1:lips:inEff}   {\ensuremath{{1.000 } } }
\vdef{default-11:SgData-APV1:lips:inEffE}   {\ensuremath{{0.077 } } }
\vdef{default-11:SgMcPU-APV1:lips:inEff}   {\ensuremath{{1.000 } } }
\vdef{default-11:SgMcPU-APV1:lips:inEffE}   {\ensuremath{{0.077 } } }
\vdef{default-11:SgMcPU-APV1:lips:inDelta}   {\ensuremath{{+0.000 } } }
\vdef{default-11:SgMcPU-APV1:lips:inDeltaE}   {\ensuremath{{0.108 } } }
\vdef{default-11:SgData-APV1:ip:loEff}   {\ensuremath{{0.769 } } }
\vdef{default-11:SgData-APV1:ip:loEffE}   {\ensuremath{{0.111 } } }
\vdef{default-11:SgData-APV1:ip:hiEff}   {\ensuremath{{0.231 } } }
\vdef{default-11:SgData-APV1:ip:hiEffE}   {\ensuremath{{0.111 } } }
\vdef{default-11:SgMcPU-APV1:ip:loEff}   {\ensuremath{{0.986 } } }
\vdef{default-11:SgMcPU-APV1:ip:loEffE}   {\ensuremath{{0.085 } } }
\vdef{default-11:SgMcPU-APV1:ip:hiEff}   {\ensuremath{{0.014 } } }
\vdef{default-11:SgMcPU-APV1:ip:hiEffE}   {\ensuremath{{0.062 } } }
\vdef{default-11:SgMcPU-APV1:ip:loDelta}   {\ensuremath{{-0.247 } } }
\vdef{default-11:SgMcPU-APV1:ip:loDeltaE}   {\ensuremath{{0.165 } } }
\vdef{default-11:SgMcPU-APV1:ip:hiDelta}   {\ensuremath{{+1.772 } } }
\vdef{default-11:SgMcPU-APV1:ip:hiDeltaE}   {\ensuremath{{0.967 } } }
\vdef{default-11:SgData-APV1:ips:loEff}   {\ensuremath{{0.714 } } }
\vdef{default-11:SgData-APV1:ips:loEffE}   {\ensuremath{{0.112 } } }
\vdef{default-11:SgData-APV1:ips:hiEff}   {\ensuremath{{0.286 } } }
\vdef{default-11:SgData-APV1:ips:hiEffE}   {\ensuremath{{0.112 } } }
\vdef{default-11:SgMcPU-APV1:ips:loEff}   {\ensuremath{{0.964 } } }
\vdef{default-11:SgMcPU-APV1:ips:loEffE}   {\ensuremath{{0.062 } } }
\vdef{default-11:SgMcPU-APV1:ips:hiEff}   {\ensuremath{{0.036 } } }
\vdef{default-11:SgMcPU-APV1:ips:hiEffE}   {\ensuremath{{0.062 } } }
\vdef{default-11:SgMcPU-APV1:ips:loDelta}   {\ensuremath{{-0.297 } } }
\vdef{default-11:SgMcPU-APV1:ips:loDeltaE}   {\ensuremath{{0.166 } } }
\vdef{default-11:SgMcPU-APV1:ips:hiDelta}   {\ensuremath{{+1.548 } } }
\vdef{default-11:SgMcPU-APV1:ips:hiDeltaE}   {\ensuremath{{0.705 } } }
\vdef{default-11:SgData-APV1:maxdoca:loEff}   {\ensuremath{{1.000 } } }
\vdef{default-11:SgData-APV1:maxdoca:loEffE}   {\ensuremath{{0.077 } } }
\vdef{default-11:SgData-APV1:maxdoca:hiEff}   {\ensuremath{{0.000 } } }
\vdef{default-11:SgData-APV1:maxdoca:hiEffE}   {\ensuremath{{0.077 } } }
\vdef{default-11:SgMcPU-APV1:maxdoca:loEff}   {\ensuremath{{1.000 } } }
\vdef{default-11:SgMcPU-APV1:maxdoca:loEffE}   {\ensuremath{{0.077 } } }
\vdef{default-11:SgMcPU-APV1:maxdoca:hiEff}   {\ensuremath{{0.000 } } }
\vdef{default-11:SgMcPU-APV1:maxdoca:hiEffE}   {\ensuremath{{0.077 } } }
\vdef{default-11:SgMcPU-APV1:maxdoca:loDelta}   {\ensuremath{{+0.000 } } }
\vdef{default-11:SgMcPU-APV1:maxdoca:loDeltaE}   {\ensuremath{{0.108 } } }
\vdef{default-11:SgMcPU-APV1:maxdoca:hiDelta}   {\ensuremath{{\mathrm{NaN} } } }
\vdef{default-11:SgMcPU-APV1:maxdoca:hiDeltaE}   {\ensuremath{{\mathrm{NaN} } } }
\vdef{default-11:SgMcPU-APV0:osiso:loEff}   {\ensuremath{{1.000 } } }
\vdef{default-11:SgMcPU-APV0:osiso:loEffE}   {\ensuremath{{0.032 } } }
\vdef{default-11:SgMcPU-APV0:osiso:hiEff}   {\ensuremath{{1.000 } } }
\vdef{default-11:SgMcPU-APV0:osiso:hiEffE}   {\ensuremath{{0.032 } } }
\vdef{default-11:SgMcPU-APV1:osiso:loEff}   {\ensuremath{{1.000 } } }
\vdef{default-11:SgMcPU-APV1:osiso:loEffE}   {\ensuremath{{0.032 } } }
\vdef{default-11:SgMcPU-APV1:osiso:hiEff}   {\ensuremath{{1.000 } } }
\vdef{default-11:SgMcPU-APV1:osiso:hiEffE}   {\ensuremath{{0.032 } } }
\vdef{default-11:SgMcPU-APV1:osiso:loDelta}   {\ensuremath{{+0.000 } } }
\vdef{default-11:SgMcPU-APV1:osiso:loDeltaE}   {\ensuremath{{0.046 } } }
\vdef{default-11:SgMcPU-APV1:osiso:hiDelta}   {\ensuremath{{+0.000 } } }
\vdef{default-11:SgMcPU-APV1:osiso:hiDeltaE}   {\ensuremath{{0.046 } } }
\vdef{default-11:SgMcPU-APV0:osreliso:loEff}   {\ensuremath{{0.143 } } }
\vdef{default-11:SgMcPU-APV0:osreliso:loEffE}   {\ensuremath{{0.067 } } }
\vdef{default-11:SgMcPU-APV0:osreliso:hiEff}   {\ensuremath{{0.857 } } }
\vdef{default-11:SgMcPU-APV0:osreliso:hiEffE}   {\ensuremath{{0.067 } } }
\vdef{default-11:SgMcPU-APV1:osreliso:loEff}   {\ensuremath{{0.238 } } }
\vdef{default-11:SgMcPU-APV1:osreliso:loEffE}   {\ensuremath{{0.078 } } }
\vdef{default-11:SgMcPU-APV1:osreliso:hiEff}   {\ensuremath{{0.762 } } }
\vdef{default-11:SgMcPU-APV1:osreliso:hiEffE}   {\ensuremath{{0.078 } } }
\vdef{default-11:SgMcPU-APV1:osreliso:loDelta}   {\ensuremath{{-0.502 } } }
\vdef{default-11:SgMcPU-APV1:osreliso:loDeltaE}   {\ensuremath{{0.536 } } }
\vdef{default-11:SgMcPU-APV1:osreliso:hiDelta}   {\ensuremath{{+0.118 } } }
\vdef{default-11:SgMcPU-APV1:osreliso:hiDeltaE}   {\ensuremath{{0.128 } } }
\vdef{default-11:SgMcPU-APV0:osmuonpt:loEff}   {\ensuremath{{0.000 } } }
\vdef{default-11:SgMcPU-APV0:osmuonpt:loEffE}   {\ensuremath{{0.163 } } }
\vdef{default-11:SgMcPU-APV0:osmuonpt:hiEff}   {\ensuremath{{1.000 } } }
\vdef{default-11:SgMcPU-APV0:osmuonpt:hiEffE}   {\ensuremath{{0.163 } } }
\vdef{default-11:SgMcPU-APV1:osmuonpt:loEff}   {\ensuremath{{0.000 } } }
\vdef{default-11:SgMcPU-APV1:osmuonpt:loEffE}   {\ensuremath{{0.163 } } }
\vdef{default-11:SgMcPU-APV1:osmuonpt:hiEff}   {\ensuremath{{1.000 } } }
\vdef{default-11:SgMcPU-APV1:osmuonpt:hiEffE}   {\ensuremath{{0.163 } } }
\vdef{default-11:SgMcPU-APV1:osmuonpt:loDelta}   {\ensuremath{{\mathrm{NaN} } } }
\vdef{default-11:SgMcPU-APV1:osmuonpt:loDeltaE}   {\ensuremath{{\mathrm{NaN} } } }
\vdef{default-11:SgMcPU-APV1:osmuonpt:hiDelta}   {\ensuremath{{+0.000 } } }
\vdef{default-11:SgMcPU-APV1:osmuonpt:hiDeltaE}   {\ensuremath{{0.231 } } }
\vdef{default-11:SgMcPU-APV0:osmuondr:loEff}   {\ensuremath{{0.000 } } }
\vdef{default-11:SgMcPU-APV0:osmuondr:loEffE}   {\ensuremath{{0.163 } } }
\vdef{default-11:SgMcPU-APV0:osmuondr:hiEff}   {\ensuremath{{1.000 } } }
\vdef{default-11:SgMcPU-APV0:osmuondr:hiEffE}   {\ensuremath{{0.163 } } }
\vdef{default-11:SgMcPU-APV1:osmuondr:loEff}   {\ensuremath{{0.000 } } }
\vdef{default-11:SgMcPU-APV1:osmuondr:loEffE}   {\ensuremath{{0.163 } } }
\vdef{default-11:SgMcPU-APV1:osmuondr:hiEff}   {\ensuremath{{1.000 } } }
\vdef{default-11:SgMcPU-APV1:osmuondr:hiEffE}   {\ensuremath{{0.163 } } }
\vdef{default-11:SgMcPU-APV1:osmuondr:loDelta}   {\ensuremath{{\mathrm{NaN} } } }
\vdef{default-11:SgMcPU-APV1:osmuondr:loDeltaE}   {\ensuremath{{\mathrm{NaN} } } }
\vdef{default-11:SgMcPU-APV1:osmuondr:hiDelta}   {\ensuremath{{+0.000 } } }
\vdef{default-11:SgMcPU-APV1:osmuondr:hiDeltaE}   {\ensuremath{{0.231 } } }
\vdef{default-11:SgMcPU-APV0:hlt:loEff}   {\ensuremath{{0.161 } } }
\vdef{default-11:SgMcPU-APV0:hlt:loEffE}   {\ensuremath{{0.066 } } }
\vdef{default-11:SgMcPU-APV0:hlt:hiEff}   {\ensuremath{{0.839 } } }
\vdef{default-11:SgMcPU-APV0:hlt:hiEffE}   {\ensuremath{{0.066 } } }
\vdef{default-11:SgMcPU-APV1:hlt:loEff}   {\ensuremath{{0.356 } } }
\vdef{default-11:SgMcPU-APV1:hlt:loEffE}   {\ensuremath{{0.082 } } }
\vdef{default-11:SgMcPU-APV1:hlt:hiEff}   {\ensuremath{{0.644 } } }
\vdef{default-11:SgMcPU-APV1:hlt:hiEffE}   {\ensuremath{{0.084 } } }
\vdef{default-11:SgMcPU-APV1:hlt:loDelta}   {\ensuremath{{-0.752 } } }
\vdef{default-11:SgMcPU-APV1:hlt:loDeltaE}   {\ensuremath{{0.405 } } }
\vdef{default-11:SgMcPU-APV1:hlt:hiDelta}   {\ensuremath{{+0.262 } } }
\vdef{default-11:SgMcPU-APV1:hlt:hiDeltaE}   {\ensuremath{{0.149 } } }
\vdef{default-11:SgMcPU-APV0:muonsid:loEff}   {\ensuremath{{0.037 } } }
\vdef{default-11:SgMcPU-APV0:muonsid:loEffE}   {\ensuremath{{0.046 } } }
\vdef{default-11:SgMcPU-APV0:muonsid:hiEff}   {\ensuremath{{0.963 } } }
\vdef{default-11:SgMcPU-APV0:muonsid:hiEffE}   {\ensuremath{{0.046 } } }
\vdef{default-11:SgMcPU-APV1:muonsid:loEff}   {\ensuremath{{0.120 } } }
\vdef{default-11:SgMcPU-APV1:muonsid:loEffE}   {\ensuremath{{0.063 } } }
\vdef{default-11:SgMcPU-APV1:muonsid:hiEff}   {\ensuremath{{0.880 } } }
\vdef{default-11:SgMcPU-APV1:muonsid:hiEffE}   {\ensuremath{{0.069 } } }
\vdef{default-11:SgMcPU-APV1:muonsid:loDelta}   {\ensuremath{{-1.059 } } }
\vdef{default-11:SgMcPU-APV1:muonsid:loDeltaE}   {\ensuremath{{0.975 } } }
\vdef{default-11:SgMcPU-APV1:muonsid:hiDelta}   {\ensuremath{{+0.090 } } }
\vdef{default-11:SgMcPU-APV1:muonsid:hiDeltaE}   {\ensuremath{{0.092 } } }
\vdef{default-11:SgMcPU-APV0:tracksqual:loEff}   {\ensuremath{{0.000 } } }
\vdef{default-11:SgMcPU-APV0:tracksqual:loEffE}   {\ensuremath{{0.034 } } }
\vdef{default-11:SgMcPU-APV0:tracksqual:hiEff}   {\ensuremath{{1.000 } } }
\vdef{default-11:SgMcPU-APV0:tracksqual:hiEffE}   {\ensuremath{{0.034 } } }
\vdef{default-11:SgMcPU-APV1:tracksqual:loEff}   {\ensuremath{{0.005 } } }
\vdef{default-11:SgMcPU-APV1:tracksqual:loEffE}   {\ensuremath{{0.034 } } }
\vdef{default-11:SgMcPU-APV1:tracksqual:hiEff}   {\ensuremath{{0.995 } } }
\vdef{default-11:SgMcPU-APV1:tracksqual:hiEffE}   {\ensuremath{{0.048 } } }
\vdef{default-11:SgMcPU-APV1:tracksqual:loDelta}   {\ensuremath{{-2.000 } } }
\vdef{default-11:SgMcPU-APV1:tracksqual:loDeltaE}   {\ensuremath{{29.361 } } }
\vdef{default-11:SgMcPU-APV1:tracksqual:hiDelta}   {\ensuremath{{+0.005 } } }
\vdef{default-11:SgMcPU-APV1:tracksqual:hiDeltaE}   {\ensuremath{{0.059 } } }
\vdef{default-11:SgMcPU-APV0:pvz:loEff}   {\ensuremath{{0.516 } } }
\vdef{default-11:SgMcPU-APV0:pvz:loEffE}   {\ensuremath{{0.086 } } }
\vdef{default-11:SgMcPU-APV0:pvz:hiEff}   {\ensuremath{{0.484 } } }
\vdef{default-11:SgMcPU-APV0:pvz:hiEffE}   {\ensuremath{{0.086 } } }
\vdef{default-11:SgMcPU-APV1:pvz:loEff}   {\ensuremath{{0.492 } } }
\vdef{default-11:SgMcPU-APV1:pvz:loEffE}   {\ensuremath{{0.086 } } }
\vdef{default-11:SgMcPU-APV1:pvz:hiEff}   {\ensuremath{{0.508 } } }
\vdef{default-11:SgMcPU-APV1:pvz:hiEffE}   {\ensuremath{{0.086 } } }
\vdef{default-11:SgMcPU-APV1:pvz:loDelta}   {\ensuremath{{+0.047 } } }
\vdef{default-11:SgMcPU-APV1:pvz:loDeltaE}   {\ensuremath{{0.240 } } }
\vdef{default-11:SgMcPU-APV1:pvz:hiDelta}   {\ensuremath{{-0.048 } } }
\vdef{default-11:SgMcPU-APV1:pvz:hiDeltaE}   {\ensuremath{{0.245 } } }
\vdef{default-11:SgMcPU-APV0:pvn:loEff}   {\ensuremath{{1.000 } } }
\vdef{default-11:SgMcPU-APV0:pvn:loEffE}   {\ensuremath{{0.029 } } }
\vdef{default-11:SgMcPU-APV0:pvn:hiEff}   {\ensuremath{{1.000 } } }
\vdef{default-11:SgMcPU-APV0:pvn:hiEffE}   {\ensuremath{{0.029 } } }
\vdef{default-11:SgMcPU-APV1:pvn:loEff}   {\ensuremath{{1.246 } } }
\vdef{default-11:SgMcPU-APV1:pvn:loEffE}   {\ensuremath{{\mathrm{NaN} } } }
\vdef{default-11:SgMcPU-APV1:pvn:hiEff}   {\ensuremath{{1.000 } } }
\vdef{default-11:SgMcPU-APV1:pvn:hiEffE}   {\ensuremath{{0.029 } } }
\vdef{default-11:SgMcPU-APV1:pvn:loDelta}   {\ensuremath{{-0.219 } } }
\vdef{default-11:SgMcPU-APV1:pvn:loDeltaE}   {\ensuremath{{\mathrm{NaN} } } }
\vdef{default-11:SgMcPU-APV1:pvn:hiDelta}   {\ensuremath{{+0.000 } } }
\vdef{default-11:SgMcPU-APV1:pvn:hiDeltaE}   {\ensuremath{{0.042 } } }
\vdef{default-11:SgMcPU-APV0:pvavew8:loEff}   {\ensuremath{{0.038 } } }
\vdef{default-11:SgMcPU-APV0:pvavew8:loEffE}   {\ensuremath{{0.048 } } }
\vdef{default-11:SgMcPU-APV0:pvavew8:hiEff}   {\ensuremath{{0.962 } } }
\vdef{default-11:SgMcPU-APV0:pvavew8:hiEffE}   {\ensuremath{{0.048 } } }
\vdef{default-11:SgMcPU-APV1:pvavew8:loEff}   {\ensuremath{{0.005 } } }
\vdef{default-11:SgMcPU-APV1:pvavew8:loEffE}   {\ensuremath{{0.034 } } }
\vdef{default-11:SgMcPU-APV1:pvavew8:hiEff}   {\ensuremath{{0.995 } } }
\vdef{default-11:SgMcPU-APV1:pvavew8:hiEffE}   {\ensuremath{{0.048 } } }
\vdef{default-11:SgMcPU-APV1:pvavew8:loDelta}   {\ensuremath{{+1.563 } } }
\vdef{default-11:SgMcPU-APV1:pvavew8:loDeltaE}   {\ensuremath{{2.885 } } }
\vdef{default-11:SgMcPU-APV1:pvavew8:hiDelta}   {\ensuremath{{-0.034 } } }
\vdef{default-11:SgMcPU-APV1:pvavew8:hiDeltaE}   {\ensuremath{{0.069 } } }
\vdef{default-11:SgMcPU-APV0:pvntrk:loEff}   {\ensuremath{{1.000 } } }
\vdef{default-11:SgMcPU-APV0:pvntrk:loEffE}   {\ensuremath{{0.029 } } }
\vdef{default-11:SgMcPU-APV0:pvntrk:hiEff}   {\ensuremath{{1.000 } } }
\vdef{default-11:SgMcPU-APV0:pvntrk:hiEffE}   {\ensuremath{{0.029 } } }
\vdef{default-11:SgMcPU-APV1:pvntrk:loEff}   {\ensuremath{{1.000 } } }
\vdef{default-11:SgMcPU-APV1:pvntrk:loEffE}   {\ensuremath{{0.030 } } }
\vdef{default-11:SgMcPU-APV1:pvntrk:hiEff}   {\ensuremath{{1.000 } } }
\vdef{default-11:SgMcPU-APV1:pvntrk:hiEffE}   {\ensuremath{{0.030 } } }
\vdef{default-11:SgMcPU-APV1:pvntrk:loDelta}   {\ensuremath{{+0.000 } } }
\vdef{default-11:SgMcPU-APV1:pvntrk:loDeltaE}   {\ensuremath{{0.042 } } }
\vdef{default-11:SgMcPU-APV1:pvntrk:hiDelta}   {\ensuremath{{+0.000 } } }
\vdef{default-11:SgMcPU-APV1:pvntrk:hiDeltaE}   {\ensuremath{{0.042 } } }
\vdef{default-11:SgMcPU-APV0:muon1pt:loEff}   {\ensuremath{{1.000 } } }
\vdef{default-11:SgMcPU-APV0:muon1pt:loEffE}   {\ensuremath{{0.034 } } }
\vdef{default-11:SgMcPU-APV0:muon1pt:hiEff}   {\ensuremath{{1.000 } } }
\vdef{default-11:SgMcPU-APV0:muon1pt:hiEffE}   {\ensuremath{{0.034 } } }
\vdef{default-11:SgMcPU-APV1:muon1pt:loEff}   {\ensuremath{{1.014 } } }
\vdef{default-11:SgMcPU-APV1:muon1pt:loEffE}   {\ensuremath{{0.034 } } }
\vdef{default-11:SgMcPU-APV1:muon1pt:hiEff}   {\ensuremath{{1.000 } } }
\vdef{default-11:SgMcPU-APV1:muon1pt:hiEffE}   {\ensuremath{{0.034 } } }
\vdef{default-11:SgMcPU-APV1:muon1pt:loDelta}   {\ensuremath{{-0.014 } } }
\vdef{default-11:SgMcPU-APV1:muon1pt:loDeltaE}   {\ensuremath{{0.048 } } }
\vdef{default-11:SgMcPU-APV1:muon1pt:hiDelta}   {\ensuremath{{+0.000 } } }
\vdef{default-11:SgMcPU-APV1:muon1pt:hiDeltaE}   {\ensuremath{{0.049 } } }
\vdef{default-11:SgMcPU-APV0:muon2pt:loEff}   {\ensuremath{{0.000 } } }
\vdef{default-11:SgMcPU-APV0:muon2pt:loEffE}   {\ensuremath{{0.034 } } }
\vdef{default-11:SgMcPU-APV0:muon2pt:hiEff}   {\ensuremath{{1.000 } } }
\vdef{default-11:SgMcPU-APV0:muon2pt:hiEffE}   {\ensuremath{{0.034 } } }
\vdef{default-11:SgMcPU-APV1:muon2pt:loEff}   {\ensuremath{{0.014 } } }
\vdef{default-11:SgMcPU-APV1:muon2pt:loEffE}   {\ensuremath{{0.034 } } }
\vdef{default-11:SgMcPU-APV1:muon2pt:hiEff}   {\ensuremath{{0.986 } } }
\vdef{default-11:SgMcPU-APV1:muon2pt:hiEffE}   {\ensuremath{{0.048 } } }
\vdef{default-11:SgMcPU-APV1:muon2pt:loDelta}   {\ensuremath{{-2.000 } } }
\vdef{default-11:SgMcPU-APV1:muon2pt:loDeltaE}   {\ensuremath{{9.879 } } }
\vdef{default-11:SgMcPU-APV1:muon2pt:hiDelta}   {\ensuremath{{+0.014 } } }
\vdef{default-11:SgMcPU-APV1:muon2pt:hiDeltaE}   {\ensuremath{{0.059 } } }
\vdef{default-11:SgMcPU-APV0:muonseta:loEff}   {\ensuremath{{0.750 } } }
\vdef{default-11:SgMcPU-APV0:muonseta:loEffE}   {\ensuremath{{0.059 } } }
\vdef{default-11:SgMcPU-APV0:muonseta:hiEff}   {\ensuremath{{0.250 } } }
\vdef{default-11:SgMcPU-APV0:muonseta:hiEffE}   {\ensuremath{{0.059 } } }
\vdef{default-11:SgMcPU-APV1:muonseta:loEff}   {\ensuremath{{0.675 } } }
\vdef{default-11:SgMcPU-APV1:muonseta:loEffE}   {\ensuremath{{0.064 } } }
\vdef{default-11:SgMcPU-APV1:muonseta:hiEff}   {\ensuremath{{0.325 } } }
\vdef{default-11:SgMcPU-APV1:muonseta:hiEffE}   {\ensuremath{{0.063 } } }
\vdef{default-11:SgMcPU-APV1:muonseta:loDelta}   {\ensuremath{{+0.106 } } }
\vdef{default-11:SgMcPU-APV1:muonseta:loDeltaE}   {\ensuremath{{0.122 } } }
\vdef{default-11:SgMcPU-APV1:muonseta:hiDelta}   {\ensuremath{{-0.262 } } }
\vdef{default-11:SgMcPU-APV1:muonseta:hiDeltaE}   {\ensuremath{{0.300 } } }
\vdef{default-11:SgMcPU-APV0:pt:loEff}   {\ensuremath{{0.000 } } }
\vdef{default-11:SgMcPU-APV0:pt:loEffE}   {\ensuremath{{0.018 } } }
\vdef{default-11:SgMcPU-APV0:pt:hiEff}   {\ensuremath{{1.000 } } }
\vdef{default-11:SgMcPU-APV0:pt:hiEffE}   {\ensuremath{{0.018 } } }
\vdef{default-11:SgMcPU-APV1:pt:loEff}   {\ensuremath{{0.000 } } }
\vdef{default-11:SgMcPU-APV1:pt:loEffE}   {\ensuremath{{0.018 } } }
\vdef{default-11:SgMcPU-APV1:pt:hiEff}   {\ensuremath{{1.000 } } }
\vdef{default-11:SgMcPU-APV1:pt:hiEffE}   {\ensuremath{{0.018 } } }
\vdef{default-11:SgMcPU-APV1:pt:loDelta}   {\ensuremath{{\mathrm{NaN} } } }
\vdef{default-11:SgMcPU-APV1:pt:loDeltaE}   {\ensuremath{{\mathrm{NaN} } } }
\vdef{default-11:SgMcPU-APV1:pt:hiDelta}   {\ensuremath{{+0.000 } } }
\vdef{default-11:SgMcPU-APV1:pt:hiDeltaE}   {\ensuremath{{0.026 } } }
\vdef{default-11:SgMcPU-APV0:p:loEff}   {\ensuremath{{1.000 } } }
\vdef{default-11:SgMcPU-APV0:p:loEffE}   {\ensuremath{{0.034 } } }
\vdef{default-11:SgMcPU-APV0:p:hiEff}   {\ensuremath{{1.000 } } }
\vdef{default-11:SgMcPU-APV0:p:hiEffE}   {\ensuremath{{0.034 } } }
\vdef{default-11:SgMcPU-APV1:p:loEff}   {\ensuremath{{1.009 } } }
\vdef{default-11:SgMcPU-APV1:p:loEffE}   {\ensuremath{{0.000 } } }
\vdef{default-11:SgMcPU-APV1:p:hiEff}   {\ensuremath{{1.000 } } }
\vdef{default-11:SgMcPU-APV1:p:hiEffE}   {\ensuremath{{0.036 } } }
\vdef{default-11:SgMcPU-APV1:p:loDelta}   {\ensuremath{{-0.009 } } }
\vdef{default-11:SgMcPU-APV1:p:loDeltaE}   {\ensuremath{{0.034 } } }
\vdef{default-11:SgMcPU-APV1:p:hiDelta}   {\ensuremath{{+0.000 } } }
\vdef{default-11:SgMcPU-APV1:p:hiDeltaE}   {\ensuremath{{0.050 } } }
\vdef{default-11:SgMcPU-APV0:eta:loEff}   {\ensuremath{{0.769 } } }
\vdef{default-11:SgMcPU-APV0:eta:loEffE}   {\ensuremath{{0.080 } } }
\vdef{default-11:SgMcPU-APV0:eta:hiEff}   {\ensuremath{{0.231 } } }
\vdef{default-11:SgMcPU-APV0:eta:hiEffE}   {\ensuremath{{0.080 } } }
\vdef{default-11:SgMcPU-APV1:eta:loEff}   {\ensuremath{{0.656 } } }
\vdef{default-11:SgMcPU-APV1:eta:loEffE}   {\ensuremath{{0.089 } } }
\vdef{default-11:SgMcPU-APV1:eta:hiEff}   {\ensuremath{{0.344 } } }
\vdef{default-11:SgMcPU-APV1:eta:hiEffE}   {\ensuremath{{0.087 } } }
\vdef{default-11:SgMcPU-APV1:eta:loDelta}   {\ensuremath{{+0.159 } } }
\vdef{default-11:SgMcPU-APV1:eta:loDeltaE}   {\ensuremath{{0.170 } } }
\vdef{default-11:SgMcPU-APV1:eta:hiDelta}   {\ensuremath{{-0.395 } } }
\vdef{default-11:SgMcPU-APV1:eta:hiDeltaE}   {\ensuremath{{0.413 } } }
\vdef{default-11:SgMcPU-APV0:bdt:loEff}   {\ensuremath{{0.968 } } }
\vdef{default-11:SgMcPU-APV0:bdt:loEffE}   {\ensuremath{{0.041 } } }
\vdef{default-11:SgMcPU-APV0:bdt:hiEff}   {\ensuremath{{0.032 } } }
\vdef{default-11:SgMcPU-APV0:bdt:hiEffE}   {\ensuremath{{0.041 } } }
\vdef{default-11:SgMcPU-APV1:bdt:loEff}   {\ensuremath{{0.888 } } }
\vdef{default-11:SgMcPU-APV1:bdt:loEffE}   {\ensuremath{{0.061 } } }
\vdef{default-11:SgMcPU-APV1:bdt:hiEff}   {\ensuremath{{0.112 } } }
\vdef{default-11:SgMcPU-APV1:bdt:hiEffE}   {\ensuremath{{0.056 } } }
\vdef{default-11:SgMcPU-APV1:bdt:loDelta}   {\ensuremath{{+0.086 } } }
\vdef{default-11:SgMcPU-APV1:bdt:loDeltaE}   {\ensuremath{{0.081 } } }
\vdef{default-11:SgMcPU-APV1:bdt:hiDelta}   {\ensuremath{{-1.108 } } }
\vdef{default-11:SgMcPU-APV1:bdt:hiDeltaE}   {\ensuremath{{0.944 } } }
\vdef{default-11:SgMcPU-APV0:fl3d:loEff}   {\ensuremath{{0.719 } } }
\vdef{default-11:SgMcPU-APV0:fl3d:loEffE}   {\ensuremath{{0.077 } } }
\vdef{default-11:SgMcPU-APV0:fl3d:hiEff}   {\ensuremath{{0.281 } } }
\vdef{default-11:SgMcPU-APV0:fl3d:hiEffE}   {\ensuremath{{0.077 } } }
\vdef{default-11:SgMcPU-APV1:fl3d:loEff}   {\ensuremath{{0.874 } } }
\vdef{default-11:SgMcPU-APV1:fl3d:loEffE}   {\ensuremath{{0.061 } } }
\vdef{default-11:SgMcPU-APV1:fl3d:hiEff}   {\ensuremath{{0.126 } } }
\vdef{default-11:SgMcPU-APV1:fl3d:hiEffE}   {\ensuremath{{0.061 } } }
\vdef{default-11:SgMcPU-APV1:fl3d:loDelta}   {\ensuremath{{-0.196 } } }
\vdef{default-11:SgMcPU-APV1:fl3d:loDeltaE}   {\ensuremath{{0.127 } } }
\vdef{default-11:SgMcPU-APV1:fl3d:hiDelta}   {\ensuremath{{+0.766 } } }
\vdef{default-11:SgMcPU-APV1:fl3d:hiDeltaE}   {\ensuremath{{0.479 } } }
\vdef{default-11:SgMcPU-APV0:fl3de:loEff}   {\ensuremath{{1.000 } } }
\vdef{default-11:SgMcPU-APV0:fl3de:loEffE}   {\ensuremath{{0.029 } } }
\vdef{default-11:SgMcPU-APV0:fl3de:hiEff}   {\ensuremath{{0.000 } } }
\vdef{default-11:SgMcPU-APV0:fl3de:hiEffE}   {\ensuremath{{0.029 } } }
\vdef{default-11:SgMcPU-APV1:fl3de:loEff}   {\ensuremath{{1.000 } } }
\vdef{default-11:SgMcPU-APV1:fl3de:loEffE}   {\ensuremath{{0.029 } } }
\vdef{default-11:SgMcPU-APV1:fl3de:hiEff}   {\ensuremath{{0.000 } } }
\vdef{default-11:SgMcPU-APV1:fl3de:hiEffE}   {\ensuremath{{0.029 } } }
\vdef{default-11:SgMcPU-APV1:fl3de:loDelta}   {\ensuremath{{+0.000 } } }
\vdef{default-11:SgMcPU-APV1:fl3de:loDeltaE}   {\ensuremath{{0.040 } } }
\vdef{default-11:SgMcPU-APV1:fl3de:hiDelta}   {\ensuremath{{\mathrm{NaN} } } }
\vdef{default-11:SgMcPU-APV1:fl3de:hiDeltaE}   {\ensuremath{{\mathrm{NaN} } } }
\vdef{default-11:SgMcPU-APV0:fls3d:loEff}   {\ensuremath{{0.062 } } }
\vdef{default-11:SgMcPU-APV0:fls3d:loEffE}   {\ensuremath{{0.048 } } }
\vdef{default-11:SgMcPU-APV0:fls3d:hiEff}   {\ensuremath{{0.938 } } }
\vdef{default-11:SgMcPU-APV0:fls3d:hiEffE}   {\ensuremath{{0.048 } } }
\vdef{default-11:SgMcPU-APV1:fls3d:loEff}   {\ensuremath{{0.074 } } }
\vdef{default-11:SgMcPU-APV1:fls3d:loEffE}   {\ensuremath{{0.048 } } }
\vdef{default-11:SgMcPU-APV1:fls3d:hiEff}   {\ensuremath{{0.926 } } }
\vdef{default-11:SgMcPU-APV1:fls3d:hiEffE}   {\ensuremath{{0.054 } } }
\vdef{default-11:SgMcPU-APV1:fls3d:loDelta}   {\ensuremath{{-0.169 } } }
\vdef{default-11:SgMcPU-APV1:fls3d:loDeltaE}   {\ensuremath{{0.996 } } }
\vdef{default-11:SgMcPU-APV1:fls3d:hiDelta}   {\ensuremath{{+0.012 } } }
\vdef{default-11:SgMcPU-APV1:fls3d:hiDeltaE}   {\ensuremath{{0.078 } } }
\vdef{default-11:SgMcPU-APV0:flsxy:loEff}   {\ensuremath{{1.000 } } }
\vdef{default-11:SgMcPU-APV0:flsxy:loEffE}   {\ensuremath{{0.029 } } }
\vdef{default-11:SgMcPU-APV0:flsxy:hiEff}   {\ensuremath{{1.000 } } }
\vdef{default-11:SgMcPU-APV0:flsxy:hiEffE}   {\ensuremath{{0.029 } } }
\vdef{default-11:SgMcPU-APV1:flsxy:loEff}   {\ensuremath{{1.012 } } }
\vdef{default-11:SgMcPU-APV1:flsxy:loEffE}   {\ensuremath{{0.000 } } }
\vdef{default-11:SgMcPU-APV1:flsxy:hiEff}   {\ensuremath{{1.000 } } }
\vdef{default-11:SgMcPU-APV1:flsxy:hiEffE}   {\ensuremath{{0.029 } } }
\vdef{default-11:SgMcPU-APV1:flsxy:loDelta}   {\ensuremath{{-0.012 } } }
\vdef{default-11:SgMcPU-APV1:flsxy:loDeltaE}   {\ensuremath{{0.029 } } }
\vdef{default-11:SgMcPU-APV1:flsxy:hiDelta}   {\ensuremath{{+0.000 } } }
\vdef{default-11:SgMcPU-APV1:flsxy:hiDeltaE}   {\ensuremath{{0.041 } } }
\vdef{default-11:SgMcPU-APV0:chi2dof:loEff}   {\ensuremath{{1.000 } } }
\vdef{default-11:SgMcPU-APV0:chi2dof:loEffE}   {\ensuremath{{0.034 } } }
\vdef{default-11:SgMcPU-APV0:chi2dof:hiEff}   {\ensuremath{{0.000 } } }
\vdef{default-11:SgMcPU-APV0:chi2dof:hiEffE}   {\ensuremath{{0.034 } } }
\vdef{default-11:SgMcPU-APV1:chi2dof:loEff}   {\ensuremath{{0.855 } } }
\vdef{default-11:SgMcPU-APV1:chi2dof:loEffE}   {\ensuremath{{0.067 } } }
\vdef{default-11:SgMcPU-APV1:chi2dof:hiEff}   {\ensuremath{{0.145 } } }
\vdef{default-11:SgMcPU-APV1:chi2dof:hiEffE}   {\ensuremath{{0.067 } } }
\vdef{default-11:SgMcPU-APV1:chi2dof:loDelta}   {\ensuremath{{+0.157 } } }
\vdef{default-11:SgMcPU-APV1:chi2dof:loDeltaE}   {\ensuremath{{0.085 } } }
\vdef{default-11:SgMcPU-APV1:chi2dof:hiDelta}   {\ensuremath{{-2.000 } } }
\vdef{default-11:SgMcPU-APV1:chi2dof:hiDeltaE}   {\ensuremath{{0.950 } } }
\vdef{default-11:SgMcPU-APV0:pchi2dof:loEff}   {\ensuremath{{0.643 } } }
\vdef{default-11:SgMcPU-APV0:pchi2dof:loEffE}   {\ensuremath{{0.087 } } }
\vdef{default-11:SgMcPU-APV0:pchi2dof:hiEff}   {\ensuremath{{0.357 } } }
\vdef{default-11:SgMcPU-APV0:pchi2dof:hiEffE}   {\ensuremath{{0.087 } } }
\vdef{default-11:SgMcPU-APV1:pchi2dof:loEff}   {\ensuremath{{0.706 } } }
\vdef{default-11:SgMcPU-APV1:pchi2dof:loEffE}   {\ensuremath{{0.084 } } }
\vdef{default-11:SgMcPU-APV1:pchi2dof:hiEff}   {\ensuremath{{0.294 } } }
\vdef{default-11:SgMcPU-APV1:pchi2dof:hiEffE}   {\ensuremath{{0.084 } } }
\vdef{default-11:SgMcPU-APV1:pchi2dof:loDelta}   {\ensuremath{{-0.093 } } }
\vdef{default-11:SgMcPU-APV1:pchi2dof:loDeltaE}   {\ensuremath{{0.180 } } }
\vdef{default-11:SgMcPU-APV1:pchi2dof:hiDelta}   {\ensuremath{{+0.194 } } }
\vdef{default-11:SgMcPU-APV1:pchi2dof:hiDeltaE}   {\ensuremath{{0.372 } } }
\vdef{default-11:SgMcPU-APV0:alpha:loEff}   {\ensuremath{{1.000 } } }
\vdef{default-11:SgMcPU-APV0:alpha:loEffE}   {\ensuremath{{0.034 } } }
\vdef{default-11:SgMcPU-APV0:alpha:hiEff}   {\ensuremath{{0.000 } } }
\vdef{default-11:SgMcPU-APV0:alpha:hiEffE}   {\ensuremath{{0.034 } } }
\vdef{default-11:SgMcPU-APV1:alpha:loEff}   {\ensuremath{{0.981 } } }
\vdef{default-11:SgMcPU-APV1:alpha:loEffE}   {\ensuremath{{0.048 } } }
\vdef{default-11:SgMcPU-APV1:alpha:hiEff}   {\ensuremath{{0.019 } } }
\vdef{default-11:SgMcPU-APV1:alpha:hiEffE}   {\ensuremath{{0.034 } } }
\vdef{default-11:SgMcPU-APV1:alpha:loDelta}   {\ensuremath{{+0.019 } } }
\vdef{default-11:SgMcPU-APV1:alpha:loDeltaE}   {\ensuremath{{0.060 } } }
\vdef{default-11:SgMcPU-APV1:alpha:hiDelta}   {\ensuremath{{-2.000 } } }
\vdef{default-11:SgMcPU-APV1:alpha:hiDeltaE}   {\ensuremath{{7.444 } } }
\vdef{default-11:SgMcPU-APV0:iso:loEff}   {\ensuremath{{0.071 } } }
\vdef{default-11:SgMcPU-APV0:iso:loEffE}   {\ensuremath{{0.054 } } }
\vdef{default-11:SgMcPU-APV0:iso:hiEff}   {\ensuremath{{0.929 } } }
\vdef{default-11:SgMcPU-APV0:iso:hiEffE}   {\ensuremath{{0.054 } } }
\vdef{default-11:SgMcPU-APV1:iso:loEff}   {\ensuremath{{0.120 } } }
\vdef{default-11:SgMcPU-APV1:iso:loEffE}   {\ensuremath{{0.063 } } }
\vdef{default-11:SgMcPU-APV1:iso:hiEff}   {\ensuremath{{0.880 } } }
\vdef{default-11:SgMcPU-APV1:iso:hiEffE}   {\ensuremath{{0.063 } } }
\vdef{default-11:SgMcPU-APV1:iso:loDelta}   {\ensuremath{{-0.510 } } }
\vdef{default-11:SgMcPU-APV1:iso:loDeltaE}   {\ensuremath{{0.858 } } }
\vdef{default-11:SgMcPU-APV1:iso:hiDelta}   {\ensuremath{{+0.054 } } }
\vdef{default-11:SgMcPU-APV1:iso:hiDeltaE}   {\ensuremath{{0.092 } } }
\vdef{default-11:SgMcPU-APV0:docatrk:loEff}   {\ensuremath{{0.071 } } }
\vdef{default-11:SgMcPU-APV0:docatrk:loEffE}   {\ensuremath{{0.054 } } }
\vdef{default-11:SgMcPU-APV0:docatrk:hiEff}   {\ensuremath{{0.929 } } }
\vdef{default-11:SgMcPU-APV0:docatrk:hiEffE}   {\ensuremath{{0.054 } } }
\vdef{default-11:SgMcPU-APV1:docatrk:loEff}   {\ensuremath{{0.142 } } }
\vdef{default-11:SgMcPU-APV1:docatrk:loEffE}   {\ensuremath{{0.063 } } }
\vdef{default-11:SgMcPU-APV1:docatrk:hiEff}   {\ensuremath{{0.858 } } }
\vdef{default-11:SgMcPU-APV1:docatrk:hiEffE}   {\ensuremath{{0.063 } } }
\vdef{default-11:SgMcPU-APV1:docatrk:loDelta}   {\ensuremath{{-0.659 } } }
\vdef{default-11:SgMcPU-APV1:docatrk:loDeltaE}   {\ensuremath{{0.780 } } }
\vdef{default-11:SgMcPU-APV1:docatrk:hiDelta}   {\ensuremath{{+0.079 } } }
\vdef{default-11:SgMcPU-APV1:docatrk:hiDeltaE}   {\ensuremath{{0.093 } } }
\vdef{default-11:SgMcPU-APV0:isotrk:loEff}   {\ensuremath{{1.000 } } }
\vdef{default-11:SgMcPU-APV0:isotrk:loEffE}   {\ensuremath{{0.032 } } }
\vdef{default-11:SgMcPU-APV0:isotrk:hiEff}   {\ensuremath{{1.000 } } }
\vdef{default-11:SgMcPU-APV0:isotrk:hiEffE}   {\ensuremath{{0.032 } } }
\vdef{default-11:SgMcPU-APV1:isotrk:loEff}   {\ensuremath{{1.000 } } }
\vdef{default-11:SgMcPU-APV1:isotrk:loEffE}   {\ensuremath{{0.032 } } }
\vdef{default-11:SgMcPU-APV1:isotrk:hiEff}   {\ensuremath{{1.000 } } }
\vdef{default-11:SgMcPU-APV1:isotrk:hiEffE}   {\ensuremath{{0.032 } } }
\vdef{default-11:SgMcPU-APV1:isotrk:loDelta}   {\ensuremath{{+0.000 } } }
\vdef{default-11:SgMcPU-APV1:isotrk:loDeltaE}   {\ensuremath{{0.046 } } }
\vdef{default-11:SgMcPU-APV1:isotrk:hiDelta}   {\ensuremath{{+0.000 } } }
\vdef{default-11:SgMcPU-APV1:isotrk:hiDeltaE}   {\ensuremath{{0.046 } } }
\vdef{default-11:SgMcPU-APV0:closetrk:loEff}   {\ensuremath{{0.963 } } }
\vdef{default-11:SgMcPU-APV0:closetrk:loEffE}   {\ensuremath{{0.046 } } }
\vdef{default-11:SgMcPU-APV0:closetrk:hiEff}   {\ensuremath{{0.037 } } }
\vdef{default-11:SgMcPU-APV0:closetrk:hiEffE}   {\ensuremath{{0.046 } } }
\vdef{default-11:SgMcPU-APV1:closetrk:loEff}   {\ensuremath{{0.972 } } }
\vdef{default-11:SgMcPU-APV1:closetrk:loEffE}   {\ensuremath{{0.046 } } }
\vdef{default-11:SgMcPU-APV1:closetrk:hiEff}   {\ensuremath{{0.028 } } }
\vdef{default-11:SgMcPU-APV1:closetrk:hiEffE}   {\ensuremath{{0.033 } } }
\vdef{default-11:SgMcPU-APV1:closetrk:loDelta}   {\ensuremath{{-0.010 } } }
\vdef{default-11:SgMcPU-APV1:closetrk:loDeltaE}   {\ensuremath{{0.068 } } }
\vdef{default-11:SgMcPU-APV1:closetrk:hiDelta}   {\ensuremath{{+0.295 } } }
\vdef{default-11:SgMcPU-APV1:closetrk:hiDeltaE}   {\ensuremath{{1.702 } } }
\vdef{default-11:SgMcPU-APV0:lip:loEff}   {\ensuremath{{1.000 } } }
\vdef{default-11:SgMcPU-APV0:lip:loEffE}   {\ensuremath{{0.034 } } }
\vdef{default-11:SgMcPU-APV0:lip:hiEff}   {\ensuremath{{0.000 } } }
\vdef{default-11:SgMcPU-APV0:lip:hiEffE}   {\ensuremath{{0.034 } } }
\vdef{default-11:SgMcPU-APV1:lip:loEff}   {\ensuremath{{1.000 } } }
\vdef{default-11:SgMcPU-APV1:lip:loEffE}   {\ensuremath{{0.034 } } }
\vdef{default-11:SgMcPU-APV1:lip:hiEff}   {\ensuremath{{0.000 } } }
\vdef{default-11:SgMcPU-APV1:lip:hiEffE}   {\ensuremath{{0.034 } } }
\vdef{default-11:SgMcPU-APV1:lip:loDelta}   {\ensuremath{{+0.000 } } }
\vdef{default-11:SgMcPU-APV1:lip:loDeltaE}   {\ensuremath{{0.049 } } }
\vdef{default-11:SgMcPU-APV1:lip:hiDelta}   {\ensuremath{{\mathrm{NaN} } } }
\vdef{default-11:SgMcPU-APV1:lip:hiDeltaE}   {\ensuremath{{\mathrm{NaN} } } }
\vdef{default-11:SgMcPU-APV0:lip:inEff}   {\ensuremath{{1.000 } } }
\vdef{default-11:SgMcPU-APV0:lip:inEffE}   {\ensuremath{{0.034 } } }
\vdef{default-11:SgMcPU-APV1:lip:inEff}   {\ensuremath{{1.000 } } }
\vdef{default-11:SgMcPU-APV1:lip:inEffE}   {\ensuremath{{0.034 } } }
\vdef{default-11:SgMcPU-APV1:lip:inDelta}   {\ensuremath{{+0.000 } } }
\vdef{default-11:SgMcPU-APV1:lip:inDeltaE}   {\ensuremath{{0.049 } } }
\vdef{default-11:SgMcPU-APV0:lips:loEff}   {\ensuremath{{1.000 } } }
\vdef{default-11:SgMcPU-APV0:lips:loEffE}   {\ensuremath{{0.034 } } }
\vdef{default-11:SgMcPU-APV0:lips:hiEff}   {\ensuremath{{0.000 } } }
\vdef{default-11:SgMcPU-APV0:lips:hiEffE}   {\ensuremath{{0.034 } } }
\vdef{default-11:SgMcPU-APV1:lips:loEff}   {\ensuremath{{1.000 } } }
\vdef{default-11:SgMcPU-APV1:lips:loEffE}   {\ensuremath{{0.036 } } }
\vdef{default-11:SgMcPU-APV1:lips:hiEff}   {\ensuremath{{0.000 } } }
\vdef{default-11:SgMcPU-APV1:lips:hiEffE}   {\ensuremath{{0.036 } } }
\vdef{default-11:SgMcPU-APV1:lips:loDelta}   {\ensuremath{{+0.000 } } }
\vdef{default-11:SgMcPU-APV1:lips:loDeltaE}   {\ensuremath{{0.050 } } }
\vdef{default-11:SgMcPU-APV1:lips:hiDelta}   {\ensuremath{{\mathrm{NaN} } } }
\vdef{default-11:SgMcPU-APV1:lips:hiDeltaE}   {\ensuremath{{\mathrm{NaN} } } }
\vdef{default-11:SgMcPU-APV0:lips:inEff}   {\ensuremath{{1.000 } } }
\vdef{default-11:SgMcPU-APV0:lips:inEffE}   {\ensuremath{{0.034 } } }
\vdef{default-11:SgMcPU-APV1:lips:inEff}   {\ensuremath{{1.000 } } }
\vdef{default-11:SgMcPU-APV1:lips:inEffE}   {\ensuremath{{0.036 } } }
\vdef{default-11:SgMcPU-APV1:lips:inDelta}   {\ensuremath{{+0.000 } } }
\vdef{default-11:SgMcPU-APV1:lips:inDeltaE}   {\ensuremath{{0.050 } } }
\vdef{default-11:SgMcPU-APV0:ip:loEff}   {\ensuremath{{1.000 } } }
\vdef{default-11:SgMcPU-APV0:ip:loEffE}   {\ensuremath{{0.034 } } }
\vdef{default-11:SgMcPU-APV0:ip:hiEff}   {\ensuremath{{0.000 } } }
\vdef{default-11:SgMcPU-APV0:ip:hiEffE}   {\ensuremath{{0.034 } } }
\vdef{default-11:SgMcPU-APV1:ip:loEff}   {\ensuremath{{0.986 } } }
\vdef{default-11:SgMcPU-APV1:ip:loEffE}   {\ensuremath{{0.048 } } }
\vdef{default-11:SgMcPU-APV1:ip:hiEff}   {\ensuremath{{0.014 } } }
\vdef{default-11:SgMcPU-APV1:ip:hiEffE}   {\ensuremath{{0.034 } } }
\vdef{default-11:SgMcPU-APV1:ip:loDelta}   {\ensuremath{{+0.014 } } }
\vdef{default-11:SgMcPU-APV1:ip:loDeltaE}   {\ensuremath{{0.059 } } }
\vdef{default-11:SgMcPU-APV1:ip:hiDelta}   {\ensuremath{{-2.000 } } }
\vdef{default-11:SgMcPU-APV1:ip:hiDeltaE}   {\ensuremath{{9.879 } } }
\vdef{default-11:SgMcPU-APV0:ips:loEff}   {\ensuremath{{0.929 } } }
\vdef{default-11:SgMcPU-APV0:ips:loEffE}   {\ensuremath{{0.054 } } }
\vdef{default-11:SgMcPU-APV0:ips:hiEff}   {\ensuremath{{0.071 } } }
\vdef{default-11:SgMcPU-APV0:ips:hiEffE}   {\ensuremath{{0.054 } } }
\vdef{default-11:SgMcPU-APV1:ips:loEff}   {\ensuremath{{0.964 } } }
\vdef{default-11:SgMcPU-APV1:ips:loEffE}   {\ensuremath{{0.054 } } }
\vdef{default-11:SgMcPU-APV1:ips:hiEff}   {\ensuremath{{0.036 } } }
\vdef{default-11:SgMcPU-APV1:ips:hiEffE}   {\ensuremath{{0.045 } } }
\vdef{default-11:SgMcPU-APV1:ips:loDelta}   {\ensuremath{{-0.037 } } }
\vdef{default-11:SgMcPU-APV1:ips:loDeltaE}   {\ensuremath{{0.081 } } }
\vdef{default-11:SgMcPU-APV1:ips:hiDelta}   {\ensuremath{{+0.651 } } }
\vdef{default-11:SgMcPU-APV1:ips:hiDeltaE}   {\ensuremath{{1.292 } } }
\vdef{default-11:SgMcPU-APV0:maxdoca:loEff}   {\ensuremath{{1.000 } } }
\vdef{default-11:SgMcPU-APV0:maxdoca:loEffE}   {\ensuremath{{0.034 } } }
\vdef{default-11:SgMcPU-APV0:maxdoca:hiEff}   {\ensuremath{{0.000 } } }
\vdef{default-11:SgMcPU-APV0:maxdoca:hiEffE}   {\ensuremath{{0.034 } } }
\vdef{default-11:SgMcPU-APV1:maxdoca:loEff}   {\ensuremath{{1.000 } } }
\vdef{default-11:SgMcPU-APV1:maxdoca:loEffE}   {\ensuremath{{0.036 } } }
\vdef{default-11:SgMcPU-APV1:maxdoca:hiEff}   {\ensuremath{{0.000 } } }
\vdef{default-11:SgMcPU-APV1:maxdoca:hiEffE}   {\ensuremath{{0.036 } } }
\vdef{default-11:SgMcPU-APV1:maxdoca:loDelta}   {\ensuremath{{+0.000 } } }
\vdef{default-11:SgMcPU-APV1:maxdoca:loDeltaE}   {\ensuremath{{0.050 } } }
\vdef{default-11:SgMcPU-APV1:maxdoca:hiDelta}   {\ensuremath{{\mathrm{NaN} } } }
\vdef{default-11:SgMcPU-APV1:maxdoca:hiDeltaE}   {\ensuremath{{\mathrm{NaN} } } }
\vdef{default-11:SgData-B:osiso:loEff}   {\ensuremath{{1.077 } } }
\vdef{default-11:SgData-B:osiso:loEffE}   {\ensuremath{{0.000 } } }
\vdef{default-11:SgData-B:osiso:hiEff}   {\ensuremath{{1.000 } } }
\vdef{default-11:SgData-B:osiso:hiEffE}   {\ensuremath{{0.062 } } }
\vdef{default-11:SgMc-B:osiso:loEff}   {\ensuremath{{1.001 } } }
\vdef{default-11:SgMc-B:osiso:loEffE}   {\ensuremath{{0.062 } } }
\vdef{default-11:SgMc-B:osiso:hiEff}   {\ensuremath{{1.000 } } }
\vdef{default-11:SgMc-B:osiso:hiEffE}   {\ensuremath{{0.062 } } }
\vdef{default-11:SgMc-B:osiso:loDelta}   {\ensuremath{{+0.073 } } }
\vdef{default-11:SgMc-B:osiso:loDeltaE}   {\ensuremath{{0.062 } } }
\vdef{default-11:SgMc-B:osiso:hiDelta}   {\ensuremath{{+0.000 } } }
\vdef{default-11:SgMc-B:osiso:hiDeltaE}   {\ensuremath{{0.088 } } }
\vdef{default-11:SgData-B:osreliso:loEff}   {\ensuremath{{0.143 } } }
\vdef{default-11:SgData-B:osreliso:loEffE}   {\ensuremath{{0.095 } } }
\vdef{default-11:SgData-B:osreliso:hiEff}   {\ensuremath{{0.857 } } }
\vdef{default-11:SgData-B:osreliso:hiEffE}   {\ensuremath{{0.095 } } }
\vdef{default-11:SgMc-B:osreliso:loEff}   {\ensuremath{{0.262 } } }
\vdef{default-11:SgMc-B:osreliso:loEffE}   {\ensuremath{{0.111 } } }
\vdef{default-11:SgMc-B:osreliso:hiEff}   {\ensuremath{{0.738 } } }
\vdef{default-11:SgMc-B:osreliso:hiEffE}   {\ensuremath{{0.111 } } }
\vdef{default-11:SgMc-B:osreliso:loDelta}   {\ensuremath{{-0.588 } } }
\vdef{default-11:SgMc-B:osreliso:loDeltaE}   {\ensuremath{{0.718 } } }
\vdef{default-11:SgMc-B:osreliso:hiDelta}   {\ensuremath{{+0.149 } } }
\vdef{default-11:SgMc-B:osreliso:hiDeltaE}   {\ensuremath{{0.185 } } }
\vdef{default-11:SgData-B:osmuonpt:loEff}   {\ensuremath{{\mathrm{NaN} } } }
\vdef{default-11:SgData-B:osmuonpt:loEffE}   {\ensuremath{{0.289 } } }
\vdef{default-11:SgData-B:osmuonpt:hiEff}   {\ensuremath{{\mathrm{NaN} } } }
\vdef{default-11:SgData-B:osmuonpt:hiEffE}   {\ensuremath{{0.289 } } }
\vdef{default-11:SgMc-B:osmuonpt:loEff}   {\ensuremath{{\mathrm{NaN} } } }
\vdef{default-11:SgMc-B:osmuonpt:loEffE}   {\ensuremath{{0.289 } } }
\vdef{default-11:SgMc-B:osmuonpt:hiEff}   {\ensuremath{{\mathrm{NaN} } } }
\vdef{default-11:SgMc-B:osmuonpt:hiEffE}   {\ensuremath{{0.289 } } }
\vdef{default-11:SgMc-B:osmuonpt:loDelta}   {\ensuremath{{\mathrm{NaN} } } }
\vdef{default-11:SgMc-B:osmuonpt:loDeltaE}   {\ensuremath{{\mathrm{NaN} } } }
\vdef{default-11:SgMc-B:osmuonpt:hiDelta}   {\ensuremath{{\mathrm{NaN} } } }
\vdef{default-11:SgMc-B:osmuonpt:hiDeltaE}   {\ensuremath{{\mathrm{NaN} } } }
\vdef{default-11:SgData-B:osmuondr:loEff}   {\ensuremath{{\mathrm{NaN} } } }
\vdef{default-11:SgData-B:osmuondr:loEffE}   {\ensuremath{{0.289 } } }
\vdef{default-11:SgData-B:osmuondr:hiEff}   {\ensuremath{{\mathrm{NaN} } } }
\vdef{default-11:SgData-B:osmuondr:hiEffE}   {\ensuremath{{0.289 } } }
\vdef{default-11:SgMc-B:osmuondr:loEff}   {\ensuremath{{\mathrm{NaN} } } }
\vdef{default-11:SgMc-B:osmuondr:loEffE}   {\ensuremath{{0.289 } } }
\vdef{default-11:SgMc-B:osmuondr:hiEff}   {\ensuremath{{\mathrm{NaN} } } }
\vdef{default-11:SgMc-B:osmuondr:hiEffE}   {\ensuremath{{0.289 } } }
\vdef{default-11:SgMc-B:osmuondr:loDelta}   {\ensuremath{{\mathrm{NaN} } } }
\vdef{default-11:SgMc-B:osmuondr:loDeltaE}   {\ensuremath{{\mathrm{NaN} } } }
\vdef{default-11:SgMc-B:osmuondr:hiDelta}   {\ensuremath{{\mathrm{NaN} } } }
\vdef{default-11:SgMc-B:osmuondr:hiDeltaE}   {\ensuremath{{\mathrm{NaN} } } }
\vdef{default-11:SgData-B:hlt:loEff}   {\ensuremath{{0.429 } } }
\vdef{default-11:SgData-B:hlt:loEffE}   {\ensuremath{{0.157 } } }
\vdef{default-11:SgData-B:hlt:hiEff}   {\ensuremath{{0.571 } } }
\vdef{default-11:SgData-B:hlt:hiEffE}   {\ensuremath{{0.157 } } }
\vdef{default-11:SgMc-B:hlt:loEff}   {\ensuremath{{0.148 } } }
\vdef{default-11:SgMc-B:hlt:loEffE}   {\ensuremath{{0.131 } } }
\vdef{default-11:SgMc-B:hlt:hiEff}   {\ensuremath{{0.852 } } }
\vdef{default-11:SgMc-B:hlt:hiEffE}   {\ensuremath{{0.149 } } }
\vdef{default-11:SgMc-B:hlt:loDelta}   {\ensuremath{{+0.972 } } }
\vdef{default-11:SgMc-B:hlt:loDeltaE}   {\ensuremath{{0.733 } } }
\vdef{default-11:SgMc-B:hlt:hiDelta}   {\ensuremath{{-0.394 } } }
\vdef{default-11:SgMc-B:hlt:hiDeltaE}   {\ensuremath{{0.313 } } }
\vdef{default-11:SgData-B:muonsid:loEff}   {\ensuremath{{0.600 } } }
\vdef{default-11:SgData-B:muonsid:loEffE}   {\ensuremath{{0.137 } } }
\vdef{default-11:SgData-B:muonsid:hiEff}   {\ensuremath{{0.400 } } }
\vdef{default-11:SgData-B:muonsid:hiEffE}   {\ensuremath{{0.137 } } }
\vdef{default-11:SgMc-B:muonsid:loEff}   {\ensuremath{{0.168 } } }
\vdef{default-11:SgMc-B:muonsid:loEffE}   {\ensuremath{{0.103 } } }
\vdef{default-11:SgMc-B:muonsid:hiEff}   {\ensuremath{{0.832 } } }
\vdef{default-11:SgMc-B:muonsid:hiEffE}   {\ensuremath{{0.120 } } }
\vdef{default-11:SgMc-B:muonsid:loDelta}   {\ensuremath{{+1.126 } } }
\vdef{default-11:SgMc-B:muonsid:loDeltaE}   {\ensuremath{{0.449 } } }
\vdef{default-11:SgMc-B:muonsid:hiDelta}   {\ensuremath{{-0.702 } } }
\vdef{default-11:SgMc-B:muonsid:hiDeltaE}   {\ensuremath{{0.325 } } }
\vdef{default-11:SgData-B:tracksqual:loEff}   {\ensuremath{{0.000 } } }
\vdef{default-11:SgData-B:tracksqual:loEffE}   {\ensuremath{{0.141 } } }
\vdef{default-11:SgData-B:tracksqual:hiEff}   {\ensuremath{{1.000 } } }
\vdef{default-11:SgData-B:tracksqual:hiEffE}   {\ensuremath{{0.141 } } }
\vdef{default-11:SgMc-B:tracksqual:loEff}   {\ensuremath{{0.001 } } }
\vdef{default-11:SgMc-B:tracksqual:loEffE}   {\ensuremath{{0.141 } } }
\vdef{default-11:SgMc-B:tracksqual:hiEff}   {\ensuremath{{0.999 } } }
\vdef{default-11:SgMc-B:tracksqual:hiEffE}   {\ensuremath{{0.178 } } }
\vdef{default-11:SgMc-B:tracksqual:loDelta}   {\ensuremath{{-2.000 } } }
\vdef{default-11:SgMc-B:tracksqual:loDeltaE}   {\ensuremath{{757.822 } } }
\vdef{default-11:SgMc-B:tracksqual:hiDelta}   {\ensuremath{{+0.001 } } }
\vdef{default-11:SgMc-B:tracksqual:hiDeltaE}   {\ensuremath{{0.227 } } }
\vdef{default-11:SgData-B:pvz:loEff}   {\ensuremath{{0.714 } } }
\vdef{default-11:SgData-B:pvz:loEffE}   {\ensuremath{{0.149 } } }
\vdef{default-11:SgData-B:pvz:hiEff}   {\ensuremath{{0.286 } } }
\vdef{default-11:SgData-B:pvz:hiEffE}   {\ensuremath{{0.149 } } }
\vdef{default-11:SgMc-B:pvz:loEff}   {\ensuremath{{0.482 } } }
\vdef{default-11:SgMc-B:pvz:loEffE}   {\ensuremath{{0.157 } } }
\vdef{default-11:SgMc-B:pvz:hiEff}   {\ensuremath{{0.518 } } }
\vdef{default-11:SgMc-B:pvz:hiEffE}   {\ensuremath{{0.157 } } }
\vdef{default-11:SgMc-B:pvz:loDelta}   {\ensuremath{{+0.388 } } }
\vdef{default-11:SgMc-B:pvz:loDeltaE}   {\ensuremath{{0.372 } } }
\vdef{default-11:SgMc-B:pvz:hiDelta}   {\ensuremath{{-0.578 } } }
\vdef{default-11:SgMc-B:pvz:hiDeltaE}   {\ensuremath{{0.553 } } }
\vdef{default-11:SgData-B:pvn:loEff}   {\ensuremath{{1.000 } } }
\vdef{default-11:SgData-B:pvn:loEffE}   {\ensuremath{{0.099 } } }
\vdef{default-11:SgData-B:pvn:hiEff}   {\ensuremath{{1.000 } } }
\vdef{default-11:SgData-B:pvn:hiEffE}   {\ensuremath{{0.099 } } }
\vdef{default-11:SgMc-B:pvn:loEff}   {\ensuremath{{1.000 } } }
\vdef{default-11:SgMc-B:pvn:loEffE}   {\ensuremath{{0.099 } } }
\vdef{default-11:SgMc-B:pvn:hiEff}   {\ensuremath{{1.000 } } }
\vdef{default-11:SgMc-B:pvn:hiEffE}   {\ensuremath{{0.099 } } }
\vdef{default-11:SgMc-B:pvn:loDelta}   {\ensuremath{{+0.000 } } }
\vdef{default-11:SgMc-B:pvn:loDeltaE}   {\ensuremath{{0.141 } } }
\vdef{default-11:SgMc-B:pvn:hiDelta}   {\ensuremath{{+0.000 } } }
\vdef{default-11:SgMc-B:pvn:hiDeltaE}   {\ensuremath{{0.141 } } }
\vdef{default-11:SgData-B:pvavew8:loEff}   {\ensuremath{{0.000 } } }
\vdef{default-11:SgData-B:pvavew8:loEffE}   {\ensuremath{{0.141 } } }
\vdef{default-11:SgData-B:pvavew8:hiEff}   {\ensuremath{{1.000 } } }
\vdef{default-11:SgData-B:pvavew8:hiEffE}   {\ensuremath{{0.141 } } }
\vdef{default-11:SgMc-B:pvavew8:loEff}   {\ensuremath{{0.004 } } }
\vdef{default-11:SgMc-B:pvavew8:loEffE}   {\ensuremath{{0.163 } } }
\vdef{default-11:SgMc-B:pvavew8:hiEff}   {\ensuremath{{0.996 } } }
\vdef{default-11:SgMc-B:pvavew8:hiEffE}   {\ensuremath{{0.163 } } }
\vdef{default-11:SgMc-B:pvavew8:loDelta}   {\ensuremath{{-2.000 } } }
\vdef{default-11:SgMc-B:pvavew8:loDeltaE}   {\ensuremath{{151.452 } } }
\vdef{default-11:SgMc-B:pvavew8:hiDelta}   {\ensuremath{{+0.004 } } }
\vdef{default-11:SgMc-B:pvavew8:hiDeltaE}   {\ensuremath{{0.216 } } }
\vdef{default-11:SgData-B:pvntrk:loEff}   {\ensuremath{{1.000 } } }
\vdef{default-11:SgData-B:pvntrk:loEffE}   {\ensuremath{{0.099 } } }
\vdef{default-11:SgData-B:pvntrk:hiEff}   {\ensuremath{{1.000 } } }
\vdef{default-11:SgData-B:pvntrk:hiEffE}   {\ensuremath{{0.099 } } }
\vdef{default-11:SgMc-B:pvntrk:loEff}   {\ensuremath{{1.000 } } }
\vdef{default-11:SgMc-B:pvntrk:loEffE}   {\ensuremath{{0.099 } } }
\vdef{default-11:SgMc-B:pvntrk:hiEff}   {\ensuremath{{1.000 } } }
\vdef{default-11:SgMc-B:pvntrk:hiEffE}   {\ensuremath{{0.099 } } }
\vdef{default-11:SgMc-B:pvntrk:loDelta}   {\ensuremath{{+0.000 } } }
\vdef{default-11:SgMc-B:pvntrk:loDeltaE}   {\ensuremath{{0.141 } } }
\vdef{default-11:SgMc-B:pvntrk:hiDelta}   {\ensuremath{{+0.000 } } }
\vdef{default-11:SgMc-B:pvntrk:hiDeltaE}   {\ensuremath{{0.141 } } }
\vdef{default-11:SgData-B:muon1pt:loEff}   {\ensuremath{{1.000 } } }
\vdef{default-11:SgData-B:muon1pt:loEffE}   {\ensuremath{{0.141 } } }
\vdef{default-11:SgData-B:muon1pt:hiEff}   {\ensuremath{{1.000 } } }
\vdef{default-11:SgData-B:muon1pt:hiEffE}   {\ensuremath{{0.141 } } }
\vdef{default-11:SgMc-B:muon1pt:loEff}   {\ensuremath{{1.017 } } }
\vdef{default-11:SgMc-B:muon1pt:loEffE}   {\ensuremath{{0.141 } } }
\vdef{default-11:SgMc-B:muon1pt:hiEff}   {\ensuremath{{1.000 } } }
\vdef{default-11:SgMc-B:muon1pt:hiEffE}   {\ensuremath{{0.141 } } }
\vdef{default-11:SgMc-B:muon1pt:loDelta}   {\ensuremath{{-0.017 } } }
\vdef{default-11:SgMc-B:muon1pt:loDeltaE}   {\ensuremath{{0.197 } } }
\vdef{default-11:SgMc-B:muon1pt:hiDelta}   {\ensuremath{{+0.000 } } }
\vdef{default-11:SgMc-B:muon1pt:hiDeltaE}   {\ensuremath{{0.199 } } }
\vdef{default-11:SgData-B:muon2pt:loEff}   {\ensuremath{{0.000 } } }
\vdef{default-11:SgData-B:muon2pt:loEffE}   {\ensuremath{{0.141 } } }
\vdef{default-11:SgData-B:muon2pt:hiEff}   {\ensuremath{{1.000 } } }
\vdef{default-11:SgData-B:muon2pt:hiEffE}   {\ensuremath{{0.141 } } }
\vdef{default-11:SgMc-B:muon2pt:loEff}   {\ensuremath{{0.041 } } }
\vdef{default-11:SgMc-B:muon2pt:loEffE}   {\ensuremath{{0.141 } } }
\vdef{default-11:SgMc-B:muon2pt:hiEff}   {\ensuremath{{0.959 } } }
\vdef{default-11:SgMc-B:muon2pt:hiEffE}   {\ensuremath{{0.178 } } }
\vdef{default-11:SgMc-B:muon2pt:loDelta}   {\ensuremath{{-2.000 } } }
\vdef{default-11:SgMc-B:muon2pt:loDeltaE}   {\ensuremath{{13.789 } } }
\vdef{default-11:SgMc-B:muon2pt:hiDelta}   {\ensuremath{{+0.042 } } }
\vdef{default-11:SgMc-B:muon2pt:hiDeltaE}   {\ensuremath{{0.233 } } }
\vdef{default-11:SgData-B:muonseta:loEff}   {\ensuremath{{0.875 } } }
\vdef{default-11:SgData-B:muonseta:loEffE}   {\ensuremath{{0.121 } } }
\vdef{default-11:SgData-B:muonseta:hiEff}   {\ensuremath{{0.125 } } }
\vdef{default-11:SgData-B:muonseta:hiEffE}   {\ensuremath{{0.121 } } }
\vdef{default-11:SgMc-B:muonseta:loEff}   {\ensuremath{{0.813 } } }
\vdef{default-11:SgMc-B:muonseta:loEffE}   {\ensuremath{{0.131 } } }
\vdef{default-11:SgMc-B:muonseta:hiEff}   {\ensuremath{{0.187 } } }
\vdef{default-11:SgMc-B:muonseta:hiEffE}   {\ensuremath{{0.131 } } }
\vdef{default-11:SgMc-B:muonseta:loDelta}   {\ensuremath{{+0.074 } } }
\vdef{default-11:SgMc-B:muonseta:loDeltaE}   {\ensuremath{{0.212 } } }
\vdef{default-11:SgMc-B:muonseta:hiDelta}   {\ensuremath{{-0.398 } } }
\vdef{default-11:SgMc-B:muonseta:hiDeltaE}   {\ensuremath{{1.146 } } }
\vdef{default-11:SgData-B:pt:loEff}   {\ensuremath{{0.000 } } }
\vdef{default-11:SgData-B:pt:loEffE}   {\ensuremath{{0.090 } } }
\vdef{default-11:SgData-B:pt:hiEff}   {\ensuremath{{1.000 } } }
\vdef{default-11:SgData-B:pt:hiEffE}   {\ensuremath{{0.090 } } }
\vdef{default-11:SgMc-B:pt:loEff}   {\ensuremath{{0.000 } } }
\vdef{default-11:SgMc-B:pt:loEffE}   {\ensuremath{{0.099 } } }
\vdef{default-11:SgMc-B:pt:hiEff}   {\ensuremath{{1.000 } } }
\vdef{default-11:SgMc-B:pt:hiEffE}   {\ensuremath{{0.099 } } }
\vdef{default-11:SgMc-B:pt:loDelta}   {\ensuremath{{\mathrm{NaN} } } }
\vdef{default-11:SgMc-B:pt:loDeltaE}   {\ensuremath{{\mathrm{NaN} } } }
\vdef{default-11:SgMc-B:pt:hiDelta}   {\ensuremath{{+0.000 } } }
\vdef{default-11:SgMc-B:pt:hiDeltaE}   {\ensuremath{{0.134 } } }
\vdef{default-11:SgData-B:p:loEff}   {\ensuremath{{1.000 } } }
\vdef{default-11:SgData-B:p:loEffE}   {\ensuremath{{0.141 } } }
\vdef{default-11:SgData-B:p:hiEff}   {\ensuremath{{1.000 } } }
\vdef{default-11:SgData-B:p:hiEffE}   {\ensuremath{{0.141 } } }
\vdef{default-11:SgMc-B:p:loEff}   {\ensuremath{{1.000 } } }
\vdef{default-11:SgMc-B:p:loEffE}   {\ensuremath{{0.141 } } }
\vdef{default-11:SgMc-B:p:hiEff}   {\ensuremath{{1.000 } } }
\vdef{default-11:SgMc-B:p:hiEffE}   {\ensuremath{{0.141 } } }
\vdef{default-11:SgMc-B:p:loDelta}   {\ensuremath{{+0.000 } } }
\vdef{default-11:SgMc-B:p:loDeltaE}   {\ensuremath{{0.199 } } }
\vdef{default-11:SgMc-B:p:hiDelta}   {\ensuremath{{+0.000 } } }
\vdef{default-11:SgMc-B:p:hiDeltaE}   {\ensuremath{{0.199 } } }
\vdef{default-11:SgData-B:eta:loEff}   {\ensuremath{{0.750 } } }
\vdef{default-11:SgData-B:eta:loEffE}   {\ensuremath{{0.178 } } }
\vdef{default-11:SgData-B:eta:hiEff}   {\ensuremath{{0.250 } } }
\vdef{default-11:SgData-B:eta:hiEffE}   {\ensuremath{{0.178 } } }
\vdef{default-11:SgMc-B:eta:loEff}   {\ensuremath{{0.804 } } }
\vdef{default-11:SgMc-B:eta:loEffE}   {\ensuremath{{0.163 } } }
\vdef{default-11:SgMc-B:eta:hiEff}   {\ensuremath{{0.196 } } }
\vdef{default-11:SgMc-B:eta:hiEffE}   {\ensuremath{{0.163 } } }
\vdef{default-11:SgMc-B:eta:loDelta}   {\ensuremath{{-0.069 } } }
\vdef{default-11:SgMc-B:eta:loDeltaE}   {\ensuremath{{0.312 } } }
\vdef{default-11:SgMc-B:eta:hiDelta}   {\ensuremath{{+0.240 } } }
\vdef{default-11:SgMc-B:eta:hiDeltaE}   {\ensuremath{{1.079 } } }
\vdef{default-11:SgData-B:bdt:loEff}   {\ensuremath{{0.714 } } }
\vdef{default-11:SgData-B:bdt:loEffE}   {\ensuremath{{0.149 } } }
\vdef{default-11:SgData-B:bdt:hiEff}   {\ensuremath{{0.286 } } }
\vdef{default-11:SgData-B:bdt:hiEffE}   {\ensuremath{{0.149 } } }
\vdef{default-11:SgMc-B:bdt:loEff}   {\ensuremath{{0.890 } } }
\vdef{default-11:SgMc-B:bdt:loEffE}   {\ensuremath{{0.131 } } }
\vdef{default-11:SgMc-B:bdt:hiEff}   {\ensuremath{{0.110 } } }
\vdef{default-11:SgMc-B:bdt:hiEffE}   {\ensuremath{{0.099 } } }
\vdef{default-11:SgMc-B:bdt:loDelta}   {\ensuremath{{-0.219 } } }
\vdef{default-11:SgMc-B:bdt:loDeltaE}   {\ensuremath{{0.253 } } }
\vdef{default-11:SgMc-B:bdt:hiDelta}   {\ensuremath{{+0.886 } } }
\vdef{default-11:SgMc-B:bdt:hiDeltaE}   {\ensuremath{{0.837 } } }
\vdef{default-11:SgData-B:fl3d:loEff}   {\ensuremath{{0.967 } } }
\vdef{default-11:SgData-B:fl3d:loEffE}   {\ensuremath{{0.042 } } }
\vdef{default-11:SgData-B:fl3d:hiEff}   {\ensuremath{{0.033 } } }
\vdef{default-11:SgData-B:fl3d:hiEffE}   {\ensuremath{{0.042 } } }
\vdef{default-11:SgMc-B:fl3d:loEff}   {\ensuremath{{0.950 } } }
\vdef{default-11:SgMc-B:fl3d:loEffE}   {\ensuremath{{0.051 } } }
\vdef{default-11:SgMc-B:fl3d:hiEff}   {\ensuremath{{0.050 } } }
\vdef{default-11:SgMc-B:fl3d:hiEffE}   {\ensuremath{{0.042 } } }
\vdef{default-11:SgMc-B:fl3d:loDelta}   {\ensuremath{{+0.018 } } }
\vdef{default-11:SgMc-B:fl3d:loDeltaE}   {\ensuremath{{0.069 } } }
\vdef{default-11:SgMc-B:fl3d:hiDelta}   {\ensuremath{{-0.405 } } }
\vdef{default-11:SgMc-B:fl3d:hiDeltaE}   {\ensuremath{{1.455 } } }
\vdef{default-11:SgData-B:fl3de:loEff}   {\ensuremath{{1.000 } } }
\vdef{default-11:SgData-B:fl3de:loEffE}   {\ensuremath{{0.033 } } }
\vdef{default-11:SgData-B:fl3de:hiEff}   {\ensuremath{{0.111 } } }
\vdef{default-11:SgData-B:fl3de:hiEffE}   {\ensuremath{{0.063 } } }
\vdef{default-11:SgMc-B:fl3de:loEff}   {\ensuremath{{1.000 } } }
\vdef{default-11:SgMc-B:fl3de:loEffE}   {\ensuremath{{0.034 } } }
\vdef{default-11:SgMc-B:fl3de:hiEff}   {\ensuremath{{0.000 } } }
\vdef{default-11:SgMc-B:fl3de:hiEffE}   {\ensuremath{{0.034 } } }
\vdef{default-11:SgMc-B:fl3de:loDelta}   {\ensuremath{{+0.000 } } }
\vdef{default-11:SgMc-B:fl3de:loDeltaE}   {\ensuremath{{0.048 } } }
\vdef{default-11:SgMc-B:fl3de:hiDelta}   {\ensuremath{{+2.000 } } }
\vdef{default-11:SgMc-B:fl3de:hiDeltaE}   {\ensuremath{{1.241 } } }
\vdef{default-11:SgData-B:fls3d:loEff}   {\ensuremath{{0.733 } } }
\vdef{default-11:SgData-B:fls3d:loEffE}   {\ensuremath{{0.078 } } }
\vdef{default-11:SgData-B:fls3d:hiEff}   {\ensuremath{{0.267 } } }
\vdef{default-11:SgData-B:fls3d:hiEffE}   {\ensuremath{{0.078 } } }
\vdef{default-11:SgMc-B:fls3d:loEff}   {\ensuremath{{0.056 } } }
\vdef{default-11:SgMc-B:fls3d:loEffE}   {\ensuremath{{0.042 } } }
\vdef{default-11:SgMc-B:fls3d:hiEff}   {\ensuremath{{0.944 } } }
\vdef{default-11:SgMc-B:fls3d:hiEffE}   {\ensuremath{{0.051 } } }
\vdef{default-11:SgMc-B:fls3d:loDelta}   {\ensuremath{{+1.714 } } }
\vdef{default-11:SgMc-B:fls3d:loDeltaE}   {\ensuremath{{0.200 } } }
\vdef{default-11:SgMc-B:fls3d:hiDelta}   {\ensuremath{{-1.119 } } }
\vdef{default-11:SgMc-B:fls3d:hiDeltaE}   {\ensuremath{{0.205 } } }
\vdef{default-11:SgData-B:flsxy:loEff}   {\ensuremath{{1.000 } } }
\vdef{default-11:SgData-B:flsxy:loEffE}   {\ensuremath{{0.030 } } }
\vdef{default-11:SgData-B:flsxy:hiEff}   {\ensuremath{{1.000 } } }
\vdef{default-11:SgData-B:flsxy:hiEffE}   {\ensuremath{{0.030 } } }
\vdef{default-11:SgMc-B:flsxy:loEff}   {\ensuremath{{1.004 } } }
\vdef{default-11:SgMc-B:flsxy:loEffE}   {\ensuremath{{0.030 } } }
\vdef{default-11:SgMc-B:flsxy:hiEff}   {\ensuremath{{1.000 } } }
\vdef{default-11:SgMc-B:flsxy:hiEffE}   {\ensuremath{{0.030 } } }
\vdef{default-11:SgMc-B:flsxy:loDelta}   {\ensuremath{{-0.004 } } }
\vdef{default-11:SgMc-B:flsxy:loDeltaE}   {\ensuremath{{0.043 } } }
\vdef{default-11:SgMc-B:flsxy:hiDelta}   {\ensuremath{{+0.000 } } }
\vdef{default-11:SgMc-B:flsxy:hiDeltaE}   {\ensuremath{{0.043 } } }
\vdef{default-11:SgData-B:chi2dof:loEff}   {\ensuremath{{0.444 } } }
\vdef{default-11:SgData-B:chi2dof:loEffE}   {\ensuremath{{0.144 } } }
\vdef{default-11:SgData-B:chi2dof:hiEff}   {\ensuremath{{0.556 } } }
\vdef{default-11:SgData-B:chi2dof:hiEffE}   {\ensuremath{{0.144 } } }
\vdef{default-11:SgMc-B:chi2dof:loEff}   {\ensuremath{{0.898 } } }
\vdef{default-11:SgMc-B:chi2dof:loEffE}   {\ensuremath{{0.090 } } }
\vdef{default-11:SgMc-B:chi2dof:hiEff}   {\ensuremath{{0.102 } } }
\vdef{default-11:SgMc-B:chi2dof:hiEffE}   {\ensuremath{{0.090 } } }
\vdef{default-11:SgMc-B:chi2dof:loDelta}   {\ensuremath{{-0.676 } } }
\vdef{default-11:SgMc-B:chi2dof:loDeltaE}   {\ensuremath{{0.300 } } }
\vdef{default-11:SgMc-B:chi2dof:hiDelta}   {\ensuremath{{+1.382 } } }
\vdef{default-11:SgMc-B:chi2dof:hiDeltaE}   {\ensuremath{{0.485 } } }
\vdef{default-11:SgData-B:pchi2dof:loEff}   {\ensuremath{{0.882 } } }
\vdef{default-11:SgData-B:pchi2dof:loEffE}   {\ensuremath{{0.082 } } }
\vdef{default-11:SgData-B:pchi2dof:hiEff}   {\ensuremath{{0.118 } } }
\vdef{default-11:SgData-B:pchi2dof:hiEffE}   {\ensuremath{{0.082 } } }
\vdef{default-11:SgMc-B:pchi2dof:loEff}   {\ensuremath{{0.705 } } }
\vdef{default-11:SgMc-B:pchi2dof:loEffE}   {\ensuremath{{0.108 } } }
\vdef{default-11:SgMc-B:pchi2dof:hiEff}   {\ensuremath{{0.295 } } }
\vdef{default-11:SgMc-B:pchi2dof:hiEffE}   {\ensuremath{{0.104 } } }
\vdef{default-11:SgMc-B:pchi2dof:loDelta}   {\ensuremath{{+0.223 } } }
\vdef{default-11:SgMc-B:pchi2dof:loDeltaE}   {\ensuremath{{0.177 } } }
\vdef{default-11:SgMc-B:pchi2dof:hiDelta}   {\ensuremath{{-0.860 } } }
\vdef{default-11:SgMc-B:pchi2dof:hiDeltaE}   {\ensuremath{{0.634 } } }
\vdef{default-11:SgData-B:alpha:loEff}   {\ensuremath{{0.571 } } }
\vdef{default-11:SgData-B:alpha:loEffE}   {\ensuremath{{0.157 } } }
\vdef{default-11:SgData-B:alpha:hiEff}   {\ensuremath{{0.429 } } }
\vdef{default-11:SgData-B:alpha:hiEffE}   {\ensuremath{{0.157 } } }
\vdef{default-11:SgMc-B:alpha:loEff}   {\ensuremath{{0.977 } } }
\vdef{default-11:SgMc-B:alpha:loEffE}   {\ensuremath{{0.110 } } }
\vdef{default-11:SgMc-B:alpha:hiEff}   {\ensuremath{{0.023 } } }
\vdef{default-11:SgMc-B:alpha:hiEffE}   {\ensuremath{{0.110 } } }
\vdef{default-11:SgMc-B:alpha:loDelta}   {\ensuremath{{-0.524 } } }
\vdef{default-11:SgMc-B:alpha:loDeltaE}   {\ensuremath{{0.277 } } }
\vdef{default-11:SgMc-B:alpha:hiDelta}   {\ensuremath{{+1.794 } } }
\vdef{default-11:SgMc-B:alpha:hiDeltaE}   {\ensuremath{{0.928 } } }
\vdef{default-11:SgData-B:iso:loEff}   {\ensuremath{{0.714 } } }
\vdef{default-11:SgData-B:iso:loEffE}   {\ensuremath{{0.112 } } }
\vdef{default-11:SgData-B:iso:hiEff}   {\ensuremath{{0.286 } } }
\vdef{default-11:SgData-B:iso:hiEffE}   {\ensuremath{{0.112 } } }
\vdef{default-11:SgMc-B:iso:loEff}   {\ensuremath{{0.101 } } }
\vdef{default-11:SgMc-B:iso:loEffE}   {\ensuremath{{0.080 } } }
\vdef{default-11:SgMc-B:iso:hiEff}   {\ensuremath{{0.899 } } }
\vdef{default-11:SgMc-B:iso:hiEffE}   {\ensuremath{{0.095 } } }
\vdef{default-11:SgMc-B:iso:loDelta}   {\ensuremath{{+1.504 } } }
\vdef{default-11:SgMc-B:iso:loDeltaE}   {\ensuremath{{0.351 } } }
\vdef{default-11:SgMc-B:iso:hiDelta}   {\ensuremath{{-1.035 } } }
\vdef{default-11:SgMc-B:iso:hiDeltaE}   {\ensuremath{{0.298 } } }
\vdef{default-11:SgData-B:docatrk:loEff}   {\ensuremath{{0.500 } } }
\vdef{default-11:SgData-B:docatrk:loEffE}   {\ensuremath{{0.151 } } }
\vdef{default-11:SgData-B:docatrk:hiEff}   {\ensuremath{{0.500 } } }
\vdef{default-11:SgData-B:docatrk:hiEffE}   {\ensuremath{{0.151 } } }
\vdef{default-11:SgMc-B:docatrk:loEff}   {\ensuremath{{0.104 } } }
\vdef{default-11:SgMc-B:docatrk:loEffE}   {\ensuremath{{0.099 } } }
\vdef{default-11:SgMc-B:docatrk:hiEff}   {\ensuremath{{0.896 } } }
\vdef{default-11:SgMc-B:docatrk:hiEffE}   {\ensuremath{{0.099 } } }
\vdef{default-11:SgMc-B:docatrk:loDelta}   {\ensuremath{{+1.311 } } }
\vdef{default-11:SgMc-B:docatrk:loDeltaE}   {\ensuremath{{0.571 } } }
\vdef{default-11:SgMc-B:docatrk:hiDelta}   {\ensuremath{{-0.567 } } }
\vdef{default-11:SgMc-B:docatrk:hiDeltaE}   {\ensuremath{{0.295 } } }
\vdef{default-11:SgData-B:isotrk:loEff}   {\ensuremath{{1.000 } } }
\vdef{default-11:SgData-B:isotrk:loEffE}   {\ensuremath{{0.059 } } }
\vdef{default-11:SgData-B:isotrk:hiEff}   {\ensuremath{{1.000 } } }
\vdef{default-11:SgData-B:isotrk:hiEffE}   {\ensuremath{{0.059 } } }
\vdef{default-11:SgMc-B:isotrk:loEff}   {\ensuremath{{1.000 } } }
\vdef{default-11:SgMc-B:isotrk:loEffE}   {\ensuremath{{0.062 } } }
\vdef{default-11:SgMc-B:isotrk:hiEff}   {\ensuremath{{1.000 } } }
\vdef{default-11:SgMc-B:isotrk:hiEffE}   {\ensuremath{{0.062 } } }
\vdef{default-11:SgMc-B:isotrk:loDelta}   {\ensuremath{{+0.000 } } }
\vdef{default-11:SgMc-B:isotrk:loDeltaE}   {\ensuremath{{0.086 } } }
\vdef{default-11:SgMc-B:isotrk:hiDelta}   {\ensuremath{{+0.000 } } }
\vdef{default-11:SgMc-B:isotrk:hiDeltaE}   {\ensuremath{{0.086 } } }
\vdef{default-11:SgData-B:closetrk:loEff}   {\ensuremath{{0.800 } } }
\vdef{default-11:SgData-B:closetrk:loEffE}   {\ensuremath{{0.160 } } }
\vdef{default-11:SgData-B:closetrk:hiEff}   {\ensuremath{{0.200 } } }
\vdef{default-11:SgData-B:closetrk:hiEffE}   {\ensuremath{{0.160 } } }
\vdef{default-11:SgMc-B:closetrk:loEff}   {\ensuremath{{0.965 } } }
\vdef{default-11:SgMc-B:closetrk:loEffE}   {\ensuremath{{0.160 } } }
\vdef{default-11:SgMc-B:closetrk:hiEff}   {\ensuremath{{0.035 } } }
\vdef{default-11:SgMc-B:closetrk:hiEffE}   {\ensuremath{{0.124 } } }
\vdef{default-11:SgMc-B:closetrk:loDelta}   {\ensuremath{{-0.187 } } }
\vdef{default-11:SgMc-B:closetrk:loDeltaE}   {\ensuremath{{0.257 } } }
\vdef{default-11:SgMc-B:closetrk:hiDelta}   {\ensuremath{{+1.402 } } }
\vdef{default-11:SgMc-B:closetrk:hiDeltaE}   {\ensuremath{{1.835 } } }
\vdef{default-11:SgData-B:lip:loEff}   {\ensuremath{{1.000 } } }
\vdef{default-11:SgData-B:lip:loEffE}   {\ensuremath{{0.141 } } }
\vdef{default-11:SgData-B:lip:hiEff}   {\ensuremath{{0.000 } } }
\vdef{default-11:SgData-B:lip:hiEffE}   {\ensuremath{{0.141 } } }
\vdef{default-11:SgMc-B:lip:loEff}   {\ensuremath{{1.000 } } }
\vdef{default-11:SgMc-B:lip:loEffE}   {\ensuremath{{0.163 } } }
\vdef{default-11:SgMc-B:lip:hiEff}   {\ensuremath{{0.000 } } }
\vdef{default-11:SgMc-B:lip:hiEffE}   {\ensuremath{{0.163 } } }
\vdef{default-11:SgMc-B:lip:loDelta}   {\ensuremath{{+0.000 } } }
\vdef{default-11:SgMc-B:lip:loDeltaE}   {\ensuremath{{0.216 } } }
\vdef{default-11:SgMc-B:lip:hiDelta}   {\ensuremath{{\mathrm{NaN} } } }
\vdef{default-11:SgMc-B:lip:hiDeltaE}   {\ensuremath{{\mathrm{NaN} } } }
\vdef{default-11:SgData-B:lip:inEff}   {\ensuremath{{1.000 } } }
\vdef{default-11:SgData-B:lip:inEffE}   {\ensuremath{{0.141 } } }
\vdef{default-11:SgMc-B:lip:inEff}   {\ensuremath{{1.000 } } }
\vdef{default-11:SgMc-B:lip:inEffE}   {\ensuremath{{0.163 } } }
\vdef{default-11:SgMc-B:lip:inDelta}   {\ensuremath{{+0.000 } } }
\vdef{default-11:SgMc-B:lip:inDeltaE}   {\ensuremath{{0.216 } } }
\vdef{default-11:SgData-B:lips:loEff}   {\ensuremath{{1.000 } } }
\vdef{default-11:SgData-B:lips:loEffE}   {\ensuremath{{0.141 } } }
\vdef{default-11:SgData-B:lips:hiEff}   {\ensuremath{{0.000 } } }
\vdef{default-11:SgData-B:lips:hiEffE}   {\ensuremath{{0.141 } } }
\vdef{default-11:SgMc-B:lips:loEff}   {\ensuremath{{1.000 } } }
\vdef{default-11:SgMc-B:lips:loEffE}   {\ensuremath{{0.141 } } }
\vdef{default-11:SgMc-B:lips:hiEff}   {\ensuremath{{0.000 } } }
\vdef{default-11:SgMc-B:lips:hiEffE}   {\ensuremath{{0.141 } } }
\vdef{default-11:SgMc-B:lips:loDelta}   {\ensuremath{{+0.000 } } }
\vdef{default-11:SgMc-B:lips:loDeltaE}   {\ensuremath{{0.199 } } }
\vdef{default-11:SgMc-B:lips:hiDelta}   {\ensuremath{{\mathrm{NaN} } } }
\vdef{default-11:SgMc-B:lips:hiDeltaE}   {\ensuremath{{\mathrm{NaN} } } }
\vdef{default-11:SgData-B:lips:inEff}   {\ensuremath{{1.000 } } }
\vdef{default-11:SgData-B:lips:inEffE}   {\ensuremath{{0.141 } } }
\vdef{default-11:SgMc-B:lips:inEff}   {\ensuremath{{1.000 } } }
\vdef{default-11:SgMc-B:lips:inEffE}   {\ensuremath{{0.141 } } }
\vdef{default-11:SgMc-B:lips:inDelta}   {\ensuremath{{+0.000 } } }
\vdef{default-11:SgMc-B:lips:inDeltaE}   {\ensuremath{{0.199 } } }
\vdef{default-11:SgData-B:ip:loEff}   {\ensuremath{{0.500 } } }
\vdef{default-11:SgData-B:ip:loEffE}   {\ensuremath{{0.151 } } }
\vdef{default-11:SgData-B:ip:hiEff}   {\ensuremath{{0.500 } } }
\vdef{default-11:SgData-B:ip:hiEffE}   {\ensuremath{{0.151 } } }
\vdef{default-11:SgMc-B:ip:loEff}   {\ensuremath{{0.962 } } }
\vdef{default-11:SgMc-B:ip:loEffE}   {\ensuremath{{0.121 } } }
\vdef{default-11:SgMc-B:ip:hiEff}   {\ensuremath{{0.038 } } }
\vdef{default-11:SgMc-B:ip:hiEffE}   {\ensuremath{{0.090 } } }
\vdef{default-11:SgMc-B:ip:loDelta}   {\ensuremath{{-0.632 } } }
\vdef{default-11:SgMc-B:ip:loDeltaE}   {\ensuremath{{0.294 } } }
\vdef{default-11:SgMc-B:ip:hiDelta}   {\ensuremath{{+1.718 } } }
\vdef{default-11:SgMc-B:ip:hiDeltaE}   {\ensuremath{{0.630 } } }
\vdef{default-11:SgData-B:ips:loEff}   {\ensuremath{{0.500 } } }
\vdef{default-11:SgData-B:ips:loEffE}   {\ensuremath{{0.151 } } }
\vdef{default-11:SgData-B:ips:hiEff}   {\ensuremath{{0.500 } } }
\vdef{default-11:SgData-B:ips:hiEffE}   {\ensuremath{{0.151 } } }
\vdef{default-11:SgMc-B:ips:loEff}   {\ensuremath{{0.974 } } }
\vdef{default-11:SgMc-B:ips:loEffE}   {\ensuremath{{0.099 } } }
\vdef{default-11:SgMc-B:ips:hiEff}   {\ensuremath{{0.026 } } }
\vdef{default-11:SgMc-B:ips:hiEffE}   {\ensuremath{{0.099 } } }
\vdef{default-11:SgMc-B:ips:loDelta}   {\ensuremath{{-0.643 } } }
\vdef{default-11:SgMc-B:ips:loDeltaE}   {\ensuremath{{0.285 } } }
\vdef{default-11:SgMc-B:ips:hiDelta}   {\ensuremath{{+1.802 } } }
\vdef{default-11:SgMc-B:ips:hiDeltaE}   {\ensuremath{{0.720 } } }
\vdef{default-11:SgData-B:maxdoca:loEff}   {\ensuremath{{1.000 } } }
\vdef{default-11:SgData-B:maxdoca:loEffE}   {\ensuremath{{0.141 } } }
\vdef{default-11:SgData-B:maxdoca:hiEff}   {\ensuremath{{0.000 } } }
\vdef{default-11:SgData-B:maxdoca:hiEffE}   {\ensuremath{{0.141 } } }
\vdef{default-11:SgMc-B:maxdoca:loEff}   {\ensuremath{{1.000 } } }
\vdef{default-11:SgMc-B:maxdoca:loEffE}   {\ensuremath{{0.141 } } }
\vdef{default-11:SgMc-B:maxdoca:hiEff}   {\ensuremath{{0.000 } } }
\vdef{default-11:SgMc-B:maxdoca:hiEffE}   {\ensuremath{{0.141 } } }
\vdef{default-11:SgMc-B:maxdoca:loDelta}   {\ensuremath{{+0.000 } } }
\vdef{default-11:SgMc-B:maxdoca:loDeltaE}   {\ensuremath{{0.199 } } }
\vdef{default-11:SgMc-B:maxdoca:hiDelta}   {\ensuremath{{\mathrm{NaN} } } }
\vdef{default-11:SgMc-B:maxdoca:hiDeltaE}   {\ensuremath{{\mathrm{NaN} } } }
\vdef{default-11:SgData-E:osiso:loEff}   {\ensuremath{{1.018 } } }
\vdef{default-11:SgData-E:osiso:loEffE}   {\ensuremath{{0.000 } } }
\vdef{default-11:SgData-E:osiso:hiEff}   {\ensuremath{{1.000 } } }
\vdef{default-11:SgData-E:osiso:hiEffE}   {\ensuremath{{0.017 } } }
\vdef{default-11:SgMc-E:osiso:loEff}   {\ensuremath{{1.000 } } }
\vdef{default-11:SgMc-E:osiso:loEffE}   {\ensuremath{{0.017 } } }
\vdef{default-11:SgMc-E:osiso:hiEff}   {\ensuremath{{1.000 } } }
\vdef{default-11:SgMc-E:osiso:hiEffE}   {\ensuremath{{0.017 } } }
\vdef{default-11:SgMc-E:osiso:loDelta}   {\ensuremath{{+0.018 } } }
\vdef{default-11:SgMc-E:osiso:loDeltaE}   {\ensuremath{{0.017 } } }
\vdef{default-11:SgMc-E:osiso:hiDelta}   {\ensuremath{{+0.000 } } }
\vdef{default-11:SgMc-E:osiso:hiDeltaE}   {\ensuremath{{0.024 } } }
\vdef{default-11:SgData-E:osreliso:loEff}   {\ensuremath{{0.189 } } }
\vdef{default-11:SgData-E:osreliso:loEffE}   {\ensuremath{{0.053 } } }
\vdef{default-11:SgData-E:osreliso:hiEff}   {\ensuremath{{0.811 } } }
\vdef{default-11:SgData-E:osreliso:hiEffE}   {\ensuremath{{0.053 } } }
\vdef{default-11:SgMc-E:osreliso:loEff}   {\ensuremath{{0.267 } } }
\vdef{default-11:SgMc-E:osreliso:loEffE}   {\ensuremath{{0.060 } } }
\vdef{default-11:SgMc-E:osreliso:hiEff}   {\ensuremath{{0.733 } } }
\vdef{default-11:SgMc-E:osreliso:hiEffE}   {\ensuremath{{0.060 } } }
\vdef{default-11:SgMc-E:osreliso:loDelta}   {\ensuremath{{-0.344 } } }
\vdef{default-11:SgMc-E:osreliso:loDeltaE}   {\ensuremath{{0.352 } } }
\vdef{default-11:SgMc-E:osreliso:hiDelta}   {\ensuremath{{+0.102 } } }
\vdef{default-11:SgMc-E:osreliso:hiDeltaE}   {\ensuremath{{0.105 } } }
\vdef{default-11:SgData-E:osmuonpt:loEff}   {\ensuremath{{0.000 } } }
\vdef{default-11:SgData-E:osmuonpt:loEffE}   {\ensuremath{{0.236 } } }
\vdef{default-11:SgData-E:osmuonpt:hiEff}   {\ensuremath{{1.000 } } }
\vdef{default-11:SgData-E:osmuonpt:hiEffE}   {\ensuremath{{0.236 } } }
\vdef{default-11:SgMc-E:osmuonpt:loEff}   {\ensuremath{{0.000 } } }
\vdef{default-11:SgMc-E:osmuonpt:loEffE}   {\ensuremath{{0.236 } } }
\vdef{default-11:SgMc-E:osmuonpt:hiEff}   {\ensuremath{{1.000 } } }
\vdef{default-11:SgMc-E:osmuonpt:hiEffE}   {\ensuremath{{0.236 } } }
\vdef{default-11:SgMc-E:osmuonpt:loDelta}   {\ensuremath{{\mathrm{NaN} } } }
\vdef{default-11:SgMc-E:osmuonpt:loDeltaE}   {\ensuremath{{\mathrm{NaN} } } }
\vdef{default-11:SgMc-E:osmuonpt:hiDelta}   {\ensuremath{{+0.000 } } }
\vdef{default-11:SgMc-E:osmuonpt:hiDeltaE}   {\ensuremath{{0.333 } } }
\vdef{default-11:SgData-E:osmuondr:loEff}   {\ensuremath{{0.000 } } }
\vdef{default-11:SgData-E:osmuondr:loEffE}   {\ensuremath{{0.236 } } }
\vdef{default-11:SgData-E:osmuondr:hiEff}   {\ensuremath{{1.000 } } }
\vdef{default-11:SgData-E:osmuondr:hiEffE}   {\ensuremath{{0.236 } } }
\vdef{default-11:SgMc-E:osmuondr:loEff}   {\ensuremath{{0.040 } } }
\vdef{default-11:SgMc-E:osmuondr:loEffE}   {\ensuremath{{0.236 } } }
\vdef{default-11:SgMc-E:osmuondr:hiEff}   {\ensuremath{{0.960 } } }
\vdef{default-11:SgMc-E:osmuondr:hiEffE}   {\ensuremath{{0.236 } } }
\vdef{default-11:SgMc-E:osmuondr:loDelta}   {\ensuremath{{-2.000 } } }
\vdef{default-11:SgMc-E:osmuondr:loDeltaE}   {\ensuremath{{23.570 } } }
\vdef{default-11:SgMc-E:osmuondr:hiDelta}   {\ensuremath{{+0.041 } } }
\vdef{default-11:SgMc-E:osmuondr:hiDeltaE}   {\ensuremath{{0.340 } } }
\vdef{default-11:SgData-E:hlt:loEff}   {\ensuremath{{0.065 } } }
\vdef{default-11:SgData-E:hlt:loEffE}   {\ensuremath{{0.049 } } }
\vdef{default-11:SgData-E:hlt:hiEff}   {\ensuremath{{0.935 } } }
\vdef{default-11:SgData-E:hlt:hiEffE}   {\ensuremath{{0.049 } } }
\vdef{default-11:SgMc-E:hlt:loEff}   {\ensuremath{{0.281 } } }
\vdef{default-11:SgMc-E:hlt:loEffE}   {\ensuremath{{0.076 } } }
\vdef{default-11:SgMc-E:hlt:hiEff}   {\ensuremath{{0.719 } } }
\vdef{default-11:SgMc-E:hlt:hiEffE}   {\ensuremath{{0.079 } } }
\vdef{default-11:SgMc-E:hlt:loDelta}   {\ensuremath{{-1.253 } } }
\vdef{default-11:SgMc-E:hlt:loDeltaE}   {\ensuremath{{0.493 } } }
\vdef{default-11:SgMc-E:hlt:hiDelta}   {\ensuremath{{+0.262 } } }
\vdef{default-11:SgMc-E:hlt:hiDeltaE}   {\ensuremath{{0.120 } } }
\vdef{default-11:SgData-E:muonsid:loEff}   {\ensuremath{{0.121 } } }
\vdef{default-11:SgData-E:muonsid:loEffE}   {\ensuremath{{0.058 } } }
\vdef{default-11:SgData-E:muonsid:hiEff}   {\ensuremath{{0.879 } } }
\vdef{default-11:SgData-E:muonsid:hiEffE}   {\ensuremath{{0.058 } } }
\vdef{default-11:SgMc-E:muonsid:loEff}   {\ensuremath{{0.105 } } }
\vdef{default-11:SgMc-E:muonsid:loEffE}   {\ensuremath{{0.053 } } }
\vdef{default-11:SgMc-E:muonsid:hiEff}   {\ensuremath{{0.895 } } }
\vdef{default-11:SgMc-E:muonsid:hiEffE}   {\ensuremath{{0.058 } } }
\vdef{default-11:SgMc-E:muonsid:loDelta}   {\ensuremath{{+0.141 } } }
\vdef{default-11:SgMc-E:muonsid:loDeltaE}   {\ensuremath{{0.693 } } }
\vdef{default-11:SgMc-E:muonsid:hiDelta}   {\ensuremath{{-0.018 } } }
\vdef{default-11:SgMc-E:muonsid:hiDeltaE}   {\ensuremath{{0.093 } } }
\vdef{default-11:SgData-E:tracksqual:loEff}   {\ensuremath{{0.000 } } }
\vdef{default-11:SgData-E:tracksqual:loEffE}   {\ensuremath{{0.031 } } }
\vdef{default-11:SgData-E:tracksqual:hiEff}   {\ensuremath{{1.000 } } }
\vdef{default-11:SgData-E:tracksqual:hiEffE}   {\ensuremath{{0.031 } } }
\vdef{default-11:SgMc-E:tracksqual:loEff}   {\ensuremath{{0.000 } } }
\vdef{default-11:SgMc-E:tracksqual:loEffE}   {\ensuremath{{0.031 } } }
\vdef{default-11:SgMc-E:tracksqual:hiEff}   {\ensuremath{{1.000 } } }
\vdef{default-11:SgMc-E:tracksqual:hiEffE}   {\ensuremath{{0.031 } } }
\vdef{default-11:SgMc-E:tracksqual:loDelta}   {\ensuremath{{\mathrm{NaN} } } }
\vdef{default-11:SgMc-E:tracksqual:loDeltaE}   {\ensuremath{{\mathrm{NaN} } } }
\vdef{default-11:SgMc-E:tracksqual:hiDelta}   {\ensuremath{{+0.000 } } }
\vdef{default-11:SgMc-E:tracksqual:hiDeltaE}   {\ensuremath{{0.044 } } }
\vdef{default-11:SgData-E:pvz:loEff}   {\ensuremath{{0.484 } } }
\vdef{default-11:SgData-E:pvz:loEffE}   {\ensuremath{{0.086 } } }
\vdef{default-11:SgData-E:pvz:hiEff}   {\ensuremath{{0.516 } } }
\vdef{default-11:SgData-E:pvz:hiEffE}   {\ensuremath{{0.086 } } }
\vdef{default-11:SgMc-E:pvz:loEff}   {\ensuremath{{0.442 } } }
\vdef{default-11:SgMc-E:pvz:loEffE}   {\ensuremath{{0.086 } } }
\vdef{default-11:SgMc-E:pvz:hiEff}   {\ensuremath{{0.558 } } }
\vdef{default-11:SgMc-E:pvz:hiEffE}   {\ensuremath{{0.086 } } }
\vdef{default-11:SgMc-E:pvz:loDelta}   {\ensuremath{{+0.090 } } }
\vdef{default-11:SgMc-E:pvz:loDeltaE}   {\ensuremath{{0.263 } } }
\vdef{default-11:SgMc-E:pvz:hiDelta}   {\ensuremath{{-0.078 } } }
\vdef{default-11:SgMc-E:pvz:hiDeltaE}   {\ensuremath{{0.227 } } }
\vdef{default-11:SgData-E:pvn:loEff}   {\ensuremath{{1.033 } } }
\vdef{default-11:SgData-E:pvn:loEffE}   {\ensuremath{{0.000 } } }
\vdef{default-11:SgData-E:pvn:hiEff}   {\ensuremath{{1.000 } } }
\vdef{default-11:SgData-E:pvn:hiEffE}   {\ensuremath{{0.030 } } }
\vdef{default-11:SgMc-E:pvn:loEff}   {\ensuremath{{1.000 } } }
\vdef{default-11:SgMc-E:pvn:loEffE}   {\ensuremath{{0.030 } } }
\vdef{default-11:SgMc-E:pvn:hiEff}   {\ensuremath{{1.000 } } }
\vdef{default-11:SgMc-E:pvn:hiEffE}   {\ensuremath{{0.030 } } }
\vdef{default-11:SgMc-E:pvn:loDelta}   {\ensuremath{{+0.033 } } }
\vdef{default-11:SgMc-E:pvn:loDeltaE}   {\ensuremath{{0.030 } } }
\vdef{default-11:SgMc-E:pvn:hiDelta}   {\ensuremath{{+0.000 } } }
\vdef{default-11:SgMc-E:pvn:hiDeltaE}   {\ensuremath{{0.043 } } }
\vdef{default-11:SgData-E:pvavew8:loEff}   {\ensuremath{{0.069 } } }
\vdef{default-11:SgData-E:pvavew8:loEffE}   {\ensuremath{{0.052 } } }
\vdef{default-11:SgData-E:pvavew8:hiEff}   {\ensuremath{{0.931 } } }
\vdef{default-11:SgData-E:pvavew8:hiEffE}   {\ensuremath{{0.052 } } }
\vdef{default-11:SgMc-E:pvavew8:loEff}   {\ensuremath{{0.004 } } }
\vdef{default-11:SgMc-E:pvavew8:loEffE}   {\ensuremath{{0.031 } } }
\vdef{default-11:SgMc-E:pvavew8:hiEff}   {\ensuremath{{0.996 } } }
\vdef{default-11:SgMc-E:pvavew8:hiEffE}   {\ensuremath{{0.043 } } }
\vdef{default-11:SgMc-E:pvavew8:loDelta}   {\ensuremath{{+1.780 } } }
\vdef{default-11:SgMc-E:pvavew8:loDeltaE}   {\ensuremath{{1.625 } } }
\vdef{default-11:SgMc-E:pvavew8:hiDelta}   {\ensuremath{{-0.067 } } }
\vdef{default-11:SgMc-E:pvavew8:hiDeltaE}   {\ensuremath{{0.071 } } }
\vdef{default-11:SgData-E:pvntrk:loEff}   {\ensuremath{{1.000 } } }
\vdef{default-11:SgData-E:pvntrk:loEffE}   {\ensuremath{{0.029 } } }
\vdef{default-11:SgData-E:pvntrk:hiEff}   {\ensuremath{{1.000 } } }
\vdef{default-11:SgData-E:pvntrk:hiEffE}   {\ensuremath{{0.029 } } }
\vdef{default-11:SgMc-E:pvntrk:loEff}   {\ensuremath{{1.000 } } }
\vdef{default-11:SgMc-E:pvntrk:loEffE}   {\ensuremath{{0.029 } } }
\vdef{default-11:SgMc-E:pvntrk:hiEff}   {\ensuremath{{1.000 } } }
\vdef{default-11:SgMc-E:pvntrk:hiEffE}   {\ensuremath{{0.029 } } }
\vdef{default-11:SgMc-E:pvntrk:loDelta}   {\ensuremath{{+0.000 } } }
\vdef{default-11:SgMc-E:pvntrk:loDeltaE}   {\ensuremath{{0.042 } } }
\vdef{default-11:SgMc-E:pvntrk:hiDelta}   {\ensuremath{{+0.000 } } }
\vdef{default-11:SgMc-E:pvntrk:hiDeltaE}   {\ensuremath{{0.042 } } }
\vdef{default-11:SgData-E:muon1pt:loEff}   {\ensuremath{{1.000 } } }
\vdef{default-11:SgData-E:muon1pt:loEffE}   {\ensuremath{{0.029 } } }
\vdef{default-11:SgData-E:muon1pt:hiEff}   {\ensuremath{{1.000 } } }
\vdef{default-11:SgData-E:muon1pt:hiEffE}   {\ensuremath{{0.029 } } }
\vdef{default-11:SgMc-E:muon1pt:loEff}   {\ensuremath{{1.007 } } }
\vdef{default-11:SgMc-E:muon1pt:loEffE}   {\ensuremath{{0.029 } } }
\vdef{default-11:SgMc-E:muon1pt:hiEff}   {\ensuremath{{1.000 } } }
\vdef{default-11:SgMc-E:muon1pt:hiEffE}   {\ensuremath{{0.029 } } }
\vdef{default-11:SgMc-E:muon1pt:loDelta}   {\ensuremath{{-0.007 } } }
\vdef{default-11:SgMc-E:muon1pt:loDeltaE}   {\ensuremath{{0.040 } } }
\vdef{default-11:SgMc-E:muon1pt:hiDelta}   {\ensuremath{{+0.000 } } }
\vdef{default-11:SgMc-E:muon1pt:hiDeltaE}   {\ensuremath{{0.040 } } }
\vdef{default-11:SgData-E:muon2pt:loEff}   {\ensuremath{{0.094 } } }
\vdef{default-11:SgData-E:muon2pt:loEffE}   {\ensuremath{{0.054 } } }
\vdef{default-11:SgData-E:muon2pt:hiEff}   {\ensuremath{{0.906 } } }
\vdef{default-11:SgData-E:muon2pt:hiEffE}   {\ensuremath{{0.054 } } }
\vdef{default-11:SgMc-E:muon2pt:loEff}   {\ensuremath{{0.160 } } }
\vdef{default-11:SgMc-E:muon2pt:loEffE}   {\ensuremath{{0.064 } } }
\vdef{default-11:SgMc-E:muon2pt:hiEff}   {\ensuremath{{0.840 } } }
\vdef{default-11:SgMc-E:muon2pt:hiEffE}   {\ensuremath{{0.068 } } }
\vdef{default-11:SgMc-E:muon2pt:loDelta}   {\ensuremath{{-0.521 } } }
\vdef{default-11:SgMc-E:muon2pt:loDeltaE}   {\ensuremath{{0.659 } } }
\vdef{default-11:SgMc-E:muon2pt:hiDelta}   {\ensuremath{{+0.076 } } }
\vdef{default-11:SgMc-E:muon2pt:hiDeltaE}   {\ensuremath{{0.101 } } }
\vdef{default-11:SgData-E:muonseta:loEff}   {\ensuremath{{0.569 } } }
\vdef{default-11:SgData-E:muonseta:loEffE}   {\ensuremath{{0.063 } } }
\vdef{default-11:SgData-E:muonseta:hiEff}   {\ensuremath{{0.431 } } }
\vdef{default-11:SgData-E:muonseta:hiEffE}   {\ensuremath{{0.063 } } }
\vdef{default-11:SgMc-E:muonseta:loEff}   {\ensuremath{{0.513 } } }
\vdef{default-11:SgMc-E:muonseta:loEffE}   {\ensuremath{{0.065 } } }
\vdef{default-11:SgMc-E:muonseta:hiEff}   {\ensuremath{{0.487 } } }
\vdef{default-11:SgMc-E:muonseta:hiEffE}   {\ensuremath{{0.065 } } }
\vdef{default-11:SgMc-E:muonseta:loDelta}   {\ensuremath{{+0.104 } } }
\vdef{default-11:SgMc-E:muonseta:loDeltaE}   {\ensuremath{{0.168 } } }
\vdef{default-11:SgMc-E:muonseta:hiDelta}   {\ensuremath{{-0.123 } } }
\vdef{default-11:SgMc-E:muonseta:hiDeltaE}   {\ensuremath{{0.197 } } }
\vdef{default-11:SgData-E:pt:loEff}   {\ensuremath{{0.000 } } }
\vdef{default-11:SgData-E:pt:loEffE}   {\ensuremath{{0.016 } } }
\vdef{default-11:SgData-E:pt:hiEff}   {\ensuremath{{1.000 } } }
\vdef{default-11:SgData-E:pt:hiEffE}   {\ensuremath{{0.016 } } }
\vdef{default-11:SgMc-E:pt:loEff}   {\ensuremath{{0.000 } } }
\vdef{default-11:SgMc-E:pt:loEffE}   {\ensuremath{{0.016 } } }
\vdef{default-11:SgMc-E:pt:hiEff}   {\ensuremath{{1.000 } } }
\vdef{default-11:SgMc-E:pt:hiEffE}   {\ensuremath{{0.016 } } }
\vdef{default-11:SgMc-E:pt:loDelta}   {\ensuremath{{\mathrm{NaN} } } }
\vdef{default-11:SgMc-E:pt:loDeltaE}   {\ensuremath{{\mathrm{NaN} } } }
\vdef{default-11:SgMc-E:pt:hiDelta}   {\ensuremath{{+0.000 } } }
\vdef{default-11:SgMc-E:pt:hiDeltaE}   {\ensuremath{{0.022 } } }
\vdef{default-11:SgData-E:p:loEff}   {\ensuremath{{1.034 } } }
\vdef{default-11:SgData-E:p:loEffE}   {\ensuremath{{0.000 } } }
\vdef{default-11:SgData-E:p:hiEff}   {\ensuremath{{1.000 } } }
\vdef{default-11:SgData-E:p:hiEffE}   {\ensuremath{{0.031 } } }
\vdef{default-11:SgMc-E:p:loEff}   {\ensuremath{{1.021 } } }
\vdef{default-11:SgMc-E:p:loEffE}   {\ensuremath{{0.000 } } }
\vdef{default-11:SgMc-E:p:hiEff}   {\ensuremath{{1.000 } } }
\vdef{default-11:SgMc-E:p:hiEffE}   {\ensuremath{{0.032 } } }
\vdef{default-11:SgMc-E:p:loDelta}   {\ensuremath{{+0.013 } } }
\vdef{default-11:SgMc-E:p:loDeltaE}   {\ensuremath{{0.000 } } }
\vdef{default-11:SgMc-E:p:hiDelta}   {\ensuremath{{+0.000 } } }
\vdef{default-11:SgMc-E:p:hiDeltaE}   {\ensuremath{{0.045 } } }
\vdef{default-11:SgData-E:eta:loEff}   {\ensuremath{{0.552 } } }
\vdef{default-11:SgData-E:eta:loEffE}   {\ensuremath{{0.088 } } }
\vdef{default-11:SgData-E:eta:hiEff}   {\ensuremath{{0.448 } } }
\vdef{default-11:SgData-E:eta:hiEffE}   {\ensuremath{{0.088 } } }
\vdef{default-11:SgMc-E:eta:loEff}   {\ensuremath{{0.495 } } }
\vdef{default-11:SgMc-E:eta:loEffE}   {\ensuremath{{0.088 } } }
\vdef{default-11:SgMc-E:eta:hiEff}   {\ensuremath{{0.505 } } }
\vdef{default-11:SgMc-E:eta:hiEffE}   {\ensuremath{{0.088 } } }
\vdef{default-11:SgMc-E:eta:loDelta}   {\ensuremath{{+0.107 } } }
\vdef{default-11:SgMc-E:eta:loDeltaE}   {\ensuremath{{0.239 } } }
\vdef{default-11:SgMc-E:eta:hiDelta}   {\ensuremath{{-0.118 } } }
\vdef{default-11:SgMc-E:eta:hiDeltaE}   {\ensuremath{{0.262 } } }
\vdef{default-11:SgData-E:bdt:loEff}   {\ensuremath{{0.968 } } }
\vdef{default-11:SgData-E:bdt:loEffE}   {\ensuremath{{0.041 } } }
\vdef{default-11:SgData-E:bdt:hiEff}   {\ensuremath{{0.032 } } }
\vdef{default-11:SgData-E:bdt:hiEffE}   {\ensuremath{{0.041 } } }
\vdef{default-11:SgMc-E:bdt:loEff}   {\ensuremath{{0.996 } } }
\vdef{default-11:SgMc-E:bdt:loEffE}   {\ensuremath{{0.041 } } }
\vdef{default-11:SgMc-E:bdt:hiEff}   {\ensuremath{{0.004 } } }
\vdef{default-11:SgMc-E:bdt:hiEffE}   {\ensuremath{{0.029 } } }
\vdef{default-11:SgMc-E:bdt:loDelta}   {\ensuremath{{-0.028 } } }
\vdef{default-11:SgMc-E:bdt:loDeltaE}   {\ensuremath{{0.059 } } }
\vdef{default-11:SgMc-E:bdt:hiDelta}   {\ensuremath{{+1.526 } } }
\vdef{default-11:SgMc-E:bdt:hiDeltaE}   {\ensuremath{{2.882 } } }
\vdef{default-11:SgData-E:fl3d:loEff}   {\ensuremath{{0.992 } } }
\vdef{default-11:SgData-E:fl3d:loEffE}   {\ensuremath{{0.005 } } }
\vdef{default-11:SgData-E:fl3d:hiEff}   {\ensuremath{{0.008 } } }
\vdef{default-11:SgData-E:fl3d:hiEffE}   {\ensuremath{{0.005 } } }
\vdef{default-11:SgMc-E:fl3d:loEff}   {\ensuremath{{0.783 } } }
\vdef{default-11:SgMc-E:fl3d:loEffE}   {\ensuremath{{0.021 } } }
\vdef{default-11:SgMc-E:fl3d:hiEff}   {\ensuremath{{0.217 } } }
\vdef{default-11:SgMc-E:fl3d:hiEffE}   {\ensuremath{{0.021 } } }
\vdef{default-11:SgMc-E:fl3d:loDelta}   {\ensuremath{{+0.235 } } }
\vdef{default-11:SgMc-E:fl3d:loDeltaE}   {\ensuremath{{0.028 } } }
\vdef{default-11:SgMc-E:fl3d:hiDelta}   {\ensuremath{{-1.855 } } }
\vdef{default-11:SgMc-E:fl3d:hiDeltaE}   {\ensuremath{{0.093 } } }
\vdef{default-11:SgData-E:fl3de:loEff}   {\ensuremath{{1.000 } } }
\vdef{default-11:SgData-E:fl3de:loEffE}   {\ensuremath{{0.003 } } }
\vdef{default-11:SgData-E:fl3de:hiEff}   {\ensuremath{{0.031 } } }
\vdef{default-11:SgData-E:fl3de:hiEffE}   {\ensuremath{{0.009 } } }
\vdef{default-11:SgMc-E:fl3de:loEff}   {\ensuremath{{1.000 } } }
\vdef{default-11:SgMc-E:fl3de:loEffE}   {\ensuremath{{0.003 } } }
\vdef{default-11:SgMc-E:fl3de:hiEff}   {\ensuremath{{0.002 } } }
\vdef{default-11:SgMc-E:fl3de:hiEffE}   {\ensuremath{{0.003 } } }
\vdef{default-11:SgMc-E:fl3de:loDelta}   {\ensuremath{{+0.000 } } }
\vdef{default-11:SgMc-E:fl3de:loDeltaE}   {\ensuremath{{0.004 } } }
\vdef{default-11:SgMc-E:fl3de:hiDelta}   {\ensuremath{{+1.808 } } }
\vdef{default-11:SgMc-E:fl3de:hiDeltaE}   {\ensuremath{{0.332 } } }
\vdef{default-11:SgData-E:fls3d:loEff}   {\ensuremath{{0.878 } } }
\vdef{default-11:SgData-E:fls3d:loEffE}   {\ensuremath{{0.017 } } }
\vdef{default-11:SgData-E:fls3d:hiEff}   {\ensuremath{{0.122 } } }
\vdef{default-11:SgData-E:fls3d:hiEffE}   {\ensuremath{{0.017 } } }
\vdef{default-11:SgMc-E:fls3d:loEff}   {\ensuremath{{0.134 } } }
\vdef{default-11:SgMc-E:fls3d:loEffE}   {\ensuremath{{0.018 } } }
\vdef{default-11:SgMc-E:fls3d:hiEff}   {\ensuremath{{0.866 } } }
\vdef{default-11:SgMc-E:fls3d:hiEffE}   {\ensuremath{{0.018 } } }
\vdef{default-11:SgMc-E:fls3d:loDelta}   {\ensuremath{{+1.470 } } }
\vdef{default-11:SgMc-E:fls3d:loDeltaE}   {\ensuremath{{0.062 } } }
\vdef{default-11:SgMc-E:fls3d:hiDelta}   {\ensuremath{{-1.505 } } }
\vdef{default-11:SgMc-E:fls3d:hiDeltaE}   {\ensuremath{{0.061 } } }
\vdef{default-11:SgData-E:flsxy:loEff}   {\ensuremath{{1.000 } } }
\vdef{default-11:SgData-E:flsxy:loEffE}   {\ensuremath{{0.003 } } }
\vdef{default-11:SgData-E:flsxy:hiEff}   {\ensuremath{{1.000 } } }
\vdef{default-11:SgData-E:flsxy:hiEffE}   {\ensuremath{{0.003 } } }
\vdef{default-11:SgMc-E:flsxy:loEff}   {\ensuremath{{1.002 } } }
\vdef{default-11:SgMc-E:flsxy:loEffE}   {\ensuremath{{0.003 } } }
\vdef{default-11:SgMc-E:flsxy:hiEff}   {\ensuremath{{1.000 } } }
\vdef{default-11:SgMc-E:flsxy:hiEffE}   {\ensuremath{{0.003 } } }
\vdef{default-11:SgMc-E:flsxy:loDelta}   {\ensuremath{{-0.002 } } }
\vdef{default-11:SgMc-E:flsxy:loDeltaE}   {\ensuremath{{0.004 } } }
\vdef{default-11:SgMc-E:flsxy:hiDelta}   {\ensuremath{{+0.000 } } }
\vdef{default-11:SgMc-E:flsxy:hiDeltaE}   {\ensuremath{{0.004 } } }
\vdef{default-11:SgData-E:chi2dof:loEff}   {\ensuremath{{0.853 } } }
\vdef{default-11:SgData-E:chi2dof:loEffE}   {\ensuremath{{0.061 } } }
\vdef{default-11:SgData-E:chi2dof:hiEff}   {\ensuremath{{0.147 } } }
\vdef{default-11:SgData-E:chi2dof:hiEffE}   {\ensuremath{{0.061 } } }
\vdef{default-11:SgMc-E:chi2dof:loEff}   {\ensuremath{{0.925 } } }
\vdef{default-11:SgMc-E:chi2dof:loEffE}   {\ensuremath{{0.047 } } }
\vdef{default-11:SgMc-E:chi2dof:hiEff}   {\ensuremath{{0.075 } } }
\vdef{default-11:SgMc-E:chi2dof:hiEffE}   {\ensuremath{{0.047 } } }
\vdef{default-11:SgMc-E:chi2dof:loDelta}   {\ensuremath{{-0.081 } } }
\vdef{default-11:SgMc-E:chi2dof:loDeltaE}   {\ensuremath{{0.088 } } }
\vdef{default-11:SgMc-E:chi2dof:hiDelta}   {\ensuremath{{+0.646 } } }
\vdef{default-11:SgMc-E:chi2dof:hiDeltaE}   {\ensuremath{{0.669 } } }
\vdef{default-11:SgData-E:pchi2dof:loEff}   {\ensuremath{{0.863 } } }
\vdef{default-11:SgData-E:pchi2dof:loEffE}   {\ensuremath{{0.041 } } }
\vdef{default-11:SgData-E:pchi2dof:hiEff}   {\ensuremath{{0.137 } } }
\vdef{default-11:SgData-E:pchi2dof:hiEffE}   {\ensuremath{{0.041 } } }
\vdef{default-11:SgMc-E:pchi2dof:loEff}   {\ensuremath{{0.702 } } }
\vdef{default-11:SgMc-E:pchi2dof:loEffE}   {\ensuremath{{0.053 } } }
\vdef{default-11:SgMc-E:pchi2dof:hiEff}   {\ensuremath{{0.298 } } }
\vdef{default-11:SgMc-E:pchi2dof:hiEffE}   {\ensuremath{{0.052 } } }
\vdef{default-11:SgMc-E:pchi2dof:loDelta}   {\ensuremath{{+0.206 } } }
\vdef{default-11:SgMc-E:pchi2dof:loDeltaE}   {\ensuremath{{0.088 } } }
\vdef{default-11:SgMc-E:pchi2dof:hiDelta}   {\ensuremath{{-0.741 } } }
\vdef{default-11:SgMc-E:pchi2dof:hiDeltaE}   {\ensuremath{{0.297 } } }
\vdef{default-11:SgData-E:alpha:loEff}   {\ensuremath{{0.784 } } }
\vdef{default-11:SgData-E:alpha:loEffE}   {\ensuremath{{0.067 } } }
\vdef{default-11:SgData-E:alpha:hiEff}   {\ensuremath{{0.216 } } }
\vdef{default-11:SgData-E:alpha:hiEffE}   {\ensuremath{{0.067 } } }
\vdef{default-11:SgMc-E:alpha:loEff}   {\ensuremath{{0.994 } } }
\vdef{default-11:SgMc-E:alpha:loEffE}   {\ensuremath{{0.035 } } }
\vdef{default-11:SgMc-E:alpha:hiEff}   {\ensuremath{{0.006 } } }
\vdef{default-11:SgMc-E:alpha:hiEffE}   {\ensuremath{{0.025 } } }
\vdef{default-11:SgMc-E:alpha:loDelta}   {\ensuremath{{-0.236 } } }
\vdef{default-11:SgMc-E:alpha:loDeltaE}   {\ensuremath{{0.091 } } }
\vdef{default-11:SgMc-E:alpha:hiDelta}   {\ensuremath{{+1.892 } } }
\vdef{default-11:SgMc-E:alpha:hiDeltaE}   {\ensuremath{{0.439 } } }
\vdef{default-11:SgData-E:iso:loEff}   {\ensuremath{{0.482 } } }
\vdef{default-11:SgData-E:iso:loEffE}   {\ensuremath{{0.065 } } }
\vdef{default-11:SgData-E:iso:hiEff}   {\ensuremath{{0.518 } } }
\vdef{default-11:SgData-E:iso:hiEffE}   {\ensuremath{{0.065 } } }
\vdef{default-11:SgMc-E:iso:loEff}   {\ensuremath{{0.112 } } }
\vdef{default-11:SgMc-E:iso:loEffE}   {\ensuremath{{0.042 } } }
\vdef{default-11:SgMc-E:iso:hiEff}   {\ensuremath{{0.888 } } }
\vdef{default-11:SgMc-E:iso:hiEffE}   {\ensuremath{{0.045 } } }
\vdef{default-11:SgMc-E:iso:loDelta}   {\ensuremath{{+1.248 } } }
\vdef{default-11:SgMc-E:iso:loDeltaE}   {\ensuremath{{0.246 } } }
\vdef{default-11:SgMc-E:iso:hiDelta}   {\ensuremath{{-0.527 } } }
\vdef{default-11:SgMc-E:iso:hiDeltaE}   {\ensuremath{{0.126 } } }
\vdef{default-11:SgData-E:docatrk:loEff}   {\ensuremath{{0.121 } } }
\vdef{default-11:SgData-E:docatrk:loEffE}   {\ensuremath{{0.058 } } }
\vdef{default-11:SgData-E:docatrk:hiEff}   {\ensuremath{{0.879 } } }
\vdef{default-11:SgData-E:docatrk:hiEffE}   {\ensuremath{{0.058 } } }
\vdef{default-11:SgMc-E:docatrk:loEff}   {\ensuremath{{0.090 } } }
\vdef{default-11:SgMc-E:docatrk:loEffE}   {\ensuremath{{0.047 } } }
\vdef{default-11:SgMc-E:docatrk:hiEff}   {\ensuremath{{0.910 } } }
\vdef{default-11:SgMc-E:docatrk:hiEffE}   {\ensuremath{{0.053 } } }
\vdef{default-11:SgMc-E:docatrk:loDelta}   {\ensuremath{{+0.294 } } }
\vdef{default-11:SgMc-E:docatrk:loDeltaE}   {\ensuremath{{0.691 } } }
\vdef{default-11:SgMc-E:docatrk:hiDelta}   {\ensuremath{{-0.035 } } }
\vdef{default-11:SgMc-E:docatrk:hiDeltaE}   {\ensuremath{{0.088 } } }
\vdef{default-11:SgData-E:isotrk:loEff}   {\ensuremath{{1.000 } } }
\vdef{default-11:SgData-E:isotrk:loEffE}   {\ensuremath{{0.017 } } }
\vdef{default-11:SgData-E:isotrk:hiEff}   {\ensuremath{{1.000 } } }
\vdef{default-11:SgData-E:isotrk:hiEffE}   {\ensuremath{{0.017 } } }
\vdef{default-11:SgMc-E:isotrk:loEff}   {\ensuremath{{1.000 } } }
\vdef{default-11:SgMc-E:isotrk:loEffE}   {\ensuremath{{0.017 } } }
\vdef{default-11:SgMc-E:isotrk:hiEff}   {\ensuremath{{1.000 } } }
\vdef{default-11:SgMc-E:isotrk:hiEffE}   {\ensuremath{{0.017 } } }
\vdef{default-11:SgMc-E:isotrk:loDelta}   {\ensuremath{{+0.000 } } }
\vdef{default-11:SgMc-E:isotrk:loDeltaE}   {\ensuremath{{0.024 } } }
\vdef{default-11:SgMc-E:isotrk:hiDelta}   {\ensuremath{{+0.000 } } }
\vdef{default-11:SgMc-E:isotrk:hiDeltaE}   {\ensuremath{{0.024 } } }
\vdef{default-11:SgData-E:closetrk:loEff}   {\ensuremath{{0.829 } } }
\vdef{default-11:SgData-E:closetrk:loEffE}   {\ensuremath{{0.064 } } }
\vdef{default-11:SgData-E:closetrk:hiEff}   {\ensuremath{{0.171 } } }
\vdef{default-11:SgData-E:closetrk:hiEffE}   {\ensuremath{{0.064 } } }
\vdef{default-11:SgMc-E:closetrk:loEff}   {\ensuremath{{0.975 } } }
\vdef{default-11:SgMc-E:closetrk:loEffE}   {\ensuremath{{0.037 } } }
\vdef{default-11:SgMc-E:closetrk:hiEff}   {\ensuremath{{0.025 } } }
\vdef{default-11:SgMc-E:closetrk:hiEffE}   {\ensuremath{{0.026 } } }
\vdef{default-11:SgMc-E:closetrk:loDelta}   {\ensuremath{{-0.163 } } }
\vdef{default-11:SgMc-E:closetrk:loDeltaE}   {\ensuremath{{0.085 } } }
\vdef{default-11:SgMc-E:closetrk:hiDelta}   {\ensuremath{{+1.500 } } }
\vdef{default-11:SgMc-E:closetrk:hiDeltaE}   {\ensuremath{{0.497 } } }
\vdef{default-11:SgData-E:lip:loEff}   {\ensuremath{{1.000 } } }
\vdef{default-11:SgData-E:lip:loEffE}   {\ensuremath{{0.031 } } }
\vdef{default-11:SgData-E:lip:hiEff}   {\ensuremath{{0.000 } } }
\vdef{default-11:SgData-E:lip:hiEffE}   {\ensuremath{{0.031 } } }
\vdef{default-11:SgMc-E:lip:loEff}   {\ensuremath{{1.000 } } }
\vdef{default-11:SgMc-E:lip:loEffE}   {\ensuremath{{0.031 } } }
\vdef{default-11:SgMc-E:lip:hiEff}   {\ensuremath{{0.000 } } }
\vdef{default-11:SgMc-E:lip:hiEffE}   {\ensuremath{{0.031 } } }
\vdef{default-11:SgMc-E:lip:loDelta}   {\ensuremath{{+0.000 } } }
\vdef{default-11:SgMc-E:lip:loDeltaE}   {\ensuremath{{0.044 } } }
\vdef{default-11:SgMc-E:lip:hiDelta}   {\ensuremath{{\mathrm{NaN} } } }
\vdef{default-11:SgMc-E:lip:hiDeltaE}   {\ensuremath{{\mathrm{NaN} } } }
\vdef{default-11:SgData-E:lip:inEff}   {\ensuremath{{1.000 } } }
\vdef{default-11:SgData-E:lip:inEffE}   {\ensuremath{{0.031 } } }
\vdef{default-11:SgMc-E:lip:inEff}   {\ensuremath{{1.000 } } }
\vdef{default-11:SgMc-E:lip:inEffE}   {\ensuremath{{0.031 } } }
\vdef{default-11:SgMc-E:lip:inDelta}   {\ensuremath{{+0.000 } } }
\vdef{default-11:SgMc-E:lip:inDeltaE}   {\ensuremath{{0.044 } } }
\vdef{default-11:SgData-E:lips:loEff}   {\ensuremath{{1.000 } } }
\vdef{default-11:SgData-E:lips:loEffE}   {\ensuremath{{0.031 } } }
\vdef{default-11:SgData-E:lips:hiEff}   {\ensuremath{{0.000 } } }
\vdef{default-11:SgData-E:lips:hiEffE}   {\ensuremath{{0.031 } } }
\vdef{default-11:SgMc-E:lips:loEff}   {\ensuremath{{1.000 } } }
\vdef{default-11:SgMc-E:lips:loEffE}   {\ensuremath{{0.031 } } }
\vdef{default-11:SgMc-E:lips:hiEff}   {\ensuremath{{0.000 } } }
\vdef{default-11:SgMc-E:lips:hiEffE}   {\ensuremath{{0.031 } } }
\vdef{default-11:SgMc-E:lips:loDelta}   {\ensuremath{{+0.000 } } }
\vdef{default-11:SgMc-E:lips:loDeltaE}   {\ensuremath{{0.044 } } }
\vdef{default-11:SgMc-E:lips:hiDelta}   {\ensuremath{{\mathrm{NaN} } } }
\vdef{default-11:SgMc-E:lips:hiDeltaE}   {\ensuremath{{\mathrm{NaN} } } }
\vdef{default-11:SgData-E:lips:inEff}   {\ensuremath{{1.000 } } }
\vdef{default-11:SgData-E:lips:inEffE}   {\ensuremath{{0.031 } } }
\vdef{default-11:SgMc-E:lips:inEff}   {\ensuremath{{1.000 } } }
\vdef{default-11:SgMc-E:lips:inEffE}   {\ensuremath{{0.031 } } }
\vdef{default-11:SgMc-E:lips:inDelta}   {\ensuremath{{+0.000 } } }
\vdef{default-11:SgMc-E:lips:inDeltaE}   {\ensuremath{{0.044 } } }
\vdef{default-11:SgData-E:ip:loEff}   {\ensuremath{{0.806 } } }
\vdef{default-11:SgData-E:ip:loEffE}   {\ensuremath{{0.065 } } }
\vdef{default-11:SgData-E:ip:hiEff}   {\ensuremath{{0.194 } } }
\vdef{default-11:SgData-E:ip:hiEffE}   {\ensuremath{{0.065 } } }
\vdef{default-11:SgMc-E:ip:loEff}   {\ensuremath{{0.956 } } }
\vdef{default-11:SgMc-E:ip:loEffE}   {\ensuremath{{0.043 } } }
\vdef{default-11:SgMc-E:ip:hiEff}   {\ensuremath{{0.044 } } }
\vdef{default-11:SgMc-E:ip:hiEffE}   {\ensuremath{{0.036 } } }
\vdef{default-11:SgMc-E:ip:loDelta}   {\ensuremath{{-0.171 } } }
\vdef{default-11:SgMc-E:ip:loDeltaE}   {\ensuremath{{0.092 } } }
\vdef{default-11:SgMc-E:ip:hiDelta}   {\ensuremath{{+1.259 } } }
\vdef{default-11:SgMc-E:ip:hiDeltaE}   {\ensuremath{{0.529 } } }
\vdef{default-11:SgData-E:ips:loEff}   {\ensuremath{{0.763 } } }
\vdef{default-11:SgData-E:ips:loEffE}   {\ensuremath{{0.068 } } }
\vdef{default-11:SgData-E:ips:hiEff}   {\ensuremath{{0.237 } } }
\vdef{default-11:SgData-E:ips:hiEffE}   {\ensuremath{{0.068 } } }
\vdef{default-11:SgMc-E:ips:loEff}   {\ensuremath{{0.983 } } }
\vdef{default-11:SgMc-E:ips:loEffE}   {\ensuremath{{0.034 } } }
\vdef{default-11:SgMc-E:ips:hiEff}   {\ensuremath{{0.017 } } }
\vdef{default-11:SgMc-E:ips:hiEffE}   {\ensuremath{{0.024 } } }
\vdef{default-11:SgMc-E:ips:loDelta}   {\ensuremath{{-0.252 } } }
\vdef{default-11:SgMc-E:ips:loDeltaE}   {\ensuremath{{0.094 } } }
\vdef{default-11:SgMc-E:ips:hiDelta}   {\ensuremath{{+1.735 } } }
\vdef{default-11:SgMc-E:ips:hiDeltaE}   {\ensuremath{{0.366 } } }
\vdef{default-11:SgData-E:maxdoca:loEff}   {\ensuremath{{1.000 } } }
\vdef{default-11:SgData-E:maxdoca:loEffE}   {\ensuremath{{0.031 } } }
\vdef{default-11:SgData-E:maxdoca:hiEff}   {\ensuremath{{0.000 } } }
\vdef{default-11:SgData-E:maxdoca:hiEffE}   {\ensuremath{{0.031 } } }
\vdef{default-11:SgMc-E:maxdoca:loEff}   {\ensuremath{{1.000 } } }
\vdef{default-11:SgMc-E:maxdoca:loEffE}   {\ensuremath{{0.032 } } }
\vdef{default-11:SgMc-E:maxdoca:hiEff}   {\ensuremath{{0.000 } } }
\vdef{default-11:SgMc-E:maxdoca:hiEffE}   {\ensuremath{{0.032 } } }
\vdef{default-11:SgMc-E:maxdoca:loDelta}   {\ensuremath{{+0.000 } } }
\vdef{default-11:SgMc-E:maxdoca:loDeltaE}   {\ensuremath{{0.045 } } }
\vdef{default-11:SgMc-E:maxdoca:hiDelta}   {\ensuremath{{\mathrm{NaN} } } }
\vdef{default-11:SgMc-E:maxdoca:hiDeltaE}   {\ensuremath{{\mathrm{NaN} } } }
\vdef{default-11:NoData-A:osiso:loEff}   {\ensuremath{{1.005 } } }
\vdef{default-11:NoData-A:osiso:loEffE}   {\ensuremath{{\mathrm{NaN} } } }
\vdef{default-11:NoData-A:osiso:hiEff}   {\ensuremath{{1.000 } } }
\vdef{default-11:NoData-A:osiso:hiEffE}   {\ensuremath{{0.000 } } }
\vdef{default-11:NoMc-A:osiso:loEff}   {\ensuremath{{1.003 } } }
\vdef{default-11:NoMc-A:osiso:loEffE}   {\ensuremath{{\mathrm{NaN} } } }
\vdef{default-11:NoMc-A:osiso:hiEff}   {\ensuremath{{1.000 } } }
\vdef{default-11:NoMc-A:osiso:hiEffE}   {\ensuremath{{0.000 } } }
\vdef{default-11:NoMc-A:osiso:loDelta}   {\ensuremath{{+0.002 } } }
\vdef{default-11:NoMc-A:osiso:loDeltaE}   {\ensuremath{{\mathrm{NaN} } } }
\vdef{default-11:NoMc-A:osiso:hiDelta}   {\ensuremath{{+0.000 } } }
\vdef{default-11:NoMc-A:osiso:hiDeltaE}   {\ensuremath{{0.000 } } }
\vdef{default-11:NoData-A:osreliso:loEff}   {\ensuremath{{0.247 } } }
\vdef{default-11:NoData-A:osreliso:loEffE}   {\ensuremath{{0.001 } } }
\vdef{default-11:NoData-A:osreliso:hiEff}   {\ensuremath{{0.753 } } }
\vdef{default-11:NoData-A:osreliso:hiEffE}   {\ensuremath{{0.001 } } }
\vdef{default-11:NoMc-A:osreliso:loEff}   {\ensuremath{{0.286 } } }
\vdef{default-11:NoMc-A:osreliso:loEffE}   {\ensuremath{{0.001 } } }
\vdef{default-11:NoMc-A:osreliso:hiEff}   {\ensuremath{{0.714 } } }
\vdef{default-11:NoMc-A:osreliso:hiEffE}   {\ensuremath{{0.001 } } }
\vdef{default-11:NoMc-A:osreliso:loDelta}   {\ensuremath{{-0.147 } } }
\vdef{default-11:NoMc-A:osreliso:loDeltaE}   {\ensuremath{{0.007 } } }
\vdef{default-11:NoMc-A:osreliso:hiDelta}   {\ensuremath{{+0.054 } } }
\vdef{default-11:NoMc-A:osreliso:hiDeltaE}   {\ensuremath{{0.003 } } }
\vdef{default-11:NoData-A:osmuonpt:loEff}   {\ensuremath{{0.000 } } }
\vdef{default-11:NoData-A:osmuonpt:loEffE}   {\ensuremath{{0.000 } } }
\vdef{default-11:NoData-A:osmuonpt:hiEff}   {\ensuremath{{1.000 } } }
\vdef{default-11:NoData-A:osmuonpt:hiEffE}   {\ensuremath{{0.000 } } }
\vdef{default-11:NoMc-A:osmuonpt:loEff}   {\ensuremath{{0.000 } } }
\vdef{default-11:NoMc-A:osmuonpt:loEffE}   {\ensuremath{{0.000 } } }
\vdef{default-11:NoMc-A:osmuonpt:hiEff}   {\ensuremath{{1.000 } } }
\vdef{default-11:NoMc-A:osmuonpt:hiEffE}   {\ensuremath{{0.000 } } }
\vdef{default-11:NoMc-A:osmuonpt:loDelta}   {\ensuremath{{\mathrm{NaN} } } }
\vdef{default-11:NoMc-A:osmuonpt:loDeltaE}   {\ensuremath{{\mathrm{NaN} } } }
\vdef{default-11:NoMc-A:osmuonpt:hiDelta}   {\ensuremath{{+0.000 } } }
\vdef{default-11:NoMc-A:osmuonpt:hiDeltaE}   {\ensuremath{{0.000 } } }
\vdef{default-11:NoData-A:osmuondr:loEff}   {\ensuremath{{0.022 } } }
\vdef{default-11:NoData-A:osmuondr:loEffE}   {\ensuremath{{0.002 } } }
\vdef{default-11:NoData-A:osmuondr:hiEff}   {\ensuremath{{0.978 } } }
\vdef{default-11:NoData-A:osmuondr:hiEffE}   {\ensuremath{{0.002 } } }
\vdef{default-11:NoMc-A:osmuondr:loEff}   {\ensuremath{{0.017 } } }
\vdef{default-11:NoMc-A:osmuondr:loEffE}   {\ensuremath{{0.002 } } }
\vdef{default-11:NoMc-A:osmuondr:hiEff}   {\ensuremath{{0.983 } } }
\vdef{default-11:NoMc-A:osmuondr:hiEffE}   {\ensuremath{{0.002 } } }
\vdef{default-11:NoMc-A:osmuondr:loDelta}   {\ensuremath{{+0.266 } } }
\vdef{default-11:NoMc-A:osmuondr:loDeltaE}   {\ensuremath{{0.162 } } }
\vdef{default-11:NoMc-A:osmuondr:hiDelta}   {\ensuremath{{-0.005 } } }
\vdef{default-11:NoMc-A:osmuondr:hiDeltaE}   {\ensuremath{{0.003 } } }
\vdef{default-11:NoData-A:hlt:loEff}   {\ensuremath{{0.057 } } }
\vdef{default-11:NoData-A:hlt:loEffE}   {\ensuremath{{0.001 } } }
\vdef{default-11:NoData-A:hlt:hiEff}   {\ensuremath{{0.943 } } }
\vdef{default-11:NoData-A:hlt:hiEffE}   {\ensuremath{{0.001 } } }
\vdef{default-11:NoMc-A:hlt:loEff}   {\ensuremath{{0.297 } } }
\vdef{default-11:NoMc-A:hlt:loEffE}   {\ensuremath{{0.001 } } }
\vdef{default-11:NoMc-A:hlt:hiEff}   {\ensuremath{{0.703 } } }
\vdef{default-11:NoMc-A:hlt:hiEffE}   {\ensuremath{{0.001 } } }
\vdef{default-11:NoMc-A:hlt:loDelta}   {\ensuremath{{-1.354 } } }
\vdef{default-11:NoMc-A:hlt:loDeltaE}   {\ensuremath{{0.007 } } }
\vdef{default-11:NoMc-A:hlt:hiDelta}   {\ensuremath{{+0.292 } } }
\vdef{default-11:NoMc-A:hlt:hiDeltaE}   {\ensuremath{{0.002 } } }
\vdef{default-11:NoData-A:muonsid:loEff}   {\ensuremath{{0.147 } } }
\vdef{default-11:NoData-A:muonsid:loEffE}   {\ensuremath{{0.001 } } }
\vdef{default-11:NoData-A:muonsid:hiEff}   {\ensuremath{{0.853 } } }
\vdef{default-11:NoData-A:muonsid:hiEffE}   {\ensuremath{{0.001 } } }
\vdef{default-11:NoMc-A:muonsid:loEff}   {\ensuremath{{0.156 } } }
\vdef{default-11:NoMc-A:muonsid:loEffE}   {\ensuremath{{0.001 } } }
\vdef{default-11:NoMc-A:muonsid:hiEff}   {\ensuremath{{0.844 } } }
\vdef{default-11:NoMc-A:muonsid:hiEffE}   {\ensuremath{{0.001 } } }
\vdef{default-11:NoMc-A:muonsid:loDelta}   {\ensuremath{{-0.062 } } }
\vdef{default-11:NoMc-A:muonsid:loDeltaE}   {\ensuremath{{0.010 } } }
\vdef{default-11:NoMc-A:muonsid:hiDelta}   {\ensuremath{{+0.011 } } }
\vdef{default-11:NoMc-A:muonsid:hiDeltaE}   {\ensuremath{{0.002 } } }
\vdef{default-11:NoData-A:tracksqual:loEff}   {\ensuremath{{0.000 } } }
\vdef{default-11:NoData-A:tracksqual:loEffE}   {\ensuremath{{0.000 } } }
\vdef{default-11:NoData-A:tracksqual:hiEff}   {\ensuremath{{1.000 } } }
\vdef{default-11:NoData-A:tracksqual:hiEffE}   {\ensuremath{{0.000 } } }
\vdef{default-11:NoMc-A:tracksqual:loEff}   {\ensuremath{{0.000 } } }
\vdef{default-11:NoMc-A:tracksqual:loEffE}   {\ensuremath{{0.000 } } }
\vdef{default-11:NoMc-A:tracksqual:hiEff}   {\ensuremath{{1.000 } } }
\vdef{default-11:NoMc-A:tracksqual:hiEffE}   {\ensuremath{{0.000 } } }
\vdef{default-11:NoMc-A:tracksqual:loDelta}   {\ensuremath{{+1.079 } } }
\vdef{default-11:NoMc-A:tracksqual:loDeltaE}   {\ensuremath{{0.222 } } }
\vdef{default-11:NoMc-A:tracksqual:hiDelta}   {\ensuremath{{-0.000 } } }
\vdef{default-11:NoMc-A:tracksqual:hiDeltaE}   {\ensuremath{{0.000 } } }
\vdef{default-11:NoData-A:pvz:loEff}   {\ensuremath{{0.513 } } }
\vdef{default-11:NoData-A:pvz:loEffE}   {\ensuremath{{0.002 } } }
\vdef{default-11:NoData-A:pvz:hiEff}   {\ensuremath{{0.487 } } }
\vdef{default-11:NoData-A:pvz:hiEffE}   {\ensuremath{{0.002 } } }
\vdef{default-11:NoMc-A:pvz:loEff}   {\ensuremath{{0.469 } } }
\vdef{default-11:NoMc-A:pvz:loEffE}   {\ensuremath{{0.002 } } }
\vdef{default-11:NoMc-A:pvz:hiEff}   {\ensuremath{{0.531 } } }
\vdef{default-11:NoMc-A:pvz:hiEffE}   {\ensuremath{{0.002 } } }
\vdef{default-11:NoMc-A:pvz:loDelta}   {\ensuremath{{+0.088 } } }
\vdef{default-11:NoMc-A:pvz:loDeltaE}   {\ensuremath{{0.005 } } }
\vdef{default-11:NoMc-A:pvz:hiDelta}   {\ensuremath{{-0.085 } } }
\vdef{default-11:NoMc-A:pvz:hiDeltaE}   {\ensuremath{{0.004 } } }
\vdef{default-11:NoData-A:pvn:loEff}   {\ensuremath{{1.008 } } }
\vdef{default-11:NoData-A:pvn:loEffE}   {\ensuremath{{\mathrm{NaN} } } }
\vdef{default-11:NoData-A:pvn:hiEff}   {\ensuremath{{1.000 } } }
\vdef{default-11:NoData-A:pvn:hiEffE}   {\ensuremath{{0.000 } } }
\vdef{default-11:NoMc-A:pvn:loEff}   {\ensuremath{{1.000 } } }
\vdef{default-11:NoMc-A:pvn:loEffE}   {\ensuremath{{0.000 } } }
\vdef{default-11:NoMc-A:pvn:hiEff}   {\ensuremath{{1.000 } } }
\vdef{default-11:NoMc-A:pvn:hiEffE}   {\ensuremath{{0.000 } } }
\vdef{default-11:NoMc-A:pvn:loDelta}   {\ensuremath{{+0.008 } } }
\vdef{default-11:NoMc-A:pvn:loDeltaE}   {\ensuremath{{\mathrm{NaN} } } }
\vdef{default-11:NoMc-A:pvn:hiDelta}   {\ensuremath{{+0.000 } } }
\vdef{default-11:NoMc-A:pvn:hiDeltaE}   {\ensuremath{{0.000 } } }
\vdef{default-11:NoData-A:pvavew8:loEff}   {\ensuremath{{0.011 } } }
\vdef{default-11:NoData-A:pvavew8:loEffE}   {\ensuremath{{0.000 } } }
\vdef{default-11:NoData-A:pvavew8:hiEff}   {\ensuremath{{0.989 } } }
\vdef{default-11:NoData-A:pvavew8:hiEffE}   {\ensuremath{{0.000 } } }
\vdef{default-11:NoMc-A:pvavew8:loEff}   {\ensuremath{{0.007 } } }
\vdef{default-11:NoMc-A:pvavew8:loEffE}   {\ensuremath{{0.000 } } }
\vdef{default-11:NoMc-A:pvavew8:hiEff}   {\ensuremath{{0.993 } } }
\vdef{default-11:NoMc-A:pvavew8:hiEffE}   {\ensuremath{{0.000 } } }
\vdef{default-11:NoMc-A:pvavew8:loDelta}   {\ensuremath{{+0.422 } } }
\vdef{default-11:NoMc-A:pvavew8:loDeltaE}   {\ensuremath{{0.047 } } }
\vdef{default-11:NoMc-A:pvavew8:hiDelta}   {\ensuremath{{-0.004 } } }
\vdef{default-11:NoMc-A:pvavew8:hiDeltaE}   {\ensuremath{{0.000 } } }
\vdef{default-11:NoData-A:pvntrk:loEff}   {\ensuremath{{1.000 } } }
\vdef{default-11:NoData-A:pvntrk:loEffE}   {\ensuremath{{0.000 } } }
\vdef{default-11:NoData-A:pvntrk:hiEff}   {\ensuremath{{1.000 } } }
\vdef{default-11:NoData-A:pvntrk:hiEffE}   {\ensuremath{{0.000 } } }
\vdef{default-11:NoMc-A:pvntrk:loEff}   {\ensuremath{{1.000 } } }
\vdef{default-11:NoMc-A:pvntrk:loEffE}   {\ensuremath{{0.000 } } }
\vdef{default-11:NoMc-A:pvntrk:hiEff}   {\ensuremath{{1.000 } } }
\vdef{default-11:NoMc-A:pvntrk:hiEffE}   {\ensuremath{{0.000 } } }
\vdef{default-11:NoMc-A:pvntrk:loDelta}   {\ensuremath{{+0.000 } } }
\vdef{default-11:NoMc-A:pvntrk:loDeltaE}   {\ensuremath{{0.000 } } }
\vdef{default-11:NoMc-A:pvntrk:hiDelta}   {\ensuremath{{+0.000 } } }
\vdef{default-11:NoMc-A:pvntrk:hiDeltaE}   {\ensuremath{{0.000 } } }
\vdef{default-11:NoData-A:muon1pt:loEff}   {\ensuremath{{1.009 } } }
\vdef{default-11:NoData-A:muon1pt:loEffE}   {\ensuremath{{\mathrm{NaN} } } }
\vdef{default-11:NoData-A:muon1pt:hiEff}   {\ensuremath{{1.000 } } }
\vdef{default-11:NoData-A:muon1pt:hiEffE}   {\ensuremath{{0.000 } } }
\vdef{default-11:NoMc-A:muon1pt:loEff}   {\ensuremath{{1.008 } } }
\vdef{default-11:NoMc-A:muon1pt:loEffE}   {\ensuremath{{\mathrm{NaN} } } }
\vdef{default-11:NoMc-A:muon1pt:hiEff}   {\ensuremath{{1.000 } } }
\vdef{default-11:NoMc-A:muon1pt:hiEffE}   {\ensuremath{{0.000 } } }
\vdef{default-11:NoMc-A:muon1pt:loDelta}   {\ensuremath{{+0.002 } } }
\vdef{default-11:NoMc-A:muon1pt:loDeltaE}   {\ensuremath{{\mathrm{NaN} } } }
\vdef{default-11:NoMc-A:muon1pt:hiDelta}   {\ensuremath{{+0.000 } } }
\vdef{default-11:NoMc-A:muon1pt:hiDeltaE}   {\ensuremath{{0.000 } } }
\vdef{default-11:NoData-A:muon2pt:loEff}   {\ensuremath{{0.074 } } }
\vdef{default-11:NoData-A:muon2pt:loEffE}   {\ensuremath{{0.001 } } }
\vdef{default-11:NoData-A:muon2pt:hiEff}   {\ensuremath{{0.926 } } }
\vdef{default-11:NoData-A:muon2pt:hiEffE}   {\ensuremath{{0.001 } } }
\vdef{default-11:NoMc-A:muon2pt:loEff}   {\ensuremath{{0.073 } } }
\vdef{default-11:NoMc-A:muon2pt:loEffE}   {\ensuremath{{0.001 } } }
\vdef{default-11:NoMc-A:muon2pt:hiEff}   {\ensuremath{{0.927 } } }
\vdef{default-11:NoMc-A:muon2pt:hiEffE}   {\ensuremath{{0.001 } } }
\vdef{default-11:NoMc-A:muon2pt:loDelta}   {\ensuremath{{+0.020 } } }
\vdef{default-11:NoMc-A:muon2pt:loDeltaE}   {\ensuremath{{0.016 } } }
\vdef{default-11:NoMc-A:muon2pt:hiDelta}   {\ensuremath{{-0.002 } } }
\vdef{default-11:NoMc-A:muon2pt:hiDeltaE}   {\ensuremath{{0.001 } } }
\vdef{default-11:NoData-A:muonseta:loEff}   {\ensuremath{{0.738 } } }
\vdef{default-11:NoData-A:muonseta:loEffE}   {\ensuremath{{0.001 } } }
\vdef{default-11:NoData-A:muonseta:hiEff}   {\ensuremath{{0.262 } } }
\vdef{default-11:NoData-A:muonseta:hiEffE}   {\ensuremath{{0.001 } } }
\vdef{default-11:NoMc-A:muonseta:loEff}   {\ensuremath{{0.732 } } }
\vdef{default-11:NoMc-A:muonseta:loEffE}   {\ensuremath{{0.001 } } }
\vdef{default-11:NoMc-A:muonseta:hiEff}   {\ensuremath{{0.268 } } }
\vdef{default-11:NoMc-A:muonseta:hiEffE}   {\ensuremath{{0.001 } } }
\vdef{default-11:NoMc-A:muonseta:loDelta}   {\ensuremath{{+0.009 } } }
\vdef{default-11:NoMc-A:muonseta:loDeltaE}   {\ensuremath{{0.002 } } }
\vdef{default-11:NoMc-A:muonseta:hiDelta}   {\ensuremath{{-0.024 } } }
\vdef{default-11:NoMc-A:muonseta:hiDeltaE}   {\ensuremath{{0.005 } } }
\vdef{default-11:NoData-A:pt:loEff}   {\ensuremath{{0.000 } } }
\vdef{default-11:NoData-A:pt:loEffE}   {\ensuremath{{0.000 } } }
\vdef{default-11:NoData-A:pt:hiEff}   {\ensuremath{{1.000 } } }
\vdef{default-11:NoData-A:pt:hiEffE}   {\ensuremath{{0.000 } } }
\vdef{default-11:NoMc-A:pt:loEff}   {\ensuremath{{0.000 } } }
\vdef{default-11:NoMc-A:pt:loEffE}   {\ensuremath{{0.000 } } }
\vdef{default-11:NoMc-A:pt:hiEff}   {\ensuremath{{1.000 } } }
\vdef{default-11:NoMc-A:pt:hiEffE}   {\ensuremath{{0.000 } } }
\vdef{default-11:NoMc-A:pt:loDelta}   {\ensuremath{{\mathrm{NaN} } } }
\vdef{default-11:NoMc-A:pt:loDeltaE}   {\ensuremath{{\mathrm{NaN} } } }
\vdef{default-11:NoMc-A:pt:hiDelta}   {\ensuremath{{+0.000 } } }
\vdef{default-11:NoMc-A:pt:hiDeltaE}   {\ensuremath{{0.000 } } }
\vdef{default-11:NoData-A:p:loEff}   {\ensuremath{{1.013 } } }
\vdef{default-11:NoData-A:p:loEffE}   {\ensuremath{{\mathrm{NaN} } } }
\vdef{default-11:NoData-A:p:hiEff}   {\ensuremath{{1.000 } } }
\vdef{default-11:NoData-A:p:hiEffE}   {\ensuremath{{0.000 } } }
\vdef{default-11:NoMc-A:p:loEff}   {\ensuremath{{1.013 } } }
\vdef{default-11:NoMc-A:p:loEffE}   {\ensuremath{{\mathrm{NaN} } } }
\vdef{default-11:NoMc-A:p:hiEff}   {\ensuremath{{1.000 } } }
\vdef{default-11:NoMc-A:p:hiEffE}   {\ensuremath{{0.000 } } }
\vdef{default-11:NoMc-A:p:loDelta}   {\ensuremath{{-0.000 } } }
\vdef{default-11:NoMc-A:p:loDeltaE}   {\ensuremath{{\mathrm{NaN} } } }
\vdef{default-11:NoMc-A:p:hiDelta}   {\ensuremath{{+0.000 } } }
\vdef{default-11:NoMc-A:p:hiDeltaE}   {\ensuremath{{0.000 } } }
\vdef{default-11:NoData-A:eta:loEff}   {\ensuremath{{0.729 } } }
\vdef{default-11:NoData-A:eta:loEffE}   {\ensuremath{{0.001 } } }
\vdef{default-11:NoData-A:eta:hiEff}   {\ensuremath{{0.271 } } }
\vdef{default-11:NoData-A:eta:hiEffE}   {\ensuremath{{0.001 } } }
\vdef{default-11:NoMc-A:eta:loEff}   {\ensuremath{{0.724 } } }
\vdef{default-11:NoMc-A:eta:loEffE}   {\ensuremath{{0.001 } } }
\vdef{default-11:NoMc-A:eta:hiEff}   {\ensuremath{{0.276 } } }
\vdef{default-11:NoMc-A:eta:hiEffE}   {\ensuremath{{0.001 } } }
\vdef{default-11:NoMc-A:eta:loDelta}   {\ensuremath{{+0.008 } } }
\vdef{default-11:NoMc-A:eta:loDeltaE}   {\ensuremath{{0.003 } } }
\vdef{default-11:NoMc-A:eta:hiDelta}   {\ensuremath{{-0.021 } } }
\vdef{default-11:NoMc-A:eta:hiDeltaE}   {\ensuremath{{0.007 } } }
\vdef{default-11:NoData-A:bdt:loEff}   {\ensuremath{{0.911 } } }
\vdef{default-11:NoData-A:bdt:loEffE}   {\ensuremath{{0.001 } } }
\vdef{default-11:NoData-A:bdt:hiEff}   {\ensuremath{{0.089 } } }
\vdef{default-11:NoData-A:bdt:hiEffE}   {\ensuremath{{0.001 } } }
\vdef{default-11:NoMc-A:bdt:loEff}   {\ensuremath{{0.914 } } }
\vdef{default-11:NoMc-A:bdt:loEffE}   {\ensuremath{{0.001 } } }
\vdef{default-11:NoMc-A:bdt:hiEff}   {\ensuremath{{0.086 } } }
\vdef{default-11:NoMc-A:bdt:hiEffE}   {\ensuremath{{0.001 } } }
\vdef{default-11:NoMc-A:bdt:loDelta}   {\ensuremath{{-0.003 } } }
\vdef{default-11:NoMc-A:bdt:loDeltaE}   {\ensuremath{{0.001 } } }
\vdef{default-11:NoMc-A:bdt:hiDelta}   {\ensuremath{{+0.036 } } }
\vdef{default-11:NoMc-A:bdt:hiDeltaE}   {\ensuremath{{0.014 } } }
\vdef{default-11:NoData-A:fl3d:loEff}   {\ensuremath{{0.832 } } }
\vdef{default-11:NoData-A:fl3d:loEffE}   {\ensuremath{{0.001 } } }
\vdef{default-11:NoData-A:fl3d:hiEff}   {\ensuremath{{0.168 } } }
\vdef{default-11:NoData-A:fl3d:hiEffE}   {\ensuremath{{0.001 } } }
\vdef{default-11:NoMc-A:fl3d:loEff}   {\ensuremath{{0.828 } } }
\vdef{default-11:NoMc-A:fl3d:loEffE}   {\ensuremath{{0.001 } } }
\vdef{default-11:NoMc-A:fl3d:hiEff}   {\ensuremath{{0.172 } } }
\vdef{default-11:NoMc-A:fl3d:hiEffE}   {\ensuremath{{0.001 } } }
\vdef{default-11:NoMc-A:fl3d:loDelta}   {\ensuremath{{+0.005 } } }
\vdef{default-11:NoMc-A:fl3d:loDeltaE}   {\ensuremath{{0.002 } } }
\vdef{default-11:NoMc-A:fl3d:hiDelta}   {\ensuremath{{-0.024 } } }
\vdef{default-11:NoMc-A:fl3d:hiDeltaE}   {\ensuremath{{0.009 } } }
\vdef{default-11:NoData-A:fl3de:loEff}   {\ensuremath{{1.000 } } }
\vdef{default-11:NoData-A:fl3de:loEffE}   {\ensuremath{{0.000 } } }
\vdef{default-11:NoData-A:fl3de:hiEff}   {\ensuremath{{0.000 } } }
\vdef{default-11:NoData-A:fl3de:hiEffE}   {\ensuremath{{0.000 } } }
\vdef{default-11:NoMc-A:fl3de:loEff}   {\ensuremath{{1.000 } } }
\vdef{default-11:NoMc-A:fl3de:loEffE}   {\ensuremath{{0.000 } } }
\vdef{default-11:NoMc-A:fl3de:hiEff}   {\ensuremath{{0.000 } } }
\vdef{default-11:NoMc-A:fl3de:hiEffE}   {\ensuremath{{0.000 } } }
\vdef{default-11:NoMc-A:fl3de:loDelta}   {\ensuremath{{+0.000 } } }
\vdef{default-11:NoMc-A:fl3de:loDeltaE}   {\ensuremath{{0.000 } } }
\vdef{default-11:NoMc-A:fl3de:hiDelta}   {\ensuremath{{-0.118 } } }
\vdef{default-11:NoMc-A:fl3de:hiDeltaE}   {\ensuremath{{0.483 } } }
\vdef{default-11:NoData-A:fls3d:loEff}   {\ensuremath{{0.074 } } }
\vdef{default-11:NoData-A:fls3d:loEffE}   {\ensuremath{{0.001 } } }
\vdef{default-11:NoData-A:fls3d:hiEff}   {\ensuremath{{0.926 } } }
\vdef{default-11:NoData-A:fls3d:hiEffE}   {\ensuremath{{0.001 } } }
\vdef{default-11:NoMc-A:fls3d:loEff}   {\ensuremath{{0.077 } } }
\vdef{default-11:NoMc-A:fls3d:loEffE}   {\ensuremath{{0.001 } } }
\vdef{default-11:NoMc-A:fls3d:hiEff}   {\ensuremath{{0.923 } } }
\vdef{default-11:NoMc-A:fls3d:hiEffE}   {\ensuremath{{0.001 } } }
\vdef{default-11:NoMc-A:fls3d:loDelta}   {\ensuremath{{-0.049 } } }
\vdef{default-11:NoMc-A:fls3d:loDeltaE}   {\ensuremath{{0.015 } } }
\vdef{default-11:NoMc-A:fls3d:hiDelta}   {\ensuremath{{+0.004 } } }
\vdef{default-11:NoMc-A:fls3d:hiDeltaE}   {\ensuremath{{0.001 } } }
\vdef{default-11:NoData-A:flsxy:loEff}   {\ensuremath{{1.012 } } }
\vdef{default-11:NoData-A:flsxy:loEffE}   {\ensuremath{{\mathrm{NaN} } } }
\vdef{default-11:NoData-A:flsxy:hiEff}   {\ensuremath{{1.000 } } }
\vdef{default-11:NoData-A:flsxy:hiEffE}   {\ensuremath{{0.000 } } }
\vdef{default-11:NoMc-A:flsxy:loEff}   {\ensuremath{{1.012 } } }
\vdef{default-11:NoMc-A:flsxy:loEffE}   {\ensuremath{{\mathrm{NaN} } } }
\vdef{default-11:NoMc-A:flsxy:hiEff}   {\ensuremath{{1.000 } } }
\vdef{default-11:NoMc-A:flsxy:hiEffE}   {\ensuremath{{0.000 } } }
\vdef{default-11:NoMc-A:flsxy:loDelta}   {\ensuremath{{+0.000 } } }
\vdef{default-11:NoMc-A:flsxy:loDeltaE}   {\ensuremath{{\mathrm{NaN} } } }
\vdef{default-11:NoMc-A:flsxy:hiDelta}   {\ensuremath{{+0.000 } } }
\vdef{default-11:NoMc-A:flsxy:hiDeltaE}   {\ensuremath{{0.000 } } }
\vdef{default-11:NoData-A:chi2dof:loEff}   {\ensuremath{{0.928 } } }
\vdef{default-11:NoData-A:chi2dof:loEffE}   {\ensuremath{{0.001 } } }
\vdef{default-11:NoData-A:chi2dof:hiEff}   {\ensuremath{{0.072 } } }
\vdef{default-11:NoData-A:chi2dof:hiEffE}   {\ensuremath{{0.001 } } }
\vdef{default-11:NoMc-A:chi2dof:loEff}   {\ensuremath{{0.936 } } }
\vdef{default-11:NoMc-A:chi2dof:loEffE}   {\ensuremath{{0.001 } } }
\vdef{default-11:NoMc-A:chi2dof:hiEff}   {\ensuremath{{0.064 } } }
\vdef{default-11:NoMc-A:chi2dof:hiEffE}   {\ensuremath{{0.001 } } }
\vdef{default-11:NoMc-A:chi2dof:loDelta}   {\ensuremath{{-0.009 } } }
\vdef{default-11:NoMc-A:chi2dof:loDeltaE}   {\ensuremath{{0.001 } } }
\vdef{default-11:NoMc-A:chi2dof:hiDelta}   {\ensuremath{{+0.129 } } }
\vdef{default-11:NoMc-A:chi2dof:hiDeltaE}   {\ensuremath{{0.016 } } }
\vdef{default-11:NoData-A:pchi2dof:loEff}   {\ensuremath{{0.642 } } }
\vdef{default-11:NoData-A:pchi2dof:loEffE}   {\ensuremath{{0.001 } } }
\vdef{default-11:NoData-A:pchi2dof:hiEff}   {\ensuremath{{0.358 } } }
\vdef{default-11:NoData-A:pchi2dof:hiEffE}   {\ensuremath{{0.001 } } }
\vdef{default-11:NoMc-A:pchi2dof:loEff}   {\ensuremath{{0.621 } } }
\vdef{default-11:NoMc-A:pchi2dof:loEffE}   {\ensuremath{{0.002 } } }
\vdef{default-11:NoMc-A:pchi2dof:hiEff}   {\ensuremath{{0.379 } } }
\vdef{default-11:NoMc-A:pchi2dof:hiEffE}   {\ensuremath{{0.002 } } }
\vdef{default-11:NoMc-A:pchi2dof:loDelta}   {\ensuremath{{+0.034 } } }
\vdef{default-11:NoMc-A:pchi2dof:loDeltaE}   {\ensuremath{{0.003 } } }
\vdef{default-11:NoMc-A:pchi2dof:hiDelta}   {\ensuremath{{-0.058 } } }
\vdef{default-11:NoMc-A:pchi2dof:hiDeltaE}   {\ensuremath{{0.006 } } }
\vdef{default-11:NoData-A:alpha:loEff}   {\ensuremath{{0.995 } } }
\vdef{default-11:NoData-A:alpha:loEffE}   {\ensuremath{{0.000 } } }
\vdef{default-11:NoData-A:alpha:hiEff}   {\ensuremath{{0.005 } } }
\vdef{default-11:NoData-A:alpha:hiEffE}   {\ensuremath{{0.000 } } }
\vdef{default-11:NoMc-A:alpha:loEff}   {\ensuremath{{0.994 } } }
\vdef{default-11:NoMc-A:alpha:loEffE}   {\ensuremath{{0.000 } } }
\vdef{default-11:NoMc-A:alpha:hiEff}   {\ensuremath{{0.006 } } }
\vdef{default-11:NoMc-A:alpha:hiEffE}   {\ensuremath{{0.000 } } }
\vdef{default-11:NoMc-A:alpha:loDelta}   {\ensuremath{{+0.000 } } }
\vdef{default-11:NoMc-A:alpha:loDeltaE}   {\ensuremath{{0.000 } } }
\vdef{default-11:NoMc-A:alpha:hiDelta}   {\ensuremath{{-0.067 } } }
\vdef{default-11:NoMc-A:alpha:hiDeltaE}   {\ensuremath{{0.061 } } }
\vdef{default-11:NoData-A:iso:loEff}   {\ensuremath{{0.129 } } }
\vdef{default-11:NoData-A:iso:loEffE}   {\ensuremath{{0.001 } } }
\vdef{default-11:NoData-A:iso:hiEff}   {\ensuremath{{0.871 } } }
\vdef{default-11:NoData-A:iso:hiEffE}   {\ensuremath{{0.001 } } }
\vdef{default-11:NoMc-A:iso:loEff}   {\ensuremath{{0.107 } } }
\vdef{default-11:NoMc-A:iso:loEffE}   {\ensuremath{{0.001 } } }
\vdef{default-11:NoMc-A:iso:hiEff}   {\ensuremath{{0.893 } } }
\vdef{default-11:NoMc-A:iso:hiEffE}   {\ensuremath{{0.001 } } }
\vdef{default-11:NoMc-A:iso:loDelta}   {\ensuremath{{+0.184 } } }
\vdef{default-11:NoMc-A:iso:loDeltaE}   {\ensuremath{{0.012 } } }
\vdef{default-11:NoMc-A:iso:hiDelta}   {\ensuremath{{-0.025 } } }
\vdef{default-11:NoMc-A:iso:hiDeltaE}   {\ensuremath{{0.002 } } }
\vdef{default-11:NoData-A:docatrk:loEff}   {\ensuremath{{0.071 } } }
\vdef{default-11:NoData-A:docatrk:loEffE}   {\ensuremath{{0.001 } } }
\vdef{default-11:NoData-A:docatrk:hiEff}   {\ensuremath{{0.929 } } }
\vdef{default-11:NoData-A:docatrk:hiEffE}   {\ensuremath{{0.001 } } }
\vdef{default-11:NoMc-A:docatrk:loEff}   {\ensuremath{{0.083 } } }
\vdef{default-11:NoMc-A:docatrk:loEffE}   {\ensuremath{{0.001 } } }
\vdef{default-11:NoMc-A:docatrk:hiEff}   {\ensuremath{{0.917 } } }
\vdef{default-11:NoMc-A:docatrk:hiEffE}   {\ensuremath{{0.001 } } }
\vdef{default-11:NoMc-A:docatrk:loDelta}   {\ensuremath{{-0.155 } } }
\vdef{default-11:NoMc-A:docatrk:loDeltaE}   {\ensuremath{{0.015 } } }
\vdef{default-11:NoMc-A:docatrk:hiDelta}   {\ensuremath{{+0.013 } } }
\vdef{default-11:NoMc-A:docatrk:hiDeltaE}   {\ensuremath{{0.001 } } }
\vdef{default-11:NoData-A:isotrk:loEff}   {\ensuremath{{1.000 } } }
\vdef{default-11:NoData-A:isotrk:loEffE}   {\ensuremath{{0.000 } } }
\vdef{default-11:NoData-A:isotrk:hiEff}   {\ensuremath{{1.000 } } }
\vdef{default-11:NoData-A:isotrk:hiEffE}   {\ensuremath{{0.000 } } }
\vdef{default-11:NoMc-A:isotrk:loEff}   {\ensuremath{{1.000 } } }
\vdef{default-11:NoMc-A:isotrk:loEffE}   {\ensuremath{{0.000 } } }
\vdef{default-11:NoMc-A:isotrk:hiEff}   {\ensuremath{{1.000 } } }
\vdef{default-11:NoMc-A:isotrk:hiEffE}   {\ensuremath{{0.000 } } }
\vdef{default-11:NoMc-A:isotrk:loDelta}   {\ensuremath{{+0.000 } } }
\vdef{default-11:NoMc-A:isotrk:loDeltaE}   {\ensuremath{{0.000 } } }
\vdef{default-11:NoMc-A:isotrk:hiDelta}   {\ensuremath{{+0.000 } } }
\vdef{default-11:NoMc-A:isotrk:hiDeltaE}   {\ensuremath{{0.000 } } }
\vdef{default-11:NoData-A:closetrk:loEff}   {\ensuremath{{0.975 } } }
\vdef{default-11:NoData-A:closetrk:loEffE}   {\ensuremath{{0.000 } } }
\vdef{default-11:NoData-A:closetrk:hiEff}   {\ensuremath{{0.025 } } }
\vdef{default-11:NoData-A:closetrk:hiEffE}   {\ensuremath{{0.000 } } }
\vdef{default-11:NoMc-A:closetrk:loEff}   {\ensuremath{{0.978 } } }
\vdef{default-11:NoMc-A:closetrk:loEffE}   {\ensuremath{{0.000 } } }
\vdef{default-11:NoMc-A:closetrk:hiEff}   {\ensuremath{{0.022 } } }
\vdef{default-11:NoMc-A:closetrk:hiEffE}   {\ensuremath{{0.000 } } }
\vdef{default-11:NoMc-A:closetrk:loDelta}   {\ensuremath{{-0.003 } } }
\vdef{default-11:NoMc-A:closetrk:loDeltaE}   {\ensuremath{{0.001 } } }
\vdef{default-11:NoMc-A:closetrk:hiDelta}   {\ensuremath{{+0.111 } } }
\vdef{default-11:NoMc-A:closetrk:hiDeltaE}   {\ensuremath{{0.029 } } }
\vdef{default-11:NoData-A:lip:loEff}   {\ensuremath{{1.000 } } }
\vdef{default-11:NoData-A:lip:loEffE}   {\ensuremath{{0.000 } } }
\vdef{default-11:NoData-A:lip:hiEff}   {\ensuremath{{0.000 } } }
\vdef{default-11:NoData-A:lip:hiEffE}   {\ensuremath{{0.000 } } }
\vdef{default-11:NoMc-A:lip:loEff}   {\ensuremath{{1.000 } } }
\vdef{default-11:NoMc-A:lip:loEffE}   {\ensuremath{{0.000 } } }
\vdef{default-11:NoMc-A:lip:hiEff}   {\ensuremath{{0.000 } } }
\vdef{default-11:NoMc-A:lip:hiEffE}   {\ensuremath{{0.000 } } }
\vdef{default-11:NoMc-A:lip:loDelta}   {\ensuremath{{+0.000 } } }
\vdef{default-11:NoMc-A:lip:loDeltaE}   {\ensuremath{{0.000 } } }
\vdef{default-11:NoMc-A:lip:hiDelta}   {\ensuremath{{\mathrm{NaN} } } }
\vdef{default-11:NoMc-A:lip:hiDeltaE}   {\ensuremath{{\mathrm{NaN} } } }
\vdef{default-11:NoData-A:lip:inEff}   {\ensuremath{{1.000 } } }
\vdef{default-11:NoData-A:lip:inEffE}   {\ensuremath{{0.000 } } }
\vdef{default-11:NoMc-A:lip:inEff}   {\ensuremath{{1.000 } } }
\vdef{default-11:NoMc-A:lip:inEffE}   {\ensuremath{{0.000 } } }
\vdef{default-11:NoMc-A:lip:inDelta}   {\ensuremath{{+0.000 } } }
\vdef{default-11:NoMc-A:lip:inDeltaE}   {\ensuremath{{0.000 } } }
\vdef{default-11:NoData-A:lips:loEff}   {\ensuremath{{1.000 } } }
\vdef{default-11:NoData-A:lips:loEffE}   {\ensuremath{{0.000 } } }
\vdef{default-11:NoData-A:lips:hiEff}   {\ensuremath{{0.000 } } }
\vdef{default-11:NoData-A:lips:hiEffE}   {\ensuremath{{0.000 } } }
\vdef{default-11:NoMc-A:lips:loEff}   {\ensuremath{{1.000 } } }
\vdef{default-11:NoMc-A:lips:loEffE}   {\ensuremath{{0.000 } } }
\vdef{default-11:NoMc-A:lips:hiEff}   {\ensuremath{{0.000 } } }
\vdef{default-11:NoMc-A:lips:hiEffE}   {\ensuremath{{0.000 } } }
\vdef{default-11:NoMc-A:lips:loDelta}   {\ensuremath{{+0.000 } } }
\vdef{default-11:NoMc-A:lips:loDeltaE}   {\ensuremath{{0.000 } } }
\vdef{default-11:NoMc-A:lips:hiDelta}   {\ensuremath{{\mathrm{NaN} } } }
\vdef{default-11:NoMc-A:lips:hiDeltaE}   {\ensuremath{{\mathrm{NaN} } } }
\vdef{default-11:NoData-A:lips:inEff}   {\ensuremath{{1.000 } } }
\vdef{default-11:NoData-A:lips:inEffE}   {\ensuremath{{0.000 } } }
\vdef{default-11:NoMc-A:lips:inEff}   {\ensuremath{{1.000 } } }
\vdef{default-11:NoMc-A:lips:inEffE}   {\ensuremath{{0.000 } } }
\vdef{default-11:NoMc-A:lips:inDelta}   {\ensuremath{{+0.000 } } }
\vdef{default-11:NoMc-A:lips:inDeltaE}   {\ensuremath{{0.000 } } }
\vdef{default-11:NoData-A:ip:loEff}   {\ensuremath{{0.972 } } }
\vdef{default-11:NoData-A:ip:loEffE}   {\ensuremath{{0.001 } } }
\vdef{default-11:NoData-A:ip:hiEff}   {\ensuremath{{0.028 } } }
\vdef{default-11:NoData-A:ip:hiEffE}   {\ensuremath{{0.001 } } }
\vdef{default-11:NoMc-A:ip:loEff}   {\ensuremath{{0.972 } } }
\vdef{default-11:NoMc-A:ip:loEffE}   {\ensuremath{{0.001 } } }
\vdef{default-11:NoMc-A:ip:hiEff}   {\ensuremath{{0.028 } } }
\vdef{default-11:NoMc-A:ip:hiEffE}   {\ensuremath{{0.001 } } }
\vdef{default-11:NoMc-A:ip:loDelta}   {\ensuremath{{+0.001 } } }
\vdef{default-11:NoMc-A:ip:loDeltaE}   {\ensuremath{{0.001 } } }
\vdef{default-11:NoMc-A:ip:hiDelta}   {\ensuremath{{-0.019 } } }
\vdef{default-11:NoMc-A:ip:hiDeltaE}   {\ensuremath{{0.027 } } }
\vdef{default-11:NoData-A:ips:loEff}   {\ensuremath{{0.944 } } }
\vdef{default-11:NoData-A:ips:loEffE}   {\ensuremath{{0.001 } } }
\vdef{default-11:NoData-A:ips:hiEff}   {\ensuremath{{0.056 } } }
\vdef{default-11:NoData-A:ips:hiEffE}   {\ensuremath{{0.001 } } }
\vdef{default-11:NoMc-A:ips:loEff}   {\ensuremath{{0.959 } } }
\vdef{default-11:NoMc-A:ips:loEffE}   {\ensuremath{{0.001 } } }
\vdef{default-11:NoMc-A:ips:hiEff}   {\ensuremath{{0.041 } } }
\vdef{default-11:NoMc-A:ips:hiEffE}   {\ensuremath{{0.001 } } }
\vdef{default-11:NoMc-A:ips:loDelta}   {\ensuremath{{-0.015 } } }
\vdef{default-11:NoMc-A:ips:loDeltaE}   {\ensuremath{{0.001 } } }
\vdef{default-11:NoMc-A:ips:hiDelta}   {\ensuremath{{+0.301 } } }
\vdef{default-11:NoMc-A:ips:hiDeltaE}   {\ensuremath{{0.020 } } }
\vdef{default-11:NoData-A:maxdoca:loEff}   {\ensuremath{{1.000 } } }
\vdef{default-11:NoData-A:maxdoca:loEffE}   {\ensuremath{{0.000 } } }
\vdef{default-11:NoData-A:maxdoca:hiEff}   {\ensuremath{{0.013 } } }
\vdef{default-11:NoData-A:maxdoca:hiEffE}   {\ensuremath{{0.000 } } }
\vdef{default-11:NoMc-A:maxdoca:loEff}   {\ensuremath{{1.000 } } }
\vdef{default-11:NoMc-A:maxdoca:loEffE}   {\ensuremath{{0.000 } } }
\vdef{default-11:NoMc-A:maxdoca:hiEff}   {\ensuremath{{0.011 } } }
\vdef{default-11:NoMc-A:maxdoca:hiEffE}   {\ensuremath{{0.000 } } }
\vdef{default-11:NoMc-A:maxdoca:loDelta}   {\ensuremath{{+0.000 } } }
\vdef{default-11:NoMc-A:maxdoca:loDeltaE}   {\ensuremath{{0.000 } } }
\vdef{default-11:NoMc-A:maxdoca:hiDelta}   {\ensuremath{{+0.201 } } }
\vdef{default-11:NoMc-A:maxdoca:hiDeltaE}   {\ensuremath{{0.041 } } }
\vdef{default-11:NoData-A:kaonpt:loEff}   {\ensuremath{{1.010 } } }
\vdef{default-11:NoData-A:kaonpt:loEffE}   {\ensuremath{{\mathrm{NaN} } } }
\vdef{default-11:NoData-A:kaonpt:hiEff}   {\ensuremath{{1.000 } } }
\vdef{default-11:NoData-A:kaonpt:hiEffE}   {\ensuremath{{0.000 } } }
\vdef{default-11:NoMc-A:kaonpt:loEff}   {\ensuremath{{1.007 } } }
\vdef{default-11:NoMc-A:kaonpt:loEffE}   {\ensuremath{{\mathrm{NaN} } } }
\vdef{default-11:NoMc-A:kaonpt:hiEff}   {\ensuremath{{1.000 } } }
\vdef{default-11:NoMc-A:kaonpt:hiEffE}   {\ensuremath{{0.000 } } }
\vdef{default-11:NoMc-A:kaonpt:loDelta}   {\ensuremath{{+0.003 } } }
\vdef{default-11:NoMc-A:kaonpt:loDeltaE}   {\ensuremath{{\mathrm{NaN} } } }
\vdef{default-11:NoMc-A:kaonpt:hiDelta}   {\ensuremath{{+0.000 } } }
\vdef{default-11:NoMc-A:kaonpt:hiDeltaE}   {\ensuremath{{0.000 } } }
\vdef{default-11:NoData-A:psipt:loEff}   {\ensuremath{{1.004 } } }
\vdef{default-11:NoData-A:psipt:loEffE}   {\ensuremath{{\mathrm{NaN} } } }
\vdef{default-11:NoData-A:psipt:hiEff}   {\ensuremath{{1.000 } } }
\vdef{default-11:NoData-A:psipt:hiEffE}   {\ensuremath{{0.000 } } }
\vdef{default-11:NoMc-A:psipt:loEff}   {\ensuremath{{1.003 } } }
\vdef{default-11:NoMc-A:psipt:loEffE}   {\ensuremath{{\mathrm{NaN} } } }
\vdef{default-11:NoMc-A:psipt:hiEff}   {\ensuremath{{1.000 } } }
\vdef{default-11:NoMc-A:psipt:hiEffE}   {\ensuremath{{0.000 } } }
\vdef{default-11:NoMc-A:psipt:loDelta}   {\ensuremath{{+0.001 } } }
\vdef{default-11:NoMc-A:psipt:loDeltaE}   {\ensuremath{{\mathrm{NaN} } } }
\vdef{default-11:NoMc-A:psipt:hiDelta}   {\ensuremath{{+0.000 } } }
\vdef{default-11:NoMc-A:psipt:hiDeltaE}   {\ensuremath{{0.000 } } }
\vdef{default-11:NoData-A:osiso:loEff}   {\ensuremath{{1.005 } } }
\vdef{default-11:NoData-A:osiso:loEffE}   {\ensuremath{{\mathrm{NaN} } } }
\vdef{default-11:NoData-A:osiso:hiEff}   {\ensuremath{{1.000 } } }
\vdef{default-11:NoData-A:osiso:hiEffE}   {\ensuremath{{0.000 } } }
\vdef{default-11:NoMcPU-A:osiso:loEff}   {\ensuremath{{1.004 } } }
\vdef{default-11:NoMcPU-A:osiso:loEffE}   {\ensuremath{{\mathrm{NaN} } } }
\vdef{default-11:NoMcPU-A:osiso:hiEff}   {\ensuremath{{1.000 } } }
\vdef{default-11:NoMcPU-A:osiso:hiEffE}   {\ensuremath{{0.000 } } }
\vdef{default-11:NoMcPU-A:osiso:loDelta}   {\ensuremath{{+0.001 } } }
\vdef{default-11:NoMcPU-A:osiso:loDeltaE}   {\ensuremath{{\mathrm{NaN} } } }
\vdef{default-11:NoMcPU-A:osiso:hiDelta}   {\ensuremath{{+0.000 } } }
\vdef{default-11:NoMcPU-A:osiso:hiDeltaE}   {\ensuremath{{0.000 } } }
\vdef{default-11:NoData-A:osreliso:loEff}   {\ensuremath{{0.247 } } }
\vdef{default-11:NoData-A:osreliso:loEffE}   {\ensuremath{{0.001 } } }
\vdef{default-11:NoData-A:osreliso:hiEff}   {\ensuremath{{0.753 } } }
\vdef{default-11:NoData-A:osreliso:hiEffE}   {\ensuremath{{0.001 } } }
\vdef{default-11:NoMcPU-A:osreliso:loEff}   {\ensuremath{{0.286 } } }
\vdef{default-11:NoMcPU-A:osreliso:loEffE}   {\ensuremath{{0.001 } } }
\vdef{default-11:NoMcPU-A:osreliso:hiEff}   {\ensuremath{{0.714 } } }
\vdef{default-11:NoMcPU-A:osreliso:hiEffE}   {\ensuremath{{0.001 } } }
\vdef{default-11:NoMcPU-A:osreliso:loDelta}   {\ensuremath{{-0.148 } } }
\vdef{default-11:NoMcPU-A:osreliso:loDeltaE}   {\ensuremath{{0.007 } } }
\vdef{default-11:NoMcPU-A:osreliso:hiDelta}   {\ensuremath{{+0.054 } } }
\vdef{default-11:NoMcPU-A:osreliso:hiDeltaE}   {\ensuremath{{0.003 } } }
\vdef{default-11:NoData-A:osmuonpt:loEff}   {\ensuremath{{0.000 } } }
\vdef{default-11:NoData-A:osmuonpt:loEffE}   {\ensuremath{{0.000 } } }
\vdef{default-11:NoData-A:osmuonpt:hiEff}   {\ensuremath{{1.000 } } }
\vdef{default-11:NoData-A:osmuonpt:hiEffE}   {\ensuremath{{0.000 } } }
\vdef{default-11:NoMcPU-A:osmuonpt:loEff}   {\ensuremath{{0.000 } } }
\vdef{default-11:NoMcPU-A:osmuonpt:loEffE}   {\ensuremath{{0.000 } } }
\vdef{default-11:NoMcPU-A:osmuonpt:hiEff}   {\ensuremath{{1.000 } } }
\vdef{default-11:NoMcPU-A:osmuonpt:hiEffE}   {\ensuremath{{0.000 } } }
\vdef{default-11:NoMcPU-A:osmuonpt:loDelta}   {\ensuremath{{\mathrm{NaN} } } }
\vdef{default-11:NoMcPU-A:osmuonpt:loDeltaE}   {\ensuremath{{\mathrm{NaN} } } }
\vdef{default-11:NoMcPU-A:osmuonpt:hiDelta}   {\ensuremath{{+0.000 } } }
\vdef{default-11:NoMcPU-A:osmuonpt:hiDeltaE}   {\ensuremath{{0.000 } } }
\vdef{default-11:NoData-A:osmuondr:loEff}   {\ensuremath{{0.022 } } }
\vdef{default-11:NoData-A:osmuondr:loEffE}   {\ensuremath{{0.002 } } }
\vdef{default-11:NoData-A:osmuondr:hiEff}   {\ensuremath{{0.978 } } }
\vdef{default-11:NoData-A:osmuondr:hiEffE}   {\ensuremath{{0.002 } } }
\vdef{default-11:NoMcPU-A:osmuondr:loEff}   {\ensuremath{{0.029 } } }
\vdef{default-11:NoMcPU-A:osmuondr:loEffE}   {\ensuremath{{0.003 } } }
\vdef{default-11:NoMcPU-A:osmuondr:hiEff}   {\ensuremath{{0.971 } } }
\vdef{default-11:NoMcPU-A:osmuondr:hiEffE}   {\ensuremath{{0.003 } } }
\vdef{default-11:NoMcPU-A:osmuondr:loDelta}   {\ensuremath{{-0.268 } } }
\vdef{default-11:NoMcPU-A:osmuondr:loDeltaE}   {\ensuremath{{0.141 } } }
\vdef{default-11:NoMcPU-A:osmuondr:hiDelta}   {\ensuremath{{+0.007 } } }
\vdef{default-11:NoMcPU-A:osmuondr:hiDeltaE}   {\ensuremath{{0.004 } } }
\vdef{default-11:NoData-A:hlt:loEff}   {\ensuremath{{0.057 } } }
\vdef{default-11:NoData-A:hlt:loEffE}   {\ensuremath{{0.001 } } }
\vdef{default-11:NoData-A:hlt:hiEff}   {\ensuremath{{0.943 } } }
\vdef{default-11:NoData-A:hlt:hiEffE}   {\ensuremath{{0.001 } } }
\vdef{default-11:NoMcPU-A:hlt:loEff}   {\ensuremath{{0.306 } } }
\vdef{default-11:NoMcPU-A:hlt:loEffE}   {\ensuremath{{0.001 } } }
\vdef{default-11:NoMcPU-A:hlt:hiEff}   {\ensuremath{{0.694 } } }
\vdef{default-11:NoMcPU-A:hlt:hiEffE}   {\ensuremath{{0.001 } } }
\vdef{default-11:NoMcPU-A:hlt:loDelta}   {\ensuremath{{-1.369 } } }
\vdef{default-11:NoMcPU-A:hlt:loDeltaE}   {\ensuremath{{0.007 } } }
\vdef{default-11:NoMcPU-A:hlt:hiDelta}   {\ensuremath{{+0.304 } } }
\vdef{default-11:NoMcPU-A:hlt:hiDeltaE}   {\ensuremath{{0.002 } } }
\vdef{default-11:NoData-A:muonsid:loEff}   {\ensuremath{{0.147 } } }
\vdef{default-11:NoData-A:muonsid:loEffE}   {\ensuremath{{0.001 } } }
\vdef{default-11:NoData-A:muonsid:hiEff}   {\ensuremath{{0.853 } } }
\vdef{default-11:NoData-A:muonsid:hiEffE}   {\ensuremath{{0.001 } } }
\vdef{default-11:NoMcPU-A:muonsid:loEff}   {\ensuremath{{0.224 } } }
\vdef{default-11:NoMcPU-A:muonsid:loEffE}   {\ensuremath{{0.001 } } }
\vdef{default-11:NoMcPU-A:muonsid:hiEff}   {\ensuremath{{0.776 } } }
\vdef{default-11:NoMcPU-A:muonsid:hiEffE}   {\ensuremath{{0.001 } } }
\vdef{default-11:NoMcPU-A:muonsid:loDelta}   {\ensuremath{{-0.416 } } }
\vdef{default-11:NoMcPU-A:muonsid:loDeltaE}   {\ensuremath{{0.009 } } }
\vdef{default-11:NoMcPU-A:muonsid:hiDelta}   {\ensuremath{{+0.095 } } }
\vdef{default-11:NoMcPU-A:muonsid:hiDeltaE}   {\ensuremath{{0.002 } } }
\vdef{default-11:NoData-A:tracksqual:loEff}   {\ensuremath{{0.000 } } }
\vdef{default-11:NoData-A:tracksqual:loEffE}   {\ensuremath{{0.000 } } }
\vdef{default-11:NoData-A:tracksqual:hiEff}   {\ensuremath{{1.000 } } }
\vdef{default-11:NoData-A:tracksqual:hiEffE}   {\ensuremath{{0.000 } } }
\vdef{default-11:NoMcPU-A:tracksqual:loEff}   {\ensuremath{{0.000 } } }
\vdef{default-11:NoMcPU-A:tracksqual:loEffE}   {\ensuremath{{0.000 } } }
\vdef{default-11:NoMcPU-A:tracksqual:hiEff}   {\ensuremath{{1.000 } } }
\vdef{default-11:NoMcPU-A:tracksqual:hiEffE}   {\ensuremath{{0.000 } } }
\vdef{default-11:NoMcPU-A:tracksqual:loDelta}   {\ensuremath{{+0.313 } } }
\vdef{default-11:NoMcPU-A:tracksqual:loDeltaE}   {\ensuremath{{0.225 } } }
\vdef{default-11:NoMcPU-A:tracksqual:hiDelta}   {\ensuremath{{-0.000 } } }
\vdef{default-11:NoMcPU-A:tracksqual:hiDeltaE}   {\ensuremath{{0.000 } } }
\vdef{default-11:NoData-A:pvz:loEff}   {\ensuremath{{0.513 } } }
\vdef{default-11:NoData-A:pvz:loEffE}   {\ensuremath{{0.002 } } }
\vdef{default-11:NoData-A:pvz:hiEff}   {\ensuremath{{0.487 } } }
\vdef{default-11:NoData-A:pvz:hiEffE}   {\ensuremath{{0.002 } } }
\vdef{default-11:NoMcPU-A:pvz:loEff}   {\ensuremath{{0.471 } } }
\vdef{default-11:NoMcPU-A:pvz:loEffE}   {\ensuremath{{0.002 } } }
\vdef{default-11:NoMcPU-A:pvz:hiEff}   {\ensuremath{{0.529 } } }
\vdef{default-11:NoMcPU-A:pvz:hiEffE}   {\ensuremath{{0.002 } } }
\vdef{default-11:NoMcPU-A:pvz:loDelta}   {\ensuremath{{+0.085 } } }
\vdef{default-11:NoMcPU-A:pvz:loDeltaE}   {\ensuremath{{0.005 } } }
\vdef{default-11:NoMcPU-A:pvz:hiDelta}   {\ensuremath{{-0.082 } } }
\vdef{default-11:NoMcPU-A:pvz:hiDeltaE}   {\ensuremath{{0.004 } } }
\vdef{default-11:NoData-A:pvn:loEff}   {\ensuremath{{1.008 } } }
\vdef{default-11:NoData-A:pvn:loEffE}   {\ensuremath{{\mathrm{NaN} } } }
\vdef{default-11:NoData-A:pvn:hiEff}   {\ensuremath{{1.000 } } }
\vdef{default-11:NoData-A:pvn:hiEffE}   {\ensuremath{{0.000 } } }
\vdef{default-11:NoMcPU-A:pvn:loEff}   {\ensuremath{{1.063 } } }
\vdef{default-11:NoMcPU-A:pvn:loEffE}   {\ensuremath{{\mathrm{NaN} } } }
\vdef{default-11:NoMcPU-A:pvn:hiEff}   {\ensuremath{{1.000 } } }
\vdef{default-11:NoMcPU-A:pvn:hiEffE}   {\ensuremath{{0.000 } } }
\vdef{default-11:NoMcPU-A:pvn:loDelta}   {\ensuremath{{-0.053 } } }
\vdef{default-11:NoMcPU-A:pvn:loDeltaE}   {\ensuremath{{\mathrm{NaN} } } }
\vdef{default-11:NoMcPU-A:pvn:hiDelta}   {\ensuremath{{+0.000 } } }
\vdef{default-11:NoMcPU-A:pvn:hiDeltaE}   {\ensuremath{{0.000 } } }
\vdef{default-11:NoData-A:pvavew8:loEff}   {\ensuremath{{0.011 } } }
\vdef{default-11:NoData-A:pvavew8:loEffE}   {\ensuremath{{0.000 } } }
\vdef{default-11:NoData-A:pvavew8:hiEff}   {\ensuremath{{0.989 } } }
\vdef{default-11:NoData-A:pvavew8:hiEffE}   {\ensuremath{{0.000 } } }
\vdef{default-11:NoMcPU-A:pvavew8:loEff}   {\ensuremath{{0.007 } } }
\vdef{default-11:NoMcPU-A:pvavew8:loEffE}   {\ensuremath{{0.000 } } }
\vdef{default-11:NoMcPU-A:pvavew8:hiEff}   {\ensuremath{{0.993 } } }
\vdef{default-11:NoMcPU-A:pvavew8:hiEffE}   {\ensuremath{{0.000 } } }
\vdef{default-11:NoMcPU-A:pvavew8:loDelta}   {\ensuremath{{+0.488 } } }
\vdef{default-11:NoMcPU-A:pvavew8:loDeltaE}   {\ensuremath{{0.047 } } }
\vdef{default-11:NoMcPU-A:pvavew8:hiDelta}   {\ensuremath{{-0.004 } } }
\vdef{default-11:NoMcPU-A:pvavew8:hiDeltaE}   {\ensuremath{{0.000 } } }
\vdef{default-11:NoData-A:pvntrk:loEff}   {\ensuremath{{1.000 } } }
\vdef{default-11:NoData-A:pvntrk:loEffE}   {\ensuremath{{0.000 } } }
\vdef{default-11:NoData-A:pvntrk:hiEff}   {\ensuremath{{1.000 } } }
\vdef{default-11:NoData-A:pvntrk:hiEffE}   {\ensuremath{{0.000 } } }
\vdef{default-11:NoMcPU-A:pvntrk:loEff}   {\ensuremath{{1.000 } } }
\vdef{default-11:NoMcPU-A:pvntrk:loEffE}   {\ensuremath{{0.000 } } }
\vdef{default-11:NoMcPU-A:pvntrk:hiEff}   {\ensuremath{{1.000 } } }
\vdef{default-11:NoMcPU-A:pvntrk:hiEffE}   {\ensuremath{{0.000 } } }
\vdef{default-11:NoMcPU-A:pvntrk:loDelta}   {\ensuremath{{+0.000 } } }
\vdef{default-11:NoMcPU-A:pvntrk:loDeltaE}   {\ensuremath{{0.000 } } }
\vdef{default-11:NoMcPU-A:pvntrk:hiDelta}   {\ensuremath{{+0.000 } } }
\vdef{default-11:NoMcPU-A:pvntrk:hiDeltaE}   {\ensuremath{{0.000 } } }
\vdef{default-11:NoData-A:muon1pt:loEff}   {\ensuremath{{1.009 } } }
\vdef{default-11:NoData-A:muon1pt:loEffE}   {\ensuremath{{\mathrm{NaN} } } }
\vdef{default-11:NoData-A:muon1pt:hiEff}   {\ensuremath{{1.000 } } }
\vdef{default-11:NoData-A:muon1pt:hiEffE}   {\ensuremath{{0.000 } } }
\vdef{default-11:NoMcPU-A:muon1pt:loEff}   {\ensuremath{{1.010 } } }
\vdef{default-11:NoMcPU-A:muon1pt:loEffE}   {\ensuremath{{\mathrm{NaN} } } }
\vdef{default-11:NoMcPU-A:muon1pt:hiEff}   {\ensuremath{{1.000 } } }
\vdef{default-11:NoMcPU-A:muon1pt:hiEffE}   {\ensuremath{{0.000 } } }
\vdef{default-11:NoMcPU-A:muon1pt:loDelta}   {\ensuremath{{-0.000 } } }
\vdef{default-11:NoMcPU-A:muon1pt:loDeltaE}   {\ensuremath{{\mathrm{NaN} } } }
\vdef{default-11:NoMcPU-A:muon1pt:hiDelta}   {\ensuremath{{+0.000 } } }
\vdef{default-11:NoMcPU-A:muon1pt:hiDeltaE}   {\ensuremath{{0.000 } } }
\vdef{default-11:NoData-A:muon2pt:loEff}   {\ensuremath{{0.074 } } }
\vdef{default-11:NoData-A:muon2pt:loEffE}   {\ensuremath{{0.001 } } }
\vdef{default-11:NoData-A:muon2pt:hiEff}   {\ensuremath{{0.926 } } }
\vdef{default-11:NoData-A:muon2pt:hiEffE}   {\ensuremath{{0.001 } } }
\vdef{default-11:NoMcPU-A:muon2pt:loEff}   {\ensuremath{{0.004 } } }
\vdef{default-11:NoMcPU-A:muon2pt:loEffE}   {\ensuremath{{0.000 } } }
\vdef{default-11:NoMcPU-A:muon2pt:hiEff}   {\ensuremath{{0.996 } } }
\vdef{default-11:NoMcPU-A:muon2pt:hiEffE}   {\ensuremath{{0.000 } } }
\vdef{default-11:NoMcPU-A:muon2pt:loDelta}   {\ensuremath{{+1.816 } } }
\vdef{default-11:NoMcPU-A:muon2pt:loDeltaE}   {\ensuremath{{0.009 } } }
\vdef{default-11:NoMcPU-A:muon2pt:hiDelta}   {\ensuremath{{-0.073 } } }
\vdef{default-11:NoMcPU-A:muon2pt:hiDeltaE}   {\ensuremath{{0.001 } } }
\vdef{default-11:NoData-A:muonseta:loEff}   {\ensuremath{{0.738 } } }
\vdef{default-11:NoData-A:muonseta:loEffE}   {\ensuremath{{0.001 } } }
\vdef{default-11:NoData-A:muonseta:hiEff}   {\ensuremath{{0.262 } } }
\vdef{default-11:NoData-A:muonseta:hiEffE}   {\ensuremath{{0.001 } } }
\vdef{default-11:NoMcPU-A:muonseta:loEff}   {\ensuremath{{0.843 } } }
\vdef{default-11:NoMcPU-A:muonseta:loEffE}   {\ensuremath{{0.001 } } }
\vdef{default-11:NoMcPU-A:muonseta:hiEff}   {\ensuremath{{0.157 } } }
\vdef{default-11:NoMcPU-A:muonseta:hiEffE}   {\ensuremath{{0.001 } } }
\vdef{default-11:NoMcPU-A:muonseta:loDelta}   {\ensuremath{{-0.133 } } }
\vdef{default-11:NoMcPU-A:muonseta:loDeltaE}   {\ensuremath{{0.002 } } }
\vdef{default-11:NoMcPU-A:muonseta:hiDelta}   {\ensuremath{{+0.503 } } }
\vdef{default-11:NoMcPU-A:muonseta:hiDeltaE}   {\ensuremath{{0.006 } } }
\vdef{default-11:NoData-A:pt:loEff}   {\ensuremath{{0.000 } } }
\vdef{default-11:NoData-A:pt:loEffE}   {\ensuremath{{0.000 } } }
\vdef{default-11:NoData-A:pt:hiEff}   {\ensuremath{{1.000 } } }
\vdef{default-11:NoData-A:pt:hiEffE}   {\ensuremath{{0.000 } } }
\vdef{default-11:NoMcPU-A:pt:loEff}   {\ensuremath{{0.000 } } }
\vdef{default-11:NoMcPU-A:pt:loEffE}   {\ensuremath{{0.000 } } }
\vdef{default-11:NoMcPU-A:pt:hiEff}   {\ensuremath{{1.000 } } }
\vdef{default-11:NoMcPU-A:pt:hiEffE}   {\ensuremath{{0.000 } } }
\vdef{default-11:NoMcPU-A:pt:loDelta}   {\ensuremath{{\mathrm{NaN} } } }
\vdef{default-11:NoMcPU-A:pt:loDeltaE}   {\ensuremath{{\mathrm{NaN} } } }
\vdef{default-11:NoMcPU-A:pt:hiDelta}   {\ensuremath{{+0.000 } } }
\vdef{default-11:NoMcPU-A:pt:hiDeltaE}   {\ensuremath{{0.000 } } }
\vdef{default-11:NoData-A:p:loEff}   {\ensuremath{{1.013 } } }
\vdef{default-11:NoData-A:p:loEffE}   {\ensuremath{{\mathrm{NaN} } } }
\vdef{default-11:NoData-A:p:hiEff}   {\ensuremath{{1.000 } } }
\vdef{default-11:NoData-A:p:hiEffE}   {\ensuremath{{0.000 } } }
\vdef{default-11:NoMcPU-A:p:loEff}   {\ensuremath{{1.004 } } }
\vdef{default-11:NoMcPU-A:p:loEffE}   {\ensuremath{{\mathrm{NaN} } } }
\vdef{default-11:NoMcPU-A:p:hiEff}   {\ensuremath{{1.000 } } }
\vdef{default-11:NoMcPU-A:p:hiEffE}   {\ensuremath{{0.000 } } }
\vdef{default-11:NoMcPU-A:p:loDelta}   {\ensuremath{{+0.008 } } }
\vdef{default-11:NoMcPU-A:p:loDeltaE}   {\ensuremath{{\mathrm{NaN} } } }
\vdef{default-11:NoMcPU-A:p:hiDelta}   {\ensuremath{{+0.000 } } }
\vdef{default-11:NoMcPU-A:p:hiDeltaE}   {\ensuremath{{0.000 } } }
\vdef{default-11:NoData-A:eta:loEff}   {\ensuremath{{0.729 } } }
\vdef{default-11:NoData-A:eta:loEffE}   {\ensuremath{{0.001 } } }
\vdef{default-11:NoData-A:eta:hiEff}   {\ensuremath{{0.271 } } }
\vdef{default-11:NoData-A:eta:hiEffE}   {\ensuremath{{0.001 } } }
\vdef{default-11:NoMcPU-A:eta:loEff}   {\ensuremath{{0.841 } } }
\vdef{default-11:NoMcPU-A:eta:loEffE}   {\ensuremath{{0.001 } } }
\vdef{default-11:NoMcPU-A:eta:hiEff}   {\ensuremath{{0.159 } } }
\vdef{default-11:NoMcPU-A:eta:hiEffE}   {\ensuremath{{0.001 } } }
\vdef{default-11:NoMcPU-A:eta:loDelta}   {\ensuremath{{-0.142 } } }
\vdef{default-11:NoMcPU-A:eta:loDeltaE}   {\ensuremath{{0.002 } } }
\vdef{default-11:NoMcPU-A:eta:hiDelta}   {\ensuremath{{+0.522 } } }
\vdef{default-11:NoMcPU-A:eta:hiDeltaE}   {\ensuremath{{0.009 } } }
\vdef{default-11:NoData-A:bdt:loEff}   {\ensuremath{{0.911 } } }
\vdef{default-11:NoData-A:bdt:loEffE}   {\ensuremath{{0.001 } } }
\vdef{default-11:NoData-A:bdt:hiEff}   {\ensuremath{{0.089 } } }
\vdef{default-11:NoData-A:bdt:hiEffE}   {\ensuremath{{0.001 } } }
\vdef{default-11:NoMcPU-A:bdt:loEff}   {\ensuremath{{0.871 } } }
\vdef{default-11:NoMcPU-A:bdt:loEffE}   {\ensuremath{{0.001 } } }
\vdef{default-11:NoMcPU-A:bdt:hiEff}   {\ensuremath{{0.129 } } }
\vdef{default-11:NoMcPU-A:bdt:hiEffE}   {\ensuremath{{0.001 } } }
\vdef{default-11:NoMcPU-A:bdt:loDelta}   {\ensuremath{{+0.045 } } }
\vdef{default-11:NoMcPU-A:bdt:loDeltaE}   {\ensuremath{{0.002 } } }
\vdef{default-11:NoMcPU-A:bdt:hiDelta}   {\ensuremath{{-0.370 } } }
\vdef{default-11:NoMcPU-A:bdt:hiDeltaE}   {\ensuremath{{0.013 } } }
\vdef{default-11:NoData-A:fl3d:loEff}   {\ensuremath{{0.832 } } }
\vdef{default-11:NoData-A:fl3d:loEffE}   {\ensuremath{{0.001 } } }
\vdef{default-11:NoData-A:fl3d:hiEff}   {\ensuremath{{0.168 } } }
\vdef{default-11:NoData-A:fl3d:hiEffE}   {\ensuremath{{0.001 } } }
\vdef{default-11:NoMcPU-A:fl3d:loEff}   {\ensuremath{{0.888 } } }
\vdef{default-11:NoMcPU-A:fl3d:loEffE}   {\ensuremath{{0.001 } } }
\vdef{default-11:NoMcPU-A:fl3d:hiEff}   {\ensuremath{{0.112 } } }
\vdef{default-11:NoMcPU-A:fl3d:hiEffE}   {\ensuremath{{0.001 } } }
\vdef{default-11:NoMcPU-A:fl3d:loDelta}   {\ensuremath{{-0.065 } } }
\vdef{default-11:NoMcPU-A:fl3d:loDeltaE}   {\ensuremath{{0.002 } } }
\vdef{default-11:NoMcPU-A:fl3d:hiDelta}   {\ensuremath{{+0.397 } } }
\vdef{default-11:NoMcPU-A:fl3d:hiDeltaE}   {\ensuremath{{0.010 } } }
\vdef{default-11:NoData-A:fl3de:loEff}   {\ensuremath{{1.000 } } }
\vdef{default-11:NoData-A:fl3de:loEffE}   {\ensuremath{{0.000 } } }
\vdef{default-11:NoData-A:fl3de:hiEff}   {\ensuremath{{0.000 } } }
\vdef{default-11:NoData-A:fl3de:hiEffE}   {\ensuremath{{0.000 } } }
\vdef{default-11:NoMcPU-A:fl3de:loEff}   {\ensuremath{{1.000 } } }
\vdef{default-11:NoMcPU-A:fl3de:loEffE}   {\ensuremath{{0.000 } } }
\vdef{default-11:NoMcPU-A:fl3de:hiEff}   {\ensuremath{{0.000 } } }
\vdef{default-11:NoMcPU-A:fl3de:hiEffE}   {\ensuremath{{0.000 } } }
\vdef{default-11:NoMcPU-A:fl3de:loDelta}   {\ensuremath{{+0.000 } } }
\vdef{default-11:NoMcPU-A:fl3de:loDeltaE}   {\ensuremath{{0.000 } } }
\vdef{default-11:NoMcPU-A:fl3de:hiDelta}   {\ensuremath{{+0.550 } } }
\vdef{default-11:NoMcPU-A:fl3de:hiDeltaE}   {\ensuremath{{0.539 } } }
\vdef{default-11:NoData-A:fls3d:loEff}   {\ensuremath{{0.074 } } }
\vdef{default-11:NoData-A:fls3d:loEffE}   {\ensuremath{{0.001 } } }
\vdef{default-11:NoData-A:fls3d:hiEff}   {\ensuremath{{0.926 } } }
\vdef{default-11:NoData-A:fls3d:hiEffE}   {\ensuremath{{0.001 } } }
\vdef{default-11:NoMcPU-A:fls3d:loEff}   {\ensuremath{{0.062 } } }
\vdef{default-11:NoMcPU-A:fls3d:loEffE}   {\ensuremath{{0.001 } } }
\vdef{default-11:NoMcPU-A:fls3d:hiEff}   {\ensuremath{{0.938 } } }
\vdef{default-11:NoMcPU-A:fls3d:hiEffE}   {\ensuremath{{0.001 } } }
\vdef{default-11:NoMcPU-A:fls3d:loDelta}   {\ensuremath{{+0.176 } } }
\vdef{default-11:NoMcPU-A:fls3d:loDeltaE}   {\ensuremath{{0.016 } } }
\vdef{default-11:NoMcPU-A:fls3d:hiDelta}   {\ensuremath{{-0.013 } } }
\vdef{default-11:NoMcPU-A:fls3d:hiDeltaE}   {\ensuremath{{0.001 } } }
\vdef{default-11:NoData-A:flsxy:loEff}   {\ensuremath{{1.012 } } }
\vdef{default-11:NoData-A:flsxy:loEffE}   {\ensuremath{{\mathrm{NaN} } } }
\vdef{default-11:NoData-A:flsxy:hiEff}   {\ensuremath{{1.000 } } }
\vdef{default-11:NoData-A:flsxy:hiEffE}   {\ensuremath{{0.000 } } }
\vdef{default-11:NoMcPU-A:flsxy:loEff}   {\ensuremath{{1.013 } } }
\vdef{default-11:NoMcPU-A:flsxy:loEffE}   {\ensuremath{{\mathrm{NaN} } } }
\vdef{default-11:NoMcPU-A:flsxy:hiEff}   {\ensuremath{{1.000 } } }
\vdef{default-11:NoMcPU-A:flsxy:hiEffE}   {\ensuremath{{0.000 } } }
\vdef{default-11:NoMcPU-A:flsxy:loDelta}   {\ensuremath{{-0.001 } } }
\vdef{default-11:NoMcPU-A:flsxy:loDeltaE}   {\ensuremath{{\mathrm{NaN} } } }
\vdef{default-11:NoMcPU-A:flsxy:hiDelta}   {\ensuremath{{+0.000 } } }
\vdef{default-11:NoMcPU-A:flsxy:hiDeltaE}   {\ensuremath{{0.000 } } }
\vdef{default-11:NoData-A:chi2dof:loEff}   {\ensuremath{{0.928 } } }
\vdef{default-11:NoData-A:chi2dof:loEffE}   {\ensuremath{{0.001 } } }
\vdef{default-11:NoData-A:chi2dof:hiEff}   {\ensuremath{{0.072 } } }
\vdef{default-11:NoData-A:chi2dof:hiEffE}   {\ensuremath{{0.001 } } }
\vdef{default-11:NoMcPU-A:chi2dof:loEff}   {\ensuremath{{0.945 } } }
\vdef{default-11:NoMcPU-A:chi2dof:loEffE}   {\ensuremath{{0.001 } } }
\vdef{default-11:NoMcPU-A:chi2dof:hiEff}   {\ensuremath{{0.055 } } }
\vdef{default-11:NoMcPU-A:chi2dof:hiEffE}   {\ensuremath{{0.001 } } }
\vdef{default-11:NoMcPU-A:chi2dof:loDelta}   {\ensuremath{{-0.019 } } }
\vdef{default-11:NoMcPU-A:chi2dof:loDeltaE}   {\ensuremath{{0.001 } } }
\vdef{default-11:NoMcPU-A:chi2dof:hiDelta}   {\ensuremath{{+0.282 } } }
\vdef{default-11:NoMcPU-A:chi2dof:hiDeltaE}   {\ensuremath{{0.017 } } }
\vdef{default-11:NoData-A:pchi2dof:loEff}   {\ensuremath{{0.642 } } }
\vdef{default-11:NoData-A:pchi2dof:loEffE}   {\ensuremath{{0.001 } } }
\vdef{default-11:NoData-A:pchi2dof:hiEff}   {\ensuremath{{0.358 } } }
\vdef{default-11:NoData-A:pchi2dof:hiEffE}   {\ensuremath{{0.001 } } }
\vdef{default-11:NoMcPU-A:pchi2dof:loEff}   {\ensuremath{{0.612 } } }
\vdef{default-11:NoMcPU-A:pchi2dof:loEffE}   {\ensuremath{{0.002 } } }
\vdef{default-11:NoMcPU-A:pchi2dof:hiEff}   {\ensuremath{{0.388 } } }
\vdef{default-11:NoMcPU-A:pchi2dof:hiEffE}   {\ensuremath{{0.002 } } }
\vdef{default-11:NoMcPU-A:pchi2dof:loDelta}   {\ensuremath{{+0.048 } } }
\vdef{default-11:NoMcPU-A:pchi2dof:loDeltaE}   {\ensuremath{{0.003 } } }
\vdef{default-11:NoMcPU-A:pchi2dof:hiDelta}   {\ensuremath{{-0.081 } } }
\vdef{default-11:NoMcPU-A:pchi2dof:hiDeltaE}   {\ensuremath{{0.006 } } }
\vdef{default-11:NoData-A:alpha:loEff}   {\ensuremath{{0.995 } } }
\vdef{default-11:NoData-A:alpha:loEffE}   {\ensuremath{{0.000 } } }
\vdef{default-11:NoData-A:alpha:hiEff}   {\ensuremath{{0.005 } } }
\vdef{default-11:NoData-A:alpha:hiEffE}   {\ensuremath{{0.000 } } }
\vdef{default-11:NoMcPU-A:alpha:loEff}   {\ensuremath{{0.993 } } }
\vdef{default-11:NoMcPU-A:alpha:loEffE}   {\ensuremath{{0.000 } } }
\vdef{default-11:NoMcPU-A:alpha:hiEff}   {\ensuremath{{0.007 } } }
\vdef{default-11:NoMcPU-A:alpha:hiEffE}   {\ensuremath{{0.000 } } }
\vdef{default-11:NoMcPU-A:alpha:loDelta}   {\ensuremath{{+0.002 } } }
\vdef{default-11:NoMcPU-A:alpha:loDeltaE}   {\ensuremath{{0.000 } } }
\vdef{default-11:NoMcPU-A:alpha:hiDelta}   {\ensuremath{{-0.307 } } }
\vdef{default-11:NoMcPU-A:alpha:hiDeltaE}   {\ensuremath{{0.056 } } }
\vdef{default-11:NoData-A:iso:loEff}   {\ensuremath{{0.129 } } }
\vdef{default-11:NoData-A:iso:loEffE}   {\ensuremath{{0.001 } } }
\vdef{default-11:NoData-A:iso:hiEff}   {\ensuremath{{0.871 } } }
\vdef{default-11:NoData-A:iso:hiEffE}   {\ensuremath{{0.001 } } }
\vdef{default-11:NoMcPU-A:iso:loEff}   {\ensuremath{{0.108 } } }
\vdef{default-11:NoMcPU-A:iso:loEffE}   {\ensuremath{{0.001 } } }
\vdef{default-11:NoMcPU-A:iso:hiEff}   {\ensuremath{{0.892 } } }
\vdef{default-11:NoMcPU-A:iso:hiEffE}   {\ensuremath{{0.001 } } }
\vdef{default-11:NoMcPU-A:iso:loDelta}   {\ensuremath{{+0.171 } } }
\vdef{default-11:NoMcPU-A:iso:loDeltaE}   {\ensuremath{{0.012 } } }
\vdef{default-11:NoMcPU-A:iso:hiDelta}   {\ensuremath{{-0.023 } } }
\vdef{default-11:NoMcPU-A:iso:hiDeltaE}   {\ensuremath{{0.002 } } }
\vdef{default-11:NoData-A:docatrk:loEff}   {\ensuremath{{0.071 } } }
\vdef{default-11:NoData-A:docatrk:loEffE}   {\ensuremath{{0.001 } } }
\vdef{default-11:NoData-A:docatrk:hiEff}   {\ensuremath{{0.929 } } }
\vdef{default-11:NoData-A:docatrk:hiEffE}   {\ensuremath{{0.001 } } }
\vdef{default-11:NoMcPU-A:docatrk:loEff}   {\ensuremath{{0.084 } } }
\vdef{default-11:NoMcPU-A:docatrk:loEffE}   {\ensuremath{{0.001 } } }
\vdef{default-11:NoMcPU-A:docatrk:hiEff}   {\ensuremath{{0.916 } } }
\vdef{default-11:NoMcPU-A:docatrk:hiEffE}   {\ensuremath{{0.001 } } }
\vdef{default-11:NoMcPU-A:docatrk:loDelta}   {\ensuremath{{-0.168 } } }
\vdef{default-11:NoMcPU-A:docatrk:loDeltaE}   {\ensuremath{{0.015 } } }
\vdef{default-11:NoMcPU-A:docatrk:hiDelta}   {\ensuremath{{+0.014 } } }
\vdef{default-11:NoMcPU-A:docatrk:hiDeltaE}   {\ensuremath{{0.001 } } }
\vdef{default-11:NoData-A:isotrk:loEff}   {\ensuremath{{1.000 } } }
\vdef{default-11:NoData-A:isotrk:loEffE}   {\ensuremath{{0.000 } } }
\vdef{default-11:NoData-A:isotrk:hiEff}   {\ensuremath{{1.000 } } }
\vdef{default-11:NoData-A:isotrk:hiEffE}   {\ensuremath{{0.000 } } }
\vdef{default-11:NoMcPU-A:isotrk:loEff}   {\ensuremath{{1.000 } } }
\vdef{default-11:NoMcPU-A:isotrk:loEffE}   {\ensuremath{{0.000 } } }
\vdef{default-11:NoMcPU-A:isotrk:hiEff}   {\ensuremath{{1.000 } } }
\vdef{default-11:NoMcPU-A:isotrk:hiEffE}   {\ensuremath{{0.000 } } }
\vdef{default-11:NoMcPU-A:isotrk:loDelta}   {\ensuremath{{+0.000 } } }
\vdef{default-11:NoMcPU-A:isotrk:loDeltaE}   {\ensuremath{{0.000 } } }
\vdef{default-11:NoMcPU-A:isotrk:hiDelta}   {\ensuremath{{+0.000 } } }
\vdef{default-11:NoMcPU-A:isotrk:hiDeltaE}   {\ensuremath{{0.000 } } }
\vdef{default-11:NoData-A:closetrk:loEff}   {\ensuremath{{0.975 } } }
\vdef{default-11:NoData-A:closetrk:loEffE}   {\ensuremath{{0.000 } } }
\vdef{default-11:NoData-A:closetrk:hiEff}   {\ensuremath{{0.025 } } }
\vdef{default-11:NoData-A:closetrk:hiEffE}   {\ensuremath{{0.000 } } }
\vdef{default-11:NoMcPU-A:closetrk:loEff}   {\ensuremath{{0.973 } } }
\vdef{default-11:NoMcPU-A:closetrk:loEffE}   {\ensuremath{{0.001 } } }
\vdef{default-11:NoMcPU-A:closetrk:hiEff}   {\ensuremath{{0.027 } } }
\vdef{default-11:NoMcPU-A:closetrk:hiEffE}   {\ensuremath{{0.001 } } }
\vdef{default-11:NoMcPU-A:closetrk:loDelta}   {\ensuremath{{+0.002 } } }
\vdef{default-11:NoMcPU-A:closetrk:loDeltaE}   {\ensuremath{{0.001 } } }
\vdef{default-11:NoMcPU-A:closetrk:hiDelta}   {\ensuremath{{-0.081 } } }
\vdef{default-11:NoMcPU-A:closetrk:hiDeltaE}   {\ensuremath{{0.028 } } }
\vdef{default-11:NoData-A:lip:loEff}   {\ensuremath{{1.000 } } }
\vdef{default-11:NoData-A:lip:loEffE}   {\ensuremath{{0.000 } } }
\vdef{default-11:NoData-A:lip:hiEff}   {\ensuremath{{0.000 } } }
\vdef{default-11:NoData-A:lip:hiEffE}   {\ensuremath{{0.000 } } }
\vdef{default-11:NoMcPU-A:lip:loEff}   {\ensuremath{{1.000 } } }
\vdef{default-11:NoMcPU-A:lip:loEffE}   {\ensuremath{{0.000 } } }
\vdef{default-11:NoMcPU-A:lip:hiEff}   {\ensuremath{{0.000 } } }
\vdef{default-11:NoMcPU-A:lip:hiEffE}   {\ensuremath{{0.000 } } }
\vdef{default-11:NoMcPU-A:lip:loDelta}   {\ensuremath{{+0.000 } } }
\vdef{default-11:NoMcPU-A:lip:loDeltaE}   {\ensuremath{{0.000 } } }
\vdef{default-11:NoMcPU-A:lip:hiDelta}   {\ensuremath{{\mathrm{NaN} } } }
\vdef{default-11:NoMcPU-A:lip:hiDeltaE}   {\ensuremath{{\mathrm{NaN} } } }
\vdef{default-11:NoData-A:lip:inEff}   {\ensuremath{{1.000 } } }
\vdef{default-11:NoData-A:lip:inEffE}   {\ensuremath{{0.000 } } }
\vdef{default-11:NoMcPU-A:lip:inEff}   {\ensuremath{{1.000 } } }
\vdef{default-11:NoMcPU-A:lip:inEffE}   {\ensuremath{{0.000 } } }
\vdef{default-11:NoMcPU-A:lip:inDelta}   {\ensuremath{{+0.000 } } }
\vdef{default-11:NoMcPU-A:lip:inDeltaE}   {\ensuremath{{0.000 } } }
\vdef{default-11:NoData-A:lips:loEff}   {\ensuremath{{1.000 } } }
\vdef{default-11:NoData-A:lips:loEffE}   {\ensuremath{{0.000 } } }
\vdef{default-11:NoData-A:lips:hiEff}   {\ensuremath{{0.000 } } }
\vdef{default-11:NoData-A:lips:hiEffE}   {\ensuremath{{0.000 } } }
\vdef{default-11:NoMcPU-A:lips:loEff}   {\ensuremath{{1.000 } } }
\vdef{default-11:NoMcPU-A:lips:loEffE}   {\ensuremath{{0.000 } } }
\vdef{default-11:NoMcPU-A:lips:hiEff}   {\ensuremath{{0.000 } } }
\vdef{default-11:NoMcPU-A:lips:hiEffE}   {\ensuremath{{0.000 } } }
\vdef{default-11:NoMcPU-A:lips:loDelta}   {\ensuremath{{+0.000 } } }
\vdef{default-11:NoMcPU-A:lips:loDeltaE}   {\ensuremath{{0.000 } } }
\vdef{default-11:NoMcPU-A:lips:hiDelta}   {\ensuremath{{\mathrm{NaN} } } }
\vdef{default-11:NoMcPU-A:lips:hiDeltaE}   {\ensuremath{{\mathrm{NaN} } } }
\vdef{default-11:NoData-A:lips:inEff}   {\ensuremath{{1.000 } } }
\vdef{default-11:NoData-A:lips:inEffE}   {\ensuremath{{0.000 } } }
\vdef{default-11:NoMcPU-A:lips:inEff}   {\ensuremath{{1.000 } } }
\vdef{default-11:NoMcPU-A:lips:inEffE}   {\ensuremath{{0.000 } } }
\vdef{default-11:NoMcPU-A:lips:inDelta}   {\ensuremath{{+0.000 } } }
\vdef{default-11:NoMcPU-A:lips:inDeltaE}   {\ensuremath{{0.000 } } }
\vdef{default-11:NoData-A:ip:loEff}   {\ensuremath{{0.972 } } }
\vdef{default-11:NoData-A:ip:loEffE}   {\ensuremath{{0.001 } } }
\vdef{default-11:NoData-A:ip:hiEff}   {\ensuremath{{0.028 } } }
\vdef{default-11:NoData-A:ip:hiEffE}   {\ensuremath{{0.001 } } }
\vdef{default-11:NoMcPU-A:ip:loEff}   {\ensuremath{{0.970 } } }
\vdef{default-11:NoMcPU-A:ip:loEffE}   {\ensuremath{{0.001 } } }
\vdef{default-11:NoMcPU-A:ip:hiEff}   {\ensuremath{{0.030 } } }
\vdef{default-11:NoMcPU-A:ip:hiEffE}   {\ensuremath{{0.001 } } }
\vdef{default-11:NoMcPU-A:ip:loDelta}   {\ensuremath{{+0.002 } } }
\vdef{default-11:NoMcPU-A:ip:loDeltaE}   {\ensuremath{{0.001 } } }
\vdef{default-11:NoMcPU-A:ip:hiDelta}   {\ensuremath{{-0.080 } } }
\vdef{default-11:NoMcPU-A:ip:hiDeltaE}   {\ensuremath{{0.026 } } }
\vdef{default-11:NoData-A:ips:loEff}   {\ensuremath{{0.944 } } }
\vdef{default-11:NoData-A:ips:loEffE}   {\ensuremath{{0.001 } } }
\vdef{default-11:NoData-A:ips:hiEff}   {\ensuremath{{0.056 } } }
\vdef{default-11:NoData-A:ips:hiEffE}   {\ensuremath{{0.001 } } }
\vdef{default-11:NoMcPU-A:ips:loEff}   {\ensuremath{{0.951 } } }
\vdef{default-11:NoMcPU-A:ips:loEffE}   {\ensuremath{{0.001 } } }
\vdef{default-11:NoMcPU-A:ips:hiEff}   {\ensuremath{{0.049 } } }
\vdef{default-11:NoMcPU-A:ips:hiEffE}   {\ensuremath{{0.001 } } }
\vdef{default-11:NoMcPU-A:ips:loDelta}   {\ensuremath{{-0.007 } } }
\vdef{default-11:NoMcPU-A:ips:loDeltaE}   {\ensuremath{{0.001 } } }
\vdef{default-11:NoMcPU-A:ips:hiDelta}   {\ensuremath{{+0.124 } } }
\vdef{default-11:NoMcPU-A:ips:hiDeltaE}   {\ensuremath{{0.019 } } }
\vdef{default-11:NoData-A:maxdoca:loEff}   {\ensuremath{{1.000 } } }
\vdef{default-11:NoData-A:maxdoca:loEffE}   {\ensuremath{{0.000 } } }
\vdef{default-11:NoData-A:maxdoca:hiEff}   {\ensuremath{{0.013 } } }
\vdef{default-11:NoData-A:maxdoca:hiEffE}   {\ensuremath{{0.000 } } }
\vdef{default-11:NoMcPU-A:maxdoca:loEff}   {\ensuremath{{1.000 } } }
\vdef{default-11:NoMcPU-A:maxdoca:loEffE}   {\ensuremath{{0.000 } } }
\vdef{default-11:NoMcPU-A:maxdoca:hiEff}   {\ensuremath{{0.014 } } }
\vdef{default-11:NoMcPU-A:maxdoca:hiEffE}   {\ensuremath{{0.000 } } }
\vdef{default-11:NoMcPU-A:maxdoca:loDelta}   {\ensuremath{{+0.000 } } }
\vdef{default-11:NoMcPU-A:maxdoca:loDeltaE}   {\ensuremath{{0.000 } } }
\vdef{default-11:NoMcPU-A:maxdoca:hiDelta}   {\ensuremath{{-0.078 } } }
\vdef{default-11:NoMcPU-A:maxdoca:hiDeltaE}   {\ensuremath{{0.039 } } }
\vdef{default-11:NoData-A:kaonpt:loEff}   {\ensuremath{{1.010 } } }
\vdef{default-11:NoData-A:kaonpt:loEffE}   {\ensuremath{{\mathrm{NaN} } } }
\vdef{default-11:NoData-A:kaonpt:hiEff}   {\ensuremath{{1.000 } } }
\vdef{default-11:NoData-A:kaonpt:hiEffE}   {\ensuremath{{0.000 } } }
\vdef{default-11:NoMcPU-A:kaonpt:loEff}   {\ensuremath{{1.009 } } }
\vdef{default-11:NoMcPU-A:kaonpt:loEffE}   {\ensuremath{{\mathrm{NaN} } } }
\vdef{default-11:NoMcPU-A:kaonpt:hiEff}   {\ensuremath{{1.000 } } }
\vdef{default-11:NoMcPU-A:kaonpt:hiEffE}   {\ensuremath{{0.000 } } }
\vdef{default-11:NoMcPU-A:kaonpt:loDelta}   {\ensuremath{{+0.001 } } }
\vdef{default-11:NoMcPU-A:kaonpt:loDeltaE}   {\ensuremath{{\mathrm{NaN} } } }
\vdef{default-11:NoMcPU-A:kaonpt:hiDelta}   {\ensuremath{{+0.000 } } }
\vdef{default-11:NoMcPU-A:kaonpt:hiDeltaE}   {\ensuremath{{0.000 } } }
\vdef{default-11:NoData-A:psipt:loEff}   {\ensuremath{{1.004 } } }
\vdef{default-11:NoData-A:psipt:loEffE}   {\ensuremath{{\mathrm{NaN} } } }
\vdef{default-11:NoData-A:psipt:hiEff}   {\ensuremath{{1.000 } } }
\vdef{default-11:NoData-A:psipt:hiEffE}   {\ensuremath{{0.000 } } }
\vdef{default-11:NoMcPU-A:psipt:loEff}   {\ensuremath{{1.003 } } }
\vdef{default-11:NoMcPU-A:psipt:loEffE}   {\ensuremath{{\mathrm{NaN} } } }
\vdef{default-11:NoMcPU-A:psipt:hiEff}   {\ensuremath{{1.000 } } }
\vdef{default-11:NoMcPU-A:psipt:hiEffE}   {\ensuremath{{0.000 } } }
\vdef{default-11:NoMcPU-A:psipt:loDelta}   {\ensuremath{{+0.000 } } }
\vdef{default-11:NoMcPU-A:psipt:loDeltaE}   {\ensuremath{{\mathrm{NaN} } } }
\vdef{default-11:NoMcPU-A:psipt:hiDelta}   {\ensuremath{{+0.000 } } }
\vdef{default-11:NoMcPU-A:psipt:hiDeltaE}   {\ensuremath{{0.000 } } }
\vdef{default-11:NoData-APV0:osiso:loEff}   {\ensuremath{{1.005 } } }
\vdef{default-11:NoData-APV0:osiso:loEffE}   {\ensuremath{{\mathrm{NaN} } } }
\vdef{default-11:NoData-APV0:osiso:hiEff}   {\ensuremath{{1.000 } } }
\vdef{default-11:NoData-APV0:osiso:hiEffE}   {\ensuremath{{0.000 } } }
\vdef{default-11:NoMcPU-APV0:osiso:loEff}   {\ensuremath{{1.005 } } }
\vdef{default-11:NoMcPU-APV0:osiso:loEffE}   {\ensuremath{{\mathrm{NaN} } } }
\vdef{default-11:NoMcPU-APV0:osiso:hiEff}   {\ensuremath{{1.000 } } }
\vdef{default-11:NoMcPU-APV0:osiso:hiEffE}   {\ensuremath{{0.000 } } }
\vdef{default-11:NoMcPU-APV0:osiso:loDelta}   {\ensuremath{{-0.000 } } }
\vdef{default-11:NoMcPU-APV0:osiso:loDeltaE}   {\ensuremath{{\mathrm{NaN} } } }
\vdef{default-11:NoMcPU-APV0:osiso:hiDelta}   {\ensuremath{{+0.000 } } }
\vdef{default-11:NoMcPU-APV0:osiso:hiDeltaE}   {\ensuremath{{0.000 } } }
\vdef{default-11:NoData-APV0:osreliso:loEff}   {\ensuremath{{0.243 } } }
\vdef{default-11:NoData-APV0:osreliso:loEffE}   {\ensuremath{{0.002 } } }
\vdef{default-11:NoData-APV0:osreliso:hiEff}   {\ensuremath{{0.757 } } }
\vdef{default-11:NoData-APV0:osreliso:hiEffE}   {\ensuremath{{0.002 } } }
\vdef{default-11:NoMcPU-APV0:osreliso:loEff}   {\ensuremath{{0.291 } } }
\vdef{default-11:NoMcPU-APV0:osreliso:loEffE}   {\ensuremath{{0.003 } } }
\vdef{default-11:NoMcPU-APV0:osreliso:hiEff}   {\ensuremath{{0.709 } } }
\vdef{default-11:NoMcPU-APV0:osreliso:hiEffE}   {\ensuremath{{0.003 } } }
\vdef{default-11:NoMcPU-APV0:osreliso:loDelta}   {\ensuremath{{-0.180 } } }
\vdef{default-11:NoMcPU-APV0:osreliso:loDeltaE}   {\ensuremath{{0.013 } } }
\vdef{default-11:NoMcPU-APV0:osreliso:hiDelta}   {\ensuremath{{+0.065 } } }
\vdef{default-11:NoMcPU-APV0:osreliso:hiDeltaE}   {\ensuremath{{0.005 } } }
\vdef{default-11:NoData-APV0:osmuonpt:loEff}   {\ensuremath{{0.000 } } }
\vdef{default-11:NoData-APV0:osmuonpt:loEffE}   {\ensuremath{{0.001 } } }
\vdef{default-11:NoData-APV0:osmuonpt:hiEff}   {\ensuremath{{1.000 } } }
\vdef{default-11:NoData-APV0:osmuonpt:hiEffE}   {\ensuremath{{0.001 } } }
\vdef{default-11:NoMcPU-APV0:osmuonpt:loEff}   {\ensuremath{{0.000 } } }
\vdef{default-11:NoMcPU-APV0:osmuonpt:loEffE}   {\ensuremath{{0.001 } } }
\vdef{default-11:NoMcPU-APV0:osmuonpt:hiEff}   {\ensuremath{{1.000 } } }
\vdef{default-11:NoMcPU-APV0:osmuonpt:hiEffE}   {\ensuremath{{0.001 } } }
\vdef{default-11:NoMcPU-APV0:osmuonpt:loDelta}   {\ensuremath{{\mathrm{NaN} } } }
\vdef{default-11:NoMcPU-APV0:osmuonpt:loDeltaE}   {\ensuremath{{\mathrm{NaN} } } }
\vdef{default-11:NoMcPU-APV0:osmuonpt:hiDelta}   {\ensuremath{{+0.000 } } }
\vdef{default-11:NoMcPU-APV0:osmuonpt:hiDeltaE}   {\ensuremath{{0.001 } } }
\vdef{default-11:NoData-APV0:osmuondr:loEff}   {\ensuremath{{0.026 } } }
\vdef{default-11:NoData-APV0:osmuondr:loEffE}   {\ensuremath{{0.005 } } }
\vdef{default-11:NoData-APV0:osmuondr:hiEff}   {\ensuremath{{0.974 } } }
\vdef{default-11:NoData-APV0:osmuondr:hiEffE}   {\ensuremath{{0.005 } } }
\vdef{default-11:NoMcPU-APV0:osmuondr:loEff}   {\ensuremath{{0.009 } } }
\vdef{default-11:NoMcPU-APV0:osmuondr:loEffE}   {\ensuremath{{0.003 } } }
\vdef{default-11:NoMcPU-APV0:osmuondr:hiEff}   {\ensuremath{{0.991 } } }
\vdef{default-11:NoMcPU-APV0:osmuondr:hiEffE}   {\ensuremath{{0.003 } } }
\vdef{default-11:NoMcPU-APV0:osmuondr:loDelta}   {\ensuremath{{+0.989 } } }
\vdef{default-11:NoMcPU-APV0:osmuondr:loDeltaE}   {\ensuremath{{0.282 } } }
\vdef{default-11:NoMcPU-APV0:osmuondr:hiDelta}   {\ensuremath{{-0.017 } } }
\vdef{default-11:NoMcPU-APV0:osmuondr:hiDeltaE}   {\ensuremath{{0.006 } } }
\vdef{default-11:NoData-APV0:hlt:loEff}   {\ensuremath{{0.043 } } }
\vdef{default-11:NoData-APV0:hlt:loEffE}   {\ensuremath{{0.001 } } }
\vdef{default-11:NoData-APV0:hlt:hiEff}   {\ensuremath{{0.957 } } }
\vdef{default-11:NoData-APV0:hlt:hiEffE}   {\ensuremath{{0.001 } } }
\vdef{default-11:NoMcPU-APV0:hlt:loEff}   {\ensuremath{{0.284 } } }
\vdef{default-11:NoMcPU-APV0:hlt:loEffE}   {\ensuremath{{0.003 } } }
\vdef{default-11:NoMcPU-APV0:hlt:hiEff}   {\ensuremath{{0.716 } } }
\vdef{default-11:NoMcPU-APV0:hlt:hiEffE}   {\ensuremath{{0.003 } } }
\vdef{default-11:NoMcPU-APV0:hlt:loDelta}   {\ensuremath{{-1.477 } } }
\vdef{default-11:NoMcPU-APV0:hlt:loDeltaE}   {\ensuremath{{0.013 } } }
\vdef{default-11:NoMcPU-APV0:hlt:hiDelta}   {\ensuremath{{+0.289 } } }
\vdef{default-11:NoMcPU-APV0:hlt:hiDeltaE}   {\ensuremath{{0.004 } } }
\vdef{default-11:NoData-APV0:muonsid:loEff}   {\ensuremath{{0.142 } } }
\vdef{default-11:NoData-APV0:muonsid:loEffE}   {\ensuremath{{0.002 } } }
\vdef{default-11:NoData-APV0:muonsid:hiEff}   {\ensuremath{{0.858 } } }
\vdef{default-11:NoData-APV0:muonsid:hiEffE}   {\ensuremath{{0.002 } } }
\vdef{default-11:NoMcPU-APV0:muonsid:loEff}   {\ensuremath{{0.229 } } }
\vdef{default-11:NoMcPU-APV0:muonsid:loEffE}   {\ensuremath{{0.002 } } }
\vdef{default-11:NoMcPU-APV0:muonsid:hiEff}   {\ensuremath{{0.771 } } }
\vdef{default-11:NoMcPU-APV0:muonsid:hiEffE}   {\ensuremath{{0.002 } } }
\vdef{default-11:NoMcPU-APV0:muonsid:loDelta}   {\ensuremath{{-0.466 } } }
\vdef{default-11:NoMcPU-APV0:muonsid:loDeltaE}   {\ensuremath{{0.016 } } }
\vdef{default-11:NoMcPU-APV0:muonsid:hiDelta}   {\ensuremath{{+0.106 } } }
\vdef{default-11:NoMcPU-APV0:muonsid:hiDeltaE}   {\ensuremath{{0.004 } } }
\vdef{default-11:NoData-APV0:tracksqual:loEff}   {\ensuremath{{0.001 } } }
\vdef{default-11:NoData-APV0:tracksqual:loEffE}   {\ensuremath{{0.000 } } }
\vdef{default-11:NoData-APV0:tracksqual:hiEff}   {\ensuremath{{0.999 } } }
\vdef{default-11:NoData-APV0:tracksqual:hiEffE}   {\ensuremath{{0.000 } } }
\vdef{default-11:NoMcPU-APV0:tracksqual:loEff}   {\ensuremath{{0.000 } } }
\vdef{default-11:NoMcPU-APV0:tracksqual:loEffE}   {\ensuremath{{0.000 } } }
\vdef{default-11:NoMcPU-APV0:tracksqual:hiEff}   {\ensuremath{{1.000 } } }
\vdef{default-11:NoMcPU-APV0:tracksqual:hiEffE}   {\ensuremath{{0.000 } } }
\vdef{default-11:NoMcPU-APV0:tracksqual:loDelta}   {\ensuremath{{+0.345 } } }
\vdef{default-11:NoMcPU-APV0:tracksqual:loDeltaE}   {\ensuremath{{0.387 } } }
\vdef{default-11:NoMcPU-APV0:tracksqual:hiDelta}   {\ensuremath{{-0.000 } } }
\vdef{default-11:NoMcPU-APV0:tracksqual:hiDeltaE}   {\ensuremath{{0.000 } } }
\vdef{default-11:NoData-APV0:pvz:loEff}   {\ensuremath{{0.516 } } }
\vdef{default-11:NoData-APV0:pvz:loEffE}   {\ensuremath{{0.003 } } }
\vdef{default-11:NoData-APV0:pvz:hiEff}   {\ensuremath{{0.484 } } }
\vdef{default-11:NoData-APV0:pvz:hiEffE}   {\ensuremath{{0.003 } } }
\vdef{default-11:NoMcPU-APV0:pvz:loEff}   {\ensuremath{{0.470 } } }
\vdef{default-11:NoMcPU-APV0:pvz:loEffE}   {\ensuremath{{0.003 } } }
\vdef{default-11:NoMcPU-APV0:pvz:hiEff}   {\ensuremath{{0.530 } } }
\vdef{default-11:NoMcPU-APV0:pvz:hiEffE}   {\ensuremath{{0.003 } } }
\vdef{default-11:NoMcPU-APV0:pvz:loDelta}   {\ensuremath{{+0.094 } } }
\vdef{default-11:NoMcPU-APV0:pvz:loDeltaE}   {\ensuremath{{0.008 } } }
\vdef{default-11:NoMcPU-APV0:pvz:hiDelta}   {\ensuremath{{-0.092 } } }
\vdef{default-11:NoMcPU-APV0:pvz:hiDeltaE}   {\ensuremath{{0.008 } } }
\vdef{default-11:NoData-APV0:pvn:loEff}   {\ensuremath{{1.000 } } }
\vdef{default-11:NoData-APV0:pvn:loEffE}   {\ensuremath{{0.000 } } }
\vdef{default-11:NoData-APV0:pvn:hiEff}   {\ensuremath{{1.000 } } }
\vdef{default-11:NoData-APV0:pvn:hiEffE}   {\ensuremath{{0.000 } } }
\vdef{default-11:NoMcPU-APV0:pvn:loEff}   {\ensuremath{{1.000 } } }
\vdef{default-11:NoMcPU-APV0:pvn:loEffE}   {\ensuremath{{0.000 } } }
\vdef{default-11:NoMcPU-APV0:pvn:hiEff}   {\ensuremath{{1.000 } } }
\vdef{default-11:NoMcPU-APV0:pvn:hiEffE}   {\ensuremath{{0.000 } } }
\vdef{default-11:NoMcPU-APV0:pvn:loDelta}   {\ensuremath{{+0.000 } } }
\vdef{default-11:NoMcPU-APV0:pvn:loDeltaE}   {\ensuremath{{0.000 } } }
\vdef{default-11:NoMcPU-APV0:pvn:hiDelta}   {\ensuremath{{+0.000 } } }
\vdef{default-11:NoMcPU-APV0:pvn:hiDeltaE}   {\ensuremath{{0.000 } } }
\vdef{default-11:NoData-APV0:pvavew8:loEff}   {\ensuremath{{0.013 } } }
\vdef{default-11:NoData-APV0:pvavew8:loEffE}   {\ensuremath{{0.001 } } }
\vdef{default-11:NoData-APV0:pvavew8:hiEff}   {\ensuremath{{0.987 } } }
\vdef{default-11:NoData-APV0:pvavew8:hiEffE}   {\ensuremath{{0.001 } } }
\vdef{default-11:NoMcPU-APV0:pvavew8:loEff}   {\ensuremath{{0.005 } } }
\vdef{default-11:NoMcPU-APV0:pvavew8:loEffE}   {\ensuremath{{0.000 } } }
\vdef{default-11:NoMcPU-APV0:pvavew8:hiEff}   {\ensuremath{{0.995 } } }
\vdef{default-11:NoMcPU-APV0:pvavew8:hiEffE}   {\ensuremath{{0.000 } } }
\vdef{default-11:NoMcPU-APV0:pvavew8:loDelta}   {\ensuremath{{+0.932 } } }
\vdef{default-11:NoMcPU-APV0:pvavew8:loDeltaE}   {\ensuremath{{0.080 } } }
\vdef{default-11:NoMcPU-APV0:pvavew8:hiDelta}   {\ensuremath{{-0.008 } } }
\vdef{default-11:NoMcPU-APV0:pvavew8:hiDeltaE}   {\ensuremath{{0.001 } } }
\vdef{default-11:NoData-APV0:pvntrk:loEff}   {\ensuremath{{1.000 } } }
\vdef{default-11:NoData-APV0:pvntrk:loEffE}   {\ensuremath{{0.000 } } }
\vdef{default-11:NoData-APV0:pvntrk:hiEff}   {\ensuremath{{1.000 } } }
\vdef{default-11:NoData-APV0:pvntrk:hiEffE}   {\ensuremath{{0.000 } } }
\vdef{default-11:NoMcPU-APV0:pvntrk:loEff}   {\ensuremath{{1.000 } } }
\vdef{default-11:NoMcPU-APV0:pvntrk:loEffE}   {\ensuremath{{0.000 } } }
\vdef{default-11:NoMcPU-APV0:pvntrk:hiEff}   {\ensuremath{{1.000 } } }
\vdef{default-11:NoMcPU-APV0:pvntrk:hiEffE}   {\ensuremath{{0.000 } } }
\vdef{default-11:NoMcPU-APV0:pvntrk:loDelta}   {\ensuremath{{+0.000 } } }
\vdef{default-11:NoMcPU-APV0:pvntrk:loDeltaE}   {\ensuremath{{0.000 } } }
\vdef{default-11:NoMcPU-APV0:pvntrk:hiDelta}   {\ensuremath{{+0.000 } } }
\vdef{default-11:NoMcPU-APV0:pvntrk:hiDeltaE}   {\ensuremath{{0.000 } } }
\vdef{default-11:NoData-APV0:muon1pt:loEff}   {\ensuremath{{1.009 } } }
\vdef{default-11:NoData-APV0:muon1pt:loEffE}   {\ensuremath{{\mathrm{NaN} } } }
\vdef{default-11:NoData-APV0:muon1pt:hiEff}   {\ensuremath{{1.000 } } }
\vdef{default-11:NoData-APV0:muon1pt:hiEffE}   {\ensuremath{{0.000 } } }
\vdef{default-11:NoMcPU-APV0:muon1pt:loEff}   {\ensuremath{{1.010 } } }
\vdef{default-11:NoMcPU-APV0:muon1pt:loEffE}   {\ensuremath{{\mathrm{NaN} } } }
\vdef{default-11:NoMcPU-APV0:muon1pt:hiEff}   {\ensuremath{{1.000 } } }
\vdef{default-11:NoMcPU-APV0:muon1pt:hiEffE}   {\ensuremath{{0.000 } } }
\vdef{default-11:NoMcPU-APV0:muon1pt:loDelta}   {\ensuremath{{-0.001 } } }
\vdef{default-11:NoMcPU-APV0:muon1pt:loDeltaE}   {\ensuremath{{\mathrm{NaN} } } }
\vdef{default-11:NoMcPU-APV0:muon1pt:hiDelta}   {\ensuremath{{+0.000 } } }
\vdef{default-11:NoMcPU-APV0:muon1pt:hiDeltaE}   {\ensuremath{{0.000 } } }
\vdef{default-11:NoData-APV0:muon2pt:loEff}   {\ensuremath{{0.118 } } }
\vdef{default-11:NoData-APV0:muon2pt:loEffE}   {\ensuremath{{0.002 } } }
\vdef{default-11:NoData-APV0:muon2pt:hiEff}   {\ensuremath{{0.882 } } }
\vdef{default-11:NoData-APV0:muon2pt:hiEffE}   {\ensuremath{{0.002 } } }
\vdef{default-11:NoMcPU-APV0:muon2pt:loEff}   {\ensuremath{{0.003 } } }
\vdef{default-11:NoMcPU-APV0:muon2pt:loEffE}   {\ensuremath{{0.000 } } }
\vdef{default-11:NoMcPU-APV0:muon2pt:hiEff}   {\ensuremath{{0.997 } } }
\vdef{default-11:NoMcPU-APV0:muon2pt:hiEffE}   {\ensuremath{{0.000 } } }
\vdef{default-11:NoMcPU-APV0:muon2pt:loDelta}   {\ensuremath{{+1.891 } } }
\vdef{default-11:NoMcPU-APV0:muon2pt:loDeltaE}   {\ensuremath{{0.010 } } }
\vdef{default-11:NoMcPU-APV0:muon2pt:hiDelta}   {\ensuremath{{-0.122 } } }
\vdef{default-11:NoMcPU-APV0:muon2pt:hiDeltaE}   {\ensuremath{{0.002 } } }
\vdef{default-11:NoData-APV0:muonseta:loEff}   {\ensuremath{{0.730 } } }
\vdef{default-11:NoData-APV0:muonseta:loEffE}   {\ensuremath{{0.002 } } }
\vdef{default-11:NoData-APV0:muonseta:hiEff}   {\ensuremath{{0.270 } } }
\vdef{default-11:NoData-APV0:muonseta:hiEffE}   {\ensuremath{{0.002 } } }
\vdef{default-11:NoMcPU-APV0:muonseta:loEff}   {\ensuremath{{0.848 } } }
\vdef{default-11:NoMcPU-APV0:muonseta:loEffE}   {\ensuremath{{0.002 } } }
\vdef{default-11:NoMcPU-APV0:muonseta:hiEff}   {\ensuremath{{0.152 } } }
\vdef{default-11:NoMcPU-APV0:muonseta:hiEffE}   {\ensuremath{{0.002 } } }
\vdef{default-11:NoMcPU-APV0:muonseta:loDelta}   {\ensuremath{{-0.150 } } }
\vdef{default-11:NoMcPU-APV0:muonseta:loDeltaE}   {\ensuremath{{0.003 } } }
\vdef{default-11:NoMcPU-APV0:muonseta:hiDelta}   {\ensuremath{{+0.562 } } }
\vdef{default-11:NoMcPU-APV0:muonseta:hiDeltaE}   {\ensuremath{{0.011 } } }
\vdef{default-11:NoData-APV0:pt:loEff}   {\ensuremath{{0.000 } } }
\vdef{default-11:NoData-APV0:pt:loEffE}   {\ensuremath{{0.000 } } }
\vdef{default-11:NoData-APV0:pt:hiEff}   {\ensuremath{{1.000 } } }
\vdef{default-11:NoData-APV0:pt:hiEffE}   {\ensuremath{{0.000 } } }
\vdef{default-11:NoMcPU-APV0:pt:loEff}   {\ensuremath{{0.000 } } }
\vdef{default-11:NoMcPU-APV0:pt:loEffE}   {\ensuremath{{0.000 } } }
\vdef{default-11:NoMcPU-APV0:pt:hiEff}   {\ensuremath{{1.000 } } }
\vdef{default-11:NoMcPU-APV0:pt:hiEffE}   {\ensuremath{{0.000 } } }
\vdef{default-11:NoMcPU-APV0:pt:loDelta}   {\ensuremath{{\mathrm{NaN} } } }
\vdef{default-11:NoMcPU-APV0:pt:loDeltaE}   {\ensuremath{{\mathrm{NaN} } } }
\vdef{default-11:NoMcPU-APV0:pt:hiDelta}   {\ensuremath{{+0.000 } } }
\vdef{default-11:NoMcPU-APV0:pt:hiDeltaE}   {\ensuremath{{0.000 } } }
\vdef{default-11:NoData-APV0:p:loEff}   {\ensuremath{{1.012 } } }
\vdef{default-11:NoData-APV0:p:loEffE}   {\ensuremath{{\mathrm{NaN} } } }
\vdef{default-11:NoData-APV0:p:hiEff}   {\ensuremath{{1.000 } } }
\vdef{default-11:NoData-APV0:p:hiEffE}   {\ensuremath{{0.000 } } }
\vdef{default-11:NoMcPU-APV0:p:loEff}   {\ensuremath{{1.003 } } }
\vdef{default-11:NoMcPU-APV0:p:loEffE}   {\ensuremath{{\mathrm{NaN} } } }
\vdef{default-11:NoMcPU-APV0:p:hiEff}   {\ensuremath{{1.000 } } }
\vdef{default-11:NoMcPU-APV0:p:hiEffE}   {\ensuremath{{0.000 } } }
\vdef{default-11:NoMcPU-APV0:p:loDelta}   {\ensuremath{{+0.009 } } }
\vdef{default-11:NoMcPU-APV0:p:loDeltaE}   {\ensuremath{{\mathrm{NaN} } } }
\vdef{default-11:NoMcPU-APV0:p:hiDelta}   {\ensuremath{{+0.000 } } }
\vdef{default-11:NoMcPU-APV0:p:hiDeltaE}   {\ensuremath{{0.000 } } }
\vdef{default-11:NoData-APV0:eta:loEff}   {\ensuremath{{0.722 } } }
\vdef{default-11:NoData-APV0:eta:loEffE}   {\ensuremath{{0.003 } } }
\vdef{default-11:NoData-APV0:eta:hiEff}   {\ensuremath{{0.278 } } }
\vdef{default-11:NoData-APV0:eta:hiEffE}   {\ensuremath{{0.003 } } }
\vdef{default-11:NoMcPU-APV0:eta:loEff}   {\ensuremath{{0.846 } } }
\vdef{default-11:NoMcPU-APV0:eta:loEffE}   {\ensuremath{{0.002 } } }
\vdef{default-11:NoMcPU-APV0:eta:hiEff}   {\ensuremath{{0.154 } } }
\vdef{default-11:NoMcPU-APV0:eta:hiEffE}   {\ensuremath{{0.002 } } }
\vdef{default-11:NoMcPU-APV0:eta:loDelta}   {\ensuremath{{-0.158 } } }
\vdef{default-11:NoMcPU-APV0:eta:loDeltaE}   {\ensuremath{{0.004 } } }
\vdef{default-11:NoMcPU-APV0:eta:hiDelta}   {\ensuremath{{+0.572 } } }
\vdef{default-11:NoMcPU-APV0:eta:hiDeltaE}   {\ensuremath{{0.016 } } }
\vdef{default-11:NoData-APV0:bdt:loEff}   {\ensuremath{{0.914 } } }
\vdef{default-11:NoData-APV0:bdt:loEffE}   {\ensuremath{{0.002 } } }
\vdef{default-11:NoData-APV0:bdt:hiEff}   {\ensuremath{{0.086 } } }
\vdef{default-11:NoData-APV0:bdt:hiEffE}   {\ensuremath{{0.002 } } }
\vdef{default-11:NoMcPU-APV0:bdt:loEff}   {\ensuremath{{0.863 } } }
\vdef{default-11:NoMcPU-APV0:bdt:loEffE}   {\ensuremath{{0.002 } } }
\vdef{default-11:NoMcPU-APV0:bdt:hiEff}   {\ensuremath{{0.137 } } }
\vdef{default-11:NoMcPU-APV0:bdt:hiEffE}   {\ensuremath{{0.002 } } }
\vdef{default-11:NoMcPU-APV0:bdt:loDelta}   {\ensuremath{{+0.056 } } }
\vdef{default-11:NoMcPU-APV0:bdt:loDeltaE}   {\ensuremath{{0.003 } } }
\vdef{default-11:NoMcPU-APV0:bdt:hiDelta}   {\ensuremath{{-0.450 } } }
\vdef{default-11:NoMcPU-APV0:bdt:hiDeltaE}   {\ensuremath{{0.023 } } }
\vdef{default-11:NoData-APV0:fl3d:loEff}   {\ensuremath{{0.826 } } }
\vdef{default-11:NoData-APV0:fl3d:loEffE}   {\ensuremath{{0.002 } } }
\vdef{default-11:NoData-APV0:fl3d:hiEff}   {\ensuremath{{0.174 } } }
\vdef{default-11:NoData-APV0:fl3d:hiEffE}   {\ensuremath{{0.002 } } }
\vdef{default-11:NoMcPU-APV0:fl3d:loEff}   {\ensuremath{{0.891 } } }
\vdef{default-11:NoMcPU-APV0:fl3d:loEffE}   {\ensuremath{{0.002 } } }
\vdef{default-11:NoMcPU-APV0:fl3d:hiEff}   {\ensuremath{{0.109 } } }
\vdef{default-11:NoMcPU-APV0:fl3d:hiEffE}   {\ensuremath{{0.002 } } }
\vdef{default-11:NoMcPU-APV0:fl3d:loDelta}   {\ensuremath{{-0.075 } } }
\vdef{default-11:NoMcPU-APV0:fl3d:loDeltaE}   {\ensuremath{{0.003 } } }
\vdef{default-11:NoMcPU-APV0:fl3d:hiDelta}   {\ensuremath{{+0.455 } } }
\vdef{default-11:NoMcPU-APV0:fl3d:hiDeltaE}   {\ensuremath{{0.019 } } }
\vdef{default-11:NoData-APV0:fl3de:loEff}   {\ensuremath{{1.000 } } }
\vdef{default-11:NoData-APV0:fl3de:loEffE}   {\ensuremath{{0.000 } } }
\vdef{default-11:NoData-APV0:fl3de:hiEff}   {\ensuremath{{0.000 } } }
\vdef{default-11:NoData-APV0:fl3de:hiEffE}   {\ensuremath{{0.000 } } }
\vdef{default-11:NoMcPU-APV0:fl3de:loEff}   {\ensuremath{{1.000 } } }
\vdef{default-11:NoMcPU-APV0:fl3de:loEffE}   {\ensuremath{{0.000 } } }
\vdef{default-11:NoMcPU-APV0:fl3de:hiEff}   {\ensuremath{{0.000 } } }
\vdef{default-11:NoMcPU-APV0:fl3de:hiEffE}   {\ensuremath{{0.000 } } }
\vdef{default-11:NoMcPU-APV0:fl3de:loDelta}   {\ensuremath{{+0.000 } } }
\vdef{default-11:NoMcPU-APV0:fl3de:loDeltaE}   {\ensuremath{{0.000 } } }
\vdef{default-11:NoMcPU-APV0:fl3de:hiDelta}   {\ensuremath{{+2.000 } } }
\vdef{default-11:NoMcPU-APV0:fl3de:hiDeltaE}   {\ensuremath{{1.527 } } }
\vdef{default-11:NoData-APV0:fls3d:loEff}   {\ensuremath{{0.077 } } }
\vdef{default-11:NoData-APV0:fls3d:loEffE}   {\ensuremath{{0.001 } } }
\vdef{default-11:NoData-APV0:fls3d:hiEff}   {\ensuremath{{0.923 } } }
\vdef{default-11:NoData-APV0:fls3d:hiEffE}   {\ensuremath{{0.001 } } }
\vdef{default-11:NoMcPU-APV0:fls3d:loEff}   {\ensuremath{{0.063 } } }
\vdef{default-11:NoMcPU-APV0:fls3d:loEffE}   {\ensuremath{{0.001 } } }
\vdef{default-11:NoMcPU-APV0:fls3d:hiEff}   {\ensuremath{{0.937 } } }
\vdef{default-11:NoMcPU-APV0:fls3d:hiEffE}   {\ensuremath{{0.001 } } }
\vdef{default-11:NoMcPU-APV0:fls3d:loDelta}   {\ensuremath{{+0.193 } } }
\vdef{default-11:NoMcPU-APV0:fls3d:loDeltaE}   {\ensuremath{{0.028 } } }
\vdef{default-11:NoMcPU-APV0:fls3d:hiDelta}   {\ensuremath{{-0.015 } } }
\vdef{default-11:NoMcPU-APV0:fls3d:hiDeltaE}   {\ensuremath{{0.002 } } }
\vdef{default-11:NoData-APV0:flsxy:loEff}   {\ensuremath{{1.012 } } }
\vdef{default-11:NoData-APV0:flsxy:loEffE}   {\ensuremath{{\mathrm{NaN} } } }
\vdef{default-11:NoData-APV0:flsxy:hiEff}   {\ensuremath{{1.000 } } }
\vdef{default-11:NoData-APV0:flsxy:hiEffE}   {\ensuremath{{0.000 } } }
\vdef{default-11:NoMcPU-APV0:flsxy:loEff}   {\ensuremath{{1.011 } } }
\vdef{default-11:NoMcPU-APV0:flsxy:loEffE}   {\ensuremath{{\mathrm{NaN} } } }
\vdef{default-11:NoMcPU-APV0:flsxy:hiEff}   {\ensuremath{{1.000 } } }
\vdef{default-11:NoMcPU-APV0:flsxy:hiEffE}   {\ensuremath{{0.000 } } }
\vdef{default-11:NoMcPU-APV0:flsxy:loDelta}   {\ensuremath{{+0.001 } } }
\vdef{default-11:NoMcPU-APV0:flsxy:loDeltaE}   {\ensuremath{{\mathrm{NaN} } } }
\vdef{default-11:NoMcPU-APV0:flsxy:hiDelta}   {\ensuremath{{+0.000 } } }
\vdef{default-11:NoMcPU-APV0:flsxy:hiDeltaE}   {\ensuremath{{0.000 } } }
\vdef{default-11:NoData-APV0:chi2dof:loEff}   {\ensuremath{{0.925 } } }
\vdef{default-11:NoData-APV0:chi2dof:loEffE}   {\ensuremath{{0.002 } } }
\vdef{default-11:NoData-APV0:chi2dof:hiEff}   {\ensuremath{{0.075 } } }
\vdef{default-11:NoData-APV0:chi2dof:hiEffE}   {\ensuremath{{0.002 } } }
\vdef{default-11:NoMcPU-APV0:chi2dof:loEff}   {\ensuremath{{0.948 } } }
\vdef{default-11:NoMcPU-APV0:chi2dof:loEffE}   {\ensuremath{{0.001 } } }
\vdef{default-11:NoMcPU-APV0:chi2dof:hiEff}   {\ensuremath{{0.052 } } }
\vdef{default-11:NoMcPU-APV0:chi2dof:hiEffE}   {\ensuremath{{0.001 } } }
\vdef{default-11:NoMcPU-APV0:chi2dof:loDelta}   {\ensuremath{{-0.024 } } }
\vdef{default-11:NoMcPU-APV0:chi2dof:loDeltaE}   {\ensuremath{{0.002 } } }
\vdef{default-11:NoMcPU-APV0:chi2dof:hiDelta}   {\ensuremath{{+0.357 } } }
\vdef{default-11:NoMcPU-APV0:chi2dof:hiDeltaE}   {\ensuremath{{0.031 } } }
\vdef{default-11:NoData-APV0:pchi2dof:loEff}   {\ensuremath{{0.647 } } }
\vdef{default-11:NoData-APV0:pchi2dof:loEffE}   {\ensuremath{{0.003 } } }
\vdef{default-11:NoData-APV0:pchi2dof:hiEff}   {\ensuremath{{0.353 } } }
\vdef{default-11:NoData-APV0:pchi2dof:hiEffE}   {\ensuremath{{0.003 } } }
\vdef{default-11:NoMcPU-APV0:pchi2dof:loEff}   {\ensuremath{{0.597 } } }
\vdef{default-11:NoMcPU-APV0:pchi2dof:loEffE}   {\ensuremath{{0.003 } } }
\vdef{default-11:NoMcPU-APV0:pchi2dof:hiEff}   {\ensuremath{{0.403 } } }
\vdef{default-11:NoMcPU-APV0:pchi2dof:hiEffE}   {\ensuremath{{0.003 } } }
\vdef{default-11:NoMcPU-APV0:pchi2dof:loDelta}   {\ensuremath{{+0.079 } } }
\vdef{default-11:NoMcPU-APV0:pchi2dof:loDeltaE}   {\ensuremath{{0.006 } } }
\vdef{default-11:NoMcPU-APV0:pchi2dof:hiDelta}   {\ensuremath{{-0.131 } } }
\vdef{default-11:NoMcPU-APV0:pchi2dof:hiDeltaE}   {\ensuremath{{0.010 } } }
\vdef{default-11:NoData-APV0:alpha:loEff}   {\ensuremath{{0.994 } } }
\vdef{default-11:NoData-APV0:alpha:loEffE}   {\ensuremath{{0.000 } } }
\vdef{default-11:NoData-APV0:alpha:hiEff}   {\ensuremath{{0.006 } } }
\vdef{default-11:NoData-APV0:alpha:hiEffE}   {\ensuremath{{0.000 } } }
\vdef{default-11:NoMcPU-APV0:alpha:loEff}   {\ensuremath{{0.993 } } }
\vdef{default-11:NoMcPU-APV0:alpha:loEffE}   {\ensuremath{{0.001 } } }
\vdef{default-11:NoMcPU-APV0:alpha:hiEff}   {\ensuremath{{0.007 } } }
\vdef{default-11:NoMcPU-APV0:alpha:hiEffE}   {\ensuremath{{0.001 } } }
\vdef{default-11:NoMcPU-APV0:alpha:loDelta}   {\ensuremath{{+0.002 } } }
\vdef{default-11:NoMcPU-APV0:alpha:loDeltaE}   {\ensuremath{{0.001 } } }
\vdef{default-11:NoMcPU-APV0:alpha:hiDelta}   {\ensuremath{{-0.265 } } }
\vdef{default-11:NoMcPU-APV0:alpha:hiDeltaE}   {\ensuremath{{0.103 } } }
\vdef{default-11:NoData-APV0:iso:loEff}   {\ensuremath{{0.127 } } }
\vdef{default-11:NoData-APV0:iso:loEffE}   {\ensuremath{{0.002 } } }
\vdef{default-11:NoData-APV0:iso:hiEff}   {\ensuremath{{0.873 } } }
\vdef{default-11:NoData-APV0:iso:hiEffE}   {\ensuremath{{0.002 } } }
\vdef{default-11:NoMcPU-APV0:iso:loEff}   {\ensuremath{{0.110 } } }
\vdef{default-11:NoMcPU-APV0:iso:loEffE}   {\ensuremath{{0.002 } } }
\vdef{default-11:NoMcPU-APV0:iso:hiEff}   {\ensuremath{{0.890 } } }
\vdef{default-11:NoMcPU-APV0:iso:hiEffE}   {\ensuremath{{0.002 } } }
\vdef{default-11:NoMcPU-APV0:iso:loDelta}   {\ensuremath{{+0.149 } } }
\vdef{default-11:NoMcPU-APV0:iso:loDeltaE}   {\ensuremath{{0.021 } } }
\vdef{default-11:NoMcPU-APV0:iso:hiDelta}   {\ensuremath{{-0.020 } } }
\vdef{default-11:NoMcPU-APV0:iso:hiDeltaE}   {\ensuremath{{0.003 } } }
\vdef{default-11:NoData-APV0:docatrk:loEff}   {\ensuremath{{0.066 } } }
\vdef{default-11:NoData-APV0:docatrk:loEffE}   {\ensuremath{{0.001 } } }
\vdef{default-11:NoData-APV0:docatrk:hiEff}   {\ensuremath{{0.934 } } }
\vdef{default-11:NoData-APV0:docatrk:hiEffE}   {\ensuremath{{0.001 } } }
\vdef{default-11:NoMcPU-APV0:docatrk:loEff}   {\ensuremath{{0.080 } } }
\vdef{default-11:NoMcPU-APV0:docatrk:loEffE}   {\ensuremath{{0.002 } } }
\vdef{default-11:NoMcPU-APV0:docatrk:hiEff}   {\ensuremath{{0.920 } } }
\vdef{default-11:NoMcPU-APV0:docatrk:hiEffE}   {\ensuremath{{0.002 } } }
\vdef{default-11:NoMcPU-APV0:docatrk:loDelta}   {\ensuremath{{-0.180 } } }
\vdef{default-11:NoMcPU-APV0:docatrk:loDeltaE}   {\ensuremath{{0.029 } } }
\vdef{default-11:NoMcPU-APV0:docatrk:hiDelta}   {\ensuremath{{+0.014 } } }
\vdef{default-11:NoMcPU-APV0:docatrk:hiDeltaE}   {\ensuremath{{0.002 } } }
\vdef{default-11:NoData-APV0:isotrk:loEff}   {\ensuremath{{1.000 } } }
\vdef{default-11:NoData-APV0:isotrk:loEffE}   {\ensuremath{{0.000 } } }
\vdef{default-11:NoData-APV0:isotrk:hiEff}   {\ensuremath{{1.000 } } }
\vdef{default-11:NoData-APV0:isotrk:hiEffE}   {\ensuremath{{0.000 } } }
\vdef{default-11:NoMcPU-APV0:isotrk:loEff}   {\ensuremath{{1.000 } } }
\vdef{default-11:NoMcPU-APV0:isotrk:loEffE}   {\ensuremath{{0.000 } } }
\vdef{default-11:NoMcPU-APV0:isotrk:hiEff}   {\ensuremath{{1.000 } } }
\vdef{default-11:NoMcPU-APV0:isotrk:hiEffE}   {\ensuremath{{0.000 } } }
\vdef{default-11:NoMcPU-APV0:isotrk:loDelta}   {\ensuremath{{+0.000 } } }
\vdef{default-11:NoMcPU-APV0:isotrk:loDeltaE}   {\ensuremath{{0.000 } } }
\vdef{default-11:NoMcPU-APV0:isotrk:hiDelta}   {\ensuremath{{+0.000 } } }
\vdef{default-11:NoMcPU-APV0:isotrk:hiDeltaE}   {\ensuremath{{0.000 } } }
\vdef{default-11:NoData-APV0:closetrk:loEff}   {\ensuremath{{0.978 } } }
\vdef{default-11:NoData-APV0:closetrk:loEffE}   {\ensuremath{{0.001 } } }
\vdef{default-11:NoData-APV0:closetrk:hiEff}   {\ensuremath{{0.022 } } }
\vdef{default-11:NoData-APV0:closetrk:hiEffE}   {\ensuremath{{0.001 } } }
\vdef{default-11:NoMcPU-APV0:closetrk:loEff}   {\ensuremath{{0.975 } } }
\vdef{default-11:NoMcPU-APV0:closetrk:loEffE}   {\ensuremath{{0.001 } } }
\vdef{default-11:NoMcPU-APV0:closetrk:hiEff}   {\ensuremath{{0.025 } } }
\vdef{default-11:NoMcPU-APV0:closetrk:hiEffE}   {\ensuremath{{0.001 } } }
\vdef{default-11:NoMcPU-APV0:closetrk:loDelta}   {\ensuremath{{+0.003 } } }
\vdef{default-11:NoMcPU-APV0:closetrk:loDeltaE}   {\ensuremath{{0.001 } } }
\vdef{default-11:NoMcPU-APV0:closetrk:hiDelta}   {\ensuremath{{-0.133 } } }
\vdef{default-11:NoMcPU-APV0:closetrk:hiDeltaE}   {\ensuremath{{0.054 } } }
\vdef{default-11:NoData-APV0:lip:loEff}   {\ensuremath{{1.000 } } }
\vdef{default-11:NoData-APV0:lip:loEffE}   {\ensuremath{{0.000 } } }
\vdef{default-11:NoData-APV0:lip:hiEff}   {\ensuremath{{0.000 } } }
\vdef{default-11:NoData-APV0:lip:hiEffE}   {\ensuremath{{0.000 } } }
\vdef{default-11:NoMcPU-APV0:lip:loEff}   {\ensuremath{{1.000 } } }
\vdef{default-11:NoMcPU-APV0:lip:loEffE}   {\ensuremath{{0.000 } } }
\vdef{default-11:NoMcPU-APV0:lip:hiEff}   {\ensuremath{{0.000 } } }
\vdef{default-11:NoMcPU-APV0:lip:hiEffE}   {\ensuremath{{0.000 } } }
\vdef{default-11:NoMcPU-APV0:lip:loDelta}   {\ensuremath{{+0.000 } } }
\vdef{default-11:NoMcPU-APV0:lip:loDeltaE}   {\ensuremath{{0.000 } } }
\vdef{default-11:NoMcPU-APV0:lip:hiDelta}   {\ensuremath{{\mathrm{NaN} } } }
\vdef{default-11:NoMcPU-APV0:lip:hiDeltaE}   {\ensuremath{{\mathrm{NaN} } } }
\vdef{default-11:NoData-APV0:lip:inEff}   {\ensuremath{{1.000 } } }
\vdef{default-11:NoData-APV0:lip:inEffE}   {\ensuremath{{0.000 } } }
\vdef{default-11:NoMcPU-APV0:lip:inEff}   {\ensuremath{{1.000 } } }
\vdef{default-11:NoMcPU-APV0:lip:inEffE}   {\ensuremath{{0.000 } } }
\vdef{default-11:NoMcPU-APV0:lip:inDelta}   {\ensuremath{{+0.000 } } }
\vdef{default-11:NoMcPU-APV0:lip:inDeltaE}   {\ensuremath{{0.000 } } }
\vdef{default-11:NoData-APV0:lips:loEff}   {\ensuremath{{1.000 } } }
\vdef{default-11:NoData-APV0:lips:loEffE}   {\ensuremath{{0.000 } } }
\vdef{default-11:NoData-APV0:lips:hiEff}   {\ensuremath{{0.000 } } }
\vdef{default-11:NoData-APV0:lips:hiEffE}   {\ensuremath{{0.000 } } }
\vdef{default-11:NoMcPU-APV0:lips:loEff}   {\ensuremath{{1.000 } } }
\vdef{default-11:NoMcPU-APV0:lips:loEffE}   {\ensuremath{{0.000 } } }
\vdef{default-11:NoMcPU-APV0:lips:hiEff}   {\ensuremath{{0.000 } } }
\vdef{default-11:NoMcPU-APV0:lips:hiEffE}   {\ensuremath{{0.000 } } }
\vdef{default-11:NoMcPU-APV0:lips:loDelta}   {\ensuremath{{+0.000 } } }
\vdef{default-11:NoMcPU-APV0:lips:loDeltaE}   {\ensuremath{{0.000 } } }
\vdef{default-11:NoMcPU-APV0:lips:hiDelta}   {\ensuremath{{\mathrm{NaN} } } }
\vdef{default-11:NoMcPU-APV0:lips:hiDeltaE}   {\ensuremath{{\mathrm{NaN} } } }
\vdef{default-11:NoData-APV0:lips:inEff}   {\ensuremath{{1.000 } } }
\vdef{default-11:NoData-APV0:lips:inEffE}   {\ensuremath{{0.000 } } }
\vdef{default-11:NoMcPU-APV0:lips:inEff}   {\ensuremath{{1.000 } } }
\vdef{default-11:NoMcPU-APV0:lips:inEffE}   {\ensuremath{{0.000 } } }
\vdef{default-11:NoMcPU-APV0:lips:inDelta}   {\ensuremath{{+0.000 } } }
\vdef{default-11:NoMcPU-APV0:lips:inDeltaE}   {\ensuremath{{0.000 } } }
\vdef{default-11:NoData-APV0:ip:loEff}   {\ensuremath{{0.972 } } }
\vdef{default-11:NoData-APV0:ip:loEffE}   {\ensuremath{{0.001 } } }
\vdef{default-11:NoData-APV0:ip:hiEff}   {\ensuremath{{0.028 } } }
\vdef{default-11:NoData-APV0:ip:hiEffE}   {\ensuremath{{0.001 } } }
\vdef{default-11:NoMcPU-APV0:ip:loEff}   {\ensuremath{{0.969 } } }
\vdef{default-11:NoMcPU-APV0:ip:loEffE}   {\ensuremath{{0.001 } } }
\vdef{default-11:NoMcPU-APV0:ip:hiEff}   {\ensuremath{{0.031 } } }
\vdef{default-11:NoMcPU-APV0:ip:hiEffE}   {\ensuremath{{0.001 } } }
\vdef{default-11:NoMcPU-APV0:ip:loDelta}   {\ensuremath{{+0.003 } } }
\vdef{default-11:NoMcPU-APV0:ip:loDeltaE}   {\ensuremath{{0.001 } } }
\vdef{default-11:NoMcPU-APV0:ip:hiDelta}   {\ensuremath{{-0.103 } } }
\vdef{default-11:NoMcPU-APV0:ip:hiDeltaE}   {\ensuremath{{0.048 } } }
\vdef{default-11:NoData-APV0:ips:loEff}   {\ensuremath{{0.947 } } }
\vdef{default-11:NoData-APV0:ips:loEffE}   {\ensuremath{{0.001 } } }
\vdef{default-11:NoData-APV0:ips:hiEff}   {\ensuremath{{0.053 } } }
\vdef{default-11:NoData-APV0:ips:hiEffE}   {\ensuremath{{0.001 } } }
\vdef{default-11:NoMcPU-APV0:ips:loEff}   {\ensuremath{{0.957 } } }
\vdef{default-11:NoMcPU-APV0:ips:loEffE}   {\ensuremath{{0.001 } } }
\vdef{default-11:NoMcPU-APV0:ips:hiEff}   {\ensuremath{{0.043 } } }
\vdef{default-11:NoMcPU-APV0:ips:hiEffE}   {\ensuremath{{0.001 } } }
\vdef{default-11:NoMcPU-APV0:ips:loDelta}   {\ensuremath{{-0.011 } } }
\vdef{default-11:NoMcPU-APV0:ips:loDeltaE}   {\ensuremath{{0.002 } } }
\vdef{default-11:NoMcPU-APV0:ips:hiDelta}   {\ensuremath{{+0.212 } } }
\vdef{default-11:NoMcPU-APV0:ips:hiDeltaE}   {\ensuremath{{0.036 } } }
\vdef{default-11:NoData-APV0:maxdoca:loEff}   {\ensuremath{{1.000 } } }
\vdef{default-11:NoData-APV0:maxdoca:loEffE}   {\ensuremath{{0.000 } } }
\vdef{default-11:NoData-APV0:maxdoca:hiEff}   {\ensuremath{{0.014 } } }
\vdef{default-11:NoData-APV0:maxdoca:hiEffE}   {\ensuremath{{0.001 } } }
\vdef{default-11:NoMcPU-APV0:maxdoca:loEff}   {\ensuremath{{1.000 } } }
\vdef{default-11:NoMcPU-APV0:maxdoca:loEffE}   {\ensuremath{{0.000 } } }
\vdef{default-11:NoMcPU-APV0:maxdoca:hiEff}   {\ensuremath{{0.014 } } }
\vdef{default-11:NoMcPU-APV0:maxdoca:hiEffE}   {\ensuremath{{0.001 } } }
\vdef{default-11:NoMcPU-APV0:maxdoca:loDelta}   {\ensuremath{{+0.000 } } }
\vdef{default-11:NoMcPU-APV0:maxdoca:loDeltaE}   {\ensuremath{{0.000 } } }
\vdef{default-11:NoMcPU-APV0:maxdoca:hiDelta}   {\ensuremath{{-0.001 } } }
\vdef{default-11:NoMcPU-APV0:maxdoca:hiDeltaE}   {\ensuremath{{0.072 } } }
\vdef{default-11:NoData-APV0:kaonpt:loEff}   {\ensuremath{{1.010 } } }
\vdef{default-11:NoData-APV0:kaonpt:loEffE}   {\ensuremath{{\mathrm{NaN} } } }
\vdef{default-11:NoData-APV0:kaonpt:hiEff}   {\ensuremath{{1.000 } } }
\vdef{default-11:NoData-APV0:kaonpt:hiEffE}   {\ensuremath{{0.000 } } }
\vdef{default-11:NoMcPU-APV0:kaonpt:loEff}   {\ensuremath{{1.010 } } }
\vdef{default-11:NoMcPU-APV0:kaonpt:loEffE}   {\ensuremath{{\mathrm{NaN} } } }
\vdef{default-11:NoMcPU-APV0:kaonpt:hiEff}   {\ensuremath{{1.000 } } }
\vdef{default-11:NoMcPU-APV0:kaonpt:hiEffE}   {\ensuremath{{0.000 } } }
\vdef{default-11:NoMcPU-APV0:kaonpt:loDelta}   {\ensuremath{{-0.000 } } }
\vdef{default-11:NoMcPU-APV0:kaonpt:loDeltaE}   {\ensuremath{{\mathrm{NaN} } } }
\vdef{default-11:NoMcPU-APV0:kaonpt:hiDelta}   {\ensuremath{{+0.000 } } }
\vdef{default-11:NoMcPU-APV0:kaonpt:hiDeltaE}   {\ensuremath{{0.000 } } }
\vdef{default-11:NoData-APV0:psipt:loEff}   {\ensuremath{{1.004 } } }
\vdef{default-11:NoData-APV0:psipt:loEffE}   {\ensuremath{{\mathrm{NaN} } } }
\vdef{default-11:NoData-APV0:psipt:hiEff}   {\ensuremath{{1.000 } } }
\vdef{default-11:NoData-APV0:psipt:hiEffE}   {\ensuremath{{0.000 } } }
\vdef{default-11:NoMcPU-APV0:psipt:loEff}   {\ensuremath{{1.004 } } }
\vdef{default-11:NoMcPU-APV0:psipt:loEffE}   {\ensuremath{{\mathrm{NaN} } } }
\vdef{default-11:NoMcPU-APV0:psipt:hiEff}   {\ensuremath{{1.000 } } }
\vdef{default-11:NoMcPU-APV0:psipt:hiEffE}   {\ensuremath{{0.000 } } }
\vdef{default-11:NoMcPU-APV0:psipt:loDelta}   {\ensuremath{{+0.000 } } }
\vdef{default-11:NoMcPU-APV0:psipt:loDeltaE}   {\ensuremath{{\mathrm{NaN} } } }
\vdef{default-11:NoMcPU-APV0:psipt:hiDelta}   {\ensuremath{{+0.000 } } }
\vdef{default-11:NoMcPU-APV0:psipt:hiDeltaE}   {\ensuremath{{0.000 } } }
\vdef{default-11:NoData-APV1:osiso:loEff}   {\ensuremath{{1.005 } } }
\vdef{default-11:NoData-APV1:osiso:loEffE}   {\ensuremath{{\mathrm{NaN} } } }
\vdef{default-11:NoData-APV1:osiso:hiEff}   {\ensuremath{{1.000 } } }
\vdef{default-11:NoData-APV1:osiso:hiEffE}   {\ensuremath{{0.000 } } }
\vdef{default-11:NoMcPU-APV1:osiso:loEff}   {\ensuremath{{1.003 } } }
\vdef{default-11:NoMcPU-APV1:osiso:loEffE}   {\ensuremath{{\mathrm{NaN} } } }
\vdef{default-11:NoMcPU-APV1:osiso:hiEff}   {\ensuremath{{1.000 } } }
\vdef{default-11:NoMcPU-APV1:osiso:hiEffE}   {\ensuremath{{0.000 } } }
\vdef{default-11:NoMcPU-APV1:osiso:loDelta}   {\ensuremath{{+0.002 } } }
\vdef{default-11:NoMcPU-APV1:osiso:loDeltaE}   {\ensuremath{{\mathrm{NaN} } } }
\vdef{default-11:NoMcPU-APV1:osiso:hiDelta}   {\ensuremath{{+0.000 } } }
\vdef{default-11:NoMcPU-APV1:osiso:hiDeltaE}   {\ensuremath{{0.000 } } }
\vdef{default-11:NoData-APV1:osreliso:loEff}   {\ensuremath{{0.248 } } }
\vdef{default-11:NoData-APV1:osreliso:loEffE}   {\ensuremath{{0.003 } } }
\vdef{default-11:NoData-APV1:osreliso:hiEff}   {\ensuremath{{0.752 } } }
\vdef{default-11:NoData-APV1:osreliso:hiEffE}   {\ensuremath{{0.003 } } }
\vdef{default-11:NoMcPU-APV1:osreliso:loEff}   {\ensuremath{{0.287 } } }
\vdef{default-11:NoMcPU-APV1:osreliso:loEffE}   {\ensuremath{{0.003 } } }
\vdef{default-11:NoMcPU-APV1:osreliso:hiEff}   {\ensuremath{{0.713 } } }
\vdef{default-11:NoMcPU-APV1:osreliso:hiEffE}   {\ensuremath{{0.003 } } }
\vdef{default-11:NoMcPU-APV1:osreliso:loDelta}   {\ensuremath{{-0.147 } } }
\vdef{default-11:NoMcPU-APV1:osreliso:loDeltaE}   {\ensuremath{{0.015 } } }
\vdef{default-11:NoMcPU-APV1:osreliso:hiDelta}   {\ensuremath{{+0.054 } } }
\vdef{default-11:NoMcPU-APV1:osreliso:hiDeltaE}   {\ensuremath{{0.006 } } }
\vdef{default-11:NoData-APV1:osmuonpt:loEff}   {\ensuremath{{0.000 } } }
\vdef{default-11:NoData-APV1:osmuonpt:loEffE}   {\ensuremath{{0.001 } } }
\vdef{default-11:NoData-APV1:osmuonpt:hiEff}   {\ensuremath{{1.000 } } }
\vdef{default-11:NoData-APV1:osmuonpt:hiEffE}   {\ensuremath{{0.001 } } }
\vdef{default-11:NoMcPU-APV1:osmuonpt:loEff}   {\ensuremath{{0.000 } } }
\vdef{default-11:NoMcPU-APV1:osmuonpt:loEffE}   {\ensuremath{{0.001 } } }
\vdef{default-11:NoMcPU-APV1:osmuonpt:hiEff}   {\ensuremath{{1.000 } } }
\vdef{default-11:NoMcPU-APV1:osmuonpt:hiEffE}   {\ensuremath{{0.001 } } }
\vdef{default-11:NoMcPU-APV1:osmuonpt:loDelta}   {\ensuremath{{\mathrm{NaN} } } }
\vdef{default-11:NoMcPU-APV1:osmuonpt:loDeltaE}   {\ensuremath{{\mathrm{NaN} } } }
\vdef{default-11:NoMcPU-APV1:osmuonpt:hiDelta}   {\ensuremath{{+0.000 } } }
\vdef{default-11:NoMcPU-APV1:osmuonpt:hiDeltaE}   {\ensuremath{{0.002 } } }
\vdef{default-11:NoData-APV1:osmuondr:loEff}   {\ensuremath{{0.028 } } }
\vdef{default-11:NoData-APV1:osmuondr:loEffE}   {\ensuremath{{0.006 } } }
\vdef{default-11:NoData-APV1:osmuondr:hiEff}   {\ensuremath{{0.972 } } }
\vdef{default-11:NoData-APV1:osmuondr:hiEffE}   {\ensuremath{{0.006 } } }
\vdef{default-11:NoMcPU-APV1:osmuondr:loEff}   {\ensuremath{{0.015 } } }
\vdef{default-11:NoMcPU-APV1:osmuondr:loEffE}   {\ensuremath{{0.004 } } }
\vdef{default-11:NoMcPU-APV1:osmuondr:hiEff}   {\ensuremath{{0.985 } } }
\vdef{default-11:NoMcPU-APV1:osmuondr:hiEffE}   {\ensuremath{{0.004 } } }
\vdef{default-11:NoMcPU-APV1:osmuondr:loDelta}   {\ensuremath{{+0.636 } } }
\vdef{default-11:NoMcPU-APV1:osmuondr:loDeltaE}   {\ensuremath{{0.322 } } }
\vdef{default-11:NoMcPU-APV1:osmuondr:hiDelta}   {\ensuremath{{-0.014 } } }
\vdef{default-11:NoMcPU-APV1:osmuondr:hiDeltaE}   {\ensuremath{{0.007 } } }
\vdef{default-11:NoData-APV1:hlt:loEff}   {\ensuremath{{0.075 } } }
\vdef{default-11:NoData-APV1:hlt:loEffE}   {\ensuremath{{0.002 } } }
\vdef{default-11:NoData-APV1:hlt:hiEff}   {\ensuremath{{0.925 } } }
\vdef{default-11:NoData-APV1:hlt:hiEffE}   {\ensuremath{{0.002 } } }
\vdef{default-11:NoMcPU-APV1:hlt:loEff}   {\ensuremath{{0.327 } } }
\vdef{default-11:NoMcPU-APV1:hlt:loEffE}   {\ensuremath{{0.003 } } }
\vdef{default-11:NoMcPU-APV1:hlt:hiEff}   {\ensuremath{{0.673 } } }
\vdef{default-11:NoMcPU-APV1:hlt:hiEffE}   {\ensuremath{{0.003 } } }
\vdef{default-11:NoMcPU-APV1:hlt:loDelta}   {\ensuremath{{-1.251 } } }
\vdef{default-11:NoMcPU-APV1:hlt:loDeltaE}   {\ensuremath{{0.015 } } }
\vdef{default-11:NoMcPU-APV1:hlt:hiDelta}   {\ensuremath{{+0.315 } } }
\vdef{default-11:NoMcPU-APV1:hlt:hiDeltaE}   {\ensuremath{{0.005 } } }
\vdef{default-11:NoData-APV1:muonsid:loEff}   {\ensuremath{{0.154 } } }
\vdef{default-11:NoData-APV1:muonsid:loEffE}   {\ensuremath{{0.002 } } }
\vdef{default-11:NoData-APV1:muonsid:hiEff}   {\ensuremath{{0.846 } } }
\vdef{default-11:NoData-APV1:muonsid:hiEffE}   {\ensuremath{{0.002 } } }
\vdef{default-11:NoMcPU-APV1:muonsid:loEff}   {\ensuremath{{0.218 } } }
\vdef{default-11:NoMcPU-APV1:muonsid:loEffE}   {\ensuremath{{0.003 } } }
\vdef{default-11:NoMcPU-APV1:muonsid:hiEff}   {\ensuremath{{0.782 } } }
\vdef{default-11:NoMcPU-APV1:muonsid:hiEffE}   {\ensuremath{{0.003 } } }
\vdef{default-11:NoMcPU-APV1:muonsid:loDelta}   {\ensuremath{{-0.344 } } }
\vdef{default-11:NoMcPU-APV1:muonsid:loDeltaE}   {\ensuremath{{0.019 } } }
\vdef{default-11:NoMcPU-APV1:muonsid:hiDelta}   {\ensuremath{{+0.079 } } }
\vdef{default-11:NoMcPU-APV1:muonsid:hiDeltaE}   {\ensuremath{{0.004 } } }
\vdef{default-11:NoData-APV1:tracksqual:loEff}   {\ensuremath{{0.001 } } }
\vdef{default-11:NoData-APV1:tracksqual:loEffE}   {\ensuremath{{0.000 } } }
\vdef{default-11:NoData-APV1:tracksqual:hiEff}   {\ensuremath{{0.999 } } }
\vdef{default-11:NoData-APV1:tracksqual:hiEffE}   {\ensuremath{{0.000 } } }
\vdef{default-11:NoMcPU-APV1:tracksqual:loEff}   {\ensuremath{{0.000 } } }
\vdef{default-11:NoMcPU-APV1:tracksqual:loEffE}   {\ensuremath{{0.000 } } }
\vdef{default-11:NoMcPU-APV1:tracksqual:hiEff}   {\ensuremath{{1.000 } } }
\vdef{default-11:NoMcPU-APV1:tracksqual:hiEffE}   {\ensuremath{{0.000 } } }
\vdef{default-11:NoMcPU-APV1:tracksqual:loDelta}   {\ensuremath{{+0.271 } } }
\vdef{default-11:NoMcPU-APV1:tracksqual:loDeltaE}   {\ensuremath{{0.458 } } }
\vdef{default-11:NoMcPU-APV1:tracksqual:hiDelta}   {\ensuremath{{-0.000 } } }
\vdef{default-11:NoMcPU-APV1:tracksqual:hiDeltaE}   {\ensuremath{{0.000 } } }
\vdef{default-11:NoData-APV1:pvz:loEff}   {\ensuremath{{0.506 } } }
\vdef{default-11:NoData-APV1:pvz:loEffE}   {\ensuremath{{0.003 } } }
\vdef{default-11:NoData-APV1:pvz:hiEff}   {\ensuremath{{0.494 } } }
\vdef{default-11:NoData-APV1:pvz:hiEffE}   {\ensuremath{{0.003 } } }
\vdef{default-11:NoMcPU-APV1:pvz:loEff}   {\ensuremath{{0.477 } } }
\vdef{default-11:NoMcPU-APV1:pvz:loEffE}   {\ensuremath{{0.003 } } }
\vdef{default-11:NoMcPU-APV1:pvz:hiEff}   {\ensuremath{{0.523 } } }
\vdef{default-11:NoMcPU-APV1:pvz:hiEffE}   {\ensuremath{{0.003 } } }
\vdef{default-11:NoMcPU-APV1:pvz:loDelta}   {\ensuremath{{+0.058 } } }
\vdef{default-11:NoMcPU-APV1:pvz:loDeltaE}   {\ensuremath{{0.010 } } }
\vdef{default-11:NoMcPU-APV1:pvz:hiDelta}   {\ensuremath{{-0.056 } } }
\vdef{default-11:NoMcPU-APV1:pvz:hiDeltaE}   {\ensuremath{{0.009 } } }
\vdef{default-11:NoData-APV1:pvn:loEff}   {\ensuremath{{1.035 } } }
\vdef{default-11:NoData-APV1:pvn:loEffE}   {\ensuremath{{\mathrm{NaN} } } }
\vdef{default-11:NoData-APV1:pvn:hiEff}   {\ensuremath{{1.000 } } }
\vdef{default-11:NoData-APV1:pvn:hiEffE}   {\ensuremath{{0.000 } } }
\vdef{default-11:NoMcPU-APV1:pvn:loEff}   {\ensuremath{{1.185 } } }
\vdef{default-11:NoMcPU-APV1:pvn:loEffE}   {\ensuremath{{\mathrm{NaN} } } }
\vdef{default-11:NoMcPU-APV1:pvn:hiEff}   {\ensuremath{{1.000 } } }
\vdef{default-11:NoMcPU-APV1:pvn:hiEffE}   {\ensuremath{{0.000 } } }
\vdef{default-11:NoMcPU-APV1:pvn:loDelta}   {\ensuremath{{-0.136 } } }
\vdef{default-11:NoMcPU-APV1:pvn:loDeltaE}   {\ensuremath{{\mathrm{NaN} } } }
\vdef{default-11:NoMcPU-APV1:pvn:hiDelta}   {\ensuremath{{+0.000 } } }
\vdef{default-11:NoMcPU-APV1:pvn:hiDeltaE}   {\ensuremath{{0.000 } } }
\vdef{default-11:NoData-APV1:pvavew8:loEff}   {\ensuremath{{0.011 } } }
\vdef{default-11:NoData-APV1:pvavew8:loEffE}   {\ensuremath{{0.001 } } }
\vdef{default-11:NoData-APV1:pvavew8:hiEff}   {\ensuremath{{0.989 } } }
\vdef{default-11:NoData-APV1:pvavew8:hiEffE}   {\ensuremath{{0.001 } } }
\vdef{default-11:NoMcPU-APV1:pvavew8:loEff}   {\ensuremath{{0.008 } } }
\vdef{default-11:NoMcPU-APV1:pvavew8:loEffE}   {\ensuremath{{0.001 } } }
\vdef{default-11:NoMcPU-APV1:pvavew8:hiEff}   {\ensuremath{{0.992 } } }
\vdef{default-11:NoMcPU-APV1:pvavew8:hiEffE}   {\ensuremath{{0.001 } } }
\vdef{default-11:NoMcPU-APV1:pvavew8:loDelta}   {\ensuremath{{+0.218 } } }
\vdef{default-11:NoMcPU-APV1:pvavew8:loDeltaE}   {\ensuremath{{0.099 } } }
\vdef{default-11:NoMcPU-APV1:pvavew8:hiDelta}   {\ensuremath{{-0.002 } } }
\vdef{default-11:NoMcPU-APV1:pvavew8:hiDeltaE}   {\ensuremath{{0.001 } } }
\vdef{default-11:NoData-APV1:pvntrk:loEff}   {\ensuremath{{1.000 } } }
\vdef{default-11:NoData-APV1:pvntrk:loEffE}   {\ensuremath{{0.000 } } }
\vdef{default-11:NoData-APV1:pvntrk:hiEff}   {\ensuremath{{1.000 } } }
\vdef{default-11:NoData-APV1:pvntrk:hiEffE}   {\ensuremath{{0.000 } } }
\vdef{default-11:NoMcPU-APV1:pvntrk:loEff}   {\ensuremath{{1.000 } } }
\vdef{default-11:NoMcPU-APV1:pvntrk:loEffE}   {\ensuremath{{0.000 } } }
\vdef{default-11:NoMcPU-APV1:pvntrk:hiEff}   {\ensuremath{{1.000 } } }
\vdef{default-11:NoMcPU-APV1:pvntrk:hiEffE}   {\ensuremath{{0.000 } } }
\vdef{default-11:NoMcPU-APV1:pvntrk:loDelta}   {\ensuremath{{+0.000 } } }
\vdef{default-11:NoMcPU-APV1:pvntrk:loDeltaE}   {\ensuremath{{0.000 } } }
\vdef{default-11:NoMcPU-APV1:pvntrk:hiDelta}   {\ensuremath{{+0.000 } } }
\vdef{default-11:NoMcPU-APV1:pvntrk:hiDeltaE}   {\ensuremath{{0.000 } } }
\vdef{default-11:NoData-APV1:muon1pt:loEff}   {\ensuremath{{1.010 } } }
\vdef{default-11:NoData-APV1:muon1pt:loEffE}   {\ensuremath{{\mathrm{NaN} } } }
\vdef{default-11:NoData-APV1:muon1pt:hiEff}   {\ensuremath{{1.000 } } }
\vdef{default-11:NoData-APV1:muon1pt:hiEffE}   {\ensuremath{{0.000 } } }
\vdef{default-11:NoMcPU-APV1:muon1pt:loEff}   {\ensuremath{{1.009 } } }
\vdef{default-11:NoMcPU-APV1:muon1pt:loEffE}   {\ensuremath{{\mathrm{NaN} } } }
\vdef{default-11:NoMcPU-APV1:muon1pt:hiEff}   {\ensuremath{{1.000 } } }
\vdef{default-11:NoMcPU-APV1:muon1pt:hiEffE}   {\ensuremath{{0.000 } } }
\vdef{default-11:NoMcPU-APV1:muon1pt:loDelta}   {\ensuremath{{+0.001 } } }
\vdef{default-11:NoMcPU-APV1:muon1pt:loDeltaE}   {\ensuremath{{\mathrm{NaN} } } }
\vdef{default-11:NoMcPU-APV1:muon1pt:hiDelta}   {\ensuremath{{+0.000 } } }
\vdef{default-11:NoMcPU-APV1:muon1pt:hiDeltaE}   {\ensuremath{{0.000 } } }
\vdef{default-11:NoData-APV1:muon2pt:loEff}   {\ensuremath{{0.020 } } }
\vdef{default-11:NoData-APV1:muon2pt:loEffE}   {\ensuremath{{0.001 } } }
\vdef{default-11:NoData-APV1:muon2pt:hiEff}   {\ensuremath{{0.980 } } }
\vdef{default-11:NoData-APV1:muon2pt:hiEffE}   {\ensuremath{{0.001 } } }
\vdef{default-11:NoMcPU-APV1:muon2pt:loEff}   {\ensuremath{{0.003 } } }
\vdef{default-11:NoMcPU-APV1:muon2pt:loEffE}   {\ensuremath{{0.000 } } }
\vdef{default-11:NoMcPU-APV1:muon2pt:hiEff}   {\ensuremath{{0.997 } } }
\vdef{default-11:NoMcPU-APV1:muon2pt:hiEffE}   {\ensuremath{{0.000 } } }
\vdef{default-11:NoMcPU-APV1:muon2pt:loDelta}   {\ensuremath{{+1.428 } } }
\vdef{default-11:NoMcPU-APV1:muon2pt:loDeltaE}   {\ensuremath{{0.063 } } }
\vdef{default-11:NoMcPU-APV1:muon2pt:hiDelta}   {\ensuremath{{-0.016 } } }
\vdef{default-11:NoMcPU-APV1:muon2pt:hiDeltaE}   {\ensuremath{{0.001 } } }
\vdef{default-11:NoData-APV1:muonseta:loEff}   {\ensuremath{{0.751 } } }
\vdef{default-11:NoData-APV1:muonseta:loEffE}   {\ensuremath{{0.002 } } }
\vdef{default-11:NoData-APV1:muonseta:hiEff}   {\ensuremath{{0.249 } } }
\vdef{default-11:NoData-APV1:muonseta:hiEffE}   {\ensuremath{{0.002 } } }
\vdef{default-11:NoMcPU-APV1:muonseta:loEff}   {\ensuremath{{0.842 } } }
\vdef{default-11:NoMcPU-APV1:muonseta:loEffE}   {\ensuremath{{0.002 } } }
\vdef{default-11:NoMcPU-APV1:muonseta:hiEff}   {\ensuremath{{0.158 } } }
\vdef{default-11:NoMcPU-APV1:muonseta:hiEffE}   {\ensuremath{{0.002 } } }
\vdef{default-11:NoMcPU-APV1:muonseta:loDelta}   {\ensuremath{{-0.114 } } }
\vdef{default-11:NoMcPU-APV1:muonseta:loDeltaE}   {\ensuremath{{0.004 } } }
\vdef{default-11:NoMcPU-APV1:muonseta:hiDelta}   {\ensuremath{{+0.449 } } }
\vdef{default-11:NoMcPU-APV1:muonseta:hiDeltaE}   {\ensuremath{{0.013 } } }
\vdef{default-11:NoData-APV1:pt:loEff}   {\ensuremath{{0.000 } } }
\vdef{default-11:NoData-APV1:pt:loEffE}   {\ensuremath{{0.000 } } }
\vdef{default-11:NoData-APV1:pt:hiEff}   {\ensuremath{{1.000 } } }
\vdef{default-11:NoData-APV1:pt:hiEffE}   {\ensuremath{{0.000 } } }
\vdef{default-11:NoMcPU-APV1:pt:loEff}   {\ensuremath{{0.000 } } }
\vdef{default-11:NoMcPU-APV1:pt:loEffE}   {\ensuremath{{0.000 } } }
\vdef{default-11:NoMcPU-APV1:pt:hiEff}   {\ensuremath{{1.000 } } }
\vdef{default-11:NoMcPU-APV1:pt:hiEffE}   {\ensuremath{{0.000 } } }
\vdef{default-11:NoMcPU-APV1:pt:loDelta}   {\ensuremath{{\mathrm{NaN} } } }
\vdef{default-11:NoMcPU-APV1:pt:loDeltaE}   {\ensuremath{{\mathrm{NaN} } } }
\vdef{default-11:NoMcPU-APV1:pt:hiDelta}   {\ensuremath{{+0.000 } } }
\vdef{default-11:NoMcPU-APV1:pt:hiDeltaE}   {\ensuremath{{0.000 } } }
\vdef{default-11:NoData-APV1:p:loEff}   {\ensuremath{{1.011 } } }
\vdef{default-11:NoData-APV1:p:loEffE}   {\ensuremath{{\mathrm{NaN} } } }
\vdef{default-11:NoData-APV1:p:hiEff}   {\ensuremath{{1.000 } } }
\vdef{default-11:NoData-APV1:p:hiEffE}   {\ensuremath{{0.000 } } }
\vdef{default-11:NoMcPU-APV1:p:loEff}   {\ensuremath{{1.004 } } }
\vdef{default-11:NoMcPU-APV1:p:loEffE}   {\ensuremath{{\mathrm{NaN} } } }
\vdef{default-11:NoMcPU-APV1:p:hiEff}   {\ensuremath{{1.000 } } }
\vdef{default-11:NoMcPU-APV1:p:hiEffE}   {\ensuremath{{0.000 } } }
\vdef{default-11:NoMcPU-APV1:p:loDelta}   {\ensuremath{{+0.007 } } }
\vdef{default-11:NoMcPU-APV1:p:loDeltaE}   {\ensuremath{{\mathrm{NaN} } } }
\vdef{default-11:NoMcPU-APV1:p:hiDelta}   {\ensuremath{{+0.000 } } }
\vdef{default-11:NoMcPU-APV1:p:hiDeltaE}   {\ensuremath{{0.000 } } }
\vdef{default-11:NoData-APV1:eta:loEff}   {\ensuremath{{0.741 } } }
\vdef{default-11:NoData-APV1:eta:loEffE}   {\ensuremath{{0.003 } } }
\vdef{default-11:NoData-APV1:eta:hiEff}   {\ensuremath{{0.259 } } }
\vdef{default-11:NoData-APV1:eta:hiEffE}   {\ensuremath{{0.003 } } }
\vdef{default-11:NoMcPU-APV1:eta:loEff}   {\ensuremath{{0.843 } } }
\vdef{default-11:NoMcPU-APV1:eta:loEffE}   {\ensuremath{{0.003 } } }
\vdef{default-11:NoMcPU-APV1:eta:hiEff}   {\ensuremath{{0.157 } } }
\vdef{default-11:NoMcPU-APV1:eta:hiEffE}   {\ensuremath{{0.003 } } }
\vdef{default-11:NoMcPU-APV1:eta:loDelta}   {\ensuremath{{-0.129 } } }
\vdef{default-11:NoMcPU-APV1:eta:loDeltaE}   {\ensuremath{{0.005 } } }
\vdef{default-11:NoMcPU-APV1:eta:hiDelta}   {\ensuremath{{+0.492 } } }
\vdef{default-11:NoMcPU-APV1:eta:hiDeltaE}   {\ensuremath{{0.019 } } }
\vdef{default-11:NoData-APV1:bdt:loEff}   {\ensuremath{{0.907 } } }
\vdef{default-11:NoData-APV1:bdt:loEffE}   {\ensuremath{{0.002 } } }
\vdef{default-11:NoData-APV1:bdt:hiEff}   {\ensuremath{{0.093 } } }
\vdef{default-11:NoData-APV1:bdt:hiEffE}   {\ensuremath{{0.002 } } }
\vdef{default-11:NoMcPU-APV1:bdt:loEff}   {\ensuremath{{0.873 } } }
\vdef{default-11:NoMcPU-APV1:bdt:loEffE}   {\ensuremath{{0.002 } } }
\vdef{default-11:NoMcPU-APV1:bdt:hiEff}   {\ensuremath{{0.127 } } }
\vdef{default-11:NoMcPU-APV1:bdt:hiEffE}   {\ensuremath{{0.002 } } }
\vdef{default-11:NoMcPU-APV1:bdt:loDelta}   {\ensuremath{{+0.038 } } }
\vdef{default-11:NoMcPU-APV1:bdt:loDeltaE}   {\ensuremath{{0.003 } } }
\vdef{default-11:NoMcPU-APV1:bdt:hiDelta}   {\ensuremath{{-0.309 } } }
\vdef{default-11:NoMcPU-APV1:bdt:hiDeltaE}   {\ensuremath{{0.026 } } }
\vdef{default-11:NoData-APV1:fl3d:loEff}   {\ensuremath{{0.837 } } }
\vdef{default-11:NoData-APV1:fl3d:loEffE}   {\ensuremath{{0.002 } } }
\vdef{default-11:NoData-APV1:fl3d:hiEff}   {\ensuremath{{0.163 } } }
\vdef{default-11:NoData-APV1:fl3d:hiEffE}   {\ensuremath{{0.002 } } }
\vdef{default-11:NoMcPU-APV1:fl3d:loEff}   {\ensuremath{{0.884 } } }
\vdef{default-11:NoMcPU-APV1:fl3d:loEffE}   {\ensuremath{{0.002 } } }
\vdef{default-11:NoMcPU-APV1:fl3d:hiEff}   {\ensuremath{{0.116 } } }
\vdef{default-11:NoMcPU-APV1:fl3d:hiEffE}   {\ensuremath{{0.002 } } }
\vdef{default-11:NoMcPU-APV1:fl3d:loDelta}   {\ensuremath{{-0.055 } } }
\vdef{default-11:NoMcPU-APV1:fl3d:loDeltaE}   {\ensuremath{{0.004 } } }
\vdef{default-11:NoMcPU-APV1:fl3d:hiDelta}   {\ensuremath{{+0.337 } } }
\vdef{default-11:NoMcPU-APV1:fl3d:hiDeltaE}   {\ensuremath{{0.022 } } }
\vdef{default-11:NoData-APV1:fl3de:loEff}   {\ensuremath{{1.000 } } }
\vdef{default-11:NoData-APV1:fl3de:loEffE}   {\ensuremath{{0.000 } } }
\vdef{default-11:NoData-APV1:fl3de:hiEff}   {\ensuremath{{0.000 } } }
\vdef{default-11:NoData-APV1:fl3de:hiEffE}   {\ensuremath{{0.000 } } }
\vdef{default-11:NoMcPU-APV1:fl3de:loEff}   {\ensuremath{{1.000 } } }
\vdef{default-11:NoMcPU-APV1:fl3de:loEffE}   {\ensuremath{{0.000 } } }
\vdef{default-11:NoMcPU-APV1:fl3de:hiEff}   {\ensuremath{{0.000 } } }
\vdef{default-11:NoMcPU-APV1:fl3de:hiEffE}   {\ensuremath{{0.000 } } }
\vdef{default-11:NoMcPU-APV1:fl3de:loDelta}   {\ensuremath{{+0.000 } } }
\vdef{default-11:NoMcPU-APV1:fl3de:loDeltaE}   {\ensuremath{{0.000 } } }
\vdef{default-11:NoMcPU-APV1:fl3de:hiDelta}   {\ensuremath{{+0.181 } } }
\vdef{default-11:NoMcPU-APV1:fl3de:hiDeltaE}   {\ensuremath{{0.823 } } }
\vdef{default-11:NoData-APV1:fls3d:loEff}   {\ensuremath{{0.069 } } }
\vdef{default-11:NoData-APV1:fls3d:loEffE}   {\ensuremath{{0.002 } } }
\vdef{default-11:NoData-APV1:fls3d:hiEff}   {\ensuremath{{0.931 } } }
\vdef{default-11:NoData-APV1:fls3d:hiEffE}   {\ensuremath{{0.002 } } }
\vdef{default-11:NoMcPU-APV1:fls3d:loEff}   {\ensuremath{{0.063 } } }
\vdef{default-11:NoMcPU-APV1:fls3d:loEffE}   {\ensuremath{{0.002 } } }
\vdef{default-11:NoMcPU-APV1:fls3d:hiEff}   {\ensuremath{{0.937 } } }
\vdef{default-11:NoMcPU-APV1:fls3d:hiEffE}   {\ensuremath{{0.002 } } }
\vdef{default-11:NoMcPU-APV1:fls3d:loDelta}   {\ensuremath{{+0.094 } } }
\vdef{default-11:NoMcPU-APV1:fls3d:loDeltaE}   {\ensuremath{{0.034 } } }
\vdef{default-11:NoMcPU-APV1:fls3d:hiDelta}   {\ensuremath{{-0.007 } } }
\vdef{default-11:NoMcPU-APV1:fls3d:hiDeltaE}   {\ensuremath{{0.002 } } }
\vdef{default-11:NoData-APV1:flsxy:loEff}   {\ensuremath{{1.012 } } }
\vdef{default-11:NoData-APV1:flsxy:loEffE}   {\ensuremath{{\mathrm{NaN} } } }
\vdef{default-11:NoData-APV1:flsxy:hiEff}   {\ensuremath{{1.000 } } }
\vdef{default-11:NoData-APV1:flsxy:hiEffE}   {\ensuremath{{0.000 } } }
\vdef{default-11:NoMcPU-APV1:flsxy:loEff}   {\ensuremath{{1.015 } } }
\vdef{default-11:NoMcPU-APV1:flsxy:loEffE}   {\ensuremath{{\mathrm{NaN} } } }
\vdef{default-11:NoMcPU-APV1:flsxy:hiEff}   {\ensuremath{{1.000 } } }
\vdef{default-11:NoMcPU-APV1:flsxy:hiEffE}   {\ensuremath{{0.000 } } }
\vdef{default-11:NoMcPU-APV1:flsxy:loDelta}   {\ensuremath{{-0.003 } } }
\vdef{default-11:NoMcPU-APV1:flsxy:loDeltaE}   {\ensuremath{{\mathrm{NaN} } } }
\vdef{default-11:NoMcPU-APV1:flsxy:hiDelta}   {\ensuremath{{+0.000 } } }
\vdef{default-11:NoMcPU-APV1:flsxy:hiDeltaE}   {\ensuremath{{0.000 } } }
\vdef{default-11:NoData-APV1:chi2dof:loEff}   {\ensuremath{{0.929 } } }
\vdef{default-11:NoData-APV1:chi2dof:loEffE}   {\ensuremath{{0.002 } } }
\vdef{default-11:NoData-APV1:chi2dof:hiEff}   {\ensuremath{{0.071 } } }
\vdef{default-11:NoData-APV1:chi2dof:hiEffE}   {\ensuremath{{0.002 } } }
\vdef{default-11:NoMcPU-APV1:chi2dof:loEff}   {\ensuremath{{0.943 } } }
\vdef{default-11:NoMcPU-APV1:chi2dof:loEffE}   {\ensuremath{{0.002 } } }
\vdef{default-11:NoMcPU-APV1:chi2dof:hiEff}   {\ensuremath{{0.057 } } }
\vdef{default-11:NoMcPU-APV1:chi2dof:hiEffE}   {\ensuremath{{0.002 } } }
\vdef{default-11:NoMcPU-APV1:chi2dof:loDelta}   {\ensuremath{{-0.016 } } }
\vdef{default-11:NoMcPU-APV1:chi2dof:loDeltaE}   {\ensuremath{{0.002 } } }
\vdef{default-11:NoMcPU-APV1:chi2dof:hiDelta}   {\ensuremath{{+0.231 } } }
\vdef{default-11:NoMcPU-APV1:chi2dof:hiDeltaE}   {\ensuremath{{0.036 } } }
\vdef{default-11:NoData-APV1:pchi2dof:loEff}   {\ensuremath{{0.639 } } }
\vdef{default-11:NoData-APV1:pchi2dof:loEffE}   {\ensuremath{{0.003 } } }
\vdef{default-11:NoData-APV1:pchi2dof:hiEff}   {\ensuremath{{0.361 } } }
\vdef{default-11:NoData-APV1:pchi2dof:hiEffE}   {\ensuremath{{0.003 } } }
\vdef{default-11:NoMcPU-APV1:pchi2dof:loEff}   {\ensuremath{{0.614 } } }
\vdef{default-11:NoMcPU-APV1:pchi2dof:loEffE}   {\ensuremath{{0.003 } } }
\vdef{default-11:NoMcPU-APV1:pchi2dof:hiEff}   {\ensuremath{{0.386 } } }
\vdef{default-11:NoMcPU-APV1:pchi2dof:hiEffE}   {\ensuremath{{0.003 } } }
\vdef{default-11:NoMcPU-APV1:pchi2dof:loDelta}   {\ensuremath{{+0.039 } } }
\vdef{default-11:NoMcPU-APV1:pchi2dof:loDeltaE}   {\ensuremath{{0.007 } } }
\vdef{default-11:NoMcPU-APV1:pchi2dof:hiDelta}   {\ensuremath{{-0.065 } } }
\vdef{default-11:NoMcPU-APV1:pchi2dof:hiDeltaE}   {\ensuremath{{0.012 } } }
\vdef{default-11:NoData-APV1:alpha:loEff}   {\ensuremath{{0.995 } } }
\vdef{default-11:NoData-APV1:alpha:loEffE}   {\ensuremath{{0.000 } } }
\vdef{default-11:NoData-APV1:alpha:hiEff}   {\ensuremath{{0.005 } } }
\vdef{default-11:NoData-APV1:alpha:hiEffE}   {\ensuremath{{0.000 } } }
\vdef{default-11:NoMcPU-APV1:alpha:loEff}   {\ensuremath{{0.993 } } }
\vdef{default-11:NoMcPU-APV1:alpha:loEffE}   {\ensuremath{{0.001 } } }
\vdef{default-11:NoMcPU-APV1:alpha:hiEff}   {\ensuremath{{0.007 } } }
\vdef{default-11:NoMcPU-APV1:alpha:hiEffE}   {\ensuremath{{0.001 } } }
\vdef{default-11:NoMcPU-APV1:alpha:loDelta}   {\ensuremath{{+0.002 } } }
\vdef{default-11:NoMcPU-APV1:alpha:loDeltaE}   {\ensuremath{{0.001 } } }
\vdef{default-11:NoMcPU-APV1:alpha:hiDelta}   {\ensuremath{{-0.393 } } }
\vdef{default-11:NoMcPU-APV1:alpha:hiDeltaE}   {\ensuremath{{0.121 } } }
\vdef{default-11:NoData-APV1:iso:loEff}   {\ensuremath{{0.129 } } }
\vdef{default-11:NoData-APV1:iso:loEffE}   {\ensuremath{{0.002 } } }
\vdef{default-11:NoData-APV1:iso:hiEff}   {\ensuremath{{0.871 } } }
\vdef{default-11:NoData-APV1:iso:hiEffE}   {\ensuremath{{0.002 } } }
\vdef{default-11:NoMcPU-APV1:iso:loEff}   {\ensuremath{{0.110 } } }
\vdef{default-11:NoMcPU-APV1:iso:loEffE}   {\ensuremath{{0.002 } } }
\vdef{default-11:NoMcPU-APV1:iso:hiEff}   {\ensuremath{{0.890 } } }
\vdef{default-11:NoMcPU-APV1:iso:hiEffE}   {\ensuremath{{0.002 } } }
\vdef{default-11:NoMcPU-APV1:iso:loDelta}   {\ensuremath{{+0.158 } } }
\vdef{default-11:NoMcPU-APV1:iso:loDeltaE}   {\ensuremath{{0.025 } } }
\vdef{default-11:NoMcPU-APV1:iso:hiDelta}   {\ensuremath{{-0.022 } } }
\vdef{default-11:NoMcPU-APV1:iso:hiDeltaE}   {\ensuremath{{0.003 } } }
\vdef{default-11:NoData-APV1:docatrk:loEff}   {\ensuremath{{0.072 } } }
\vdef{default-11:NoData-APV1:docatrk:loEffE}   {\ensuremath{{0.002 } } }
\vdef{default-11:NoData-APV1:docatrk:hiEff}   {\ensuremath{{0.928 } } }
\vdef{default-11:NoData-APV1:docatrk:hiEffE}   {\ensuremath{{0.002 } } }
\vdef{default-11:NoMcPU-APV1:docatrk:loEff}   {\ensuremath{{0.085 } } }
\vdef{default-11:NoMcPU-APV1:docatrk:loEffE}   {\ensuremath{{0.002 } } }
\vdef{default-11:NoMcPU-APV1:docatrk:hiEff}   {\ensuremath{{0.915 } } }
\vdef{default-11:NoMcPU-APV1:docatrk:hiEffE}   {\ensuremath{{0.002 } } }
\vdef{default-11:NoMcPU-APV1:docatrk:loDelta}   {\ensuremath{{-0.171 } } }
\vdef{default-11:NoMcPU-APV1:docatrk:loDeltaE}   {\ensuremath{{0.033 } } }
\vdef{default-11:NoMcPU-APV1:docatrk:hiDelta}   {\ensuremath{{+0.014 } } }
\vdef{default-11:NoMcPU-APV1:docatrk:hiDeltaE}   {\ensuremath{{0.003 } } }
\vdef{default-11:NoData-APV1:isotrk:loEff}   {\ensuremath{{1.000 } } }
\vdef{default-11:NoData-APV1:isotrk:loEffE}   {\ensuremath{{0.000 } } }
\vdef{default-11:NoData-APV1:isotrk:hiEff}   {\ensuremath{{1.000 } } }
\vdef{default-11:NoData-APV1:isotrk:hiEffE}   {\ensuremath{{0.000 } } }
\vdef{default-11:NoMcPU-APV1:isotrk:loEff}   {\ensuremath{{1.000 } } }
\vdef{default-11:NoMcPU-APV1:isotrk:loEffE}   {\ensuremath{{0.000 } } }
\vdef{default-11:NoMcPU-APV1:isotrk:hiEff}   {\ensuremath{{1.000 } } }
\vdef{default-11:NoMcPU-APV1:isotrk:hiEffE}   {\ensuremath{{0.000 } } }
\vdef{default-11:NoMcPU-APV1:isotrk:loDelta}   {\ensuremath{{+0.000 } } }
\vdef{default-11:NoMcPU-APV1:isotrk:loDeltaE}   {\ensuremath{{0.000 } } }
\vdef{default-11:NoMcPU-APV1:isotrk:hiDelta}   {\ensuremath{{+0.000 } } }
\vdef{default-11:NoMcPU-APV1:isotrk:hiDeltaE}   {\ensuremath{{0.000 } } }
\vdef{default-11:NoData-APV1:closetrk:loEff}   {\ensuremath{{0.971 } } }
\vdef{default-11:NoData-APV1:closetrk:loEffE}   {\ensuremath{{0.001 } } }
\vdef{default-11:NoData-APV1:closetrk:hiEff}   {\ensuremath{{0.029 } } }
\vdef{default-11:NoData-APV1:closetrk:hiEffE}   {\ensuremath{{0.001 } } }
\vdef{default-11:NoMcPU-APV1:closetrk:loEff}   {\ensuremath{{0.969 } } }
\vdef{default-11:NoMcPU-APV1:closetrk:loEffE}   {\ensuremath{{0.001 } } }
\vdef{default-11:NoMcPU-APV1:closetrk:hiEff}   {\ensuremath{{0.031 } } }
\vdef{default-11:NoMcPU-APV1:closetrk:hiEffE}   {\ensuremath{{0.001 } } }
\vdef{default-11:NoMcPU-APV1:closetrk:loDelta}   {\ensuremath{{+0.002 } } }
\vdef{default-11:NoMcPU-APV1:closetrk:loDeltaE}   {\ensuremath{{0.002 } } }
\vdef{default-11:NoMcPU-APV1:closetrk:hiDelta}   {\ensuremath{{-0.074 } } }
\vdef{default-11:NoMcPU-APV1:closetrk:hiDeltaE}   {\ensuremath{{0.054 } } }
\vdef{default-11:NoData-APV1:lip:loEff}   {\ensuremath{{1.000 } } }
\vdef{default-11:NoData-APV1:lip:loEffE}   {\ensuremath{{0.000 } } }
\vdef{default-11:NoData-APV1:lip:hiEff}   {\ensuremath{{0.000 } } }
\vdef{default-11:NoData-APV1:lip:hiEffE}   {\ensuremath{{0.000 } } }
\vdef{default-11:NoMcPU-APV1:lip:loEff}   {\ensuremath{{1.000 } } }
\vdef{default-11:NoMcPU-APV1:lip:loEffE}   {\ensuremath{{0.000 } } }
\vdef{default-11:NoMcPU-APV1:lip:hiEff}   {\ensuremath{{0.000 } } }
\vdef{default-11:NoMcPU-APV1:lip:hiEffE}   {\ensuremath{{0.000 } } }
\vdef{default-11:NoMcPU-APV1:lip:loDelta}   {\ensuremath{{+0.000 } } }
\vdef{default-11:NoMcPU-APV1:lip:loDeltaE}   {\ensuremath{{0.000 } } }
\vdef{default-11:NoMcPU-APV1:lip:hiDelta}   {\ensuremath{{\mathrm{NaN} } } }
\vdef{default-11:NoMcPU-APV1:lip:hiDeltaE}   {\ensuremath{{\mathrm{NaN} } } }
\vdef{default-11:NoData-APV1:lip:inEff}   {\ensuremath{{1.000 } } }
\vdef{default-11:NoData-APV1:lip:inEffE}   {\ensuremath{{0.000 } } }
\vdef{default-11:NoMcPU-APV1:lip:inEff}   {\ensuremath{{1.000 } } }
\vdef{default-11:NoMcPU-APV1:lip:inEffE}   {\ensuremath{{0.000 } } }
\vdef{default-11:NoMcPU-APV1:lip:inDelta}   {\ensuremath{{+0.000 } } }
\vdef{default-11:NoMcPU-APV1:lip:inDeltaE}   {\ensuremath{{0.000 } } }
\vdef{default-11:NoData-APV1:lips:loEff}   {\ensuremath{{1.000 } } }
\vdef{default-11:NoData-APV1:lips:loEffE}   {\ensuremath{{0.000 } } }
\vdef{default-11:NoData-APV1:lips:hiEff}   {\ensuremath{{0.000 } } }
\vdef{default-11:NoData-APV1:lips:hiEffE}   {\ensuremath{{0.000 } } }
\vdef{default-11:NoMcPU-APV1:lips:loEff}   {\ensuremath{{1.000 } } }
\vdef{default-11:NoMcPU-APV1:lips:loEffE}   {\ensuremath{{0.000 } } }
\vdef{default-11:NoMcPU-APV1:lips:hiEff}   {\ensuremath{{0.000 } } }
\vdef{default-11:NoMcPU-APV1:lips:hiEffE}   {\ensuremath{{0.000 } } }
\vdef{default-11:NoMcPU-APV1:lips:loDelta}   {\ensuremath{{+0.000 } } }
\vdef{default-11:NoMcPU-APV1:lips:loDeltaE}   {\ensuremath{{0.000 } } }
\vdef{default-11:NoMcPU-APV1:lips:hiDelta}   {\ensuremath{{\mathrm{NaN} } } }
\vdef{default-11:NoMcPU-APV1:lips:hiDeltaE}   {\ensuremath{{\mathrm{NaN} } } }
\vdef{default-11:NoData-APV1:lips:inEff}   {\ensuremath{{1.000 } } }
\vdef{default-11:NoData-APV1:lips:inEffE}   {\ensuremath{{0.000 } } }
\vdef{default-11:NoMcPU-APV1:lips:inEff}   {\ensuremath{{1.000 } } }
\vdef{default-11:NoMcPU-APV1:lips:inEffE}   {\ensuremath{{0.000 } } }
\vdef{default-11:NoMcPU-APV1:lips:inDelta}   {\ensuremath{{+0.000 } } }
\vdef{default-11:NoMcPU-APV1:lips:inDeltaE}   {\ensuremath{{0.000 } } }
\vdef{default-11:NoData-APV1:ip:loEff}   {\ensuremath{{0.971 } } }
\vdef{default-11:NoData-APV1:ip:loEffE}   {\ensuremath{{0.001 } } }
\vdef{default-11:NoData-APV1:ip:hiEff}   {\ensuremath{{0.029 } } }
\vdef{default-11:NoData-APV1:ip:hiEffE}   {\ensuremath{{0.001 } } }
\vdef{default-11:NoMcPU-APV1:ip:loEff}   {\ensuremath{{0.970 } } }
\vdef{default-11:NoMcPU-APV1:ip:loEffE}   {\ensuremath{{0.001 } } }
\vdef{default-11:NoMcPU-APV1:ip:hiEff}   {\ensuremath{{0.030 } } }
\vdef{default-11:NoMcPU-APV1:ip:hiEffE}   {\ensuremath{{0.001 } } }
\vdef{default-11:NoMcPU-APV1:ip:loDelta}   {\ensuremath{{+0.000 } } }
\vdef{default-11:NoMcPU-APV1:ip:loDeltaE}   {\ensuremath{{0.002 } } }
\vdef{default-11:NoMcPU-APV1:ip:hiDelta}   {\ensuremath{{-0.004 } } }
\vdef{default-11:NoMcPU-APV1:ip:hiDeltaE}   {\ensuremath{{0.055 } } }
\vdef{default-11:NoData-APV1:ips:loEff}   {\ensuremath{{0.941 } } }
\vdef{default-11:NoData-APV1:ips:loEffE}   {\ensuremath{{0.002 } } }
\vdef{default-11:NoData-APV1:ips:hiEff}   {\ensuremath{{0.059 } } }
\vdef{default-11:NoData-APV1:ips:hiEffE}   {\ensuremath{{0.002 } } }
\vdef{default-11:NoMcPU-APV1:ips:loEff}   {\ensuremath{{0.948 } } }
\vdef{default-11:NoMcPU-APV1:ips:loEffE}   {\ensuremath{{0.001 } } }
\vdef{default-11:NoMcPU-APV1:ips:hiEff}   {\ensuremath{{0.052 } } }
\vdef{default-11:NoMcPU-APV1:ips:hiEffE}   {\ensuremath{{0.001 } } }
\vdef{default-11:NoMcPU-APV1:ips:loDelta}   {\ensuremath{{-0.007 } } }
\vdef{default-11:NoMcPU-APV1:ips:loDeltaE}   {\ensuremath{{0.002 } } }
\vdef{default-11:NoMcPU-APV1:ips:hiDelta}   {\ensuremath{{+0.121 } } }
\vdef{default-11:NoMcPU-APV1:ips:hiDeltaE}   {\ensuremath{{0.039 } } }
\vdef{default-11:NoData-APV1:maxdoca:loEff}   {\ensuremath{{1.000 } } }
\vdef{default-11:NoData-APV1:maxdoca:loEffE}   {\ensuremath{{0.000 } } }
\vdef{default-11:NoData-APV1:maxdoca:hiEff}   {\ensuremath{{0.013 } } }
\vdef{default-11:NoData-APV1:maxdoca:hiEffE}   {\ensuremath{{0.001 } } }
\vdef{default-11:NoMcPU-APV1:maxdoca:loEff}   {\ensuremath{{1.000 } } }
\vdef{default-11:NoMcPU-APV1:maxdoca:loEffE}   {\ensuremath{{0.000 } } }
\vdef{default-11:NoMcPU-APV1:maxdoca:hiEff}   {\ensuremath{{0.014 } } }
\vdef{default-11:NoMcPU-APV1:maxdoca:hiEffE}   {\ensuremath{{0.001 } } }
\vdef{default-11:NoMcPU-APV1:maxdoca:loDelta}   {\ensuremath{{+0.000 } } }
\vdef{default-11:NoMcPU-APV1:maxdoca:loDeltaE}   {\ensuremath{{0.000 } } }
\vdef{default-11:NoMcPU-APV1:maxdoca:hiDelta}   {\ensuremath{{-0.062 } } }
\vdef{default-11:NoMcPU-APV1:maxdoca:hiDeltaE}   {\ensuremath{{0.083 } } }
\vdef{default-11:NoData-APV1:kaonpt:loEff}   {\ensuremath{{1.010 } } }
\vdef{default-11:NoData-APV1:kaonpt:loEffE}   {\ensuremath{{\mathrm{NaN} } } }
\vdef{default-11:NoData-APV1:kaonpt:hiEff}   {\ensuremath{{1.000 } } }
\vdef{default-11:NoData-APV1:kaonpt:hiEffE}   {\ensuremath{{0.000 } } }
\vdef{default-11:NoMcPU-APV1:kaonpt:loEff}   {\ensuremath{{1.009 } } }
\vdef{default-11:NoMcPU-APV1:kaonpt:loEffE}   {\ensuremath{{\mathrm{NaN} } } }
\vdef{default-11:NoMcPU-APV1:kaonpt:hiEff}   {\ensuremath{{1.000 } } }
\vdef{default-11:NoMcPU-APV1:kaonpt:hiEffE}   {\ensuremath{{0.000 } } }
\vdef{default-11:NoMcPU-APV1:kaonpt:loDelta}   {\ensuremath{{+0.000 } } }
\vdef{default-11:NoMcPU-APV1:kaonpt:loDeltaE}   {\ensuremath{{\mathrm{NaN} } } }
\vdef{default-11:NoMcPU-APV1:kaonpt:hiDelta}   {\ensuremath{{+0.000 } } }
\vdef{default-11:NoMcPU-APV1:kaonpt:hiDeltaE}   {\ensuremath{{0.000 } } }
\vdef{default-11:NoData-APV1:psipt:loEff}   {\ensuremath{{1.003 } } }
\vdef{default-11:NoData-APV1:psipt:loEffE}   {\ensuremath{{\mathrm{NaN} } } }
\vdef{default-11:NoData-APV1:psipt:hiEff}   {\ensuremath{{1.000 } } }
\vdef{default-11:NoData-APV1:psipt:hiEffE}   {\ensuremath{{0.000 } } }
\vdef{default-11:NoMcPU-APV1:psipt:loEff}   {\ensuremath{{1.003 } } }
\vdef{default-11:NoMcPU-APV1:psipt:loEffE}   {\ensuremath{{\mathrm{NaN} } } }
\vdef{default-11:NoMcPU-APV1:psipt:hiEff}   {\ensuremath{{1.000 } } }
\vdef{default-11:NoMcPU-APV1:psipt:hiEffE}   {\ensuremath{{0.000 } } }
\vdef{default-11:NoMcPU-APV1:psipt:loDelta}   {\ensuremath{{-0.000 } } }
\vdef{default-11:NoMcPU-APV1:psipt:loDeltaE}   {\ensuremath{{\mathrm{NaN} } } }
\vdef{default-11:NoMcPU-APV1:psipt:hiDelta}   {\ensuremath{{+0.000 } } }
\vdef{default-11:NoMcPU-APV1:psipt:hiDeltaE}   {\ensuremath{{0.000 } } }
\vdef{default-11:NoMcPU-APV0:osiso:loEff}   {\ensuremath{{1.005 } } }
\vdef{default-11:NoMcPU-APV0:osiso:loEffE}   {\ensuremath{{\mathrm{NaN} } } }
\vdef{default-11:NoMcPU-APV0:osiso:hiEff}   {\ensuremath{{1.000 } } }
\vdef{default-11:NoMcPU-APV0:osiso:hiEffE}   {\ensuremath{{0.000 } } }
\vdef{default-11:NoMcPU-APV1:osiso:loEff}   {\ensuremath{{1.003 } } }
\vdef{default-11:NoMcPU-APV1:osiso:loEffE}   {\ensuremath{{\mathrm{NaN} } } }
\vdef{default-11:NoMcPU-APV1:osiso:hiEff}   {\ensuremath{{1.000 } } }
\vdef{default-11:NoMcPU-APV1:osiso:hiEffE}   {\ensuremath{{0.000 } } }
\vdef{default-11:NoMcPU-APV1:osiso:loDelta}   {\ensuremath{{+0.001 } } }
\vdef{default-11:NoMcPU-APV1:osiso:loDeltaE}   {\ensuremath{{\mathrm{NaN} } } }
\vdef{default-11:NoMcPU-APV1:osiso:hiDelta}   {\ensuremath{{+0.000 } } }
\vdef{default-11:NoMcPU-APV1:osiso:hiDeltaE}   {\ensuremath{{0.000 } } }
\vdef{default-11:NoMcPU-APV0:osreliso:loEff}   {\ensuremath{{0.291 } } }
\vdef{default-11:NoMcPU-APV0:osreliso:loEffE}   {\ensuremath{{0.003 } } }
\vdef{default-11:NoMcPU-APV0:osreliso:hiEff}   {\ensuremath{{0.709 } } }
\vdef{default-11:NoMcPU-APV0:osreliso:hiEffE}   {\ensuremath{{0.003 } } }
\vdef{default-11:NoMcPU-APV1:osreliso:loEff}   {\ensuremath{{0.287 } } }
\vdef{default-11:NoMcPU-APV1:osreliso:loEffE}   {\ensuremath{{0.003 } } }
\vdef{default-11:NoMcPU-APV1:osreliso:hiEff}   {\ensuremath{{0.713 } } }
\vdef{default-11:NoMcPU-APV1:osreliso:hiEffE}   {\ensuremath{{0.003 } } }
\vdef{default-11:NoMcPU-APV1:osreliso:loDelta}   {\ensuremath{{+0.013 } } }
\vdef{default-11:NoMcPU-APV1:osreliso:loDeltaE}   {\ensuremath{{0.012 } } }
\vdef{default-11:NoMcPU-APV1:osreliso:hiDelta}   {\ensuremath{{-0.005 } } }
\vdef{default-11:NoMcPU-APV1:osreliso:hiDeltaE}   {\ensuremath{{0.005 } } }
\vdef{default-11:NoMcPU-APV0:osmuonpt:loEff}   {\ensuremath{{0.000 } } }
\vdef{default-11:NoMcPU-APV0:osmuonpt:loEffE}   {\ensuremath{{0.001 } } }
\vdef{default-11:NoMcPU-APV0:osmuonpt:hiEff}   {\ensuremath{{1.000 } } }
\vdef{default-11:NoMcPU-APV0:osmuonpt:hiEffE}   {\ensuremath{{0.001 } } }
\vdef{default-11:NoMcPU-APV1:osmuonpt:loEff}   {\ensuremath{{0.000 } } }
\vdef{default-11:NoMcPU-APV1:osmuonpt:loEffE}   {\ensuremath{{0.001 } } }
\vdef{default-11:NoMcPU-APV1:osmuonpt:hiEff}   {\ensuremath{{1.000 } } }
\vdef{default-11:NoMcPU-APV1:osmuonpt:hiEffE}   {\ensuremath{{0.001 } } }
\vdef{default-11:NoMcPU-APV1:osmuonpt:loDelta}   {\ensuremath{{\mathrm{NaN} } } }
\vdef{default-11:NoMcPU-APV1:osmuonpt:loDeltaE}   {\ensuremath{{\mathrm{NaN} } } }
\vdef{default-11:NoMcPU-APV1:osmuonpt:hiDelta}   {\ensuremath{{+0.000 } } }
\vdef{default-11:NoMcPU-APV1:osmuonpt:hiDeltaE}   {\ensuremath{{0.001 } } }
\vdef{default-11:NoMcPU-APV0:osmuondr:loEff}   {\ensuremath{{0.009 } } }
\vdef{default-11:NoMcPU-APV0:osmuondr:loEffE}   {\ensuremath{{0.003 } } }
\vdef{default-11:NoMcPU-APV0:osmuondr:hiEff}   {\ensuremath{{0.991 } } }
\vdef{default-11:NoMcPU-APV0:osmuondr:hiEffE}   {\ensuremath{{0.003 } } }
\vdef{default-11:NoMcPU-APV1:osmuondr:loEff}   {\ensuremath{{0.015 } } }
\vdef{default-11:NoMcPU-APV1:osmuondr:loEffE}   {\ensuremath{{0.004 } } }
\vdef{default-11:NoMcPU-APV1:osmuondr:hiEff}   {\ensuremath{{0.985 } } }
\vdef{default-11:NoMcPU-APV1:osmuondr:hiEffE}   {\ensuremath{{0.004 } } }
\vdef{default-11:NoMcPU-APV1:osmuondr:loDelta}   {\ensuremath{{-0.514 } } }
\vdef{default-11:NoMcPU-APV1:osmuondr:loDeltaE}   {\ensuremath{{0.381 } } }
\vdef{default-11:NoMcPU-APV1:osmuondr:hiDelta}   {\ensuremath{{+0.006 } } }
\vdef{default-11:NoMcPU-APV1:osmuondr:hiDeltaE}   {\ensuremath{{0.005 } } }
\vdef{default-11:NoMcPU-APV0:hlt:loEff}   {\ensuremath{{0.284 } } }
\vdef{default-11:NoMcPU-APV0:hlt:loEffE}   {\ensuremath{{0.003 } } }
\vdef{default-11:NoMcPU-APV0:hlt:hiEff}   {\ensuremath{{0.716 } } }
\vdef{default-11:NoMcPU-APV0:hlt:hiEffE}   {\ensuremath{{0.003 } } }
\vdef{default-11:NoMcPU-APV1:hlt:loEff}   {\ensuremath{{0.327 } } }
\vdef{default-11:NoMcPU-APV1:hlt:loEffE}   {\ensuremath{{0.003 } } }
\vdef{default-11:NoMcPU-APV1:hlt:hiEff}   {\ensuremath{{0.673 } } }
\vdef{default-11:NoMcPU-APV1:hlt:hiEffE}   {\ensuremath{{0.003 } } }
\vdef{default-11:NoMcPU-APV1:hlt:loDelta}   {\ensuremath{{-0.140 } } }
\vdef{default-11:NoMcPU-APV1:hlt:loDeltaE}   {\ensuremath{{0.012 } } }
\vdef{default-11:NoMcPU-APV1:hlt:hiDelta}   {\ensuremath{{+0.062 } } }
\vdef{default-11:NoMcPU-APV1:hlt:hiDeltaE}   {\ensuremath{{0.005 } } }
\vdef{default-11:NoMcPU-APV0:muonsid:loEff}   {\ensuremath{{0.229 } } }
\vdef{default-11:NoMcPU-APV0:muonsid:loEffE}   {\ensuremath{{0.002 } } }
\vdef{default-11:NoMcPU-APV0:muonsid:hiEff}   {\ensuremath{{0.771 } } }
\vdef{default-11:NoMcPU-APV0:muonsid:hiEffE}   {\ensuremath{{0.002 } } }
\vdef{default-11:NoMcPU-APV1:muonsid:loEff}   {\ensuremath{{0.218 } } }
\vdef{default-11:NoMcPU-APV1:muonsid:loEffE}   {\ensuremath{{0.002 } } }
\vdef{default-11:NoMcPU-APV1:muonsid:hiEff}   {\ensuremath{{0.782 } } }
\vdef{default-11:NoMcPU-APV1:muonsid:hiEffE}   {\ensuremath{{0.002 } } }
\vdef{default-11:NoMcPU-APV1:muonsid:loDelta}   {\ensuremath{{+0.048 } } }
\vdef{default-11:NoMcPU-APV1:muonsid:loDeltaE}   {\ensuremath{{0.015 } } }
\vdef{default-11:NoMcPU-APV1:muonsid:hiDelta}   {\ensuremath{{-0.014 } } }
\vdef{default-11:NoMcPU-APV1:muonsid:hiDeltaE}   {\ensuremath{{0.004 } } }
\vdef{default-11:NoMcPU-APV0:tracksqual:loEff}   {\ensuremath{{0.000 } } }
\vdef{default-11:NoMcPU-APV0:tracksqual:loEffE}   {\ensuremath{{0.000 } } }
\vdef{default-11:NoMcPU-APV0:tracksqual:hiEff}   {\ensuremath{{1.000 } } }
\vdef{default-11:NoMcPU-APV0:tracksqual:hiEffE}   {\ensuremath{{0.000 } } }
\vdef{default-11:NoMcPU-APV1:tracksqual:loEff}   {\ensuremath{{0.000 } } }
\vdef{default-11:NoMcPU-APV1:tracksqual:loEffE}   {\ensuremath{{0.000 } } }
\vdef{default-11:NoMcPU-APV1:tracksqual:hiEff}   {\ensuremath{{1.000 } } }
\vdef{default-11:NoMcPU-APV1:tracksqual:hiEffE}   {\ensuremath{{0.000 } } }
\vdef{default-11:NoMcPU-APV1:tracksqual:loDelta}   {\ensuremath{{-0.054 } } }
\vdef{default-11:NoMcPU-APV1:tracksqual:loDeltaE}   {\ensuremath{{0.427 } } }
\vdef{default-11:NoMcPU-APV1:tracksqual:hiDelta}   {\ensuremath{{+0.000 } } }
\vdef{default-11:NoMcPU-APV1:tracksqual:hiDeltaE}   {\ensuremath{{0.000 } } }
\vdef{default-11:NoMcPU-APV0:pvz:loEff}   {\ensuremath{{0.470 } } }
\vdef{default-11:NoMcPU-APV0:pvz:loEffE}   {\ensuremath{{0.003 } } }
\vdef{default-11:NoMcPU-APV0:pvz:hiEff}   {\ensuremath{{0.530 } } }
\vdef{default-11:NoMcPU-APV0:pvz:hiEffE}   {\ensuremath{{0.003 } } }
\vdef{default-11:NoMcPU-APV1:pvz:loEff}   {\ensuremath{{0.477 } } }
\vdef{default-11:NoMcPU-APV1:pvz:loEffE}   {\ensuremath{{0.003 } } }
\vdef{default-11:NoMcPU-APV1:pvz:hiEff}   {\ensuremath{{0.523 } } }
\vdef{default-11:NoMcPU-APV1:pvz:hiEffE}   {\ensuremath{{0.003 } } }
\vdef{default-11:NoMcPU-APV1:pvz:loDelta}   {\ensuremath{{-0.016 } } }
\vdef{default-11:NoMcPU-APV1:pvz:loDeltaE}   {\ensuremath{{0.009 } } }
\vdef{default-11:NoMcPU-APV1:pvz:hiDelta}   {\ensuremath{{+0.014 } } }
\vdef{default-11:NoMcPU-APV1:pvz:hiDeltaE}   {\ensuremath{{0.008 } } }
\vdef{default-11:NoMcPU-APV0:pvn:loEff}   {\ensuremath{{1.000 } } }
\vdef{default-11:NoMcPU-APV0:pvn:loEffE}   {\ensuremath{{0.000 } } }
\vdef{default-11:NoMcPU-APV0:pvn:hiEff}   {\ensuremath{{1.000 } } }
\vdef{default-11:NoMcPU-APV0:pvn:hiEffE}   {\ensuremath{{0.000 } } }
\vdef{default-11:NoMcPU-APV1:pvn:loEff}   {\ensuremath{{1.185 } } }
\vdef{default-11:NoMcPU-APV1:pvn:loEffE}   {\ensuremath{{\mathrm{NaN} } } }
\vdef{default-11:NoMcPU-APV1:pvn:hiEff}   {\ensuremath{{1.000 } } }
\vdef{default-11:NoMcPU-APV1:pvn:hiEffE}   {\ensuremath{{0.000 } } }
\vdef{default-11:NoMcPU-APV1:pvn:loDelta}   {\ensuremath{{-0.170 } } }
\vdef{default-11:NoMcPU-APV1:pvn:loDeltaE}   {\ensuremath{{\mathrm{NaN} } } }
\vdef{default-11:NoMcPU-APV1:pvn:hiDelta}   {\ensuremath{{+0.000 } } }
\vdef{default-11:NoMcPU-APV1:pvn:hiDeltaE}   {\ensuremath{{0.000 } } }
\vdef{default-11:NoMcPU-APV0:pvavew8:loEff}   {\ensuremath{{0.005 } } }
\vdef{default-11:NoMcPU-APV0:pvavew8:loEffE}   {\ensuremath{{0.000 } } }
\vdef{default-11:NoMcPU-APV0:pvavew8:hiEff}   {\ensuremath{{0.995 } } }
\vdef{default-11:NoMcPU-APV0:pvavew8:hiEffE}   {\ensuremath{{0.000 } } }
\vdef{default-11:NoMcPU-APV1:pvavew8:loEff}   {\ensuremath{{0.008 } } }
\vdef{default-11:NoMcPU-APV1:pvavew8:loEffE}   {\ensuremath{{0.001 } } }
\vdef{default-11:NoMcPU-APV1:pvavew8:hiEff}   {\ensuremath{{0.992 } } }
\vdef{default-11:NoMcPU-APV1:pvavew8:hiEffE}   {\ensuremath{{0.001 } } }
\vdef{default-11:NoMcPU-APV1:pvavew8:loDelta}   {\ensuremath{{-0.580 } } }
\vdef{default-11:NoMcPU-APV1:pvavew8:loDeltaE}   {\ensuremath{{0.100 } } }
\vdef{default-11:NoMcPU-APV1:pvavew8:hiDelta}   {\ensuremath{{+0.004 } } }
\vdef{default-11:NoMcPU-APV1:pvavew8:hiDeltaE}   {\ensuremath{{0.001 } } }
\vdef{default-11:NoMcPU-APV0:pvntrk:loEff}   {\ensuremath{{1.000 } } }
\vdef{default-11:NoMcPU-APV0:pvntrk:loEffE}   {\ensuremath{{0.000 } } }
\vdef{default-11:NoMcPU-APV0:pvntrk:hiEff}   {\ensuremath{{1.000 } } }
\vdef{default-11:NoMcPU-APV0:pvntrk:hiEffE}   {\ensuremath{{0.000 } } }
\vdef{default-11:NoMcPU-APV1:pvntrk:loEff}   {\ensuremath{{1.000 } } }
\vdef{default-11:NoMcPU-APV1:pvntrk:loEffE}   {\ensuremath{{0.000 } } }
\vdef{default-11:NoMcPU-APV1:pvntrk:hiEff}   {\ensuremath{{1.000 } } }
\vdef{default-11:NoMcPU-APV1:pvntrk:hiEffE}   {\ensuremath{{0.000 } } }
\vdef{default-11:NoMcPU-APV1:pvntrk:loDelta}   {\ensuremath{{+0.000 } } }
\vdef{default-11:NoMcPU-APV1:pvntrk:loDeltaE}   {\ensuremath{{0.000 } } }
\vdef{default-11:NoMcPU-APV1:pvntrk:hiDelta}   {\ensuremath{{+0.000 } } }
\vdef{default-11:NoMcPU-APV1:pvntrk:hiDeltaE}   {\ensuremath{{0.000 } } }
\vdef{default-11:NoMcPU-APV0:muon1pt:loEff}   {\ensuremath{{1.010 } } }
\vdef{default-11:NoMcPU-APV0:muon1pt:loEffE}   {\ensuremath{{\mathrm{NaN} } } }
\vdef{default-11:NoMcPU-APV0:muon1pt:hiEff}   {\ensuremath{{1.000 } } }
\vdef{default-11:NoMcPU-APV0:muon1pt:hiEffE}   {\ensuremath{{0.000 } } }
\vdef{default-11:NoMcPU-APV1:muon1pt:loEff}   {\ensuremath{{1.009 } } }
\vdef{default-11:NoMcPU-APV1:muon1pt:loEffE}   {\ensuremath{{\mathrm{NaN} } } }
\vdef{default-11:NoMcPU-APV1:muon1pt:hiEff}   {\ensuremath{{1.000 } } }
\vdef{default-11:NoMcPU-APV1:muon1pt:hiEffE}   {\ensuremath{{0.000 } } }
\vdef{default-11:NoMcPU-APV1:muon1pt:loDelta}   {\ensuremath{{+0.001 } } }
\vdef{default-11:NoMcPU-APV1:muon1pt:loDeltaE}   {\ensuremath{{\mathrm{NaN} } } }
\vdef{default-11:NoMcPU-APV1:muon1pt:hiDelta}   {\ensuremath{{+0.000 } } }
\vdef{default-11:NoMcPU-APV1:muon1pt:hiDeltaE}   {\ensuremath{{0.000 } } }
\vdef{default-11:NoMcPU-APV0:muon2pt:loEff}   {\ensuremath{{0.003 } } }
\vdef{default-11:NoMcPU-APV0:muon2pt:loEffE}   {\ensuremath{{0.000 } } }
\vdef{default-11:NoMcPU-APV0:muon2pt:hiEff}   {\ensuremath{{0.997 } } }
\vdef{default-11:NoMcPU-APV0:muon2pt:hiEffE}   {\ensuremath{{0.000 } } }
\vdef{default-11:NoMcPU-APV1:muon2pt:loEff}   {\ensuremath{{0.003 } } }
\vdef{default-11:NoMcPU-APV1:muon2pt:loEffE}   {\ensuremath{{0.000 } } }
\vdef{default-11:NoMcPU-APV1:muon2pt:hiEff}   {\ensuremath{{0.997 } } }
\vdef{default-11:NoMcPU-APV1:muon2pt:hiEffE}   {\ensuremath{{0.000 } } }
\vdef{default-11:NoMcPU-APV1:muon2pt:loDelta}   {\ensuremath{{+0.007 } } }
\vdef{default-11:NoMcPU-APV1:muon2pt:loDeltaE}   {\ensuremath{{0.139 } } }
\vdef{default-11:NoMcPU-APV1:muon2pt:hiDelta}   {\ensuremath{{-0.000 } } }
\vdef{default-11:NoMcPU-APV1:muon2pt:hiDeltaE}   {\ensuremath{{0.000 } } }
\vdef{default-11:NoMcPU-APV0:muonseta:loEff}   {\ensuremath{{0.848 } } }
\vdef{default-11:NoMcPU-APV0:muonseta:loEffE}   {\ensuremath{{0.002 } } }
\vdef{default-11:NoMcPU-APV0:muonseta:hiEff}   {\ensuremath{{0.152 } } }
\vdef{default-11:NoMcPU-APV0:muonseta:hiEffE}   {\ensuremath{{0.002 } } }
\vdef{default-11:NoMcPU-APV1:muonseta:loEff}   {\ensuremath{{0.842 } } }
\vdef{default-11:NoMcPU-APV1:muonseta:loEffE}   {\ensuremath{{0.002 } } }
\vdef{default-11:NoMcPU-APV1:muonseta:hiEff}   {\ensuremath{{0.158 } } }
\vdef{default-11:NoMcPU-APV1:muonseta:hiEffE}   {\ensuremath{{0.002 } } }
\vdef{default-11:NoMcPU-APV1:muonseta:loDelta}   {\ensuremath{{+0.007 } } }
\vdef{default-11:NoMcPU-APV1:muonseta:loDeltaE}   {\ensuremath{{0.003 } } }
\vdef{default-11:NoMcPU-APV1:muonseta:hiDelta}   {\ensuremath{{-0.038 } } }
\vdef{default-11:NoMcPU-APV1:muonseta:hiDeltaE}   {\ensuremath{{0.014 } } }
\vdef{default-11:NoMcPU-APV0:pt:loEff}   {\ensuremath{{0.000 } } }
\vdef{default-11:NoMcPU-APV0:pt:loEffE}   {\ensuremath{{0.000 } } }
\vdef{default-11:NoMcPU-APV0:pt:hiEff}   {\ensuremath{{1.000 } } }
\vdef{default-11:NoMcPU-APV0:pt:hiEffE}   {\ensuremath{{0.000 } } }
\vdef{default-11:NoMcPU-APV1:pt:loEff}   {\ensuremath{{0.000 } } }
\vdef{default-11:NoMcPU-APV1:pt:loEffE}   {\ensuremath{{0.000 } } }
\vdef{default-11:NoMcPU-APV1:pt:hiEff}   {\ensuremath{{1.000 } } }
\vdef{default-11:NoMcPU-APV1:pt:hiEffE}   {\ensuremath{{0.000 } } }
\vdef{default-11:NoMcPU-APV1:pt:loDelta}   {\ensuremath{{\mathrm{NaN} } } }
\vdef{default-11:NoMcPU-APV1:pt:loDeltaE}   {\ensuremath{{\mathrm{NaN} } } }
\vdef{default-11:NoMcPU-APV1:pt:hiDelta}   {\ensuremath{{+0.000 } } }
\vdef{default-11:NoMcPU-APV1:pt:hiDeltaE}   {\ensuremath{{0.000 } } }
\vdef{default-11:NoMcPU-APV0:p:loEff}   {\ensuremath{{1.003 } } }
\vdef{default-11:NoMcPU-APV0:p:loEffE}   {\ensuremath{{\mathrm{NaN} } } }
\vdef{default-11:NoMcPU-APV0:p:hiEff}   {\ensuremath{{1.000 } } }
\vdef{default-11:NoMcPU-APV0:p:hiEffE}   {\ensuremath{{0.000 } } }
\vdef{default-11:NoMcPU-APV1:p:loEff}   {\ensuremath{{1.004 } } }
\vdef{default-11:NoMcPU-APV1:p:loEffE}   {\ensuremath{{\mathrm{NaN} } } }
\vdef{default-11:NoMcPU-APV1:p:hiEff}   {\ensuremath{{1.000 } } }
\vdef{default-11:NoMcPU-APV1:p:hiEffE}   {\ensuremath{{0.000 } } }
\vdef{default-11:NoMcPU-APV1:p:loDelta}   {\ensuremath{{-0.001 } } }
\vdef{default-11:NoMcPU-APV1:p:loDeltaE}   {\ensuremath{{\mathrm{NaN} } } }
\vdef{default-11:NoMcPU-APV1:p:hiDelta}   {\ensuremath{{+0.000 } } }
\vdef{default-11:NoMcPU-APV1:p:hiDeltaE}   {\ensuremath{{0.000 } } }
\vdef{default-11:NoMcPU-APV0:eta:loEff}   {\ensuremath{{0.846 } } }
\vdef{default-11:NoMcPU-APV0:eta:loEffE}   {\ensuremath{{0.002 } } }
\vdef{default-11:NoMcPU-APV0:eta:hiEff}   {\ensuremath{{0.154 } } }
\vdef{default-11:NoMcPU-APV0:eta:hiEffE}   {\ensuremath{{0.002 } } }
\vdef{default-11:NoMcPU-APV1:eta:loEff}   {\ensuremath{{0.843 } } }
\vdef{default-11:NoMcPU-APV1:eta:loEffE}   {\ensuremath{{0.002 } } }
\vdef{default-11:NoMcPU-APV1:eta:hiEff}   {\ensuremath{{0.157 } } }
\vdef{default-11:NoMcPU-APV1:eta:hiEffE}   {\ensuremath{{0.002 } } }
\vdef{default-11:NoMcPU-APV1:eta:loDelta}   {\ensuremath{{+0.003 } } }
\vdef{default-11:NoMcPU-APV1:eta:loDeltaE}   {\ensuremath{{0.004 } } }
\vdef{default-11:NoMcPU-APV1:eta:hiDelta}   {\ensuremath{{-0.017 } } }
\vdef{default-11:NoMcPU-APV1:eta:hiDeltaE}   {\ensuremath{{0.020 } } }
\vdef{default-11:NoMcPU-APV0:bdt:loEff}   {\ensuremath{{0.863 } } }
\vdef{default-11:NoMcPU-APV0:bdt:loEffE}   {\ensuremath{{0.002 } } }
\vdef{default-11:NoMcPU-APV0:bdt:hiEff}   {\ensuremath{{0.137 } } }
\vdef{default-11:NoMcPU-APV0:bdt:hiEffE}   {\ensuremath{{0.002 } } }
\vdef{default-11:NoMcPU-APV1:bdt:loEff}   {\ensuremath{{0.873 } } }
\vdef{default-11:NoMcPU-APV1:bdt:loEffE}   {\ensuremath{{0.002 } } }
\vdef{default-11:NoMcPU-APV1:bdt:hiEff}   {\ensuremath{{0.127 } } }
\vdef{default-11:NoMcPU-APV1:bdt:hiEffE}   {\ensuremath{{0.002 } } }
\vdef{default-11:NoMcPU-APV1:bdt:loDelta}   {\ensuremath{{-0.011 } } }
\vdef{default-11:NoMcPU-APV1:bdt:loDeltaE}   {\ensuremath{{0.003 } } }
\vdef{default-11:NoMcPU-APV1:bdt:hiDelta}   {\ensuremath{{+0.073 } } }
\vdef{default-11:NoMcPU-APV1:bdt:hiDeltaE}   {\ensuremath{{0.021 } } }
\vdef{default-11:NoMcPU-APV0:fl3d:loEff}   {\ensuremath{{0.891 } } }
\vdef{default-11:NoMcPU-APV0:fl3d:loEffE}   {\ensuremath{{0.002 } } }
\vdef{default-11:NoMcPU-APV0:fl3d:hiEff}   {\ensuremath{{0.109 } } }
\vdef{default-11:NoMcPU-APV0:fl3d:hiEffE}   {\ensuremath{{0.002 } } }
\vdef{default-11:NoMcPU-APV1:fl3d:loEff}   {\ensuremath{{0.884 } } }
\vdef{default-11:NoMcPU-APV1:fl3d:loEffE}   {\ensuremath{{0.002 } } }
\vdef{default-11:NoMcPU-APV1:fl3d:hiEff}   {\ensuremath{{0.116 } } }
\vdef{default-11:NoMcPU-APV1:fl3d:hiEffE}   {\ensuremath{{0.002 } } }
\vdef{default-11:NoMcPU-APV1:fl3d:loDelta}   {\ensuremath{{+0.008 } } }
\vdef{default-11:NoMcPU-APV1:fl3d:loDeltaE}   {\ensuremath{{0.003 } } }
\vdef{default-11:NoMcPU-APV1:fl3d:hiDelta}   {\ensuremath{{-0.061 } } }
\vdef{default-11:NoMcPU-APV1:fl3d:hiDeltaE}   {\ensuremath{{0.022 } } }
\vdef{default-11:NoMcPU-APV0:fl3de:loEff}   {\ensuremath{{1.000 } } }
\vdef{default-11:NoMcPU-APV0:fl3de:loEffE}   {\ensuremath{{0.000 } } }
\vdef{default-11:NoMcPU-APV0:fl3de:hiEff}   {\ensuremath{{0.000 } } }
\vdef{default-11:NoMcPU-APV0:fl3de:hiEffE}   {\ensuremath{{0.000 } } }
\vdef{default-11:NoMcPU-APV1:fl3de:loEff}   {\ensuremath{{1.000 } } }
\vdef{default-11:NoMcPU-APV1:fl3de:loEffE}   {\ensuremath{{0.000 } } }
\vdef{default-11:NoMcPU-APV1:fl3de:hiEff}   {\ensuremath{{0.000 } } }
\vdef{default-11:NoMcPU-APV1:fl3de:hiEffE}   {\ensuremath{{0.000 } } }
\vdef{default-11:NoMcPU-APV1:fl3de:loDelta}   {\ensuremath{{+0.000 } } }
\vdef{default-11:NoMcPU-APV1:fl3de:loDeltaE}   {\ensuremath{{0.000 } } }
\vdef{default-11:NoMcPU-APV1:fl3de:hiDelta}   {\ensuremath{{-2.000 } } }
\vdef{default-11:NoMcPU-APV1:fl3de:hiDeltaE}   {\ensuremath{{1.014 } } }
\vdef{default-11:NoMcPU-APV0:fls3d:loEff}   {\ensuremath{{0.063 } } }
\vdef{default-11:NoMcPU-APV0:fls3d:loEffE}   {\ensuremath{{0.001 } } }
\vdef{default-11:NoMcPU-APV0:fls3d:hiEff}   {\ensuremath{{0.937 } } }
\vdef{default-11:NoMcPU-APV0:fls3d:hiEffE}   {\ensuremath{{0.001 } } }
\vdef{default-11:NoMcPU-APV1:fls3d:loEff}   {\ensuremath{{0.063 } } }
\vdef{default-11:NoMcPU-APV1:fls3d:loEffE}   {\ensuremath{{0.001 } } }
\vdef{default-11:NoMcPU-APV1:fls3d:hiEff}   {\ensuremath{{0.937 } } }
\vdef{default-11:NoMcPU-APV1:fls3d:hiEffE}   {\ensuremath{{0.001 } } }
\vdef{default-11:NoMcPU-APV1:fls3d:loDelta}   {\ensuremath{{+0.009 } } }
\vdef{default-11:NoMcPU-APV1:fls3d:loDeltaE}   {\ensuremath{{0.030 } } }
\vdef{default-11:NoMcPU-APV1:fls3d:hiDelta}   {\ensuremath{{-0.001 } } }
\vdef{default-11:NoMcPU-APV1:fls3d:hiDeltaE}   {\ensuremath{{0.002 } } }
\vdef{default-11:NoMcPU-APV0:flsxy:loEff}   {\ensuremath{{1.011 } } }
\vdef{default-11:NoMcPU-APV0:flsxy:loEffE}   {\ensuremath{{\mathrm{NaN} } } }
\vdef{default-11:NoMcPU-APV0:flsxy:hiEff}   {\ensuremath{{1.000 } } }
\vdef{default-11:NoMcPU-APV0:flsxy:hiEffE}   {\ensuremath{{0.000 } } }
\vdef{default-11:NoMcPU-APV1:flsxy:loEff}   {\ensuremath{{1.015 } } }
\vdef{default-11:NoMcPU-APV1:flsxy:loEffE}   {\ensuremath{{\mathrm{NaN} } } }
\vdef{default-11:NoMcPU-APV1:flsxy:hiEff}   {\ensuremath{{1.000 } } }
\vdef{default-11:NoMcPU-APV1:flsxy:hiEffE}   {\ensuremath{{0.000 } } }
\vdef{default-11:NoMcPU-APV1:flsxy:loDelta}   {\ensuremath{{-0.005 } } }
\vdef{default-11:NoMcPU-APV1:flsxy:loDeltaE}   {\ensuremath{{\mathrm{NaN} } } }
\vdef{default-11:NoMcPU-APV1:flsxy:hiDelta}   {\ensuremath{{+0.000 } } }
\vdef{default-11:NoMcPU-APV1:flsxy:hiDeltaE}   {\ensuremath{{0.000 } } }
\vdef{default-11:NoMcPU-APV0:chi2dof:loEff}   {\ensuremath{{0.948 } } }
\vdef{default-11:NoMcPU-APV0:chi2dof:loEffE}   {\ensuremath{{0.001 } } }
\vdef{default-11:NoMcPU-APV0:chi2dof:hiEff}   {\ensuremath{{0.052 } } }
\vdef{default-11:NoMcPU-APV0:chi2dof:hiEffE}   {\ensuremath{{0.001 } } }
\vdef{default-11:NoMcPU-APV1:chi2dof:loEff}   {\ensuremath{{0.943 } } }
\vdef{default-11:NoMcPU-APV1:chi2dof:loEffE}   {\ensuremath{{0.001 } } }
\vdef{default-11:NoMcPU-APV1:chi2dof:hiEff}   {\ensuremath{{0.057 } } }
\vdef{default-11:NoMcPU-APV1:chi2dof:hiEffE}   {\ensuremath{{0.001 } } }
\vdef{default-11:NoMcPU-APV1:chi2dof:loDelta}   {\ensuremath{{+0.005 } } }
\vdef{default-11:NoMcPU-APV1:chi2dof:loDeltaE}   {\ensuremath{{0.002 } } }
\vdef{default-11:NoMcPU-APV1:chi2dof:hiDelta}   {\ensuremath{{-0.081 } } }
\vdef{default-11:NoMcPU-APV1:chi2dof:hiDeltaE}   {\ensuremath{{0.034 } } }
\vdef{default-11:NoMcPU-APV0:pchi2dof:loEff}   {\ensuremath{{0.597 } } }
\vdef{default-11:NoMcPU-APV0:pchi2dof:loEffE}   {\ensuremath{{0.003 } } }
\vdef{default-11:NoMcPU-APV0:pchi2dof:hiEff}   {\ensuremath{{0.403 } } }
\vdef{default-11:NoMcPU-APV0:pchi2dof:hiEffE}   {\ensuremath{{0.003 } } }
\vdef{default-11:NoMcPU-APV1:pchi2dof:loEff}   {\ensuremath{{0.614 } } }
\vdef{default-11:NoMcPU-APV1:pchi2dof:loEffE}   {\ensuremath{{0.003 } } }
\vdef{default-11:NoMcPU-APV1:pchi2dof:hiEff}   {\ensuremath{{0.386 } } }
\vdef{default-11:NoMcPU-APV1:pchi2dof:hiEffE}   {\ensuremath{{0.003 } } }
\vdef{default-11:NoMcPU-APV1:pchi2dof:loDelta}   {\ensuremath{{-0.028 } } }
\vdef{default-11:NoMcPU-APV1:pchi2dof:loDeltaE}   {\ensuremath{{0.007 } } }
\vdef{default-11:NoMcPU-APV1:pchi2dof:hiDelta}   {\ensuremath{{+0.043 } } }
\vdef{default-11:NoMcPU-APV1:pchi2dof:hiDeltaE}   {\ensuremath{{0.010 } } }
\vdef{default-11:NoMcPU-APV0:alpha:loEff}   {\ensuremath{{0.993 } } }
\vdef{default-11:NoMcPU-APV0:alpha:loEffE}   {\ensuremath{{0.001 } } }
\vdef{default-11:NoMcPU-APV0:alpha:hiEff}   {\ensuremath{{0.007 } } }
\vdef{default-11:NoMcPU-APV0:alpha:hiEffE}   {\ensuremath{{0.001 } } }
\vdef{default-11:NoMcPU-APV1:alpha:loEff}   {\ensuremath{{0.993 } } }
\vdef{default-11:NoMcPU-APV1:alpha:loEffE}   {\ensuremath{{0.001 } } }
\vdef{default-11:NoMcPU-APV1:alpha:hiEff}   {\ensuremath{{0.007 } } }
\vdef{default-11:NoMcPU-APV1:alpha:hiEffE}   {\ensuremath{{0.001 } } }
\vdef{default-11:NoMcPU-APV1:alpha:loDelta}   {\ensuremath{{+0.000 } } }
\vdef{default-11:NoMcPU-APV1:alpha:loDeltaE}   {\ensuremath{{0.001 } } }
\vdef{default-11:NoMcPU-APV1:alpha:hiDelta}   {\ensuremath{{-0.018 } } }
\vdef{default-11:NoMcPU-APV1:alpha:hiDeltaE}   {\ensuremath{{0.098 } } }
\vdef{default-11:NoMcPU-APV0:iso:loEff}   {\ensuremath{{0.110 } } }
\vdef{default-11:NoMcPU-APV0:iso:loEffE}   {\ensuremath{{0.002 } } }
\vdef{default-11:NoMcPU-APV0:iso:hiEff}   {\ensuremath{{0.890 } } }
\vdef{default-11:NoMcPU-APV0:iso:hiEffE}   {\ensuremath{{0.002 } } }
\vdef{default-11:NoMcPU-APV1:iso:loEff}   {\ensuremath{{0.110 } } }
\vdef{default-11:NoMcPU-APV1:iso:loEffE}   {\ensuremath{{0.002 } } }
\vdef{default-11:NoMcPU-APV1:iso:hiEff}   {\ensuremath{{0.890 } } }
\vdef{default-11:NoMcPU-APV1:iso:hiEffE}   {\ensuremath{{0.002 } } }
\vdef{default-11:NoMcPU-APV1:iso:loDelta}   {\ensuremath{{-0.005 } } }
\vdef{default-11:NoMcPU-APV1:iso:loDeltaE}   {\ensuremath{{0.022 } } }
\vdef{default-11:NoMcPU-APV1:iso:hiDelta}   {\ensuremath{{+0.001 } } }
\vdef{default-11:NoMcPU-APV1:iso:hiDeltaE}   {\ensuremath{{0.003 } } }
\vdef{default-11:NoMcPU-APV0:docatrk:loEff}   {\ensuremath{{0.080 } } }
\vdef{default-11:NoMcPU-APV0:docatrk:loEffE}   {\ensuremath{{0.002 } } }
\vdef{default-11:NoMcPU-APV0:docatrk:hiEff}   {\ensuremath{{0.920 } } }
\vdef{default-11:NoMcPU-APV0:docatrk:hiEffE}   {\ensuremath{{0.002 } } }
\vdef{default-11:NoMcPU-APV1:docatrk:loEff}   {\ensuremath{{0.085 } } }
\vdef{default-11:NoMcPU-APV1:docatrk:loEffE}   {\ensuremath{{0.002 } } }
\vdef{default-11:NoMcPU-APV1:docatrk:hiEff}   {\ensuremath{{0.915 } } }
\vdef{default-11:NoMcPU-APV1:docatrk:hiEffE}   {\ensuremath{{0.002 } } }
\vdef{default-11:NoMcPU-APV1:docatrk:loDelta}   {\ensuremath{{-0.065 } } }
\vdef{default-11:NoMcPU-APV1:docatrk:loDeltaE}   {\ensuremath{{0.028 } } }
\vdef{default-11:NoMcPU-APV1:docatrk:hiDelta}   {\ensuremath{{+0.006 } } }
\vdef{default-11:NoMcPU-APV1:docatrk:hiDeltaE}   {\ensuremath{{0.002 } } }
\vdef{default-11:NoMcPU-APV0:isotrk:loEff}   {\ensuremath{{1.000 } } }
\vdef{default-11:NoMcPU-APV0:isotrk:loEffE}   {\ensuremath{{0.000 } } }
\vdef{default-11:NoMcPU-APV0:isotrk:hiEff}   {\ensuremath{{1.000 } } }
\vdef{default-11:NoMcPU-APV0:isotrk:hiEffE}   {\ensuremath{{0.000 } } }
\vdef{default-11:NoMcPU-APV1:isotrk:loEff}   {\ensuremath{{1.000 } } }
\vdef{default-11:NoMcPU-APV1:isotrk:loEffE}   {\ensuremath{{0.000 } } }
\vdef{default-11:NoMcPU-APV1:isotrk:hiEff}   {\ensuremath{{1.000 } } }
\vdef{default-11:NoMcPU-APV1:isotrk:hiEffE}   {\ensuremath{{0.000 } } }
\vdef{default-11:NoMcPU-APV1:isotrk:loDelta}   {\ensuremath{{+0.000 } } }
\vdef{default-11:NoMcPU-APV1:isotrk:loDeltaE}   {\ensuremath{{0.000 } } }
\vdef{default-11:NoMcPU-APV1:isotrk:hiDelta}   {\ensuremath{{+0.000 } } }
\vdef{default-11:NoMcPU-APV1:isotrk:hiDeltaE}   {\ensuremath{{0.000 } } }
\vdef{default-11:NoMcPU-APV0:closetrk:loEff}   {\ensuremath{{0.975 } } }
\vdef{default-11:NoMcPU-APV0:closetrk:loEffE}   {\ensuremath{{0.001 } } }
\vdef{default-11:NoMcPU-APV0:closetrk:hiEff}   {\ensuremath{{0.025 } } }
\vdef{default-11:NoMcPU-APV0:closetrk:hiEffE}   {\ensuremath{{0.001 } } }
\vdef{default-11:NoMcPU-APV1:closetrk:loEff}   {\ensuremath{{0.969 } } }
\vdef{default-11:NoMcPU-APV1:closetrk:loEffE}   {\ensuremath{{0.001 } } }
\vdef{default-11:NoMcPU-APV1:closetrk:hiEff}   {\ensuremath{{0.031 } } }
\vdef{default-11:NoMcPU-APV1:closetrk:hiEffE}   {\ensuremath{{0.001 } } }
\vdef{default-11:NoMcPU-APV1:closetrk:loDelta}   {\ensuremath{{+0.006 } } }
\vdef{default-11:NoMcPU-APV1:closetrk:loDeltaE}   {\ensuremath{{0.001 } } }
\vdef{default-11:NoMcPU-APV1:closetrk:hiDelta}   {\ensuremath{{-0.223 } } }
\vdef{default-11:NoMcPU-APV1:closetrk:hiDeltaE}   {\ensuremath{{0.049 } } }
\vdef{default-11:NoMcPU-APV0:lip:loEff}   {\ensuremath{{1.000 } } }
\vdef{default-11:NoMcPU-APV0:lip:loEffE}   {\ensuremath{{0.000 } } }
\vdef{default-11:NoMcPU-APV0:lip:hiEff}   {\ensuremath{{0.000 } } }
\vdef{default-11:NoMcPU-APV0:lip:hiEffE}   {\ensuremath{{0.000 } } }
\vdef{default-11:NoMcPU-APV1:lip:loEff}   {\ensuremath{{1.000 } } }
\vdef{default-11:NoMcPU-APV1:lip:loEffE}   {\ensuremath{{0.000 } } }
\vdef{default-11:NoMcPU-APV1:lip:hiEff}   {\ensuremath{{0.000 } } }
\vdef{default-11:NoMcPU-APV1:lip:hiEffE}   {\ensuremath{{0.000 } } }
\vdef{default-11:NoMcPU-APV1:lip:loDelta}   {\ensuremath{{+0.000 } } }
\vdef{default-11:NoMcPU-APV1:lip:loDeltaE}   {\ensuremath{{0.000 } } }
\vdef{default-11:NoMcPU-APV1:lip:hiDelta}   {\ensuremath{{\mathrm{NaN} } } }
\vdef{default-11:NoMcPU-APV1:lip:hiDeltaE}   {\ensuremath{{\mathrm{NaN} } } }
\vdef{default-11:NoMcPU-APV0:lip:inEff}   {\ensuremath{{1.000 } } }
\vdef{default-11:NoMcPU-APV0:lip:inEffE}   {\ensuremath{{0.000 } } }
\vdef{default-11:NoMcPU-APV1:lip:inEff}   {\ensuremath{{1.000 } } }
\vdef{default-11:NoMcPU-APV1:lip:inEffE}   {\ensuremath{{0.000 } } }
\vdef{default-11:NoMcPU-APV1:lip:inDelta}   {\ensuremath{{+0.000 } } }
\vdef{default-11:NoMcPU-APV1:lip:inDeltaE}   {\ensuremath{{0.000 } } }
\vdef{default-11:NoMcPU-APV0:lips:loEff}   {\ensuremath{{1.000 } } }
\vdef{default-11:NoMcPU-APV0:lips:loEffE}   {\ensuremath{{0.000 } } }
\vdef{default-11:NoMcPU-APV0:lips:hiEff}   {\ensuremath{{0.000 } } }
\vdef{default-11:NoMcPU-APV0:lips:hiEffE}   {\ensuremath{{0.000 } } }
\vdef{default-11:NoMcPU-APV1:lips:loEff}   {\ensuremath{{1.000 } } }
\vdef{default-11:NoMcPU-APV1:lips:loEffE}   {\ensuremath{{0.000 } } }
\vdef{default-11:NoMcPU-APV1:lips:hiEff}   {\ensuremath{{0.000 } } }
\vdef{default-11:NoMcPU-APV1:lips:hiEffE}   {\ensuremath{{0.000 } } }
\vdef{default-11:NoMcPU-APV1:lips:loDelta}   {\ensuremath{{+0.000 } } }
\vdef{default-11:NoMcPU-APV1:lips:loDeltaE}   {\ensuremath{{0.000 } } }
\vdef{default-11:NoMcPU-APV1:lips:hiDelta}   {\ensuremath{{\mathrm{NaN} } } }
\vdef{default-11:NoMcPU-APV1:lips:hiDeltaE}   {\ensuremath{{\mathrm{NaN} } } }
\vdef{default-11:NoMcPU-APV0:lips:inEff}   {\ensuremath{{1.000 } } }
\vdef{default-11:NoMcPU-APV0:lips:inEffE}   {\ensuremath{{0.000 } } }
\vdef{default-11:NoMcPU-APV1:lips:inEff}   {\ensuremath{{1.000 } } }
\vdef{default-11:NoMcPU-APV1:lips:inEffE}   {\ensuremath{{0.000 } } }
\vdef{default-11:NoMcPU-APV1:lips:inDelta}   {\ensuremath{{+0.000 } } }
\vdef{default-11:NoMcPU-APV1:lips:inDeltaE}   {\ensuremath{{0.000 } } }
\vdef{default-11:NoMcPU-APV0:ip:loEff}   {\ensuremath{{0.969 } } }
\vdef{default-11:NoMcPU-APV0:ip:loEffE}   {\ensuremath{{0.001 } } }
\vdef{default-11:NoMcPU-APV0:ip:hiEff}   {\ensuremath{{0.031 } } }
\vdef{default-11:NoMcPU-APV0:ip:hiEffE}   {\ensuremath{{0.001 } } }
\vdef{default-11:NoMcPU-APV1:ip:loEff}   {\ensuremath{{0.970 } } }
\vdef{default-11:NoMcPU-APV1:ip:loEffE}   {\ensuremath{{0.001 } } }
\vdef{default-11:NoMcPU-APV1:ip:hiEff}   {\ensuremath{{0.030 } } }
\vdef{default-11:NoMcPU-APV1:ip:hiEffE}   {\ensuremath{{0.001 } } }
\vdef{default-11:NoMcPU-APV1:ip:loDelta}   {\ensuremath{{-0.001 } } }
\vdef{default-11:NoMcPU-APV1:ip:loDeltaE}   {\ensuremath{{0.001 } } }
\vdef{default-11:NoMcPU-APV1:ip:hiDelta}   {\ensuremath{{+0.039 } } }
\vdef{default-11:NoMcPU-APV1:ip:hiDeltaE}   {\ensuremath{{0.047 } } }
\vdef{default-11:NoMcPU-APV0:ips:loEff}   {\ensuremath{{0.957 } } }
\vdef{default-11:NoMcPU-APV0:ips:loEffE}   {\ensuremath{{0.001 } } }
\vdef{default-11:NoMcPU-APV0:ips:hiEff}   {\ensuremath{{0.043 } } }
\vdef{default-11:NoMcPU-APV0:ips:hiEffE}   {\ensuremath{{0.001 } } }
\vdef{default-11:NoMcPU-APV1:ips:loEff}   {\ensuremath{{0.948 } } }
\vdef{default-11:NoMcPU-APV1:ips:loEffE}   {\ensuremath{{0.001 } } }
\vdef{default-11:NoMcPU-APV1:ips:hiEff}   {\ensuremath{{0.052 } } }
\vdef{default-11:NoMcPU-APV1:ips:hiEffE}   {\ensuremath{{0.001 } } }
\vdef{default-11:NoMcPU-APV1:ips:loDelta}   {\ensuremath{{+0.010 } } }
\vdef{default-11:NoMcPU-APV1:ips:loDeltaE}   {\ensuremath{{0.002 } } }
\vdef{default-11:NoMcPU-APV1:ips:hiDelta}   {\ensuremath{{-0.201 } } }
\vdef{default-11:NoMcPU-APV1:ips:hiDeltaE}   {\ensuremath{{0.037 } } }
\vdef{default-11:NoMcPU-APV0:maxdoca:loEff}   {\ensuremath{{1.000 } } }
\vdef{default-11:NoMcPU-APV0:maxdoca:loEffE}   {\ensuremath{{0.000 } } }
\vdef{default-11:NoMcPU-APV0:maxdoca:hiEff}   {\ensuremath{{0.014 } } }
\vdef{default-11:NoMcPU-APV0:maxdoca:hiEffE}   {\ensuremath{{0.001 } } }
\vdef{default-11:NoMcPU-APV1:maxdoca:loEff}   {\ensuremath{{1.000 } } }
\vdef{default-11:NoMcPU-APV1:maxdoca:loEffE}   {\ensuremath{{0.000 } } }
\vdef{default-11:NoMcPU-APV1:maxdoca:hiEff}   {\ensuremath{{0.014 } } }
\vdef{default-11:NoMcPU-APV1:maxdoca:hiEffE}   {\ensuremath{{0.001 } } }
\vdef{default-11:NoMcPU-APV1:maxdoca:loDelta}   {\ensuremath{{+0.000 } } }
\vdef{default-11:NoMcPU-APV1:maxdoca:loDeltaE}   {\ensuremath{{0.000 } } }
\vdef{default-11:NoMcPU-APV1:maxdoca:hiDelta}   {\ensuremath{{-0.035 } } }
\vdef{default-11:NoMcPU-APV1:maxdoca:hiDeltaE}   {\ensuremath{{0.071 } } }
\vdef{default-11:NoMcPU-APV0:kaonpt:loEff}   {\ensuremath{{1.010 } } }
\vdef{default-11:NoMcPU-APV0:kaonpt:loEffE}   {\ensuremath{{\mathrm{NaN} } } }
\vdef{default-11:NoMcPU-APV0:kaonpt:hiEff}   {\ensuremath{{1.000 } } }
\vdef{default-11:NoMcPU-APV0:kaonpt:hiEffE}   {\ensuremath{{0.000 } } }
\vdef{default-11:NoMcPU-APV1:kaonpt:loEff}   {\ensuremath{{1.009 } } }
\vdef{default-11:NoMcPU-APV1:kaonpt:loEffE}   {\ensuremath{{\mathrm{NaN} } } }
\vdef{default-11:NoMcPU-APV1:kaonpt:hiEff}   {\ensuremath{{1.000 } } }
\vdef{default-11:NoMcPU-APV1:kaonpt:hiEffE}   {\ensuremath{{0.000 } } }
\vdef{default-11:NoMcPU-APV1:kaonpt:loDelta}   {\ensuremath{{+0.001 } } }
\vdef{default-11:NoMcPU-APV1:kaonpt:loDeltaE}   {\ensuremath{{\mathrm{NaN} } } }
\vdef{default-11:NoMcPU-APV1:kaonpt:hiDelta}   {\ensuremath{{+0.000 } } }
\vdef{default-11:NoMcPU-APV1:kaonpt:hiDeltaE}   {\ensuremath{{0.000 } } }
\vdef{default-11:NoMcPU-APV0:psipt:loEff}   {\ensuremath{{1.004 } } }
\vdef{default-11:NoMcPU-APV0:psipt:loEffE}   {\ensuremath{{\mathrm{NaN} } } }
\vdef{default-11:NoMcPU-APV0:psipt:hiEff}   {\ensuremath{{1.000 } } }
\vdef{default-11:NoMcPU-APV0:psipt:hiEffE}   {\ensuremath{{0.000 } } }
\vdef{default-11:NoMcPU-APV1:psipt:loEff}   {\ensuremath{{1.003 } } }
\vdef{default-11:NoMcPU-APV1:psipt:loEffE}   {\ensuremath{{\mathrm{NaN} } } }
\vdef{default-11:NoMcPU-APV1:psipt:hiEff}   {\ensuremath{{1.000 } } }
\vdef{default-11:NoMcPU-APV1:psipt:hiEffE}   {\ensuremath{{0.000 } } }
\vdef{default-11:NoMcPU-APV1:psipt:loDelta}   {\ensuremath{{+0.001 } } }
\vdef{default-11:NoMcPU-APV1:psipt:loDeltaE}   {\ensuremath{{\mathrm{NaN} } } }
\vdef{default-11:NoMcPU-APV1:psipt:hiDelta}   {\ensuremath{{+0.000 } } }
\vdef{default-11:NoMcPU-APV1:psipt:hiDeltaE}   {\ensuremath{{0.000 } } }
\vdef{default-11:NoData-B:osiso:loEff}   {\ensuremath{{1.005 } } }
\vdef{default-11:NoData-B:osiso:loEffE}   {\ensuremath{{\mathrm{NaN} } } }
\vdef{default-11:NoData-B:osiso:hiEff}   {\ensuremath{{1.000 } } }
\vdef{default-11:NoData-B:osiso:hiEffE}   {\ensuremath{{0.000 } } }
\vdef{default-11:NoMc-B:osiso:loEff}   {\ensuremath{{1.003 } } }
\vdef{default-11:NoMc-B:osiso:loEffE}   {\ensuremath{{\mathrm{NaN} } } }
\vdef{default-11:NoMc-B:osiso:hiEff}   {\ensuremath{{1.000 } } }
\vdef{default-11:NoMc-B:osiso:hiEffE}   {\ensuremath{{0.000 } } }
\vdef{default-11:NoMc-B:osiso:loDelta}   {\ensuremath{{+0.002 } } }
\vdef{default-11:NoMc-B:osiso:loDeltaE}   {\ensuremath{{\mathrm{NaN} } } }
\vdef{default-11:NoMc-B:osiso:hiDelta}   {\ensuremath{{+0.000 } } }
\vdef{default-11:NoMc-B:osiso:hiDeltaE}   {\ensuremath{{0.000 } } }
\vdef{default-11:NoData-B:osreliso:loEff}   {\ensuremath{{0.247 } } }
\vdef{default-11:NoData-B:osreliso:loEffE}   {\ensuremath{{0.002 } } }
\vdef{default-11:NoData-B:osreliso:hiEff}   {\ensuremath{{0.753 } } }
\vdef{default-11:NoData-B:osreliso:hiEffE}   {\ensuremath{{0.002 } } }
\vdef{default-11:NoMc-B:osreliso:loEff}   {\ensuremath{{0.284 } } }
\vdef{default-11:NoMc-B:osreliso:loEffE}   {\ensuremath{{0.002 } } }
\vdef{default-11:NoMc-B:osreliso:hiEff}   {\ensuremath{{0.716 } } }
\vdef{default-11:NoMc-B:osreliso:hiEffE}   {\ensuremath{{0.002 } } }
\vdef{default-11:NoMc-B:osreliso:loDelta}   {\ensuremath{{-0.140 } } }
\vdef{default-11:NoMc-B:osreliso:loDeltaE}   {\ensuremath{{0.008 } } }
\vdef{default-11:NoMc-B:osreliso:hiDelta}   {\ensuremath{{+0.051 } } }
\vdef{default-11:NoMc-B:osreliso:hiDeltaE}   {\ensuremath{{0.003 } } }
\vdef{default-11:NoData-B:osmuonpt:loEff}   {\ensuremath{{0.000 } } }
\vdef{default-11:NoData-B:osmuonpt:loEffE}   {\ensuremath{{0.000 } } }
\vdef{default-11:NoData-B:osmuonpt:hiEff}   {\ensuremath{{1.000 } } }
\vdef{default-11:NoData-B:osmuonpt:hiEffE}   {\ensuremath{{0.000 } } }
\vdef{default-11:NoMc-B:osmuonpt:loEff}   {\ensuremath{{0.000 } } }
\vdef{default-11:NoMc-B:osmuonpt:loEffE}   {\ensuremath{{0.000 } } }
\vdef{default-11:NoMc-B:osmuonpt:hiEff}   {\ensuremath{{1.000 } } }
\vdef{default-11:NoMc-B:osmuonpt:hiEffE}   {\ensuremath{{0.000 } } }
\vdef{default-11:NoMc-B:osmuonpt:loDelta}   {\ensuremath{{\mathrm{NaN} } } }
\vdef{default-11:NoMc-B:osmuonpt:loDeltaE}   {\ensuremath{{\mathrm{NaN} } } }
\vdef{default-11:NoMc-B:osmuonpt:hiDelta}   {\ensuremath{{+0.000 } } }
\vdef{default-11:NoMc-B:osmuonpt:hiDeltaE}   {\ensuremath{{0.000 } } }
\vdef{default-11:NoData-B:osmuondr:loEff}   {\ensuremath{{0.018 } } }
\vdef{default-11:NoData-B:osmuondr:loEffE}   {\ensuremath{{0.002 } } }
\vdef{default-11:NoData-B:osmuondr:hiEff}   {\ensuremath{{0.982 } } }
\vdef{default-11:NoData-B:osmuondr:hiEffE}   {\ensuremath{{0.002 } } }
\vdef{default-11:NoMc-B:osmuondr:loEff}   {\ensuremath{{0.016 } } }
\vdef{default-11:NoMc-B:osmuondr:loEffE}   {\ensuremath{{0.002 } } }
\vdef{default-11:NoMc-B:osmuondr:hiEff}   {\ensuremath{{0.984 } } }
\vdef{default-11:NoMc-B:osmuondr:hiEffE}   {\ensuremath{{0.002 } } }
\vdef{default-11:NoMc-B:osmuondr:loDelta}   {\ensuremath{{+0.100 } } }
\vdef{default-11:NoMc-B:osmuondr:loDeltaE}   {\ensuremath{{0.202 } } }
\vdef{default-11:NoMc-B:osmuondr:hiDelta}   {\ensuremath{{-0.002 } } }
\vdef{default-11:NoMc-B:osmuondr:hiDeltaE}   {\ensuremath{{0.003 } } }
\vdef{default-11:NoData-B:hlt:loEff}   {\ensuremath{{0.065 } } }
\vdef{default-11:NoData-B:hlt:loEffE}   {\ensuremath{{0.001 } } }
\vdef{default-11:NoData-B:hlt:hiEff}   {\ensuremath{{0.935 } } }
\vdef{default-11:NoData-B:hlt:hiEffE}   {\ensuremath{{0.001 } } }
\vdef{default-11:NoMc-B:hlt:loEff}   {\ensuremath{{0.233 } } }
\vdef{default-11:NoMc-B:hlt:loEffE}   {\ensuremath{{0.002 } } }
\vdef{default-11:NoMc-B:hlt:hiEff}   {\ensuremath{{0.767 } } }
\vdef{default-11:NoMc-B:hlt:hiEffE}   {\ensuremath{{0.002 } } }
\vdef{default-11:NoMc-B:hlt:loDelta}   {\ensuremath{{-1.131 } } }
\vdef{default-11:NoMc-B:hlt:loDeltaE}   {\ensuremath{{0.010 } } }
\vdef{default-11:NoMc-B:hlt:hiDelta}   {\ensuremath{{+0.198 } } }
\vdef{default-11:NoMc-B:hlt:hiDeltaE}   {\ensuremath{{0.002 } } }
\vdef{default-11:NoData-B:muonsid:loEff}   {\ensuremath{{0.147 } } }
\vdef{default-11:NoData-B:muonsid:loEffE}   {\ensuremath{{0.001 } } }
\vdef{default-11:NoData-B:muonsid:hiEff}   {\ensuremath{{0.853 } } }
\vdef{default-11:NoData-B:muonsid:hiEffE}   {\ensuremath{{0.001 } } }
\vdef{default-11:NoMc-B:muonsid:loEff}   {\ensuremath{{0.162 } } }
\vdef{default-11:NoMc-B:muonsid:loEffE}   {\ensuremath{{0.001 } } }
\vdef{default-11:NoMc-B:muonsid:hiEff}   {\ensuremath{{0.838 } } }
\vdef{default-11:NoMc-B:muonsid:hiEffE}   {\ensuremath{{0.001 } } }
\vdef{default-11:NoMc-B:muonsid:loDelta}   {\ensuremath{{-0.102 } } }
\vdef{default-11:NoMc-B:muonsid:loDeltaE}   {\ensuremath{{0.011 } } }
\vdef{default-11:NoMc-B:muonsid:hiDelta}   {\ensuremath{{+0.019 } } }
\vdef{default-11:NoMc-B:muonsid:hiDeltaE}   {\ensuremath{{0.002 } } }
\vdef{default-11:NoData-B:tracksqual:loEff}   {\ensuremath{{0.000 } } }
\vdef{default-11:NoData-B:tracksqual:loEffE}   {\ensuremath{{0.000 } } }
\vdef{default-11:NoData-B:tracksqual:hiEff}   {\ensuremath{{1.000 } } }
\vdef{default-11:NoData-B:tracksqual:hiEffE}   {\ensuremath{{0.000 } } }
\vdef{default-11:NoMc-B:tracksqual:loEff}   {\ensuremath{{0.000 } } }
\vdef{default-11:NoMc-B:tracksqual:loEffE}   {\ensuremath{{0.000 } } }
\vdef{default-11:NoMc-B:tracksqual:hiEff}   {\ensuremath{{1.000 } } }
\vdef{default-11:NoMc-B:tracksqual:hiEffE}   {\ensuremath{{0.000 } } }
\vdef{default-11:NoMc-B:tracksqual:loDelta}   {\ensuremath{{+0.664 } } }
\vdef{default-11:NoMc-B:tracksqual:loDeltaE}   {\ensuremath{{0.318 } } }
\vdef{default-11:NoMc-B:tracksqual:hiDelta}   {\ensuremath{{-0.000 } } }
\vdef{default-11:NoMc-B:tracksqual:hiDeltaE}   {\ensuremath{{0.000 } } }
\vdef{default-11:NoData-B:pvz:loEff}   {\ensuremath{{0.515 } } }
\vdef{default-11:NoData-B:pvz:loEffE}   {\ensuremath{{0.002 } } }
\vdef{default-11:NoData-B:pvz:hiEff}   {\ensuremath{{0.485 } } }
\vdef{default-11:NoData-B:pvz:hiEffE}   {\ensuremath{{0.002 } } }
\vdef{default-11:NoMc-B:pvz:loEff}   {\ensuremath{{0.471 } } }
\vdef{default-11:NoMc-B:pvz:loEffE}   {\ensuremath{{0.002 } } }
\vdef{default-11:NoMc-B:pvz:hiEff}   {\ensuremath{{0.529 } } }
\vdef{default-11:NoMc-B:pvz:hiEffE}   {\ensuremath{{0.002 } } }
\vdef{default-11:NoMc-B:pvz:loDelta}   {\ensuremath{{+0.089 } } }
\vdef{default-11:NoMc-B:pvz:loDeltaE}   {\ensuremath{{0.005 } } }
\vdef{default-11:NoMc-B:pvz:hiDelta}   {\ensuremath{{-0.087 } } }
\vdef{default-11:NoMc-B:pvz:hiDeltaE}   {\ensuremath{{0.005 } } }
\vdef{default-11:NoData-B:pvn:loEff}   {\ensuremath{{1.008 } } }
\vdef{default-11:NoData-B:pvn:loEffE}   {\ensuremath{{\mathrm{NaN} } } }
\vdef{default-11:NoData-B:pvn:hiEff}   {\ensuremath{{1.000 } } }
\vdef{default-11:NoData-B:pvn:hiEffE}   {\ensuremath{{0.000 } } }
\vdef{default-11:NoMc-B:pvn:loEff}   {\ensuremath{{1.000 } } }
\vdef{default-11:NoMc-B:pvn:loEffE}   {\ensuremath{{0.000 } } }
\vdef{default-11:NoMc-B:pvn:hiEff}   {\ensuremath{{1.000 } } }
\vdef{default-11:NoMc-B:pvn:hiEffE}   {\ensuremath{{0.000 } } }
\vdef{default-11:NoMc-B:pvn:loDelta}   {\ensuremath{{+0.008 } } }
\vdef{default-11:NoMc-B:pvn:loDeltaE}   {\ensuremath{{\mathrm{NaN} } } }
\vdef{default-11:NoMc-B:pvn:hiDelta}   {\ensuremath{{+0.000 } } }
\vdef{default-11:NoMc-B:pvn:hiDeltaE}   {\ensuremath{{0.000 } } }
\vdef{default-11:NoData-B:pvavew8:loEff}   {\ensuremath{{0.010 } } }
\vdef{default-11:NoData-B:pvavew8:loEffE}   {\ensuremath{{0.000 } } }
\vdef{default-11:NoData-B:pvavew8:hiEff}   {\ensuremath{{0.990 } } }
\vdef{default-11:NoData-B:pvavew8:hiEffE}   {\ensuremath{{0.000 } } }
\vdef{default-11:NoMc-B:pvavew8:loEff}   {\ensuremath{{0.006 } } }
\vdef{default-11:NoMc-B:pvavew8:loEffE}   {\ensuremath{{0.000 } } }
\vdef{default-11:NoMc-B:pvavew8:hiEff}   {\ensuremath{{0.994 } } }
\vdef{default-11:NoMc-B:pvavew8:hiEffE}   {\ensuremath{{0.000 } } }
\vdef{default-11:NoMc-B:pvavew8:loDelta}   {\ensuremath{{+0.451 } } }
\vdef{default-11:NoMc-B:pvavew8:loDeltaE}   {\ensuremath{{0.056 } } }
\vdef{default-11:NoMc-B:pvavew8:hiDelta}   {\ensuremath{{-0.004 } } }
\vdef{default-11:NoMc-B:pvavew8:hiDeltaE}   {\ensuremath{{0.000 } } }
\vdef{default-11:NoData-B:pvntrk:loEff}   {\ensuremath{{1.000 } } }
\vdef{default-11:NoData-B:pvntrk:loEffE}   {\ensuremath{{0.000 } } }
\vdef{default-11:NoData-B:pvntrk:hiEff}   {\ensuremath{{1.000 } } }
\vdef{default-11:NoData-B:pvntrk:hiEffE}   {\ensuremath{{0.000 } } }
\vdef{default-11:NoMc-B:pvntrk:loEff}   {\ensuremath{{1.000 } } }
\vdef{default-11:NoMc-B:pvntrk:loEffE}   {\ensuremath{{0.000 } } }
\vdef{default-11:NoMc-B:pvntrk:hiEff}   {\ensuremath{{1.000 } } }
\vdef{default-11:NoMc-B:pvntrk:hiEffE}   {\ensuremath{{0.000 } } }
\vdef{default-11:NoMc-B:pvntrk:loDelta}   {\ensuremath{{+0.000 } } }
\vdef{default-11:NoMc-B:pvntrk:loDeltaE}   {\ensuremath{{0.000 } } }
\vdef{default-11:NoMc-B:pvntrk:hiDelta}   {\ensuremath{{+0.000 } } }
\vdef{default-11:NoMc-B:pvntrk:hiDeltaE}   {\ensuremath{{0.000 } } }
\vdef{default-11:NoData-B:muon1pt:loEff}   {\ensuremath{{1.012 } } }
\vdef{default-11:NoData-B:muon1pt:loEffE}   {\ensuremath{{\mathrm{NaN} } } }
\vdef{default-11:NoData-B:muon1pt:hiEff}   {\ensuremath{{1.000 } } }
\vdef{default-11:NoData-B:muon1pt:hiEffE}   {\ensuremath{{0.000 } } }
\vdef{default-11:NoMc-B:muon1pt:loEff}   {\ensuremath{{1.009 } } }
\vdef{default-11:NoMc-B:muon1pt:loEffE}   {\ensuremath{{\mathrm{NaN} } } }
\vdef{default-11:NoMc-B:muon1pt:hiEff}   {\ensuremath{{1.000 } } }
\vdef{default-11:NoMc-B:muon1pt:hiEffE}   {\ensuremath{{0.000 } } }
\vdef{default-11:NoMc-B:muon1pt:loDelta}   {\ensuremath{{+0.002 } } }
\vdef{default-11:NoMc-B:muon1pt:loDeltaE}   {\ensuremath{{\mathrm{NaN} } } }
\vdef{default-11:NoMc-B:muon1pt:hiDelta}   {\ensuremath{{+0.000 } } }
\vdef{default-11:NoMc-B:muon1pt:hiDeltaE}   {\ensuremath{{0.000 } } }
\vdef{default-11:NoData-B:muon2pt:loEff}   {\ensuremath{{0.046 } } }
\vdef{default-11:NoData-B:muon2pt:loEffE}   {\ensuremath{{0.001 } } }
\vdef{default-11:NoData-B:muon2pt:hiEff}   {\ensuremath{{0.954 } } }
\vdef{default-11:NoData-B:muon2pt:hiEffE}   {\ensuremath{{0.001 } } }
\vdef{default-11:NoMc-B:muon2pt:loEff}   {\ensuremath{{0.044 } } }
\vdef{default-11:NoMc-B:muon2pt:loEffE}   {\ensuremath{{0.001 } } }
\vdef{default-11:NoMc-B:muon2pt:hiEff}   {\ensuremath{{0.956 } } }
\vdef{default-11:NoMc-B:muon2pt:hiEffE}   {\ensuremath{{0.001 } } }
\vdef{default-11:NoMc-B:muon2pt:loDelta}   {\ensuremath{{+0.030 } } }
\vdef{default-11:NoMc-B:muon2pt:loDeltaE}   {\ensuremath{{0.024 } } }
\vdef{default-11:NoMc-B:muon2pt:hiDelta}   {\ensuremath{{-0.001 } } }
\vdef{default-11:NoMc-B:muon2pt:hiDeltaE}   {\ensuremath{{0.001 } } }
\vdef{default-11:NoData-B:muonseta:loEff}   {\ensuremath{{0.814 } } }
\vdef{default-11:NoData-B:muonseta:loEffE}   {\ensuremath{{0.001 } } }
\vdef{default-11:NoData-B:muonseta:hiEff}   {\ensuremath{{0.186 } } }
\vdef{default-11:NoData-B:muonseta:hiEffE}   {\ensuremath{{0.001 } } }
\vdef{default-11:NoMc-B:muonseta:loEff}   {\ensuremath{{0.812 } } }
\vdef{default-11:NoMc-B:muonseta:loEffE}   {\ensuremath{{0.001 } } }
\vdef{default-11:NoMc-B:muonseta:hiEff}   {\ensuremath{{0.188 } } }
\vdef{default-11:NoMc-B:muonseta:hiEffE}   {\ensuremath{{0.001 } } }
\vdef{default-11:NoMc-B:muonseta:loDelta}   {\ensuremath{{+0.002 } } }
\vdef{default-11:NoMc-B:muonseta:loDeltaE}   {\ensuremath{{0.002 } } }
\vdef{default-11:NoMc-B:muonseta:hiDelta}   {\ensuremath{{-0.008 } } }
\vdef{default-11:NoMc-B:muonseta:hiDeltaE}   {\ensuremath{{0.008 } } }
\vdef{default-11:NoData-B:pt:loEff}   {\ensuremath{{0.000 } } }
\vdef{default-11:NoData-B:pt:loEffE}   {\ensuremath{{0.000 } } }
\vdef{default-11:NoData-B:pt:hiEff}   {\ensuremath{{1.000 } } }
\vdef{default-11:NoData-B:pt:hiEffE}   {\ensuremath{{0.000 } } }
\vdef{default-11:NoMc-B:pt:loEff}   {\ensuremath{{0.000 } } }
\vdef{default-11:NoMc-B:pt:loEffE}   {\ensuremath{{0.000 } } }
\vdef{default-11:NoMc-B:pt:hiEff}   {\ensuremath{{1.000 } } }
\vdef{default-11:NoMc-B:pt:hiEffE}   {\ensuremath{{0.000 } } }
\vdef{default-11:NoMc-B:pt:loDelta}   {\ensuremath{{\mathrm{NaN} } } }
\vdef{default-11:NoMc-B:pt:loDeltaE}   {\ensuremath{{\mathrm{NaN} } } }
\vdef{default-11:NoMc-B:pt:hiDelta}   {\ensuremath{{+0.000 } } }
\vdef{default-11:NoMc-B:pt:hiDeltaE}   {\ensuremath{{0.000 } } }
\vdef{default-11:NoData-B:p:loEff}   {\ensuremath{{1.003 } } }
\vdef{default-11:NoData-B:p:loEffE}   {\ensuremath{{\mathrm{NaN} } } }
\vdef{default-11:NoData-B:p:hiEff}   {\ensuremath{{1.000 } } }
\vdef{default-11:NoData-B:p:hiEffE}   {\ensuremath{{0.000 } } }
\vdef{default-11:NoMc-B:p:loEff}   {\ensuremath{{1.002 } } }
\vdef{default-11:NoMc-B:p:loEffE}   {\ensuremath{{\mathrm{NaN} } } }
\vdef{default-11:NoMc-B:p:hiEff}   {\ensuremath{{1.000 } } }
\vdef{default-11:NoMc-B:p:hiEffE}   {\ensuremath{{0.000 } } }
\vdef{default-11:NoMc-B:p:loDelta}   {\ensuremath{{+0.001 } } }
\vdef{default-11:NoMc-B:p:loDeltaE}   {\ensuremath{{\mathrm{NaN} } } }
\vdef{default-11:NoMc-B:p:hiDelta}   {\ensuremath{{+0.000 } } }
\vdef{default-11:NoMc-B:p:hiDeltaE}   {\ensuremath{{0.000 } } }
\vdef{default-11:NoData-B:eta:loEff}   {\ensuremath{{0.802 } } }
\vdef{default-11:NoData-B:eta:loEffE}   {\ensuremath{{0.001 } } }
\vdef{default-11:NoData-B:eta:hiEff}   {\ensuremath{{0.198 } } }
\vdef{default-11:NoData-B:eta:hiEffE}   {\ensuremath{{0.001 } } }
\vdef{default-11:NoMc-B:eta:loEff}   {\ensuremath{{0.801 } } }
\vdef{default-11:NoMc-B:eta:loEffE}   {\ensuremath{{0.001 } } }
\vdef{default-11:NoMc-B:eta:hiEff}   {\ensuremath{{0.199 } } }
\vdef{default-11:NoMc-B:eta:hiEffE}   {\ensuremath{{0.001 } } }
\vdef{default-11:NoMc-B:eta:loDelta}   {\ensuremath{{+0.001 } } }
\vdef{default-11:NoMc-B:eta:loDeltaE}   {\ensuremath{{0.003 } } }
\vdef{default-11:NoMc-B:eta:hiDelta}   {\ensuremath{{-0.004 } } }
\vdef{default-11:NoMc-B:eta:hiDeltaE}   {\ensuremath{{0.011 } } }
\vdef{default-11:NoData-B:bdt:loEff}   {\ensuremath{{0.883 } } }
\vdef{default-11:NoData-B:bdt:loEffE}   {\ensuremath{{0.001 } } }
\vdef{default-11:NoData-B:bdt:hiEff}   {\ensuremath{{0.117 } } }
\vdef{default-11:NoData-B:bdt:hiEffE}   {\ensuremath{{0.001 } } }
\vdef{default-11:NoMc-B:bdt:loEff}   {\ensuremath{{0.873 } } }
\vdef{default-11:NoMc-B:bdt:loEffE}   {\ensuremath{{0.001 } } }
\vdef{default-11:NoMc-B:bdt:hiEff}   {\ensuremath{{0.127 } } }
\vdef{default-11:NoMc-B:bdt:hiEffE}   {\ensuremath{{0.001 } } }
\vdef{default-11:NoMc-B:bdt:loDelta}   {\ensuremath{{+0.011 } } }
\vdef{default-11:NoMc-B:bdt:loDeltaE}   {\ensuremath{{0.002 } } }
\vdef{default-11:NoMc-B:bdt:hiDelta}   {\ensuremath{{-0.082 } } }
\vdef{default-11:NoMc-B:bdt:hiDeltaE}   {\ensuremath{{0.014 } } }
\vdef{default-11:NoData-B:fl3d:loEff}   {\ensuremath{{0.893 } } }
\vdef{default-11:NoData-B:fl3d:loEffE}   {\ensuremath{{0.001 } } }
\vdef{default-11:NoData-B:fl3d:hiEff}   {\ensuremath{{0.107 } } }
\vdef{default-11:NoData-B:fl3d:hiEffE}   {\ensuremath{{0.001 } } }
\vdef{default-11:NoMc-B:fl3d:loEff}   {\ensuremath{{0.895 } } }
\vdef{default-11:NoMc-B:fl3d:loEffE}   {\ensuremath{{0.001 } } }
\vdef{default-11:NoMc-B:fl3d:hiEff}   {\ensuremath{{0.105 } } }
\vdef{default-11:NoMc-B:fl3d:hiEffE}   {\ensuremath{{0.001 } } }
\vdef{default-11:NoMc-B:fl3d:loDelta}   {\ensuremath{{-0.002 } } }
\vdef{default-11:NoMc-B:fl3d:loDeltaE}   {\ensuremath{{0.002 } } }
\vdef{default-11:NoMc-B:fl3d:hiDelta}   {\ensuremath{{+0.015 } } }
\vdef{default-11:NoMc-B:fl3d:hiDeltaE}   {\ensuremath{{0.014 } } }
\vdef{default-11:NoData-B:fl3de:loEff}   {\ensuremath{{1.000 } } }
\vdef{default-11:NoData-B:fl3de:loEffE}   {\ensuremath{{0.000 } } }
\vdef{default-11:NoData-B:fl3de:hiEff}   {\ensuremath{{0.000 } } }
\vdef{default-11:NoData-B:fl3de:hiEffE}   {\ensuremath{{0.000 } } }
\vdef{default-11:NoMc-B:fl3de:loEff}   {\ensuremath{{1.000 } } }
\vdef{default-11:NoMc-B:fl3de:loEffE}   {\ensuremath{{0.000 } } }
\vdef{default-11:NoMc-B:fl3de:hiEff}   {\ensuremath{{0.000 } } }
\vdef{default-11:NoMc-B:fl3de:hiEffE}   {\ensuremath{{0.000 } } }
\vdef{default-11:NoMc-B:fl3de:loDelta}   {\ensuremath{{+0.000 } } }
\vdef{default-11:NoMc-B:fl3de:loDeltaE}   {\ensuremath{{0.000 } } }
\vdef{default-11:NoMc-B:fl3de:hiDelta}   {\ensuremath{{-2.000 } } }
\vdef{default-11:NoMc-B:fl3de:hiDeltaE}   {\ensuremath{{4.893 } } }
\vdef{default-11:NoData-B:fls3d:loEff}   {\ensuremath{{0.055 } } }
\vdef{default-11:NoData-B:fls3d:loEffE}   {\ensuremath{{0.001 } } }
\vdef{default-11:NoData-B:fls3d:hiEff}   {\ensuremath{{0.945 } } }
\vdef{default-11:NoData-B:fls3d:hiEffE}   {\ensuremath{{0.001 } } }
\vdef{default-11:NoMc-B:fls3d:loEff}   {\ensuremath{{0.057 } } }
\vdef{default-11:NoMc-B:fls3d:loEffE}   {\ensuremath{{0.001 } } }
\vdef{default-11:NoMc-B:fls3d:hiEff}   {\ensuremath{{0.943 } } }
\vdef{default-11:NoMc-B:fls3d:hiEffE}   {\ensuremath{{0.001 } } }
\vdef{default-11:NoMc-B:fls3d:loDelta}   {\ensuremath{{-0.034 } } }
\vdef{default-11:NoMc-B:fls3d:loDeltaE}   {\ensuremath{{0.021 } } }
\vdef{default-11:NoMc-B:fls3d:hiDelta}   {\ensuremath{{+0.002 } } }
\vdef{default-11:NoMc-B:fls3d:hiDeltaE}   {\ensuremath{{0.001 } } }
\vdef{default-11:NoData-B:flsxy:loEff}   {\ensuremath{{1.013 } } }
\vdef{default-11:NoData-B:flsxy:loEffE}   {\ensuremath{{\mathrm{NaN} } } }
\vdef{default-11:NoData-B:flsxy:hiEff}   {\ensuremath{{1.000 } } }
\vdef{default-11:NoData-B:flsxy:hiEffE}   {\ensuremath{{0.000 } } }
\vdef{default-11:NoMc-B:flsxy:loEff}   {\ensuremath{{1.013 } } }
\vdef{default-11:NoMc-B:flsxy:loEffE}   {\ensuremath{{\mathrm{NaN} } } }
\vdef{default-11:NoMc-B:flsxy:hiEff}   {\ensuremath{{1.000 } } }
\vdef{default-11:NoMc-B:flsxy:hiEffE}   {\ensuremath{{0.000 } } }
\vdef{default-11:NoMc-B:flsxy:loDelta}   {\ensuremath{{+0.001 } } }
\vdef{default-11:NoMc-B:flsxy:loDeltaE}   {\ensuremath{{\mathrm{NaN} } } }
\vdef{default-11:NoMc-B:flsxy:hiDelta}   {\ensuremath{{+0.000 } } }
\vdef{default-11:NoMc-B:flsxy:hiDeltaE}   {\ensuremath{{0.000 } } }
\vdef{default-11:NoData-B:chi2dof:loEff}   {\ensuremath{{0.932 } } }
\vdef{default-11:NoData-B:chi2dof:loEffE}   {\ensuremath{{0.001 } } }
\vdef{default-11:NoData-B:chi2dof:hiEff}   {\ensuremath{{0.068 } } }
\vdef{default-11:NoData-B:chi2dof:hiEffE}   {\ensuremath{{0.001 } } }
\vdef{default-11:NoMc-B:chi2dof:loEff}   {\ensuremath{{0.938 } } }
\vdef{default-11:NoMc-B:chi2dof:loEffE}   {\ensuremath{{0.001 } } }
\vdef{default-11:NoMc-B:chi2dof:hiEff}   {\ensuremath{{0.062 } } }
\vdef{default-11:NoMc-B:chi2dof:hiEffE}   {\ensuremath{{0.001 } } }
\vdef{default-11:NoMc-B:chi2dof:loDelta}   {\ensuremath{{-0.006 } } }
\vdef{default-11:NoMc-B:chi2dof:loDeltaE}   {\ensuremath{{0.001 } } }
\vdef{default-11:NoMc-B:chi2dof:hiDelta}   {\ensuremath{{+0.091 } } }
\vdef{default-11:NoMc-B:chi2dof:hiDeltaE}   {\ensuremath{{0.019 } } }
\vdef{default-11:NoData-B:pchi2dof:loEff}   {\ensuremath{{0.637 } } }
\vdef{default-11:NoData-B:pchi2dof:loEffE}   {\ensuremath{{0.002 } } }
\vdef{default-11:NoData-B:pchi2dof:hiEff}   {\ensuremath{{0.363 } } }
\vdef{default-11:NoData-B:pchi2dof:hiEffE}   {\ensuremath{{0.002 } } }
\vdef{default-11:NoMc-B:pchi2dof:loEff}   {\ensuremath{{0.624 } } }
\vdef{default-11:NoMc-B:pchi2dof:loEffE}   {\ensuremath{{0.002 } } }
\vdef{default-11:NoMc-B:pchi2dof:hiEff}   {\ensuremath{{0.376 } } }
\vdef{default-11:NoMc-B:pchi2dof:hiEffE}   {\ensuremath{{0.002 } } }
\vdef{default-11:NoMc-B:pchi2dof:loDelta}   {\ensuremath{{+0.020 } } }
\vdef{default-11:NoMc-B:pchi2dof:loDeltaE}   {\ensuremath{{0.004 } } }
\vdef{default-11:NoMc-B:pchi2dof:hiDelta}   {\ensuremath{{-0.034 } } }
\vdef{default-11:NoMc-B:pchi2dof:hiDeltaE}   {\ensuremath{{0.007 } } }
\vdef{default-11:NoData-B:alpha:loEff}   {\ensuremath{{0.993 } } }
\vdef{default-11:NoData-B:alpha:loEffE}   {\ensuremath{{0.000 } } }
\vdef{default-11:NoData-B:alpha:hiEff}   {\ensuremath{{0.007 } } }
\vdef{default-11:NoData-B:alpha:hiEffE}   {\ensuremath{{0.000 } } }
\vdef{default-11:NoMc-B:alpha:loEff}   {\ensuremath{{0.992 } } }
\vdef{default-11:NoMc-B:alpha:loEffE}   {\ensuremath{{0.000 } } }
\vdef{default-11:NoMc-B:alpha:hiEff}   {\ensuremath{{0.008 } } }
\vdef{default-11:NoMc-B:alpha:hiEffE}   {\ensuremath{{0.000 } } }
\vdef{default-11:NoMc-B:alpha:loDelta}   {\ensuremath{{+0.001 } } }
\vdef{default-11:NoMc-B:alpha:loDeltaE}   {\ensuremath{{0.000 } } }
\vdef{default-11:NoMc-B:alpha:hiDelta}   {\ensuremath{{-0.105 } } }
\vdef{default-11:NoMc-B:alpha:hiDeltaE}   {\ensuremath{{0.062 } } }
\vdef{default-11:NoData-B:iso:loEff}   {\ensuremath{{0.134 } } }
\vdef{default-11:NoData-B:iso:loEffE}   {\ensuremath{{0.001 } } }
\vdef{default-11:NoData-B:iso:hiEff}   {\ensuremath{{0.866 } } }
\vdef{default-11:NoData-B:iso:hiEffE}   {\ensuremath{{0.001 } } }
\vdef{default-11:NoMc-B:iso:loEff}   {\ensuremath{{0.112 } } }
\vdef{default-11:NoMc-B:iso:loEffE}   {\ensuremath{{0.001 } } }
\vdef{default-11:NoMc-B:iso:hiEff}   {\ensuremath{{0.888 } } }
\vdef{default-11:NoMc-B:iso:hiEffE}   {\ensuremath{{0.001 } } }
\vdef{default-11:NoMc-B:iso:loDelta}   {\ensuremath{{+0.174 } } }
\vdef{default-11:NoMc-B:iso:loDeltaE}   {\ensuremath{{0.013 } } }
\vdef{default-11:NoMc-B:iso:hiDelta}   {\ensuremath{{-0.025 } } }
\vdef{default-11:NoMc-B:iso:hiDeltaE}   {\ensuremath{{0.002 } } }
\vdef{default-11:NoData-B:docatrk:loEff}   {\ensuremath{{0.071 } } }
\vdef{default-11:NoData-B:docatrk:loEffE}   {\ensuremath{{0.001 } } }
\vdef{default-11:NoData-B:docatrk:hiEff}   {\ensuremath{{0.929 } } }
\vdef{default-11:NoData-B:docatrk:hiEffE}   {\ensuremath{{0.001 } } }
\vdef{default-11:NoMc-B:docatrk:loEff}   {\ensuremath{{0.085 } } }
\vdef{default-11:NoMc-B:docatrk:loEffE}   {\ensuremath{{0.001 } } }
\vdef{default-11:NoMc-B:docatrk:hiEff}   {\ensuremath{{0.915 } } }
\vdef{default-11:NoMc-B:docatrk:hiEffE}   {\ensuremath{{0.001 } } }
\vdef{default-11:NoMc-B:docatrk:loDelta}   {\ensuremath{{-0.185 } } }
\vdef{default-11:NoMc-B:docatrk:loDeltaE}   {\ensuremath{{0.018 } } }
\vdef{default-11:NoMc-B:docatrk:hiDelta}   {\ensuremath{{+0.016 } } }
\vdef{default-11:NoMc-B:docatrk:hiDeltaE}   {\ensuremath{{0.002 } } }
\vdef{default-11:NoData-B:isotrk:loEff}   {\ensuremath{{1.000 } } }
\vdef{default-11:NoData-B:isotrk:loEffE}   {\ensuremath{{0.000 } } }
\vdef{default-11:NoData-B:isotrk:hiEff}   {\ensuremath{{1.000 } } }
\vdef{default-11:NoData-B:isotrk:hiEffE}   {\ensuremath{{0.000 } } }
\vdef{default-11:NoMc-B:isotrk:loEff}   {\ensuremath{{1.000 } } }
\vdef{default-11:NoMc-B:isotrk:loEffE}   {\ensuremath{{0.000 } } }
\vdef{default-11:NoMc-B:isotrk:hiEff}   {\ensuremath{{1.000 } } }
\vdef{default-11:NoMc-B:isotrk:hiEffE}   {\ensuremath{{0.000 } } }
\vdef{default-11:NoMc-B:isotrk:loDelta}   {\ensuremath{{+0.000 } } }
\vdef{default-11:NoMc-B:isotrk:loDeltaE}   {\ensuremath{{0.000 } } }
\vdef{default-11:NoMc-B:isotrk:hiDelta}   {\ensuremath{{+0.000 } } }
\vdef{default-11:NoMc-B:isotrk:hiDeltaE}   {\ensuremath{{0.000 } } }
\vdef{default-11:NoData-B:closetrk:loEff}   {\ensuremath{{0.975 } } }
\vdef{default-11:NoData-B:closetrk:loEffE}   {\ensuremath{{0.001 } } }
\vdef{default-11:NoData-B:closetrk:hiEff}   {\ensuremath{{0.025 } } }
\vdef{default-11:NoData-B:closetrk:hiEffE}   {\ensuremath{{0.001 } } }
\vdef{default-11:NoMc-B:closetrk:loEff}   {\ensuremath{{0.977 } } }
\vdef{default-11:NoMc-B:closetrk:loEffE}   {\ensuremath{{0.001 } } }
\vdef{default-11:NoMc-B:closetrk:hiEff}   {\ensuremath{{0.023 } } }
\vdef{default-11:NoMc-B:closetrk:hiEffE}   {\ensuremath{{0.001 } } }
\vdef{default-11:NoMc-B:closetrk:loDelta}   {\ensuremath{{-0.002 } } }
\vdef{default-11:NoMc-B:closetrk:loDeltaE}   {\ensuremath{{0.001 } } }
\vdef{default-11:NoMc-B:closetrk:hiDelta}   {\ensuremath{{+0.098 } } }
\vdef{default-11:NoMc-B:closetrk:hiDeltaE}   {\ensuremath{{0.033 } } }
\vdef{default-11:NoData-B:lip:loEff}   {\ensuremath{{1.000 } } }
\vdef{default-11:NoData-B:lip:loEffE}   {\ensuremath{{0.000 } } }
\vdef{default-11:NoData-B:lip:hiEff}   {\ensuremath{{0.000 } } }
\vdef{default-11:NoData-B:lip:hiEffE}   {\ensuremath{{0.000 } } }
\vdef{default-11:NoMc-B:lip:loEff}   {\ensuremath{{1.000 } } }
\vdef{default-11:NoMc-B:lip:loEffE}   {\ensuremath{{0.000 } } }
\vdef{default-11:NoMc-B:lip:hiEff}   {\ensuremath{{0.000 } } }
\vdef{default-11:NoMc-B:lip:hiEffE}   {\ensuremath{{0.000 } } }
\vdef{default-11:NoMc-B:lip:loDelta}   {\ensuremath{{+0.000 } } }
\vdef{default-11:NoMc-B:lip:loDeltaE}   {\ensuremath{{0.000 } } }
\vdef{default-11:NoMc-B:lip:hiDelta}   {\ensuremath{{\mathrm{NaN} } } }
\vdef{default-11:NoMc-B:lip:hiDeltaE}   {\ensuremath{{\mathrm{NaN} } } }
\vdef{default-11:NoData-B:lip:inEff}   {\ensuremath{{1.000 } } }
\vdef{default-11:NoData-B:lip:inEffE}   {\ensuremath{{0.000 } } }
\vdef{default-11:NoMc-B:lip:inEff}   {\ensuremath{{1.000 } } }
\vdef{default-11:NoMc-B:lip:inEffE}   {\ensuremath{{0.000 } } }
\vdef{default-11:NoMc-B:lip:inDelta}   {\ensuremath{{+0.000 } } }
\vdef{default-11:NoMc-B:lip:inDeltaE}   {\ensuremath{{0.000 } } }
\vdef{default-11:NoData-B:lips:loEff}   {\ensuremath{{1.000 } } }
\vdef{default-11:NoData-B:lips:loEffE}   {\ensuremath{{0.000 } } }
\vdef{default-11:NoData-B:lips:hiEff}   {\ensuremath{{0.000 } } }
\vdef{default-11:NoData-B:lips:hiEffE}   {\ensuremath{{0.000 } } }
\vdef{default-11:NoMc-B:lips:loEff}   {\ensuremath{{1.000 } } }
\vdef{default-11:NoMc-B:lips:loEffE}   {\ensuremath{{0.000 } } }
\vdef{default-11:NoMc-B:lips:hiEff}   {\ensuremath{{0.000 } } }
\vdef{default-11:NoMc-B:lips:hiEffE}   {\ensuremath{{0.000 } } }
\vdef{default-11:NoMc-B:lips:loDelta}   {\ensuremath{{+0.000 } } }
\vdef{default-11:NoMc-B:lips:loDeltaE}   {\ensuremath{{0.000 } } }
\vdef{default-11:NoMc-B:lips:hiDelta}   {\ensuremath{{\mathrm{NaN} } } }
\vdef{default-11:NoMc-B:lips:hiDeltaE}   {\ensuremath{{\mathrm{NaN} } } }
\vdef{default-11:NoData-B:lips:inEff}   {\ensuremath{{1.000 } } }
\vdef{default-11:NoData-B:lips:inEffE}   {\ensuremath{{0.000 } } }
\vdef{default-11:NoMc-B:lips:inEff}   {\ensuremath{{1.000 } } }
\vdef{default-11:NoMc-B:lips:inEffE}   {\ensuremath{{0.000 } } }
\vdef{default-11:NoMc-B:lips:inDelta}   {\ensuremath{{+0.000 } } }
\vdef{default-11:NoMc-B:lips:inDeltaE}   {\ensuremath{{0.000 } } }
\vdef{default-11:NoData-B:ip:loEff}   {\ensuremath{{0.971 } } }
\vdef{default-11:NoData-B:ip:loEffE}   {\ensuremath{{0.001 } } }
\vdef{default-11:NoData-B:ip:hiEff}   {\ensuremath{{0.029 } } }
\vdef{default-11:NoData-B:ip:hiEffE}   {\ensuremath{{0.001 } } }
\vdef{default-11:NoMc-B:ip:loEff}   {\ensuremath{{0.970 } } }
\vdef{default-11:NoMc-B:ip:loEffE}   {\ensuremath{{0.001 } } }
\vdef{default-11:NoMc-B:ip:hiEff}   {\ensuremath{{0.030 } } }
\vdef{default-11:NoMc-B:ip:hiEffE}   {\ensuremath{{0.001 } } }
\vdef{default-11:NoMc-B:ip:loDelta}   {\ensuremath{{+0.001 } } }
\vdef{default-11:NoMc-B:ip:loDeltaE}   {\ensuremath{{0.001 } } }
\vdef{default-11:NoMc-B:ip:hiDelta}   {\ensuremath{{-0.026 } } }
\vdef{default-11:NoMc-B:ip:hiDeltaE}   {\ensuremath{{0.030 } } }
\vdef{default-11:NoData-B:ips:loEff}   {\ensuremath{{0.937 } } }
\vdef{default-11:NoData-B:ips:loEffE}   {\ensuremath{{0.001 } } }
\vdef{default-11:NoData-B:ips:hiEff}   {\ensuremath{{0.063 } } }
\vdef{default-11:NoData-B:ips:hiEffE}   {\ensuremath{{0.001 } } }
\vdef{default-11:NoMc-B:ips:loEff}   {\ensuremath{{0.953 } } }
\vdef{default-11:NoMc-B:ips:loEffE}   {\ensuremath{{0.001 } } }
\vdef{default-11:NoMc-B:ips:hiEff}   {\ensuremath{{0.047 } } }
\vdef{default-11:NoMc-B:ips:hiEffE}   {\ensuremath{{0.001 } } }
\vdef{default-11:NoMc-B:ips:loDelta}   {\ensuremath{{-0.017 } } }
\vdef{default-11:NoMc-B:ips:loDeltaE}   {\ensuremath{{0.001 } } }
\vdef{default-11:NoMc-B:ips:hiDelta}   {\ensuremath{{+0.292 } } }
\vdef{default-11:NoMc-B:ips:hiDeltaE}   {\ensuremath{{0.021 } } }
\vdef{default-11:NoData-B:maxdoca:loEff}   {\ensuremath{{1.000 } } }
\vdef{default-11:NoData-B:maxdoca:loEffE}   {\ensuremath{{0.000 } } }
\vdef{default-11:NoData-B:maxdoca:hiEff}   {\ensuremath{{0.012 } } }
\vdef{default-11:NoData-B:maxdoca:hiEffE}   {\ensuremath{{0.000 } } }
\vdef{default-11:NoMc-B:maxdoca:loEff}   {\ensuremath{{1.000 } } }
\vdef{default-11:NoMc-B:maxdoca:loEffE}   {\ensuremath{{0.000 } } }
\vdef{default-11:NoMc-B:maxdoca:hiEff}   {\ensuremath{{0.010 } } }
\vdef{default-11:NoMc-B:maxdoca:hiEffE}   {\ensuremath{{0.000 } } }
\vdef{default-11:NoMc-B:maxdoca:loDelta}   {\ensuremath{{+0.000 } } }
\vdef{default-11:NoMc-B:maxdoca:loDeltaE}   {\ensuremath{{0.000 } } }
\vdef{default-11:NoMc-B:maxdoca:hiDelta}   {\ensuremath{{+0.173 } } }
\vdef{default-11:NoMc-B:maxdoca:hiDeltaE}   {\ensuremath{{0.050 } } }
\vdef{default-11:NoData-B:kaonpt:loEff}   {\ensuremath{{1.012 } } }
\vdef{default-11:NoData-B:kaonpt:loEffE}   {\ensuremath{{\mathrm{NaN} } } }
\vdef{default-11:NoData-B:kaonpt:hiEff}   {\ensuremath{{1.000 } } }
\vdef{default-11:NoData-B:kaonpt:hiEffE}   {\ensuremath{{0.000 } } }
\vdef{default-11:NoMc-B:kaonpt:loEff}   {\ensuremath{{1.008 } } }
\vdef{default-11:NoMc-B:kaonpt:loEffE}   {\ensuremath{{\mathrm{NaN} } } }
\vdef{default-11:NoMc-B:kaonpt:hiEff}   {\ensuremath{{1.000 } } }
\vdef{default-11:NoMc-B:kaonpt:hiEffE}   {\ensuremath{{0.000 } } }
\vdef{default-11:NoMc-B:kaonpt:loDelta}   {\ensuremath{{+0.003 } } }
\vdef{default-11:NoMc-B:kaonpt:loDeltaE}   {\ensuremath{{\mathrm{NaN} } } }
\vdef{default-11:NoMc-B:kaonpt:hiDelta}   {\ensuremath{{+0.000 } } }
\vdef{default-11:NoMc-B:kaonpt:hiDeltaE}   {\ensuremath{{0.000 } } }
\vdef{default-11:NoData-B:psipt:loEff}   {\ensuremath{{1.005 } } }
\vdef{default-11:NoData-B:psipt:loEffE}   {\ensuremath{{\mathrm{NaN} } } }
\vdef{default-11:NoData-B:psipt:hiEff}   {\ensuremath{{1.000 } } }
\vdef{default-11:NoData-B:psipt:hiEffE}   {\ensuremath{{0.000 } } }
\vdef{default-11:NoMc-B:psipt:loEff}   {\ensuremath{{1.003 } } }
\vdef{default-11:NoMc-B:psipt:loEffE}   {\ensuremath{{\mathrm{NaN} } } }
\vdef{default-11:NoMc-B:psipt:hiEff}   {\ensuremath{{1.000 } } }
\vdef{default-11:NoMc-B:psipt:hiEffE}   {\ensuremath{{0.000 } } }
\vdef{default-11:NoMc-B:psipt:loDelta}   {\ensuremath{{+0.001 } } }
\vdef{default-11:NoMc-B:psipt:loDeltaE}   {\ensuremath{{\mathrm{NaN} } } }
\vdef{default-11:NoMc-B:psipt:hiDelta}   {\ensuremath{{+0.000 } } }
\vdef{default-11:NoMc-B:psipt:hiDeltaE}   {\ensuremath{{0.000 } } }
\vdef{default-11:NoData-E:osiso:loEff}   {\ensuremath{{1.003 } } }
\vdef{default-11:NoData-E:osiso:loEffE}   {\ensuremath{{\mathrm{NaN} } } }
\vdef{default-11:NoData-E:osiso:hiEff}   {\ensuremath{{1.000 } } }
\vdef{default-11:NoData-E:osiso:hiEffE}   {\ensuremath{{0.000 } } }
\vdef{default-11:NoMc-E:osiso:loEff}   {\ensuremath{{1.002 } } }
\vdef{default-11:NoMc-E:osiso:loEffE}   {\ensuremath{{\mathrm{NaN} } } }
\vdef{default-11:NoMc-E:osiso:hiEff}   {\ensuremath{{1.000 } } }
\vdef{default-11:NoMc-E:osiso:hiEffE}   {\ensuremath{{0.000 } } }
\vdef{default-11:NoMc-E:osiso:loDelta}   {\ensuremath{{+0.000 } } }
\vdef{default-11:NoMc-E:osiso:loDeltaE}   {\ensuremath{{\mathrm{NaN} } } }
\vdef{default-11:NoMc-E:osiso:hiDelta}   {\ensuremath{{+0.000 } } }
\vdef{default-11:NoMc-E:osiso:hiDeltaE}   {\ensuremath{{0.000 } } }
\vdef{default-11:NoData-E:osreliso:loEff}   {\ensuremath{{0.247 } } }
\vdef{default-11:NoData-E:osreliso:loEffE}   {\ensuremath{{0.003 } } }
\vdef{default-11:NoData-E:osreliso:hiEff}   {\ensuremath{{0.753 } } }
\vdef{default-11:NoData-E:osreliso:hiEffE}   {\ensuremath{{0.003 } } }
\vdef{default-11:NoMc-E:osreliso:loEff}   {\ensuremath{{0.292 } } }
\vdef{default-11:NoMc-E:osreliso:loEffE}   {\ensuremath{{0.003 } } }
\vdef{default-11:NoMc-E:osreliso:hiEff}   {\ensuremath{{0.708 } } }
\vdef{default-11:NoMc-E:osreliso:hiEffE}   {\ensuremath{{0.003 } } }
\vdef{default-11:NoMc-E:osreliso:loDelta}   {\ensuremath{{-0.167 } } }
\vdef{default-11:NoMc-E:osreliso:loDeltaE}   {\ensuremath{{0.014 } } }
\vdef{default-11:NoMc-E:osreliso:hiDelta}   {\ensuremath{{+0.062 } } }
\vdef{default-11:NoMc-E:osreliso:hiDeltaE}   {\ensuremath{{0.005 } } }
\vdef{default-11:NoData-E:osmuonpt:loEff}   {\ensuremath{{0.000 } } }
\vdef{default-11:NoData-E:osmuonpt:loEffE}   {\ensuremath{{0.001 } } }
\vdef{default-11:NoData-E:osmuonpt:hiEff}   {\ensuremath{{1.000 } } }
\vdef{default-11:NoData-E:osmuonpt:hiEffE}   {\ensuremath{{0.001 } } }
\vdef{default-11:NoMc-E:osmuonpt:loEff}   {\ensuremath{{0.000 } } }
\vdef{default-11:NoMc-E:osmuonpt:loEffE}   {\ensuremath{{0.001 } } }
\vdef{default-11:NoMc-E:osmuonpt:hiEff}   {\ensuremath{{1.000 } } }
\vdef{default-11:NoMc-E:osmuonpt:hiEffE}   {\ensuremath{{0.001 } } }
\vdef{default-11:NoMc-E:osmuonpt:loDelta}   {\ensuremath{{\mathrm{NaN} } } }
\vdef{default-11:NoMc-E:osmuonpt:loDeltaE}   {\ensuremath{{\mathrm{NaN} } } }
\vdef{default-11:NoMc-E:osmuonpt:hiDelta}   {\ensuremath{{+0.000 } } }
\vdef{default-11:NoMc-E:osmuonpt:hiDeltaE}   {\ensuremath{{0.002 } } }
\vdef{default-11:NoData-E:osmuondr:loEff}   {\ensuremath{{0.036 } } }
\vdef{default-11:NoData-E:osmuondr:loEffE}   {\ensuremath{{0.006 } } }
\vdef{default-11:NoData-E:osmuondr:hiEff}   {\ensuremath{{0.964 } } }
\vdef{default-11:NoData-E:osmuondr:hiEffE}   {\ensuremath{{0.006 } } }
\vdef{default-11:NoMc-E:osmuondr:loEff}   {\ensuremath{{0.020 } } }
\vdef{default-11:NoMc-E:osmuondr:loEffE}   {\ensuremath{{0.005 } } }
\vdef{default-11:NoMc-E:osmuondr:hiEff}   {\ensuremath{{0.980 } } }
\vdef{default-11:NoMc-E:osmuondr:hiEffE}   {\ensuremath{{0.005 } } }
\vdef{default-11:NoMc-E:osmuondr:loDelta}   {\ensuremath{{+0.588 } } }
\vdef{default-11:NoMc-E:osmuondr:loDeltaE}   {\ensuremath{{0.266 } } }
\vdef{default-11:NoMc-E:osmuondr:hiDelta}   {\ensuremath{{-0.017 } } }
\vdef{default-11:NoMc-E:osmuondr:hiDeltaE}   {\ensuremath{{0.008 } } }
\vdef{default-11:NoData-E:hlt:loEff}   {\ensuremath{{0.034 } } }
\vdef{default-11:NoData-E:hlt:loEffE}   {\ensuremath{{0.001 } } }
\vdef{default-11:NoData-E:hlt:hiEff}   {\ensuremath{{0.966 } } }
\vdef{default-11:NoData-E:hlt:hiEffE}   {\ensuremath{{0.001 } } }
\vdef{default-11:NoMc-E:hlt:loEff}   {\ensuremath{{0.429 } } }
\vdef{default-11:NoMc-E:hlt:loEffE}   {\ensuremath{{0.003 } } }
\vdef{default-11:NoMc-E:hlt:hiEff}   {\ensuremath{{0.571 } } }
\vdef{default-11:NoMc-E:hlt:hiEffE}   {\ensuremath{{0.003 } } }
\vdef{default-11:NoMc-E:hlt:loDelta}   {\ensuremath{{-1.707 } } }
\vdef{default-11:NoMc-E:hlt:loDeltaE}   {\ensuremath{{0.009 } } }
\vdef{default-11:NoMc-E:hlt:hiDelta}   {\ensuremath{{+0.514 } } }
\vdef{default-11:NoMc-E:hlt:hiDeltaE}   {\ensuremath{{0.005 } } }
\vdef{default-11:NoData-E:muonsid:loEff}   {\ensuremath{{0.148 } } }
\vdef{default-11:NoData-E:muonsid:loEffE}   {\ensuremath{{0.002 } } }
\vdef{default-11:NoData-E:muonsid:hiEff}   {\ensuremath{{0.852 } } }
\vdef{default-11:NoData-E:muonsid:hiEffE}   {\ensuremath{{0.002 } } }
\vdef{default-11:NoMc-E:muonsid:loEff}   {\ensuremath{{0.139 } } }
\vdef{default-11:NoMc-E:muonsid:loEffE}   {\ensuremath{{0.002 } } }
\vdef{default-11:NoMc-E:muonsid:hiEff}   {\ensuremath{{0.861 } } }
\vdef{default-11:NoMc-E:muonsid:hiEffE}   {\ensuremath{{0.002 } } }
\vdef{default-11:NoMc-E:muonsid:loDelta}   {\ensuremath{{+0.061 } } }
\vdef{default-11:NoMc-E:muonsid:loDeltaE}   {\ensuremath{{0.021 } } }
\vdef{default-11:NoMc-E:muonsid:hiDelta}   {\ensuremath{{-0.010 } } }
\vdef{default-11:NoMc-E:muonsid:hiDeltaE}   {\ensuremath{{0.003 } } }
\vdef{default-11:NoData-E:tracksqual:loEff}   {\ensuremath{{0.001 } } }
\vdef{default-11:NoData-E:tracksqual:loEffE}   {\ensuremath{{0.000 } } }
\vdef{default-11:NoData-E:tracksqual:hiEff}   {\ensuremath{{0.999 } } }
\vdef{default-11:NoData-E:tracksqual:hiEffE}   {\ensuremath{{0.000 } } }
\vdef{default-11:NoMc-E:tracksqual:loEff}   {\ensuremath{{0.000 } } }
\vdef{default-11:NoMc-E:tracksqual:loEffE}   {\ensuremath{{0.000 } } }
\vdef{default-11:NoMc-E:tracksqual:hiEff}   {\ensuremath{{1.000 } } }
\vdef{default-11:NoMc-E:tracksqual:hiEffE}   {\ensuremath{{0.000 } } }
\vdef{default-11:NoMc-E:tracksqual:loDelta}   {\ensuremath{{+1.734 } } }
\vdef{default-11:NoMc-E:tracksqual:loDeltaE}   {\ensuremath{{0.241 } } }
\vdef{default-11:NoMc-E:tracksqual:hiDelta}   {\ensuremath{{-0.001 } } }
\vdef{default-11:NoMc-E:tracksqual:hiDeltaE}   {\ensuremath{{0.000 } } }
\vdef{default-11:NoData-E:pvz:loEff}   {\ensuremath{{0.506 } } }
\vdef{default-11:NoData-E:pvz:loEffE}   {\ensuremath{{0.003 } } }
\vdef{default-11:NoData-E:pvz:hiEff}   {\ensuremath{{0.494 } } }
\vdef{default-11:NoData-E:pvz:hiEffE}   {\ensuremath{{0.003 } } }
\vdef{default-11:NoMc-E:pvz:loEff}   {\ensuremath{{0.466 } } }
\vdef{default-11:NoMc-E:pvz:loEffE}   {\ensuremath{{0.003 } } }
\vdef{default-11:NoMc-E:pvz:hiEff}   {\ensuremath{{0.534 } } }
\vdef{default-11:NoMc-E:pvz:hiEffE}   {\ensuremath{{0.003 } } }
\vdef{default-11:NoMc-E:pvz:loDelta}   {\ensuremath{{+0.082 } } }
\vdef{default-11:NoMc-E:pvz:loDeltaE}   {\ensuremath{{0.009 } } }
\vdef{default-11:NoMc-E:pvz:hiDelta}   {\ensuremath{{-0.077 } } }
\vdef{default-11:NoMc-E:pvz:hiDeltaE}   {\ensuremath{{0.009 } } }
\vdef{default-11:NoData-E:pvn:loEff}   {\ensuremath{{1.007 } } }
\vdef{default-11:NoData-E:pvn:loEffE}   {\ensuremath{{\mathrm{NaN} } } }
\vdef{default-11:NoData-E:pvn:hiEff}   {\ensuremath{{1.000 } } }
\vdef{default-11:NoData-E:pvn:hiEffE}   {\ensuremath{{0.000 } } }
\vdef{default-11:NoMc-E:pvn:loEff}   {\ensuremath{{1.000 } } }
\vdef{default-11:NoMc-E:pvn:loEffE}   {\ensuremath{{0.000 } } }
\vdef{default-11:NoMc-E:pvn:hiEff}   {\ensuremath{{1.000 } } }
\vdef{default-11:NoMc-E:pvn:hiEffE}   {\ensuremath{{0.000 } } }
\vdef{default-11:NoMc-E:pvn:loDelta}   {\ensuremath{{+0.007 } } }
\vdef{default-11:NoMc-E:pvn:loDeltaE}   {\ensuremath{{\mathrm{NaN} } } }
\vdef{default-11:NoMc-E:pvn:hiDelta}   {\ensuremath{{+0.000 } } }
\vdef{default-11:NoMc-E:pvn:hiDeltaE}   {\ensuremath{{0.000 } } }
\vdef{default-11:NoData-E:pvavew8:loEff}   {\ensuremath{{0.013 } } }
\vdef{default-11:NoData-E:pvavew8:loEffE}   {\ensuremath{{0.001 } } }
\vdef{default-11:NoData-E:pvavew8:hiEff}   {\ensuremath{{0.987 } } }
\vdef{default-11:NoData-E:pvavew8:hiEffE}   {\ensuremath{{0.001 } } }
\vdef{default-11:NoMc-E:pvavew8:loEff}   {\ensuremath{{0.009 } } }
\vdef{default-11:NoMc-E:pvavew8:loEffE}   {\ensuremath{{0.001 } } }
\vdef{default-11:NoMc-E:pvavew8:hiEff}   {\ensuremath{{0.991 } } }
\vdef{default-11:NoMc-E:pvavew8:hiEffE}   {\ensuremath{{0.001 } } }
\vdef{default-11:NoMc-E:pvavew8:loDelta}   {\ensuremath{{+0.373 } } }
\vdef{default-11:NoMc-E:pvavew8:loDeltaE}   {\ensuremath{{0.086 } } }
\vdef{default-11:NoMc-E:pvavew8:hiDelta}   {\ensuremath{{-0.004 } } }
\vdef{default-11:NoMc-E:pvavew8:hiDeltaE}   {\ensuremath{{0.001 } } }
\vdef{default-11:NoData-E:pvntrk:loEff}   {\ensuremath{{1.000 } } }
\vdef{default-11:NoData-E:pvntrk:loEffE}   {\ensuremath{{0.000 } } }
\vdef{default-11:NoData-E:pvntrk:hiEff}   {\ensuremath{{1.000 } } }
\vdef{default-11:NoData-E:pvntrk:hiEffE}   {\ensuremath{{0.000 } } }
\vdef{default-11:NoMc-E:pvntrk:loEff}   {\ensuremath{{1.000 } } }
\vdef{default-11:NoMc-E:pvntrk:loEffE}   {\ensuremath{{0.000 } } }
\vdef{default-11:NoMc-E:pvntrk:hiEff}   {\ensuremath{{1.000 } } }
\vdef{default-11:NoMc-E:pvntrk:hiEffE}   {\ensuremath{{0.000 } } }
\vdef{default-11:NoMc-E:pvntrk:loDelta}   {\ensuremath{{+0.000 } } }
\vdef{default-11:NoMc-E:pvntrk:loDeltaE}   {\ensuremath{{0.000 } } }
\vdef{default-11:NoMc-E:pvntrk:hiDelta}   {\ensuremath{{+0.000 } } }
\vdef{default-11:NoMc-E:pvntrk:hiDeltaE}   {\ensuremath{{0.000 } } }
\vdef{default-11:NoData-E:muon1pt:loEff}   {\ensuremath{{1.003 } } }
\vdef{default-11:NoData-E:muon1pt:loEffE}   {\ensuremath{{\mathrm{NaN} } } }
\vdef{default-11:NoData-E:muon1pt:hiEff}   {\ensuremath{{1.000 } } }
\vdef{default-11:NoData-E:muon1pt:hiEffE}   {\ensuremath{{0.000 } } }
\vdef{default-11:NoMc-E:muon1pt:loEff}   {\ensuremath{{1.004 } } }
\vdef{default-11:NoMc-E:muon1pt:loEffE}   {\ensuremath{{\mathrm{NaN} } } }
\vdef{default-11:NoMc-E:muon1pt:hiEff}   {\ensuremath{{1.000 } } }
\vdef{default-11:NoMc-E:muon1pt:hiEffE}   {\ensuremath{{0.000 } } }
\vdef{default-11:NoMc-E:muon1pt:loDelta}   {\ensuremath{{-0.000 } } }
\vdef{default-11:NoMc-E:muon1pt:loDeltaE}   {\ensuremath{{\mathrm{NaN} } } }
\vdef{default-11:NoMc-E:muon1pt:hiDelta}   {\ensuremath{{+0.000 } } }
\vdef{default-11:NoMc-E:muon1pt:hiDeltaE}   {\ensuremath{{0.000 } } }
\vdef{default-11:NoData-E:muon2pt:loEff}   {\ensuremath{{0.151 } } }
\vdef{default-11:NoData-E:muon2pt:loEffE}   {\ensuremath{{0.002 } } }
\vdef{default-11:NoData-E:muon2pt:hiEff}   {\ensuremath{{0.849 } } }
\vdef{default-11:NoData-E:muon2pt:hiEffE}   {\ensuremath{{0.002 } } }
\vdef{default-11:NoMc-E:muon2pt:loEff}   {\ensuremath{{0.143 } } }
\vdef{default-11:NoMc-E:muon2pt:loEffE}   {\ensuremath{{0.002 } } }
\vdef{default-11:NoMc-E:muon2pt:hiEff}   {\ensuremath{{0.857 } } }
\vdef{default-11:NoMc-E:muon2pt:hiEffE}   {\ensuremath{{0.002 } } }
\vdef{default-11:NoMc-E:muon2pt:loDelta}   {\ensuremath{{+0.054 } } }
\vdef{default-11:NoMc-E:muon2pt:loDeltaE}   {\ensuremath{{0.020 } } }
\vdef{default-11:NoMc-E:muon2pt:hiDelta}   {\ensuremath{{-0.009 } } }
\vdef{default-11:NoMc-E:muon2pt:hiDeltaE}   {\ensuremath{{0.004 } } }
\vdef{default-11:NoData-E:muonseta:loEff}   {\ensuremath{{0.510 } } }
\vdef{default-11:NoData-E:muonseta:loEffE}   {\ensuremath{{0.002 } } }
\vdef{default-11:NoData-E:muonseta:hiEff}   {\ensuremath{{0.490 } } }
\vdef{default-11:NoData-E:muonseta:hiEffE}   {\ensuremath{{0.002 } } }
\vdef{default-11:NoMc-E:muonseta:loEff}   {\ensuremath{{0.510 } } }
\vdef{default-11:NoMc-E:muonseta:loEffE}   {\ensuremath{{0.002 } } }
\vdef{default-11:NoMc-E:muonseta:hiEff}   {\ensuremath{{0.490 } } }
\vdef{default-11:NoMc-E:muonseta:hiEffE}   {\ensuremath{{0.002 } } }
\vdef{default-11:NoMc-E:muonseta:loDelta}   {\ensuremath{{-0.001 } } }
\vdef{default-11:NoMc-E:muonseta:loDeltaE}   {\ensuremath{{0.006 } } }
\vdef{default-11:NoMc-E:muonseta:hiDelta}   {\ensuremath{{+0.001 } } }
\vdef{default-11:NoMc-E:muonseta:hiDeltaE}   {\ensuremath{{0.007 } } }
\vdef{default-11:NoData-E:pt:loEff}   {\ensuremath{{0.000 } } }
\vdef{default-11:NoData-E:pt:loEffE}   {\ensuremath{{0.000 } } }
\vdef{default-11:NoData-E:pt:hiEff}   {\ensuremath{{1.000 } } }
\vdef{default-11:NoData-E:pt:hiEffE}   {\ensuremath{{0.000 } } }
\vdef{default-11:NoMc-E:pt:loEff}   {\ensuremath{{0.000 } } }
\vdef{default-11:NoMc-E:pt:loEffE}   {\ensuremath{{0.000 } } }
\vdef{default-11:NoMc-E:pt:hiEff}   {\ensuremath{{1.000 } } }
\vdef{default-11:NoMc-E:pt:hiEffE}   {\ensuremath{{0.000 } } }
\vdef{default-11:NoMc-E:pt:loDelta}   {\ensuremath{{\mathrm{NaN} } } }
\vdef{default-11:NoMc-E:pt:loDeltaE}   {\ensuremath{{\mathrm{NaN} } } }
\vdef{default-11:NoMc-E:pt:hiDelta}   {\ensuremath{{+0.000 } } }
\vdef{default-11:NoMc-E:pt:hiDeltaE}   {\ensuremath{{0.000 } } }
\vdef{default-11:NoData-E:p:loEff}   {\ensuremath{{1.042 } } }
\vdef{default-11:NoData-E:p:loEffE}   {\ensuremath{{\mathrm{NaN} } } }
\vdef{default-11:NoData-E:p:hiEff}   {\ensuremath{{1.000 } } }
\vdef{default-11:NoData-E:p:hiEffE}   {\ensuremath{{0.000 } } }
\vdef{default-11:NoMc-E:p:loEff}   {\ensuremath{{1.043 } } }
\vdef{default-11:NoMc-E:p:loEffE}   {\ensuremath{{\mathrm{NaN} } } }
\vdef{default-11:NoMc-E:p:hiEff}   {\ensuremath{{1.000 } } }
\vdef{default-11:NoMc-E:p:hiEffE}   {\ensuremath{{0.000 } } }
\vdef{default-11:NoMc-E:p:loDelta}   {\ensuremath{{-0.001 } } }
\vdef{default-11:NoMc-E:p:loDeltaE}   {\ensuremath{{\mathrm{NaN} } } }
\vdef{default-11:NoMc-E:p:hiDelta}   {\ensuremath{{+0.000 } } }
\vdef{default-11:NoMc-E:p:hiDeltaE}   {\ensuremath{{0.000 } } }
\vdef{default-11:NoData-E:eta:loEff}   {\ensuremath{{0.510 } } }
\vdef{default-11:NoData-E:eta:loEffE}   {\ensuremath{{0.003 } } }
\vdef{default-11:NoData-E:eta:hiEff}   {\ensuremath{{0.490 } } }
\vdef{default-11:NoData-E:eta:hiEffE}   {\ensuremath{{0.003 } } }
\vdef{default-11:NoMc-E:eta:loEff}   {\ensuremath{{0.510 } } }
\vdef{default-11:NoMc-E:eta:loEffE}   {\ensuremath{{0.003 } } }
\vdef{default-11:NoMc-E:eta:hiEff}   {\ensuremath{{0.490 } } }
\vdef{default-11:NoMc-E:eta:hiEffE}   {\ensuremath{{0.003 } } }
\vdef{default-11:NoMc-E:eta:loDelta}   {\ensuremath{{-0.001 } } }
\vdef{default-11:NoMc-E:eta:loDeltaE}   {\ensuremath{{0.009 } } }
\vdef{default-11:NoMc-E:eta:hiDelta}   {\ensuremath{{+0.001 } } }
\vdef{default-11:NoMc-E:eta:hiDeltaE}   {\ensuremath{{0.009 } } }
\vdef{default-11:NoData-E:bdt:loEff}   {\ensuremath{{1.000 } } }
\vdef{default-11:NoData-E:bdt:loEffE}   {\ensuremath{{0.000 } } }
\vdef{default-11:NoData-E:bdt:hiEff}   {\ensuremath{{0.000 } } }
\vdef{default-11:NoData-E:bdt:hiEffE}   {\ensuremath{{0.000 } } }
\vdef{default-11:NoMc-E:bdt:loEff}   {\ensuremath{{1.000 } } }
\vdef{default-11:NoMc-E:bdt:loEffE}   {\ensuremath{{0.000 } } }
\vdef{default-11:NoMc-E:bdt:hiEff}   {\ensuremath{{0.000 } } }
\vdef{default-11:NoMc-E:bdt:hiEffE}   {\ensuremath{{0.000 } } }
\vdef{default-11:NoMc-E:bdt:loDelta}   {\ensuremath{{+0.000 } } }
\vdef{default-11:NoMc-E:bdt:loDeltaE}   {\ensuremath{{0.000 } } }
\vdef{default-11:NoMc-E:bdt:hiDelta}   {\ensuremath{{-0.112 } } }
\vdef{default-11:NoMc-E:bdt:hiDeltaE}   {\ensuremath{{0.442 } } }
\vdef{default-11:NoData-E:fl3d:loEff}   {\ensuremath{{0.664 } } }
\vdef{default-11:NoData-E:fl3d:loEffE}   {\ensuremath{{0.003 } } }
\vdef{default-11:NoData-E:fl3d:hiEff}   {\ensuremath{{0.336 } } }
\vdef{default-11:NoData-E:fl3d:hiEffE}   {\ensuremath{{0.003 } } }
\vdef{default-11:NoMc-E:fl3d:loEff}   {\ensuremath{{0.662 } } }
\vdef{default-11:NoMc-E:fl3d:loEffE}   {\ensuremath{{0.003 } } }
\vdef{default-11:NoMc-E:fl3d:hiEff}   {\ensuremath{{0.338 } } }
\vdef{default-11:NoMc-E:fl3d:hiEffE}   {\ensuremath{{0.003 } } }
\vdef{default-11:NoMc-E:fl3d:loDelta}   {\ensuremath{{+0.004 } } }
\vdef{default-11:NoMc-E:fl3d:loDeltaE}   {\ensuremath{{0.006 } } }
\vdef{default-11:NoMc-E:fl3d:hiDelta}   {\ensuremath{{-0.007 } } }
\vdef{default-11:NoMc-E:fl3d:hiDeltaE}   {\ensuremath{{0.012 } } }
\vdef{default-11:NoData-E:fl3de:loEff}   {\ensuremath{{1.000 } } }
\vdef{default-11:NoData-E:fl3de:loEffE}   {\ensuremath{{0.000 } } }
\vdef{default-11:NoData-E:fl3de:hiEff}   {\ensuremath{{0.000 } } }
\vdef{default-11:NoData-E:fl3de:hiEffE}   {\ensuremath{{0.000 } } }
\vdef{default-11:NoMc-E:fl3de:loEff}   {\ensuremath{{1.000 } } }
\vdef{default-11:NoMc-E:fl3de:loEffE}   {\ensuremath{{0.000 } } }
\vdef{default-11:NoMc-E:fl3de:hiEff}   {\ensuremath{{0.000 } } }
\vdef{default-11:NoMc-E:fl3de:hiEffE}   {\ensuremath{{0.000 } } }
\vdef{default-11:NoMc-E:fl3de:loDelta}   {\ensuremath{{+0.000 } } }
\vdef{default-11:NoMc-E:fl3de:loDeltaE}   {\ensuremath{{0.000 } } }
\vdef{default-11:NoMc-E:fl3de:hiDelta}   {\ensuremath{{+0.040 } } }
\vdef{default-11:NoMc-E:fl3de:hiDeltaE}   {\ensuremath{{0.511 } } }
\vdef{default-11:NoData-E:fls3d:loEff}   {\ensuremath{{0.125 } } }
\vdef{default-11:NoData-E:fls3d:loEffE}   {\ensuremath{{0.002 } } }
\vdef{default-11:NoData-E:fls3d:hiEff}   {\ensuremath{{0.875 } } }
\vdef{default-11:NoData-E:fls3d:hiEffE}   {\ensuremath{{0.002 } } }
\vdef{default-11:NoMc-E:fls3d:loEff}   {\ensuremath{{0.128 } } }
\vdef{default-11:NoMc-E:fls3d:loEffE}   {\ensuremath{{0.002 } } }
\vdef{default-11:NoMc-E:fls3d:hiEff}   {\ensuremath{{0.872 } } }
\vdef{default-11:NoMc-E:fls3d:hiEffE}   {\ensuremath{{0.002 } } }
\vdef{default-11:NoMc-E:fls3d:loDelta}   {\ensuremath{{-0.027 } } }
\vdef{default-11:NoMc-E:fls3d:loDeltaE}   {\ensuremath{{0.022 } } }
\vdef{default-11:NoMc-E:fls3d:hiDelta}   {\ensuremath{{+0.004 } } }
\vdef{default-11:NoMc-E:fls3d:hiDeltaE}   {\ensuremath{{0.003 } } }
\vdef{default-11:NoData-E:flsxy:loEff}   {\ensuremath{{1.008 } } }
\vdef{default-11:NoData-E:flsxy:loEffE}   {\ensuremath{{\mathrm{NaN} } } }
\vdef{default-11:NoData-E:flsxy:hiEff}   {\ensuremath{{1.000 } } }
\vdef{default-11:NoData-E:flsxy:hiEffE}   {\ensuremath{{0.000 } } }
\vdef{default-11:NoMc-E:flsxy:loEff}   {\ensuremath{{1.009 } } }
\vdef{default-11:NoMc-E:flsxy:loEffE}   {\ensuremath{{\mathrm{NaN} } } }
\vdef{default-11:NoMc-E:flsxy:hiEff}   {\ensuremath{{1.000 } } }
\vdef{default-11:NoMc-E:flsxy:hiEffE}   {\ensuremath{{0.000 } } }
\vdef{default-11:NoMc-E:flsxy:loDelta}   {\ensuremath{{-0.001 } } }
\vdef{default-11:NoMc-E:flsxy:loDeltaE}   {\ensuremath{{\mathrm{NaN} } } }
\vdef{default-11:NoMc-E:flsxy:hiDelta}   {\ensuremath{{+0.000 } } }
\vdef{default-11:NoMc-E:flsxy:hiDeltaE}   {\ensuremath{{0.000 } } }
\vdef{default-11:NoData-E:chi2dof:loEff}   {\ensuremath{{0.914 } } }
\vdef{default-11:NoData-E:chi2dof:loEffE}   {\ensuremath{{0.002 } } }
\vdef{default-11:NoData-E:chi2dof:hiEff}   {\ensuremath{{0.086 } } }
\vdef{default-11:NoData-E:chi2dof:hiEffE}   {\ensuremath{{0.002 } } }
\vdef{default-11:NoMc-E:chi2dof:loEff}   {\ensuremath{{0.932 } } }
\vdef{default-11:NoMc-E:chi2dof:loEffE}   {\ensuremath{{0.002 } } }
\vdef{default-11:NoMc-E:chi2dof:hiEff}   {\ensuremath{{0.068 } } }
\vdef{default-11:NoMc-E:chi2dof:hiEffE}   {\ensuremath{{0.002 } } }
\vdef{default-11:NoMc-E:chi2dof:loDelta}   {\ensuremath{{-0.019 } } }
\vdef{default-11:NoMc-E:chi2dof:loDeltaE}   {\ensuremath{{0.003 } } }
\vdef{default-11:NoMc-E:chi2dof:hiDelta}   {\ensuremath{{+0.228 } } }
\vdef{default-11:NoMc-E:chi2dof:hiDeltaE}   {\ensuremath{{0.030 } } }
\vdef{default-11:NoData-E:pchi2dof:loEff}   {\ensuremath{{0.659 } } }
\vdef{default-11:NoData-E:pchi2dof:loEffE}   {\ensuremath{{0.003 } } }
\vdef{default-11:NoData-E:pchi2dof:hiEff}   {\ensuremath{{0.341 } } }
\vdef{default-11:NoData-E:pchi2dof:hiEffE}   {\ensuremath{{0.003 } } }
\vdef{default-11:NoMc-E:pchi2dof:loEff}   {\ensuremath{{0.612 } } }
\vdef{default-11:NoMc-E:pchi2dof:loEffE}   {\ensuremath{{0.003 } } }
\vdef{default-11:NoMc-E:pchi2dof:hiEff}   {\ensuremath{{0.388 } } }
\vdef{default-11:NoMc-E:pchi2dof:hiEffE}   {\ensuremath{{0.003 } } }
\vdef{default-11:NoMc-E:pchi2dof:loDelta}   {\ensuremath{{+0.073 } } }
\vdef{default-11:NoMc-E:pchi2dof:loDeltaE}   {\ensuremath{{0.007 } } }
\vdef{default-11:NoMc-E:pchi2dof:hiDelta}   {\ensuremath{{-0.127 } } }
\vdef{default-11:NoMc-E:pchi2dof:hiDeltaE}   {\ensuremath{{0.012 } } }
\vdef{default-11:NoData-E:alpha:loEff}   {\ensuremath{{0.999 } } }
\vdef{default-11:NoData-E:alpha:loEffE}   {\ensuremath{{0.000 } } }
\vdef{default-11:NoData-E:alpha:hiEff}   {\ensuremath{{0.001 } } }
\vdef{default-11:NoData-E:alpha:hiEffE}   {\ensuremath{{0.000 } } }
\vdef{default-11:NoMc-E:alpha:loEff}   {\ensuremath{{0.999 } } }
\vdef{default-11:NoMc-E:alpha:loEffE}   {\ensuremath{{0.000 } } }
\vdef{default-11:NoMc-E:alpha:hiEff}   {\ensuremath{{0.001 } } }
\vdef{default-11:NoMc-E:alpha:hiEffE}   {\ensuremath{{0.000 } } }
\vdef{default-11:NoMc-E:alpha:loDelta}   {\ensuremath{{-0.000 } } }
\vdef{default-11:NoMc-E:alpha:loDeltaE}   {\ensuremath{{0.000 } } }
\vdef{default-11:NoMc-E:alpha:hiDelta}   {\ensuremath{{+0.218 } } }
\vdef{default-11:NoMc-E:alpha:hiDeltaE}   {\ensuremath{{0.255 } } }
\vdef{default-11:NoData-E:iso:loEff}   {\ensuremath{{0.112 } } }
\vdef{default-11:NoData-E:iso:loEffE}   {\ensuremath{{0.002 } } }
\vdef{default-11:NoData-E:iso:hiEff}   {\ensuremath{{0.888 } } }
\vdef{default-11:NoData-E:iso:hiEffE}   {\ensuremath{{0.002 } } }
\vdef{default-11:NoMc-E:iso:loEff}   {\ensuremath{{0.091 } } }
\vdef{default-11:NoMc-E:iso:loEffE}   {\ensuremath{{0.002 } } }
\vdef{default-11:NoMc-E:iso:hiEff}   {\ensuremath{{0.909 } } }
\vdef{default-11:NoMc-E:iso:hiEffE}   {\ensuremath{{0.002 } } }
\vdef{default-11:NoMc-E:iso:loDelta}   {\ensuremath{{+0.203 } } }
\vdef{default-11:NoMc-E:iso:loDeltaE}   {\ensuremath{{0.026 } } }
\vdef{default-11:NoMc-E:iso:hiDelta}   {\ensuremath{{-0.023 } } }
\vdef{default-11:NoMc-E:iso:hiDeltaE}   {\ensuremath{{0.003 } } }
\vdef{default-11:NoData-E:docatrk:loEff}   {\ensuremath{{0.070 } } }
\vdef{default-11:NoData-E:docatrk:loEffE}   {\ensuremath{{0.002 } } }
\vdef{default-11:NoData-E:docatrk:hiEff}   {\ensuremath{{0.930 } } }
\vdef{default-11:NoData-E:docatrk:hiEffE}   {\ensuremath{{0.002 } } }
\vdef{default-11:NoMc-E:docatrk:loEff}   {\ensuremath{{0.075 } } }
\vdef{default-11:NoMc-E:docatrk:loEffE}   {\ensuremath{{0.002 } } }
\vdef{default-11:NoMc-E:docatrk:hiEff}   {\ensuremath{{0.925 } } }
\vdef{default-11:NoMc-E:docatrk:hiEffE}   {\ensuremath{{0.002 } } }
\vdef{default-11:NoMc-E:docatrk:loDelta}   {\ensuremath{{-0.068 } } }
\vdef{default-11:NoMc-E:docatrk:loDeltaE}   {\ensuremath{{0.032 } } }
\vdef{default-11:NoMc-E:docatrk:hiDelta}   {\ensuremath{{+0.005 } } }
\vdef{default-11:NoMc-E:docatrk:hiDeltaE}   {\ensuremath{{0.003 } } }
\vdef{default-11:NoData-E:isotrk:loEff}   {\ensuremath{{1.000 } } }
\vdef{default-11:NoData-E:isotrk:loEffE}   {\ensuremath{{0.000 } } }
\vdef{default-11:NoData-E:isotrk:hiEff}   {\ensuremath{{1.000 } } }
\vdef{default-11:NoData-E:isotrk:hiEffE}   {\ensuremath{{0.000 } } }
\vdef{default-11:NoMc-E:isotrk:loEff}   {\ensuremath{{1.000 } } }
\vdef{default-11:NoMc-E:isotrk:loEffE}   {\ensuremath{{0.000 } } }
\vdef{default-11:NoMc-E:isotrk:hiEff}   {\ensuremath{{1.000 } } }
\vdef{default-11:NoMc-E:isotrk:hiEffE}   {\ensuremath{{0.000 } } }
\vdef{default-11:NoMc-E:isotrk:loDelta}   {\ensuremath{{+0.000 } } }
\vdef{default-11:NoMc-E:isotrk:loDeltaE}   {\ensuremath{{0.000 } } }
\vdef{default-11:NoMc-E:isotrk:hiDelta}   {\ensuremath{{+0.000 } } }
\vdef{default-11:NoMc-E:isotrk:hiDeltaE}   {\ensuremath{{0.000 } } }
\vdef{default-11:NoData-E:closetrk:loEff}   {\ensuremath{{0.978 } } }
\vdef{default-11:NoData-E:closetrk:loEffE}   {\ensuremath{{0.001 } } }
\vdef{default-11:NoData-E:closetrk:hiEff}   {\ensuremath{{0.022 } } }
\vdef{default-11:NoData-E:closetrk:hiEffE}   {\ensuremath{{0.001 } } }
\vdef{default-11:NoMc-E:closetrk:loEff}   {\ensuremath{{0.981 } } }
\vdef{default-11:NoMc-E:closetrk:loEffE}   {\ensuremath{{0.001 } } }
\vdef{default-11:NoMc-E:closetrk:hiEff}   {\ensuremath{{0.019 } } }
\vdef{default-11:NoMc-E:closetrk:hiEffE}   {\ensuremath{{0.001 } } }
\vdef{default-11:NoMc-E:closetrk:loDelta}   {\ensuremath{{-0.003 } } }
\vdef{default-11:NoMc-E:closetrk:loDeltaE}   {\ensuremath{{0.001 } } }
\vdef{default-11:NoMc-E:closetrk:hiDelta}   {\ensuremath{{+0.143 } } }
\vdef{default-11:NoMc-E:closetrk:hiDeltaE}   {\ensuremath{{0.063 } } }
\vdef{default-11:NoData-E:lip:loEff}   {\ensuremath{{1.000 } } }
\vdef{default-11:NoData-E:lip:loEffE}   {\ensuremath{{0.000 } } }
\vdef{default-11:NoData-E:lip:hiEff}   {\ensuremath{{0.000 } } }
\vdef{default-11:NoData-E:lip:hiEffE}   {\ensuremath{{0.000 } } }
\vdef{default-11:NoMc-E:lip:loEff}   {\ensuremath{{1.000 } } }
\vdef{default-11:NoMc-E:lip:loEffE}   {\ensuremath{{0.000 } } }
\vdef{default-11:NoMc-E:lip:hiEff}   {\ensuremath{{0.000 } } }
\vdef{default-11:NoMc-E:lip:hiEffE}   {\ensuremath{{0.000 } } }
\vdef{default-11:NoMc-E:lip:loDelta}   {\ensuremath{{+0.000 } } }
\vdef{default-11:NoMc-E:lip:loDeltaE}   {\ensuremath{{0.000 } } }
\vdef{default-11:NoMc-E:lip:hiDelta}   {\ensuremath{{\mathrm{NaN} } } }
\vdef{default-11:NoMc-E:lip:hiDeltaE}   {\ensuremath{{\mathrm{NaN} } } }
\vdef{default-11:NoData-E:lip:inEff}   {\ensuremath{{1.000 } } }
\vdef{default-11:NoData-E:lip:inEffE}   {\ensuremath{{0.000 } } }
\vdef{default-11:NoMc-E:lip:inEff}   {\ensuremath{{1.000 } } }
\vdef{default-11:NoMc-E:lip:inEffE}   {\ensuremath{{0.000 } } }
\vdef{default-11:NoMc-E:lip:inDelta}   {\ensuremath{{+0.000 } } }
\vdef{default-11:NoMc-E:lip:inDeltaE}   {\ensuremath{{0.000 } } }
\vdef{default-11:NoData-E:lips:loEff}   {\ensuremath{{1.000 } } }
\vdef{default-11:NoData-E:lips:loEffE}   {\ensuremath{{0.000 } } }
\vdef{default-11:NoData-E:lips:hiEff}   {\ensuremath{{0.000 } } }
\vdef{default-11:NoData-E:lips:hiEffE}   {\ensuremath{{0.000 } } }
\vdef{default-11:NoMc-E:lips:loEff}   {\ensuremath{{1.000 } } }
\vdef{default-11:NoMc-E:lips:loEffE}   {\ensuremath{{0.000 } } }
\vdef{default-11:NoMc-E:lips:hiEff}   {\ensuremath{{0.000 } } }
\vdef{default-11:NoMc-E:lips:hiEffE}   {\ensuremath{{0.000 } } }
\vdef{default-11:NoMc-E:lips:loDelta}   {\ensuremath{{+0.000 } } }
\vdef{default-11:NoMc-E:lips:loDeltaE}   {\ensuremath{{0.000 } } }
\vdef{default-11:NoMc-E:lips:hiDelta}   {\ensuremath{{\mathrm{NaN} } } }
\vdef{default-11:NoMc-E:lips:hiDeltaE}   {\ensuremath{{\mathrm{NaN} } } }
\vdef{default-11:NoData-E:lips:inEff}   {\ensuremath{{1.000 } } }
\vdef{default-11:NoData-E:lips:inEffE}   {\ensuremath{{0.000 } } }
\vdef{default-11:NoMc-E:lips:inEff}   {\ensuremath{{1.000 } } }
\vdef{default-11:NoMc-E:lips:inEffE}   {\ensuremath{{0.000 } } }
\vdef{default-11:NoMc-E:lips:inDelta}   {\ensuremath{{+0.000 } } }
\vdef{default-11:NoMc-E:lips:inDeltaE}   {\ensuremath{{0.000 } } }
\vdef{default-11:NoData-E:ip:loEff}   {\ensuremath{{0.975 } } }
\vdef{default-11:NoData-E:ip:loEffE}   {\ensuremath{{0.001 } } }
\vdef{default-11:NoData-E:ip:hiEff}   {\ensuremath{{0.025 } } }
\vdef{default-11:NoData-E:ip:hiEffE}   {\ensuremath{{0.001 } } }
\vdef{default-11:NoMc-E:ip:loEff}   {\ensuremath{{0.975 } } }
\vdef{default-11:NoMc-E:ip:loEffE}   {\ensuremath{{0.001 } } }
\vdef{default-11:NoMc-E:ip:hiEff}   {\ensuremath{{0.025 } } }
\vdef{default-11:NoMc-E:ip:hiEffE}   {\ensuremath{{0.001 } } }
\vdef{default-11:NoMc-E:ip:loDelta}   {\ensuremath{{+0.000 } } }
\vdef{default-11:NoMc-E:ip:loDeltaE}   {\ensuremath{{0.001 } } }
\vdef{default-11:NoMc-E:ip:hiDelta}   {\ensuremath{{-0.006 } } }
\vdef{default-11:NoMc-E:ip:hiDeltaE}   {\ensuremath{{0.057 } } }
\vdef{default-11:NoData-E:ips:loEff}   {\ensuremath{{0.967 } } }
\vdef{default-11:NoData-E:ips:loEffE}   {\ensuremath{{0.001 } } }
\vdef{default-11:NoData-E:ips:hiEff}   {\ensuremath{{0.033 } } }
\vdef{default-11:NoData-E:ips:hiEffE}   {\ensuremath{{0.001 } } }
\vdef{default-11:NoMc-E:ips:loEff}   {\ensuremath{{0.975 } } }
\vdef{default-11:NoMc-E:ips:loEffE}   {\ensuremath{{0.001 } } }
\vdef{default-11:NoMc-E:ips:hiEff}   {\ensuremath{{0.025 } } }
\vdef{default-11:NoMc-E:ips:hiEffE}   {\ensuremath{{0.001 } } }
\vdef{default-11:NoMc-E:ips:loDelta}   {\ensuremath{{-0.008 } } }
\vdef{default-11:NoMc-E:ips:loDeltaE}   {\ensuremath{{0.002 } } }
\vdef{default-11:NoMc-E:ips:hiDelta}   {\ensuremath{{+0.282 } } }
\vdef{default-11:NoMc-E:ips:hiDeltaE}   {\ensuremath{{0.052 } } }
\vdef{default-11:NoData-E:maxdoca:loEff}   {\ensuremath{{1.000 } } }
\vdef{default-11:NoData-E:maxdoca:loEffE}   {\ensuremath{{0.000 } } }
\vdef{default-11:NoData-E:maxdoca:hiEff}   {\ensuremath{{0.018 } } }
\vdef{default-11:NoData-E:maxdoca:hiEffE}   {\ensuremath{{0.001 } } }
\vdef{default-11:NoMc-E:maxdoca:loEff}   {\ensuremath{{1.000 } } }
\vdef{default-11:NoMc-E:maxdoca:loEffE}   {\ensuremath{{0.000 } } }
\vdef{default-11:NoMc-E:maxdoca:hiEff}   {\ensuremath{{0.013 } } }
\vdef{default-11:NoMc-E:maxdoca:hiEffE}   {\ensuremath{{0.001 } } }
\vdef{default-11:NoMc-E:maxdoca:loDelta}   {\ensuremath{{+0.000 } } }
\vdef{default-11:NoMc-E:maxdoca:loDeltaE}   {\ensuremath{{0.000 } } }
\vdef{default-11:NoMc-E:maxdoca:hiDelta}   {\ensuremath{{+0.278 } } }
\vdef{default-11:NoMc-E:maxdoca:hiDeltaE}   {\ensuremath{{0.073 } } }
\vdef{default-11:NoData-E:kaonpt:loEff}   {\ensuremath{{1.005 } } }
\vdef{default-11:NoData-E:kaonpt:loEffE}   {\ensuremath{{\mathrm{NaN} } } }
\vdef{default-11:NoData-E:kaonpt:hiEff}   {\ensuremath{{1.000 } } }
\vdef{default-11:NoData-E:kaonpt:hiEffE}   {\ensuremath{{0.000 } } }
\vdef{default-11:NoMc-E:kaonpt:loEff}   {\ensuremath{{1.005 } } }
\vdef{default-11:NoMc-E:kaonpt:loEffE}   {\ensuremath{{\mathrm{NaN} } } }
\vdef{default-11:NoMc-E:kaonpt:hiEff}   {\ensuremath{{1.000 } } }
\vdef{default-11:NoMc-E:kaonpt:hiEffE}   {\ensuremath{{0.000 } } }
\vdef{default-11:NoMc-E:kaonpt:loDelta}   {\ensuremath{{+0.000 } } }
\vdef{default-11:NoMc-E:kaonpt:loDeltaE}   {\ensuremath{{\mathrm{NaN} } } }
\vdef{default-11:NoMc-E:kaonpt:hiDelta}   {\ensuremath{{+0.000 } } }
\vdef{default-11:NoMc-E:kaonpt:hiDeltaE}   {\ensuremath{{0.000 } } }
\vdef{default-11:NoData-E:psipt:loEff}   {\ensuremath{{1.001 } } }
\vdef{default-11:NoData-E:psipt:loEffE}   {\ensuremath{{\mathrm{NaN} } } }
\vdef{default-11:NoData-E:psipt:hiEff}   {\ensuremath{{1.000 } } }
\vdef{default-11:NoData-E:psipt:hiEffE}   {\ensuremath{{0.000 } } }
\vdef{default-11:NoMc-E:psipt:loEff}   {\ensuremath{{1.001 } } }
\vdef{default-11:NoMc-E:psipt:loEffE}   {\ensuremath{{\mathrm{NaN} } } }
\vdef{default-11:NoMc-E:psipt:hiEff}   {\ensuremath{{1.000 } } }
\vdef{default-11:NoMc-E:psipt:hiEffE}   {\ensuremath{{0.000 } } }
\vdef{default-11:NoMc-E:psipt:loDelta}   {\ensuremath{{-0.000 } } }
\vdef{default-11:NoMc-E:psipt:loDeltaE}   {\ensuremath{{\mathrm{NaN} } } }
\vdef{default-11:NoMc-E:psipt:hiDelta}   {\ensuremath{{+0.000 } } }
\vdef{default-11:NoMc-E:psipt:hiDeltaE}   {\ensuremath{{0.000 } } }
\vdef{default-11:NoMc2e33-A:osiso:loEff}   {\ensuremath{{1.003 } } }
\vdef{default-11:NoMc2e33-A:osiso:loEffE}   {\ensuremath{{\mathrm{NaN} } } }
\vdef{default-11:NoMc2e33-A:osiso:hiEff}   {\ensuremath{{1.000 } } }
\vdef{default-11:NoMc2e33-A:osiso:hiEffE}   {\ensuremath{{0.000 } } }
\vdef{default-11:NoMcCMS-A:osiso:loEff}   {\ensuremath{{1.003 } } }
\vdef{default-11:NoMcCMS-A:osiso:loEffE}   {\ensuremath{{\mathrm{NaN} } } }
\vdef{default-11:NoMcCMS-A:osiso:hiEff}   {\ensuremath{{1.000 } } }
\vdef{default-11:NoMcCMS-A:osiso:hiEffE}   {\ensuremath{{0.000 } } }
\vdef{default-11:NoMcCMS-A:osiso:loDelta}   {\ensuremath{{+0.000 } } }
\vdef{default-11:NoMcCMS-A:osiso:loDeltaE}   {\ensuremath{{\mathrm{NaN} } } }
\vdef{default-11:NoMcCMS-A:osiso:hiDelta}   {\ensuremath{{+0.000 } } }
\vdef{default-11:NoMcCMS-A:osiso:hiDeltaE}   {\ensuremath{{0.000 } } }
\vdef{default-11:NoMc2e33-A:osreliso:loEff}   {\ensuremath{{0.289 } } }
\vdef{default-11:NoMc2e33-A:osreliso:loEffE}   {\ensuremath{{0.001 } } }
\vdef{default-11:NoMc2e33-A:osreliso:hiEff}   {\ensuremath{{0.711 } } }
\vdef{default-11:NoMc2e33-A:osreliso:hiEffE}   {\ensuremath{{0.001 } } }
\vdef{default-11:NoMcCMS-A:osreliso:loEff}   {\ensuremath{{0.282 } } }
\vdef{default-11:NoMcCMS-A:osreliso:loEffE}   {\ensuremath{{0.001 } } }
\vdef{default-11:NoMcCMS-A:osreliso:hiEff}   {\ensuremath{{0.718 } } }
\vdef{default-11:NoMcCMS-A:osreliso:hiEffE}   {\ensuremath{{0.001 } } }
\vdef{default-11:NoMcCMS-A:osreliso:loDelta}   {\ensuremath{{+0.024 } } }
\vdef{default-11:NoMcCMS-A:osreliso:loDeltaE}   {\ensuremath{{0.006 } } }
\vdef{default-11:NoMcCMS-A:osreliso:hiDelta}   {\ensuremath{{-0.009 } } }
\vdef{default-11:NoMcCMS-A:osreliso:hiDeltaE}   {\ensuremath{{0.002 } } }
\vdef{default-11:NoMc2e33-A:osmuonpt:loEff}   {\ensuremath{{0.000 } } }
\vdef{default-11:NoMc2e33-A:osmuonpt:loEffE}   {\ensuremath{{0.000 } } }
\vdef{default-11:NoMc2e33-A:osmuonpt:hiEff}   {\ensuremath{{1.000 } } }
\vdef{default-11:NoMc2e33-A:osmuonpt:hiEffE}   {\ensuremath{{0.000 } } }
\vdef{default-11:NoMcCMS-A:osmuonpt:loEff}   {\ensuremath{{0.000 } } }
\vdef{default-11:NoMcCMS-A:osmuonpt:loEffE}   {\ensuremath{{0.000 } } }
\vdef{default-11:NoMcCMS-A:osmuonpt:hiEff}   {\ensuremath{{1.000 } } }
\vdef{default-11:NoMcCMS-A:osmuonpt:hiEffE}   {\ensuremath{{0.000 } } }
\vdef{default-11:NoMcCMS-A:osmuonpt:loDelta}   {\ensuremath{{\mathrm{NaN} } } }
\vdef{default-11:NoMcCMS-A:osmuonpt:loDeltaE}   {\ensuremath{{\mathrm{NaN} } } }
\vdef{default-11:NoMcCMS-A:osmuonpt:hiDelta}   {\ensuremath{{+0.000 } } }
\vdef{default-11:NoMcCMS-A:osmuonpt:hiDeltaE}   {\ensuremath{{0.000 } } }
\vdef{default-11:NoMc2e33-A:osmuondr:loEff}   {\ensuremath{{0.018 } } }
\vdef{default-11:NoMc2e33-A:osmuondr:loEffE}   {\ensuremath{{0.002 } } }
\vdef{default-11:NoMc2e33-A:osmuondr:hiEff}   {\ensuremath{{0.982 } } }
\vdef{default-11:NoMc2e33-A:osmuondr:hiEffE}   {\ensuremath{{0.002 } } }
\vdef{default-11:NoMcCMS-A:osmuondr:loEff}   {\ensuremath{{0.012 } } }
\vdef{default-11:NoMcCMS-A:osmuondr:loEffE}   {\ensuremath{{0.002 } } }
\vdef{default-11:NoMcCMS-A:osmuondr:hiEff}   {\ensuremath{{0.988 } } }
\vdef{default-11:NoMcCMS-A:osmuondr:hiEffE}   {\ensuremath{{0.002 } } }
\vdef{default-11:NoMcCMS-A:osmuondr:loDelta}   {\ensuremath{{+0.420 } } }
\vdef{default-11:NoMcCMS-A:osmuondr:loDeltaE}   {\ensuremath{{0.168 } } }
\vdef{default-11:NoMcCMS-A:osmuondr:hiDelta}   {\ensuremath{{-0.006 } } }
\vdef{default-11:NoMcCMS-A:osmuondr:hiDeltaE}   {\ensuremath{{0.003 } } }
\vdef{default-11:NoMc2e33-A:hlt:loEff}   {\ensuremath{{0.277 } } }
\vdef{default-11:NoMc2e33-A:hlt:loEffE}   {\ensuremath{{0.001 } } }
\vdef{default-11:NoMc2e33-A:hlt:hiEff}   {\ensuremath{{0.723 } } }
\vdef{default-11:NoMc2e33-A:hlt:hiEffE}   {\ensuremath{{0.001 } } }
\vdef{default-11:NoMcCMS-A:hlt:loEff}   {\ensuremath{{0.267 } } }
\vdef{default-11:NoMcCMS-A:hlt:loEffE}   {\ensuremath{{0.001 } } }
\vdef{default-11:NoMcCMS-A:hlt:hiEff}   {\ensuremath{{0.733 } } }
\vdef{default-11:NoMcCMS-A:hlt:hiEffE}   {\ensuremath{{0.001 } } }
\vdef{default-11:NoMcCMS-A:hlt:loDelta}   {\ensuremath{{+0.036 } } }
\vdef{default-11:NoMcCMS-A:hlt:loDeltaE}   {\ensuremath{{0.005 } } }
\vdef{default-11:NoMcCMS-A:hlt:hiDelta}   {\ensuremath{{-0.013 } } }
\vdef{default-11:NoMcCMS-A:hlt:hiDeltaE}   {\ensuremath{{0.002 } } }
\vdef{default-11:NoMc2e33-A:muonsid:loEff}   {\ensuremath{{0.158 } } }
\vdef{default-11:NoMc2e33-A:muonsid:loEffE}   {\ensuremath{{0.001 } } }
\vdef{default-11:NoMc2e33-A:muonsid:hiEff}   {\ensuremath{{0.842 } } }
\vdef{default-11:NoMc2e33-A:muonsid:hiEffE}   {\ensuremath{{0.001 } } }
\vdef{default-11:NoMcCMS-A:muonsid:loEff}   {\ensuremath{{0.220 } } }
\vdef{default-11:NoMcCMS-A:muonsid:loEffE}   {\ensuremath{{0.001 } } }
\vdef{default-11:NoMcCMS-A:muonsid:hiEff}   {\ensuremath{{0.780 } } }
\vdef{default-11:NoMcCMS-A:muonsid:hiEffE}   {\ensuremath{{0.001 } } }
\vdef{default-11:NoMcCMS-A:muonsid:loDelta}   {\ensuremath{{-0.326 } } }
\vdef{default-11:NoMcCMS-A:muonsid:loDeltaE}   {\ensuremath{{0.007 } } }
\vdef{default-11:NoMcCMS-A:muonsid:hiDelta}   {\ensuremath{{+0.076 } } }
\vdef{default-11:NoMcCMS-A:muonsid:hiDeltaE}   {\ensuremath{{0.002 } } }
\vdef{default-11:NoMc2e33-A:tracksqual:loEff}   {\ensuremath{{0.000 } } }
\vdef{default-11:NoMc2e33-A:tracksqual:loEffE}   {\ensuremath{{0.000 } } }
\vdef{default-11:NoMc2e33-A:tracksqual:hiEff}   {\ensuremath{{1.000 } } }
\vdef{default-11:NoMc2e33-A:tracksqual:hiEffE}   {\ensuremath{{0.000 } } }
\vdef{default-11:NoMcCMS-A:tracksqual:loEff}   {\ensuremath{{0.000 } } }
\vdef{default-11:NoMcCMS-A:tracksqual:loEffE}   {\ensuremath{{0.000 } } }
\vdef{default-11:NoMcCMS-A:tracksqual:hiEff}   {\ensuremath{{1.000 } } }
\vdef{default-11:NoMcCMS-A:tracksqual:hiEffE}   {\ensuremath{{0.000 } } }
\vdef{default-11:NoMcCMS-A:tracksqual:loDelta}   {\ensuremath{{-0.395 } } }
\vdef{default-11:NoMcCMS-A:tracksqual:loDeltaE}   {\ensuremath{{0.276 } } }
\vdef{default-11:NoMcCMS-A:tracksqual:hiDelta}   {\ensuremath{{+0.000 } } }
\vdef{default-11:NoMcCMS-A:tracksqual:hiDeltaE}   {\ensuremath{{0.000 } } }
\vdef{default-11:NoMc2e33-A:pvz:loEff}   {\ensuremath{{0.469 } } }
\vdef{default-11:NoMc2e33-A:pvz:loEffE}   {\ensuremath{{0.001 } } }
\vdef{default-11:NoMc2e33-A:pvz:hiEff}   {\ensuremath{{0.531 } } }
\vdef{default-11:NoMc2e33-A:pvz:hiEffE}   {\ensuremath{{0.001 } } }
\vdef{default-11:NoMcCMS-A:pvz:loEff}   {\ensuremath{{0.470 } } }
\vdef{default-11:NoMcCMS-A:pvz:loEffE}   {\ensuremath{{0.001 } } }
\vdef{default-11:NoMcCMS-A:pvz:hiEff}   {\ensuremath{{0.530 } } }
\vdef{default-11:NoMcCMS-A:pvz:hiEffE}   {\ensuremath{{0.001 } } }
\vdef{default-11:NoMcCMS-A:pvz:loDelta}   {\ensuremath{{-0.001 } } }
\vdef{default-11:NoMcCMS-A:pvz:loDeltaE}   {\ensuremath{{0.004 } } }
\vdef{default-11:NoMcCMS-A:pvz:hiDelta}   {\ensuremath{{+0.001 } } }
\vdef{default-11:NoMcCMS-A:pvz:hiDeltaE}   {\ensuremath{{0.003 } } }
\vdef{default-11:NoMc2e33-A:pvn:loEff}   {\ensuremath{{1.000 } } }
\vdef{default-11:NoMc2e33-A:pvn:loEffE}   {\ensuremath{{0.000 } } }
\vdef{default-11:NoMc2e33-A:pvn:hiEff}   {\ensuremath{{1.000 } } }
\vdef{default-11:NoMc2e33-A:pvn:hiEffE}   {\ensuremath{{0.000 } } }
\vdef{default-11:NoMcCMS-A:pvn:loEff}   {\ensuremath{{1.065 } } }
\vdef{default-11:NoMcCMS-A:pvn:loEffE}   {\ensuremath{{\mathrm{NaN} } } }
\vdef{default-11:NoMcCMS-A:pvn:hiEff}   {\ensuremath{{1.000 } } }
\vdef{default-11:NoMcCMS-A:pvn:hiEffE}   {\ensuremath{{0.000 } } }
\vdef{default-11:NoMcCMS-A:pvn:loDelta}   {\ensuremath{{-0.063 } } }
\vdef{default-11:NoMcCMS-A:pvn:loDeltaE}   {\ensuremath{{\mathrm{NaN} } } }
\vdef{default-11:NoMcCMS-A:pvn:hiDelta}   {\ensuremath{{+0.000 } } }
\vdef{default-11:NoMcCMS-A:pvn:hiDeltaE}   {\ensuremath{{0.000 } } }
\vdef{default-11:NoMc2e33-A:pvavew8:loEff}   {\ensuremath{{0.004 } } }
\vdef{default-11:NoMc2e33-A:pvavew8:loEffE}   {\ensuremath{{0.000 } } }
\vdef{default-11:NoMc2e33-A:pvavew8:hiEff}   {\ensuremath{{0.996 } } }
\vdef{default-11:NoMc2e33-A:pvavew8:hiEffE}   {\ensuremath{{0.000 } } }
\vdef{default-11:NoMcCMS-A:pvavew8:loEff}   {\ensuremath{{0.006 } } }
\vdef{default-11:NoMcCMS-A:pvavew8:loEffE}   {\ensuremath{{0.000 } } }
\vdef{default-11:NoMcCMS-A:pvavew8:hiEff}   {\ensuremath{{0.994 } } }
\vdef{default-11:NoMcCMS-A:pvavew8:hiEffE}   {\ensuremath{{0.000 } } }
\vdef{default-11:NoMcCMS-A:pvavew8:loDelta}   {\ensuremath{{-0.358 } } }
\vdef{default-11:NoMcCMS-A:pvavew8:loDeltaE}   {\ensuremath{{0.055 } } }
\vdef{default-11:NoMcCMS-A:pvavew8:hiDelta}   {\ensuremath{{+0.002 } } }
\vdef{default-11:NoMcCMS-A:pvavew8:hiDeltaE}   {\ensuremath{{0.000 } } }
\vdef{default-11:NoMc2e33-A:pvntrk:loEff}   {\ensuremath{{1.000 } } }
\vdef{default-11:NoMc2e33-A:pvntrk:loEffE}   {\ensuremath{{0.000 } } }
\vdef{default-11:NoMc2e33-A:pvntrk:hiEff}   {\ensuremath{{1.000 } } }
\vdef{default-11:NoMc2e33-A:pvntrk:hiEffE}   {\ensuremath{{0.000 } } }
\vdef{default-11:NoMcCMS-A:pvntrk:loEff}   {\ensuremath{{1.000 } } }
\vdef{default-11:NoMcCMS-A:pvntrk:loEffE}   {\ensuremath{{0.000 } } }
\vdef{default-11:NoMcCMS-A:pvntrk:hiEff}   {\ensuremath{{1.000 } } }
\vdef{default-11:NoMcCMS-A:pvntrk:hiEffE}   {\ensuremath{{0.000 } } }
\vdef{default-11:NoMcCMS-A:pvntrk:loDelta}   {\ensuremath{{+0.000 } } }
\vdef{default-11:NoMcCMS-A:pvntrk:loDeltaE}   {\ensuremath{{0.000 } } }
\vdef{default-11:NoMcCMS-A:pvntrk:hiDelta}   {\ensuremath{{+0.000 } } }
\vdef{default-11:NoMcCMS-A:pvntrk:hiDeltaE}   {\ensuremath{{0.000 } } }
\vdef{default-11:NoMc2e33-A:muon1pt:loEff}   {\ensuremath{{1.007 } } }
\vdef{default-11:NoMc2e33-A:muon1pt:loEffE}   {\ensuremath{{\mathrm{NaN} } } }
\vdef{default-11:NoMc2e33-A:muon1pt:hiEff}   {\ensuremath{{1.000 } } }
\vdef{default-11:NoMc2e33-A:muon1pt:hiEffE}   {\ensuremath{{0.000 } } }
\vdef{default-11:NoMcCMS-A:muon1pt:loEff}   {\ensuremath{{1.009 } } }
\vdef{default-11:NoMcCMS-A:muon1pt:loEffE}   {\ensuremath{{\mathrm{NaN} } } }
\vdef{default-11:NoMcCMS-A:muon1pt:hiEff}   {\ensuremath{{1.000 } } }
\vdef{default-11:NoMcCMS-A:muon1pt:hiEffE}   {\ensuremath{{0.000 } } }
\vdef{default-11:NoMcCMS-A:muon1pt:loDelta}   {\ensuremath{{-0.002 } } }
\vdef{default-11:NoMcCMS-A:muon1pt:loDeltaE}   {\ensuremath{{\mathrm{NaN} } } }
\vdef{default-11:NoMcCMS-A:muon1pt:hiDelta}   {\ensuremath{{+0.000 } } }
\vdef{default-11:NoMcCMS-A:muon1pt:hiDeltaE}   {\ensuremath{{0.000 } } }
\vdef{default-11:NoMc2e33-A:muon2pt:loEff}   {\ensuremath{{0.136 } } }
\vdef{default-11:NoMc2e33-A:muon2pt:loEffE}   {\ensuremath{{0.001 } } }
\vdef{default-11:NoMc2e33-A:muon2pt:hiEff}   {\ensuremath{{0.864 } } }
\vdef{default-11:NoMc2e33-A:muon2pt:hiEffE}   {\ensuremath{{0.001 } } }
\vdef{default-11:NoMcCMS-A:muon2pt:loEff}   {\ensuremath{{0.090 } } }
\vdef{default-11:NoMcCMS-A:muon2pt:loEffE}   {\ensuremath{{0.001 } } }
\vdef{default-11:NoMcCMS-A:muon2pt:hiEff}   {\ensuremath{{0.910 } } }
\vdef{default-11:NoMcCMS-A:muon2pt:hiEffE}   {\ensuremath{{0.001 } } }
\vdef{default-11:NoMcCMS-A:muon2pt:loDelta}   {\ensuremath{{+0.408 } } }
\vdef{default-11:NoMcCMS-A:muon2pt:loDeltaE}   {\ensuremath{{0.010 } } }
\vdef{default-11:NoMcCMS-A:muon2pt:hiDelta}   {\ensuremath{{-0.052 } } }
\vdef{default-11:NoMcCMS-A:muon2pt:hiDeltaE}   {\ensuremath{{0.001 } } }
\vdef{default-11:NoMc2e33-A:muonseta:loEff}   {\ensuremath{{0.728 } } }
\vdef{default-11:NoMc2e33-A:muonseta:loEffE}   {\ensuremath{{0.001 } } }
\vdef{default-11:NoMc2e33-A:muonseta:hiEff}   {\ensuremath{{0.272 } } }
\vdef{default-11:NoMc2e33-A:muonseta:hiEffE}   {\ensuremath{{0.001 } } }
\vdef{default-11:NoMcCMS-A:muonseta:loEff}   {\ensuremath{{0.844 } } }
\vdef{default-11:NoMcCMS-A:muonseta:loEffE}   {\ensuremath{{0.001 } } }
\vdef{default-11:NoMcCMS-A:muonseta:hiEff}   {\ensuremath{{0.156 } } }
\vdef{default-11:NoMcCMS-A:muonseta:hiEffE}   {\ensuremath{{0.001 } } }
\vdef{default-11:NoMcCMS-A:muonseta:loDelta}   {\ensuremath{{-0.148 } } }
\vdef{default-11:NoMcCMS-A:muonseta:loDeltaE}   {\ensuremath{{0.001 } } }
\vdef{default-11:NoMcCMS-A:muonseta:hiDelta}   {\ensuremath{{+0.542 } } }
\vdef{default-11:NoMcCMS-A:muonseta:hiDeltaE}   {\ensuremath{{0.005 } } }
\vdef{default-11:NoMc2e33-A:pt:loEff}   {\ensuremath{{0.000 } } }
\vdef{default-11:NoMc2e33-A:pt:loEffE}   {\ensuremath{{0.000 } } }
\vdef{default-11:NoMc2e33-A:pt:hiEff}   {\ensuremath{{1.000 } } }
\vdef{default-11:NoMc2e33-A:pt:hiEffE}   {\ensuremath{{0.000 } } }
\vdef{default-11:NoMcCMS-A:pt:loEff}   {\ensuremath{{0.000 } } }
\vdef{default-11:NoMcCMS-A:pt:loEffE}   {\ensuremath{{0.000 } } }
\vdef{default-11:NoMcCMS-A:pt:hiEff}   {\ensuremath{{1.000 } } }
\vdef{default-11:NoMcCMS-A:pt:hiEffE}   {\ensuremath{{0.000 } } }
\vdef{default-11:NoMcCMS-A:pt:loDelta}   {\ensuremath{{\mathrm{NaN} } } }
\vdef{default-11:NoMcCMS-A:pt:loDeltaE}   {\ensuremath{{\mathrm{NaN} } } }
\vdef{default-11:NoMcCMS-A:pt:hiDelta}   {\ensuremath{{+0.000 } } }
\vdef{default-11:NoMcCMS-A:pt:hiDeltaE}   {\ensuremath{{0.000 } } }
\vdef{default-11:NoMc2e33-A:p:loEff}   {\ensuremath{{1.013 } } }
\vdef{default-11:NoMc2e33-A:p:loEffE}   {\ensuremath{{\mathrm{NaN} } } }
\vdef{default-11:NoMc2e33-A:p:hiEff}   {\ensuremath{{1.000 } } }
\vdef{default-11:NoMc2e33-A:p:hiEffE}   {\ensuremath{{0.000 } } }
\vdef{default-11:NoMcCMS-A:p:loEff}   {\ensuremath{{1.004 } } }
\vdef{default-11:NoMcCMS-A:p:loEffE}   {\ensuremath{{\mathrm{NaN} } } }
\vdef{default-11:NoMcCMS-A:p:hiEff}   {\ensuremath{{1.000 } } }
\vdef{default-11:NoMcCMS-A:p:hiEffE}   {\ensuremath{{0.000 } } }
\vdef{default-11:NoMcCMS-A:p:loDelta}   {\ensuremath{{+0.008 } } }
\vdef{default-11:NoMcCMS-A:p:loDeltaE}   {\ensuremath{{\mathrm{NaN} } } }
\vdef{default-11:NoMcCMS-A:p:hiDelta}   {\ensuremath{{+0.000 } } }
\vdef{default-11:NoMcCMS-A:p:hiDeltaE}   {\ensuremath{{0.000 } } }
\vdef{default-11:NoMc2e33-A:eta:loEff}   {\ensuremath{{0.720 } } }
\vdef{default-11:NoMc2e33-A:eta:loEffE}   {\ensuremath{{0.001 } } }
\vdef{default-11:NoMc2e33-A:eta:hiEff}   {\ensuremath{{0.280 } } }
\vdef{default-11:NoMc2e33-A:eta:hiEffE}   {\ensuremath{{0.001 } } }
\vdef{default-11:NoMcCMS-A:eta:loEff}   {\ensuremath{{0.841 } } }
\vdef{default-11:NoMcCMS-A:eta:loEffE}   {\ensuremath{{0.001 } } }
\vdef{default-11:NoMcCMS-A:eta:hiEff}   {\ensuremath{{0.159 } } }
\vdef{default-11:NoMcCMS-A:eta:hiEffE}   {\ensuremath{{0.001 } } }
\vdef{default-11:NoMcCMS-A:eta:loDelta}   {\ensuremath{{-0.155 } } }
\vdef{default-11:NoMcCMS-A:eta:loDeltaE}   {\ensuremath{{0.002 } } }
\vdef{default-11:NoMcCMS-A:eta:hiDelta}   {\ensuremath{{+0.551 } } }
\vdef{default-11:NoMcCMS-A:eta:hiDeltaE}   {\ensuremath{{0.007 } } }
\vdef{default-11:NoMc2e33-A:bdt:loEff}   {\ensuremath{{0.914 } } }
\vdef{default-11:NoMc2e33-A:bdt:loEffE}   {\ensuremath{{0.001 } } }
\vdef{default-11:NoMc2e33-A:bdt:hiEff}   {\ensuremath{{0.086 } } }
\vdef{default-11:NoMc2e33-A:bdt:hiEffE}   {\ensuremath{{0.001 } } }
\vdef{default-11:NoMcCMS-A:bdt:loEff}   {\ensuremath{{0.871 } } }
\vdef{default-11:NoMcCMS-A:bdt:loEffE}   {\ensuremath{{0.001 } } }
\vdef{default-11:NoMcCMS-A:bdt:hiEff}   {\ensuremath{{0.129 } } }
\vdef{default-11:NoMcCMS-A:bdt:hiEffE}   {\ensuremath{{0.001 } } }
\vdef{default-11:NoMcCMS-A:bdt:loDelta}   {\ensuremath{{+0.049 } } }
\vdef{default-11:NoMcCMS-A:bdt:loDeltaE}   {\ensuremath{{0.001 } } }
\vdef{default-11:NoMcCMS-A:bdt:hiDelta}   {\ensuremath{{-0.404 } } }
\vdef{default-11:NoMcCMS-A:bdt:hiDeltaE}   {\ensuremath{{0.009 } } }
\vdef{default-11:NoMc2e33-A:fl3d:loEff}   {\ensuremath{{0.826 } } }
\vdef{default-11:NoMc2e33-A:fl3d:loEffE}   {\ensuremath{{0.001 } } }
\vdef{default-11:NoMc2e33-A:fl3d:hiEff}   {\ensuremath{{0.174 } } }
\vdef{default-11:NoMc2e33-A:fl3d:hiEffE}   {\ensuremath{{0.001 } } }
\vdef{default-11:NoMcCMS-A:fl3d:loEff}   {\ensuremath{{0.881 } } }
\vdef{default-11:NoMcCMS-A:fl3d:loEffE}   {\ensuremath{{0.001 } } }
\vdef{default-11:NoMcCMS-A:fl3d:hiEff}   {\ensuremath{{0.119 } } }
\vdef{default-11:NoMcCMS-A:fl3d:hiEffE}   {\ensuremath{{0.001 } } }
\vdef{default-11:NoMcCMS-A:fl3d:loDelta}   {\ensuremath{{-0.065 } } }
\vdef{default-11:NoMcCMS-A:fl3d:loDeltaE}   {\ensuremath{{0.002 } } }
\vdef{default-11:NoMcCMS-A:fl3d:hiDelta}   {\ensuremath{{+0.376 } } }
\vdef{default-11:NoMcCMS-A:fl3d:hiDeltaE}   {\ensuremath{{0.009 } } }
\vdef{default-11:NoMc2e33-A:fl3de:loEff}   {\ensuremath{{1.000 } } }
\vdef{default-11:NoMc2e33-A:fl3de:loEffE}   {\ensuremath{{0.000 } } }
\vdef{default-11:NoMc2e33-A:fl3de:hiEff}   {\ensuremath{{0.000 } } }
\vdef{default-11:NoMc2e33-A:fl3de:hiEffE}   {\ensuremath{{0.000 } } }
\vdef{default-11:NoMcCMS-A:fl3de:loEff}   {\ensuremath{{1.000 } } }
\vdef{default-11:NoMcCMS-A:fl3de:loEffE}   {\ensuremath{{0.000 } } }
\vdef{default-11:NoMcCMS-A:fl3de:hiEff}   {\ensuremath{{0.000 } } }
\vdef{default-11:NoMcCMS-A:fl3de:hiEffE}   {\ensuremath{{0.000 } } }
\vdef{default-11:NoMcCMS-A:fl3de:loDelta}   {\ensuremath{{+0.000 } } }
\vdef{default-11:NoMcCMS-A:fl3de:loDeltaE}   {\ensuremath{{0.000 } } }
\vdef{default-11:NoMcCMS-A:fl3de:hiDelta}   {\ensuremath{{+0.083 } } }
\vdef{default-11:NoMcCMS-A:fl3de:hiDeltaE}   {\ensuremath{{0.575 } } }
\vdef{default-11:NoMc2e33-A:fls3d:loEff}   {\ensuremath{{0.078 } } }
\vdef{default-11:NoMc2e33-A:fls3d:loEffE}   {\ensuremath{{0.001 } } }
\vdef{default-11:NoMc2e33-A:fls3d:hiEff}   {\ensuremath{{0.922 } } }
\vdef{default-11:NoMc2e33-A:fls3d:hiEffE}   {\ensuremath{{0.001 } } }
\vdef{default-11:NoMcCMS-A:fls3d:loEff}   {\ensuremath{{0.060 } } }
\vdef{default-11:NoMcCMS-A:fls3d:loEffE}   {\ensuremath{{0.001 } } }
\vdef{default-11:NoMcCMS-A:fls3d:hiEff}   {\ensuremath{{0.940 } } }
\vdef{default-11:NoMcCMS-A:fls3d:hiEffE}   {\ensuremath{{0.001 } } }
\vdef{default-11:NoMcCMS-A:fls3d:loDelta}   {\ensuremath{{+0.263 } } }
\vdef{default-11:NoMcCMS-A:fls3d:loDeltaE}   {\ensuremath{{0.013 } } }
\vdef{default-11:NoMcCMS-A:fls3d:hiDelta}   {\ensuremath{{-0.019 } } }
\vdef{default-11:NoMcCMS-A:fls3d:hiDeltaE}   {\ensuremath{{0.001 } } }
\vdef{default-11:NoMc2e33-A:flsxy:loEff}   {\ensuremath{{1.012 } } }
\vdef{default-11:NoMc2e33-A:flsxy:loEffE}   {\ensuremath{{\mathrm{NaN} } } }
\vdef{default-11:NoMc2e33-A:flsxy:hiEff}   {\ensuremath{{1.000 } } }
\vdef{default-11:NoMc2e33-A:flsxy:hiEffE}   {\ensuremath{{0.000 } } }
\vdef{default-11:NoMcCMS-A:flsxy:loEff}   {\ensuremath{{1.013 } } }
\vdef{default-11:NoMcCMS-A:flsxy:loEffE}   {\ensuremath{{\mathrm{NaN} } } }
\vdef{default-11:NoMcCMS-A:flsxy:hiEff}   {\ensuremath{{1.000 } } }
\vdef{default-11:NoMcCMS-A:flsxy:hiEffE}   {\ensuremath{{0.000 } } }
\vdef{default-11:NoMcCMS-A:flsxy:loDelta}   {\ensuremath{{-0.001 } } }
\vdef{default-11:NoMcCMS-A:flsxy:loDeltaE}   {\ensuremath{{\mathrm{NaN} } } }
\vdef{default-11:NoMcCMS-A:flsxy:hiDelta}   {\ensuremath{{+0.000 } } }
\vdef{default-11:NoMcCMS-A:flsxy:hiDeltaE}   {\ensuremath{{0.000 } } }
\vdef{default-11:NoMc2e33-A:chi2dof:loEff}   {\ensuremath{{0.940 } } }
\vdef{default-11:NoMc2e33-A:chi2dof:loEffE}   {\ensuremath{{0.001 } } }
\vdef{default-11:NoMc2e33-A:chi2dof:hiEff}   {\ensuremath{{0.060 } } }
\vdef{default-11:NoMc2e33-A:chi2dof:hiEffE}   {\ensuremath{{0.001 } } }
\vdef{default-11:NoMcCMS-A:chi2dof:loEff}   {\ensuremath{{0.939 } } }
\vdef{default-11:NoMcCMS-A:chi2dof:loEffE}   {\ensuremath{{0.001 } } }
\vdef{default-11:NoMcCMS-A:chi2dof:hiEff}   {\ensuremath{{0.061 } } }
\vdef{default-11:NoMcCMS-A:chi2dof:hiEffE}   {\ensuremath{{0.001 } } }
\vdef{default-11:NoMcCMS-A:chi2dof:loDelta}   {\ensuremath{{+0.000 } } }
\vdef{default-11:NoMcCMS-A:chi2dof:loDeltaE}   {\ensuremath{{0.001 } } }
\vdef{default-11:NoMcCMS-A:chi2dof:hiDelta}   {\ensuremath{{-0.006 } } }
\vdef{default-11:NoMcCMS-A:chi2dof:hiDeltaE}   {\ensuremath{{0.015 } } }
\vdef{default-11:NoMc2e33-A:pchi2dof:loEff}   {\ensuremath{{0.621 } } }
\vdef{default-11:NoMc2e33-A:pchi2dof:loEffE}   {\ensuremath{{0.001 } } }
\vdef{default-11:NoMc2e33-A:pchi2dof:hiEff}   {\ensuremath{{0.379 } } }
\vdef{default-11:NoMc2e33-A:pchi2dof:hiEffE}   {\ensuremath{{0.001 } } }
\vdef{default-11:NoMcCMS-A:pchi2dof:loEff}   {\ensuremath{{0.629 } } }
\vdef{default-11:NoMcCMS-A:pchi2dof:loEffE}   {\ensuremath{{0.001 } } }
\vdef{default-11:NoMcCMS-A:pchi2dof:hiEff}   {\ensuremath{{0.371 } } }
\vdef{default-11:NoMcCMS-A:pchi2dof:hiEffE}   {\ensuremath{{0.001 } } }
\vdef{default-11:NoMcCMS-A:pchi2dof:loDelta}   {\ensuremath{{-0.013 } } }
\vdef{default-11:NoMcCMS-A:pchi2dof:loDeltaE}   {\ensuremath{{0.003 } } }
\vdef{default-11:NoMcCMS-A:pchi2dof:hiDelta}   {\ensuremath{{+0.022 } } }
\vdef{default-11:NoMcCMS-A:pchi2dof:hiDeltaE}   {\ensuremath{{0.005 } } }
\vdef{default-11:NoMc2e33-A:alpha:loEff}   {\ensuremath{{0.994 } } }
\vdef{default-11:NoMc2e33-A:alpha:loEffE}   {\ensuremath{{0.000 } } }
\vdef{default-11:NoMc2e33-A:alpha:hiEff}   {\ensuremath{{0.006 } } }
\vdef{default-11:NoMc2e33-A:alpha:hiEffE}   {\ensuremath{{0.000 } } }
\vdef{default-11:NoMcCMS-A:alpha:loEff}   {\ensuremath{{0.993 } } }
\vdef{default-11:NoMcCMS-A:alpha:loEffE}   {\ensuremath{{0.000 } } }
\vdef{default-11:NoMcCMS-A:alpha:hiEff}   {\ensuremath{{0.007 } } }
\vdef{default-11:NoMcCMS-A:alpha:hiEffE}   {\ensuremath{{0.000 } } }
\vdef{default-11:NoMcCMS-A:alpha:loDelta}   {\ensuremath{{+0.001 } } }
\vdef{default-11:NoMcCMS-A:alpha:loDeltaE}   {\ensuremath{{0.000 } } }
\vdef{default-11:NoMcCMS-A:alpha:hiDelta}   {\ensuremath{{-0.128 } } }
\vdef{default-11:NoMcCMS-A:alpha:hiDeltaE}   {\ensuremath{{0.049 } } }
\vdef{default-11:NoMc2e33-A:iso:loEff}   {\ensuremath{{0.107 } } }
\vdef{default-11:NoMc2e33-A:iso:loEffE}   {\ensuremath{{0.001 } } }
\vdef{default-11:NoMc2e33-A:iso:hiEff}   {\ensuremath{{0.893 } } }
\vdef{default-11:NoMc2e33-A:iso:hiEffE}   {\ensuremath{{0.001 } } }
\vdef{default-11:NoMcCMS-A:iso:loEff}   {\ensuremath{{0.111 } } }
\vdef{default-11:NoMcCMS-A:iso:loEffE}   {\ensuremath{{0.001 } } }
\vdef{default-11:NoMcCMS-A:iso:hiEff}   {\ensuremath{{0.889 } } }
\vdef{default-11:NoMcCMS-A:iso:hiEffE}   {\ensuremath{{0.001 } } }
\vdef{default-11:NoMcCMS-A:iso:loDelta}   {\ensuremath{{-0.039 } } }
\vdef{default-11:NoMcCMS-A:iso:loDeltaE}   {\ensuremath{{0.011 } } }
\vdef{default-11:NoMcCMS-A:iso:hiDelta}   {\ensuremath{{+0.005 } } }
\vdef{default-11:NoMcCMS-A:iso:hiDeltaE}   {\ensuremath{{0.001 } } }
\vdef{default-11:NoMc2e33-A:docatrk:loEff}   {\ensuremath{{0.082 } } }
\vdef{default-11:NoMc2e33-A:docatrk:loEffE}   {\ensuremath{{0.001 } } }
\vdef{default-11:NoMc2e33-A:docatrk:hiEff}   {\ensuremath{{0.918 } } }
\vdef{default-11:NoMc2e33-A:docatrk:hiEffE}   {\ensuremath{{0.001 } } }
\vdef{default-11:NoMcCMS-A:docatrk:loEff}   {\ensuremath{{0.087 } } }
\vdef{default-11:NoMcCMS-A:docatrk:loEffE}   {\ensuremath{{0.001 } } }
\vdef{default-11:NoMcCMS-A:docatrk:hiEff}   {\ensuremath{{0.913 } } }
\vdef{default-11:NoMcCMS-A:docatrk:hiEffE}   {\ensuremath{{0.001 } } }
\vdef{default-11:NoMcCMS-A:docatrk:loDelta}   {\ensuremath{{-0.059 } } }
\vdef{default-11:NoMcCMS-A:docatrk:loDeltaE}   {\ensuremath{{0.013 } } }
\vdef{default-11:NoMcCMS-A:docatrk:hiDelta}   {\ensuremath{{+0.005 } } }
\vdef{default-11:NoMcCMS-A:docatrk:hiDeltaE}   {\ensuremath{{0.001 } } }
\vdef{default-11:NoMc2e33-A:isotrk:loEff}   {\ensuremath{{1.000 } } }
\vdef{default-11:NoMc2e33-A:isotrk:loEffE}   {\ensuremath{{0.000 } } }
\vdef{default-11:NoMc2e33-A:isotrk:hiEff}   {\ensuremath{{1.000 } } }
\vdef{default-11:NoMc2e33-A:isotrk:hiEffE}   {\ensuremath{{0.000 } } }
\vdef{default-11:NoMcCMS-A:isotrk:loEff}   {\ensuremath{{1.000 } } }
\vdef{default-11:NoMcCMS-A:isotrk:loEffE}   {\ensuremath{{0.000 } } }
\vdef{default-11:NoMcCMS-A:isotrk:hiEff}   {\ensuremath{{1.000 } } }
\vdef{default-11:NoMcCMS-A:isotrk:hiEffE}   {\ensuremath{{0.000 } } }
\vdef{default-11:NoMcCMS-A:isotrk:loDelta}   {\ensuremath{{+0.000 } } }
\vdef{default-11:NoMcCMS-A:isotrk:loDeltaE}   {\ensuremath{{0.000 } } }
\vdef{default-11:NoMcCMS-A:isotrk:hiDelta}   {\ensuremath{{+0.000 } } }
\vdef{default-11:NoMcCMS-A:isotrk:hiDeltaE}   {\ensuremath{{0.000 } } }
\vdef{default-11:NoMc2e33-A:closetrk:loEff}   {\ensuremath{{0.978 } } }
\vdef{default-11:NoMc2e33-A:closetrk:loEffE}   {\ensuremath{{0.000 } } }
\vdef{default-11:NoMc2e33-A:closetrk:hiEff}   {\ensuremath{{0.022 } } }
\vdef{default-11:NoMc2e33-A:closetrk:hiEffE}   {\ensuremath{{0.000 } } }
\vdef{default-11:NoMcCMS-A:closetrk:loEff}   {\ensuremath{{0.974 } } }
\vdef{default-11:NoMcCMS-A:closetrk:loEffE}   {\ensuremath{{0.000 } } }
\vdef{default-11:NoMcCMS-A:closetrk:hiEff}   {\ensuremath{{0.026 } } }
\vdef{default-11:NoMcCMS-A:closetrk:hiEffE}   {\ensuremath{{0.000 } } }
\vdef{default-11:NoMcCMS-A:closetrk:loDelta}   {\ensuremath{{+0.004 } } }
\vdef{default-11:NoMcCMS-A:closetrk:loDeltaE}   {\ensuremath{{0.001 } } }
\vdef{default-11:NoMcCMS-A:closetrk:hiDelta}   {\ensuremath{{-0.165 } } }
\vdef{default-11:NoMcCMS-A:closetrk:hiDeltaE}   {\ensuremath{{0.025 } } }
\vdef{default-11:NoMc2e33-A:lip:loEff}   {\ensuremath{{1.000 } } }
\vdef{default-11:NoMc2e33-A:lip:loEffE}   {\ensuremath{{0.000 } } }
\vdef{default-11:NoMc2e33-A:lip:hiEff}   {\ensuremath{{0.000 } } }
\vdef{default-11:NoMc2e33-A:lip:hiEffE}   {\ensuremath{{0.000 } } }
\vdef{default-11:NoMcCMS-A:lip:loEff}   {\ensuremath{{1.000 } } }
\vdef{default-11:NoMcCMS-A:lip:loEffE}   {\ensuremath{{0.000 } } }
\vdef{default-11:NoMcCMS-A:lip:hiEff}   {\ensuremath{{0.000 } } }
\vdef{default-11:NoMcCMS-A:lip:hiEffE}   {\ensuremath{{0.000 } } }
\vdef{default-11:NoMcCMS-A:lip:loDelta}   {\ensuremath{{+0.000 } } }
\vdef{default-11:NoMcCMS-A:lip:loDeltaE}   {\ensuremath{{0.000 } } }
\vdef{default-11:NoMcCMS-A:lip:hiDelta}   {\ensuremath{{\mathrm{NaN} } } }
\vdef{default-11:NoMcCMS-A:lip:hiDeltaE}   {\ensuremath{{\mathrm{NaN} } } }
\vdef{default-11:NoMc2e33-A:lip:inEff}   {\ensuremath{{1.000 } } }
\vdef{default-11:NoMc2e33-A:lip:inEffE}   {\ensuremath{{0.000 } } }
\vdef{default-11:NoMcCMS-A:lip:inEff}   {\ensuremath{{1.000 } } }
\vdef{default-11:NoMcCMS-A:lip:inEffE}   {\ensuremath{{0.000 } } }
\vdef{default-11:NoMcCMS-A:lip:inDelta}   {\ensuremath{{+0.000 } } }
\vdef{default-11:NoMcCMS-A:lip:inDeltaE}   {\ensuremath{{0.000 } } }
\vdef{default-11:NoMc2e33-A:lips:loEff}   {\ensuremath{{1.000 } } }
\vdef{default-11:NoMc2e33-A:lips:loEffE}   {\ensuremath{{0.000 } } }
\vdef{default-11:NoMc2e33-A:lips:hiEff}   {\ensuremath{{0.000 } } }
\vdef{default-11:NoMc2e33-A:lips:hiEffE}   {\ensuremath{{0.000 } } }
\vdef{default-11:NoMcCMS-A:lips:loEff}   {\ensuremath{{1.000 } } }
\vdef{default-11:NoMcCMS-A:lips:loEffE}   {\ensuremath{{0.000 } } }
\vdef{default-11:NoMcCMS-A:lips:hiEff}   {\ensuremath{{0.000 } } }
\vdef{default-11:NoMcCMS-A:lips:hiEffE}   {\ensuremath{{0.000 } } }
\vdef{default-11:NoMcCMS-A:lips:loDelta}   {\ensuremath{{+0.000 } } }
\vdef{default-11:NoMcCMS-A:lips:loDeltaE}   {\ensuremath{{0.000 } } }
\vdef{default-11:NoMcCMS-A:lips:hiDelta}   {\ensuremath{{\mathrm{NaN} } } }
\vdef{default-11:NoMcCMS-A:lips:hiDeltaE}   {\ensuremath{{\mathrm{NaN} } } }
\vdef{default-11:NoMc2e33-A:lips:inEff}   {\ensuremath{{1.000 } } }
\vdef{default-11:NoMc2e33-A:lips:inEffE}   {\ensuremath{{0.000 } } }
\vdef{default-11:NoMcCMS-A:lips:inEff}   {\ensuremath{{1.000 } } }
\vdef{default-11:NoMcCMS-A:lips:inEffE}   {\ensuremath{{0.000 } } }
\vdef{default-11:NoMcCMS-A:lips:inDelta}   {\ensuremath{{+0.000 } } }
\vdef{default-11:NoMcCMS-A:lips:inDeltaE}   {\ensuremath{{0.000 } } }
\vdef{default-11:NoMc2e33-A:ip:loEff}   {\ensuremath{{0.972 } } }
\vdef{default-11:NoMc2e33-A:ip:loEffE}   {\ensuremath{{0.000 } } }
\vdef{default-11:NoMc2e33-A:ip:hiEff}   {\ensuremath{{0.028 } } }
\vdef{default-11:NoMc2e33-A:ip:hiEffE}   {\ensuremath{{0.000 } } }
\vdef{default-11:NoMcCMS-A:ip:loEff}   {\ensuremath{{0.970 } } }
\vdef{default-11:NoMcCMS-A:ip:loEffE}   {\ensuremath{{0.000 } } }
\vdef{default-11:NoMcCMS-A:ip:hiEff}   {\ensuremath{{0.030 } } }
\vdef{default-11:NoMcCMS-A:ip:hiEffE}   {\ensuremath{{0.000 } } }
\vdef{default-11:NoMcCMS-A:ip:loDelta}   {\ensuremath{{+0.002 } } }
\vdef{default-11:NoMcCMS-A:ip:loDeltaE}   {\ensuremath{{0.001 } } }
\vdef{default-11:NoMcCMS-A:ip:hiDelta}   {\ensuremath{{-0.073 } } }
\vdef{default-11:NoMcCMS-A:ip:hiDeltaE}   {\ensuremath{{0.022 } } }
\vdef{default-11:NoMc2e33-A:ips:loEff}   {\ensuremath{{0.959 } } }
\vdef{default-11:NoMc2e33-A:ips:loEffE}   {\ensuremath{{0.001 } } }
\vdef{default-11:NoMc2e33-A:ips:hiEff}   {\ensuremath{{0.041 } } }
\vdef{default-11:NoMc2e33-A:ips:hiEffE}   {\ensuremath{{0.001 } } }
\vdef{default-11:NoMcCMS-A:ips:loEff}   {\ensuremath{{0.951 } } }
\vdef{default-11:NoMcCMS-A:ips:loEffE}   {\ensuremath{{0.001 } } }
\vdef{default-11:NoMcCMS-A:ips:hiEff}   {\ensuremath{{0.049 } } }
\vdef{default-11:NoMcCMS-A:ips:hiEffE}   {\ensuremath{{0.001 } } }
\vdef{default-11:NoMcCMS-A:ips:loDelta}   {\ensuremath{{+0.008 } } }
\vdef{default-11:NoMcCMS-A:ips:loDeltaE}   {\ensuremath{{0.001 } } }
\vdef{default-11:NoMcCMS-A:ips:hiDelta}   {\ensuremath{{-0.170 } } }
\vdef{default-11:NoMcCMS-A:ips:hiDeltaE}   {\ensuremath{{0.018 } } }
\vdef{default-11:NoMc2e33-A:maxdoca:loEff}   {\ensuremath{{1.000 } } }
\vdef{default-11:NoMc2e33-A:maxdoca:loEffE}   {\ensuremath{{0.000 } } }
\vdef{default-11:NoMc2e33-A:maxdoca:hiEff}   {\ensuremath{{0.011 } } }
\vdef{default-11:NoMc2e33-A:maxdoca:hiEffE}   {\ensuremath{{0.000 } } }
\vdef{default-11:NoMcCMS-A:maxdoca:loEff}   {\ensuremath{{1.000 } } }
\vdef{default-11:NoMcCMS-A:maxdoca:loEffE}   {\ensuremath{{0.000 } } }
\vdef{default-11:NoMcCMS-A:maxdoca:hiEff}   {\ensuremath{{0.009 } } }
\vdef{default-11:NoMcCMS-A:maxdoca:hiEffE}   {\ensuremath{{0.000 } } }
\vdef{default-11:NoMcCMS-A:maxdoca:loDelta}   {\ensuremath{{+0.000 } } }
\vdef{default-11:NoMcCMS-A:maxdoca:loDeltaE}   {\ensuremath{{0.000 } } }
\vdef{default-11:NoMcCMS-A:maxdoca:hiDelta}   {\ensuremath{{+0.233 } } }
\vdef{default-11:NoMcCMS-A:maxdoca:hiDeltaE}   {\ensuremath{{0.040 } } }
\vdef{default-11:NoMc2e33-A:kaonpt:loEff}   {\ensuremath{{1.007 } } }
\vdef{default-11:NoMc2e33-A:kaonpt:loEffE}   {\ensuremath{{\mathrm{NaN} } } }
\vdef{default-11:NoMc2e33-A:kaonpt:hiEff}   {\ensuremath{{1.000 } } }
\vdef{default-11:NoMc2e33-A:kaonpt:hiEffE}   {\ensuremath{{0.000 } } }
\vdef{default-11:NoMcCMS-A:kaonpt:loEff}   {\ensuremath{{1.008 } } }
\vdef{default-11:NoMcCMS-A:kaonpt:loEffE}   {\ensuremath{{\mathrm{NaN} } } }
\vdef{default-11:NoMcCMS-A:kaonpt:hiEff}   {\ensuremath{{1.000 } } }
\vdef{default-11:NoMcCMS-A:kaonpt:hiEffE}   {\ensuremath{{0.000 } } }
\vdef{default-11:NoMcCMS-A:kaonpt:loDelta}   {\ensuremath{{-0.001 } } }
\vdef{default-11:NoMcCMS-A:kaonpt:loDeltaE}   {\ensuremath{{\mathrm{NaN} } } }
\vdef{default-11:NoMcCMS-A:kaonpt:hiDelta}   {\ensuremath{{+0.000 } } }
\vdef{default-11:NoMcCMS-A:kaonpt:hiDeltaE}   {\ensuremath{{0.000 } } }
\vdef{default-11:NoMc2e33-A:psipt:loEff}   {\ensuremath{{1.003 } } }
\vdef{default-11:NoMc2e33-A:psipt:loEffE}   {\ensuremath{{\mathrm{NaN} } } }
\vdef{default-11:NoMc2e33-A:psipt:hiEff}   {\ensuremath{{1.000 } } }
\vdef{default-11:NoMc2e33-A:psipt:hiEffE}   {\ensuremath{{0.000 } } }
\vdef{default-11:NoMcCMS-A:psipt:loEff}   {\ensuremath{{1.003 } } }
\vdef{default-11:NoMcCMS-A:psipt:loEffE}   {\ensuremath{{\mathrm{NaN} } } }
\vdef{default-11:NoMcCMS-A:psipt:hiEff}   {\ensuremath{{1.000 } } }
\vdef{default-11:NoMcCMS-A:psipt:hiEffE}   {\ensuremath{{0.000 } } }
\vdef{default-11:NoMcCMS-A:psipt:loDelta}   {\ensuremath{{-0.000 } } }
\vdef{default-11:NoMcCMS-A:psipt:loDeltaE}   {\ensuremath{{\mathrm{NaN} } } }
\vdef{default-11:NoMcCMS-A:psipt:hiDelta}   {\ensuremath{{+0.000 } } }
\vdef{default-11:NoMcCMS-A:psipt:hiDeltaE}   {\ensuremath{{0.000 } } }
\vdef{default-11:NoData-AR3:osiso:loEff}   {\ensuremath{{1.003 } } }
\vdef{default-11:NoData-AR3:osiso:loEffE}   {\ensuremath{{\mathrm{NaN} } } }
\vdef{default-11:NoData-AR3:osiso:hiEff}   {\ensuremath{{1.000 } } }
\vdef{default-11:NoData-AR3:osiso:hiEffE}   {\ensuremath{{0.000 } } }
\vdef{default-11:NoMc2e33-A:osiso:loEff}   {\ensuremath{{1.003 } } }
\vdef{default-11:NoMc2e33-A:osiso:loEffE}   {\ensuremath{{\mathrm{NaN} } } }
\vdef{default-11:NoMc2e33-A:osiso:hiEff}   {\ensuremath{{1.000 } } }
\vdef{default-11:NoMc2e33-A:osiso:hiEffE}   {\ensuremath{{0.000 } } }
\vdef{default-11:NoMc2e33-A:osiso:loDelta}   {\ensuremath{{+0.000 } } }
\vdef{default-11:NoMc2e33-A:osiso:loDeltaE}   {\ensuremath{{\mathrm{NaN} } } }
\vdef{default-11:NoMc2e33-A:osiso:hiDelta}   {\ensuremath{{+0.000 } } }
\vdef{default-11:NoMc2e33-A:osiso:hiDeltaE}   {\ensuremath{{0.000 } } }
\vdef{default-11:NoData-AR3:osreliso:loEff}   {\ensuremath{{0.254 } } }
\vdef{default-11:NoData-AR3:osreliso:loEffE}   {\ensuremath{{0.003 } } }
\vdef{default-11:NoData-AR3:osreliso:hiEff}   {\ensuremath{{0.746 } } }
\vdef{default-11:NoData-AR3:osreliso:hiEffE}   {\ensuremath{{0.003 } } }
\vdef{default-11:NoMc2e33-A:osreliso:loEff}   {\ensuremath{{0.289 } } }
\vdef{default-11:NoMc2e33-A:osreliso:loEffE}   {\ensuremath{{0.003 } } }
\vdef{default-11:NoMc2e33-A:osreliso:hiEff}   {\ensuremath{{0.711 } } }
\vdef{default-11:NoMc2e33-A:osreliso:hiEffE}   {\ensuremath{{0.003 } } }
\vdef{default-11:NoMc2e33-A:osreliso:loDelta}   {\ensuremath{{-0.126 } } }
\vdef{default-11:NoMc2e33-A:osreliso:loDeltaE}   {\ensuremath{{0.016 } } }
\vdef{default-11:NoMc2e33-A:osreliso:hiDelta}   {\ensuremath{{+0.047 } } }
\vdef{default-11:NoMc2e33-A:osreliso:hiDeltaE}   {\ensuremath{{0.006 } } }
\vdef{default-11:NoData-AR3:osmuonpt:loEff}   {\ensuremath{{0.000 } } }
\vdef{default-11:NoData-AR3:osmuonpt:loEffE}   {\ensuremath{{0.001 } } }
\vdef{default-11:NoData-AR3:osmuonpt:hiEff}   {\ensuremath{{1.000 } } }
\vdef{default-11:NoData-AR3:osmuonpt:hiEffE}   {\ensuremath{{0.001 } } }
\vdef{default-11:NoMc2e33-A:osmuonpt:loEff}   {\ensuremath{{0.000 } } }
\vdef{default-11:NoMc2e33-A:osmuonpt:loEffE}   {\ensuremath{{0.001 } } }
\vdef{default-11:NoMc2e33-A:osmuonpt:hiEff}   {\ensuremath{{1.000 } } }
\vdef{default-11:NoMc2e33-A:osmuonpt:hiEffE}   {\ensuremath{{0.001 } } }
\vdef{default-11:NoMc2e33-A:osmuonpt:loDelta}   {\ensuremath{{\mathrm{NaN} } } }
\vdef{default-11:NoMc2e33-A:osmuonpt:loDeltaE}   {\ensuremath{{\mathrm{NaN} } } }
\vdef{default-11:NoMc2e33-A:osmuonpt:hiDelta}   {\ensuremath{{+0.000 } } }
\vdef{default-11:NoMc2e33-A:osmuonpt:hiDeltaE}   {\ensuremath{{0.002 } } }
\vdef{default-11:NoData-AR3:osmuondr:loEff}   {\ensuremath{{0.022 } } }
\vdef{default-11:NoData-AR3:osmuondr:loEffE}   {\ensuremath{{0.005 } } }
\vdef{default-11:NoData-AR3:osmuondr:hiEff}   {\ensuremath{{0.978 } } }
\vdef{default-11:NoData-AR3:osmuondr:hiEffE}   {\ensuremath{{0.006 } } }
\vdef{default-11:NoMc2e33-A:osmuondr:loEff}   {\ensuremath{{0.018 } } }
\vdef{default-11:NoMc2e33-A:osmuondr:loEffE}   {\ensuremath{{0.005 } } }
\vdef{default-11:NoMc2e33-A:osmuondr:hiEff}   {\ensuremath{{0.982 } } }
\vdef{default-11:NoMc2e33-A:osmuondr:hiEffE}   {\ensuremath{{0.005 } } }
\vdef{default-11:NoMc2e33-A:osmuondr:loDelta}   {\ensuremath{{+0.209 } } }
\vdef{default-11:NoMc2e33-A:osmuondr:loDeltaE}   {\ensuremath{{0.361 } } }
\vdef{default-11:NoMc2e33-A:osmuondr:hiDelta}   {\ensuremath{{-0.004 } } }
\vdef{default-11:NoMc2e33-A:osmuondr:hiDeltaE}   {\ensuremath{{0.008 } } }
\vdef{default-11:NoData-AR3:hlt:loEff}   {\ensuremath{{0.049 } } }
\vdef{default-11:NoData-AR3:hlt:loEffE}   {\ensuremath{{0.002 } } }
\vdef{default-11:NoData-AR3:hlt:hiEff}   {\ensuremath{{0.951 } } }
\vdef{default-11:NoData-AR3:hlt:hiEffE}   {\ensuremath{{0.002 } } }
\vdef{default-11:NoMc2e33-A:hlt:loEff}   {\ensuremath{{0.277 } } }
\vdef{default-11:NoMc2e33-A:hlt:loEffE}   {\ensuremath{{0.003 } } }
\vdef{default-11:NoMc2e33-A:hlt:hiEff}   {\ensuremath{{0.723 } } }
\vdef{default-11:NoMc2e33-A:hlt:hiEffE}   {\ensuremath{{0.003 } } }
\vdef{default-11:NoMc2e33-A:hlt:loDelta}   {\ensuremath{{-1.401 } } }
\vdef{default-11:NoMc2e33-A:hlt:loDeltaE}   {\ensuremath{{0.017 } } }
\vdef{default-11:NoMc2e33-A:hlt:hiDelta}   {\ensuremath{{+0.273 } } }
\vdef{default-11:NoMc2e33-A:hlt:hiDeltaE}   {\ensuremath{{0.005 } } }
\vdef{default-11:NoData-AR3:muonsid:loEff}   {\ensuremath{{0.140 } } }
\vdef{default-11:NoData-AR3:muonsid:loEffE}   {\ensuremath{{0.002 } } }
\vdef{default-11:NoData-AR3:muonsid:hiEff}   {\ensuremath{{0.860 } } }
\vdef{default-11:NoData-AR3:muonsid:hiEffE}   {\ensuremath{{0.002 } } }
\vdef{default-11:NoMc2e33-A:muonsid:loEff}   {\ensuremath{{0.158 } } }
\vdef{default-11:NoMc2e33-A:muonsid:loEffE}   {\ensuremath{{0.002 } } }
\vdef{default-11:NoMc2e33-A:muonsid:hiEff}   {\ensuremath{{0.842 } } }
\vdef{default-11:NoMc2e33-A:muonsid:hiEffE}   {\ensuremath{{0.002 } } }
\vdef{default-11:NoMc2e33-A:muonsid:loDelta}   {\ensuremath{{-0.120 } } }
\vdef{default-11:NoMc2e33-A:muonsid:loDeltaE}   {\ensuremath{{0.023 } } }
\vdef{default-11:NoMc2e33-A:muonsid:hiDelta}   {\ensuremath{{+0.021 } } }
\vdef{default-11:NoMc2e33-A:muonsid:hiDeltaE}   {\ensuremath{{0.004 } } }
\vdef{default-11:NoData-AR3:tracksqual:loEff}   {\ensuremath{{0.000 } } }
\vdef{default-11:NoData-AR3:tracksqual:loEffE}   {\ensuremath{{0.000 } } }
\vdef{default-11:NoData-AR3:tracksqual:hiEff}   {\ensuremath{{1.000 } } }
\vdef{default-11:NoData-AR3:tracksqual:hiEffE}   {\ensuremath{{0.000 } } }
\vdef{default-11:NoMc2e33-A:tracksqual:loEff}   {\ensuremath{{0.000 } } }
\vdef{default-11:NoMc2e33-A:tracksqual:loEffE}   {\ensuremath{{0.000 } } }
\vdef{default-11:NoMc2e33-A:tracksqual:hiEff}   {\ensuremath{{1.000 } } }
\vdef{default-11:NoMc2e33-A:tracksqual:hiEffE}   {\ensuremath{{0.000 } } }
\vdef{default-11:NoMc2e33-A:tracksqual:loDelta}   {\ensuremath{{+0.952 } } }
\vdef{default-11:NoMc2e33-A:tracksqual:loDeltaE}   {\ensuremath{{0.582 } } }
\vdef{default-11:NoMc2e33-A:tracksqual:hiDelta}   {\ensuremath{{-0.000 } } }
\vdef{default-11:NoMc2e33-A:tracksqual:hiDeltaE}   {\ensuremath{{0.000 } } }
\vdef{default-11:NoData-AR3:pvz:loEff}   {\ensuremath{{0.508 } } }
\vdef{default-11:NoData-AR3:pvz:loEffE}   {\ensuremath{{0.004 } } }
\vdef{default-11:NoData-AR3:pvz:hiEff}   {\ensuremath{{0.492 } } }
\vdef{default-11:NoData-AR3:pvz:hiEffE}   {\ensuremath{{0.004 } } }
\vdef{default-11:NoMc2e33-A:pvz:loEff}   {\ensuremath{{0.469 } } }
\vdef{default-11:NoMc2e33-A:pvz:loEffE}   {\ensuremath{{0.004 } } }
\vdef{default-11:NoMc2e33-A:pvz:hiEff}   {\ensuremath{{0.531 } } }
\vdef{default-11:NoMc2e33-A:pvz:hiEffE}   {\ensuremath{{0.004 } } }
\vdef{default-11:NoMc2e33-A:pvz:loDelta}   {\ensuremath{{+0.080 } } }
\vdef{default-11:NoMc2e33-A:pvz:loDeltaE}   {\ensuremath{{0.010 } } }
\vdef{default-11:NoMc2e33-A:pvz:hiDelta}   {\ensuremath{{-0.076 } } }
\vdef{default-11:NoMc2e33-A:pvz:hiDeltaE}   {\ensuremath{{0.010 } } }
\vdef{default-11:NoData-AR3:pvn:loEff}   {\ensuremath{{1.000 } } }
\vdef{default-11:NoData-AR3:pvn:loEffE}   {\ensuremath{{\mathrm{NaN} } } }
\vdef{default-11:NoData-AR3:pvn:hiEff}   {\ensuremath{{1.000 } } }
\vdef{default-11:NoData-AR3:pvn:hiEffE}   {\ensuremath{{0.000 } } }
\vdef{default-11:NoMc2e33-A:pvn:loEff}   {\ensuremath{{1.000 } } }
\vdef{default-11:NoMc2e33-A:pvn:loEffE}   {\ensuremath{{0.000 } } }
\vdef{default-11:NoMc2e33-A:pvn:hiEff}   {\ensuremath{{1.000 } } }
\vdef{default-11:NoMc2e33-A:pvn:hiEffE}   {\ensuremath{{0.000 } } }
\vdef{default-11:NoMc2e33-A:pvn:loDelta}   {\ensuremath{{+0.000 } } }
\vdef{default-11:NoMc2e33-A:pvn:loDeltaE}   {\ensuremath{{\mathrm{NaN} } } }
\vdef{default-11:NoMc2e33-A:pvn:hiDelta}   {\ensuremath{{+0.000 } } }
\vdef{default-11:NoMc2e33-A:pvn:hiDeltaE}   {\ensuremath{{0.000 } } }
\vdef{default-11:NoData-AR3:pvavew8:loEff}   {\ensuremath{{0.013 } } }
\vdef{default-11:NoData-AR3:pvavew8:loEffE}   {\ensuremath{{0.001 } } }
\vdef{default-11:NoData-AR3:pvavew8:hiEff}   {\ensuremath{{0.987 } } }
\vdef{default-11:NoData-AR3:pvavew8:hiEffE}   {\ensuremath{{0.001 } } }
\vdef{default-11:NoMc2e33-A:pvavew8:loEff}   {\ensuremath{{0.004 } } }
\vdef{default-11:NoMc2e33-A:pvavew8:loEffE}   {\ensuremath{{0.000 } } }
\vdef{default-11:NoMc2e33-A:pvavew8:hiEff}   {\ensuremath{{0.996 } } }
\vdef{default-11:NoMc2e33-A:pvavew8:hiEffE}   {\ensuremath{{0.000 } } }
\vdef{default-11:NoMc2e33-A:pvavew8:loDelta}   {\ensuremath{{+1.035 } } }
\vdef{default-11:NoMc2e33-A:pvavew8:loDeltaE}   {\ensuremath{{0.098 } } }
\vdef{default-11:NoMc2e33-A:pvavew8:hiDelta}   {\ensuremath{{-0.009 } } }
\vdef{default-11:NoMc2e33-A:pvavew8:hiDeltaE}   {\ensuremath{{0.001 } } }
\vdef{default-11:NoData-AR3:pvntrk:loEff}   {\ensuremath{{1.000 } } }
\vdef{default-11:NoData-AR3:pvntrk:loEffE}   {\ensuremath{{0.000 } } }
\vdef{default-11:NoData-AR3:pvntrk:hiEff}   {\ensuremath{{1.000 } } }
\vdef{default-11:NoData-AR3:pvntrk:hiEffE}   {\ensuremath{{0.000 } } }
\vdef{default-11:NoMc2e33-A:pvntrk:loEff}   {\ensuremath{{1.000 } } }
\vdef{default-11:NoMc2e33-A:pvntrk:loEffE}   {\ensuremath{{0.000 } } }
\vdef{default-11:NoMc2e33-A:pvntrk:hiEff}   {\ensuremath{{1.000 } } }
\vdef{default-11:NoMc2e33-A:pvntrk:hiEffE}   {\ensuremath{{0.000 } } }
\vdef{default-11:NoMc2e33-A:pvntrk:loDelta}   {\ensuremath{{+0.000 } } }
\vdef{default-11:NoMc2e33-A:pvntrk:loDeltaE}   {\ensuremath{{0.000 } } }
\vdef{default-11:NoMc2e33-A:pvntrk:hiDelta}   {\ensuremath{{+0.000 } } }
\vdef{default-11:NoMc2e33-A:pvntrk:hiDeltaE}   {\ensuremath{{0.000 } } }
\vdef{default-11:NoData-AR3:muon1pt:loEff}   {\ensuremath{{1.008 } } }
\vdef{default-11:NoData-AR3:muon1pt:loEffE}   {\ensuremath{{\mathrm{NaN} } } }
\vdef{default-11:NoData-AR3:muon1pt:hiEff}   {\ensuremath{{1.000 } } }
\vdef{default-11:NoData-AR3:muon1pt:hiEffE}   {\ensuremath{{0.000 } } }
\vdef{default-11:NoMc2e33-A:muon1pt:loEff}   {\ensuremath{{1.007 } } }
\vdef{default-11:NoMc2e33-A:muon1pt:loEffE}   {\ensuremath{{\mathrm{NaN} } } }
\vdef{default-11:NoMc2e33-A:muon1pt:hiEff}   {\ensuremath{{1.000 } } }
\vdef{default-11:NoMc2e33-A:muon1pt:hiEffE}   {\ensuremath{{0.000 } } }
\vdef{default-11:NoMc2e33-A:muon1pt:loDelta}   {\ensuremath{{+0.001 } } }
\vdef{default-11:NoMc2e33-A:muon1pt:loDeltaE}   {\ensuremath{{\mathrm{NaN} } } }
\vdef{default-11:NoMc2e33-A:muon1pt:hiDelta}   {\ensuremath{{+0.000 } } }
\vdef{default-11:NoMc2e33-A:muon1pt:hiDeltaE}   {\ensuremath{{0.000 } } }
\vdef{default-11:NoData-AR3:muon2pt:loEff}   {\ensuremath{{0.134 } } }
\vdef{default-11:NoData-AR3:muon2pt:loEffE}   {\ensuremath{{0.002 } } }
\vdef{default-11:NoData-AR3:muon2pt:hiEff}   {\ensuremath{{0.866 } } }
\vdef{default-11:NoData-AR3:muon2pt:hiEffE}   {\ensuremath{{0.002 } } }
\vdef{default-11:NoMc2e33-A:muon2pt:loEff}   {\ensuremath{{0.136 } } }
\vdef{default-11:NoMc2e33-A:muon2pt:loEffE}   {\ensuremath{{0.002 } } }
\vdef{default-11:NoMc2e33-A:muon2pt:hiEff}   {\ensuremath{{0.864 } } }
\vdef{default-11:NoMc2e33-A:muon2pt:hiEffE}   {\ensuremath{{0.002 } } }
\vdef{default-11:NoMc2e33-A:muon2pt:loDelta}   {\ensuremath{{-0.010 } } }
\vdef{default-11:NoMc2e33-A:muon2pt:loDeltaE}   {\ensuremath{{0.025 } } }
\vdef{default-11:NoMc2e33-A:muon2pt:hiDelta}   {\ensuremath{{+0.002 } } }
\vdef{default-11:NoMc2e33-A:muon2pt:hiDeltaE}   {\ensuremath{{0.004 } } }
\vdef{default-11:NoData-AR3:muonseta:loEff}   {\ensuremath{{0.733 } } }
\vdef{default-11:NoData-AR3:muonseta:loEffE}   {\ensuremath{{0.002 } } }
\vdef{default-11:NoData-AR3:muonseta:hiEff}   {\ensuremath{{0.267 } } }
\vdef{default-11:NoData-AR3:muonseta:hiEffE}   {\ensuremath{{0.002 } } }
\vdef{default-11:NoMc2e33-A:muonseta:loEff}   {\ensuremath{{0.728 } } }
\vdef{default-11:NoMc2e33-A:muonseta:loEffE}   {\ensuremath{{0.002 } } }
\vdef{default-11:NoMc2e33-A:muonseta:hiEff}   {\ensuremath{{0.272 } } }
\vdef{default-11:NoMc2e33-A:muonseta:hiEffE}   {\ensuremath{{0.002 } } }
\vdef{default-11:NoMc2e33-A:muonseta:loDelta}   {\ensuremath{{+0.007 } } }
\vdef{default-11:NoMc2e33-A:muonseta:loDeltaE}   {\ensuremath{{0.004 } } }
\vdef{default-11:NoMc2e33-A:muonseta:hiDelta}   {\ensuremath{{-0.020 } } }
\vdef{default-11:NoMc2e33-A:muonseta:hiDeltaE}   {\ensuremath{{0.012 } } }
\vdef{default-11:NoData-AR3:pt:loEff}   {\ensuremath{{0.000 } } }
\vdef{default-11:NoData-AR3:pt:loEffE}   {\ensuremath{{0.000 } } }
\vdef{default-11:NoData-AR3:pt:hiEff}   {\ensuremath{{1.000 } } }
\vdef{default-11:NoData-AR3:pt:hiEffE}   {\ensuremath{{0.000 } } }
\vdef{default-11:NoMc2e33-A:pt:loEff}   {\ensuremath{{0.000 } } }
\vdef{default-11:NoMc2e33-A:pt:loEffE}   {\ensuremath{{0.000 } } }
\vdef{default-11:NoMc2e33-A:pt:hiEff}   {\ensuremath{{1.000 } } }
\vdef{default-11:NoMc2e33-A:pt:hiEffE}   {\ensuremath{{0.000 } } }
\vdef{default-11:NoMc2e33-A:pt:loDelta}   {\ensuremath{{\mathrm{NaN} } } }
\vdef{default-11:NoMc2e33-A:pt:loDeltaE}   {\ensuremath{{\mathrm{NaN} } } }
\vdef{default-11:NoMc2e33-A:pt:hiDelta}   {\ensuremath{{+0.000 } } }
\vdef{default-11:NoMc2e33-A:pt:hiDeltaE}   {\ensuremath{{0.000 } } }
\vdef{default-11:NoData-AR3:p:loEff}   {\ensuremath{{1.011 } } }
\vdef{default-11:NoData-AR3:p:loEffE}   {\ensuremath{{\mathrm{NaN} } } }
\vdef{default-11:NoData-AR3:p:hiEff}   {\ensuremath{{1.000 } } }
\vdef{default-11:NoData-AR3:p:hiEffE}   {\ensuremath{{0.000 } } }
\vdef{default-11:NoMc2e33-A:p:loEff}   {\ensuremath{{1.013 } } }
\vdef{default-11:NoMc2e33-A:p:loEffE}   {\ensuremath{{\mathrm{NaN} } } }
\vdef{default-11:NoMc2e33-A:p:hiEff}   {\ensuremath{{1.000 } } }
\vdef{default-11:NoMc2e33-A:p:hiEffE}   {\ensuremath{{0.000 } } }
\vdef{default-11:NoMc2e33-A:p:loDelta}   {\ensuremath{{-0.001 } } }
\vdef{default-11:NoMc2e33-A:p:loDeltaE}   {\ensuremath{{\mathrm{NaN} } } }
\vdef{default-11:NoMc2e33-A:p:hiDelta}   {\ensuremath{{+0.000 } } }
\vdef{default-11:NoMc2e33-A:p:hiDeltaE}   {\ensuremath{{0.000 } } }
\vdef{default-11:NoData-AR3:eta:loEff}   {\ensuremath{{0.726 } } }
\vdef{default-11:NoData-AR3:eta:loEffE}   {\ensuremath{{0.003 } } }
\vdef{default-11:NoData-AR3:eta:hiEff}   {\ensuremath{{0.274 } } }
\vdef{default-11:NoData-AR3:eta:hiEffE}   {\ensuremath{{0.003 } } }
\vdef{default-11:NoMc2e33-A:eta:loEff}   {\ensuremath{{0.720 } } }
\vdef{default-11:NoMc2e33-A:eta:loEffE}   {\ensuremath{{0.003 } } }
\vdef{default-11:NoMc2e33-A:eta:hiEff}   {\ensuremath{{0.280 } } }
\vdef{default-11:NoMc2e33-A:eta:hiEffE}   {\ensuremath{{0.003 } } }
\vdef{default-11:NoMc2e33-A:eta:loDelta}   {\ensuremath{{+0.008 } } }
\vdef{default-11:NoMc2e33-A:eta:loDeltaE}   {\ensuremath{{0.006 } } }
\vdef{default-11:NoMc2e33-A:eta:hiDelta}   {\ensuremath{{-0.021 } } }
\vdef{default-11:NoMc2e33-A:eta:hiDeltaE}   {\ensuremath{{0.017 } } }
\vdef{default-11:NoData-AR3:bdt:loEff}   {\ensuremath{{0.911 } } }
\vdef{default-11:NoData-AR3:bdt:loEffE}   {\ensuremath{{0.002 } } }
\vdef{default-11:NoData-AR3:bdt:hiEff}   {\ensuremath{{0.089 } } }
\vdef{default-11:NoData-AR3:bdt:hiEffE}   {\ensuremath{{0.002 } } }
\vdef{default-11:NoMc2e33-A:bdt:loEff}   {\ensuremath{{0.914 } } }
\vdef{default-11:NoMc2e33-A:bdt:loEffE}   {\ensuremath{{0.002 } } }
\vdef{default-11:NoMc2e33-A:bdt:hiEff}   {\ensuremath{{0.086 } } }
\vdef{default-11:NoMc2e33-A:bdt:hiEffE}   {\ensuremath{{0.002 } } }
\vdef{default-11:NoMc2e33-A:bdt:loDelta}   {\ensuremath{{-0.004 } } }
\vdef{default-11:NoMc2e33-A:bdt:loDeltaE}   {\ensuremath{{0.003 } } }
\vdef{default-11:NoMc2e33-A:bdt:hiDelta}   {\ensuremath{{+0.038 } } }
\vdef{default-11:NoMc2e33-A:bdt:hiDeltaE}   {\ensuremath{{0.033 } } }
\vdef{default-11:NoData-AR3:fl3d:loEff}   {\ensuremath{{0.834 } } }
\vdef{default-11:NoData-AR3:fl3d:loEffE}   {\ensuremath{{0.003 } } }
\vdef{default-11:NoData-AR3:fl3d:hiEff}   {\ensuremath{{0.166 } } }
\vdef{default-11:NoData-AR3:fl3d:hiEffE}   {\ensuremath{{0.003 } } }
\vdef{default-11:NoMc2e33-A:fl3d:loEff}   {\ensuremath{{0.826 } } }
\vdef{default-11:NoMc2e33-A:fl3d:loEffE}   {\ensuremath{{0.003 } } }
\vdef{default-11:NoMc2e33-A:fl3d:hiEff}   {\ensuremath{{0.174 } } }
\vdef{default-11:NoMc2e33-A:fl3d:hiEffE}   {\ensuremath{{0.003 } } }
\vdef{default-11:NoMc2e33-A:fl3d:loDelta}   {\ensuremath{{+0.010 } } }
\vdef{default-11:NoMc2e33-A:fl3d:loDeltaE}   {\ensuremath{{0.004 } } }
\vdef{default-11:NoMc2e33-A:fl3d:hiDelta}   {\ensuremath{{-0.050 } } }
\vdef{default-11:NoMc2e33-A:fl3d:hiDeltaE}   {\ensuremath{{0.021 } } }
\vdef{default-11:NoData-AR3:fl3de:loEff}   {\ensuremath{{1.000 } } }
\vdef{default-11:NoData-AR3:fl3de:loEffE}   {\ensuremath{{0.000 } } }
\vdef{default-11:NoData-AR3:fl3de:hiEff}   {\ensuremath{{-0.000 } } }
\vdef{default-11:NoData-AR3:fl3de:hiEffE}   {\ensuremath{{\mathrm{NaN} } } }
\vdef{default-11:NoMc2e33-A:fl3de:loEff}   {\ensuremath{{1.000 } } }
\vdef{default-11:NoMc2e33-A:fl3de:loEffE}   {\ensuremath{{0.000 } } }
\vdef{default-11:NoMc2e33-A:fl3de:hiEff}   {\ensuremath{{0.000 } } }
\vdef{default-11:NoMc2e33-A:fl3de:hiEffE}   {\ensuremath{{0.000 } } }
\vdef{default-11:NoMc2e33-A:fl3de:loDelta}   {\ensuremath{{+0.000 } } }
\vdef{default-11:NoMc2e33-A:fl3de:loDeltaE}   {\ensuremath{{0.000 } } }
\vdef{default-11:NoMc2e33-A:fl3de:hiDelta}   {\ensuremath{{+3.948 } } }
\vdef{default-11:NoMc2e33-A:fl3de:hiDeltaE}   {\ensuremath{{\mathrm{NaN} } } }
\vdef{default-11:NoData-AR3:fls3d:loEff}   {\ensuremath{{0.074 } } }
\vdef{default-11:NoData-AR3:fls3d:loEffE}   {\ensuremath{{0.002 } } }
\vdef{default-11:NoData-AR3:fls3d:hiEff}   {\ensuremath{{0.926 } } }
\vdef{default-11:NoData-AR3:fls3d:hiEffE}   {\ensuremath{{0.002 } } }
\vdef{default-11:NoMc2e33-A:fls3d:loEff}   {\ensuremath{{0.078 } } }
\vdef{default-11:NoMc2e33-A:fls3d:loEffE}   {\ensuremath{{0.002 } } }
\vdef{default-11:NoMc2e33-A:fls3d:hiEff}   {\ensuremath{{0.922 } } }
\vdef{default-11:NoMc2e33-A:fls3d:hiEffE}   {\ensuremath{{0.002 } } }
\vdef{default-11:NoMc2e33-A:fls3d:loDelta}   {\ensuremath{{-0.051 } } }
\vdef{default-11:NoMc2e33-A:fls3d:loDeltaE}   {\ensuremath{{0.034 } } }
\vdef{default-11:NoMc2e33-A:fls3d:hiDelta}   {\ensuremath{{+0.004 } } }
\vdef{default-11:NoMc2e33-A:fls3d:hiDeltaE}   {\ensuremath{{0.003 } } }
\vdef{default-11:NoData-AR3:flsxy:loEff}   {\ensuremath{{1.012 } } }
\vdef{default-11:NoData-AR3:flsxy:loEffE}   {\ensuremath{{\mathrm{NaN} } } }
\vdef{default-11:NoData-AR3:flsxy:hiEff}   {\ensuremath{{1.000 } } }
\vdef{default-11:NoData-AR3:flsxy:hiEffE}   {\ensuremath{{0.000 } } }
\vdef{default-11:NoMc2e33-A:flsxy:loEff}   {\ensuremath{{1.012 } } }
\vdef{default-11:NoMc2e33-A:flsxy:loEffE}   {\ensuremath{{\mathrm{NaN} } } }
\vdef{default-11:NoMc2e33-A:flsxy:hiEff}   {\ensuremath{{1.000 } } }
\vdef{default-11:NoMc2e33-A:flsxy:hiEffE}   {\ensuremath{{0.000 } } }
\vdef{default-11:NoMc2e33-A:flsxy:loDelta}   {\ensuremath{{+0.000 } } }
\vdef{default-11:NoMc2e33-A:flsxy:loDeltaE}   {\ensuremath{{\mathrm{NaN} } } }
\vdef{default-11:NoMc2e33-A:flsxy:hiDelta}   {\ensuremath{{+0.000 } } }
\vdef{default-11:NoMc2e33-A:flsxy:hiDeltaE}   {\ensuremath{{0.000 } } }
\vdef{default-11:NoData-AR3:chi2dof:loEff}   {\ensuremath{{0.935 } } }
\vdef{default-11:NoData-AR3:chi2dof:loEffE}   {\ensuremath{{0.002 } } }
\vdef{default-11:NoData-AR3:chi2dof:hiEff}   {\ensuremath{{0.065 } } }
\vdef{default-11:NoData-AR3:chi2dof:hiEffE}   {\ensuremath{{0.002 } } }
\vdef{default-11:NoMc2e33-A:chi2dof:loEff}   {\ensuremath{{0.940 } } }
\vdef{default-11:NoMc2e33-A:chi2dof:loEffE}   {\ensuremath{{0.002 } } }
\vdef{default-11:NoMc2e33-A:chi2dof:hiEff}   {\ensuremath{{0.060 } } }
\vdef{default-11:NoMc2e33-A:chi2dof:hiEffE}   {\ensuremath{{0.002 } } }
\vdef{default-11:NoMc2e33-A:chi2dof:loDelta}   {\ensuremath{{-0.005 } } }
\vdef{default-11:NoMc2e33-A:chi2dof:loDeltaE}   {\ensuremath{{0.003 } } }
\vdef{default-11:NoMc2e33-A:chi2dof:hiDelta}   {\ensuremath{{+0.071 } } }
\vdef{default-11:NoMc2e33-A:chi2dof:hiDeltaE}   {\ensuremath{{0.039 } } }
\vdef{default-11:NoData-AR3:pchi2dof:loEff}   {\ensuremath{{0.634 } } }
\vdef{default-11:NoData-AR3:pchi2dof:loEffE}   {\ensuremath{{0.003 } } }
\vdef{default-11:NoData-AR3:pchi2dof:hiEff}   {\ensuremath{{0.366 } } }
\vdef{default-11:NoData-AR3:pchi2dof:hiEffE}   {\ensuremath{{0.003 } } }
\vdef{default-11:NoMc2e33-A:pchi2dof:loEff}   {\ensuremath{{0.621 } } }
\vdef{default-11:NoMc2e33-A:pchi2dof:loEffE}   {\ensuremath{{0.003 } } }
\vdef{default-11:NoMc2e33-A:pchi2dof:hiEff}   {\ensuremath{{0.379 } } }
\vdef{default-11:NoMc2e33-A:pchi2dof:hiEffE}   {\ensuremath{{0.003 } } }
\vdef{default-11:NoMc2e33-A:pchi2dof:loDelta}   {\ensuremath{{+0.021 } } }
\vdef{default-11:NoMc2e33-A:pchi2dof:loDeltaE}   {\ensuremath{{0.008 } } }
\vdef{default-11:NoMc2e33-A:pchi2dof:hiDelta}   {\ensuremath{{-0.036 } } }
\vdef{default-11:NoMc2e33-A:pchi2dof:hiDeltaE}   {\ensuremath{{0.013 } } }
\vdef{default-11:NoData-AR3:alpha:loEff}   {\ensuremath{{0.994 } } }
\vdef{default-11:NoData-AR3:alpha:loEffE}   {\ensuremath{{0.001 } } }
\vdef{default-11:NoData-AR3:alpha:hiEff}   {\ensuremath{{0.006 } } }
\vdef{default-11:NoData-AR3:alpha:hiEffE}   {\ensuremath{{0.001 } } }
\vdef{default-11:NoMc2e33-A:alpha:loEff}   {\ensuremath{{0.994 } } }
\vdef{default-11:NoMc2e33-A:alpha:loEffE}   {\ensuremath{{0.001 } } }
\vdef{default-11:NoMc2e33-A:alpha:hiEff}   {\ensuremath{{0.006 } } }
\vdef{default-11:NoMc2e33-A:alpha:hiEffE}   {\ensuremath{{0.001 } } }
\vdef{default-11:NoMc2e33-A:alpha:loDelta}   {\ensuremath{{+0.000 } } }
\vdef{default-11:NoMc2e33-A:alpha:loDeltaE}   {\ensuremath{{0.001 } } }
\vdef{default-11:NoMc2e33-A:alpha:hiDelta}   {\ensuremath{{-0.029 } } }
\vdef{default-11:NoMc2e33-A:alpha:hiDeltaE}   {\ensuremath{{0.136 } } }
\vdef{default-11:NoData-AR3:iso:loEff}   {\ensuremath{{0.130 } } }
\vdef{default-11:NoData-AR3:iso:loEffE}   {\ensuremath{{0.002 } } }
\vdef{default-11:NoData-AR3:iso:hiEff}   {\ensuremath{{0.870 } } }
\vdef{default-11:NoData-AR3:iso:hiEffE}   {\ensuremath{{0.002 } } }
\vdef{default-11:NoMc2e33-A:iso:loEff}   {\ensuremath{{0.107 } } }
\vdef{default-11:NoMc2e33-A:iso:loEffE}   {\ensuremath{{0.002 } } }
\vdef{default-11:NoMc2e33-A:iso:hiEff}   {\ensuremath{{0.893 } } }
\vdef{default-11:NoMc2e33-A:iso:hiEffE}   {\ensuremath{{0.002 } } }
\vdef{default-11:NoMc2e33-A:iso:loDelta}   {\ensuremath{{+0.199 } } }
\vdef{default-11:NoMc2e33-A:iso:loDeltaE}   {\ensuremath{{0.026 } } }
\vdef{default-11:NoMc2e33-A:iso:hiDelta}   {\ensuremath{{-0.027 } } }
\vdef{default-11:NoMc2e33-A:iso:hiDeltaE}   {\ensuremath{{0.004 } } }
\vdef{default-11:NoData-AR3:docatrk:loEff}   {\ensuremath{{0.072 } } }
\vdef{default-11:NoData-AR3:docatrk:loEffE}   {\ensuremath{{0.002 } } }
\vdef{default-11:NoData-AR3:docatrk:hiEff}   {\ensuremath{{0.928 } } }
\vdef{default-11:NoData-AR3:docatrk:hiEffE}   {\ensuremath{{0.002 } } }
\vdef{default-11:NoMc2e33-A:docatrk:loEff}   {\ensuremath{{0.082 } } }
\vdef{default-11:NoMc2e33-A:docatrk:loEffE}   {\ensuremath{{0.002 } } }
\vdef{default-11:NoMc2e33-A:docatrk:hiEff}   {\ensuremath{{0.918 } } }
\vdef{default-11:NoMc2e33-A:docatrk:hiEffE}   {\ensuremath{{0.002 } } }
\vdef{default-11:NoMc2e33-A:docatrk:loDelta}   {\ensuremath{{-0.135 } } }
\vdef{default-11:NoMc2e33-A:docatrk:loDeltaE}   {\ensuremath{{0.035 } } }
\vdef{default-11:NoMc2e33-A:docatrk:hiDelta}   {\ensuremath{{+0.011 } } }
\vdef{default-11:NoMc2e33-A:docatrk:hiDeltaE}   {\ensuremath{{0.003 } } }
\vdef{default-11:NoData-AR3:isotrk:loEff}   {\ensuremath{{1.000 } } }
\vdef{default-11:NoData-AR3:isotrk:loEffE}   {\ensuremath{{0.000 } } }
\vdef{default-11:NoData-AR3:isotrk:hiEff}   {\ensuremath{{1.000 } } }
\vdef{default-11:NoData-AR3:isotrk:hiEffE}   {\ensuremath{{0.000 } } }
\vdef{default-11:NoMc2e33-A:isotrk:loEff}   {\ensuremath{{1.000 } } }
\vdef{default-11:NoMc2e33-A:isotrk:loEffE}   {\ensuremath{{0.000 } } }
\vdef{default-11:NoMc2e33-A:isotrk:hiEff}   {\ensuremath{{1.000 } } }
\vdef{default-11:NoMc2e33-A:isotrk:hiEffE}   {\ensuremath{{0.000 } } }
\vdef{default-11:NoMc2e33-A:isotrk:loDelta}   {\ensuremath{{+0.000 } } }
\vdef{default-11:NoMc2e33-A:isotrk:loDeltaE}   {\ensuremath{{0.000 } } }
\vdef{default-11:NoMc2e33-A:isotrk:hiDelta}   {\ensuremath{{+0.000 } } }
\vdef{default-11:NoMc2e33-A:isotrk:hiDeltaE}   {\ensuremath{{0.000 } } }
\vdef{default-11:NoData-AR3:closetrk:loEff}   {\ensuremath{{0.977 } } }
\vdef{default-11:NoData-AR3:closetrk:loEffE}   {\ensuremath{{0.001 } } }
\vdef{default-11:NoData-AR3:closetrk:hiEff}   {\ensuremath{{0.023 } } }
\vdef{default-11:NoData-AR3:closetrk:hiEffE}   {\ensuremath{{0.001 } } }
\vdef{default-11:NoMc2e33-A:closetrk:loEff}   {\ensuremath{{0.978 } } }
\vdef{default-11:NoMc2e33-A:closetrk:loEffE}   {\ensuremath{{0.001 } } }
\vdef{default-11:NoMc2e33-A:closetrk:hiEff}   {\ensuremath{{0.022 } } }
\vdef{default-11:NoMc2e33-A:closetrk:hiEffE}   {\ensuremath{{0.001 } } }
\vdef{default-11:NoMc2e33-A:closetrk:loDelta}   {\ensuremath{{-0.001 } } }
\vdef{default-11:NoMc2e33-A:closetrk:loDeltaE}   {\ensuremath{{0.002 } } }
\vdef{default-11:NoMc2e33-A:closetrk:hiDelta}   {\ensuremath{{+0.031 } } }
\vdef{default-11:NoMc2e33-A:closetrk:hiDeltaE}   {\ensuremath{{0.068 } } }
\vdef{default-11:NoData-AR3:lip:loEff}   {\ensuremath{{1.000 } } }
\vdef{default-11:NoData-AR3:lip:loEffE}   {\ensuremath{{0.000 } } }
\vdef{default-11:NoData-AR3:lip:hiEff}   {\ensuremath{{0.000 } } }
\vdef{default-11:NoData-AR3:lip:hiEffE}   {\ensuremath{{0.000 } } }
\vdef{default-11:NoMc2e33-A:lip:loEff}   {\ensuremath{{1.000 } } }
\vdef{default-11:NoMc2e33-A:lip:loEffE}   {\ensuremath{{0.000 } } }
\vdef{default-11:NoMc2e33-A:lip:hiEff}   {\ensuremath{{0.000 } } }
\vdef{default-11:NoMc2e33-A:lip:hiEffE}   {\ensuremath{{0.000 } } }
\vdef{default-11:NoMc2e33-A:lip:loDelta}   {\ensuremath{{+0.000 } } }
\vdef{default-11:NoMc2e33-A:lip:loDeltaE}   {\ensuremath{{0.000 } } }
\vdef{default-11:NoMc2e33-A:lip:hiDelta}   {\ensuremath{{\mathrm{NaN} } } }
\vdef{default-11:NoMc2e33-A:lip:hiDeltaE}   {\ensuremath{{\mathrm{NaN} } } }
\vdef{default-11:NoData-AR3:lip:inEff}   {\ensuremath{{1.000 } } }
\vdef{default-11:NoData-AR3:lip:inEffE}   {\ensuremath{{0.000 } } }
\vdef{default-11:NoMc2e33-A:lip:inEff}   {\ensuremath{{1.000 } } }
\vdef{default-11:NoMc2e33-A:lip:inEffE}   {\ensuremath{{0.000 } } }
\vdef{default-11:NoMc2e33-A:lip:inDelta}   {\ensuremath{{+0.000 } } }
\vdef{default-11:NoMc2e33-A:lip:inDeltaE}   {\ensuremath{{0.000 } } }
\vdef{default-11:NoData-AR3:lips:loEff}   {\ensuremath{{1.000 } } }
\vdef{default-11:NoData-AR3:lips:loEffE}   {\ensuremath{{0.000 } } }
\vdef{default-11:NoData-AR3:lips:hiEff}   {\ensuremath{{0.000 } } }
\vdef{default-11:NoData-AR3:lips:hiEffE}   {\ensuremath{{0.000 } } }
\vdef{default-11:NoMc2e33-A:lips:loEff}   {\ensuremath{{1.000 } } }
\vdef{default-11:NoMc2e33-A:lips:loEffE}   {\ensuremath{{0.000 } } }
\vdef{default-11:NoMc2e33-A:lips:hiEff}   {\ensuremath{{0.000 } } }
\vdef{default-11:NoMc2e33-A:lips:hiEffE}   {\ensuremath{{0.000 } } }
\vdef{default-11:NoMc2e33-A:lips:loDelta}   {\ensuremath{{+0.000 } } }
\vdef{default-11:NoMc2e33-A:lips:loDeltaE}   {\ensuremath{{0.000 } } }
\vdef{default-11:NoMc2e33-A:lips:hiDelta}   {\ensuremath{{\mathrm{NaN} } } }
\vdef{default-11:NoMc2e33-A:lips:hiDeltaE}   {\ensuremath{{\mathrm{NaN} } } }
\vdef{default-11:NoData-AR3:lips:inEff}   {\ensuremath{{1.000 } } }
\vdef{default-11:NoData-AR3:lips:inEffE}   {\ensuremath{{0.000 } } }
\vdef{default-11:NoMc2e33-A:lips:inEff}   {\ensuremath{{1.000 } } }
\vdef{default-11:NoMc2e33-A:lips:inEffE}   {\ensuremath{{0.000 } } }
\vdef{default-11:NoMc2e33-A:lips:inDelta}   {\ensuremath{{+0.000 } } }
\vdef{default-11:NoMc2e33-A:lips:inDeltaE}   {\ensuremath{{0.000 } } }
\vdef{default-11:NoData-AR3:ip:loEff}   {\ensuremath{{0.971 } } }
\vdef{default-11:NoData-AR3:ip:loEffE}   {\ensuremath{{0.001 } } }
\vdef{default-11:NoData-AR3:ip:hiEff}   {\ensuremath{{0.029 } } }
\vdef{default-11:NoData-AR3:ip:hiEffE}   {\ensuremath{{0.001 } } }
\vdef{default-11:NoMc2e33-A:ip:loEff}   {\ensuremath{{0.972 } } }
\vdef{default-11:NoMc2e33-A:ip:loEffE}   {\ensuremath{{0.001 } } }
\vdef{default-11:NoMc2e33-A:ip:hiEff}   {\ensuremath{{0.028 } } }
\vdef{default-11:NoMc2e33-A:ip:hiEffE}   {\ensuremath{{0.001 } } }
\vdef{default-11:NoMc2e33-A:ip:loDelta}   {\ensuremath{{-0.001 } } }
\vdef{default-11:NoMc2e33-A:ip:loDeltaE}   {\ensuremath{{0.002 } } }
\vdef{default-11:NoMc2e33-A:ip:hiDelta}   {\ensuremath{{+0.021 } } }
\vdef{default-11:NoMc2e33-A:ip:hiDeltaE}   {\ensuremath{{0.060 } } }
\vdef{default-11:NoData-AR3:ips:loEff}   {\ensuremath{{0.948 } } }
\vdef{default-11:NoData-AR3:ips:loEffE}   {\ensuremath{{0.002 } } }
\vdef{default-11:NoData-AR3:ips:hiEff}   {\ensuremath{{0.052 } } }
\vdef{default-11:NoData-AR3:ips:hiEffE}   {\ensuremath{{0.002 } } }
\vdef{default-11:NoMc2e33-A:ips:loEff}   {\ensuremath{{0.959 } } }
\vdef{default-11:NoMc2e33-A:ips:loEffE}   {\ensuremath{{0.001 } } }
\vdef{default-11:NoMc2e33-A:ips:hiEff}   {\ensuremath{{0.041 } } }
\vdef{default-11:NoMc2e33-A:ips:hiEffE}   {\ensuremath{{0.001 } } }
\vdef{default-11:NoMc2e33-A:ips:loDelta}   {\ensuremath{{-0.011 } } }
\vdef{default-11:NoMc2e33-A:ips:loDeltaE}   {\ensuremath{{0.002 } } }
\vdef{default-11:NoMc2e33-A:ips:hiDelta}   {\ensuremath{{+0.230 } } }
\vdef{default-11:NoMc2e33-A:ips:hiDeltaE}   {\ensuremath{{0.046 } } }
\vdef{default-11:NoData-AR3:maxdoca:loEff}   {\ensuremath{{1.000 } } }
\vdef{default-11:NoData-AR3:maxdoca:loEffE}   {\ensuremath{{0.000 } } }
\vdef{default-11:NoData-AR3:maxdoca:hiEff}   {\ensuremath{{0.014 } } }
\vdef{default-11:NoData-AR3:maxdoca:hiEffE}   {\ensuremath{{0.001 } } }
\vdef{default-11:NoMc2e33-A:maxdoca:loEff}   {\ensuremath{{1.000 } } }
\vdef{default-11:NoMc2e33-A:maxdoca:loEffE}   {\ensuremath{{0.000 } } }
\vdef{default-11:NoMc2e33-A:maxdoca:hiEff}   {\ensuremath{{0.011 } } }
\vdef{default-11:NoMc2e33-A:maxdoca:hiEffE}   {\ensuremath{{0.001 } } }
\vdef{default-11:NoMc2e33-A:maxdoca:loDelta}   {\ensuremath{{+0.000 } } }
\vdef{default-11:NoMc2e33-A:maxdoca:loDeltaE}   {\ensuremath{{0.000 } } }
\vdef{default-11:NoMc2e33-A:maxdoca:hiDelta}   {\ensuremath{{+0.270 } } }
\vdef{default-11:NoMc2e33-A:maxdoca:hiDeltaE}   {\ensuremath{{0.092 } } }
\vdef{default-11:NoData-AR3:kaonpt:loEff}   {\ensuremath{{1.009 } } }
\vdef{default-11:NoData-AR3:kaonpt:loEffE}   {\ensuremath{{\mathrm{NaN} } } }
\vdef{default-11:NoData-AR3:kaonpt:hiEff}   {\ensuremath{{1.000 } } }
\vdef{default-11:NoData-AR3:kaonpt:hiEffE}   {\ensuremath{{0.000 } } }
\vdef{default-11:NoMc2e33-A:kaonpt:loEff}   {\ensuremath{{1.007 } } }
\vdef{default-11:NoMc2e33-A:kaonpt:loEffE}   {\ensuremath{{\mathrm{NaN} } } }
\vdef{default-11:NoMc2e33-A:kaonpt:hiEff}   {\ensuremath{{1.000 } } }
\vdef{default-11:NoMc2e33-A:kaonpt:hiEffE}   {\ensuremath{{0.000 } } }
\vdef{default-11:NoMc2e33-A:kaonpt:loDelta}   {\ensuremath{{+0.002 } } }
\vdef{default-11:NoMc2e33-A:kaonpt:loDeltaE}   {\ensuremath{{\mathrm{NaN} } } }
\vdef{default-11:NoMc2e33-A:kaonpt:hiDelta}   {\ensuremath{{+0.000 } } }
\vdef{default-11:NoMc2e33-A:kaonpt:hiDeltaE}   {\ensuremath{{0.000 } } }
\vdef{default-11:NoData-AR3:psipt:loEff}   {\ensuremath{{1.004 } } }
\vdef{default-11:NoData-AR3:psipt:loEffE}   {\ensuremath{{\mathrm{NaN} } } }
\vdef{default-11:NoData-AR3:psipt:hiEff}   {\ensuremath{{1.000 } } }
\vdef{default-11:NoData-AR3:psipt:hiEffE}   {\ensuremath{{0.000 } } }
\vdef{default-11:NoMc2e33-A:psipt:loEff}   {\ensuremath{{1.003 } } }
\vdef{default-11:NoMc2e33-A:psipt:loEffE}   {\ensuremath{{\mathrm{NaN} } } }
\vdef{default-11:NoMc2e33-A:psipt:hiEff}   {\ensuremath{{1.000 } } }
\vdef{default-11:NoMc2e33-A:psipt:hiEffE}   {\ensuremath{{0.000 } } }
\vdef{default-11:NoMc2e33-A:psipt:loDelta}   {\ensuremath{{+0.001 } } }
\vdef{default-11:NoMc2e33-A:psipt:loDeltaE}   {\ensuremath{{\mathrm{NaN} } } }
\vdef{default-11:NoMc2e33-A:psipt:hiDelta}   {\ensuremath{{+0.000 } } }
\vdef{default-11:NoMc2e33-A:psipt:hiDeltaE}   {\ensuremath{{0.000 } } }
\vdef{default-11:NoData-AR3:osiso:loEff}   {\ensuremath{{1.003 } } }
\vdef{default-11:NoData-AR3:osiso:loEffE}   {\ensuremath{{\mathrm{NaN} } } }
\vdef{default-11:NoData-AR3:osiso:hiEff}   {\ensuremath{{1.000 } } }
\vdef{default-11:NoData-AR3:osiso:hiEffE}   {\ensuremath{{0.000 } } }
\vdef{default-11:NoMcCMS-A:osiso:loEff}   {\ensuremath{{1.003 } } }
\vdef{default-11:NoMcCMS-A:osiso:loEffE}   {\ensuremath{{\mathrm{NaN} } } }
\vdef{default-11:NoMcCMS-A:osiso:hiEff}   {\ensuremath{{1.000 } } }
\vdef{default-11:NoMcCMS-A:osiso:hiEffE}   {\ensuremath{{0.000 } } }
\vdef{default-11:NoMcCMS-A:osiso:loDelta}   {\ensuremath{{+0.001 } } }
\vdef{default-11:NoMcCMS-A:osiso:loDeltaE}   {\ensuremath{{\mathrm{NaN} } } }
\vdef{default-11:NoMcCMS-A:osiso:hiDelta}   {\ensuremath{{+0.000 } } }
\vdef{default-11:NoMcCMS-A:osiso:hiDeltaE}   {\ensuremath{{0.000 } } }
\vdef{default-11:NoData-AR3:osreliso:loEff}   {\ensuremath{{0.254 } } }
\vdef{default-11:NoData-AR3:osreliso:loEffE}   {\ensuremath{{0.003 } } }
\vdef{default-11:NoData-AR3:osreliso:hiEff}   {\ensuremath{{0.746 } } }
\vdef{default-11:NoData-AR3:osreliso:hiEffE}   {\ensuremath{{0.003 } } }
\vdef{default-11:NoMcCMS-A:osreliso:loEff}   {\ensuremath{{0.282 } } }
\vdef{default-11:NoMcCMS-A:osreliso:loEffE}   {\ensuremath{{0.003 } } }
\vdef{default-11:NoMcCMS-A:osreliso:hiEff}   {\ensuremath{{0.718 } } }
\vdef{default-11:NoMcCMS-A:osreliso:hiEffE}   {\ensuremath{{0.003 } } }
\vdef{default-11:NoMcCMS-A:osreliso:loDelta}   {\ensuremath{{-0.102 } } }
\vdef{default-11:NoMcCMS-A:osreliso:loDeltaE}   {\ensuremath{{0.016 } } }
\vdef{default-11:NoMcCMS-A:osreliso:hiDelta}   {\ensuremath{{+0.037 } } }
\vdef{default-11:NoMcCMS-A:osreliso:hiDeltaE}   {\ensuremath{{0.006 } } }
\vdef{default-11:NoData-AR3:osmuonpt:loEff}   {\ensuremath{{0.000 } } }
\vdef{default-11:NoData-AR3:osmuonpt:loEffE}   {\ensuremath{{0.001 } } }
\vdef{default-11:NoData-AR3:osmuonpt:hiEff}   {\ensuremath{{1.000 } } }
\vdef{default-11:NoData-AR3:osmuonpt:hiEffE}   {\ensuremath{{0.001 } } }
\vdef{default-11:NoMcCMS-A:osmuonpt:loEff}   {\ensuremath{{0.000 } } }
\vdef{default-11:NoMcCMS-A:osmuonpt:loEffE}   {\ensuremath{{0.001 } } }
\vdef{default-11:NoMcCMS-A:osmuonpt:hiEff}   {\ensuremath{{1.000 } } }
\vdef{default-11:NoMcCMS-A:osmuonpt:hiEffE}   {\ensuremath{{0.001 } } }
\vdef{default-11:NoMcCMS-A:osmuonpt:loDelta}   {\ensuremath{{\mathrm{NaN} } } }
\vdef{default-11:NoMcCMS-A:osmuonpt:loDeltaE}   {\ensuremath{{\mathrm{NaN} } } }
\vdef{default-11:NoMcCMS-A:osmuonpt:hiDelta}   {\ensuremath{{+0.000 } } }
\vdef{default-11:NoMcCMS-A:osmuonpt:hiDeltaE}   {\ensuremath{{0.002 } } }
\vdef{default-11:NoData-AR3:osmuondr:loEff}   {\ensuremath{{0.022 } } }
\vdef{default-11:NoData-AR3:osmuondr:loEffE}   {\ensuremath{{0.005 } } }
\vdef{default-11:NoData-AR3:osmuondr:hiEff}   {\ensuremath{{0.978 } } }
\vdef{default-11:NoData-AR3:osmuondr:hiEffE}   {\ensuremath{{0.006 } } }
\vdef{default-11:NoMcCMS-A:osmuondr:loEff}   {\ensuremath{{0.012 } } }
\vdef{default-11:NoMcCMS-A:osmuondr:loEffE}   {\ensuremath{{0.004 } } }
\vdef{default-11:NoMcCMS-A:osmuondr:hiEff}   {\ensuremath{{0.988 } } }
\vdef{default-11:NoMcCMS-A:osmuondr:hiEffE}   {\ensuremath{{0.004 } } }
\vdef{default-11:NoMcCMS-A:osmuondr:loDelta}   {\ensuremath{{+0.615 } } }
\vdef{default-11:NoMcCMS-A:osmuondr:loDeltaE}   {\ensuremath{{0.376 } } }
\vdef{default-11:NoMcCMS-A:osmuondr:hiDelta}   {\ensuremath{{-0.011 } } }
\vdef{default-11:NoMcCMS-A:osmuondr:hiDeltaE}   {\ensuremath{{0.007 } } }
\vdef{default-11:NoData-AR3:hlt:loEff}   {\ensuremath{{0.049 } } }
\vdef{default-11:NoData-AR3:hlt:loEffE}   {\ensuremath{{0.002 } } }
\vdef{default-11:NoData-AR3:hlt:hiEff}   {\ensuremath{{0.951 } } }
\vdef{default-11:NoData-AR3:hlt:hiEffE}   {\ensuremath{{0.002 } } }
\vdef{default-11:NoMcCMS-A:hlt:loEff}   {\ensuremath{{0.267 } } }
\vdef{default-11:NoMcCMS-A:hlt:loEffE}   {\ensuremath{{0.003 } } }
\vdef{default-11:NoMcCMS-A:hlt:hiEff}   {\ensuremath{{0.733 } } }
\vdef{default-11:NoMcCMS-A:hlt:hiEffE}   {\ensuremath{{0.003 } } }
\vdef{default-11:NoMcCMS-A:hlt:loDelta}   {\ensuremath{{-1.383 } } }
\vdef{default-11:NoMcCMS-A:hlt:loDeltaE}   {\ensuremath{{0.018 } } }
\vdef{default-11:NoMcCMS-A:hlt:hiDelta}   {\ensuremath{{+0.259 } } }
\vdef{default-11:NoMcCMS-A:hlt:hiDeltaE}   {\ensuremath{{0.005 } } }
\vdef{default-11:NoData-AR3:muonsid:loEff}   {\ensuremath{{0.140 } } }
\vdef{default-11:NoData-AR3:muonsid:loEffE}   {\ensuremath{{0.002 } } }
\vdef{default-11:NoData-AR3:muonsid:hiEff}   {\ensuremath{{0.860 } } }
\vdef{default-11:NoData-AR3:muonsid:hiEffE}   {\ensuremath{{0.002 } } }
\vdef{default-11:NoMcCMS-A:muonsid:loEff}   {\ensuremath{{0.220 } } }
\vdef{default-11:NoMcCMS-A:muonsid:loEffE}   {\ensuremath{{0.003 } } }
\vdef{default-11:NoMcCMS-A:muonsid:hiEff}   {\ensuremath{{0.780 } } }
\vdef{default-11:NoMcCMS-A:muonsid:hiEffE}   {\ensuremath{{0.003 } } }
\vdef{default-11:NoMcCMS-A:muonsid:loDelta}   {\ensuremath{{-0.443 } } }
\vdef{default-11:NoMcCMS-A:muonsid:loDeltaE}   {\ensuremath{{0.020 } } }
\vdef{default-11:NoMcCMS-A:muonsid:hiDelta}   {\ensuremath{{+0.097 } } }
\vdef{default-11:NoMcCMS-A:muonsid:hiDeltaE}   {\ensuremath{{0.005 } } }
\vdef{default-11:NoData-AR3:tracksqual:loEff}   {\ensuremath{{0.000 } } }
\vdef{default-11:NoData-AR3:tracksqual:loEffE}   {\ensuremath{{0.000 } } }
\vdef{default-11:NoData-AR3:tracksqual:hiEff}   {\ensuremath{{1.000 } } }
\vdef{default-11:NoData-AR3:tracksqual:hiEffE}   {\ensuremath{{0.000 } } }
\vdef{default-11:NoMcCMS-A:tracksqual:loEff}   {\ensuremath{{0.000 } } }
\vdef{default-11:NoMcCMS-A:tracksqual:loEffE}   {\ensuremath{{0.000 } } }
\vdef{default-11:NoMcCMS-A:tracksqual:hiEff}   {\ensuremath{{1.000 } } }
\vdef{default-11:NoMcCMS-A:tracksqual:hiEffE}   {\ensuremath{{0.000 } } }
\vdef{default-11:NoMcCMS-A:tracksqual:loDelta}   {\ensuremath{{+0.615 } } }
\vdef{default-11:NoMcCMS-A:tracksqual:loDeltaE}   {\ensuremath{{0.553 } } }
\vdef{default-11:NoMcCMS-A:tracksqual:hiDelta}   {\ensuremath{{-0.000 } } }
\vdef{default-11:NoMcCMS-A:tracksqual:hiDeltaE}   {\ensuremath{{0.000 } } }
\vdef{default-11:NoData-AR3:pvz:loEff}   {\ensuremath{{0.508 } } }
\vdef{default-11:NoData-AR3:pvz:loEffE}   {\ensuremath{{0.004 } } }
\vdef{default-11:NoData-AR3:pvz:hiEff}   {\ensuremath{{0.492 } } }
\vdef{default-11:NoData-AR3:pvz:hiEffE}   {\ensuremath{{0.004 } } }
\vdef{default-11:NoMcCMS-A:pvz:loEff}   {\ensuremath{{0.470 } } }
\vdef{default-11:NoMcCMS-A:pvz:loEffE}   {\ensuremath{{0.004 } } }
\vdef{default-11:NoMcCMS-A:pvz:hiEff}   {\ensuremath{{0.530 } } }
\vdef{default-11:NoMcCMS-A:pvz:hiEffE}   {\ensuremath{{0.004 } } }
\vdef{default-11:NoMcCMS-A:pvz:loDelta}   {\ensuremath{{+0.079 } } }
\vdef{default-11:NoMcCMS-A:pvz:loDeltaE}   {\ensuremath{{0.010 } } }
\vdef{default-11:NoMcCMS-A:pvz:hiDelta}   {\ensuremath{{-0.075 } } }
\vdef{default-11:NoMcCMS-A:pvz:hiDeltaE}   {\ensuremath{{0.010 } } }
\vdef{default-11:NoData-AR3:pvn:loEff}   {\ensuremath{{1.000 } } }
\vdef{default-11:NoData-AR3:pvn:loEffE}   {\ensuremath{{\mathrm{NaN} } } }
\vdef{default-11:NoData-AR3:pvn:hiEff}   {\ensuremath{{1.000 } } }
\vdef{default-11:NoData-AR3:pvn:hiEffE}   {\ensuremath{{0.000 } } }
\vdef{default-11:NoMcCMS-A:pvn:loEff}   {\ensuremath{{1.065 } } }
\vdef{default-11:NoMcCMS-A:pvn:loEffE}   {\ensuremath{{\mathrm{NaN} } } }
\vdef{default-11:NoMcCMS-A:pvn:hiEff}   {\ensuremath{{1.000 } } }
\vdef{default-11:NoMcCMS-A:pvn:hiEffE}   {\ensuremath{{0.000 } } }
\vdef{default-11:NoMcCMS-A:pvn:loDelta}   {\ensuremath{{-0.063 } } }
\vdef{default-11:NoMcCMS-A:pvn:loDeltaE}   {\ensuremath{{\mathrm{NaN} } } }
\vdef{default-11:NoMcCMS-A:pvn:hiDelta}   {\ensuremath{{+0.000 } } }
\vdef{default-11:NoMcCMS-A:pvn:hiDeltaE}   {\ensuremath{{0.000 } } }
\vdef{default-11:NoData-AR3:pvavew8:loEff}   {\ensuremath{{0.013 } } }
\vdef{default-11:NoData-AR3:pvavew8:loEffE}   {\ensuremath{{0.001 } } }
\vdef{default-11:NoData-AR3:pvavew8:hiEff}   {\ensuremath{{0.987 } } }
\vdef{default-11:NoData-AR3:pvavew8:hiEffE}   {\ensuremath{{0.001 } } }
\vdef{default-11:NoMcCMS-A:pvavew8:loEff}   {\ensuremath{{0.006 } } }
\vdef{default-11:NoMcCMS-A:pvavew8:loEffE}   {\ensuremath{{0.001 } } }
\vdef{default-11:NoMcCMS-A:pvavew8:hiEff}   {\ensuremath{{0.994 } } }
\vdef{default-11:NoMcCMS-A:pvavew8:hiEffE}   {\ensuremath{{0.001 } } }
\vdef{default-11:NoMcCMS-A:pvavew8:loDelta}   {\ensuremath{{+0.746 } } }
\vdef{default-11:NoMcCMS-A:pvavew8:loDeltaE}   {\ensuremath{{0.101 } } }
\vdef{default-11:NoMcCMS-A:pvavew8:hiDelta}   {\ensuremath{{-0.007 } } }
\vdef{default-11:NoMcCMS-A:pvavew8:hiDeltaE}   {\ensuremath{{0.001 } } }
\vdef{default-11:NoData-AR3:pvntrk:loEff}   {\ensuremath{{1.000 } } }
\vdef{default-11:NoData-AR3:pvntrk:loEffE}   {\ensuremath{{0.000 } } }
\vdef{default-11:NoData-AR3:pvntrk:hiEff}   {\ensuremath{{1.000 } } }
\vdef{default-11:NoData-AR3:pvntrk:hiEffE}   {\ensuremath{{0.000 } } }
\vdef{default-11:NoMcCMS-A:pvntrk:loEff}   {\ensuremath{{1.000 } } }
\vdef{default-11:NoMcCMS-A:pvntrk:loEffE}   {\ensuremath{{0.000 } } }
\vdef{default-11:NoMcCMS-A:pvntrk:hiEff}   {\ensuremath{{1.000 } } }
\vdef{default-11:NoMcCMS-A:pvntrk:hiEffE}   {\ensuremath{{0.000 } } }
\vdef{default-11:NoMcCMS-A:pvntrk:loDelta}   {\ensuremath{{+0.000 } } }
\vdef{default-11:NoMcCMS-A:pvntrk:loDeltaE}   {\ensuremath{{0.000 } } }
\vdef{default-11:NoMcCMS-A:pvntrk:hiDelta}   {\ensuremath{{+0.000 } } }
\vdef{default-11:NoMcCMS-A:pvntrk:hiDeltaE}   {\ensuremath{{0.000 } } }
\vdef{default-11:NoData-AR3:muon1pt:loEff}   {\ensuremath{{1.008 } } }
\vdef{default-11:NoData-AR3:muon1pt:loEffE}   {\ensuremath{{\mathrm{NaN} } } }
\vdef{default-11:NoData-AR3:muon1pt:hiEff}   {\ensuremath{{1.000 } } }
\vdef{default-11:NoData-AR3:muon1pt:hiEffE}   {\ensuremath{{0.000 } } }
\vdef{default-11:NoMcCMS-A:muon1pt:loEff}   {\ensuremath{{1.009 } } }
\vdef{default-11:NoMcCMS-A:muon1pt:loEffE}   {\ensuremath{{\mathrm{NaN} } } }
\vdef{default-11:NoMcCMS-A:muon1pt:hiEff}   {\ensuremath{{1.000 } } }
\vdef{default-11:NoMcCMS-A:muon1pt:hiEffE}   {\ensuremath{{0.000 } } }
\vdef{default-11:NoMcCMS-A:muon1pt:loDelta}   {\ensuremath{{-0.001 } } }
\vdef{default-11:NoMcCMS-A:muon1pt:loDeltaE}   {\ensuremath{{\mathrm{NaN} } } }
\vdef{default-11:NoMcCMS-A:muon1pt:hiDelta}   {\ensuremath{{+0.000 } } }
\vdef{default-11:NoMcCMS-A:muon1pt:hiDeltaE}   {\ensuremath{{0.000 } } }
\vdef{default-11:NoData-AR3:muon2pt:loEff}   {\ensuremath{{0.134 } } }
\vdef{default-11:NoData-AR3:muon2pt:loEffE}   {\ensuremath{{0.002 } } }
\vdef{default-11:NoData-AR3:muon2pt:hiEff}   {\ensuremath{{0.866 } } }
\vdef{default-11:NoData-AR3:muon2pt:hiEffE}   {\ensuremath{{0.002 } } }
\vdef{default-11:NoMcCMS-A:muon2pt:loEff}   {\ensuremath{{0.090 } } }
\vdef{default-11:NoMcCMS-A:muon2pt:loEffE}   {\ensuremath{{0.002 } } }
\vdef{default-11:NoMcCMS-A:muon2pt:hiEff}   {\ensuremath{{0.910 } } }
\vdef{default-11:NoMcCMS-A:muon2pt:hiEffE}   {\ensuremath{{0.002 } } }
\vdef{default-11:NoMcCMS-A:muon2pt:loDelta}   {\ensuremath{{+0.398 } } }
\vdef{default-11:NoMcCMS-A:muon2pt:loDeltaE}   {\ensuremath{{0.027 } } }
\vdef{default-11:NoMcCMS-A:muon2pt:hiDelta}   {\ensuremath{{-0.050 } } }
\vdef{default-11:NoMcCMS-A:muon2pt:hiDeltaE}   {\ensuremath{{0.003 } } }
\vdef{default-11:NoData-AR3:muonseta:loEff}   {\ensuremath{{0.733 } } }
\vdef{default-11:NoData-AR3:muonseta:loEffE}   {\ensuremath{{0.002 } } }
\vdef{default-11:NoData-AR3:muonseta:hiEff}   {\ensuremath{{0.267 } } }
\vdef{default-11:NoData-AR3:muonseta:hiEffE}   {\ensuremath{{0.002 } } }
\vdef{default-11:NoMcCMS-A:muonseta:loEff}   {\ensuremath{{0.844 } } }
\vdef{default-11:NoMcCMS-A:muonseta:loEffE}   {\ensuremath{{0.002 } } }
\vdef{default-11:NoMcCMS-A:muonseta:hiEff}   {\ensuremath{{0.156 } } }
\vdef{default-11:NoMcCMS-A:muonseta:hiEffE}   {\ensuremath{{0.002 } } }
\vdef{default-11:NoMcCMS-A:muonseta:loDelta}   {\ensuremath{{-0.140 } } }
\vdef{default-11:NoMcCMS-A:muonseta:loDeltaE}   {\ensuremath{{0.004 } } }
\vdef{default-11:NoMcCMS-A:muonseta:hiDelta}   {\ensuremath{{+0.523 } } }
\vdef{default-11:NoMcCMS-A:muonseta:hiDeltaE}   {\ensuremath{{0.014 } } }
\vdef{default-11:NoData-AR3:pt:loEff}   {\ensuremath{{0.000 } } }
\vdef{default-11:NoData-AR3:pt:loEffE}   {\ensuremath{{0.000 } } }
\vdef{default-11:NoData-AR3:pt:hiEff}   {\ensuremath{{1.000 } } }
\vdef{default-11:NoData-AR3:pt:hiEffE}   {\ensuremath{{0.000 } } }
\vdef{default-11:NoMcCMS-A:pt:loEff}   {\ensuremath{{0.000 } } }
\vdef{default-11:NoMcCMS-A:pt:loEffE}   {\ensuremath{{0.000 } } }
\vdef{default-11:NoMcCMS-A:pt:hiEff}   {\ensuremath{{1.000 } } }
\vdef{default-11:NoMcCMS-A:pt:hiEffE}   {\ensuremath{{0.000 } } }
\vdef{default-11:NoMcCMS-A:pt:loDelta}   {\ensuremath{{\mathrm{NaN} } } }
\vdef{default-11:NoMcCMS-A:pt:loDeltaE}   {\ensuremath{{\mathrm{NaN} } } }
\vdef{default-11:NoMcCMS-A:pt:hiDelta}   {\ensuremath{{+0.000 } } }
\vdef{default-11:NoMcCMS-A:pt:hiDeltaE}   {\ensuremath{{0.000 } } }
\vdef{default-11:NoData-AR3:p:loEff}   {\ensuremath{{1.011 } } }
\vdef{default-11:NoData-AR3:p:loEffE}   {\ensuremath{{\mathrm{NaN} } } }
\vdef{default-11:NoData-AR3:p:hiEff}   {\ensuremath{{1.000 } } }
\vdef{default-11:NoData-AR3:p:hiEffE}   {\ensuremath{{0.000 } } }
\vdef{default-11:NoMcCMS-A:p:loEff}   {\ensuremath{{1.004 } } }
\vdef{default-11:NoMcCMS-A:p:loEffE}   {\ensuremath{{\mathrm{NaN} } } }
\vdef{default-11:NoMcCMS-A:p:hiEff}   {\ensuremath{{1.000 } } }
\vdef{default-11:NoMcCMS-A:p:hiEffE}   {\ensuremath{{0.000 } } }
\vdef{default-11:NoMcCMS-A:p:loDelta}   {\ensuremath{{+0.007 } } }
\vdef{default-11:NoMcCMS-A:p:loDeltaE}   {\ensuremath{{\mathrm{NaN} } } }
\vdef{default-11:NoMcCMS-A:p:hiDelta}   {\ensuremath{{+0.000 } } }
\vdef{default-11:NoMcCMS-A:p:hiDeltaE}   {\ensuremath{{0.000 } } }
\vdef{default-11:NoData-AR3:eta:loEff}   {\ensuremath{{0.726 } } }
\vdef{default-11:NoData-AR3:eta:loEffE}   {\ensuremath{{0.003 } } }
\vdef{default-11:NoData-AR3:eta:hiEff}   {\ensuremath{{0.274 } } }
\vdef{default-11:NoData-AR3:eta:hiEffE}   {\ensuremath{{0.003 } } }
\vdef{default-11:NoMcCMS-A:eta:loEff}   {\ensuremath{{0.841 } } }
\vdef{default-11:NoMcCMS-A:eta:loEffE}   {\ensuremath{{0.003 } } }
\vdef{default-11:NoMcCMS-A:eta:hiEff}   {\ensuremath{{0.159 } } }
\vdef{default-11:NoMcCMS-A:eta:hiEffE}   {\ensuremath{{0.003 } } }
\vdef{default-11:NoMcCMS-A:eta:loDelta}   {\ensuremath{{-0.147 } } }
\vdef{default-11:NoMcCMS-A:eta:loDeltaE}   {\ensuremath{{0.006 } } }
\vdef{default-11:NoMcCMS-A:eta:hiDelta}   {\ensuremath{{+0.532 } } }
\vdef{default-11:NoMcCMS-A:eta:hiDeltaE}   {\ensuremath{{0.019 } } }
\vdef{default-11:NoData-AR3:bdt:loEff}   {\ensuremath{{0.911 } } }
\vdef{default-11:NoData-AR3:bdt:loEffE}   {\ensuremath{{0.002 } } }
\vdef{default-11:NoData-AR3:bdt:hiEff}   {\ensuremath{{0.089 } } }
\vdef{default-11:NoData-AR3:bdt:hiEffE}   {\ensuremath{{0.002 } } }
\vdef{default-11:NoMcCMS-A:bdt:loEff}   {\ensuremath{{0.871 } } }
\vdef{default-11:NoMcCMS-A:bdt:loEffE}   {\ensuremath{{0.002 } } }
\vdef{default-11:NoMcCMS-A:bdt:hiEff}   {\ensuremath{{0.129 } } }
\vdef{default-11:NoMcCMS-A:bdt:hiEffE}   {\ensuremath{{0.002 } } }
\vdef{default-11:NoMcCMS-A:bdt:loDelta}   {\ensuremath{{+0.045 } } }
\vdef{default-11:NoMcCMS-A:bdt:loDeltaE}   {\ensuremath{{0.004 } } }
\vdef{default-11:NoMcCMS-A:bdt:hiDelta}   {\ensuremath{{-0.367 } } }
\vdef{default-11:NoMcCMS-A:bdt:hiDeltaE}   {\ensuremath{{0.028 } } }
\vdef{default-11:NoData-AR3:fl3d:loEff}   {\ensuremath{{0.834 } } }
\vdef{default-11:NoData-AR3:fl3d:loEffE}   {\ensuremath{{0.003 } } }
\vdef{default-11:NoData-AR3:fl3d:hiEff}   {\ensuremath{{0.166 } } }
\vdef{default-11:NoData-AR3:fl3d:hiEffE}   {\ensuremath{{0.003 } } }
\vdef{default-11:NoMcCMS-A:fl3d:loEff}   {\ensuremath{{0.881 } } }
\vdef{default-11:NoMcCMS-A:fl3d:loEffE}   {\ensuremath{{0.002 } } }
\vdef{default-11:NoMcCMS-A:fl3d:hiEff}   {\ensuremath{{0.119 } } }
\vdef{default-11:NoMcCMS-A:fl3d:hiEffE}   {\ensuremath{{0.002 } } }
\vdef{default-11:NoMcCMS-A:fl3d:loDelta}   {\ensuremath{{-0.054 } } }
\vdef{default-11:NoMcCMS-A:fl3d:loDeltaE}   {\ensuremath{{0.004 } } }
\vdef{default-11:NoMcCMS-A:fl3d:hiDelta}   {\ensuremath{{+0.328 } } }
\vdef{default-11:NoMcCMS-A:fl3d:hiDeltaE}   {\ensuremath{{0.023 } } }
\vdef{default-11:NoData-AR3:fl3de:loEff}   {\ensuremath{{1.000 } } }
\vdef{default-11:NoData-AR3:fl3de:loEffE}   {\ensuremath{{0.000 } } }
\vdef{default-11:NoData-AR3:fl3de:hiEff}   {\ensuremath{{-0.000 } } }
\vdef{default-11:NoData-AR3:fl3de:hiEffE}   {\ensuremath{{\mathrm{NaN} } } }
\vdef{default-11:NoMcCMS-A:fl3de:loEff}   {\ensuremath{{1.000 } } }
\vdef{default-11:NoMcCMS-A:fl3de:loEffE}   {\ensuremath{{0.000 } } }
\vdef{default-11:NoMcCMS-A:fl3de:hiEff}   {\ensuremath{{0.000 } } }
\vdef{default-11:NoMcCMS-A:fl3de:hiEffE}   {\ensuremath{{0.000 } } }
\vdef{default-11:NoMcCMS-A:fl3de:loDelta}   {\ensuremath{{+0.000 } } }
\vdef{default-11:NoMcCMS-A:fl3de:loDeltaE}   {\ensuremath{{0.000 } } }
\vdef{default-11:NoMcCMS-A:fl3de:hiDelta}   {\ensuremath{{+3.726 } } }
\vdef{default-11:NoMcCMS-A:fl3de:hiDeltaE}   {\ensuremath{{\mathrm{NaN} } } }
\vdef{default-11:NoData-AR3:fls3d:loEff}   {\ensuremath{{0.074 } } }
\vdef{default-11:NoData-AR3:fls3d:loEffE}   {\ensuremath{{0.002 } } }
\vdef{default-11:NoData-AR3:fls3d:hiEff}   {\ensuremath{{0.926 } } }
\vdef{default-11:NoData-AR3:fls3d:hiEffE}   {\ensuremath{{0.002 } } }
\vdef{default-11:NoMcCMS-A:fls3d:loEff}   {\ensuremath{{0.060 } } }
\vdef{default-11:NoMcCMS-A:fls3d:loEffE}   {\ensuremath{{0.002 } } }
\vdef{default-11:NoMcCMS-A:fls3d:hiEff}   {\ensuremath{{0.940 } } }
\vdef{default-11:NoMcCMS-A:fls3d:hiEffE}   {\ensuremath{{0.002 } } }
\vdef{default-11:NoMcCMS-A:fls3d:loDelta}   {\ensuremath{{+0.212 } } }
\vdef{default-11:NoMcCMS-A:fls3d:loDeltaE}   {\ensuremath{{0.036 } } }
\vdef{default-11:NoMcCMS-A:fls3d:hiDelta}   {\ensuremath{{-0.015 } } }
\vdef{default-11:NoMcCMS-A:fls3d:hiDeltaE}   {\ensuremath{{0.003 } } }
\vdef{default-11:NoData-AR3:flsxy:loEff}   {\ensuremath{{1.012 } } }
\vdef{default-11:NoData-AR3:flsxy:loEffE}   {\ensuremath{{\mathrm{NaN} } } }
\vdef{default-11:NoData-AR3:flsxy:hiEff}   {\ensuremath{{1.000 } } }
\vdef{default-11:NoData-AR3:flsxy:hiEffE}   {\ensuremath{{0.000 } } }
\vdef{default-11:NoMcCMS-A:flsxy:loEff}   {\ensuremath{{1.013 } } }
\vdef{default-11:NoMcCMS-A:flsxy:loEffE}   {\ensuremath{{\mathrm{NaN} } } }
\vdef{default-11:NoMcCMS-A:flsxy:hiEff}   {\ensuremath{{1.000 } } }
\vdef{default-11:NoMcCMS-A:flsxy:hiEffE}   {\ensuremath{{0.000 } } }
\vdef{default-11:NoMcCMS-A:flsxy:loDelta}   {\ensuremath{{-0.001 } } }
\vdef{default-11:NoMcCMS-A:flsxy:loDeltaE}   {\ensuremath{{\mathrm{NaN} } } }
\vdef{default-11:NoMcCMS-A:flsxy:hiDelta}   {\ensuremath{{+0.000 } } }
\vdef{default-11:NoMcCMS-A:flsxy:hiDeltaE}   {\ensuremath{{0.000 } } }
\vdef{default-11:NoData-AR3:chi2dof:loEff}   {\ensuremath{{0.935 } } }
\vdef{default-11:NoData-AR3:chi2dof:loEffE}   {\ensuremath{{0.002 } } }
\vdef{default-11:NoData-AR3:chi2dof:hiEff}   {\ensuremath{{0.065 } } }
\vdef{default-11:NoData-AR3:chi2dof:hiEffE}   {\ensuremath{{0.002 } } }
\vdef{default-11:NoMcCMS-A:chi2dof:loEff}   {\ensuremath{{0.939 } } }
\vdef{default-11:NoMcCMS-A:chi2dof:loEffE}   {\ensuremath{{0.002 } } }
\vdef{default-11:NoMcCMS-A:chi2dof:hiEff}   {\ensuremath{{0.061 } } }
\vdef{default-11:NoMcCMS-A:chi2dof:hiEffE}   {\ensuremath{{0.002 } } }
\vdef{default-11:NoMcCMS-A:chi2dof:loDelta}   {\ensuremath{{-0.004 } } }
\vdef{default-11:NoMcCMS-A:chi2dof:loDeltaE}   {\ensuremath{{0.003 } } }
\vdef{default-11:NoMcCMS-A:chi2dof:hiDelta}   {\ensuremath{{+0.065 } } }
\vdef{default-11:NoMcCMS-A:chi2dof:hiDeltaE}   {\ensuremath{{0.039 } } }
\vdef{default-11:NoData-AR3:pchi2dof:loEff}   {\ensuremath{{0.634 } } }
\vdef{default-11:NoData-AR3:pchi2dof:loEffE}   {\ensuremath{{0.003 } } }
\vdef{default-11:NoData-AR3:pchi2dof:hiEff}   {\ensuremath{{0.366 } } }
\vdef{default-11:NoData-AR3:pchi2dof:hiEffE}   {\ensuremath{{0.003 } } }
\vdef{default-11:NoMcCMS-A:pchi2dof:loEff}   {\ensuremath{{0.629 } } }
\vdef{default-11:NoMcCMS-A:pchi2dof:loEffE}   {\ensuremath{{0.003 } } }
\vdef{default-11:NoMcCMS-A:pchi2dof:hiEff}   {\ensuremath{{0.371 } } }
\vdef{default-11:NoMcCMS-A:pchi2dof:hiEffE}   {\ensuremath{{0.003 } } }
\vdef{default-11:NoMcCMS-A:pchi2dof:loDelta}   {\ensuremath{{+0.008 } } }
\vdef{default-11:NoMcCMS-A:pchi2dof:loDeltaE}   {\ensuremath{{0.008 } } }
\vdef{default-11:NoMcCMS-A:pchi2dof:hiDelta}   {\ensuremath{{-0.014 } } }
\vdef{default-11:NoMcCMS-A:pchi2dof:hiDeltaE}   {\ensuremath{{0.013 } } }
\vdef{default-11:NoData-AR3:alpha:loEff}   {\ensuremath{{0.994 } } }
\vdef{default-11:NoData-AR3:alpha:loEffE}   {\ensuremath{{0.001 } } }
\vdef{default-11:NoData-AR3:alpha:hiEff}   {\ensuremath{{0.006 } } }
\vdef{default-11:NoData-AR3:alpha:hiEffE}   {\ensuremath{{0.001 } } }
\vdef{default-11:NoMcCMS-A:alpha:loEff}   {\ensuremath{{0.993 } } }
\vdef{default-11:NoMcCMS-A:alpha:loEffE}   {\ensuremath{{0.001 } } }
\vdef{default-11:NoMcCMS-A:alpha:hiEff}   {\ensuremath{{0.007 } } }
\vdef{default-11:NoMcCMS-A:alpha:hiEffE}   {\ensuremath{{0.001 } } }
\vdef{default-11:NoMcCMS-A:alpha:loDelta}   {\ensuremath{{+0.001 } } }
\vdef{default-11:NoMcCMS-A:alpha:loDeltaE}   {\ensuremath{{0.001 } } }
\vdef{default-11:NoMcCMS-A:alpha:hiDelta}   {\ensuremath{{-0.157 } } }
\vdef{default-11:NoMcCMS-A:alpha:hiDeltaE}   {\ensuremath{{0.131 } } }
\vdef{default-11:NoData-AR3:iso:loEff}   {\ensuremath{{0.130 } } }
\vdef{default-11:NoData-AR3:iso:loEffE}   {\ensuremath{{0.002 } } }
\vdef{default-11:NoData-AR3:iso:hiEff}   {\ensuremath{{0.870 } } }
\vdef{default-11:NoData-AR3:iso:hiEffE}   {\ensuremath{{0.002 } } }
\vdef{default-11:NoMcCMS-A:iso:loEff}   {\ensuremath{{0.111 } } }
\vdef{default-11:NoMcCMS-A:iso:loEffE}   {\ensuremath{{0.002 } } }
\vdef{default-11:NoMcCMS-A:iso:hiEff}   {\ensuremath{{0.889 } } }
\vdef{default-11:NoMcCMS-A:iso:hiEffE}   {\ensuremath{{0.002 } } }
\vdef{default-11:NoMcCMS-A:iso:loDelta}   {\ensuremath{{+0.161 } } }
\vdef{default-11:NoMcCMS-A:iso:loDeltaE}   {\ensuremath{{0.026 } } }
\vdef{default-11:NoMcCMS-A:iso:hiDelta}   {\ensuremath{{-0.022 } } }
\vdef{default-11:NoMcCMS-A:iso:hiDeltaE}   {\ensuremath{{0.004 } } }
\vdef{default-11:NoData-AR3:docatrk:loEff}   {\ensuremath{{0.072 } } }
\vdef{default-11:NoData-AR3:docatrk:loEffE}   {\ensuremath{{0.002 } } }
\vdef{default-11:NoData-AR3:docatrk:hiEff}   {\ensuremath{{0.928 } } }
\vdef{default-11:NoData-AR3:docatrk:hiEffE}   {\ensuremath{{0.002 } } }
\vdef{default-11:NoMcCMS-A:docatrk:loEff}   {\ensuremath{{0.087 } } }
\vdef{default-11:NoMcCMS-A:docatrk:loEffE}   {\ensuremath{{0.002 } } }
\vdef{default-11:NoMcCMS-A:docatrk:hiEff}   {\ensuremath{{0.913 } } }
\vdef{default-11:NoMcCMS-A:docatrk:hiEffE}   {\ensuremath{{0.002 } } }
\vdef{default-11:NoMcCMS-A:docatrk:loDelta}   {\ensuremath{{-0.193 } } }
\vdef{default-11:NoMcCMS-A:docatrk:loDeltaE}   {\ensuremath{{0.034 } } }
\vdef{default-11:NoMcCMS-A:docatrk:hiDelta}   {\ensuremath{{+0.017 } } }
\vdef{default-11:NoMcCMS-A:docatrk:hiDeltaE}   {\ensuremath{{0.003 } } }
\vdef{default-11:NoData-AR3:isotrk:loEff}   {\ensuremath{{1.000 } } }
\vdef{default-11:NoData-AR3:isotrk:loEffE}   {\ensuremath{{0.000 } } }
\vdef{default-11:NoData-AR3:isotrk:hiEff}   {\ensuremath{{1.000 } } }
\vdef{default-11:NoData-AR3:isotrk:hiEffE}   {\ensuremath{{0.000 } } }
\vdef{default-11:NoMcCMS-A:isotrk:loEff}   {\ensuremath{{1.000 } } }
\vdef{default-11:NoMcCMS-A:isotrk:loEffE}   {\ensuremath{{0.000 } } }
\vdef{default-11:NoMcCMS-A:isotrk:hiEff}   {\ensuremath{{1.000 } } }
\vdef{default-11:NoMcCMS-A:isotrk:hiEffE}   {\ensuremath{{0.000 } } }
\vdef{default-11:NoMcCMS-A:isotrk:loDelta}   {\ensuremath{{+0.000 } } }
\vdef{default-11:NoMcCMS-A:isotrk:loDeltaE}   {\ensuremath{{0.000 } } }
\vdef{default-11:NoMcCMS-A:isotrk:hiDelta}   {\ensuremath{{+0.000 } } }
\vdef{default-11:NoMcCMS-A:isotrk:hiDeltaE}   {\ensuremath{{0.000 } } }
\vdef{default-11:NoData-AR3:closetrk:loEff}   {\ensuremath{{0.977 } } }
\vdef{default-11:NoData-AR3:closetrk:loEffE}   {\ensuremath{{0.001 } } }
\vdef{default-11:NoData-AR3:closetrk:hiEff}   {\ensuremath{{0.023 } } }
\vdef{default-11:NoData-AR3:closetrk:hiEffE}   {\ensuremath{{0.001 } } }
\vdef{default-11:NoMcCMS-A:closetrk:loEff}   {\ensuremath{{0.974 } } }
\vdef{default-11:NoMcCMS-A:closetrk:loEffE}   {\ensuremath{{0.001 } } }
\vdef{default-11:NoMcCMS-A:closetrk:hiEff}   {\ensuremath{{0.026 } } }
\vdef{default-11:NoMcCMS-A:closetrk:hiEffE}   {\ensuremath{{0.001 } } }
\vdef{default-11:NoMcCMS-A:closetrk:loDelta}   {\ensuremath{{+0.003 } } }
\vdef{default-11:NoMcCMS-A:closetrk:loDeltaE}   {\ensuremath{{0.002 } } }
\vdef{default-11:NoMcCMS-A:closetrk:hiDelta}   {\ensuremath{{-0.133 } } }
\vdef{default-11:NoMcCMS-A:closetrk:hiDeltaE}   {\ensuremath{{0.065 } } }
\vdef{default-11:NoData-AR3:lip:loEff}   {\ensuremath{{1.000 } } }
\vdef{default-11:NoData-AR3:lip:loEffE}   {\ensuremath{{0.000 } } }
\vdef{default-11:NoData-AR3:lip:hiEff}   {\ensuremath{{0.000 } } }
\vdef{default-11:NoData-AR3:lip:hiEffE}   {\ensuremath{{0.000 } } }
\vdef{default-11:NoMcCMS-A:lip:loEff}   {\ensuremath{{1.000 } } }
\vdef{default-11:NoMcCMS-A:lip:loEffE}   {\ensuremath{{0.000 } } }
\vdef{default-11:NoMcCMS-A:lip:hiEff}   {\ensuremath{{0.000 } } }
\vdef{default-11:NoMcCMS-A:lip:hiEffE}   {\ensuremath{{0.000 } } }
\vdef{default-11:NoMcCMS-A:lip:loDelta}   {\ensuremath{{+0.000 } } }
\vdef{default-11:NoMcCMS-A:lip:loDeltaE}   {\ensuremath{{0.000 } } }
\vdef{default-11:NoMcCMS-A:lip:hiDelta}   {\ensuremath{{\mathrm{NaN} } } }
\vdef{default-11:NoMcCMS-A:lip:hiDeltaE}   {\ensuremath{{\mathrm{NaN} } } }
\vdef{default-11:NoData-AR3:lip:inEff}   {\ensuremath{{1.000 } } }
\vdef{default-11:NoData-AR3:lip:inEffE}   {\ensuremath{{0.000 } } }
\vdef{default-11:NoMcCMS-A:lip:inEff}   {\ensuremath{{1.000 } } }
\vdef{default-11:NoMcCMS-A:lip:inEffE}   {\ensuremath{{0.000 } } }
\vdef{default-11:NoMcCMS-A:lip:inDelta}   {\ensuremath{{+0.000 } } }
\vdef{default-11:NoMcCMS-A:lip:inDeltaE}   {\ensuremath{{0.000 } } }
\vdef{default-11:NoData-AR3:lips:loEff}   {\ensuremath{{1.000 } } }
\vdef{default-11:NoData-AR3:lips:loEffE}   {\ensuremath{{0.000 } } }
\vdef{default-11:NoData-AR3:lips:hiEff}   {\ensuremath{{0.000 } } }
\vdef{default-11:NoData-AR3:lips:hiEffE}   {\ensuremath{{0.000 } } }
\vdef{default-11:NoMcCMS-A:lips:loEff}   {\ensuremath{{1.000 } } }
\vdef{default-11:NoMcCMS-A:lips:loEffE}   {\ensuremath{{0.000 } } }
\vdef{default-11:NoMcCMS-A:lips:hiEff}   {\ensuremath{{0.000 } } }
\vdef{default-11:NoMcCMS-A:lips:hiEffE}   {\ensuremath{{0.000 } } }
\vdef{default-11:NoMcCMS-A:lips:loDelta}   {\ensuremath{{+0.000 } } }
\vdef{default-11:NoMcCMS-A:lips:loDeltaE}   {\ensuremath{{0.000 } } }
\vdef{default-11:NoMcCMS-A:lips:hiDelta}   {\ensuremath{{\mathrm{NaN} } } }
\vdef{default-11:NoMcCMS-A:lips:hiDeltaE}   {\ensuremath{{\mathrm{NaN} } } }
\vdef{default-11:NoData-AR3:lips:inEff}   {\ensuremath{{1.000 } } }
\vdef{default-11:NoData-AR3:lips:inEffE}   {\ensuremath{{0.000 } } }
\vdef{default-11:NoMcCMS-A:lips:inEff}   {\ensuremath{{1.000 } } }
\vdef{default-11:NoMcCMS-A:lips:inEffE}   {\ensuremath{{0.000 } } }
\vdef{default-11:NoMcCMS-A:lips:inDelta}   {\ensuremath{{+0.000 } } }
\vdef{default-11:NoMcCMS-A:lips:inDeltaE}   {\ensuremath{{0.000 } } }
\vdef{default-11:NoData-AR3:ip:loEff}   {\ensuremath{{0.971 } } }
\vdef{default-11:NoData-AR3:ip:loEffE}   {\ensuremath{{0.001 } } }
\vdef{default-11:NoData-AR3:ip:hiEff}   {\ensuremath{{0.029 } } }
\vdef{default-11:NoData-AR3:ip:hiEffE}   {\ensuremath{{0.001 } } }
\vdef{default-11:NoMcCMS-A:ip:loEff}   {\ensuremath{{0.970 } } }
\vdef{default-11:NoMcCMS-A:ip:loEffE}   {\ensuremath{{0.001 } } }
\vdef{default-11:NoMcCMS-A:ip:hiEff}   {\ensuremath{{0.030 } } }
\vdef{default-11:NoMcCMS-A:ip:hiEffE}   {\ensuremath{{0.001 } } }
\vdef{default-11:NoMcCMS-A:ip:loDelta}   {\ensuremath{{+0.002 } } }
\vdef{default-11:NoMcCMS-A:ip:loDeltaE}   {\ensuremath{{0.002 } } }
\vdef{default-11:NoMcCMS-A:ip:hiDelta}   {\ensuremath{{-0.051 } } }
\vdef{default-11:NoMcCMS-A:ip:hiDeltaE}   {\ensuremath{{0.059 } } }
\vdef{default-11:NoData-AR3:ips:loEff}   {\ensuremath{{0.948 } } }
\vdef{default-11:NoData-AR3:ips:loEffE}   {\ensuremath{{0.002 } } }
\vdef{default-11:NoData-AR3:ips:hiEff}   {\ensuremath{{0.052 } } }
\vdef{default-11:NoData-AR3:ips:hiEffE}   {\ensuremath{{0.002 } } }
\vdef{default-11:NoMcCMS-A:ips:loEff}   {\ensuremath{{0.951 } } }
\vdef{default-11:NoMcCMS-A:ips:loEffE}   {\ensuremath{{0.002 } } }
\vdef{default-11:NoMcCMS-A:ips:hiEff}   {\ensuremath{{0.049 } } }
\vdef{default-11:NoMcCMS-A:ips:hiEffE}   {\ensuremath{{0.002 } } }
\vdef{default-11:NoMcCMS-A:ips:loDelta}   {\ensuremath{{-0.003 } } }
\vdef{default-11:NoMcCMS-A:ips:loDeltaE}   {\ensuremath{{0.002 } } }
\vdef{default-11:NoMcCMS-A:ips:hiDelta}   {\ensuremath{{+0.061 } } }
\vdef{default-11:NoMcCMS-A:ips:hiDeltaE}   {\ensuremath{{0.044 } } }
\vdef{default-11:NoData-AR3:maxdoca:loEff}   {\ensuremath{{1.000 } } }
\vdef{default-11:NoData-AR3:maxdoca:loEffE}   {\ensuremath{{0.000 } } }
\vdef{default-11:NoData-AR3:maxdoca:hiEff}   {\ensuremath{{0.014 } } }
\vdef{default-11:NoData-AR3:maxdoca:hiEffE}   {\ensuremath{{0.001 } } }
\vdef{default-11:NoMcCMS-A:maxdoca:loEff}   {\ensuremath{{1.000 } } }
\vdef{default-11:NoMcCMS-A:maxdoca:loEffE}   {\ensuremath{{0.000 } } }
\vdef{default-11:NoMcCMS-A:maxdoca:hiEff}   {\ensuremath{{0.009 } } }
\vdef{default-11:NoMcCMS-A:maxdoca:hiEffE}   {\ensuremath{{0.001 } } }
\vdef{default-11:NoMcCMS-A:maxdoca:loDelta}   {\ensuremath{{+0.000 } } }
\vdef{default-11:NoMcCMS-A:maxdoca:loDeltaE}   {\ensuremath{{0.000 } } }
\vdef{default-11:NoMcCMS-A:maxdoca:hiDelta}   {\ensuremath{{+0.495 } } }
\vdef{default-11:NoMcCMS-A:maxdoca:hiDeltaE}   {\ensuremath{{0.094 } } }
\vdef{default-11:NoData-AR3:kaonpt:loEff}   {\ensuremath{{1.009 } } }
\vdef{default-11:NoData-AR3:kaonpt:loEffE}   {\ensuremath{{\mathrm{NaN} } } }
\vdef{default-11:NoData-AR3:kaonpt:hiEff}   {\ensuremath{{1.000 } } }
\vdef{default-11:NoData-AR3:kaonpt:hiEffE}   {\ensuremath{{0.000 } } }
\vdef{default-11:NoMcCMS-A:kaonpt:loEff}   {\ensuremath{{1.008 } } }
\vdef{default-11:NoMcCMS-A:kaonpt:loEffE}   {\ensuremath{{\mathrm{NaN} } } }
\vdef{default-11:NoMcCMS-A:kaonpt:hiEff}   {\ensuremath{{1.000 } } }
\vdef{default-11:NoMcCMS-A:kaonpt:hiEffE}   {\ensuremath{{0.000 } } }
\vdef{default-11:NoMcCMS-A:kaonpt:loDelta}   {\ensuremath{{+0.001 } } }
\vdef{default-11:NoMcCMS-A:kaonpt:loDeltaE}   {\ensuremath{{\mathrm{NaN} } } }
\vdef{default-11:NoMcCMS-A:kaonpt:hiDelta}   {\ensuremath{{+0.000 } } }
\vdef{default-11:NoMcCMS-A:kaonpt:hiDeltaE}   {\ensuremath{{0.000 } } }
\vdef{default-11:NoData-AR3:psipt:loEff}   {\ensuremath{{1.004 } } }
\vdef{default-11:NoData-AR3:psipt:loEffE}   {\ensuremath{{\mathrm{NaN} } } }
\vdef{default-11:NoData-AR3:psipt:hiEff}   {\ensuremath{{1.000 } } }
\vdef{default-11:NoData-AR3:psipt:hiEffE}   {\ensuremath{{0.000 } } }
\vdef{default-11:NoMcCMS-A:psipt:loEff}   {\ensuremath{{1.003 } } }
\vdef{default-11:NoMcCMS-A:psipt:loEffE}   {\ensuremath{{\mathrm{NaN} } } }
\vdef{default-11:NoMcCMS-A:psipt:hiEff}   {\ensuremath{{1.000 } } }
\vdef{default-11:NoMcCMS-A:psipt:hiEffE}   {\ensuremath{{0.000 } } }
\vdef{default-11:NoMcCMS-A:psipt:loDelta}   {\ensuremath{{+0.001 } } }
\vdef{default-11:NoMcCMS-A:psipt:loDeltaE}   {\ensuremath{{\mathrm{NaN} } } }
\vdef{default-11:NoMcCMS-A:psipt:hiDelta}   {\ensuremath{{+0.000 } } }
\vdef{default-11:NoMcCMS-A:psipt:hiDeltaE}   {\ensuremath{{0.000 } } }
\vdef{default-11:NoMc3e33-A:osiso:loEff}   {\ensuremath{{1.003 } } }
\vdef{default-11:NoMc3e33-A:osiso:loEffE}   {\ensuremath{{\mathrm{NaN} } } }
\vdef{default-11:NoMc3e33-A:osiso:hiEff}   {\ensuremath{{1.000 } } }
\vdef{default-11:NoMc3e33-A:osiso:hiEffE}   {\ensuremath{{0.000 } } }
\vdef{default-11:NoMcCMS-A:osiso:loEff}   {\ensuremath{{1.003 } } }
\vdef{default-11:NoMcCMS-A:osiso:loEffE}   {\ensuremath{{\mathrm{NaN} } } }
\vdef{default-11:NoMcCMS-A:osiso:hiEff}   {\ensuremath{{1.000 } } }
\vdef{default-11:NoMcCMS-A:osiso:hiEffE}   {\ensuremath{{0.000 } } }
\vdef{default-11:NoMcCMS-A:osiso:loDelta}   {\ensuremath{{+0.000 } } }
\vdef{default-11:NoMcCMS-A:osiso:loDeltaE}   {\ensuremath{{\mathrm{NaN} } } }
\vdef{default-11:NoMcCMS-A:osiso:hiDelta}   {\ensuremath{{+0.000 } } }
\vdef{default-11:NoMcCMS-A:osiso:hiDeltaE}   {\ensuremath{{0.000 } } }
\vdef{default-11:NoMc3e33-A:osreliso:loEff}   {\ensuremath{{0.288 } } }
\vdef{default-11:NoMc3e33-A:osreliso:loEffE}   {\ensuremath{{0.001 } } }
\vdef{default-11:NoMc3e33-A:osreliso:hiEff}   {\ensuremath{{0.712 } } }
\vdef{default-11:NoMc3e33-A:osreliso:hiEffE}   {\ensuremath{{0.001 } } }
\vdef{default-11:NoMcCMS-A:osreliso:loEff}   {\ensuremath{{0.282 } } }
\vdef{default-11:NoMcCMS-A:osreliso:loEffE}   {\ensuremath{{0.001 } } }
\vdef{default-11:NoMcCMS-A:osreliso:hiEff}   {\ensuremath{{0.718 } } }
\vdef{default-11:NoMcCMS-A:osreliso:hiEffE}   {\ensuremath{{0.001 } } }
\vdef{default-11:NoMcCMS-A:osreliso:loDelta}   {\ensuremath{{+0.022 } } }
\vdef{default-11:NoMcCMS-A:osreliso:loDeltaE}   {\ensuremath{{0.006 } } }
\vdef{default-11:NoMcCMS-A:osreliso:hiDelta}   {\ensuremath{{-0.009 } } }
\vdef{default-11:NoMcCMS-A:osreliso:hiDeltaE}   {\ensuremath{{0.002 } } }
\vdef{default-11:NoMc3e33-A:osmuonpt:loEff}   {\ensuremath{{0.000 } } }
\vdef{default-11:NoMc3e33-A:osmuonpt:loEffE}   {\ensuremath{{0.000 } } }
\vdef{default-11:NoMc3e33-A:osmuonpt:hiEff}   {\ensuremath{{1.000 } } }
\vdef{default-11:NoMc3e33-A:osmuonpt:hiEffE}   {\ensuremath{{0.000 } } }
\vdef{default-11:NoMcCMS-A:osmuonpt:loEff}   {\ensuremath{{0.000 } } }
\vdef{default-11:NoMcCMS-A:osmuonpt:loEffE}   {\ensuremath{{0.000 } } }
\vdef{default-11:NoMcCMS-A:osmuonpt:hiEff}   {\ensuremath{{1.000 } } }
\vdef{default-11:NoMcCMS-A:osmuonpt:hiEffE}   {\ensuremath{{0.000 } } }
\vdef{default-11:NoMcCMS-A:osmuonpt:loDelta}   {\ensuremath{{\mathrm{NaN} } } }
\vdef{default-11:NoMcCMS-A:osmuonpt:loDeltaE}   {\ensuremath{{\mathrm{NaN} } } }
\vdef{default-11:NoMcCMS-A:osmuonpt:hiDelta}   {\ensuremath{{+0.000 } } }
\vdef{default-11:NoMcCMS-A:osmuonpt:hiDeltaE}   {\ensuremath{{0.000 } } }
\vdef{default-11:NoMc3e33-A:osmuondr:loEff}   {\ensuremath{{0.016 } } }
\vdef{default-11:NoMc3e33-A:osmuondr:loEffE}   {\ensuremath{{0.002 } } }
\vdef{default-11:NoMc3e33-A:osmuondr:hiEff}   {\ensuremath{{0.984 } } }
\vdef{default-11:NoMc3e33-A:osmuondr:hiEffE}   {\ensuremath{{0.002 } } }
\vdef{default-11:NoMcCMS-A:osmuondr:loEff}   {\ensuremath{{0.012 } } }
\vdef{default-11:NoMcCMS-A:osmuondr:loEffE}   {\ensuremath{{0.002 } } }
\vdef{default-11:NoMcCMS-A:osmuondr:hiEff}   {\ensuremath{{0.988 } } }
\vdef{default-11:NoMcCMS-A:osmuondr:hiEffE}   {\ensuremath{{0.002 } } }
\vdef{default-11:NoMcCMS-A:osmuondr:loDelta}   {\ensuremath{{+0.321 } } }
\vdef{default-11:NoMcCMS-A:osmuondr:loDeltaE}   {\ensuremath{{0.190 } } }
\vdef{default-11:NoMcCMS-A:osmuondr:hiDelta}   {\ensuremath{{-0.005 } } }
\vdef{default-11:NoMcCMS-A:osmuondr:hiDeltaE}   {\ensuremath{{0.003 } } }
\vdef{default-11:NoMc3e33-A:hlt:loEff}   {\ensuremath{{0.336 } } }
\vdef{default-11:NoMc3e33-A:hlt:loEffE}   {\ensuremath{{0.001 } } }
\vdef{default-11:NoMc3e33-A:hlt:hiEff}   {\ensuremath{{0.664 } } }
\vdef{default-11:NoMc3e33-A:hlt:hiEffE}   {\ensuremath{{0.001 } } }
\vdef{default-11:NoMcCMS-A:hlt:loEff}   {\ensuremath{{0.267 } } }
\vdef{default-11:NoMcCMS-A:hlt:loEffE}   {\ensuremath{{0.001 } } }
\vdef{default-11:NoMcCMS-A:hlt:hiEff}   {\ensuremath{{0.733 } } }
\vdef{default-11:NoMcCMS-A:hlt:hiEffE}   {\ensuremath{{0.001 } } }
\vdef{default-11:NoMcCMS-A:hlt:loDelta}   {\ensuremath{{+0.229 } } }
\vdef{default-11:NoMcCMS-A:hlt:loDeltaE}   {\ensuremath{{0.005 } } }
\vdef{default-11:NoMcCMS-A:hlt:hiDelta}   {\ensuremath{{-0.099 } } }
\vdef{default-11:NoMcCMS-A:hlt:hiDeltaE}   {\ensuremath{{0.002 } } }
\vdef{default-11:NoMc3e33-A:muonsid:loEff}   {\ensuremath{{0.157 } } }
\vdef{default-11:NoMc3e33-A:muonsid:loEffE}   {\ensuremath{{0.001 } } }
\vdef{default-11:NoMc3e33-A:muonsid:hiEff}   {\ensuremath{{0.843 } } }
\vdef{default-11:NoMc3e33-A:muonsid:hiEffE}   {\ensuremath{{0.001 } } }
\vdef{default-11:NoMcCMS-A:muonsid:loEff}   {\ensuremath{{0.220 } } }
\vdef{default-11:NoMcCMS-A:muonsid:loEffE}   {\ensuremath{{0.001 } } }
\vdef{default-11:NoMcCMS-A:muonsid:hiEff}   {\ensuremath{{0.780 } } }
\vdef{default-11:NoMcCMS-A:muonsid:hiEffE}   {\ensuremath{{0.001 } } }
\vdef{default-11:NoMcCMS-A:muonsid:loDelta}   {\ensuremath{{-0.334 } } }
\vdef{default-11:NoMcCMS-A:muonsid:loDeltaE}   {\ensuremath{{0.008 } } }
\vdef{default-11:NoMcCMS-A:muonsid:hiDelta}   {\ensuremath{{+0.078 } } }
\vdef{default-11:NoMcCMS-A:muonsid:hiDeltaE}   {\ensuremath{{0.002 } } }
\vdef{default-11:NoMc3e33-A:tracksqual:loEff}   {\ensuremath{{0.000 } } }
\vdef{default-11:NoMc3e33-A:tracksqual:loEffE}   {\ensuremath{{0.000 } } }
\vdef{default-11:NoMc3e33-A:tracksqual:hiEff}   {\ensuremath{{1.000 } } }
\vdef{default-11:NoMc3e33-A:tracksqual:hiEffE}   {\ensuremath{{0.000 } } }
\vdef{default-11:NoMcCMS-A:tracksqual:loEff}   {\ensuremath{{0.000 } } }
\vdef{default-11:NoMcCMS-A:tracksqual:loEffE}   {\ensuremath{{0.000 } } }
\vdef{default-11:NoMcCMS-A:tracksqual:hiEff}   {\ensuremath{{1.000 } } }
\vdef{default-11:NoMcCMS-A:tracksqual:hiEffE}   {\ensuremath{{0.000 } } }
\vdef{default-11:NoMcCMS-A:tracksqual:loDelta}   {\ensuremath{{-0.361 } } }
\vdef{default-11:NoMcCMS-A:tracksqual:loDeltaE}   {\ensuremath{{0.286 } } }
\vdef{default-11:NoMcCMS-A:tracksqual:hiDelta}   {\ensuremath{{+0.000 } } }
\vdef{default-11:NoMcCMS-A:tracksqual:hiDeltaE}   {\ensuremath{{0.000 } } }
\vdef{default-11:NoMc3e33-A:pvz:loEff}   {\ensuremath{{0.469 } } }
\vdef{default-11:NoMc3e33-A:pvz:loEffE}   {\ensuremath{{0.001 } } }
\vdef{default-11:NoMc3e33-A:pvz:hiEff}   {\ensuremath{{0.531 } } }
\vdef{default-11:NoMc3e33-A:pvz:hiEffE}   {\ensuremath{{0.001 } } }
\vdef{default-11:NoMcCMS-A:pvz:loEff}   {\ensuremath{{0.470 } } }
\vdef{default-11:NoMcCMS-A:pvz:loEffE}   {\ensuremath{{0.001 } } }
\vdef{default-11:NoMcCMS-A:pvz:hiEff}   {\ensuremath{{0.530 } } }
\vdef{default-11:NoMcCMS-A:pvz:hiEffE}   {\ensuremath{{0.001 } } }
\vdef{default-11:NoMcCMS-A:pvz:loDelta}   {\ensuremath{{-0.001 } } }
\vdef{default-11:NoMcCMS-A:pvz:loDeltaE}   {\ensuremath{{0.004 } } }
\vdef{default-11:NoMcCMS-A:pvz:hiDelta}   {\ensuremath{{+0.001 } } }
\vdef{default-11:NoMcCMS-A:pvz:hiDeltaE}   {\ensuremath{{0.003 } } }
\vdef{default-11:NoMc3e33-A:pvn:loEff}   {\ensuremath{{1.000 } } }
\vdef{default-11:NoMc3e33-A:pvn:loEffE}   {\ensuremath{{0.000 } } }
\vdef{default-11:NoMc3e33-A:pvn:hiEff}   {\ensuremath{{1.000 } } }
\vdef{default-11:NoMc3e33-A:pvn:hiEffE}   {\ensuremath{{0.000 } } }
\vdef{default-11:NoMcCMS-A:pvn:loEff}   {\ensuremath{{1.065 } } }
\vdef{default-11:NoMcCMS-A:pvn:loEffE}   {\ensuremath{{\mathrm{NaN} } } }
\vdef{default-11:NoMcCMS-A:pvn:hiEff}   {\ensuremath{{1.000 } } }
\vdef{default-11:NoMcCMS-A:pvn:hiEffE}   {\ensuremath{{0.000 } } }
\vdef{default-11:NoMcCMS-A:pvn:loDelta}   {\ensuremath{{-0.063 } } }
\vdef{default-11:NoMcCMS-A:pvn:loDeltaE}   {\ensuremath{{\mathrm{NaN} } } }
\vdef{default-11:NoMcCMS-A:pvn:hiDelta}   {\ensuremath{{+0.000 } } }
\vdef{default-11:NoMcCMS-A:pvn:hiDeltaE}   {\ensuremath{{0.000 } } }
\vdef{default-11:NoMc3e33-A:pvavew8:loEff}   {\ensuremath{{0.004 } } }
\vdef{default-11:NoMc3e33-A:pvavew8:loEffE}   {\ensuremath{{0.000 } } }
\vdef{default-11:NoMc3e33-A:pvavew8:hiEff}   {\ensuremath{{0.996 } } }
\vdef{default-11:NoMc3e33-A:pvavew8:hiEffE}   {\ensuremath{{0.000 } } }
\vdef{default-11:NoMcCMS-A:pvavew8:loEff}   {\ensuremath{{0.006 } } }
\vdef{default-11:NoMcCMS-A:pvavew8:loEffE}   {\ensuremath{{0.000 } } }
\vdef{default-11:NoMcCMS-A:pvavew8:hiEff}   {\ensuremath{{0.994 } } }
\vdef{default-11:NoMcCMS-A:pvavew8:hiEffE}   {\ensuremath{{0.000 } } }
\vdef{default-11:NoMcCMS-A:pvavew8:loDelta}   {\ensuremath{{-0.345 } } }
\vdef{default-11:NoMcCMS-A:pvavew8:loDeltaE}   {\ensuremath{{0.058 } } }
\vdef{default-11:NoMcCMS-A:pvavew8:hiDelta}   {\ensuremath{{+0.002 } } }
\vdef{default-11:NoMcCMS-A:pvavew8:hiDeltaE}   {\ensuremath{{0.000 } } }
\vdef{default-11:NoMc3e33-A:pvntrk:loEff}   {\ensuremath{{1.000 } } }
\vdef{default-11:NoMc3e33-A:pvntrk:loEffE}   {\ensuremath{{0.000 } } }
\vdef{default-11:NoMc3e33-A:pvntrk:hiEff}   {\ensuremath{{1.000 } } }
\vdef{default-11:NoMc3e33-A:pvntrk:hiEffE}   {\ensuremath{{0.000 } } }
\vdef{default-11:NoMcCMS-A:pvntrk:loEff}   {\ensuremath{{1.000 } } }
\vdef{default-11:NoMcCMS-A:pvntrk:loEffE}   {\ensuremath{{0.000 } } }
\vdef{default-11:NoMcCMS-A:pvntrk:hiEff}   {\ensuremath{{1.000 } } }
\vdef{default-11:NoMcCMS-A:pvntrk:hiEffE}   {\ensuremath{{0.000 } } }
\vdef{default-11:NoMcCMS-A:pvntrk:loDelta}   {\ensuremath{{+0.000 } } }
\vdef{default-11:NoMcCMS-A:pvntrk:loDeltaE}   {\ensuremath{{0.000 } } }
\vdef{default-11:NoMcCMS-A:pvntrk:hiDelta}   {\ensuremath{{+0.000 } } }
\vdef{default-11:NoMcCMS-A:pvntrk:hiDeltaE}   {\ensuremath{{0.000 } } }
\vdef{default-11:NoMc3e33-A:muon1pt:loEff}   {\ensuremath{{1.008 } } }
\vdef{default-11:NoMc3e33-A:muon1pt:loEffE}   {\ensuremath{{\mathrm{NaN} } } }
\vdef{default-11:NoMc3e33-A:muon1pt:hiEff}   {\ensuremath{{1.000 } } }
\vdef{default-11:NoMc3e33-A:muon1pt:hiEffE}   {\ensuremath{{0.000 } } }
\vdef{default-11:NoMcCMS-A:muon1pt:loEff}   {\ensuremath{{1.009 } } }
\vdef{default-11:NoMcCMS-A:muon1pt:loEffE}   {\ensuremath{{\mathrm{NaN} } } }
\vdef{default-11:NoMcCMS-A:muon1pt:hiEff}   {\ensuremath{{1.000 } } }
\vdef{default-11:NoMcCMS-A:muon1pt:hiEffE}   {\ensuremath{{0.000 } } }
\vdef{default-11:NoMcCMS-A:muon1pt:loDelta}   {\ensuremath{{-0.001 } } }
\vdef{default-11:NoMcCMS-A:muon1pt:loDeltaE}   {\ensuremath{{\mathrm{NaN} } } }
\vdef{default-11:NoMcCMS-A:muon1pt:hiDelta}   {\ensuremath{{+0.000 } } }
\vdef{default-11:NoMcCMS-A:muon1pt:hiDeltaE}   {\ensuremath{{0.000 } } }
\vdef{default-11:NoMc3e33-A:muon2pt:loEff}   {\ensuremath{{0.005 } } }
\vdef{default-11:NoMc3e33-A:muon2pt:loEffE}   {\ensuremath{{0.000 } } }
\vdef{default-11:NoMc3e33-A:muon2pt:hiEff}   {\ensuremath{{0.995 } } }
\vdef{default-11:NoMc3e33-A:muon2pt:hiEffE}   {\ensuremath{{0.000 } } }
\vdef{default-11:NoMcCMS-A:muon2pt:loEff}   {\ensuremath{{0.090 } } }
\vdef{default-11:NoMcCMS-A:muon2pt:loEffE}   {\ensuremath{{0.001 } } }
\vdef{default-11:NoMcCMS-A:muon2pt:hiEff}   {\ensuremath{{0.910 } } }
\vdef{default-11:NoMcCMS-A:muon2pt:hiEffE}   {\ensuremath{{0.001 } } }
\vdef{default-11:NoMcCMS-A:muon2pt:loDelta}   {\ensuremath{{-1.772 } } }
\vdef{default-11:NoMcCMS-A:muon2pt:loDeltaE}   {\ensuremath{{0.009 } } }
\vdef{default-11:NoMcCMS-A:muon2pt:hiDelta}   {\ensuremath{{+0.088 } } }
\vdef{default-11:NoMcCMS-A:muon2pt:hiDeltaE}   {\ensuremath{{0.001 } } }
\vdef{default-11:NoMc3e33-A:muonseta:loEff}   {\ensuremath{{0.738 } } }
\vdef{default-11:NoMc3e33-A:muonseta:loEffE}   {\ensuremath{{0.001 } } }
\vdef{default-11:NoMc3e33-A:muonseta:hiEff}   {\ensuremath{{0.262 } } }
\vdef{default-11:NoMc3e33-A:muonseta:hiEffE}   {\ensuremath{{0.001 } } }
\vdef{default-11:NoMcCMS-A:muonseta:loEff}   {\ensuremath{{0.844 } } }
\vdef{default-11:NoMcCMS-A:muonseta:loEffE}   {\ensuremath{{0.001 } } }
\vdef{default-11:NoMcCMS-A:muonseta:hiEff}   {\ensuremath{{0.156 } } }
\vdef{default-11:NoMcCMS-A:muonseta:hiEffE}   {\ensuremath{{0.001 } } }
\vdef{default-11:NoMcCMS-A:muonseta:loDelta}   {\ensuremath{{-0.135 } } }
\vdef{default-11:NoMcCMS-A:muonseta:loDeltaE}   {\ensuremath{{0.002 } } }
\vdef{default-11:NoMcCMS-A:muonseta:hiDelta}   {\ensuremath{{+0.508 } } }
\vdef{default-11:NoMcCMS-A:muonseta:hiDeltaE}   {\ensuremath{{0.006 } } }
\vdef{default-11:NoMc3e33-A:pt:loEff}   {\ensuremath{{0.000 } } }
\vdef{default-11:NoMc3e33-A:pt:loEffE}   {\ensuremath{{0.000 } } }
\vdef{default-11:NoMc3e33-A:pt:hiEff}   {\ensuremath{{1.000 } } }
\vdef{default-11:NoMc3e33-A:pt:hiEffE}   {\ensuremath{{0.000 } } }
\vdef{default-11:NoMcCMS-A:pt:loEff}   {\ensuremath{{0.000 } } }
\vdef{default-11:NoMcCMS-A:pt:loEffE}   {\ensuremath{{0.000 } } }
\vdef{default-11:NoMcCMS-A:pt:hiEff}   {\ensuremath{{1.000 } } }
\vdef{default-11:NoMcCMS-A:pt:hiEffE}   {\ensuremath{{0.000 } } }
\vdef{default-11:NoMcCMS-A:pt:loDelta}   {\ensuremath{{\mathrm{NaN} } } }
\vdef{default-11:NoMcCMS-A:pt:loDeltaE}   {\ensuremath{{\mathrm{NaN} } } }
\vdef{default-11:NoMcCMS-A:pt:hiDelta}   {\ensuremath{{+0.000 } } }
\vdef{default-11:NoMcCMS-A:pt:hiDeltaE}   {\ensuremath{{0.000 } } }
\vdef{default-11:NoMc3e33-A:p:loEff}   {\ensuremath{{1.011 } } }
\vdef{default-11:NoMc3e33-A:p:loEffE}   {\ensuremath{{\mathrm{NaN} } } }
\vdef{default-11:NoMc3e33-A:p:hiEff}   {\ensuremath{{1.000 } } }
\vdef{default-11:NoMc3e33-A:p:hiEffE}   {\ensuremath{{0.000 } } }
\vdef{default-11:NoMcCMS-A:p:loEff}   {\ensuremath{{1.004 } } }
\vdef{default-11:NoMcCMS-A:p:loEffE}   {\ensuremath{{\mathrm{NaN} } } }
\vdef{default-11:NoMcCMS-A:p:hiEff}   {\ensuremath{{1.000 } } }
\vdef{default-11:NoMcCMS-A:p:hiEffE}   {\ensuremath{{0.000 } } }
\vdef{default-11:NoMcCMS-A:p:loDelta}   {\ensuremath{{+0.007 } } }
\vdef{default-11:NoMcCMS-A:p:loDeltaE}   {\ensuremath{{\mathrm{NaN} } } }
\vdef{default-11:NoMcCMS-A:p:hiDelta}   {\ensuremath{{+0.000 } } }
\vdef{default-11:NoMcCMS-A:p:hiDeltaE}   {\ensuremath{{0.000 } } }
\vdef{default-11:NoMc3e33-A:eta:loEff}   {\ensuremath{{0.729 } } }
\vdef{default-11:NoMc3e33-A:eta:loEffE}   {\ensuremath{{0.001 } } }
\vdef{default-11:NoMc3e33-A:eta:hiEff}   {\ensuremath{{0.271 } } }
\vdef{default-11:NoMc3e33-A:eta:hiEffE}   {\ensuremath{{0.001 } } }
\vdef{default-11:NoMcCMS-A:eta:loEff}   {\ensuremath{{0.841 } } }
\vdef{default-11:NoMcCMS-A:eta:loEffE}   {\ensuremath{{0.001 } } }
\vdef{default-11:NoMcCMS-A:eta:hiEff}   {\ensuremath{{0.159 } } }
\vdef{default-11:NoMcCMS-A:eta:hiEffE}   {\ensuremath{{0.001 } } }
\vdef{default-11:NoMcCMS-A:eta:loDelta}   {\ensuremath{{-0.142 } } }
\vdef{default-11:NoMcCMS-A:eta:loDeltaE}   {\ensuremath{{0.002 } } }
\vdef{default-11:NoMcCMS-A:eta:hiDelta}   {\ensuremath{{+0.519 } } }
\vdef{default-11:NoMcCMS-A:eta:hiDeltaE}   {\ensuremath{{0.008 } } }
\vdef{default-11:NoMc3e33-A:bdt:loEff}   {\ensuremath{{0.914 } } }
\vdef{default-11:NoMc3e33-A:bdt:loEffE}   {\ensuremath{{0.001 } } }
\vdef{default-11:NoMc3e33-A:bdt:hiEff}   {\ensuremath{{0.086 } } }
\vdef{default-11:NoMc3e33-A:bdt:hiEffE}   {\ensuremath{{0.001 } } }
\vdef{default-11:NoMcCMS-A:bdt:loEff}   {\ensuremath{{0.871 } } }
\vdef{default-11:NoMcCMS-A:bdt:loEffE}   {\ensuremath{{0.001 } } }
\vdef{default-11:NoMcCMS-A:bdt:hiEff}   {\ensuremath{{0.129 } } }
\vdef{default-11:NoMcCMS-A:bdt:hiEffE}   {\ensuremath{{0.001 } } }
\vdef{default-11:NoMcCMS-A:bdt:loDelta}   {\ensuremath{{+0.049 } } }
\vdef{default-11:NoMcCMS-A:bdt:loDeltaE}   {\ensuremath{{0.001 } } }
\vdef{default-11:NoMcCMS-A:bdt:hiDelta}   {\ensuremath{{-0.404 } } }
\vdef{default-11:NoMcCMS-A:bdt:hiDeltaE}   {\ensuremath{{0.009 } } }
\vdef{default-11:NoMc3e33-A:fl3d:loEff}   {\ensuremath{{0.835 } } }
\vdef{default-11:NoMc3e33-A:fl3d:loEffE}   {\ensuremath{{0.001 } } }
\vdef{default-11:NoMc3e33-A:fl3d:hiEff}   {\ensuremath{{0.165 } } }
\vdef{default-11:NoMc3e33-A:fl3d:hiEffE}   {\ensuremath{{0.001 } } }
\vdef{default-11:NoMcCMS-A:fl3d:loEff}   {\ensuremath{{0.881 } } }
\vdef{default-11:NoMcCMS-A:fl3d:loEffE}   {\ensuremath{{0.001 } } }
\vdef{default-11:NoMcCMS-A:fl3d:hiEff}   {\ensuremath{{0.119 } } }
\vdef{default-11:NoMcCMS-A:fl3d:hiEffE}   {\ensuremath{{0.001 } } }
\vdef{default-11:NoMcCMS-A:fl3d:loDelta}   {\ensuremath{{-0.053 } } }
\vdef{default-11:NoMcCMS-A:fl3d:loDeltaE}   {\ensuremath{{0.002 } } }
\vdef{default-11:NoMcCMS-A:fl3d:hiDelta}   {\ensuremath{{+0.322 } } }
\vdef{default-11:NoMcCMS-A:fl3d:hiDeltaE}   {\ensuremath{{0.009 } } }
\vdef{default-11:NoMc3e33-A:fl3de:loEff}   {\ensuremath{{1.000 } } }
\vdef{default-11:NoMc3e33-A:fl3de:loEffE}   {\ensuremath{{0.000 } } }
\vdef{default-11:NoMc3e33-A:fl3de:hiEff}   {\ensuremath{{0.000 } } }
\vdef{default-11:NoMc3e33-A:fl3de:hiEffE}   {\ensuremath{{0.000 } } }
\vdef{default-11:NoMcCMS-A:fl3de:loEff}   {\ensuremath{{1.000 } } }
\vdef{default-11:NoMcCMS-A:fl3de:loEffE}   {\ensuremath{{0.000 } } }
\vdef{default-11:NoMcCMS-A:fl3de:hiEff}   {\ensuremath{{0.000 } } }
\vdef{default-11:NoMcCMS-A:fl3de:hiEffE}   {\ensuremath{{0.000 } } }
\vdef{default-11:NoMcCMS-A:fl3de:loDelta}   {\ensuremath{{+0.000 } } }
\vdef{default-11:NoMcCMS-A:fl3de:loDeltaE}   {\ensuremath{{0.000 } } }
\vdef{default-11:NoMcCMS-A:fl3de:hiDelta}   {\ensuremath{{-0.651 } } }
\vdef{default-11:NoMcCMS-A:fl3de:hiDeltaE}   {\ensuremath{{0.702 } } }
\vdef{default-11:NoMc3e33-A:fls3d:loEff}   {\ensuremath{{0.074 } } }
\vdef{default-11:NoMc3e33-A:fls3d:loEffE}   {\ensuremath{{0.001 } } }
\vdef{default-11:NoMc3e33-A:fls3d:hiEff}   {\ensuremath{{0.926 } } }
\vdef{default-11:NoMc3e33-A:fls3d:hiEffE}   {\ensuremath{{0.001 } } }
\vdef{default-11:NoMcCMS-A:fls3d:loEff}   {\ensuremath{{0.060 } } }
\vdef{default-11:NoMcCMS-A:fls3d:loEffE}   {\ensuremath{{0.001 } } }
\vdef{default-11:NoMcCMS-A:fls3d:hiEff}   {\ensuremath{{0.940 } } }
\vdef{default-11:NoMcCMS-A:fls3d:hiEffE}   {\ensuremath{{0.001 } } }
\vdef{default-11:NoMcCMS-A:fls3d:loDelta}   {\ensuremath{{+0.216 } } }
\vdef{default-11:NoMcCMS-A:fls3d:loDeltaE}   {\ensuremath{{0.014 } } }
\vdef{default-11:NoMcCMS-A:fls3d:hiDelta}   {\ensuremath{{-0.015 } } }
\vdef{default-11:NoMcCMS-A:fls3d:hiDeltaE}   {\ensuremath{{0.001 } } }
\vdef{default-11:NoMc3e33-A:flsxy:loEff}   {\ensuremath{{1.012 } } }
\vdef{default-11:NoMc3e33-A:flsxy:loEffE}   {\ensuremath{{\mathrm{NaN} } } }
\vdef{default-11:NoMc3e33-A:flsxy:hiEff}   {\ensuremath{{1.000 } } }
\vdef{default-11:NoMc3e33-A:flsxy:hiEffE}   {\ensuremath{{0.000 } } }
\vdef{default-11:NoMcCMS-A:flsxy:loEff}   {\ensuremath{{1.013 } } }
\vdef{default-11:NoMcCMS-A:flsxy:loEffE}   {\ensuremath{{\mathrm{NaN} } } }
\vdef{default-11:NoMcCMS-A:flsxy:hiEff}   {\ensuremath{{1.000 } } }
\vdef{default-11:NoMcCMS-A:flsxy:hiEffE}   {\ensuremath{{0.000 } } }
\vdef{default-11:NoMcCMS-A:flsxy:loDelta}   {\ensuremath{{-0.001 } } }
\vdef{default-11:NoMcCMS-A:flsxy:loDeltaE}   {\ensuremath{{\mathrm{NaN} } } }
\vdef{default-11:NoMcCMS-A:flsxy:hiDelta}   {\ensuremath{{+0.000 } } }
\vdef{default-11:NoMcCMS-A:flsxy:hiDeltaE}   {\ensuremath{{0.000 } } }
\vdef{default-11:NoMc3e33-A:chi2dof:loEff}   {\ensuremath{{0.944 } } }
\vdef{default-11:NoMc3e33-A:chi2dof:loEffE}   {\ensuremath{{0.001 } } }
\vdef{default-11:NoMc3e33-A:chi2dof:hiEff}   {\ensuremath{{0.056 } } }
\vdef{default-11:NoMc3e33-A:chi2dof:hiEffE}   {\ensuremath{{0.001 } } }
\vdef{default-11:NoMcCMS-A:chi2dof:loEff}   {\ensuremath{{0.939 } } }
\vdef{default-11:NoMcCMS-A:chi2dof:loEffE}   {\ensuremath{{0.001 } } }
\vdef{default-11:NoMcCMS-A:chi2dof:hiEff}   {\ensuremath{{0.061 } } }
\vdef{default-11:NoMcCMS-A:chi2dof:hiEffE}   {\ensuremath{{0.001 } } }
\vdef{default-11:NoMcCMS-A:chi2dof:loDelta}   {\ensuremath{{+0.005 } } }
\vdef{default-11:NoMcCMS-A:chi2dof:loDeltaE}   {\ensuremath{{0.001 } } }
\vdef{default-11:NoMcCMS-A:chi2dof:hiDelta}   {\ensuremath{{-0.078 } } }
\vdef{default-11:NoMcCMS-A:chi2dof:hiDeltaE}   {\ensuremath{{0.016 } } }
\vdef{default-11:NoMc3e33-A:pchi2dof:loEff}   {\ensuremath{{0.607 } } }
\vdef{default-11:NoMc3e33-A:pchi2dof:loEffE}   {\ensuremath{{0.001 } } }
\vdef{default-11:NoMc3e33-A:pchi2dof:hiEff}   {\ensuremath{{0.393 } } }
\vdef{default-11:NoMc3e33-A:pchi2dof:hiEffE}   {\ensuremath{{0.001 } } }
\vdef{default-11:NoMcCMS-A:pchi2dof:loEff}   {\ensuremath{{0.629 } } }
\vdef{default-11:NoMcCMS-A:pchi2dof:loEffE}   {\ensuremath{{0.001 } } }
\vdef{default-11:NoMcCMS-A:pchi2dof:hiEff}   {\ensuremath{{0.371 } } }
\vdef{default-11:NoMcCMS-A:pchi2dof:hiEffE}   {\ensuremath{{0.001 } } }
\vdef{default-11:NoMcCMS-A:pchi2dof:loDelta}   {\ensuremath{{-0.036 } } }
\vdef{default-11:NoMcCMS-A:pchi2dof:loDeltaE}   {\ensuremath{{0.003 } } }
\vdef{default-11:NoMcCMS-A:pchi2dof:hiDelta}   {\ensuremath{{+0.058 } } }
\vdef{default-11:NoMcCMS-A:pchi2dof:hiDeltaE}   {\ensuremath{{0.005 } } }
\vdef{default-11:NoMc3e33-A:alpha:loEff}   {\ensuremath{{0.994 } } }
\vdef{default-11:NoMc3e33-A:alpha:loEffE}   {\ensuremath{{0.000 } } }
\vdef{default-11:NoMc3e33-A:alpha:hiEff}   {\ensuremath{{0.006 } } }
\vdef{default-11:NoMc3e33-A:alpha:hiEffE}   {\ensuremath{{0.000 } } }
\vdef{default-11:NoMcCMS-A:alpha:loEff}   {\ensuremath{{0.993 } } }
\vdef{default-11:NoMcCMS-A:alpha:loEffE}   {\ensuremath{{0.000 } } }
\vdef{default-11:NoMcCMS-A:alpha:hiEff}   {\ensuremath{{0.007 } } }
\vdef{default-11:NoMcCMS-A:alpha:hiEffE}   {\ensuremath{{0.000 } } }
\vdef{default-11:NoMcCMS-A:alpha:loDelta}   {\ensuremath{{+0.001 } } }
\vdef{default-11:NoMcCMS-A:alpha:loDeltaE}   {\ensuremath{{0.000 } } }
\vdef{default-11:NoMcCMS-A:alpha:hiDelta}   {\ensuremath{{-0.110 } } }
\vdef{default-11:NoMcCMS-A:alpha:hiDeltaE}   {\ensuremath{{0.051 } } }
\vdef{default-11:NoMc3e33-A:iso:loEff}   {\ensuremath{{0.107 } } }
\vdef{default-11:NoMc3e33-A:iso:loEffE}   {\ensuremath{{0.001 } } }
\vdef{default-11:NoMc3e33-A:iso:hiEff}   {\ensuremath{{0.893 } } }
\vdef{default-11:NoMc3e33-A:iso:hiEffE}   {\ensuremath{{0.001 } } }
\vdef{default-11:NoMcCMS-A:iso:loEff}   {\ensuremath{{0.111 } } }
\vdef{default-11:NoMcCMS-A:iso:loEffE}   {\ensuremath{{0.001 } } }
\vdef{default-11:NoMcCMS-A:iso:hiEff}   {\ensuremath{{0.889 } } }
\vdef{default-11:NoMcCMS-A:iso:hiEffE}   {\ensuremath{{0.001 } } }
\vdef{default-11:NoMcCMS-A:iso:loDelta}   {\ensuremath{{-0.032 } } }
\vdef{default-11:NoMcCMS-A:iso:loDeltaE}   {\ensuremath{{0.011 } } }
\vdef{default-11:NoMcCMS-A:iso:hiDelta}   {\ensuremath{{+0.004 } } }
\vdef{default-11:NoMcCMS-A:iso:hiDeltaE}   {\ensuremath{{0.001 } } }
\vdef{default-11:NoMc3e33-A:docatrk:loEff}   {\ensuremath{{0.083 } } }
\vdef{default-11:NoMc3e33-A:docatrk:loEffE}   {\ensuremath{{0.001 } } }
\vdef{default-11:NoMc3e33-A:docatrk:hiEff}   {\ensuremath{{0.917 } } }
\vdef{default-11:NoMc3e33-A:docatrk:hiEffE}   {\ensuremath{{0.001 } } }
\vdef{default-11:NoMcCMS-A:docatrk:loEff}   {\ensuremath{{0.087 } } }
\vdef{default-11:NoMcCMS-A:docatrk:loEffE}   {\ensuremath{{0.001 } } }
\vdef{default-11:NoMcCMS-A:docatrk:hiEff}   {\ensuremath{{0.913 } } }
\vdef{default-11:NoMcCMS-A:docatrk:hiEffE}   {\ensuremath{{0.001 } } }
\vdef{default-11:NoMcCMS-A:docatrk:loDelta}   {\ensuremath{{-0.050 } } }
\vdef{default-11:NoMcCMS-A:docatrk:loDeltaE}   {\ensuremath{{0.013 } } }
\vdef{default-11:NoMcCMS-A:docatrk:hiDelta}   {\ensuremath{{+0.005 } } }
\vdef{default-11:NoMcCMS-A:docatrk:hiDeltaE}   {\ensuremath{{0.001 } } }
\vdef{default-11:NoMc3e33-A:isotrk:loEff}   {\ensuremath{{1.000 } } }
\vdef{default-11:NoMc3e33-A:isotrk:loEffE}   {\ensuremath{{0.000 } } }
\vdef{default-11:NoMc3e33-A:isotrk:hiEff}   {\ensuremath{{1.000 } } }
\vdef{default-11:NoMc3e33-A:isotrk:hiEffE}   {\ensuremath{{0.000 } } }
\vdef{default-11:NoMcCMS-A:isotrk:loEff}   {\ensuremath{{1.000 } } }
\vdef{default-11:NoMcCMS-A:isotrk:loEffE}   {\ensuremath{{0.000 } } }
\vdef{default-11:NoMcCMS-A:isotrk:hiEff}   {\ensuremath{{1.000 } } }
\vdef{default-11:NoMcCMS-A:isotrk:hiEffE}   {\ensuremath{{0.000 } } }
\vdef{default-11:NoMcCMS-A:isotrk:loDelta}   {\ensuremath{{+0.000 } } }
\vdef{default-11:NoMcCMS-A:isotrk:loDeltaE}   {\ensuremath{{0.000 } } }
\vdef{default-11:NoMcCMS-A:isotrk:hiDelta}   {\ensuremath{{+0.000 } } }
\vdef{default-11:NoMcCMS-A:isotrk:hiDeltaE}   {\ensuremath{{0.000 } } }
\vdef{default-11:NoMc3e33-A:closetrk:loEff}   {\ensuremath{{0.978 } } }
\vdef{default-11:NoMc3e33-A:closetrk:loEffE}   {\ensuremath{{0.000 } } }
\vdef{default-11:NoMc3e33-A:closetrk:hiEff}   {\ensuremath{{0.022 } } }
\vdef{default-11:NoMc3e33-A:closetrk:hiEffE}   {\ensuremath{{0.000 } } }
\vdef{default-11:NoMcCMS-A:closetrk:loEff}   {\ensuremath{{0.974 } } }
\vdef{default-11:NoMcCMS-A:closetrk:loEffE}   {\ensuremath{{0.000 } } }
\vdef{default-11:NoMcCMS-A:closetrk:hiEff}   {\ensuremath{{0.026 } } }
\vdef{default-11:NoMcCMS-A:closetrk:hiEffE}   {\ensuremath{{0.000 } } }
\vdef{default-11:NoMcCMS-A:closetrk:loDelta}   {\ensuremath{{+0.004 } } }
\vdef{default-11:NoMcCMS-A:closetrk:loDeltaE}   {\ensuremath{{0.001 } } }
\vdef{default-11:NoMcCMS-A:closetrk:hiDelta}   {\ensuremath{{-0.158 } } }
\vdef{default-11:NoMcCMS-A:closetrk:hiDeltaE}   {\ensuremath{{0.026 } } }
\vdef{default-11:NoMc3e33-A:lip:loEff}   {\ensuremath{{1.000 } } }
\vdef{default-11:NoMc3e33-A:lip:loEffE}   {\ensuremath{{0.000 } } }
\vdef{default-11:NoMc3e33-A:lip:hiEff}   {\ensuremath{{0.000 } } }
\vdef{default-11:NoMc3e33-A:lip:hiEffE}   {\ensuremath{{0.000 } } }
\vdef{default-11:NoMcCMS-A:lip:loEff}   {\ensuremath{{1.000 } } }
\vdef{default-11:NoMcCMS-A:lip:loEffE}   {\ensuremath{{0.000 } } }
\vdef{default-11:NoMcCMS-A:lip:hiEff}   {\ensuremath{{0.000 } } }
\vdef{default-11:NoMcCMS-A:lip:hiEffE}   {\ensuremath{{0.000 } } }
\vdef{default-11:NoMcCMS-A:lip:loDelta}   {\ensuremath{{+0.000 } } }
\vdef{default-11:NoMcCMS-A:lip:loDeltaE}   {\ensuremath{{0.000 } } }
\vdef{default-11:NoMcCMS-A:lip:hiDelta}   {\ensuremath{{\mathrm{NaN} } } }
\vdef{default-11:NoMcCMS-A:lip:hiDeltaE}   {\ensuremath{{\mathrm{NaN} } } }
\vdef{default-11:NoMc3e33-A:lip:inEff}   {\ensuremath{{1.000 } } }
\vdef{default-11:NoMc3e33-A:lip:inEffE}   {\ensuremath{{0.000 } } }
\vdef{default-11:NoMcCMS-A:lip:inEff}   {\ensuremath{{1.000 } } }
\vdef{default-11:NoMcCMS-A:lip:inEffE}   {\ensuremath{{0.000 } } }
\vdef{default-11:NoMcCMS-A:lip:inDelta}   {\ensuremath{{+0.000 } } }
\vdef{default-11:NoMcCMS-A:lip:inDeltaE}   {\ensuremath{{0.000 } } }
\vdef{default-11:NoMc3e33-A:lips:loEff}   {\ensuremath{{1.000 } } }
\vdef{default-11:NoMc3e33-A:lips:loEffE}   {\ensuremath{{0.000 } } }
\vdef{default-11:NoMc3e33-A:lips:hiEff}   {\ensuremath{{0.000 } } }
\vdef{default-11:NoMc3e33-A:lips:hiEffE}   {\ensuremath{{0.000 } } }
\vdef{default-11:NoMcCMS-A:lips:loEff}   {\ensuremath{{1.000 } } }
\vdef{default-11:NoMcCMS-A:lips:loEffE}   {\ensuremath{{0.000 } } }
\vdef{default-11:NoMcCMS-A:lips:hiEff}   {\ensuremath{{0.000 } } }
\vdef{default-11:NoMcCMS-A:lips:hiEffE}   {\ensuremath{{0.000 } } }
\vdef{default-11:NoMcCMS-A:lips:loDelta}   {\ensuremath{{+0.000 } } }
\vdef{default-11:NoMcCMS-A:lips:loDeltaE}   {\ensuremath{{0.000 } } }
\vdef{default-11:NoMcCMS-A:lips:hiDelta}   {\ensuremath{{\mathrm{NaN} } } }
\vdef{default-11:NoMcCMS-A:lips:hiDeltaE}   {\ensuremath{{\mathrm{NaN} } } }
\vdef{default-11:NoMc3e33-A:lips:inEff}   {\ensuremath{{1.000 } } }
\vdef{default-11:NoMc3e33-A:lips:inEffE}   {\ensuremath{{0.000 } } }
\vdef{default-11:NoMcCMS-A:lips:inEff}   {\ensuremath{{1.000 } } }
\vdef{default-11:NoMcCMS-A:lips:inEffE}   {\ensuremath{{0.000 } } }
\vdef{default-11:NoMcCMS-A:lips:inDelta}   {\ensuremath{{+0.000 } } }
\vdef{default-11:NoMcCMS-A:lips:inDeltaE}   {\ensuremath{{0.000 } } }
\vdef{default-11:NoMc3e33-A:ip:loEff}   {\ensuremath{{0.972 } } }
\vdef{default-11:NoMc3e33-A:ip:loEffE}   {\ensuremath{{0.000 } } }
\vdef{default-11:NoMc3e33-A:ip:hiEff}   {\ensuremath{{0.028 } } }
\vdef{default-11:NoMc3e33-A:ip:hiEffE}   {\ensuremath{{0.000 } } }
\vdef{default-11:NoMcCMS-A:ip:loEff}   {\ensuremath{{0.970 } } }
\vdef{default-11:NoMcCMS-A:ip:loEffE}   {\ensuremath{{0.000 } } }
\vdef{default-11:NoMcCMS-A:ip:hiEff}   {\ensuremath{{0.030 } } }
\vdef{default-11:NoMcCMS-A:ip:hiEffE}   {\ensuremath{{0.000 } } }
\vdef{default-11:NoMcCMS-A:ip:loDelta}   {\ensuremath{{+0.002 } } }
\vdef{default-11:NoMcCMS-A:ip:loDeltaE}   {\ensuremath{{0.001 } } }
\vdef{default-11:NoMcCMS-A:ip:hiDelta}   {\ensuremath{{-0.075 } } }
\vdef{default-11:NoMcCMS-A:ip:hiDeltaE}   {\ensuremath{{0.023 } } }
\vdef{default-11:NoMc3e33-A:ips:loEff}   {\ensuremath{{0.958 } } }
\vdef{default-11:NoMc3e33-A:ips:loEffE}   {\ensuremath{{0.001 } } }
\vdef{default-11:NoMc3e33-A:ips:hiEff}   {\ensuremath{{0.042 } } }
\vdef{default-11:NoMc3e33-A:ips:hiEffE}   {\ensuremath{{0.001 } } }
\vdef{default-11:NoMcCMS-A:ips:loEff}   {\ensuremath{{0.951 } } }
\vdef{default-11:NoMcCMS-A:ips:loEffE}   {\ensuremath{{0.001 } } }
\vdef{default-11:NoMcCMS-A:ips:hiEff}   {\ensuremath{{0.049 } } }
\vdef{default-11:NoMcCMS-A:ips:hiEffE}   {\ensuremath{{0.001 } } }
\vdef{default-11:NoMcCMS-A:ips:loDelta}   {\ensuremath{{+0.007 } } }
\vdef{default-11:NoMcCMS-A:ips:loDeltaE}   {\ensuremath{{0.001 } } }
\vdef{default-11:NoMcCMS-A:ips:hiDelta}   {\ensuremath{{-0.154 } } }
\vdef{default-11:NoMcCMS-A:ips:hiDeltaE}   {\ensuremath{{0.018 } } }
\vdef{default-11:NoMc3e33-A:maxdoca:loEff}   {\ensuremath{{1.000 } } }
\vdef{default-11:NoMc3e33-A:maxdoca:loEffE}   {\ensuremath{{0.000 } } }
\vdef{default-11:NoMc3e33-A:maxdoca:hiEff}   {\ensuremath{{0.011 } } }
\vdef{default-11:NoMc3e33-A:maxdoca:hiEffE}   {\ensuremath{{0.000 } } }
\vdef{default-11:NoMcCMS-A:maxdoca:loEff}   {\ensuremath{{1.000 } } }
\vdef{default-11:NoMcCMS-A:maxdoca:loEffE}   {\ensuremath{{0.000 } } }
\vdef{default-11:NoMcCMS-A:maxdoca:hiEff}   {\ensuremath{{0.009 } } }
\vdef{default-11:NoMcCMS-A:maxdoca:hiEffE}   {\ensuremath{{0.000 } } }
\vdef{default-11:NoMcCMS-A:maxdoca:loDelta}   {\ensuremath{{+0.000 } } }
\vdef{default-11:NoMcCMS-A:maxdoca:loDeltaE}   {\ensuremath{{0.000 } } }
\vdef{default-11:NoMcCMS-A:maxdoca:hiDelta}   {\ensuremath{{+0.229 } } }
\vdef{default-11:NoMcCMS-A:maxdoca:hiDeltaE}   {\ensuremath{{0.041 } } }
\vdef{default-11:NoMc3e33-A:kaonpt:loEff}   {\ensuremath{{1.007 } } }
\vdef{default-11:NoMc3e33-A:kaonpt:loEffE}   {\ensuremath{{\mathrm{NaN} } } }
\vdef{default-11:NoMc3e33-A:kaonpt:hiEff}   {\ensuremath{{1.000 } } }
\vdef{default-11:NoMc3e33-A:kaonpt:hiEffE}   {\ensuremath{{0.000 } } }
\vdef{default-11:NoMcCMS-A:kaonpt:loEff}   {\ensuremath{{1.008 } } }
\vdef{default-11:NoMcCMS-A:kaonpt:loEffE}   {\ensuremath{{\mathrm{NaN} } } }
\vdef{default-11:NoMcCMS-A:kaonpt:hiEff}   {\ensuremath{{1.000 } } }
\vdef{default-11:NoMcCMS-A:kaonpt:hiEffE}   {\ensuremath{{0.000 } } }
\vdef{default-11:NoMcCMS-A:kaonpt:loDelta}   {\ensuremath{{-0.000 } } }
\vdef{default-11:NoMcCMS-A:kaonpt:loDeltaE}   {\ensuremath{{\mathrm{NaN} } } }
\vdef{default-11:NoMcCMS-A:kaonpt:hiDelta}   {\ensuremath{{+0.000 } } }
\vdef{default-11:NoMcCMS-A:kaonpt:hiDeltaE}   {\ensuremath{{0.000 } } }
\vdef{default-11:NoMc3e33-A:psipt:loEff}   {\ensuremath{{1.003 } } }
\vdef{default-11:NoMc3e33-A:psipt:loEffE}   {\ensuremath{{\mathrm{NaN} } } }
\vdef{default-11:NoMc3e33-A:psipt:hiEff}   {\ensuremath{{1.000 } } }
\vdef{default-11:NoMc3e33-A:psipt:hiEffE}   {\ensuremath{{0.000 } } }
\vdef{default-11:NoMcCMS-A:psipt:loEff}   {\ensuremath{{1.003 } } }
\vdef{default-11:NoMcCMS-A:psipt:loEffE}   {\ensuremath{{\mathrm{NaN} } } }
\vdef{default-11:NoMcCMS-A:psipt:hiEff}   {\ensuremath{{1.000 } } }
\vdef{default-11:NoMcCMS-A:psipt:hiEffE}   {\ensuremath{{0.000 } } }
\vdef{default-11:NoMcCMS-A:psipt:loDelta}   {\ensuremath{{+0.000 } } }
\vdef{default-11:NoMcCMS-A:psipt:loDeltaE}   {\ensuremath{{\mathrm{NaN} } } }
\vdef{default-11:NoMcCMS-A:psipt:hiDelta}   {\ensuremath{{+0.000 } } }
\vdef{default-11:NoMcCMS-A:psipt:hiDeltaE}   {\ensuremath{{0.000 } } }
\vdef{default-11:NoData-AR4:osiso:loEff}   {\ensuremath{{1.005 } } }
\vdef{default-11:NoData-AR4:osiso:loEffE}   {\ensuremath{{\mathrm{NaN} } } }
\vdef{default-11:NoData-AR4:osiso:hiEff}   {\ensuremath{{1.000 } } }
\vdef{default-11:NoData-AR4:osiso:hiEffE}   {\ensuremath{{0.000 } } }
\vdef{default-11:NoMc3e33-A:osiso:loEff}   {\ensuremath{{1.003 } } }
\vdef{default-11:NoMc3e33-A:osiso:loEffE}   {\ensuremath{{\mathrm{NaN} } } }
\vdef{default-11:NoMc3e33-A:osiso:hiEff}   {\ensuremath{{1.000 } } }
\vdef{default-11:NoMc3e33-A:osiso:hiEffE}   {\ensuremath{{0.000 } } }
\vdef{default-11:NoMc3e33-A:osiso:loDelta}   {\ensuremath{{+0.002 } } }
\vdef{default-11:NoMc3e33-A:osiso:loDeltaE}   {\ensuremath{{\mathrm{NaN} } } }
\vdef{default-11:NoMc3e33-A:osiso:hiDelta}   {\ensuremath{{+0.000 } } }
\vdef{default-11:NoMc3e33-A:osiso:hiDeltaE}   {\ensuremath{{0.000 } } }
\vdef{default-11:NoData-AR4:osreliso:loEff}   {\ensuremath{{0.247 } } }
\vdef{default-11:NoData-AR4:osreliso:loEffE}   {\ensuremath{{0.002 } } }
\vdef{default-11:NoData-AR4:osreliso:hiEff}   {\ensuremath{{0.753 } } }
\vdef{default-11:NoData-AR4:osreliso:hiEffE}   {\ensuremath{{0.002 } } }
\vdef{default-11:NoMc3e33-A:osreliso:loEff}   {\ensuremath{{0.288 } } }
\vdef{default-11:NoMc3e33-A:osreliso:loEffE}   {\ensuremath{{0.002 } } }
\vdef{default-11:NoMc3e33-A:osreliso:hiEff}   {\ensuremath{{0.712 } } }
\vdef{default-11:NoMc3e33-A:osreliso:hiEffE}   {\ensuremath{{0.002 } } }
\vdef{default-11:NoMc3e33-A:osreliso:loDelta}   {\ensuremath{{-0.152 } } }
\vdef{default-11:NoMc3e33-A:osreliso:loDeltaE}   {\ensuremath{{0.011 } } }
\vdef{default-11:NoMc3e33-A:osreliso:hiDelta}   {\ensuremath{{+0.056 } } }
\vdef{default-11:NoMc3e33-A:osreliso:hiDeltaE}   {\ensuremath{{0.004 } } }
\vdef{default-11:NoData-AR4:osmuonpt:loEff}   {\ensuremath{{0.000 } } }
\vdef{default-11:NoData-AR4:osmuonpt:loEffE}   {\ensuremath{{0.001 } } }
\vdef{default-11:NoData-AR4:osmuonpt:hiEff}   {\ensuremath{{1.000 } } }
\vdef{default-11:NoData-AR4:osmuonpt:hiEffE}   {\ensuremath{{0.001 } } }
\vdef{default-11:NoMc3e33-A:osmuonpt:loEff}   {\ensuremath{{0.000 } } }
\vdef{default-11:NoMc3e33-A:osmuonpt:loEffE}   {\ensuremath{{0.001 } } }
\vdef{default-11:NoMc3e33-A:osmuonpt:hiEff}   {\ensuremath{{1.000 } } }
\vdef{default-11:NoMc3e33-A:osmuonpt:hiEffE}   {\ensuremath{{0.001 } } }
\vdef{default-11:NoMc3e33-A:osmuonpt:loDelta}   {\ensuremath{{\mathrm{NaN} } } }
\vdef{default-11:NoMc3e33-A:osmuonpt:loDeltaE}   {\ensuremath{{\mathrm{NaN} } } }
\vdef{default-11:NoMc3e33-A:osmuonpt:hiDelta}   {\ensuremath{{+0.000 } } }
\vdef{default-11:NoMc3e33-A:osmuonpt:hiDeltaE}   {\ensuremath{{0.001 } } }
\vdef{default-11:NoData-AR4:osmuondr:loEff}   {\ensuremath{{0.015 } } }
\vdef{default-11:NoData-AR4:osmuondr:loEffE}   {\ensuremath{{0.003 } } }
\vdef{default-11:NoData-AR4:osmuondr:hiEff}   {\ensuremath{{0.985 } } }
\vdef{default-11:NoData-AR4:osmuondr:hiEffE}   {\ensuremath{{0.003 } } }
\vdef{default-11:NoMc3e33-A:osmuondr:loEff}   {\ensuremath{{0.016 } } }
\vdef{default-11:NoMc3e33-A:osmuondr:loEffE}   {\ensuremath{{0.003 } } }
\vdef{default-11:NoMc3e33-A:osmuondr:hiEff}   {\ensuremath{{0.984 } } }
\vdef{default-11:NoMc3e33-A:osmuondr:hiEffE}   {\ensuremath{{0.003 } } }
\vdef{default-11:NoMc3e33-A:osmuondr:loDelta}   {\ensuremath{{-0.067 } } }
\vdef{default-11:NoMc3e33-A:osmuondr:loDeltaE}   {\ensuremath{{0.304 } } }
\vdef{default-11:NoMc3e33-A:osmuondr:hiDelta}   {\ensuremath{{+0.001 } } }
\vdef{default-11:NoMc3e33-A:osmuondr:hiDeltaE}   {\ensuremath{{0.005 } } }
\vdef{default-11:NoData-AR4:hlt:loEff}   {\ensuremath{{0.074 } } }
\vdef{default-11:NoData-AR4:hlt:loEffE}   {\ensuremath{{0.001 } } }
\vdef{default-11:NoData-AR4:hlt:hiEff}   {\ensuremath{{0.926 } } }
\vdef{default-11:NoData-AR4:hlt:hiEffE}   {\ensuremath{{0.001 } } }
\vdef{default-11:NoMc3e33-A:hlt:loEff}   {\ensuremath{{0.336 } } }
\vdef{default-11:NoMc3e33-A:hlt:loEffE}   {\ensuremath{{0.002 } } }
\vdef{default-11:NoMc3e33-A:hlt:hiEff}   {\ensuremath{{0.664 } } }
\vdef{default-11:NoMc3e33-A:hlt:hiEffE}   {\ensuremath{{0.002 } } }
\vdef{default-11:NoMc3e33-A:hlt:loDelta}   {\ensuremath{{-1.275 } } }
\vdef{default-11:NoMc3e33-A:hlt:loDeltaE}   {\ensuremath{{0.011 } } }
\vdef{default-11:NoMc3e33-A:hlt:hiDelta}   {\ensuremath{{+0.329 } } }
\vdef{default-11:NoMc3e33-A:hlt:hiDeltaE}   {\ensuremath{{0.004 } } }
\vdef{default-11:NoData-AR4:muonsid:loEff}   {\ensuremath{{0.153 } } }
\vdef{default-11:NoData-AR4:muonsid:loEffE}   {\ensuremath{{0.002 } } }
\vdef{default-11:NoData-AR4:muonsid:hiEff}   {\ensuremath{{0.847 } } }
\vdef{default-11:NoData-AR4:muonsid:hiEffE}   {\ensuremath{{0.002 } } }
\vdef{default-11:NoMc3e33-A:muonsid:loEff}   {\ensuremath{{0.157 } } }
\vdef{default-11:NoMc3e33-A:muonsid:loEffE}   {\ensuremath{{0.002 } } }
\vdef{default-11:NoMc3e33-A:muonsid:hiEff}   {\ensuremath{{0.843 } } }
\vdef{default-11:NoMc3e33-A:muonsid:hiEffE}   {\ensuremath{{0.002 } } }
\vdef{default-11:NoMc3e33-A:muonsid:loDelta}   {\ensuremath{{-0.031 } } }
\vdef{default-11:NoMc3e33-A:muonsid:loDeltaE}   {\ensuremath{{0.016 } } }
\vdef{default-11:NoMc3e33-A:muonsid:hiDelta}   {\ensuremath{{+0.006 } } }
\vdef{default-11:NoMc3e33-A:muonsid:hiDeltaE}   {\ensuremath{{0.003 } } }
\vdef{default-11:NoData-AR4:tracksqual:loEff}   {\ensuremath{{0.000 } } }
\vdef{default-11:NoData-AR4:tracksqual:loEffE}   {\ensuremath{{0.000 } } }
\vdef{default-11:NoData-AR4:tracksqual:hiEff}   {\ensuremath{{1.000 } } }
\vdef{default-11:NoData-AR4:tracksqual:hiEffE}   {\ensuremath{{0.000 } } }
\vdef{default-11:NoMc3e33-A:tracksqual:loEff}   {\ensuremath{{0.000 } } }
\vdef{default-11:NoMc3e33-A:tracksqual:loEffE}   {\ensuremath{{0.000 } } }
\vdef{default-11:NoMc3e33-A:tracksqual:hiEff}   {\ensuremath{{1.000 } } }
\vdef{default-11:NoMc3e33-A:tracksqual:hiEffE}   {\ensuremath{{0.000 } } }
\vdef{default-11:NoMc3e33-A:tracksqual:loDelta}   {\ensuremath{{+0.850 } } }
\vdef{default-11:NoMc3e33-A:tracksqual:loDeltaE}   {\ensuremath{{0.413 } } }
\vdef{default-11:NoMc3e33-A:tracksqual:hiDelta}   {\ensuremath{{-0.000 } } }
\vdef{default-11:NoMc3e33-A:tracksqual:hiDeltaE}   {\ensuremath{{0.000 } } }
\vdef{default-11:NoData-AR4:pvz:loEff}   {\ensuremath{{0.515 } } }
\vdef{default-11:NoData-AR4:pvz:loEffE}   {\ensuremath{{0.003 } } }
\vdef{default-11:NoData-AR4:pvz:hiEff}   {\ensuremath{{0.485 } } }
\vdef{default-11:NoData-AR4:pvz:hiEffE}   {\ensuremath{{0.003 } } }
\vdef{default-11:NoMc3e33-A:pvz:loEff}   {\ensuremath{{0.469 } } }
\vdef{default-11:NoMc3e33-A:pvz:loEffE}   {\ensuremath{{0.003 } } }
\vdef{default-11:NoMc3e33-A:pvz:hiEff}   {\ensuremath{{0.531 } } }
\vdef{default-11:NoMc3e33-A:pvz:hiEffE}   {\ensuremath{{0.003 } } }
\vdef{default-11:NoMc3e33-A:pvz:loDelta}   {\ensuremath{{+0.094 } } }
\vdef{default-11:NoMc3e33-A:pvz:loDeltaE}   {\ensuremath{{0.007 } } }
\vdef{default-11:NoMc3e33-A:pvz:hiDelta}   {\ensuremath{{-0.091 } } }
\vdef{default-11:NoMc3e33-A:pvz:hiDeltaE}   {\ensuremath{{0.007 } } }
\vdef{default-11:NoData-AR4:pvn:loEff}   {\ensuremath{{1.011 } } }
\vdef{default-11:NoData-AR4:pvn:loEffE}   {\ensuremath{{\mathrm{NaN} } } }
\vdef{default-11:NoData-AR4:pvn:hiEff}   {\ensuremath{{1.000 } } }
\vdef{default-11:NoData-AR4:pvn:hiEffE}   {\ensuremath{{0.000 } } }
\vdef{default-11:NoMc3e33-A:pvn:loEff}   {\ensuremath{{1.000 } } }
\vdef{default-11:NoMc3e33-A:pvn:loEffE}   {\ensuremath{{0.000 } } }
\vdef{default-11:NoMc3e33-A:pvn:hiEff}   {\ensuremath{{1.000 } } }
\vdef{default-11:NoMc3e33-A:pvn:hiEffE}   {\ensuremath{{0.000 } } }
\vdef{default-11:NoMc3e33-A:pvn:loDelta}   {\ensuremath{{+0.011 } } }
\vdef{default-11:NoMc3e33-A:pvn:loDeltaE}   {\ensuremath{{\mathrm{NaN} } } }
\vdef{default-11:NoMc3e33-A:pvn:hiDelta}   {\ensuremath{{+0.000 } } }
\vdef{default-11:NoMc3e33-A:pvn:hiDeltaE}   {\ensuremath{{0.000 } } }
\vdef{default-11:NoData-AR4:pvavew8:loEff}   {\ensuremath{{0.007 } } }
\vdef{default-11:NoData-AR4:pvavew8:loEffE}   {\ensuremath{{0.000 } } }
\vdef{default-11:NoData-AR4:pvavew8:hiEff}   {\ensuremath{{0.993 } } }
\vdef{default-11:NoData-AR4:pvavew8:hiEffE}   {\ensuremath{{0.000 } } }
\vdef{default-11:NoMc3e33-A:pvavew8:loEff}   {\ensuremath{{0.004 } } }
\vdef{default-11:NoMc3e33-A:pvavew8:loEffE}   {\ensuremath{{0.000 } } }
\vdef{default-11:NoMc3e33-A:pvavew8:hiEff}   {\ensuremath{{0.996 } } }
\vdef{default-11:NoMc3e33-A:pvavew8:hiEffE}   {\ensuremath{{0.000 } } }
\vdef{default-11:NoMc3e33-A:pvavew8:loDelta}   {\ensuremath{{+0.485 } } }
\vdef{default-11:NoMc3e33-A:pvavew8:loDeltaE}   {\ensuremath{{0.098 } } }
\vdef{default-11:NoMc3e33-A:pvavew8:hiDelta}   {\ensuremath{{-0.003 } } }
\vdef{default-11:NoMc3e33-A:pvavew8:hiDeltaE}   {\ensuremath{{0.001 } } }
\vdef{default-11:NoData-AR4:pvntrk:loEff}   {\ensuremath{{1.000 } } }
\vdef{default-11:NoData-AR4:pvntrk:loEffE}   {\ensuremath{{0.000 } } }
\vdef{default-11:NoData-AR4:pvntrk:hiEff}   {\ensuremath{{1.000 } } }
\vdef{default-11:NoData-AR4:pvntrk:hiEffE}   {\ensuremath{{0.000 } } }
\vdef{default-11:NoMc3e33-A:pvntrk:loEff}   {\ensuremath{{1.000 } } }
\vdef{default-11:NoMc3e33-A:pvntrk:loEffE}   {\ensuremath{{0.000 } } }
\vdef{default-11:NoMc3e33-A:pvntrk:hiEff}   {\ensuremath{{1.000 } } }
\vdef{default-11:NoMc3e33-A:pvntrk:hiEffE}   {\ensuremath{{0.000 } } }
\vdef{default-11:NoMc3e33-A:pvntrk:loDelta}   {\ensuremath{{+0.000 } } }
\vdef{default-11:NoMc3e33-A:pvntrk:loDeltaE}   {\ensuremath{{0.000 } } }
\vdef{default-11:NoMc3e33-A:pvntrk:hiDelta}   {\ensuremath{{+0.000 } } }
\vdef{default-11:NoMc3e33-A:pvntrk:hiDeltaE}   {\ensuremath{{0.000 } } }
\vdef{default-11:NoData-AR4:muon1pt:loEff}   {\ensuremath{{1.011 } } }
\vdef{default-11:NoData-AR4:muon1pt:loEffE}   {\ensuremath{{\mathrm{NaN} } } }
\vdef{default-11:NoData-AR4:muon1pt:hiEff}   {\ensuremath{{1.000 } } }
\vdef{default-11:NoData-AR4:muon1pt:hiEffE}   {\ensuremath{{0.000 } } }
\vdef{default-11:NoMc3e33-A:muon1pt:loEff}   {\ensuremath{{1.008 } } }
\vdef{default-11:NoMc3e33-A:muon1pt:loEffE}   {\ensuremath{{\mathrm{NaN} } } }
\vdef{default-11:NoMc3e33-A:muon1pt:hiEff}   {\ensuremath{{1.000 } } }
\vdef{default-11:NoMc3e33-A:muon1pt:hiEffE}   {\ensuremath{{0.000 } } }
\vdef{default-11:NoMc3e33-A:muon1pt:loDelta}   {\ensuremath{{+0.003 } } }
\vdef{default-11:NoMc3e33-A:muon1pt:loDeltaE}   {\ensuremath{{\mathrm{NaN} } } }
\vdef{default-11:NoMc3e33-A:muon1pt:hiDelta}   {\ensuremath{{+0.000 } } }
\vdef{default-11:NoMc3e33-A:muon1pt:hiDeltaE}   {\ensuremath{{0.000 } } }
\vdef{default-11:NoData-AR4:muon2pt:loEff}   {\ensuremath{{0.007 } } }
\vdef{default-11:NoData-AR4:muon2pt:loEffE}   {\ensuremath{{0.000 } } }
\vdef{default-11:NoData-AR4:muon2pt:hiEff}   {\ensuremath{{0.993 } } }
\vdef{default-11:NoData-AR4:muon2pt:hiEffE}   {\ensuremath{{0.000 } } }
\vdef{default-11:NoMc3e33-A:muon2pt:loEff}   {\ensuremath{{0.005 } } }
\vdef{default-11:NoMc3e33-A:muon2pt:loEffE}   {\ensuremath{{0.000 } } }
\vdef{default-11:NoMc3e33-A:muon2pt:hiEff}   {\ensuremath{{0.995 } } }
\vdef{default-11:NoMc3e33-A:muon2pt:hiEffE}   {\ensuremath{{0.000 } } }
\vdef{default-11:NoMc3e33-A:muon2pt:loDelta}   {\ensuremath{{+0.247 } } }
\vdef{default-11:NoMc3e33-A:muon2pt:loDeltaE}   {\ensuremath{{0.093 } } }
\vdef{default-11:NoMc3e33-A:muon2pt:hiDelta}   {\ensuremath{{-0.002 } } }
\vdef{default-11:NoMc3e33-A:muon2pt:hiDeltaE}   {\ensuremath{{0.001 } } }
\vdef{default-11:NoData-AR4:muonseta:loEff}   {\ensuremath{{0.747 } } }
\vdef{default-11:NoData-AR4:muonseta:loEffE}   {\ensuremath{{0.002 } } }
\vdef{default-11:NoData-AR4:muonseta:hiEff}   {\ensuremath{{0.253 } } }
\vdef{default-11:NoData-AR4:muonseta:hiEffE}   {\ensuremath{{0.002 } } }
\vdef{default-11:NoMc3e33-A:muonseta:loEff}   {\ensuremath{{0.738 } } }
\vdef{default-11:NoMc3e33-A:muonseta:loEffE}   {\ensuremath{{0.002 } } }
\vdef{default-11:NoMc3e33-A:muonseta:hiEff}   {\ensuremath{{0.262 } } }
\vdef{default-11:NoMc3e33-A:muonseta:hiEffE}   {\ensuremath{{0.002 } } }
\vdef{default-11:NoMc3e33-A:muonseta:loDelta}   {\ensuremath{{+0.013 } } }
\vdef{default-11:NoMc3e33-A:muonseta:loDeltaE}   {\ensuremath{{0.003 } } }
\vdef{default-11:NoMc3e33-A:muonseta:hiDelta}   {\ensuremath{{-0.038 } } }
\vdef{default-11:NoMc3e33-A:muonseta:hiDeltaE}   {\ensuremath{{0.009 } } }
\vdef{default-11:NoData-AR4:pt:loEff}   {\ensuremath{{0.000 } } }
\vdef{default-11:NoData-AR4:pt:loEffE}   {\ensuremath{{0.000 } } }
\vdef{default-11:NoData-AR4:pt:hiEff}   {\ensuremath{{1.000 } } }
\vdef{default-11:NoData-AR4:pt:hiEffE}   {\ensuremath{{0.000 } } }
\vdef{default-11:NoMc3e33-A:pt:loEff}   {\ensuremath{{0.000 } } }
\vdef{default-11:NoMc3e33-A:pt:loEffE}   {\ensuremath{{0.000 } } }
\vdef{default-11:NoMc3e33-A:pt:hiEff}   {\ensuremath{{1.000 } } }
\vdef{default-11:NoMc3e33-A:pt:hiEffE}   {\ensuremath{{0.000 } } }
\vdef{default-11:NoMc3e33-A:pt:loDelta}   {\ensuremath{{\mathrm{NaN} } } }
\vdef{default-11:NoMc3e33-A:pt:loDeltaE}   {\ensuremath{{\mathrm{NaN} } } }
\vdef{default-11:NoMc3e33-A:pt:hiDelta}   {\ensuremath{{+0.000 } } }
\vdef{default-11:NoMc3e33-A:pt:hiDeltaE}   {\ensuremath{{0.000 } } }
\vdef{default-11:NoData-AR4:p:loEff}   {\ensuremath{{1.011 } } }
\vdef{default-11:NoData-AR4:p:loEffE}   {\ensuremath{{\mathrm{NaN} } } }
\vdef{default-11:NoData-AR4:p:hiEff}   {\ensuremath{{1.000 } } }
\vdef{default-11:NoData-AR4:p:hiEffE}   {\ensuremath{{0.000 } } }
\vdef{default-11:NoMc3e33-A:p:loEff}   {\ensuremath{{1.011 } } }
\vdef{default-11:NoMc3e33-A:p:loEffE}   {\ensuremath{{\mathrm{NaN} } } }
\vdef{default-11:NoMc3e33-A:p:hiEff}   {\ensuremath{{1.000 } } }
\vdef{default-11:NoMc3e33-A:p:hiEffE}   {\ensuremath{{0.000 } } }
\vdef{default-11:NoMc3e33-A:p:loDelta}   {\ensuremath{{+0.000 } } }
\vdef{default-11:NoMc3e33-A:p:loDeltaE}   {\ensuremath{{\mathrm{NaN} } } }
\vdef{default-11:NoMc3e33-A:p:hiDelta}   {\ensuremath{{+0.000 } } }
\vdef{default-11:NoMc3e33-A:p:hiDeltaE}   {\ensuremath{{0.000 } } }
\vdef{default-11:NoData-AR4:eta:loEff}   {\ensuremath{{0.739 } } }
\vdef{default-11:NoData-AR4:eta:loEffE}   {\ensuremath{{0.002 } } }
\vdef{default-11:NoData-AR4:eta:hiEff}   {\ensuremath{{0.261 } } }
\vdef{default-11:NoData-AR4:eta:hiEffE}   {\ensuremath{{0.002 } } }
\vdef{default-11:NoMc3e33-A:eta:loEff}   {\ensuremath{{0.729 } } }
\vdef{default-11:NoMc3e33-A:eta:loEffE}   {\ensuremath{{0.002 } } }
\vdef{default-11:NoMc3e33-A:eta:hiEff}   {\ensuremath{{0.271 } } }
\vdef{default-11:NoMc3e33-A:eta:hiEffE}   {\ensuremath{{0.002 } } }
\vdef{default-11:NoMc3e33-A:eta:loDelta}   {\ensuremath{{+0.013 } } }
\vdef{default-11:NoMc3e33-A:eta:loDeltaE}   {\ensuremath{{0.004 } } }
\vdef{default-11:NoMc3e33-A:eta:hiDelta}   {\ensuremath{{-0.035 } } }
\vdef{default-11:NoMc3e33-A:eta:hiDeltaE}   {\ensuremath{{0.012 } } }
\vdef{default-11:NoData-AR4:bdt:loEff}   {\ensuremath{{0.907 } } }
\vdef{default-11:NoData-AR4:bdt:loEffE}   {\ensuremath{{0.001 } } }
\vdef{default-11:NoData-AR4:bdt:hiEff}   {\ensuremath{{0.093 } } }
\vdef{default-11:NoData-AR4:bdt:hiEffE}   {\ensuremath{{0.001 } } }
\vdef{default-11:NoMc3e33-A:bdt:loEff}   {\ensuremath{{0.914 } } }
\vdef{default-11:NoMc3e33-A:bdt:loEffE}   {\ensuremath{{0.001 } } }
\vdef{default-11:NoMc3e33-A:bdt:hiEff}   {\ensuremath{{0.086 } } }
\vdef{default-11:NoMc3e33-A:bdt:hiEffE}   {\ensuremath{{0.001 } } }
\vdef{default-11:NoMc3e33-A:bdt:loDelta}   {\ensuremath{{-0.008 } } }
\vdef{default-11:NoMc3e33-A:bdt:loDeltaE}   {\ensuremath{{0.002 } } }
\vdef{default-11:NoMc3e33-A:bdt:hiDelta}   {\ensuremath{{+0.082 } } }
\vdef{default-11:NoMc3e33-A:bdt:hiDeltaE}   {\ensuremath{{0.023 } } }
\vdef{default-11:NoData-AR4:fl3d:loEff}   {\ensuremath{{0.837 } } }
\vdef{default-11:NoData-AR4:fl3d:loEffE}   {\ensuremath{{0.002 } } }
\vdef{default-11:NoData-AR4:fl3d:hiEff}   {\ensuremath{{0.163 } } }
\vdef{default-11:NoData-AR4:fl3d:hiEffE}   {\ensuremath{{0.002 } } }
\vdef{default-11:NoMc3e33-A:fl3d:loEff}   {\ensuremath{{0.835 } } }
\vdef{default-11:NoMc3e33-A:fl3d:loEffE}   {\ensuremath{{0.002 } } }
\vdef{default-11:NoMc3e33-A:fl3d:hiEff}   {\ensuremath{{0.165 } } }
\vdef{default-11:NoMc3e33-A:fl3d:hiEffE}   {\ensuremath{{0.002 } } }
\vdef{default-11:NoMc3e33-A:fl3d:loDelta}   {\ensuremath{{+0.002 } } }
\vdef{default-11:NoMc3e33-A:fl3d:loDeltaE}   {\ensuremath{{0.003 } } }
\vdef{default-11:NoMc3e33-A:fl3d:hiDelta}   {\ensuremath{{-0.010 } } }
\vdef{default-11:NoMc3e33-A:fl3d:hiDeltaE}   {\ensuremath{{0.016 } } }
\vdef{default-11:NoData-AR4:fl3de:loEff}   {\ensuremath{{1.000 } } }
\vdef{default-11:NoData-AR4:fl3de:loEffE}   {\ensuremath{{0.000 } } }
\vdef{default-11:NoData-AR4:fl3de:hiEff}   {\ensuremath{{0.000 } } }
\vdef{default-11:NoData-AR4:fl3de:hiEffE}   {\ensuremath{{0.000 } } }
\vdef{default-11:NoMc3e33-A:fl3de:loEff}   {\ensuremath{{1.000 } } }
\vdef{default-11:NoMc3e33-A:fl3de:loEffE}   {\ensuremath{{0.000 } } }
\vdef{default-11:NoMc3e33-A:fl3de:hiEff}   {\ensuremath{{0.000 } } }
\vdef{default-11:NoMc3e33-A:fl3de:hiEffE}   {\ensuremath{{0.000 } } }
\vdef{default-11:NoMc3e33-A:fl3de:loDelta}   {\ensuremath{{+0.000 } } }
\vdef{default-11:NoMc3e33-A:fl3de:loDeltaE}   {\ensuremath{{0.000 } } }
\vdef{default-11:NoMc3e33-A:fl3de:hiDelta}   {\ensuremath{{+1.375 } } }
\vdef{default-11:NoMc3e33-A:fl3de:hiDeltaE}   {\ensuremath{{0.622 } } }
\vdef{default-11:NoData-AR4:fls3d:loEff}   {\ensuremath{{0.071 } } }
\vdef{default-11:NoData-AR4:fls3d:loEffE}   {\ensuremath{{0.001 } } }
\vdef{default-11:NoData-AR4:fls3d:hiEff}   {\ensuremath{{0.929 } } }
\vdef{default-11:NoData-AR4:fls3d:hiEffE}   {\ensuremath{{0.001 } } }
\vdef{default-11:NoMc3e33-A:fls3d:loEff}   {\ensuremath{{0.074 } } }
\vdef{default-11:NoMc3e33-A:fls3d:loEffE}   {\ensuremath{{0.001 } } }
\vdef{default-11:NoMc3e33-A:fls3d:hiEff}   {\ensuremath{{0.926 } } }
\vdef{default-11:NoMc3e33-A:fls3d:hiEffE}   {\ensuremath{{0.001 } } }
\vdef{default-11:NoMc3e33-A:fls3d:loDelta}   {\ensuremath{{-0.049 } } }
\vdef{default-11:NoMc3e33-A:fls3d:loDeltaE}   {\ensuremath{{0.025 } } }
\vdef{default-11:NoMc3e33-A:fls3d:hiDelta}   {\ensuremath{{+0.004 } } }
\vdef{default-11:NoMc3e33-A:fls3d:hiDeltaE}   {\ensuremath{{0.002 } } }
\vdef{default-11:NoData-AR4:flsxy:loEff}   {\ensuremath{{1.013 } } }
\vdef{default-11:NoData-AR4:flsxy:loEffE}   {\ensuremath{{\mathrm{NaN} } } }
\vdef{default-11:NoData-AR4:flsxy:hiEff}   {\ensuremath{{1.000 } } }
\vdef{default-11:NoData-AR4:flsxy:hiEffE}   {\ensuremath{{0.000 } } }
\vdef{default-11:NoMc3e33-A:flsxy:loEff}   {\ensuremath{{1.012 } } }
\vdef{default-11:NoMc3e33-A:flsxy:loEffE}   {\ensuremath{{\mathrm{NaN} } } }
\vdef{default-11:NoMc3e33-A:flsxy:hiEff}   {\ensuremath{{1.000 } } }
\vdef{default-11:NoMc3e33-A:flsxy:hiEffE}   {\ensuremath{{0.000 } } }
\vdef{default-11:NoMc3e33-A:flsxy:loDelta}   {\ensuremath{{+0.001 } } }
\vdef{default-11:NoMc3e33-A:flsxy:loDeltaE}   {\ensuremath{{\mathrm{NaN} } } }
\vdef{default-11:NoMc3e33-A:flsxy:hiDelta}   {\ensuremath{{+0.000 } } }
\vdef{default-11:NoMc3e33-A:flsxy:hiDeltaE}   {\ensuremath{{0.000 } } }
\vdef{default-11:NoData-AR4:chi2dof:loEff}   {\ensuremath{{0.934 } } }
\vdef{default-11:NoData-AR4:chi2dof:loEffE}   {\ensuremath{{0.001 } } }
\vdef{default-11:NoData-AR4:chi2dof:hiEff}   {\ensuremath{{0.066 } } }
\vdef{default-11:NoData-AR4:chi2dof:hiEffE}   {\ensuremath{{0.001 } } }
\vdef{default-11:NoMc3e33-A:chi2dof:loEff}   {\ensuremath{{0.944 } } }
\vdef{default-11:NoMc3e33-A:chi2dof:loEffE}   {\ensuremath{{0.001 } } }
\vdef{default-11:NoMc3e33-A:chi2dof:hiEff}   {\ensuremath{{0.056 } } }
\vdef{default-11:NoMc3e33-A:chi2dof:hiEffE}   {\ensuremath{{0.001 } } }
\vdef{default-11:NoMc3e33-A:chi2dof:loDelta}   {\ensuremath{{-0.010 } } }
\vdef{default-11:NoMc3e33-A:chi2dof:loDeltaE}   {\ensuremath{{0.002 } } }
\vdef{default-11:NoMc3e33-A:chi2dof:hiDelta}   {\ensuremath{{+0.157 } } }
\vdef{default-11:NoMc3e33-A:chi2dof:hiDeltaE}   {\ensuremath{{0.028 } } }
\vdef{default-11:NoData-AR4:pchi2dof:loEff}   {\ensuremath{{0.629 } } }
\vdef{default-11:NoData-AR4:pchi2dof:loEffE}   {\ensuremath{{0.002 } } }
\vdef{default-11:NoData-AR4:pchi2dof:hiEff}   {\ensuremath{{0.371 } } }
\vdef{default-11:NoData-AR4:pchi2dof:hiEffE}   {\ensuremath{{0.002 } } }
\vdef{default-11:NoMc3e33-A:pchi2dof:loEff}   {\ensuremath{{0.607 } } }
\vdef{default-11:NoMc3e33-A:pchi2dof:loEffE}   {\ensuremath{{0.002 } } }
\vdef{default-11:NoMc3e33-A:pchi2dof:hiEff}   {\ensuremath{{0.393 } } }
\vdef{default-11:NoMc3e33-A:pchi2dof:hiEffE}   {\ensuremath{{0.002 } } }
\vdef{default-11:NoMc3e33-A:pchi2dof:loDelta}   {\ensuremath{{+0.035 } } }
\vdef{default-11:NoMc3e33-A:pchi2dof:loDeltaE}   {\ensuremath{{0.006 } } }
\vdef{default-11:NoMc3e33-A:pchi2dof:hiDelta}   {\ensuremath{{-0.056 } } }
\vdef{default-11:NoMc3e33-A:pchi2dof:hiDeltaE}   {\ensuremath{{0.009 } } }
\vdef{default-11:NoData-AR4:alpha:loEff}   {\ensuremath{{0.995 } } }
\vdef{default-11:NoData-AR4:alpha:loEffE}   {\ensuremath{{0.000 } } }
\vdef{default-11:NoData-AR4:alpha:hiEff}   {\ensuremath{{0.005 } } }
\vdef{default-11:NoData-AR4:alpha:hiEffE}   {\ensuremath{{0.000 } } }
\vdef{default-11:NoMc3e33-A:alpha:loEff}   {\ensuremath{{0.994 } } }
\vdef{default-11:NoMc3e33-A:alpha:loEffE}   {\ensuremath{{0.000 } } }
\vdef{default-11:NoMc3e33-A:alpha:hiEff}   {\ensuremath{{0.006 } } }
\vdef{default-11:NoMc3e33-A:alpha:hiEffE}   {\ensuremath{{0.000 } } }
\vdef{default-11:NoMc3e33-A:alpha:loDelta}   {\ensuremath{{+0.001 } } }
\vdef{default-11:NoMc3e33-A:alpha:loDeltaE}   {\ensuremath{{0.001 } } }
\vdef{default-11:NoMc3e33-A:alpha:hiDelta}   {\ensuremath{{-0.144 } } }
\vdef{default-11:NoMc3e33-A:alpha:hiDeltaE}   {\ensuremath{{0.098 } } }
\vdef{default-11:NoData-AR4:iso:loEff}   {\ensuremath{{0.127 } } }
\vdef{default-11:NoData-AR4:iso:loEffE}   {\ensuremath{{0.002 } } }
\vdef{default-11:NoData-AR4:iso:hiEff}   {\ensuremath{{0.873 } } }
\vdef{default-11:NoData-AR4:iso:hiEffE}   {\ensuremath{{0.002 } } }
\vdef{default-11:NoMc3e33-A:iso:loEff}   {\ensuremath{{0.107 } } }
\vdef{default-11:NoMc3e33-A:iso:loEffE}   {\ensuremath{{0.002 } } }
\vdef{default-11:NoMc3e33-A:iso:hiEff}   {\ensuremath{{0.893 } } }
\vdef{default-11:NoMc3e33-A:iso:hiEffE}   {\ensuremath{{0.002 } } }
\vdef{default-11:NoMc3e33-A:iso:loDelta}   {\ensuremath{{+0.169 } } }
\vdef{default-11:NoMc3e33-A:iso:loDeltaE}   {\ensuremath{{0.019 } } }
\vdef{default-11:NoMc3e33-A:iso:hiDelta}   {\ensuremath{{-0.022 } } }
\vdef{default-11:NoMc3e33-A:iso:hiDeltaE}   {\ensuremath{{0.003 } } }
\vdef{default-11:NoData-AR4:docatrk:loEff}   {\ensuremath{{0.073 } } }
\vdef{default-11:NoData-AR4:docatrk:loEffE}   {\ensuremath{{0.001 } } }
\vdef{default-11:NoData-AR4:docatrk:hiEff}   {\ensuremath{{0.927 } } }
\vdef{default-11:NoData-AR4:docatrk:hiEffE}   {\ensuremath{{0.001 } } }
\vdef{default-11:NoMc3e33-A:docatrk:loEff}   {\ensuremath{{0.083 } } }
\vdef{default-11:NoMc3e33-A:docatrk:loEffE}   {\ensuremath{{0.001 } } }
\vdef{default-11:NoMc3e33-A:docatrk:hiEff}   {\ensuremath{{0.917 } } }
\vdef{default-11:NoMc3e33-A:docatrk:hiEffE}   {\ensuremath{{0.001 } } }
\vdef{default-11:NoMc3e33-A:docatrk:loDelta}   {\ensuremath{{-0.125 } } }
\vdef{default-11:NoMc3e33-A:docatrk:loDeltaE}   {\ensuremath{{0.025 } } }
\vdef{default-11:NoMc3e33-A:docatrk:hiDelta}   {\ensuremath{{+0.011 } } }
\vdef{default-11:NoMc3e33-A:docatrk:hiDeltaE}   {\ensuremath{{0.002 } } }
\vdef{default-11:NoData-AR4:isotrk:loEff}   {\ensuremath{{1.000 } } }
\vdef{default-11:NoData-AR4:isotrk:loEffE}   {\ensuremath{{0.000 } } }
\vdef{default-11:NoData-AR4:isotrk:hiEff}   {\ensuremath{{1.000 } } }
\vdef{default-11:NoData-AR4:isotrk:hiEffE}   {\ensuremath{{0.000 } } }
\vdef{default-11:NoMc3e33-A:isotrk:loEff}   {\ensuremath{{1.000 } } }
\vdef{default-11:NoMc3e33-A:isotrk:loEffE}   {\ensuremath{{0.000 } } }
\vdef{default-11:NoMc3e33-A:isotrk:hiEff}   {\ensuremath{{1.000 } } }
\vdef{default-11:NoMc3e33-A:isotrk:hiEffE}   {\ensuremath{{0.000 } } }
\vdef{default-11:NoMc3e33-A:isotrk:loDelta}   {\ensuremath{{+0.000 } } }
\vdef{default-11:NoMc3e33-A:isotrk:loDeltaE}   {\ensuremath{{0.000 } } }
\vdef{default-11:NoMc3e33-A:isotrk:hiDelta}   {\ensuremath{{+0.000 } } }
\vdef{default-11:NoMc3e33-A:isotrk:hiDeltaE}   {\ensuremath{{0.000 } } }
\vdef{default-11:NoData-AR4:closetrk:loEff}   {\ensuremath{{0.974 } } }
\vdef{default-11:NoData-AR4:closetrk:loEffE}   {\ensuremath{{0.001 } } }
\vdef{default-11:NoData-AR4:closetrk:hiEff}   {\ensuremath{{0.026 } } }
\vdef{default-11:NoData-AR4:closetrk:hiEffE}   {\ensuremath{{0.001 } } }
\vdef{default-11:NoMc3e33-A:closetrk:loEff}   {\ensuremath{{0.978 } } }
\vdef{default-11:NoMc3e33-A:closetrk:loEffE}   {\ensuremath{{0.001 } } }
\vdef{default-11:NoMc3e33-A:closetrk:hiEff}   {\ensuremath{{0.022 } } }
\vdef{default-11:NoMc3e33-A:closetrk:hiEffE}   {\ensuremath{{0.001 } } }
\vdef{default-11:NoMc3e33-A:closetrk:loDelta}   {\ensuremath{{-0.004 } } }
\vdef{default-11:NoMc3e33-A:closetrk:loDeltaE}   {\ensuremath{{0.001 } } }
\vdef{default-11:NoMc3e33-A:closetrk:hiDelta}   {\ensuremath{{+0.143 } } }
\vdef{default-11:NoMc3e33-A:closetrk:hiDeltaE}   {\ensuremath{{0.046 } } }
\vdef{default-11:NoData-AR4:lip:loEff}   {\ensuremath{{1.000 } } }
\vdef{default-11:NoData-AR4:lip:loEffE}   {\ensuremath{{0.000 } } }
\vdef{default-11:NoData-AR4:lip:hiEff}   {\ensuremath{{0.000 } } }
\vdef{default-11:NoData-AR4:lip:hiEffE}   {\ensuremath{{0.000 } } }
\vdef{default-11:NoMc3e33-A:lip:loEff}   {\ensuremath{{1.000 } } }
\vdef{default-11:NoMc3e33-A:lip:loEffE}   {\ensuremath{{0.000 } } }
\vdef{default-11:NoMc3e33-A:lip:hiEff}   {\ensuremath{{0.000 } } }
\vdef{default-11:NoMc3e33-A:lip:hiEffE}   {\ensuremath{{0.000 } } }
\vdef{default-11:NoMc3e33-A:lip:loDelta}   {\ensuremath{{+0.000 } } }
\vdef{default-11:NoMc3e33-A:lip:loDeltaE}   {\ensuremath{{0.000 } } }
\vdef{default-11:NoMc3e33-A:lip:hiDelta}   {\ensuremath{{\mathrm{NaN} } } }
\vdef{default-11:NoMc3e33-A:lip:hiDeltaE}   {\ensuremath{{\mathrm{NaN} } } }
\vdef{default-11:NoData-AR4:lip:inEff}   {\ensuremath{{1.000 } } }
\vdef{default-11:NoData-AR4:lip:inEffE}   {\ensuremath{{0.000 } } }
\vdef{default-11:NoMc3e33-A:lip:inEff}   {\ensuremath{{1.000 } } }
\vdef{default-11:NoMc3e33-A:lip:inEffE}   {\ensuremath{{0.000 } } }
\vdef{default-11:NoMc3e33-A:lip:inDelta}   {\ensuremath{{+0.000 } } }
\vdef{default-11:NoMc3e33-A:lip:inDeltaE}   {\ensuremath{{0.000 } } }
\vdef{default-11:NoData-AR4:lips:loEff}   {\ensuremath{{1.000 } } }
\vdef{default-11:NoData-AR4:lips:loEffE}   {\ensuremath{{0.000 } } }
\vdef{default-11:NoData-AR4:lips:hiEff}   {\ensuremath{{0.000 } } }
\vdef{default-11:NoData-AR4:lips:hiEffE}   {\ensuremath{{0.000 } } }
\vdef{default-11:NoMc3e33-A:lips:loEff}   {\ensuremath{{1.000 } } }
\vdef{default-11:NoMc3e33-A:lips:loEffE}   {\ensuremath{{0.000 } } }
\vdef{default-11:NoMc3e33-A:lips:hiEff}   {\ensuremath{{0.000 } } }
\vdef{default-11:NoMc3e33-A:lips:hiEffE}   {\ensuremath{{0.000 } } }
\vdef{default-11:NoMc3e33-A:lips:loDelta}   {\ensuremath{{+0.000 } } }
\vdef{default-11:NoMc3e33-A:lips:loDeltaE}   {\ensuremath{{0.000 } } }
\vdef{default-11:NoMc3e33-A:lips:hiDelta}   {\ensuremath{{\mathrm{NaN} } } }
\vdef{default-11:NoMc3e33-A:lips:hiDeltaE}   {\ensuremath{{\mathrm{NaN} } } }
\vdef{default-11:NoData-AR4:lips:inEff}   {\ensuremath{{1.000 } } }
\vdef{default-11:NoData-AR4:lips:inEffE}   {\ensuremath{{0.000 } } }
\vdef{default-11:NoMc3e33-A:lips:inEff}   {\ensuremath{{1.000 } } }
\vdef{default-11:NoMc3e33-A:lips:inEffE}   {\ensuremath{{0.000 } } }
\vdef{default-11:NoMc3e33-A:lips:inDelta}   {\ensuremath{{+0.000 } } }
\vdef{default-11:NoMc3e33-A:lips:inDeltaE}   {\ensuremath{{0.000 } } }
\vdef{default-11:NoData-AR4:ip:loEff}   {\ensuremath{{0.971 } } }
\vdef{default-11:NoData-AR4:ip:loEffE}   {\ensuremath{{0.001 } } }
\vdef{default-11:NoData-AR4:ip:hiEff}   {\ensuremath{{0.029 } } }
\vdef{default-11:NoData-AR4:ip:hiEffE}   {\ensuremath{{0.001 } } }
\vdef{default-11:NoMc3e33-A:ip:loEff}   {\ensuremath{{0.972 } } }
\vdef{default-11:NoMc3e33-A:ip:loEffE}   {\ensuremath{{0.001 } } }
\vdef{default-11:NoMc3e33-A:ip:hiEff}   {\ensuremath{{0.028 } } }
\vdef{default-11:NoMc3e33-A:ip:hiEffE}   {\ensuremath{{0.001 } } }
\vdef{default-11:NoMc3e33-A:ip:loDelta}   {\ensuremath{{-0.000 } } }
\vdef{default-11:NoMc3e33-A:ip:loDeltaE}   {\ensuremath{{0.001 } } }
\vdef{default-11:NoMc3e33-A:ip:hiDelta}   {\ensuremath{{+0.010 } } }
\vdef{default-11:NoMc3e33-A:ip:hiDeltaE}   {\ensuremath{{0.043 } } }
\vdef{default-11:NoData-AR4:ips:loEff}   {\ensuremath{{0.942 } } }
\vdef{default-11:NoData-AR4:ips:loEffE}   {\ensuremath{{0.001 } } }
\vdef{default-11:NoData-AR4:ips:hiEff}   {\ensuremath{{0.058 } } }
\vdef{default-11:NoData-AR4:ips:hiEffE}   {\ensuremath{{0.001 } } }
\vdef{default-11:NoMc3e33-A:ips:loEff}   {\ensuremath{{0.958 } } }
\vdef{default-11:NoMc3e33-A:ips:loEffE}   {\ensuremath{{0.001 } } }
\vdef{default-11:NoMc3e33-A:ips:hiEff}   {\ensuremath{{0.042 } } }
\vdef{default-11:NoMc3e33-A:ips:hiEffE}   {\ensuremath{{0.001 } } }
\vdef{default-11:NoMc3e33-A:ips:loDelta}   {\ensuremath{{-0.017 } } }
\vdef{default-11:NoMc3e33-A:ips:loDeltaE}   {\ensuremath{{0.002 } } }
\vdef{default-11:NoMc3e33-A:ips:hiDelta}   {\ensuremath{{+0.322 } } }
\vdef{default-11:NoMc3e33-A:ips:hiDeltaE}   {\ensuremath{{0.031 } } }
\vdef{default-11:NoData-AR4:maxdoca:loEff}   {\ensuremath{{1.000 } } }
\vdef{default-11:NoData-AR4:maxdoca:loEffE}   {\ensuremath{{0.000 } } }
\vdef{default-11:NoData-AR4:maxdoca:hiEff}   {\ensuremath{{0.012 } } }
\vdef{default-11:NoData-AR4:maxdoca:hiEffE}   {\ensuremath{{0.001 } } }
\vdef{default-11:NoMc3e33-A:maxdoca:loEff}   {\ensuremath{{1.000 } } }
\vdef{default-11:NoMc3e33-A:maxdoca:loEffE}   {\ensuremath{{0.000 } } }
\vdef{default-11:NoMc3e33-A:maxdoca:hiEff}   {\ensuremath{{0.011 } } }
\vdef{default-11:NoMc3e33-A:maxdoca:hiEffE}   {\ensuremath{{0.001 } } }
\vdef{default-11:NoMc3e33-A:maxdoca:loDelta}   {\ensuremath{{+0.000 } } }
\vdef{default-11:NoMc3e33-A:maxdoca:loDeltaE}   {\ensuremath{{0.000 } } }
\vdef{default-11:NoMc3e33-A:maxdoca:hiDelta}   {\ensuremath{{+0.091 } } }
\vdef{default-11:NoMc3e33-A:maxdoca:hiDeltaE}   {\ensuremath{{0.069 } } }
\vdef{default-11:NoData-AR4:kaonpt:loEff}   {\ensuremath{{1.009 } } }
\vdef{default-11:NoData-AR4:kaonpt:loEffE}   {\ensuremath{{\mathrm{NaN} } } }
\vdef{default-11:NoData-AR4:kaonpt:hiEff}   {\ensuremath{{1.000 } } }
\vdef{default-11:NoData-AR4:kaonpt:hiEffE}   {\ensuremath{{0.000 } } }
\vdef{default-11:NoMc3e33-A:kaonpt:loEff}   {\ensuremath{{1.007 } } }
\vdef{default-11:NoMc3e33-A:kaonpt:loEffE}   {\ensuremath{{\mathrm{NaN} } } }
\vdef{default-11:NoMc3e33-A:kaonpt:hiEff}   {\ensuremath{{1.000 } } }
\vdef{default-11:NoMc3e33-A:kaonpt:hiEffE}   {\ensuremath{{0.000 } } }
\vdef{default-11:NoMc3e33-A:kaonpt:loDelta}   {\ensuremath{{+0.002 } } }
\vdef{default-11:NoMc3e33-A:kaonpt:loDeltaE}   {\ensuremath{{\mathrm{NaN} } } }
\vdef{default-11:NoMc3e33-A:kaonpt:hiDelta}   {\ensuremath{{+0.000 } } }
\vdef{default-11:NoMc3e33-A:kaonpt:hiDeltaE}   {\ensuremath{{0.000 } } }
\vdef{default-11:NoData-AR4:psipt:loEff}   {\ensuremath{{1.004 } } }
\vdef{default-11:NoData-AR4:psipt:loEffE}   {\ensuremath{{\mathrm{NaN} } } }
\vdef{default-11:NoData-AR4:psipt:hiEff}   {\ensuremath{{1.000 } } }
\vdef{default-11:NoData-AR4:psipt:hiEffE}   {\ensuremath{{0.000 } } }
\vdef{default-11:NoMc3e33-A:psipt:loEff}   {\ensuremath{{1.003 } } }
\vdef{default-11:NoMc3e33-A:psipt:loEffE}   {\ensuremath{{\mathrm{NaN} } } }
\vdef{default-11:NoMc3e33-A:psipt:hiEff}   {\ensuremath{{1.000 } } }
\vdef{default-11:NoMc3e33-A:psipt:hiEffE}   {\ensuremath{{0.000 } } }
\vdef{default-11:NoMc3e33-A:psipt:loDelta}   {\ensuremath{{+0.001 } } }
\vdef{default-11:NoMc3e33-A:psipt:loDeltaE}   {\ensuremath{{\mathrm{NaN} } } }
\vdef{default-11:NoMc3e33-A:psipt:hiDelta}   {\ensuremath{{+0.000 } } }
\vdef{default-11:NoMc3e33-A:psipt:hiDeltaE}   {\ensuremath{{0.000 } } }
\vdef{default-11:NoData-AR4:osiso:loEff}   {\ensuremath{{1.005 } } }
\vdef{default-11:NoData-AR4:osiso:loEffE}   {\ensuremath{{\mathrm{NaN} } } }
\vdef{default-11:NoData-AR4:osiso:hiEff}   {\ensuremath{{1.000 } } }
\vdef{default-11:NoData-AR4:osiso:hiEffE}   {\ensuremath{{0.000 } } }
\vdef{default-11:NoMcCMS-A:osiso:loEff}   {\ensuremath{{1.003 } } }
\vdef{default-11:NoMcCMS-A:osiso:loEffE}   {\ensuremath{{\mathrm{NaN} } } }
\vdef{default-11:NoMcCMS-A:osiso:hiEff}   {\ensuremath{{1.000 } } }
\vdef{default-11:NoMcCMS-A:osiso:hiEffE}   {\ensuremath{{0.000 } } }
\vdef{default-11:NoMcCMS-A:osiso:loDelta}   {\ensuremath{{+0.002 } } }
\vdef{default-11:NoMcCMS-A:osiso:loDeltaE}   {\ensuremath{{\mathrm{NaN} } } }
\vdef{default-11:NoMcCMS-A:osiso:hiDelta}   {\ensuremath{{+0.000 } } }
\vdef{default-11:NoMcCMS-A:osiso:hiDeltaE}   {\ensuremath{{0.000 } } }
\vdef{default-11:NoData-AR4:osreliso:loEff}   {\ensuremath{{0.247 } } }
\vdef{default-11:NoData-AR4:osreliso:loEffE}   {\ensuremath{{0.002 } } }
\vdef{default-11:NoData-AR4:osreliso:hiEff}   {\ensuremath{{0.753 } } }
\vdef{default-11:NoData-AR4:osreliso:hiEffE}   {\ensuremath{{0.002 } } }
\vdef{default-11:NoMcCMS-A:osreliso:loEff}   {\ensuremath{{0.282 } } }
\vdef{default-11:NoMcCMS-A:osreliso:loEffE}   {\ensuremath{{0.002 } } }
\vdef{default-11:NoMcCMS-A:osreliso:hiEff}   {\ensuremath{{0.718 } } }
\vdef{default-11:NoMcCMS-A:osreliso:hiEffE}   {\ensuremath{{0.002 } } }
\vdef{default-11:NoMcCMS-A:osreliso:loDelta}   {\ensuremath{{-0.130 } } }
\vdef{default-11:NoMcCMS-A:osreliso:loDeltaE}   {\ensuremath{{0.012 } } }
\vdef{default-11:NoMcCMS-A:osreliso:hiDelta}   {\ensuremath{{+0.047 } } }
\vdef{default-11:NoMcCMS-A:osreliso:hiDeltaE}   {\ensuremath{{0.004 } } }
\vdef{default-11:NoData-AR4:osmuonpt:loEff}   {\ensuremath{{0.000 } } }
\vdef{default-11:NoData-AR4:osmuonpt:loEffE}   {\ensuremath{{0.001 } } }
\vdef{default-11:NoData-AR4:osmuonpt:hiEff}   {\ensuremath{{1.000 } } }
\vdef{default-11:NoData-AR4:osmuonpt:hiEffE}   {\ensuremath{{0.001 } } }
\vdef{default-11:NoMcCMS-A:osmuonpt:loEff}   {\ensuremath{{0.000 } } }
\vdef{default-11:NoMcCMS-A:osmuonpt:loEffE}   {\ensuremath{{0.001 } } }
\vdef{default-11:NoMcCMS-A:osmuonpt:hiEff}   {\ensuremath{{1.000 } } }
\vdef{default-11:NoMcCMS-A:osmuonpt:hiEffE}   {\ensuremath{{0.001 } } }
\vdef{default-11:NoMcCMS-A:osmuonpt:loDelta}   {\ensuremath{{\mathrm{NaN} } } }
\vdef{default-11:NoMcCMS-A:osmuonpt:loDeltaE}   {\ensuremath{{\mathrm{NaN} } } }
\vdef{default-11:NoMcCMS-A:osmuonpt:hiDelta}   {\ensuremath{{+0.000 } } }
\vdef{default-11:NoMcCMS-A:osmuonpt:hiDeltaE}   {\ensuremath{{0.001 } } }
\vdef{default-11:NoData-AR4:osmuondr:loEff}   {\ensuremath{{0.015 } } }
\vdef{default-11:NoData-AR4:osmuondr:loEffE}   {\ensuremath{{0.003 } } }
\vdef{default-11:NoData-AR4:osmuondr:hiEff}   {\ensuremath{{0.985 } } }
\vdef{default-11:NoData-AR4:osmuondr:hiEffE}   {\ensuremath{{0.003 } } }
\vdef{default-11:NoMcCMS-A:osmuondr:loEff}   {\ensuremath{{0.012 } } }
\vdef{default-11:NoMcCMS-A:osmuondr:loEffE}   {\ensuremath{{0.003 } } }
\vdef{default-11:NoMcCMS-A:osmuondr:hiEff}   {\ensuremath{{0.988 } } }
\vdef{default-11:NoMcCMS-A:osmuondr:hiEffE}   {\ensuremath{{0.003 } } }
\vdef{default-11:NoMcCMS-A:osmuondr:loDelta}   {\ensuremath{{+0.255 } } }
\vdef{default-11:NoMcCMS-A:osmuondr:loDeltaE}   {\ensuremath{{0.328 } } }
\vdef{default-11:NoMcCMS-A:osmuondr:hiDelta}   {\ensuremath{{-0.003 } } }
\vdef{default-11:NoMcCMS-A:osmuondr:hiDeltaE}   {\ensuremath{{0.005 } } }
\vdef{default-11:NoData-AR4:hlt:loEff}   {\ensuremath{{0.074 } } }
\vdef{default-11:NoData-AR4:hlt:loEffE}   {\ensuremath{{0.001 } } }
\vdef{default-11:NoData-AR4:hlt:hiEff}   {\ensuremath{{0.926 } } }
\vdef{default-11:NoData-AR4:hlt:hiEffE}   {\ensuremath{{0.001 } } }
\vdef{default-11:NoMcCMS-A:hlt:loEff}   {\ensuremath{{0.267 } } }
\vdef{default-11:NoMcCMS-A:hlt:loEffE}   {\ensuremath{{0.002 } } }
\vdef{default-11:NoMcCMS-A:hlt:hiEff}   {\ensuremath{{0.733 } } }
\vdef{default-11:NoMcCMS-A:hlt:hiEffE}   {\ensuremath{{0.002 } } }
\vdef{default-11:NoMcCMS-A:hlt:loDelta}   {\ensuremath{{-1.128 } } }
\vdef{default-11:NoMcCMS-A:hlt:loDeltaE}   {\ensuremath{{0.013 } } }
\vdef{default-11:NoMcCMS-A:hlt:hiDelta}   {\ensuremath{{+0.232 } } }
\vdef{default-11:NoMcCMS-A:hlt:hiDeltaE}   {\ensuremath{{0.003 } } }
\vdef{default-11:NoData-AR4:muonsid:loEff}   {\ensuremath{{0.153 } } }
\vdef{default-11:NoData-AR4:muonsid:loEffE}   {\ensuremath{{0.002 } } }
\vdef{default-11:NoData-AR4:muonsid:hiEff}   {\ensuremath{{0.847 } } }
\vdef{default-11:NoData-AR4:muonsid:hiEffE}   {\ensuremath{{0.002 } } }
\vdef{default-11:NoMcCMS-A:muonsid:loEff}   {\ensuremath{{0.220 } } }
\vdef{default-11:NoMcCMS-A:muonsid:loEffE}   {\ensuremath{{0.002 } } }
\vdef{default-11:NoMcCMS-A:muonsid:hiEff}   {\ensuremath{{0.780 } } }
\vdef{default-11:NoMcCMS-A:muonsid:hiEffE}   {\ensuremath{{0.002 } } }
\vdef{default-11:NoMcCMS-A:muonsid:loDelta}   {\ensuremath{{-0.364 } } }
\vdef{default-11:NoMcCMS-A:muonsid:loDeltaE}   {\ensuremath{{0.014 } } }
\vdef{default-11:NoMcCMS-A:muonsid:hiDelta}   {\ensuremath{{+0.083 } } }
\vdef{default-11:NoMcCMS-A:muonsid:hiDeltaE}   {\ensuremath{{0.003 } } }
\vdef{default-11:NoData-AR4:tracksqual:loEff}   {\ensuremath{{0.000 } } }
\vdef{default-11:NoData-AR4:tracksqual:loEffE}   {\ensuremath{{0.000 } } }
\vdef{default-11:NoData-AR4:tracksqual:hiEff}   {\ensuremath{{1.000 } } }
\vdef{default-11:NoData-AR4:tracksqual:hiEffE}   {\ensuremath{{0.000 } } }
\vdef{default-11:NoMcCMS-A:tracksqual:loEff}   {\ensuremath{{0.000 } } }
\vdef{default-11:NoMcCMS-A:tracksqual:loEffE}   {\ensuremath{{0.000 } } }
\vdef{default-11:NoMcCMS-A:tracksqual:hiEff}   {\ensuremath{{1.000 } } }
\vdef{default-11:NoMcCMS-A:tracksqual:hiEffE}   {\ensuremath{{0.000 } } }
\vdef{default-11:NoMcCMS-A:tracksqual:loDelta}   {\ensuremath{{+0.530 } } }
\vdef{default-11:NoMcCMS-A:tracksqual:loDeltaE}   {\ensuremath{{0.398 } } }
\vdef{default-11:NoMcCMS-A:tracksqual:hiDelta}   {\ensuremath{{-0.000 } } }
\vdef{default-11:NoMcCMS-A:tracksqual:hiDeltaE}   {\ensuremath{{0.000 } } }
\vdef{default-11:NoData-AR4:pvz:loEff}   {\ensuremath{{0.515 } } }
\vdef{default-11:NoData-AR4:pvz:loEffE}   {\ensuremath{{0.003 } } }
\vdef{default-11:NoData-AR4:pvz:hiEff}   {\ensuremath{{0.485 } } }
\vdef{default-11:NoData-AR4:pvz:hiEffE}   {\ensuremath{{0.003 } } }
\vdef{default-11:NoMcCMS-A:pvz:loEff}   {\ensuremath{{0.470 } } }
\vdef{default-11:NoMcCMS-A:pvz:loEffE}   {\ensuremath{{0.003 } } }
\vdef{default-11:NoMcCMS-A:pvz:hiEff}   {\ensuremath{{0.530 } } }
\vdef{default-11:NoMcCMS-A:pvz:hiEffE}   {\ensuremath{{0.003 } } }
\vdef{default-11:NoMcCMS-A:pvz:loDelta}   {\ensuremath{{+0.092 } } }
\vdef{default-11:NoMcCMS-A:pvz:loDeltaE}   {\ensuremath{{0.007 } } }
\vdef{default-11:NoMcCMS-A:pvz:hiDelta}   {\ensuremath{{-0.090 } } }
\vdef{default-11:NoMcCMS-A:pvz:hiDeltaE}   {\ensuremath{{0.007 } } }
\vdef{default-11:NoData-AR4:pvn:loEff}   {\ensuremath{{1.011 } } }
\vdef{default-11:NoData-AR4:pvn:loEffE}   {\ensuremath{{\mathrm{NaN} } } }
\vdef{default-11:NoData-AR4:pvn:hiEff}   {\ensuremath{{1.000 } } }
\vdef{default-11:NoData-AR4:pvn:hiEffE}   {\ensuremath{{0.000 } } }
\vdef{default-11:NoMcCMS-A:pvn:loEff}   {\ensuremath{{1.065 } } }
\vdef{default-11:NoMcCMS-A:pvn:loEffE}   {\ensuremath{{\mathrm{NaN} } } }
\vdef{default-11:NoMcCMS-A:pvn:hiEff}   {\ensuremath{{1.000 } } }
\vdef{default-11:NoMcCMS-A:pvn:hiEffE}   {\ensuremath{{0.000 } } }
\vdef{default-11:NoMcCMS-A:pvn:loDelta}   {\ensuremath{{-0.052 } } }
\vdef{default-11:NoMcCMS-A:pvn:loDeltaE}   {\ensuremath{{\mathrm{NaN} } } }
\vdef{default-11:NoMcCMS-A:pvn:hiDelta}   {\ensuremath{{+0.000 } } }
\vdef{default-11:NoMcCMS-A:pvn:hiDeltaE}   {\ensuremath{{0.000 } } }
\vdef{default-11:NoData-AR4:pvavew8:loEff}   {\ensuremath{{0.007 } } }
\vdef{default-11:NoData-AR4:pvavew8:loEffE}   {\ensuremath{{0.000 } } }
\vdef{default-11:NoData-AR4:pvavew8:hiEff}   {\ensuremath{{0.993 } } }
\vdef{default-11:NoData-AR4:pvavew8:hiEffE}   {\ensuremath{{0.000 } } }
\vdef{default-11:NoMcCMS-A:pvavew8:loEff}   {\ensuremath{{0.006 } } }
\vdef{default-11:NoMcCMS-A:pvavew8:loEffE}   {\ensuremath{{0.000 } } }
\vdef{default-11:NoMcCMS-A:pvavew8:hiEff}   {\ensuremath{{0.994 } } }
\vdef{default-11:NoMcCMS-A:pvavew8:hiEffE}   {\ensuremath{{0.000 } } }
\vdef{default-11:NoMcCMS-A:pvavew8:loDelta}   {\ensuremath{{+0.146 } } }
\vdef{default-11:NoMcCMS-A:pvavew8:loDeltaE}   {\ensuremath{{0.094 } } }
\vdef{default-11:NoMcCMS-A:pvavew8:hiDelta}   {\ensuremath{{-0.001 } } }
\vdef{default-11:NoMcCMS-A:pvavew8:hiDeltaE}   {\ensuremath{{0.001 } } }
\vdef{default-11:NoData-AR4:pvntrk:loEff}   {\ensuremath{{1.000 } } }
\vdef{default-11:NoData-AR4:pvntrk:loEffE}   {\ensuremath{{0.000 } } }
\vdef{default-11:NoData-AR4:pvntrk:hiEff}   {\ensuremath{{1.000 } } }
\vdef{default-11:NoData-AR4:pvntrk:hiEffE}   {\ensuremath{{0.000 } } }
\vdef{default-11:NoMcCMS-A:pvntrk:loEff}   {\ensuremath{{1.000 } } }
\vdef{default-11:NoMcCMS-A:pvntrk:loEffE}   {\ensuremath{{0.000 } } }
\vdef{default-11:NoMcCMS-A:pvntrk:hiEff}   {\ensuremath{{1.000 } } }
\vdef{default-11:NoMcCMS-A:pvntrk:hiEffE}   {\ensuremath{{0.000 } } }
\vdef{default-11:NoMcCMS-A:pvntrk:loDelta}   {\ensuremath{{+0.000 } } }
\vdef{default-11:NoMcCMS-A:pvntrk:loDeltaE}   {\ensuremath{{0.000 } } }
\vdef{default-11:NoMcCMS-A:pvntrk:hiDelta}   {\ensuremath{{+0.000 } } }
\vdef{default-11:NoMcCMS-A:pvntrk:hiDeltaE}   {\ensuremath{{0.000 } } }
\vdef{default-11:NoData-AR4:muon1pt:loEff}   {\ensuremath{{1.011 } } }
\vdef{default-11:NoData-AR4:muon1pt:loEffE}   {\ensuremath{{\mathrm{NaN} } } }
\vdef{default-11:NoData-AR4:muon1pt:hiEff}   {\ensuremath{{1.000 } } }
\vdef{default-11:NoData-AR4:muon1pt:hiEffE}   {\ensuremath{{0.000 } } }
\vdef{default-11:NoMcCMS-A:muon1pt:loEff}   {\ensuremath{{1.009 } } }
\vdef{default-11:NoMcCMS-A:muon1pt:loEffE}   {\ensuremath{{\mathrm{NaN} } } }
\vdef{default-11:NoMcCMS-A:muon1pt:hiEff}   {\ensuremath{{1.000 } } }
\vdef{default-11:NoMcCMS-A:muon1pt:hiEffE}   {\ensuremath{{0.000 } } }
\vdef{default-11:NoMcCMS-A:muon1pt:loDelta}   {\ensuremath{{+0.002 } } }
\vdef{default-11:NoMcCMS-A:muon1pt:loDeltaE}   {\ensuremath{{\mathrm{NaN} } } }
\vdef{default-11:NoMcCMS-A:muon1pt:hiDelta}   {\ensuremath{{+0.000 } } }
\vdef{default-11:NoMcCMS-A:muon1pt:hiDeltaE}   {\ensuremath{{0.000 } } }
\vdef{default-11:NoData-AR4:muon2pt:loEff}   {\ensuremath{{0.007 } } }
\vdef{default-11:NoData-AR4:muon2pt:loEffE}   {\ensuremath{{0.000 } } }
\vdef{default-11:NoData-AR4:muon2pt:hiEff}   {\ensuremath{{0.993 } } }
\vdef{default-11:NoData-AR4:muon2pt:hiEffE}   {\ensuremath{{0.000 } } }
\vdef{default-11:NoMcCMS-A:muon2pt:loEff}   {\ensuremath{{0.090 } } }
\vdef{default-11:NoMcCMS-A:muon2pt:loEffE}   {\ensuremath{{0.001 } } }
\vdef{default-11:NoMcCMS-A:muon2pt:hiEff}   {\ensuremath{{0.910 } } }
\vdef{default-11:NoMcCMS-A:muon2pt:hiEffE}   {\ensuremath{{0.001 } } }
\vdef{default-11:NoMcCMS-A:muon2pt:loDelta}   {\ensuremath{{-1.713 } } }
\vdef{default-11:NoMcCMS-A:muon2pt:loDeltaE}   {\ensuremath{{0.017 } } }
\vdef{default-11:NoMcCMS-A:muon2pt:hiDelta}   {\ensuremath{{+0.087 } } }
\vdef{default-11:NoMcCMS-A:muon2pt:hiDeltaE}   {\ensuremath{{0.002 } } }
\vdef{default-11:NoData-AR4:muonseta:loEff}   {\ensuremath{{0.747 } } }
\vdef{default-11:NoData-AR4:muonseta:loEffE}   {\ensuremath{{0.002 } } }
\vdef{default-11:NoData-AR4:muonseta:hiEff}   {\ensuremath{{0.253 } } }
\vdef{default-11:NoData-AR4:muonseta:hiEffE}   {\ensuremath{{0.002 } } }
\vdef{default-11:NoMcCMS-A:muonseta:loEff}   {\ensuremath{{0.844 } } }
\vdef{default-11:NoMcCMS-A:muonseta:loEffE}   {\ensuremath{{0.001 } } }
\vdef{default-11:NoMcCMS-A:muonseta:hiEff}   {\ensuremath{{0.156 } } }
\vdef{default-11:NoMcCMS-A:muonseta:hiEffE}   {\ensuremath{{0.001 } } }
\vdef{default-11:NoMcCMS-A:muonseta:loDelta}   {\ensuremath{{-0.121 } } }
\vdef{default-11:NoMcCMS-A:muonseta:loDeltaE}   {\ensuremath{{0.003 } } }
\vdef{default-11:NoMcCMS-A:muonseta:hiDelta}   {\ensuremath{{+0.473 } } }
\vdef{default-11:NoMcCMS-A:muonseta:hiDeltaE}   {\ensuremath{{0.010 } } }
\vdef{default-11:NoData-AR4:pt:loEff}   {\ensuremath{{0.000 } } }
\vdef{default-11:NoData-AR4:pt:loEffE}   {\ensuremath{{0.000 } } }
\vdef{default-11:NoData-AR4:pt:hiEff}   {\ensuremath{{1.000 } } }
\vdef{default-11:NoData-AR4:pt:hiEffE}   {\ensuremath{{0.000 } } }
\vdef{default-11:NoMcCMS-A:pt:loEff}   {\ensuremath{{0.000 } } }
\vdef{default-11:NoMcCMS-A:pt:loEffE}   {\ensuremath{{0.000 } } }
\vdef{default-11:NoMcCMS-A:pt:hiEff}   {\ensuremath{{1.000 } } }
\vdef{default-11:NoMcCMS-A:pt:hiEffE}   {\ensuremath{{0.000 } } }
\vdef{default-11:NoMcCMS-A:pt:loDelta}   {\ensuremath{{\mathrm{NaN} } } }
\vdef{default-11:NoMcCMS-A:pt:loDeltaE}   {\ensuremath{{\mathrm{NaN} } } }
\vdef{default-11:NoMcCMS-A:pt:hiDelta}   {\ensuremath{{+0.000 } } }
\vdef{default-11:NoMcCMS-A:pt:hiDeltaE}   {\ensuremath{{0.000 } } }
\vdef{default-11:NoData-AR4:p:loEff}   {\ensuremath{{1.011 } } }
\vdef{default-11:NoData-AR4:p:loEffE}   {\ensuremath{{\mathrm{NaN} } } }
\vdef{default-11:NoData-AR4:p:hiEff}   {\ensuremath{{1.000 } } }
\vdef{default-11:NoData-AR4:p:hiEffE}   {\ensuremath{{0.000 } } }
\vdef{default-11:NoMcCMS-A:p:loEff}   {\ensuremath{{1.004 } } }
\vdef{default-11:NoMcCMS-A:p:loEffE}   {\ensuremath{{\mathrm{NaN} } } }
\vdef{default-11:NoMcCMS-A:p:hiEff}   {\ensuremath{{1.000 } } }
\vdef{default-11:NoMcCMS-A:p:hiEffE}   {\ensuremath{{0.000 } } }
\vdef{default-11:NoMcCMS-A:p:loDelta}   {\ensuremath{{+0.007 } } }
\vdef{default-11:NoMcCMS-A:p:loDeltaE}   {\ensuremath{{\mathrm{NaN} } } }
\vdef{default-11:NoMcCMS-A:p:hiDelta}   {\ensuremath{{+0.000 } } }
\vdef{default-11:NoMcCMS-A:p:hiDeltaE}   {\ensuremath{{0.000 } } }
\vdef{default-11:NoData-AR4:eta:loEff}   {\ensuremath{{0.739 } } }
\vdef{default-11:NoData-AR4:eta:loEffE}   {\ensuremath{{0.002 } } }
\vdef{default-11:NoData-AR4:eta:hiEff}   {\ensuremath{{0.261 } } }
\vdef{default-11:NoData-AR4:eta:hiEffE}   {\ensuremath{{0.002 } } }
\vdef{default-11:NoMcCMS-A:eta:loEff}   {\ensuremath{{0.841 } } }
\vdef{default-11:NoMcCMS-A:eta:loEffE}   {\ensuremath{{0.002 } } }
\vdef{default-11:NoMcCMS-A:eta:hiEff}   {\ensuremath{{0.159 } } }
\vdef{default-11:NoMcCMS-A:eta:hiEffE}   {\ensuremath{{0.002 } } }
\vdef{default-11:NoMcCMS-A:eta:loDelta}   {\ensuremath{{-0.129 } } }
\vdef{default-11:NoMcCMS-A:eta:loDeltaE}   {\ensuremath{{0.004 } } }
\vdef{default-11:NoMcCMS-A:eta:hiDelta}   {\ensuremath{{+0.486 } } }
\vdef{default-11:NoMcCMS-A:eta:hiDeltaE}   {\ensuremath{{0.014 } } }
\vdef{default-11:NoData-AR4:bdt:loEff}   {\ensuremath{{0.907 } } }
\vdef{default-11:NoData-AR4:bdt:loEffE}   {\ensuremath{{0.001 } } }
\vdef{default-11:NoData-AR4:bdt:hiEff}   {\ensuremath{{0.093 } } }
\vdef{default-11:NoData-AR4:bdt:hiEffE}   {\ensuremath{{0.001 } } }
\vdef{default-11:NoMcCMS-A:bdt:loEff}   {\ensuremath{{0.871 } } }
\vdef{default-11:NoMcCMS-A:bdt:loEffE}   {\ensuremath{{0.002 } } }
\vdef{default-11:NoMcCMS-A:bdt:hiEff}   {\ensuremath{{0.129 } } }
\vdef{default-11:NoMcCMS-A:bdt:hiEffE}   {\ensuremath{{0.002 } } }
\vdef{default-11:NoMcCMS-A:bdt:loDelta}   {\ensuremath{{+0.041 } } }
\vdef{default-11:NoMcCMS-A:bdt:loDeltaE}   {\ensuremath{{0.003 } } }
\vdef{default-11:NoMcCMS-A:bdt:hiDelta}   {\ensuremath{{-0.325 } } }
\vdef{default-11:NoMcCMS-A:bdt:hiDeltaE}   {\ensuremath{{0.020 } } }
\vdef{default-11:NoData-AR4:fl3d:loEff}   {\ensuremath{{0.837 } } }
\vdef{default-11:NoData-AR4:fl3d:loEffE}   {\ensuremath{{0.002 } } }
\vdef{default-11:NoData-AR4:fl3d:hiEff}   {\ensuremath{{0.163 } } }
\vdef{default-11:NoData-AR4:fl3d:hiEffE}   {\ensuremath{{0.002 } } }
\vdef{default-11:NoMcCMS-A:fl3d:loEff}   {\ensuremath{{0.881 } } }
\vdef{default-11:NoMcCMS-A:fl3d:loEffE}   {\ensuremath{{0.002 } } }
\vdef{default-11:NoMcCMS-A:fl3d:hiEff}   {\ensuremath{{0.119 } } }
\vdef{default-11:NoMcCMS-A:fl3d:hiEffE}   {\ensuremath{{0.002 } } }
\vdef{default-11:NoMcCMS-A:fl3d:loDelta}   {\ensuremath{{-0.051 } } }
\vdef{default-11:NoMcCMS-A:fl3d:loDeltaE}   {\ensuremath{{0.003 } } }
\vdef{default-11:NoMcCMS-A:fl3d:hiDelta}   {\ensuremath{{+0.312 } } }
\vdef{default-11:NoMcCMS-A:fl3d:hiDeltaE}   {\ensuremath{{0.017 } } }
\vdef{default-11:NoData-AR4:fl3de:loEff}   {\ensuremath{{1.000 } } }
\vdef{default-11:NoData-AR4:fl3de:loEffE}   {\ensuremath{{0.000 } } }
\vdef{default-11:NoData-AR4:fl3de:hiEff}   {\ensuremath{{0.000 } } }
\vdef{default-11:NoData-AR4:fl3de:hiEffE}   {\ensuremath{{0.000 } } }
\vdef{default-11:NoMcCMS-A:fl3de:loEff}   {\ensuremath{{1.000 } } }
\vdef{default-11:NoMcCMS-A:fl3de:loEffE}   {\ensuremath{{0.000 } } }
\vdef{default-11:NoMcCMS-A:fl3de:hiEff}   {\ensuremath{{0.000 } } }
\vdef{default-11:NoMcCMS-A:fl3de:hiEffE}   {\ensuremath{{0.000 } } }
\vdef{default-11:NoMcCMS-A:fl3de:loDelta}   {\ensuremath{{+0.000 } } }
\vdef{default-11:NoMcCMS-A:fl3de:loDeltaE}   {\ensuremath{{0.000 } } }
\vdef{default-11:NoMcCMS-A:fl3de:hiDelta}   {\ensuremath{{+0.933 } } }
\vdef{default-11:NoMcCMS-A:fl3de:hiDeltaE}   {\ensuremath{{0.708 } } }
\vdef{default-11:NoData-AR4:fls3d:loEff}   {\ensuremath{{0.071 } } }
\vdef{default-11:NoData-AR4:fls3d:loEffE}   {\ensuremath{{0.001 } } }
\vdef{default-11:NoData-AR4:fls3d:hiEff}   {\ensuremath{{0.929 } } }
\vdef{default-11:NoData-AR4:fls3d:hiEffE}   {\ensuremath{{0.001 } } }
\vdef{default-11:NoMcCMS-A:fls3d:loEff}   {\ensuremath{{0.060 } } }
\vdef{default-11:NoMcCMS-A:fls3d:loEffE}   {\ensuremath{{0.001 } } }
\vdef{default-11:NoMcCMS-A:fls3d:hiEff}   {\ensuremath{{0.940 } } }
\vdef{default-11:NoMcCMS-A:fls3d:hiEffE}   {\ensuremath{{0.001 } } }
\vdef{default-11:NoMcCMS-A:fls3d:loDelta}   {\ensuremath{{+0.167 } } }
\vdef{default-11:NoMcCMS-A:fls3d:loDeltaE}   {\ensuremath{{0.026 } } }
\vdef{default-11:NoMcCMS-A:fls3d:hiDelta}   {\ensuremath{{-0.012 } } }
\vdef{default-11:NoMcCMS-A:fls3d:hiDeltaE}   {\ensuremath{{0.002 } } }
\vdef{default-11:NoData-AR4:flsxy:loEff}   {\ensuremath{{1.013 } } }
\vdef{default-11:NoData-AR4:flsxy:loEffE}   {\ensuremath{{\mathrm{NaN} } } }
\vdef{default-11:NoData-AR4:flsxy:hiEff}   {\ensuremath{{1.000 } } }
\vdef{default-11:NoData-AR4:flsxy:hiEffE}   {\ensuremath{{0.000 } } }
\vdef{default-11:NoMcCMS-A:flsxy:loEff}   {\ensuremath{{1.013 } } }
\vdef{default-11:NoMcCMS-A:flsxy:loEffE}   {\ensuremath{{\mathrm{NaN} } } }
\vdef{default-11:NoMcCMS-A:flsxy:hiEff}   {\ensuremath{{1.000 } } }
\vdef{default-11:NoMcCMS-A:flsxy:hiEffE}   {\ensuremath{{0.000 } } }
\vdef{default-11:NoMcCMS-A:flsxy:loDelta}   {\ensuremath{{+0.000 } } }
\vdef{default-11:NoMcCMS-A:flsxy:loDeltaE}   {\ensuremath{{\mathrm{NaN} } } }
\vdef{default-11:NoMcCMS-A:flsxy:hiDelta}   {\ensuremath{{+0.000 } } }
\vdef{default-11:NoMcCMS-A:flsxy:hiDeltaE}   {\ensuremath{{0.000 } } }
\vdef{default-11:NoData-AR4:chi2dof:loEff}   {\ensuremath{{0.934 } } }
\vdef{default-11:NoData-AR4:chi2dof:loEffE}   {\ensuremath{{0.001 } } }
\vdef{default-11:NoData-AR4:chi2dof:hiEff}   {\ensuremath{{0.066 } } }
\vdef{default-11:NoData-AR4:chi2dof:hiEffE}   {\ensuremath{{0.001 } } }
\vdef{default-11:NoMcCMS-A:chi2dof:loEff}   {\ensuremath{{0.939 } } }
\vdef{default-11:NoMcCMS-A:chi2dof:loEffE}   {\ensuremath{{0.001 } } }
\vdef{default-11:NoMcCMS-A:chi2dof:hiEff}   {\ensuremath{{0.061 } } }
\vdef{default-11:NoMcCMS-A:chi2dof:hiEffE}   {\ensuremath{{0.001 } } }
\vdef{default-11:NoMcCMS-A:chi2dof:loDelta}   {\ensuremath{{-0.005 } } }
\vdef{default-11:NoMcCMS-A:chi2dof:loDeltaE}   {\ensuremath{{0.002 } } }
\vdef{default-11:NoMcCMS-A:chi2dof:hiDelta}   {\ensuremath{{+0.079 } } }
\vdef{default-11:NoMcCMS-A:chi2dof:hiDeltaE}   {\ensuremath{{0.028 } } }
\vdef{default-11:NoData-AR4:pchi2dof:loEff}   {\ensuremath{{0.629 } } }
\vdef{default-11:NoData-AR4:pchi2dof:loEffE}   {\ensuremath{{0.002 } } }
\vdef{default-11:NoData-AR4:pchi2dof:hiEff}   {\ensuremath{{0.371 } } }
\vdef{default-11:NoData-AR4:pchi2dof:hiEffE}   {\ensuremath{{0.002 } } }
\vdef{default-11:NoMcCMS-A:pchi2dof:loEff}   {\ensuremath{{0.629 } } }
\vdef{default-11:NoMcCMS-A:pchi2dof:loEffE}   {\ensuremath{{0.002 } } }
\vdef{default-11:NoMcCMS-A:pchi2dof:hiEff}   {\ensuremath{{0.371 } } }
\vdef{default-11:NoMcCMS-A:pchi2dof:hiEffE}   {\ensuremath{{0.002 } } }
\vdef{default-11:NoMcCMS-A:pchi2dof:loDelta}   {\ensuremath{{-0.001 } } }
\vdef{default-11:NoMcCMS-A:pchi2dof:loDeltaE}   {\ensuremath{{0.005 } } }
\vdef{default-11:NoMcCMS-A:pchi2dof:hiDelta}   {\ensuremath{{+0.002 } } }
\vdef{default-11:NoMcCMS-A:pchi2dof:hiDeltaE}   {\ensuremath{{0.009 } } }
\vdef{default-11:NoData-AR4:alpha:loEff}   {\ensuremath{{0.995 } } }
\vdef{default-11:NoData-AR4:alpha:loEffE}   {\ensuremath{{0.000 } } }
\vdef{default-11:NoData-AR4:alpha:hiEff}   {\ensuremath{{0.005 } } }
\vdef{default-11:NoData-AR4:alpha:hiEffE}   {\ensuremath{{0.000 } } }
\vdef{default-11:NoMcCMS-A:alpha:loEff}   {\ensuremath{{0.993 } } }
\vdef{default-11:NoMcCMS-A:alpha:loEffE}   {\ensuremath{{0.000 } } }
\vdef{default-11:NoMcCMS-A:alpha:hiEff}   {\ensuremath{{0.007 } } }
\vdef{default-11:NoMcCMS-A:alpha:hiEffE}   {\ensuremath{{0.000 } } }
\vdef{default-11:NoMcCMS-A:alpha:loDelta}   {\ensuremath{{+0.002 } } }
\vdef{default-11:NoMcCMS-A:alpha:loDeltaE}   {\ensuremath{{0.001 } } }
\vdef{default-11:NoMcCMS-A:alpha:hiDelta}   {\ensuremath{{-0.254 } } }
\vdef{default-11:NoMcCMS-A:alpha:hiDeltaE}   {\ensuremath{{0.094 } } }
\vdef{default-11:NoData-AR4:iso:loEff}   {\ensuremath{{0.127 } } }
\vdef{default-11:NoData-AR4:iso:loEffE}   {\ensuremath{{0.002 } } }
\vdef{default-11:NoData-AR4:iso:hiEff}   {\ensuremath{{0.873 } } }
\vdef{default-11:NoData-AR4:iso:hiEffE}   {\ensuremath{{0.002 } } }
\vdef{default-11:NoMcCMS-A:iso:loEff}   {\ensuremath{{0.111 } } }
\vdef{default-11:NoMcCMS-A:iso:loEffE}   {\ensuremath{{0.002 } } }
\vdef{default-11:NoMcCMS-A:iso:hiEff}   {\ensuremath{{0.889 } } }
\vdef{default-11:NoMcCMS-A:iso:hiEffE}   {\ensuremath{{0.002 } } }
\vdef{default-11:NoMcCMS-A:iso:loDelta}   {\ensuremath{{+0.137 } } }
\vdef{default-11:NoMcCMS-A:iso:loDeltaE}   {\ensuremath{{0.019 } } }
\vdef{default-11:NoMcCMS-A:iso:hiDelta}   {\ensuremath{{-0.018 } } }
\vdef{default-11:NoMcCMS-A:iso:hiDeltaE}   {\ensuremath{{0.003 } } }
\vdef{default-11:NoData-AR4:docatrk:loEff}   {\ensuremath{{0.073 } } }
\vdef{default-11:NoData-AR4:docatrk:loEffE}   {\ensuremath{{0.001 } } }
\vdef{default-11:NoData-AR4:docatrk:hiEff}   {\ensuremath{{0.927 } } }
\vdef{default-11:NoData-AR4:docatrk:hiEffE}   {\ensuremath{{0.001 } } }
\vdef{default-11:NoMcCMS-A:docatrk:loEff}   {\ensuremath{{0.087 } } }
\vdef{default-11:NoMcCMS-A:docatrk:loEffE}   {\ensuremath{{0.001 } } }
\vdef{default-11:NoMcCMS-A:docatrk:hiEff}   {\ensuremath{{0.913 } } }
\vdef{default-11:NoMcCMS-A:docatrk:hiEffE}   {\ensuremath{{0.001 } } }
\vdef{default-11:NoMcCMS-A:docatrk:loDelta}   {\ensuremath{{-0.174 } } }
\vdef{default-11:NoMcCMS-A:docatrk:loDeltaE}   {\ensuremath{{0.024 } } }
\vdef{default-11:NoMcCMS-A:docatrk:hiDelta}   {\ensuremath{{+0.015 } } }
\vdef{default-11:NoMcCMS-A:docatrk:hiDeltaE}   {\ensuremath{{0.002 } } }
\vdef{default-11:NoData-AR4:isotrk:loEff}   {\ensuremath{{1.000 } } }
\vdef{default-11:NoData-AR4:isotrk:loEffE}   {\ensuremath{{0.000 } } }
\vdef{default-11:NoData-AR4:isotrk:hiEff}   {\ensuremath{{1.000 } } }
\vdef{default-11:NoData-AR4:isotrk:hiEffE}   {\ensuremath{{0.000 } } }
\vdef{default-11:NoMcCMS-A:isotrk:loEff}   {\ensuremath{{1.000 } } }
\vdef{default-11:NoMcCMS-A:isotrk:loEffE}   {\ensuremath{{0.000 } } }
\vdef{default-11:NoMcCMS-A:isotrk:hiEff}   {\ensuremath{{1.000 } } }
\vdef{default-11:NoMcCMS-A:isotrk:hiEffE}   {\ensuremath{{0.000 } } }
\vdef{default-11:NoMcCMS-A:isotrk:loDelta}   {\ensuremath{{+0.000 } } }
\vdef{default-11:NoMcCMS-A:isotrk:loDeltaE}   {\ensuremath{{0.000 } } }
\vdef{default-11:NoMcCMS-A:isotrk:hiDelta}   {\ensuremath{{+0.000 } } }
\vdef{default-11:NoMcCMS-A:isotrk:hiDeltaE}   {\ensuremath{{0.000 } } }
\vdef{default-11:NoData-AR4:closetrk:loEff}   {\ensuremath{{0.974 } } }
\vdef{default-11:NoData-AR4:closetrk:loEffE}   {\ensuremath{{0.001 } } }
\vdef{default-11:NoData-AR4:closetrk:hiEff}   {\ensuremath{{0.026 } } }
\vdef{default-11:NoData-AR4:closetrk:hiEffE}   {\ensuremath{{0.001 } } }
\vdef{default-11:NoMcCMS-A:closetrk:loEff}   {\ensuremath{{0.974 } } }
\vdef{default-11:NoMcCMS-A:closetrk:loEffE}   {\ensuremath{{0.001 } } }
\vdef{default-11:NoMcCMS-A:closetrk:hiEff}   {\ensuremath{{0.026 } } }
\vdef{default-11:NoMcCMS-A:closetrk:hiEffE}   {\ensuremath{{0.001 } } }
\vdef{default-11:NoMcCMS-A:closetrk:loDelta}   {\ensuremath{{+0.000 } } }
\vdef{default-11:NoMcCMS-A:closetrk:loDeltaE}   {\ensuremath{{0.001 } } }
\vdef{default-11:NoMcCMS-A:closetrk:hiDelta}   {\ensuremath{{-0.014 } } }
\vdef{default-11:NoMcCMS-A:closetrk:hiDeltaE}   {\ensuremath{{0.045 } } }
\vdef{default-11:NoData-AR4:lip:loEff}   {\ensuremath{{1.000 } } }
\vdef{default-11:NoData-AR4:lip:loEffE}   {\ensuremath{{0.000 } } }
\vdef{default-11:NoData-AR4:lip:hiEff}   {\ensuremath{{0.000 } } }
\vdef{default-11:NoData-AR4:lip:hiEffE}   {\ensuremath{{0.000 } } }
\vdef{default-11:NoMcCMS-A:lip:loEff}   {\ensuremath{{1.000 } } }
\vdef{default-11:NoMcCMS-A:lip:loEffE}   {\ensuremath{{0.000 } } }
\vdef{default-11:NoMcCMS-A:lip:hiEff}   {\ensuremath{{0.000 } } }
\vdef{default-11:NoMcCMS-A:lip:hiEffE}   {\ensuremath{{0.000 } } }
\vdef{default-11:NoMcCMS-A:lip:loDelta}   {\ensuremath{{+0.000 } } }
\vdef{default-11:NoMcCMS-A:lip:loDeltaE}   {\ensuremath{{0.000 } } }
\vdef{default-11:NoMcCMS-A:lip:hiDelta}   {\ensuremath{{\mathrm{NaN} } } }
\vdef{default-11:NoMcCMS-A:lip:hiDeltaE}   {\ensuremath{{\mathrm{NaN} } } }
\vdef{default-11:NoData-AR4:lip:inEff}   {\ensuremath{{1.000 } } }
\vdef{default-11:NoData-AR4:lip:inEffE}   {\ensuremath{{0.000 } } }
\vdef{default-11:NoMcCMS-A:lip:inEff}   {\ensuremath{{1.000 } } }
\vdef{default-11:NoMcCMS-A:lip:inEffE}   {\ensuremath{{0.000 } } }
\vdef{default-11:NoMcCMS-A:lip:inDelta}   {\ensuremath{{+0.000 } } }
\vdef{default-11:NoMcCMS-A:lip:inDeltaE}   {\ensuremath{{0.000 } } }
\vdef{default-11:NoData-AR4:lips:loEff}   {\ensuremath{{1.000 } } }
\vdef{default-11:NoData-AR4:lips:loEffE}   {\ensuremath{{0.000 } } }
\vdef{default-11:NoData-AR4:lips:hiEff}   {\ensuremath{{0.000 } } }
\vdef{default-11:NoData-AR4:lips:hiEffE}   {\ensuremath{{0.000 } } }
\vdef{default-11:NoMcCMS-A:lips:loEff}   {\ensuremath{{1.000 } } }
\vdef{default-11:NoMcCMS-A:lips:loEffE}   {\ensuremath{{0.000 } } }
\vdef{default-11:NoMcCMS-A:lips:hiEff}   {\ensuremath{{0.000 } } }
\vdef{default-11:NoMcCMS-A:lips:hiEffE}   {\ensuremath{{0.000 } } }
\vdef{default-11:NoMcCMS-A:lips:loDelta}   {\ensuremath{{+0.000 } } }
\vdef{default-11:NoMcCMS-A:lips:loDeltaE}   {\ensuremath{{0.000 } } }
\vdef{default-11:NoMcCMS-A:lips:hiDelta}   {\ensuremath{{\mathrm{NaN} } } }
\vdef{default-11:NoMcCMS-A:lips:hiDeltaE}   {\ensuremath{{\mathrm{NaN} } } }
\vdef{default-11:NoData-AR4:lips:inEff}   {\ensuremath{{1.000 } } }
\vdef{default-11:NoData-AR4:lips:inEffE}   {\ensuremath{{0.000 } } }
\vdef{default-11:NoMcCMS-A:lips:inEff}   {\ensuremath{{1.000 } } }
\vdef{default-11:NoMcCMS-A:lips:inEffE}   {\ensuremath{{0.000 } } }
\vdef{default-11:NoMcCMS-A:lips:inDelta}   {\ensuremath{{+0.000 } } }
\vdef{default-11:NoMcCMS-A:lips:inDeltaE}   {\ensuremath{{0.000 } } }
\vdef{default-11:NoData-AR4:ip:loEff}   {\ensuremath{{0.971 } } }
\vdef{default-11:NoData-AR4:ip:loEffE}   {\ensuremath{{0.001 } } }
\vdef{default-11:NoData-AR4:ip:hiEff}   {\ensuremath{{0.029 } } }
\vdef{default-11:NoData-AR4:ip:hiEffE}   {\ensuremath{{0.001 } } }
\vdef{default-11:NoMcCMS-A:ip:loEff}   {\ensuremath{{0.970 } } }
\vdef{default-11:NoMcCMS-A:ip:loEffE}   {\ensuremath{{0.001 } } }
\vdef{default-11:NoMcCMS-A:ip:hiEff}   {\ensuremath{{0.030 } } }
\vdef{default-11:NoMcCMS-A:ip:hiEffE}   {\ensuremath{{0.001 } } }
\vdef{default-11:NoMcCMS-A:ip:loDelta}   {\ensuremath{{+0.002 } } }
\vdef{default-11:NoMcCMS-A:ip:loDeltaE}   {\ensuremath{{0.001 } } }
\vdef{default-11:NoMcCMS-A:ip:hiDelta}   {\ensuremath{{-0.065 } } }
\vdef{default-11:NoMcCMS-A:ip:hiDeltaE}   {\ensuremath{{0.042 } } }
\vdef{default-11:NoData-AR4:ips:loEff}   {\ensuremath{{0.942 } } }
\vdef{default-11:NoData-AR4:ips:loEffE}   {\ensuremath{{0.001 } } }
\vdef{default-11:NoData-AR4:ips:hiEff}   {\ensuremath{{0.058 } } }
\vdef{default-11:NoData-AR4:ips:hiEffE}   {\ensuremath{{0.001 } } }
\vdef{default-11:NoMcCMS-A:ips:loEff}   {\ensuremath{{0.951 } } }
\vdef{default-11:NoMcCMS-A:ips:loEffE}   {\ensuremath{{0.001 } } }
\vdef{default-11:NoMcCMS-A:ips:hiEff}   {\ensuremath{{0.049 } } }
\vdef{default-11:NoMcCMS-A:ips:hiEffE}   {\ensuremath{{0.001 } } }
\vdef{default-11:NoMcCMS-A:ips:loDelta}   {\ensuremath{{-0.010 } } }
\vdef{default-11:NoMcCMS-A:ips:loDeltaE}   {\ensuremath{{0.002 } } }
\vdef{default-11:NoMcCMS-A:ips:hiDelta}   {\ensuremath{{+0.170 } } }
\vdef{default-11:NoMcCMS-A:ips:hiDeltaE}   {\ensuremath{{0.030 } } }
\vdef{default-11:NoData-AR4:maxdoca:loEff}   {\ensuremath{{1.000 } } }
\vdef{default-11:NoData-AR4:maxdoca:loEffE}   {\ensuremath{{0.000 } } }
\vdef{default-11:NoData-AR4:maxdoca:hiEff}   {\ensuremath{{0.012 } } }
\vdef{default-11:NoData-AR4:maxdoca:hiEffE}   {\ensuremath{{0.001 } } }
\vdef{default-11:NoMcCMS-A:maxdoca:loEff}   {\ensuremath{{1.000 } } }
\vdef{default-11:NoMcCMS-A:maxdoca:loEffE}   {\ensuremath{{0.000 } } }
\vdef{default-11:NoMcCMS-A:maxdoca:hiEff}   {\ensuremath{{0.009 } } }
\vdef{default-11:NoMcCMS-A:maxdoca:hiEffE}   {\ensuremath{{0.000 } } }
\vdef{default-11:NoMcCMS-A:maxdoca:loDelta}   {\ensuremath{{+0.000 } } }
\vdef{default-11:NoMcCMS-A:maxdoca:loDeltaE}   {\ensuremath{{0.000 } } }
\vdef{default-11:NoMcCMS-A:maxdoca:hiDelta}   {\ensuremath{{+0.319 } } }
\vdef{default-11:NoMcCMS-A:maxdoca:hiDeltaE}   {\ensuremath{{0.072 } } }
\vdef{default-11:NoData-AR4:kaonpt:loEff}   {\ensuremath{{1.009 } } }
\vdef{default-11:NoData-AR4:kaonpt:loEffE}   {\ensuremath{{\mathrm{NaN} } } }
\vdef{default-11:NoData-AR4:kaonpt:hiEff}   {\ensuremath{{1.000 } } }
\vdef{default-11:NoData-AR4:kaonpt:hiEffE}   {\ensuremath{{0.000 } } }
\vdef{default-11:NoMcCMS-A:kaonpt:loEff}   {\ensuremath{{1.008 } } }
\vdef{default-11:NoMcCMS-A:kaonpt:loEffE}   {\ensuremath{{\mathrm{NaN} } } }
\vdef{default-11:NoMcCMS-A:kaonpt:hiEff}   {\ensuremath{{1.000 } } }
\vdef{default-11:NoMcCMS-A:kaonpt:hiEffE}   {\ensuremath{{0.000 } } }
\vdef{default-11:NoMcCMS-A:kaonpt:loDelta}   {\ensuremath{{+0.001 } } }
\vdef{default-11:NoMcCMS-A:kaonpt:loDeltaE}   {\ensuremath{{\mathrm{NaN} } } }
\vdef{default-11:NoMcCMS-A:kaonpt:hiDelta}   {\ensuremath{{+0.000 } } }
\vdef{default-11:NoMcCMS-A:kaonpt:hiDeltaE}   {\ensuremath{{0.000 } } }
\vdef{default-11:NoData-AR4:psipt:loEff}   {\ensuremath{{1.004 } } }
\vdef{default-11:NoData-AR4:psipt:loEffE}   {\ensuremath{{\mathrm{NaN} } } }
\vdef{default-11:NoData-AR4:psipt:hiEff}   {\ensuremath{{1.000 } } }
\vdef{default-11:NoData-AR4:psipt:hiEffE}   {\ensuremath{{0.000 } } }
\vdef{default-11:NoMcCMS-A:psipt:loEff}   {\ensuremath{{1.003 } } }
\vdef{default-11:NoMcCMS-A:psipt:loEffE}   {\ensuremath{{\mathrm{NaN} } } }
\vdef{default-11:NoMcCMS-A:psipt:hiEff}   {\ensuremath{{1.000 } } }
\vdef{default-11:NoMcCMS-A:psipt:hiEffE}   {\ensuremath{{0.000 } } }
\vdef{default-11:NoMcCMS-A:psipt:loDelta}   {\ensuremath{{+0.001 } } }
\vdef{default-11:NoMcCMS-A:psipt:loDeltaE}   {\ensuremath{{\mathrm{NaN} } } }
\vdef{default-11:NoMcCMS-A:psipt:hiDelta}   {\ensuremath{{+0.000 } } }
\vdef{default-11:NoMcCMS-A:psipt:hiDeltaE}   {\ensuremath{{0.000 } } }
\vdef{default-11:NoData-AR5:osiso:loEff}   {\ensuremath{{1.005 } } }
\vdef{default-11:NoData-AR5:osiso:loEffE}   {\ensuremath{{\mathrm{NaN} } } }
\vdef{default-11:NoData-AR5:osiso:hiEff}   {\ensuremath{{1.000 } } }
\vdef{default-11:NoData-AR5:osiso:hiEffE}   {\ensuremath{{0.000 } } }
\vdef{default-11:NoMc3e33-A:osiso:loEff}   {\ensuremath{{1.003 } } }
\vdef{default-11:NoMc3e33-A:osiso:loEffE}   {\ensuremath{{\mathrm{NaN} } } }
\vdef{default-11:NoMc3e33-A:osiso:hiEff}   {\ensuremath{{1.000 } } }
\vdef{default-11:NoMc3e33-A:osiso:hiEffE}   {\ensuremath{{0.000 } } }
\vdef{default-11:NoMc3e33-A:osiso:loDelta}   {\ensuremath{{+0.002 } } }
\vdef{default-11:NoMc3e33-A:osiso:loDeltaE}   {\ensuremath{{\mathrm{NaN} } } }
\vdef{default-11:NoMc3e33-A:osiso:hiDelta}   {\ensuremath{{+0.000 } } }
\vdef{default-11:NoMc3e33-A:osiso:hiDeltaE}   {\ensuremath{{0.000 } } }
\vdef{default-11:NoData-AR5:osreliso:loEff}   {\ensuremath{{0.252 } } }
\vdef{default-11:NoData-AR5:osreliso:loEffE}   {\ensuremath{{0.003 } } }
\vdef{default-11:NoData-AR5:osreliso:hiEff}   {\ensuremath{{0.748 } } }
\vdef{default-11:NoData-AR5:osreliso:hiEffE}   {\ensuremath{{0.003 } } }
\vdef{default-11:NoMc3e33-A:osreliso:loEff}   {\ensuremath{{0.288 } } }
\vdef{default-11:NoMc3e33-A:osreliso:loEffE}   {\ensuremath{{0.004 } } }
\vdef{default-11:NoMc3e33-A:osreliso:hiEff}   {\ensuremath{{0.712 } } }
\vdef{default-11:NoMc3e33-A:osreliso:hiEffE}   {\ensuremath{{0.004 } } }
\vdef{default-11:NoMc3e33-A:osreliso:loDelta}   {\ensuremath{{-0.133 } } }
\vdef{default-11:NoMc3e33-A:osreliso:loDeltaE}   {\ensuremath{{0.019 } } }
\vdef{default-11:NoMc3e33-A:osreliso:hiDelta}   {\ensuremath{{+0.049 } } }
\vdef{default-11:NoMc3e33-A:osreliso:hiDeltaE}   {\ensuremath{{0.007 } } }
\vdef{default-11:NoData-AR5:osmuonpt:loEff}   {\ensuremath{{0.000 } } }
\vdef{default-11:NoData-AR5:osmuonpt:loEffE}   {\ensuremath{{0.002 } } }
\vdef{default-11:NoData-AR5:osmuonpt:hiEff}   {\ensuremath{{1.000 } } }
\vdef{default-11:NoData-AR5:osmuonpt:hiEffE}   {\ensuremath{{0.002 } } }
\vdef{default-11:NoMc3e33-A:osmuonpt:loEff}   {\ensuremath{{0.000 } } }
\vdef{default-11:NoMc3e33-A:osmuonpt:loEffE}   {\ensuremath{{0.002 } } }
\vdef{default-11:NoMc3e33-A:osmuonpt:hiEff}   {\ensuremath{{1.000 } } }
\vdef{default-11:NoMc3e33-A:osmuonpt:hiEffE}   {\ensuremath{{0.002 } } }
\vdef{default-11:NoMc3e33-A:osmuonpt:loDelta}   {\ensuremath{{\mathrm{NaN} } } }
\vdef{default-11:NoMc3e33-A:osmuonpt:loDeltaE}   {\ensuremath{{\mathrm{NaN} } } }
\vdef{default-11:NoMc3e33-A:osmuonpt:hiDelta}   {\ensuremath{{+0.000 } } }
\vdef{default-11:NoMc3e33-A:osmuonpt:hiDeltaE}   {\ensuremath{{0.003 } } }
\vdef{default-11:NoData-AR5:osmuondr:loEff}   {\ensuremath{{0.030 } } }
\vdef{default-11:NoData-AR5:osmuondr:loEffE}   {\ensuremath{{0.008 } } }
\vdef{default-11:NoData-AR5:osmuondr:hiEff}   {\ensuremath{{0.970 } } }
\vdef{default-11:NoData-AR5:osmuondr:hiEffE}   {\ensuremath{{0.008 } } }
\vdef{default-11:NoMc3e33-A:osmuondr:loEff}   {\ensuremath{{0.016 } } }
\vdef{default-11:NoMc3e33-A:osmuondr:loEffE}   {\ensuremath{{0.006 } } }
\vdef{default-11:NoMc3e33-A:osmuondr:hiEff}   {\ensuremath{{0.984 } } }
\vdef{default-11:NoMc3e33-A:osmuondr:hiEffE}   {\ensuremath{{0.006 } } }
\vdef{default-11:NoMc3e33-A:osmuondr:loDelta}   {\ensuremath{{+0.602 } } }
\vdef{default-11:NoMc3e33-A:osmuondr:loDeltaE}   {\ensuremath{{0.389 } } }
\vdef{default-11:NoMc3e33-A:osmuondr:hiDelta}   {\ensuremath{{-0.014 } } }
\vdef{default-11:NoMc3e33-A:osmuondr:hiDeltaE}   {\ensuremath{{0.010 } } }
\vdef{default-11:NoData-AR5:hlt:loEff}   {\ensuremath{{0.078 } } }
\vdef{default-11:NoData-AR5:hlt:loEffE}   {\ensuremath{{0.002 } } }
\vdef{default-11:NoData-AR5:hlt:hiEff}   {\ensuremath{{0.922 } } }
\vdef{default-11:NoData-AR5:hlt:hiEffE}   {\ensuremath{{0.002 } } }
\vdef{default-11:NoMc3e33-A:hlt:loEff}   {\ensuremath{{0.336 } } }
\vdef{default-11:NoMc3e33-A:hlt:loEffE}   {\ensuremath{{0.004 } } }
\vdef{default-11:NoMc3e33-A:hlt:hiEff}   {\ensuremath{{0.664 } } }
\vdef{default-11:NoMc3e33-A:hlt:hiEffE}   {\ensuremath{{0.004 } } }
\vdef{default-11:NoMc3e33-A:hlt:loDelta}   {\ensuremath{{-1.247 } } }
\vdef{default-11:NoMc3e33-A:hlt:loDeltaE}   {\ensuremath{{0.019 } } }
\vdef{default-11:NoMc3e33-A:hlt:hiDelta}   {\ensuremath{{+0.326 } } }
\vdef{default-11:NoMc3e33-A:hlt:hiDeltaE}   {\ensuremath{{0.006 } } }
\vdef{default-11:NoData-AR5:muonsid:loEff}   {\ensuremath{{0.150 } } }
\vdef{default-11:NoData-AR5:muonsid:loEffE}   {\ensuremath{{0.003 } } }
\vdef{default-11:NoData-AR5:muonsid:hiEff}   {\ensuremath{{0.850 } } }
\vdef{default-11:NoData-AR5:muonsid:hiEffE}   {\ensuremath{{0.003 } } }
\vdef{default-11:NoMc3e33-A:muonsid:loEff}   {\ensuremath{{0.157 } } }
\vdef{default-11:NoMc3e33-A:muonsid:loEffE}   {\ensuremath{{0.003 } } }
\vdef{default-11:NoMc3e33-A:muonsid:hiEff}   {\ensuremath{{0.843 } } }
\vdef{default-11:NoMc3e33-A:muonsid:hiEffE}   {\ensuremath{{0.003 } } }
\vdef{default-11:NoMc3e33-A:muonsid:loDelta}   {\ensuremath{{-0.049 } } }
\vdef{default-11:NoMc3e33-A:muonsid:loDeltaE}   {\ensuremath{{0.026 } } }
\vdef{default-11:NoMc3e33-A:muonsid:hiDelta}   {\ensuremath{{+0.009 } } }
\vdef{default-11:NoMc3e33-A:muonsid:hiDeltaE}   {\ensuremath{{0.005 } } }
\vdef{default-11:NoData-AR5:tracksqual:loEff}   {\ensuremath{{0.001 } } }
\vdef{default-11:NoData-AR5:tracksqual:loEffE}   {\ensuremath{{0.000 } } }
\vdef{default-11:NoData-AR5:tracksqual:hiEff}   {\ensuremath{{0.999 } } }
\vdef{default-11:NoData-AR5:tracksqual:hiEffE}   {\ensuremath{{0.000 } } }
\vdef{default-11:NoMc3e33-A:tracksqual:loEff}   {\ensuremath{{0.000 } } }
\vdef{default-11:NoMc3e33-A:tracksqual:loEffE}   {\ensuremath{{0.000 } } }
\vdef{default-11:NoMc3e33-A:tracksqual:hiEff}   {\ensuremath{{1.000 } } }
\vdef{default-11:NoMc3e33-A:tracksqual:hiEffE}   {\ensuremath{{0.000 } } }
\vdef{default-11:NoMc3e33-A:tracksqual:loDelta}   {\ensuremath{{+1.062 } } }
\vdef{default-11:NoMc3e33-A:tracksqual:loDeltaE}   {\ensuremath{{0.607 } } }
\vdef{default-11:NoMc3e33-A:tracksqual:hiDelta}   {\ensuremath{{-0.000 } } }
\vdef{default-11:NoMc3e33-A:tracksqual:hiDeltaE}   {\ensuremath{{0.000 } } }
\vdef{default-11:NoData-AR5:pvz:loEff}   {\ensuremath{{0.506 } } }
\vdef{default-11:NoData-AR5:pvz:loEffE}   {\ensuremath{{0.004 } } }
\vdef{default-11:NoData-AR5:pvz:hiEff}   {\ensuremath{{0.494 } } }
\vdef{default-11:NoData-AR5:pvz:hiEffE}   {\ensuremath{{0.004 } } }
\vdef{default-11:NoMc3e33-A:pvz:loEff}   {\ensuremath{{0.469 } } }
\vdef{default-11:NoMc3e33-A:pvz:loEffE}   {\ensuremath{{0.004 } } }
\vdef{default-11:NoMc3e33-A:pvz:hiEff}   {\ensuremath{{0.531 } } }
\vdef{default-11:NoMc3e33-A:pvz:hiEffE}   {\ensuremath{{0.004 } } }
\vdef{default-11:NoMc3e33-A:pvz:loDelta}   {\ensuremath{{+0.077 } } }
\vdef{default-11:NoMc3e33-A:pvz:loDeltaE}   {\ensuremath{{0.012 } } }
\vdef{default-11:NoMc3e33-A:pvz:hiDelta}   {\ensuremath{{-0.073 } } }
\vdef{default-11:NoMc3e33-A:pvz:hiDeltaE}   {\ensuremath{{0.011 } } }
\vdef{default-11:NoData-AR5:pvn:loEff}   {\ensuremath{{1.022 } } }
\vdef{default-11:NoData-AR5:pvn:loEffE}   {\ensuremath{{\mathrm{NaN} } } }
\vdef{default-11:NoData-AR5:pvn:hiEff}   {\ensuremath{{1.000 } } }
\vdef{default-11:NoData-AR5:pvn:hiEffE}   {\ensuremath{{0.000 } } }
\vdef{default-11:NoMc3e33-A:pvn:loEff}   {\ensuremath{{1.000 } } }
\vdef{default-11:NoMc3e33-A:pvn:loEffE}   {\ensuremath{{0.000 } } }
\vdef{default-11:NoMc3e33-A:pvn:hiEff}   {\ensuremath{{1.000 } } }
\vdef{default-11:NoMc3e33-A:pvn:hiEffE}   {\ensuremath{{0.000 } } }
\vdef{default-11:NoMc3e33-A:pvn:loDelta}   {\ensuremath{{+0.022 } } }
\vdef{default-11:NoMc3e33-A:pvn:loDeltaE}   {\ensuremath{{\mathrm{NaN} } } }
\vdef{default-11:NoMc3e33-A:pvn:hiDelta}   {\ensuremath{{+0.000 } } }
\vdef{default-11:NoMc3e33-A:pvn:hiDeltaE}   {\ensuremath{{0.000 } } }
\vdef{default-11:NoData-AR5:pvavew8:loEff}   {\ensuremath{{0.008 } } }
\vdef{default-11:NoData-AR5:pvavew8:loEffE}   {\ensuremath{{0.001 } } }
\vdef{default-11:NoData-AR5:pvavew8:hiEff}   {\ensuremath{{0.992 } } }
\vdef{default-11:NoData-AR5:pvavew8:hiEffE}   {\ensuremath{{0.001 } } }
\vdef{default-11:NoMc3e33-A:pvavew8:loEff}   {\ensuremath{{0.004 } } }
\vdef{default-11:NoMc3e33-A:pvavew8:loEffE}   {\ensuremath{{0.001 } } }
\vdef{default-11:NoMc3e33-A:pvavew8:hiEff}   {\ensuremath{{0.996 } } }
\vdef{default-11:NoMc3e33-A:pvavew8:hiEffE}   {\ensuremath{{0.001 } } }
\vdef{default-11:NoMc3e33-A:pvavew8:loDelta}   {\ensuremath{{+0.678 } } }
\vdef{default-11:NoMc3e33-A:pvavew8:loDeltaE}   {\ensuremath{{0.146 } } }
\vdef{default-11:NoMc3e33-A:pvavew8:hiDelta}   {\ensuremath{{-0.004 } } }
\vdef{default-11:NoMc3e33-A:pvavew8:hiDeltaE}   {\ensuremath{{0.001 } } }
\vdef{default-11:NoData-AR5:pvntrk:loEff}   {\ensuremath{{1.000 } } }
\vdef{default-11:NoData-AR5:pvntrk:loEffE}   {\ensuremath{{0.000 } } }
\vdef{default-11:NoData-AR5:pvntrk:hiEff}   {\ensuremath{{1.000 } } }
\vdef{default-11:NoData-AR5:pvntrk:hiEffE}   {\ensuremath{{0.000 } } }
\vdef{default-11:NoMc3e33-A:pvntrk:loEff}   {\ensuremath{{1.000 } } }
\vdef{default-11:NoMc3e33-A:pvntrk:loEffE}   {\ensuremath{{0.000 } } }
\vdef{default-11:NoMc3e33-A:pvntrk:hiEff}   {\ensuremath{{1.000 } } }
\vdef{default-11:NoMc3e33-A:pvntrk:hiEffE}   {\ensuremath{{0.000 } } }
\vdef{default-11:NoMc3e33-A:pvntrk:loDelta}   {\ensuremath{{+0.000 } } }
\vdef{default-11:NoMc3e33-A:pvntrk:loDeltaE}   {\ensuremath{{0.000 } } }
\vdef{default-11:NoMc3e33-A:pvntrk:hiDelta}   {\ensuremath{{+0.000 } } }
\vdef{default-11:NoMc3e33-A:pvntrk:hiDeltaE}   {\ensuremath{{0.000 } } }
\vdef{default-11:NoData-AR5:muon1pt:loEff}   {\ensuremath{{1.009 } } }
\vdef{default-11:NoData-AR5:muon1pt:loEffE}   {\ensuremath{{\mathrm{NaN} } } }
\vdef{default-11:NoData-AR5:muon1pt:hiEff}   {\ensuremath{{1.000 } } }
\vdef{default-11:NoData-AR5:muon1pt:hiEffE}   {\ensuremath{{0.000 } } }
\vdef{default-11:NoMc3e33-A:muon1pt:loEff}   {\ensuremath{{1.008 } } }
\vdef{default-11:NoMc3e33-A:muon1pt:loEffE}   {\ensuremath{{\mathrm{NaN} } } }
\vdef{default-11:NoMc3e33-A:muon1pt:hiEff}   {\ensuremath{{1.000 } } }
\vdef{default-11:NoMc3e33-A:muon1pt:hiEffE}   {\ensuremath{{0.000 } } }
\vdef{default-11:NoMc3e33-A:muon1pt:loDelta}   {\ensuremath{{+0.001 } } }
\vdef{default-11:NoMc3e33-A:muon1pt:loDeltaE}   {\ensuremath{{\mathrm{NaN} } } }
\vdef{default-11:NoMc3e33-A:muon1pt:hiDelta}   {\ensuremath{{+0.000 } } }
\vdef{default-11:NoMc3e33-A:muon1pt:hiDeltaE}   {\ensuremath{{0.000 } } }
\vdef{default-11:NoData-AR5:muon2pt:loEff}   {\ensuremath{{0.008 } } }
\vdef{default-11:NoData-AR5:muon2pt:loEffE}   {\ensuremath{{0.001 } } }
\vdef{default-11:NoData-AR5:muon2pt:hiEff}   {\ensuremath{{0.992 } } }
\vdef{default-11:NoData-AR5:muon2pt:hiEffE}   {\ensuremath{{0.001 } } }
\vdef{default-11:NoMc3e33-A:muon2pt:loEff}   {\ensuremath{{0.005 } } }
\vdef{default-11:NoMc3e33-A:muon2pt:loEffE}   {\ensuremath{{0.001 } } }
\vdef{default-11:NoMc3e33-A:muon2pt:hiEff}   {\ensuremath{{0.995 } } }
\vdef{default-11:NoMc3e33-A:muon2pt:hiEffE}   {\ensuremath{{0.001 } } }
\vdef{default-11:NoMc3e33-A:muon2pt:loDelta}   {\ensuremath{{+0.348 } } }
\vdef{default-11:NoMc3e33-A:muon2pt:loDeltaE}   {\ensuremath{{0.147 } } }
\vdef{default-11:NoMc3e33-A:muon2pt:hiDelta}   {\ensuremath{{-0.002 } } }
\vdef{default-11:NoMc3e33-A:muon2pt:hiDeltaE}   {\ensuremath{{0.001 } } }
\vdef{default-11:NoData-AR5:muonseta:loEff}   {\ensuremath{{0.746 } } }
\vdef{default-11:NoData-AR5:muonseta:loEffE}   {\ensuremath{{0.003 } } }
\vdef{default-11:NoData-AR5:muonseta:hiEff}   {\ensuremath{{0.254 } } }
\vdef{default-11:NoData-AR5:muonseta:hiEffE}   {\ensuremath{{0.003 } } }
\vdef{default-11:NoMc3e33-A:muonseta:loEff}   {\ensuremath{{0.738 } } }
\vdef{default-11:NoMc3e33-A:muonseta:loEffE}   {\ensuremath{{0.003 } } }
\vdef{default-11:NoMc3e33-A:muonseta:hiEff}   {\ensuremath{{0.262 } } }
\vdef{default-11:NoMc3e33-A:muonseta:hiEffE}   {\ensuremath{{0.003 } } }
\vdef{default-11:NoMc3e33-A:muonseta:loDelta}   {\ensuremath{{+0.011 } } }
\vdef{default-11:NoMc3e33-A:muonseta:loDeltaE}   {\ensuremath{{0.005 } } }
\vdef{default-11:NoMc3e33-A:muonseta:hiDelta}   {\ensuremath{{-0.032 } } }
\vdef{default-11:NoMc3e33-A:muonseta:hiDeltaE}   {\ensuremath{{0.015 } } }
\vdef{default-11:NoData-AR5:pt:loEff}   {\ensuremath{{0.000 } } }
\vdef{default-11:NoData-AR5:pt:loEffE}   {\ensuremath{{0.000 } } }
\vdef{default-11:NoData-AR5:pt:hiEff}   {\ensuremath{{1.000 } } }
\vdef{default-11:NoData-AR5:pt:hiEffE}   {\ensuremath{{0.000 } } }
\vdef{default-11:NoMc3e33-A:pt:loEff}   {\ensuremath{{0.000 } } }
\vdef{default-11:NoMc3e33-A:pt:loEffE}   {\ensuremath{{0.000 } } }
\vdef{default-11:NoMc3e33-A:pt:hiEff}   {\ensuremath{{1.000 } } }
\vdef{default-11:NoMc3e33-A:pt:hiEffE}   {\ensuremath{{0.000 } } }
\vdef{default-11:NoMc3e33-A:pt:loDelta}   {\ensuremath{{\mathrm{NaN} } } }
\vdef{default-11:NoMc3e33-A:pt:loDeltaE}   {\ensuremath{{\mathrm{NaN} } } }
\vdef{default-11:NoMc3e33-A:pt:hiDelta}   {\ensuremath{{+0.000 } } }
\vdef{default-11:NoMc3e33-A:pt:hiDeltaE}   {\ensuremath{{0.000 } } }
\vdef{default-11:NoData-AR5:p:loEff}   {\ensuremath{{1.010 } } }
\vdef{default-11:NoData-AR5:p:loEffE}   {\ensuremath{{\mathrm{NaN} } } }
\vdef{default-11:NoData-AR5:p:hiEff}   {\ensuremath{{1.000 } } }
\vdef{default-11:NoData-AR5:p:hiEffE}   {\ensuremath{{0.000 } } }
\vdef{default-11:NoMc3e33-A:p:loEff}   {\ensuremath{{1.011 } } }
\vdef{default-11:NoMc3e33-A:p:loEffE}   {\ensuremath{{\mathrm{NaN} } } }
\vdef{default-11:NoMc3e33-A:p:hiEff}   {\ensuremath{{1.000 } } }
\vdef{default-11:NoMc3e33-A:p:hiEffE}   {\ensuremath{{0.000 } } }
\vdef{default-11:NoMc3e33-A:p:loDelta}   {\ensuremath{{-0.000 } } }
\vdef{default-11:NoMc3e33-A:p:loDeltaE}   {\ensuremath{{\mathrm{NaN} } } }
\vdef{default-11:NoMc3e33-A:p:hiDelta}   {\ensuremath{{+0.000 } } }
\vdef{default-11:NoMc3e33-A:p:hiDeltaE}   {\ensuremath{{0.000 } } }
\vdef{default-11:NoData-AR5:eta:loEff}   {\ensuremath{{0.735 } } }
\vdef{default-11:NoData-AR5:eta:loEffE}   {\ensuremath{{0.004 } } }
\vdef{default-11:NoData-AR5:eta:hiEff}   {\ensuremath{{0.265 } } }
\vdef{default-11:NoData-AR5:eta:hiEffE}   {\ensuremath{{0.004 } } }
\vdef{default-11:NoMc3e33-A:eta:loEff}   {\ensuremath{{0.729 } } }
\vdef{default-11:NoMc3e33-A:eta:loEffE}   {\ensuremath{{0.004 } } }
\vdef{default-11:NoMc3e33-A:eta:hiEff}   {\ensuremath{{0.271 } } }
\vdef{default-11:NoMc3e33-A:eta:hiEffE}   {\ensuremath{{0.004 } } }
\vdef{default-11:NoMc3e33-A:eta:loDelta}   {\ensuremath{{+0.008 } } }
\vdef{default-11:NoMc3e33-A:eta:loDeltaE}   {\ensuremath{{0.007 } } }
\vdef{default-11:NoMc3e33-A:eta:hiDelta}   {\ensuremath{{-0.023 } } }
\vdef{default-11:NoMc3e33-A:eta:hiDeltaE}   {\ensuremath{{0.020 } } }
\vdef{default-11:NoData-AR5:bdt:loEff}   {\ensuremath{{0.910 } } }
\vdef{default-11:NoData-AR5:bdt:loEffE}   {\ensuremath{{0.002 } } }
\vdef{default-11:NoData-AR5:bdt:hiEff}   {\ensuremath{{0.090 } } }
\vdef{default-11:NoData-AR5:bdt:hiEffE}   {\ensuremath{{0.002 } } }
\vdef{default-11:NoMc3e33-A:bdt:loEff}   {\ensuremath{{0.914 } } }
\vdef{default-11:NoMc3e33-A:bdt:loEffE}   {\ensuremath{{0.002 } } }
\vdef{default-11:NoMc3e33-A:bdt:hiEff}   {\ensuremath{{0.086 } } }
\vdef{default-11:NoMc3e33-A:bdt:hiEffE}   {\ensuremath{{0.002 } } }
\vdef{default-11:NoMc3e33-A:bdt:loDelta}   {\ensuremath{{-0.004 } } }
\vdef{default-11:NoMc3e33-A:bdt:loDeltaE}   {\ensuremath{{0.004 } } }
\vdef{default-11:NoMc3e33-A:bdt:hiDelta}   {\ensuremath{{+0.042 } } }
\vdef{default-11:NoMc3e33-A:bdt:hiDeltaE}   {\ensuremath{{0.038 } } }
\vdef{default-11:NoData-AR5:fl3d:loEff}   {\ensuremath{{0.843 } } }
\vdef{default-11:NoData-AR5:fl3d:loEffE}   {\ensuremath{{0.003 } } }
\vdef{default-11:NoData-AR5:fl3d:hiEff}   {\ensuremath{{0.157 } } }
\vdef{default-11:NoData-AR5:fl3d:hiEffE}   {\ensuremath{{0.003 } } }
\vdef{default-11:NoMc3e33-A:fl3d:loEff}   {\ensuremath{{0.835 } } }
\vdef{default-11:NoMc3e33-A:fl3d:loEffE}   {\ensuremath{{0.003 } } }
\vdef{default-11:NoMc3e33-A:fl3d:hiEff}   {\ensuremath{{0.165 } } }
\vdef{default-11:NoMc3e33-A:fl3d:hiEffE}   {\ensuremath{{0.003 } } }
\vdef{default-11:NoMc3e33-A:fl3d:loDelta}   {\ensuremath{{+0.009 } } }
\vdef{default-11:NoMc3e33-A:fl3d:loDeltaE}   {\ensuremath{{0.005 } } }
\vdef{default-11:NoMc3e33-A:fl3d:hiDelta}   {\ensuremath{{-0.049 } } }
\vdef{default-11:NoMc3e33-A:fl3d:hiDeltaE}   {\ensuremath{{0.026 } } }
\vdef{default-11:NoData-AR5:fl3de:loEff}   {\ensuremath{{1.000 } } }
\vdef{default-11:NoData-AR5:fl3de:loEffE}   {\ensuremath{{0.000 } } }
\vdef{default-11:NoData-AR5:fl3de:hiEff}   {\ensuremath{{0.000 } } }
\vdef{default-11:NoData-AR5:fl3de:hiEffE}   {\ensuremath{{0.000 } } }
\vdef{default-11:NoMc3e33-A:fl3de:loEff}   {\ensuremath{{1.000 } } }
\vdef{default-11:NoMc3e33-A:fl3de:loEffE}   {\ensuremath{{0.000 } } }
\vdef{default-11:NoMc3e33-A:fl3de:hiEff}   {\ensuremath{{0.000 } } }
\vdef{default-11:NoMc3e33-A:fl3de:hiEffE}   {\ensuremath{{0.000 } } }
\vdef{default-11:NoMc3e33-A:fl3de:loDelta}   {\ensuremath{{+0.000 } } }
\vdef{default-11:NoMc3e33-A:fl3de:loDeltaE}   {\ensuremath{{0.000 } } }
\vdef{default-11:NoMc3e33-A:fl3de:hiDelta}   {\ensuremath{{+1.136 } } }
\vdef{default-11:NoMc3e33-A:fl3de:hiDeltaE}   {\ensuremath{{2.131 } } }
\vdef{default-11:NoData-AR5:fls3d:loEff}   {\ensuremath{{0.070 } } }
\vdef{default-11:NoData-AR5:fls3d:loEffE}   {\ensuremath{{0.002 } } }
\vdef{default-11:NoData-AR5:fls3d:hiEff}   {\ensuremath{{0.930 } } }
\vdef{default-11:NoData-AR5:fls3d:hiEffE}   {\ensuremath{{0.002 } } }
\vdef{default-11:NoMc3e33-A:fls3d:loEff}   {\ensuremath{{0.074 } } }
\vdef{default-11:NoMc3e33-A:fls3d:loEffE}   {\ensuremath{{0.002 } } }
\vdef{default-11:NoMc3e33-A:fls3d:hiEff}   {\ensuremath{{0.926 } } }
\vdef{default-11:NoMc3e33-A:fls3d:hiEffE}   {\ensuremath{{0.002 } } }
\vdef{default-11:NoMc3e33-A:fls3d:loDelta}   {\ensuremath{{-0.064 } } }
\vdef{default-11:NoMc3e33-A:fls3d:loDeltaE}   {\ensuremath{{0.041 } } }
\vdef{default-11:NoMc3e33-A:fls3d:hiDelta}   {\ensuremath{{+0.005 } } }
\vdef{default-11:NoMc3e33-A:fls3d:hiDeltaE}   {\ensuremath{{0.003 } } }
\vdef{default-11:NoData-AR5:flsxy:loEff}   {\ensuremath{{1.012 } } }
\vdef{default-11:NoData-AR5:flsxy:loEffE}   {\ensuremath{{\mathrm{NaN} } } }
\vdef{default-11:NoData-AR5:flsxy:hiEff}   {\ensuremath{{1.000 } } }
\vdef{default-11:NoData-AR5:flsxy:hiEffE}   {\ensuremath{{0.000 } } }
\vdef{default-11:NoMc3e33-A:flsxy:loEff}   {\ensuremath{{1.012 } } }
\vdef{default-11:NoMc3e33-A:flsxy:loEffE}   {\ensuremath{{\mathrm{NaN} } } }
\vdef{default-11:NoMc3e33-A:flsxy:hiEff}   {\ensuremath{{1.000 } } }
\vdef{default-11:NoMc3e33-A:flsxy:hiEffE}   {\ensuremath{{0.000 } } }
\vdef{default-11:NoMc3e33-A:flsxy:loDelta}   {\ensuremath{{+0.000 } } }
\vdef{default-11:NoMc3e33-A:flsxy:loDeltaE}   {\ensuremath{{\mathrm{NaN} } } }
\vdef{default-11:NoMc3e33-A:flsxy:hiDelta}   {\ensuremath{{+0.000 } } }
\vdef{default-11:NoMc3e33-A:flsxy:hiDeltaE}   {\ensuremath{{0.000 } } }
\vdef{default-11:NoData-AR5:chi2dof:loEff}   {\ensuremath{{0.925 } } }
\vdef{default-11:NoData-AR5:chi2dof:loEffE}   {\ensuremath{{0.002 } } }
\vdef{default-11:NoData-AR5:chi2dof:hiEff}   {\ensuremath{{0.075 } } }
\vdef{default-11:NoData-AR5:chi2dof:hiEffE}   {\ensuremath{{0.002 } } }
\vdef{default-11:NoMc3e33-A:chi2dof:loEff}   {\ensuremath{{0.944 } } }
\vdef{default-11:NoMc3e33-A:chi2dof:loEffE}   {\ensuremath{{0.002 } } }
\vdef{default-11:NoMc3e33-A:chi2dof:hiEff}   {\ensuremath{{0.056 } } }
\vdef{default-11:NoMc3e33-A:chi2dof:hiEffE}   {\ensuremath{{0.002 } } }
\vdef{default-11:NoMc3e33-A:chi2dof:loDelta}   {\ensuremath{{-0.020 } } }
\vdef{default-11:NoMc3e33-A:chi2dof:loDeltaE}   {\ensuremath{{0.003 } } }
\vdef{default-11:NoMc3e33-A:chi2dof:hiDelta}   {\ensuremath{{+0.287 } } }
\vdef{default-11:NoMc3e33-A:chi2dof:hiDeltaE}   {\ensuremath{{0.044 } } }
\vdef{default-11:NoData-AR5:pchi2dof:loEff}   {\ensuremath{{0.647 } } }
\vdef{default-11:NoData-AR5:pchi2dof:loEffE}   {\ensuremath{{0.004 } } }
\vdef{default-11:NoData-AR5:pchi2dof:hiEff}   {\ensuremath{{0.353 } } }
\vdef{default-11:NoData-AR5:pchi2dof:hiEffE}   {\ensuremath{{0.004 } } }
\vdef{default-11:NoMc3e33-A:pchi2dof:loEff}   {\ensuremath{{0.607 } } }
\vdef{default-11:NoMc3e33-A:pchi2dof:loEffE}   {\ensuremath{{0.004 } } }
\vdef{default-11:NoMc3e33-A:pchi2dof:hiEff}   {\ensuremath{{0.393 } } }
\vdef{default-11:NoMc3e33-A:pchi2dof:hiEffE}   {\ensuremath{{0.004 } } }
\vdef{default-11:NoMc3e33-A:pchi2dof:loDelta}   {\ensuremath{{+0.064 } } }
\vdef{default-11:NoMc3e33-A:pchi2dof:loDeltaE}   {\ensuremath{{0.009 } } }
\vdef{default-11:NoMc3e33-A:pchi2dof:hiDelta}   {\ensuremath{{-0.107 } } }
\vdef{default-11:NoMc3e33-A:pchi2dof:hiDeltaE}   {\ensuremath{{0.015 } } }
\vdef{default-11:NoData-AR5:alpha:loEff}   {\ensuremath{{0.994 } } }
\vdef{default-11:NoData-AR5:alpha:loEffE}   {\ensuremath{{0.001 } } }
\vdef{default-11:NoData-AR5:alpha:hiEff}   {\ensuremath{{0.006 } } }
\vdef{default-11:NoData-AR5:alpha:hiEffE}   {\ensuremath{{0.001 } } }
\vdef{default-11:NoMc3e33-A:alpha:loEff}   {\ensuremath{{0.994 } } }
\vdef{default-11:NoMc3e33-A:alpha:loEffE}   {\ensuremath{{0.001 } } }
\vdef{default-11:NoMc3e33-A:alpha:hiEff}   {\ensuremath{{0.006 } } }
\vdef{default-11:NoMc3e33-A:alpha:hiEffE}   {\ensuremath{{0.001 } } }
\vdef{default-11:NoMc3e33-A:alpha:loDelta}   {\ensuremath{{-0.000 } } }
\vdef{default-11:NoMc3e33-A:alpha:loDeltaE}   {\ensuremath{{0.001 } } }
\vdef{default-11:NoMc3e33-A:alpha:hiDelta}   {\ensuremath{{+0.063 } } }
\vdef{default-11:NoMc3e33-A:alpha:hiDeltaE}   {\ensuremath{{0.153 } } }
\vdef{default-11:NoData-AR5:iso:loEff}   {\ensuremath{{0.129 } } }
\vdef{default-11:NoData-AR5:iso:loEffE}   {\ensuremath{{0.003 } } }
\vdef{default-11:NoData-AR5:iso:hiEff}   {\ensuremath{{0.871 } } }
\vdef{default-11:NoData-AR5:iso:hiEffE}   {\ensuremath{{0.003 } } }
\vdef{default-11:NoMc3e33-A:iso:loEff}   {\ensuremath{{0.107 } } }
\vdef{default-11:NoMc3e33-A:iso:loEffE}   {\ensuremath{{0.002 } } }
\vdef{default-11:NoMc3e33-A:iso:hiEff}   {\ensuremath{{0.893 } } }
\vdef{default-11:NoMc3e33-A:iso:hiEffE}   {\ensuremath{{0.002 } } }
\vdef{default-11:NoMc3e33-A:iso:loDelta}   {\ensuremath{{+0.183 } } }
\vdef{default-11:NoMc3e33-A:iso:loDeltaE}   {\ensuremath{{0.031 } } }
\vdef{default-11:NoMc3e33-A:iso:hiDelta}   {\ensuremath{{-0.025 } } }
\vdef{default-11:NoMc3e33-A:iso:hiDeltaE}   {\ensuremath{{0.004 } } }
\vdef{default-11:NoData-AR5:docatrk:loEff}   {\ensuremath{{0.074 } } }
\vdef{default-11:NoData-AR5:docatrk:loEffE}   {\ensuremath{{0.002 } } }
\vdef{default-11:NoData-AR5:docatrk:hiEff}   {\ensuremath{{0.926 } } }
\vdef{default-11:NoData-AR5:docatrk:hiEffE}   {\ensuremath{{0.002 } } }
\vdef{default-11:NoMc3e33-A:docatrk:loEff}   {\ensuremath{{0.083 } } }
\vdef{default-11:NoMc3e33-A:docatrk:loEffE}   {\ensuremath{{0.002 } } }
\vdef{default-11:NoMc3e33-A:docatrk:hiEff}   {\ensuremath{{0.917 } } }
\vdef{default-11:NoMc3e33-A:docatrk:hiEffE}   {\ensuremath{{0.002 } } }
\vdef{default-11:NoMc3e33-A:docatrk:loDelta}   {\ensuremath{{-0.121 } } }
\vdef{default-11:NoMc3e33-A:docatrk:loDeltaE}   {\ensuremath{{0.040 } } }
\vdef{default-11:NoMc3e33-A:docatrk:hiDelta}   {\ensuremath{{+0.010 } } }
\vdef{default-11:NoMc3e33-A:docatrk:hiDeltaE}   {\ensuremath{{0.003 } } }
\vdef{default-11:NoData-AR5:isotrk:loEff}   {\ensuremath{{1.000 } } }
\vdef{default-11:NoData-AR5:isotrk:loEffE}   {\ensuremath{{0.000 } } }
\vdef{default-11:NoData-AR5:isotrk:hiEff}   {\ensuremath{{1.000 } } }
\vdef{default-11:NoData-AR5:isotrk:hiEffE}   {\ensuremath{{0.000 } } }
\vdef{default-11:NoMc3e33-A:isotrk:loEff}   {\ensuremath{{1.000 } } }
\vdef{default-11:NoMc3e33-A:isotrk:loEffE}   {\ensuremath{{0.000 } } }
\vdef{default-11:NoMc3e33-A:isotrk:hiEff}   {\ensuremath{{1.000 } } }
\vdef{default-11:NoMc3e33-A:isotrk:hiEffE}   {\ensuremath{{0.000 } } }
\vdef{default-11:NoMc3e33-A:isotrk:loDelta}   {\ensuremath{{+0.000 } } }
\vdef{default-11:NoMc3e33-A:isotrk:loDeltaE}   {\ensuremath{{0.000 } } }
\vdef{default-11:NoMc3e33-A:isotrk:hiDelta}   {\ensuremath{{+0.000 } } }
\vdef{default-11:NoMc3e33-A:isotrk:hiDeltaE}   {\ensuremath{{0.000 } } }
\vdef{default-11:NoData-AR5:closetrk:loEff}   {\ensuremath{{0.972 } } }
\vdef{default-11:NoData-AR5:closetrk:loEffE}   {\ensuremath{{0.001 } } }
\vdef{default-11:NoData-AR5:closetrk:hiEff}   {\ensuremath{{0.028 } } }
\vdef{default-11:NoData-AR5:closetrk:hiEffE}   {\ensuremath{{0.001 } } }
\vdef{default-11:NoMc3e33-A:closetrk:loEff}   {\ensuremath{{0.978 } } }
\vdef{default-11:NoMc3e33-A:closetrk:loEffE}   {\ensuremath{{0.001 } } }
\vdef{default-11:NoMc3e33-A:closetrk:hiEff}   {\ensuremath{{0.022 } } }
\vdef{default-11:NoMc3e33-A:closetrk:hiEffE}   {\ensuremath{{0.001 } } }
\vdef{default-11:NoMc3e33-A:closetrk:loDelta}   {\ensuremath{{-0.006 } } }
\vdef{default-11:NoMc3e33-A:closetrk:loDeltaE}   {\ensuremath{{0.002 } } }
\vdef{default-11:NoMc3e33-A:closetrk:hiDelta}   {\ensuremath{{+0.219 } } }
\vdef{default-11:NoMc3e33-A:closetrk:hiDeltaE}   {\ensuremath{{0.074 } } }
\vdef{default-11:NoData-AR5:lip:loEff}   {\ensuremath{{1.000 } } }
\vdef{default-11:NoData-AR5:lip:loEffE}   {\ensuremath{{0.000 } } }
\vdef{default-11:NoData-AR5:lip:hiEff}   {\ensuremath{{0.000 } } }
\vdef{default-11:NoData-AR5:lip:hiEffE}   {\ensuremath{{0.000 } } }
\vdef{default-11:NoMc3e33-A:lip:loEff}   {\ensuremath{{1.000 } } }
\vdef{default-11:NoMc3e33-A:lip:loEffE}   {\ensuremath{{0.000 } } }
\vdef{default-11:NoMc3e33-A:lip:hiEff}   {\ensuremath{{0.000 } } }
\vdef{default-11:NoMc3e33-A:lip:hiEffE}   {\ensuremath{{0.000 } } }
\vdef{default-11:NoMc3e33-A:lip:loDelta}   {\ensuremath{{+0.000 } } }
\vdef{default-11:NoMc3e33-A:lip:loDeltaE}   {\ensuremath{{0.000 } } }
\vdef{default-11:NoMc3e33-A:lip:hiDelta}   {\ensuremath{{\mathrm{NaN} } } }
\vdef{default-11:NoMc3e33-A:lip:hiDeltaE}   {\ensuremath{{\mathrm{NaN} } } }
\vdef{default-11:NoData-AR5:lip:inEff}   {\ensuremath{{1.000 } } }
\vdef{default-11:NoData-AR5:lip:inEffE}   {\ensuremath{{0.000 } } }
\vdef{default-11:NoMc3e33-A:lip:inEff}   {\ensuremath{{1.000 } } }
\vdef{default-11:NoMc3e33-A:lip:inEffE}   {\ensuremath{{0.000 } } }
\vdef{default-11:NoMc3e33-A:lip:inDelta}   {\ensuremath{{+0.000 } } }
\vdef{default-11:NoMc3e33-A:lip:inDeltaE}   {\ensuremath{{0.000 } } }
\vdef{default-11:NoData-AR5:lips:loEff}   {\ensuremath{{1.000 } } }
\vdef{default-11:NoData-AR5:lips:loEffE}   {\ensuremath{{0.000 } } }
\vdef{default-11:NoData-AR5:lips:hiEff}   {\ensuremath{{0.000 } } }
\vdef{default-11:NoData-AR5:lips:hiEffE}   {\ensuremath{{0.000 } } }
\vdef{default-11:NoMc3e33-A:lips:loEff}   {\ensuremath{{1.000 } } }
\vdef{default-11:NoMc3e33-A:lips:loEffE}   {\ensuremath{{0.000 } } }
\vdef{default-11:NoMc3e33-A:lips:hiEff}   {\ensuremath{{0.000 } } }
\vdef{default-11:NoMc3e33-A:lips:hiEffE}   {\ensuremath{{0.000 } } }
\vdef{default-11:NoMc3e33-A:lips:loDelta}   {\ensuremath{{+0.000 } } }
\vdef{default-11:NoMc3e33-A:lips:loDeltaE}   {\ensuremath{{0.000 } } }
\vdef{default-11:NoMc3e33-A:lips:hiDelta}   {\ensuremath{{\mathrm{NaN} } } }
\vdef{default-11:NoMc3e33-A:lips:hiDeltaE}   {\ensuremath{{\mathrm{NaN} } } }
\vdef{default-11:NoData-AR5:lips:inEff}   {\ensuremath{{1.000 } } }
\vdef{default-11:NoData-AR5:lips:inEffE}   {\ensuremath{{0.000 } } }
\vdef{default-11:NoMc3e33-A:lips:inEff}   {\ensuremath{{1.000 } } }
\vdef{default-11:NoMc3e33-A:lips:inEffE}   {\ensuremath{{0.000 } } }
\vdef{default-11:NoMc3e33-A:lips:inDelta}   {\ensuremath{{+0.000 } } }
\vdef{default-11:NoMc3e33-A:lips:inDeltaE}   {\ensuremath{{0.000 } } }
\vdef{default-11:NoData-AR5:ip:loEff}   {\ensuremath{{0.971 } } }
\vdef{default-11:NoData-AR5:ip:loEffE}   {\ensuremath{{0.001 } } }
\vdef{default-11:NoData-AR5:ip:hiEff}   {\ensuremath{{0.029 } } }
\vdef{default-11:NoData-AR5:ip:hiEffE}   {\ensuremath{{0.001 } } }
\vdef{default-11:NoMc3e33-A:ip:loEff}   {\ensuremath{{0.972 } } }
\vdef{default-11:NoMc3e33-A:ip:loEffE}   {\ensuremath{{0.001 } } }
\vdef{default-11:NoMc3e33-A:ip:hiEff}   {\ensuremath{{0.028 } } }
\vdef{default-11:NoMc3e33-A:ip:hiEffE}   {\ensuremath{{0.001 } } }
\vdef{default-11:NoMc3e33-A:ip:loDelta}   {\ensuremath{{-0.001 } } }
\vdef{default-11:NoMc3e33-A:ip:loDeltaE}   {\ensuremath{{0.002 } } }
\vdef{default-11:NoMc3e33-A:ip:hiDelta}   {\ensuremath{{+0.029 } } }
\vdef{default-11:NoMc3e33-A:ip:hiDeltaE}   {\ensuremath{{0.070 } } }
\vdef{default-11:NoData-AR5:ips:loEff}   {\ensuremath{{0.939 } } }
\vdef{default-11:NoData-AR5:ips:loEffE}   {\ensuremath{{0.002 } } }
\vdef{default-11:NoData-AR5:ips:hiEff}   {\ensuremath{{0.061 } } }
\vdef{default-11:NoData-AR5:ips:hiEffE}   {\ensuremath{{0.002 } } }
\vdef{default-11:NoMc3e33-A:ips:loEff}   {\ensuremath{{0.958 } } }
\vdef{default-11:NoMc3e33-A:ips:loEffE}   {\ensuremath{{0.002 } } }
\vdef{default-11:NoMc3e33-A:ips:hiEff}   {\ensuremath{{0.042 } } }
\vdef{default-11:NoMc3e33-A:ips:hiEffE}   {\ensuremath{{0.002 } } }
\vdef{default-11:NoMc3e33-A:ips:loDelta}   {\ensuremath{{-0.020 } } }
\vdef{default-11:NoMc3e33-A:ips:loDeltaE}   {\ensuremath{{0.003 } } }
\vdef{default-11:NoMc3e33-A:ips:hiDelta}   {\ensuremath{{+0.369 } } }
\vdef{default-11:NoMc3e33-A:ips:hiDeltaE}   {\ensuremath{{0.050 } } }
\vdef{default-11:NoData-AR5:maxdoca:loEff}   {\ensuremath{{1.000 } } }
\vdef{default-11:NoData-AR5:maxdoca:loEffE}   {\ensuremath{{0.000 } } }
\vdef{default-11:NoData-AR5:maxdoca:hiEff}   {\ensuremath{{0.014 } } }
\vdef{default-11:NoData-AR5:maxdoca:hiEffE}   {\ensuremath{{0.001 } } }
\vdef{default-11:NoMc3e33-A:maxdoca:loEff}   {\ensuremath{{1.000 } } }
\vdef{default-11:NoMc3e33-A:maxdoca:loEffE}   {\ensuremath{{0.000 } } }
\vdef{default-11:NoMc3e33-A:maxdoca:hiEff}   {\ensuremath{{0.011 } } }
\vdef{default-11:NoMc3e33-A:maxdoca:hiEffE}   {\ensuremath{{0.001 } } }
\vdef{default-11:NoMc3e33-A:maxdoca:loDelta}   {\ensuremath{{+0.000 } } }
\vdef{default-11:NoMc3e33-A:maxdoca:loDeltaE}   {\ensuremath{{0.000 } } }
\vdef{default-11:NoMc3e33-A:maxdoca:hiDelta}   {\ensuremath{{+0.261 } } }
\vdef{default-11:NoMc3e33-A:maxdoca:hiDeltaE}   {\ensuremath{{0.108 } } }
\vdef{default-11:NoData-AR5:kaonpt:loEff}   {\ensuremath{{1.011 } } }
\vdef{default-11:NoData-AR5:kaonpt:loEffE}   {\ensuremath{{\mathrm{NaN} } } }
\vdef{default-11:NoData-AR5:kaonpt:hiEff}   {\ensuremath{{1.000 } } }
\vdef{default-11:NoData-AR5:kaonpt:hiEffE}   {\ensuremath{{0.000 } } }
\vdef{default-11:NoMc3e33-A:kaonpt:loEff}   {\ensuremath{{1.007 } } }
\vdef{default-11:NoMc3e33-A:kaonpt:loEffE}   {\ensuremath{{\mathrm{NaN} } } }
\vdef{default-11:NoMc3e33-A:kaonpt:hiEff}   {\ensuremath{{1.000 } } }
\vdef{default-11:NoMc3e33-A:kaonpt:hiEffE}   {\ensuremath{{0.000 } } }
\vdef{default-11:NoMc3e33-A:kaonpt:loDelta}   {\ensuremath{{+0.003 } } }
\vdef{default-11:NoMc3e33-A:kaonpt:loDeltaE}   {\ensuremath{{\mathrm{NaN} } } }
\vdef{default-11:NoMc3e33-A:kaonpt:hiDelta}   {\ensuremath{{+0.000 } } }
\vdef{default-11:NoMc3e33-A:kaonpt:hiDeltaE}   {\ensuremath{{0.000 } } }
\vdef{default-11:NoData-AR5:psipt:loEff}   {\ensuremath{{1.004 } } }
\vdef{default-11:NoData-AR5:psipt:loEffE}   {\ensuremath{{\mathrm{NaN} } } }
\vdef{default-11:NoData-AR5:psipt:hiEff}   {\ensuremath{{1.000 } } }
\vdef{default-11:NoData-AR5:psipt:hiEffE}   {\ensuremath{{0.000 } } }
\vdef{default-11:NoMc3e33-A:psipt:loEff}   {\ensuremath{{1.003 } } }
\vdef{default-11:NoMc3e33-A:psipt:loEffE}   {\ensuremath{{\mathrm{NaN} } } }
\vdef{default-11:NoMc3e33-A:psipt:hiEff}   {\ensuremath{{1.000 } } }
\vdef{default-11:NoMc3e33-A:psipt:hiEffE}   {\ensuremath{{0.000 } } }
\vdef{default-11:NoMc3e33-A:psipt:loDelta}   {\ensuremath{{+0.001 } } }
\vdef{default-11:NoMc3e33-A:psipt:loDeltaE}   {\ensuremath{{\mathrm{NaN} } } }
\vdef{default-11:NoMc3e33-A:psipt:hiDelta}   {\ensuremath{{+0.000 } } }
\vdef{default-11:NoMc3e33-A:psipt:hiDeltaE}   {\ensuremath{{0.000 } } }
\vdef{default-11:CsData-A:osiso:loEff}   {\ensuremath{{1.006 } } }
\vdef{default-11:CsData-A:osiso:loEffE}   {\ensuremath{{\mathrm{NaN} } } }
\vdef{default-11:CsData-A:osiso:hiEff}   {\ensuremath{{1.000 } } }
\vdef{default-11:CsData-A:osiso:hiEffE}   {\ensuremath{{0.000 } } }
\vdef{default-11:CsMc-A:osiso:loEff}   {\ensuremath{{1.004 } } }
\vdef{default-11:CsMc-A:osiso:loEffE}   {\ensuremath{{\mathrm{NaN} } } }
\vdef{default-11:CsMc-A:osiso:hiEff}   {\ensuremath{{1.000 } } }
\vdef{default-11:CsMc-A:osiso:hiEffE}   {\ensuremath{{0.000 } } }
\vdef{default-11:CsMc-A:osiso:loDelta}   {\ensuremath{{+0.002 } } }
\vdef{default-11:CsMc-A:osiso:loDeltaE}   {\ensuremath{{\mathrm{NaN} } } }
\vdef{default-11:CsMc-A:osiso:hiDelta}   {\ensuremath{{+0.000 } } }
\vdef{default-11:CsMc-A:osiso:hiDeltaE}   {\ensuremath{{0.000 } } }
\vdef{default-11:CsData-A:osreliso:loEff}   {\ensuremath{{0.272 } } }
\vdef{default-11:CsData-A:osreliso:loEffE}   {\ensuremath{{0.005 } } }
\vdef{default-11:CsData-A:osreliso:hiEff}   {\ensuremath{{0.728 } } }
\vdef{default-11:CsData-A:osreliso:hiEffE}   {\ensuremath{{0.005 } } }
\vdef{default-11:CsMc-A:osreliso:loEff}   {\ensuremath{{0.304 } } }
\vdef{default-11:CsMc-A:osreliso:loEffE}   {\ensuremath{{0.005 } } }
\vdef{default-11:CsMc-A:osreliso:hiEff}   {\ensuremath{{0.696 } } }
\vdef{default-11:CsMc-A:osreliso:hiEffE}   {\ensuremath{{0.005 } } }
\vdef{default-11:CsMc-A:osreliso:loDelta}   {\ensuremath{{-0.114 } } }
\vdef{default-11:CsMc-A:osreliso:loDeltaE}   {\ensuremath{{0.023 } } }
\vdef{default-11:CsMc-A:osreliso:hiDelta}   {\ensuremath{{+0.046 } } }
\vdef{default-11:CsMc-A:osreliso:hiDeltaE}   {\ensuremath{{0.009 } } }
\vdef{default-11:CsData-A:osmuonpt:loEff}   {\ensuremath{{0.000 } } }
\vdef{default-11:CsData-A:osmuonpt:loEffE}   {\ensuremath{{0.003 } } }
\vdef{default-11:CsData-A:osmuonpt:hiEff}   {\ensuremath{{1.000 } } }
\vdef{default-11:CsData-A:osmuonpt:hiEffE}   {\ensuremath{{0.003 } } }
\vdef{default-11:CsMc-A:osmuonpt:loEff}   {\ensuremath{{0.000 } } }
\vdef{default-11:CsMc-A:osmuonpt:loEffE}   {\ensuremath{{0.003 } } }
\vdef{default-11:CsMc-A:osmuonpt:hiEff}   {\ensuremath{{1.000 } } }
\vdef{default-11:CsMc-A:osmuonpt:hiEffE}   {\ensuremath{{0.003 } } }
\vdef{default-11:CsMc-A:osmuonpt:loDelta}   {\ensuremath{{\mathrm{NaN} } } }
\vdef{default-11:CsMc-A:osmuonpt:loDeltaE}   {\ensuremath{{\mathrm{NaN} } } }
\vdef{default-11:CsMc-A:osmuonpt:hiDelta}   {\ensuremath{{+0.000 } } }
\vdef{default-11:CsMc-A:osmuonpt:hiDeltaE}   {\ensuremath{{0.004 } } }
\vdef{default-11:CsData-A:osmuondr:loEff}   {\ensuremath{{0.009 } } }
\vdef{default-11:CsData-A:osmuondr:loEffE}   {\ensuremath{{0.005 } } }
\vdef{default-11:CsData-A:osmuondr:hiEff}   {\ensuremath{{0.991 } } }
\vdef{default-11:CsData-A:osmuondr:hiEffE}   {\ensuremath{{0.005 } } }
\vdef{default-11:CsMc-A:osmuondr:loEff}   {\ensuremath{{0.023 } } }
\vdef{default-11:CsMc-A:osmuondr:loEffE}   {\ensuremath{{0.008 } } }
\vdef{default-11:CsMc-A:osmuondr:hiEff}   {\ensuremath{{0.977 } } }
\vdef{default-11:CsMc-A:osmuondr:hiEffE}   {\ensuremath{{0.008 } } }
\vdef{default-11:CsMc-A:osmuondr:loDelta}   {\ensuremath{{-0.853 } } }
\vdef{default-11:CsMc-A:osmuondr:loDeltaE}   {\ensuremath{{0.535 } } }
\vdef{default-11:CsMc-A:osmuondr:hiDelta}   {\ensuremath{{+0.014 } } }
\vdef{default-11:CsMc-A:osmuondr:hiDeltaE}   {\ensuremath{{0.009 } } }
\vdef{default-11:CsData-A:hlt:loEff}   {\ensuremath{{0.065 } } }
\vdef{default-11:CsData-A:hlt:loEffE}   {\ensuremath{{0.003 } } }
\vdef{default-11:CsData-A:hlt:hiEff}   {\ensuremath{{0.935 } } }
\vdef{default-11:CsData-A:hlt:hiEffE}   {\ensuremath{{0.003 } } }
\vdef{default-11:CsMc-A:hlt:loEff}   {\ensuremath{{0.305 } } }
\vdef{default-11:CsMc-A:hlt:loEffE}   {\ensuremath{{0.005 } } }
\vdef{default-11:CsMc-A:hlt:hiEff}   {\ensuremath{{0.695 } } }
\vdef{default-11:CsMc-A:hlt:hiEffE}   {\ensuremath{{0.005 } } }
\vdef{default-11:CsMc-A:hlt:loDelta}   {\ensuremath{{-1.300 } } }
\vdef{default-11:CsMc-A:hlt:loDeltaE}   {\ensuremath{{0.025 } } }
\vdef{default-11:CsMc-A:hlt:hiDelta}   {\ensuremath{{+0.294 } } }
\vdef{default-11:CsMc-A:hlt:hiDeltaE}   {\ensuremath{{0.007 } } }
\vdef{default-11:CsData-A:muonsid:loEff}   {\ensuremath{{0.149 } } }
\vdef{default-11:CsData-A:muonsid:loEffE}   {\ensuremath{{0.004 } } }
\vdef{default-11:CsData-A:muonsid:hiEff}   {\ensuremath{{0.851 } } }
\vdef{default-11:CsData-A:muonsid:hiEffE}   {\ensuremath{{0.004 } } }
\vdef{default-11:CsMc-A:muonsid:loEff}   {\ensuremath{{0.146 } } }
\vdef{default-11:CsMc-A:muonsid:loEffE}   {\ensuremath{{0.004 } } }
\vdef{default-11:CsMc-A:muonsid:hiEff}   {\ensuremath{{0.854 } } }
\vdef{default-11:CsMc-A:muonsid:hiEffE}   {\ensuremath{{0.004 } } }
\vdef{default-11:CsMc-A:muonsid:loDelta}   {\ensuremath{{+0.018 } } }
\vdef{default-11:CsMc-A:muonsid:loDeltaE}   {\ensuremath{{0.034 } } }
\vdef{default-11:CsMc-A:muonsid:hiDelta}   {\ensuremath{{-0.003 } } }
\vdef{default-11:CsMc-A:muonsid:hiDeltaE}   {\ensuremath{{0.006 } } }
\vdef{default-11:CsData-A:tracksqual:loEff}   {\ensuremath{{0.001 } } }
\vdef{default-11:CsData-A:tracksqual:loEffE}   {\ensuremath{{0.000 } } }
\vdef{default-11:CsData-A:tracksqual:hiEff}   {\ensuremath{{0.999 } } }
\vdef{default-11:CsData-A:tracksqual:hiEffE}   {\ensuremath{{0.000 } } }
\vdef{default-11:CsMc-A:tracksqual:loEff}   {\ensuremath{{0.000 } } }
\vdef{default-11:CsMc-A:tracksqual:loEffE}   {\ensuremath{{0.000 } } }
\vdef{default-11:CsMc-A:tracksqual:hiEff}   {\ensuremath{{1.000 } } }
\vdef{default-11:CsMc-A:tracksqual:hiEffE}   {\ensuremath{{0.000 } } }
\vdef{default-11:CsMc-A:tracksqual:loDelta}   {\ensuremath{{+1.479 } } }
\vdef{default-11:CsMc-A:tracksqual:loDeltaE}   {\ensuremath{{0.520 } } }
\vdef{default-11:CsMc-A:tracksqual:hiDelta}   {\ensuremath{{-0.001 } } }
\vdef{default-11:CsMc-A:tracksqual:hiDeltaE}   {\ensuremath{{0.000 } } }
\vdef{default-11:CsData-A:pvz:loEff}   {\ensuremath{{0.513 } } }
\vdef{default-11:CsData-A:pvz:loEffE}   {\ensuremath{{0.005 } } }
\vdef{default-11:CsData-A:pvz:hiEff}   {\ensuremath{{0.487 } } }
\vdef{default-11:CsData-A:pvz:hiEffE}   {\ensuremath{{0.005 } } }
\vdef{default-11:CsMc-A:pvz:loEff}   {\ensuremath{{0.471 } } }
\vdef{default-11:CsMc-A:pvz:loEffE}   {\ensuremath{{0.005 } } }
\vdef{default-11:CsMc-A:pvz:hiEff}   {\ensuremath{{0.529 } } }
\vdef{default-11:CsMc-A:pvz:hiEffE}   {\ensuremath{{0.005 } } }
\vdef{default-11:CsMc-A:pvz:loDelta}   {\ensuremath{{+0.085 } } }
\vdef{default-11:CsMc-A:pvz:loDeltaE}   {\ensuremath{{0.015 } } }
\vdef{default-11:CsMc-A:pvz:hiDelta}   {\ensuremath{{-0.082 } } }
\vdef{default-11:CsMc-A:pvz:hiDeltaE}   {\ensuremath{{0.015 } } }
\vdef{default-11:CsData-A:pvn:loEff}   {\ensuremath{{1.007 } } }
\vdef{default-11:CsData-A:pvn:loEffE}   {\ensuremath{{\mathrm{NaN} } } }
\vdef{default-11:CsData-A:pvn:hiEff}   {\ensuremath{{1.000 } } }
\vdef{default-11:CsData-A:pvn:hiEffE}   {\ensuremath{{0.000 } } }
\vdef{default-11:CsMc-A:pvn:loEff}   {\ensuremath{{1.000 } } }
\vdef{default-11:CsMc-A:pvn:loEffE}   {\ensuremath{{0.000 } } }
\vdef{default-11:CsMc-A:pvn:hiEff}   {\ensuremath{{1.000 } } }
\vdef{default-11:CsMc-A:pvn:hiEffE}   {\ensuremath{{0.000 } } }
\vdef{default-11:CsMc-A:pvn:loDelta}   {\ensuremath{{+0.007 } } }
\vdef{default-11:CsMc-A:pvn:loDeltaE}   {\ensuremath{{\mathrm{NaN} } } }
\vdef{default-11:CsMc-A:pvn:hiDelta}   {\ensuremath{{+0.000 } } }
\vdef{default-11:CsMc-A:pvn:hiDeltaE}   {\ensuremath{{0.000 } } }
\vdef{default-11:CsData-A:pvavew8:loEff}   {\ensuremath{{0.014 } } }
\vdef{default-11:CsData-A:pvavew8:loEffE}   {\ensuremath{{0.001 } } }
\vdef{default-11:CsData-A:pvavew8:hiEff}   {\ensuremath{{0.986 } } }
\vdef{default-11:CsData-A:pvavew8:hiEffE}   {\ensuremath{{0.001 } } }
\vdef{default-11:CsMc-A:pvavew8:loEff}   {\ensuremath{{0.011 } } }
\vdef{default-11:CsMc-A:pvavew8:loEffE}   {\ensuremath{{0.001 } } }
\vdef{default-11:CsMc-A:pvavew8:hiEff}   {\ensuremath{{0.989 } } }
\vdef{default-11:CsMc-A:pvavew8:hiEffE}   {\ensuremath{{0.001 } } }
\vdef{default-11:CsMc-A:pvavew8:loDelta}   {\ensuremath{{+0.214 } } }
\vdef{default-11:CsMc-A:pvavew8:loDeltaE}   {\ensuremath{{0.134 } } }
\vdef{default-11:CsMc-A:pvavew8:hiDelta}   {\ensuremath{{-0.003 } } }
\vdef{default-11:CsMc-A:pvavew8:hiDeltaE}   {\ensuremath{{0.002 } } }
\vdef{default-11:CsData-A:pvntrk:loEff}   {\ensuremath{{1.000 } } }
\vdef{default-11:CsData-A:pvntrk:loEffE}   {\ensuremath{{0.000 } } }
\vdef{default-11:CsData-A:pvntrk:hiEff}   {\ensuremath{{1.000 } } }
\vdef{default-11:CsData-A:pvntrk:hiEffE}   {\ensuremath{{0.000 } } }
\vdef{default-11:CsMc-A:pvntrk:loEff}   {\ensuremath{{1.000 } } }
\vdef{default-11:CsMc-A:pvntrk:loEffE}   {\ensuremath{{0.000 } } }
\vdef{default-11:CsMc-A:pvntrk:hiEff}   {\ensuremath{{1.000 } } }
\vdef{default-11:CsMc-A:pvntrk:hiEffE}   {\ensuremath{{0.000 } } }
\vdef{default-11:CsMc-A:pvntrk:loDelta}   {\ensuremath{{+0.000 } } }
\vdef{default-11:CsMc-A:pvntrk:loDeltaE}   {\ensuremath{{0.000 } } }
\vdef{default-11:CsMc-A:pvntrk:hiDelta}   {\ensuremath{{+0.000 } } }
\vdef{default-11:CsMc-A:pvntrk:hiDeltaE}   {\ensuremath{{0.000 } } }
\vdef{default-11:CsData-A:muon1pt:loEff}   {\ensuremath{{1.013 } } }
\vdef{default-11:CsData-A:muon1pt:loEffE}   {\ensuremath{{\mathrm{NaN} } } }
\vdef{default-11:CsData-A:muon1pt:hiEff}   {\ensuremath{{1.000 } } }
\vdef{default-11:CsData-A:muon1pt:hiEffE}   {\ensuremath{{0.000 } } }
\vdef{default-11:CsMc-A:muon1pt:loEff}   {\ensuremath{{1.010 } } }
\vdef{default-11:CsMc-A:muon1pt:loEffE}   {\ensuremath{{\mathrm{NaN} } } }
\vdef{default-11:CsMc-A:muon1pt:hiEff}   {\ensuremath{{1.000 } } }
\vdef{default-11:CsMc-A:muon1pt:hiEffE}   {\ensuremath{{0.000 } } }
\vdef{default-11:CsMc-A:muon1pt:loDelta}   {\ensuremath{{+0.003 } } }
\vdef{default-11:CsMc-A:muon1pt:loDeltaE}   {\ensuremath{{\mathrm{NaN} } } }
\vdef{default-11:CsMc-A:muon1pt:hiDelta}   {\ensuremath{{+0.000 } } }
\vdef{default-11:CsMc-A:muon1pt:hiDeltaE}   {\ensuremath{{0.000 } } }
\vdef{default-11:CsData-A:muon2pt:loEff}   {\ensuremath{{0.074 } } }
\vdef{default-11:CsData-A:muon2pt:loEffE}   {\ensuremath{{0.003 } } }
\vdef{default-11:CsData-A:muon2pt:hiEff}   {\ensuremath{{0.926 } } }
\vdef{default-11:CsData-A:muon2pt:hiEffE}   {\ensuremath{{0.003 } } }
\vdef{default-11:CsMc-A:muon2pt:loEff}   {\ensuremath{{0.068 } } }
\vdef{default-11:CsMc-A:muon2pt:loEffE}   {\ensuremath{{0.003 } } }
\vdef{default-11:CsMc-A:muon2pt:hiEff}   {\ensuremath{{0.932 } } }
\vdef{default-11:CsMc-A:muon2pt:hiEffE}   {\ensuremath{{0.003 } } }
\vdef{default-11:CsMc-A:muon2pt:loDelta}   {\ensuremath{{+0.074 } } }
\vdef{default-11:CsMc-A:muon2pt:loDeltaE}   {\ensuremath{{0.053 } } }
\vdef{default-11:CsMc-A:muon2pt:hiDelta}   {\ensuremath{{-0.006 } } }
\vdef{default-11:CsMc-A:muon2pt:hiDeltaE}   {\ensuremath{{0.004 } } }
\vdef{default-11:CsData-A:muonseta:loEff}   {\ensuremath{{0.724 } } }
\vdef{default-11:CsData-A:muonseta:loEffE}   {\ensuremath{{0.003 } } }
\vdef{default-11:CsData-A:muonseta:hiEff}   {\ensuremath{{0.276 } } }
\vdef{default-11:CsData-A:muonseta:hiEffE}   {\ensuremath{{0.003 } } }
\vdef{default-11:CsMc-A:muonseta:loEff}   {\ensuremath{{0.732 } } }
\vdef{default-11:CsMc-A:muonseta:loEffE}   {\ensuremath{{0.003 } } }
\vdef{default-11:CsMc-A:muonseta:hiEff}   {\ensuremath{{0.268 } } }
\vdef{default-11:CsMc-A:muonseta:hiEffE}   {\ensuremath{{0.003 } } }
\vdef{default-11:CsMc-A:muonseta:loDelta}   {\ensuremath{{-0.011 } } }
\vdef{default-11:CsMc-A:muonseta:loDeltaE}   {\ensuremath{{0.007 } } }
\vdef{default-11:CsMc-A:muonseta:hiDelta}   {\ensuremath{{+0.028 } } }
\vdef{default-11:CsMc-A:muonseta:hiDeltaE}   {\ensuremath{{0.018 } } }
\vdef{default-11:CsData-A:pt:loEff}   {\ensuremath{{0.000 } } }
\vdef{default-11:CsData-A:pt:loEffE}   {\ensuremath{{0.000 } } }
\vdef{default-11:CsData-A:pt:hiEff}   {\ensuremath{{1.000 } } }
\vdef{default-11:CsData-A:pt:hiEffE}   {\ensuremath{{0.000 } } }
\vdef{default-11:CsMc-A:pt:loEff}   {\ensuremath{{0.000 } } }
\vdef{default-11:CsMc-A:pt:loEffE}   {\ensuremath{{0.000 } } }
\vdef{default-11:CsMc-A:pt:hiEff}   {\ensuremath{{1.000 } } }
\vdef{default-11:CsMc-A:pt:hiEffE}   {\ensuremath{{0.000 } } }
\vdef{default-11:CsMc-A:pt:loDelta}   {\ensuremath{{\mathrm{NaN} } } }
\vdef{default-11:CsMc-A:pt:loDeltaE}   {\ensuremath{{\mathrm{NaN} } } }
\vdef{default-11:CsMc-A:pt:hiDelta}   {\ensuremath{{+0.000 } } }
\vdef{default-11:CsMc-A:pt:hiDeltaE}   {\ensuremath{{0.000 } } }
\vdef{default-11:CsData-A:p:loEff}   {\ensuremath{{1.023 } } }
\vdef{default-11:CsData-A:p:loEffE}   {\ensuremath{{\mathrm{NaN} } } }
\vdef{default-11:CsData-A:p:hiEff}   {\ensuremath{{1.000 } } }
\vdef{default-11:CsData-A:p:hiEffE}   {\ensuremath{{0.000 } } }
\vdef{default-11:CsMc-A:p:loEff}   {\ensuremath{{1.017 } } }
\vdef{default-11:CsMc-A:p:loEffE}   {\ensuremath{{\mathrm{NaN} } } }
\vdef{default-11:CsMc-A:p:hiEff}   {\ensuremath{{1.000 } } }
\vdef{default-11:CsMc-A:p:hiEffE}   {\ensuremath{{0.000 } } }
\vdef{default-11:CsMc-A:p:loDelta}   {\ensuremath{{+0.005 } } }
\vdef{default-11:CsMc-A:p:loDeltaE}   {\ensuremath{{\mathrm{NaN} } } }
\vdef{default-11:CsMc-A:p:hiDelta}   {\ensuremath{{+0.000 } } }
\vdef{default-11:CsMc-A:p:hiDeltaE}   {\ensuremath{{0.000 } } }
\vdef{default-11:CsData-A:eta:loEff}   {\ensuremath{{0.719 } } }
\vdef{default-11:CsData-A:eta:loEffE}   {\ensuremath{{0.005 } } }
\vdef{default-11:CsData-A:eta:hiEff}   {\ensuremath{{0.281 } } }
\vdef{default-11:CsData-A:eta:hiEffE}   {\ensuremath{{0.005 } } }
\vdef{default-11:CsMc-A:eta:loEff}   {\ensuremath{{0.723 } } }
\vdef{default-11:CsMc-A:eta:loEffE}   {\ensuremath{{0.005 } } }
\vdef{default-11:CsMc-A:eta:hiEff}   {\ensuremath{{0.277 } } }
\vdef{default-11:CsMc-A:eta:hiEffE}   {\ensuremath{{0.005 } } }
\vdef{default-11:CsMc-A:eta:loDelta}   {\ensuremath{{-0.006 } } }
\vdef{default-11:CsMc-A:eta:loDeltaE}   {\ensuremath{{0.010 } } }
\vdef{default-11:CsMc-A:eta:hiDelta}   {\ensuremath{{+0.015 } } }
\vdef{default-11:CsMc-A:eta:hiDeltaE}   {\ensuremath{{0.025 } } }
\vdef{default-11:CsData-A:bdt:loEff}   {\ensuremath{{0.896 } } }
\vdef{default-11:CsData-A:bdt:loEffE}   {\ensuremath{{0.003 } } }
\vdef{default-11:CsData-A:bdt:hiEff}   {\ensuremath{{0.104 } } }
\vdef{default-11:CsData-A:bdt:hiEffE}   {\ensuremath{{0.003 } } }
\vdef{default-11:CsMc-A:bdt:loEff}   {\ensuremath{{0.908 } } }
\vdef{default-11:CsMc-A:bdt:loEffE}   {\ensuremath{{0.003 } } }
\vdef{default-11:CsMc-A:bdt:hiEff}   {\ensuremath{{0.092 } } }
\vdef{default-11:CsMc-A:bdt:hiEffE}   {\ensuremath{{0.003 } } }
\vdef{default-11:CsMc-A:bdt:loDelta}   {\ensuremath{{-0.013 } } }
\vdef{default-11:CsMc-A:bdt:loDeltaE}   {\ensuremath{{0.005 } } }
\vdef{default-11:CsMc-A:bdt:hiDelta}   {\ensuremath{{+0.121 } } }
\vdef{default-11:CsMc-A:bdt:hiDeltaE}   {\ensuremath{{0.045 } } }
\vdef{default-11:CsData-A:fl3d:loEff}   {\ensuremath{{0.823 } } }
\vdef{default-11:CsData-A:fl3d:loEffE}   {\ensuremath{{0.004 } } }
\vdef{default-11:CsData-A:fl3d:hiEff}   {\ensuremath{{0.177 } } }
\vdef{default-11:CsData-A:fl3d:hiEffE}   {\ensuremath{{0.004 } } }
\vdef{default-11:CsMc-A:fl3d:loEff}   {\ensuremath{{0.835 } } }
\vdef{default-11:CsMc-A:fl3d:loEffE}   {\ensuremath{{0.004 } } }
\vdef{default-11:CsMc-A:fl3d:hiEff}   {\ensuremath{{0.165 } } }
\vdef{default-11:CsMc-A:fl3d:hiEffE}   {\ensuremath{{0.004 } } }
\vdef{default-11:CsMc-A:fl3d:loDelta}   {\ensuremath{{-0.015 } } }
\vdef{default-11:CsMc-A:fl3d:loDeltaE}   {\ensuremath{{0.006 } } }
\vdef{default-11:CsMc-A:fl3d:hiDelta}   {\ensuremath{{+0.073 } } }
\vdef{default-11:CsMc-A:fl3d:hiDeltaE}   {\ensuremath{{0.031 } } }
\vdef{default-11:CsData-A:fl3de:loEff}   {\ensuremath{{1.000 } } }
\vdef{default-11:CsData-A:fl3de:loEffE}   {\ensuremath{{0.000 } } }
\vdef{default-11:CsData-A:fl3de:hiEff}   {\ensuremath{{0.000 } } }
\vdef{default-11:CsData-A:fl3de:hiEffE}   {\ensuremath{{0.000 } } }
\vdef{default-11:CsMc-A:fl3de:loEff}   {\ensuremath{{1.000 } } }
\vdef{default-11:CsMc-A:fl3de:loEffE}   {\ensuremath{{0.000 } } }
\vdef{default-11:CsMc-A:fl3de:hiEff}   {\ensuremath{{0.000 } } }
\vdef{default-11:CsMc-A:fl3de:hiEffE}   {\ensuremath{{0.000 } } }
\vdef{default-11:CsMc-A:fl3de:loDelta}   {\ensuremath{{+0.000 } } }
\vdef{default-11:CsMc-A:fl3de:loDeltaE}   {\ensuremath{{0.000 } } }
\vdef{default-11:CsMc-A:fl3de:hiDelta}   {\ensuremath{{+0.462 } } }
\vdef{default-11:CsMc-A:fl3de:hiDeltaE}   {\ensuremath{{1.361 } } }
\vdef{default-11:CsData-A:fls3d:loEff}   {\ensuremath{{0.095 } } }
\vdef{default-11:CsData-A:fls3d:loEffE}   {\ensuremath{{0.003 } } }
\vdef{default-11:CsData-A:fls3d:hiEff}   {\ensuremath{{0.905 } } }
\vdef{default-11:CsData-A:fls3d:hiEffE}   {\ensuremath{{0.003 } } }
\vdef{default-11:CsMc-A:fls3d:loEff}   {\ensuremath{{0.097 } } }
\vdef{default-11:CsMc-A:fls3d:loEffE}   {\ensuremath{{0.003 } } }
\vdef{default-11:CsMc-A:fls3d:hiEff}   {\ensuremath{{0.903 } } }
\vdef{default-11:CsMc-A:fls3d:hiEffE}   {\ensuremath{{0.003 } } }
\vdef{default-11:CsMc-A:fls3d:loDelta}   {\ensuremath{{-0.023 } } }
\vdef{default-11:CsMc-A:fls3d:loDeltaE}   {\ensuremath{{0.043 } } }
\vdef{default-11:CsMc-A:fls3d:hiDelta}   {\ensuremath{{+0.002 } } }
\vdef{default-11:CsMc-A:fls3d:hiDeltaE}   {\ensuremath{{0.005 } } }
\vdef{default-11:CsData-A:flsxy:loEff}   {\ensuremath{{1.010 } } }
\vdef{default-11:CsData-A:flsxy:loEffE}   {\ensuremath{{\mathrm{NaN} } } }
\vdef{default-11:CsData-A:flsxy:hiEff}   {\ensuremath{{1.000 } } }
\vdef{default-11:CsData-A:flsxy:hiEffE}   {\ensuremath{{0.000 } } }
\vdef{default-11:CsMc-A:flsxy:loEff}   {\ensuremath{{1.007 } } }
\vdef{default-11:CsMc-A:flsxy:loEffE}   {\ensuremath{{\mathrm{NaN} } } }
\vdef{default-11:CsMc-A:flsxy:hiEff}   {\ensuremath{{1.000 } } }
\vdef{default-11:CsMc-A:flsxy:hiEffE}   {\ensuremath{{0.000 } } }
\vdef{default-11:CsMc-A:flsxy:loDelta}   {\ensuremath{{+0.003 } } }
\vdef{default-11:CsMc-A:flsxy:loDeltaE}   {\ensuremath{{\mathrm{NaN} } } }
\vdef{default-11:CsMc-A:flsxy:hiDelta}   {\ensuremath{{+0.000 } } }
\vdef{default-11:CsMc-A:flsxy:hiDeltaE}   {\ensuremath{{0.000 } } }
\vdef{default-11:CsData-A:chi2dof:loEff}   {\ensuremath{{0.929 } } }
\vdef{default-11:CsData-A:chi2dof:loEffE}   {\ensuremath{{0.003 } } }
\vdef{default-11:CsData-A:chi2dof:hiEff}   {\ensuremath{{0.071 } } }
\vdef{default-11:CsData-A:chi2dof:hiEffE}   {\ensuremath{{0.003 } } }
\vdef{default-11:CsMc-A:chi2dof:loEff}   {\ensuremath{{0.944 } } }
\vdef{default-11:CsMc-A:chi2dof:loEffE}   {\ensuremath{{0.002 } } }
\vdef{default-11:CsMc-A:chi2dof:hiEff}   {\ensuremath{{0.056 } } }
\vdef{default-11:CsMc-A:chi2dof:hiEffE}   {\ensuremath{{0.002 } } }
\vdef{default-11:CsMc-A:chi2dof:loDelta}   {\ensuremath{{-0.016 } } }
\vdef{default-11:CsMc-A:chi2dof:loDeltaE}   {\ensuremath{{0.004 } } }
\vdef{default-11:CsMc-A:chi2dof:hiDelta}   {\ensuremath{{+0.239 } } }
\vdef{default-11:CsMc-A:chi2dof:hiDeltaE}   {\ensuremath{{0.057 } } }
\vdef{default-11:CsData-A:pchi2dof:loEff}   {\ensuremath{{0.624 } } }
\vdef{default-11:CsData-A:pchi2dof:loEffE}   {\ensuremath{{0.005 } } }
\vdef{default-11:CsData-A:pchi2dof:hiEff}   {\ensuremath{{0.376 } } }
\vdef{default-11:CsData-A:pchi2dof:hiEffE}   {\ensuremath{{0.005 } } }
\vdef{default-11:CsMc-A:pchi2dof:loEff}   {\ensuremath{{0.589 } } }
\vdef{default-11:CsMc-A:pchi2dof:loEffE}   {\ensuremath{{0.005 } } }
\vdef{default-11:CsMc-A:pchi2dof:hiEff}   {\ensuremath{{0.411 } } }
\vdef{default-11:CsMc-A:pchi2dof:hiEffE}   {\ensuremath{{0.005 } } }
\vdef{default-11:CsMc-A:pchi2dof:loDelta}   {\ensuremath{{+0.058 } } }
\vdef{default-11:CsMc-A:pchi2dof:loDeltaE}   {\ensuremath{{0.012 } } }
\vdef{default-11:CsMc-A:pchi2dof:hiDelta}   {\ensuremath{{-0.089 } } }
\vdef{default-11:CsMc-A:pchi2dof:hiDeltaE}   {\ensuremath{{0.018 } } }
\vdef{default-11:CsData-A:alpha:loEff}   {\ensuremath{{0.995 } } }
\vdef{default-11:CsData-A:alpha:loEffE}   {\ensuremath{{0.001 } } }
\vdef{default-11:CsData-A:alpha:hiEff}   {\ensuremath{{0.005 } } }
\vdef{default-11:CsData-A:alpha:hiEffE}   {\ensuremath{{0.001 } } }
\vdef{default-11:CsMc-A:alpha:loEff}   {\ensuremath{{0.996 } } }
\vdef{default-11:CsMc-A:alpha:loEffE}   {\ensuremath{{0.001 } } }
\vdef{default-11:CsMc-A:alpha:hiEff}   {\ensuremath{{0.004 } } }
\vdef{default-11:CsMc-A:alpha:hiEffE}   {\ensuremath{{0.001 } } }
\vdef{default-11:CsMc-A:alpha:loDelta}   {\ensuremath{{-0.001 } } }
\vdef{default-11:CsMc-A:alpha:loDeltaE}   {\ensuremath{{0.001 } } }
\vdef{default-11:CsMc-A:alpha:hiDelta}   {\ensuremath{{+0.191 } } }
\vdef{default-11:CsMc-A:alpha:hiDeltaE}   {\ensuremath{{0.230 } } }
\vdef{default-11:CsData-A:iso:loEff}   {\ensuremath{{0.093 } } }
\vdef{default-11:CsData-A:iso:loEffE}   {\ensuremath{{0.003 } } }
\vdef{default-11:CsData-A:iso:hiEff}   {\ensuremath{{0.907 } } }
\vdef{default-11:CsData-A:iso:hiEffE}   {\ensuremath{{0.003 } } }
\vdef{default-11:CsMc-A:iso:loEff}   {\ensuremath{{0.085 } } }
\vdef{default-11:CsMc-A:iso:loEffE}   {\ensuremath{{0.003 } } }
\vdef{default-11:CsMc-A:iso:hiEff}   {\ensuremath{{0.915 } } }
\vdef{default-11:CsMc-A:iso:hiEffE}   {\ensuremath{{0.003 } } }
\vdef{default-11:CsMc-A:iso:loDelta}   {\ensuremath{{+0.093 } } }
\vdef{default-11:CsMc-A:iso:loDeltaE}   {\ensuremath{{0.046 } } }
\vdef{default-11:CsMc-A:iso:hiDelta}   {\ensuremath{{-0.009 } } }
\vdef{default-11:CsMc-A:iso:hiDeltaE}   {\ensuremath{{0.005 } } }
\vdef{default-11:CsData-A:docatrk:loEff}   {\ensuremath{{0.067 } } }
\vdef{default-11:CsData-A:docatrk:loEffE}   {\ensuremath{{0.003 } } }
\vdef{default-11:CsData-A:docatrk:hiEff}   {\ensuremath{{0.933 } } }
\vdef{default-11:CsData-A:docatrk:hiEffE}   {\ensuremath{{0.003 } } }
\vdef{default-11:CsMc-A:docatrk:loEff}   {\ensuremath{{0.080 } } }
\vdef{default-11:CsMc-A:docatrk:loEffE}   {\ensuremath{{0.003 } } }
\vdef{default-11:CsMc-A:docatrk:hiEff}   {\ensuremath{{0.920 } } }
\vdef{default-11:CsMc-A:docatrk:hiEffE}   {\ensuremath{{0.003 } } }
\vdef{default-11:CsMc-A:docatrk:loDelta}   {\ensuremath{{-0.182 } } }
\vdef{default-11:CsMc-A:docatrk:loDeltaE}   {\ensuremath{{0.053 } } }
\vdef{default-11:CsMc-A:docatrk:hiDelta}   {\ensuremath{{+0.014 } } }
\vdef{default-11:CsMc-A:docatrk:hiDeltaE}   {\ensuremath{{0.004 } } }
\vdef{default-11:CsData-A:isotrk:loEff}   {\ensuremath{{1.000 } } }
\vdef{default-11:CsData-A:isotrk:loEffE}   {\ensuremath{{0.000 } } }
\vdef{default-11:CsData-A:isotrk:hiEff}   {\ensuremath{{1.000 } } }
\vdef{default-11:CsData-A:isotrk:hiEffE}   {\ensuremath{{0.000 } } }
\vdef{default-11:CsMc-A:isotrk:loEff}   {\ensuremath{{1.000 } } }
\vdef{default-11:CsMc-A:isotrk:loEffE}   {\ensuremath{{0.000 } } }
\vdef{default-11:CsMc-A:isotrk:hiEff}   {\ensuremath{{1.000 } } }
\vdef{default-11:CsMc-A:isotrk:hiEffE}   {\ensuremath{{0.000 } } }
\vdef{default-11:CsMc-A:isotrk:loDelta}   {\ensuremath{{+0.000 } } }
\vdef{default-11:CsMc-A:isotrk:loDeltaE}   {\ensuremath{{0.000 } } }
\vdef{default-11:CsMc-A:isotrk:hiDelta}   {\ensuremath{{+0.000 } } }
\vdef{default-11:CsMc-A:isotrk:hiDeltaE}   {\ensuremath{{0.000 } } }
\vdef{default-11:CsData-A:closetrk:loEff}   {\ensuremath{{0.982 } } }
\vdef{default-11:CsData-A:closetrk:loEffE}   {\ensuremath{{0.001 } } }
\vdef{default-11:CsData-A:closetrk:hiEff}   {\ensuremath{{0.018 } } }
\vdef{default-11:CsData-A:closetrk:hiEffE}   {\ensuremath{{0.001 } } }
\vdef{default-11:CsMc-A:closetrk:loEff}   {\ensuremath{{0.981 } } }
\vdef{default-11:CsMc-A:closetrk:loEffE}   {\ensuremath{{0.001 } } }
\vdef{default-11:CsMc-A:closetrk:hiEff}   {\ensuremath{{0.019 } } }
\vdef{default-11:CsMc-A:closetrk:hiEffE}   {\ensuremath{{0.001 } } }
\vdef{default-11:CsMc-A:closetrk:loDelta}   {\ensuremath{{+0.001 } } }
\vdef{default-11:CsMc-A:closetrk:loDeltaE}   {\ensuremath{{0.002 } } }
\vdef{default-11:CsMc-A:closetrk:hiDelta}   {\ensuremath{{-0.076 } } }
\vdef{default-11:CsMc-A:closetrk:hiDeltaE}   {\ensuremath{{0.110 } } }
\vdef{default-11:CsData-A:lip:loEff}   {\ensuremath{{1.000 } } }
\vdef{default-11:CsData-A:lip:loEffE}   {\ensuremath{{0.000 } } }
\vdef{default-11:CsData-A:lip:hiEff}   {\ensuremath{{0.000 } } }
\vdef{default-11:CsData-A:lip:hiEffE}   {\ensuremath{{0.000 } } }
\vdef{default-11:CsMc-A:lip:loEff}   {\ensuremath{{1.000 } } }
\vdef{default-11:CsMc-A:lip:loEffE}   {\ensuremath{{0.000 } } }
\vdef{default-11:CsMc-A:lip:hiEff}   {\ensuremath{{0.000 } } }
\vdef{default-11:CsMc-A:lip:hiEffE}   {\ensuremath{{0.000 } } }
\vdef{default-11:CsMc-A:lip:loDelta}   {\ensuremath{{+0.000 } } }
\vdef{default-11:CsMc-A:lip:loDeltaE}   {\ensuremath{{0.000 } } }
\vdef{default-11:CsMc-A:lip:hiDelta}   {\ensuremath{{\mathrm{NaN} } } }
\vdef{default-11:CsMc-A:lip:hiDeltaE}   {\ensuremath{{\mathrm{NaN} } } }
\vdef{default-11:CsData-A:lip:inEff}   {\ensuremath{{1.000 } } }
\vdef{default-11:CsData-A:lip:inEffE}   {\ensuremath{{0.000 } } }
\vdef{default-11:CsMc-A:lip:inEff}   {\ensuremath{{1.000 } } }
\vdef{default-11:CsMc-A:lip:inEffE}   {\ensuremath{{0.000 } } }
\vdef{default-11:CsMc-A:lip:inDelta}   {\ensuremath{{+0.000 } } }
\vdef{default-11:CsMc-A:lip:inDeltaE}   {\ensuremath{{0.000 } } }
\vdef{default-11:CsData-A:lips:loEff}   {\ensuremath{{1.000 } } }
\vdef{default-11:CsData-A:lips:loEffE}   {\ensuremath{{0.000 } } }
\vdef{default-11:CsData-A:lips:hiEff}   {\ensuremath{{0.000 } } }
\vdef{default-11:CsData-A:lips:hiEffE}   {\ensuremath{{0.000 } } }
\vdef{default-11:CsMc-A:lips:loEff}   {\ensuremath{{1.000 } } }
\vdef{default-11:CsMc-A:lips:loEffE}   {\ensuremath{{0.000 } } }
\vdef{default-11:CsMc-A:lips:hiEff}   {\ensuremath{{0.000 } } }
\vdef{default-11:CsMc-A:lips:hiEffE}   {\ensuremath{{0.000 } } }
\vdef{default-11:CsMc-A:lips:loDelta}   {\ensuremath{{+0.000 } } }
\vdef{default-11:CsMc-A:lips:loDeltaE}   {\ensuremath{{0.000 } } }
\vdef{default-11:CsMc-A:lips:hiDelta}   {\ensuremath{{\mathrm{NaN} } } }
\vdef{default-11:CsMc-A:lips:hiDeltaE}   {\ensuremath{{\mathrm{NaN} } } }
\vdef{default-11:CsData-A:lips:inEff}   {\ensuremath{{1.000 } } }
\vdef{default-11:CsData-A:lips:inEffE}   {\ensuremath{{0.000 } } }
\vdef{default-11:CsMc-A:lips:inEff}   {\ensuremath{{1.000 } } }
\vdef{default-11:CsMc-A:lips:inEffE}   {\ensuremath{{0.000 } } }
\vdef{default-11:CsMc-A:lips:inDelta}   {\ensuremath{{+0.000 } } }
\vdef{default-11:CsMc-A:lips:inDeltaE}   {\ensuremath{{0.000 } } }
\vdef{default-11:CsData-A:ip:loEff}   {\ensuremath{{0.974 } } }
\vdef{default-11:CsData-A:ip:loEffE}   {\ensuremath{{0.002 } } }
\vdef{default-11:CsData-A:ip:hiEff}   {\ensuremath{{0.026 } } }
\vdef{default-11:CsData-A:ip:hiEffE}   {\ensuremath{{0.002 } } }
\vdef{default-11:CsMc-A:ip:loEff}   {\ensuremath{{0.974 } } }
\vdef{default-11:CsMc-A:ip:loEffE}   {\ensuremath{{0.002 } } }
\vdef{default-11:CsMc-A:ip:hiEff}   {\ensuremath{{0.026 } } }
\vdef{default-11:CsMc-A:ip:hiEffE}   {\ensuremath{{0.002 } } }
\vdef{default-11:CsMc-A:ip:loDelta}   {\ensuremath{{-0.000 } } }
\vdef{default-11:CsMc-A:ip:loDeltaE}   {\ensuremath{{0.002 } } }
\vdef{default-11:CsMc-A:ip:hiDelta}   {\ensuremath{{+0.008 } } }
\vdef{default-11:CsMc-A:ip:hiDeltaE}   {\ensuremath{{0.092 } } }
\vdef{default-11:CsData-A:ips:loEff}   {\ensuremath{{0.939 } } }
\vdef{default-11:CsData-A:ips:loEffE}   {\ensuremath{{0.003 } } }
\vdef{default-11:CsData-A:ips:hiEff}   {\ensuremath{{0.061 } } }
\vdef{default-11:CsData-A:ips:hiEffE}   {\ensuremath{{0.003 } } }
\vdef{default-11:CsMc-A:ips:loEff}   {\ensuremath{{0.951 } } }
\vdef{default-11:CsMc-A:ips:loEffE}   {\ensuremath{{0.002 } } }
\vdef{default-11:CsMc-A:ips:hiEff}   {\ensuremath{{0.049 } } }
\vdef{default-11:CsMc-A:ips:hiEffE}   {\ensuremath{{0.002 } } }
\vdef{default-11:CsMc-A:ips:loDelta}   {\ensuremath{{-0.013 } } }
\vdef{default-11:CsMc-A:ips:loDeltaE}   {\ensuremath{{0.004 } } }
\vdef{default-11:CsMc-A:ips:hiDelta}   {\ensuremath{{+0.224 } } }
\vdef{default-11:CsMc-A:ips:hiDeltaE}   {\ensuremath{{0.061 } } }
\vdef{default-11:CsData-A:maxdoca:loEff}   {\ensuremath{{1.000 } } }
\vdef{default-11:CsData-A:maxdoca:loEffE}   {\ensuremath{{0.000 } } }
\vdef{default-11:CsData-A:maxdoca:hiEff}   {\ensuremath{{0.034 } } }
\vdef{default-11:CsData-A:maxdoca:hiEffE}   {\ensuremath{{0.002 } } }
\vdef{default-11:CsMc-A:maxdoca:loEff}   {\ensuremath{{1.000 } } }
\vdef{default-11:CsMc-A:maxdoca:loEffE}   {\ensuremath{{0.000 } } }
\vdef{default-11:CsMc-A:maxdoca:hiEff}   {\ensuremath{{0.021 } } }
\vdef{default-11:CsMc-A:maxdoca:hiEffE}   {\ensuremath{{0.002 } } }
\vdef{default-11:CsMc-A:maxdoca:loDelta}   {\ensuremath{{+0.000 } } }
\vdef{default-11:CsMc-A:maxdoca:loDeltaE}   {\ensuremath{{0.000 } } }
\vdef{default-11:CsMc-A:maxdoca:hiDelta}   {\ensuremath{{+0.460 } } }
\vdef{default-11:CsMc-A:maxdoca:hiDeltaE}   {\ensuremath{{0.090 } } }
\vdef{default-11:CsData-A:kaonspt:loEff}   {\ensuremath{{1.000 } } }
\vdef{default-11:CsData-A:kaonspt:loEffE}   {\ensuremath{{\mathrm{NaN} } } }
\vdef{default-11:CsData-A:kaonspt:hiEff}   {\ensuremath{{1.000 } } }
\vdef{default-11:CsData-A:kaonspt:hiEffE}   {\ensuremath{{0.000 } } }
\vdef{default-11:CsMc-A:kaonspt:loEff}   {\ensuremath{{1.000 } } }
\vdef{default-11:CsMc-A:kaonspt:loEffE}   {\ensuremath{{0.000 } } }
\vdef{default-11:CsMc-A:kaonspt:hiEff}   {\ensuremath{{1.000 } } }
\vdef{default-11:CsMc-A:kaonspt:hiEffE}   {\ensuremath{{0.000 } } }
\vdef{default-11:CsMc-A:kaonspt:loDelta}   {\ensuremath{{+0.000 } } }
\vdef{default-11:CsMc-A:kaonspt:loDeltaE}   {\ensuremath{{\mathrm{NaN} } } }
\vdef{default-11:CsMc-A:kaonspt:hiDelta}   {\ensuremath{{+0.000 } } }
\vdef{default-11:CsMc-A:kaonspt:hiDeltaE}   {\ensuremath{{0.000 } } }
\vdef{default-11:CsData-A:psipt:loEff}   {\ensuremath{{1.006 } } }
\vdef{default-11:CsData-A:psipt:loEffE}   {\ensuremath{{\mathrm{NaN} } } }
\vdef{default-11:CsData-A:psipt:hiEff}   {\ensuremath{{1.000 } } }
\vdef{default-11:CsData-A:psipt:hiEffE}   {\ensuremath{{0.000 } } }
\vdef{default-11:CsMc-A:psipt:loEff}   {\ensuremath{{1.003 } } }
\vdef{default-11:CsMc-A:psipt:loEffE}   {\ensuremath{{\mathrm{NaN} } } }
\vdef{default-11:CsMc-A:psipt:hiEff}   {\ensuremath{{1.000 } } }
\vdef{default-11:CsMc-A:psipt:hiEffE}   {\ensuremath{{0.000 } } }
\vdef{default-11:CsMc-A:psipt:loDelta}   {\ensuremath{{+0.002 } } }
\vdef{default-11:CsMc-A:psipt:loDeltaE}   {\ensuremath{{\mathrm{NaN} } } }
\vdef{default-11:CsMc-A:psipt:hiDelta}   {\ensuremath{{+0.000 } } }
\vdef{default-11:CsMc-A:psipt:hiDeltaE}   {\ensuremath{{0.000 } } }
\vdef{default-11:CsData-A:phipt:loEff}   {\ensuremath{{1.015 } } }
\vdef{default-11:CsData-A:phipt:loEffE}   {\ensuremath{{\mathrm{NaN} } } }
\vdef{default-11:CsData-A:phipt:hiEff}   {\ensuremath{{1.000 } } }
\vdef{default-11:CsData-A:phipt:hiEffE}   {\ensuremath{{0.000 } } }
\vdef{default-11:CsMc-A:phipt:loEff}   {\ensuremath{{1.008 } } }
\vdef{default-11:CsMc-A:phipt:loEffE}   {\ensuremath{{\mathrm{NaN} } } }
\vdef{default-11:CsMc-A:phipt:hiEff}   {\ensuremath{{1.000 } } }
\vdef{default-11:CsMc-A:phipt:hiEffE}   {\ensuremath{{0.000 } } }
\vdef{default-11:CsMc-A:phipt:loDelta}   {\ensuremath{{+0.008 } } }
\vdef{default-11:CsMc-A:phipt:loDeltaE}   {\ensuremath{{\mathrm{NaN} } } }
\vdef{default-11:CsMc-A:phipt:hiDelta}   {\ensuremath{{+0.000 } } }
\vdef{default-11:CsMc-A:phipt:hiDeltaE}   {\ensuremath{{0.000 } } }
\vdef{default-11:CsData-A:deltar:loEff}   {\ensuremath{{1.000 } } }
\vdef{default-11:CsData-A:deltar:loEffE}   {\ensuremath{{0.000 } } }
\vdef{default-11:CsData-A:deltar:hiEff}   {\ensuremath{{0.000 } } }
\vdef{default-11:CsData-A:deltar:hiEffE}   {\ensuremath{{0.000 } } }
\vdef{default-11:CsMc-A:deltar:loEff}   {\ensuremath{{1.000 } } }
\vdef{default-11:CsMc-A:deltar:loEffE}   {\ensuremath{{0.000 } } }
\vdef{default-11:CsMc-A:deltar:hiEff}   {\ensuremath{{0.000 } } }
\vdef{default-11:CsMc-A:deltar:hiEffE}   {\ensuremath{{0.000 } } }
\vdef{default-11:CsMc-A:deltar:loDelta}   {\ensuremath{{+0.000 } } }
\vdef{default-11:CsMc-A:deltar:loDeltaE}   {\ensuremath{{0.000 } } }
\vdef{default-11:CsMc-A:deltar:hiDelta}   {\ensuremath{{\mathrm{NaN} } } }
\vdef{default-11:CsMc-A:deltar:hiDeltaE}   {\ensuremath{{\mathrm{NaN} } } }
\vdef{default-11:CsData-A:mkk:loEff}   {\ensuremath{{1.159 } } }
\vdef{default-11:CsData-A:mkk:loEffE}   {\ensuremath{{\mathrm{NaN} } } }
\vdef{default-11:CsData-A:mkk:hiEff}   {\ensuremath{{1.000 } } }
\vdef{default-11:CsData-A:mkk:hiEffE}   {\ensuremath{{0.000 } } }
\vdef{default-11:CsMc-A:mkk:loEff}   {\ensuremath{{1.000 } } }
\vdef{default-11:CsMc-A:mkk:loEffE}   {\ensuremath{{0.000 } } }
\vdef{default-11:CsMc-A:mkk:hiEff}   {\ensuremath{{1.000 } } }
\vdef{default-11:CsMc-A:mkk:hiEffE}   {\ensuremath{{0.000 } } }
\vdef{default-11:CsMc-A:mkk:loDelta}   {\ensuremath{{+0.147 } } }
\vdef{default-11:CsMc-A:mkk:loDeltaE}   {\ensuremath{{\mathrm{NaN} } } }
\vdef{default-11:CsMc-A:mkk:hiDelta}   {\ensuremath{{+0.000 } } }
\vdef{default-11:CsMc-A:mkk:hiDeltaE}   {\ensuremath{{0.000 } } }
\vdef{default-11:CsData-A:osiso:loEff}   {\ensuremath{{1.006 } } }
\vdef{default-11:CsData-A:osiso:loEffE}   {\ensuremath{{\mathrm{NaN} } } }
\vdef{default-11:CsData-A:osiso:hiEff}   {\ensuremath{{1.000 } } }
\vdef{default-11:CsData-A:osiso:hiEffE}   {\ensuremath{{0.000 } } }
\vdef{default-11:CsMcPU-A:osiso:loEff}   {\ensuremath{{1.002 } } }
\vdef{default-11:CsMcPU-A:osiso:loEffE}   {\ensuremath{{\mathrm{NaN} } } }
\vdef{default-11:CsMcPU-A:osiso:hiEff}   {\ensuremath{{1.000 } } }
\vdef{default-11:CsMcPU-A:osiso:hiEffE}   {\ensuremath{{0.000 } } }
\vdef{default-11:CsMcPU-A:osiso:loDelta}   {\ensuremath{{+0.003 } } }
\vdef{default-11:CsMcPU-A:osiso:loDeltaE}   {\ensuremath{{\mathrm{NaN} } } }
\vdef{default-11:CsMcPU-A:osiso:hiDelta}   {\ensuremath{{+0.000 } } }
\vdef{default-11:CsMcPU-A:osiso:hiDeltaE}   {\ensuremath{{0.000 } } }
\vdef{default-11:CsData-A:osreliso:loEff}   {\ensuremath{{0.272 } } }
\vdef{default-11:CsData-A:osreliso:loEffE}   {\ensuremath{{0.005 } } }
\vdef{default-11:CsData-A:osreliso:hiEff}   {\ensuremath{{0.728 } } }
\vdef{default-11:CsData-A:osreliso:hiEffE}   {\ensuremath{{0.005 } } }
\vdef{default-11:CsMcPU-A:osreliso:loEff}   {\ensuremath{{0.294 } } }
\vdef{default-11:CsMcPU-A:osreliso:loEffE}   {\ensuremath{{0.005 } } }
\vdef{default-11:CsMcPU-A:osreliso:hiEff}   {\ensuremath{{0.706 } } }
\vdef{default-11:CsMcPU-A:osreliso:hiEffE}   {\ensuremath{{0.005 } } }
\vdef{default-11:CsMcPU-A:osreliso:loDelta}   {\ensuremath{{-0.078 } } }
\vdef{default-11:CsMcPU-A:osreliso:loDeltaE}   {\ensuremath{{0.023 } } }
\vdef{default-11:CsMcPU-A:osreliso:hiDelta}   {\ensuremath{{+0.031 } } }
\vdef{default-11:CsMcPU-A:osreliso:hiDeltaE}   {\ensuremath{{0.009 } } }
\vdef{default-11:CsData-A:osmuonpt:loEff}   {\ensuremath{{0.000 } } }
\vdef{default-11:CsData-A:osmuonpt:loEffE}   {\ensuremath{{0.003 } } }
\vdef{default-11:CsData-A:osmuonpt:hiEff}   {\ensuremath{{1.000 } } }
\vdef{default-11:CsData-A:osmuonpt:hiEffE}   {\ensuremath{{0.003 } } }
\vdef{default-11:CsMcPU-A:osmuonpt:loEff}   {\ensuremath{{0.000 } } }
\vdef{default-11:CsMcPU-A:osmuonpt:loEffE}   {\ensuremath{{0.003 } } }
\vdef{default-11:CsMcPU-A:osmuonpt:hiEff}   {\ensuremath{{1.000 } } }
\vdef{default-11:CsMcPU-A:osmuonpt:hiEffE}   {\ensuremath{{0.003 } } }
\vdef{default-11:CsMcPU-A:osmuonpt:loDelta}   {\ensuremath{{\mathrm{NaN} } } }
\vdef{default-11:CsMcPU-A:osmuonpt:loDeltaE}   {\ensuremath{{\mathrm{NaN} } } }
\vdef{default-11:CsMcPU-A:osmuonpt:hiDelta}   {\ensuremath{{+0.000 } } }
\vdef{default-11:CsMcPU-A:osmuonpt:hiDeltaE}   {\ensuremath{{0.004 } } }
\vdef{default-11:CsData-A:osmuondr:loEff}   {\ensuremath{{0.009 } } }
\vdef{default-11:CsData-A:osmuondr:loEffE}   {\ensuremath{{0.005 } } }
\vdef{default-11:CsData-A:osmuondr:hiEff}   {\ensuremath{{0.991 } } }
\vdef{default-11:CsData-A:osmuondr:hiEffE}   {\ensuremath{{0.005 } } }
\vdef{default-11:CsMcPU-A:osmuondr:loEff}   {\ensuremath{{0.037 } } }
\vdef{default-11:CsMcPU-A:osmuondr:loEffE}   {\ensuremath{{0.010 } } }
\vdef{default-11:CsMcPU-A:osmuondr:hiEff}   {\ensuremath{{0.963 } } }
\vdef{default-11:CsMcPU-A:osmuondr:hiEffE}   {\ensuremath{{0.010 } } }
\vdef{default-11:CsMcPU-A:osmuondr:loDelta}   {\ensuremath{{-1.202 } } }
\vdef{default-11:CsMcPU-A:osmuondr:loDeltaE}   {\ensuremath{{0.397 } } }
\vdef{default-11:CsMcPU-A:osmuondr:hiDelta}   {\ensuremath{{+0.028 } } }
\vdef{default-11:CsMcPU-A:osmuondr:hiDeltaE}   {\ensuremath{{0.011 } } }
\vdef{default-11:CsData-A:hlt:loEff}   {\ensuremath{{0.065 } } }
\vdef{default-11:CsData-A:hlt:loEffE}   {\ensuremath{{0.003 } } }
\vdef{default-11:CsData-A:hlt:hiEff}   {\ensuremath{{0.935 } } }
\vdef{default-11:CsData-A:hlt:hiEffE}   {\ensuremath{{0.003 } } }
\vdef{default-11:CsMcPU-A:hlt:loEff}   {\ensuremath{{0.322 } } }
\vdef{default-11:CsMcPU-A:hlt:loEffE}   {\ensuremath{{0.005 } } }
\vdef{default-11:CsMcPU-A:hlt:hiEff}   {\ensuremath{{0.678 } } }
\vdef{default-11:CsMcPU-A:hlt:hiEffE}   {\ensuremath{{0.005 } } }
\vdef{default-11:CsMcPU-A:hlt:loDelta}   {\ensuremath{{-1.332 } } }
\vdef{default-11:CsMcPU-A:hlt:loDeltaE}   {\ensuremath{{0.024 } } }
\vdef{default-11:CsMcPU-A:hlt:hiDelta}   {\ensuremath{{+0.319 } } }
\vdef{default-11:CsMcPU-A:hlt:hiDeltaE}   {\ensuremath{{0.008 } } }
\vdef{default-11:CsData-A:muonsid:loEff}   {\ensuremath{{0.149 } } }
\vdef{default-11:CsData-A:muonsid:loEffE}   {\ensuremath{{0.004 } } }
\vdef{default-11:CsData-A:muonsid:hiEff}   {\ensuremath{{0.851 } } }
\vdef{default-11:CsData-A:muonsid:hiEffE}   {\ensuremath{{0.004 } } }
\vdef{default-11:CsMcPU-A:muonsid:loEff}   {\ensuremath{{0.230 } } }
\vdef{default-11:CsMcPU-A:muonsid:loEffE}   {\ensuremath{{0.004 } } }
\vdef{default-11:CsMcPU-A:muonsid:hiEff}   {\ensuremath{{0.770 } } }
\vdef{default-11:CsMcPU-A:muonsid:hiEffE}   {\ensuremath{{0.004 } } }
\vdef{default-11:CsMcPU-A:muonsid:loDelta}   {\ensuremath{{-0.427 } } }
\vdef{default-11:CsMcPU-A:muonsid:loDeltaE}   {\ensuremath{{0.029 } } }
\vdef{default-11:CsMcPU-A:muonsid:hiDelta}   {\ensuremath{{+0.100 } } }
\vdef{default-11:CsMcPU-A:muonsid:hiDeltaE}   {\ensuremath{{0.007 } } }
\vdef{default-11:CsData-A:tracksqual:loEff}   {\ensuremath{{0.001 } } }
\vdef{default-11:CsData-A:tracksqual:loEffE}   {\ensuremath{{0.000 } } }
\vdef{default-11:CsData-A:tracksqual:hiEff}   {\ensuremath{{0.999 } } }
\vdef{default-11:CsData-A:tracksqual:hiEffE}   {\ensuremath{{0.000 } } }
\vdef{default-11:CsMcPU-A:tracksqual:loEff}   {\ensuremath{{0.000 } } }
\vdef{default-11:CsMcPU-A:tracksqual:loEffE}   {\ensuremath{{0.000 } } }
\vdef{default-11:CsMcPU-A:tracksqual:hiEff}   {\ensuremath{{1.000 } } }
\vdef{default-11:CsMcPU-A:tracksqual:hiEffE}   {\ensuremath{{0.000 } } }
\vdef{default-11:CsMcPU-A:tracksqual:loDelta}   {\ensuremath{{+2.000 } } }
\vdef{default-11:CsMcPU-A:tracksqual:loDeltaE}   {\ensuremath{{0.639 } } }
\vdef{default-11:CsMcPU-A:tracksqual:hiDelta}   {\ensuremath{{-0.001 } } }
\vdef{default-11:CsMcPU-A:tracksqual:hiDeltaE}   {\ensuremath{{0.000 } } }
\vdef{default-11:CsData-A:pvz:loEff}   {\ensuremath{{0.513 } } }
\vdef{default-11:CsData-A:pvz:loEffE}   {\ensuremath{{0.005 } } }
\vdef{default-11:CsData-A:pvz:hiEff}   {\ensuremath{{0.487 } } }
\vdef{default-11:CsData-A:pvz:hiEffE}   {\ensuremath{{0.005 } } }
\vdef{default-11:CsMcPU-A:pvz:loEff}   {\ensuremath{{0.477 } } }
\vdef{default-11:CsMcPU-A:pvz:loEffE}   {\ensuremath{{0.005 } } }
\vdef{default-11:CsMcPU-A:pvz:hiEff}   {\ensuremath{{0.523 } } }
\vdef{default-11:CsMcPU-A:pvz:hiEffE}   {\ensuremath{{0.005 } } }
\vdef{default-11:CsMcPU-A:pvz:loDelta}   {\ensuremath{{+0.073 } } }
\vdef{default-11:CsMcPU-A:pvz:loDeltaE}   {\ensuremath{{0.015 } } }
\vdef{default-11:CsMcPU-A:pvz:hiDelta}   {\ensuremath{{-0.071 } } }
\vdef{default-11:CsMcPU-A:pvz:hiDeltaE}   {\ensuremath{{0.015 } } }
\vdef{default-11:CsData-A:pvn:loEff}   {\ensuremath{{1.007 } } }
\vdef{default-11:CsData-A:pvn:loEffE}   {\ensuremath{{\mathrm{NaN} } } }
\vdef{default-11:CsData-A:pvn:hiEff}   {\ensuremath{{1.000 } } }
\vdef{default-11:CsData-A:pvn:hiEffE}   {\ensuremath{{0.000 } } }
\vdef{default-11:CsMcPU-A:pvn:loEff}   {\ensuremath{{1.062 } } }
\vdef{default-11:CsMcPU-A:pvn:loEffE}   {\ensuremath{{\mathrm{NaN} } } }
\vdef{default-11:CsMcPU-A:pvn:hiEff}   {\ensuremath{{1.000 } } }
\vdef{default-11:CsMcPU-A:pvn:hiEffE}   {\ensuremath{{0.000 } } }
\vdef{default-11:CsMcPU-A:pvn:loDelta}   {\ensuremath{{-0.053 } } }
\vdef{default-11:CsMcPU-A:pvn:loDeltaE}   {\ensuremath{{\mathrm{NaN} } } }
\vdef{default-11:CsMcPU-A:pvn:hiDelta}   {\ensuremath{{+0.000 } } }
\vdef{default-11:CsMcPU-A:pvn:hiDeltaE}   {\ensuremath{{0.000 } } }
\vdef{default-11:CsData-A:pvavew8:loEff}   {\ensuremath{{0.014 } } }
\vdef{default-11:CsData-A:pvavew8:loEffE}   {\ensuremath{{0.001 } } }
\vdef{default-11:CsData-A:pvavew8:hiEff}   {\ensuremath{{0.986 } } }
\vdef{default-11:CsData-A:pvavew8:hiEffE}   {\ensuremath{{0.001 } } }
\vdef{default-11:CsMcPU-A:pvavew8:loEff}   {\ensuremath{{0.010 } } }
\vdef{default-11:CsMcPU-A:pvavew8:loEffE}   {\ensuremath{{0.001 } } }
\vdef{default-11:CsMcPU-A:pvavew8:hiEff}   {\ensuremath{{0.990 } } }
\vdef{default-11:CsMcPU-A:pvavew8:hiEffE}   {\ensuremath{{0.001 } } }
\vdef{default-11:CsMcPU-A:pvavew8:loDelta}   {\ensuremath{{+0.332 } } }
\vdef{default-11:CsMcPU-A:pvavew8:loDeltaE}   {\ensuremath{{0.136 } } }
\vdef{default-11:CsMcPU-A:pvavew8:hiDelta}   {\ensuremath{{-0.004 } } }
\vdef{default-11:CsMcPU-A:pvavew8:hiDeltaE}   {\ensuremath{{0.002 } } }
\vdef{default-11:CsData-A:pvntrk:loEff}   {\ensuremath{{1.000 } } }
\vdef{default-11:CsData-A:pvntrk:loEffE}   {\ensuremath{{0.000 } } }
\vdef{default-11:CsData-A:pvntrk:hiEff}   {\ensuremath{{1.000 } } }
\vdef{default-11:CsData-A:pvntrk:hiEffE}   {\ensuremath{{0.000 } } }
\vdef{default-11:CsMcPU-A:pvntrk:loEff}   {\ensuremath{{1.000 } } }
\vdef{default-11:CsMcPU-A:pvntrk:loEffE}   {\ensuremath{{0.000 } } }
\vdef{default-11:CsMcPU-A:pvntrk:hiEff}   {\ensuremath{{1.000 } } }
\vdef{default-11:CsMcPU-A:pvntrk:hiEffE}   {\ensuremath{{0.000 } } }
\vdef{default-11:CsMcPU-A:pvntrk:loDelta}   {\ensuremath{{+0.000 } } }
\vdef{default-11:CsMcPU-A:pvntrk:loDeltaE}   {\ensuremath{{0.000 } } }
\vdef{default-11:CsMcPU-A:pvntrk:hiDelta}   {\ensuremath{{+0.000 } } }
\vdef{default-11:CsMcPU-A:pvntrk:hiDeltaE}   {\ensuremath{{0.000 } } }
\vdef{default-11:CsData-A:muon1pt:loEff}   {\ensuremath{{1.013 } } }
\vdef{default-11:CsData-A:muon1pt:loEffE}   {\ensuremath{{\mathrm{NaN} } } }
\vdef{default-11:CsData-A:muon1pt:hiEff}   {\ensuremath{{1.000 } } }
\vdef{default-11:CsData-A:muon1pt:hiEffE}   {\ensuremath{{0.000 } } }
\vdef{default-11:CsMcPU-A:muon1pt:loEff}   {\ensuremath{{1.014 } } }
\vdef{default-11:CsMcPU-A:muon1pt:loEffE}   {\ensuremath{{\mathrm{NaN} } } }
\vdef{default-11:CsMcPU-A:muon1pt:hiEff}   {\ensuremath{{1.000 } } }
\vdef{default-11:CsMcPU-A:muon1pt:hiEffE}   {\ensuremath{{0.000 } } }
\vdef{default-11:CsMcPU-A:muon1pt:loDelta}   {\ensuremath{{-0.001 } } }
\vdef{default-11:CsMcPU-A:muon1pt:loDeltaE}   {\ensuremath{{\mathrm{NaN} } } }
\vdef{default-11:CsMcPU-A:muon1pt:hiDelta}   {\ensuremath{{+0.000 } } }
\vdef{default-11:CsMcPU-A:muon1pt:hiDeltaE}   {\ensuremath{{0.000 } } }
\vdef{default-11:CsData-A:muon2pt:loEff}   {\ensuremath{{0.074 } } }
\vdef{default-11:CsData-A:muon2pt:loEffE}   {\ensuremath{{0.003 } } }
\vdef{default-11:CsData-A:muon2pt:hiEff}   {\ensuremath{{0.926 } } }
\vdef{default-11:CsData-A:muon2pt:hiEffE}   {\ensuremath{{0.003 } } }
\vdef{default-11:CsMcPU-A:muon2pt:loEff}   {\ensuremath{{0.002 } } }
\vdef{default-11:CsMcPU-A:muon2pt:loEffE}   {\ensuremath{{0.000 } } }
\vdef{default-11:CsMcPU-A:muon2pt:hiEff}   {\ensuremath{{0.998 } } }
\vdef{default-11:CsMcPU-A:muon2pt:hiEffE}   {\ensuremath{{0.000 } } }
\vdef{default-11:CsMcPU-A:muon2pt:loDelta}   {\ensuremath{{+1.893 } } }
\vdef{default-11:CsMcPU-A:muon2pt:loDeltaE}   {\ensuremath{{0.025 } } }
\vdef{default-11:CsMcPU-A:muon2pt:hiDelta}   {\ensuremath{{-0.074 } } }
\vdef{default-11:CsMcPU-A:muon2pt:hiDeltaE}   {\ensuremath{{0.003 } } }
\vdef{default-11:CsData-A:muonseta:loEff}   {\ensuremath{{0.724 } } }
\vdef{default-11:CsData-A:muonseta:loEffE}   {\ensuremath{{0.003 } } }
\vdef{default-11:CsData-A:muonseta:hiEff}   {\ensuremath{{0.276 } } }
\vdef{default-11:CsData-A:muonseta:hiEffE}   {\ensuremath{{0.003 } } }
\vdef{default-11:CsMcPU-A:muonseta:loEff}   {\ensuremath{{0.857 } } }
\vdef{default-11:CsMcPU-A:muonseta:loEffE}   {\ensuremath{{0.003 } } }
\vdef{default-11:CsMcPU-A:muonseta:hiEff}   {\ensuremath{{0.143 } } }
\vdef{default-11:CsMcPU-A:muonseta:hiEffE}   {\ensuremath{{0.003 } } }
\vdef{default-11:CsMcPU-A:muonseta:loDelta}   {\ensuremath{{-0.168 } } }
\vdef{default-11:CsMcPU-A:muonseta:loDeltaE}   {\ensuremath{{0.006 } } }
\vdef{default-11:CsMcPU-A:muonseta:hiDelta}   {\ensuremath{{+0.632 } } }
\vdef{default-11:CsMcPU-A:muonseta:hiDeltaE}   {\ensuremath{{0.020 } } }
\vdef{default-11:CsData-A:pt:loEff}   {\ensuremath{{0.000 } } }
\vdef{default-11:CsData-A:pt:loEffE}   {\ensuremath{{0.000 } } }
\vdef{default-11:CsData-A:pt:hiEff}   {\ensuremath{{1.000 } } }
\vdef{default-11:CsData-A:pt:hiEffE}   {\ensuremath{{0.000 } } }
\vdef{default-11:CsMcPU-A:pt:loEff}   {\ensuremath{{0.000 } } }
\vdef{default-11:CsMcPU-A:pt:loEffE}   {\ensuremath{{0.000 } } }
\vdef{default-11:CsMcPU-A:pt:hiEff}   {\ensuremath{{1.000 } } }
\vdef{default-11:CsMcPU-A:pt:hiEffE}   {\ensuremath{{0.000 } } }
\vdef{default-11:CsMcPU-A:pt:loDelta}   {\ensuremath{{\mathrm{NaN} } } }
\vdef{default-11:CsMcPU-A:pt:loDeltaE}   {\ensuremath{{\mathrm{NaN} } } }
\vdef{default-11:CsMcPU-A:pt:hiDelta}   {\ensuremath{{+0.000 } } }
\vdef{default-11:CsMcPU-A:pt:hiDeltaE}   {\ensuremath{{0.000 } } }
\vdef{default-11:CsData-A:p:loEff}   {\ensuremath{{1.023 } } }
\vdef{default-11:CsData-A:p:loEffE}   {\ensuremath{{\mathrm{NaN} } } }
\vdef{default-11:CsData-A:p:hiEff}   {\ensuremath{{1.000 } } }
\vdef{default-11:CsData-A:p:hiEffE}   {\ensuremath{{0.000 } } }
\vdef{default-11:CsMcPU-A:p:loEff}   {\ensuremath{{1.008 } } }
\vdef{default-11:CsMcPU-A:p:loEffE}   {\ensuremath{{\mathrm{NaN} } } }
\vdef{default-11:CsMcPU-A:p:hiEff}   {\ensuremath{{1.000 } } }
\vdef{default-11:CsMcPU-A:p:hiEffE}   {\ensuremath{{0.000 } } }
\vdef{default-11:CsMcPU-A:p:loDelta}   {\ensuremath{{+0.015 } } }
\vdef{default-11:CsMcPU-A:p:loDeltaE}   {\ensuremath{{\mathrm{NaN} } } }
\vdef{default-11:CsMcPU-A:p:hiDelta}   {\ensuremath{{+0.000 } } }
\vdef{default-11:CsMcPU-A:p:hiDeltaE}   {\ensuremath{{0.000 } } }
\vdef{default-11:CsData-A:eta:loEff}   {\ensuremath{{0.719 } } }
\vdef{default-11:CsData-A:eta:loEffE}   {\ensuremath{{0.005 } } }
\vdef{default-11:CsData-A:eta:hiEff}   {\ensuremath{{0.281 } } }
\vdef{default-11:CsData-A:eta:hiEffE}   {\ensuremath{{0.005 } } }
\vdef{default-11:CsMcPU-A:eta:loEff}   {\ensuremath{{0.853 } } }
\vdef{default-11:CsMcPU-A:eta:loEffE}   {\ensuremath{{0.004 } } }
\vdef{default-11:CsMcPU-A:eta:hiEff}   {\ensuremath{{0.147 } } }
\vdef{default-11:CsMcPU-A:eta:hiEffE}   {\ensuremath{{0.004 } } }
\vdef{default-11:CsMcPU-A:eta:loDelta}   {\ensuremath{{-0.171 } } }
\vdef{default-11:CsMcPU-A:eta:loDeltaE}   {\ensuremath{{0.008 } } }
\vdef{default-11:CsMcPU-A:eta:hiDelta}   {\ensuremath{{+0.628 } } }
\vdef{default-11:CsMcPU-A:eta:hiDeltaE}   {\ensuremath{{0.028 } } }
\vdef{default-11:CsData-A:bdt:loEff}   {\ensuremath{{0.896 } } }
\vdef{default-11:CsData-A:bdt:loEffE}   {\ensuremath{{0.003 } } }
\vdef{default-11:CsData-A:bdt:hiEff}   {\ensuremath{{0.104 } } }
\vdef{default-11:CsData-A:bdt:hiEffE}   {\ensuremath{{0.003 } } }
\vdef{default-11:CsMcPU-A:bdt:loEff}   {\ensuremath{{0.857 } } }
\vdef{default-11:CsMcPU-A:bdt:loEffE}   {\ensuremath{{0.004 } } }
\vdef{default-11:CsMcPU-A:bdt:hiEff}   {\ensuremath{{0.143 } } }
\vdef{default-11:CsMcPU-A:bdt:hiEffE}   {\ensuremath{{0.004 } } }
\vdef{default-11:CsMcPU-A:bdt:loDelta}   {\ensuremath{{+0.044 } } }
\vdef{default-11:CsMcPU-A:bdt:loDeltaE}   {\ensuremath{{0.006 } } }
\vdef{default-11:CsMcPU-A:bdt:hiDelta}   {\ensuremath{{-0.314 } } }
\vdef{default-11:CsMcPU-A:bdt:hiDeltaE}   {\ensuremath{{0.039 } } }
\vdef{default-11:CsData-A:fl3d:loEff}   {\ensuremath{{0.823 } } }
\vdef{default-11:CsData-A:fl3d:loEffE}   {\ensuremath{{0.004 } } }
\vdef{default-11:CsData-A:fl3d:hiEff}   {\ensuremath{{0.177 } } }
\vdef{default-11:CsData-A:fl3d:hiEffE}   {\ensuremath{{0.004 } } }
\vdef{default-11:CsMcPU-A:fl3d:loEff}   {\ensuremath{{0.895 } } }
\vdef{default-11:CsMcPU-A:fl3d:loEffE}   {\ensuremath{{0.003 } } }
\vdef{default-11:CsMcPU-A:fl3d:hiEff}   {\ensuremath{{0.105 } } }
\vdef{default-11:CsMcPU-A:fl3d:hiEffE}   {\ensuremath{{0.003 } } }
\vdef{default-11:CsMcPU-A:fl3d:loDelta}   {\ensuremath{{-0.084 } } }
\vdef{default-11:CsMcPU-A:fl3d:loDeltaE}   {\ensuremath{{0.006 } } }
\vdef{default-11:CsMcPU-A:fl3d:hiDelta}   {\ensuremath{{+0.514 } } }
\vdef{default-11:CsMcPU-A:fl3d:hiDeltaE}   {\ensuremath{{0.033 } } }
\vdef{default-11:CsData-A:fl3de:loEff}   {\ensuremath{{1.000 } } }
\vdef{default-11:CsData-A:fl3de:loEffE}   {\ensuremath{{0.000 } } }
\vdef{default-11:CsData-A:fl3de:hiEff}   {\ensuremath{{0.000 } } }
\vdef{default-11:CsData-A:fl3de:hiEffE}   {\ensuremath{{0.000 } } }
\vdef{default-11:CsMcPU-A:fl3de:loEff}   {\ensuremath{{1.000 } } }
\vdef{default-11:CsMcPU-A:fl3de:loEffE}   {\ensuremath{{0.000 } } }
\vdef{default-11:CsMcPU-A:fl3de:hiEff}   {\ensuremath{{0.000 } } }
\vdef{default-11:CsMcPU-A:fl3de:hiEffE}   {\ensuremath{{0.000 } } }
\vdef{default-11:CsMcPU-A:fl3de:loDelta}   {\ensuremath{{+0.000 } } }
\vdef{default-11:CsMcPU-A:fl3de:loDeltaE}   {\ensuremath{{0.000 } } }
\vdef{default-11:CsMcPU-A:fl3de:hiDelta}   {\ensuremath{{-0.518 } } }
\vdef{default-11:CsMcPU-A:fl3de:hiDeltaE}   {\ensuremath{{1.095 } } }
\vdef{default-11:CsData-A:fls3d:loEff}   {\ensuremath{{0.095 } } }
\vdef{default-11:CsData-A:fls3d:loEffE}   {\ensuremath{{0.003 } } }
\vdef{default-11:CsData-A:fls3d:hiEff}   {\ensuremath{{0.905 } } }
\vdef{default-11:CsData-A:fls3d:hiEffE}   {\ensuremath{{0.003 } } }
\vdef{default-11:CsMcPU-A:fls3d:loEff}   {\ensuremath{{0.072 } } }
\vdef{default-11:CsMcPU-A:fls3d:loEffE}   {\ensuremath{{0.003 } } }
\vdef{default-11:CsMcPU-A:fls3d:hiEff}   {\ensuremath{{0.928 } } }
\vdef{default-11:CsMcPU-A:fls3d:hiEffE}   {\ensuremath{{0.003 } } }
\vdef{default-11:CsMcPU-A:fls3d:loDelta}   {\ensuremath{{+0.274 } } }
\vdef{default-11:CsMcPU-A:fls3d:loDeltaE}   {\ensuremath{{0.046 } } }
\vdef{default-11:CsMcPU-A:fls3d:hiDelta}   {\ensuremath{{-0.025 } } }
\vdef{default-11:CsMcPU-A:fls3d:hiDeltaE}   {\ensuremath{{0.004 } } }
\vdef{default-11:CsData-A:flsxy:loEff}   {\ensuremath{{1.010 } } }
\vdef{default-11:CsData-A:flsxy:loEffE}   {\ensuremath{{\mathrm{NaN} } } }
\vdef{default-11:CsData-A:flsxy:hiEff}   {\ensuremath{{1.000 } } }
\vdef{default-11:CsData-A:flsxy:hiEffE}   {\ensuremath{{0.000 } } }
\vdef{default-11:CsMcPU-A:flsxy:loEff}   {\ensuremath{{1.009 } } }
\vdef{default-11:CsMcPU-A:flsxy:loEffE}   {\ensuremath{{\mathrm{NaN} } } }
\vdef{default-11:CsMcPU-A:flsxy:hiEff}   {\ensuremath{{1.000 } } }
\vdef{default-11:CsMcPU-A:flsxy:hiEffE}   {\ensuremath{{0.000 } } }
\vdef{default-11:CsMcPU-A:flsxy:loDelta}   {\ensuremath{{+0.001 } } }
\vdef{default-11:CsMcPU-A:flsxy:loDeltaE}   {\ensuremath{{\mathrm{NaN} } } }
\vdef{default-11:CsMcPU-A:flsxy:hiDelta}   {\ensuremath{{+0.000 } } }
\vdef{default-11:CsMcPU-A:flsxy:hiDeltaE}   {\ensuremath{{0.000 } } }
\vdef{default-11:CsData-A:chi2dof:loEff}   {\ensuremath{{0.929 } } }
\vdef{default-11:CsData-A:chi2dof:loEffE}   {\ensuremath{{0.003 } } }
\vdef{default-11:CsData-A:chi2dof:hiEff}   {\ensuremath{{0.071 } } }
\vdef{default-11:CsData-A:chi2dof:hiEffE}   {\ensuremath{{0.003 } } }
\vdef{default-11:CsMcPU-A:chi2dof:loEff}   {\ensuremath{{0.951 } } }
\vdef{default-11:CsMcPU-A:chi2dof:loEffE}   {\ensuremath{{0.002 } } }
\vdef{default-11:CsMcPU-A:chi2dof:hiEff}   {\ensuremath{{0.049 } } }
\vdef{default-11:CsMcPU-A:chi2dof:hiEffE}   {\ensuremath{{0.002 } } }
\vdef{default-11:CsMcPU-A:chi2dof:loDelta}   {\ensuremath{{-0.023 } } }
\vdef{default-11:CsMcPU-A:chi2dof:loDeltaE}   {\ensuremath{{0.004 } } }
\vdef{default-11:CsMcPU-A:chi2dof:hiDelta}   {\ensuremath{{+0.359 } } }
\vdef{default-11:CsMcPU-A:chi2dof:hiDeltaE}   {\ensuremath{{0.058 } } }
\vdef{default-11:CsData-A:pchi2dof:loEff}   {\ensuremath{{0.624 } } }
\vdef{default-11:CsData-A:pchi2dof:loEffE}   {\ensuremath{{0.005 } } }
\vdef{default-11:CsData-A:pchi2dof:hiEff}   {\ensuremath{{0.376 } } }
\vdef{default-11:CsData-A:pchi2dof:hiEffE}   {\ensuremath{{0.005 } } }
\vdef{default-11:CsMcPU-A:pchi2dof:loEff}   {\ensuremath{{0.577 } } }
\vdef{default-11:CsMcPU-A:pchi2dof:loEffE}   {\ensuremath{{0.005 } } }
\vdef{default-11:CsMcPU-A:pchi2dof:hiEff}   {\ensuremath{{0.423 } } }
\vdef{default-11:CsMcPU-A:pchi2dof:hiEffE}   {\ensuremath{{0.005 } } }
\vdef{default-11:CsMcPU-A:pchi2dof:loDelta}   {\ensuremath{{+0.078 } } }
\vdef{default-11:CsMcPU-A:pchi2dof:loDeltaE}   {\ensuremath{{0.012 } } }
\vdef{default-11:CsMcPU-A:pchi2dof:hiDelta}   {\ensuremath{{-0.118 } } }
\vdef{default-11:CsMcPU-A:pchi2dof:hiDeltaE}   {\ensuremath{{0.018 } } }
\vdef{default-11:CsData-A:alpha:loEff}   {\ensuremath{{0.995 } } }
\vdef{default-11:CsData-A:alpha:loEffE}   {\ensuremath{{0.001 } } }
\vdef{default-11:CsData-A:alpha:hiEff}   {\ensuremath{{0.005 } } }
\vdef{default-11:CsData-A:alpha:hiEffE}   {\ensuremath{{0.001 } } }
\vdef{default-11:CsMcPU-A:alpha:loEff}   {\ensuremath{{0.991 } } }
\vdef{default-11:CsMcPU-A:alpha:loEffE}   {\ensuremath{{0.001 } } }
\vdef{default-11:CsMcPU-A:alpha:hiEff}   {\ensuremath{{0.009 } } }
\vdef{default-11:CsMcPU-A:alpha:hiEffE}   {\ensuremath{{0.001 } } }
\vdef{default-11:CsMcPU-A:alpha:loDelta}   {\ensuremath{{+0.004 } } }
\vdef{default-11:CsMcPU-A:alpha:loDeltaE}   {\ensuremath{{0.001 } } }
\vdef{default-11:CsMcPU-A:alpha:hiDelta}   {\ensuremath{{-0.573 } } }
\vdef{default-11:CsMcPU-A:alpha:hiDeltaE}   {\ensuremath{{0.179 } } }
\vdef{default-11:CsData-A:iso:loEff}   {\ensuremath{{0.093 } } }
\vdef{default-11:CsData-A:iso:loEffE}   {\ensuremath{{0.003 } } }
\vdef{default-11:CsData-A:iso:hiEff}   {\ensuremath{{0.907 } } }
\vdef{default-11:CsData-A:iso:hiEffE}   {\ensuremath{{0.003 } } }
\vdef{default-11:CsMcPU-A:iso:loEff}   {\ensuremath{{0.097 } } }
\vdef{default-11:CsMcPU-A:iso:loEffE}   {\ensuremath{{0.003 } } }
\vdef{default-11:CsMcPU-A:iso:hiEff}   {\ensuremath{{0.903 } } }
\vdef{default-11:CsMcPU-A:iso:hiEffE}   {\ensuremath{{0.003 } } }
\vdef{default-11:CsMcPU-A:iso:loDelta}   {\ensuremath{{-0.041 } } }
\vdef{default-11:CsMcPU-A:iso:loDeltaE}   {\ensuremath{{0.045 } } }
\vdef{default-11:CsMcPU-A:iso:hiDelta}   {\ensuremath{{+0.004 } } }
\vdef{default-11:CsMcPU-A:iso:hiDeltaE}   {\ensuremath{{0.005 } } }
\vdef{default-11:CsData-A:docatrk:loEff}   {\ensuremath{{0.067 } } }
\vdef{default-11:CsData-A:docatrk:loEffE}   {\ensuremath{{0.003 } } }
\vdef{default-11:CsData-A:docatrk:hiEff}   {\ensuremath{{0.933 } } }
\vdef{default-11:CsData-A:docatrk:hiEffE}   {\ensuremath{{0.003 } } }
\vdef{default-11:CsMcPU-A:docatrk:loEff}   {\ensuremath{{0.084 } } }
\vdef{default-11:CsMcPU-A:docatrk:loEffE}   {\ensuremath{{0.003 } } }
\vdef{default-11:CsMcPU-A:docatrk:hiEff}   {\ensuremath{{0.916 } } }
\vdef{default-11:CsMcPU-A:docatrk:hiEffE}   {\ensuremath{{0.003 } } }
\vdef{default-11:CsMcPU-A:docatrk:loDelta}   {\ensuremath{{-0.228 } } }
\vdef{default-11:CsMcPU-A:docatrk:loDeltaE}   {\ensuremath{{0.052 } } }
\vdef{default-11:CsMcPU-A:docatrk:hiDelta}   {\ensuremath{{+0.019 } } }
\vdef{default-11:CsMcPU-A:docatrk:hiDeltaE}   {\ensuremath{{0.004 } } }
\vdef{default-11:CsData-A:isotrk:loEff}   {\ensuremath{{1.000 } } }
\vdef{default-11:CsData-A:isotrk:loEffE}   {\ensuremath{{0.000 } } }
\vdef{default-11:CsData-A:isotrk:hiEff}   {\ensuremath{{1.000 } } }
\vdef{default-11:CsData-A:isotrk:hiEffE}   {\ensuremath{{0.000 } } }
\vdef{default-11:CsMcPU-A:isotrk:loEff}   {\ensuremath{{1.000 } } }
\vdef{default-11:CsMcPU-A:isotrk:loEffE}   {\ensuremath{{0.000 } } }
\vdef{default-11:CsMcPU-A:isotrk:hiEff}   {\ensuremath{{1.000 } } }
\vdef{default-11:CsMcPU-A:isotrk:hiEffE}   {\ensuremath{{0.000 } } }
\vdef{default-11:CsMcPU-A:isotrk:loDelta}   {\ensuremath{{+0.000 } } }
\vdef{default-11:CsMcPU-A:isotrk:loDeltaE}   {\ensuremath{{0.000 } } }
\vdef{default-11:CsMcPU-A:isotrk:hiDelta}   {\ensuremath{{+0.000 } } }
\vdef{default-11:CsMcPU-A:isotrk:hiDeltaE}   {\ensuremath{{0.000 } } }
\vdef{default-11:CsData-A:closetrk:loEff}   {\ensuremath{{0.982 } } }
\vdef{default-11:CsData-A:closetrk:loEffE}   {\ensuremath{{0.001 } } }
\vdef{default-11:CsData-A:closetrk:hiEff}   {\ensuremath{{0.018 } } }
\vdef{default-11:CsData-A:closetrk:hiEffE}   {\ensuremath{{0.001 } } }
\vdef{default-11:CsMcPU-A:closetrk:loEff}   {\ensuremath{{0.975 } } }
\vdef{default-11:CsMcPU-A:closetrk:loEffE}   {\ensuremath{{0.002 } } }
\vdef{default-11:CsMcPU-A:closetrk:hiEff}   {\ensuremath{{0.025 } } }
\vdef{default-11:CsMcPU-A:closetrk:hiEffE}   {\ensuremath{{0.002 } } }
\vdef{default-11:CsMcPU-A:closetrk:loDelta}   {\ensuremath{{+0.007 } } }
\vdef{default-11:CsMcPU-A:closetrk:loDeltaE}   {\ensuremath{{0.002 } } }
\vdef{default-11:CsMcPU-A:closetrk:hiDelta}   {\ensuremath{{-0.315 } } }
\vdef{default-11:CsMcPU-A:closetrk:hiDeltaE}   {\ensuremath{{0.101 } } }
\vdef{default-11:CsData-A:lip:loEff}   {\ensuremath{{1.000 } } }
\vdef{default-11:CsData-A:lip:loEffE}   {\ensuremath{{0.000 } } }
\vdef{default-11:CsData-A:lip:hiEff}   {\ensuremath{{0.000 } } }
\vdef{default-11:CsData-A:lip:hiEffE}   {\ensuremath{{0.000 } } }
\vdef{default-11:CsMcPU-A:lip:loEff}   {\ensuremath{{1.000 } } }
\vdef{default-11:CsMcPU-A:lip:loEffE}   {\ensuremath{{0.000 } } }
\vdef{default-11:CsMcPU-A:lip:hiEff}   {\ensuremath{{0.000 } } }
\vdef{default-11:CsMcPU-A:lip:hiEffE}   {\ensuremath{{0.000 } } }
\vdef{default-11:CsMcPU-A:lip:loDelta}   {\ensuremath{{+0.000 } } }
\vdef{default-11:CsMcPU-A:lip:loDeltaE}   {\ensuremath{{0.000 } } }
\vdef{default-11:CsMcPU-A:lip:hiDelta}   {\ensuremath{{\mathrm{NaN} } } }
\vdef{default-11:CsMcPU-A:lip:hiDeltaE}   {\ensuremath{{\mathrm{NaN} } } }
\vdef{default-11:CsData-A:lip:inEff}   {\ensuremath{{1.000 } } }
\vdef{default-11:CsData-A:lip:inEffE}   {\ensuremath{{0.000 } } }
\vdef{default-11:CsMcPU-A:lip:inEff}   {\ensuremath{{1.000 } } }
\vdef{default-11:CsMcPU-A:lip:inEffE}   {\ensuremath{{0.000 } } }
\vdef{default-11:CsMcPU-A:lip:inDelta}   {\ensuremath{{+0.000 } } }
\vdef{default-11:CsMcPU-A:lip:inDeltaE}   {\ensuremath{{0.000 } } }
\vdef{default-11:CsData-A:lips:loEff}   {\ensuremath{{1.000 } } }
\vdef{default-11:CsData-A:lips:loEffE}   {\ensuremath{{0.000 } } }
\vdef{default-11:CsData-A:lips:hiEff}   {\ensuremath{{0.000 } } }
\vdef{default-11:CsData-A:lips:hiEffE}   {\ensuremath{{0.000 } } }
\vdef{default-11:CsMcPU-A:lips:loEff}   {\ensuremath{{1.000 } } }
\vdef{default-11:CsMcPU-A:lips:loEffE}   {\ensuremath{{0.000 } } }
\vdef{default-11:CsMcPU-A:lips:hiEff}   {\ensuremath{{0.000 } } }
\vdef{default-11:CsMcPU-A:lips:hiEffE}   {\ensuremath{{0.000 } } }
\vdef{default-11:CsMcPU-A:lips:loDelta}   {\ensuremath{{+0.000 } } }
\vdef{default-11:CsMcPU-A:lips:loDeltaE}   {\ensuremath{{0.000 } } }
\vdef{default-11:CsMcPU-A:lips:hiDelta}   {\ensuremath{{\mathrm{NaN} } } }
\vdef{default-11:CsMcPU-A:lips:hiDeltaE}   {\ensuremath{{\mathrm{NaN} } } }
\vdef{default-11:CsData-A:lips:inEff}   {\ensuremath{{1.000 } } }
\vdef{default-11:CsData-A:lips:inEffE}   {\ensuremath{{0.000 } } }
\vdef{default-11:CsMcPU-A:lips:inEff}   {\ensuremath{{1.000 } } }
\vdef{default-11:CsMcPU-A:lips:inEffE}   {\ensuremath{{0.000 } } }
\vdef{default-11:CsMcPU-A:lips:inDelta}   {\ensuremath{{+0.000 } } }
\vdef{default-11:CsMcPU-A:lips:inDeltaE}   {\ensuremath{{0.000 } } }
\vdef{default-11:CsData-A:ip:loEff}   {\ensuremath{{0.974 } } }
\vdef{default-11:CsData-A:ip:loEffE}   {\ensuremath{{0.002 } } }
\vdef{default-11:CsData-A:ip:hiEff}   {\ensuremath{{0.026 } } }
\vdef{default-11:CsData-A:ip:hiEffE}   {\ensuremath{{0.002 } } }
\vdef{default-11:CsMcPU-A:ip:loEff}   {\ensuremath{{0.968 } } }
\vdef{default-11:CsMcPU-A:ip:loEffE}   {\ensuremath{{0.002 } } }
\vdef{default-11:CsMcPU-A:ip:hiEff}   {\ensuremath{{0.032 } } }
\vdef{default-11:CsMcPU-A:ip:hiEffE}   {\ensuremath{{0.002 } } }
\vdef{default-11:CsMcPU-A:ip:loDelta}   {\ensuremath{{+0.006 } } }
\vdef{default-11:CsMcPU-A:ip:loDeltaE}   {\ensuremath{{0.003 } } }
\vdef{default-11:CsMcPU-A:ip:hiDelta}   {\ensuremath{{-0.207 } } }
\vdef{default-11:CsMcPU-A:ip:hiDeltaE}   {\ensuremath{{0.087 } } }
\vdef{default-11:CsData-A:ips:loEff}   {\ensuremath{{0.939 } } }
\vdef{default-11:CsData-A:ips:loEffE}   {\ensuremath{{0.003 } } }
\vdef{default-11:CsData-A:ips:hiEff}   {\ensuremath{{0.061 } } }
\vdef{default-11:CsData-A:ips:hiEffE}   {\ensuremath{{0.003 } } }
\vdef{default-11:CsMcPU-A:ips:loEff}   {\ensuremath{{0.938 } } }
\vdef{default-11:CsMcPU-A:ips:loEffE}   {\ensuremath{{0.003 } } }
\vdef{default-11:CsMcPU-A:ips:hiEff}   {\ensuremath{{0.062 } } }
\vdef{default-11:CsMcPU-A:ips:hiEffE}   {\ensuremath{{0.003 } } }
\vdef{default-11:CsMcPU-A:ips:loDelta}   {\ensuremath{{+0.001 } } }
\vdef{default-11:CsMcPU-A:ips:loDeltaE}   {\ensuremath{{0.004 } } }
\vdef{default-11:CsMcPU-A:ips:hiDelta}   {\ensuremath{{-0.014 } } }
\vdef{default-11:CsMcPU-A:ips:hiDeltaE}   {\ensuremath{{0.058 } } }
\vdef{default-11:CsData-A:maxdoca:loEff}   {\ensuremath{{1.000 } } }
\vdef{default-11:CsData-A:maxdoca:loEffE}   {\ensuremath{{0.000 } } }
\vdef{default-11:CsData-A:maxdoca:hiEff}   {\ensuremath{{0.034 } } }
\vdef{default-11:CsData-A:maxdoca:hiEffE}   {\ensuremath{{0.002 } } }
\vdef{default-11:CsMcPU-A:maxdoca:loEff}   {\ensuremath{{1.000 } } }
\vdef{default-11:CsMcPU-A:maxdoca:loEffE}   {\ensuremath{{0.000 } } }
\vdef{default-11:CsMcPU-A:maxdoca:hiEff}   {\ensuremath{{0.021 } } }
\vdef{default-11:CsMcPU-A:maxdoca:hiEffE}   {\ensuremath{{0.002 } } }
\vdef{default-11:CsMcPU-A:maxdoca:loDelta}   {\ensuremath{{+0.000 } } }
\vdef{default-11:CsMcPU-A:maxdoca:loDeltaE}   {\ensuremath{{0.000 } } }
\vdef{default-11:CsMcPU-A:maxdoca:hiDelta}   {\ensuremath{{+0.465 } } }
\vdef{default-11:CsMcPU-A:maxdoca:hiDeltaE}   {\ensuremath{{0.090 } } }
\vdef{default-11:CsData-A:kaonspt:loEff}   {\ensuremath{{1.000 } } }
\vdef{default-11:CsData-A:kaonspt:loEffE}   {\ensuremath{{\mathrm{NaN} } } }
\vdef{default-11:CsData-A:kaonspt:hiEff}   {\ensuremath{{1.000 } } }
\vdef{default-11:CsData-A:kaonspt:hiEffE}   {\ensuremath{{0.000 } } }
\vdef{default-11:CsMcPU-A:kaonspt:loEff}   {\ensuremath{{1.000 } } }
\vdef{default-11:CsMcPU-A:kaonspt:loEffE}   {\ensuremath{{\mathrm{NaN} } } }
\vdef{default-11:CsMcPU-A:kaonspt:hiEff}   {\ensuremath{{1.000 } } }
\vdef{default-11:CsMcPU-A:kaonspt:hiEffE}   {\ensuremath{{0.000 } } }
\vdef{default-11:CsMcPU-A:kaonspt:loDelta}   {\ensuremath{{+0.000 } } }
\vdef{default-11:CsMcPU-A:kaonspt:loDeltaE}   {\ensuremath{{\mathrm{NaN} } } }
\vdef{default-11:CsMcPU-A:kaonspt:hiDelta}   {\ensuremath{{+0.000 } } }
\vdef{default-11:CsMcPU-A:kaonspt:hiDeltaE}   {\ensuremath{{0.000 } } }
\vdef{default-11:CsData-A:psipt:loEff}   {\ensuremath{{1.006 } } }
\vdef{default-11:CsData-A:psipt:loEffE}   {\ensuremath{{\mathrm{NaN} } } }
\vdef{default-11:CsData-A:psipt:hiEff}   {\ensuremath{{1.000 } } }
\vdef{default-11:CsData-A:psipt:hiEffE}   {\ensuremath{{0.000 } } }
\vdef{default-11:CsMcPU-A:psipt:loEff}   {\ensuremath{{1.004 } } }
\vdef{default-11:CsMcPU-A:psipt:loEffE}   {\ensuremath{{\mathrm{NaN} } } }
\vdef{default-11:CsMcPU-A:psipt:hiEff}   {\ensuremath{{1.000 } } }
\vdef{default-11:CsMcPU-A:psipt:hiEffE}   {\ensuremath{{0.000 } } }
\vdef{default-11:CsMcPU-A:psipt:loDelta}   {\ensuremath{{+0.001 } } }
\vdef{default-11:CsMcPU-A:psipt:loDeltaE}   {\ensuremath{{\mathrm{NaN} } } }
\vdef{default-11:CsMcPU-A:psipt:hiDelta}   {\ensuremath{{+0.000 } } }
\vdef{default-11:CsMcPU-A:psipt:hiDeltaE}   {\ensuremath{{0.000 } } }
\vdef{default-11:CsData-A:phipt:loEff}   {\ensuremath{{1.015 } } }
\vdef{default-11:CsData-A:phipt:loEffE}   {\ensuremath{{\mathrm{NaN} } } }
\vdef{default-11:CsData-A:phipt:hiEff}   {\ensuremath{{1.000 } } }
\vdef{default-11:CsData-A:phipt:hiEffE}   {\ensuremath{{0.000 } } }
\vdef{default-11:CsMcPU-A:phipt:loEff}   {\ensuremath{{1.008 } } }
\vdef{default-11:CsMcPU-A:phipt:loEffE}   {\ensuremath{{\mathrm{NaN} } } }
\vdef{default-11:CsMcPU-A:phipt:hiEff}   {\ensuremath{{1.000 } } }
\vdef{default-11:CsMcPU-A:phipt:hiEffE}   {\ensuremath{{0.000 } } }
\vdef{default-11:CsMcPU-A:phipt:loDelta}   {\ensuremath{{+0.007 } } }
\vdef{default-11:CsMcPU-A:phipt:loDeltaE}   {\ensuremath{{\mathrm{NaN} } } }
\vdef{default-11:CsMcPU-A:phipt:hiDelta}   {\ensuremath{{+0.000 } } }
\vdef{default-11:CsMcPU-A:phipt:hiDeltaE}   {\ensuremath{{0.000 } } }
\vdef{default-11:CsData-A:deltar:loEff}   {\ensuremath{{1.000 } } }
\vdef{default-11:CsData-A:deltar:loEffE}   {\ensuremath{{0.000 } } }
\vdef{default-11:CsData-A:deltar:hiEff}   {\ensuremath{{0.000 } } }
\vdef{default-11:CsData-A:deltar:hiEffE}   {\ensuremath{{0.000 } } }
\vdef{default-11:CsMcPU-A:deltar:loEff}   {\ensuremath{{0.999 } } }
\vdef{default-11:CsMcPU-A:deltar:loEffE}   {\ensuremath{{0.000 } } }
\vdef{default-11:CsMcPU-A:deltar:hiEff}   {\ensuremath{{0.001 } } }
\vdef{default-11:CsMcPU-A:deltar:hiEffE}   {\ensuremath{{0.000 } } }
\vdef{default-11:CsMcPU-A:deltar:loDelta}   {\ensuremath{{+0.001 } } }
\vdef{default-11:CsMcPU-A:deltar:loDeltaE}   {\ensuremath{{0.000 } } }
\vdef{default-11:CsMcPU-A:deltar:hiDelta}   {\ensuremath{{-2.000 } } }
\vdef{default-11:CsMcPU-A:deltar:hiDeltaE}   {\ensuremath{{0.419 } } }
\vdef{default-11:CsData-A:mkk:loEff}   {\ensuremath{{1.159 } } }
\vdef{default-11:CsData-A:mkk:loEffE}   {\ensuremath{{\mathrm{NaN} } } }
\vdef{default-11:CsData-A:mkk:hiEff}   {\ensuremath{{1.000 } } }
\vdef{default-11:CsData-A:mkk:hiEffE}   {\ensuremath{{0.000 } } }
\vdef{default-11:CsMcPU-A:mkk:loEff}   {\ensuremath{{1.000 } } }
\vdef{default-11:CsMcPU-A:mkk:loEffE}   {\ensuremath{{0.000 } } }
\vdef{default-11:CsMcPU-A:mkk:hiEff}   {\ensuremath{{1.000 } } }
\vdef{default-11:CsMcPU-A:mkk:hiEffE}   {\ensuremath{{0.000 } } }
\vdef{default-11:CsMcPU-A:mkk:loDelta}   {\ensuremath{{+0.147 } } }
\vdef{default-11:CsMcPU-A:mkk:loDeltaE}   {\ensuremath{{\mathrm{NaN} } } }
\vdef{default-11:CsMcPU-A:mkk:hiDelta}   {\ensuremath{{+0.000 } } }
\vdef{default-11:CsMcPU-A:mkk:hiDeltaE}   {\ensuremath{{0.000 } } }
\vdef{default-11:CsData-APV0:osiso:loEff}   {\ensuremath{{1.006 } } }
\vdef{default-11:CsData-APV0:osiso:loEffE}   {\ensuremath{{\mathrm{NaN} } } }
\vdef{default-11:CsData-APV0:osiso:hiEff}   {\ensuremath{{1.000 } } }
\vdef{default-11:CsData-APV0:osiso:hiEffE}   {\ensuremath{{0.000 } } }
\vdef{default-11:CsMcPU-APV0:osiso:loEff}   {\ensuremath{{1.002 } } }
\vdef{default-11:CsMcPU-APV0:osiso:loEffE}   {\ensuremath{{\mathrm{NaN} } } }
\vdef{default-11:CsMcPU-APV0:osiso:hiEff}   {\ensuremath{{1.000 } } }
\vdef{default-11:CsMcPU-APV0:osiso:hiEffE}   {\ensuremath{{0.000 } } }
\vdef{default-11:CsMcPU-APV0:osiso:loDelta}   {\ensuremath{{+0.004 } } }
\vdef{default-11:CsMcPU-APV0:osiso:loDeltaE}   {\ensuremath{{\mathrm{NaN} } } }
\vdef{default-11:CsMcPU-APV0:osiso:hiDelta}   {\ensuremath{{+0.000 } } }
\vdef{default-11:CsMcPU-APV0:osiso:hiDeltaE}   {\ensuremath{{0.001 } } }
\vdef{default-11:CsData-APV0:osreliso:loEff}   {\ensuremath{{0.260 } } }
\vdef{default-11:CsData-APV0:osreliso:loEffE}   {\ensuremath{{0.008 } } }
\vdef{default-11:CsData-APV0:osreliso:hiEff}   {\ensuremath{{0.740 } } }
\vdef{default-11:CsData-APV0:osreliso:hiEffE}   {\ensuremath{{0.008 } } }
\vdef{default-11:CsMcPU-APV0:osreliso:loEff}   {\ensuremath{{0.287 } } }
\vdef{default-11:CsMcPU-APV0:osreliso:loEffE}   {\ensuremath{{0.009 } } }
\vdef{default-11:CsMcPU-APV0:osreliso:hiEff}   {\ensuremath{{0.713 } } }
\vdef{default-11:CsMcPU-APV0:osreliso:hiEffE}   {\ensuremath{{0.009 } } }
\vdef{default-11:CsMcPU-APV0:osreliso:loDelta}   {\ensuremath{{-0.097 } } }
\vdef{default-11:CsMcPU-APV0:osreliso:loDeltaE}   {\ensuremath{{0.044 } } }
\vdef{default-11:CsMcPU-APV0:osreliso:hiDelta}   {\ensuremath{{+0.037 } } }
\vdef{default-11:CsMcPU-APV0:osreliso:hiDeltaE}   {\ensuremath{{0.017 } } }
\vdef{default-11:CsData-APV0:osmuonpt:loEff}   {\ensuremath{{0.000 } } }
\vdef{default-11:CsData-APV0:osmuonpt:loEffE}   {\ensuremath{{0.009 } } }
\vdef{default-11:CsData-APV0:osmuonpt:hiEff}   {\ensuremath{{1.000 } } }
\vdef{default-11:CsData-APV0:osmuonpt:hiEffE}   {\ensuremath{{0.009 } } }
\vdef{default-11:CsMcPU-APV0:osmuonpt:loEff}   {\ensuremath{{0.000 } } }
\vdef{default-11:CsMcPU-APV0:osmuonpt:loEffE}   {\ensuremath{{0.009 } } }
\vdef{default-11:CsMcPU-APV0:osmuonpt:hiEff}   {\ensuremath{{1.000 } } }
\vdef{default-11:CsMcPU-APV0:osmuonpt:hiEffE}   {\ensuremath{{0.009 } } }
\vdef{default-11:CsMcPU-APV0:osmuonpt:loDelta}   {\ensuremath{{\mathrm{NaN} } } }
\vdef{default-11:CsMcPU-APV0:osmuonpt:loDeltaE}   {\ensuremath{{\mathrm{NaN} } } }
\vdef{default-11:CsMcPU-APV0:osmuonpt:hiDelta}   {\ensuremath{{+0.000 } } }
\vdef{default-11:CsMcPU-APV0:osmuonpt:hiDeltaE}   {\ensuremath{{0.012 } } }
\vdef{default-11:CsData-APV0:osmuondr:loEff}   {\ensuremath{{0.020 } } }
\vdef{default-11:CsData-APV0:osmuondr:loEffE}   {\ensuremath{{0.015 } } }
\vdef{default-11:CsData-APV0:osmuondr:hiEff}   {\ensuremath{{0.980 } } }
\vdef{default-11:CsData-APV0:osmuondr:hiEffE}   {\ensuremath{{0.015 } } }
\vdef{default-11:CsMcPU-APV0:osmuondr:loEff}   {\ensuremath{{0.053 } } }
\vdef{default-11:CsMcPU-APV0:osmuondr:loEffE}   {\ensuremath{{0.022 } } }
\vdef{default-11:CsMcPU-APV0:osmuondr:hiEff}   {\ensuremath{{0.947 } } }
\vdef{default-11:CsMcPU-APV0:osmuondr:hiEffE}   {\ensuremath{{0.022 } } }
\vdef{default-11:CsMcPU-APV0:osmuondr:loDelta}   {\ensuremath{{-0.903 } } }
\vdef{default-11:CsMcPU-APV0:osmuondr:loDeltaE}   {\ensuremath{{0.675 } } }
\vdef{default-11:CsMcPU-APV0:osmuondr:hiDelta}   {\ensuremath{{+0.034 } } }
\vdef{default-11:CsMcPU-APV0:osmuondr:hiDeltaE}   {\ensuremath{{0.028 } } }
\vdef{default-11:CsData-APV0:hlt:loEff}   {\ensuremath{{0.049 } } }
\vdef{default-11:CsData-APV0:hlt:loEffE}   {\ensuremath{{0.004 } } }
\vdef{default-11:CsData-APV0:hlt:hiEff}   {\ensuremath{{0.951 } } }
\vdef{default-11:CsData-APV0:hlt:hiEffE}   {\ensuremath{{0.004 } } }
\vdef{default-11:CsMcPU-APV0:hlt:loEff}   {\ensuremath{{0.281 } } }
\vdef{default-11:CsMcPU-APV0:hlt:loEffE}   {\ensuremath{{0.009 } } }
\vdef{default-11:CsMcPU-APV0:hlt:hiEff}   {\ensuremath{{0.719 } } }
\vdef{default-11:CsMcPU-APV0:hlt:hiEffE}   {\ensuremath{{0.009 } } }
\vdef{default-11:CsMcPU-APV0:hlt:loDelta}   {\ensuremath{{-1.400 } } }
\vdef{default-11:CsMcPU-APV0:hlt:loDeltaE}   {\ensuremath{{0.047 } } }
\vdef{default-11:CsMcPU-APV0:hlt:hiDelta}   {\ensuremath{{+0.277 } } }
\vdef{default-11:CsMcPU-APV0:hlt:hiDeltaE}   {\ensuremath{{0.013 } } }
\vdef{default-11:CsData-APV0:muonsid:loEff}   {\ensuremath{{0.150 } } }
\vdef{default-11:CsData-APV0:muonsid:loEffE}   {\ensuremath{{0.007 } } }
\vdef{default-11:CsData-APV0:muonsid:hiEff}   {\ensuremath{{0.850 } } }
\vdef{default-11:CsData-APV0:muonsid:hiEffE}   {\ensuremath{{0.007 } } }
\vdef{default-11:CsMcPU-APV0:muonsid:loEff}   {\ensuremath{{0.230 } } }
\vdef{default-11:CsMcPU-APV0:muonsid:loEffE}   {\ensuremath{{0.008 } } }
\vdef{default-11:CsMcPU-APV0:muonsid:hiEff}   {\ensuremath{{0.770 } } }
\vdef{default-11:CsMcPU-APV0:muonsid:hiEffE}   {\ensuremath{{0.008 } } }
\vdef{default-11:CsMcPU-APV0:muonsid:loDelta}   {\ensuremath{{-0.424 } } }
\vdef{default-11:CsMcPU-APV0:muonsid:loDeltaE}   {\ensuremath{{0.053 } } }
\vdef{default-11:CsMcPU-APV0:muonsid:hiDelta}   {\ensuremath{{+0.099 } } }
\vdef{default-11:CsMcPU-APV0:muonsid:hiDeltaE}   {\ensuremath{{0.013 } } }
\vdef{default-11:CsData-APV0:tracksqual:loEff}   {\ensuremath{{0.002 } } }
\vdef{default-11:CsData-APV0:tracksqual:loEffE}   {\ensuremath{{0.001 } } }
\vdef{default-11:CsData-APV0:tracksqual:hiEff}   {\ensuremath{{0.998 } } }
\vdef{default-11:CsData-APV0:tracksqual:hiEffE}   {\ensuremath{{0.001 } } }
\vdef{default-11:CsMcPU-APV0:tracksqual:loEff}   {\ensuremath{{0.000 } } }
\vdef{default-11:CsMcPU-APV0:tracksqual:loEffE}   {\ensuremath{{0.000 } } }
\vdef{default-11:CsMcPU-APV0:tracksqual:hiEff}   {\ensuremath{{1.000 } } }
\vdef{default-11:CsMcPU-APV0:tracksqual:hiEffE}   {\ensuremath{{0.000 } } }
\vdef{default-11:CsMcPU-APV0:tracksqual:loDelta}   {\ensuremath{{+2.000 } } }
\vdef{default-11:CsMcPU-APV0:tracksqual:loDeltaE}   {\ensuremath{{0.999 } } }
\vdef{default-11:CsMcPU-APV0:tracksqual:hiDelta}   {\ensuremath{{-0.002 } } }
\vdef{default-11:CsMcPU-APV0:tracksqual:hiDeltaE}   {\ensuremath{{0.001 } } }
\vdef{default-11:CsData-APV0:pvz:loEff}   {\ensuremath{{0.523 } } }
\vdef{default-11:CsData-APV0:pvz:loEffE}   {\ensuremath{{0.010 } } }
\vdef{default-11:CsData-APV0:pvz:hiEff}   {\ensuremath{{0.477 } } }
\vdef{default-11:CsData-APV0:pvz:hiEffE}   {\ensuremath{{0.010 } } }
\vdef{default-11:CsMcPU-APV0:pvz:loEff}   {\ensuremath{{0.481 } } }
\vdef{default-11:CsMcPU-APV0:pvz:loEffE}   {\ensuremath{{0.010 } } }
\vdef{default-11:CsMcPU-APV0:pvz:hiEff}   {\ensuremath{{0.519 } } }
\vdef{default-11:CsMcPU-APV0:pvz:hiEffE}   {\ensuremath{{0.010 } } }
\vdef{default-11:CsMcPU-APV0:pvz:loDelta}   {\ensuremath{{+0.084 } } }
\vdef{default-11:CsMcPU-APV0:pvz:loDeltaE}   {\ensuremath{{0.028 } } }
\vdef{default-11:CsMcPU-APV0:pvz:hiDelta}   {\ensuremath{{-0.085 } } }
\vdef{default-11:CsMcPU-APV0:pvz:hiDeltaE}   {\ensuremath{{0.028 } } }
\vdef{default-11:CsData-APV0:pvn:loEff}   {\ensuremath{{1.000 } } }
\vdef{default-11:CsData-APV0:pvn:loEffE}   {\ensuremath{{0.000 } } }
\vdef{default-11:CsData-APV0:pvn:hiEff}   {\ensuremath{{1.000 } } }
\vdef{default-11:CsData-APV0:pvn:hiEffE}   {\ensuremath{{0.000 } } }
\vdef{default-11:CsMcPU-APV0:pvn:loEff}   {\ensuremath{{1.000 } } }
\vdef{default-11:CsMcPU-APV0:pvn:loEffE}   {\ensuremath{{0.000 } } }
\vdef{default-11:CsMcPU-APV0:pvn:hiEff}   {\ensuremath{{1.000 } } }
\vdef{default-11:CsMcPU-APV0:pvn:hiEffE}   {\ensuremath{{0.000 } } }
\vdef{default-11:CsMcPU-APV0:pvn:loDelta}   {\ensuremath{{+0.000 } } }
\vdef{default-11:CsMcPU-APV0:pvn:loDeltaE}   {\ensuremath{{0.001 } } }
\vdef{default-11:CsMcPU-APV0:pvn:hiDelta}   {\ensuremath{{+0.000 } } }
\vdef{default-11:CsMcPU-APV0:pvn:hiDeltaE}   {\ensuremath{{0.001 } } }
\vdef{default-11:CsData-APV0:pvavew8:loEff}   {\ensuremath{{0.014 } } }
\vdef{default-11:CsData-APV0:pvavew8:loEffE}   {\ensuremath{{0.002 } } }
\vdef{default-11:CsData-APV0:pvavew8:hiEff}   {\ensuremath{{0.986 } } }
\vdef{default-11:CsData-APV0:pvavew8:hiEffE}   {\ensuremath{{0.002 } } }
\vdef{default-11:CsMcPU-APV0:pvavew8:loEff}   {\ensuremath{{0.007 } } }
\vdef{default-11:CsMcPU-APV0:pvavew8:loEffE}   {\ensuremath{{0.002 } } }
\vdef{default-11:CsMcPU-APV0:pvavew8:hiEff}   {\ensuremath{{0.993 } } }
\vdef{default-11:CsMcPU-APV0:pvavew8:hiEffE}   {\ensuremath{{0.002 } } }
\vdef{default-11:CsMcPU-APV0:pvavew8:loDelta}   {\ensuremath{{+0.671 } } }
\vdef{default-11:CsMcPU-APV0:pvavew8:loDeltaE}   {\ensuremath{{0.267 } } }
\vdef{default-11:CsMcPU-APV0:pvavew8:hiDelta}   {\ensuremath{{-0.007 } } }
\vdef{default-11:CsMcPU-APV0:pvavew8:hiDeltaE}   {\ensuremath{{0.003 } } }
\vdef{default-11:CsData-APV0:pvntrk:loEff}   {\ensuremath{{1.000 } } }
\vdef{default-11:CsData-APV0:pvntrk:loEffE}   {\ensuremath{{0.000 } } }
\vdef{default-11:CsData-APV0:pvntrk:hiEff}   {\ensuremath{{1.000 } } }
\vdef{default-11:CsData-APV0:pvntrk:hiEffE}   {\ensuremath{{0.000 } } }
\vdef{default-11:CsMcPU-APV0:pvntrk:loEff}   {\ensuremath{{1.000 } } }
\vdef{default-11:CsMcPU-APV0:pvntrk:loEffE}   {\ensuremath{{0.000 } } }
\vdef{default-11:CsMcPU-APV0:pvntrk:hiEff}   {\ensuremath{{1.000 } } }
\vdef{default-11:CsMcPU-APV0:pvntrk:hiEffE}   {\ensuremath{{0.000 } } }
\vdef{default-11:CsMcPU-APV0:pvntrk:loDelta}   {\ensuremath{{+0.000 } } }
\vdef{default-11:CsMcPU-APV0:pvntrk:loDeltaE}   {\ensuremath{{0.001 } } }
\vdef{default-11:CsMcPU-APV0:pvntrk:hiDelta}   {\ensuremath{{+0.000 } } }
\vdef{default-11:CsMcPU-APV0:pvntrk:hiDeltaE}   {\ensuremath{{0.001 } } }
\vdef{default-11:CsData-APV0:muon1pt:loEff}   {\ensuremath{{1.014 } } }
\vdef{default-11:CsData-APV0:muon1pt:loEffE}   {\ensuremath{{\mathrm{NaN} } } }
\vdef{default-11:CsData-APV0:muon1pt:hiEff}   {\ensuremath{{1.000 } } }
\vdef{default-11:CsData-APV0:muon1pt:hiEffE}   {\ensuremath{{0.000 } } }
\vdef{default-11:CsMcPU-APV0:muon1pt:loEff}   {\ensuremath{{1.014 } } }
\vdef{default-11:CsMcPU-APV0:muon1pt:loEffE}   {\ensuremath{{\mathrm{NaN} } } }
\vdef{default-11:CsMcPU-APV0:muon1pt:hiEff}   {\ensuremath{{1.000 } } }
\vdef{default-11:CsMcPU-APV0:muon1pt:hiEffE}   {\ensuremath{{0.000 } } }
\vdef{default-11:CsMcPU-APV0:muon1pt:loDelta}   {\ensuremath{{+0.000 } } }
\vdef{default-11:CsMcPU-APV0:muon1pt:loDeltaE}   {\ensuremath{{\mathrm{NaN} } } }
\vdef{default-11:CsMcPU-APV0:muon1pt:hiDelta}   {\ensuremath{{+0.000 } } }
\vdef{default-11:CsMcPU-APV0:muon1pt:hiDeltaE}   {\ensuremath{{0.001 } } }
\vdef{default-11:CsData-APV0:muon2pt:loEff}   {\ensuremath{{0.114 } } }
\vdef{default-11:CsData-APV0:muon2pt:loEffE}   {\ensuremath{{0.006 } } }
\vdef{default-11:CsData-APV0:muon2pt:hiEff}   {\ensuremath{{0.886 } } }
\vdef{default-11:CsData-APV0:muon2pt:hiEffE}   {\ensuremath{{0.006 } } }
\vdef{default-11:CsMcPU-APV0:muon2pt:loEff}   {\ensuremath{{0.002 } } }
\vdef{default-11:CsMcPU-APV0:muon2pt:loEffE}   {\ensuremath{{0.001 } } }
\vdef{default-11:CsMcPU-APV0:muon2pt:hiEff}   {\ensuremath{{0.998 } } }
\vdef{default-11:CsMcPU-APV0:muon2pt:hiEffE}   {\ensuremath{{0.001 } } }
\vdef{default-11:CsMcPU-APV0:muon2pt:loDelta}   {\ensuremath{{+1.923 } } }
\vdef{default-11:CsMcPU-APV0:muon2pt:loDeltaE}   {\ensuremath{{0.032 } } }
\vdef{default-11:CsMcPU-APV0:muon2pt:hiDelta}   {\ensuremath{{-0.119 } } }
\vdef{default-11:CsMcPU-APV0:muon2pt:hiDeltaE}   {\ensuremath{{0.007 } } }
\vdef{default-11:CsData-APV0:muonseta:loEff}   {\ensuremath{{0.725 } } }
\vdef{default-11:CsData-APV0:muonseta:loEffE}   {\ensuremath{{0.006 } } }
\vdef{default-11:CsData-APV0:muonseta:hiEff}   {\ensuremath{{0.275 } } }
\vdef{default-11:CsData-APV0:muonseta:hiEffE}   {\ensuremath{{0.006 } } }
\vdef{default-11:CsMcPU-APV0:muonseta:loEff}   {\ensuremath{{0.851 } } }
\vdef{default-11:CsMcPU-APV0:muonseta:loEffE}   {\ensuremath{{0.005 } } }
\vdef{default-11:CsMcPU-APV0:muonseta:hiEff}   {\ensuremath{{0.149 } } }
\vdef{default-11:CsMcPU-APV0:muonseta:hiEffE}   {\ensuremath{{0.005 } } }
\vdef{default-11:CsMcPU-APV0:muonseta:loDelta}   {\ensuremath{{-0.159 } } }
\vdef{default-11:CsMcPU-APV0:muonseta:loDeltaE}   {\ensuremath{{0.011 } } }
\vdef{default-11:CsMcPU-APV0:muonseta:hiDelta}   {\ensuremath{{+0.592 } } }
\vdef{default-11:CsMcPU-APV0:muonseta:hiDeltaE}   {\ensuremath{{0.038 } } }
\vdef{default-11:CsData-APV0:pt:loEff}   {\ensuremath{{0.000 } } }
\vdef{default-11:CsData-APV0:pt:loEffE}   {\ensuremath{{0.000 } } }
\vdef{default-11:CsData-APV0:pt:hiEff}   {\ensuremath{{1.000 } } }
\vdef{default-11:CsData-APV0:pt:hiEffE}   {\ensuremath{{0.000 } } }
\vdef{default-11:CsMcPU-APV0:pt:loEff}   {\ensuremath{{0.000 } } }
\vdef{default-11:CsMcPU-APV0:pt:loEffE}   {\ensuremath{{0.000 } } }
\vdef{default-11:CsMcPU-APV0:pt:hiEff}   {\ensuremath{{1.000 } } }
\vdef{default-11:CsMcPU-APV0:pt:hiEffE}   {\ensuremath{{0.000 } } }
\vdef{default-11:CsMcPU-APV0:pt:loDelta}   {\ensuremath{{\mathrm{NaN} } } }
\vdef{default-11:CsMcPU-APV0:pt:loDeltaE}   {\ensuremath{{\mathrm{NaN} } } }
\vdef{default-11:CsMcPU-APV0:pt:hiDelta}   {\ensuremath{{+0.000 } } }
\vdef{default-11:CsMcPU-APV0:pt:hiDeltaE}   {\ensuremath{{0.000 } } }
\vdef{default-11:CsData-APV0:p:loEff}   {\ensuremath{{1.028 } } }
\vdef{default-11:CsData-APV0:p:loEffE}   {\ensuremath{{\mathrm{NaN} } } }
\vdef{default-11:CsData-APV0:p:hiEff}   {\ensuremath{{1.000 } } }
\vdef{default-11:CsData-APV0:p:hiEffE}   {\ensuremath{{0.000 } } }
\vdef{default-11:CsMcPU-APV0:p:loEff}   {\ensuremath{{1.008 } } }
\vdef{default-11:CsMcPU-APV0:p:loEffE}   {\ensuremath{{\mathrm{NaN} } } }
\vdef{default-11:CsMcPU-APV0:p:hiEff}   {\ensuremath{{1.000 } } }
\vdef{default-11:CsMcPU-APV0:p:hiEffE}   {\ensuremath{{0.000 } } }
\vdef{default-11:CsMcPU-APV0:p:loDelta}   {\ensuremath{{+0.020 } } }
\vdef{default-11:CsMcPU-APV0:p:loDeltaE}   {\ensuremath{{\mathrm{NaN} } } }
\vdef{default-11:CsMcPU-APV0:p:hiDelta}   {\ensuremath{{+0.000 } } }
\vdef{default-11:CsMcPU-APV0:p:hiDeltaE}   {\ensuremath{{0.001 } } }
\vdef{default-11:CsData-APV0:eta:loEff}   {\ensuremath{{0.717 } } }
\vdef{default-11:CsData-APV0:eta:loEffE}   {\ensuremath{{0.009 } } }
\vdef{default-11:CsData-APV0:eta:hiEff}   {\ensuremath{{0.283 } } }
\vdef{default-11:CsData-APV0:eta:hiEffE}   {\ensuremath{{0.009 } } }
\vdef{default-11:CsMcPU-APV0:eta:loEff}   {\ensuremath{{0.855 } } }
\vdef{default-11:CsMcPU-APV0:eta:loEffE}   {\ensuremath{{0.007 } } }
\vdef{default-11:CsMcPU-APV0:eta:hiEff}   {\ensuremath{{0.145 } } }
\vdef{default-11:CsMcPU-APV0:eta:hiEffE}   {\ensuremath{{0.007 } } }
\vdef{default-11:CsMcPU-APV0:eta:loDelta}   {\ensuremath{{-0.175 } } }
\vdef{default-11:CsMcPU-APV0:eta:loDeltaE}   {\ensuremath{{0.015 } } }
\vdef{default-11:CsMcPU-APV0:eta:hiDelta}   {\ensuremath{{+0.642 } } }
\vdef{default-11:CsMcPU-APV0:eta:hiDeltaE}   {\ensuremath{{0.052 } } }
\vdef{default-11:CsData-APV0:bdt:loEff}   {\ensuremath{{0.894 } } }
\vdef{default-11:CsData-APV0:bdt:loEffE}   {\ensuremath{{0.006 } } }
\vdef{default-11:CsData-APV0:bdt:hiEff}   {\ensuremath{{0.106 } } }
\vdef{default-11:CsData-APV0:bdt:hiEffE}   {\ensuremath{{0.006 } } }
\vdef{default-11:CsMcPU-APV0:bdt:loEff}   {\ensuremath{{0.873 } } }
\vdef{default-11:CsMcPU-APV0:bdt:loEffE}   {\ensuremath{{0.007 } } }
\vdef{default-11:CsMcPU-APV0:bdt:hiEff}   {\ensuremath{{0.127 } } }
\vdef{default-11:CsMcPU-APV0:bdt:hiEffE}   {\ensuremath{{0.007 } } }
\vdef{default-11:CsMcPU-APV0:bdt:loDelta}   {\ensuremath{{+0.024 } } }
\vdef{default-11:CsMcPU-APV0:bdt:loDeltaE}   {\ensuremath{{0.010 } } }
\vdef{default-11:CsMcPU-APV0:bdt:hiDelta}   {\ensuremath{{-0.186 } } }
\vdef{default-11:CsMcPU-APV0:bdt:hiDeltaE}   {\ensuremath{{0.076 } } }
\vdef{default-11:CsData-APV0:fl3d:loEff}   {\ensuremath{{0.817 } } }
\vdef{default-11:CsData-APV0:fl3d:loEffE}   {\ensuremath{{0.007 } } }
\vdef{default-11:CsData-APV0:fl3d:hiEff}   {\ensuremath{{0.183 } } }
\vdef{default-11:CsData-APV0:fl3d:hiEffE}   {\ensuremath{{0.007 } } }
\vdef{default-11:CsMcPU-APV0:fl3d:loEff}   {\ensuremath{{0.896 } } }
\vdef{default-11:CsMcPU-APV0:fl3d:loEffE}   {\ensuremath{{0.006 } } }
\vdef{default-11:CsMcPU-APV0:fl3d:hiEff}   {\ensuremath{{0.104 } } }
\vdef{default-11:CsMcPU-APV0:fl3d:hiEffE}   {\ensuremath{{0.006 } } }
\vdef{default-11:CsMcPU-APV0:fl3d:loDelta}   {\ensuremath{{-0.092 } } }
\vdef{default-11:CsMcPU-APV0:fl3d:loDeltaE}   {\ensuremath{{0.011 } } }
\vdef{default-11:CsMcPU-APV0:fl3d:hiDelta}   {\ensuremath{{+0.549 } } }
\vdef{default-11:CsMcPU-APV0:fl3d:hiDeltaE}   {\ensuremath{{0.061 } } }
\vdef{default-11:CsData-APV0:fl3de:loEff}   {\ensuremath{{1.000 } } }
\vdef{default-11:CsData-APV0:fl3de:loEffE}   {\ensuremath{{0.000 } } }
\vdef{default-11:CsData-APV0:fl3de:hiEff}   {\ensuremath{{-0.000 } } }
\vdef{default-11:CsData-APV0:fl3de:hiEffE}   {\ensuremath{{0.000 } } }
\vdef{default-11:CsMcPU-APV0:fl3de:loEff}   {\ensuremath{{1.000 } } }
\vdef{default-11:CsMcPU-APV0:fl3de:loEffE}   {\ensuremath{{0.000 } } }
\vdef{default-11:CsMcPU-APV0:fl3de:hiEff}   {\ensuremath{{0.000 } } }
\vdef{default-11:CsMcPU-APV0:fl3de:hiEffE}   {\ensuremath{{0.000 } } }
\vdef{default-11:CsMcPU-APV0:fl3de:loDelta}   {\ensuremath{{+0.000 } } }
\vdef{default-11:CsMcPU-APV0:fl3de:loDeltaE}   {\ensuremath{{0.000 } } }
\vdef{default-11:CsMcPU-APV0:fl3de:hiDelta}   {\ensuremath{{+2.000 } } }
\vdef{default-11:CsMcPU-APV0:fl3de:hiDeltaE}   {\ensuremath{{5.338 } } }
\vdef{default-11:CsData-APV0:fls3d:loEff}   {\ensuremath{{0.114 } } }
\vdef{default-11:CsData-APV0:fls3d:loEffE}   {\ensuremath{{0.006 } } }
\vdef{default-11:CsData-APV0:fls3d:hiEff}   {\ensuremath{{0.886 } } }
\vdef{default-11:CsData-APV0:fls3d:hiEffE}   {\ensuremath{{0.006 } } }
\vdef{default-11:CsMcPU-APV0:fls3d:loEff}   {\ensuremath{{0.082 } } }
\vdef{default-11:CsMcPU-APV0:fls3d:loEffE}   {\ensuremath{{0.005 } } }
\vdef{default-11:CsMcPU-APV0:fls3d:hiEff}   {\ensuremath{{0.918 } } }
\vdef{default-11:CsMcPU-APV0:fls3d:hiEffE}   {\ensuremath{{0.005 } } }
\vdef{default-11:CsMcPU-APV0:fls3d:loDelta}   {\ensuremath{{+0.325 } } }
\vdef{default-11:CsMcPU-APV0:fls3d:loDeltaE}   {\ensuremath{{0.078 } } }
\vdef{default-11:CsMcPU-APV0:fls3d:hiDelta}   {\ensuremath{{-0.035 } } }
\vdef{default-11:CsMcPU-APV0:fls3d:hiDeltaE}   {\ensuremath{{0.009 } } }
\vdef{default-11:CsData-APV0:flsxy:loEff}   {\ensuremath{{1.010 } } }
\vdef{default-11:CsData-APV0:flsxy:loEffE}   {\ensuremath{{\mathrm{NaN} } } }
\vdef{default-11:CsData-APV0:flsxy:hiEff}   {\ensuremath{{1.000 } } }
\vdef{default-11:CsData-APV0:flsxy:hiEffE}   {\ensuremath{{0.000 } } }
\vdef{default-11:CsMcPU-APV0:flsxy:loEff}   {\ensuremath{{1.006 } } }
\vdef{default-11:CsMcPU-APV0:flsxy:loEffE}   {\ensuremath{{\mathrm{NaN} } } }
\vdef{default-11:CsMcPU-APV0:flsxy:hiEff}   {\ensuremath{{1.000 } } }
\vdef{default-11:CsMcPU-APV0:flsxy:hiEffE}   {\ensuremath{{0.000 } } }
\vdef{default-11:CsMcPU-APV0:flsxy:loDelta}   {\ensuremath{{+0.004 } } }
\vdef{default-11:CsMcPU-APV0:flsxy:loDeltaE}   {\ensuremath{{\mathrm{NaN} } } }
\vdef{default-11:CsMcPU-APV0:flsxy:hiDelta}   {\ensuremath{{+0.000 } } }
\vdef{default-11:CsMcPU-APV0:flsxy:hiDeltaE}   {\ensuremath{{0.000 } } }
\vdef{default-11:CsData-APV0:chi2dof:loEff}   {\ensuremath{{0.924 } } }
\vdef{default-11:CsData-APV0:chi2dof:loEffE}   {\ensuremath{{0.005 } } }
\vdef{default-11:CsData-APV0:chi2dof:hiEff}   {\ensuremath{{0.076 } } }
\vdef{default-11:CsData-APV0:chi2dof:hiEffE}   {\ensuremath{{0.005 } } }
\vdef{default-11:CsMcPU-APV0:chi2dof:loEff}   {\ensuremath{{0.954 } } }
\vdef{default-11:CsMcPU-APV0:chi2dof:loEffE}   {\ensuremath{{0.004 } } }
\vdef{default-11:CsMcPU-APV0:chi2dof:hiEff}   {\ensuremath{{0.046 } } }
\vdef{default-11:CsMcPU-APV0:chi2dof:hiEffE}   {\ensuremath{{0.004 } } }
\vdef{default-11:CsMcPU-APV0:chi2dof:loDelta}   {\ensuremath{{-0.032 } } }
\vdef{default-11:CsMcPU-APV0:chi2dof:loDeltaE}   {\ensuremath{{0.007 } } }
\vdef{default-11:CsMcPU-APV0:chi2dof:hiDelta}   {\ensuremath{{+0.486 } } }
\vdef{default-11:CsMcPU-APV0:chi2dof:hiDeltaE}   {\ensuremath{{0.106 } } }
\vdef{default-11:CsData-APV0:pchi2dof:loEff}   {\ensuremath{{0.627 } } }
\vdef{default-11:CsData-APV0:pchi2dof:loEffE}   {\ensuremath{{0.009 } } }
\vdef{default-11:CsData-APV0:pchi2dof:hiEff}   {\ensuremath{{0.373 } } }
\vdef{default-11:CsData-APV0:pchi2dof:hiEffE}   {\ensuremath{{0.009 } } }
\vdef{default-11:CsMcPU-APV0:pchi2dof:loEff}   {\ensuremath{{0.567 } } }
\vdef{default-11:CsMcPU-APV0:pchi2dof:loEffE}   {\ensuremath{{0.010 } } }
\vdef{default-11:CsMcPU-APV0:pchi2dof:hiEff}   {\ensuremath{{0.433 } } }
\vdef{default-11:CsMcPU-APV0:pchi2dof:hiEffE}   {\ensuremath{{0.010 } } }
\vdef{default-11:CsMcPU-APV0:pchi2dof:loDelta}   {\ensuremath{{+0.099 } } }
\vdef{default-11:CsMcPU-APV0:pchi2dof:loDeltaE}   {\ensuremath{{0.022 } } }
\vdef{default-11:CsMcPU-APV0:pchi2dof:hiDelta}   {\ensuremath{{-0.147 } } }
\vdef{default-11:CsMcPU-APV0:pchi2dof:hiDeltaE}   {\ensuremath{{0.033 } } }
\vdef{default-11:CsData-APV0:alpha:loEff}   {\ensuremath{{0.993 } } }
\vdef{default-11:CsData-APV0:alpha:loEffE}   {\ensuremath{{0.002 } } }
\vdef{default-11:CsData-APV0:alpha:hiEff}   {\ensuremath{{0.007 } } }
\vdef{default-11:CsData-APV0:alpha:hiEffE}   {\ensuremath{{0.002 } } }
\vdef{default-11:CsMcPU-APV0:alpha:loEff}   {\ensuremath{{0.992 } } }
\vdef{default-11:CsMcPU-APV0:alpha:loEffE}   {\ensuremath{{0.002 } } }
\vdef{default-11:CsMcPU-APV0:alpha:hiEff}   {\ensuremath{{0.008 } } }
\vdef{default-11:CsMcPU-APV0:alpha:hiEffE}   {\ensuremath{{0.002 } } }
\vdef{default-11:CsMcPU-APV0:alpha:loDelta}   {\ensuremath{{+0.001 } } }
\vdef{default-11:CsMcPU-APV0:alpha:loDeltaE}   {\ensuremath{{0.003 } } }
\vdef{default-11:CsMcPU-APV0:alpha:hiDelta}   {\ensuremath{{-0.184 } } }
\vdef{default-11:CsMcPU-APV0:alpha:hiDeltaE}   {\ensuremath{{0.340 } } }
\vdef{default-11:CsData-APV0:iso:loEff}   {\ensuremath{{0.100 } } }
\vdef{default-11:CsData-APV0:iso:loEffE}   {\ensuremath{{0.006 } } }
\vdef{default-11:CsData-APV0:iso:hiEff}   {\ensuremath{{0.900 } } }
\vdef{default-11:CsData-APV0:iso:hiEffE}   {\ensuremath{{0.006 } } }
\vdef{default-11:CsMcPU-APV0:iso:loEff}   {\ensuremath{{0.098 } } }
\vdef{default-11:CsMcPU-APV0:iso:loEffE}   {\ensuremath{{0.006 } } }
\vdef{default-11:CsMcPU-APV0:iso:hiEff}   {\ensuremath{{0.902 } } }
\vdef{default-11:CsMcPU-APV0:iso:hiEffE}   {\ensuremath{{0.006 } } }
\vdef{default-11:CsMcPU-APV0:iso:loDelta}   {\ensuremath{{+0.023 } } }
\vdef{default-11:CsMcPU-APV0:iso:loDeltaE}   {\ensuremath{{0.082 } } }
\vdef{default-11:CsMcPU-APV0:iso:hiDelta}   {\ensuremath{{-0.002 } } }
\vdef{default-11:CsMcPU-APV0:iso:hiDeltaE}   {\ensuremath{{0.009 } } }
\vdef{default-11:CsData-APV0:docatrk:loEff}   {\ensuremath{{0.067 } } }
\vdef{default-11:CsData-APV0:docatrk:loEffE}   {\ensuremath{{0.005 } } }
\vdef{default-11:CsData-APV0:docatrk:hiEff}   {\ensuremath{{0.933 } } }
\vdef{default-11:CsData-APV0:docatrk:hiEffE}   {\ensuremath{{0.005 } } }
\vdef{default-11:CsMcPU-APV0:docatrk:loEff}   {\ensuremath{{0.082 } } }
\vdef{default-11:CsMcPU-APV0:docatrk:loEffE}   {\ensuremath{{0.005 } } }
\vdef{default-11:CsMcPU-APV0:docatrk:hiEff}   {\ensuremath{{0.918 } } }
\vdef{default-11:CsMcPU-APV0:docatrk:hiEffE}   {\ensuremath{{0.005 } } }
\vdef{default-11:CsMcPU-APV0:docatrk:loDelta}   {\ensuremath{{-0.201 } } }
\vdef{default-11:CsMcPU-APV0:docatrk:loDeltaE}   {\ensuremath{{0.098 } } }
\vdef{default-11:CsMcPU-APV0:docatrk:hiDelta}   {\ensuremath{{+0.016 } } }
\vdef{default-11:CsMcPU-APV0:docatrk:hiDeltaE}   {\ensuremath{{0.008 } } }
\vdef{default-11:CsData-APV0:isotrk:loEff}   {\ensuremath{{1.000 } } }
\vdef{default-11:CsData-APV0:isotrk:loEffE}   {\ensuremath{{0.000 } } }
\vdef{default-11:CsData-APV0:isotrk:hiEff}   {\ensuremath{{1.000 } } }
\vdef{default-11:CsData-APV0:isotrk:hiEffE}   {\ensuremath{{0.000 } } }
\vdef{default-11:CsMcPU-APV0:isotrk:loEff}   {\ensuremath{{1.000 } } }
\vdef{default-11:CsMcPU-APV0:isotrk:loEffE}   {\ensuremath{{0.000 } } }
\vdef{default-11:CsMcPU-APV0:isotrk:hiEff}   {\ensuremath{{1.000 } } }
\vdef{default-11:CsMcPU-APV0:isotrk:hiEffE}   {\ensuremath{{0.000 } } }
\vdef{default-11:CsMcPU-APV0:isotrk:loDelta}   {\ensuremath{{+0.000 } } }
\vdef{default-11:CsMcPU-APV0:isotrk:loDeltaE}   {\ensuremath{{0.001 } } }
\vdef{default-11:CsMcPU-APV0:isotrk:hiDelta}   {\ensuremath{{+0.000 } } }
\vdef{default-11:CsMcPU-APV0:isotrk:hiDeltaE}   {\ensuremath{{0.001 } } }
\vdef{default-11:CsData-APV0:closetrk:loEff}   {\ensuremath{{0.987 } } }
\vdef{default-11:CsData-APV0:closetrk:loEffE}   {\ensuremath{{0.002 } } }
\vdef{default-11:CsData-APV0:closetrk:hiEff}   {\ensuremath{{0.013 } } }
\vdef{default-11:CsData-APV0:closetrk:hiEffE}   {\ensuremath{{0.002 } } }
\vdef{default-11:CsMcPU-APV0:closetrk:loEff}   {\ensuremath{{0.978 } } }
\vdef{default-11:CsMcPU-APV0:closetrk:loEffE}   {\ensuremath{{0.003 } } }
\vdef{default-11:CsMcPU-APV0:closetrk:hiEff}   {\ensuremath{{0.022 } } }
\vdef{default-11:CsMcPU-APV0:closetrk:hiEffE}   {\ensuremath{{0.003 } } }
\vdef{default-11:CsMcPU-APV0:closetrk:loDelta}   {\ensuremath{{+0.009 } } }
\vdef{default-11:CsMcPU-APV0:closetrk:loDeltaE}   {\ensuremath{{0.004 } } }
\vdef{default-11:CsMcPU-APV0:closetrk:hiDelta}   {\ensuremath{{-0.535 } } }
\vdef{default-11:CsMcPU-APV0:closetrk:hiDeltaE}   {\ensuremath{{0.206 } } }
\vdef{default-11:CsData-APV0:lip:loEff}   {\ensuremath{{1.000 } } }
\vdef{default-11:CsData-APV0:lip:loEffE}   {\ensuremath{{0.000 } } }
\vdef{default-11:CsData-APV0:lip:hiEff}   {\ensuremath{{0.000 } } }
\vdef{default-11:CsData-APV0:lip:hiEffE}   {\ensuremath{{0.000 } } }
\vdef{default-11:CsMcPU-APV0:lip:loEff}   {\ensuremath{{1.000 } } }
\vdef{default-11:CsMcPU-APV0:lip:loEffE}   {\ensuremath{{0.000 } } }
\vdef{default-11:CsMcPU-APV0:lip:hiEff}   {\ensuremath{{0.000 } } }
\vdef{default-11:CsMcPU-APV0:lip:hiEffE}   {\ensuremath{{0.000 } } }
\vdef{default-11:CsMcPU-APV0:lip:loDelta}   {\ensuremath{{+0.000 } } }
\vdef{default-11:CsMcPU-APV0:lip:loDeltaE}   {\ensuremath{{0.001 } } }
\vdef{default-11:CsMcPU-APV0:lip:hiDelta}   {\ensuremath{{\mathrm{NaN} } } }
\vdef{default-11:CsMcPU-APV0:lip:hiDeltaE}   {\ensuremath{{\mathrm{NaN} } } }
\vdef{default-11:CsData-APV0:lip:inEff}   {\ensuremath{{1.000 } } }
\vdef{default-11:CsData-APV0:lip:inEffE}   {\ensuremath{{0.000 } } }
\vdef{default-11:CsMcPU-APV0:lip:inEff}   {\ensuremath{{1.000 } } }
\vdef{default-11:CsMcPU-APV0:lip:inEffE}   {\ensuremath{{0.000 } } }
\vdef{default-11:CsMcPU-APV0:lip:inDelta}   {\ensuremath{{+0.000 } } }
\vdef{default-11:CsMcPU-APV0:lip:inDeltaE}   {\ensuremath{{0.001 } } }
\vdef{default-11:CsData-APV0:lips:loEff}   {\ensuremath{{1.000 } } }
\vdef{default-11:CsData-APV0:lips:loEffE}   {\ensuremath{{0.000 } } }
\vdef{default-11:CsData-APV0:lips:hiEff}   {\ensuremath{{0.000 } } }
\vdef{default-11:CsData-APV0:lips:hiEffE}   {\ensuremath{{0.000 } } }
\vdef{default-11:CsMcPU-APV0:lips:loEff}   {\ensuremath{{1.000 } } }
\vdef{default-11:CsMcPU-APV0:lips:loEffE}   {\ensuremath{{0.000 } } }
\vdef{default-11:CsMcPU-APV0:lips:hiEff}   {\ensuremath{{0.000 } } }
\vdef{default-11:CsMcPU-APV0:lips:hiEffE}   {\ensuremath{{0.000 } } }
\vdef{default-11:CsMcPU-APV0:lips:loDelta}   {\ensuremath{{+0.000 } } }
\vdef{default-11:CsMcPU-APV0:lips:loDeltaE}   {\ensuremath{{0.001 } } }
\vdef{default-11:CsMcPU-APV0:lips:hiDelta}   {\ensuremath{{\mathrm{NaN} } } }
\vdef{default-11:CsMcPU-APV0:lips:hiDeltaE}   {\ensuremath{{\mathrm{NaN} } } }
\vdef{default-11:CsData-APV0:lips:inEff}   {\ensuremath{{1.000 } } }
\vdef{default-11:CsData-APV0:lips:inEffE}   {\ensuremath{{0.000 } } }
\vdef{default-11:CsMcPU-APV0:lips:inEff}   {\ensuremath{{1.000 } } }
\vdef{default-11:CsMcPU-APV0:lips:inEffE}   {\ensuremath{{0.000 } } }
\vdef{default-11:CsMcPU-APV0:lips:inDelta}   {\ensuremath{{+0.000 } } }
\vdef{default-11:CsMcPU-APV0:lips:inDeltaE}   {\ensuremath{{0.001 } } }
\vdef{default-11:CsData-APV0:ip:loEff}   {\ensuremath{{0.972 } } }
\vdef{default-11:CsData-APV0:ip:loEffE}   {\ensuremath{{0.003 } } }
\vdef{default-11:CsData-APV0:ip:hiEff}   {\ensuremath{{0.028 } } }
\vdef{default-11:CsData-APV0:ip:hiEffE}   {\ensuremath{{0.003 } } }
\vdef{default-11:CsMcPU-APV0:ip:loEff}   {\ensuremath{{0.974 } } }
\vdef{default-11:CsMcPU-APV0:ip:loEffE}   {\ensuremath{{0.003 } } }
\vdef{default-11:CsMcPU-APV0:ip:hiEff}   {\ensuremath{{0.026 } } }
\vdef{default-11:CsMcPU-APV0:ip:hiEffE}   {\ensuremath{{0.003 } } }
\vdef{default-11:CsMcPU-APV0:ip:loDelta}   {\ensuremath{{-0.002 } } }
\vdef{default-11:CsMcPU-APV0:ip:loDeltaE}   {\ensuremath{{0.005 } } }
\vdef{default-11:CsMcPU-APV0:ip:hiDelta}   {\ensuremath{{+0.067 } } }
\vdef{default-11:CsMcPU-APV0:ip:hiDeltaE}   {\ensuremath{{0.168 } } }
\vdef{default-11:CsData-APV0:ips:loEff}   {\ensuremath{{0.941 } } }
\vdef{default-11:CsData-APV0:ips:loEffE}   {\ensuremath{{0.005 } } }
\vdef{default-11:CsData-APV0:ips:hiEff}   {\ensuremath{{0.059 } } }
\vdef{default-11:CsData-APV0:ips:hiEffE}   {\ensuremath{{0.005 } } }
\vdef{default-11:CsMcPU-APV0:ips:loEff}   {\ensuremath{{0.931 } } }
\vdef{default-11:CsMcPU-APV0:ips:loEffE}   {\ensuremath{{0.005 } } }
\vdef{default-11:CsMcPU-APV0:ips:hiEff}   {\ensuremath{{0.069 } } }
\vdef{default-11:CsMcPU-APV0:ips:hiEffE}   {\ensuremath{{0.005 } } }
\vdef{default-11:CsMcPU-APV0:ips:loDelta}   {\ensuremath{{+0.011 } } }
\vdef{default-11:CsMcPU-APV0:ips:loDeltaE}   {\ensuremath{{0.007 } } }
\vdef{default-11:CsMcPU-APV0:ips:hiDelta}   {\ensuremath{{-0.162 } } }
\vdef{default-11:CsMcPU-APV0:ips:hiDeltaE}   {\ensuremath{{0.105 } } }
\vdef{default-11:CsData-APV0:maxdoca:loEff}   {\ensuremath{{1.000 } } }
\vdef{default-11:CsData-APV0:maxdoca:loEffE}   {\ensuremath{{0.000 } } }
\vdef{default-11:CsData-APV0:maxdoca:hiEff}   {\ensuremath{{0.028 } } }
\vdef{default-11:CsData-APV0:maxdoca:hiEffE}   {\ensuremath{{0.003 } } }
\vdef{default-11:CsMcPU-APV0:maxdoca:loEff}   {\ensuremath{{1.000 } } }
\vdef{default-11:CsMcPU-APV0:maxdoca:loEffE}   {\ensuremath{{0.000 } } }
\vdef{default-11:CsMcPU-APV0:maxdoca:hiEff}   {\ensuremath{{0.024 } } }
\vdef{default-11:CsMcPU-APV0:maxdoca:hiEffE}   {\ensuremath{{0.003 } } }
\vdef{default-11:CsMcPU-APV0:maxdoca:loDelta}   {\ensuremath{{+0.000 } } }
\vdef{default-11:CsMcPU-APV0:maxdoca:loDeltaE}   {\ensuremath{{0.001 } } }
\vdef{default-11:CsMcPU-APV0:maxdoca:hiDelta}   {\ensuremath{{+0.145 } } }
\vdef{default-11:CsMcPU-APV0:maxdoca:hiDeltaE}   {\ensuremath{{0.176 } } }
\vdef{default-11:CsData-APV0:kaonspt:loEff}   {\ensuremath{{1.001 } } }
\vdef{default-11:CsData-APV0:kaonspt:loEffE}   {\ensuremath{{\mathrm{NaN} } } }
\vdef{default-11:CsData-APV0:kaonspt:hiEff}   {\ensuremath{{1.000 } } }
\vdef{default-11:CsData-APV0:kaonspt:hiEffE}   {\ensuremath{{0.000 } } }
\vdef{default-11:CsMcPU-APV0:kaonspt:loEff}   {\ensuremath{{1.000 } } }
\vdef{default-11:CsMcPU-APV0:kaonspt:loEffE}   {\ensuremath{{0.000 } } }
\vdef{default-11:CsMcPU-APV0:kaonspt:hiEff}   {\ensuremath{{1.000 } } }
\vdef{default-11:CsMcPU-APV0:kaonspt:hiEffE}   {\ensuremath{{0.000 } } }
\vdef{default-11:CsMcPU-APV0:kaonspt:loDelta}   {\ensuremath{{+0.001 } } }
\vdef{default-11:CsMcPU-APV0:kaonspt:loDeltaE}   {\ensuremath{{\mathrm{NaN} } } }
\vdef{default-11:CsMcPU-APV0:kaonspt:hiDelta}   {\ensuremath{{+0.000 } } }
\vdef{default-11:CsMcPU-APV0:kaonspt:hiDeltaE}   {\ensuremath{{0.000 } } }
\vdef{default-11:CsData-APV0:psipt:loEff}   {\ensuremath{{1.006 } } }
\vdef{default-11:CsData-APV0:psipt:loEffE}   {\ensuremath{{\mathrm{NaN} } } }
\vdef{default-11:CsData-APV0:psipt:hiEff}   {\ensuremath{{1.000 } } }
\vdef{default-11:CsData-APV0:psipt:hiEffE}   {\ensuremath{{0.000 } } }
\vdef{default-11:CsMcPU-APV0:psipt:loEff}   {\ensuremath{{1.006 } } }
\vdef{default-11:CsMcPU-APV0:psipt:loEffE}   {\ensuremath{{\mathrm{NaN} } } }
\vdef{default-11:CsMcPU-APV0:psipt:hiEff}   {\ensuremath{{1.000 } } }
\vdef{default-11:CsMcPU-APV0:psipt:hiEffE}   {\ensuremath{{0.000 } } }
\vdef{default-11:CsMcPU-APV0:psipt:loDelta}   {\ensuremath{{+0.000 } } }
\vdef{default-11:CsMcPU-APV0:psipt:loDeltaE}   {\ensuremath{{\mathrm{NaN} } } }
\vdef{default-11:CsMcPU-APV0:psipt:hiDelta}   {\ensuremath{{+0.000 } } }
\vdef{default-11:CsMcPU-APV0:psipt:hiDeltaE}   {\ensuremath{{0.001 } } }
\vdef{default-11:CsData-APV0:phipt:loEff}   {\ensuremath{{1.020 } } }
\vdef{default-11:CsData-APV0:phipt:loEffE}   {\ensuremath{{\mathrm{NaN} } } }
\vdef{default-11:CsData-APV0:phipt:hiEff}   {\ensuremath{{1.000 } } }
\vdef{default-11:CsData-APV0:phipt:hiEffE}   {\ensuremath{{0.000 } } }
\vdef{default-11:CsMcPU-APV0:phipt:loEff}   {\ensuremath{{1.008 } } }
\vdef{default-11:CsMcPU-APV0:phipt:loEffE}   {\ensuremath{{\mathrm{NaN} } } }
\vdef{default-11:CsMcPU-APV0:phipt:hiEff}   {\ensuremath{{1.000 } } }
\vdef{default-11:CsMcPU-APV0:phipt:hiEffE}   {\ensuremath{{0.000 } } }
\vdef{default-11:CsMcPU-APV0:phipt:loDelta}   {\ensuremath{{+0.011 } } }
\vdef{default-11:CsMcPU-APV0:phipt:loDeltaE}   {\ensuremath{{\mathrm{NaN} } } }
\vdef{default-11:CsMcPU-APV0:phipt:hiDelta}   {\ensuremath{{+0.000 } } }
\vdef{default-11:CsMcPU-APV0:phipt:hiDeltaE}   {\ensuremath{{0.001 } } }
\vdef{default-11:CsData-APV0:deltar:loEff}   {\ensuremath{{1.000 } } }
\vdef{default-11:CsData-APV0:deltar:loEffE}   {\ensuremath{{0.000 } } }
\vdef{default-11:CsData-APV0:deltar:hiEff}   {\ensuremath{{0.000 } } }
\vdef{default-11:CsData-APV0:deltar:hiEffE}   {\ensuremath{{0.000 } } }
\vdef{default-11:CsMcPU-APV0:deltar:loEff}   {\ensuremath{{0.999 } } }
\vdef{default-11:CsMcPU-APV0:deltar:loEffE}   {\ensuremath{{0.001 } } }
\vdef{default-11:CsMcPU-APV0:deltar:hiEff}   {\ensuremath{{0.001 } } }
\vdef{default-11:CsMcPU-APV0:deltar:hiEffE}   {\ensuremath{{0.001 } } }
\vdef{default-11:CsMcPU-APV0:deltar:loDelta}   {\ensuremath{{+0.001 } } }
\vdef{default-11:CsMcPU-APV0:deltar:loDeltaE}   {\ensuremath{{0.001 } } }
\vdef{default-11:CsMcPU-APV0:deltar:hiDelta}   {\ensuremath{{-2.000 } } }
\vdef{default-11:CsMcPU-APV0:deltar:hiDeltaE}   {\ensuremath{{1.477 } } }
\vdef{default-11:CsData-APV0:mkk:loEff}   {\ensuremath{{1.164 } } }
\vdef{default-11:CsData-APV0:mkk:loEffE}   {\ensuremath{{\mathrm{NaN} } } }
\vdef{default-11:CsData-APV0:mkk:hiEff}   {\ensuremath{{1.000 } } }
\vdef{default-11:CsData-APV0:mkk:hiEffE}   {\ensuremath{{0.000 } } }
\vdef{default-11:CsMcPU-APV0:mkk:loEff}   {\ensuremath{{1.000 } } }
\vdef{default-11:CsMcPU-APV0:mkk:loEffE}   {\ensuremath{{0.000 } } }
\vdef{default-11:CsMcPU-APV0:mkk:hiEff}   {\ensuremath{{1.000 } } }
\vdef{default-11:CsMcPU-APV0:mkk:hiEffE}   {\ensuremath{{0.000 } } }
\vdef{default-11:CsMcPU-APV0:mkk:loDelta}   {\ensuremath{{+0.152 } } }
\vdef{default-11:CsMcPU-APV0:mkk:loDeltaE}   {\ensuremath{{\mathrm{NaN} } } }
\vdef{default-11:CsMcPU-APV0:mkk:hiDelta}   {\ensuremath{{+0.000 } } }
\vdef{default-11:CsMcPU-APV0:mkk:hiDeltaE}   {\ensuremath{{0.000 } } }
\vdef{default-11:CsData-APV1:osiso:loEff}   {\ensuremath{{1.007 } } }
\vdef{default-11:CsData-APV1:osiso:loEffE}   {\ensuremath{{\mathrm{NaN} } } }
\vdef{default-11:CsData-APV1:osiso:hiEff}   {\ensuremath{{1.000 } } }
\vdef{default-11:CsData-APV1:osiso:hiEffE}   {\ensuremath{{0.000 } } }
\vdef{default-11:CsMcPU-APV1:osiso:loEff}   {\ensuremath{{1.004 } } }
\vdef{default-11:CsMcPU-APV1:osiso:loEffE}   {\ensuremath{{\mathrm{NaN} } } }
\vdef{default-11:CsMcPU-APV1:osiso:hiEff}   {\ensuremath{{1.000 } } }
\vdef{default-11:CsMcPU-APV1:osiso:hiEffE}   {\ensuremath{{0.000 } } }
\vdef{default-11:CsMcPU-APV1:osiso:loDelta}   {\ensuremath{{+0.003 } } }
\vdef{default-11:CsMcPU-APV1:osiso:loDeltaE}   {\ensuremath{{\mathrm{NaN} } } }
\vdef{default-11:CsMcPU-APV1:osiso:hiDelta}   {\ensuremath{{+0.000 } } }
\vdef{default-11:CsMcPU-APV1:osiso:hiDeltaE}   {\ensuremath{{0.001 } } }
\vdef{default-11:CsData-APV1:osreliso:loEff}   {\ensuremath{{0.280 } } }
\vdef{default-11:CsData-APV1:osreliso:loEffE}   {\ensuremath{{0.010 } } }
\vdef{default-11:CsData-APV1:osreliso:hiEff}   {\ensuremath{{0.720 } } }
\vdef{default-11:CsData-APV1:osreliso:hiEffE}   {\ensuremath{{0.010 } } }
\vdef{default-11:CsMcPU-APV1:osreliso:loEff}   {\ensuremath{{0.293 } } }
\vdef{default-11:CsMcPU-APV1:osreliso:loEffE}   {\ensuremath{{0.010 } } }
\vdef{default-11:CsMcPU-APV1:osreliso:hiEff}   {\ensuremath{{0.707 } } }
\vdef{default-11:CsMcPU-APV1:osreliso:hiEffE}   {\ensuremath{{0.010 } } }
\vdef{default-11:CsMcPU-APV1:osreliso:loDelta}   {\ensuremath{{-0.044 } } }
\vdef{default-11:CsMcPU-APV1:osreliso:loDeltaE}   {\ensuremath{{0.049 } } }
\vdef{default-11:CsMcPU-APV1:osreliso:hiDelta}   {\ensuremath{{+0.018 } } }
\vdef{default-11:CsMcPU-APV1:osreliso:hiDeltaE}   {\ensuremath{{0.020 } } }
\vdef{default-11:CsData-APV1:osmuonpt:loEff}   {\ensuremath{{0.000 } } }
\vdef{default-11:CsData-APV1:osmuonpt:loEffE}   {\ensuremath{{0.012 } } }
\vdef{default-11:CsData-APV1:osmuonpt:hiEff}   {\ensuremath{{1.000 } } }
\vdef{default-11:CsData-APV1:osmuonpt:hiEffE}   {\ensuremath{{0.012 } } }
\vdef{default-11:CsMcPU-APV1:osmuonpt:loEff}   {\ensuremath{{0.000 } } }
\vdef{default-11:CsMcPU-APV1:osmuonpt:loEffE}   {\ensuremath{{0.012 } } }
\vdef{default-11:CsMcPU-APV1:osmuonpt:hiEff}   {\ensuremath{{1.000 } } }
\vdef{default-11:CsMcPU-APV1:osmuonpt:hiEffE}   {\ensuremath{{0.012 } } }
\vdef{default-11:CsMcPU-APV1:osmuonpt:loDelta}   {\ensuremath{{\mathrm{NaN} } } }
\vdef{default-11:CsMcPU-APV1:osmuonpt:loDeltaE}   {\ensuremath{{\mathrm{NaN} } } }
\vdef{default-11:CsMcPU-APV1:osmuonpt:hiDelta}   {\ensuremath{{+0.000 } } }
\vdef{default-11:CsMcPU-APV1:osmuonpt:hiDeltaE}   {\ensuremath{{0.017 } } }
\vdef{default-11:CsData-APV1:osmuondr:loEff}   {\ensuremath{{0.012 } } }
\vdef{default-11:CsData-APV1:osmuondr:loEffE}   {\ensuremath{{0.017 } } }
\vdef{default-11:CsData-APV1:osmuondr:hiEff}   {\ensuremath{{0.988 } } }
\vdef{default-11:CsData-APV1:osmuondr:hiEffE}   {\ensuremath{{0.017 } } }
\vdef{default-11:CsMcPU-APV1:osmuondr:loEff}   {\ensuremath{{0.000 } } }
\vdef{default-11:CsMcPU-APV1:osmuondr:loEffE}   {\ensuremath{{0.012 } } }
\vdef{default-11:CsMcPU-APV1:osmuondr:hiEff}   {\ensuremath{{1.000 } } }
\vdef{default-11:CsMcPU-APV1:osmuondr:hiEffE}   {\ensuremath{{0.012 } } }
\vdef{default-11:CsMcPU-APV1:osmuondr:loDelta}   {\ensuremath{{+2.000 } } }
\vdef{default-11:CsMcPU-APV1:osmuondr:loDeltaE}   {\ensuremath{{3.898 } } }
\vdef{default-11:CsMcPU-APV1:osmuondr:hiDelta}   {\ensuremath{{-0.012 } } }
\vdef{default-11:CsMcPU-APV1:osmuondr:hiDeltaE}   {\ensuremath{{0.021 } } }
\vdef{default-11:CsData-APV1:hlt:loEff}   {\ensuremath{{0.091 } } }
\vdef{default-11:CsData-APV1:hlt:loEffE}   {\ensuremath{{0.006 } } }
\vdef{default-11:CsData-APV1:hlt:hiEff}   {\ensuremath{{0.909 } } }
\vdef{default-11:CsData-APV1:hlt:hiEffE}   {\ensuremath{{0.006 } } }
\vdef{default-11:CsMcPU-APV1:hlt:loEff}   {\ensuremath{{0.357 } } }
\vdef{default-11:CsMcPU-APV1:hlt:loEffE}   {\ensuremath{{0.010 } } }
\vdef{default-11:CsMcPU-APV1:hlt:hiEff}   {\ensuremath{{0.643 } } }
\vdef{default-11:CsMcPU-APV1:hlt:hiEffE}   {\ensuremath{{0.010 } } }
\vdef{default-11:CsMcPU-APV1:hlt:loDelta}   {\ensuremath{{-1.188 } } }
\vdef{default-11:CsMcPU-APV1:hlt:loDeltaE}   {\ensuremath{{0.049 } } }
\vdef{default-11:CsMcPU-APV1:hlt:hiDelta}   {\ensuremath{{+0.343 } } }
\vdef{default-11:CsMcPU-APV1:hlt:hiDeltaE}   {\ensuremath{{0.017 } } }
\vdef{default-11:CsData-APV1:muonsid:loEff}   {\ensuremath{{0.161 } } }
\vdef{default-11:CsData-APV1:muonsid:loEffE}   {\ensuremath{{0.008 } } }
\vdef{default-11:CsData-APV1:muonsid:hiEff}   {\ensuremath{{0.839 } } }
\vdef{default-11:CsData-APV1:muonsid:hiEffE}   {\ensuremath{{0.008 } } }
\vdef{default-11:CsMcPU-APV1:muonsid:loEff}   {\ensuremath{{0.223 } } }
\vdef{default-11:CsMcPU-APV1:muonsid:loEffE}   {\ensuremath{{0.009 } } }
\vdef{default-11:CsMcPU-APV1:muonsid:hiEff}   {\ensuremath{{0.777 } } }
\vdef{default-11:CsMcPU-APV1:muonsid:hiEffE}   {\ensuremath{{0.009 } } }
\vdef{default-11:CsMcPU-APV1:muonsid:loDelta}   {\ensuremath{{-0.323 } } }
\vdef{default-11:CsMcPU-APV1:muonsid:loDeltaE}   {\ensuremath{{0.061 } } }
\vdef{default-11:CsMcPU-APV1:muonsid:hiDelta}   {\ensuremath{{+0.077 } } }
\vdef{default-11:CsMcPU-APV1:muonsid:hiDeltaE}   {\ensuremath{{0.015 } } }
\vdef{default-11:CsData-APV1:tracksqual:loEff}   {\ensuremath{{0.001 } } }
\vdef{default-11:CsData-APV1:tracksqual:loEffE}   {\ensuremath{{0.001 } } }
\vdef{default-11:CsData-APV1:tracksqual:hiEff}   {\ensuremath{{0.999 } } }
\vdef{default-11:CsData-APV1:tracksqual:hiEffE}   {\ensuremath{{0.001 } } }
\vdef{default-11:CsMcPU-APV1:tracksqual:loEff}   {\ensuremath{{0.000 } } }
\vdef{default-11:CsMcPU-APV1:tracksqual:loEffE}   {\ensuremath{{0.001 } } }
\vdef{default-11:CsMcPU-APV1:tracksqual:hiEff}   {\ensuremath{{1.000 } } }
\vdef{default-11:CsMcPU-APV1:tracksqual:hiEffE}   {\ensuremath{{0.001 } } }
\vdef{default-11:CsMcPU-APV1:tracksqual:loDelta}   {\ensuremath{{+2.000 } } }
\vdef{default-11:CsMcPU-APV1:tracksqual:loDeltaE}   {\ensuremath{{3.995 } } }
\vdef{default-11:CsMcPU-APV1:tracksqual:hiDelta}   {\ensuremath{{-0.001 } } }
\vdef{default-11:CsMcPU-APV1:tracksqual:hiDeltaE}   {\ensuremath{{0.001 } } }
\vdef{default-11:CsData-APV1:pvz:loEff}   {\ensuremath{{0.518 } } }
\vdef{default-11:CsData-APV1:pvz:loEffE}   {\ensuremath{{0.011 } } }
\vdef{default-11:CsData-APV1:pvz:hiEff}   {\ensuremath{{0.482 } } }
\vdef{default-11:CsData-APV1:pvz:hiEffE}   {\ensuremath{{0.011 } } }
\vdef{default-11:CsMcPU-APV1:pvz:loEff}   {\ensuremath{{0.464 } } }
\vdef{default-11:CsMcPU-APV1:pvz:loEffE}   {\ensuremath{{0.011 } } }
\vdef{default-11:CsMcPU-APV1:pvz:hiEff}   {\ensuremath{{0.536 } } }
\vdef{default-11:CsMcPU-APV1:pvz:hiEffE}   {\ensuremath{{0.011 } } }
\vdef{default-11:CsMcPU-APV1:pvz:loDelta}   {\ensuremath{{+0.110 } } }
\vdef{default-11:CsMcPU-APV1:pvz:loDeltaE}   {\ensuremath{{0.032 } } }
\vdef{default-11:CsMcPU-APV1:pvz:hiDelta}   {\ensuremath{{-0.106 } } }
\vdef{default-11:CsMcPU-APV1:pvz:hiDeltaE}   {\ensuremath{{0.030 } } }
\vdef{default-11:CsData-APV1:pvn:loEff}   {\ensuremath{{1.034 } } }
\vdef{default-11:CsData-APV1:pvn:loEffE}   {\ensuremath{{\mathrm{NaN} } } }
\vdef{default-11:CsData-APV1:pvn:hiEff}   {\ensuremath{{1.000 } } }
\vdef{default-11:CsData-APV1:pvn:hiEffE}   {\ensuremath{{0.000 } } }
\vdef{default-11:CsMcPU-APV1:pvn:loEff}   {\ensuremath{{1.184 } } }
\vdef{default-11:CsMcPU-APV1:pvn:loEffE}   {\ensuremath{{\mathrm{NaN} } } }
\vdef{default-11:CsMcPU-APV1:pvn:hiEff}   {\ensuremath{{1.000 } } }
\vdef{default-11:CsMcPU-APV1:pvn:hiEffE}   {\ensuremath{{0.000 } } }
\vdef{default-11:CsMcPU-APV1:pvn:loDelta}   {\ensuremath{{-0.135 } } }
\vdef{default-11:CsMcPU-APV1:pvn:loDeltaE}   {\ensuremath{{\mathrm{NaN} } } }
\vdef{default-11:CsMcPU-APV1:pvn:hiDelta}   {\ensuremath{{+0.000 } } }
\vdef{default-11:CsMcPU-APV1:pvn:hiDeltaE}   {\ensuremath{{0.001 } } }
\vdef{default-11:CsData-APV1:pvavew8:loEff}   {\ensuremath{{0.015 } } }
\vdef{default-11:CsData-APV1:pvavew8:loEffE}   {\ensuremath{{0.003 } } }
\vdef{default-11:CsData-APV1:pvavew8:hiEff}   {\ensuremath{{0.985 } } }
\vdef{default-11:CsData-APV1:pvavew8:hiEffE}   {\ensuremath{{0.003 } } }
\vdef{default-11:CsMcPU-APV1:pvavew8:loEff}   {\ensuremath{{0.014 } } }
\vdef{default-11:CsMcPU-APV1:pvavew8:loEffE}   {\ensuremath{{0.003 } } }
\vdef{default-11:CsMcPU-APV1:pvavew8:hiEff}   {\ensuremath{{0.986 } } }
\vdef{default-11:CsMcPU-APV1:pvavew8:hiEffE}   {\ensuremath{{0.003 } } }
\vdef{default-11:CsMcPU-APV1:pvavew8:loDelta}   {\ensuremath{{+0.070 } } }
\vdef{default-11:CsMcPU-APV1:pvavew8:loDeltaE}   {\ensuremath{{0.271 } } }
\vdef{default-11:CsMcPU-APV1:pvavew8:hiDelta}   {\ensuremath{{-0.001 } } }
\vdef{default-11:CsMcPU-APV1:pvavew8:hiDeltaE}   {\ensuremath{{0.004 } } }
\vdef{default-11:CsData-APV1:pvntrk:loEff}   {\ensuremath{{1.000 } } }
\vdef{default-11:CsData-APV1:pvntrk:loEffE}   {\ensuremath{{0.000 } } }
\vdef{default-11:CsData-APV1:pvntrk:hiEff}   {\ensuremath{{1.000 } } }
\vdef{default-11:CsData-APV1:pvntrk:hiEffE}   {\ensuremath{{0.000 } } }
\vdef{default-11:CsMcPU-APV1:pvntrk:loEff}   {\ensuremath{{1.000 } } }
\vdef{default-11:CsMcPU-APV1:pvntrk:loEffE}   {\ensuremath{{0.000 } } }
\vdef{default-11:CsMcPU-APV1:pvntrk:hiEff}   {\ensuremath{{1.000 } } }
\vdef{default-11:CsMcPU-APV1:pvntrk:hiEffE}   {\ensuremath{{0.000 } } }
\vdef{default-11:CsMcPU-APV1:pvntrk:loDelta}   {\ensuremath{{+0.000 } } }
\vdef{default-11:CsMcPU-APV1:pvntrk:loDeltaE}   {\ensuremath{{0.001 } } }
\vdef{default-11:CsMcPU-APV1:pvntrk:hiDelta}   {\ensuremath{{+0.000 } } }
\vdef{default-11:CsMcPU-APV1:pvntrk:hiDeltaE}   {\ensuremath{{0.001 } } }
\vdef{default-11:CsData-APV1:muon1pt:loEff}   {\ensuremath{{1.010 } } }
\vdef{default-11:CsData-APV1:muon1pt:loEffE}   {\ensuremath{{\mathrm{NaN} } } }
\vdef{default-11:CsData-APV1:muon1pt:hiEff}   {\ensuremath{{1.000 } } }
\vdef{default-11:CsData-APV1:muon1pt:hiEffE}   {\ensuremath{{0.001 } } }
\vdef{default-11:CsMcPU-APV1:muon1pt:loEff}   {\ensuremath{{1.014 } } }
\vdef{default-11:CsMcPU-APV1:muon1pt:loEffE}   {\ensuremath{{\mathrm{NaN} } } }
\vdef{default-11:CsMcPU-APV1:muon1pt:hiEff}   {\ensuremath{{1.000 } } }
\vdef{default-11:CsMcPU-APV1:muon1pt:hiEffE}   {\ensuremath{{0.001 } } }
\vdef{default-11:CsMcPU-APV1:muon1pt:loDelta}   {\ensuremath{{-0.004 } } }
\vdef{default-11:CsMcPU-APV1:muon1pt:loDeltaE}   {\ensuremath{{\mathrm{NaN} } } }
\vdef{default-11:CsMcPU-APV1:muon1pt:hiDelta}   {\ensuremath{{+0.000 } } }
\vdef{default-11:CsMcPU-APV1:muon1pt:hiDeltaE}   {\ensuremath{{0.001 } } }
\vdef{default-11:CsData-APV1:muon2pt:loEff}   {\ensuremath{{0.019 } } }
\vdef{default-11:CsData-APV1:muon2pt:loEffE}   {\ensuremath{{0.003 } } }
\vdef{default-11:CsData-APV1:muon2pt:hiEff}   {\ensuremath{{0.981 } } }
\vdef{default-11:CsData-APV1:muon2pt:hiEffE}   {\ensuremath{{0.003 } } }
\vdef{default-11:CsMcPU-APV1:muon2pt:loEff}   {\ensuremath{{0.002 } } }
\vdef{default-11:CsMcPU-APV1:muon2pt:loEffE}   {\ensuremath{{0.001 } } }
\vdef{default-11:CsMcPU-APV1:muon2pt:hiEff}   {\ensuremath{{0.998 } } }
\vdef{default-11:CsMcPU-APV1:muon2pt:hiEffE}   {\ensuremath{{0.001 } } }
\vdef{default-11:CsMcPU-APV1:muon2pt:loDelta}   {\ensuremath{{+1.684 } } }
\vdef{default-11:CsMcPU-APV1:muon2pt:loDeltaE}   {\ensuremath{{0.194 } } }
\vdef{default-11:CsMcPU-APV1:muon2pt:hiDelta}   {\ensuremath{{-0.017 } } }
\vdef{default-11:CsMcPU-APV1:muon2pt:hiDeltaE}   {\ensuremath{{0.003 } } }
\vdef{default-11:CsData-APV1:muonseta:loEff}   {\ensuremath{{0.738 } } }
\vdef{default-11:CsData-APV1:muonseta:loEffE}   {\ensuremath{{0.007 } } }
\vdef{default-11:CsData-APV1:muonseta:hiEff}   {\ensuremath{{0.262 } } }
\vdef{default-11:CsData-APV1:muonseta:hiEffE}   {\ensuremath{{0.007 } } }
\vdef{default-11:CsMcPU-APV1:muonseta:loEff}   {\ensuremath{{0.851 } } }
\vdef{default-11:CsMcPU-APV1:muonseta:loEffE}   {\ensuremath{{0.006 } } }
\vdef{default-11:CsMcPU-APV1:muonseta:hiEff}   {\ensuremath{{0.149 } } }
\vdef{default-11:CsMcPU-APV1:muonseta:hiEffE}   {\ensuremath{{0.006 } } }
\vdef{default-11:CsMcPU-APV1:muonseta:loDelta}   {\ensuremath{{-0.142 } } }
\vdef{default-11:CsMcPU-APV1:muonseta:loDeltaE}   {\ensuremath{{0.012 } } }
\vdef{default-11:CsMcPU-APV1:muonseta:hiDelta}   {\ensuremath{{+0.549 } } }
\vdef{default-11:CsMcPU-APV1:muonseta:hiDeltaE}   {\ensuremath{{0.044 } } }
\vdef{default-11:CsData-APV1:pt:loEff}   {\ensuremath{{0.000 } } }
\vdef{default-11:CsData-APV1:pt:loEffE}   {\ensuremath{{0.000 } } }
\vdef{default-11:CsData-APV1:pt:hiEff}   {\ensuremath{{1.000 } } }
\vdef{default-11:CsData-APV1:pt:hiEffE}   {\ensuremath{{0.000 } } }
\vdef{default-11:CsMcPU-APV1:pt:loEff}   {\ensuremath{{0.000 } } }
\vdef{default-11:CsMcPU-APV1:pt:loEffE}   {\ensuremath{{0.000 } } }
\vdef{default-11:CsMcPU-APV1:pt:hiEff}   {\ensuremath{{1.000 } } }
\vdef{default-11:CsMcPU-APV1:pt:hiEffE}   {\ensuremath{{0.000 } } }
\vdef{default-11:CsMcPU-APV1:pt:loDelta}   {\ensuremath{{\mathrm{NaN} } } }
\vdef{default-11:CsMcPU-APV1:pt:loDeltaE}   {\ensuremath{{\mathrm{NaN} } } }
\vdef{default-11:CsMcPU-APV1:pt:hiDelta}   {\ensuremath{{+0.000 } } }
\vdef{default-11:CsMcPU-APV1:pt:hiDeltaE}   {\ensuremath{{0.000 } } }
\vdef{default-11:CsData-APV1:p:loEff}   {\ensuremath{{1.019 } } }
\vdef{default-11:CsData-APV1:p:loEffE}   {\ensuremath{{\mathrm{NaN} } } }
\vdef{default-11:CsData-APV1:p:hiEff}   {\ensuremath{{1.000 } } }
\vdef{default-11:CsData-APV1:p:hiEffE}   {\ensuremath{{0.001 } } }
\vdef{default-11:CsMcPU-APV1:p:loEff}   {\ensuremath{{1.006 } } }
\vdef{default-11:CsMcPU-APV1:p:loEffE}   {\ensuremath{{\mathrm{NaN} } } }
\vdef{default-11:CsMcPU-APV1:p:hiEff}   {\ensuremath{{1.000 } } }
\vdef{default-11:CsMcPU-APV1:p:hiEffE}   {\ensuremath{{0.001 } } }
\vdef{default-11:CsMcPU-APV1:p:loDelta}   {\ensuremath{{+0.013 } } }
\vdef{default-11:CsMcPU-APV1:p:loDeltaE}   {\ensuremath{{\mathrm{NaN} } } }
\vdef{default-11:CsMcPU-APV1:p:hiDelta}   {\ensuremath{{+0.000 } } }
\vdef{default-11:CsMcPU-APV1:p:hiDeltaE}   {\ensuremath{{0.001 } } }
\vdef{default-11:CsData-APV1:eta:loEff}   {\ensuremath{{0.731 } } }
\vdef{default-11:CsData-APV1:eta:loEffE}   {\ensuremath{{0.010 } } }
\vdef{default-11:CsData-APV1:eta:hiEff}   {\ensuremath{{0.269 } } }
\vdef{default-11:CsData-APV1:eta:hiEffE}   {\ensuremath{{0.010 } } }
\vdef{default-11:CsMcPU-APV1:eta:loEff}   {\ensuremath{{0.847 } } }
\vdef{default-11:CsMcPU-APV1:eta:loEffE}   {\ensuremath{{0.008 } } }
\vdef{default-11:CsMcPU-APV1:eta:hiEff}   {\ensuremath{{0.153 } } }
\vdef{default-11:CsMcPU-APV1:eta:hiEffE}   {\ensuremath{{0.008 } } }
\vdef{default-11:CsMcPU-APV1:eta:loDelta}   {\ensuremath{{-0.147 } } }
\vdef{default-11:CsMcPU-APV1:eta:loDeltaE}   {\ensuremath{{0.017 } } }
\vdef{default-11:CsMcPU-APV1:eta:hiDelta}   {\ensuremath{{+0.550 } } }
\vdef{default-11:CsMcPU-APV1:eta:hiDeltaE}   {\ensuremath{{0.061 } } }
\vdef{default-11:CsData-APV1:bdt:loEff}   {\ensuremath{{0.897 } } }
\vdef{default-11:CsData-APV1:bdt:loEffE}   {\ensuremath{{0.007 } } }
\vdef{default-11:CsData-APV1:bdt:hiEff}   {\ensuremath{{0.103 } } }
\vdef{default-11:CsData-APV1:bdt:hiEffE}   {\ensuremath{{0.007 } } }
\vdef{default-11:CsMcPU-APV1:bdt:loEff}   {\ensuremath{{0.857 } } }
\vdef{default-11:CsMcPU-APV1:bdt:loEffE}   {\ensuremath{{0.008 } } }
\vdef{default-11:CsMcPU-APV1:bdt:hiEff}   {\ensuremath{{0.143 } } }
\vdef{default-11:CsMcPU-APV1:bdt:hiEffE}   {\ensuremath{{0.008 } } }
\vdef{default-11:CsMcPU-APV1:bdt:loDelta}   {\ensuremath{{+0.046 } } }
\vdef{default-11:CsMcPU-APV1:bdt:loDeltaE}   {\ensuremath{{0.012 } } }
\vdef{default-11:CsMcPU-APV1:bdt:hiDelta}   {\ensuremath{{-0.325 } } }
\vdef{default-11:CsMcPU-APV1:bdt:hiDeltaE}   {\ensuremath{{0.082 } } }
\vdef{default-11:CsData-APV1:fl3d:loEff}   {\ensuremath{{0.830 } } }
\vdef{default-11:CsData-APV1:fl3d:loEffE}   {\ensuremath{{0.008 } } }
\vdef{default-11:CsData-APV1:fl3d:hiEff}   {\ensuremath{{0.170 } } }
\vdef{default-11:CsData-APV1:fl3d:hiEffE}   {\ensuremath{{0.008 } } }
\vdef{default-11:CsMcPU-APV1:fl3d:loEff}   {\ensuremath{{0.899 } } }
\vdef{default-11:CsMcPU-APV1:fl3d:loEffE}   {\ensuremath{{0.006 } } }
\vdef{default-11:CsMcPU-APV1:fl3d:hiEff}   {\ensuremath{{0.101 } } }
\vdef{default-11:CsMcPU-APV1:fl3d:hiEffE}   {\ensuremath{{0.006 } } }
\vdef{default-11:CsMcPU-APV1:fl3d:loDelta}   {\ensuremath{{-0.080 } } }
\vdef{default-11:CsMcPU-APV1:fl3d:loDeltaE}   {\ensuremath{{0.012 } } }
\vdef{default-11:CsMcPU-APV1:fl3d:hiDelta}   {\ensuremath{{+0.508 } } }
\vdef{default-11:CsMcPU-APV1:fl3d:hiDeltaE}   {\ensuremath{{0.073 } } }
\vdef{default-11:CsData-APV1:fl3de:loEff}   {\ensuremath{{1.000 } } }
\vdef{default-11:CsData-APV1:fl3de:loEffE}   {\ensuremath{{0.000 } } }
\vdef{default-11:CsData-APV1:fl3de:hiEff}   {\ensuremath{{0.000 } } }
\vdef{default-11:CsData-APV1:fl3de:hiEffE}   {\ensuremath{{0.001 } } }
\vdef{default-11:CsMcPU-APV1:fl3de:loEff}   {\ensuremath{{1.000 } } }
\vdef{default-11:CsMcPU-APV1:fl3de:loEffE}   {\ensuremath{{0.000 } } }
\vdef{default-11:CsMcPU-APV1:fl3de:hiEff}   {\ensuremath{{0.001 } } }
\vdef{default-11:CsMcPU-APV1:fl3de:hiEffE}   {\ensuremath{{0.001 } } }
\vdef{default-11:CsMcPU-APV1:fl3de:loDelta}   {\ensuremath{{+0.000 } } }
\vdef{default-11:CsMcPU-APV1:fl3de:loDeltaE}   {\ensuremath{{0.001 } } }
\vdef{default-11:CsMcPU-APV1:fl3de:hiDelta}   {\ensuremath{{-0.426 } } }
\vdef{default-11:CsMcPU-APV1:fl3de:hiDeltaE}   {\ensuremath{{1.607 } } }
\vdef{default-11:CsData-APV1:fls3d:loEff}   {\ensuremath{{0.087 } } }
\vdef{default-11:CsData-APV1:fls3d:loEffE}   {\ensuremath{{0.006 } } }
\vdef{default-11:CsData-APV1:fls3d:hiEff}   {\ensuremath{{0.913 } } }
\vdef{default-11:CsData-APV1:fls3d:hiEffE}   {\ensuremath{{0.006 } } }
\vdef{default-11:CsMcPU-APV1:fls3d:loEff}   {\ensuremath{{0.069 } } }
\vdef{default-11:CsMcPU-APV1:fls3d:loEffE}   {\ensuremath{{0.005 } } }
\vdef{default-11:CsMcPU-APV1:fls3d:hiEff}   {\ensuremath{{0.931 } } }
\vdef{default-11:CsMcPU-APV1:fls3d:hiEffE}   {\ensuremath{{0.005 } } }
\vdef{default-11:CsMcPU-APV1:fls3d:loDelta}   {\ensuremath{{+0.225 } } }
\vdef{default-11:CsMcPU-APV1:fls3d:loDeltaE}   {\ensuremath{{0.102 } } }
\vdef{default-11:CsMcPU-APV1:fls3d:hiDelta}   {\ensuremath{{-0.019 } } }
\vdef{default-11:CsMcPU-APV1:fls3d:hiDeltaE}   {\ensuremath{{0.009 } } }
\vdef{default-11:CsData-APV1:flsxy:loEff}   {\ensuremath{{1.008 } } }
\vdef{default-11:CsData-APV1:flsxy:loEffE}   {\ensuremath{{\mathrm{NaN} } } }
\vdef{default-11:CsData-APV1:flsxy:hiEff}   {\ensuremath{{1.000 } } }
\vdef{default-11:CsData-APV1:flsxy:hiEffE}   {\ensuremath{{0.000 } } }
\vdef{default-11:CsMcPU-APV1:flsxy:loEff}   {\ensuremath{{1.011 } } }
\vdef{default-11:CsMcPU-APV1:flsxy:loEffE}   {\ensuremath{{\mathrm{NaN} } } }
\vdef{default-11:CsMcPU-APV1:flsxy:hiEff}   {\ensuremath{{1.000 } } }
\vdef{default-11:CsMcPU-APV1:flsxy:hiEffE}   {\ensuremath{{0.000 } } }
\vdef{default-11:CsMcPU-APV1:flsxy:loDelta}   {\ensuremath{{-0.003 } } }
\vdef{default-11:CsMcPU-APV1:flsxy:loDeltaE}   {\ensuremath{{\mathrm{NaN} } } }
\vdef{default-11:CsMcPU-APV1:flsxy:hiDelta}   {\ensuremath{{+0.000 } } }
\vdef{default-11:CsMcPU-APV1:flsxy:hiDeltaE}   {\ensuremath{{0.001 } } }
\vdef{default-11:CsData-APV1:chi2dof:loEff}   {\ensuremath{{0.930 } } }
\vdef{default-11:CsData-APV1:chi2dof:loEffE}   {\ensuremath{{0.006 } } }
\vdef{default-11:CsData-APV1:chi2dof:hiEff}   {\ensuremath{{0.070 } } }
\vdef{default-11:CsData-APV1:chi2dof:hiEffE}   {\ensuremath{{0.006 } } }
\vdef{default-11:CsMcPU-APV1:chi2dof:loEff}   {\ensuremath{{0.942 } } }
\vdef{default-11:CsMcPU-APV1:chi2dof:loEffE}   {\ensuremath{{0.005 } } }
\vdef{default-11:CsMcPU-APV1:chi2dof:hiEff}   {\ensuremath{{0.058 } } }
\vdef{default-11:CsMcPU-APV1:chi2dof:hiEffE}   {\ensuremath{{0.005 } } }
\vdef{default-11:CsMcPU-APV1:chi2dof:loDelta}   {\ensuremath{{-0.013 } } }
\vdef{default-11:CsMcPU-APV1:chi2dof:loDeltaE}   {\ensuremath{{0.008 } } }
\vdef{default-11:CsMcPU-APV1:chi2dof:hiDelta}   {\ensuremath{{+0.191 } } }
\vdef{default-11:CsMcPU-APV1:chi2dof:hiDeltaE}   {\ensuremath{{0.121 } } }
\vdef{default-11:CsData-APV1:pchi2dof:loEff}   {\ensuremath{{0.628 } } }
\vdef{default-11:CsData-APV1:pchi2dof:loEffE}   {\ensuremath{{0.011 } } }
\vdef{default-11:CsData-APV1:pchi2dof:hiEff}   {\ensuremath{{0.372 } } }
\vdef{default-11:CsData-APV1:pchi2dof:hiEffE}   {\ensuremath{{0.011 } } }
\vdef{default-11:CsMcPU-APV1:pchi2dof:loEff}   {\ensuremath{{0.588 } } }
\vdef{default-11:CsMcPU-APV1:pchi2dof:loEffE}   {\ensuremath{{0.011 } } }
\vdef{default-11:CsMcPU-APV1:pchi2dof:hiEff}   {\ensuremath{{0.412 } } }
\vdef{default-11:CsMcPU-APV1:pchi2dof:hiEffE}   {\ensuremath{{0.011 } } }
\vdef{default-11:CsMcPU-APV1:pchi2dof:loDelta}   {\ensuremath{{+0.066 } } }
\vdef{default-11:CsMcPU-APV1:pchi2dof:loDeltaE}   {\ensuremath{{0.025 } } }
\vdef{default-11:CsMcPU-APV1:pchi2dof:hiDelta}   {\ensuremath{{-0.103 } } }
\vdef{default-11:CsMcPU-APV1:pchi2dof:hiDeltaE}   {\ensuremath{{0.038 } } }
\vdef{default-11:CsData-APV1:alpha:loEff}   {\ensuremath{{0.996 } } }
\vdef{default-11:CsData-APV1:alpha:loEffE}   {\ensuremath{{0.001 } } }
\vdef{default-11:CsData-APV1:alpha:hiEff}   {\ensuremath{{0.004 } } }
\vdef{default-11:CsData-APV1:alpha:hiEffE}   {\ensuremath{{0.001 } } }
\vdef{default-11:CsMcPU-APV1:alpha:loEff}   {\ensuremath{{0.990 } } }
\vdef{default-11:CsMcPU-APV1:alpha:loEffE}   {\ensuremath{{0.002 } } }
\vdef{default-11:CsMcPU-APV1:alpha:hiEff}   {\ensuremath{{0.010 } } }
\vdef{default-11:CsMcPU-APV1:alpha:hiEffE}   {\ensuremath{{0.002 } } }
\vdef{default-11:CsMcPU-APV1:alpha:loDelta}   {\ensuremath{{+0.006 } } }
\vdef{default-11:CsMcPU-APV1:alpha:loDeltaE}   {\ensuremath{{0.003 } } }
\vdef{default-11:CsMcPU-APV1:alpha:hiDelta}   {\ensuremath{{-0.888 } } }
\vdef{default-11:CsMcPU-APV1:alpha:hiDeltaE}   {\ensuremath{{0.376 } } }
\vdef{default-11:CsData-APV1:iso:loEff}   {\ensuremath{{0.094 } } }
\vdef{default-11:CsData-APV1:iso:loEffE}   {\ensuremath{{0.006 } } }
\vdef{default-11:CsData-APV1:iso:hiEff}   {\ensuremath{{0.906 } } }
\vdef{default-11:CsData-APV1:iso:hiEffE}   {\ensuremath{{0.006 } } }
\vdef{default-11:CsMcPU-APV1:iso:loEff}   {\ensuremath{{0.104 } } }
\vdef{default-11:CsMcPU-APV1:iso:loEffE}   {\ensuremath{{0.007 } } }
\vdef{default-11:CsMcPU-APV1:iso:hiEff}   {\ensuremath{{0.896 } } }
\vdef{default-11:CsMcPU-APV1:iso:hiEffE}   {\ensuremath{{0.007 } } }
\vdef{default-11:CsMcPU-APV1:iso:loDelta}   {\ensuremath{{-0.104 } } }
\vdef{default-11:CsMcPU-APV1:iso:loDeltaE}   {\ensuremath{{0.093 } } }
\vdef{default-11:CsMcPU-APV1:iso:hiDelta}   {\ensuremath{{+0.011 } } }
\vdef{default-11:CsMcPU-APV1:iso:hiDeltaE}   {\ensuremath{{0.010 } } }
\vdef{default-11:CsData-APV1:docatrk:loEff}   {\ensuremath{{0.082 } } }
\vdef{default-11:CsData-APV1:docatrk:loEffE}   {\ensuremath{{0.006 } } }
\vdef{default-11:CsData-APV1:docatrk:hiEff}   {\ensuremath{{0.918 } } }
\vdef{default-11:CsData-APV1:docatrk:hiEffE}   {\ensuremath{{0.006 } } }
\vdef{default-11:CsMcPU-APV1:docatrk:loEff}   {\ensuremath{{0.092 } } }
\vdef{default-11:CsMcPU-APV1:docatrk:loEffE}   {\ensuremath{{0.006 } } }
\vdef{default-11:CsMcPU-APV1:docatrk:hiEff}   {\ensuremath{{0.908 } } }
\vdef{default-11:CsMcPU-APV1:docatrk:hiEffE}   {\ensuremath{{0.006 } } }
\vdef{default-11:CsMcPU-APV1:docatrk:loDelta}   {\ensuremath{{-0.114 } } }
\vdef{default-11:CsMcPU-APV1:docatrk:loDeltaE}   {\ensuremath{{0.102 } } }
\vdef{default-11:CsMcPU-APV1:docatrk:hiDelta}   {\ensuremath{{+0.011 } } }
\vdef{default-11:CsMcPU-APV1:docatrk:hiDeltaE}   {\ensuremath{{0.010 } } }
\vdef{default-11:CsData-APV1:isotrk:loEff}   {\ensuremath{{1.000 } } }
\vdef{default-11:CsData-APV1:isotrk:loEffE}   {\ensuremath{{0.000 } } }
\vdef{default-11:CsData-APV1:isotrk:hiEff}   {\ensuremath{{1.000 } } }
\vdef{default-11:CsData-APV1:isotrk:hiEffE}   {\ensuremath{{0.000 } } }
\vdef{default-11:CsMcPU-APV1:isotrk:loEff}   {\ensuremath{{1.000 } } }
\vdef{default-11:CsMcPU-APV1:isotrk:loEffE}   {\ensuremath{{0.000 } } }
\vdef{default-11:CsMcPU-APV1:isotrk:hiEff}   {\ensuremath{{1.000 } } }
\vdef{default-11:CsMcPU-APV1:isotrk:hiEffE}   {\ensuremath{{0.000 } } }
\vdef{default-11:CsMcPU-APV1:isotrk:loDelta}   {\ensuremath{{+0.000 } } }
\vdef{default-11:CsMcPU-APV1:isotrk:loDeltaE}   {\ensuremath{{0.001 } } }
\vdef{default-11:CsMcPU-APV1:isotrk:hiDelta}   {\ensuremath{{+0.000 } } }
\vdef{default-11:CsMcPU-APV1:isotrk:hiDeltaE}   {\ensuremath{{0.001 } } }
\vdef{default-11:CsData-APV1:closetrk:loEff}   {\ensuremath{{0.978 } } }
\vdef{default-11:CsData-APV1:closetrk:loEffE}   {\ensuremath{{0.003 } } }
\vdef{default-11:CsData-APV1:closetrk:hiEff}   {\ensuremath{{0.022 } } }
\vdef{default-11:CsData-APV1:closetrk:hiEffE}   {\ensuremath{{0.003 } } }
\vdef{default-11:CsMcPU-APV1:closetrk:loEff}   {\ensuremath{{0.972 } } }
\vdef{default-11:CsMcPU-APV1:closetrk:loEffE}   {\ensuremath{{0.004 } } }
\vdef{default-11:CsMcPU-APV1:closetrk:hiEff}   {\ensuremath{{0.028 } } }
\vdef{default-11:CsMcPU-APV1:closetrk:hiEffE}   {\ensuremath{{0.004 } } }
\vdef{default-11:CsMcPU-APV1:closetrk:loDelta}   {\ensuremath{{+0.006 } } }
\vdef{default-11:CsMcPU-APV1:closetrk:loDeltaE}   {\ensuremath{{0.005 } } }
\vdef{default-11:CsMcPU-APV1:closetrk:hiDelta}   {\ensuremath{{-0.230 } } }
\vdef{default-11:CsMcPU-APV1:closetrk:hiDeltaE}   {\ensuremath{{0.199 } } }
\vdef{default-11:CsData-APV1:lip:loEff}   {\ensuremath{{1.000 } } }
\vdef{default-11:CsData-APV1:lip:loEffE}   {\ensuremath{{0.001 } } }
\vdef{default-11:CsData-APV1:lip:hiEff}   {\ensuremath{{0.000 } } }
\vdef{default-11:CsData-APV1:lip:hiEffE}   {\ensuremath{{0.001 } } }
\vdef{default-11:CsMcPU-APV1:lip:loEff}   {\ensuremath{{1.000 } } }
\vdef{default-11:CsMcPU-APV1:lip:loEffE}   {\ensuremath{{0.001 } } }
\vdef{default-11:CsMcPU-APV1:lip:hiEff}   {\ensuremath{{0.000 } } }
\vdef{default-11:CsMcPU-APV1:lip:hiEffE}   {\ensuremath{{0.001 } } }
\vdef{default-11:CsMcPU-APV1:lip:loDelta}   {\ensuremath{{+0.000 } } }
\vdef{default-11:CsMcPU-APV1:lip:loDeltaE}   {\ensuremath{{0.001 } } }
\vdef{default-11:CsMcPU-APV1:lip:hiDelta}   {\ensuremath{{\mathrm{NaN} } } }
\vdef{default-11:CsMcPU-APV1:lip:hiDeltaE}   {\ensuremath{{\mathrm{NaN} } } }
\vdef{default-11:CsData-APV1:lip:inEff}   {\ensuremath{{1.000 } } }
\vdef{default-11:CsData-APV1:lip:inEffE}   {\ensuremath{{0.001 } } }
\vdef{default-11:CsMcPU-APV1:lip:inEff}   {\ensuremath{{1.000 } } }
\vdef{default-11:CsMcPU-APV1:lip:inEffE}   {\ensuremath{{0.001 } } }
\vdef{default-11:CsMcPU-APV1:lip:inDelta}   {\ensuremath{{+0.000 } } }
\vdef{default-11:CsMcPU-APV1:lip:inDeltaE}   {\ensuremath{{0.001 } } }
\vdef{default-11:CsData-APV1:lips:loEff}   {\ensuremath{{1.000 } } }
\vdef{default-11:CsData-APV1:lips:loEffE}   {\ensuremath{{0.001 } } }
\vdef{default-11:CsData-APV1:lips:hiEff}   {\ensuremath{{0.000 } } }
\vdef{default-11:CsData-APV1:lips:hiEffE}   {\ensuremath{{0.001 } } }
\vdef{default-11:CsMcPU-APV1:lips:loEff}   {\ensuremath{{1.000 } } }
\vdef{default-11:CsMcPU-APV1:lips:loEffE}   {\ensuremath{{0.001 } } }
\vdef{default-11:CsMcPU-APV1:lips:hiEff}   {\ensuremath{{0.000 } } }
\vdef{default-11:CsMcPU-APV1:lips:hiEffE}   {\ensuremath{{0.001 } } }
\vdef{default-11:CsMcPU-APV1:lips:loDelta}   {\ensuremath{{+0.000 } } }
\vdef{default-11:CsMcPU-APV1:lips:loDeltaE}   {\ensuremath{{0.001 } } }
\vdef{default-11:CsMcPU-APV1:lips:hiDelta}   {\ensuremath{{\mathrm{NaN} } } }
\vdef{default-11:CsMcPU-APV1:lips:hiDeltaE}   {\ensuremath{{\mathrm{NaN} } } }
\vdef{default-11:CsData-APV1:lips:inEff}   {\ensuremath{{1.000 } } }
\vdef{default-11:CsData-APV1:lips:inEffE}   {\ensuremath{{0.001 } } }
\vdef{default-11:CsMcPU-APV1:lips:inEff}   {\ensuremath{{1.000 } } }
\vdef{default-11:CsMcPU-APV1:lips:inEffE}   {\ensuremath{{0.001 } } }
\vdef{default-11:CsMcPU-APV1:lips:inDelta}   {\ensuremath{{+0.000 } } }
\vdef{default-11:CsMcPU-APV1:lips:inDeltaE}   {\ensuremath{{0.001 } } }
\vdef{default-11:CsData-APV1:ip:loEff}   {\ensuremath{{0.970 } } }
\vdef{default-11:CsData-APV1:ip:loEffE}   {\ensuremath{{0.004 } } }
\vdef{default-11:CsData-APV1:ip:hiEff}   {\ensuremath{{0.030 } } }
\vdef{default-11:CsData-APV1:ip:hiEffE}   {\ensuremath{{0.004 } } }
\vdef{default-11:CsMcPU-APV1:ip:loEff}   {\ensuremath{{0.964 } } }
\vdef{default-11:CsMcPU-APV1:ip:loEffE}   {\ensuremath{{0.004 } } }
\vdef{default-11:CsMcPU-APV1:ip:hiEff}   {\ensuremath{{0.036 } } }
\vdef{default-11:CsMcPU-APV1:ip:hiEffE}   {\ensuremath{{0.004 } } }
\vdef{default-11:CsMcPU-APV1:ip:loDelta}   {\ensuremath{{+0.006 } } }
\vdef{default-11:CsMcPU-APV1:ip:loDeltaE}   {\ensuremath{{0.006 } } }
\vdef{default-11:CsMcPU-APV1:ip:hiDelta}   {\ensuremath{{-0.181 } } }
\vdef{default-11:CsMcPU-APV1:ip:hiDeltaE}   {\ensuremath{{0.174 } } }
\vdef{default-11:CsData-APV1:ips:loEff}   {\ensuremath{{0.937 } } }
\vdef{default-11:CsData-APV1:ips:loEffE}   {\ensuremath{{0.005 } } }
\vdef{default-11:CsData-APV1:ips:hiEff}   {\ensuremath{{0.063 } } }
\vdef{default-11:CsData-APV1:ips:hiEffE}   {\ensuremath{{0.005 } } }
\vdef{default-11:CsMcPU-APV1:ips:loEff}   {\ensuremath{{0.928 } } }
\vdef{default-11:CsMcPU-APV1:ips:loEffE}   {\ensuremath{{0.006 } } }
\vdef{default-11:CsMcPU-APV1:ips:hiEff}   {\ensuremath{{0.072 } } }
\vdef{default-11:CsMcPU-APV1:ips:hiEffE}   {\ensuremath{{0.006 } } }
\vdef{default-11:CsMcPU-APV1:ips:loDelta}   {\ensuremath{{+0.010 } } }
\vdef{default-11:CsMcPU-APV1:ips:loDeltaE}   {\ensuremath{{0.008 } } }
\vdef{default-11:CsMcPU-APV1:ips:hiDelta}   {\ensuremath{{-0.136 } } }
\vdef{default-11:CsMcPU-APV1:ips:hiDeltaE}   {\ensuremath{{0.116 } } }
\vdef{default-11:CsData-APV1:maxdoca:loEff}   {\ensuremath{{1.000 } } }
\vdef{default-11:CsData-APV1:maxdoca:loEffE}   {\ensuremath{{0.001 } } }
\vdef{default-11:CsData-APV1:maxdoca:hiEff}   {\ensuremath{{0.036 } } }
\vdef{default-11:CsData-APV1:maxdoca:hiEffE}   {\ensuremath{{0.004 } } }
\vdef{default-11:CsMcPU-APV1:maxdoca:loEff}   {\ensuremath{{1.000 } } }
\vdef{default-11:CsMcPU-APV1:maxdoca:loEffE}   {\ensuremath{{0.001 } } }
\vdef{default-11:CsMcPU-APV1:maxdoca:hiEff}   {\ensuremath{{0.017 } } }
\vdef{default-11:CsMcPU-APV1:maxdoca:hiEffE}   {\ensuremath{{0.003 } } }
\vdef{default-11:CsMcPU-APV1:maxdoca:loDelta}   {\ensuremath{{+0.000 } } }
\vdef{default-11:CsMcPU-APV1:maxdoca:loDeltaE}   {\ensuremath{{0.001 } } }
\vdef{default-11:CsMcPU-APV1:maxdoca:hiDelta}   {\ensuremath{{+0.706 } } }
\vdef{default-11:CsMcPU-APV1:maxdoca:hiDeltaE}   {\ensuremath{{0.188 } } }
\vdef{default-11:CsData-APV1:kaonspt:loEff}   {\ensuremath{{1.000 } } }
\vdef{default-11:CsData-APV1:kaonspt:loEffE}   {\ensuremath{{0.000 } } }
\vdef{default-11:CsData-APV1:kaonspt:hiEff}   {\ensuremath{{1.000 } } }
\vdef{default-11:CsData-APV1:kaonspt:hiEffE}   {\ensuremath{{0.000 } } }
\vdef{default-11:CsMcPU-APV1:kaonspt:loEff}   {\ensuremath{{1.001 } } }
\vdef{default-11:CsMcPU-APV1:kaonspt:loEffE}   {\ensuremath{{\mathrm{NaN} } } }
\vdef{default-11:CsMcPU-APV1:kaonspt:hiEff}   {\ensuremath{{1.000 } } }
\vdef{default-11:CsMcPU-APV1:kaonspt:hiEffE}   {\ensuremath{{0.000 } } }
\vdef{default-11:CsMcPU-APV1:kaonspt:loDelta}   {\ensuremath{{-0.001 } } }
\vdef{default-11:CsMcPU-APV1:kaonspt:loDeltaE}   {\ensuremath{{\mathrm{NaN} } } }
\vdef{default-11:CsMcPU-APV1:kaonspt:hiDelta}   {\ensuremath{{+0.000 } } }
\vdef{default-11:CsMcPU-APV1:kaonspt:hiDeltaE}   {\ensuremath{{0.000 } } }
\vdef{default-11:CsData-APV1:psipt:loEff}   {\ensuremath{{1.006 } } }
\vdef{default-11:CsData-APV1:psipt:loEffE}   {\ensuremath{{\mathrm{NaN} } } }
\vdef{default-11:CsData-APV1:psipt:hiEff}   {\ensuremath{{1.000 } } }
\vdef{default-11:CsData-APV1:psipt:hiEffE}   {\ensuremath{{0.001 } } }
\vdef{default-11:CsMcPU-APV1:psipt:loEff}   {\ensuremath{{1.005 } } }
\vdef{default-11:CsMcPU-APV1:psipt:loEffE}   {\ensuremath{{\mathrm{NaN} } } }
\vdef{default-11:CsMcPU-APV1:psipt:hiEff}   {\ensuremath{{1.000 } } }
\vdef{default-11:CsMcPU-APV1:psipt:hiEffE}   {\ensuremath{{0.001 } } }
\vdef{default-11:CsMcPU-APV1:psipt:loDelta}   {\ensuremath{{+0.001 } } }
\vdef{default-11:CsMcPU-APV1:psipt:loDeltaE}   {\ensuremath{{\mathrm{NaN} } } }
\vdef{default-11:CsMcPU-APV1:psipt:hiDelta}   {\ensuremath{{+0.000 } } }
\vdef{default-11:CsMcPU-APV1:psipt:hiDeltaE}   {\ensuremath{{0.001 } } }
\vdef{default-11:CsData-APV1:phipt:loEff}   {\ensuremath{{1.015 } } }
\vdef{default-11:CsData-APV1:phipt:loEffE}   {\ensuremath{{\mathrm{NaN} } } }
\vdef{default-11:CsData-APV1:phipt:hiEff}   {\ensuremath{{1.000 } } }
\vdef{default-11:CsData-APV1:phipt:hiEffE}   {\ensuremath{{0.001 } } }
\vdef{default-11:CsMcPU-APV1:phipt:loEff}   {\ensuremath{{1.011 } } }
\vdef{default-11:CsMcPU-APV1:phipt:loEffE}   {\ensuremath{{\mathrm{NaN} } } }
\vdef{default-11:CsMcPU-APV1:phipt:hiEff}   {\ensuremath{{1.000 } } }
\vdef{default-11:CsMcPU-APV1:phipt:hiEffE}   {\ensuremath{{0.001 } } }
\vdef{default-11:CsMcPU-APV1:phipt:loDelta}   {\ensuremath{{+0.004 } } }
\vdef{default-11:CsMcPU-APV1:phipt:loDeltaE}   {\ensuremath{{\mathrm{NaN} } } }
\vdef{default-11:CsMcPU-APV1:phipt:hiDelta}   {\ensuremath{{+0.000 } } }
\vdef{default-11:CsMcPU-APV1:phipt:hiDeltaE}   {\ensuremath{{0.001 } } }
\vdef{default-11:CsData-APV1:deltar:loEff}   {\ensuremath{{1.000 } } }
\vdef{default-11:CsData-APV1:deltar:loEffE}   {\ensuremath{{0.000 } } }
\vdef{default-11:CsData-APV1:deltar:hiEff}   {\ensuremath{{0.000 } } }
\vdef{default-11:CsData-APV1:deltar:hiEffE}   {\ensuremath{{0.000 } } }
\vdef{default-11:CsMcPU-APV1:deltar:loEff}   {\ensuremath{{0.999 } } }
\vdef{default-11:CsMcPU-APV1:deltar:loEffE}   {\ensuremath{{0.001 } } }
\vdef{default-11:CsMcPU-APV1:deltar:hiEff}   {\ensuremath{{0.001 } } }
\vdef{default-11:CsMcPU-APV1:deltar:hiEffE}   {\ensuremath{{0.001 } } }
\vdef{default-11:CsMcPU-APV1:deltar:loDelta}   {\ensuremath{{+0.001 } } }
\vdef{default-11:CsMcPU-APV1:deltar:loDeltaE}   {\ensuremath{{0.001 } } }
\vdef{default-11:CsMcPU-APV1:deltar:hiDelta}   {\ensuremath{{-2.000 } } }
\vdef{default-11:CsMcPU-APV1:deltar:hiDeltaE}   {\ensuremath{{2.717 } } }
\vdef{default-11:CsData-APV1:mkk:loEff}   {\ensuremath{{1.161 } } }
\vdef{default-11:CsData-APV1:mkk:loEffE}   {\ensuremath{{\mathrm{NaN} } } }
\vdef{default-11:CsData-APV1:mkk:hiEff}   {\ensuremath{{1.000 } } }
\vdef{default-11:CsData-APV1:mkk:hiEffE}   {\ensuremath{{0.000 } } }
\vdef{default-11:CsMcPU-APV1:mkk:loEff}   {\ensuremath{{1.000 } } }
\vdef{default-11:CsMcPU-APV1:mkk:loEffE}   {\ensuremath{{0.000 } } }
\vdef{default-11:CsMcPU-APV1:mkk:hiEff}   {\ensuremath{{1.000 } } }
\vdef{default-11:CsMcPU-APV1:mkk:hiEffE}   {\ensuremath{{0.000 } } }
\vdef{default-11:CsMcPU-APV1:mkk:loDelta}   {\ensuremath{{+0.149 } } }
\vdef{default-11:CsMcPU-APV1:mkk:loDeltaE}   {\ensuremath{{\mathrm{NaN} } } }
\vdef{default-11:CsMcPU-APV1:mkk:hiDelta}   {\ensuremath{{+0.000 } } }
\vdef{default-11:CsMcPU-APV1:mkk:hiDeltaE}   {\ensuremath{{0.000 } } }
\vdef{default-11:CsMcPU-APV0:osiso:loEff}   {\ensuremath{{1.002 } } }
\vdef{default-11:CsMcPU-APV0:osiso:loEffE}   {\ensuremath{{\mathrm{NaN} } } }
\vdef{default-11:CsMcPU-APV0:osiso:hiEff}   {\ensuremath{{1.000 } } }
\vdef{default-11:CsMcPU-APV0:osiso:hiEffE}   {\ensuremath{{0.000 } } }
\vdef{default-11:CsMcPU-APV1:osiso:loEff}   {\ensuremath{{1.004 } } }
\vdef{default-11:CsMcPU-APV1:osiso:loEffE}   {\ensuremath{{\mathrm{NaN} } } }
\vdef{default-11:CsMcPU-APV1:osiso:hiEff}   {\ensuremath{{1.000 } } }
\vdef{default-11:CsMcPU-APV1:osiso:hiEffE}   {\ensuremath{{0.000 } } }
\vdef{default-11:CsMcPU-APV1:osiso:loDelta}   {\ensuremath{{-0.002 } } }
\vdef{default-11:CsMcPU-APV1:osiso:loDeltaE}   {\ensuremath{{\mathrm{NaN} } } }
\vdef{default-11:CsMcPU-APV1:osiso:hiDelta}   {\ensuremath{{+0.000 } } }
\vdef{default-11:CsMcPU-APV1:osiso:hiDeltaE}   {\ensuremath{{0.001 } } }
\vdef{default-11:CsMcPU-APV0:osreliso:loEff}   {\ensuremath{{0.287 } } }
\vdef{default-11:CsMcPU-APV0:osreliso:loEffE}   {\ensuremath{{0.009 } } }
\vdef{default-11:CsMcPU-APV0:osreliso:hiEff}   {\ensuremath{{0.713 } } }
\vdef{default-11:CsMcPU-APV0:osreliso:hiEffE}   {\ensuremath{{0.009 } } }
\vdef{default-11:CsMcPU-APV1:osreliso:loEff}   {\ensuremath{{0.293 } } }
\vdef{default-11:CsMcPU-APV1:osreliso:loEffE}   {\ensuremath{{0.009 } } }
\vdef{default-11:CsMcPU-APV1:osreliso:hiEff}   {\ensuremath{{0.707 } } }
\vdef{default-11:CsMcPU-APV1:osreliso:hiEffE}   {\ensuremath{{0.009 } } }
\vdef{default-11:CsMcPU-APV1:osreliso:loDelta}   {\ensuremath{{-0.021 } } }
\vdef{default-11:CsMcPU-APV1:osreliso:loDeltaE}   {\ensuremath{{0.042 } } }
\vdef{default-11:CsMcPU-APV1:osreliso:hiDelta}   {\ensuremath{{+0.009 } } }
\vdef{default-11:CsMcPU-APV1:osreliso:hiDeltaE}   {\ensuremath{{0.017 } } }
\vdef{default-11:CsMcPU-APV0:osmuonpt:loEff}   {\ensuremath{{0.000 } } }
\vdef{default-11:CsMcPU-APV0:osmuonpt:loEffE}   {\ensuremath{{0.009 } } }
\vdef{default-11:CsMcPU-APV0:osmuonpt:hiEff}   {\ensuremath{{1.000 } } }
\vdef{default-11:CsMcPU-APV0:osmuonpt:hiEffE}   {\ensuremath{{0.009 } } }
\vdef{default-11:CsMcPU-APV1:osmuonpt:loEff}   {\ensuremath{{0.000 } } }
\vdef{default-11:CsMcPU-APV1:osmuonpt:loEffE}   {\ensuremath{{0.009 } } }
\vdef{default-11:CsMcPU-APV1:osmuonpt:hiEff}   {\ensuremath{{1.000 } } }
\vdef{default-11:CsMcPU-APV1:osmuonpt:hiEffE}   {\ensuremath{{0.009 } } }
\vdef{default-11:CsMcPU-APV1:osmuonpt:loDelta}   {\ensuremath{{\mathrm{NaN} } } }
\vdef{default-11:CsMcPU-APV1:osmuonpt:loDeltaE}   {\ensuremath{{\mathrm{NaN} } } }
\vdef{default-11:CsMcPU-APV1:osmuonpt:hiDelta}   {\ensuremath{{+0.000 } } }
\vdef{default-11:CsMcPU-APV1:osmuonpt:hiDeltaE}   {\ensuremath{{0.012 } } }
\vdef{default-11:CsMcPU-APV0:osmuondr:loEff}   {\ensuremath{{0.053 } } }
\vdef{default-11:CsMcPU-APV0:osmuondr:loEffE}   {\ensuremath{{0.022 } } }
\vdef{default-11:CsMcPU-APV0:osmuondr:hiEff}   {\ensuremath{{0.947 } } }
\vdef{default-11:CsMcPU-APV0:osmuondr:hiEffE}   {\ensuremath{{0.022 } } }
\vdef{default-11:CsMcPU-APV1:osmuondr:loEff}   {\ensuremath{{0.000 } } }
\vdef{default-11:CsMcPU-APV1:osmuondr:loEffE}   {\ensuremath{{0.009 } } }
\vdef{default-11:CsMcPU-APV1:osmuondr:hiEff}   {\ensuremath{{1.000 } } }
\vdef{default-11:CsMcPU-APV1:osmuondr:hiEffE}   {\ensuremath{{0.009 } } }
\vdef{default-11:CsMcPU-APV1:osmuondr:loDelta}   {\ensuremath{{+2.000 } } }
\vdef{default-11:CsMcPU-APV1:osmuondr:loDeltaE}   {\ensuremath{{0.650 } } }
\vdef{default-11:CsMcPU-APV1:osmuondr:hiDelta}   {\ensuremath{{-0.054 } } }
\vdef{default-11:CsMcPU-APV1:osmuondr:hiDeltaE}   {\ensuremath{{0.025 } } }
\vdef{default-11:CsMcPU-APV0:hlt:loEff}   {\ensuremath{{0.281 } } }
\vdef{default-11:CsMcPU-APV0:hlt:loEffE}   {\ensuremath{{0.009 } } }
\vdef{default-11:CsMcPU-APV0:hlt:hiEff}   {\ensuremath{{0.719 } } }
\vdef{default-11:CsMcPU-APV0:hlt:hiEffE}   {\ensuremath{{0.009 } } }
\vdef{default-11:CsMcPU-APV1:hlt:loEff}   {\ensuremath{{0.357 } } }
\vdef{default-11:CsMcPU-APV1:hlt:loEffE}   {\ensuremath{{0.009 } } }
\vdef{default-11:CsMcPU-APV1:hlt:hiEff}   {\ensuremath{{0.643 } } }
\vdef{default-11:CsMcPU-APV1:hlt:hiEffE}   {\ensuremath{{0.009 } } }
\vdef{default-11:CsMcPU-APV1:hlt:loDelta}   {\ensuremath{{-0.240 } } }
\vdef{default-11:CsMcPU-APV1:hlt:loDeltaE}   {\ensuremath{{0.040 } } }
\vdef{default-11:CsMcPU-APV1:hlt:hiDelta}   {\ensuremath{{+0.113 } } }
\vdef{default-11:CsMcPU-APV1:hlt:hiDeltaE}   {\ensuremath{{0.019 } } }
\vdef{default-11:CsMcPU-APV0:muonsid:loEff}   {\ensuremath{{0.230 } } }
\vdef{default-11:CsMcPU-APV0:muonsid:loEffE}   {\ensuremath{{0.008 } } }
\vdef{default-11:CsMcPU-APV0:muonsid:hiEff}   {\ensuremath{{0.770 } } }
\vdef{default-11:CsMcPU-APV0:muonsid:hiEffE}   {\ensuremath{{0.008 } } }
\vdef{default-11:CsMcPU-APV1:muonsid:loEff}   {\ensuremath{{0.223 } } }
\vdef{default-11:CsMcPU-APV1:muonsid:loEffE}   {\ensuremath{{0.008 } } }
\vdef{default-11:CsMcPU-APV1:muonsid:hiEff}   {\ensuremath{{0.777 } } }
\vdef{default-11:CsMcPU-APV1:muonsid:hiEffE}   {\ensuremath{{0.008 } } }
\vdef{default-11:CsMcPU-APV1:muonsid:loDelta}   {\ensuremath{{+0.033 } } }
\vdef{default-11:CsMcPU-APV1:muonsid:loDeltaE}   {\ensuremath{{0.049 } } }
\vdef{default-11:CsMcPU-APV1:muonsid:hiDelta}   {\ensuremath{{-0.010 } } }
\vdef{default-11:CsMcPU-APV1:muonsid:hiDeltaE}   {\ensuremath{{0.014 } } }
\vdef{default-11:CsMcPU-APV0:tracksqual:loEff}   {\ensuremath{{0.000 } } }
\vdef{default-11:CsMcPU-APV0:tracksqual:loEffE}   {\ensuremath{{0.000 } } }
\vdef{default-11:CsMcPU-APV0:tracksqual:hiEff}   {\ensuremath{{1.000 } } }
\vdef{default-11:CsMcPU-APV0:tracksqual:hiEffE}   {\ensuremath{{0.000 } } }
\vdef{default-11:CsMcPU-APV1:tracksqual:loEff}   {\ensuremath{{0.000 } } }
\vdef{default-11:CsMcPU-APV1:tracksqual:loEffE}   {\ensuremath{{0.000 } } }
\vdef{default-11:CsMcPU-APV1:tracksqual:hiEff}   {\ensuremath{{1.000 } } }
\vdef{default-11:CsMcPU-APV1:tracksqual:hiEffE}   {\ensuremath{{0.000 } } }
\vdef{default-11:CsMcPU-APV1:tracksqual:loDelta}   {\ensuremath{{\mathrm{NaN} } } }
\vdef{default-11:CsMcPU-APV1:tracksqual:loDeltaE}   {\ensuremath{{\mathrm{NaN} } } }
\vdef{default-11:CsMcPU-APV1:tracksqual:hiDelta}   {\ensuremath{{+0.000 } } }
\vdef{default-11:CsMcPU-APV1:tracksqual:hiDeltaE}   {\ensuremath{{0.001 } } }
\vdef{default-11:CsMcPU-APV0:pvz:loEff}   {\ensuremath{{0.481 } } }
\vdef{default-11:CsMcPU-APV0:pvz:loEffE}   {\ensuremath{{0.010 } } }
\vdef{default-11:CsMcPU-APV0:pvz:hiEff}   {\ensuremath{{0.519 } } }
\vdef{default-11:CsMcPU-APV0:pvz:hiEffE}   {\ensuremath{{0.010 } } }
\vdef{default-11:CsMcPU-APV1:pvz:loEff}   {\ensuremath{{0.464 } } }
\vdef{default-11:CsMcPU-APV1:pvz:loEffE}   {\ensuremath{{0.010 } } }
\vdef{default-11:CsMcPU-APV1:pvz:hiEff}   {\ensuremath{{0.536 } } }
\vdef{default-11:CsMcPU-APV1:pvz:hiEffE}   {\ensuremath{{0.010 } } }
\vdef{default-11:CsMcPU-APV1:pvz:loDelta}   {\ensuremath{{+0.036 } } }
\vdef{default-11:CsMcPU-APV1:pvz:loDeltaE}   {\ensuremath{{0.029 } } }
\vdef{default-11:CsMcPU-APV1:pvz:hiDelta}   {\ensuremath{{-0.032 } } }
\vdef{default-11:CsMcPU-APV1:pvz:hiDeltaE}   {\ensuremath{{0.026 } } }
\vdef{default-11:CsMcPU-APV0:pvn:loEff}   {\ensuremath{{1.000 } } }
\vdef{default-11:CsMcPU-APV0:pvn:loEffE}   {\ensuremath{{0.000 } } }
\vdef{default-11:CsMcPU-APV0:pvn:hiEff}   {\ensuremath{{1.000 } } }
\vdef{default-11:CsMcPU-APV0:pvn:hiEffE}   {\ensuremath{{0.000 } } }
\vdef{default-11:CsMcPU-APV1:pvn:loEff}   {\ensuremath{{1.184 } } }
\vdef{default-11:CsMcPU-APV1:pvn:loEffE}   {\ensuremath{{\mathrm{NaN} } } }
\vdef{default-11:CsMcPU-APV1:pvn:hiEff}   {\ensuremath{{1.000 } } }
\vdef{default-11:CsMcPU-APV1:pvn:hiEffE}   {\ensuremath{{0.000 } } }
\vdef{default-11:CsMcPU-APV1:pvn:loDelta}   {\ensuremath{{-0.168 } } }
\vdef{default-11:CsMcPU-APV1:pvn:loDeltaE}   {\ensuremath{{\mathrm{NaN} } } }
\vdef{default-11:CsMcPU-APV1:pvn:hiDelta}   {\ensuremath{{+0.000 } } }
\vdef{default-11:CsMcPU-APV1:pvn:hiDeltaE}   {\ensuremath{{0.001 } } }
\vdef{default-11:CsMcPU-APV0:pvavew8:loEff}   {\ensuremath{{0.007 } } }
\vdef{default-11:CsMcPU-APV0:pvavew8:loEffE}   {\ensuremath{{0.002 } } }
\vdef{default-11:CsMcPU-APV0:pvavew8:hiEff}   {\ensuremath{{0.993 } } }
\vdef{default-11:CsMcPU-APV0:pvavew8:hiEffE}   {\ensuremath{{0.002 } } }
\vdef{default-11:CsMcPU-APV1:pvavew8:loEff}   {\ensuremath{{0.014 } } }
\vdef{default-11:CsMcPU-APV1:pvavew8:loEffE}   {\ensuremath{{0.002 } } }
\vdef{default-11:CsMcPU-APV1:pvavew8:hiEff}   {\ensuremath{{0.986 } } }
\vdef{default-11:CsMcPU-APV1:pvavew8:hiEffE}   {\ensuremath{{0.002 } } }
\vdef{default-11:CsMcPU-APV1:pvavew8:loDelta}   {\ensuremath{{-0.676 } } }
\vdef{default-11:CsMcPU-APV1:pvavew8:loDeltaE}   {\ensuremath{{0.266 } } }
\vdef{default-11:CsMcPU-APV1:pvavew8:hiDelta}   {\ensuremath{{+0.007 } } }
\vdef{default-11:CsMcPU-APV1:pvavew8:hiDeltaE}   {\ensuremath{{0.003 } } }
\vdef{default-11:CsMcPU-APV0:pvntrk:loEff}   {\ensuremath{{1.000 } } }
\vdef{default-11:CsMcPU-APV0:pvntrk:loEffE}   {\ensuremath{{0.000 } } }
\vdef{default-11:CsMcPU-APV0:pvntrk:hiEff}   {\ensuremath{{1.000 } } }
\vdef{default-11:CsMcPU-APV0:pvntrk:hiEffE}   {\ensuremath{{0.000 } } }
\vdef{default-11:CsMcPU-APV1:pvntrk:loEff}   {\ensuremath{{1.000 } } }
\vdef{default-11:CsMcPU-APV1:pvntrk:loEffE}   {\ensuremath{{0.000 } } }
\vdef{default-11:CsMcPU-APV1:pvntrk:hiEff}   {\ensuremath{{1.000 } } }
\vdef{default-11:CsMcPU-APV1:pvntrk:hiEffE}   {\ensuremath{{0.000 } } }
\vdef{default-11:CsMcPU-APV1:pvntrk:loDelta}   {\ensuremath{{+0.000 } } }
\vdef{default-11:CsMcPU-APV1:pvntrk:loDeltaE}   {\ensuremath{{0.001 } } }
\vdef{default-11:CsMcPU-APV1:pvntrk:hiDelta}   {\ensuremath{{+0.000 } } }
\vdef{default-11:CsMcPU-APV1:pvntrk:hiDeltaE}   {\ensuremath{{0.001 } } }
\vdef{default-11:CsMcPU-APV0:muon1pt:loEff}   {\ensuremath{{1.014 } } }
\vdef{default-11:CsMcPU-APV0:muon1pt:loEffE}   {\ensuremath{{\mathrm{NaN} } } }
\vdef{default-11:CsMcPU-APV0:muon1pt:hiEff}   {\ensuremath{{1.000 } } }
\vdef{default-11:CsMcPU-APV0:muon1pt:hiEffE}   {\ensuremath{{0.000 } } }
\vdef{default-11:CsMcPU-APV1:muon1pt:loEff}   {\ensuremath{{1.014 } } }
\vdef{default-11:CsMcPU-APV1:muon1pt:loEffE}   {\ensuremath{{\mathrm{NaN} } } }
\vdef{default-11:CsMcPU-APV1:muon1pt:hiEff}   {\ensuremath{{1.000 } } }
\vdef{default-11:CsMcPU-APV1:muon1pt:hiEffE}   {\ensuremath{{0.000 } } }
\vdef{default-11:CsMcPU-APV1:muon1pt:loDelta}   {\ensuremath{{-0.000 } } }
\vdef{default-11:CsMcPU-APV1:muon1pt:loDeltaE}   {\ensuremath{{\mathrm{NaN} } } }
\vdef{default-11:CsMcPU-APV1:muon1pt:hiDelta}   {\ensuremath{{+0.000 } } }
\vdef{default-11:CsMcPU-APV1:muon1pt:hiDeltaE}   {\ensuremath{{0.001 } } }
\vdef{default-11:CsMcPU-APV0:muon2pt:loEff}   {\ensuremath{{0.002 } } }
\vdef{default-11:CsMcPU-APV0:muon2pt:loEffE}   {\ensuremath{{0.001 } } }
\vdef{default-11:CsMcPU-APV0:muon2pt:hiEff}   {\ensuremath{{0.998 } } }
\vdef{default-11:CsMcPU-APV0:muon2pt:hiEffE}   {\ensuremath{{0.001 } } }
\vdef{default-11:CsMcPU-APV1:muon2pt:loEff}   {\ensuremath{{0.002 } } }
\vdef{default-11:CsMcPU-APV1:muon2pt:loEffE}   {\ensuremath{{0.001 } } }
\vdef{default-11:CsMcPU-APV1:muon2pt:hiEff}   {\ensuremath{{0.998 } } }
\vdef{default-11:CsMcPU-APV1:muon2pt:hiEffE}   {\ensuremath{{0.001 } } }
\vdef{default-11:CsMcPU-APV1:muon2pt:loDelta}   {\ensuremath{{+0.332 } } }
\vdef{default-11:CsMcPU-APV1:muon2pt:loDeltaE}   {\ensuremath{{0.636 } } }
\vdef{default-11:CsMcPU-APV1:muon2pt:hiDelta}   {\ensuremath{{-0.001 } } }
\vdef{default-11:CsMcPU-APV1:muon2pt:hiDeltaE}   {\ensuremath{{0.001 } } }
\vdef{default-11:CsMcPU-APV0:muonseta:loEff}   {\ensuremath{{0.851 } } }
\vdef{default-11:CsMcPU-APV0:muonseta:loEffE}   {\ensuremath{{0.005 } } }
\vdef{default-11:CsMcPU-APV0:muonseta:hiEff}   {\ensuremath{{0.149 } } }
\vdef{default-11:CsMcPU-APV0:muonseta:hiEffE}   {\ensuremath{{0.005 } } }
\vdef{default-11:CsMcPU-APV1:muonseta:loEff}   {\ensuremath{{0.851 } } }
\vdef{default-11:CsMcPU-APV1:muonseta:loEffE}   {\ensuremath{{0.005 } } }
\vdef{default-11:CsMcPU-APV1:muonseta:hiEff}   {\ensuremath{{0.149 } } }
\vdef{default-11:CsMcPU-APV1:muonseta:hiEffE}   {\ensuremath{{0.005 } } }
\vdef{default-11:CsMcPU-APV1:muonseta:loDelta}   {\ensuremath{{-0.001 } } }
\vdef{default-11:CsMcPU-APV1:muonseta:loDeltaE}   {\ensuremath{{0.008 } } }
\vdef{default-11:CsMcPU-APV1:muonseta:hiDelta}   {\ensuremath{{+0.004 } } }
\vdef{default-11:CsMcPU-APV1:muonseta:hiDeltaE}   {\ensuremath{{0.048 } } }
\vdef{default-11:CsMcPU-APV0:pt:loEff}   {\ensuremath{{0.000 } } }
\vdef{default-11:CsMcPU-APV0:pt:loEffE}   {\ensuremath{{0.000 } } }
\vdef{default-11:CsMcPU-APV0:pt:hiEff}   {\ensuremath{{1.000 } } }
\vdef{default-11:CsMcPU-APV0:pt:hiEffE}   {\ensuremath{{0.000 } } }
\vdef{default-11:CsMcPU-APV1:pt:loEff}   {\ensuremath{{0.000 } } }
\vdef{default-11:CsMcPU-APV1:pt:loEffE}   {\ensuremath{{0.000 } } }
\vdef{default-11:CsMcPU-APV1:pt:hiEff}   {\ensuremath{{1.000 } } }
\vdef{default-11:CsMcPU-APV1:pt:hiEffE}   {\ensuremath{{0.000 } } }
\vdef{default-11:CsMcPU-APV1:pt:loDelta}   {\ensuremath{{\mathrm{NaN} } } }
\vdef{default-11:CsMcPU-APV1:pt:loDeltaE}   {\ensuremath{{\mathrm{NaN} } } }
\vdef{default-11:CsMcPU-APV1:pt:hiDelta}   {\ensuremath{{+0.000 } } }
\vdef{default-11:CsMcPU-APV1:pt:hiDeltaE}   {\ensuremath{{0.000 } } }
\vdef{default-11:CsMcPU-APV0:p:loEff}   {\ensuremath{{1.008 } } }
\vdef{default-11:CsMcPU-APV0:p:loEffE}   {\ensuremath{{\mathrm{NaN} } } }
\vdef{default-11:CsMcPU-APV0:p:hiEff}   {\ensuremath{{1.000 } } }
\vdef{default-11:CsMcPU-APV0:p:hiEffE}   {\ensuremath{{0.000 } } }
\vdef{default-11:CsMcPU-APV1:p:loEff}   {\ensuremath{{1.006 } } }
\vdef{default-11:CsMcPU-APV1:p:loEffE}   {\ensuremath{{\mathrm{NaN} } } }
\vdef{default-11:CsMcPU-APV1:p:hiEff}   {\ensuremath{{1.000 } } }
\vdef{default-11:CsMcPU-APV1:p:hiEffE}   {\ensuremath{{0.000 } } }
\vdef{default-11:CsMcPU-APV1:p:loDelta}   {\ensuremath{{+0.002 } } }
\vdef{default-11:CsMcPU-APV1:p:loDeltaE}   {\ensuremath{{\mathrm{NaN} } } }
\vdef{default-11:CsMcPU-APV1:p:hiDelta}   {\ensuremath{{+0.000 } } }
\vdef{default-11:CsMcPU-APV1:p:hiDeltaE}   {\ensuremath{{0.001 } } }
\vdef{default-11:CsMcPU-APV0:eta:loEff}   {\ensuremath{{0.855 } } }
\vdef{default-11:CsMcPU-APV0:eta:loEffE}   {\ensuremath{{0.007 } } }
\vdef{default-11:CsMcPU-APV0:eta:hiEff}   {\ensuremath{{0.145 } } }
\vdef{default-11:CsMcPU-APV0:eta:hiEffE}   {\ensuremath{{0.007 } } }
\vdef{default-11:CsMcPU-APV1:eta:loEff}   {\ensuremath{{0.847 } } }
\vdef{default-11:CsMcPU-APV1:eta:loEffE}   {\ensuremath{{0.007 } } }
\vdef{default-11:CsMcPU-APV1:eta:hiEff}   {\ensuremath{{0.153 } } }
\vdef{default-11:CsMcPU-APV1:eta:hiEffE}   {\ensuremath{{0.007 } } }
\vdef{default-11:CsMcPU-APV1:eta:loDelta}   {\ensuremath{{+0.009 } } }
\vdef{default-11:CsMcPU-APV1:eta:loDeltaE}   {\ensuremath{{0.012 } } }
\vdef{default-11:CsMcPU-APV1:eta:hiDelta}   {\ensuremath{{-0.050 } } }
\vdef{default-11:CsMcPU-APV1:eta:hiDeltaE}   {\ensuremath{{0.068 } } }
\vdef{default-11:CsMcPU-APV0:bdt:loEff}   {\ensuremath{{0.873 } } }
\vdef{default-11:CsMcPU-APV0:bdt:loEffE}   {\ensuremath{{0.007 } } }
\vdef{default-11:CsMcPU-APV0:bdt:hiEff}   {\ensuremath{{0.127 } } }
\vdef{default-11:CsMcPU-APV0:bdt:hiEffE}   {\ensuremath{{0.007 } } }
\vdef{default-11:CsMcPU-APV1:bdt:loEff}   {\ensuremath{{0.857 } } }
\vdef{default-11:CsMcPU-APV1:bdt:loEffE}   {\ensuremath{{0.007 } } }
\vdef{default-11:CsMcPU-APV1:bdt:hiEff}   {\ensuremath{{0.143 } } }
\vdef{default-11:CsMcPU-APV1:bdt:hiEffE}   {\ensuremath{{0.007 } } }
\vdef{default-11:CsMcPU-APV1:bdt:loDelta}   {\ensuremath{{+0.018 } } }
\vdef{default-11:CsMcPU-APV1:bdt:loDeltaE}   {\ensuremath{{0.011 } } }
\vdef{default-11:CsMcPU-APV1:bdt:hiDelta}   {\ensuremath{{-0.117 } } }
\vdef{default-11:CsMcPU-APV1:bdt:hiDeltaE}   {\ensuremath{{0.070 } } }
\vdef{default-11:CsMcPU-APV0:fl3d:loEff}   {\ensuremath{{0.896 } } }
\vdef{default-11:CsMcPU-APV0:fl3d:loEffE}   {\ensuremath{{0.006 } } }
\vdef{default-11:CsMcPU-APV0:fl3d:hiEff}   {\ensuremath{{0.104 } } }
\vdef{default-11:CsMcPU-APV0:fl3d:hiEffE}   {\ensuremath{{0.006 } } }
\vdef{default-11:CsMcPU-APV1:fl3d:loEff}   {\ensuremath{{0.899 } } }
\vdef{default-11:CsMcPU-APV1:fl3d:loEffE}   {\ensuremath{{0.006 } } }
\vdef{default-11:CsMcPU-APV1:fl3d:hiEff}   {\ensuremath{{0.101 } } }
\vdef{default-11:CsMcPU-APV1:fl3d:hiEffE}   {\ensuremath{{0.005 } } }
\vdef{default-11:CsMcPU-APV1:fl3d:loDelta}   {\ensuremath{{-0.003 } } }
\vdef{default-11:CsMcPU-APV1:fl3d:loDeltaE}   {\ensuremath{{0.009 } } }
\vdef{default-11:CsMcPU-APV1:fl3d:hiDelta}   {\ensuremath{{+0.028 } } }
\vdef{default-11:CsMcPU-APV1:fl3d:hiDeltaE}   {\ensuremath{{0.076 } } }
\vdef{default-11:CsMcPU-APV0:fl3de:loEff}   {\ensuremath{{1.000 } } }
\vdef{default-11:CsMcPU-APV0:fl3de:loEffE}   {\ensuremath{{0.000 } } }
\vdef{default-11:CsMcPU-APV0:fl3de:hiEff}   {\ensuremath{{0.000 } } }
\vdef{default-11:CsMcPU-APV0:fl3de:hiEffE}   {\ensuremath{{0.000 } } }
\vdef{default-11:CsMcPU-APV1:fl3de:loEff}   {\ensuremath{{1.000 } } }
\vdef{default-11:CsMcPU-APV1:fl3de:loEffE}   {\ensuremath{{0.000 } } }
\vdef{default-11:CsMcPU-APV1:fl3de:hiEff}   {\ensuremath{{0.001 } } }
\vdef{default-11:CsMcPU-APV1:fl3de:hiEffE}   {\ensuremath{{0.001 } } }
\vdef{default-11:CsMcPU-APV1:fl3de:loDelta}   {\ensuremath{{+0.000 } } }
\vdef{default-11:CsMcPU-APV1:fl3de:loDeltaE}   {\ensuremath{{0.000 } } }
\vdef{default-11:CsMcPU-APV1:fl3de:hiDelta}   {\ensuremath{{-2.000 } } }
\vdef{default-11:CsMcPU-APV1:fl3de:hiDeltaE}   {\ensuremath{{1.932 } } }
\vdef{default-11:CsMcPU-APV0:fls3d:loEff}   {\ensuremath{{0.082 } } }
\vdef{default-11:CsMcPU-APV0:fls3d:loEffE}   {\ensuremath{{0.005 } } }
\vdef{default-11:CsMcPU-APV0:fls3d:hiEff}   {\ensuremath{{0.918 } } }
\vdef{default-11:CsMcPU-APV0:fls3d:hiEffE}   {\ensuremath{{0.005 } } }
\vdef{default-11:CsMcPU-APV1:fls3d:loEff}   {\ensuremath{{0.069 } } }
\vdef{default-11:CsMcPU-APV1:fls3d:loEffE}   {\ensuremath{{0.005 } } }
\vdef{default-11:CsMcPU-APV1:fls3d:hiEff}   {\ensuremath{{0.931 } } }
\vdef{default-11:CsMcPU-APV1:fls3d:hiEffE}   {\ensuremath{{0.005 } } }
\vdef{default-11:CsMcPU-APV1:fls3d:loDelta}   {\ensuremath{{+0.168 } } }
\vdef{default-11:CsMcPU-APV1:fls3d:loDeltaE}   {\ensuremath{{0.090 } } }
\vdef{default-11:CsMcPU-APV1:fls3d:hiDelta}   {\ensuremath{{-0.014 } } }
\vdef{default-11:CsMcPU-APV1:fls3d:hiDeltaE}   {\ensuremath{{0.007 } } }
\vdef{default-11:CsMcPU-APV0:flsxy:loEff}   {\ensuremath{{1.006 } } }
\vdef{default-11:CsMcPU-APV0:flsxy:loEffE}   {\ensuremath{{\mathrm{NaN} } } }
\vdef{default-11:CsMcPU-APV0:flsxy:hiEff}   {\ensuremath{{1.000 } } }
\vdef{default-11:CsMcPU-APV0:flsxy:hiEffE}   {\ensuremath{{0.000 } } }
\vdef{default-11:CsMcPU-APV1:flsxy:loEff}   {\ensuremath{{1.011 } } }
\vdef{default-11:CsMcPU-APV1:flsxy:loEffE}   {\ensuremath{{\mathrm{NaN} } } }
\vdef{default-11:CsMcPU-APV1:flsxy:hiEff}   {\ensuremath{{1.000 } } }
\vdef{default-11:CsMcPU-APV1:flsxy:hiEffE}   {\ensuremath{{0.000 } } }
\vdef{default-11:CsMcPU-APV1:flsxy:loDelta}   {\ensuremath{{-0.005 } } }
\vdef{default-11:CsMcPU-APV1:flsxy:loDeltaE}   {\ensuremath{{\mathrm{NaN} } } }
\vdef{default-11:CsMcPU-APV1:flsxy:hiDelta}   {\ensuremath{{+0.000 } } }
\vdef{default-11:CsMcPU-APV1:flsxy:hiDeltaE}   {\ensuremath{{0.000 } } }
\vdef{default-11:CsMcPU-APV0:chi2dof:loEff}   {\ensuremath{{0.954 } } }
\vdef{default-11:CsMcPU-APV0:chi2dof:loEffE}   {\ensuremath{{0.004 } } }
\vdef{default-11:CsMcPU-APV0:chi2dof:hiEff}   {\ensuremath{{0.046 } } }
\vdef{default-11:CsMcPU-APV0:chi2dof:hiEffE}   {\ensuremath{{0.004 } } }
\vdef{default-11:CsMcPU-APV1:chi2dof:loEff}   {\ensuremath{{0.942 } } }
\vdef{default-11:CsMcPU-APV1:chi2dof:loEffE}   {\ensuremath{{0.005 } } }
\vdef{default-11:CsMcPU-APV1:chi2dof:hiEff}   {\ensuremath{{0.058 } } }
\vdef{default-11:CsMcPU-APV1:chi2dof:hiEffE}   {\ensuremath{{0.005 } } }
\vdef{default-11:CsMcPU-APV1:chi2dof:loDelta}   {\ensuremath{{+0.012 } } }
\vdef{default-11:CsMcPU-APV1:chi2dof:loDeltaE}   {\ensuremath{{0.007 } } }
\vdef{default-11:CsMcPU-APV1:chi2dof:hiDelta}   {\ensuremath{{-0.219 } } }
\vdef{default-11:CsMcPU-APV1:chi2dof:hiDeltaE}   {\ensuremath{{0.118 } } }
\vdef{default-11:CsMcPU-APV0:pchi2dof:loEff}   {\ensuremath{{0.567 } } }
\vdef{default-11:CsMcPU-APV0:pchi2dof:loEffE}   {\ensuremath{{0.010 } } }
\vdef{default-11:CsMcPU-APV0:pchi2dof:hiEff}   {\ensuremath{{0.433 } } }
\vdef{default-11:CsMcPU-APV0:pchi2dof:hiEffE}   {\ensuremath{{0.010 } } }
\vdef{default-11:CsMcPU-APV1:pchi2dof:loEff}   {\ensuremath{{0.588 } } }
\vdef{default-11:CsMcPU-APV1:pchi2dof:loEffE}   {\ensuremath{{0.009 } } }
\vdef{default-11:CsMcPU-APV1:pchi2dof:hiEff}   {\ensuremath{{0.412 } } }
\vdef{default-11:CsMcPU-APV1:pchi2dof:hiEffE}   {\ensuremath{{0.009 } } }
\vdef{default-11:CsMcPU-APV1:pchi2dof:loDelta}   {\ensuremath{{-0.035 } } }
\vdef{default-11:CsMcPU-APV1:pchi2dof:loDeltaE}   {\ensuremath{{0.023 } } }
\vdef{default-11:CsMcPU-APV1:pchi2dof:hiDelta}   {\ensuremath{{+0.048 } } }
\vdef{default-11:CsMcPU-APV1:pchi2dof:hiDeltaE}   {\ensuremath{{0.032 } } }
\vdef{default-11:CsMcPU-APV0:alpha:loEff}   {\ensuremath{{0.992 } } }
\vdef{default-11:CsMcPU-APV0:alpha:loEffE}   {\ensuremath{{0.002 } } }
\vdef{default-11:CsMcPU-APV0:alpha:hiEff}   {\ensuremath{{0.008 } } }
\vdef{default-11:CsMcPU-APV0:alpha:hiEffE}   {\ensuremath{{0.002 } } }
\vdef{default-11:CsMcPU-APV1:alpha:loEff}   {\ensuremath{{0.990 } } }
\vdef{default-11:CsMcPU-APV1:alpha:loEffE}   {\ensuremath{{0.002 } } }
\vdef{default-11:CsMcPU-APV1:alpha:hiEff}   {\ensuremath{{0.010 } } }
\vdef{default-11:CsMcPU-APV1:alpha:hiEffE}   {\ensuremath{{0.002 } } }
\vdef{default-11:CsMcPU-APV1:alpha:loDelta}   {\ensuremath{{+0.002 } } }
\vdef{default-11:CsMcPU-APV1:alpha:loDeltaE}   {\ensuremath{{0.003 } } }
\vdef{default-11:CsMcPU-APV1:alpha:hiDelta}   {\ensuremath{{-0.199 } } }
\vdef{default-11:CsMcPU-APV1:alpha:hiDeltaE}   {\ensuremath{{0.305 } } }
\vdef{default-11:CsMcPU-APV0:iso:loEff}   {\ensuremath{{0.098 } } }
\vdef{default-11:CsMcPU-APV0:iso:loEffE}   {\ensuremath{{0.006 } } }
\vdef{default-11:CsMcPU-APV0:iso:hiEff}   {\ensuremath{{0.902 } } }
\vdef{default-11:CsMcPU-APV0:iso:hiEffE}   {\ensuremath{{0.006 } } }
\vdef{default-11:CsMcPU-APV1:iso:loEff}   {\ensuremath{{0.104 } } }
\vdef{default-11:CsMcPU-APV1:iso:loEffE}   {\ensuremath{{0.006 } } }
\vdef{default-11:CsMcPU-APV1:iso:hiEff}   {\ensuremath{{0.896 } } }
\vdef{default-11:CsMcPU-APV1:iso:hiEffE}   {\ensuremath{{0.006 } } }
\vdef{default-11:CsMcPU-APV1:iso:loDelta}   {\ensuremath{{-0.061 } } }
\vdef{default-11:CsMcPU-APV1:iso:loDeltaE}   {\ensuremath{{0.081 } } }
\vdef{default-11:CsMcPU-APV1:iso:hiDelta}   {\ensuremath{{+0.007 } } }
\vdef{default-11:CsMcPU-APV1:iso:hiDeltaE}   {\ensuremath{{0.009 } } }
\vdef{default-11:CsMcPU-APV0:docatrk:loEff}   {\ensuremath{{0.082 } } }
\vdef{default-11:CsMcPU-APV0:docatrk:loEffE}   {\ensuremath{{0.005 } } }
\vdef{default-11:CsMcPU-APV0:docatrk:hiEff}   {\ensuremath{{0.918 } } }
\vdef{default-11:CsMcPU-APV0:docatrk:hiEffE}   {\ensuremath{{0.005 } } }
\vdef{default-11:CsMcPU-APV1:docatrk:loEff}   {\ensuremath{{0.092 } } }
\vdef{default-11:CsMcPU-APV1:docatrk:loEffE}   {\ensuremath{{0.006 } } }
\vdef{default-11:CsMcPU-APV1:docatrk:hiEff}   {\ensuremath{{0.908 } } }
\vdef{default-11:CsMcPU-APV1:docatrk:hiEffE}   {\ensuremath{{0.006 } } }
\vdef{default-11:CsMcPU-APV1:docatrk:loDelta}   {\ensuremath{{-0.107 } } }
\vdef{default-11:CsMcPU-APV1:docatrk:loDeltaE}   {\ensuremath{{0.091 } } }
\vdef{default-11:CsMcPU-APV1:docatrk:hiDelta}   {\ensuremath{{+0.010 } } }
\vdef{default-11:CsMcPU-APV1:docatrk:hiDeltaE}   {\ensuremath{{0.009 } } }
\vdef{default-11:CsMcPU-APV0:isotrk:loEff}   {\ensuremath{{1.000 } } }
\vdef{default-11:CsMcPU-APV0:isotrk:loEffE}   {\ensuremath{{0.000 } } }
\vdef{default-11:CsMcPU-APV0:isotrk:hiEff}   {\ensuremath{{1.000 } } }
\vdef{default-11:CsMcPU-APV0:isotrk:hiEffE}   {\ensuremath{{0.000 } } }
\vdef{default-11:CsMcPU-APV1:isotrk:loEff}   {\ensuremath{{1.000 } } }
\vdef{default-11:CsMcPU-APV1:isotrk:loEffE}   {\ensuremath{{0.000 } } }
\vdef{default-11:CsMcPU-APV1:isotrk:hiEff}   {\ensuremath{{1.000 } } }
\vdef{default-11:CsMcPU-APV1:isotrk:hiEffE}   {\ensuremath{{0.000 } } }
\vdef{default-11:CsMcPU-APV1:isotrk:loDelta}   {\ensuremath{{+0.000 } } }
\vdef{default-11:CsMcPU-APV1:isotrk:loDeltaE}   {\ensuremath{{0.001 } } }
\vdef{default-11:CsMcPU-APV1:isotrk:hiDelta}   {\ensuremath{{+0.000 } } }
\vdef{default-11:CsMcPU-APV1:isotrk:hiDeltaE}   {\ensuremath{{0.001 } } }
\vdef{default-11:CsMcPU-APV0:closetrk:loEff}   {\ensuremath{{0.978 } } }
\vdef{default-11:CsMcPU-APV0:closetrk:loEffE}   {\ensuremath{{0.003 } } }
\vdef{default-11:CsMcPU-APV0:closetrk:hiEff}   {\ensuremath{{0.022 } } }
\vdef{default-11:CsMcPU-APV0:closetrk:hiEffE}   {\ensuremath{{0.003 } } }
\vdef{default-11:CsMcPU-APV1:closetrk:loEff}   {\ensuremath{{0.972 } } }
\vdef{default-11:CsMcPU-APV1:closetrk:loEffE}   {\ensuremath{{0.003 } } }
\vdef{default-11:CsMcPU-APV1:closetrk:hiEff}   {\ensuremath{{0.028 } } }
\vdef{default-11:CsMcPU-APV1:closetrk:hiEffE}   {\ensuremath{{0.003 } } }
\vdef{default-11:CsMcPU-APV1:closetrk:loDelta}   {\ensuremath{{+0.006 } } }
\vdef{default-11:CsMcPU-APV1:closetrk:loDeltaE}   {\ensuremath{{0.005 } } }
\vdef{default-11:CsMcPU-APV1:closetrk:hiDelta}   {\ensuremath{{-0.243 } } }
\vdef{default-11:CsMcPU-APV1:closetrk:hiDeltaE}   {\ensuremath{{0.176 } } }
\vdef{default-11:CsMcPU-APV0:lip:loEff}   {\ensuremath{{1.000 } } }
\vdef{default-11:CsMcPU-APV0:lip:loEffE}   {\ensuremath{{0.000 } } }
\vdef{default-11:CsMcPU-APV0:lip:hiEff}   {\ensuremath{{0.000 } } }
\vdef{default-11:CsMcPU-APV0:lip:hiEffE}   {\ensuremath{{0.000 } } }
\vdef{default-11:CsMcPU-APV1:lip:loEff}   {\ensuremath{{1.000 } } }
\vdef{default-11:CsMcPU-APV1:lip:loEffE}   {\ensuremath{{0.000 } } }
\vdef{default-11:CsMcPU-APV1:lip:hiEff}   {\ensuremath{{0.000 } } }
\vdef{default-11:CsMcPU-APV1:lip:hiEffE}   {\ensuremath{{0.000 } } }
\vdef{default-11:CsMcPU-APV1:lip:loDelta}   {\ensuremath{{+0.000 } } }
\vdef{default-11:CsMcPU-APV1:lip:loDeltaE}   {\ensuremath{{0.001 } } }
\vdef{default-11:CsMcPU-APV1:lip:hiDelta}   {\ensuremath{{\mathrm{NaN} } } }
\vdef{default-11:CsMcPU-APV1:lip:hiDeltaE}   {\ensuremath{{\mathrm{NaN} } } }
\vdef{default-11:CsMcPU-APV0:lip:inEff}   {\ensuremath{{1.000 } } }
\vdef{default-11:CsMcPU-APV0:lip:inEffE}   {\ensuremath{{0.000 } } }
\vdef{default-11:CsMcPU-APV1:lip:inEff}   {\ensuremath{{1.000 } } }
\vdef{default-11:CsMcPU-APV1:lip:inEffE}   {\ensuremath{{0.000 } } }
\vdef{default-11:CsMcPU-APV1:lip:inDelta}   {\ensuremath{{+0.000 } } }
\vdef{default-11:CsMcPU-APV1:lip:inDeltaE}   {\ensuremath{{0.001 } } }
\vdef{default-11:CsMcPU-APV0:lips:loEff}   {\ensuremath{{1.000 } } }
\vdef{default-11:CsMcPU-APV0:lips:loEffE}   {\ensuremath{{0.000 } } }
\vdef{default-11:CsMcPU-APV0:lips:hiEff}   {\ensuremath{{0.000 } } }
\vdef{default-11:CsMcPU-APV0:lips:hiEffE}   {\ensuremath{{0.000 } } }
\vdef{default-11:CsMcPU-APV1:lips:loEff}   {\ensuremath{{1.000 } } }
\vdef{default-11:CsMcPU-APV1:lips:loEffE}   {\ensuremath{{0.000 } } }
\vdef{default-11:CsMcPU-APV1:lips:hiEff}   {\ensuremath{{0.000 } } }
\vdef{default-11:CsMcPU-APV1:lips:hiEffE}   {\ensuremath{{0.000 } } }
\vdef{default-11:CsMcPU-APV1:lips:loDelta}   {\ensuremath{{+0.000 } } }
\vdef{default-11:CsMcPU-APV1:lips:loDeltaE}   {\ensuremath{{0.001 } } }
\vdef{default-11:CsMcPU-APV1:lips:hiDelta}   {\ensuremath{{\mathrm{NaN} } } }
\vdef{default-11:CsMcPU-APV1:lips:hiDeltaE}   {\ensuremath{{\mathrm{NaN} } } }
\vdef{default-11:CsMcPU-APV0:lips:inEff}   {\ensuremath{{1.000 } } }
\vdef{default-11:CsMcPU-APV0:lips:inEffE}   {\ensuremath{{0.000 } } }
\vdef{default-11:CsMcPU-APV1:lips:inEff}   {\ensuremath{{1.000 } } }
\vdef{default-11:CsMcPU-APV1:lips:inEffE}   {\ensuremath{{0.000 } } }
\vdef{default-11:CsMcPU-APV1:lips:inDelta}   {\ensuremath{{+0.000 } } }
\vdef{default-11:CsMcPU-APV1:lips:inDeltaE}   {\ensuremath{{0.001 } } }
\vdef{default-11:CsMcPU-APV0:ip:loEff}   {\ensuremath{{0.974 } } }
\vdef{default-11:CsMcPU-APV0:ip:loEffE}   {\ensuremath{{0.003 } } }
\vdef{default-11:CsMcPU-APV0:ip:hiEff}   {\ensuremath{{0.026 } } }
\vdef{default-11:CsMcPU-APV0:ip:hiEffE}   {\ensuremath{{0.003 } } }
\vdef{default-11:CsMcPU-APV1:ip:loEff}   {\ensuremath{{0.964 } } }
\vdef{default-11:CsMcPU-APV1:ip:loEffE}   {\ensuremath{{0.004 } } }
\vdef{default-11:CsMcPU-APV1:ip:hiEff}   {\ensuremath{{0.036 } } }
\vdef{default-11:CsMcPU-APV1:ip:hiEffE}   {\ensuremath{{0.004 } } }
\vdef{default-11:CsMcPU-APV1:ip:loDelta}   {\ensuremath{{+0.010 } } }
\vdef{default-11:CsMcPU-APV1:ip:loDeltaE}   {\ensuremath{{0.005 } } }
\vdef{default-11:CsMcPU-APV1:ip:hiDelta}   {\ensuremath{{-0.301 } } }
\vdef{default-11:CsMcPU-APV1:ip:hiDeltaE}   {\ensuremath{{0.156 } } }
\vdef{default-11:CsMcPU-APV0:ips:loEff}   {\ensuremath{{0.931 } } }
\vdef{default-11:CsMcPU-APV0:ips:loEffE}   {\ensuremath{{0.005 } } }
\vdef{default-11:CsMcPU-APV0:ips:hiEff}   {\ensuremath{{0.069 } } }
\vdef{default-11:CsMcPU-APV0:ips:hiEffE}   {\ensuremath{{0.005 } } }
\vdef{default-11:CsMcPU-APV1:ips:loEff}   {\ensuremath{{0.928 } } }
\vdef{default-11:CsMcPU-APV1:ips:loEffE}   {\ensuremath{{0.005 } } }
\vdef{default-11:CsMcPU-APV1:ips:hiEff}   {\ensuremath{{0.072 } } }
\vdef{default-11:CsMcPU-APV1:ips:hiEffE}   {\ensuremath{{0.005 } } }
\vdef{default-11:CsMcPU-APV1:ips:loDelta}   {\ensuremath{{+0.003 } } }
\vdef{default-11:CsMcPU-APV1:ips:loDeltaE}   {\ensuremath{{0.008 } } }
\vdef{default-11:CsMcPU-APV1:ips:hiDelta}   {\ensuremath{{-0.044 } } }
\vdef{default-11:CsMcPU-APV1:ips:hiDeltaE}   {\ensuremath{{0.100 } } }
\vdef{default-11:CsMcPU-APV0:maxdoca:loEff}   {\ensuremath{{1.000 } } }
\vdef{default-11:CsMcPU-APV0:maxdoca:loEffE}   {\ensuremath{{0.000 } } }
\vdef{default-11:CsMcPU-APV0:maxdoca:hiEff}   {\ensuremath{{0.024 } } }
\vdef{default-11:CsMcPU-APV0:maxdoca:hiEffE}   {\ensuremath{{0.003 } } }
\vdef{default-11:CsMcPU-APV1:maxdoca:loEff}   {\ensuremath{{1.000 } } }
\vdef{default-11:CsMcPU-APV1:maxdoca:loEffE}   {\ensuremath{{0.000 } } }
\vdef{default-11:CsMcPU-APV1:maxdoca:hiEff}   {\ensuremath{{0.017 } } }
\vdef{default-11:CsMcPU-APV1:maxdoca:hiEffE}   {\ensuremath{{0.003 } } }
\vdef{default-11:CsMcPU-APV1:maxdoca:loDelta}   {\ensuremath{{+0.000 } } }
\vdef{default-11:CsMcPU-APV1:maxdoca:loDeltaE}   {\ensuremath{{0.001 } } }
\vdef{default-11:CsMcPU-APV1:maxdoca:hiDelta}   {\ensuremath{{+0.341 } } }
\vdef{default-11:CsMcPU-APV1:maxdoca:hiDeltaE}   {\ensuremath{{0.196 } } }
\vdef{default-11:CsMcPU-APV0:kaonspt:loEff}   {\ensuremath{{1.000 } } }
\vdef{default-11:CsMcPU-APV0:kaonspt:loEffE}   {\ensuremath{{0.000 } } }
\vdef{default-11:CsMcPU-APV0:kaonspt:hiEff}   {\ensuremath{{1.000 } } }
\vdef{default-11:CsMcPU-APV0:kaonspt:hiEffE}   {\ensuremath{{0.000 } } }
\vdef{default-11:CsMcPU-APV1:kaonspt:loEff}   {\ensuremath{{1.001 } } }
\vdef{default-11:CsMcPU-APV1:kaonspt:loEffE}   {\ensuremath{{\mathrm{NaN} } } }
\vdef{default-11:CsMcPU-APV1:kaonspt:hiEff}   {\ensuremath{{1.000 } } }
\vdef{default-11:CsMcPU-APV1:kaonspt:hiEffE}   {\ensuremath{{0.000 } } }
\vdef{default-11:CsMcPU-APV1:kaonspt:loDelta}   {\ensuremath{{-0.001 } } }
\vdef{default-11:CsMcPU-APV1:kaonspt:loDeltaE}   {\ensuremath{{\mathrm{NaN} } } }
\vdef{default-11:CsMcPU-APV1:kaonspt:hiDelta}   {\ensuremath{{+0.000 } } }
\vdef{default-11:CsMcPU-APV1:kaonspt:hiDeltaE}   {\ensuremath{{0.000 } } }
\vdef{default-11:CsMcPU-APV0:psipt:loEff}   {\ensuremath{{1.006 } } }
\vdef{default-11:CsMcPU-APV0:psipt:loEffE}   {\ensuremath{{\mathrm{NaN} } } }
\vdef{default-11:CsMcPU-APV0:psipt:hiEff}   {\ensuremath{{1.000 } } }
\vdef{default-11:CsMcPU-APV0:psipt:hiEffE}   {\ensuremath{{0.000 } } }
\vdef{default-11:CsMcPU-APV1:psipt:loEff}   {\ensuremath{{1.005 } } }
\vdef{default-11:CsMcPU-APV1:psipt:loEffE}   {\ensuremath{{\mathrm{NaN} } } }
\vdef{default-11:CsMcPU-APV1:psipt:hiEff}   {\ensuremath{{1.000 } } }
\vdef{default-11:CsMcPU-APV1:psipt:hiEffE}   {\ensuremath{{0.000 } } }
\vdef{default-11:CsMcPU-APV1:psipt:loDelta}   {\ensuremath{{+0.001 } } }
\vdef{default-11:CsMcPU-APV1:psipt:loDeltaE}   {\ensuremath{{\mathrm{NaN} } } }
\vdef{default-11:CsMcPU-APV1:psipt:hiDelta}   {\ensuremath{{+0.000 } } }
\vdef{default-11:CsMcPU-APV1:psipt:hiDeltaE}   {\ensuremath{{0.001 } } }
\vdef{default-11:CsMcPU-APV0:phipt:loEff}   {\ensuremath{{1.008 } } }
\vdef{default-11:CsMcPU-APV0:phipt:loEffE}   {\ensuremath{{\mathrm{NaN} } } }
\vdef{default-11:CsMcPU-APV0:phipt:hiEff}   {\ensuremath{{1.000 } } }
\vdef{default-11:CsMcPU-APV0:phipt:hiEffE}   {\ensuremath{{0.000 } } }
\vdef{default-11:CsMcPU-APV1:phipt:loEff}   {\ensuremath{{1.011 } } }
\vdef{default-11:CsMcPU-APV1:phipt:loEffE}   {\ensuremath{{\mathrm{NaN} } } }
\vdef{default-11:CsMcPU-APV1:phipt:hiEff}   {\ensuremath{{1.000 } } }
\vdef{default-11:CsMcPU-APV1:phipt:hiEffE}   {\ensuremath{{0.000 } } }
\vdef{default-11:CsMcPU-APV1:phipt:loDelta}   {\ensuremath{{-0.003 } } }
\vdef{default-11:CsMcPU-APV1:phipt:loDeltaE}   {\ensuremath{{\mathrm{NaN} } } }
\vdef{default-11:CsMcPU-APV1:phipt:hiDelta}   {\ensuremath{{+0.000 } } }
\vdef{default-11:CsMcPU-APV1:phipt:hiDeltaE}   {\ensuremath{{0.001 } } }
\vdef{default-11:CsMcPU-APV0:deltar:loEff}   {\ensuremath{{0.999 } } }
\vdef{default-11:CsMcPU-APV0:deltar:loEffE}   {\ensuremath{{0.001 } } }
\vdef{default-11:CsMcPU-APV0:deltar:hiEff}   {\ensuremath{{0.001 } } }
\vdef{default-11:CsMcPU-APV0:deltar:hiEffE}   {\ensuremath{{0.001 } } }
\vdef{default-11:CsMcPU-APV1:deltar:loEff}   {\ensuremath{{0.999 } } }
\vdef{default-11:CsMcPU-APV1:deltar:loEffE}   {\ensuremath{{0.001 } } }
\vdef{default-11:CsMcPU-APV1:deltar:hiEff}   {\ensuremath{{0.001 } } }
\vdef{default-11:CsMcPU-APV1:deltar:hiEffE}   {\ensuremath{{0.001 } } }
\vdef{default-11:CsMcPU-APV1:deltar:loDelta}   {\ensuremath{{-0.000 } } }
\vdef{default-11:CsMcPU-APV1:deltar:loDeltaE}   {\ensuremath{{0.001 } } }
\vdef{default-11:CsMcPU-APV1:deltar:hiDelta}   {\ensuremath{{+0.336 } } }
\vdef{default-11:CsMcPU-APV1:deltar:hiDeltaE}   {\ensuremath{{0.945 } } }
\vdef{default-11:CsMcPU-APV0:mkk:loEff}   {\ensuremath{{1.000 } } }
\vdef{default-11:CsMcPU-APV0:mkk:loEffE}   {\ensuremath{{0.000 } } }
\vdef{default-11:CsMcPU-APV0:mkk:hiEff}   {\ensuremath{{1.000 } } }
\vdef{default-11:CsMcPU-APV0:mkk:hiEffE}   {\ensuremath{{0.000 } } }
\vdef{default-11:CsMcPU-APV1:mkk:loEff}   {\ensuremath{{1.000 } } }
\vdef{default-11:CsMcPU-APV1:mkk:loEffE}   {\ensuremath{{0.000 } } }
\vdef{default-11:CsMcPU-APV1:mkk:hiEff}   {\ensuremath{{1.000 } } }
\vdef{default-11:CsMcPU-APV1:mkk:hiEffE}   {\ensuremath{{0.000 } } }
\vdef{default-11:CsMcPU-APV1:mkk:loDelta}   {\ensuremath{{+0.000 } } }
\vdef{default-11:CsMcPU-APV1:mkk:loDeltaE}   {\ensuremath{{0.000 } } }
\vdef{default-11:CsMcPU-APV1:mkk:hiDelta}   {\ensuremath{{+0.000 } } }
\vdef{default-11:CsMcPU-APV1:mkk:hiDeltaE}   {\ensuremath{{0.000 } } }
\vdef{default-11:CsData-B:osiso:loEff}   {\ensuremath{{1.006 } } }
\vdef{default-11:CsData-B:osiso:loEffE}   {\ensuremath{{\mathrm{NaN} } } }
\vdef{default-11:CsData-B:osiso:hiEff}   {\ensuremath{{1.000 } } }
\vdef{default-11:CsData-B:osiso:hiEffE}   {\ensuremath{{0.000 } } }
\vdef{default-11:CsMc-B:osiso:loEff}   {\ensuremath{{1.004 } } }
\vdef{default-11:CsMc-B:osiso:loEffE}   {\ensuremath{{\mathrm{NaN} } } }
\vdef{default-11:CsMc-B:osiso:hiEff}   {\ensuremath{{1.000 } } }
\vdef{default-11:CsMc-B:osiso:hiEffE}   {\ensuremath{{0.000 } } }
\vdef{default-11:CsMc-B:osiso:loDelta}   {\ensuremath{{+0.001 } } }
\vdef{default-11:CsMc-B:osiso:loDeltaE}   {\ensuremath{{\mathrm{NaN} } } }
\vdef{default-11:CsMc-B:osiso:hiDelta}   {\ensuremath{{+0.000 } } }
\vdef{default-11:CsMc-B:osiso:hiDeltaE}   {\ensuremath{{0.000 } } }
\vdef{default-11:CsData-B:osreliso:loEff}   {\ensuremath{{0.278 } } }
\vdef{default-11:CsData-B:osreliso:loEffE}   {\ensuremath{{0.005 } } }
\vdef{default-11:CsData-B:osreliso:hiEff}   {\ensuremath{{0.722 } } }
\vdef{default-11:CsData-B:osreliso:hiEffE}   {\ensuremath{{0.005 } } }
\vdef{default-11:CsMc-B:osreliso:loEff}   {\ensuremath{{0.301 } } }
\vdef{default-11:CsMc-B:osreliso:loEffE}   {\ensuremath{{0.005 } } }
\vdef{default-11:CsMc-B:osreliso:hiEff}   {\ensuremath{{0.699 } } }
\vdef{default-11:CsMc-B:osreliso:hiEffE}   {\ensuremath{{0.005 } } }
\vdef{default-11:CsMc-B:osreliso:loDelta}   {\ensuremath{{-0.081 } } }
\vdef{default-11:CsMc-B:osreliso:loDeltaE}   {\ensuremath{{0.026 } } }
\vdef{default-11:CsMc-B:osreliso:hiDelta}   {\ensuremath{{+0.033 } } }
\vdef{default-11:CsMc-B:osreliso:hiDeltaE}   {\ensuremath{{0.011 } } }
\vdef{default-11:CsData-B:osmuonpt:loEff}   {\ensuremath{{0.000 } } }
\vdef{default-11:CsData-B:osmuonpt:loEffE}   {\ensuremath{{0.004 } } }
\vdef{default-11:CsData-B:osmuonpt:hiEff}   {\ensuremath{{1.000 } } }
\vdef{default-11:CsData-B:osmuonpt:hiEffE}   {\ensuremath{{0.004 } } }
\vdef{default-11:CsMc-B:osmuonpt:loEff}   {\ensuremath{{0.000 } } }
\vdef{default-11:CsMc-B:osmuonpt:loEffE}   {\ensuremath{{0.004 } } }
\vdef{default-11:CsMc-B:osmuonpt:hiEff}   {\ensuremath{{1.000 } } }
\vdef{default-11:CsMc-B:osmuonpt:hiEffE}   {\ensuremath{{0.004 } } }
\vdef{default-11:CsMc-B:osmuonpt:loDelta}   {\ensuremath{{\mathrm{NaN} } } }
\vdef{default-11:CsMc-B:osmuonpt:loDeltaE}   {\ensuremath{{\mathrm{NaN} } } }
\vdef{default-11:CsMc-B:osmuonpt:hiDelta}   {\ensuremath{{+0.000 } } }
\vdef{default-11:CsMc-B:osmuonpt:hiDeltaE}   {\ensuremath{{0.005 } } }
\vdef{default-11:CsData-B:osmuondr:loEff}   {\ensuremath{{0.008 } } }
\vdef{default-11:CsData-B:osmuondr:loEffE}   {\ensuremath{{0.006 } } }
\vdef{default-11:CsData-B:osmuondr:hiEff}   {\ensuremath{{0.992 } } }
\vdef{default-11:CsData-B:osmuondr:hiEffE}   {\ensuremath{{0.006 } } }
\vdef{default-11:CsMc-B:osmuondr:loEff}   {\ensuremath{{0.013 } } }
\vdef{default-11:CsMc-B:osmuondr:loEffE}   {\ensuremath{{0.007 } } }
\vdef{default-11:CsMc-B:osmuondr:hiEff}   {\ensuremath{{0.987 } } }
\vdef{default-11:CsMc-B:osmuondr:hiEffE}   {\ensuremath{{0.007 } } }
\vdef{default-11:CsMc-B:osmuondr:loDelta}   {\ensuremath{{-0.464 } } }
\vdef{default-11:CsMc-B:osmuondr:loDeltaE}   {\ensuremath{{0.883 } } }
\vdef{default-11:CsMc-B:osmuondr:hiDelta}   {\ensuremath{{+0.005 } } }
\vdef{default-11:CsMc-B:osmuondr:hiDeltaE}   {\ensuremath{{0.010 } } }
\vdef{default-11:CsData-B:hlt:loEff}   {\ensuremath{{0.072 } } }
\vdef{default-11:CsData-B:hlt:loEffE}   {\ensuremath{{0.003 } } }
\vdef{default-11:CsData-B:hlt:hiEff}   {\ensuremath{{0.928 } } }
\vdef{default-11:CsData-B:hlt:hiEffE}   {\ensuremath{{0.003 } } }
\vdef{default-11:CsMc-B:hlt:loEff}   {\ensuremath{{0.238 } } }
\vdef{default-11:CsMc-B:hlt:loEffE}   {\ensuremath{{0.005 } } }
\vdef{default-11:CsMc-B:hlt:hiEff}   {\ensuremath{{0.762 } } }
\vdef{default-11:CsMc-B:hlt:hiEffE}   {\ensuremath{{0.005 } } }
\vdef{default-11:CsMc-B:hlt:loDelta}   {\ensuremath{{-1.069 } } }
\vdef{default-11:CsMc-B:hlt:loDeltaE}   {\ensuremath{{0.035 } } }
\vdef{default-11:CsMc-B:hlt:hiDelta}   {\ensuremath{{+0.196 } } }
\vdef{default-11:CsMc-B:hlt:hiDeltaE}   {\ensuremath{{0.007 } } }
\vdef{default-11:CsData-B:muonsid:loEff}   {\ensuremath{{0.151 } } }
\vdef{default-11:CsData-B:muonsid:loEffE}   {\ensuremath{{0.004 } } }
\vdef{default-11:CsData-B:muonsid:hiEff}   {\ensuremath{{0.849 } } }
\vdef{default-11:CsData-B:muonsid:hiEffE}   {\ensuremath{{0.004 } } }
\vdef{default-11:CsMc-B:muonsid:loEff}   {\ensuremath{{0.149 } } }
\vdef{default-11:CsMc-B:muonsid:loEffE}   {\ensuremath{{0.004 } } }
\vdef{default-11:CsMc-B:muonsid:hiEff}   {\ensuremath{{0.851 } } }
\vdef{default-11:CsMc-B:muonsid:hiEffE}   {\ensuremath{{0.004 } } }
\vdef{default-11:CsMc-B:muonsid:loDelta}   {\ensuremath{{+0.010 } } }
\vdef{default-11:CsMc-B:muonsid:loDeltaE}   {\ensuremath{{0.039 } } }
\vdef{default-11:CsMc-B:muonsid:hiDelta}   {\ensuremath{{-0.002 } } }
\vdef{default-11:CsMc-B:muonsid:hiDeltaE}   {\ensuremath{{0.007 } } }
\vdef{default-11:CsData-B:tracksqual:loEff}   {\ensuremath{{0.001 } } }
\vdef{default-11:CsData-B:tracksqual:loEffE}   {\ensuremath{{0.000 } } }
\vdef{default-11:CsData-B:tracksqual:hiEff}   {\ensuremath{{0.999 } } }
\vdef{default-11:CsData-B:tracksqual:hiEffE}   {\ensuremath{{0.000 } } }
\vdef{default-11:CsMc-B:tracksqual:loEff}   {\ensuremath{{0.000 } } }
\vdef{default-11:CsMc-B:tracksqual:loEffE}   {\ensuremath{{0.000 } } }
\vdef{default-11:CsMc-B:tracksqual:hiEff}   {\ensuremath{{1.000 } } }
\vdef{default-11:CsMc-B:tracksqual:hiEffE}   {\ensuremath{{0.000 } } }
\vdef{default-11:CsMc-B:tracksqual:loDelta}   {\ensuremath{{+1.104 } } }
\vdef{default-11:CsMc-B:tracksqual:loDeltaE}   {\ensuremath{{0.854 } } }
\vdef{default-11:CsMc-B:tracksqual:hiDelta}   {\ensuremath{{-0.000 } } }
\vdef{default-11:CsMc-B:tracksqual:hiDeltaE}   {\ensuremath{{0.000 } } }
\vdef{default-11:CsData-B:pvz:loEff}   {\ensuremath{{0.519 } } }
\vdef{default-11:CsData-B:pvz:loEffE}   {\ensuremath{{0.006 } } }
\vdef{default-11:CsData-B:pvz:hiEff}   {\ensuremath{{0.481 } } }
\vdef{default-11:CsData-B:pvz:hiEffE}   {\ensuremath{{0.006 } } }
\vdef{default-11:CsMc-B:pvz:loEff}   {\ensuremath{{0.479 } } }
\vdef{default-11:CsMc-B:pvz:loEffE}   {\ensuremath{{0.006 } } }
\vdef{default-11:CsMc-B:pvz:hiEff}   {\ensuremath{{0.521 } } }
\vdef{default-11:CsMc-B:pvz:hiEffE}   {\ensuremath{{0.006 } } }
\vdef{default-11:CsMc-B:pvz:loDelta}   {\ensuremath{{+0.079 } } }
\vdef{default-11:CsMc-B:pvz:loDeltaE}   {\ensuremath{{0.017 } } }
\vdef{default-11:CsMc-B:pvz:hiDelta}   {\ensuremath{{-0.079 } } }
\vdef{default-11:CsMc-B:pvz:hiDeltaE}   {\ensuremath{{0.017 } } }
\vdef{default-11:CsData-B:pvn:loEff}   {\ensuremath{{1.008 } } }
\vdef{default-11:CsData-B:pvn:loEffE}   {\ensuremath{{\mathrm{NaN} } } }
\vdef{default-11:CsData-B:pvn:hiEff}   {\ensuremath{{1.000 } } }
\vdef{default-11:CsData-B:pvn:hiEffE}   {\ensuremath{{0.000 } } }
\vdef{default-11:CsMc-B:pvn:loEff}   {\ensuremath{{1.000 } } }
\vdef{default-11:CsMc-B:pvn:loEffE}   {\ensuremath{{0.000 } } }
\vdef{default-11:CsMc-B:pvn:hiEff}   {\ensuremath{{1.000 } } }
\vdef{default-11:CsMc-B:pvn:hiEffE}   {\ensuremath{{0.000 } } }
\vdef{default-11:CsMc-B:pvn:loDelta}   {\ensuremath{{+0.008 } } }
\vdef{default-11:CsMc-B:pvn:loDeltaE}   {\ensuremath{{\mathrm{NaN} } } }
\vdef{default-11:CsMc-B:pvn:hiDelta}   {\ensuremath{{+0.000 } } }
\vdef{default-11:CsMc-B:pvn:hiDeltaE}   {\ensuremath{{0.000 } } }
\vdef{default-11:CsData-B:pvavew8:loEff}   {\ensuremath{{0.012 } } }
\vdef{default-11:CsData-B:pvavew8:loEffE}   {\ensuremath{{0.001 } } }
\vdef{default-11:CsData-B:pvavew8:hiEff}   {\ensuremath{{0.988 } } }
\vdef{default-11:CsData-B:pvavew8:hiEffE}   {\ensuremath{{0.001 } } }
\vdef{default-11:CsMc-B:pvavew8:loEff}   {\ensuremath{{0.011 } } }
\vdef{default-11:CsMc-B:pvavew8:loEffE}   {\ensuremath{{0.001 } } }
\vdef{default-11:CsMc-B:pvavew8:hiEff}   {\ensuremath{{0.989 } } }
\vdef{default-11:CsMc-B:pvavew8:hiEffE}   {\ensuremath{{0.001 } } }
\vdef{default-11:CsMc-B:pvavew8:loDelta}   {\ensuremath{{+0.091 } } }
\vdef{default-11:CsMc-B:pvavew8:loDeltaE}   {\ensuremath{{0.168 } } }
\vdef{default-11:CsMc-B:pvavew8:hiDelta}   {\ensuremath{{-0.001 } } }
\vdef{default-11:CsMc-B:pvavew8:hiDeltaE}   {\ensuremath{{0.002 } } }
\vdef{default-11:CsData-B:pvntrk:loEff}   {\ensuremath{{1.000 } } }
\vdef{default-11:CsData-B:pvntrk:loEffE}   {\ensuremath{{0.000 } } }
\vdef{default-11:CsData-B:pvntrk:hiEff}   {\ensuremath{{1.000 } } }
\vdef{default-11:CsData-B:pvntrk:hiEffE}   {\ensuremath{{0.000 } } }
\vdef{default-11:CsMc-B:pvntrk:loEff}   {\ensuremath{{1.000 } } }
\vdef{default-11:CsMc-B:pvntrk:loEffE}   {\ensuremath{{0.000 } } }
\vdef{default-11:CsMc-B:pvntrk:hiEff}   {\ensuremath{{1.000 } } }
\vdef{default-11:CsMc-B:pvntrk:hiEffE}   {\ensuremath{{0.000 } } }
\vdef{default-11:CsMc-B:pvntrk:loDelta}   {\ensuremath{{+0.000 } } }
\vdef{default-11:CsMc-B:pvntrk:loDeltaE}   {\ensuremath{{0.000 } } }
\vdef{default-11:CsMc-B:pvntrk:hiDelta}   {\ensuremath{{+0.000 } } }
\vdef{default-11:CsMc-B:pvntrk:hiDeltaE}   {\ensuremath{{0.000 } } }
\vdef{default-11:CsData-B:muon1pt:loEff}   {\ensuremath{{1.017 } } }
\vdef{default-11:CsData-B:muon1pt:loEffE}   {\ensuremath{{\mathrm{NaN} } } }
\vdef{default-11:CsData-B:muon1pt:hiEff}   {\ensuremath{{1.000 } } }
\vdef{default-11:CsData-B:muon1pt:hiEffE}   {\ensuremath{{0.000 } } }
\vdef{default-11:CsMc-B:muon1pt:loEff}   {\ensuremath{{1.013 } } }
\vdef{default-11:CsMc-B:muon1pt:loEffE}   {\ensuremath{{\mathrm{NaN} } } }
\vdef{default-11:CsMc-B:muon1pt:hiEff}   {\ensuremath{{1.000 } } }
\vdef{default-11:CsMc-B:muon1pt:hiEffE}   {\ensuremath{{0.000 } } }
\vdef{default-11:CsMc-B:muon1pt:loDelta}   {\ensuremath{{+0.004 } } }
\vdef{default-11:CsMc-B:muon1pt:loDeltaE}   {\ensuremath{{\mathrm{NaN} } } }
\vdef{default-11:CsMc-B:muon1pt:hiDelta}   {\ensuremath{{+0.000 } } }
\vdef{default-11:CsMc-B:muon1pt:hiDeltaE}   {\ensuremath{{0.000 } } }
\vdef{default-11:CsData-B:muon2pt:loEff}   {\ensuremath{{0.044 } } }
\vdef{default-11:CsData-B:muon2pt:loEffE}   {\ensuremath{{0.003 } } }
\vdef{default-11:CsData-B:muon2pt:hiEff}   {\ensuremath{{0.956 } } }
\vdef{default-11:CsData-B:muon2pt:hiEffE}   {\ensuremath{{0.003 } } }
\vdef{default-11:CsMc-B:muon2pt:loEff}   {\ensuremath{{0.041 } } }
\vdef{default-11:CsMc-B:muon2pt:loEffE}   {\ensuremath{{0.002 } } }
\vdef{default-11:CsMc-B:muon2pt:hiEff}   {\ensuremath{{0.959 } } }
\vdef{default-11:CsMc-B:muon2pt:hiEffE}   {\ensuremath{{0.002 } } }
\vdef{default-11:CsMc-B:muon2pt:loDelta}   {\ensuremath{{+0.074 } } }
\vdef{default-11:CsMc-B:muon2pt:loDeltaE}   {\ensuremath{{0.083 } } }
\vdef{default-11:CsMc-B:muon2pt:hiDelta}   {\ensuremath{{-0.003 } } }
\vdef{default-11:CsMc-B:muon2pt:hiDeltaE}   {\ensuremath{{0.004 } } }
\vdef{default-11:CsData-B:muonseta:loEff}   {\ensuremath{{0.805 } } }
\vdef{default-11:CsData-B:muonseta:loEffE}   {\ensuremath{{0.004 } } }
\vdef{default-11:CsData-B:muonseta:hiEff}   {\ensuremath{{0.195 } } }
\vdef{default-11:CsData-B:muonseta:hiEffE}   {\ensuremath{{0.004 } } }
\vdef{default-11:CsMc-B:muonseta:loEff}   {\ensuremath{{0.807 } } }
\vdef{default-11:CsMc-B:muonseta:loEffE}   {\ensuremath{{0.004 } } }
\vdef{default-11:CsMc-B:muonseta:hiEff}   {\ensuremath{{0.193 } } }
\vdef{default-11:CsMc-B:muonseta:hiEffE}   {\ensuremath{{0.004 } } }
\vdef{default-11:CsMc-B:muonseta:loDelta}   {\ensuremath{{-0.002 } } }
\vdef{default-11:CsMc-B:muonseta:loDeltaE}   {\ensuremath{{0.006 } } }
\vdef{default-11:CsMc-B:muonseta:hiDelta}   {\ensuremath{{+0.009 } } }
\vdef{default-11:CsMc-B:muonseta:hiDeltaE}   {\ensuremath{{0.026 } } }
\vdef{default-11:CsData-B:pt:loEff}   {\ensuremath{{0.000 } } }
\vdef{default-11:CsData-B:pt:loEffE}   {\ensuremath{{0.000 } } }
\vdef{default-11:CsData-B:pt:hiEff}   {\ensuremath{{1.000 } } }
\vdef{default-11:CsData-B:pt:hiEffE}   {\ensuremath{{0.000 } } }
\vdef{default-11:CsMc-B:pt:loEff}   {\ensuremath{{0.000 } } }
\vdef{default-11:CsMc-B:pt:loEffE}   {\ensuremath{{0.000 } } }
\vdef{default-11:CsMc-B:pt:hiEff}   {\ensuremath{{1.000 } } }
\vdef{default-11:CsMc-B:pt:hiEffE}   {\ensuremath{{0.000 } } }
\vdef{default-11:CsMc-B:pt:loDelta}   {\ensuremath{{\mathrm{NaN} } } }
\vdef{default-11:CsMc-B:pt:loDeltaE}   {\ensuremath{{\mathrm{NaN} } } }
\vdef{default-11:CsMc-B:pt:hiDelta}   {\ensuremath{{+0.000 } } }
\vdef{default-11:CsMc-B:pt:hiDeltaE}   {\ensuremath{{0.000 } } }
\vdef{default-11:CsData-B:p:loEff}   {\ensuremath{{1.005 } } }
\vdef{default-11:CsData-B:p:loEffE}   {\ensuremath{{\mathrm{NaN} } } }
\vdef{default-11:CsData-B:p:hiEff}   {\ensuremath{{1.000 } } }
\vdef{default-11:CsData-B:p:hiEffE}   {\ensuremath{{0.000 } } }
\vdef{default-11:CsMc-B:p:loEff}   {\ensuremath{{1.003 } } }
\vdef{default-11:CsMc-B:p:loEffE}   {\ensuremath{{\mathrm{NaN} } } }
\vdef{default-11:CsMc-B:p:hiEff}   {\ensuremath{{1.000 } } }
\vdef{default-11:CsMc-B:p:hiEffE}   {\ensuremath{{0.000 } } }
\vdef{default-11:CsMc-B:p:loDelta}   {\ensuremath{{+0.002 } } }
\vdef{default-11:CsMc-B:p:loDeltaE}   {\ensuremath{{\mathrm{NaN} } } }
\vdef{default-11:CsMc-B:p:hiDelta}   {\ensuremath{{+0.000 } } }
\vdef{default-11:CsMc-B:p:hiDeltaE}   {\ensuremath{{0.000 } } }
\vdef{default-11:CsData-B:eta:loEff}   {\ensuremath{{0.799 } } }
\vdef{default-11:CsData-B:eta:loEffE}   {\ensuremath{{0.005 } } }
\vdef{default-11:CsData-B:eta:hiEff}   {\ensuremath{{0.201 } } }
\vdef{default-11:CsData-B:eta:hiEffE}   {\ensuremath{{0.005 } } }
\vdef{default-11:CsMc-B:eta:loEff}   {\ensuremath{{0.795 } } }
\vdef{default-11:CsMc-B:eta:loEffE}   {\ensuremath{{0.005 } } }
\vdef{default-11:CsMc-B:eta:hiEff}   {\ensuremath{{0.205 } } }
\vdef{default-11:CsMc-B:eta:hiEffE}   {\ensuremath{{0.005 } } }
\vdef{default-11:CsMc-B:eta:loDelta}   {\ensuremath{{+0.004 } } }
\vdef{default-11:CsMc-B:eta:loDeltaE}   {\ensuremath{{0.009 } } }
\vdef{default-11:CsMc-B:eta:hiDelta}   {\ensuremath{{-0.017 } } }
\vdef{default-11:CsMc-B:eta:hiDeltaE}   {\ensuremath{{0.035 } } }
\vdef{default-11:CsData-B:bdt:loEff}   {\ensuremath{{0.860 } } }
\vdef{default-11:CsData-B:bdt:loEffE}   {\ensuremath{{0.004 } } }
\vdef{default-11:CsData-B:bdt:hiEff}   {\ensuremath{{0.140 } } }
\vdef{default-11:CsData-B:bdt:hiEffE}   {\ensuremath{{0.004 } } }
\vdef{default-11:CsMc-B:bdt:loEff}   {\ensuremath{{0.863 } } }
\vdef{default-11:CsMc-B:bdt:loEffE}   {\ensuremath{{0.004 } } }
\vdef{default-11:CsMc-B:bdt:hiEff}   {\ensuremath{{0.137 } } }
\vdef{default-11:CsMc-B:bdt:hiEffE}   {\ensuremath{{0.004 } } }
\vdef{default-11:CsMc-B:bdt:loDelta}   {\ensuremath{{-0.003 } } }
\vdef{default-11:CsMc-B:bdt:loDeltaE}   {\ensuremath{{0.007 } } }
\vdef{default-11:CsMc-B:bdt:hiDelta}   {\ensuremath{{+0.020 } } }
\vdef{default-11:CsMc-B:bdt:hiDeltaE}   {\ensuremath{{0.043 } } }
\vdef{default-11:CsData-B:fl3d:loEff}   {\ensuremath{{0.890 } } }
\vdef{default-11:CsData-B:fl3d:loEffE}   {\ensuremath{{0.004 } } }
\vdef{default-11:CsData-B:fl3d:hiEff}   {\ensuremath{{0.110 } } }
\vdef{default-11:CsData-B:fl3d:hiEffE}   {\ensuremath{{0.004 } } }
\vdef{default-11:CsMc-B:fl3d:loEff}   {\ensuremath{{0.901 } } }
\vdef{default-11:CsMc-B:fl3d:loEffE}   {\ensuremath{{0.003 } } }
\vdef{default-11:CsMc-B:fl3d:hiEff}   {\ensuremath{{0.099 } } }
\vdef{default-11:CsMc-B:fl3d:hiEffE}   {\ensuremath{{0.003 } } }
\vdef{default-11:CsMc-B:fl3d:loDelta}   {\ensuremath{{-0.012 } } }
\vdef{default-11:CsMc-B:fl3d:loDeltaE}   {\ensuremath{{0.006 } } }
\vdef{default-11:CsMc-B:fl3d:hiDelta}   {\ensuremath{{+0.099 } } }
\vdef{default-11:CsMc-B:fl3d:hiDeltaE}   {\ensuremath{{0.048 } } }
\vdef{default-11:CsData-B:fl3de:loEff}   {\ensuremath{{1.000 } } }
\vdef{default-11:CsData-B:fl3de:loEffE}   {\ensuremath{{0.000 } } }
\vdef{default-11:CsData-B:fl3de:hiEff}   {\ensuremath{{0.000 } } }
\vdef{default-11:CsData-B:fl3de:hiEffE}   {\ensuremath{{0.000 } } }
\vdef{default-11:CsMc-B:fl3de:loEff}   {\ensuremath{{1.000 } } }
\vdef{default-11:CsMc-B:fl3de:loEffE}   {\ensuremath{{0.000 } } }
\vdef{default-11:CsMc-B:fl3de:hiEff}   {\ensuremath{{0.000 } } }
\vdef{default-11:CsMc-B:fl3de:hiEffE}   {\ensuremath{{0.000 } } }
\vdef{default-11:CsMc-B:fl3de:loDelta}   {\ensuremath{{+0.000 } } }
\vdef{default-11:CsMc-B:fl3de:loDeltaE}   {\ensuremath{{0.000 } } }
\vdef{default-11:CsMc-B:fl3de:hiDelta}   {\ensuremath{{\mathrm{NaN} } } }
\vdef{default-11:CsMc-B:fl3de:hiDeltaE}   {\ensuremath{{\mathrm{NaN} } } }
\vdef{default-11:CsData-B:fls3d:loEff}   {\ensuremath{{0.060 } } }
\vdef{default-11:CsData-B:fls3d:loEffE}   {\ensuremath{{0.003 } } }
\vdef{default-11:CsData-B:fls3d:hiEff}   {\ensuremath{{0.940 } } }
\vdef{default-11:CsData-B:fls3d:hiEffE}   {\ensuremath{{0.003 } } }
\vdef{default-11:CsMc-B:fls3d:loEff}   {\ensuremath{{0.065 } } }
\vdef{default-11:CsMc-B:fls3d:loEffE}   {\ensuremath{{0.003 } } }
\vdef{default-11:CsMc-B:fls3d:hiEff}   {\ensuremath{{0.935 } } }
\vdef{default-11:CsMc-B:fls3d:hiEffE}   {\ensuremath{{0.003 } } }
\vdef{default-11:CsMc-B:fls3d:loDelta}   {\ensuremath{{-0.074 } } }
\vdef{default-11:CsMc-B:fls3d:loDeltaE}   {\ensuremath{{0.064 } } }
\vdef{default-11:CsMc-B:fls3d:hiDelta}   {\ensuremath{{+0.005 } } }
\vdef{default-11:CsMc-B:fls3d:hiDeltaE}   {\ensuremath{{0.004 } } }
\vdef{default-11:CsData-B:flsxy:loEff}   {\ensuremath{{1.011 } } }
\vdef{default-11:CsData-B:flsxy:loEffE}   {\ensuremath{{\mathrm{NaN} } } }
\vdef{default-11:CsData-B:flsxy:hiEff}   {\ensuremath{{1.000 } } }
\vdef{default-11:CsData-B:flsxy:hiEffE}   {\ensuremath{{0.000 } } }
\vdef{default-11:CsMc-B:flsxy:loEff}   {\ensuremath{{1.008 } } }
\vdef{default-11:CsMc-B:flsxy:loEffE}   {\ensuremath{{\mathrm{NaN} } } }
\vdef{default-11:CsMc-B:flsxy:hiEff}   {\ensuremath{{1.000 } } }
\vdef{default-11:CsMc-B:flsxy:hiEffE}   {\ensuremath{{0.000 } } }
\vdef{default-11:CsMc-B:flsxy:loDelta}   {\ensuremath{{+0.003 } } }
\vdef{default-11:CsMc-B:flsxy:loDeltaE}   {\ensuremath{{\mathrm{NaN} } } }
\vdef{default-11:CsMc-B:flsxy:hiDelta}   {\ensuremath{{+0.000 } } }
\vdef{default-11:CsMc-B:flsxy:hiDeltaE}   {\ensuremath{{0.000 } } }
\vdef{default-11:CsData-B:chi2dof:loEff}   {\ensuremath{{0.934 } } }
\vdef{default-11:CsData-B:chi2dof:loEffE}   {\ensuremath{{0.003 } } }
\vdef{default-11:CsData-B:chi2dof:hiEff}   {\ensuremath{{0.066 } } }
\vdef{default-11:CsData-B:chi2dof:hiEffE}   {\ensuremath{{0.003 } } }
\vdef{default-11:CsMc-B:chi2dof:loEff}   {\ensuremath{{0.946 } } }
\vdef{default-11:CsMc-B:chi2dof:loEffE}   {\ensuremath{{0.003 } } }
\vdef{default-11:CsMc-B:chi2dof:hiEff}   {\ensuremath{{0.054 } } }
\vdef{default-11:CsMc-B:chi2dof:hiEffE}   {\ensuremath{{0.003 } } }
\vdef{default-11:CsMc-B:chi2dof:loDelta}   {\ensuremath{{-0.013 } } }
\vdef{default-11:CsMc-B:chi2dof:loDeltaE}   {\ensuremath{{0.004 } } }
\vdef{default-11:CsMc-B:chi2dof:hiDelta}   {\ensuremath{{+0.206 } } }
\vdef{default-11:CsMc-B:chi2dof:hiDeltaE}   {\ensuremath{{0.068 } } }
\vdef{default-11:CsData-B:pchi2dof:loEff}   {\ensuremath{{0.626 } } }
\vdef{default-11:CsData-B:pchi2dof:loEffE}   {\ensuremath{{0.006 } } }
\vdef{default-11:CsData-B:pchi2dof:hiEff}   {\ensuremath{{0.374 } } }
\vdef{default-11:CsData-B:pchi2dof:hiEffE}   {\ensuremath{{0.006 } } }
\vdef{default-11:CsMc-B:pchi2dof:loEff}   {\ensuremath{{0.592 } } }
\vdef{default-11:CsMc-B:pchi2dof:loEffE}   {\ensuremath{{0.006 } } }
\vdef{default-11:CsMc-B:pchi2dof:hiEff}   {\ensuremath{{0.408 } } }
\vdef{default-11:CsMc-B:pchi2dof:hiEffE}   {\ensuremath{{0.006 } } }
\vdef{default-11:CsMc-B:pchi2dof:loDelta}   {\ensuremath{{+0.056 } } }
\vdef{default-11:CsMc-B:pchi2dof:loDeltaE}   {\ensuremath{{0.014 } } }
\vdef{default-11:CsMc-B:pchi2dof:hiDelta}   {\ensuremath{{-0.087 } } }
\vdef{default-11:CsMc-B:pchi2dof:hiDeltaE}   {\ensuremath{{0.021 } } }
\vdef{default-11:CsData-B:alpha:loEff}   {\ensuremath{{0.994 } } }
\vdef{default-11:CsData-B:alpha:loEffE}   {\ensuremath{{0.001 } } }
\vdef{default-11:CsData-B:alpha:hiEff}   {\ensuremath{{0.006 } } }
\vdef{default-11:CsData-B:alpha:hiEffE}   {\ensuremath{{0.001 } } }
\vdef{default-11:CsMc-B:alpha:loEff}   {\ensuremath{{0.995 } } }
\vdef{default-11:CsMc-B:alpha:loEffE}   {\ensuremath{{0.001 } } }
\vdef{default-11:CsMc-B:alpha:hiEff}   {\ensuremath{{0.005 } } }
\vdef{default-11:CsMc-B:alpha:hiEffE}   {\ensuremath{{0.001 } } }
\vdef{default-11:CsMc-B:alpha:loDelta}   {\ensuremath{{-0.001 } } }
\vdef{default-11:CsMc-B:alpha:loDeltaE}   {\ensuremath{{0.001 } } }
\vdef{default-11:CsMc-B:alpha:hiDelta}   {\ensuremath{{+0.100 } } }
\vdef{default-11:CsMc-B:alpha:hiDeltaE}   {\ensuremath{{0.242 } } }
\vdef{default-11:CsData-B:iso:loEff}   {\ensuremath{{0.097 } } }
\vdef{default-11:CsData-B:iso:loEffE}   {\ensuremath{{0.004 } } }
\vdef{default-11:CsData-B:iso:hiEff}   {\ensuremath{{0.903 } } }
\vdef{default-11:CsData-B:iso:hiEffE}   {\ensuremath{{0.004 } } }
\vdef{default-11:CsMc-B:iso:loEff}   {\ensuremath{{0.089 } } }
\vdef{default-11:CsMc-B:iso:loEffE}   {\ensuremath{{0.003 } } }
\vdef{default-11:CsMc-B:iso:hiEff}   {\ensuremath{{0.911 } } }
\vdef{default-11:CsMc-B:iso:hiEffE}   {\ensuremath{{0.003 } } }
\vdef{default-11:CsMc-B:iso:loDelta}   {\ensuremath{{+0.079 } } }
\vdef{default-11:CsMc-B:iso:loDeltaE}   {\ensuremath{{0.053 } } }
\vdef{default-11:CsMc-B:iso:hiDelta}   {\ensuremath{{-0.008 } } }
\vdef{default-11:CsMc-B:iso:hiDeltaE}   {\ensuremath{{0.005 } } }
\vdef{default-11:CsData-B:docatrk:loEff}   {\ensuremath{{0.066 } } }
\vdef{default-11:CsData-B:docatrk:loEffE}   {\ensuremath{{0.003 } } }
\vdef{default-11:CsData-B:docatrk:hiEff}   {\ensuremath{{0.934 } } }
\vdef{default-11:CsData-B:docatrk:hiEffE}   {\ensuremath{{0.003 } } }
\vdef{default-11:CsMc-B:docatrk:loEff}   {\ensuremath{{0.083 } } }
\vdef{default-11:CsMc-B:docatrk:loEffE}   {\ensuremath{{0.003 } } }
\vdef{default-11:CsMc-B:docatrk:hiEff}   {\ensuremath{{0.917 } } }
\vdef{default-11:CsMc-B:docatrk:hiEffE}   {\ensuremath{{0.003 } } }
\vdef{default-11:CsMc-B:docatrk:loDelta}   {\ensuremath{{-0.230 } } }
\vdef{default-11:CsMc-B:docatrk:loDeltaE}   {\ensuremath{{0.061 } } }
\vdef{default-11:CsMc-B:docatrk:hiDelta}   {\ensuremath{{+0.019 } } }
\vdef{default-11:CsMc-B:docatrk:hiDeltaE}   {\ensuremath{{0.005 } } }
\vdef{default-11:CsData-B:isotrk:loEff}   {\ensuremath{{1.000 } } }
\vdef{default-11:CsData-B:isotrk:loEffE}   {\ensuremath{{0.000 } } }
\vdef{default-11:CsData-B:isotrk:hiEff}   {\ensuremath{{1.000 } } }
\vdef{default-11:CsData-B:isotrk:hiEffE}   {\ensuremath{{0.000 } } }
\vdef{default-11:CsMc-B:isotrk:loEff}   {\ensuremath{{1.000 } } }
\vdef{default-11:CsMc-B:isotrk:loEffE}   {\ensuremath{{0.000 } } }
\vdef{default-11:CsMc-B:isotrk:hiEff}   {\ensuremath{{1.000 } } }
\vdef{default-11:CsMc-B:isotrk:hiEffE}   {\ensuremath{{0.000 } } }
\vdef{default-11:CsMc-B:isotrk:loDelta}   {\ensuremath{{+0.000 } } }
\vdef{default-11:CsMc-B:isotrk:loDeltaE}   {\ensuremath{{0.000 } } }
\vdef{default-11:CsMc-B:isotrk:hiDelta}   {\ensuremath{{+0.000 } } }
\vdef{default-11:CsMc-B:isotrk:hiDeltaE}   {\ensuremath{{0.000 } } }
\vdef{default-11:CsData-B:closetrk:loEff}   {\ensuremath{{0.981 } } }
\vdef{default-11:CsData-B:closetrk:loEffE}   {\ensuremath{{0.002 } } }
\vdef{default-11:CsData-B:closetrk:hiEff}   {\ensuremath{{0.019 } } }
\vdef{default-11:CsData-B:closetrk:hiEffE}   {\ensuremath{{0.002 } } }
\vdef{default-11:CsMc-B:closetrk:loEff}   {\ensuremath{{0.980 } } }
\vdef{default-11:CsMc-B:closetrk:loEffE}   {\ensuremath{{0.002 } } }
\vdef{default-11:CsMc-B:closetrk:hiEff}   {\ensuremath{{0.020 } } }
\vdef{default-11:CsMc-B:closetrk:hiEffE}   {\ensuremath{{0.002 } } }
\vdef{default-11:CsMc-B:closetrk:loDelta}   {\ensuremath{{+0.001 } } }
\vdef{default-11:CsMc-B:closetrk:loDeltaE}   {\ensuremath{{0.002 } } }
\vdef{default-11:CsMc-B:closetrk:hiDelta}   {\ensuremath{{-0.069 } } }
\vdef{default-11:CsMc-B:closetrk:hiDeltaE}   {\ensuremath{{0.125 } } }
\vdef{default-11:CsData-B:lip:loEff}   {\ensuremath{{1.000 } } }
\vdef{default-11:CsData-B:lip:loEffE}   {\ensuremath{{0.000 } } }
\vdef{default-11:CsData-B:lip:hiEff}   {\ensuremath{{0.000 } } }
\vdef{default-11:CsData-B:lip:hiEffE}   {\ensuremath{{0.000 } } }
\vdef{default-11:CsMc-B:lip:loEff}   {\ensuremath{{1.000 } } }
\vdef{default-11:CsMc-B:lip:loEffE}   {\ensuremath{{0.000 } } }
\vdef{default-11:CsMc-B:lip:hiEff}   {\ensuremath{{0.000 } } }
\vdef{default-11:CsMc-B:lip:hiEffE}   {\ensuremath{{0.000 } } }
\vdef{default-11:CsMc-B:lip:loDelta}   {\ensuremath{{+0.000 } } }
\vdef{default-11:CsMc-B:lip:loDeltaE}   {\ensuremath{{0.000 } } }
\vdef{default-11:CsMc-B:lip:hiDelta}   {\ensuremath{{\mathrm{NaN} } } }
\vdef{default-11:CsMc-B:lip:hiDeltaE}   {\ensuremath{{\mathrm{NaN} } } }
\vdef{default-11:CsData-B:lip:inEff}   {\ensuremath{{1.000 } } }
\vdef{default-11:CsData-B:lip:inEffE}   {\ensuremath{{0.000 } } }
\vdef{default-11:CsMc-B:lip:inEff}   {\ensuremath{{1.000 } } }
\vdef{default-11:CsMc-B:lip:inEffE}   {\ensuremath{{0.000 } } }
\vdef{default-11:CsMc-B:lip:inDelta}   {\ensuremath{{+0.000 } } }
\vdef{default-11:CsMc-B:lip:inDeltaE}   {\ensuremath{{0.000 } } }
\vdef{default-11:CsData-B:lips:loEff}   {\ensuremath{{1.000 } } }
\vdef{default-11:CsData-B:lips:loEffE}   {\ensuremath{{0.000 } } }
\vdef{default-11:CsData-B:lips:hiEff}   {\ensuremath{{0.000 } } }
\vdef{default-11:CsData-B:lips:hiEffE}   {\ensuremath{{0.000 } } }
\vdef{default-11:CsMc-B:lips:loEff}   {\ensuremath{{1.000 } } }
\vdef{default-11:CsMc-B:lips:loEffE}   {\ensuremath{{0.000 } } }
\vdef{default-11:CsMc-B:lips:hiEff}   {\ensuremath{{0.000 } } }
\vdef{default-11:CsMc-B:lips:hiEffE}   {\ensuremath{{0.000 } } }
\vdef{default-11:CsMc-B:lips:loDelta}   {\ensuremath{{+0.000 } } }
\vdef{default-11:CsMc-B:lips:loDeltaE}   {\ensuremath{{0.000 } } }
\vdef{default-11:CsMc-B:lips:hiDelta}   {\ensuremath{{\mathrm{NaN} } } }
\vdef{default-11:CsMc-B:lips:hiDeltaE}   {\ensuremath{{\mathrm{NaN} } } }
\vdef{default-11:CsData-B:lips:inEff}   {\ensuremath{{1.000 } } }
\vdef{default-11:CsData-B:lips:inEffE}   {\ensuremath{{0.000 } } }
\vdef{default-11:CsMc-B:lips:inEff}   {\ensuremath{{1.000 } } }
\vdef{default-11:CsMc-B:lips:inEffE}   {\ensuremath{{0.000 } } }
\vdef{default-11:CsMc-B:lips:inDelta}   {\ensuremath{{+0.000 } } }
\vdef{default-11:CsMc-B:lips:inDeltaE}   {\ensuremath{{0.000 } } }
\vdef{default-11:CsData-B:ip:loEff}   {\ensuremath{{0.972 } } }
\vdef{default-11:CsData-B:ip:loEffE}   {\ensuremath{{0.002 } } }
\vdef{default-11:CsData-B:ip:hiEff}   {\ensuremath{{0.028 } } }
\vdef{default-11:CsData-B:ip:hiEffE}   {\ensuremath{{0.002 } } }
\vdef{default-11:CsMc-B:ip:loEff}   {\ensuremath{{0.973 } } }
\vdef{default-11:CsMc-B:ip:loEffE}   {\ensuremath{{0.002 } } }
\vdef{default-11:CsMc-B:ip:hiEff}   {\ensuremath{{0.027 } } }
\vdef{default-11:CsMc-B:ip:hiEffE}   {\ensuremath{{0.002 } } }
\vdef{default-11:CsMc-B:ip:loDelta}   {\ensuremath{{-0.001 } } }
\vdef{default-11:CsMc-B:ip:loDeltaE}   {\ensuremath{{0.003 } } }
\vdef{default-11:CsMc-B:ip:hiDelta}   {\ensuremath{{+0.052 } } }
\vdef{default-11:CsMc-B:ip:hiDeltaE}   {\ensuremath{{0.105 } } }
\vdef{default-11:CsData-B:ips:loEff}   {\ensuremath{{0.932 } } }
\vdef{default-11:CsData-B:ips:loEffE}   {\ensuremath{{0.003 } } }
\vdef{default-11:CsData-B:ips:hiEff}   {\ensuremath{{0.068 } } }
\vdef{default-11:CsData-B:ips:hiEffE}   {\ensuremath{{0.003 } } }
\vdef{default-11:CsMc-B:ips:loEff}   {\ensuremath{{0.945 } } }
\vdef{default-11:CsMc-B:ips:loEffE}   {\ensuremath{{0.003 } } }
\vdef{default-11:CsMc-B:ips:hiEff}   {\ensuremath{{0.055 } } }
\vdef{default-11:CsMc-B:ips:hiEffE}   {\ensuremath{{0.003 } } }
\vdef{default-11:CsMc-B:ips:loDelta}   {\ensuremath{{-0.013 } } }
\vdef{default-11:CsMc-B:ips:loDeltaE}   {\ensuremath{{0.004 } } }
\vdef{default-11:CsMc-B:ips:hiDelta}   {\ensuremath{{+0.202 } } }
\vdef{default-11:CsMc-B:ips:hiDeltaE}   {\ensuremath{{0.067 } } }
\vdef{default-11:CsData-B:maxdoca:loEff}   {\ensuremath{{1.000 } } }
\vdef{default-11:CsData-B:maxdoca:loEffE}   {\ensuremath{{0.000 } } }
\vdef{default-11:CsData-B:maxdoca:hiEff}   {\ensuremath{{0.025 } } }
\vdef{default-11:CsData-B:maxdoca:hiEffE}   {\ensuremath{{0.002 } } }
\vdef{default-11:CsMc-B:maxdoca:loEff}   {\ensuremath{{1.000 } } }
\vdef{default-11:CsMc-B:maxdoca:loEffE}   {\ensuremath{{0.000 } } }
\vdef{default-11:CsMc-B:maxdoca:hiEff}   {\ensuremath{{0.015 } } }
\vdef{default-11:CsMc-B:maxdoca:hiEffE}   {\ensuremath{{0.002 } } }
\vdef{default-11:CsMc-B:maxdoca:loDelta}   {\ensuremath{{+0.000 } } }
\vdef{default-11:CsMc-B:maxdoca:loDeltaE}   {\ensuremath{{0.000 } } }
\vdef{default-11:CsMc-B:maxdoca:hiDelta}   {\ensuremath{{+0.503 } } }
\vdef{default-11:CsMc-B:maxdoca:hiDeltaE}   {\ensuremath{{0.123 } } }
\vdef{default-11:CsData-B:kaonspt:loEff}   {\ensuremath{{1.000 } } }
\vdef{default-11:CsData-B:kaonspt:loEffE}   {\ensuremath{{\mathrm{NaN} } } }
\vdef{default-11:CsData-B:kaonspt:hiEff}   {\ensuremath{{1.000 } } }
\vdef{default-11:CsData-B:kaonspt:hiEffE}   {\ensuremath{{0.000 } } }
\vdef{default-11:CsMc-B:kaonspt:loEff}   {\ensuremath{{1.000 } } }
\vdef{default-11:CsMc-B:kaonspt:loEffE}   {\ensuremath{{\mathrm{NaN} } } }
\vdef{default-11:CsMc-B:kaonspt:hiEff}   {\ensuremath{{1.000 } } }
\vdef{default-11:CsMc-B:kaonspt:hiEffE}   {\ensuremath{{0.000 } } }
\vdef{default-11:CsMc-B:kaonspt:loDelta}   {\ensuremath{{+0.000 } } }
\vdef{default-11:CsMc-B:kaonspt:loDeltaE}   {\ensuremath{{\mathrm{NaN} } } }
\vdef{default-11:CsMc-B:kaonspt:hiDelta}   {\ensuremath{{+0.000 } } }
\vdef{default-11:CsMc-B:kaonspt:hiDeltaE}   {\ensuremath{{0.000 } } }
\vdef{default-11:CsData-B:psipt:loEff}   {\ensuremath{{1.007 } } }
\vdef{default-11:CsData-B:psipt:loEffE}   {\ensuremath{{\mathrm{NaN} } } }
\vdef{default-11:CsData-B:psipt:hiEff}   {\ensuremath{{1.000 } } }
\vdef{default-11:CsData-B:psipt:hiEffE}   {\ensuremath{{0.000 } } }
\vdef{default-11:CsMc-B:psipt:loEff}   {\ensuremath{{1.004 } } }
\vdef{default-11:CsMc-B:psipt:loEffE}   {\ensuremath{{\mathrm{NaN} } } }
\vdef{default-11:CsMc-B:psipt:hiEff}   {\ensuremath{{1.000 } } }
\vdef{default-11:CsMc-B:psipt:hiEffE}   {\ensuremath{{0.000 } } }
\vdef{default-11:CsMc-B:psipt:loDelta}   {\ensuremath{{+0.002 } } }
\vdef{default-11:CsMc-B:psipt:loDeltaE}   {\ensuremath{{\mathrm{NaN} } } }
\vdef{default-11:CsMc-B:psipt:hiDelta}   {\ensuremath{{+0.000 } } }
\vdef{default-11:CsMc-B:psipt:hiDeltaE}   {\ensuremath{{0.000 } } }
\vdef{default-11:CsData-B:phipt:loEff}   {\ensuremath{{1.018 } } }
\vdef{default-11:CsData-B:phipt:loEffE}   {\ensuremath{{\mathrm{NaN} } } }
\vdef{default-11:CsData-B:phipt:hiEff}   {\ensuremath{{1.000 } } }
\vdef{default-11:CsData-B:phipt:hiEffE}   {\ensuremath{{0.000 } } }
\vdef{default-11:CsMc-B:phipt:loEff}   {\ensuremath{{1.009 } } }
\vdef{default-11:CsMc-B:phipt:loEffE}   {\ensuremath{{\mathrm{NaN} } } }
\vdef{default-11:CsMc-B:phipt:hiEff}   {\ensuremath{{1.000 } } }
\vdef{default-11:CsMc-B:phipt:hiEffE}   {\ensuremath{{0.000 } } }
\vdef{default-11:CsMc-B:phipt:loDelta}   {\ensuremath{{+0.009 } } }
\vdef{default-11:CsMc-B:phipt:loDeltaE}   {\ensuremath{{\mathrm{NaN} } } }
\vdef{default-11:CsMc-B:phipt:hiDelta}   {\ensuremath{{+0.000 } } }
\vdef{default-11:CsMc-B:phipt:hiDeltaE}   {\ensuremath{{0.000 } } }
\vdef{default-11:CsData-B:deltar:loEff}   {\ensuremath{{1.000 } } }
\vdef{default-11:CsData-B:deltar:loEffE}   {\ensuremath{{0.000 } } }
\vdef{default-11:CsData-B:deltar:hiEff}   {\ensuremath{{0.000 } } }
\vdef{default-11:CsData-B:deltar:hiEffE}   {\ensuremath{{0.000 } } }
\vdef{default-11:CsMc-B:deltar:loEff}   {\ensuremath{{1.000 } } }
\vdef{default-11:CsMc-B:deltar:loEffE}   {\ensuremath{{0.000 } } }
\vdef{default-11:CsMc-B:deltar:hiEff}   {\ensuremath{{0.000 } } }
\vdef{default-11:CsMc-B:deltar:hiEffE}   {\ensuremath{{0.000 } } }
\vdef{default-11:CsMc-B:deltar:loDelta}   {\ensuremath{{+0.000 } } }
\vdef{default-11:CsMc-B:deltar:loDeltaE}   {\ensuremath{{0.000 } } }
\vdef{default-11:CsMc-B:deltar:hiDelta}   {\ensuremath{{\mathrm{NaN} } } }
\vdef{default-11:CsMc-B:deltar:hiDeltaE}   {\ensuremath{{\mathrm{NaN} } } }
\vdef{default-11:CsData-B:mkk:loEff}   {\ensuremath{{1.162 } } }
\vdef{default-11:CsData-B:mkk:loEffE}   {\ensuremath{{\mathrm{NaN} } } }
\vdef{default-11:CsData-B:mkk:hiEff}   {\ensuremath{{1.000 } } }
\vdef{default-11:CsData-B:mkk:hiEffE}   {\ensuremath{{0.000 } } }
\vdef{default-11:CsMc-B:mkk:loEff}   {\ensuremath{{1.000 } } }
\vdef{default-11:CsMc-B:mkk:loEffE}   {\ensuremath{{0.000 } } }
\vdef{default-11:CsMc-B:mkk:hiEff}   {\ensuremath{{1.000 } } }
\vdef{default-11:CsMc-B:mkk:hiEffE}   {\ensuremath{{0.000 } } }
\vdef{default-11:CsMc-B:mkk:loDelta}   {\ensuremath{{+0.150 } } }
\vdef{default-11:CsMc-B:mkk:loDeltaE}   {\ensuremath{{\mathrm{NaN} } } }
\vdef{default-11:CsMc-B:mkk:hiDelta}   {\ensuremath{{+0.000 } } }
\vdef{default-11:CsMc-B:mkk:hiDeltaE}   {\ensuremath{{0.000 } } }
\vdef{default-11:CsData-E:osiso:loEff}   {\ensuremath{{1.006 } } }
\vdef{default-11:CsData-E:osiso:loEffE}   {\ensuremath{{\mathrm{NaN} } } }
\vdef{default-11:CsData-E:osiso:hiEff}   {\ensuremath{{1.000 } } }
\vdef{default-11:CsData-E:osiso:hiEffE}   {\ensuremath{{0.000 } } }
\vdef{default-11:CsMc-E:osiso:loEff}   {\ensuremath{{1.003 } } }
\vdef{default-11:CsMc-E:osiso:loEffE}   {\ensuremath{{\mathrm{NaN} } } }
\vdef{default-11:CsMc-E:osiso:hiEff}   {\ensuremath{{1.000 } } }
\vdef{default-11:CsMc-E:osiso:hiEffE}   {\ensuremath{{0.000 } } }
\vdef{default-11:CsMc-E:osiso:loDelta}   {\ensuremath{{+0.003 } } }
\vdef{default-11:CsMc-E:osiso:loDeltaE}   {\ensuremath{{\mathrm{NaN} } } }
\vdef{default-11:CsMc-E:osiso:hiDelta}   {\ensuremath{{+0.000 } } }
\vdef{default-11:CsMc-E:osiso:hiDeltaE}   {\ensuremath{{0.001 } } }
\vdef{default-11:CsData-E:osreliso:loEff}   {\ensuremath{{0.254 } } }
\vdef{default-11:CsData-E:osreliso:loEffE}   {\ensuremath{{0.009 } } }
\vdef{default-11:CsData-E:osreliso:hiEff}   {\ensuremath{{0.746 } } }
\vdef{default-11:CsData-E:osreliso:hiEffE}   {\ensuremath{{0.009 } } }
\vdef{default-11:CsMc-E:osreliso:loEff}   {\ensuremath{{0.314 } } }
\vdef{default-11:CsMc-E:osreliso:loEffE}   {\ensuremath{{0.009 } } }
\vdef{default-11:CsMc-E:osreliso:hiEff}   {\ensuremath{{0.686 } } }
\vdef{default-11:CsMc-E:osreliso:hiEffE}   {\ensuremath{{0.009 } } }
\vdef{default-11:CsMc-E:osreliso:loDelta}   {\ensuremath{{-0.210 } } }
\vdef{default-11:CsMc-E:osreliso:loDeltaE}   {\ensuremath{{0.045 } } }
\vdef{default-11:CsMc-E:osreliso:hiDelta}   {\ensuremath{{+0.083 } } }
\vdef{default-11:CsMc-E:osreliso:hiDeltaE}   {\ensuremath{{0.018 } } }
\vdef{default-11:CsData-E:osmuonpt:loEff}   {\ensuremath{{0.000 } } }
\vdef{default-11:CsData-E:osmuonpt:loEffE}   {\ensuremath{{0.009 } } }
\vdef{default-11:CsData-E:osmuonpt:hiEff}   {\ensuremath{{1.000 } } }
\vdef{default-11:CsData-E:osmuonpt:hiEffE}   {\ensuremath{{0.009 } } }
\vdef{default-11:CsMc-E:osmuonpt:loEff}   {\ensuremath{{0.000 } } }
\vdef{default-11:CsMc-E:osmuonpt:loEffE}   {\ensuremath{{0.009 } } }
\vdef{default-11:CsMc-E:osmuonpt:hiEff}   {\ensuremath{{1.000 } } }
\vdef{default-11:CsMc-E:osmuonpt:hiEffE}   {\ensuremath{{0.009 } } }
\vdef{default-11:CsMc-E:osmuonpt:loDelta}   {\ensuremath{{\mathrm{NaN} } } }
\vdef{default-11:CsMc-E:osmuonpt:loDeltaE}   {\ensuremath{{\mathrm{NaN} } } }
\vdef{default-11:CsMc-E:osmuonpt:hiDelta}   {\ensuremath{{+0.000 } } }
\vdef{default-11:CsMc-E:osmuonpt:hiDeltaE}   {\ensuremath{{0.012 } } }
\vdef{default-11:CsData-E:osmuondr:loEff}   {\ensuremath{{0.012 } } }
\vdef{default-11:CsData-E:osmuondr:loEffE}   {\ensuremath{{0.012 } } }
\vdef{default-11:CsData-E:osmuondr:hiEff}   {\ensuremath{{0.988 } } }
\vdef{default-11:CsData-E:osmuondr:hiEffE}   {\ensuremath{{0.012 } } }
\vdef{default-11:CsMc-E:osmuondr:loEff}   {\ensuremath{{0.052 } } }
\vdef{default-11:CsMc-E:osmuondr:loEffE}   {\ensuremath{{0.021 } } }
\vdef{default-11:CsMc-E:osmuondr:hiEff}   {\ensuremath{{0.948 } } }
\vdef{default-11:CsMc-E:osmuondr:hiEffE}   {\ensuremath{{0.023 } } }
\vdef{default-11:CsMc-E:osmuondr:loDelta}   {\ensuremath{{-1.272 } } }
\vdef{default-11:CsMc-E:osmuondr:loDeltaE}   {\ensuremath{{0.675 } } }
\vdef{default-11:CsMc-E:osmuondr:hiDelta}   {\ensuremath{{+0.042 } } }
\vdef{default-11:CsMc-E:osmuondr:hiDeltaE}   {\ensuremath{{0.027 } } }
\vdef{default-11:CsData-E:hlt:loEff}   {\ensuremath{{0.042 } } }
\vdef{default-11:CsData-E:hlt:loEffE}   {\ensuremath{{0.004 } } }
\vdef{default-11:CsData-E:hlt:hiEff}   {\ensuremath{{0.958 } } }
\vdef{default-11:CsData-E:hlt:hiEffE}   {\ensuremath{{0.004 } } }
\vdef{default-11:CsMc-E:hlt:loEff}   {\ensuremath{{0.441 } } }
\vdef{default-11:CsMc-E:hlt:loEffE}   {\ensuremath{{0.010 } } }
\vdef{default-11:CsMc-E:hlt:hiEff}   {\ensuremath{{0.559 } } }
\vdef{default-11:CsMc-E:hlt:hiEffE}   {\ensuremath{{0.010 } } }
\vdef{default-11:CsMc-E:hlt:loDelta}   {\ensuremath{{-1.655 } } }
\vdef{default-11:CsMc-E:hlt:loDeltaE}   {\ensuremath{{0.032 } } }
\vdef{default-11:CsMc-E:hlt:hiDelta}   {\ensuremath{{+0.526 } } }
\vdef{default-11:CsMc-E:hlt:hiDeltaE}   {\ensuremath{{0.017 } } }
\vdef{default-11:CsData-E:muonsid:loEff}   {\ensuremath{{0.144 } } }
\vdef{default-11:CsData-E:muonsid:loEffE}   {\ensuremath{{0.007 } } }
\vdef{default-11:CsData-E:muonsid:hiEff}   {\ensuremath{{0.856 } } }
\vdef{default-11:CsData-E:muonsid:hiEffE}   {\ensuremath{{0.007 } } }
\vdef{default-11:CsMc-E:muonsid:loEff}   {\ensuremath{{0.138 } } }
\vdef{default-11:CsMc-E:muonsid:loEffE}   {\ensuremath{{0.007 } } }
\vdef{default-11:CsMc-E:muonsid:hiEff}   {\ensuremath{{0.862 } } }
\vdef{default-11:CsMc-E:muonsid:hiEffE}   {\ensuremath{{0.007 } } }
\vdef{default-11:CsMc-E:muonsid:loDelta}   {\ensuremath{{+0.040 } } }
\vdef{default-11:CsMc-E:muonsid:loDeltaE}   {\ensuremath{{0.068 } } }
\vdef{default-11:CsMc-E:muonsid:hiDelta}   {\ensuremath{{-0.007 } } }
\vdef{default-11:CsMc-E:muonsid:hiDeltaE}   {\ensuremath{{0.011 } } }
\vdef{default-11:CsData-E:tracksqual:loEff}   {\ensuremath{{0.001 } } }
\vdef{default-11:CsData-E:tracksqual:loEffE}   {\ensuremath{{0.001 } } }
\vdef{default-11:CsData-E:tracksqual:hiEff}   {\ensuremath{{0.999 } } }
\vdef{default-11:CsData-E:tracksqual:hiEffE}   {\ensuremath{{0.001 } } }
\vdef{default-11:CsMc-E:tracksqual:loEff}   {\ensuremath{{0.000 } } }
\vdef{default-11:CsMc-E:tracksqual:loEffE}   {\ensuremath{{0.000 } } }
\vdef{default-11:CsMc-E:tracksqual:hiEff}   {\ensuremath{{1.000 } } }
\vdef{default-11:CsMc-E:tracksqual:hiEffE}   {\ensuremath{{0.000 } } }
\vdef{default-11:CsMc-E:tracksqual:loDelta}   {\ensuremath{{+2.000 } } }
\vdef{default-11:CsMc-E:tracksqual:loDeltaE}   {\ensuremath{{1.332 } } }
\vdef{default-11:CsMc-E:tracksqual:hiDelta}   {\ensuremath{{-0.001 } } }
\vdef{default-11:CsMc-E:tracksqual:hiDeltaE}   {\ensuremath{{0.001 } } }
\vdef{default-11:CsData-E:pvz:loEff}   {\ensuremath{{0.496 } } }
\vdef{default-11:CsData-E:pvz:loEffE}   {\ensuremath{{0.010 } } }
\vdef{default-11:CsData-E:pvz:hiEff}   {\ensuremath{{0.504 } } }
\vdef{default-11:CsData-E:pvz:hiEffE}   {\ensuremath{{0.010 } } }
\vdef{default-11:CsMc-E:pvz:loEff}   {\ensuremath{{0.455 } } }
\vdef{default-11:CsMc-E:pvz:loEffE}   {\ensuremath{{0.010 } } }
\vdef{default-11:CsMc-E:pvz:hiEff}   {\ensuremath{{0.545 } } }
\vdef{default-11:CsMc-E:pvz:hiEffE}   {\ensuremath{{0.010 } } }
\vdef{default-11:CsMc-E:pvz:loDelta}   {\ensuremath{{+0.087 } } }
\vdef{default-11:CsMc-E:pvz:loDeltaE}   {\ensuremath{{0.031 } } }
\vdef{default-11:CsMc-E:pvz:hiDelta}   {\ensuremath{{-0.079 } } }
\vdef{default-11:CsMc-E:pvz:hiDeltaE}   {\ensuremath{{0.028 } } }
\vdef{default-11:CsData-E:pvn:loEff}   {\ensuremath{{1.007 } } }
\vdef{default-11:CsData-E:pvn:loEffE}   {\ensuremath{{\mathrm{NaN} } } }
\vdef{default-11:CsData-E:pvn:hiEff}   {\ensuremath{{1.000 } } }
\vdef{default-11:CsData-E:pvn:hiEffE}   {\ensuremath{{0.000 } } }
\vdef{default-11:CsMc-E:pvn:loEff}   {\ensuremath{{1.000 } } }
\vdef{default-11:CsMc-E:pvn:loEffE}   {\ensuremath{{0.000 } } }
\vdef{default-11:CsMc-E:pvn:hiEff}   {\ensuremath{{1.000 } } }
\vdef{default-11:CsMc-E:pvn:hiEffE}   {\ensuremath{{0.000 } } }
\vdef{default-11:CsMc-E:pvn:loDelta}   {\ensuremath{{+0.007 } } }
\vdef{default-11:CsMc-E:pvn:loDeltaE}   {\ensuremath{{\mathrm{NaN} } } }
\vdef{default-11:CsMc-E:pvn:hiDelta}   {\ensuremath{{+0.000 } } }
\vdef{default-11:CsMc-E:pvn:hiDeltaE}   {\ensuremath{{0.001 } } }
\vdef{default-11:CsData-E:pvavew8:loEff}   {\ensuremath{{0.021 } } }
\vdef{default-11:CsData-E:pvavew8:loEffE}   {\ensuremath{{0.003 } } }
\vdef{default-11:CsData-E:pvavew8:hiEff}   {\ensuremath{{0.979 } } }
\vdef{default-11:CsData-E:pvavew8:hiEffE}   {\ensuremath{{0.003 } } }
\vdef{default-11:CsMc-E:pvavew8:loEff}   {\ensuremath{{0.013 } } }
\vdef{default-11:CsMc-E:pvavew8:loEffE}   {\ensuremath{{0.002 } } }
\vdef{default-11:CsMc-E:pvavew8:hiEff}   {\ensuremath{{0.987 } } }
\vdef{default-11:CsMc-E:pvavew8:hiEffE}   {\ensuremath{{0.002 } } }
\vdef{default-11:CsMc-E:pvavew8:loDelta}   {\ensuremath{{+0.446 } } }
\vdef{default-11:CsMc-E:pvavew8:loDeltaE}   {\ensuremath{{0.222 } } }
\vdef{default-11:CsMc-E:pvavew8:hiDelta}   {\ensuremath{{-0.008 } } }
\vdef{default-11:CsMc-E:pvavew8:hiDeltaE}   {\ensuremath{{0.004 } } }
\vdef{default-11:CsData-E:pvntrk:loEff}   {\ensuremath{{1.000 } } }
\vdef{default-11:CsData-E:pvntrk:loEffE}   {\ensuremath{{0.000 } } }
\vdef{default-11:CsData-E:pvntrk:hiEff}   {\ensuremath{{1.000 } } }
\vdef{default-11:CsData-E:pvntrk:hiEffE}   {\ensuremath{{0.000 } } }
\vdef{default-11:CsMc-E:pvntrk:loEff}   {\ensuremath{{1.000 } } }
\vdef{default-11:CsMc-E:pvntrk:loEffE}   {\ensuremath{{0.000 } } }
\vdef{default-11:CsMc-E:pvntrk:hiEff}   {\ensuremath{{1.000 } } }
\vdef{default-11:CsMc-E:pvntrk:hiEffE}   {\ensuremath{{0.000 } } }
\vdef{default-11:CsMc-E:pvntrk:loDelta}   {\ensuremath{{+0.000 } } }
\vdef{default-11:CsMc-E:pvntrk:loDeltaE}   {\ensuremath{{0.001 } } }
\vdef{default-11:CsMc-E:pvntrk:hiDelta}   {\ensuremath{{+0.000 } } }
\vdef{default-11:CsMc-E:pvntrk:hiDeltaE}   {\ensuremath{{0.001 } } }
\vdef{default-11:CsData-E:muon1pt:loEff}   {\ensuremath{{1.003 } } }
\vdef{default-11:CsData-E:muon1pt:loEffE}   {\ensuremath{{\mathrm{NaN} } } }
\vdef{default-11:CsData-E:muon1pt:hiEff}   {\ensuremath{{1.000 } } }
\vdef{default-11:CsData-E:muon1pt:hiEffE}   {\ensuremath{{0.000 } } }
\vdef{default-11:CsMc-E:muon1pt:loEff}   {\ensuremath{{1.003 } } }
\vdef{default-11:CsMc-E:muon1pt:loEffE}   {\ensuremath{{\mathrm{NaN} } } }
\vdef{default-11:CsMc-E:muon1pt:hiEff}   {\ensuremath{{1.000 } } }
\vdef{default-11:CsMc-E:muon1pt:hiEffE}   {\ensuremath{{0.000 } } }
\vdef{default-11:CsMc-E:muon1pt:loDelta}   {\ensuremath{{+0.000 } } }
\vdef{default-11:CsMc-E:muon1pt:loDeltaE}   {\ensuremath{{\mathrm{NaN} } } }
\vdef{default-11:CsMc-E:muon1pt:hiDelta}   {\ensuremath{{+0.000 } } }
\vdef{default-11:CsMc-E:muon1pt:hiDeltaE}   {\ensuremath{{0.001 } } }
\vdef{default-11:CsData-E:muon2pt:loEff}   {\ensuremath{{0.148 } } }
\vdef{default-11:CsData-E:muon2pt:loEffE}   {\ensuremath{{0.007 } } }
\vdef{default-11:CsData-E:muon2pt:hiEff}   {\ensuremath{{0.852 } } }
\vdef{default-11:CsData-E:muon2pt:hiEffE}   {\ensuremath{{0.007 } } }
\vdef{default-11:CsMc-E:muon2pt:loEff}   {\ensuremath{{0.138 } } }
\vdef{default-11:CsMc-E:muon2pt:loEffE}   {\ensuremath{{0.007 } } }
\vdef{default-11:CsMc-E:muon2pt:hiEff}   {\ensuremath{{0.862 } } }
\vdef{default-11:CsMc-E:muon2pt:hiEffE}   {\ensuremath{{0.007 } } }
\vdef{default-11:CsMc-E:muon2pt:loDelta}   {\ensuremath{{+0.073 } } }
\vdef{default-11:CsMc-E:muon2pt:loDeltaE}   {\ensuremath{{0.067 } } }
\vdef{default-11:CsMc-E:muon2pt:hiDelta}   {\ensuremath{{-0.012 } } }
\vdef{default-11:CsMc-E:muon2pt:hiDeltaE}   {\ensuremath{{0.011 } } }
\vdef{default-11:CsData-E:muonseta:loEff}   {\ensuremath{{0.495 } } }
\vdef{default-11:CsData-E:muonseta:loEffE}   {\ensuremath{{0.007 } } }
\vdef{default-11:CsData-E:muonseta:hiEff}   {\ensuremath{{0.505 } } }
\vdef{default-11:CsData-E:muonseta:hiEffE}   {\ensuremath{{0.007 } } }
\vdef{default-11:CsMc-E:muonseta:loEff}   {\ensuremath{{0.522 } } }
\vdef{default-11:CsMc-E:muonseta:loEffE}   {\ensuremath{{0.007 } } }
\vdef{default-11:CsMc-E:muonseta:hiEff}   {\ensuremath{{0.478 } } }
\vdef{default-11:CsMc-E:muonseta:hiEffE}   {\ensuremath{{0.007 } } }
\vdef{default-11:CsMc-E:muonseta:loDelta}   {\ensuremath{{-0.053 } } }
\vdef{default-11:CsMc-E:muonseta:loDeltaE}   {\ensuremath{{0.021 } } }
\vdef{default-11:CsMc-E:muonseta:hiDelta}   {\ensuremath{{+0.055 } } }
\vdef{default-11:CsMc-E:muonseta:hiDeltaE}   {\ensuremath{{0.021 } } }
\vdef{default-11:CsData-E:pt:loEff}   {\ensuremath{{0.000 } } }
\vdef{default-11:CsData-E:pt:loEffE}   {\ensuremath{{0.000 } } }
\vdef{default-11:CsData-E:pt:hiEff}   {\ensuremath{{1.000 } } }
\vdef{default-11:CsData-E:pt:hiEffE}   {\ensuremath{{0.000 } } }
\vdef{default-11:CsMc-E:pt:loEff}   {\ensuremath{{0.000 } } }
\vdef{default-11:CsMc-E:pt:loEffE}   {\ensuremath{{0.000 } } }
\vdef{default-11:CsMc-E:pt:hiEff}   {\ensuremath{{1.000 } } }
\vdef{default-11:CsMc-E:pt:hiEffE}   {\ensuremath{{0.000 } } }
\vdef{default-11:CsMc-E:pt:loDelta}   {\ensuremath{{\mathrm{NaN} } } }
\vdef{default-11:CsMc-E:pt:loDeltaE}   {\ensuremath{{\mathrm{NaN} } } }
\vdef{default-11:CsMc-E:pt:hiDelta}   {\ensuremath{{+0.000 } } }
\vdef{default-11:CsMc-E:pt:hiDeltaE}   {\ensuremath{{0.000 } } }
\vdef{default-11:CsData-E:p:loEff}   {\ensuremath{{1.075 } } }
\vdef{default-11:CsData-E:p:loEffE}   {\ensuremath{{\mathrm{NaN} } } }
\vdef{default-11:CsData-E:p:hiEff}   {\ensuremath{{1.000 } } }
\vdef{default-11:CsData-E:p:hiEffE}   {\ensuremath{{0.000 } } }
\vdef{default-11:CsMc-E:p:loEff}   {\ensuremath{{1.059 } } }
\vdef{default-11:CsMc-E:p:loEffE}   {\ensuremath{{\mathrm{NaN} } } }
\vdef{default-11:CsMc-E:p:hiEff}   {\ensuremath{{1.000 } } }
\vdef{default-11:CsMc-E:p:hiEffE}   {\ensuremath{{0.000 } } }
\vdef{default-11:CsMc-E:p:loDelta}   {\ensuremath{{+0.015 } } }
\vdef{default-11:CsMc-E:p:loDeltaE}   {\ensuremath{{\mathrm{NaN} } } }
\vdef{default-11:CsMc-E:p:hiDelta}   {\ensuremath{{+0.000 } } }
\vdef{default-11:CsMc-E:p:hiDeltaE}   {\ensuremath{{0.001 } } }
\vdef{default-11:CsData-E:eta:loEff}   {\ensuremath{{0.494 } } }
\vdef{default-11:CsData-E:eta:loEffE}   {\ensuremath{{0.011 } } }
\vdef{default-11:CsData-E:eta:hiEff}   {\ensuremath{{0.506 } } }
\vdef{default-11:CsData-E:eta:hiEffE}   {\ensuremath{{0.011 } } }
\vdef{default-11:CsMc-E:eta:loEff}   {\ensuremath{{0.522 } } }
\vdef{default-11:CsMc-E:eta:loEffE}   {\ensuremath{{0.011 } } }
\vdef{default-11:CsMc-E:eta:hiEff}   {\ensuremath{{0.478 } } }
\vdef{default-11:CsMc-E:eta:hiEffE}   {\ensuremath{{0.011 } } }
\vdef{default-11:CsMc-E:eta:loDelta}   {\ensuremath{{-0.054 } } }
\vdef{default-11:CsMc-E:eta:loDeltaE}   {\ensuremath{{0.029 } } }
\vdef{default-11:CsMc-E:eta:hiDelta}   {\ensuremath{{+0.056 } } }
\vdef{default-11:CsMc-E:eta:hiDeltaE}   {\ensuremath{{0.030 } } }
\vdef{default-11:CsData-E:bdt:loEff}   {\ensuremath{{1.000 } } }
\vdef{default-11:CsData-E:bdt:loEffE}   {\ensuremath{{0.000 } } }
\vdef{default-11:CsData-E:bdt:hiEff}   {\ensuremath{{0.000 } } }
\vdef{default-11:CsData-E:bdt:hiEffE}   {\ensuremath{{0.000 } } }
\vdef{default-11:CsMc-E:bdt:loEff}   {\ensuremath{{1.000 } } }
\vdef{default-11:CsMc-E:bdt:loEffE}   {\ensuremath{{0.001 } } }
\vdef{default-11:CsMc-E:bdt:hiEff}   {\ensuremath{{0.000 } } }
\vdef{default-11:CsMc-E:bdt:hiEffE}   {\ensuremath{{0.000 } } }
\vdef{default-11:CsMc-E:bdt:loDelta}   {\ensuremath{{+0.000 } } }
\vdef{default-11:CsMc-E:bdt:loDeltaE}   {\ensuremath{{0.001 } } }
\vdef{default-11:CsMc-E:bdt:hiDelta}   {\ensuremath{{-2.000 } } }
\vdef{default-11:CsMc-E:bdt:hiDeltaE}   {\ensuremath{{4.845 } } }
\vdef{default-11:CsData-E:fl3d:loEff}   {\ensuremath{{0.658 } } }
\vdef{default-11:CsData-E:fl3d:loEffE}   {\ensuremath{{0.009 } } }
\vdef{default-11:CsData-E:fl3d:hiEff}   {\ensuremath{{0.342 } } }
\vdef{default-11:CsData-E:fl3d:hiEffE}   {\ensuremath{{0.009 } } }
\vdef{default-11:CsMc-E:fl3d:loEff}   {\ensuremath{{0.679 } } }
\vdef{default-11:CsMc-E:fl3d:loEffE}   {\ensuremath{{0.009 } } }
\vdef{default-11:CsMc-E:fl3d:hiEff}   {\ensuremath{{0.321 } } }
\vdef{default-11:CsMc-E:fl3d:hiEffE}   {\ensuremath{{0.009 } } }
\vdef{default-11:CsMc-E:fl3d:loDelta}   {\ensuremath{{-0.032 } } }
\vdef{default-11:CsMc-E:fl3d:loDeltaE}   {\ensuremath{{0.018 } } }
\vdef{default-11:CsMc-E:fl3d:hiDelta}   {\ensuremath{{+0.064 } } }
\vdef{default-11:CsMc-E:fl3d:hiDeltaE}   {\ensuremath{{0.037 } } }
\vdef{default-11:CsData-E:fl3de:loEff}   {\ensuremath{{1.000 } } }
\vdef{default-11:CsData-E:fl3de:loEffE}   {\ensuremath{{0.000 } } }
\vdef{default-11:CsData-E:fl3de:hiEff}   {\ensuremath{{0.000 } } }
\vdef{default-11:CsData-E:fl3de:hiEffE}   {\ensuremath{{0.000 } } }
\vdef{default-11:CsMc-E:fl3de:loEff}   {\ensuremath{{1.000 } } }
\vdef{default-11:CsMc-E:fl3de:loEffE}   {\ensuremath{{0.000 } } }
\vdef{default-11:CsMc-E:fl3de:hiEff}   {\ensuremath{{0.000 } } }
\vdef{default-11:CsMc-E:fl3de:hiEffE}   {\ensuremath{{0.000 } } }
\vdef{default-11:CsMc-E:fl3de:loDelta}   {\ensuremath{{+0.000 } } }
\vdef{default-11:CsMc-E:fl3de:loDeltaE}   {\ensuremath{{0.000 } } }
\vdef{default-11:CsMc-E:fl3de:hiDelta}   {\ensuremath{{+0.479 } } }
\vdef{default-11:CsMc-E:fl3de:hiDeltaE}   {\ensuremath{{1.365 } } }
\vdef{default-11:CsData-E:fls3d:loEff}   {\ensuremath{{0.181 } } }
\vdef{default-11:CsData-E:fls3d:loEffE}   {\ensuremath{{0.007 } } }
\vdef{default-11:CsData-E:fls3d:hiEff}   {\ensuremath{{0.819 } } }
\vdef{default-11:CsData-E:fls3d:hiEffE}   {\ensuremath{{0.007 } } }
\vdef{default-11:CsMc-E:fls3d:loEff}   {\ensuremath{{0.174 } } }
\vdef{default-11:CsMc-E:fls3d:loEffE}   {\ensuremath{{0.007 } } }
\vdef{default-11:CsMc-E:fls3d:hiEff}   {\ensuremath{{0.826 } } }
\vdef{default-11:CsMc-E:fls3d:hiEffE}   {\ensuremath{{0.007 } } }
\vdef{default-11:CsMc-E:fls3d:loDelta}   {\ensuremath{{+0.036 } } }
\vdef{default-11:CsMc-E:fls3d:loDeltaE}   {\ensuremath{{0.056 } } }
\vdef{default-11:CsMc-E:fls3d:hiDelta}   {\ensuremath{{-0.008 } } }
\vdef{default-11:CsMc-E:fls3d:hiDeltaE}   {\ensuremath{{0.012 } } }
\vdef{default-11:CsData-E:flsxy:loEff}   {\ensuremath{{1.007 } } }
\vdef{default-11:CsData-E:flsxy:loEffE}   {\ensuremath{{\mathrm{NaN} } } }
\vdef{default-11:CsData-E:flsxy:hiEff}   {\ensuremath{{1.000 } } }
\vdef{default-11:CsData-E:flsxy:hiEffE}   {\ensuremath{{0.000 } } }
\vdef{default-11:CsMc-E:flsxy:loEff}   {\ensuremath{{1.004 } } }
\vdef{default-11:CsMc-E:flsxy:loEffE}   {\ensuremath{{\mathrm{NaN} } } }
\vdef{default-11:CsMc-E:flsxy:hiEff}   {\ensuremath{{1.000 } } }
\vdef{default-11:CsMc-E:flsxy:hiEffE}   {\ensuremath{{0.000 } } }
\vdef{default-11:CsMc-E:flsxy:loDelta}   {\ensuremath{{+0.003 } } }
\vdef{default-11:CsMc-E:flsxy:loDeltaE}   {\ensuremath{{\mathrm{NaN} } } }
\vdef{default-11:CsMc-E:flsxy:hiDelta}   {\ensuremath{{+0.000 } } }
\vdef{default-11:CsMc-E:flsxy:hiDeltaE}   {\ensuremath{{0.000 } } }
\vdef{default-11:CsData-E:chi2dof:loEff}   {\ensuremath{{0.915 } } }
\vdef{default-11:CsData-E:chi2dof:loEffE}   {\ensuremath{{0.006 } } }
\vdef{default-11:CsData-E:chi2dof:hiEff}   {\ensuremath{{0.085 } } }
\vdef{default-11:CsData-E:chi2dof:hiEffE}   {\ensuremath{{0.006 } } }
\vdef{default-11:CsMc-E:chi2dof:loEff}   {\ensuremath{{0.939 } } }
\vdef{default-11:CsMc-E:chi2dof:loEffE}   {\ensuremath{{0.005 } } }
\vdef{default-11:CsMc-E:chi2dof:hiEff}   {\ensuremath{{0.061 } } }
\vdef{default-11:CsMc-E:chi2dof:hiEffE}   {\ensuremath{{0.005 } } }
\vdef{default-11:CsMc-E:chi2dof:loDelta}   {\ensuremath{{-0.026 } } }
\vdef{default-11:CsMc-E:chi2dof:loDeltaE}   {\ensuremath{{0.008 } } }
\vdef{default-11:CsMc-E:chi2dof:hiDelta}   {\ensuremath{{+0.325 } } }
\vdef{default-11:CsMc-E:chi2dof:hiDeltaE}   {\ensuremath{{0.102 } } }
\vdef{default-11:CsData-E:pchi2dof:loEff}   {\ensuremath{{0.620 } } }
\vdef{default-11:CsData-E:pchi2dof:loEffE}   {\ensuremath{{0.010 } } }
\vdef{default-11:CsData-E:pchi2dof:hiEff}   {\ensuremath{{0.380 } } }
\vdef{default-11:CsData-E:pchi2dof:hiEffE}   {\ensuremath{{0.010 } } }
\vdef{default-11:CsMc-E:pchi2dof:loEff}   {\ensuremath{{0.580 } } }
\vdef{default-11:CsMc-E:pchi2dof:loEffE}   {\ensuremath{{0.010 } } }
\vdef{default-11:CsMc-E:pchi2dof:hiEff}   {\ensuremath{{0.420 } } }
\vdef{default-11:CsMc-E:pchi2dof:hiEffE}   {\ensuremath{{0.010 } } }
\vdef{default-11:CsMc-E:pchi2dof:loDelta}   {\ensuremath{{+0.066 } } }
\vdef{default-11:CsMc-E:pchi2dof:loDeltaE}   {\ensuremath{{0.023 } } }
\vdef{default-11:CsMc-E:pchi2dof:hiDelta}   {\ensuremath{{-0.099 } } }
\vdef{default-11:CsMc-E:pchi2dof:hiDeltaE}   {\ensuremath{{0.034 } } }
\vdef{default-11:CsData-E:alpha:loEff}   {\ensuremath{{0.998 } } }
\vdef{default-11:CsData-E:alpha:loEffE}   {\ensuremath{{0.001 } } }
\vdef{default-11:CsData-E:alpha:hiEff}   {\ensuremath{{0.002 } } }
\vdef{default-11:CsData-E:alpha:hiEffE}   {\ensuremath{{0.001 } } }
\vdef{default-11:CsMc-E:alpha:loEff}   {\ensuremath{{0.999 } } }
\vdef{default-11:CsMc-E:alpha:loEffE}   {\ensuremath{{0.001 } } }
\vdef{default-11:CsMc-E:alpha:hiEff}   {\ensuremath{{0.001 } } }
\vdef{default-11:CsMc-E:alpha:hiEffE}   {\ensuremath{{0.001 } } }
\vdef{default-11:CsMc-E:alpha:loDelta}   {\ensuremath{{-0.002 } } }
\vdef{default-11:CsMc-E:alpha:loDeltaE}   {\ensuremath{{0.001 } } }
\vdef{default-11:CsMc-E:alpha:hiDelta}   {\ensuremath{{+1.118 } } }
\vdef{default-11:CsMc-E:alpha:hiDeltaE}   {\ensuremath{{0.765 } } }
\vdef{default-11:CsData-E:iso:loEff}   {\ensuremath{{0.084 } } }
\vdef{default-11:CsData-E:iso:loEffE}   {\ensuremath{{0.006 } } }
\vdef{default-11:CsData-E:iso:hiEff}   {\ensuremath{{0.916 } } }
\vdef{default-11:CsData-E:iso:hiEffE}   {\ensuremath{{0.006 } } }
\vdef{default-11:CsMc-E:iso:loEff}   {\ensuremath{{0.072 } } }
\vdef{default-11:CsMc-E:iso:loEffE}   {\ensuremath{{0.005 } } }
\vdef{default-11:CsMc-E:iso:hiEff}   {\ensuremath{{0.928 } } }
\vdef{default-11:CsMc-E:iso:hiEffE}   {\ensuremath{{0.005 } } }
\vdef{default-11:CsMc-E:iso:loDelta}   {\ensuremath{{+0.147 } } }
\vdef{default-11:CsMc-E:iso:loDeltaE}   {\ensuremath{{0.098 } } }
\vdef{default-11:CsMc-E:iso:hiDelta}   {\ensuremath{{-0.012 } } }
\vdef{default-11:CsMc-E:iso:hiDeltaE}   {\ensuremath{{0.008 } } }
\vdef{default-11:CsData-E:docatrk:loEff}   {\ensuremath{{0.071 } } }
\vdef{default-11:CsData-E:docatrk:loEffE}   {\ensuremath{{0.005 } } }
\vdef{default-11:CsData-E:docatrk:hiEff}   {\ensuremath{{0.929 } } }
\vdef{default-11:CsData-E:docatrk:hiEffE}   {\ensuremath{{0.005 } } }
\vdef{default-11:CsMc-E:docatrk:loEff}   {\ensuremath{{0.072 } } }
\vdef{default-11:CsMc-E:docatrk:loEffE}   {\ensuremath{{0.005 } } }
\vdef{default-11:CsMc-E:docatrk:hiEff}   {\ensuremath{{0.928 } } }
\vdef{default-11:CsMc-E:docatrk:hiEffE}   {\ensuremath{{0.005 } } }
\vdef{default-11:CsMc-E:docatrk:loDelta}   {\ensuremath{{-0.015 } } }
\vdef{default-11:CsMc-E:docatrk:loDeltaE}   {\ensuremath{{0.105 } } }
\vdef{default-11:CsMc-E:docatrk:hiDelta}   {\ensuremath{{+0.001 } } }
\vdef{default-11:CsMc-E:docatrk:hiDeltaE}   {\ensuremath{{0.008 } } }
\vdef{default-11:CsData-E:isotrk:loEff}   {\ensuremath{{1.000 } } }
\vdef{default-11:CsData-E:isotrk:loEffE}   {\ensuremath{{0.000 } } }
\vdef{default-11:CsData-E:isotrk:hiEff}   {\ensuremath{{1.000 } } }
\vdef{default-11:CsData-E:isotrk:hiEffE}   {\ensuremath{{0.000 } } }
\vdef{default-11:CsMc-E:isotrk:loEff}   {\ensuremath{{1.000 } } }
\vdef{default-11:CsMc-E:isotrk:loEffE}   {\ensuremath{{0.000 } } }
\vdef{default-11:CsMc-E:isotrk:hiEff}   {\ensuremath{{1.000 } } }
\vdef{default-11:CsMc-E:isotrk:hiEffE}   {\ensuremath{{0.000 } } }
\vdef{default-11:CsMc-E:isotrk:loDelta}   {\ensuremath{{+0.000 } } }
\vdef{default-11:CsMc-E:isotrk:loDeltaE}   {\ensuremath{{0.001 } } }
\vdef{default-11:CsMc-E:isotrk:hiDelta}   {\ensuremath{{+0.000 } } }
\vdef{default-11:CsMc-E:isotrk:hiDeltaE}   {\ensuremath{{0.001 } } }
\vdef{default-11:CsData-E:closetrk:loEff}   {\ensuremath{{0.984 } } }
\vdef{default-11:CsData-E:closetrk:loEffE}   {\ensuremath{{0.003 } } }
\vdef{default-11:CsData-E:closetrk:hiEff}   {\ensuremath{{0.016 } } }
\vdef{default-11:CsData-E:closetrk:hiEffE}   {\ensuremath{{0.003 } } }
\vdef{default-11:CsMc-E:closetrk:loEff}   {\ensuremath{{0.983 } } }
\vdef{default-11:CsMc-E:closetrk:loEffE}   {\ensuremath{{0.003 } } }
\vdef{default-11:CsMc-E:closetrk:hiEff}   {\ensuremath{{0.017 } } }
\vdef{default-11:CsMc-E:closetrk:hiEffE}   {\ensuremath{{0.003 } } }
\vdef{default-11:CsMc-E:closetrk:loDelta}   {\ensuremath{{+0.002 } } }
\vdef{default-11:CsMc-E:closetrk:loDeltaE}   {\ensuremath{{0.004 } } }
\vdef{default-11:CsMc-E:closetrk:hiDelta}   {\ensuremath{{-0.096 } } }
\vdef{default-11:CsMc-E:closetrk:hiDeltaE}   {\ensuremath{{0.228 } } }
\vdef{default-11:CsData-E:lip:loEff}   {\ensuremath{{1.000 } } }
\vdef{default-11:CsData-E:lip:loEffE}   {\ensuremath{{0.000 } } }
\vdef{default-11:CsData-E:lip:hiEff}   {\ensuremath{{0.000 } } }
\vdef{default-11:CsData-E:lip:hiEffE}   {\ensuremath{{0.000 } } }
\vdef{default-11:CsMc-E:lip:loEff}   {\ensuremath{{1.000 } } }
\vdef{default-11:CsMc-E:lip:loEffE}   {\ensuremath{{0.000 } } }
\vdef{default-11:CsMc-E:lip:hiEff}   {\ensuremath{{0.000 } } }
\vdef{default-11:CsMc-E:lip:hiEffE}   {\ensuremath{{0.000 } } }
\vdef{default-11:CsMc-E:lip:loDelta}   {\ensuremath{{+0.000 } } }
\vdef{default-11:CsMc-E:lip:loDeltaE}   {\ensuremath{{0.001 } } }
\vdef{default-11:CsMc-E:lip:hiDelta}   {\ensuremath{{\mathrm{NaN} } } }
\vdef{default-11:CsMc-E:lip:hiDeltaE}   {\ensuremath{{\mathrm{NaN} } } }
\vdef{default-11:CsData-E:lip:inEff}   {\ensuremath{{1.000 } } }
\vdef{default-11:CsData-E:lip:inEffE}   {\ensuremath{{0.000 } } }
\vdef{default-11:CsMc-E:lip:inEff}   {\ensuremath{{1.000 } } }
\vdef{default-11:CsMc-E:lip:inEffE}   {\ensuremath{{0.000 } } }
\vdef{default-11:CsMc-E:lip:inDelta}   {\ensuremath{{+0.000 } } }
\vdef{default-11:CsMc-E:lip:inDeltaE}   {\ensuremath{{0.001 } } }
\vdef{default-11:CsData-E:lips:loEff}   {\ensuremath{{1.000 } } }
\vdef{default-11:CsData-E:lips:loEffE}   {\ensuremath{{0.000 } } }
\vdef{default-11:CsData-E:lips:hiEff}   {\ensuremath{{0.000 } } }
\vdef{default-11:CsData-E:lips:hiEffE}   {\ensuremath{{0.000 } } }
\vdef{default-11:CsMc-E:lips:loEff}   {\ensuremath{{1.000 } } }
\vdef{default-11:CsMc-E:lips:loEffE}   {\ensuremath{{0.000 } } }
\vdef{default-11:CsMc-E:lips:hiEff}   {\ensuremath{{0.000 } } }
\vdef{default-11:CsMc-E:lips:hiEffE}   {\ensuremath{{0.000 } } }
\vdef{default-11:CsMc-E:lips:loDelta}   {\ensuremath{{+0.000 } } }
\vdef{default-11:CsMc-E:lips:loDeltaE}   {\ensuremath{{0.001 } } }
\vdef{default-11:CsMc-E:lips:hiDelta}   {\ensuremath{{\mathrm{NaN} } } }
\vdef{default-11:CsMc-E:lips:hiDeltaE}   {\ensuremath{{\mathrm{NaN} } } }
\vdef{default-11:CsData-E:lips:inEff}   {\ensuremath{{1.000 } } }
\vdef{default-11:CsData-E:lips:inEffE}   {\ensuremath{{0.000 } } }
\vdef{default-11:CsMc-E:lips:inEff}   {\ensuremath{{1.000 } } }
\vdef{default-11:CsMc-E:lips:inEffE}   {\ensuremath{{0.000 } } }
\vdef{default-11:CsMc-E:lips:inDelta}   {\ensuremath{{+0.000 } } }
\vdef{default-11:CsMc-E:lips:inDeltaE}   {\ensuremath{{0.001 } } }
\vdef{default-11:CsData-E:ip:loEff}   {\ensuremath{{0.979 } } }
\vdef{default-11:CsData-E:ip:loEffE}   {\ensuremath{{0.003 } } }
\vdef{default-11:CsData-E:ip:hiEff}   {\ensuremath{{0.021 } } }
\vdef{default-11:CsData-E:ip:hiEffE}   {\ensuremath{{0.003 } } }
\vdef{default-11:CsMc-E:ip:loEff}   {\ensuremath{{0.976 } } }
\vdef{default-11:CsMc-E:ip:loEffE}   {\ensuremath{{0.003 } } }
\vdef{default-11:CsMc-E:ip:hiEff}   {\ensuremath{{0.024 } } }
\vdef{default-11:CsMc-E:ip:hiEffE}   {\ensuremath{{0.003 } } }
\vdef{default-11:CsMc-E:ip:loDelta}   {\ensuremath{{+0.003 } } }
\vdef{default-11:CsMc-E:ip:loDeltaE}   {\ensuremath{{0.004 } } }
\vdef{default-11:CsMc-E:ip:hiDelta}   {\ensuremath{{-0.136 } } }
\vdef{default-11:CsMc-E:ip:hiDeltaE}   {\ensuremath{{0.195 } } }
\vdef{default-11:CsData-E:ips:loEff}   {\ensuremath{{0.958 } } }
\vdef{default-11:CsData-E:ips:loEffE}   {\ensuremath{{0.004 } } }
\vdef{default-11:CsData-E:ips:hiEff}   {\ensuremath{{0.042 } } }
\vdef{default-11:CsData-E:ips:hiEffE}   {\ensuremath{{0.004 } } }
\vdef{default-11:CsMc-E:ips:loEff}   {\ensuremath{{0.970 } } }
\vdef{default-11:CsMc-E:ips:loEffE}   {\ensuremath{{0.004 } } }
\vdef{default-11:CsMc-E:ips:hiEff}   {\ensuremath{{0.030 } } }
\vdef{default-11:CsMc-E:ips:hiEffE}   {\ensuremath{{0.004 } } }
\vdef{default-11:CsMc-E:ips:loDelta}   {\ensuremath{{-0.013 } } }
\vdef{default-11:CsMc-E:ips:loDeltaE}   {\ensuremath{{0.006 } } }
\vdef{default-11:CsMc-E:ips:hiDelta}   {\ensuremath{{+0.340 } } }
\vdef{default-11:CsMc-E:ips:hiDeltaE}   {\ensuremath{{0.149 } } }
\vdef{default-11:CsData-E:maxdoca:loEff}   {\ensuremath{{1.000 } } }
\vdef{default-11:CsData-E:maxdoca:loEffE}   {\ensuremath{{0.000 } } }
\vdef{default-11:CsData-E:maxdoca:hiEff}   {\ensuremath{{0.061 } } }
\vdef{default-11:CsData-E:maxdoca:hiEffE}   {\ensuremath{{0.005 } } }
\vdef{default-11:CsMc-E:maxdoca:loEff}   {\ensuremath{{1.000 } } }
\vdef{default-11:CsMc-E:maxdoca:loEffE}   {\ensuremath{{0.000 } } }
\vdef{default-11:CsMc-E:maxdoca:hiEff}   {\ensuremath{{0.039 } } }
\vdef{default-11:CsMc-E:maxdoca:hiEffE}   {\ensuremath{{0.004 } } }
\vdef{default-11:CsMc-E:maxdoca:loDelta}   {\ensuremath{{+0.000 } } }
\vdef{default-11:CsMc-E:maxdoca:loDeltaE}   {\ensuremath{{0.001 } } }
\vdef{default-11:CsMc-E:maxdoca:hiDelta}   {\ensuremath{{+0.446 } } }
\vdef{default-11:CsMc-E:maxdoca:hiDeltaE}   {\ensuremath{{0.130 } } }
\vdef{default-11:CsData-E:kaonspt:loEff}   {\ensuremath{{1.000 } } }
\vdef{default-11:CsData-E:kaonspt:loEffE}   {\ensuremath{{0.000 } } }
\vdef{default-11:CsData-E:kaonspt:hiEff}   {\ensuremath{{1.000 } } }
\vdef{default-11:CsData-E:kaonspt:hiEffE}   {\ensuremath{{0.000 } } }
\vdef{default-11:CsMc-E:kaonspt:loEff}   {\ensuremath{{1.000 } } }
\vdef{default-11:CsMc-E:kaonspt:loEffE}   {\ensuremath{{0.000 } } }
\vdef{default-11:CsMc-E:kaonspt:hiEff}   {\ensuremath{{1.000 } } }
\vdef{default-11:CsMc-E:kaonspt:hiEffE}   {\ensuremath{{0.000 } } }
\vdef{default-11:CsMc-E:kaonspt:loDelta}   {\ensuremath{{+0.000 } } }
\vdef{default-11:CsMc-E:kaonspt:loDeltaE}   {\ensuremath{{0.000 } } }
\vdef{default-11:CsMc-E:kaonspt:hiDelta}   {\ensuremath{{+0.000 } } }
\vdef{default-11:CsMc-E:kaonspt:hiDeltaE}   {\ensuremath{{0.000 } } }
\vdef{default-11:CsData-E:psipt:loEff}   {\ensuremath{{1.002 } } }
\vdef{default-11:CsData-E:psipt:loEffE}   {\ensuremath{{\mathrm{NaN} } } }
\vdef{default-11:CsData-E:psipt:hiEff}   {\ensuremath{{1.000 } } }
\vdef{default-11:CsData-E:psipt:hiEffE}   {\ensuremath{{0.000 } } }
\vdef{default-11:CsMc-E:psipt:loEff}   {\ensuremath{{1.000 } } }
\vdef{default-11:CsMc-E:psipt:loEffE}   {\ensuremath{{0.000 } } }
\vdef{default-11:CsMc-E:psipt:hiEff}   {\ensuremath{{1.000 } } }
\vdef{default-11:CsMc-E:psipt:hiEffE}   {\ensuremath{{0.000 } } }
\vdef{default-11:CsMc-E:psipt:loDelta}   {\ensuremath{{+0.001 } } }
\vdef{default-11:CsMc-E:psipt:loDeltaE}   {\ensuremath{{\mathrm{NaN} } } }
\vdef{default-11:CsMc-E:psipt:hiDelta}   {\ensuremath{{+0.000 } } }
\vdef{default-11:CsMc-E:psipt:hiDeltaE}   {\ensuremath{{0.001 } } }
\vdef{default-11:CsData-E:phipt:loEff}   {\ensuremath{{1.008 } } }
\vdef{default-11:CsData-E:phipt:loEffE}   {\ensuremath{{\mathrm{NaN} } } }
\vdef{default-11:CsData-E:phipt:hiEff}   {\ensuremath{{1.000 } } }
\vdef{default-11:CsData-E:phipt:hiEffE}   {\ensuremath{{0.000 } } }
\vdef{default-11:CsMc-E:phipt:loEff}   {\ensuremath{{1.004 } } }
\vdef{default-11:CsMc-E:phipt:loEffE}   {\ensuremath{{\mathrm{NaN} } } }
\vdef{default-11:CsMc-E:phipt:hiEff}   {\ensuremath{{1.000 } } }
\vdef{default-11:CsMc-E:phipt:hiEffE}   {\ensuremath{{0.000 } } }
\vdef{default-11:CsMc-E:phipt:loDelta}   {\ensuremath{{+0.004 } } }
\vdef{default-11:CsMc-E:phipt:loDeltaE}   {\ensuremath{{\mathrm{NaN} } } }
\vdef{default-11:CsMc-E:phipt:hiDelta}   {\ensuremath{{+0.000 } } }
\vdef{default-11:CsMc-E:phipt:hiDeltaE}   {\ensuremath{{0.001 } } }
\vdef{default-11:CsData-E:deltar:loEff}   {\ensuremath{{1.000 } } }
\vdef{default-11:CsData-E:deltar:loEffE}   {\ensuremath{{0.000 } } }
\vdef{default-11:CsData-E:deltar:hiEff}   {\ensuremath{{0.000 } } }
\vdef{default-11:CsData-E:deltar:hiEffE}   {\ensuremath{{0.000 } } }
\vdef{default-11:CsMc-E:deltar:loEff}   {\ensuremath{{1.000 } } }
\vdef{default-11:CsMc-E:deltar:loEffE}   {\ensuremath{{0.000 } } }
\vdef{default-11:CsMc-E:deltar:hiEff}   {\ensuremath{{0.000 } } }
\vdef{default-11:CsMc-E:deltar:hiEffE}   {\ensuremath{{0.000 } } }
\vdef{default-11:CsMc-E:deltar:loDelta}   {\ensuremath{{+0.000 } } }
\vdef{default-11:CsMc-E:deltar:loDeltaE}   {\ensuremath{{0.001 } } }
\vdef{default-11:CsMc-E:deltar:hiDelta}   {\ensuremath{{\mathrm{NaN} } } }
\vdef{default-11:CsMc-E:deltar:hiDeltaE}   {\ensuremath{{\mathrm{NaN} } } }
\vdef{default-11:CsData-E:mkk:loEff}   {\ensuremath{{1.152 } } }
\vdef{default-11:CsData-E:mkk:loEffE}   {\ensuremath{{\mathrm{NaN} } } }
\vdef{default-11:CsData-E:mkk:hiEff}   {\ensuremath{{1.000 } } }
\vdef{default-11:CsData-E:mkk:hiEffE}   {\ensuremath{{0.000 } } }
\vdef{default-11:CsMc-E:mkk:loEff}   {\ensuremath{{1.000 } } }
\vdef{default-11:CsMc-E:mkk:loEffE}   {\ensuremath{{0.000 } } }
\vdef{default-11:CsMc-E:mkk:hiEff}   {\ensuremath{{1.000 } } }
\vdef{default-11:CsMc-E:mkk:hiEffE}   {\ensuremath{{0.000 } } }
\vdef{default-11:CsMc-E:mkk:loDelta}   {\ensuremath{{+0.141 } } }
\vdef{default-11:CsMc-E:mkk:loDeltaE}   {\ensuremath{{\mathrm{NaN} } } }
\vdef{default-11:CsMc-E:mkk:hiDelta}   {\ensuremath{{+0.000 } } }
\vdef{default-11:CsMc-E:mkk:hiDeltaE}   {\ensuremath{{0.000 } } }

\vdef{default-11anaBarrel:MuidSg0-pT11pT11:val}   {\ensuremath{{0.712 } } }
\vdef{default-11anaBarrel:MuidNo0-pT11pT11:val}   {\ensuremath{{0.769 } } }
\vdef{default-11anaBarrel:TNPMCMuidSg0-pT11pT11:val}   {\ensuremath{{0.789 } } }
\vdef{default-11anaBarrel:TNPMCMuidNo0-pT11pT11:val}   {\ensuremath{{0.773 } } }
\vdef{default-11anaBarrel:TNPMuidSg0-pT11pT11:val}   {\ensuremath{{0.793 } } }
\vdef{default-11anaBarrel:TNPMuidNo0-pT11pT11:val}   {\ensuremath{{0.786 } } }
\vdef{default-11anaBarrel:rhoMuidSg0-pT11pT11:val}   {\ensuremath{{0.902 } } }
\vdef{default-11anaBarrel:rhoMuidNo0-pT11pT11:val}   {\ensuremath{{0.994 } } }
\vdef{default-11anaBarrel:rMcMuid0-pT11pT11:val}   {\ensuremath{{0.925 } } }
\vdef{default-11anaBarrel:rMcMuid0-pT11pT11:err}   {\ensuremath{{0.012 } } }
\vdef{default-11anaBarrel:rTNPMCMuid0-pT11pT11:val}   {\ensuremath{{1.021 } } }
\vdef{default-11anaBarrel:rTNPMCMuid0-pT11pT11:err}   {\ensuremath{{0.003 } } }
\vdef{default-11anaBarrel:rTNPMuid0-pT11pT11:val}   {\ensuremath{{1.009 } } }
\vdef{default-11anaBarrel:rTNPMuid0-pT11pT11:err}   {\ensuremath{{0.002 } } }
\vdef{default-11anaBarrel:rRhoTNPMCMuid0-pT11pT11:val}   {\ensuremath{{0.925 } } }
\vdef{default-11anaBarrel:rRhoTNPMCMuid0-pT11pT11:err}   {\ensuremath{{0.003 } } }
\vdef{default-11anaBarrel:rRhoTNPMuid0-pT11pT11:val}   {\ensuremath{{0.915 } } }
\vdef{default-11anaBarrel:rRhoTNPMuid0-pT11pT11:err}   {\ensuremath{{0.002 } } }
\vdef{default-11anaBarrel:TrigSg0-pT11pT11:val}   {\ensuremath{{0.844 } } }
\vdef{default-11anaBarrel:TrigNo0-pT11pT11:val}   {\ensuremath{{0.767 } } }
\vdef{default-11anaBarrel:TNPMCTrigSg0-pT11pT11:val}   {\ensuremath{{0.844 } } }
\vdef{default-11anaBarrel:TNPMCTrigNo0-pT11pT11:val}   {\ensuremath{{0.829 } } }
\vdef{default-11anaBarrel:TNPTrigSg0-pT11pT11:val}   {\ensuremath{{0.799 } } }
\vdef{default-11anaBarrel:TNPTrigNo0-pT11pT11:val}   {\ensuremath{{0.786 } } }
\vdef{default-11anaBarrel:rhoTrigSg0-pT11pT11:val}   {\ensuremath{{0.999 } } }
\vdef{default-11anaBarrel:rhoTrigNo0-pT11pT11:val}   {\ensuremath{{0.925 } } }
\vdef{default-11anaBarrel:rMcTrig0-pT11pT11:val}   {\ensuremath{{1.101 } } }
\vdef{default-11anaBarrel:rMcTrig0-pT11pT11:err}   {\ensuremath{{0.011 } } }
\vdef{default-11anaBarrel:rTNPMCTrig0-pT11pT11:val}   {\ensuremath{{1.018 } } }
\vdef{default-11anaBarrel:rTNPMCTrig0-pT11pT11:err}   {\ensuremath{{0.003 } } }
\vdef{default-11anaBarrel:rTNPTrig0-pT11pT11:val}   {\ensuremath{{1.017 } } }
\vdef{default-11anaBarrel:rTNPTrig0-pT11pT11:err}   {\ensuremath{{0.003 } } }
\vdef{default-11anaBarrel:rRhoTNPMCTrig0-pT11pT11:val}   {\ensuremath{{1.101 } } }
\vdef{default-11anaBarrel:rRhoTNPMCTrig0-pT11pT11:err}   {\ensuremath{{0.003 } } }
\vdef{default-11anaBarrel:rRhoTNPTrig0-pT11pT11:val}   {\ensuremath{{1.099 } } }
\vdef{default-11anaBarrel:rRhoTNPTrig0-pT11pT11:err}   {\ensuremath{{0.003 } } }
\vdef{default-11anaBarrel:MuidSg1-pT11pT11:val}   {\ensuremath{{0.849 } } }
\vdef{default-11anaBarrel:MuidNo1-pT11pT11:val}   {\ensuremath{{0.778 } } }
\vdef{default-11anaBarrel:TNPMCMuidSg1-pT11pT11:val}   {\ensuremath{{0.830 } } }
\vdef{default-11anaBarrel:TNPMCMuidNo1-pT11pT11:val}   {\ensuremath{{0.828 } } }
\vdef{default-11anaBarrel:TNPMuidSg1-pT11pT11:val}   {\ensuremath{{0.781 } } }
\vdef{default-11anaBarrel:TNPMuidNo1-pT11pT11:val}   {\ensuremath{{0.781 } } }
\vdef{default-11anaBarrel:rhoMuidSg1-pT11pT11:val}   {\ensuremath{{1.022 } } }
\vdef{default-11anaBarrel:rhoMuidNo1-pT11pT11:val}   {\ensuremath{{0.941 } } }
\vdef{default-11anaBarrel:rMcMuid1-pT11pT11:val}   {\ensuremath{{1.091 } } }
\vdef{default-11anaBarrel:rMcMuid1-pT11pT11:err}   {\ensuremath{{0.012 } } }
\vdef{default-11anaBarrel:rTNPMCMuid1-pT11pT11:val}   {\ensuremath{{1.003 } } }
\vdef{default-11anaBarrel:rTNPMCMuid1-pT11pT11:err}   {\ensuremath{{0.002 } } }
\vdef{default-11anaBarrel:rTNPMuid1-pT11pT11:val}   {\ensuremath{{1.000 } } }
\vdef{default-11anaBarrel:rTNPMuid1-pT11pT11:err}   {\ensuremath{{0.002 } } }
\vdef{default-11anaBarrel:rRhoTNPMCMuid1-pT11pT11:val}   {\ensuremath{{1.091 } } }
\vdef{default-11anaBarrel:rRhoTNPMCMuid1-pT11pT11:err}   {\ensuremath{{0.002 } } }
\vdef{default-11anaBarrel:rRhoTNPMuid1-pT11pT11:val}   {\ensuremath{{1.087 } } }
\vdef{default-11anaBarrel:rRhoTNPMuid1-pT11pT11:err}   {\ensuremath{{0.002 } } }
\vdef{default-11anaBarrel:TrigSg1-pT11pT11:val}   {\ensuremath{{0.731 } } }
\vdef{default-11anaBarrel:TrigNo1-pT11pT11:val}   {\ensuremath{{0.591 } } }
\vdef{default-11anaBarrel:TNPMCTrigSg1-pT11pT11:val}   {\ensuremath{{0.755 } } }
\vdef{default-11anaBarrel:TNPMCTrigNo1-pT11pT11:val}   {\ensuremath{{0.733 } } }
\vdef{default-11anaBarrel:TNPTrigSg1-pT11pT11:val}   {\ensuremath{{0.765 } } }
\vdef{default-11anaBarrel:TNPTrigNo1-pT11pT11:val}   {\ensuremath{{0.748 } } }
\vdef{default-11anaBarrel:rhoTrigSg1-pT11pT11:val}   {\ensuremath{{0.969 } } }
\vdef{default-11anaBarrel:rhoTrigNo1-pT11pT11:val}   {\ensuremath{{0.807 } } }
\vdef{default-11anaBarrel:rMcTrig1-pT11pT11:val}   {\ensuremath{{1.237 } } }
\vdef{default-11anaBarrel:rMcTrig1-pT11pT11:err}   {\ensuremath{{0.021 } } }
\vdef{default-11anaBarrel:rTNPMCTrig1-pT11pT11:val}   {\ensuremath{{1.029 } } }
\vdef{default-11anaBarrel:rTNPMCTrig1-pT11pT11:err}   {\ensuremath{{0.005 } } }
\vdef{default-11anaBarrel:rTNPTrig1-pT11pT11:val}   {\ensuremath{{1.023 } } }
\vdef{default-11anaBarrel:rTNPTrig1-pT11pT11:err}   {\ensuremath{{0.004 } } }
\vdef{default-11anaBarrel:rRhoTNPMCTrig1-pT11pT11:val}   {\ensuremath{{1.237 } } }
\vdef{default-11anaBarrel:rRhoTNPMCTrig1-pT11pT11:err}   {\ensuremath{{0.006 } } }
\vdef{default-11anaBarrel:rRhoTNPTrig1-pT11pT11:val}   {\ensuremath{{1.230 } } }
\vdef{default-11anaBarrel:rRhoTNPTrig1-pT11pT11:err}   {\ensuremath{{0.005 } } }
\vdef{default-11anaEndcap:MuidSg0-pT11pT11:val}   {\ensuremath{{0.712 } } }
\vdef{default-11anaEndcap:MuidNo0-pT11pT11:val}   {\ensuremath{{0.769 } } }
\vdef{default-11anaEndcap:TNPMCMuidSg0-pT11pT11:val}   {\ensuremath{{0.789 } } }
\vdef{default-11anaEndcap:TNPMCMuidNo0-pT11pT11:val}   {\ensuremath{{0.773 } } }
\vdef{default-11anaEndcap:TNPMuidSg0-pT11pT11:val}   {\ensuremath{{0.793 } } }
\vdef{default-11anaEndcap:TNPMuidNo0-pT11pT11:val}   {\ensuremath{{0.786 } } }
\vdef{default-11anaEndcap:rhoMuidSg0-pT11pT11:val}   {\ensuremath{{0.902 } } }
\vdef{default-11anaEndcap:rhoMuidNo0-pT11pT11:val}   {\ensuremath{{0.994 } } }
\vdef{default-11anaEndcap:rMcMuid0-pT11pT11:val}   {\ensuremath{{0.925 } } }
\vdef{default-11anaEndcap:rMcMuid0-pT11pT11:err}   {\ensuremath{{0.012 } } }
\vdef{default-11anaEndcap:rTNPMCMuid0-pT11pT11:val}   {\ensuremath{{1.021 } } }
\vdef{default-11anaEndcap:rTNPMCMuid0-pT11pT11:err}   {\ensuremath{{0.003 } } }
\vdef{default-11anaEndcap:rTNPMuid0-pT11pT11:val}   {\ensuremath{{1.009 } } }
\vdef{default-11anaEndcap:rTNPMuid0-pT11pT11:err}   {\ensuremath{{0.002 } } }
\vdef{default-11anaEndcap:rRhoTNPMCMuid0-pT11pT11:val}   {\ensuremath{{0.925 } } }
\vdef{default-11anaEndcap:rRhoTNPMCMuid0-pT11pT11:err}   {\ensuremath{{0.003 } } }
\vdef{default-11anaEndcap:rRhoTNPMuid0-pT11pT11:val}   {\ensuremath{{0.915 } } }
\vdef{default-11anaEndcap:rRhoTNPMuid0-pT11pT11:err}   {\ensuremath{{0.002 } } }
\vdef{default-11anaEndcap:TrigSg0-pT11pT11:val}   {\ensuremath{{0.844 } } }
\vdef{default-11anaEndcap:TrigNo0-pT11pT11:val}   {\ensuremath{{0.767 } } }
\vdef{default-11anaEndcap:TNPMCTrigSg0-pT11pT11:val}   {\ensuremath{{0.844 } } }
\vdef{default-11anaEndcap:TNPMCTrigNo0-pT11pT11:val}   {\ensuremath{{0.829 } } }
\vdef{default-11anaEndcap:TNPTrigSg0-pT11pT11:val}   {\ensuremath{{0.799 } } }
\vdef{default-11anaEndcap:TNPTrigNo0-pT11pT11:val}   {\ensuremath{{0.786 } } }
\vdef{default-11anaEndcap:rhoTrigSg0-pT11pT11:val}   {\ensuremath{{0.999 } } }
\vdef{default-11anaEndcap:rhoTrigNo0-pT11pT11:val}   {\ensuremath{{0.925 } } }
\vdef{default-11anaEndcap:rMcTrig0-pT11pT11:val}   {\ensuremath{{1.101 } } }
\vdef{default-11anaEndcap:rMcTrig0-pT11pT11:err}   {\ensuremath{{0.011 } } }
\vdef{default-11anaEndcap:rTNPMCTrig0-pT11pT11:val}   {\ensuremath{{1.018 } } }
\vdef{default-11anaEndcap:rTNPMCTrig0-pT11pT11:err}   {\ensuremath{{0.003 } } }
\vdef{default-11anaEndcap:rTNPTrig0-pT11pT11:val}   {\ensuremath{{1.017 } } }
\vdef{default-11anaEndcap:rTNPTrig0-pT11pT11:err}   {\ensuremath{{0.003 } } }
\vdef{default-11anaEndcap:rRhoTNPMCTrig0-pT11pT11:val}   {\ensuremath{{1.101 } } }
\vdef{default-11anaEndcap:rRhoTNPMCTrig0-pT11pT11:err}   {\ensuremath{{0.003 } } }
\vdef{default-11anaEndcap:rRhoTNPTrig0-pT11pT11:val}   {\ensuremath{{1.099 } } }
\vdef{default-11anaEndcap:rRhoTNPTrig0-pT11pT11:err}   {\ensuremath{{0.003 } } }
\vdef{default-11anaEndcap:MuidSg1-pT11pT11:val}   {\ensuremath{{0.849 } } }
\vdef{default-11anaEndcap:MuidNo1-pT11pT11:val}   {\ensuremath{{0.778 } } }
\vdef{default-11anaEndcap:TNPMCMuidSg1-pT11pT11:val}   {\ensuremath{{0.830 } } }
\vdef{default-11anaEndcap:TNPMCMuidNo1-pT11pT11:val}   {\ensuremath{{0.828 } } }
\vdef{default-11anaEndcap:TNPMuidSg1-pT11pT11:val}   {\ensuremath{{0.781 } } }
\vdef{default-11anaEndcap:TNPMuidNo1-pT11pT11:val}   {\ensuremath{{0.781 } } }
\vdef{default-11anaEndcap:rhoMuidSg1-pT11pT11:val}   {\ensuremath{{1.022 } } }
\vdef{default-11anaEndcap:rhoMuidNo1-pT11pT11:val}   {\ensuremath{{0.941 } } }
\vdef{default-11anaEndcap:rMcMuid1-pT11pT11:val}   {\ensuremath{{1.091 } } }
\vdef{default-11anaEndcap:rMcMuid1-pT11pT11:err}   {\ensuremath{{0.012 } } }
\vdef{default-11anaEndcap:rTNPMCMuid1-pT11pT11:val}   {\ensuremath{{1.003 } } }
\vdef{default-11anaEndcap:rTNPMCMuid1-pT11pT11:err}   {\ensuremath{{0.002 } } }
\vdef{default-11anaEndcap:rTNPMuid1-pT11pT11:val}   {\ensuremath{{1.000 } } }
\vdef{default-11anaEndcap:rTNPMuid1-pT11pT11:err}   {\ensuremath{{0.002 } } }
\vdef{default-11anaEndcap:rRhoTNPMCMuid1-pT11pT11:val}   {\ensuremath{{1.091 } } }
\vdef{default-11anaEndcap:rRhoTNPMCMuid1-pT11pT11:err}   {\ensuremath{{0.002 } } }
\vdef{default-11anaEndcap:rRhoTNPMuid1-pT11pT11:val}   {\ensuremath{{1.087 } } }
\vdef{default-11anaEndcap:rRhoTNPMuid1-pT11pT11:err}   {\ensuremath{{0.002 } } }
\vdef{default-11anaEndcap:TrigSg1-pT11pT11:val}   {\ensuremath{{0.731 } } }
\vdef{default-11anaEndcap:TrigNo1-pT11pT11:val}   {\ensuremath{{0.591 } } }
\vdef{default-11anaEndcap:TNPMCTrigSg1-pT11pT11:val}   {\ensuremath{{0.755 } } }
\vdef{default-11anaEndcap:TNPMCTrigNo1-pT11pT11:val}   {\ensuremath{{0.733 } } }
\vdef{default-11anaEndcap:TNPTrigSg1-pT11pT11:val}   {\ensuremath{{0.765 } } }
\vdef{default-11anaEndcap:TNPTrigNo1-pT11pT11:val}   {\ensuremath{{0.748 } } }
\vdef{default-11anaEndcap:rhoTrigSg1-pT11pT11:val}   {\ensuremath{{0.969 } } }
\vdef{default-11anaEndcap:rhoTrigNo1-pT11pT11:val}   {\ensuremath{{0.807 } } }
\vdef{default-11anaEndcap:rMcTrig1-pT11pT11:val}   {\ensuremath{{1.237 } } }
\vdef{default-11anaEndcap:rMcTrig1-pT11pT11:err}   {\ensuremath{{0.021 } } }
\vdef{default-11anaEndcap:rTNPMCTrig1-pT11pT11:val}   {\ensuremath{{1.029 } } }
\vdef{default-11anaEndcap:rTNPMCTrig1-pT11pT11:err}   {\ensuremath{{0.005 } } }
\vdef{default-11anaEndcap:rTNPTrig1-pT11pT11:val}   {\ensuremath{{1.023 } } }
\vdef{default-11anaEndcap:rTNPTrig1-pT11pT11:err}   {\ensuremath{{0.004 } } }
\vdef{default-11anaEndcap:rRhoTNPMCTrig1-pT11pT11:val}   {\ensuremath{{1.237 } } }
\vdef{default-11anaEndcap:rRhoTNPMCTrig1-pT11pT11:err}   {\ensuremath{{0.006 } } }
\vdef{default-11anaEndcap:rRhoTNPTrig1-pT11pT11:val}   {\ensuremath{{1.230 } } }
\vdef{default-11anaEndcap:rRhoTNPTrig1-pT11pT11:err}   {\ensuremath{{0.005 } } }
\vdef{default-11woCowboyVeto:MuidSg0-pT40pT40:val}   {\ensuremath{{0.710 } } }
\vdef{default-11woCowboyVeto:MuidNo0-pT40pT40:val}   {\ensuremath{{0.763 } } }
\vdef{default-11woCowboyVeto:TNPMCMuidSg0-pT40pT40:val}   {\ensuremath{{0.784 } } }
\vdef{default-11woCowboyVeto:TNPMCMuidNo0-pT40pT40:val}   {\ensuremath{{0.763 } } }
\vdef{default-11woCowboyVeto:TNPMuidSg0-pT40pT40:val}   {\ensuremath{{0.791 } } }
\vdef{default-11woCowboyVeto:TNPMuidNo0-pT40pT40:val}   {\ensuremath{{0.782 } } }
\vdef{default-11woCowboyVeto:rhoMuidSg0-pT40pT40:val}   {\ensuremath{{0.906 } } }
\vdef{default-11woCowboyVeto:rhoMuidNo0-pT40pT40:val}   {\ensuremath{{1.000 } } }
\vdef{default-11woCowboyVeto:rMcMuid0-pT40pT40:val}   {\ensuremath{{0.931 } } }
\vdef{default-11woCowboyVeto:rMcMuid0-pT40pT40:err}   {\ensuremath{{0.012 } } }
\vdef{default-11woCowboyVeto:rTNPMCMuid0-pT40pT40:val}   {\ensuremath{{1.027 } } }
\vdef{default-11woCowboyVeto:rTNPMCMuid0-pT40pT40:err}   {\ensuremath{{0.003 } } }
\vdef{default-11woCowboyVeto:rTNPMuid0-pT40pT40:val}   {\ensuremath{{1.012 } } }
\vdef{default-11woCowboyVeto:rTNPMuid0-pT40pT40:err}   {\ensuremath{{0.002 } } }
\vdef{default-11woCowboyVeto:rRhoTNPMCMuid0-pT40pT40:val}   {\ensuremath{{0.931 } } }
\vdef{default-11woCowboyVeto:rRhoTNPMCMuid0-pT40pT40:err}   {\ensuremath{{0.003 } } }
\vdef{default-11woCowboyVeto:rRhoTNPMuid0-pT40pT40:val}   {\ensuremath{{0.917 } } }
\vdef{default-11woCowboyVeto:rRhoTNPMuid0-pT40pT40:err}   {\ensuremath{{0.002 } } }
\vdef{default-11woCowboyVeto:TrigSg0-pT40pT40:val}   {\ensuremath{{0.841 } } }
\vdef{default-11woCowboyVeto:TrigNo0-pT40pT40:val}   {\ensuremath{{0.762 } } }
\vdef{default-11woCowboyVeto:TNPMCTrigSg0-pT40pT40:val}   {\ensuremath{{0.840 } } }
\vdef{default-11woCowboyVeto:TNPMCTrigNo0-pT40pT40:val}   {\ensuremath{{0.821 } } }
\vdef{default-11woCowboyVeto:TNPTrigSg0-pT40pT40:val}   {\ensuremath{{0.796 } } }
\vdef{default-11woCowboyVeto:TNPTrigNo0-pT40pT40:val}   {\ensuremath{{0.779 } } }
\vdef{default-11woCowboyVeto:rhoTrigSg0-pT40pT40:val}   {\ensuremath{{1.001 } } }
\vdef{default-11woCowboyVeto:rhoTrigNo0-pT40pT40:val}   {\ensuremath{{0.928 } } }
\vdef{default-11woCowboyVeto:rMcTrig0-pT40pT40:val}   {\ensuremath{{1.104 } } }
\vdef{default-11woCowboyVeto:rMcTrig0-pT40pT40:err}   {\ensuremath{{0.011 } } }
\vdef{default-11woCowboyVeto:rTNPMCTrig0-pT40pT40:val}   {\ensuremath{{1.023 } } }
\vdef{default-11woCowboyVeto:rTNPMCTrig0-pT40pT40:err}   {\ensuremath{{0.003 } } }
\vdef{default-11woCowboyVeto:rTNPTrig0-pT40pT40:val}   {\ensuremath{{1.022 } } }
\vdef{default-11woCowboyVeto:rTNPTrig0-pT40pT40:err}   {\ensuremath{{0.003 } } }
\vdef{default-11woCowboyVeto:rRhoTNPMCTrig0-pT40pT40:val}   {\ensuremath{{1.104 } } }
\vdef{default-11woCowboyVeto:rRhoTNPMCTrig0-pT40pT40:err}   {\ensuremath{{0.003 } } }
\vdef{default-11woCowboyVeto:rRhoTNPTrig0-pT40pT40:val}   {\ensuremath{{1.102 } } }
\vdef{default-11woCowboyVeto:rRhoTNPTrig0-pT40pT40:err}   {\ensuremath{{0.003 } } }
\vdef{default-11woCowboyVeto:MuidSg1-pT40pT40:val}   {\ensuremath{{0.848 } } }
\vdef{default-11woCowboyVeto:MuidNo1-pT40pT40:val}   {\ensuremath{{0.777 } } }
\vdef{default-11woCowboyVeto:TNPMCMuidSg1-pT40pT40:val}   {\ensuremath{{0.830 } } }
\vdef{default-11woCowboyVeto:TNPMCMuidNo1-pT40pT40:val}   {\ensuremath{{0.824 } } }
\vdef{default-11woCowboyVeto:TNPMuidSg1-pT40pT40:val}   {\ensuremath{{0.781 } } }
\vdef{default-11woCowboyVeto:TNPMuidNo1-pT40pT40:val}   {\ensuremath{{0.780 } } }
\vdef{default-11woCowboyVeto:rhoMuidSg1-pT40pT40:val}   {\ensuremath{{1.021 } } }
\vdef{default-11woCowboyVeto:rhoMuidNo1-pT40pT40:val}   {\ensuremath{{0.943 } } }
\vdef{default-11woCowboyVeto:rMcMuid1-pT40pT40:val}   {\ensuremath{{1.091 } } }
\vdef{default-11woCowboyVeto:rMcMuid1-pT40pT40:err}   {\ensuremath{{0.012 } } }
\vdef{default-11woCowboyVeto:rTNPMCMuid1-pT40pT40:val}   {\ensuremath{{1.007 } } }
\vdef{default-11woCowboyVeto:rTNPMCMuid1-pT40pT40:err}   {\ensuremath{{0.002 } } }
\vdef{default-11woCowboyVeto:rTNPMuid1-pT40pT40:val}   {\ensuremath{{1.001 } } }
\vdef{default-11woCowboyVeto:rTNPMuid1-pT40pT40:err}   {\ensuremath{{0.002 } } }
\vdef{default-11woCowboyVeto:rRhoTNPMCMuid1-pT40pT40:val}   {\ensuremath{{1.091 } } }
\vdef{default-11woCowboyVeto:rRhoTNPMCMuid1-pT40pT40:err}   {\ensuremath{{0.002 } } }
\vdef{default-11woCowboyVeto:rRhoTNPMuid1-pT40pT40:val}   {\ensuremath{{1.085 } } }
\vdef{default-11woCowboyVeto:rRhoTNPMuid1-pT40pT40:err}   {\ensuremath{{0.002 } } }
\vdef{default-11woCowboyVeto:TrigSg1-pT40pT40:val}   {\ensuremath{{0.731 } } }
\vdef{default-11woCowboyVeto:TrigNo1-pT40pT40:val}   {\ensuremath{{0.588 } } }
\vdef{default-11woCowboyVeto:TNPMCTrigSg1-pT40pT40:val}   {\ensuremath{{0.754 } } }
\vdef{default-11woCowboyVeto:TNPMCTrigNo1-pT40pT40:val}   {\ensuremath{{0.729 } } }
\vdef{default-11woCowboyVeto:TNPTrigSg1-pT40pT40:val}   {\ensuremath{{0.765 } } }
\vdef{default-11woCowboyVeto:TNPTrigNo1-pT40pT40:val}   {\ensuremath{{0.746 } } }
\vdef{default-11woCowboyVeto:rhoTrigSg1-pT40pT40:val}   {\ensuremath{{0.970 } } }
\vdef{default-11woCowboyVeto:rhoTrigNo1-pT40pT40:val}   {\ensuremath{{0.806 } } }
\vdef{default-11woCowboyVeto:rMcTrig1-pT40pT40:val}   {\ensuremath{{1.244 } } }
\vdef{default-11woCowboyVeto:rMcTrig1-pT40pT40:err}   {\ensuremath{{0.021 } } }
\vdef{default-11woCowboyVeto:rTNPMCTrig1-pT40pT40:val}   {\ensuremath{{1.034 } } }
\vdef{default-11woCowboyVeto:rTNPMCTrig1-pT40pT40:err}   {\ensuremath{{0.005 } } }
\vdef{default-11woCowboyVeto:rTNPTrig1-pT40pT40:val}   {\ensuremath{{1.026 } } }
\vdef{default-11woCowboyVeto:rTNPTrig1-pT40pT40:err}   {\ensuremath{{0.004 } } }
\vdef{default-11woCowboyVeto:rRhoTNPMCTrig1-pT40pT40:val}   {\ensuremath{{1.244 } } }
\vdef{default-11woCowboyVeto:rRhoTNPMCTrig1-pT40pT40:err}   {\ensuremath{{0.006 } } }
\vdef{default-11woCowboyVeto:rRhoTNPTrig1-pT40pT40:val}   {\ensuremath{{1.234 } } }
\vdef{default-11woCowboyVeto:rRhoTNPTrig1-pT40pT40:err}   {\ensuremath{{0.005 } } }
\vdef{default-11woCowboyVeto:MuidSg0-pT50pT50:val}   {\ensuremath{{0.771 } } }
\vdef{default-11woCowboyVeto:MuidNo0-pT50pT50:val}   {\ensuremath{{0.844 } } }
\vdef{default-11woCowboyVeto:TNPMCMuidSg0-pT50pT50:val}   {\ensuremath{{0.869 } } }
\vdef{default-11woCowboyVeto:TNPMCMuidNo0-pT50pT50:val}   {\ensuremath{{0.866 } } }
\vdef{default-11woCowboyVeto:TNPMuidSg0-pT50pT50:val}   {\ensuremath{{0.840 } } }
\vdef{default-11woCowboyVeto:TNPMuidNo0-pT50pT50:val}   {\ensuremath{{0.842 } } }
\vdef{default-11woCowboyVeto:rhoMuidSg0-pT50pT50:val}   {\ensuremath{{0.887 } } }
\vdef{default-11woCowboyVeto:rhoMuidNo0-pT50pT50:val}   {\ensuremath{{0.974 } } }
\vdef{default-11woCowboyVeto:rMcMuid0-pT50pT50:val}   {\ensuremath{{0.914 } } }
\vdef{default-11woCowboyVeto:rMcMuid0-pT50pT50:err}   {\ensuremath{{0.012 } } }
\vdef{default-11woCowboyVeto:rTNPMCMuid0-pT50pT50:val}   {\ensuremath{{1.003 } } }
\vdef{default-11woCowboyVeto:rTNPMCMuid0-pT50pT50:err}   {\ensuremath{{0.001 } } }
\vdef{default-11woCowboyVeto:rTNPMuid0-pT50pT50:val}   {\ensuremath{{0.998 } } }
\vdef{default-11woCowboyVeto:rTNPMuid0-pT50pT50:err}   {\ensuremath{{0.001 } } }
\vdef{default-11woCowboyVeto:rRhoTNPMCMuid0-pT50pT50:val}   {\ensuremath{{0.914 } } }
\vdef{default-11woCowboyVeto:rRhoTNPMCMuid0-pT50pT50:err}   {\ensuremath{{0.001 } } }
\vdef{default-11woCowboyVeto:rRhoTNPMuid0-pT50pT50:val}   {\ensuremath{{0.909 } } }
\vdef{default-11woCowboyVeto:rRhoTNPMuid0-pT50pT50:err}   {\ensuremath{{0.001 } } }
\vdef{default-11woCowboyVeto:TrigSg0-pT50pT50:val}   {\ensuremath{{0.897 } } }
\vdef{default-11woCowboyVeto:TrigNo0-pT50pT50:val}   {\ensuremath{{0.820 } } }
\vdef{default-11woCowboyVeto:TNPMCTrigSg0-pT50pT50:val}   {\ensuremath{{0.915 } } }
\vdef{default-11woCowboyVeto:TNPMCTrigNo0-pT50pT50:val}   {\ensuremath{{0.913 } } }
\vdef{default-11woCowboyVeto:TNPTrigSg0-pT50pT50:val}   {\ensuremath{{0.864 } } }
\vdef{default-11woCowboyVeto:TNPTrigNo0-pT50pT50:val}   {\ensuremath{{0.864 } } }
\vdef{default-11woCowboyVeto:rhoTrigSg0-pT50pT50:val}   {\ensuremath{{0.980 } } }
\vdef{default-11woCowboyVeto:rhoTrigNo0-pT50pT50:val}   {\ensuremath{{0.899 } } }
\vdef{default-11woCowboyVeto:rMcTrig0-pT50pT50:val}   {\ensuremath{{1.093 } } }
\vdef{default-11woCowboyVeto:rMcTrig0-pT50pT50:err}   {\ensuremath{{0.011 } } }
\vdef{default-11woCowboyVeto:rTNPMCTrig0-pT50pT50:val}   {\ensuremath{{1.002 } } }
\vdef{default-11woCowboyVeto:rTNPMCTrig0-pT50pT50:err}   {\ensuremath{{0.001 } } }
\vdef{default-11woCowboyVeto:rTNPTrig0-pT50pT50:val}   {\ensuremath{{1.000 } } }
\vdef{default-11woCowboyVeto:rTNPTrig0-pT50pT50:err}   {\ensuremath{{0.001 } } }
\vdef{default-11woCowboyVeto:rRhoTNPMCTrig0-pT50pT50:val}   {\ensuremath{{1.093 } } }
\vdef{default-11woCowboyVeto:rRhoTNPMCTrig0-pT50pT50:err}   {\ensuremath{{0.001 } } }
\vdef{default-11woCowboyVeto:rRhoTNPTrig0-pT50pT50:val}   {\ensuremath{{1.090 } } }
\vdef{default-11woCowboyVeto:rRhoTNPTrig0-pT50pT50:err}   {\ensuremath{{0.001 } } }
\vdef{default-11woCowboyVeto:MuidSg1-pT50pT50:val}   {\ensuremath{{0.876 } } }
\vdef{default-11woCowboyVeto:MuidNo1-pT50pT50:val}   {\ensuremath{{0.784 } } }
\vdef{default-11woCowboyVeto:TNPMCMuidSg1-pT50pT50:val}   {\ensuremath{{0.853 } } }
\vdef{default-11woCowboyVeto:TNPMCMuidNo1-pT50pT50:val}   {\ensuremath{{0.854 } } }
\vdef{default-11woCowboyVeto:TNPMuidSg1-pT50pT50:val}   {\ensuremath{{0.788 } } }
\vdef{default-11woCowboyVeto:TNPMuidNo1-pT50pT50:val}   {\ensuremath{{0.789 } } }
\vdef{default-11woCowboyVeto:rhoMuidSg1-pT50pT50:val}   {\ensuremath{{1.026 } } }
\vdef{default-11woCowboyVeto:rhoMuidNo1-pT50pT50:val}   {\ensuremath{{0.919 } } }
\vdef{default-11woCowboyVeto:rMcMuid1-pT50pT50:val}   {\ensuremath{{1.116 } } }
\vdef{default-11woCowboyVeto:rMcMuid1-pT50pT50:err}   {\ensuremath{{0.014 } } }
\vdef{default-11woCowboyVeto:rTNPMCMuid1-pT50pT50:val}   {\ensuremath{{1.000 } } }
\vdef{default-11woCowboyVeto:rTNPMCMuid1-pT50pT50:err}   {\ensuremath{{0.002 } } }
\vdef{default-11woCowboyVeto:rTNPMuid1-pT50pT50:val}   {\ensuremath{{0.999 } } }
\vdef{default-11woCowboyVeto:rTNPMuid1-pT50pT50:err}   {\ensuremath{{0.003 } } }
\vdef{default-11woCowboyVeto:rRhoTNPMCMuid1-pT50pT50:val}   {\ensuremath{{1.116 } } }
\vdef{default-11woCowboyVeto:rRhoTNPMCMuid1-pT50pT50:err}   {\ensuremath{{0.002 } } }
\vdef{default-11woCowboyVeto:rRhoTNPMuid1-pT50pT50:val}   {\ensuremath{{1.115 } } }
\vdef{default-11woCowboyVeto:rRhoTNPMuid1-pT50pT50:err}   {\ensuremath{{0.003 } } }
\vdef{default-11woCowboyVeto:TrigSg1-pT50pT50:val}   {\ensuremath{{0.742 } } }
\vdef{default-11woCowboyVeto:TrigNo1-pT50pT50:val}   {\ensuremath{{0.611 } } }
\vdef{default-11woCowboyVeto:TNPMCTrigSg1-pT50pT50:val}   {\ensuremath{{0.771 } } }
\vdef{default-11woCowboyVeto:TNPMCTrigNo1-pT50pT50:val}   {\ensuremath{{0.762 } } }
\vdef{default-11woCowboyVeto:TNPTrigSg1-pT50pT50:val}   {\ensuremath{{0.771 } } }
\vdef{default-11woCowboyVeto:TNPTrigNo1-pT50pT50:val}   {\ensuremath{{0.762 } } }
\vdef{default-11woCowboyVeto:rhoTrigSg1-pT50pT50:val}   {\ensuremath{{0.963 } } }
\vdef{default-11woCowboyVeto:rhoTrigNo1-pT50pT50:val}   {\ensuremath{{0.802 } } }
\vdef{default-11woCowboyVeto:rMcTrig1-pT50pT50:val}   {\ensuremath{{1.215 } } }
\vdef{default-11woCowboyVeto:rMcTrig1-pT50pT50:err}   {\ensuremath{{0.026 } } }
\vdef{default-11woCowboyVeto:rTNPMCTrig1-pT50pT50:val}   {\ensuremath{{1.011 } } }
\vdef{default-11woCowboyVeto:rTNPMCTrig1-pT50pT50:err}   {\ensuremath{{0.007 } } }
\vdef{default-11woCowboyVeto:rTNPTrig1-pT50pT50:val}   {\ensuremath{{1.012 } } }
\vdef{default-11woCowboyVeto:rTNPTrig1-pT50pT50:err}   {\ensuremath{{0.006 } } }
\vdef{default-11woCowboyVeto:rRhoTNPMCTrig1-pT50pT50:val}   {\ensuremath{{1.215 } } }
\vdef{default-11woCowboyVeto:rRhoTNPMCTrig1-pT50pT50:err}   {\ensuremath{{0.008 } } }
\vdef{default-11woCowboyVeto:rRhoTNPTrig1-pT50pT50:val}   {\ensuremath{{1.216 } } }
\vdef{default-11woCowboyVeto:rRhoTNPTrig1-pT50pT50:err}   {\ensuremath{{0.007 } } }
\vdef{default-11woCowboyVeto:MuidSg0-pT60pT60:val}   {\ensuremath{{0.776 } } }
\vdef{default-11woCowboyVeto:MuidNo0-pT60pT60:val}   {\ensuremath{{0.856 } } }
\vdef{default-11woCowboyVeto:TNPMCMuidSg0-pT60pT60:val}   {\ensuremath{{0.886 } } }
\vdef{default-11woCowboyVeto:TNPMCMuidNo0-pT60pT60:val}   {\ensuremath{{0.887 } } }
\vdef{default-11woCowboyVeto:TNPMuidSg0-pT60pT60:val}   {\ensuremath{{0.844 } } }
\vdef{default-11woCowboyVeto:TNPMuidNo0-pT60pT60:val}   {\ensuremath{{0.846 } } }
\vdef{default-11woCowboyVeto:rhoMuidSg0-pT60pT60:val}   {\ensuremath{{0.876 } } }
\vdef{default-11woCowboyVeto:rhoMuidNo0-pT60pT60:val}   {\ensuremath{{0.965 } } }
\vdef{default-11woCowboyVeto:rMcMuid0-pT60pT60:val}   {\ensuremath{{0.907 } } }
\vdef{default-11woCowboyVeto:rMcMuid0-pT60pT60:err}   {\ensuremath{{0.016 } } }
\vdef{default-11woCowboyVeto:rTNPMCMuid0-pT60pT60:val}   {\ensuremath{{1.000 } } }
\vdef{default-11woCowboyVeto:rTNPMCMuid0-pT60pT60:err}   {\ensuremath{{0.001 } } }
\vdef{default-11woCowboyVeto:rTNPMuid0-pT60pT60:val}   {\ensuremath{{0.997 } } }
\vdef{default-11woCowboyVeto:rTNPMuid0-pT60pT60:err}   {\ensuremath{{0.001 } } }
\vdef{default-11woCowboyVeto:rRhoTNPMCMuid0-pT60pT60:val}   {\ensuremath{{0.907 } } }
\vdef{default-11woCowboyVeto:rRhoTNPMCMuid0-pT60pT60:err}   {\ensuremath{{0.001 } } }
\vdef{default-11woCowboyVeto:rRhoTNPMuid0-pT60pT60:val}   {\ensuremath{{0.905 } } }
\vdef{default-11woCowboyVeto:rRhoTNPMuid0-pT60pT60:err}   {\ensuremath{{0.002 } } }
\vdef{default-11woCowboyVeto:TrigSg0-pT60pT60:val}   {\ensuremath{{0.912 } } }
\vdef{default-11woCowboyVeto:TrigNo0-pT60pT60:val}   {\ensuremath{{0.824 } } }
\vdef{default-11woCowboyVeto:TNPMCTrigSg0-pT60pT60:val}   {\ensuremath{{0.932 } } }
\vdef{default-11woCowboyVeto:TNPMCTrigNo0-pT60pT60:val}   {\ensuremath{{0.932 } } }
\vdef{default-11woCowboyVeto:TNPTrigSg0-pT60pT60:val}   {\ensuremath{{0.874 } } }
\vdef{default-11woCowboyVeto:TNPTrigNo0-pT60pT60:val}   {\ensuremath{{0.877 } } }
\vdef{default-11woCowboyVeto:rhoTrigSg0-pT60pT60:val}   {\ensuremath{{0.979 } } }
\vdef{default-11woCowboyVeto:rhoTrigNo0-pT60pT60:val}   {\ensuremath{{0.883 } } }
\vdef{default-11woCowboyVeto:rMcTrig0-pT60pT60:val}   {\ensuremath{{1.108 } } }
\vdef{default-11woCowboyVeto:rMcTrig0-pT60pT60:err}   {\ensuremath{{0.013 } } }
\vdef{default-11woCowboyVeto:rTNPMCTrig0-pT60pT60:val}   {\ensuremath{{0.999 } } }
\vdef{default-11woCowboyVeto:rTNPMCTrig0-pT60pT60:err}   {\ensuremath{{0.001 } } }
\vdef{default-11woCowboyVeto:rTNPTrig0-pT60pT60:val}   {\ensuremath{{0.996 } } }
\vdef{default-11woCowboyVeto:rTNPTrig0-pT60pT60:err}   {\ensuremath{{0.001 } } }
\vdef{default-11woCowboyVeto:rRhoTNPMCTrig0-pT60pT60:val}   {\ensuremath{{1.108 } } }
\vdef{default-11woCowboyVeto:rRhoTNPMCTrig0-pT60pT60:err}   {\ensuremath{{0.001 } } }
\vdef{default-11woCowboyVeto:rRhoTNPTrig0-pT60pT60:val}   {\ensuremath{{1.104 } } }
\vdef{default-11woCowboyVeto:rRhoTNPTrig0-pT60pT60:err}   {\ensuremath{{0.001 } } }
\vdef{default-11woCowboyVeto:MuidSg1-pT60pT60:val}   {\ensuremath{{0.877 } } }
\vdef{default-11woCowboyVeto:MuidNo1-pT60pT60:val}   {\ensuremath{{0.776 } } }
\vdef{default-11woCowboyVeto:TNPMCMuidSg1-pT60pT60:val}   {\ensuremath{{0.859 } } }
\vdef{default-11woCowboyVeto:TNPMCMuidNo1-pT60pT60:val}   {\ensuremath{{0.861 } } }
\vdef{default-11woCowboyVeto:TNPMuidSg1-pT60pT60:val}   {\ensuremath{{0.771 } } }
\vdef{default-11woCowboyVeto:TNPMuidNo1-pT60pT60:val}   {\ensuremath{{0.774 } } }
\vdef{default-11woCowboyVeto:rhoMuidSg1-pT60pT60:val}   {\ensuremath{{1.020 } } }
\vdef{default-11woCowboyVeto:rhoMuidNo1-pT60pT60:val}   {\ensuremath{{0.901 } } }
\vdef{default-11woCowboyVeto:rMcMuid1-pT60pT60:val}   {\ensuremath{{1.130 } } }
\vdef{default-11woCowboyVeto:rMcMuid1-pT60pT60:err}   {\ensuremath{{0.020 } } }
\vdef{default-11woCowboyVeto:rTNPMCMuid1-pT60pT60:val}   {\ensuremath{{0.998 } } }
\vdef{default-11woCowboyVeto:rTNPMCMuid1-pT60pT60:err}   {\ensuremath{{0.002 } } }
\vdef{default-11woCowboyVeto:rTNPMuid1-pT60pT60:val}   {\ensuremath{{0.996 } } }
\vdef{default-11woCowboyVeto:rTNPMuid1-pT60pT60:err}   {\ensuremath{{0.004 } } }
\vdef{default-11woCowboyVeto:rRhoTNPMCMuid1-pT60pT60:val}   {\ensuremath{{1.130 } } }
\vdef{default-11woCowboyVeto:rRhoTNPMCMuid1-pT60pT60:err}   {\ensuremath{{0.003 } } }
\vdef{default-11woCowboyVeto:rRhoTNPMuid1-pT60pT60:val}   {\ensuremath{{1.128 } } }
\vdef{default-11woCowboyVeto:rRhoTNPMuid1-pT60pT60:err}   {\ensuremath{{0.004 } } }
\vdef{default-11woCowboyVeto:TrigSg1-pT60pT60:val}   {\ensuremath{{0.717 } } }
\vdef{default-11woCowboyVeto:TrigNo1-pT60pT60:val}   {\ensuremath{{0.612 } } }
\vdef{default-11woCowboyVeto:TNPMCTrigSg1-pT60pT60:val}   {\ensuremath{{0.766 } } }
\vdef{default-11woCowboyVeto:TNPMCTrigNo1-pT60pT60:val}   {\ensuremath{{0.768 } } }
\vdef{default-11woCowboyVeto:TNPTrigSg1-pT60pT60:val}   {\ensuremath{{0.760 } } }
\vdef{default-11woCowboyVeto:TNPTrigNo1-pT60pT60:val}   {\ensuremath{{0.758 } } }
\vdef{default-11woCowboyVeto:rhoTrigSg1-pT60pT60:val}   {\ensuremath{{0.935 } } }
\vdef{default-11woCowboyVeto:rhoTrigNo1-pT60pT60:val}   {\ensuremath{{0.797 } } }
\vdef{default-11woCowboyVeto:rMcTrig1-pT60pT60:val}   {\ensuremath{{1.171 } } }
\vdef{default-11woCowboyVeto:rMcTrig1-pT60pT60:err}   {\ensuremath{{0.036 } } }
\vdef{default-11woCowboyVeto:rTNPMCTrig1-pT60pT60:val}   {\ensuremath{{0.998 } } }
\vdef{default-11woCowboyVeto:rTNPMCTrig1-pT60pT60:err}   {\ensuremath{{0.010 } } }
\vdef{default-11woCowboyVeto:rTNPTrig1-pT60pT60:val}   {\ensuremath{{1.002 } } }
\vdef{default-11woCowboyVeto:rTNPTrig1-pT60pT60:err}   {\ensuremath{{0.009 } } }
\vdef{default-11woCowboyVeto:rRhoTNPMCTrig1-pT60pT60:val}   {\ensuremath{{1.171 } } }
\vdef{default-11woCowboyVeto:rRhoTNPMCTrig1-pT60pT60:err}   {\ensuremath{{0.013 } } }
\vdef{default-11woCowboyVeto:rRhoTNPTrig1-pT60pT60:val}   {\ensuremath{{1.176 } } }
\vdef{default-11woCowboyVeto:rRhoTNPTrig1-pT60pT60:err}   {\ensuremath{{0.011 } } }
\vdef{default-11woCowboyVeto:MuidSg0-pT70pT70:val}   {\ensuremath{{0.755 } } }
\vdef{default-11woCowboyVeto:MuidNo0-pT70pT70:val}   {\ensuremath{{0.843 } } }
\vdef{default-11woCowboyVeto:TNPMCMuidSg0-pT70pT70:val}   {\ensuremath{{0.888 } } }
\vdef{default-11woCowboyVeto:TNPMCMuidNo0-pT70pT70:val}   {\ensuremath{{0.888 } } }
\vdef{default-11woCowboyVeto:TNPMuidSg0-pT70pT70:val}   {\ensuremath{{0.844 } } }
\vdef{default-11woCowboyVeto:TNPMuidNo0-pT70pT70:val}   {\ensuremath{{0.845 } } }
\vdef{default-11woCowboyVeto:rhoMuidSg0-pT70pT70:val}   {\ensuremath{{0.850 } } }
\vdef{default-11woCowboyVeto:rhoMuidNo0-pT70pT70:val}   {\ensuremath{{0.949 } } }
\vdef{default-11woCowboyVeto:rMcMuid0-pT70pT70:val}   {\ensuremath{{0.895 } } }
\vdef{default-11woCowboyVeto:rMcMuid0-pT70pT70:err}   {\ensuremath{{0.020 } } }
\vdef{default-11woCowboyVeto:rTNPMCMuid0-pT70pT70:val}   {\ensuremath{{1.000 } } }
\vdef{default-11woCowboyVeto:rTNPMCMuid0-pT70pT70:err}   {\ensuremath{{0.001 } } }
\vdef{default-11woCowboyVeto:rTNPMuid0-pT70pT70:val}   {\ensuremath{{0.998 } } }
\vdef{default-11woCowboyVeto:rTNPMuid0-pT70pT70:err}   {\ensuremath{{0.002 } } }
\vdef{default-11woCowboyVeto:rRhoTNPMCMuid0-pT70pT70:val}   {\ensuremath{{0.895 } } }
\vdef{default-11woCowboyVeto:rRhoTNPMCMuid0-pT70pT70:err}   {\ensuremath{{0.001 } } }
\vdef{default-11woCowboyVeto:rRhoTNPMuid0-pT70pT70:val}   {\ensuremath{{0.894 } } }
\vdef{default-11woCowboyVeto:rRhoTNPMuid0-pT70pT70:err}   {\ensuremath{{0.002 } } }
\vdef{default-11woCowboyVeto:TrigSg0-pT70pT70:val}   {\ensuremath{{0.912 } } }
\vdef{default-11woCowboyVeto:TrigNo0-pT70pT70:val}   {\ensuremath{{0.820 } } }
\vdef{default-11woCowboyVeto:TNPMCTrigSg0-pT70pT70:val}   {\ensuremath{{0.936 } } }
\vdef{default-11woCowboyVeto:TNPMCTrigNo0-pT70pT70:val}   {\ensuremath{{0.938 } } }
\vdef{default-11woCowboyVeto:TNPTrigSg0-pT70pT70:val}   {\ensuremath{{0.870 } } }
\vdef{default-11woCowboyVeto:TNPTrigNo0-pT70pT70:val}   {\ensuremath{{0.874 } } }
\vdef{default-11woCowboyVeto:rhoTrigSg0-pT70pT70:val}   {\ensuremath{{0.974 } } }
\vdef{default-11woCowboyVeto:rhoTrigNo0-pT70pT70:val}   {\ensuremath{{0.874 } } }
\vdef{default-11woCowboyVeto:rMcTrig0-pT70pT70:val}   {\ensuremath{{1.112 } } }
\vdef{default-11woCowboyVeto:rMcTrig0-pT70pT70:err}   {\ensuremath{{0.016 } } }
\vdef{default-11woCowboyVeto:rTNPMCTrig0-pT70pT70:val}   {\ensuremath{{0.998 } } }
\vdef{default-11woCowboyVeto:rTNPMCTrig0-pT70pT70:err}   {\ensuremath{{0.001 } } }
\vdef{default-11woCowboyVeto:rTNPTrig0-pT70pT70:val}   {\ensuremath{{0.995 } } }
\vdef{default-11woCowboyVeto:rTNPTrig0-pT70pT70:err}   {\ensuremath{{0.002 } } }
\vdef{default-11woCowboyVeto:rRhoTNPMCTrig0-pT70pT70:val}   {\ensuremath{{1.112 } } }
\vdef{default-11woCowboyVeto:rRhoTNPMCTrig0-pT70pT70:err}   {\ensuremath{{0.001 } } }
\vdef{default-11woCowboyVeto:rRhoTNPTrig0-pT70pT70:val}   {\ensuremath{{1.110 } } }
\vdef{default-11woCowboyVeto:rRhoTNPTrig0-pT70pT70:err}   {\ensuremath{{0.002 } } }
\vdef{default-11woCowboyVeto:MuidSg1-pT70pT70:val}   {\ensuremath{{0.875 } } }
\vdef{default-11woCowboyVeto:MuidNo1-pT70pT70:val}   {\ensuremath{{0.751 } } }
\vdef{default-11woCowboyVeto:TNPMCMuidSg1-pT70pT70:val}   {\ensuremath{{0.858 } } }
\vdef{default-11woCowboyVeto:TNPMCMuidNo1-pT70pT70:val}   {\ensuremath{{0.858 } } }
\vdef{default-11woCowboyVeto:TNPMuidSg1-pT70pT70:val}   {\ensuremath{{0.766 } } }
\vdef{default-11woCowboyVeto:TNPMuidNo1-pT70pT70:val}   {\ensuremath{{0.770 } } }
\vdef{default-11woCowboyVeto:rhoMuidSg1-pT70pT70:val}   {\ensuremath{{1.019 } } }
\vdef{default-11woCowboyVeto:rhoMuidNo1-pT70pT70:val}   {\ensuremath{{0.875 } } }
\vdef{default-11woCowboyVeto:rMcMuid1-pT70pT70:val}   {\ensuremath{{1.165 } } }
\vdef{default-11woCowboyVeto:rMcMuid1-pT70pT70:err}   {\ensuremath{{0.027 } } }
\vdef{default-11woCowboyVeto:rTNPMCMuid1-pT70pT70:val}   {\ensuremath{{1.001 } } }
\vdef{default-11woCowboyVeto:rTNPMCMuid1-pT70pT70:err}   {\ensuremath{{0.004 } } }
\vdef{default-11woCowboyVeto:rTNPMuid1-pT70pT70:val}   {\ensuremath{{0.995 } } }
\vdef{default-11woCowboyVeto:rTNPMuid1-pT70pT70:err}   {\ensuremath{{0.005 } } }
\vdef{default-11woCowboyVeto:rRhoTNPMCMuid1-pT70pT70:val}   {\ensuremath{{1.165 } } }
\vdef{default-11woCowboyVeto:rRhoTNPMCMuid1-pT70pT70:err}   {\ensuremath{{0.004 } } }
\vdef{default-11woCowboyVeto:rRhoTNPMuid1-pT70pT70:val}   {\ensuremath{{1.159 } } }
\vdef{default-11woCowboyVeto:rRhoTNPMuid1-pT70pT70:err}   {\ensuremath{{0.006 } } }
\vdef{default-11woCowboyVeto:TrigSg1-pT70pT70:val}   {\ensuremath{{0.722 } } }
\vdef{default-11woCowboyVeto:TrigNo1-pT70pT70:val}   {\ensuremath{{0.612 } } }
\vdef{default-11woCowboyVeto:TNPMCTrigSg1-pT70pT70:val}   {\ensuremath{{0.748 } } }
\vdef{default-11woCowboyVeto:TNPMCTrigNo1-pT70pT70:val}   {\ensuremath{{0.772 } } }
\vdef{default-11woCowboyVeto:TNPTrigSg1-pT70pT70:val}   {\ensuremath{{0.736 } } }
\vdef{default-11woCowboyVeto:TNPTrigNo1-pT70pT70:val}   {\ensuremath{{0.754 } } }
\vdef{default-11woCowboyVeto:rhoTrigSg1-pT70pT70:val}   {\ensuremath{{0.965 } } }
\vdef{default-11woCowboyVeto:rhoTrigNo1-pT70pT70:val}   {\ensuremath{{0.793 } } }
\vdef{default-11woCowboyVeto:rMcTrig1-pT70pT70:val}   {\ensuremath{{1.180 } } }
\vdef{default-11woCowboyVeto:rMcTrig1-pT70pT70:err}   {\ensuremath{{0.048 } } }
\vdef{default-11woCowboyVeto:rTNPMCTrig1-pT70pT70:val}   {\ensuremath{{0.969 } } }
\vdef{default-11woCowboyVeto:rTNPMCTrig1-pT70pT70:err}   {\ensuremath{{0.015 } } }
\vdef{default-11woCowboyVeto:rTNPTrig1-pT70pT70:val}   {\ensuremath{{0.976 } } }
\vdef{default-11woCowboyVeto:rTNPTrig1-pT70pT70:err}   {\ensuremath{{0.013 } } }
\vdef{default-11woCowboyVeto:rRhoTNPMCTrig1-pT70pT70:val}   {\ensuremath{{1.180 } } }
\vdef{default-11woCowboyVeto:rRhoTNPMCTrig1-pT70pT70:err}   {\ensuremath{{0.019 } } }
\vdef{default-11woCowboyVeto:rRhoTNPTrig1-pT70pT70:val}   {\ensuremath{{1.188 } } }
\vdef{default-11woCowboyVeto:rRhoTNPTrig1-pT70pT70:err}   {\ensuremath{{0.016 } } }
\vdef{default-11woCowboyVeto:MuidSg0-pT80pT80:val}   {\ensuremath{{0.761 } } }
\vdef{default-11woCowboyVeto:MuidNo0-pT80pT80:val}   {\ensuremath{{0.820 } } }
\vdef{default-11woCowboyVeto:TNPMCMuidSg0-pT80pT80:val}   {\ensuremath{{0.886 } } }
\vdef{default-11woCowboyVeto:TNPMCMuidNo0-pT80pT80:val}   {\ensuremath{{0.886 } } }
\vdef{default-11woCowboyVeto:TNPMuidSg0-pT80pT80:val}   {\ensuremath{{0.829 } } }
\vdef{default-11woCowboyVeto:TNPMuidNo0-pT80pT80:val}   {\ensuremath{{0.830 } } }
\vdef{default-11woCowboyVeto:rhoMuidSg0-pT80pT80:val}   {\ensuremath{{0.860 } } }
\vdef{default-11woCowboyVeto:rhoMuidNo0-pT80pT80:val}   {\ensuremath{{0.926 } } }
\vdef{default-11woCowboyVeto:rMcMuid0-pT80pT80:val}   {\ensuremath{{0.928 } } }
\vdef{default-11woCowboyVeto:rMcMuid0-pT80pT80:err}   {\ensuremath{{0.026 } } }
\vdef{default-11woCowboyVeto:rTNPMCMuid0-pT80pT80:val}   {\ensuremath{{1.000 } } }
\vdef{default-11woCowboyVeto:rTNPMCMuid0-pT80pT80:err}   {\ensuremath{{0.001 } } }
\vdef{default-11woCowboyVeto:rTNPMuid0-pT80pT80:val}   {\ensuremath{{0.998 } } }
\vdef{default-11woCowboyVeto:rTNPMuid0-pT80pT80:err}   {\ensuremath{{0.002 } } }
\vdef{default-11woCowboyVeto:rRhoTNPMCMuid0-pT80pT80:val}   {\ensuremath{{0.928 } } }
\vdef{default-11woCowboyVeto:rRhoTNPMCMuid0-pT80pT80:err}   {\ensuremath{{0.001 } } }
\vdef{default-11woCowboyVeto:rRhoTNPMuid0-pT80pT80:val}   {\ensuremath{{0.927 } } }
\vdef{default-11woCowboyVeto:rRhoTNPMuid0-pT80pT80:err}   {\ensuremath{{0.003 } } }
\vdef{default-11woCowboyVeto:TrigSg0-pT80pT80:val}   {\ensuremath{{0.911 } } }
\vdef{default-11woCowboyVeto:TrigNo0-pT80pT80:val}   {\ensuremath{{0.813 } } }
\vdef{default-11woCowboyVeto:TNPMCTrigSg0-pT80pT80:val}   {\ensuremath{{0.938 } } }
\vdef{default-11woCowboyVeto:TNPMCTrigNo0-pT80pT80:val}   {\ensuremath{{0.941 } } }
\vdef{default-11woCowboyVeto:TNPTrigSg0-pT80pT80:val}   {\ensuremath{{0.866 } } }
\vdef{default-11woCowboyVeto:TNPTrigNo0-pT80pT80:val}   {\ensuremath{{0.870 } } }
\vdef{default-11woCowboyVeto:rhoTrigSg0-pT80pT80:val}   {\ensuremath{{0.971 } } }
\vdef{default-11woCowboyVeto:rhoTrigNo0-pT80pT80:val}   {\ensuremath{{0.864 } } }
\vdef{default-11woCowboyVeto:rMcTrig0-pT80pT80:val}   {\ensuremath{{1.120 } } }
\vdef{default-11woCowboyVeto:rMcTrig0-pT80pT80:err}   {\ensuremath{{0.021 } } }
\vdef{default-11woCowboyVeto:rTNPMCTrig0-pT80pT80:val}   {\ensuremath{{0.997 } } }
\vdef{default-11woCowboyVeto:rTNPMCTrig0-pT80pT80:err}   {\ensuremath{{0.002 } } }
\vdef{default-11woCowboyVeto:rTNPTrig0-pT80pT80:val}   {\ensuremath{{0.995 } } }
\vdef{default-11woCowboyVeto:rTNPTrig0-pT80pT80:err}   {\ensuremath{{0.002 } } }
\vdef{default-11woCowboyVeto:rRhoTNPMCTrig0-pT80pT80:val}   {\ensuremath{{1.120 } } }
\vdef{default-11woCowboyVeto:rRhoTNPMCTrig0-pT80pT80:err}   {\ensuremath{{0.002 } } }
\vdef{default-11woCowboyVeto:rRhoTNPTrig0-pT80pT80:val}   {\ensuremath{{1.118 } } }
\vdef{default-11woCowboyVeto:rRhoTNPTrig0-pT80pT80:err}   {\ensuremath{{0.002 } } }
\vdef{default-11woCowboyVeto:MuidSg1-pT80pT80:val}   {\ensuremath{{0.851 } } }
\vdef{default-11woCowboyVeto:MuidNo1-pT80pT80:val}   {\ensuremath{{0.725 } } }
\vdef{default-11woCowboyVeto:TNPMCMuidSg1-pT80pT80:val}   {\ensuremath{{0.851 } } }
\vdef{default-11woCowboyVeto:TNPMCMuidNo1-pT80pT80:val}   {\ensuremath{{0.849 } } }
\vdef{default-11woCowboyVeto:TNPMuidSg1-pT80pT80:val}   {\ensuremath{{0.747 } } }
\vdef{default-11woCowboyVeto:TNPMuidNo1-pT80pT80:val}   {\ensuremath{{0.746 } } }
\vdef{default-11woCowboyVeto:rhoMuidSg1-pT80pT80:val}   {\ensuremath{{1.000 } } }
\vdef{default-11woCowboyVeto:rhoMuidNo1-pT80pT80:val}   {\ensuremath{{0.854 } } }
\vdef{default-11woCowboyVeto:rMcMuid1-pT80pT80:val}   {\ensuremath{{1.174 } } }
\vdef{default-11woCowboyVeto:rMcMuid1-pT80pT80:err}   {\ensuremath{{0.038 } } }
\vdef{default-11woCowboyVeto:rTNPMCMuid1-pT80pT80:val}   {\ensuremath{{1.003 } } }
\vdef{default-11woCowboyVeto:rTNPMCMuid1-pT80pT80:err}   {\ensuremath{{0.005 } } }
\vdef{default-11woCowboyVeto:rTNPMuid1-pT80pT80:val}   {\ensuremath{{1.001 } } }
\vdef{default-11woCowboyVeto:rTNPMuid1-pT80pT80:err}   {\ensuremath{{0.007 } } }
\vdef{default-11woCowboyVeto:rRhoTNPMCMuid1-pT80pT80:val}   {\ensuremath{{1.174 } } }
\vdef{default-11woCowboyVeto:rRhoTNPMCMuid1-pT80pT80:err}   {\ensuremath{{0.006 } } }
\vdef{default-11woCowboyVeto:rRhoTNPMuid1-pT80pT80:val}   {\ensuremath{{1.172 } } }
\vdef{default-11woCowboyVeto:rRhoTNPMuid1-pT80pT80:err}   {\ensuremath{{0.008 } } }
\vdef{default-11woCowboyVeto:TrigSg1-pT80pT80:val}   {\ensuremath{{0.727 } } }
\vdef{default-11woCowboyVeto:TrigNo1-pT80pT80:val}   {\ensuremath{{0.599 } } }
\vdef{default-11woCowboyVeto:TNPMCTrigSg1-pT80pT80:val}   {\ensuremath{{0.740 } } }
\vdef{default-11woCowboyVeto:TNPMCTrigNo1-pT80pT80:val}   {\ensuremath{{0.776 } } }
\vdef{default-11woCowboyVeto:TNPTrigSg1-pT80pT80:val}   {\ensuremath{{0.716 } } }
\vdef{default-11woCowboyVeto:TNPTrigNo1-pT80pT80:val}   {\ensuremath{{0.745 } } }
\vdef{default-11woCowboyVeto:rhoTrigSg1-pT80pT80:val}   {\ensuremath{{0.982 } } }
\vdef{default-11woCowboyVeto:rhoTrigNo1-pT80pT80:val}   {\ensuremath{{0.772 } } }
\vdef{default-11woCowboyVeto:rMcTrig1-pT80pT80:val}   {\ensuremath{{1.214 } } }
\vdef{default-11woCowboyVeto:rMcTrig1-pT80pT80:err}   {\ensuremath{{0.062 } } }
\vdef{default-11woCowboyVeto:rTNPMCTrig1-pT80pT80:val}   {\ensuremath{{0.955 } } }
\vdef{default-11woCowboyVeto:rTNPMCTrig1-pT80pT80:err}   {\ensuremath{{0.020 } } }
\vdef{default-11woCowboyVeto:rTNPTrig1-pT80pT80:val}   {\ensuremath{{0.961 } } }
\vdef{default-11woCowboyVeto:rTNPTrig1-pT80pT80:err}   {\ensuremath{{0.017 } } }
\vdef{default-11woCowboyVeto:rRhoTNPMCTrig1-pT80pT80:val}   {\ensuremath{{1.214 } } }
\vdef{default-11woCowboyVeto:rRhoTNPMCTrig1-pT80pT80:err}   {\ensuremath{{0.026 } } }
\vdef{default-11woCowboyVeto:rRhoTNPTrig1-pT80pT80:val}   {\ensuremath{{1.222 } } }
\vdef{default-11woCowboyVeto:rRhoTNPTrig1-pT80pT80:err}   {\ensuremath{{0.023 } } }
\vdef{default-11woCowboyVeto:MuidSg0-pT90pT90:val}   {\ensuremath{{0.744 } } }
\vdef{default-11woCowboyVeto:MuidNo0-pT90pT90:val}   {\ensuremath{{0.794 } } }
\vdef{default-11woCowboyVeto:TNPMCMuidSg0-pT90pT90:val}   {\ensuremath{{0.883 } } }
\vdef{default-11woCowboyVeto:TNPMCMuidNo0-pT90pT90:val}   {\ensuremath{{0.884 } } }
\vdef{default-11woCowboyVeto:TNPMuidSg0-pT90pT90:val}   {\ensuremath{{0.817 } } }
\vdef{default-11woCowboyVeto:TNPMuidNo0-pT90pT90:val}   {\ensuremath{{0.819 } } }
\vdef{default-11woCowboyVeto:rhoMuidSg0-pT90pT90:val}   {\ensuremath{{0.842 } } }
\vdef{default-11woCowboyVeto:rhoMuidNo0-pT90pT90:val}   {\ensuremath{{0.897 } } }
\vdef{default-11woCowboyVeto:rMcMuid0-pT90pT90:val}   {\ensuremath{{0.937 } } }
\vdef{default-11woCowboyVeto:rMcMuid0-pT90pT90:err}   {\ensuremath{{0.035 } } }
\vdef{default-11woCowboyVeto:rTNPMCMuid0-pT90pT90:val}   {\ensuremath{{0.999 } } }
\vdef{default-11woCowboyVeto:rTNPMCMuid0-pT90pT90:err}   {\ensuremath{{0.002 } } }
\vdef{default-11woCowboyVeto:rTNPMuid0-pT90pT90:val}   {\ensuremath{{0.998 } } }
\vdef{default-11woCowboyVeto:rTNPMuid0-pT90pT90:err}   {\ensuremath{{0.003 } } }
\vdef{default-11woCowboyVeto:rRhoTNPMCMuid0-pT90pT90:val}   {\ensuremath{{0.937 } } }
\vdef{default-11woCowboyVeto:rRhoTNPMCMuid0-pT90pT90:err}   {\ensuremath{{0.002 } } }
\vdef{default-11woCowboyVeto:rRhoTNPMuid0-pT90pT90:val}   {\ensuremath{{0.936 } } }
\vdef{default-11woCowboyVeto:rRhoTNPMuid0-pT90pT90:err}   {\ensuremath{{0.003 } } }
\vdef{default-11woCowboyVeto:TrigSg0-pT90pT90:val}   {\ensuremath{{0.919 } } }
\vdef{default-11woCowboyVeto:TrigNo0-pT90pT90:val}   {\ensuremath{{0.809 } } }
\vdef{default-11woCowboyVeto:TNPMCTrigSg0-pT90pT90:val}   {\ensuremath{{0.937 } } }
\vdef{default-11woCowboyVeto:TNPMCTrigNo0-pT90pT90:val}   {\ensuremath{{0.940 } } }
\vdef{default-11woCowboyVeto:TNPTrigSg0-pT90pT90:val}   {\ensuremath{{0.859 } } }
\vdef{default-11woCowboyVeto:TNPTrigNo0-pT90pT90:val}   {\ensuremath{{0.865 } } }
\vdef{default-11woCowboyVeto:rhoTrigSg0-pT90pT90:val}   {\ensuremath{{0.981 } } }
\vdef{default-11woCowboyVeto:rhoTrigNo0-pT90pT90:val}   {\ensuremath{{0.861 } } }
\vdef{default-11woCowboyVeto:rMcTrig0-pT90pT90:val}   {\ensuremath{{1.136 } } }
\vdef{default-11woCowboyVeto:rMcTrig0-pT90pT90:err}   {\ensuremath{{0.026 } } }
\vdef{default-11woCowboyVeto:rTNPMCTrig0-pT90pT90:val}   {\ensuremath{{0.996 } } }
\vdef{default-11woCowboyVeto:rTNPMCTrig0-pT90pT90:err}   {\ensuremath{{0.002 } } }
\vdef{default-11woCowboyVeto:rTNPTrig0-pT90pT90:val}   {\ensuremath{{0.994 } } }
\vdef{default-11woCowboyVeto:rTNPTrig0-pT90pT90:err}   {\ensuremath{{0.003 } } }
\vdef{default-11woCowboyVeto:rRhoTNPMCTrig0-pT90pT90:val}   {\ensuremath{{1.136 } } }
\vdef{default-11woCowboyVeto:rRhoTNPMCTrig0-pT90pT90:err}   {\ensuremath{{0.003 } } }
\vdef{default-11woCowboyVeto:rRhoTNPTrig0-pT90pT90:val}   {\ensuremath{{1.133 } } }
\vdef{default-11woCowboyVeto:rRhoTNPTrig0-pT90pT90:err}   {\ensuremath{{0.003 } } }
\vdef{default-11woCowboyVeto:MuidSg1-pT90pT90:val}   {\ensuremath{{0.864 } } }
\vdef{default-11woCowboyVeto:MuidNo1-pT90pT90:val}   {\ensuremath{{0.694 } } }
\vdef{default-11woCowboyVeto:TNPMCMuidSg1-pT90pT90:val}   {\ensuremath{{0.837 } } }
\vdef{default-11woCowboyVeto:TNPMCMuidNo1-pT90pT90:val}   {\ensuremath{{0.836 } } }
\vdef{default-11woCowboyVeto:TNPMuidSg1-pT90pT90:val}   {\ensuremath{{0.730 } } }
\vdef{default-11woCowboyVeto:TNPMuidNo1-pT90pT90:val}   {\ensuremath{{0.728 } } }
\vdef{default-11woCowboyVeto:rhoMuidSg1-pT90pT90:val}   {\ensuremath{{1.031 } } }
\vdef{default-11woCowboyVeto:rhoMuidNo1-pT90pT90:val}   {\ensuremath{{0.829 } } }
\vdef{default-11woCowboyVeto:rMcMuid1-pT90pT90:val}   {\ensuremath{{1.245 } } }
\vdef{default-11woCowboyVeto:rMcMuid1-pT90pT90:err}   {\ensuremath{{0.049 } } }
\vdef{default-11woCowboyVeto:rTNPMCMuid1-pT90pT90:val}   {\ensuremath{{1.001 } } }
\vdef{default-11woCowboyVeto:rTNPMCMuid1-pT90pT90:err}   {\ensuremath{{0.007 } } }
\vdef{default-11woCowboyVeto:rTNPMuid1-pT90pT90:val}   {\ensuremath{{1.003 } } }
\vdef{default-11woCowboyVeto:rTNPMuid1-pT90pT90:err}   {\ensuremath{{0.010 } } }
\vdef{default-11woCowboyVeto:rRhoTNPMCMuid1-pT90pT90:val}   {\ensuremath{{1.245 } } }
\vdef{default-11woCowboyVeto:rRhoTNPMCMuid1-pT90pT90:err}   {\ensuremath{{0.009 } } }
\vdef{default-11woCowboyVeto:rRhoTNPMuid1-pT90pT90:val}   {\ensuremath{{1.247 } } }
\vdef{default-11woCowboyVeto:rRhoTNPMuid1-pT90pT90:err}   {\ensuremath{{0.012 } } }
\vdef{default-11woCowboyVeto:TrigSg1-pT90pT90:val}   {\ensuremath{{0.705 } } }
\vdef{default-11woCowboyVeto:TrigNo1-pT90pT90:val}   {\ensuremath{{0.588 } } }
\vdef{default-11woCowboyVeto:TNPMCTrigSg1-pT90pT90:val}   {\ensuremath{{0.760 } } }
\vdef{default-11woCowboyVeto:TNPMCTrigNo1-pT90pT90:val}   {\ensuremath{{0.779 } } }
\vdef{default-11woCowboyVeto:TNPTrigSg1-pT90pT90:val}   {\ensuremath{{0.721 } } }
\vdef{default-11woCowboyVeto:TNPTrigNo1-pT90pT90:val}   {\ensuremath{{0.741 } } }
\vdef{default-11woCowboyVeto:rhoTrigSg1-pT90pT90:val}   {\ensuremath{{0.928 } } }
\vdef{default-11woCowboyVeto:rhoTrigNo1-pT90pT90:val}   {\ensuremath{{0.755 } } }
\vdef{default-11woCowboyVeto:rMcTrig1-pT90pT90:val}   {\ensuremath{{1.199 } } }
\vdef{default-11woCowboyVeto:rMcTrig1-pT90pT90:err}   {\ensuremath{{0.081 } } }
\vdef{default-11woCowboyVeto:rTNPMCTrig1-pT90pT90:val}   {\ensuremath{{0.975 } } }
\vdef{default-11woCowboyVeto:rTNPMCTrig1-pT90pT90:err}   {\ensuremath{{0.026 } } }
\vdef{default-11woCowboyVeto:rTNPTrig1-pT90pT90:val}   {\ensuremath{{0.973 } } }
\vdef{default-11woCowboyVeto:rTNPTrig1-pT90pT90:err}   {\ensuremath{{0.022 } } }
\vdef{default-11woCowboyVeto:rRhoTNPMCTrig1-pT90pT90:val}   {\ensuremath{{1.199 } } }
\vdef{default-11woCowboyVeto:rRhoTNPMCTrig1-pT90pT90:err}   {\ensuremath{{0.035 } } }
\vdef{default-11woCowboyVeto:rRhoTNPTrig1-pT90pT90:val}   {\ensuremath{{1.197 } } }
\vdef{default-11woCowboyVeto:rRhoTNPTrig1-pT90pT90:err}   {\ensuremath{{0.030 } } }
\vdef{default-11wCowboyVeto:MuidSg0-pT40pT40:val}   {\ensuremath{{0.710 } } }
\vdef{default-11wCowboyVeto:MuidNo0-pT40pT40:val}   {\ensuremath{{0.778 } } }
\vdef{default-11wCowboyVeto:TNPMCMuidSg0-pT40pT40:val}   {\ensuremath{{0.784 } } }
\vdef{default-11wCowboyVeto:TNPMCMuidNo0-pT40pT40:val}   {\ensuremath{{0.764 } } }
\vdef{default-11wCowboyVeto:TNPMuidSg0-pT40pT40:val}   {\ensuremath{{0.791 } } }
\vdef{default-11wCowboyVeto:TNPMuidNo0-pT40pT40:val}   {\ensuremath{{0.782 } } }
\vdef{default-11wCowboyVeto:rhoMuidSg0-pT40pT40:val}   {\ensuremath{{0.906 } } }
\vdef{default-11wCowboyVeto:rhoMuidNo0-pT40pT40:val}   {\ensuremath{{1.019 } } }
\vdef{default-11wCowboyVeto:rMcMuid0-pT40pT40:val}   {\ensuremath{{0.913 } } }
\vdef{default-11wCowboyVeto:rMcMuid0-pT40pT40:err}   {\ensuremath{{0.011 } } }
\vdef{default-11wCowboyVeto:rTNPMCMuid0-pT40pT40:val}   {\ensuremath{{1.027 } } }
\vdef{default-11wCowboyVeto:rTNPMCMuid0-pT40pT40:err}   {\ensuremath{{0.003 } } }
\vdef{default-11wCowboyVeto:rTNPMuid0-pT40pT40:val}   {\ensuremath{{1.012 } } }
\vdef{default-11wCowboyVeto:rTNPMuid0-pT40pT40:err}   {\ensuremath{{0.002 } } }
\vdef{default-11wCowboyVeto:rRhoTNPMCMuid0-pT40pT40:val}   {\ensuremath{{0.913 } } }
\vdef{default-11wCowboyVeto:rRhoTNPMCMuid0-pT40pT40:err}   {\ensuremath{{0.003 } } }
\vdef{default-11wCowboyVeto:rRhoTNPMuid0-pT40pT40:val}   {\ensuremath{{0.900 } } }
\vdef{default-11wCowboyVeto:rRhoTNPMuid0-pT40pT40:err}   {\ensuremath{{0.002 } } }
\vdef{default-11wCowboyVeto:TrigSg0-pT40pT40:val}   {\ensuremath{{0.841 } } }
\vdef{default-11wCowboyVeto:TrigNo0-pT40pT40:val}   {\ensuremath{{0.782 } } }
\vdef{default-11wCowboyVeto:TNPMCTrigSg0-pT40pT40:val}   {\ensuremath{{0.840 } } }
\vdef{default-11wCowboyVeto:TNPMCTrigNo0-pT40pT40:val}   {\ensuremath{{0.822 } } }
\vdef{default-11wCowboyVeto:TNPTrigSg0-pT40pT40:val}   {\ensuremath{{0.796 } } }
\vdef{default-11wCowboyVeto:TNPTrigNo0-pT40pT40:val}   {\ensuremath{{0.779 } } }
\vdef{default-11wCowboyVeto:rhoTrigSg0-pT40pT40:val}   {\ensuremath{{1.001 } } }
\vdef{default-11wCowboyVeto:rhoTrigNo0-pT40pT40:val}   {\ensuremath{{0.952 } } }
\vdef{default-11wCowboyVeto:rMcTrig0-pT40pT40:val}   {\ensuremath{{1.076 } } }
\vdef{default-11wCowboyVeto:rMcTrig0-pT40pT40:err}   {\ensuremath{{0.011 } } }
\vdef{default-11wCowboyVeto:rTNPMCTrig0-pT40pT40:val}   {\ensuremath{{1.023 } } }
\vdef{default-11wCowboyVeto:rTNPMCTrig0-pT40pT40:err}   {\ensuremath{{0.003 } } }
\vdef{default-11wCowboyVeto:rTNPTrig0-pT40pT40:val}   {\ensuremath{{1.021 } } }
\vdef{default-11wCowboyVeto:rTNPTrig0-pT40pT40:err}   {\ensuremath{{0.003 } } }
\vdef{default-11wCowboyVeto:rRhoTNPMCTrig0-pT40pT40:val}   {\ensuremath{{1.076 } } }
\vdef{default-11wCowboyVeto:rRhoTNPMCTrig0-pT40pT40:err}   {\ensuremath{{0.003 } } }
\vdef{default-11wCowboyVeto:rRhoTNPTrig0-pT40pT40:val}   {\ensuremath{{1.074 } } }
\vdef{default-11wCowboyVeto:rRhoTNPTrig0-pT40pT40:err}   {\ensuremath{{0.003 } } }
\vdef{default-11wCowboyVeto:MuidSg1-pT40pT40:val}   {\ensuremath{{0.848 } } }
\vdef{default-11wCowboyVeto:MuidNo1-pT40pT40:val}   {\ensuremath{{0.871 } } }
\vdef{default-11wCowboyVeto:TNPMCMuidSg1-pT40pT40:val}   {\ensuremath{{0.830 } } }
\vdef{default-11wCowboyVeto:TNPMCMuidNo1-pT40pT40:val}   {\ensuremath{{0.824 } } }
\vdef{default-11wCowboyVeto:TNPMuidSg1-pT40pT40:val}   {\ensuremath{{0.781 } } }
\vdef{default-11wCowboyVeto:TNPMuidNo1-pT40pT40:val}   {\ensuremath{{0.780 } } }
\vdef{default-11wCowboyVeto:rhoMuidSg1-pT40pT40:val}   {\ensuremath{{1.021 } } }
\vdef{default-11wCowboyVeto:rhoMuidNo1-pT40pT40:val}   {\ensuremath{{1.058 } } }
\vdef{default-11wCowboyVeto:rMcMuid1-pT40pT40:val}   {\ensuremath{{0.973 } } }
\vdef{default-11wCowboyVeto:rMcMuid1-pT40pT40:err}   {\ensuremath{{0.011 } } }
\vdef{default-11wCowboyVeto:rTNPMCMuid1-pT40pT40:val}   {\ensuremath{{1.008 } } }
\vdef{default-11wCowboyVeto:rTNPMCMuid1-pT40pT40:err}   {\ensuremath{{0.002 } } }
\vdef{default-11wCowboyVeto:rTNPMuid1-pT40pT40:val}   {\ensuremath{{1.002 } } }
\vdef{default-11wCowboyVeto:rTNPMuid1-pT40pT40:err}   {\ensuremath{{0.002 } } }
\vdef{default-11wCowboyVeto:rRhoTNPMCMuid1-pT40pT40:val}   {\ensuremath{{0.973 } } }
\vdef{default-11wCowboyVeto:rRhoTNPMCMuid1-pT40pT40:err}   {\ensuremath{{0.002 } } }
\vdef{default-11wCowboyVeto:rRhoTNPMuid1-pT40pT40:val}   {\ensuremath{{0.967 } } }
\vdef{default-11wCowboyVeto:rRhoTNPMuid1-pT40pT40:err}   {\ensuremath{{0.002 } } }
\vdef{default-11wCowboyVeto:TrigSg1-pT40pT40:val}   {\ensuremath{{0.731 } } }
\vdef{default-11wCowboyVeto:TrigNo1-pT40pT40:val}   {\ensuremath{{0.648 } } }
\vdef{default-11wCowboyVeto:TNPMCTrigSg1-pT40pT40:val}   {\ensuremath{{0.754 } } }
\vdef{default-11wCowboyVeto:TNPMCTrigNo1-pT40pT40:val}   {\ensuremath{{0.729 } } }
\vdef{default-11wCowboyVeto:TNPTrigSg1-pT40pT40:val}   {\ensuremath{{0.765 } } }
\vdef{default-11wCowboyVeto:TNPTrigNo1-pT40pT40:val}   {\ensuremath{{0.745 } } }
\vdef{default-11wCowboyVeto:rhoTrigSg1-pT40pT40:val}   {\ensuremath{{0.970 } } }
\vdef{default-11wCowboyVeto:rhoTrigNo1-pT40pT40:val}   {\ensuremath{{0.889 } } }
\vdef{default-11wCowboyVeto:rMcTrig1-pT40pT40:val}   {\ensuremath{{1.128 } } }
\vdef{default-11wCowboyVeto:rMcTrig1-pT40pT40:err}   {\ensuremath{{0.019 } } }
\vdef{default-11wCowboyVeto:rTNPMCTrig1-pT40pT40:val}   {\ensuremath{{1.034 } } }
\vdef{default-11wCowboyVeto:rTNPMCTrig1-pT40pT40:err}   {\ensuremath{{0.005 } } }
\vdef{default-11wCowboyVeto:rTNPTrig1-pT40pT40:val}   {\ensuremath{{1.027 } } }
\vdef{default-11wCowboyVeto:rTNPTrig1-pT40pT40:err}   {\ensuremath{{0.004 } } }
\vdef{default-11wCowboyVeto:rRhoTNPMCTrig1-pT40pT40:val}   {\ensuremath{{1.128 } } }
\vdef{default-11wCowboyVeto:rRhoTNPMCTrig1-pT40pT40:err}   {\ensuremath{{0.006 } } }
\vdef{default-11wCowboyVeto:rRhoTNPTrig1-pT40pT40:val}   {\ensuremath{{1.120 } } }
\vdef{default-11wCowboyVeto:rRhoTNPTrig1-pT40pT40:err}   {\ensuremath{{0.005 } } }
\vdef{default-11wCowboyVeto:MuidSg0-pT50pT50:val}   {\ensuremath{{0.771 } } }
\vdef{default-11wCowboyVeto:MuidNo0-pT50pT50:val}   {\ensuremath{{0.870 } } }
\vdef{default-11wCowboyVeto:TNPMCMuidSg0-pT50pT50:val}   {\ensuremath{{0.869 } } }
\vdef{default-11wCowboyVeto:TNPMCMuidNo0-pT50pT50:val}   {\ensuremath{{0.866 } } }
\vdef{default-11wCowboyVeto:TNPMuidSg0-pT50pT50:val}   {\ensuremath{{0.840 } } }
\vdef{default-11wCowboyVeto:TNPMuidNo0-pT50pT50:val}   {\ensuremath{{0.842 } } }
\vdef{default-11wCowboyVeto:rhoMuidSg0-pT50pT50:val}   {\ensuremath{{0.887 } } }
\vdef{default-11wCowboyVeto:rhoMuidNo0-pT50pT50:val}   {\ensuremath{{1.004 } } }
\vdef{default-11wCowboyVeto:rMcMuid0-pT50pT50:val}   {\ensuremath{{0.886 } } }
\vdef{default-11wCowboyVeto:rMcMuid0-pT50pT50:err}   {\ensuremath{{0.012 } } }
\vdef{default-11wCowboyVeto:rTNPMCMuid0-pT50pT50:val}   {\ensuremath{{1.003 } } }
\vdef{default-11wCowboyVeto:rTNPMCMuid0-pT50pT50:err}   {\ensuremath{{0.001 } } }
\vdef{default-11wCowboyVeto:rTNPMuid0-pT50pT50:val}   {\ensuremath{{0.998 } } }
\vdef{default-11wCowboyVeto:rTNPMuid0-pT50pT50:err}   {\ensuremath{{0.001 } } }
\vdef{default-11wCowboyVeto:rRhoTNPMCMuid0-pT50pT50:val}   {\ensuremath{{0.886 } } }
\vdef{default-11wCowboyVeto:rRhoTNPMCMuid0-pT50pT50:err}   {\ensuremath{{0.001 } } }
\vdef{default-11wCowboyVeto:rRhoTNPMuid0-pT50pT50:val}   {\ensuremath{{0.882 } } }
\vdef{default-11wCowboyVeto:rRhoTNPMuid0-pT50pT50:err}   {\ensuremath{{0.001 } } }
\vdef{default-11wCowboyVeto:TrigSg0-pT50pT50:val}   {\ensuremath{{0.897 } } }
\vdef{default-11wCowboyVeto:TrigNo0-pT50pT50:val}   {\ensuremath{{0.850 } } }
\vdef{default-11wCowboyVeto:TNPMCTrigSg0-pT50pT50:val}   {\ensuremath{{0.915 } } }
\vdef{default-11wCowboyVeto:TNPMCTrigNo0-pT50pT50:val}   {\ensuremath{{0.913 } } }
\vdef{default-11wCowboyVeto:TNPTrigSg0-pT50pT50:val}   {\ensuremath{{0.864 } } }
\vdef{default-11wCowboyVeto:TNPTrigNo0-pT50pT50:val}   {\ensuremath{{0.864 } } }
\vdef{default-11wCowboyVeto:rhoTrigSg0-pT50pT50:val}   {\ensuremath{{0.980 } } }
\vdef{default-11wCowboyVeto:rhoTrigNo0-pT50pT50:val}   {\ensuremath{{0.931 } } }
\vdef{default-11wCowboyVeto:rMcTrig0-pT50pT50:val}   {\ensuremath{{1.055 } } }
\vdef{default-11wCowboyVeto:rMcTrig0-pT50pT50:err}   {\ensuremath{{0.010 } } }
\vdef{default-11wCowboyVeto:rTNPMCTrig0-pT50pT50:val}   {\ensuremath{{1.002 } } }
\vdef{default-11wCowboyVeto:rTNPMCTrig0-pT50pT50:err}   {\ensuremath{{0.001 } } }
\vdef{default-11wCowboyVeto:rTNPTrig0-pT50pT50:val}   {\ensuremath{{1.000 } } }
\vdef{default-11wCowboyVeto:rTNPTrig0-pT50pT50:err}   {\ensuremath{{0.001 } } }
\vdef{default-11wCowboyVeto:rRhoTNPMCTrig0-pT50pT50:val}   {\ensuremath{{1.055 } } }
\vdef{default-11wCowboyVeto:rRhoTNPMCTrig0-pT50pT50:err}   {\ensuremath{{0.001 } } }
\vdef{default-11wCowboyVeto:rRhoTNPTrig0-pT50pT50:val}   {\ensuremath{{1.052 } } }
\vdef{default-11wCowboyVeto:rRhoTNPTrig0-pT50pT50:err}   {\ensuremath{{0.001 } } }
\vdef{default-11wCowboyVeto:MuidSg1-pT50pT50:val}   {\ensuremath{{0.876 } } }
\vdef{default-11wCowboyVeto:MuidNo1-pT50pT50:val}   {\ensuremath{{0.902 } } }
\vdef{default-11wCowboyVeto:TNPMCMuidSg1-pT50pT50:val}   {\ensuremath{{0.853 } } }
\vdef{default-11wCowboyVeto:TNPMCMuidNo1-pT50pT50:val}   {\ensuremath{{0.854 } } }
\vdef{default-11wCowboyVeto:TNPMuidSg1-pT50pT50:val}   {\ensuremath{{0.788 } } }
\vdef{default-11wCowboyVeto:TNPMuidNo1-pT50pT50:val}   {\ensuremath{{0.789 } } }
\vdef{default-11wCowboyVeto:rhoMuidSg1-pT50pT50:val}   {\ensuremath{{1.026 } } }
\vdef{default-11wCowboyVeto:rhoMuidNo1-pT50pT50:val}   {\ensuremath{{1.057 } } }
\vdef{default-11wCowboyVeto:rMcMuid1-pT50pT50:val}   {\ensuremath{{0.971 } } }
\vdef{default-11wCowboyVeto:rMcMuid1-pT50pT50:err}   {\ensuremath{{0.012 } } }
\vdef{default-11wCowboyVeto:rTNPMCMuid1-pT50pT50:val}   {\ensuremath{{1.000 } } }
\vdef{default-11wCowboyVeto:rTNPMCMuid1-pT50pT50:err}   {\ensuremath{{0.002 } } }
\vdef{default-11wCowboyVeto:rTNPMuid1-pT50pT50:val}   {\ensuremath{{0.999 } } }
\vdef{default-11wCowboyVeto:rTNPMuid1-pT50pT50:err}   {\ensuremath{{0.003 } } }
\vdef{default-11wCowboyVeto:rRhoTNPMCMuid1-pT50pT50:val}   {\ensuremath{{0.971 } } }
\vdef{default-11wCowboyVeto:rRhoTNPMCMuid1-pT50pT50:err}   {\ensuremath{{0.002 } } }
\vdef{default-11wCowboyVeto:rRhoTNPMuid1-pT50pT50:val}   {\ensuremath{{0.970 } } }
\vdef{default-11wCowboyVeto:rRhoTNPMuid1-pT50pT50:err}   {\ensuremath{{0.002 } } }
\vdef{default-11wCowboyVeto:TrigSg1-pT50pT50:val}   {\ensuremath{{0.742 } } }
\vdef{default-11wCowboyVeto:TrigNo1-pT50pT50:val}   {\ensuremath{{0.684 } } }
\vdef{default-11wCowboyVeto:TNPMCTrigSg1-pT50pT50:val}   {\ensuremath{{0.771 } } }
\vdef{default-11wCowboyVeto:TNPMCTrigNo1-pT50pT50:val}   {\ensuremath{{0.762 } } }
\vdef{default-11wCowboyVeto:TNPTrigSg1-pT50pT50:val}   {\ensuremath{{0.771 } } }
\vdef{default-11wCowboyVeto:TNPTrigNo1-pT50pT50:val}   {\ensuremath{{0.762 } } }
\vdef{default-11wCowboyVeto:rhoTrigSg1-pT50pT50:val}   {\ensuremath{{0.963 } } }
\vdef{default-11wCowboyVeto:rhoTrigNo1-pT50pT50:val}   {\ensuremath{{0.898 } } }
\vdef{default-11wCowboyVeto:rMcTrig1-pT50pT50:val}   {\ensuremath{{1.085 } } }
\vdef{default-11wCowboyVeto:rMcTrig1-pT50pT50:err}   {\ensuremath{{0.023 } } }
\vdef{default-11wCowboyVeto:rTNPMCTrig1-pT50pT50:val}   {\ensuremath{{1.011 } } }
\vdef{default-11wCowboyVeto:rTNPMCTrig1-pT50pT50:err}   {\ensuremath{{0.007 } } }
\vdef{default-11wCowboyVeto:rTNPTrig1-pT50pT50:val}   {\ensuremath{{1.013 } } }
\vdef{default-11wCowboyVeto:rTNPTrig1-pT50pT50:err}   {\ensuremath{{0.006 } } }
\vdef{default-11wCowboyVeto:rRhoTNPMCTrig1-pT50pT50:val}   {\ensuremath{{1.085 } } }
\vdef{default-11wCowboyVeto:rRhoTNPMCTrig1-pT50pT50:err}   {\ensuremath{{0.008 } } }
\vdef{default-11wCowboyVeto:rRhoTNPTrig1-pT50pT50:val}   {\ensuremath{{1.086 } } }
\vdef{default-11wCowboyVeto:rRhoTNPTrig1-pT50pT50:err}   {\ensuremath{{0.007 } } }
\vdef{default-11wCowboyVeto:MuidSg0-pT60pT60:val}   {\ensuremath{{0.776 } } }
\vdef{default-11wCowboyVeto:MuidNo0-pT60pT60:val}   {\ensuremath{{0.902 } } }
\vdef{default-11wCowboyVeto:TNPMCMuidSg0-pT60pT60:val}   {\ensuremath{{0.886 } } }
\vdef{default-11wCowboyVeto:TNPMCMuidNo0-pT60pT60:val}   {\ensuremath{{0.886 } } }
\vdef{default-11wCowboyVeto:TNPMuidSg0-pT60pT60:val}   {\ensuremath{{0.844 } } }
\vdef{default-11wCowboyVeto:TNPMuidNo0-pT60pT60:val}   {\ensuremath{{0.846 } } }
\vdef{default-11wCowboyVeto:rhoMuidSg0-pT60pT60:val}   {\ensuremath{{0.876 } } }
\vdef{default-11wCowboyVeto:rhoMuidNo0-pT60pT60:val}   {\ensuremath{{1.017 } } }
\vdef{default-11wCowboyVeto:rMcMuid0-pT60pT60:val}   {\ensuremath{{0.861 } } }
\vdef{default-11wCowboyVeto:rMcMuid0-pT60pT60:err}   {\ensuremath{{0.015 } } }
\vdef{default-11wCowboyVeto:rTNPMCMuid0-pT60pT60:val}   {\ensuremath{{1.000 } } }
\vdef{default-11wCowboyVeto:rTNPMCMuid0-pT60pT60:err}   {\ensuremath{{0.001 } } }
\vdef{default-11wCowboyVeto:rTNPMuid0-pT60pT60:val}   {\ensuremath{{0.997 } } }
\vdef{default-11wCowboyVeto:rTNPMuid0-pT60pT60:err}   {\ensuremath{{0.001 } } }
\vdef{default-11wCowboyVeto:rRhoTNPMCMuid0-pT60pT60:val}   {\ensuremath{{0.861 } } }
\vdef{default-11wCowboyVeto:rRhoTNPMCMuid0-pT60pT60:err}   {\ensuremath{{0.001 } } }
\vdef{default-11wCowboyVeto:rRhoTNPMuid0-pT60pT60:val}   {\ensuremath{{0.859 } } }
\vdef{default-11wCowboyVeto:rRhoTNPMuid0-pT60pT60:err}   {\ensuremath{{0.001 } } }
\vdef{default-11wCowboyVeto:TrigSg0-pT60pT60:val}   {\ensuremath{{0.912 } } }
\vdef{default-11wCowboyVeto:TrigNo0-pT60pT60:val}   {\ensuremath{{0.866 } } }
\vdef{default-11wCowboyVeto:TNPMCTrigSg0-pT60pT60:val}   {\ensuremath{{0.932 } } }
\vdef{default-11wCowboyVeto:TNPMCTrigNo0-pT60pT60:val}   {\ensuremath{{0.932 } } }
\vdef{default-11wCowboyVeto:TNPTrigSg0-pT60pT60:val}   {\ensuremath{{0.874 } } }
\vdef{default-11wCowboyVeto:TNPTrigNo0-pT60pT60:val}   {\ensuremath{{0.877 } } }
\vdef{default-11wCowboyVeto:rhoTrigSg0-pT60pT60:val}   {\ensuremath{{0.979 } } }
\vdef{default-11wCowboyVeto:rhoTrigNo0-pT60pT60:val}   {\ensuremath{{0.929 } } }
\vdef{default-11wCowboyVeto:rMcTrig0-pT60pT60:val}   {\ensuremath{{1.053 } } }
\vdef{default-11wCowboyVeto:rMcTrig0-pT60pT60:err}   {\ensuremath{{0.012 } } }
\vdef{default-11wCowboyVeto:rTNPMCTrig0-pT60pT60:val}   {\ensuremath{{0.999 } } }
\vdef{default-11wCowboyVeto:rTNPMCTrig0-pT60pT60:err}   {\ensuremath{{0.001 } } }
\vdef{default-11wCowboyVeto:rTNPTrig0-pT60pT60:val}   {\ensuremath{{0.996 } } }
\vdef{default-11wCowboyVeto:rTNPTrig0-pT60pT60:err}   {\ensuremath{{0.001 } } }
\vdef{default-11wCowboyVeto:rRhoTNPMCTrig0-pT60pT60:val}   {\ensuremath{{1.053 } } }
\vdef{default-11wCowboyVeto:rRhoTNPMCTrig0-pT60pT60:err}   {\ensuremath{{0.001 } } }
\vdef{default-11wCowboyVeto:rRhoTNPTrig0-pT60pT60:val}   {\ensuremath{{1.050 } } }
\vdef{default-11wCowboyVeto:rRhoTNPTrig0-pT60pT60:err}   {\ensuremath{{0.001 } } }
\vdef{default-11wCowboyVeto:MuidSg1-pT60pT60:val}   {\ensuremath{{0.877 } } }
\vdef{default-11wCowboyVeto:MuidNo1-pT60pT60:val}   {\ensuremath{{0.913 } } }
\vdef{default-11wCowboyVeto:TNPMCMuidSg1-pT60pT60:val}   {\ensuremath{{0.859 } } }
\vdef{default-11wCowboyVeto:TNPMCMuidNo1-pT60pT60:val}   {\ensuremath{{0.861 } } }
\vdef{default-11wCowboyVeto:TNPMuidSg1-pT60pT60:val}   {\ensuremath{{0.771 } } }
\vdef{default-11wCowboyVeto:TNPMuidNo1-pT60pT60:val}   {\ensuremath{{0.773 } } }
\vdef{default-11wCowboyVeto:rhoMuidSg1-pT60pT60:val}   {\ensuremath{{1.020 } } }
\vdef{default-11wCowboyVeto:rhoMuidNo1-pT60pT60:val}   {\ensuremath{{1.060 } } }
\vdef{default-11wCowboyVeto:rMcMuid1-pT60pT60:val}   {\ensuremath{{0.961 } } }
\vdef{default-11wCowboyVeto:rMcMuid1-pT60pT60:err}   {\ensuremath{{0.017 } } }
\vdef{default-11wCowboyVeto:rTNPMCMuid1-pT60pT60:val}   {\ensuremath{{0.998 } } }
\vdef{default-11wCowboyVeto:rTNPMCMuid1-pT60pT60:err}   {\ensuremath{{0.003 } } }
\vdef{default-11wCowboyVeto:rTNPMuid1-pT60pT60:val}   {\ensuremath{{0.997 } } }
\vdef{default-11wCowboyVeto:rTNPMuid1-pT60pT60:err}   {\ensuremath{{0.004 } } }
\vdef{default-11wCowboyVeto:rRhoTNPMCMuid1-pT60pT60:val}   {\ensuremath{{0.961 } } }
\vdef{default-11wCowboyVeto:rRhoTNPMCMuid1-pT60pT60:err}   {\ensuremath{{0.002 } } }
\vdef{default-11wCowboyVeto:rRhoTNPMuid1-pT60pT60:val}   {\ensuremath{{0.960 } } }
\vdef{default-11wCowboyVeto:rRhoTNPMuid1-pT60pT60:err}   {\ensuremath{{0.003 } } }
\vdef{default-11wCowboyVeto:TrigSg1-pT60pT60:val}   {\ensuremath{{0.717 } } }
\vdef{default-11wCowboyVeto:TrigNo1-pT60pT60:val}   {\ensuremath{{0.696 } } }
\vdef{default-11wCowboyVeto:TNPMCTrigSg1-pT60pT60:val}   {\ensuremath{{0.766 } } }
\vdef{default-11wCowboyVeto:TNPMCTrigNo1-pT60pT60:val}   {\ensuremath{{0.770 } } }
\vdef{default-11wCowboyVeto:TNPTrigSg1-pT60pT60:val}   {\ensuremath{{0.760 } } }
\vdef{default-11wCowboyVeto:TNPTrigNo1-pT60pT60:val}   {\ensuremath{{0.760 } } }
\vdef{default-11wCowboyVeto:rhoTrigSg1-pT60pT60:val}   {\ensuremath{{0.935 } } }
\vdef{default-11wCowboyVeto:rhoTrigNo1-pT60pT60:val}   {\ensuremath{{0.904 } } }
\vdef{default-11wCowboyVeto:rMcTrig1-pT60pT60:val}   {\ensuremath{{1.029 } } }
\vdef{default-11wCowboyVeto:rMcTrig1-pT60pT60:err}   {\ensuremath{{0.032 } } }
\vdef{default-11wCowboyVeto:rTNPMCTrig1-pT60pT60:val}   {\ensuremath{{0.995 } } }
\vdef{default-11wCowboyVeto:rTNPMCTrig1-pT60pT60:err}   {\ensuremath{{0.011 } } }
\vdef{default-11wCowboyVeto:rTNPTrig1-pT60pT60:val}   {\ensuremath{{1.000 } } }
\vdef{default-11wCowboyVeto:rTNPTrig1-pT60pT60:err}   {\ensuremath{{0.009 } } }
\vdef{default-11wCowboyVeto:rRhoTNPMCTrig1-pT60pT60:val}   {\ensuremath{{1.029 } } }
\vdef{default-11wCowboyVeto:rRhoTNPMCTrig1-pT60pT60:err}   {\ensuremath{{0.012 } } }
\vdef{default-11wCowboyVeto:rRhoTNPTrig1-pT60pT60:val}   {\ensuremath{{1.034 } } }
\vdef{default-11wCowboyVeto:rRhoTNPTrig1-pT60pT60:err}   {\ensuremath{{0.010 } } }
\vdef{default-11wCowboyVeto:MuidSg0-pT70pT70:val}   {\ensuremath{{0.755 } } }
\vdef{default-11wCowboyVeto:MuidNo0-pT70pT70:val}   {\ensuremath{{0.913 } } }
\vdef{default-11wCowboyVeto:TNPMCMuidSg0-pT70pT70:val}   {\ensuremath{{0.888 } } }
\vdef{default-11wCowboyVeto:TNPMCMuidNo0-pT70pT70:val}   {\ensuremath{{0.888 } } }
\vdef{default-11wCowboyVeto:TNPMuidSg0-pT70pT70:val}   {\ensuremath{{0.844 } } }
\vdef{default-11wCowboyVeto:TNPMuidNo0-pT70pT70:val}   {\ensuremath{{0.845 } } }
\vdef{default-11wCowboyVeto:rhoMuidSg0-pT70pT70:val}   {\ensuremath{{0.850 } } }
\vdef{default-11wCowboyVeto:rhoMuidNo0-pT70pT70:val}   {\ensuremath{{1.028 } } }
\vdef{default-11wCowboyVeto:rMcMuid0-pT70pT70:val}   {\ensuremath{{0.827 } } }
\vdef{default-11wCowboyVeto:rMcMuid0-pT70pT70:err}   {\ensuremath{{0.019 } } }
\vdef{default-11wCowboyVeto:rTNPMCMuid0-pT70pT70:val}   {\ensuremath{{1.000 } } }
\vdef{default-11wCowboyVeto:rTNPMCMuid0-pT70pT70:err}   {\ensuremath{{0.001 } } }
\vdef{default-11wCowboyVeto:rTNPMuid0-pT70pT70:val}   {\ensuremath{{0.998 } } }
\vdef{default-11wCowboyVeto:rTNPMuid0-pT70pT70:err}   {\ensuremath{{0.002 } } }
\vdef{default-11wCowboyVeto:rRhoTNPMCMuid0-pT70pT70:val}   {\ensuremath{{0.827 } } }
\vdef{default-11wCowboyVeto:rRhoTNPMCMuid0-pT70pT70:err}   {\ensuremath{{0.001 } } }
\vdef{default-11wCowboyVeto:rRhoTNPMuid0-pT70pT70:val}   {\ensuremath{{0.825 } } }
\vdef{default-11wCowboyVeto:rRhoTNPMuid0-pT70pT70:err}   {\ensuremath{{0.002 } } }
\vdef{default-11wCowboyVeto:TrigSg0-pT70pT70:val}   {\ensuremath{{0.912 } } }
\vdef{default-11wCowboyVeto:TrigNo0-pT70pT70:val}   {\ensuremath{{0.873 } } }
\vdef{default-11wCowboyVeto:TNPMCTrigSg0-pT70pT70:val}   {\ensuremath{{0.936 } } }
\vdef{default-11wCowboyVeto:TNPMCTrigNo0-pT70pT70:val}   {\ensuremath{{0.938 } } }
\vdef{default-11wCowboyVeto:TNPTrigSg0-pT70pT70:val}   {\ensuremath{{0.870 } } }
\vdef{default-11wCowboyVeto:TNPTrigNo0-pT70pT70:val}   {\ensuremath{{0.874 } } }
\vdef{default-11wCowboyVeto:rhoTrigSg0-pT70pT70:val}   {\ensuremath{{0.974 } } }
\vdef{default-11wCowboyVeto:rhoTrigNo0-pT70pT70:val}   {\ensuremath{{0.930 } } }
\vdef{default-11wCowboyVeto:rMcTrig0-pT70pT70:val}   {\ensuremath{{1.045 } } }
\vdef{default-11wCowboyVeto:rMcTrig0-pT70pT70:err}   {\ensuremath{{0.015 } } }
\vdef{default-11wCowboyVeto:rTNPMCTrig0-pT70pT70:val}   {\ensuremath{{0.998 } } }
\vdef{default-11wCowboyVeto:rTNPMCTrig0-pT70pT70:err}   {\ensuremath{{0.001 } } }
\vdef{default-11wCowboyVeto:rTNPTrig0-pT70pT70:val}   {\ensuremath{{0.995 } } }
\vdef{default-11wCowboyVeto:rTNPTrig0-pT70pT70:err}   {\ensuremath{{0.002 } } }
\vdef{default-11wCowboyVeto:rRhoTNPMCTrig0-pT70pT70:val}   {\ensuremath{{1.045 } } }
\vdef{default-11wCowboyVeto:rRhoTNPMCTrig0-pT70pT70:err}   {\ensuremath{{0.001 } } }
\vdef{default-11wCowboyVeto:rRhoTNPTrig0-pT70pT70:val}   {\ensuremath{{1.043 } } }
\vdef{default-11wCowboyVeto:rRhoTNPTrig0-pT70pT70:err}   {\ensuremath{{0.002 } } }
\vdef{default-11wCowboyVeto:MuidSg1-pT70pT70:val}   {\ensuremath{{0.875 } } }
\vdef{default-11wCowboyVeto:MuidNo1-pT70pT70:val}   {\ensuremath{{0.914 } } }
\vdef{default-11wCowboyVeto:TNPMCMuidSg1-pT70pT70:val}   {\ensuremath{{0.858 } } }
\vdef{default-11wCowboyVeto:TNPMCMuidNo1-pT70pT70:val}   {\ensuremath{{0.858 } } }
\vdef{default-11wCowboyVeto:TNPMuidSg1-pT70pT70:val}   {\ensuremath{{0.766 } } }
\vdef{default-11wCowboyVeto:TNPMuidNo1-pT70pT70:val}   {\ensuremath{{0.769 } } }
\vdef{default-11wCowboyVeto:rhoMuidSg1-pT70pT70:val}   {\ensuremath{{1.019 } } }
\vdef{default-11wCowboyVeto:rhoMuidNo1-pT70pT70:val}   {\ensuremath{{1.066 } } }
\vdef{default-11wCowboyVeto:rMcMuid1-pT70pT70:val}   {\ensuremath{{0.957 } } }
\vdef{default-11wCowboyVeto:rMcMuid1-pT70pT70:err}   {\ensuremath{{0.022 } } }
\vdef{default-11wCowboyVeto:rTNPMCMuid1-pT70pT70:val}   {\ensuremath{{1.001 } } }
\vdef{default-11wCowboyVeto:rTNPMCMuid1-pT70pT70:err}   {\ensuremath{{0.004 } } }
\vdef{default-11wCowboyVeto:rTNPMuid1-pT70pT70:val}   {\ensuremath{{0.996 } } }
\vdef{default-11wCowboyVeto:rTNPMuid1-pT70pT70:err}   {\ensuremath{{0.005 } } }
\vdef{default-11wCowboyVeto:rRhoTNPMCMuid1-pT70pT70:val}   {\ensuremath{{0.957 } } }
\vdef{default-11wCowboyVeto:rRhoTNPMCMuid1-pT70pT70:err}   {\ensuremath{{0.003 } } }
\vdef{default-11wCowboyVeto:rRhoTNPMuid1-pT70pT70:val}   {\ensuremath{{0.953 } } }
\vdef{default-11wCowboyVeto:rRhoTNPMuid1-pT70pT70:err}   {\ensuremath{{0.005 } } }
\vdef{default-11wCowboyVeto:TrigSg1-pT70pT70:val}   {\ensuremath{{0.722 } } }
\vdef{default-11wCowboyVeto:TrigNo1-pT70pT70:val}   {\ensuremath{{0.699 } } }
\vdef{default-11wCowboyVeto:TNPMCTrigSg1-pT70pT70:val}   {\ensuremath{{0.748 } } }
\vdef{default-11wCowboyVeto:TNPMCTrigNo1-pT70pT70:val}   {\ensuremath{{0.772 } } }
\vdef{default-11wCowboyVeto:TNPTrigSg1-pT70pT70:val}   {\ensuremath{{0.736 } } }
\vdef{default-11wCowboyVeto:TNPTrigNo1-pT70pT70:val}   {\ensuremath{{0.754 } } }
\vdef{default-11wCowboyVeto:rhoTrigSg1-pT70pT70:val}   {\ensuremath{{0.965 } } }
\vdef{default-11wCowboyVeto:rhoTrigNo1-pT70pT70:val}   {\ensuremath{{0.905 } } }
\vdef{default-11wCowboyVeto:rMcTrig1-pT70pT70:val}   {\ensuremath{{1.034 } } }
\vdef{default-11wCowboyVeto:rMcTrig1-pT70pT70:err}   {\ensuremath{{0.042 } } }
\vdef{default-11wCowboyVeto:rTNPMCTrig1-pT70pT70:val}   {\ensuremath{{0.969 } } }
\vdef{default-11wCowboyVeto:rTNPMCTrig1-pT70pT70:err}   {\ensuremath{{0.015 } } }
\vdef{default-11wCowboyVeto:rTNPTrig1-pT70pT70:val}   {\ensuremath{{0.977 } } }
\vdef{default-11wCowboyVeto:rTNPTrig1-pT70pT70:err}   {\ensuremath{{0.013 } } }
\vdef{default-11wCowboyVeto:rRhoTNPMCTrig1-pT70pT70:val}   {\ensuremath{{1.034 } } }
\vdef{default-11wCowboyVeto:rRhoTNPMCTrig1-pT70pT70:err}   {\ensuremath{{0.017 } } }
\vdef{default-11wCowboyVeto:rRhoTNPTrig1-pT70pT70:val}   {\ensuremath{{1.041 } } }
\vdef{default-11wCowboyVeto:rRhoTNPTrig1-pT70pT70:err}   {\ensuremath{{0.015 } } }
\vdef{default-11wCowboyVeto:MuidSg0-pT80pT80:val}   {\ensuremath{{0.761 } } }
\vdef{default-11wCowboyVeto:MuidNo0-pT80pT80:val}   {\ensuremath{{0.917 } } }
\vdef{default-11wCowboyVeto:TNPMCMuidSg0-pT80pT80:val}   {\ensuremath{{0.886 } } }
\vdef{default-11wCowboyVeto:TNPMCMuidNo0-pT80pT80:val}   {\ensuremath{{0.886 } } }
\vdef{default-11wCowboyVeto:TNPMuidSg0-pT80pT80:val}   {\ensuremath{{0.829 } } }
\vdef{default-11wCowboyVeto:TNPMuidNo0-pT80pT80:val}   {\ensuremath{{0.830 } } }
\vdef{default-11wCowboyVeto:rhoMuidSg0-pT80pT80:val}   {\ensuremath{{0.860 } } }
\vdef{default-11wCowboyVeto:rhoMuidNo0-pT80pT80:val}   {\ensuremath{{1.035 } } }
\vdef{default-11wCowboyVeto:rMcMuid0-pT80pT80:val}   {\ensuremath{{0.830 } } }
\vdef{default-11wCowboyVeto:rMcMuid0-pT80pT80:err}   {\ensuremath{{0.023 } } }
\vdef{default-11wCowboyVeto:rTNPMCMuid0-pT80pT80:val}   {\ensuremath{{1.000 } } }
\vdef{default-11wCowboyVeto:rTNPMCMuid0-pT80pT80:err}   {\ensuremath{{0.001 } } }
\vdef{default-11wCowboyVeto:rTNPMuid0-pT80pT80:val}   {\ensuremath{{0.998 } } }
\vdef{default-11wCowboyVeto:rTNPMuid0-pT80pT80:err}   {\ensuremath{{0.002 } } }
\vdef{default-11wCowboyVeto:rRhoTNPMCMuid0-pT80pT80:val}   {\ensuremath{{0.830 } } }
\vdef{default-11wCowboyVeto:rRhoTNPMCMuid0-pT80pT80:err}   {\ensuremath{{0.001 } } }
\vdef{default-11wCowboyVeto:rRhoTNPMuid0-pT80pT80:val}   {\ensuremath{{0.829 } } }
\vdef{default-11wCowboyVeto:rRhoTNPMuid0-pT80pT80:err}   {\ensuremath{{0.002 } } }
\vdef{default-11wCowboyVeto:TrigSg0-pT80pT80:val}   {\ensuremath{{0.911 } } }
\vdef{default-11wCowboyVeto:TrigNo0-pT80pT80:val}   {\ensuremath{{0.873 } } }
\vdef{default-11wCowboyVeto:TNPMCTrigSg0-pT80pT80:val}   {\ensuremath{{0.938 } } }
\vdef{default-11wCowboyVeto:TNPMCTrigNo0-pT80pT80:val}   {\ensuremath{{0.941 } } }
\vdef{default-11wCowboyVeto:TNPTrigSg0-pT80pT80:val}   {\ensuremath{{0.866 } } }
\vdef{default-11wCowboyVeto:TNPTrigNo0-pT80pT80:val}   {\ensuremath{{0.870 } } }
\vdef{default-11wCowboyVeto:rhoTrigSg0-pT80pT80:val}   {\ensuremath{{0.971 } } }
\vdef{default-11wCowboyVeto:rhoTrigNo0-pT80pT80:val}   {\ensuremath{{0.927 } } }
\vdef{default-11wCowboyVeto:rMcTrig0-pT80pT80:val}   {\ensuremath{{1.044 } } }
\vdef{default-11wCowboyVeto:rMcTrig0-pT80pT80:err}   {\ensuremath{{0.019 } } }
\vdef{default-11wCowboyVeto:rTNPMCTrig0-pT80pT80:val}   {\ensuremath{{0.997 } } }
\vdef{default-11wCowboyVeto:rTNPMCTrig0-pT80pT80:err}   {\ensuremath{{0.002 } } }
\vdef{default-11wCowboyVeto:rTNPTrig0-pT80pT80:val}   {\ensuremath{{0.995 } } }
\vdef{default-11wCowboyVeto:rTNPTrig0-pT80pT80:err}   {\ensuremath{{0.002 } } }
\vdef{default-11wCowboyVeto:rRhoTNPMCTrig0-pT80pT80:val}   {\ensuremath{{1.044 } } }
\vdef{default-11wCowboyVeto:rRhoTNPMCTrig0-pT80pT80:err}   {\ensuremath{{0.002 } } }
\vdef{default-11wCowboyVeto:rRhoTNPTrig0-pT80pT80:val}   {\ensuremath{{1.042 } } }
\vdef{default-11wCowboyVeto:rRhoTNPTrig0-pT80pT80:err}   {\ensuremath{{0.002 } } }
\vdef{default-11wCowboyVeto:MuidSg1-pT80pT80:val}   {\ensuremath{{0.851 } } }
\vdef{default-11wCowboyVeto:MuidNo1-pT80pT80:val}   {\ensuremath{{0.906 } } }
\vdef{default-11wCowboyVeto:TNPMCMuidSg1-pT80pT80:val}   {\ensuremath{{0.851 } } }
\vdef{default-11wCowboyVeto:TNPMCMuidNo1-pT80pT80:val}   {\ensuremath{{0.849 } } }
\vdef{default-11wCowboyVeto:TNPMuidSg1-pT80pT80:val}   {\ensuremath{{0.747 } } }
\vdef{default-11wCowboyVeto:TNPMuidNo1-pT80pT80:val}   {\ensuremath{{0.746 } } }
\vdef{default-11wCowboyVeto:rhoMuidSg1-pT80pT80:val}   {\ensuremath{{1.000 } } }
\vdef{default-11wCowboyVeto:rhoMuidNo1-pT80pT80:val}   {\ensuremath{{1.067 } } }
\vdef{default-11wCowboyVeto:rMcMuid1-pT80pT80:val}   {\ensuremath{{0.940 } } }
\vdef{default-11wCowboyVeto:rMcMuid1-pT80pT80:err}   {\ensuremath{{0.030 } } }
\vdef{default-11wCowboyVeto:rTNPMCMuid1-pT80pT80:val}   {\ensuremath{{1.003 } } }
\vdef{default-11wCowboyVeto:rTNPMCMuid1-pT80pT80:err}   {\ensuremath{{0.005 } } }
\vdef{default-11wCowboyVeto:rTNPMuid1-pT80pT80:val}   {\ensuremath{{1.002 } } }
\vdef{default-11wCowboyVeto:rTNPMuid1-pT80pT80:err}   {\ensuremath{{0.007 } } }
\vdef{default-11wCowboyVeto:rRhoTNPMCMuid1-pT80pT80:val}   {\ensuremath{{0.940 } } }
\vdef{default-11wCowboyVeto:rRhoTNPMCMuid1-pT80pT80:err}   {\ensuremath{{0.005 } } }
\vdef{default-11wCowboyVeto:rRhoTNPMuid1-pT80pT80:val}   {\ensuremath{{0.939 } } }
\vdef{default-11wCowboyVeto:rRhoTNPMuid1-pT80pT80:err}   {\ensuremath{{0.006 } } }
\vdef{default-11wCowboyVeto:TrigSg1-pT80pT80:val}   {\ensuremath{{0.727 } } }
\vdef{default-11wCowboyVeto:TrigNo1-pT80pT80:val}   {\ensuremath{{0.694 } } }
\vdef{default-11wCowboyVeto:TNPMCTrigSg1-pT80pT80:val}   {\ensuremath{{0.740 } } }
\vdef{default-11wCowboyVeto:TNPMCTrigNo1-pT80pT80:val}   {\ensuremath{{0.777 } } }
\vdef{default-11wCowboyVeto:TNPTrigSg1-pT80pT80:val}   {\ensuremath{{0.716 } } }
\vdef{default-11wCowboyVeto:TNPTrigNo1-pT80pT80:val}   {\ensuremath{{0.746 } } }
\vdef{default-11wCowboyVeto:rhoTrigSg1-pT80pT80:val}   {\ensuremath{{0.982 } } }
\vdef{default-11wCowboyVeto:rhoTrigNo1-pT80pT80:val}   {\ensuremath{{0.893 } } }
\vdef{default-11wCowboyVeto:rMcTrig1-pT80pT80:val}   {\ensuremath{{1.048 } } }
\vdef{default-11wCowboyVeto:rMcTrig1-pT80pT80:err}   {\ensuremath{{0.053 } } }
\vdef{default-11wCowboyVeto:rTNPMCTrig1-pT80pT80:val}   {\ensuremath{{0.953 } } }
\vdef{default-11wCowboyVeto:rTNPMCTrig1-pT80pT80:err}   {\ensuremath{{0.020 } } }
\vdef{default-11wCowboyVeto:rTNPTrig1-pT80pT80:val}   {\ensuremath{{0.960 } } }
\vdef{default-11wCowboyVeto:rTNPTrig1-pT80pT80:err}   {\ensuremath{{0.018 } } }
\vdef{default-11wCowboyVeto:rRhoTNPMCTrig1-pT80pT80:val}   {\ensuremath{{1.048 } } }
\vdef{default-11wCowboyVeto:rRhoTNPMCTrig1-pT80pT80:err}   {\ensuremath{{0.023 } } }
\vdef{default-11wCowboyVeto:rRhoTNPTrig1-pT80pT80:val}   {\ensuremath{{1.056 } } }
\vdef{default-11wCowboyVeto:rRhoTNPTrig1-pT80pT80:err}   {\ensuremath{{0.020 } } }
\vdef{default-11wCowboyVeto:MuidSg0-pT90pT90:val}   {\ensuremath{{0.744 } } }
\vdef{default-11wCowboyVeto:MuidNo0-pT90pT90:val}   {\ensuremath{{0.917 } } }
\vdef{default-11wCowboyVeto:TNPMCMuidSg0-pT90pT90:val}   {\ensuremath{{0.883 } } }
\vdef{default-11wCowboyVeto:TNPMCMuidNo0-pT90pT90:val}   {\ensuremath{{0.885 } } }
\vdef{default-11wCowboyVeto:TNPMuidSg0-pT90pT90:val}   {\ensuremath{{0.817 } } }
\vdef{default-11wCowboyVeto:TNPMuidNo0-pT90pT90:val}   {\ensuremath{{0.819 } } }
\vdef{default-11wCowboyVeto:rhoMuidSg0-pT90pT90:val}   {\ensuremath{{0.842 } } }
\vdef{default-11wCowboyVeto:rhoMuidNo0-pT90pT90:val}   {\ensuremath{{1.037 } } }
\vdef{default-11wCowboyVeto:rMcMuid0-pT90pT90:val}   {\ensuremath{{0.811 } } }
\vdef{default-11wCowboyVeto:rMcMuid0-pT90pT90:err}   {\ensuremath{{0.030 } } }
\vdef{default-11wCowboyVeto:rTNPMCMuid0-pT90pT90:val}   {\ensuremath{{0.999 } } }
\vdef{default-11wCowboyVeto:rTNPMCMuid0-pT90pT90:err}   {\ensuremath{{0.002 } } }
\vdef{default-11wCowboyVeto:rTNPMuid0-pT90pT90:val}   {\ensuremath{{0.997 } } }
\vdef{default-11wCowboyVeto:rTNPMuid0-pT90pT90:err}   {\ensuremath{{0.003 } } }
\vdef{default-11wCowboyVeto:rRhoTNPMCMuid0-pT90pT90:val}   {\ensuremath{{0.811 } } }
\vdef{default-11wCowboyVeto:rRhoTNPMCMuid0-pT90pT90:err}   {\ensuremath{{0.002 } } }
\vdef{default-11wCowboyVeto:rRhoTNPMuid0-pT90pT90:val}   {\ensuremath{{0.810 } } }
\vdef{default-11wCowboyVeto:rRhoTNPMuid0-pT90pT90:err}   {\ensuremath{{0.003 } } }
\vdef{default-11wCowboyVeto:TrigSg0-pT90pT90:val}   {\ensuremath{{0.919 } } }
\vdef{default-11wCowboyVeto:TrigNo0-pT90pT90:val}   {\ensuremath{{0.871 } } }
\vdef{default-11wCowboyVeto:TNPMCTrigSg0-pT90pT90:val}   {\ensuremath{{0.937 } } }
\vdef{default-11wCowboyVeto:TNPMCTrigNo0-pT90pT90:val}   {\ensuremath{{0.940 } } }
\vdef{default-11wCowboyVeto:TNPTrigSg0-pT90pT90:val}   {\ensuremath{{0.859 } } }
\vdef{default-11wCowboyVeto:TNPTrigNo0-pT90pT90:val}   {\ensuremath{{0.865 } } }
\vdef{default-11wCowboyVeto:rhoTrigSg0-pT90pT90:val}   {\ensuremath{{0.981 } } }
\vdef{default-11wCowboyVeto:rhoTrigNo0-pT90pT90:val}   {\ensuremath{{0.926 } } }
\vdef{default-11wCowboyVeto:rMcTrig0-pT90pT90:val}   {\ensuremath{{1.055 } } }
\vdef{default-11wCowboyVeto:rMcTrig0-pT90pT90:err}   {\ensuremath{{0.024 } } }
\vdef{default-11wCowboyVeto:rTNPMCTrig0-pT90pT90:val}   {\ensuremath{{0.996 } } }
\vdef{default-11wCowboyVeto:rTNPMCTrig0-pT90pT90:err}   {\ensuremath{{0.002 } } }
\vdef{default-11wCowboyVeto:rTNPTrig0-pT90pT90:val}   {\ensuremath{{0.993 } } }
\vdef{default-11wCowboyVeto:rTNPTrig0-pT90pT90:err}   {\ensuremath{{0.003 } } }
\vdef{default-11wCowboyVeto:rRhoTNPMCTrig0-pT90pT90:val}   {\ensuremath{{1.055 } } }
\vdef{default-11wCowboyVeto:rRhoTNPMCTrig0-pT90pT90:err}   {\ensuremath{{0.002 } } }
\vdef{default-11wCowboyVeto:rRhoTNPTrig0-pT90pT90:val}   {\ensuremath{{1.052 } } }
\vdef{default-11wCowboyVeto:rRhoTNPTrig0-pT90pT90:err}   {\ensuremath{{0.003 } } }
\vdef{default-11wCowboyVeto:MuidSg1-pT90pT90:val}   {\ensuremath{{0.864 } } }
\vdef{default-11wCowboyVeto:MuidNo1-pT90pT90:val}   {\ensuremath{{0.888 } } }
\vdef{default-11wCowboyVeto:TNPMCMuidSg1-pT90pT90:val}   {\ensuremath{{0.837 } } }
\vdef{default-11wCowboyVeto:TNPMCMuidNo1-pT90pT90:val}   {\ensuremath{{0.836 } } }
\vdef{default-11wCowboyVeto:TNPMuidSg1-pT90pT90:val}   {\ensuremath{{0.730 } } }
\vdef{default-11wCowboyVeto:TNPMuidNo1-pT90pT90:val}   {\ensuremath{{0.727 } } }
\vdef{default-11wCowboyVeto:rhoMuidSg1-pT90pT90:val}   {\ensuremath{{1.031 } } }
\vdef{default-11wCowboyVeto:rhoMuidNo1-pT90pT90:val}   {\ensuremath{{1.061 } } }
\vdef{default-11wCowboyVeto:rMcMuid1-pT90pT90:val}   {\ensuremath{{0.973 } } }
\vdef{default-11wCowboyVeto:rMcMuid1-pT90pT90:err}   {\ensuremath{{0.038 } } }
\vdef{default-11wCowboyVeto:rTNPMCMuid1-pT90pT90:val}   {\ensuremath{{1.001 } } }
\vdef{default-11wCowboyVeto:rTNPMCMuid1-pT90pT90:err}   {\ensuremath{{0.007 } } }
\vdef{default-11wCowboyVeto:rTNPMuid1-pT90pT90:val}   {\ensuremath{{1.005 } } }
\vdef{default-11wCowboyVeto:rTNPMuid1-pT90pT90:err}   {\ensuremath{{0.010 } } }
\vdef{default-11wCowboyVeto:rRhoTNPMCMuid1-pT90pT90:val}   {\ensuremath{{0.973 } } }
\vdef{default-11wCowboyVeto:rRhoTNPMCMuid1-pT90pT90:err}   {\ensuremath{{0.007 } } }
\vdef{default-11wCowboyVeto:rRhoTNPMuid1-pT90pT90:val}   {\ensuremath{{0.976 } } }
\vdef{default-11wCowboyVeto:rRhoTNPMuid1-pT90pT90:err}   {\ensuremath{{0.009 } } }
\vdef{default-11wCowboyVeto:TrigSg1-pT90pT90:val}   {\ensuremath{{0.705 } } }
\vdef{default-11wCowboyVeto:TrigNo1-pT90pT90:val}   {\ensuremath{{0.692 } } }
\vdef{default-11wCowboyVeto:TNPMCTrigSg1-pT90pT90:val}   {\ensuremath{{0.760 } } }
\vdef{default-11wCowboyVeto:TNPMCTrigNo1-pT90pT90:val}   {\ensuremath{{0.785 } } }
\vdef{default-11wCowboyVeto:TNPTrigSg1-pT90pT90:val}   {\ensuremath{{0.721 } } }
\vdef{default-11wCowboyVeto:TNPTrigNo1-pT90pT90:val}   {\ensuremath{{0.746 } } }
\vdef{default-11wCowboyVeto:rhoTrigSg1-pT90pT90:val}   {\ensuremath{{0.928 } } }
\vdef{default-11wCowboyVeto:rhoTrigNo1-pT90pT90:val}   {\ensuremath{{0.881 } } }
\vdef{default-11wCowboyVeto:rMcTrig1-pT90pT90:val}   {\ensuremath{{1.020 } } }
\vdef{default-11wCowboyVeto:rMcTrig1-pT90pT90:err}   {\ensuremath{{0.069 } } }
\vdef{default-11wCowboyVeto:rTNPMCTrig1-pT90pT90:val}   {\ensuremath{{0.967 } } }
\vdef{default-11wCowboyVeto:rTNPMCTrig1-pT90pT90:err}   {\ensuremath{{0.026 } } }
\vdef{default-11wCowboyVeto:rTNPTrig1-pT90pT90:val}   {\ensuremath{{0.967 } } }
\vdef{default-11wCowboyVeto:rTNPTrig1-pT90pT90:err}   {\ensuremath{{0.023 } } }
\vdef{default-11wCowboyVeto:rRhoTNPMCTrig1-pT90pT90:val}   {\ensuremath{{1.020 } } }
\vdef{default-11wCowboyVeto:rRhoTNPMCTrig1-pT90pT90:err}   {\ensuremath{{0.030 } } }
\vdef{default-11wCowboyVeto:rRhoTNPTrig1-pT90pT90:val}   {\ensuremath{{1.019 } } }
\vdef{default-11wCowboyVeto:rRhoTNPTrig1-pT90pT90:err}   {\ensuremath{{0.026 } } }
\vdef{default-11:SgMc3e33:trigEff0:val}   {\ensuremath{{0.926 } } }
\vdef{default-11:SgMc3e33:trigEff0:err}   {\ensuremath{{0.019 } } }
\vdef{default-11:SgMc3e33:trigEff1:val}   {\ensuremath{{0.663 } } }
\vdef{default-11:SgMc3e33:trigEff1:err}   {\ensuremath{{0.048 } } }
\vdef{default-11:SgMc2e33:trigEff0:val}   {\ensuremath{{0.932 } } }
\vdef{default-11:SgMc2e33:trigEff0:err}   {\ensuremath{{0.019 } } }
\vdef{default-11:SgMc2e33:trigEff1:val}   {\ensuremath{{0.737 } } }
\vdef{default-11:SgMc2e33:trigEff1:err}   {\ensuremath{{0.045 } } }
\vdef{default-11:SgMc1e33:trigEff0:val}   {\ensuremath{{0.923 } } }
\vdef{default-11:SgMc1e33:trigEff0:err}   {\ensuremath{{0.020 } } }
\vdef{default-11:SgMc1e33:trigEff1:val}   {\ensuremath{{0.747 } } }
\vdef{default-11:SgMc1e33:trigEff1:err}   {\ensuremath{{0.044 } } }
\vdef{default-11:NoMc3e33:trigEff0:val}   {\ensuremath{{0.775 } } }
\vdef{default-11:NoMc3e33:trigEff0:err}   {\ensuremath{{0.004 } } }
\vdef{default-11:NoMc3e33:trigEff1:val}   {\ensuremath{{0.523 } } }
\vdef{default-11:NoMc3e33:trigEff1:err}   {\ensuremath{{0.010 } } }
\vdef{default-11:NoMc2e33:trigEff0:val}   {\ensuremath{{0.811 } } }
\vdef{default-11:NoMc2e33:trigEff0:err}   {\ensuremath{{0.004 } } }
\vdef{default-11:NoMc2e33:trigEff1:val}   {\ensuremath{{0.602 } } }
\vdef{default-11:NoMc2e33:trigEff1:err}   {\ensuremath{{0.009 } } }
\vdef{default-11:NoMc1e33:trigEff0:val}   {\ensuremath{{0.852 } } }
\vdef{default-11:NoMc1e33:trigEff0:err}   {\ensuremath{{0.003 } } }
\vdef{default-11:NoMc1e33:trigEff1:val}   {\ensuremath{{0.659 } } }
\vdef{default-11:NoMc1e33:trigEff1:err}   {\ensuremath{{0.009 } } }
\vdef{default-11:CsMc3e33:trigEff0:val}   {\ensuremath{{0.765 } } }
\vdef{default-11:CsMc3e33:trigEff0:err}   {\ensuremath{{0.009 } } }
\vdef{default-11:CsMc3e33:trigEff1:val}   {\ensuremath{{0.547 } } }
\vdef{default-11:CsMc3e33:trigEff1:err}   {\ensuremath{{0.023 } } }
\vdef{default-11:CsMc2e33:trigEff0:val}   {\ensuremath{{0.801 } } }
\vdef{default-11:CsMc2e33:trigEff0:err}   {\ensuremath{{0.009 } } }
\vdef{default-11:CsMc2e33:trigEff1:val}   {\ensuremath{{0.633 } } }
\vdef{default-11:CsMc2e33:trigEff1:err}   {\ensuremath{{0.023 } } }
\vdef{default-11:CsMc1e33:trigEff0:val}   {\ensuremath{{0.851 } } }
\vdef{default-11:CsMc1e33:trigEff0:err}   {\ensuremath{{0.008 } } }
\vdef{default-11:CsMc1e33:trigEff1:val}   {\ensuremath{{0.708 } } }
\vdef{default-11:CsMc1e33:trigEff1:err}   {\ensuremath{{0.021 } } }

\vdef{default-11:bdt:0}     {\ensuremath{{-0.030 } } }
\vdef{default-11:mBdLo:0}   {\ensuremath{{5.200 } } }
\vdef{default-11:mBdHi:0}   {\ensuremath{{5.300 } } }
\vdef{default-11:mBsLo:0}   {\ensuremath{{5.300 } } }
\vdef{default-11:mBsHi:0}   {\ensuremath{{5.450 } } }
\vdef{default-11:etaMin:0}   {\ensuremath{{0.0 } } }
\vdef{default-11:etaMax:0}   {\ensuremath{{1.4 } } }
\vdef{default-11:pt:0}   {\ensuremath{{6.5 } } }
\vdef{default-11:m1pt:0}   {\ensuremath{{4.5 } } }
\vdef{default-11:m2pt:0}   {\ensuremath{{4.0 } } }
\vdef{default-11:m1eta:0}   {\ensuremath{{1.4 } } }
\vdef{default-11:m2eta:0}   {\ensuremath{{1.4 } } }
\vdef{default-11:iso:0}   {\ensuremath{{0.80 } } }
\vdef{default-11:chi2dof:0}   {\ensuremath{{2.2 } } }
\vdef{default-11:alpha:0}   {\ensuremath{{0.050 } } }
\vdef{default-11:fls3d:0}   {\ensuremath{{13.0 } } }
\vdef{default-11:docatrk:0}   {\ensuremath{{0.015 } } }
\vdef{default-11:closetrk:0}   {\ensuremath{{2 } } }
\vdef{default-11:pvlip:0}   {\ensuremath{{100.008 } } }
\vdef{default-11:pvlips:0}   {\ensuremath{{102.000 } } }
\vdef{default-11:pvlip2:0}   {\ensuremath{{-0.050 } } }
\vdef{default-11:pvlips2:0}   {\ensuremath{{-2.000 } } }
\vdef{default-11:maxdoca:0}   {\ensuremath{{3.000 } } }
\vdef{default-11:pvip:0}   {\ensuremath{{0.008 } } }
\vdef{default-11:pvips:0}   {\ensuremath{{2.000 } } }
\vdef{default-11:doApplyCowboyVeto:0}   {no }
\vdef{default-11:fDoApplyCowboyVetoAlsoInSignal:0}   {no }
\vdef{default-11:bdt:1}     {\ensuremath{{-0.080 } } }
\vdef{default-11:mBdLo:1}   {\ensuremath{{5.200 } } }
\vdef{default-11:mBdHi:1}   {\ensuremath{{5.300 } } }
\vdef{default-11:mBsLo:1}   {\ensuremath{{5.300 } } }
\vdef{default-11:mBsHi:1}   {\ensuremath{{5.450 } } }
\vdef{default-11:etaMin:1}   {\ensuremath{{1.4 } } }
\vdef{default-11:etaMax:1}   {\ensuremath{{2.4 } } }
\vdef{default-11:pt:1}   {\ensuremath{{8.5 } } }
\vdef{default-11:m1pt:1}   {\ensuremath{{4.5 } } }
\vdef{default-11:m2pt:1}   {\ensuremath{{4.2 } } }
\vdef{default-11:m1eta:1}   {\ensuremath{{2.4 } } }
\vdef{default-11:m2eta:1}   {\ensuremath{{2.4 } } }
\vdef{default-11:iso:1}   {\ensuremath{{0.80 } } }
\vdef{default-11:chi2dof:1}   {\ensuremath{{1.8 } } }
\vdef{default-11:alpha:1}   {\ensuremath{{0.030 } } }
\vdef{default-11:fls3d:1}   {\ensuremath{{15.0 } } }
\vdef{default-11:docatrk:1}   {\ensuremath{{0.015 } } }
\vdef{default-11:closetrk:1}   {\ensuremath{{2 } } }
\vdef{default-11:pvlip:1}   {\ensuremath{{100.008 } } }
\vdef{default-11:pvlips:1}   {\ensuremath{{102.000 } } }
\vdef{default-11:pvlip2:1}   {\ensuremath{{-0.050 } } }
\vdef{default-11:pvlips2:1}   {\ensuremath{{-2.000 } } }
\vdef{default-11:maxdoca:1}   {\ensuremath{{3.000 } } }
\vdef{default-11:pvip:1}   {\ensuremath{{0.008 } } }
\vdef{default-11:pvips:1}   {\ensuremath{{2.000 } } }
\vdef{default-11:doApplyCowboyVeto:1}   {no }
\vdef{default-11:fDoApplyCowboyVetoAlsoInSignal:1}   {no }
\vdef{default-11:bgBd2KK:bsRare0}   {\ensuremath{{0.000012 } } }
\vdef{default-11:bgBd2KK:bsRare0E}  {\ensuremath{{0.000009 } } }
\vdef{default-11:bgBd2KK:bdRare0}   {\ensuremath{{0.000240 } } }
\vdef{default-11:bgBd2KK:bdRare0E}  {\ensuremath{{0.000184 } } }
\vdef{default-11:bgBd2KK:loSideband0:val}   {\ensuremath{{0.000 } } }
\vdef{default-11:bgBd2KK:loSideband0:err}   {\ensuremath{{0.000 } } }
\vdef{default-11:bgBd2KK:bsRare1}   {\ensuremath{{0.000014 } } }
\vdef{default-11:bgBd2KK:bsRare1E}  {\ensuremath{{0.000011 } } }
\vdef{default-11:bgBd2KK:bdRare1}   {\ensuremath{{0.000084 } } }
\vdef{default-11:bgBd2KK:bdRare1E}  {\ensuremath{{0.000064 } } }
\vdef{default-11:bgBd2KK:loSideband1:val}   {\ensuremath{{0.000 } } }
\vdef{default-11:bgBd2KK:loSideband1:err}   {\ensuremath{{0.000 } } }
\vdef{default-11:bgBd2KPi:bsRare0}   {\ensuremath{{0.006180 } } }
\vdef{default-11:bgBd2KPi:bsRare0E}  {\ensuremath{{0.001479 } } }
\vdef{default-11:bgBd2KPi:bdRare0}   {\ensuremath{{0.058643 } } }
\vdef{default-11:bgBd2KPi:bdRare0E}  {\ensuremath{{0.014005 } } }
\vdef{default-11:bgBd2KPi:loSideband0:val}   {\ensuremath{{0.024 } } }
\vdef{default-11:bgBd2KPi:loSideband0:err}   {\ensuremath{{0.007 } } }
\vdef{default-11:bgBd2KPi:bsRare1}   {\ensuremath{{0.005002 } } }
\vdef{default-11:bgBd2KPi:bsRare1E}  {\ensuremath{{0.001196 } } }
\vdef{default-11:bgBd2KPi:bdRare1}   {\ensuremath{{0.015481 } } }
\vdef{default-11:bgBd2KPi:bdRare1E}  {\ensuremath{{0.003692 } } }
\vdef{default-11:bgBd2KPi:loSideband1:val}   {\ensuremath{{0.013 } } }
\vdef{default-11:bgBd2KPi:loSideband1:err}   {\ensuremath{{0.004 } } }
\vdef{default-11:bgBd2PiMuNu:bsRare0}   {\ensuremath{{0.008697 } } }
\vdef{default-11:bgBd2PiMuNu:bsRare0E}  {\ensuremath{{0.000678 } } }
\vdef{default-11:bgBd2PiMuNu:bdRare0}   {\ensuremath{{0.098914 } } }
\vdef{default-11:bgBd2PiMuNu:bdRare0E}  {\ensuremath{{0.010843 } } }
\vdef{default-11:bgBd2PiMuNu:loSideband0:val}   {\ensuremath{{1.646 } } }
\vdef{default-11:bgBd2PiMuNu:loSideband0:err}   {\ensuremath{{0.494 } } }
\vdef{default-11:bgBd2PiMuNu:bsRare1}   {\ensuremath{{0.008934 } } }
\vdef{default-11:bgBd2PiMuNu:bsRare1E}  {\ensuremath{{0.001059 } } }
\vdef{default-11:bgBd2PiMuNu:bdRare1}   {\ensuremath{{0.046938 } } }
\vdef{default-11:bgBd2PiMuNu:bdRare1E}  {\ensuremath{{0.008470 } } }
\vdef{default-11:bgBd2PiMuNu:loSideband1:val}   {\ensuremath{{0.670 } } }
\vdef{default-11:bgBd2PiMuNu:loSideband1:err}   {\ensuremath{{0.201 } } }
\vdef{default-11:bgBd2PiPi:bsRare0}   {\ensuremath{{0.015087 } } }
\vdef{default-11:bgBd2PiPi:bsRare0E}  {\ensuremath{{0.001520 } } }
\vdef{default-11:bgBd2PiPi:bdRare0}   {\ensuremath{{0.113944 } } }
\vdef{default-11:bgBd2PiPi:bdRare0E}  {\ensuremath{{0.003576 } } }
\vdef{default-11:bgBd2PiPi:loSideband0:val}   {\ensuremath{{0.002 } } }
\vdef{default-11:bgBd2PiPi:loSideband0:err}   {\ensuremath{{0.001 } } }
\vdef{default-11:bgBd2PiPi:bsRare1}   {\ensuremath{{0.011643 } } }
\vdef{default-11:bgBd2PiPi:bsRare1E}  {\ensuremath{{0.000644 } } }
\vdef{default-11:bgBd2PiPi:bdRare1}   {\ensuremath{{0.050848 } } }
\vdef{default-11:bgBd2PiPi:bdRare1E}  {\ensuremath{{0.000930 } } }
\vdef{default-11:bgBd2PiPi:loSideband1:val}   {\ensuremath{{0.002 } } }
\vdef{default-11:bgBd2PiPi:loSideband1:err}   {\ensuremath{{0.001 } } }
\vdef{default-11:bgBs2KK:bsRare0}   {\ensuremath{{0.023993 } } }
\vdef{default-11:bgBs2KK:bsRare0E}  {\ensuremath{{0.002672 } } }
\vdef{default-11:bgBs2KK:bdRare0}   {\ensuremath{{0.135641 } } }
\vdef{default-11:bgBs2KK:bdRare0E}  {\ensuremath{{0.006509 } } }
\vdef{default-11:bgBs2KK:loSideband0:val}   {\ensuremath{{0.003 } } }
\vdef{default-11:bgBs2KK:loSideband0:err}   {\ensuremath{{0.001 } } }
\vdef{default-11:bgBs2KK:bsRare1}   {\ensuremath{{0.015499 } } }
\vdef{default-11:bgBs2KK:bsRare1E}  {\ensuremath{{0.001157 } } }
\vdef{default-11:bgBs2KK:bdRare1}   {\ensuremath{{0.057038 } } }
\vdef{default-11:bgBs2KK:bdRare1E}  {\ensuremath{{0.001857 } } }
\vdef{default-11:bgBs2KK:loSideband1:val}   {\ensuremath{{0.002 } } }
\vdef{default-11:bgBs2KK:loSideband1:err}   {\ensuremath{{0.001 } } }
\vdef{default-11:bgBs2KMuNu:bsRare0}   {\ensuremath{{0.023993 } } }
\vdef{default-11:bgBs2KMuNu:bsRare0E}  {\ensuremath{{0.000000 } } }
\vdef{default-11:bgBs2KMuNu:bdRare0}   {\ensuremath{{0.151685 } } }
\vdef{default-11:bgBs2KMuNu:bdRare0E}  {\ensuremath{{0.004678 } } }
\vdef{default-11:bgBs2KMuNu:loSideband0:val}   {\ensuremath{{0.529 } } }
\vdef{default-11:bgBs2KMuNu:loSideband0:err}   {\ensuremath{{0.159 } } }
\vdef{default-11:bgBs2KMuNu:bsRare1}   {\ensuremath{{0.019676 } } }
\vdef{default-11:bgBs2KMuNu:bsRare1E}  {\ensuremath{{0.001218 } } }
\vdef{default-11:bgBs2KMuNu:bdRare1}   {\ensuremath{{0.079318 } } }
\vdef{default-11:bgBs2KMuNu:bdRare1E}  {\ensuremath{{0.006496 } } }
\vdef{default-11:bgBs2KMuNu:loSideband1:val}   {\ensuremath{{0.235 } } }
\vdef{default-11:bgBs2KMuNu:loSideband1:err}   {\ensuremath{{0.071 } } }
\vdef{default-11:bgBs2KPi:bsRare0}   {\ensuremath{{0.027909 } } }
\vdef{default-11:bgBs2KPi:bsRare0E}  {\ensuremath{{0.001333 } } }
\vdef{default-11:bgBs2KPi:bdRare0}   {\ensuremath{{0.153520 } } }
\vdef{default-11:bgBs2KPi:bdRare0E}  {\ensuremath{{0.000625 } } }
\vdef{default-11:bgBs2KPi:loSideband0:val}   {\ensuremath{{0.000 } } }
\vdef{default-11:bgBs2KPi:loSideband0:err}   {\ensuremath{{0.000 } } }
\vdef{default-11:bgBs2KPi:bsRare1}   {\ensuremath{{0.020822 } } }
\vdef{default-11:bgBs2KPi:bsRare1E}  {\ensuremath{{0.000390 } } }
\vdef{default-11:bgBs2KPi:bdRare1}   {\ensuremath{{0.080086 } } }
\vdef{default-11:bgBs2KPi:bdRare1E}  {\ensuremath{{0.000261 } } }
\vdef{default-11:bgBs2KPi:loSideband1:val}   {\ensuremath{{0.000 } } }
\vdef{default-11:bgBs2KPi:loSideband1:err}   {\ensuremath{{0.000 } } }
\vdef{default-11:bgBs2PiPi:bsRare0}   {\ensuremath{{0.029079 } } }
\vdef{default-11:bgBs2PiPi:bsRare0E}  {\ensuremath{{0.001209 } } }
\vdef{default-11:bgBs2PiPi:bdRare0}   {\ensuremath{{0.153647 } } }
\vdef{default-11:bgBs2PiPi:bdRare0E}  {\ensuremath{{0.000132 } } }
\vdef{default-11:bgBs2PiPi:loSideband0:val}   {\ensuremath{{0.000 } } }
\vdef{default-11:bgBs2PiPi:loSideband0:err}   {\ensuremath{{0.000 } } }
\vdef{default-11:bgBs2PiPi:bsRare1}   {\ensuremath{{0.021143 } } }
\vdef{default-11:bgBs2PiPi:bsRare1E}  {\ensuremath{{0.000332 } } }
\vdef{default-11:bgBs2PiPi:bdRare1}   {\ensuremath{{0.080182 } } }
\vdef{default-11:bgBs2PiPi:bdRare1E}  {\ensuremath{{0.000099 } } }
\vdef{default-11:bgBs2PiPi:loSideband1:val}   {\ensuremath{{0.000 } } }
\vdef{default-11:bgBs2PiPi:loSideband1:err}   {\ensuremath{{0.000 } } }
\vdef{default-11:bgLb2KP:bsRare0}   {\ensuremath{{0.030621 } } }
\vdef{default-11:bgLb2KP:bsRare0E}  {\ensuremath{{0.000612 } } }
\vdef{default-11:bgLb2KP:bdRare0}   {\ensuremath{{0.153841 } } }
\vdef{default-11:bgLb2KP:bdRare0E}  {\ensuremath{{0.000077 } } }
\vdef{default-11:bgLb2KP:loSideband0:val}   {\ensuremath{{0.000 } } }
\vdef{default-11:bgLb2KP:loSideband0:err}   {\ensuremath{{0.000 } } }
\vdef{default-11:bgLb2KP:bsRare1}   {\ensuremath{{0.021660 } } }
\vdef{default-11:bgLb2KP:bsRare1E}  {\ensuremath{{0.000205 } } }
\vdef{default-11:bgLb2KP:bdRare1}   {\ensuremath{{0.080296 } } }
\vdef{default-11:bgLb2KP:bdRare1E}  {\ensuremath{{0.000045 } } }
\vdef{default-11:bgLb2KP:loSideband1:val}   {\ensuremath{{0.000 } } }
\vdef{default-11:bgLb2KP:loSideband1:err}   {\ensuremath{{0.000 } } }
\vdef{default-11:bgLb2PMuNu:bsRare0}   {\ensuremath{{0.181765 } } }
\vdef{default-11:bgLb2PMuNu:bsRare0E}  {\ensuremath{{0.056775 } } }
\vdef{default-11:bgLb2PMuNu:bdRare0}   {\ensuremath{{0.331736 } } }
\vdef{default-11:bgLb2PMuNu:bdRare0E}  {\ensuremath{{0.066823 } } }
\vdef{default-11:bgLb2PMuNu:loSideband0:val}   {\ensuremath{{0.808 } } }
\vdef{default-11:bgLb2PMuNu:loSideband0:err}   {\ensuremath{{0.242 } } }
\vdef{default-11:bgLb2PMuNu:bsRare1}   {\ensuremath{{0.082258 } } }
\vdef{default-11:bgLb2PMuNu:bsRare1E}  {\ensuremath{{0.022763 } } }
\vdef{default-11:bgLb2PMuNu:bdRare1}   {\ensuremath{{0.149253 } } }
\vdef{default-11:bgLb2PMuNu:bdRare1E}  {\ensuremath{{0.025902 } } }
\vdef{default-11:bgLb2PMuNu:loSideband1:val}   {\ensuremath{{0.335 } } }
\vdef{default-11:bgLb2PMuNu:loSideband1:err}   {\ensuremath{{0.101 } } }
\vdef{default-11:bgLb2PiP:bsRare0}   {\ensuremath{{0.182322 } } }
\vdef{default-11:bgLb2PiP:bsRare0E}  {\ensuremath{{0.000225 } } }
\vdef{default-11:bgLb2PiP:bdRare0}   {\ensuremath{{0.331808 } } }
\vdef{default-11:bgLb2PiP:bdRare0E}  {\ensuremath{{0.000029 } } }
\vdef{default-11:bgLb2PiP:loSideband0:val}   {\ensuremath{{0.000 } } }
\vdef{default-11:bgLb2PiP:loSideband0:err}   {\ensuremath{{0.000 } } }
\vdef{default-11:bgLb2PiP:bsRare1}   {\ensuremath{{0.082482 } } }
\vdef{default-11:bgLb2PiP:bsRare1E}  {\ensuremath{{0.000091 } } }
\vdef{default-11:bgLb2PiP:bdRare1}   {\ensuremath{{0.149304 } } }
\vdef{default-11:bgLb2PiP:bdRare1E}  {\ensuremath{{0.000021 } } }
\vdef{default-11:bgLb2PiP:loSideband1:val}   {\ensuremath{{0.000 } } }
\vdef{default-11:bgLb2PiP:loSideband1:err}   {\ensuremath{{0.000 } } }
\vdef{default-11:bsRare0}   {\ensuremath{{0.182 } } }
\vdef{default-11:bsRare0E}  {\ensuremath{{0.057 } } }
\vdef{default-11:bsRare1}   {\ensuremath{{0.082 } } }
\vdef{default-11:bsRare1E}  {\ensuremath{{0.023 } } }
\vdef{default-11:bdRare0}   {\ensuremath{{0.332 } } }
\vdef{default-11:bdRare0E}  {\ensuremath{{0.070 } } }
\vdef{default-11:bdRare1}   {\ensuremath{{0.149 } } }
\vdef{default-11:bdRare1E}  {\ensuremath{{0.028 } } }
\vdef{default-11:BsSgEvt0:0:run}   {\ensuremath{{167913 } } }
\vdef{default-11:BsSgEvt0:0:evt}   {\ensuremath{{405277425 } } }
\vdef{default-11:BsSgEvt0:0:chan}   {\ensuremath{{0 } } }
\vdef{default-11:BsSgEvt0:0:m}   {\ensuremath{{5.403 } } }
\vdef{default-11:BsSgEvt0:0:pt}   {\ensuremath{{13.676 } } }
\vdef{default-11:BsSgEvt0:0:phi}   {\ensuremath{{-2.753 } } }
\vdef{default-11:BsSgEvt0:0:eta}   {\ensuremath{{0.828 } } }
\vdef{default-11:BsSgEvt0:0:channel}   {barrel }
\vdef{default-11:BsSgEvt0:0:cowboy}   {\ensuremath{{1 } } }
\vdef{default-11:BsSgEvt0:0:m1pt}   {\ensuremath{{7.213 } } }
\vdef{default-11:BsSgEvt0:0:m2pt}   {\ensuremath{{6.441 } } }
\vdef{default-11:BsSgEvt0:0:m1eta}   {\ensuremath{{0.419 } } }
\vdef{default-11:BsSgEvt0:0:m2eta}   {\ensuremath{{1.181 } } }
\vdef{default-11:BsSgEvt0:0:m1phi}   {\ensuremath{{-2.699 } } }
\vdef{default-11:BsSgEvt0:0:m2phi}   {\ensuremath{{-2.814 } } }
\vdef{default-11:BsSgEvt0:0:m1q}   {\ensuremath{{1 } } }
\vdef{default-11:BsSgEvt0:0:m2q}   {\ensuremath{{-1 } } }
\vdef{default-11:BsSgEvt0:0:iso}   {\ensuremath{{0.888 } } }
\vdef{default-11:BsSgEvt0:0:alpha}   {\ensuremath{{0.0175 } } }
\vdef{default-11:BsSgEvt0:0:chi2}   {\ensuremath{{ 1.96 } } }
\vdef{default-11:BsSgEvt0:0:dof}   {\ensuremath{{1 } } }
\vdef{default-11:BsSgEvt0:0:fls3d}   {\ensuremath{{38.53 } } }
\vdef{default-11:BsSgEvt0:0:fl3d}   {\ensuremath{{0.3000 } } }
\vdef{default-11:BsSgEvt0:0:fl3dE}   {\ensuremath{{0.0078 } } }
\vdef{default-11:BsSgEvt0:0:docatrk}   {\ensuremath{{0.0283 } } }
\vdef{default-11:BsSgEvt0:0:closetrk}   {\ensuremath{{1 } } }
\vdef{default-11:BsSgEvt0:0:lip}   {\ensuremath{{0.0037 } } }
\vdef{default-11:BsSgEvt0:0:lipE}   {\ensuremath{{0.0029 } } }
\vdef{default-11:BsSgEvt0:0:tip}   {\ensuremath{{0.0017 } } }
\vdef{default-11:BsSgEvt0:0:tipE}   {\ensuremath{{0.0024 } } }
\vdef{default-11:BsSgEvt0:0:pvlip}   {\ensuremath{{0.0037 } } }
\vdef{default-11:BsSgEvt0:0:pvlips}   {\ensuremath{{1.2537 } } }
\vdef{default-11:BsSgEvt0:0:pvip}   {\ensuremath{{0.0040 } } }
\vdef{default-11:BsSgEvt0:0:pvips}   {\ensuremath{{1.4250 } } }
\vdef{default-11:BsSgEvt0:0:maxdoca}   {\ensuremath{{0.0054 } } }
\vdef{default-11:BsSgEvt0:0:pvw8}   {\ensuremath{{0.9138 } } }
\vdef{default-11:BsSgEvt0:0:bdt}   {\ensuremath{{-0.2294 } } }
\vdef{default-11:BsSgEvt0:0:m1pix}   {\ensuremath{{3 } } }
\vdef{default-11:BsSgEvt0:0:m2pix}   {\ensuremath{{3 } } }
\vdef{default-11:BsSgEvt0:0:m1bpix}   {\ensuremath{{3 } } }
\vdef{default-11:BsSgEvt0:0:m2bpix}   {\ensuremath{{3 } } }
\vdef{default-11:BsSgEvt0:0:m1bpixl1}   {\ensuremath{{1 } } }
\vdef{default-11:BsSgEvt0:0:m2bpixl1}   {\ensuremath{{1 } } }
\vdef{default-11:SgEvt0:0:run}   {\ensuremath{{167281 } } }
\vdef{default-11:SgEvt0:0:evt}   {\ensuremath{{642317961 } } }
\vdef{default-11:SgEvt0:0:chan}   {\ensuremath{{0 } } }
\vdef{default-11:SgEvt0:0:m}   {\ensuremath{{4.959 } } }
\vdef{default-11:SgEvt0:0:pt}   {\ensuremath{{27.971 } } }
\vdef{default-11:SgEvt0:0:phi}   {\ensuremath{{-1.144 } } }
\vdef{default-11:SgEvt0:0:eta}   {\ensuremath{{-1.112 } } }
\vdef{default-11:SgEvt0:0:channel}   {barrel }
\vdef{default-11:SgEvt0:0:cowboy}   {\ensuremath{{1 } } }
\vdef{default-11:SgEvt0:0:m1pt}   {\ensuremath{{16.354 } } }
\vdef{default-11:SgEvt0:0:m2pt}   {\ensuremath{{11.634 } } }
\vdef{default-11:SgEvt0:0:m1eta}   {\ensuremath{{-0.976 } } }
\vdef{default-11:SgEvt0:0:m2eta}   {\ensuremath{{-1.273 } } }
\vdef{default-11:SgEvt0:0:m1phi}   {\ensuremath{{-1.227 } } }
\vdef{default-11:SgEvt0:0:m2phi}   {\ensuremath{{-1.028 } } }
\vdef{default-11:SgEvt0:0:m1q}   {\ensuremath{{-1 } } }
\vdef{default-11:SgEvt0:0:m2q}   {\ensuremath{{1 } } }
\vdef{default-11:SgEvt0:0:iso}   {\ensuremath{{0.965 } } }
\vdef{default-11:SgEvt0:0:alpha}   {\ensuremath{{0.0021 } } }
\vdef{default-11:SgEvt0:0:chi2}   {\ensuremath{{ 1.62 } } }
\vdef{default-11:SgEvt0:0:dof}   {\ensuremath{{1 } } }
\vdef{default-11:SgEvt0:0:fls3d}   {\ensuremath{{68.01 } } }
\vdef{default-11:SgEvt0:0:fl3d}   {\ensuremath{{1.0590 } } }
\vdef{default-11:SgEvt0:0:fl3dE}   {\ensuremath{{0.0156 } } }
\vdef{default-11:SgEvt0:0:docatrk}   {\ensuremath{{0.1825 } } }
\vdef{default-11:SgEvt0:0:closetrk}   {\ensuremath{{0 } } }
\vdef{default-11:SgEvt0:0:lip}   {\ensuremath{{0.0011 } } }
\vdef{default-11:SgEvt0:0:lipE}   {\ensuremath{{0.0032 } } }
\vdef{default-11:SgEvt0:0:tip}   {\ensuremath{{0.0011 } } }
\vdef{default-11:SgEvt0:0:tipE}   {\ensuremath{{0.0033 } } }
\vdef{default-11:SgEvt0:0:pvlip}   {\ensuremath{{0.0011 } } }
\vdef{default-11:SgEvt0:0:pvlips}   {\ensuremath{{0.3644 } } }
\vdef{default-11:SgEvt0:0:pvip}   {\ensuremath{{0.0016 } } }
\vdef{default-11:SgEvt0:0:pvips}   {\ensuremath{{0.4846 } } }
\vdef{default-11:SgEvt0:0:maxdoca}   {\ensuremath{{0.0035 } } }
\vdef{default-11:SgEvt0:0:pvw8}   {\ensuremath{{0.9370 } } }
\vdef{default-11:SgEvt0:0:bdt}   {\ensuremath{{-0.1354 } } }
\vdef{default-11:SgEvt0:0:m1pix}   {\ensuremath{{3 } } }
\vdef{default-11:SgEvt0:0:m2pix}   {\ensuremath{{3 } } }
\vdef{default-11:SgEvt0:0:m1bpix}   {\ensuremath{{3 } } }
\vdef{default-11:SgEvt0:0:m2bpix}   {\ensuremath{{3 } } }
\vdef{default-11:SgEvt0:0:m1bpixl1}   {\ensuremath{{1 } } }
\vdef{default-11:SgEvt0:0:m2bpixl1}   {\ensuremath{{1 } } }
\vdef{default-11:SgEvt1:0:run}   {\ensuremath{{167281 } } }
\vdef{default-11:SgEvt1:0:evt}   {\ensuremath{{408306037 } } }
\vdef{default-11:SgEvt1:0:chan}   {\ensuremath{{1 } } }
\vdef{default-11:SgEvt1:0:m}   {\ensuremath{{4.988 } } }
\vdef{default-11:SgEvt1:0:pt}   {\ensuremath{{10.625 } } }
\vdef{default-11:SgEvt1:0:phi}   {\ensuremath{{-1.176 } } }
\vdef{default-11:SgEvt1:0:eta}   {\ensuremath{{1.203 } } }
\vdef{default-11:SgEvt1:0:channel}   {endcap }
\vdef{default-11:SgEvt1:0:cowboy}   {\ensuremath{{1 } } }
\vdef{default-11:SgEvt1:0:m1pt}   {\ensuremath{{5.793 } } }
\vdef{default-11:SgEvt1:0:m2pt}   {\ensuremath{{5.177 } } }
\vdef{default-11:SgEvt1:0:m1eta}   {\ensuremath{{0.771 } } }
\vdef{default-11:SgEvt1:0:m2eta}   {\ensuremath{{1.512 } } }
\vdef{default-11:SgEvt1:0:m1phi}   {\ensuremath{{-1.416 } } }
\vdef{default-11:SgEvt1:0:m2phi}   {\ensuremath{{-0.906 } } }
\vdef{default-11:SgEvt1:0:m1q}   {\ensuremath{{-1 } } }
\vdef{default-11:SgEvt1:0:m2q}   {\ensuremath{{1 } } }
\vdef{default-11:SgEvt1:0:iso}   {\ensuremath{{1.000 } } }
\vdef{default-11:SgEvt1:0:alpha}   {\ensuremath{{0.0112 } } }
\vdef{default-11:SgEvt1:0:chi2}   {\ensuremath{{ 0.08 } } }
\vdef{default-11:SgEvt1:0:dof}   {\ensuremath{{1 } } }
\vdef{default-11:SgEvt1:0:fls3d}   {\ensuremath{{25.45 } } }
\vdef{default-11:SgEvt1:0:fl3d}   {\ensuremath{{0.2983 } } }
\vdef{default-11:SgEvt1:0:fl3dE}   {\ensuremath{{0.0117 } } }
\vdef{default-11:SgEvt1:0:docatrk}   {\ensuremath{{0.0842 } } }
\vdef{default-11:SgEvt1:0:closetrk}   {\ensuremath{{0 } } }
\vdef{default-11:SgEvt1:0:lip}   {\ensuremath{{0.0001 } } }
\vdef{default-11:SgEvt1:0:lipE}   {\ensuremath{{0.0056 } } }
\vdef{default-11:SgEvt1:0:tip}   {\ensuremath{{0.0033 } } }
\vdef{default-11:SgEvt1:0:tipE}   {\ensuremath{{0.0047 } } }
\vdef{default-11:SgEvt1:0:pvlip}   {\ensuremath{{0.0001 } } }
\vdef{default-11:SgEvt1:0:pvlips}   {\ensuremath{{0.0202 } } }
\vdef{default-11:SgEvt1:0:pvip}   {\ensuremath{{0.0033 } } }
\vdef{default-11:SgEvt1:0:pvips}   {\ensuremath{{0.7174 } } }
\vdef{default-11:SgEvt1:0:maxdoca}   {\ensuremath{{0.0014 } } }
\vdef{default-11:SgEvt1:0:pvw8}   {\ensuremath{{0.7814 } } }
\vdef{default-11:SgEvt1:0:bdt}   {\ensuremath{{0.1607 } } }
\vdef{default-11:SgEvt1:0:m1pix}   {\ensuremath{{3 } } }
\vdef{default-11:SgEvt1:0:m2pix}   {\ensuremath{{3 } } }
\vdef{default-11:SgEvt1:0:m1bpix}   {\ensuremath{{3 } } }
\vdef{default-11:SgEvt1:0:m2bpix}   {\ensuremath{{3 } } }
\vdef{default-11:SgEvt1:0:m1bpixl1}   {\ensuremath{{1 } } }
\vdef{default-11:SgEvt1:0:m2bpixl1}   {\ensuremath{{1 } } }
\vdef{default-11:BsSgEvt0:1:run}   {\ensuremath{{172268 } } }
\vdef{default-11:BsSgEvt0:1:evt}   {\ensuremath{{69388599 } } }
\vdef{default-11:BsSgEvt0:1:chan}   {\ensuremath{{0 } } }
\vdef{default-11:BsSgEvt0:1:m}   {\ensuremath{{5.325 } } }
\vdef{default-11:BsSgEvt0:1:pt}   {\ensuremath{{36.115 } } }
\vdef{default-11:BsSgEvt0:1:phi}   {\ensuremath{{-1.714 } } }
\vdef{default-11:BsSgEvt0:1:eta}   {\ensuremath{{0.049 } } }
\vdef{default-11:BsSgEvt0:1:channel}   {barrel }
\vdef{default-11:BsSgEvt0:1:cowboy}   {\ensuremath{{1 } } }
\vdef{default-11:BsSgEvt0:1:m1pt}   {\ensuremath{{28.910 } } }
\vdef{default-11:BsSgEvt0:1:m2pt}   {\ensuremath{{7.361 } } }
\vdef{default-11:BsSgEvt0:1:m1eta}   {\ensuremath{{0.112 } } }
\vdef{default-11:BsSgEvt0:1:m2eta}   {\ensuremath{{-0.198 } } }
\vdef{default-11:BsSgEvt0:1:m1phi}   {\ensuremath{{-1.675 } } }
\vdef{default-11:BsSgEvt0:1:m2phi}   {\ensuremath{{-1.868 } } }
\vdef{default-11:BsSgEvt0:1:m1q}   {\ensuremath{{1 } } }
\vdef{default-11:BsSgEvt0:1:m2q}   {\ensuremath{{-1 } } }
\vdef{default-11:BsSgEvt0:1:iso}   {\ensuremath{{0.870 } } }
\vdef{default-11:BsSgEvt0:1:alpha}   {\ensuremath{{0.0149 } } }
\vdef{default-11:BsSgEvt0:1:chi2}   {\ensuremath{{ 0.92 } } }
\vdef{default-11:BsSgEvt0:1:dof}   {\ensuremath{{1 } } }
\vdef{default-11:BsSgEvt0:1:fls3d}   {\ensuremath{{30.24 } } }
\vdef{default-11:BsSgEvt0:1:fl3d}   {\ensuremath{{0.3296 } } }
\vdef{default-11:BsSgEvt0:1:fl3dE}   {\ensuremath{{0.0109 } } }
\vdef{default-11:BsSgEvt0:1:docatrk}   {\ensuremath{{0.0645 } } }
\vdef{default-11:BsSgEvt0:1:closetrk}   {\ensuremath{{0 } } }
\vdef{default-11:BsSgEvt0:1:lip}   {\ensuremath{{0.0047 } } }
\vdef{default-11:BsSgEvt0:1:lipE}   {\ensuremath{{0.0032 } } }
\vdef{default-11:BsSgEvt0:1:tip}   {\ensuremath{{0.0013 } } }
\vdef{default-11:BsSgEvt0:1:tipE}   {\ensuremath{{0.0035 } } }
\vdef{default-11:BsSgEvt0:1:pvlip}   {\ensuremath{{0.0047 } } }
\vdef{default-11:BsSgEvt0:1:pvlips}   {\ensuremath{{1.4832 } } }
\vdef{default-11:BsSgEvt0:1:pvip}   {\ensuremath{{0.0049 } } }
\vdef{default-11:BsSgEvt0:1:pvips}   {\ensuremath{{1.5237 } } }
\vdef{default-11:BsSgEvt0:1:maxdoca}   {\ensuremath{{0.0034 } } }
\vdef{default-11:BsSgEvt0:1:pvw8}   {\ensuremath{{0.8360 } } }
\vdef{default-11:BsSgEvt0:1:bdt}   {\ensuremath{{0.4904 } } }
\vdef{default-11:BsSgEvt0:1:m1pix}   {\ensuremath{{3 } } }
\vdef{default-11:BsSgEvt0:1:m2pix}   {\ensuremath{{3 } } }
\vdef{default-11:BsSgEvt0:1:m1bpix}   {\ensuremath{{3 } } }
\vdef{default-11:BsSgEvt0:1:m2bpix}   {\ensuremath{{3 } } }
\vdef{default-11:BsSgEvt0:1:m1bpixl1}   {\ensuremath{{1 } } }
\vdef{default-11:BsSgEvt0:1:m2bpixl1}   {\ensuremath{{1 } } }
\vdef{default-11:SgEvt1:1:run}   {\ensuremath{{172033 } } }
\vdef{default-11:SgEvt1:1:evt}   {\ensuremath{{830147524 } } }
\vdef{default-11:SgEvt1:1:chan}   {\ensuremath{{1 } } }
\vdef{default-11:SgEvt1:1:m}   {\ensuremath{{5.896 } } }
\vdef{default-11:SgEvt1:1:pt}   {\ensuremath{{21.839 } } }
\vdef{default-11:SgEvt1:1:phi}   {\ensuremath{{-0.520 } } }
\vdef{default-11:SgEvt1:1:eta}   {\ensuremath{{-1.138 } } }
\vdef{default-11:SgEvt1:1:channel}   {endcap }
\vdef{default-11:SgEvt1:1:cowboy}   {\ensuremath{{0 } } }
\vdef{default-11:SgEvt1:1:m1pt}   {\ensuremath{{15.275 } } }
\vdef{default-11:SgEvt1:1:m2pt}   {\ensuremath{{6.551 } } }
\vdef{default-11:SgEvt1:1:m1eta}   {\ensuremath{{-0.933 } } }
\vdef{default-11:SgEvt1:1:m2eta}   {\ensuremath{{-1.511 } } }
\vdef{default-11:SgEvt1:1:m1phi}   {\ensuremath{{-0.505 } } }
\vdef{default-11:SgEvt1:1:m2phi}   {\ensuremath{{-0.555 } } }
\vdef{default-11:SgEvt1:1:m1q}   {\ensuremath{{-1 } } }
\vdef{default-11:SgEvt1:1:m2q}   {\ensuremath{{1 } } }
\vdef{default-11:SgEvt1:1:iso}   {\ensuremath{{1.000 } } }
\vdef{default-11:SgEvt1:1:alpha}   {\ensuremath{{0.0137 } } }
\vdef{default-11:SgEvt1:1:chi2}   {\ensuremath{{ 0.06 } } }
\vdef{default-11:SgEvt1:1:dof}   {\ensuremath{{1 } } }
\vdef{default-11:SgEvt1:1:fls3d}   {\ensuremath{{35.05 } } }
\vdef{default-11:SgEvt1:1:fl3d}   {\ensuremath{{0.5235 } } }
\vdef{default-11:SgEvt1:1:fl3dE}   {\ensuremath{{0.0149 } } }
\vdef{default-11:SgEvt1:1:docatrk}   {\ensuremath{{0.1003 } } }
\vdef{default-11:SgEvt1:1:closetrk}   {\ensuremath{{0 } } }
\vdef{default-11:SgEvt1:1:lip}   {\ensuremath{{0.0024 } } }
\vdef{default-11:SgEvt1:1:lipE}   {\ensuremath{{0.0062 } } }
\vdef{default-11:SgEvt1:1:tip}   {\ensuremath{{0.0058 } } }
\vdef{default-11:SgEvt1:1:tipE}   {\ensuremath{{0.0054 } } }
\vdef{default-11:SgEvt1:1:pvlip}   {\ensuremath{{0.0024 } } }
\vdef{default-11:SgEvt1:1:pvlips}   {\ensuremath{{0.3899 } } }
\vdef{default-11:SgEvt1:1:pvip}   {\ensuremath{{0.0063 } } }
\vdef{default-11:SgEvt1:1:pvips}   {\ensuremath{{1.1343 } } }
\vdef{default-11:SgEvt1:1:maxdoca}   {\ensuremath{{0.0012 } } }
\vdef{default-11:SgEvt1:1:pvw8}   {\ensuremath{{0.8369 } } }
\vdef{default-11:SgEvt1:1:bdt}   {\ensuremath{{0.0080 } } }
\vdef{default-11:SgEvt1:1:m1pix}   {\ensuremath{{2 } } }
\vdef{default-11:SgEvt1:1:m2pix}   {\ensuremath{{3 } } }
\vdef{default-11:SgEvt1:1:m1bpix}   {\ensuremath{{2 } } }
\vdef{default-11:SgEvt1:1:m2bpix}   {\ensuremath{{3 } } }
\vdef{default-11:SgEvt1:1:m1bpixl1}   {\ensuremath{{1 } } }
\vdef{default-11:SgEvt1:1:m2bpixl1}   {\ensuremath{{1 } } }
\vdef{default-11:BsSgEvt1:0:run}   {\ensuremath{{171446 } } }
\vdef{default-11:BsSgEvt1:0:evt}   {\ensuremath{{393652751 } } }
\vdef{default-11:BsSgEvt1:0:chan}   {\ensuremath{{1 } } }
\vdef{default-11:BsSgEvt1:0:m}   {\ensuremath{{5.333 } } }
\vdef{default-11:BsSgEvt1:0:pt}   {\ensuremath{{10.104 } } }
\vdef{default-11:BsSgEvt1:0:phi}   {\ensuremath{{-2.317 } } }
\vdef{default-11:BsSgEvt1:0:eta}   {\ensuremath{{1.431 } } }
\vdef{default-11:BsSgEvt1:0:channel}   {endcap }
\vdef{default-11:BsSgEvt1:0:cowboy}   {\ensuremath{{1 } } }
\vdef{default-11:BsSgEvt1:0:m1pt}   {\ensuremath{{6.789 } } }
\vdef{default-11:BsSgEvt1:0:m2pt}   {\ensuremath{{4.538 } } }
\vdef{default-11:BsSgEvt1:0:m1eta}   {\ensuremath{{1.216 } } }
\vdef{default-11:BsSgEvt1:0:m2eta}   {\ensuremath{{1.483 } } }
\vdef{default-11:BsSgEvt1:0:m1phi}   {\ensuremath{{-1.941 } } }
\vdef{default-11:BsSgEvt1:0:m2phi}   {\ensuremath{{-2.900 } } }
\vdef{default-11:BsSgEvt1:0:m1q}   {\ensuremath{{1 } } }
\vdef{default-11:BsSgEvt1:0:m2q}   {\ensuremath{{-1 } } }
\vdef{default-11:BsSgEvt1:0:iso}   {\ensuremath{{1.000 } } }
\vdef{default-11:BsSgEvt1:0:alpha}   {\ensuremath{{0.0103 } } }
\vdef{default-11:BsSgEvt1:0:chi2}   {\ensuremath{{ 0.13 } } }
\vdef{default-11:BsSgEvt1:0:dof}   {\ensuremath{{1 } } }
\vdef{default-11:BsSgEvt1:0:fls3d}   {\ensuremath{{37.81 } } }
\vdef{default-11:BsSgEvt1:0:fl3d}   {\ensuremath{{0.5463 } } }
\vdef{default-11:BsSgEvt1:0:fl3dE}   {\ensuremath{{0.0144 } } }
\vdef{default-11:BsSgEvt1:0:docatrk}   {\ensuremath{{0.0795 } } }
\vdef{default-11:BsSgEvt1:0:closetrk}   {\ensuremath{{0 } } }
\vdef{default-11:BsSgEvt1:0:lip}   {\ensuremath{{0.0010 } } }
\vdef{default-11:BsSgEvt1:0:lipE}   {\ensuremath{{0.0040 } } }
\vdef{default-11:BsSgEvt1:0:tip}   {\ensuremath{{0.0052 } } }
\vdef{default-11:BsSgEvt1:0:tipE}   {\ensuremath{{0.0039 } } }
\vdef{default-11:BsSgEvt1:0:pvlip}   {\ensuremath{{0.0010 } } }
\vdef{default-11:BsSgEvt1:0:pvlips}   {\ensuremath{{0.2391 } } }
\vdef{default-11:BsSgEvt1:0:pvip}   {\ensuremath{{0.0053 } } }
\vdef{default-11:BsSgEvt1:0:pvips}   {\ensuremath{{1.3586 } } }
\vdef{default-11:BsSgEvt1:0:maxdoca}   {\ensuremath{{0.0021 } } }
\vdef{default-11:BsSgEvt1:0:pvw8}   {\ensuremath{{0.8778 } } }
\vdef{default-11:BsSgEvt1:0:bdt}   {\ensuremath{{0.1607 } } }
\vdef{default-11:BsSgEvt1:0:m1pix}   {\ensuremath{{3 } } }
\vdef{default-11:BsSgEvt1:0:m2pix}   {\ensuremath{{3 } } }
\vdef{default-11:BsSgEvt1:0:m1bpix}   {\ensuremath{{3 } } }
\vdef{default-11:BsSgEvt1:0:m2bpix}   {\ensuremath{{3 } } }
\vdef{default-11:BsSgEvt1:0:m1bpixl1}   {\ensuremath{{1 } } }
\vdef{default-11:BsSgEvt1:0:m2bpixl1}   {\ensuremath{{1 } } }
\vdef{default-11:SgEvt0:1:run}   {\ensuremath{{171446 } } }
\vdef{default-11:SgEvt0:1:evt}   {\ensuremath{{292436855 } } }
\vdef{default-11:SgEvt0:1:chan}   {\ensuremath{{0 } } }
\vdef{default-11:SgEvt0:1:m}   {\ensuremath{{5.542 } } }
\vdef{default-11:SgEvt0:1:pt}   {\ensuremath{{13.409 } } }
\vdef{default-11:SgEvt0:1:phi}   {\ensuremath{{2.379 } } }
\vdef{default-11:SgEvt0:1:eta}   {\ensuremath{{1.031 } } }
\vdef{default-11:SgEvt0:1:channel}   {barrel }
\vdef{default-11:SgEvt0:1:cowboy}   {\ensuremath{{0 } } }
\vdef{default-11:SgEvt0:1:m1pt}   {\ensuremath{{7.749 } } }
\vdef{default-11:SgEvt0:1:m2pt}   {\ensuremath{{5.732 } } }
\vdef{default-11:SgEvt0:1:m1eta}   {\ensuremath{{1.315 } } }
\vdef{default-11:SgEvt0:1:m2eta}   {\ensuremath{{0.513 } } }
\vdef{default-11:SgEvt0:1:m1phi}   {\ensuremath{{2.433 } } }
\vdef{default-11:SgEvt0:1:m2phi}   {\ensuremath{{2.306 } } }
\vdef{default-11:SgEvt0:1:m1q}   {\ensuremath{{-1 } } }
\vdef{default-11:SgEvt0:1:m2q}   {\ensuremath{{1 } } }
\vdef{default-11:SgEvt0:1:iso}   {\ensuremath{{0.899 } } }
\vdef{default-11:SgEvt0:1:alpha}   {\ensuremath{{0.0320 } } }
\vdef{default-11:SgEvt0:1:chi2}   {\ensuremath{{ 1.07 } } }
\vdef{default-11:SgEvt0:1:dof}   {\ensuremath{{1 } } }
\vdef{default-11:SgEvt0:1:fls3d}   {\ensuremath{{19.13 } } }
\vdef{default-11:SgEvt0:1:fl3d}   {\ensuremath{{0.1330 } } }
\vdef{default-11:SgEvt0:1:fl3dE}   {\ensuremath{{0.0069 } } }
\vdef{default-11:SgEvt0:1:docatrk}   {\ensuremath{{0.0283 } } }
\vdef{default-11:SgEvt0:1:closetrk}   {\ensuremath{{1 } } }
\vdef{default-11:SgEvt0:1:lip}   {\ensuremath{{0.0021 } } }
\vdef{default-11:SgEvt0:1:lipE}   {\ensuremath{{0.0023 } } }
\vdef{default-11:SgEvt0:1:tip}   {\ensuremath{{0.0026 } } }
\vdef{default-11:SgEvt0:1:tipE}   {\ensuremath{{0.0025 } } }
\vdef{default-11:SgEvt0:1:pvlip}   {\ensuremath{{0.0021 } } }
\vdef{default-11:SgEvt0:1:pvlips}   {\ensuremath{{0.9351 } } }
\vdef{default-11:SgEvt0:1:pvip}   {\ensuremath{{0.0034 } } }
\vdef{default-11:SgEvt0:1:pvips}   {\ensuremath{{1.3949 } } }
\vdef{default-11:SgEvt0:1:maxdoca}   {\ensuremath{{0.0041 } } }
\vdef{default-11:SgEvt0:1:pvw8}   {\ensuremath{{0.8568 } } }
\vdef{default-11:SgEvt0:1:bdt}   {\ensuremath{{-0.3761 } } }
\vdef{default-11:SgEvt0:1:m1pix}   {\ensuremath{{3 } } }
\vdef{default-11:SgEvt0:1:m2pix}   {\ensuremath{{3 } } }
\vdef{default-11:SgEvt0:1:m1bpix}   {\ensuremath{{3 } } }
\vdef{default-11:SgEvt0:1:m2bpix}   {\ensuremath{{3 } } }
\vdef{default-11:SgEvt0:1:m1bpixl1}   {\ensuremath{{1 } } }
\vdef{default-11:SgEvt0:1:m2bpixl1}   {\ensuremath{{1 } } }
\vdef{default-11:SgEvt0:2:run}   {\ensuremath{{171106 } } }
\vdef{default-11:SgEvt0:2:evt}   {\ensuremath{{158689598 } } }
\vdef{default-11:SgEvt0:2:chan}   {\ensuremath{{0 } } }
\vdef{default-11:SgEvt0:2:m}   {\ensuremath{{5.010 } } }
\vdef{default-11:SgEvt0:2:pt}   {\ensuremath{{15.832 } } }
\vdef{default-11:SgEvt0:2:phi}   {\ensuremath{{-2.976 } } }
\vdef{default-11:SgEvt0:2:eta}   {\ensuremath{{-0.449 } } }
\vdef{default-11:SgEvt0:2:channel}   {barrel }
\vdef{default-11:SgEvt0:2:cowboy}   {\ensuremath{{0 } } }
\vdef{default-11:SgEvt0:2:m1pt}   {\ensuremath{{11.241 } } }
\vdef{default-11:SgEvt0:2:m2pt}   {\ensuremath{{4.678 } } }
\vdef{default-11:SgEvt0:2:m1eta}   {\ensuremath{{-0.621 } } }
\vdef{default-11:SgEvt0:2:m2eta}   {\ensuremath{{0.018 } } }
\vdef{default-11:SgEvt0:2:m1phi}   {\ensuremath{{-3.044 } } }
\vdef{default-11:SgEvt0:2:m2phi}   {\ensuremath{{-2.812 } } }
\vdef{default-11:SgEvt0:2:m1q}   {\ensuremath{{1 } } }
\vdef{default-11:SgEvt0:2:m2q}   {\ensuremath{{-1 } } }
\vdef{default-11:SgEvt0:2:iso}   {\ensuremath{{0.942 } } }
\vdef{default-11:SgEvt0:2:alpha}   {\ensuremath{{0.0175 } } }
\vdef{default-11:SgEvt0:2:chi2}   {\ensuremath{{ 0.01 } } }
\vdef{default-11:SgEvt0:2:dof}   {\ensuremath{{1 } } }
\vdef{default-11:SgEvt0:2:fls3d}   {\ensuremath{{39.71 } } }
\vdef{default-11:SgEvt0:2:fl3d}   {\ensuremath{{0.4003 } } }
\vdef{default-11:SgEvt0:2:fl3dE}   {\ensuremath{{0.0101 } } }
\vdef{default-11:SgEvt0:2:docatrk}   {\ensuremath{{0.0177 } } }
\vdef{default-11:SgEvt0:2:closetrk}   {\ensuremath{{1 } } }
\vdef{default-11:SgEvt0:2:lip}   {\ensuremath{{0.0049 } } }
\vdef{default-11:SgEvt0:2:lipE}   {\ensuremath{{0.0035 } } }
\vdef{default-11:SgEvt0:2:tip}   {\ensuremath{{0.0045 } } }
\vdef{default-11:SgEvt0:2:tipE}   {\ensuremath{{0.0032 } } }
\vdef{default-11:SgEvt0:2:pvlip}   {\ensuremath{{0.0049 } } }
\vdef{default-11:SgEvt0:2:pvlips}   {\ensuremath{{1.3937 } } }
\vdef{default-11:SgEvt0:2:pvip}   {\ensuremath{{0.0066 } } }
\vdef{default-11:SgEvt0:2:pvips}   {\ensuremath{{1.9713 } } }
\vdef{default-11:SgEvt0:2:maxdoca}   {\ensuremath{{0.0004 } } }
\vdef{default-11:SgEvt0:2:pvw8}   {\ensuremath{{0.8640 } } }
\vdef{default-11:SgEvt0:2:bdt}   {\ensuremath{{-0.0497 } } }
\vdef{default-11:SgEvt0:2:m1pix}   {\ensuremath{{3 } } }
\vdef{default-11:SgEvt0:2:m2pix}   {\ensuremath{{3 } } }
\vdef{default-11:SgEvt0:2:m1bpix}   {\ensuremath{{3 } } }
\vdef{default-11:SgEvt0:2:m2bpix}   {\ensuremath{{3 } } }
\vdef{default-11:SgEvt0:2:m1bpixl1}   {\ensuremath{{2 } } }
\vdef{default-11:SgEvt0:2:m2bpixl1}   {\ensuremath{{1 } } }
\vdef{default-11:BdSgEvt0:0:run}   {\ensuremath{{173389 } } }
\vdef{default-11:BdSgEvt0:0:evt}   {\ensuremath{{173713433 } } }
\vdef{default-11:BdSgEvt0:0:chan}   {\ensuremath{{0 } } }
\vdef{default-11:BdSgEvt0:0:m}   {\ensuremath{{5.263 } } }
\vdef{default-11:BdSgEvt0:0:pt}   {\ensuremath{{17.605 } } }
\vdef{default-11:BdSgEvt0:0:phi}   {\ensuremath{{-1.667 } } }
\vdef{default-11:BdSgEvt0:0:eta}   {\ensuremath{{0.862 } } }
\vdef{default-11:BdSgEvt0:0:channel}   {barrel }
\vdef{default-11:BdSgEvt0:0:cowboy}   {\ensuremath{{0 } } }
\vdef{default-11:BdSgEvt0:0:m1pt}   {\ensuremath{{13.428 } } }
\vdef{default-11:BdSgEvt0:0:m2pt}   {\ensuremath{{4.941 } } }
\vdef{default-11:BdSgEvt0:0:m1eta}   {\ensuremath{{0.841 } } }
\vdef{default-11:BdSgEvt0:0:m2eta}   {\ensuremath{{0.808 } } }
\vdef{default-11:BdSgEvt0:0:m1phi}   {\ensuremath{{-1.495 } } }
\vdef{default-11:BdSgEvt0:0:m2phi}   {\ensuremath{{-2.151 } } }
\vdef{default-11:BdSgEvt0:0:m1q}   {\ensuremath{{-1 } } }
\vdef{default-11:BdSgEvt0:0:m2q}   {\ensuremath{{1 } } }
\vdef{default-11:BdSgEvt0:0:iso}   {\ensuremath{{1.000 } } }
\vdef{default-11:BdSgEvt0:0:alpha}   {\ensuremath{{0.0419 } } }
\vdef{default-11:BdSgEvt0:0:chi2}   {\ensuremath{{ 1.12 } } }
\vdef{default-11:BdSgEvt0:0:dof}   {\ensuremath{{1 } } }
\vdef{default-11:BdSgEvt0:0:fls3d}   {\ensuremath{{14.76 } } }
\vdef{default-11:BdSgEvt0:0:fl3d}   {\ensuremath{{0.1837 } } }
\vdef{default-11:BdSgEvt0:0:fl3dE}   {\ensuremath{{0.0124 } } }
\vdef{default-11:BdSgEvt0:0:docatrk}   {\ensuremath{{0.0791 } } }
\vdef{default-11:BdSgEvt0:0:closetrk}   {\ensuremath{{0 } } }
\vdef{default-11:BdSgEvt0:0:lip}   {\ensuremath{{0.0049 } } }
\vdef{default-11:BdSgEvt0:0:lipE}   {\ensuremath{{0.0066 } } }
\vdef{default-11:BdSgEvt0:0:tip}   {\ensuremath{{0.0036 } } }
\vdef{default-11:BdSgEvt0:0:tipE}   {\ensuremath{{0.0052 } } }
\vdef{default-11:BdSgEvt0:0:pvlip}   {\ensuremath{{0.0049 } } }
\vdef{default-11:BdSgEvt0:0:pvlips}   {\ensuremath{{0.7350 } } }
\vdef{default-11:BdSgEvt0:0:pvip}   {\ensuremath{{0.0061 } } }
\vdef{default-11:BdSgEvt0:0:pvips}   {\ensuremath{{0.9853 } } }
\vdef{default-11:BdSgEvt0:0:maxdoca}   {\ensuremath{{0.0058 } } }
\vdef{default-11:BdSgEvt0:0:pvw8}   {\ensuremath{{0.8792 } } }
\vdef{default-11:BdSgEvt0:0:bdt}   {\ensuremath{{0.0041 } } }
\vdef{default-11:BdSgEvt0:0:m1pix}   {\ensuremath{{3 } } }
\vdef{default-11:BdSgEvt0:0:m2pix}   {\ensuremath{{3 } } }
\vdef{default-11:BdSgEvt0:0:m1bpix}   {\ensuremath{{3 } } }
\vdef{default-11:BdSgEvt0:0:m2bpix}   {\ensuremath{{3 } } }
\vdef{default-11:BdSgEvt0:0:m1bpixl1}   {\ensuremath{{1 } } }
\vdef{default-11:BdSgEvt0:0:m2bpixl1}   {\ensuremath{{1 } } }
\vdef{default-11:SgEvt1:2:run}   {\ensuremath{{172952 } } }
\vdef{default-11:SgEvt1:2:evt}   {\ensuremath{{1015831381 } } }
\vdef{default-11:SgEvt1:2:chan}   {\ensuremath{{1 } } }
\vdef{default-11:SgEvt1:2:m}   {\ensuremath{{5.579 } } }
\vdef{default-11:SgEvt1:2:pt}   {\ensuremath{{12.403 } } }
\vdef{default-11:SgEvt1:2:phi}   {\ensuremath{{-2.581 } } }
\vdef{default-11:SgEvt1:2:eta}   {\ensuremath{{-1.942 } } }
\vdef{default-11:SgEvt1:2:channel}   {endcap }
\vdef{default-11:SgEvt1:2:cowboy}   {\ensuremath{{0 } } }
\vdef{default-11:SgEvt1:2:m1pt}   {\ensuremath{{8.339 } } }
\vdef{default-11:SgEvt1:2:m2pt}   {\ensuremath{{5.259 } } }
\vdef{default-11:SgEvt1:2:m1eta}   {\ensuremath{{-1.875 } } }
\vdef{default-11:SgEvt1:2:m2eta}   {\ensuremath{{-1.821 } } }
\vdef{default-11:SgEvt1:2:m1phi}   {\ensuremath{{-2.252 } } }
\vdef{default-11:SgEvt1:2:m2phi}   {\ensuremath{{-3.118 } } }
\vdef{default-11:SgEvt1:2:m1q}   {\ensuremath{{-1 } } }
\vdef{default-11:SgEvt1:2:m2q}   {\ensuremath{{1 } } }
\vdef{default-11:SgEvt1:2:iso}   {\ensuremath{{0.924 } } }
\vdef{default-11:SgEvt1:2:alpha}   {\ensuremath{{0.0097 } } }
\vdef{default-11:SgEvt1:2:chi2}   {\ensuremath{{ 1.34 } } }
\vdef{default-11:SgEvt1:2:dof}   {\ensuremath{{1 } } }
\vdef{default-11:SgEvt1:2:fls3d}   {\ensuremath{{16.01 } } }
\vdef{default-11:SgEvt1:2:fl3d}   {\ensuremath{{0.3859 } } }
\vdef{default-11:SgEvt1:2:fl3dE}   {\ensuremath{{0.0241 } } }
\vdef{default-11:SgEvt1:2:docatrk}   {\ensuremath{{0.0352 } } }
\vdef{default-11:SgEvt1:2:closetrk}   {\ensuremath{{0 } } }
\vdef{default-11:SgEvt1:2:lip}   {\ensuremath{{-0.0003 } } }
\vdef{default-11:SgEvt1:2:lipE}   {\ensuremath{{0.0022 } } }
\vdef{default-11:SgEvt1:2:tip}   {\ensuremath{{0.0035 } } }
\vdef{default-11:SgEvt1:2:tipE}   {\ensuremath{{0.0042 } } }
\vdef{default-11:SgEvt1:2:pvlip}   {\ensuremath{{-0.0003 } } }
\vdef{default-11:SgEvt1:2:pvlips}   {\ensuremath{{-0.1579 } } }
\vdef{default-11:SgEvt1:2:pvip}   {\ensuremath{{0.0036 } } }
\vdef{default-11:SgEvt1:2:pvips}   {\ensuremath{{0.8530 } } }
\vdef{default-11:SgEvt1:2:maxdoca}   {\ensuremath{{0.0067 } } }
\vdef{default-11:SgEvt1:2:pvw8}   {\ensuremath{{0.8912 } } }
\vdef{default-11:SgEvt1:2:bdt}   {\ensuremath{{-0.0809 } } }
\vdef{default-11:SgEvt1:2:m1pix}   {\ensuremath{{3 } } }
\vdef{default-11:SgEvt1:2:m2pix}   {\ensuremath{{3 } } }
\vdef{default-11:SgEvt1:2:m1bpix}   {\ensuremath{{2 } } }
\vdef{default-11:SgEvt1:2:m2bpix}   {\ensuremath{{2 } } }
\vdef{default-11:SgEvt1:2:m1bpixl1}   {\ensuremath{{1 } } }
\vdef{default-11:SgEvt1:2:m2bpixl1}   {\ensuremath{{1 } } }
\vdef{default-11:SgEvt1:3:run}   {\ensuremath{{172952 } } }
\vdef{default-11:SgEvt1:3:evt}   {\ensuremath{{386613368 } } }
\vdef{default-11:SgEvt1:3:chan}   {\ensuremath{{1 } } }
\vdef{default-11:SgEvt1:3:m}   {\ensuremath{{5.102 } } }
\vdef{default-11:SgEvt1:3:pt}   {\ensuremath{{16.373 } } }
\vdef{default-11:SgEvt1:3:phi}   {\ensuremath{{-2.521 } } }
\vdef{default-11:SgEvt1:3:eta}   {\ensuremath{{-1.960 } } }
\vdef{default-11:SgEvt1:3:channel}   {endcap }
\vdef{default-11:SgEvt1:3:cowboy}   {\ensuremath{{1 } } }
\vdef{default-11:SgEvt1:3:m1pt}   {\ensuremath{{10.728 } } }
\vdef{default-11:SgEvt1:3:m2pt}   {\ensuremath{{6.052 } } }
\vdef{default-11:SgEvt1:3:m1eta}   {\ensuremath{{-1.762 } } }
\vdef{default-11:SgEvt1:3:m2eta}   {\ensuremath{{-2.185 } } }
\vdef{default-11:SgEvt1:3:m1phi}   {\ensuremath{{-2.352 } } }
\vdef{default-11:SgEvt1:3:m2phi}   {\ensuremath{{-2.823 } } }
\vdef{default-11:SgEvt1:3:m1q}   {\ensuremath{{1 } } }
\vdef{default-11:SgEvt1:3:m2q}   {\ensuremath{{-1 } } }
\vdef{default-11:SgEvt1:3:iso}   {\ensuremath{{0.878 } } }
\vdef{default-11:SgEvt1:3:alpha}   {\ensuremath{{0.0225 } } }
\vdef{default-11:SgEvt1:3:chi2}   {\ensuremath{{ 0.08 } } }
\vdef{default-11:SgEvt1:3:dof}   {\ensuremath{{1 } } }
\vdef{default-11:SgEvt1:3:fls3d}   {\ensuremath{{15.22 } } }
\vdef{default-11:SgEvt1:3:fl3d}   {\ensuremath{{0.6627 } } }
\vdef{default-11:SgEvt1:3:fl3dE}   {\ensuremath{{0.0435 } } }
\vdef{default-11:SgEvt1:3:docatrk}   {\ensuremath{{0.0239 } } }
\vdef{default-11:SgEvt1:3:closetrk}   {\ensuremath{{1 } } }
\vdef{default-11:SgEvt1:3:lip}   {\ensuremath{{0.0041 } } }
\vdef{default-11:SgEvt1:3:lipE}   {\ensuremath{{0.0034 } } }
\vdef{default-11:SgEvt1:3:tip}   {\ensuremath{{0.0025 } } }
\vdef{default-11:SgEvt1:3:tipE}   {\ensuremath{{0.0039 } } }
\vdef{default-11:SgEvt1:3:pvlip}   {\ensuremath{{0.0041 } } }
\vdef{default-11:SgEvt1:3:pvlips}   {\ensuremath{{1.1824 } } }
\vdef{default-11:SgEvt1:3:pvip}   {\ensuremath{{0.0047 } } }
\vdef{default-11:SgEvt1:3:pvips}   {\ensuremath{{1.3286 } } }
\vdef{default-11:SgEvt1:3:maxdoca}   {\ensuremath{{0.0021 } } }
\vdef{default-11:SgEvt1:3:pvw8}   {\ensuremath{{0.9129 } } }
\vdef{default-11:SgEvt1:3:bdt}   {\ensuremath{{-0.3662 } } }
\vdef{default-11:SgEvt1:3:m1pix}   {\ensuremath{{1 } } }
\vdef{default-11:SgEvt1:3:m2pix}   {\ensuremath{{2 } } }
\vdef{default-11:SgEvt1:3:m1bpix}   {\ensuremath{{1 } } }
\vdef{default-11:SgEvt1:3:m2bpix}   {\ensuremath{{1 } } }
\vdef{default-11:SgEvt1:3:m1bpixl1}   {\ensuremath{{0 } } }
\vdef{default-11:SgEvt1:3:m2bpixl1}   {\ensuremath{{1 } } }
\vdef{default-11:BsSgEvt1:1:run}   {\ensuremath{{179889 } } }
\vdef{default-11:BsSgEvt1:1:evt}   {\ensuremath{{533479508 } } }
\vdef{default-11:BsSgEvt1:1:chan}   {\ensuremath{{1 } } }
\vdef{default-11:BsSgEvt1:1:m}   {\ensuremath{{5.311 } } }
\vdef{default-11:BsSgEvt1:1:pt}   {\ensuremath{{14.914 } } }
\vdef{default-11:BsSgEvt1:1:phi}   {\ensuremath{{0.571 } } }
\vdef{default-11:BsSgEvt1:1:eta}   {\ensuremath{{-1.532 } } }
\vdef{default-11:BsSgEvt1:1:channel}   {endcap }
\vdef{default-11:BsSgEvt1:1:cowboy}   {\ensuremath{{0 } } }
\vdef{default-11:BsSgEvt1:1:m1pt}   {\ensuremath{{8.050 } } }
\vdef{default-11:BsSgEvt1:1:m2pt}   {\ensuremath{{7.620 } } }
\vdef{default-11:BsSgEvt1:1:m1eta}   {\ensuremath{{-1.616 } } }
\vdef{default-11:BsSgEvt1:1:m2eta}   {\ensuremath{{-1.333 } } }
\vdef{default-11:BsSgEvt1:1:m1phi}   {\ensuremath{{0.875 } } }
\vdef{default-11:BsSgEvt1:1:m2phi}   {\ensuremath{{0.250 } } }
\vdef{default-11:BsSgEvt1:1:m1q}   {\ensuremath{{-1 } } }
\vdef{default-11:BsSgEvt1:1:m2q}   {\ensuremath{{1 } } }
\vdef{default-11:BsSgEvt1:1:iso}   {\ensuremath{{1.000 } } }
\vdef{default-11:BsSgEvt1:1:alpha}   {\ensuremath{{0.0158 } } }
\vdef{default-11:BsSgEvt1:1:chi2}   {\ensuremath{{ 0.05 } } }
\vdef{default-11:BsSgEvt1:1:dof}   {\ensuremath{{1 } } }
\vdef{default-11:BsSgEvt1:1:fls3d}   {\ensuremath{{18.99 } } }
\vdef{default-11:BsSgEvt1:1:fl3d}   {\ensuremath{{0.2519 } } }
\vdef{default-11:BsSgEvt1:1:fl3dE}   {\ensuremath{{0.0133 } } }
\vdef{default-11:BsSgEvt1:1:docatrk}   {\ensuremath{{0.0486 } } }
\vdef{default-11:BsSgEvt1:1:closetrk}   {\ensuremath{{0 } } }
\vdef{default-11:BsSgEvt1:1:lip}   {\ensuremath{{-0.0014 } } }
\vdef{default-11:BsSgEvt1:1:lipE}   {\ensuremath{{0.0034 } } }
\vdef{default-11:BsSgEvt1:1:tip}   {\ensuremath{{0.0021 } } }
\vdef{default-11:BsSgEvt1:1:tipE}   {\ensuremath{{0.0032 } } }
\vdef{default-11:BsSgEvt1:1:pvlip}   {\ensuremath{{-0.0014 } } }
\vdef{default-11:BsSgEvt1:1:pvlips}   {\ensuremath{{-0.4098 } } }
\vdef{default-11:BsSgEvt1:1:pvip}   {\ensuremath{{0.0025 } } }
\vdef{default-11:BsSgEvt1:1:pvips}   {\ensuremath{{0.7730 } } }
\vdef{default-11:BsSgEvt1:1:maxdoca}   {\ensuremath{{0.0008 } } }
\vdef{default-11:BsSgEvt1:1:pvw8}   {\ensuremath{{0.9004 } } }
\vdef{default-11:BsSgEvt1:1:bdt}   {\ensuremath{{0.1044 } } }
\vdef{default-11:BsSgEvt1:1:m1pix}   {\ensuremath{{3 } } }
\vdef{default-11:BsSgEvt1:1:m2pix}   {\ensuremath{{3 } } }
\vdef{default-11:BsSgEvt1:1:m1bpix}   {\ensuremath{{3 } } }
\vdef{default-11:BsSgEvt1:1:m2bpix}   {\ensuremath{{3 } } }
\vdef{default-11:BsSgEvt1:1:m1bpixl1}   {\ensuremath{{1 } } }
\vdef{default-11:BsSgEvt1:1:m2bpixl1}   {\ensuremath{{2 } } }
\vdef{default-11:SgEvt0:3:run}   {\ensuremath{{179452 } } }
\vdef{default-11:SgEvt0:3:evt}   {\ensuremath{{495629806 } } }
\vdef{default-11:SgEvt0:3:chan}   {\ensuremath{{0 } } }
\vdef{default-11:SgEvt0:3:m}   {\ensuremath{{5.555 } } }
\vdef{default-11:SgEvt0:3:pt}   {\ensuremath{{9.469 } } }
\vdef{default-11:SgEvt0:3:phi}   {\ensuremath{{2.642 } } }
\vdef{default-11:SgEvt0:3:eta}   {\ensuremath{{1.436 } } }
\vdef{default-11:SgEvt0:3:channel}   {barrel }
\vdef{default-11:SgEvt0:3:cowboy}   {\ensuremath{{0 } } }
\vdef{default-11:SgEvt0:3:m1pt}   {\ensuremath{{6.770 } } }
\vdef{default-11:SgEvt0:3:m2pt}   {\ensuremath{{4.201 } } }
\vdef{default-11:SgEvt0:3:m1eta}   {\ensuremath{{1.334 } } }
\vdef{default-11:SgEvt0:3:m2eta}   {\ensuremath{{1.261 } } }
\vdef{default-11:SgEvt0:3:m1phi}   {\ensuremath{{3.048 } } }
\vdef{default-11:SgEvt0:3:m2phi}   {\ensuremath{{1.956 } } }
\vdef{default-11:SgEvt0:3:m1q}   {\ensuremath{{-1 } } }
\vdef{default-11:SgEvt0:3:m2q}   {\ensuremath{{1 } } }
\vdef{default-11:SgEvt0:3:iso}   {\ensuremath{{1.000 } } }
\vdef{default-11:SgEvt0:3:alpha}   {\ensuremath{{0.0394 } } }
\vdef{default-11:SgEvt0:3:chi2}   {\ensuremath{{ 1.23 } } }
\vdef{default-11:SgEvt0:3:dof}   {\ensuremath{{1 } } }
\vdef{default-11:SgEvt0:3:fls3d}   {\ensuremath{{13.52 } } }
\vdef{default-11:SgEvt0:3:fl3d}   {\ensuremath{{0.1299 } } }
\vdef{default-11:SgEvt0:3:fl3dE}   {\ensuremath{{0.0096 } } }
\vdef{default-11:SgEvt0:3:docatrk}   {\ensuremath{{0.0526 } } }
\vdef{default-11:SgEvt0:3:closetrk}   {\ensuremath{{0 } } }
\vdef{default-11:SgEvt0:3:lip}   {\ensuremath{{-0.0015 } } }
\vdef{default-11:SgEvt0:3:lipE}   {\ensuremath{{0.0040 } } }
\vdef{default-11:SgEvt0:3:tip}   {\ensuremath{{0.0039 } } }
\vdef{default-11:SgEvt0:3:tipE}   {\ensuremath{{0.0048 } } }
\vdef{default-11:SgEvt0:3:pvlip}   {\ensuremath{{-0.0015 } } }
\vdef{default-11:SgEvt0:3:pvlips}   {\ensuremath{{-0.3814 } } }
\vdef{default-11:SgEvt0:3:pvip}   {\ensuremath{{0.0042 } } }
\vdef{default-11:SgEvt0:3:pvips}   {\ensuremath{{0.8897 } } }
\vdef{default-11:SgEvt0:3:maxdoca}   {\ensuremath{{0.0054 } } }
\vdef{default-11:SgEvt0:3:pvw8}   {\ensuremath{{0.8141 } } }
\vdef{default-11:SgEvt0:3:bdt}   {\ensuremath{{-0.3534 } } }
\vdef{default-11:SgEvt0:3:m1pix}   {\ensuremath{{2 } } }
\vdef{default-11:SgEvt0:3:m2pix}   {\ensuremath{{2 } } }
\vdef{default-11:SgEvt0:3:m1bpix}   {\ensuremath{{1 } } }
\vdef{default-11:SgEvt0:3:m2bpix}   {\ensuremath{{1 } } }
\vdef{default-11:SgEvt0:3:m1bpixl1}   {\ensuremath{{1 } } }
\vdef{default-11:SgEvt0:3:m2bpixl1}   {\ensuremath{{1 } } }
\vdef{default-11:SgEvt1:4:run}   {\ensuremath{{178871 } } }
\vdef{default-11:SgEvt1:4:evt}   {\ensuremath{{24104557 } } }
\vdef{default-11:SgEvt1:4:chan}   {\ensuremath{{1 } } }
\vdef{default-11:SgEvt1:4:m}   {\ensuremath{{5.885 } } }
\vdef{default-11:SgEvt1:4:pt}   {\ensuremath{{16.150 } } }
\vdef{default-11:SgEvt1:4:phi}   {\ensuremath{{0.975 } } }
\vdef{default-11:SgEvt1:4:eta}   {\ensuremath{{-1.873 } } }
\vdef{default-11:SgEvt1:4:channel}   {endcap }
\vdef{default-11:SgEvt1:4:cowboy}   {\ensuremath{{1 } } }
\vdef{default-11:SgEvt1:4:m1pt}   {\ensuremath{{9.525 } } }
\vdef{default-11:SgEvt1:4:m2pt}   {\ensuremath{{6.580 } } }
\vdef{default-11:SgEvt1:4:m1eta}   {\ensuremath{{-1.516 } } }
\vdef{default-11:SgEvt1:4:m2eta}   {\ensuremath{{-2.237 } } }
\vdef{default-11:SgEvt1:4:m1phi}   {\ensuremath{{1.003 } } }
\vdef{default-11:SgEvt1:4:m2phi}   {\ensuremath{{0.934 } } }
\vdef{default-11:SgEvt1:4:m1q}   {\ensuremath{{1 } } }
\vdef{default-11:SgEvt1:4:m2q}   {\ensuremath{{-1 } } }
\vdef{default-11:SgEvt1:4:iso}   {\ensuremath{{1.000 } } }
\vdef{default-11:SgEvt1:4:alpha}   {\ensuremath{{0.0106 } } }
\vdef{default-11:SgEvt1:4:chi2}   {\ensuremath{{ 0.62 } } }
\vdef{default-11:SgEvt1:4:dof}   {\ensuremath{{1 } } }
\vdef{default-11:SgEvt1:4:fls3d}   {\ensuremath{{38.30 } } }
\vdef{default-11:SgEvt1:4:fl3d}   {\ensuremath{{0.6969 } } }
\vdef{default-11:SgEvt1:4:fl3dE}   {\ensuremath{{0.0182 } } }
\vdef{default-11:SgEvt1:4:docatrk}   {\ensuremath{{0.0950 } } }
\vdef{default-11:SgEvt1:4:closetrk}   {\ensuremath{{0 } } }
\vdef{default-11:SgEvt1:4:lip}   {\ensuremath{{0.0019 } } }
\vdef{default-11:SgEvt1:4:lipE}   {\ensuremath{{0.0024 } } }
\vdef{default-11:SgEvt1:4:tip}   {\ensuremath{{0.0038 } } }
\vdef{default-11:SgEvt1:4:tipE}   {\ensuremath{{0.0029 } } }
\vdef{default-11:SgEvt1:4:pvlip}   {\ensuremath{{0.0019 } } }
\vdef{default-11:SgEvt1:4:pvlips}   {\ensuremath{{0.7960 } } }
\vdef{default-11:SgEvt1:4:pvip}   {\ensuremath{{0.0042 } } }
\vdef{default-11:SgEvt1:4:pvips}   {\ensuremath{{1.4860 } } }
\vdef{default-11:SgEvt1:4:maxdoca}   {\ensuremath{{0.0038 } } }
\vdef{default-11:SgEvt1:4:pvw8}   {\ensuremath{{0.9219 } } }
\vdef{default-11:SgEvt1:4:bdt}   {\ensuremath{{0.0624 } } }
\vdef{default-11:SgEvt1:4:m1pix}   {\ensuremath{{3 } } }
\vdef{default-11:SgEvt1:4:m2pix}   {\ensuremath{{3 } } }
\vdef{default-11:SgEvt1:4:m1bpix}   {\ensuremath{{3 } } }
\vdef{default-11:SgEvt1:4:m2bpix}   {\ensuremath{{1 } } }
\vdef{default-11:SgEvt1:4:m1bpixl1}   {\ensuremath{{1 } } }
\vdef{default-11:SgEvt1:4:m2bpixl1}   {\ensuremath{{1 } } }
\vdef{default-11:BdSgEvt0:1:run}   {\ensuremath{{178840 } } }
\vdef{default-11:BdSgEvt0:1:evt}   {\ensuremath{{859654315 } } }
\vdef{default-11:BdSgEvt0:1:chan}   {\ensuremath{{0 } } }
\vdef{default-11:BdSgEvt0:1:m}   {\ensuremath{{5.273 } } }
\vdef{default-11:BdSgEvt0:1:pt}   {\ensuremath{{12.959 } } }
\vdef{default-11:BdSgEvt0:1:phi}   {\ensuremath{{-1.993 } } }
\vdef{default-11:BdSgEvt0:1:eta}   {\ensuremath{{0.993 } } }
\vdef{default-11:BdSgEvt0:1:channel}   {barrel }
\vdef{default-11:BdSgEvt0:1:cowboy}   {\ensuremath{{0 } } }
\vdef{default-11:BdSgEvt0:1:m1pt}   {\ensuremath{{7.587 } } }
\vdef{default-11:BdSgEvt0:1:m2pt}   {\ensuremath{{6.286 } } }
\vdef{default-11:BdSgEvt0:1:m1eta}   {\ensuremath{{1.056 } } }
\vdef{default-11:BdSgEvt0:1:m2eta}   {\ensuremath{{0.792 } } }
\vdef{default-11:BdSgEvt0:1:m1phi}   {\ensuremath{{-1.663 } } }
\vdef{default-11:BdSgEvt0:1:m2phi}   {\ensuremath{{-2.394 } } }
\vdef{default-11:BdSgEvt0:1:m1q}   {\ensuremath{{-1 } } }
\vdef{default-11:BdSgEvt0:1:m2q}   {\ensuremath{{1 } } }
\vdef{default-11:BdSgEvt0:1:iso}   {\ensuremath{{0.925 } } }
\vdef{default-11:BdSgEvt0:1:alpha}   {\ensuremath{{0.0044 } } }
\vdef{default-11:BdSgEvt0:1:chi2}   {\ensuremath{{ 0.28 } } }
\vdef{default-11:BdSgEvt0:1:dof}   {\ensuremath{{1 } } }
\vdef{default-11:BdSgEvt0:1:fls3d}   {\ensuremath{{46.54 } } }
\vdef{default-11:BdSgEvt0:1:fl3d}   {\ensuremath{{0.4705 } } }
\vdef{default-11:BdSgEvt0:1:fl3dE}   {\ensuremath{{0.0101 } } }
\vdef{default-11:BdSgEvt0:1:docatrk}   {\ensuremath{{0.0619 } } }
\vdef{default-11:BdSgEvt0:1:closetrk}   {\ensuremath{{0 } } }
\vdef{default-11:BdSgEvt0:1:lip}   {\ensuremath{{0.0013 } } }
\vdef{default-11:BdSgEvt0:1:lipE}   {\ensuremath{{0.0053 } } }
\vdef{default-11:BdSgEvt0:1:tip}   {\ensuremath{{0.0006 } } }
\vdef{default-11:BdSgEvt0:1:tipE}   {\ensuremath{{0.0049 } } }
\vdef{default-11:BdSgEvt0:1:pvlip}   {\ensuremath{{0.0013 } } }
\vdef{default-11:BdSgEvt0:1:pvlips}   {\ensuremath{{0.2410 } } }
\vdef{default-11:BdSgEvt0:1:pvip}   {\ensuremath{{0.0014 } } }
\vdef{default-11:BdSgEvt0:1:pvips}   {\ensuremath{{0.2660 } } }
\vdef{default-11:BdSgEvt0:1:maxdoca}   {\ensuremath{{0.0025 } } }
\vdef{default-11:BdSgEvt0:1:pvw8}   {\ensuremath{{0.9211 } } }
\vdef{default-11:BdSgEvt0:1:bdt}   {\ensuremath{{0.4245 } } }
\vdef{default-11:BdSgEvt0:1:m1pix}   {\ensuremath{{3 } } }
\vdef{default-11:BdSgEvt0:1:m2pix}   {\ensuremath{{3 } } }
\vdef{default-11:BdSgEvt0:1:m1bpix}   {\ensuremath{{3 } } }
\vdef{default-11:BdSgEvt0:1:m2bpix}   {\ensuremath{{3 } } }
\vdef{default-11:BdSgEvt0:1:m1bpixl1}   {\ensuremath{{1 } } }
\vdef{default-11:BdSgEvt0:1:m2bpixl1}   {\ensuremath{{1 } } }
\vdef{default-11:SgEvt0:4:run}   {\ensuremath{{178708 } } }
\vdef{default-11:SgEvt0:4:evt}   {\ensuremath{{214563111 } } }
\vdef{default-11:SgEvt0:4:chan}   {\ensuremath{{0 } } }
\vdef{default-11:SgEvt0:4:m}   {\ensuremath{{4.984 } } }
\vdef{default-11:SgEvt0:4:pt}   {\ensuremath{{23.709 } } }
\vdef{default-11:SgEvt0:4:phi}   {\ensuremath{{-2.593 } } }
\vdef{default-11:SgEvt0:4:eta}   {\ensuremath{{-0.494 } } }
\vdef{default-11:SgEvt0:4:channel}   {barrel }
\vdef{default-11:SgEvt0:4:cowboy}   {\ensuremath{{1 } } }
\vdef{default-11:SgEvt0:4:m1pt}   {\ensuremath{{15.039 } } }
\vdef{default-11:SgEvt0:4:m2pt}   {\ensuremath{{8.808 } } }
\vdef{default-11:SgEvt0:4:m1eta}   {\ensuremath{{-0.334 } } }
\vdef{default-11:SgEvt0:4:m2eta}   {\ensuremath{{-0.738 } } }
\vdef{default-11:SgEvt0:4:m1phi}   {\ensuremath{{-2.537 } } }
\vdef{default-11:SgEvt0:4:m2phi}   {\ensuremath{{-2.688 } } }
\vdef{default-11:SgEvt0:4:m1q}   {\ensuremath{{1 } } }
\vdef{default-11:SgEvt0:4:m2q}   {\ensuremath{{-1 } } }
\vdef{default-11:SgEvt0:4:iso}   {\ensuremath{{0.921 } } }
\vdef{default-11:SgEvt0:4:alpha}   {\ensuremath{{0.0033 } } }
\vdef{default-11:SgEvt0:4:chi2}   {\ensuremath{{ 2.19 } } }
\vdef{default-11:SgEvt0:4:dof}   {\ensuremath{{1 } } }
\vdef{default-11:SgEvt0:4:fls3d}   {\ensuremath{{32.27 } } }
\vdef{default-11:SgEvt0:4:fl3d}   {\ensuremath{{0.3267 } } }
\vdef{default-11:SgEvt0:4:fl3dE}   {\ensuremath{{0.0101 } } }
\vdef{default-11:SgEvt0:4:docatrk}   {\ensuremath{{0.0351 } } }
\vdef{default-11:SgEvt0:4:closetrk}   {\ensuremath{{0 } } }
\vdef{default-11:SgEvt0:4:lip}   {\ensuremath{{-0.0010 } } }
\vdef{default-11:SgEvt0:4:lipE}   {\ensuremath{{0.0052 } } }
\vdef{default-11:SgEvt0:4:tip}   {\ensuremath{{0.0001 } } }
\vdef{default-11:SgEvt0:4:tipE}   {\ensuremath{{0.0029 } } }
\vdef{default-11:SgEvt0:4:pvlip}   {\ensuremath{{-0.0010 } } }
\vdef{default-11:SgEvt0:4:pvlips}   {\ensuremath{{-0.1834 } } }
\vdef{default-11:SgEvt0:4:pvip}   {\ensuremath{{0.0010 } } }
\vdef{default-11:SgEvt0:4:pvips}   {\ensuremath{{0.1850 } } }
\vdef{default-11:SgEvt0:4:maxdoca}   {\ensuremath{{0.0043 } } }
\vdef{default-11:SgEvt0:4:pvw8}   {\ensuremath{{0.9372 } } }
\vdef{default-11:SgEvt0:4:bdt}   {\ensuremath{{0.0708 } } }
\vdef{default-11:SgEvt0:4:m1pix}   {\ensuremath{{3 } } }
\vdef{default-11:SgEvt0:4:m2pix}   {\ensuremath{{3 } } }
\vdef{default-11:SgEvt0:4:m1bpix}   {\ensuremath{{3 } } }
\vdef{default-11:SgEvt0:4:m2bpix}   {\ensuremath{{3 } } }
\vdef{default-11:SgEvt0:4:m1bpixl1}   {\ensuremath{{1 } } }
\vdef{default-11:SgEvt0:4:m2bpixl1}   {\ensuremath{{1 } } }
\vdef{default-11:SgEvt0:5:run}   {\ensuremath{{178100 } } }
\vdef{default-11:SgEvt0:5:evt}   {\ensuremath{{258643059 } } }
\vdef{default-11:SgEvt0:5:chan}   {\ensuremath{{0 } } }
\vdef{default-11:SgEvt0:5:m}   {\ensuremath{{4.907 } } }
\vdef{default-11:SgEvt0:5:pt}   {\ensuremath{{31.104 } } }
\vdef{default-11:SgEvt0:5:phi}   {\ensuremath{{-2.359 } } }
\vdef{default-11:SgEvt0:5:eta}   {\ensuremath{{-0.768 } } }
\vdef{default-11:SgEvt0:5:channel}   {barrel }
\vdef{default-11:SgEvt0:5:cowboy}   {\ensuremath{{0 } } }
\vdef{default-11:SgEvt0:5:m1pt}   {\ensuremath{{24.493 } } }
\vdef{default-11:SgEvt0:5:m2pt}   {\ensuremath{{6.863 } } }
\vdef{default-11:SgEvt0:5:m1eta}   {\ensuremath{{-0.810 } } }
\vdef{default-11:SgEvt0:5:m2eta}   {\ensuremath{{-0.583 } } }
\vdef{default-11:SgEvt0:5:m1phi}   {\ensuremath{{-2.294 } } }
\vdef{default-11:SgEvt0:5:m2phi}   {\ensuremath{{-2.597 } } }
\vdef{default-11:SgEvt0:5:m1q}   {\ensuremath{{-1 } } }
\vdef{default-11:SgEvt0:5:m2q}   {\ensuremath{{1 } } }
\vdef{default-11:SgEvt0:5:iso}   {\ensuremath{{0.810 } } }
\vdef{default-11:SgEvt0:5:alpha}   {\ensuremath{{0.0318 } } }
\vdef{default-11:SgEvt0:5:chi2}   {\ensuremath{{ 0.04 } } }
\vdef{default-11:SgEvt0:5:dof}   {\ensuremath{{1 } } }
\vdef{default-11:SgEvt0:5:fls3d}   {\ensuremath{{17.04 } } }
\vdef{default-11:SgEvt0:5:fl3d}   {\ensuremath{{0.1997 } } }
\vdef{default-11:SgEvt0:5:fl3dE}   {\ensuremath{{0.0117 } } }
\vdef{default-11:SgEvt0:5:docatrk}   {\ensuremath{{0.0291 } } }
\vdef{default-11:SgEvt0:5:closetrk}   {\ensuremath{{0 } } }
\vdef{default-11:SgEvt0:5:lip}   {\ensuremath{{0.0048 } } }
\vdef{default-11:SgEvt0:5:lipE}   {\ensuremath{{0.0028 } } }
\vdef{default-11:SgEvt0:5:tip}   {\ensuremath{{0.0005 } } }
\vdef{default-11:SgEvt0:5:tipE}   {\ensuremath{{0.0019 } } }
\vdef{default-11:SgEvt0:5:pvlip}   {\ensuremath{{0.0048 } } }
\vdef{default-11:SgEvt0:5:pvlips}   {\ensuremath{{1.7562 } } }
\vdef{default-11:SgEvt0:5:pvip}   {\ensuremath{{0.0049 } } }
\vdef{default-11:SgEvt0:5:pvips}   {\ensuremath{{1.7716 } } }
\vdef{default-11:SgEvt0:5:maxdoca}   {\ensuremath{{0.0007 } } }
\vdef{default-11:SgEvt0:5:pvw8}   {\ensuremath{{0.9202 } } }
\vdef{default-11:SgEvt0:5:bdt}   {\ensuremath{{-0.2665 } } }
\vdef{default-11:SgEvt0:5:m1pix}   {\ensuremath{{3 } } }
\vdef{default-11:SgEvt0:5:m2pix}   {\ensuremath{{3 } } }
\vdef{default-11:SgEvt0:5:m1bpix}   {\ensuremath{{3 } } }
\vdef{default-11:SgEvt0:5:m2bpix}   {\ensuremath{{3 } } }
\vdef{default-11:SgEvt0:5:m1bpixl1}   {\ensuremath{{1 } } }
\vdef{default-11:SgEvt0:5:m2bpixl1}   {\ensuremath{{1 } } }
\vdef{default-11:BsSgEvt1:2:run}   {\ensuremath{{177139 } } }
\vdef{default-11:BsSgEvt1:2:evt}   {\ensuremath{{59599661 } } }
\vdef{default-11:BsSgEvt1:2:chan}   {\ensuremath{{1 } } }
\vdef{default-11:BsSgEvt1:2:m}   {\ensuremath{{5.339 } } }
\vdef{default-11:BsSgEvt1:2:pt}   {\ensuremath{{19.643 } } }
\vdef{default-11:BsSgEvt1:2:phi}   {\ensuremath{{2.475 } } }
\vdef{default-11:BsSgEvt1:2:eta}   {\ensuremath{{-2.119 } } }
\vdef{default-11:BsSgEvt1:2:channel}   {endcap }
\vdef{default-11:BsSgEvt1:2:cowboy}   {\ensuremath{{0 } } }
\vdef{default-11:BsSgEvt1:2:m1pt}   {\ensuremath{{14.857 } } }
\vdef{default-11:BsSgEvt1:2:m2pt}   {\ensuremath{{5.391 } } }
\vdef{default-11:BsSgEvt1:2:m1eta}   {\ensuremath{{-2.015 } } }
\vdef{default-11:BsSgEvt1:2:m2eta}   {\ensuremath{{-2.275 } } }
\vdef{default-11:BsSgEvt1:2:m1phi}   {\ensuremath{{2.334 } } }
\vdef{default-11:BsSgEvt1:2:m2phi}   {\ensuremath{{2.876 } } }
\vdef{default-11:BsSgEvt1:2:m1q}   {\ensuremath{{1 } } }
\vdef{default-11:BsSgEvt1:2:m2q}   {\ensuremath{{-1 } } }
\vdef{default-11:BsSgEvt1:2:iso}   {\ensuremath{{1.000 } } }
\vdef{default-11:BsSgEvt1:2:alpha}   {\ensuremath{{0.0047 } } }
\vdef{default-11:BsSgEvt1:2:chi2}   {\ensuremath{{ 0.28 } } }
\vdef{default-11:BsSgEvt1:2:dof}   {\ensuremath{{1 } } }
\vdef{default-11:BsSgEvt1:2:fls3d}   {\ensuremath{{33.20 } } }
\vdef{default-11:BsSgEvt1:2:fl3d}   {\ensuremath{{1.4587 } } }
\vdef{default-11:BsSgEvt1:2:fl3dE}   {\ensuremath{{0.0439 } } }
\vdef{default-11:BsSgEvt1:2:docatrk}   {\ensuremath{{0.1915 } } }
\vdef{default-11:BsSgEvt1:2:closetrk}   {\ensuremath{{0 } } }
\vdef{default-11:BsSgEvt1:2:lip}   {\ensuremath{{0.0006 } } }
\vdef{default-11:BsSgEvt1:2:lipE}   {\ensuremath{{0.0024 } } }
\vdef{default-11:BsSgEvt1:2:tip}   {\ensuremath{{0.0065 } } }
\vdef{default-11:BsSgEvt1:2:tipE}   {\ensuremath{{0.0035 } } }
\vdef{default-11:BsSgEvt1:2:pvlip}   {\ensuremath{{0.0006 } } }
\vdef{default-11:BsSgEvt1:2:pvlips}   {\ensuremath{{0.2435 } } }
\vdef{default-11:BsSgEvt1:2:pvip}   {\ensuremath{{0.0065 } } }
\vdef{default-11:BsSgEvt1:2:pvips}   {\ensuremath{{1.8554 } } }
\vdef{default-11:BsSgEvt1:2:maxdoca}   {\ensuremath{{0.0026 } } }
\vdef{default-11:BsSgEvt1:2:pvw8}   {\ensuremath{{0.9105 } } }
\vdef{default-11:BsSgEvt1:2:bdt}   {\ensuremath{{0.4679 } } }
\vdef{default-11:BsSgEvt1:2:m1pix}   {\ensuremath{{3 } } }
\vdef{default-11:BsSgEvt1:2:m2pix}   {\ensuremath{{4 } } }
\vdef{default-11:BsSgEvt1:2:m1bpix}   {\ensuremath{{2 } } }
\vdef{default-11:BsSgEvt1:2:m2bpix}   {\ensuremath{{2 } } }
\vdef{default-11:BsSgEvt1:2:m1bpixl1}   {\ensuremath{{1 } } }
\vdef{default-11:BsSgEvt1:2:m2bpixl1}   {\ensuremath{{1 } } }
\vdef{default-11:BsSgEvt1:3:run}   {\ensuremath{{177096 } } }
\vdef{default-11:BsSgEvt1:3:evt}   {\ensuremath{{17449956 } } }
\vdef{default-11:BsSgEvt1:3:chan}   {\ensuremath{{1 } } }
\vdef{default-11:BsSgEvt1:3:m}   {\ensuremath{{5.341 } } }
\vdef{default-11:BsSgEvt1:3:pt}   {\ensuremath{{17.403 } } }
\vdef{default-11:BsSgEvt1:3:phi}   {\ensuremath{{-0.484 } } }
\vdef{default-11:BsSgEvt1:3:eta}   {\ensuremath{{-2.030 } } }
\vdef{default-11:BsSgEvt1:3:channel}   {endcap }
\vdef{default-11:BsSgEvt1:3:cowboy}   {\ensuremath{{1 } } }
\vdef{default-11:BsSgEvt1:3:m1pt}   {\ensuremath{{11.975 } } }
\vdef{default-11:BsSgEvt1:3:m2pt}   {\ensuremath{{5.474 } } }
\vdef{default-11:BsSgEvt1:3:m1eta}   {\ensuremath{{-2.188 } } }
\vdef{default-11:BsSgEvt1:3:m2eta}   {\ensuremath{{-1.558 } } }
\vdef{default-11:BsSgEvt1:3:m1phi}   {\ensuremath{{-0.533 } } }
\vdef{default-11:BsSgEvt1:3:m2phi}   {\ensuremath{{-0.378 } } }
\vdef{default-11:BsSgEvt1:3:m1q}   {\ensuremath{{-1 } } }
\vdef{default-11:BsSgEvt1:3:m2q}   {\ensuremath{{1 } } }
\vdef{default-11:BsSgEvt1:3:iso}   {\ensuremath{{0.827 } } }
\vdef{default-11:BsSgEvt1:3:alpha}   {\ensuremath{{0.0109 } } }
\vdef{default-11:BsSgEvt1:3:chi2}   {\ensuremath{{ 0.00 } } }
\vdef{default-11:BsSgEvt1:3:dof}   {\ensuremath{{1 } } }
\vdef{default-11:BsSgEvt1:3:fls3d}   {\ensuremath{{28.62 } } }
\vdef{default-11:BsSgEvt1:3:fl3d}   {\ensuremath{{0.5828 } } }
\vdef{default-11:BsSgEvt1:3:fl3dE}   {\ensuremath{{0.0204 } } }
\vdef{default-11:BsSgEvt1:3:docatrk}   {\ensuremath{{0.0319 } } }
\vdef{default-11:BsSgEvt1:3:closetrk}   {\ensuremath{{0 } } }
\vdef{default-11:BsSgEvt1:3:lip}   {\ensuremath{{0.0007 } } }
\vdef{default-11:BsSgEvt1:3:lipE}   {\ensuremath{{0.0027 } } }
\vdef{default-11:BsSgEvt1:3:tip}   {\ensuremath{{0.0058 } } }
\vdef{default-11:BsSgEvt1:3:tipE}   {\ensuremath{{0.0030 } } }
\vdef{default-11:BsSgEvt1:3:pvlip}   {\ensuremath{{0.0007 } } }
\vdef{default-11:BsSgEvt1:3:pvlips}   {\ensuremath{{0.2533 } } }
\vdef{default-11:BsSgEvt1:3:pvip}   {\ensuremath{{0.0058 } } }
\vdef{default-11:BsSgEvt1:3:pvips}   {\ensuremath{{1.9661 } } }
\vdef{default-11:BsSgEvt1:3:maxdoca}   {\ensuremath{{0.0000 } } }
\vdef{default-11:BsSgEvt1:3:pvw8}   {\ensuremath{{0.8932 } } }
\vdef{default-11:BsSgEvt1:3:bdt}   {\ensuremath{{-0.1501 } } }
\vdef{default-11:BsSgEvt1:3:m1pix}   {\ensuremath{{3 } } }
\vdef{default-11:BsSgEvt1:3:m2pix}   {\ensuremath{{3 } } }
\vdef{default-11:BsSgEvt1:3:m1bpix}   {\ensuremath{{1 } } }
\vdef{default-11:BsSgEvt1:3:m2bpix}   {\ensuremath{{3 } } }
\vdef{default-11:BsSgEvt1:3:m1bpixl1}   {\ensuremath{{1 } } }
\vdef{default-11:BsSgEvt1:3:m2bpixl1}   {\ensuremath{{1 } } }
\vdef{default-11:SgEvt1:5:run}   {\ensuremath{{176929 } } }
\vdef{default-11:SgEvt1:5:evt}   {\ensuremath{{212806156 } } }
\vdef{default-11:SgEvt1:5:chan}   {\ensuremath{{1 } } }
\vdef{default-11:SgEvt1:5:m}   {\ensuremath{{5.115 } } }
\vdef{default-11:SgEvt1:5:pt}   {\ensuremath{{12.601 } } }
\vdef{default-11:SgEvt1:5:phi}   {\ensuremath{{2.312 } } }
\vdef{default-11:SgEvt1:5:eta}   {\ensuremath{{1.809 } } }
\vdef{default-11:SgEvt1:5:channel}   {endcap }
\vdef{default-11:SgEvt1:5:cowboy}   {\ensuremath{{0 } } }
\vdef{default-11:SgEvt1:5:m1pt}   {\ensuremath{{6.762 } } }
\vdef{default-11:SgEvt1:5:m2pt}   {\ensuremath{{5.865 } } }
\vdef{default-11:SgEvt1:5:m1eta}   {\ensuremath{{2.089 } } }
\vdef{default-11:SgEvt1:5:m2eta}   {\ensuremath{{1.334 } } }
\vdef{default-11:SgEvt1:5:m1phi}   {\ensuremath{{2.203 } } }
\vdef{default-11:SgEvt1:5:m2phi}   {\ensuremath{{2.437 } } }
\vdef{default-11:SgEvt1:5:m1q}   {\ensuremath{{1 } } }
\vdef{default-11:SgEvt1:5:m2q}   {\ensuremath{{-1 } } }
\vdef{default-11:SgEvt1:5:iso}   {\ensuremath{{0.894 } } }
\vdef{default-11:SgEvt1:5:alpha}   {\ensuremath{{0.0275 } } }
\vdef{default-11:SgEvt1:5:chi2}   {\ensuremath{{ 1.67 } } }
\vdef{default-11:SgEvt1:5:dof}   {\ensuremath{{1 } } }
\vdef{default-11:SgEvt1:5:fls3d}   {\ensuremath{{17.90 } } }
\vdef{default-11:SgEvt1:5:fl3d}   {\ensuremath{{0.3241 } } }
\vdef{default-11:SgEvt1:5:fl3dE}   {\ensuremath{{0.0181 } } }
\vdef{default-11:SgEvt1:5:docatrk}   {\ensuremath{{0.0171 } } }
\vdef{default-11:SgEvt1:5:closetrk}   {\ensuremath{{1 } } }
\vdef{default-11:SgEvt1:5:lip}   {\ensuremath{{0.0028 } } }
\vdef{default-11:SgEvt1:5:lipE}   {\ensuremath{{0.0023 } } }
\vdef{default-11:SgEvt1:5:tip}   {\ensuremath{{0.0021 } } }
\vdef{default-11:SgEvt1:5:tipE}   {\ensuremath{{0.0033 } } }
\vdef{default-11:SgEvt1:5:pvlip}   {\ensuremath{{0.0028 } } }
\vdef{default-11:SgEvt1:5:pvlips}   {\ensuremath{{1.2024 } } }
\vdef{default-11:SgEvt1:5:pvip}   {\ensuremath{{0.0034 } } }
\vdef{default-11:SgEvt1:5:pvips}   {\ensuremath{{1.2776 } } }
\vdef{default-11:SgEvt1:5:maxdoca}   {\ensuremath{{0.0066 } } }
\vdef{default-11:SgEvt1:5:pvw8}   {\ensuremath{{0.9202 } } }
\vdef{default-11:SgEvt1:5:bdt}   {\ensuremath{{-0.3881 } } }
\vdef{default-11:SgEvt1:5:m1pix}   {\ensuremath{{3 } } }
\vdef{default-11:SgEvt1:5:m2pix}   {\ensuremath{{2 } } }
\vdef{default-11:SgEvt1:5:m1bpix}   {\ensuremath{{1 } } }
\vdef{default-11:SgEvt1:5:m2bpix}   {\ensuremath{{2 } } }
\vdef{default-11:SgEvt1:5:m1bpixl1}   {\ensuremath{{1 } } }
\vdef{default-11:SgEvt1:5:m2bpixl1}   {\ensuremath{{1 } } }
\vdef{default-11:SgEvt1:6:run}   {\ensuremath{{176701 } } }
\vdef{default-11:SgEvt1:6:evt}   {\ensuremath{{115159408 } } }
\vdef{default-11:SgEvt1:6:chan}   {\ensuremath{{1 } } }
\vdef{default-11:SgEvt1:6:m}   {\ensuremath{{5.789 } } }
\vdef{default-11:SgEvt1:6:pt}   {\ensuremath{{12.939 } } }
\vdef{default-11:SgEvt1:6:phi}   {\ensuremath{{0.906 } } }
\vdef{default-11:SgEvt1:6:eta}   {\ensuremath{{-1.955 } } }
\vdef{default-11:SgEvt1:6:channel}   {endcap }
\vdef{default-11:SgEvt1:6:cowboy}   {\ensuremath{{0 } } }
\vdef{default-11:SgEvt1:6:m1pt}   {\ensuremath{{7.006 } } }
\vdef{default-11:SgEvt1:6:m2pt}   {\ensuremath{{6.117 } } }
\vdef{default-11:SgEvt1:6:m1eta}   {\ensuremath{{-1.489 } } }
\vdef{default-11:SgEvt1:6:m2eta}   {\ensuremath{{-2.300 } } }
\vdef{default-11:SgEvt1:6:m1phi}   {\ensuremath{{0.766 } } }
\vdef{default-11:SgEvt1:6:m2phi}   {\ensuremath{{1.067 } } }
\vdef{default-11:SgEvt1:6:m1q}   {\ensuremath{{1 } } }
\vdef{default-11:SgEvt1:6:m2q}   {\ensuremath{{-1 } } }
\vdef{default-11:SgEvt1:6:iso}   {\ensuremath{{0.914 } } }
\vdef{default-11:SgEvt1:6:alpha}   {\ensuremath{{0.0213 } } }
\vdef{default-11:SgEvt1:6:chi2}   {\ensuremath{{ 0.44 } } }
\vdef{default-11:SgEvt1:6:dof}   {\ensuremath{{1 } } }
\vdef{default-11:SgEvt1:6:fls3d}   {\ensuremath{{23.78 } } }
\vdef{default-11:SgEvt1:6:fl3d}   {\ensuremath{{0.4307 } } }
\vdef{default-11:SgEvt1:6:fl3dE}   {\ensuremath{{0.0181 } } }
\vdef{default-11:SgEvt1:6:docatrk}   {\ensuremath{{0.0236 } } }
\vdef{default-11:SgEvt1:6:closetrk}   {\ensuremath{{1 } } }
\vdef{default-11:SgEvt1:6:lip}   {\ensuremath{{-0.0024 } } }
\vdef{default-11:SgEvt1:6:lipE}   {\ensuremath{{0.0031 } } }
\vdef{default-11:SgEvt1:6:tip}   {\ensuremath{{0.0029 } } }
\vdef{default-11:SgEvt1:6:tipE}   {\ensuremath{{0.0034 } } }
\vdef{default-11:SgEvt1:6:pvlip}   {\ensuremath{{-0.0024 } } }
\vdef{default-11:SgEvt1:6:pvlips}   {\ensuremath{{-0.7803 } } }
\vdef{default-11:SgEvt1:6:pvip}   {\ensuremath{{0.0038 } } }
\vdef{default-11:SgEvt1:6:pvips}   {\ensuremath{{1.1683 } } }
\vdef{default-11:SgEvt1:6:maxdoca}   {\ensuremath{{0.0032 } } }
\vdef{default-11:SgEvt1:6:pvw8}   {\ensuremath{{0.9018 } } }
\vdef{default-11:SgEvt1:6:bdt}   {\ensuremath{{-0.3172 } } }
\vdef{default-11:SgEvt1:6:m1pix}   {\ensuremath{{3 } } }
\vdef{default-11:SgEvt1:6:m2pix}   {\ensuremath{{3 } } }
\vdef{default-11:SgEvt1:6:m1bpix}   {\ensuremath{{2 } } }
\vdef{default-11:SgEvt1:6:m2bpix}   {\ensuremath{{1 } } }
\vdef{default-11:SgEvt1:6:m1bpixl1}   {\ensuremath{{1 } } }
\vdef{default-11:SgEvt1:6:m2bpixl1}   {\ensuremath{{1 } } }
\vdef{default-11:N-EFF-TOT-BPLUS0:val}   {\ensuremath{{0.00110 } } }
\vdef{default-11:N-EFF-TOT-BPLUS0:err}   {\ensuremath{{0.000004 } } }
\vdef{default-11:N-EFF-TOT-BPLUS0:tot}   {\ensuremath{{0.00009 } } }
\vdef{default-11:N-EFF-TOT-BPLUS0:all}   {\ensuremath{{(1.10 \pm 0.09)\times 10^{-3}} } }
\vdef{default-11:N-ACC-BPLUS0:val}   {\ensuremath{{0.162 } } }
\vdef{default-11:N-ACC-BPLUS0:err}   {\ensuremath{{0.000 } } }
\vdef{default-11:N-ACC-BPLUS0:tot}   {\ensuremath{{0.006 } } }
\vdef{default-11:N-ACC-BPLUS0:all}   {\ensuremath{{(16.19 \pm 0.57)\times 10^{-2}} } }
\vdef{default-11:N-EFF-MU-PID-BPLUS0:val}   {\ensuremath{{0.786 } } }
\vdef{default-11:N-EFF-MU-PID-BPLUS0:err}   {\ensuremath{{0.000 } } }
\vdef{default-11:N-EFF-MU-PID-BPLUS0:tot}   {\ensuremath{{0.031 } } }
\vdef{default-11:N-EFF-MU-PID-BPLUS0:all}   {\ensuremath{{(78.61 \pm 3.14)\times 10^{-2}} } }
\vdef{default-11:N-EFF-MU-PIDMC-BPLUS0:val}   {\ensuremath{{0.773 } } }
\vdef{default-11:N-EFF-MU-PIDMC-BPLUS0:err}   {\ensuremath{{0.000 } } }
\vdef{default-11:N-EFF-MU-PIDMC-BPLUS0:tot}   {\ensuremath{{0.031 } } }
\vdef{default-11:N-EFF-MU-PIDMC-BPLUS0:all}   {\ensuremath{{(77.35 \pm 3.09)\times 10^{-2}} } }
\vdef{default-11:N-EFF-MU-MC-BPLUS0:val}   {\ensuremath{{0.769 } } }
\vdef{default-11:N-EFF-MU-MC-BPLUS0:err}   {\ensuremath{{0.001 } } }
\vdef{default-11:N-EFF-MU-MC-BPLUS0:tot}   {\ensuremath{{0.031 } } }
\vdef{default-11:N-EFF-MU-MC-BPLUS0:all}   {\ensuremath{{(76.92 \pm 3.08)\times 10^{-2}} } }
\vdef{default-11:N-EFF-TRIG-PID-BPLUS0:val}   {\ensuremath{{0.786 } } }
\vdef{default-11:N-EFF-TRIG-PID-BPLUS0:err}   {\ensuremath{{0.000 } } }
\vdef{default-11:N-EFF-TRIG-PID-BPLUS0:tot}   {\ensuremath{{0.024 } } }
\vdef{default-11:N-EFF-TRIG-PID-BPLUS0:all}   {\ensuremath{{(78.61 \pm 2.36)\times 10^{-2}} } }
\vdef{default-11:N-EFF-TRIG-PIDMC-BPLUS0:val}   {\ensuremath{{0.829 } } }
\vdef{default-11:N-EFF-TRIG-PIDMC-BPLUS0:err}   {\ensuremath{{0.000 } } }
\vdef{default-11:N-EFF-TRIG-PIDMC-BPLUS0:tot}   {\ensuremath{{0.025 } } }
\vdef{default-11:N-EFF-TRIG-PIDMC-BPLUS0:all}   {\ensuremath{{(82.91 \pm 2.49)\times 10^{-2}} } }
\vdef{default-11:N-EFF-TRIG-MC-BPLUS0:val}   {\ensuremath{{0.767 } } }
\vdef{default-11:N-EFF-TRIG-MC-BPLUS0:err}   {\ensuremath{{0.001 } } }
\vdef{default-11:N-EFF-TRIG-MC-BPLUS0:tot}   {\ensuremath{{0.023 } } }
\vdef{default-11:N-EFF-TRIG-MC-BPLUS0:all}   {\ensuremath{{(76.66 \pm 2.30)\times 10^{-2}} } }
\vdef{default-11:N-EFF-CAND-BPLUS0:val}   {\ensuremath{{0.980 } } }
\vdef{default-11:N-EFF-CAND-BPLUS0:err}   {\ensuremath{{0.003 } } }
\vdef{default-11:N-EFF-CAND-BPLUS0:tot}   {\ensuremath{{0.010 } } }
\vdef{default-11:N-EFF-CAND-BPLUS0:all}   {\ensuremath{{(98.00 \pm 1.01)\times 10^{-2}} } }
\vdef{default-11:N-EFF-ANA-BPLUS0:val}   {\ensuremath{{0.0118 } } }
\vdef{default-11:N-EFF-ANA-BPLUS0:err}   {\ensuremath{{0.0004 } } }
\vdef{default-11:N-EFF-ANA-BPLUS0:tot}   {\ensuremath{{0.0008 } } }
\vdef{default-11:N-EFF-ANA-BPLUS0:all}   {\ensuremath{{(1.18 \pm 0.08)\times 10^{-2}} } }
\vdef{default-11:N-OBS-BPLUS0:val}   {\ensuremath{{82712 } } }
\vdef{default-11:N-OBS-BPLUS0:err}   {\ensuremath{{298 } } }
\vdef{default-11:N-OBS-BPLUS0:tot}   {\ensuremath{{4146 } } }
\vdef{default-11:N-OBS-BPLUS0:all}   {\ensuremath{{82712 } } }
\vdef{default-11:N-OBS-CBPLUS0:val}   {\ensuremath{{81754 } } }
\vdef{default-11:N-OBS-CBPLUS0:err}   {\ensuremath{{894 } } }
\vdef{default-11:N-EFF-TOT-BPLUS1:val}   {\ensuremath{{0.00032 } } }
\vdef{default-11:N-EFF-TOT-BPLUS1:err}   {\ensuremath{{0.000002 } } }
\vdef{default-11:N-EFF-TOT-BPLUS1:tot}   {\ensuremath{{0.00004 } } }
\vdef{default-11:N-EFF-TOT-BPLUS1:all}   {\ensuremath{{(0.32 \pm 0.04)\times 10^{-3}} } }
\vdef{default-11:N-ACC-BPLUS1:val}   {\ensuremath{{0.111 } } }
\vdef{default-11:N-ACC-BPLUS1:err}   {\ensuremath{{0.000 } } }
\vdef{default-11:N-ACC-BPLUS1:tot}   {\ensuremath{{0.006 } } }
\vdef{default-11:N-ACC-BPLUS1:all}   {\ensuremath{{(11.05 \pm 0.55)\times 10^{-2}} } }
\vdef{default-11:N-EFF-MU-PID-BPLUS1:val}   {\ensuremath{{0.783 } } }
\vdef{default-11:N-EFF-MU-PID-BPLUS1:err}   {\ensuremath{{0.000 } } }
\vdef{default-11:N-EFF-MU-PID-BPLUS1:tot}   {\ensuremath{{0.063 } } }
\vdef{default-11:N-EFF-MU-PID-BPLUS1:all}   {\ensuremath{{(78.25 \pm 6.26)\times 10^{-2}} } }
\vdef{default-11:N-EFF-MU-PIDMC-BPLUS1:val}   {\ensuremath{{0.834 } } }
\vdef{default-11:N-EFF-MU-PIDMC-BPLUS1:err}   {\ensuremath{{0.000 } } }
\vdef{default-11:N-EFF-MU-PIDMC-BPLUS1:tot}   {\ensuremath{{0.067 } } }
\vdef{default-11:N-EFF-MU-PIDMC-BPLUS1:all}   {\ensuremath{{(83.36 \pm 6.67)\times 10^{-2}} } }
\vdef{default-11:N-EFF-MU-MC-BPLUS1:val}   {\ensuremath{{0.782 } } }
\vdef{default-11:N-EFF-MU-MC-BPLUS1:err}   {\ensuremath{{0.002 } } }
\vdef{default-11:N-EFF-MU-MC-BPLUS1:tot}   {\ensuremath{{0.063 } } }
\vdef{default-11:N-EFF-MU-MC-BPLUS1:all}   {\ensuremath{{(78.25 \pm 6.26)\times 10^{-2}} } }
\vdef{default-11:N-EFF-TRIG-PID-BPLUS1:val}   {\ensuremath{{0.751 } } }
\vdef{default-11:N-EFF-TRIG-PID-BPLUS1:err}   {\ensuremath{{0.001 } } }
\vdef{default-11:N-EFF-TRIG-PID-BPLUS1:tot}   {\ensuremath{{0.045 } } }
\vdef{default-11:N-EFF-TRIG-PID-BPLUS1:all}   {\ensuremath{{(75.13 \pm 4.51)\times 10^{-2}} } }
\vdef{default-11:N-EFF-TRIG-PIDMC-BPLUS1:val}   {\ensuremath{{0.740 } } }
\vdef{default-11:N-EFF-TRIG-PIDMC-BPLUS1:err}   {\ensuremath{{0.001 } } }
\vdef{default-11:N-EFF-TRIG-PIDMC-BPLUS1:tot}   {\ensuremath{{0.044 } } }
\vdef{default-11:N-EFF-TRIG-PIDMC-BPLUS1:all}   {\ensuremath{{(74.01 \pm 4.44)\times 10^{-2}} } }
\vdef{default-11:N-EFF-TRIG-MC-BPLUS1:val}   {\ensuremath{{0.599 } } }
\vdef{default-11:N-EFF-TRIG-MC-BPLUS1:err}   {\ensuremath{{0.002 } } }
\vdef{default-11:N-EFF-TRIG-MC-BPLUS1:tot}   {\ensuremath{{0.036 } } }
\vdef{default-11:N-EFF-TRIG-MC-BPLUS1:all}   {\ensuremath{{(59.86 \pm 3.60)\times 10^{-2}} } }
\vdef{default-11:N-EFF-CAND-BPLUS1:val}   {\ensuremath{{0.980 } } }
\vdef{default-11:N-EFF-CAND-BPLUS1:err}   {\ensuremath{{0.005 } } }
\vdef{default-11:N-EFF-CAND-BPLUS1:tot}   {\ensuremath{{0.011 } } }
\vdef{default-11:N-EFF-CAND-BPLUS1:all}   {\ensuremath{{(98.00 \pm 1.08)\times 10^{-2}} } }
\vdef{default-11:N-EFF-ANA-BPLUS1:val}   {\ensuremath{{0.0063 } } }
\vdef{default-11:N-EFF-ANA-BPLUS1:err}   {\ensuremath{{0.0005 } } }
\vdef{default-11:N-EFF-ANA-BPLUS1:tot}   {\ensuremath{{0.0006 } } }
\vdef{default-11:N-EFF-ANA-BPLUS1:all}   {\ensuremath{{(0.63 \pm 0.06)\times 10^{-2}} } }
\vdef{default-11:N-OBS-BPLUS1:val}   {\ensuremath{{23809 } } }
\vdef{default-11:N-OBS-BPLUS1:err}   {\ensuremath{{173 } } }
\vdef{default-11:N-OBS-BPLUS1:tot}   {\ensuremath{{1203 } } }
\vdef{default-11:N-OBS-BPLUS1:all}   {\ensuremath{{23809 } } }
\vdef{default-11:N-OBS-CBPLUS1:val}   {\ensuremath{{22751 } } }
\vdef{default-11:N-OBS-CBPLUS1:err}   {\ensuremath{{158 } } }
\vdef{default-11:N-EXP2-SIG-BSMM0:val}   {\ensuremath{{ 2.70 } } }
\vdef{default-11:N-EXP2-SIG-BSMM0:err}   {\ensuremath{{ 0.41 } } }
\vdef{default-11:N-EXP2-SIG-BDMM0:val}   {\ensuremath{{0.236 } } }
\vdef{default-11:N-EXP2-SIG-BDMM0:err}   {\ensuremath{{0.024 } } }
\vdef{default-11:N-OBS-BKG0:val}   {\ensuremath{{6 } } }
\vdef{default-11:N-EXP-BSMM0:val}   {\ensuremath{{ 0.59 } } }
\vdef{default-11:N-EXP-BSMM0:err}   {\ensuremath{{ 0.50 } } }
\vdef{default-11:N-EXP-BDMM0:val}   {\ensuremath{{ 0.40 } } }
\vdef{default-11:N-EXP-BDMM0:err}   {\ensuremath{{ 0.34 } } }
\vdef{default-11:N-LOW-BD0:val}   {\ensuremath{{5.200 } } }
\vdef{default-11:N-HIGH-BD0:val}   {\ensuremath{{5.300 } } }
\vdef{default-11:N-LOW-BS0:val}   {\ensuremath{{5.300 } } }
\vdef{default-11:N-HIGH-BS0:val}   {\ensuremath{{5.450 } } }
\vdef{default-11:N-PSS0:val}   {\ensuremath{{0.863 } } }
\vdef{default-11:N-PSS0:err}   {\ensuremath{{0.009 } } }
\vdef{default-11:N-PSS0:tot}   {\ensuremath{{0.044 } } }
\vdef{default-11:N-PSD0:val}   {\ensuremath{{0.292 } } }
\vdef{default-11:N-PSD0:err}   {\ensuremath{{0.017 } } }
\vdef{default-11:N-PSD0:tot}   {\ensuremath{{0.022 } } }
\vdef{default-11:N-PDS0:val}   {\ensuremath{{0.071 } } }
\vdef{default-11:N-PDS0:err}   {\ensuremath{{0.007 } } }
\vdef{default-11:N-PDS0:tot}   {\ensuremath{{0.007 } } }
\vdef{default-11:N-PDD0:val}   {\ensuremath{{0.639 } } }
\vdef{default-11:N-PDD0:err}   {\ensuremath{{0.018 } } }
\vdef{default-11:N-PDD0:tot}   {\ensuremath{{0.037 } } }
\vdef{default-11:N-EFF-TOT-BSMM0:val}   {\ensuremath{{0.0029 } } }
\vdef{default-11:N-EFF-TOT-BSMM0:err}   {\ensuremath{{0.0001 } } }
\vdef{default-11:N-EFF-TOT-BSMM0:tot}   {\ensuremath{{0.0002 } } }
\vdef{default-11:N-EFF-TOT-BSMM0:all}   {\ensuremath{{(0.29 \pm 0.02)\times 10^{-2}} } }
\vdef{default-11:N-ACC-BSMM0:val}   {\ensuremath{{0.248 } } }
\vdef{default-11:N-ACC-BSMM0:err}   {\ensuremath{{0.001 } } }
\vdef{default-11:N-ACC-BSMM0:tot}   {\ensuremath{{0.009 } } }
\vdef{default-11:N-ACC-BSMM0:all}   {\ensuremath{{(24.76 \pm 0.87)\times 10^{-2}} } }
\vdef{default-11:N-EFF-MU-PID-BSMM0:val}   {\ensuremath{{0.793 } } }
\vdef{default-11:N-EFF-MU-PID-BSMM0:err}   {\ensuremath{{0.002 } } }
\vdef{default-11:N-EFF-MU-PID-BSMM0:tot}   {\ensuremath{{0.032 } } }
\vdef{default-11:N-EFF-MU-PID-BSMM0:all}   {\ensuremath{{(79.32 \pm 3.18)\times 10^{-2}} } }
\vdef{default-11:N-EFF-MU-PIDMC-BSMM0:val}   {\ensuremath{{0.789 } } }
\vdef{default-11:N-EFF-MU-PIDMC-BSMM0:err}   {\ensuremath{{0.002 } } }
\vdef{default-11:N-EFF-MU-PIDMC-BSMM0:tot}   {\ensuremath{{0.032 } } }
\vdef{default-11:N-EFF-MU-PIDMC-BSMM0:all}   {\ensuremath{{(78.94 \pm 3.16)\times 10^{-2}} } }
\vdef{default-11:N-EFF-MU-MC-BSMM0:val}   {\ensuremath{{0.712 } } }
\vdef{default-11:N-EFF-MU-MC-BSMM0:err}   {\ensuremath{{0.009 } } }
\vdef{default-11:N-EFF-MU-MC-BSMM0:tot}   {\ensuremath{{0.030 } } }
\vdef{default-11:N-EFF-MU-MC-BSMM0:all}   {\ensuremath{{(71.17 \pm 2.98)\times 10^{-2}} } }
\vdef{default-11:N-EFF-TRIG-PID-BSMM0:val}   {\ensuremath{{0.799 } } }
\vdef{default-11:N-EFF-TRIG-PID-BSMM0:err}   {\ensuremath{{0.002 } } }
\vdef{default-11:N-EFF-TRIG-PID-BSMM0:tot}   {\ensuremath{{0.024 } } }
\vdef{default-11:N-EFF-TRIG-PID-BSMM0:all}   {\ensuremath{{(79.93 \pm 2.41)\times 10^{-2}} } }
\vdef{default-11:N-EFF-TRIG-PIDMC-BSMM0:val}   {\ensuremath{{0.844 } } }
\vdef{default-11:N-EFF-TRIG-PIDMC-BSMM0:err}   {\ensuremath{{0.002 } } }
\vdef{default-11:N-EFF-TRIG-PIDMC-BSMM0:tot}   {\ensuremath{{0.025 } } }
\vdef{default-11:N-EFF-TRIG-PIDMC-BSMM0:all}   {\ensuremath{{(84.44 \pm 2.54)\times 10^{-2}} } }
\vdef{default-11:N-EFF-TRIG-MC-BSMM0:val}   {\ensuremath{{0.844 } } }
\vdef{default-11:N-EFF-TRIG-MC-BSMM0:err}   {\ensuremath{{0.008 } } }
\vdef{default-11:N-EFF-TRIG-MC-BSMM0:tot}   {\ensuremath{{0.027 } } }
\vdef{default-11:N-EFF-TRIG-MC-BSMM0:all}   {\ensuremath{{(84.39 \pm 2.67)\times 10^{-2}} } }
\vdef{default-11:N-EFF-CAND-BSMM0:val}   {\ensuremath{{0.980 } } }
\vdef{default-11:N-EFF-CAND-BSMM0:err}   {\ensuremath{{0.002 } } }
\vdef{default-11:N-EFF-CAND-BSMM0:tot}   {\ensuremath{{0.010 } } }
\vdef{default-11:N-EFF-CAND-BSMM0:all}   {\ensuremath{{(98.00 \pm 0.99)\times 10^{-2}} } }
\vdef{default-11:N-EFF-ANA-BSMM0:val}   {\ensuremath{{0.020 } } }
\vdef{default-11:N-EFF-ANA-BSMM0:err}   {\ensuremath{{0.000 } } }
\vdef{default-11:N-EFF-ANA-BSMM0:tot}   {\ensuremath{{0.001 } } }
\vdef{default-11:N-EFF-ANA-BSMM0:all}   {\ensuremath{{(2.01 \pm 0.07)\times 10^{-2}} } }
\vdef{default-11:N-EFF-TOT-BDMM0:val}   {\ensuremath{{0.0029 } } }
\vdef{default-11:N-EFF-TOT-BDMM0:err}   {\ensuremath{{0.0001 } } }
\vdef{default-11:N-EFF-TOT-BDMM0:tot}   {\ensuremath{{0.0002 } } }
\vdef{default-11:N-EFF-TOT-BDMM0:all}   {\ensuremath{{(0.29 \pm 0.02)\times 10^{-2}} } }
\vdef{default-11:N-ACC-BDMM0:val}   {\ensuremath{{0.247 } } }
\vdef{default-11:N-ACC-BDMM0:err}   {\ensuremath{{0.001 } } }
\vdef{default-11:N-ACC-BDMM0:tot}   {\ensuremath{{0.009 } } }
\vdef{default-11:N-ACC-BDMM0:all}   {\ensuremath{{(24.65 \pm 0.87)\times 10^{-2}} } }
\vdef{default-11:N-EFF-MU-PID-BDMM0:val}   {\ensuremath{{0.789 } } }
\vdef{default-11:N-EFF-MU-PID-BDMM0:err}   {\ensuremath{{0.002 } } }
\vdef{default-11:N-EFF-MU-PID-BDMM0:tot}   {\ensuremath{{0.032 } } }
\vdef{default-11:N-EFF-MU-PID-BDMM0:all}   {\ensuremath{{(78.89 \pm 3.16)\times 10^{-2}} } }
\vdef{default-11:N-EFF-MU-PIDMC-BDMM0:val}   {\ensuremath{{0.783 } } }
\vdef{default-11:N-EFF-MU-PIDMC-BDMM0:err}   {\ensuremath{{0.003 } } }
\vdef{default-11:N-EFF-MU-PIDMC-BDMM0:tot}   {\ensuremath{{0.031 } } }
\vdef{default-11:N-EFF-MU-PIDMC-BDMM0:all}   {\ensuremath{{(78.33 \pm 3.15)\times 10^{-2}} } }
\vdef{default-11:N-EFF-MU-MC-BDMM0:val}   {\ensuremath{{0.692 } } }
\vdef{default-11:N-EFF-MU-MC-BDMM0:err}   {\ensuremath{{0.013 } } }
\vdef{default-11:N-EFF-MU-MC-BDMM0:tot}   {\ensuremath{{0.031 } } }
\vdef{default-11:N-EFF-MU-MC-BDMM0:all}   {\ensuremath{{(69.20 \pm 3.07)\times 10^{-2}} } }
\vdef{default-11:N-EFF-TRIG-PID-BDMM0:val}   {\ensuremath{{0.793 } } }
\vdef{default-11:N-EFF-TRIG-PID-BDMM0:err}   {\ensuremath{{0.003 } } }
\vdef{default-11:N-EFF-TRIG-PID-BDMM0:tot}   {\ensuremath{{0.024 } } }
\vdef{default-11:N-EFF-TRIG-PID-BDMM0:all}   {\ensuremath{{(79.26 \pm 2.40)\times 10^{-2}} } }
\vdef{default-11:N-EFF-TRIG-PIDMC-BDMM0:val}   {\ensuremath{{0.838 } } }
\vdef{default-11:N-EFF-TRIG-PIDMC-BDMM0:err}   {\ensuremath{{0.003 } } }
\vdef{default-11:N-EFF-TRIG-PIDMC-BDMM0:tot}   {\ensuremath{{0.025 } } }
\vdef{default-11:N-EFF-TRIG-PIDMC-BDMM0:all}   {\ensuremath{{(83.85 \pm 2.54)\times 10^{-2}} } }
\vdef{default-11:N-EFF-TRIG-MC-BDMM0:val}   {\ensuremath{{0.837 } } }
\vdef{default-11:N-EFF-TRIG-MC-BDMM0:err}   {\ensuremath{{0.013 } } }
\vdef{default-11:N-EFF-TRIG-MC-BDMM0:tot}   {\ensuremath{{0.028 } } }
\vdef{default-11:N-EFF-TRIG-MC-BDMM0:all}   {\ensuremath{{(83.71 \pm 2.81)\times 10^{-2}} } }
\vdef{default-11:N-EFF-CAND-BDMM0:val}   {\ensuremath{{0.980 } } }
\vdef{default-11:N-EFF-CAND-BDMM0:err}   {\ensuremath{{0.002 } } }
\vdef{default-11:N-EFF-CAND-BDMM0:tot}   {\ensuremath{{0.010 } } }
\vdef{default-11:N-EFF-CAND-BDMM0:all}   {\ensuremath{{(98.00 \pm 1.00)\times 10^{-2}} } }
\vdef{default-11:N-EFF-ANA-BDMM0:val}   {\ensuremath{{0.021 } } }
\vdef{default-11:N-EFF-ANA-BDMM0:err}   {\ensuremath{{0.001 } } }
\vdef{default-11:N-EFF-ANA-BDMM0:tot}   {\ensuremath{{0.001 } } }
\vdef{default-11:N-EFF-ANA-BDMM0:all}   {\ensuremath{{(2.11 \pm 0.09)\times 10^{-2}} } }
\vdef{default-11:N-EXP-OBS-BS0:val}   {\ensuremath{{ 3.48 } } }
\vdef{default-11:N-EXP-OBS-BS0:err}   {\ensuremath{{ 0.65 } } }
\vdef{default-11:N-EXP-OBS-BD0:val}   {\ensuremath{{ 0.96 } } }
\vdef{default-11:N-EXP-OBS-BD0:err}   {\ensuremath{{ 0.34 } } }
\vdef{default-11:N-OBS-BSMM0:val}   {\ensuremath{{2 } } }
\vdef{default-11:N-OBS-BDMM0:val}   {\ensuremath{{2 } } }
\vdef{default-11:N-OFFLO-RARE0:val}   {\ensuremath{{ 3.01 } } }
\vdef{default-11:N-OFFLO-RARE0:err}   {\ensuremath{{ 0.63 } } }
\vdef{default-11:N-OFFHI-RARE0:val}   {\ensuremath{{ 0.02 } } }
\vdef{default-11:N-OFFHI-RARE0:err}   {\ensuremath{{ 0.00 } } }
\vdef{default-11:N-PEAK-BKG-BS0:val}   {\ensuremath{{ 0.18 } } }
\vdef{default-11:N-PEAK-BKG-BS0:err}   {\ensuremath{{ 0.06 } } }
\vdef{default-11:N-PEAK-BKG-BD0:val}   {\ensuremath{{ 0.33 } } }
\vdef{default-11:N-PEAK-BKG-BD0:err}   {\ensuremath{{ 0.07 } } }
\vdef{default-11:N-TAU-BS0:val}   {\ensuremath{{ 0.20 } } }
\vdef{default-11:N-TAU-BS0:err}   {\ensuremath{{ 0.01 } } }
\vdef{default-11:N-TAU-BD0:val}   {\ensuremath{{ 0.13 } } }
\vdef{default-11:N-TAU-BD0:err}   {\ensuremath{{ 0.01 } } }
\vdef{default-11:N-EXP2-SIG-BSMM1:val}   {\ensuremath{{ 1.23 } } }
\vdef{default-11:N-EXP2-SIG-BSMM1:err}   {\ensuremath{{ 0.18 } } }
\vdef{default-11:N-EXP2-SIG-BDMM1:val}   {\ensuremath{{0.104 } } }
\vdef{default-11:N-EXP2-SIG-BDMM1:err}   {\ensuremath{{0.010 } } }
\vdef{default-11:N-OBS-BKG1:val}   {\ensuremath{{7 } } }
\vdef{default-11:N-EXP-BSMM1:val}   {\ensuremath{{ 1.14 } } }
\vdef{default-11:N-EXP-BSMM1:err}   {\ensuremath{{ 0.53 } } }
\vdef{default-11:N-EXP-BDMM1:val}   {\ensuremath{{ 0.76 } } }
\vdef{default-11:N-EXP-BDMM1:err}   {\ensuremath{{ 0.35 } } }
\vdef{default-11:N-LOW-BD1:val}   {\ensuremath{{5.200 } } }
\vdef{default-11:N-HIGH-BD1:val}   {\ensuremath{{5.300 } } }
\vdef{default-11:N-LOW-BS1:val}   {\ensuremath{{5.300 } } }
\vdef{default-11:N-HIGH-BS1:val}   {\ensuremath{{5.450 } } }
\vdef{default-11:N-PSS1:val}   {\ensuremath{{0.714 } } }
\vdef{default-11:N-PSS1:err}   {\ensuremath{{0.015 } } }
\vdef{default-11:N-PSS1:tot}   {\ensuremath{{0.039 } } }
\vdef{default-11:N-PSD1:val}   {\ensuremath{{0.319 } } }
\vdef{default-11:N-PSD1:err}   {\ensuremath{{0.024 } } }
\vdef{default-11:N-PSD1:tot}   {\ensuremath{{0.029 } } }
\vdef{default-11:N-PDS1:val}   {\ensuremath{{0.164 } } }
\vdef{default-11:N-PDS1:err}   {\ensuremath{{0.012 } } }
\vdef{default-11:N-PDS1:tot}   {\ensuremath{{0.015 } } }
\vdef{default-11:N-PDD1:val}   {\ensuremath{{0.531 } } }
\vdef{default-11:N-PDD1:err}   {\ensuremath{{0.025 } } }
\vdef{default-11:N-PDD1:tot}   {\ensuremath{{0.037 } } }
\vdef{default-11:N-EFF-TOT-BSMM1:val}   {\ensuremath{{0.0016 } } }
\vdef{default-11:N-EFF-TOT-BSMM1:err}   {\ensuremath{{0.0001 } } }
\vdef{default-11:N-EFF-TOT-BSMM1:tot}   {\ensuremath{{0.0002 } } }
\vdef{default-11:N-EFF-TOT-BSMM1:all}   {\ensuremath{{(0.16 \pm 0.02)\times 10^{-2}} } }
\vdef{default-11:N-ACC-BSMM1:val}   {\ensuremath{{0.229 } } }
\vdef{default-11:N-ACC-BSMM1:err}   {\ensuremath{{0.001 } } }
\vdef{default-11:N-ACC-BSMM1:tot}   {\ensuremath{{0.011 } } }
\vdef{default-11:N-ACC-BSMM1:all}   {\ensuremath{{(22.94 \pm 1.15)\times 10^{-2}} } }
\vdef{default-11:N-EFF-MU-PID-BSMM1:val}   {\ensuremath{{0.783 } } }
\vdef{default-11:N-EFF-MU-PID-BSMM1:err}   {\ensuremath{{0.002 } } }
\vdef{default-11:N-EFF-MU-PID-BSMM1:tot}   {\ensuremath{{0.063 } } }
\vdef{default-11:N-EFF-MU-PID-BSMM1:all}   {\ensuremath{{(78.25 \pm 6.26)\times 10^{-2}} } }
\vdef{default-11:N-EFF-MU-PIDMC-BSMM1:val}   {\ensuremath{{0.835 } } }
\vdef{default-11:N-EFF-MU-PIDMC-BSMM1:err}   {\ensuremath{{0.002 } } }
\vdef{default-11:N-EFF-MU-PIDMC-BSMM1:tot}   {\ensuremath{{0.067 } } }
\vdef{default-11:N-EFF-MU-PIDMC-BSMM1:all}   {\ensuremath{{(83.51 \pm 6.68)\times 10^{-2}} } }
\vdef{default-11:N-EFF-MU-MC-BSMM1:val}   {\ensuremath{{0.850 } } }
\vdef{default-11:N-EFF-MU-MC-BSMM1:err}   {\ensuremath{{0.009 } } }
\vdef{default-11:N-EFF-MU-MC-BSMM1:tot}   {\ensuremath{{0.069 } } }
\vdef{default-11:N-EFF-MU-MC-BSMM1:all}   {\ensuremath{{(84.97 \pm 6.86)\times 10^{-2}} } }
\vdef{default-11:N-EFF-TRIG-PID-BSMM1:val}   {\ensuremath{{0.768 } } }
\vdef{default-11:N-EFF-TRIG-PID-BSMM1:err}   {\ensuremath{{0.003 } } }
\vdef{default-11:N-EFF-TRIG-PID-BSMM1:tot}   {\ensuremath{{0.046 } } }
\vdef{default-11:N-EFF-TRIG-PID-BSMM1:all}   {\ensuremath{{(76.82 \pm 4.62)\times 10^{-2}} } }
\vdef{default-11:N-EFF-TRIG-PIDMC-BSMM1:val}   {\ensuremath{{0.760 } } }
\vdef{default-11:N-EFF-TRIG-PIDMC-BSMM1:err}   {\ensuremath{{0.004 } } }
\vdef{default-11:N-EFF-TRIG-PIDMC-BSMM1:tot}   {\ensuremath{{0.046 } } }
\vdef{default-11:N-EFF-TRIG-PIDMC-BSMM1:all}   {\ensuremath{{(76.01 \pm 4.57)\times 10^{-2}} } }
\vdef{default-11:N-EFF-TRIG-MC-BSMM1:val}   {\ensuremath{{0.735 } } }
\vdef{default-11:N-EFF-TRIG-MC-BSMM1:err}   {\ensuremath{{0.013 } } }
\vdef{default-11:N-EFF-TRIG-MC-BSMM1:tot}   {\ensuremath{{0.046 } } }
\vdef{default-11:N-EFF-TRIG-MC-BSMM1:all}   {\ensuremath{{(73.50 \pm 4.59)\times 10^{-2}} } }
\vdef{default-11:N-EFF-CAND-BSMM1:val}   {\ensuremath{{0.980 } } }
\vdef{default-11:N-EFF-CAND-BSMM1:err}   {\ensuremath{{0.002 } } }
\vdef{default-11:N-EFF-CAND-BSMM1:tot}   {\ensuremath{{0.010 } } }
\vdef{default-11:N-EFF-CAND-BSMM1:all}   {\ensuremath{{(98.00 \pm 1.00)\times 10^{-2}} } }
\vdef{default-11:N-EFF-ANA-BSMM1:val}   {\ensuremath{{0.012 } } }
\vdef{default-11:N-EFF-ANA-BSMM1:err}   {\ensuremath{{0.000 } } }
\vdef{default-11:N-EFF-ANA-BSMM1:tot}   {\ensuremath{{0.000 } } }
\vdef{default-11:N-EFF-ANA-BSMM1:all}   {\ensuremath{{(1.18 \pm 0.05)\times 10^{-2}} } }
\vdef{default-11:N-EFF-TOT-BDMM1:val}   {\ensuremath{{0.0016 } } }
\vdef{default-11:N-EFF-TOT-BDMM1:err}   {\ensuremath{{0.0001 } } }
\vdef{default-11:N-EFF-TOT-BDMM1:tot}   {\ensuremath{{0.0002 } } }
\vdef{default-11:N-EFF-TOT-BDMM1:all}   {\ensuremath{{(0.16 \pm 0.02)\times 10^{-2}} } }
\vdef{default-11:N-ACC-BDMM1:val}   {\ensuremath{{0.226 } } }
\vdef{default-11:N-ACC-BDMM1:err}   {\ensuremath{{0.001 } } }
\vdef{default-11:N-ACC-BDMM1:tot}   {\ensuremath{{0.011 } } }
\vdef{default-11:N-ACC-BDMM1:all}   {\ensuremath{{(22.58 \pm 1.13)\times 10^{-2}} } }
\vdef{default-11:N-EFF-MU-PID-BDMM1:val}   {\ensuremath{{0.781 } } }
\vdef{default-11:N-EFF-MU-PID-BDMM1:err}   {\ensuremath{{0.002 } } }
\vdef{default-11:N-EFF-MU-PID-BDMM1:tot}   {\ensuremath{{0.063 } } }
\vdef{default-11:N-EFF-MU-PID-BDMM1:all}   {\ensuremath{{(78.11 \pm 6.25)\times 10^{-2}} } }
\vdef{default-11:N-EFF-MU-PIDMC-BDMM1:val}   {\ensuremath{{0.839 } } }
\vdef{default-11:N-EFF-MU-PIDMC-BDMM1:err}   {\ensuremath{{0.002 } } }
\vdef{default-11:N-EFF-MU-PIDMC-BDMM1:tot}   {\ensuremath{{0.067 } } }
\vdef{default-11:N-EFF-MU-PIDMC-BDMM1:all}   {\ensuremath{{(83.86 \pm 6.71)\times 10^{-2}} } }
\vdef{default-11:N-EFF-MU-MC-BDMM1:val}   {\ensuremath{{0.861 } } }
\vdef{default-11:N-EFF-MU-MC-BDMM1:err}   {\ensuremath{{0.013 } } }
\vdef{default-11:N-EFF-MU-MC-BDMM1:tot}   {\ensuremath{{0.070 } } }
\vdef{default-11:N-EFF-MU-MC-BDMM1:all}   {\ensuremath{{(86.14 \pm 7.02)\times 10^{-2}} } }
\vdef{default-11:N-EFF-TRIG-PID-BDMM1:val}   {\ensuremath{{0.770 } } }
\vdef{default-11:N-EFF-TRIG-PID-BDMM1:err}   {\ensuremath{{0.005 } } }
\vdef{default-11:N-EFF-TRIG-PID-BDMM1:tot}   {\ensuremath{{0.046 } } }
\vdef{default-11:N-EFF-TRIG-PID-BDMM1:all}   {\ensuremath{{(77.01 \pm 4.64)\times 10^{-2}} } }
\vdef{default-11:N-EFF-TRIG-PIDMC-BDMM1:val}   {\ensuremath{{0.765 } } }
\vdef{default-11:N-EFF-TRIG-PIDMC-BDMM1:err}   {\ensuremath{{0.005 } } }
\vdef{default-11:N-EFF-TRIG-PIDMC-BDMM1:tot}   {\ensuremath{{0.046 } } }
\vdef{default-11:N-EFF-TRIG-PIDMC-BDMM1:all}   {\ensuremath{{(76.48 \pm 4.62)\times 10^{-2}} } }
\vdef{default-11:N-EFF-TRIG-MC-BDMM1:val}   {\ensuremath{{0.675 } } }
\vdef{default-11:N-EFF-TRIG-MC-BDMM1:err}   {\ensuremath{{0.020 } } }
\vdef{default-11:N-EFF-TRIG-MC-BDMM1:tot}   {\ensuremath{{0.045 } } }
\vdef{default-11:N-EFF-TRIG-MC-BDMM1:all}   {\ensuremath{{(67.48 \pm 4.50)\times 10^{-2}} } }
\vdef{default-11:N-EFF-CAND-BDMM1:val}   {\ensuremath{{0.980 } } }
\vdef{default-11:N-EFF-CAND-BDMM1:err}   {\ensuremath{{0.003 } } }
\vdef{default-11:N-EFF-CAND-BDMM1:tot}   {\ensuremath{{0.010 } } }
\vdef{default-11:N-EFF-CAND-BDMM1:all}   {\ensuremath{{(98.00 \pm 1.03)\times 10^{-2}} } }
\vdef{default-11:N-EFF-ANA-BDMM1:val}   {\ensuremath{{0.013 } } }
\vdef{default-11:N-EFF-ANA-BDMM1:err}   {\ensuremath{{0.000 } } }
\vdef{default-11:N-EFF-ANA-BDMM1:tot}   {\ensuremath{{0.001 } } }
\vdef{default-11:N-EFF-ANA-BDMM1:all}   {\ensuremath{{(1.26 \pm 0.06)\times 10^{-2}} } }
\vdef{default-11:N-EXP-OBS-BS1:val}   {\ensuremath{{ 2.45 } } }
\vdef{default-11:N-EXP-OBS-BS1:err}   {\ensuremath{{ 0.56 } } }
\vdef{default-11:N-EXP-OBS-BD1:val}   {\ensuremath{{ 1.02 } } }
\vdef{default-11:N-EXP-OBS-BD1:err}   {\ensuremath{{ 0.36 } } }
\vdef{default-11:N-OBS-BSMM1:val}   {\ensuremath{{4 } } }
\vdef{default-11:N-OBS-BDMM1:val}   {\ensuremath{{0 } } }
\vdef{default-11:N-OFFLO-RARE1:val}   {\ensuremath{{ 1.26 } } }
\vdef{default-11:N-OFFLO-RARE1:err}   {\ensuremath{{ 0.24 } } }
\vdef{default-11:N-OFFHI-RARE1:val}   {\ensuremath{{ 0.02 } } }
\vdef{default-11:N-OFFHI-RARE1:err}   {\ensuremath{{ 0.00 } } }
\vdef{default-11:N-PEAK-BKG-BS1:val}   {\ensuremath{{ 0.08 } } }
\vdef{default-11:N-PEAK-BKG-BS1:err}   {\ensuremath{{ 0.02 } } }
\vdef{default-11:N-PEAK-BKG-BD1:val}   {\ensuremath{{ 0.15 } } }
\vdef{default-11:N-PEAK-BKG-BD1:err}   {\ensuremath{{ 0.03 } } }
\vdef{default-11:N-TAU-BS1:val}   {\ensuremath{{ 0.20 } } }
\vdef{default-11:N-TAU-BS1:err}   {\ensuremath{{ 0.01 } } }
\vdef{default-11:N-TAU-BD1:val}   {\ensuremath{{ 0.13 } } }
\vdef{default-11:N-TAU-BD1:err}   {\ensuremath{{ 0.01 } } }
\vdef{default-11:N-CSBF-TNP-BS0:val}   {\ensuremath{{0.000025 } } }
\vdef{default-11:N-CSBF-TNP-BS0:err}   {\ensuremath{{0.000000 } } }
\vdef{default-11:N-CSBF-MC-BS0:val}   {\ensuremath{{0.000026 } } }
\vdef{default-11:N-CSBF-MC-BS0:err}   {\ensuremath{{0.000000 } } }
\vdef{default-11:N-CSBF-BS0:val}   {\ensuremath{{0.000026 } } }
\vdef{default-11:N-CSBF-BS0:err}   {\ensuremath{{0.000000 } } }
\vdef{default-11:N-CSBF-TNP-BS1:val}   {\ensuremath{{0.000025 } } }
\vdef{default-11:N-CSBF-TNP-BS1:err}   {\ensuremath{{0.000001 } } }
\vdef{default-11:N-CSBF-MC-BS1:val}   {\ensuremath{{0.000025 } } }
\vdef{default-11:N-CSBF-MC-BS1:err}   {\ensuremath{{0.000001 } } }
\vdef{default-11:N-CSBF-BS1:val}   {\ensuremath{{0.000026 } } }
\vdef{default-11:N-CSBF-BS1:err}   {\ensuremath{{0.000001 } } }
\vdef{default-11:N-EFF-TOT-BS0:val}   {\ensuremath{{0.000740 } } }
\vdef{default-11:N-EFF-TOT-BS0:err}   {\ensuremath{{0.000006 } } }
\vdef{default-11:N-ACC-BS0:val}   {\ensuremath{{0.1145 } } }
\vdef{default-11:N-ACC-BS0:err}   {\ensuremath{{0.0002 } } }
\vdef{default-11:N-EFF-MU-PID-BS0:val}   {\ensuremath{{0.7892 } } }
\vdef{default-11:N-EFF-MU-PID-BS0:err}   {\ensuremath{{0.0005 } } }
\vdef{default-11:N-EFF-MU-PIDMC-BS0:val}   {\ensuremath{{0.7796 } } }
\vdef{default-11:N-EFF-MU-PIDMC-BS0:err}   {\ensuremath{{0.0007 } } }
\vdef{default-11:N-EFF-MU-MC-BS0:val}   {\ensuremath{{0.7752 } } }
\vdef{default-11:N-EFF-MU-MC-BS0:err}   {\ensuremath{{0.0026 } } }
\vdef{default-11:N-EFF-TRIG-PID-BS0:val}   {\ensuremath{{0.7914 } } }
\vdef{default-11:N-EFF-TRIG-PID-BS0:err}   {\ensuremath{{0.0007 } } }
\vdef{default-11:N-EFF-TRIG-PIDMC-BS0:val}   {\ensuremath{{0.8355 } } }
\vdef{default-11:N-EFF-TRIG-PIDMC-BS0:err}   {\ensuremath{{0.0007 } } }
\vdef{default-11:N-EFF-TRIG-MC-BS0:val}   {\ensuremath{{0.7626 } } }
\vdef{default-11:N-EFF-TRIG-MC-BS0:err}   {\ensuremath{{0.0030 } } }
\vdef{default-11:N-EFF-CAND-BS0:val}   {\ensuremath{{0.9800 } } }
\vdef{default-11:N-EFF-CAND-BS0:err}   {\ensuremath{{0.0021 } } }
\vdef{default-11:N-EFF-ANA-BS0:val}   {\ensuremath{{0.0114 } } }
\vdef{default-11:N-EFF-ANA-BS0:err}   {\ensuremath{{0.0009 } } }
\vdef{default-11:N-OBS-BS0:val}   {\ensuremath{{6805 } } }
\vdef{default-11:N-OBS-BS0:err}   {\ensuremath{{89 } } }
\vdef{default-11:N-EFF-TOT-BS1:val}   {\ensuremath{{0.000202 } } }
\vdef{default-11:N-EFF-TOT-BS1:err}   {\ensuremath{{0.000003 } } }
\vdef{default-11:N-ACC-BS1:val}   {\ensuremath{{0.0760 } } }
\vdef{default-11:N-ACC-BS1:err}   {\ensuremath{{0.0002 } } }
\vdef{default-11:N-EFF-MU-PID-BS1:val}   {\ensuremath{{0.7804 } } }
\vdef{default-11:N-EFF-MU-PID-BS1:err}   {\ensuremath{{0.0006 } } }
\vdef{default-11:N-EFF-MU-PIDMC-BS1:val}   {\ensuremath{{0.8332 } } }
\vdef{default-11:N-EFF-MU-PIDMC-BS1:err}   {\ensuremath{{0.0006 } } }
\vdef{default-11:N-EFF-MU-MC-BS1:val}   {\ensuremath{{0.7795 } } }
\vdef{default-11:N-EFF-MU-MC-BS1:err}   {\ensuremath{{0.0044 } } }
\vdef{default-11:N-EFF-TRIG-PID-BS1:val}   {\ensuremath{{0.7538 } } }
\vdef{default-11:N-EFF-TRIG-PID-BS1:err}   {\ensuremath{{0.0023 } } }
\vdef{default-11:N-EFF-TRIG-PIDMC-BS1:val}   {\ensuremath{{0.7444 } } }
\vdef{default-11:N-EFF-TRIG-PIDMC-BS1:err}   {\ensuremath{{0.0024 } } }
\vdef{default-11:N-EFF-TRIG-MC-BS1:val}   {\ensuremath{{0.5972 } } }
\vdef{default-11:N-EFF-TRIG-MC-BS1:err}   {\ensuremath{{0.0059 } } }
\vdef{default-11:N-EFF-CAND-BS1:val}   {\ensuremath{{0.9800 } } }
\vdef{default-11:N-EFF-CAND-BS1:err}   {\ensuremath{{0.0040 } } }
\vdef{default-11:N-EFF-ANA-BS1:val}   {\ensuremath{{0.0061 } } }
\vdef{default-11:N-EFF-ANA-BS1:err}   {\ensuremath{{0.0011 } } }
\vdef{default-11:N-OBS-BS1:val}   {\ensuremath{{1821 } } }
\vdef{default-11:N-OBS-BS1:err}   {\ensuremath{{47 } } }
\vdef{default-11:ggf0-0:val}   {\ensuremath{{0.259 } } }
\vdef{default-11:ggf0-0:err}   {\ensuremath{{0.002 } } }
\vdef{default-11:fex0-0:val}   {\ensuremath{{0.240 } } }
\vdef{default-11:fex0-0:err}   {\ensuremath{{0.001 } } }
\vdef{default-11:gsp0-0:val}   {\ensuremath{{0.249 } } }
\vdef{default-11:gsp0-0:err}   {\ensuremath{{0.001 } } }
\vdef{default-11:ggf0-1:val}   {\ensuremath{{0.231 } } }
\vdef{default-11:ggf0-1:err}   {\ensuremath{{0.002 } } }
\vdef{default-11:fex0-1:val}   {\ensuremath{{0.221 } } }
\vdef{default-11:fex0-1:err}   {\ensuremath{{0.001 } } }
\vdef{default-11:gsp0-1:val}   {\ensuremath{{0.235 } } }
\vdef{default-11:gsp0-1:err}   {\ensuremath{{0.001 } } }
\vdef{default-11:ggf10-0:val}   {\ensuremath{{0.176 } } }
\vdef{default-11:ggf10-0:err}   {\ensuremath{{0.001 } } }
\vdef{default-11:fex10-0:val}   {\ensuremath{{0.160 } } }
\vdef{default-11:fex10-0:err}   {\ensuremath{{0.001 } } }
\vdef{default-11:gsp10-0:val}   {\ensuremath{{0.159 } } }
\vdef{default-11:gsp10-0:err}   {\ensuremath{{0.001 } } }
\vdef{default-11:ggf10-1:val}   {\ensuremath{{0.118 } } }
\vdef{default-11:ggf10-1:err}   {\ensuremath{{0.001 } } }
\vdef{default-11:fex10-1:val}   {\ensuremath{{0.110 } } }
\vdef{default-11:fex10-1:err}   {\ensuremath{{0.001 } } }
\vdef{default-11:gsp10-1:val}   {\ensuremath{{0.109 } } }
\vdef{default-11:gsp10-1:err}   {\ensuremath{{0.001 } } }
\vdef{default-11:ggfRatio0:val}   {\ensuremath{{1.471 } } }
\vdef{default-11:ggfRatio0:err}   {\ensuremath{{0.014 } } }
\vdef{default-11:fexRatio0:val}   {\ensuremath{{1.500 } } }
\vdef{default-11:fexRatio0:err}   {\ensuremath{{0.009 } } }
\vdef{default-11:gspRatio0:val}   {\ensuremath{{1.567 } } }
\vdef{default-11:gspRatio0:err}   {\ensuremath{{0.008 } } }
\vdef{default-11:ggfRatio1:val}   {\ensuremath{{1.959 } } }
\vdef{default-11:ggfRatio1:err}   {\ensuremath{{0.022 } } }
\vdef{default-11:fexRatio1:val}   {\ensuremath{{2.020 } } }
\vdef{default-11:fexRatio1:err}   {\ensuremath{{0.015 } } }
\vdef{default-11:gspRatio1:val}   {\ensuremath{{2.147 } } }
\vdef{default-11:gspRatio1:err}   {\ensuremath{{0.013 } } }
\vdef{bdtdefault-11:bgBd2KK:bsRare0}   {\ensuremath{{0.000014 } } }
\vdef{bdtdefault-11:bgBd2KK:bsRare0E}  {\ensuremath{{0.000011 } } }
\vdef{bdtdefault-11:bgBd2KK:bdRare0}   {\ensuremath{{0.000323 } } }
\vdef{bdtdefault-11:bgBd2KK:bdRare0E}  {\ensuremath{{0.000248 } } }
\vdef{bdtdefault-11:bgBd2KK:loSideband0:val}   {\ensuremath{{0.001 } } }
\vdef{bdtdefault-11:bgBd2KK:loSideband0:err}   {\ensuremath{{0.000 } } }
\vdef{bdtdefault-11:bgBd2KK:bsRare1}   {\ensuremath{{0.000022 } } }
\vdef{bdtdefault-11:bgBd2KK:bsRare1E}  {\ensuremath{{0.000017 } } }
\vdef{bdtdefault-11:bgBd2KK:bdRare1}   {\ensuremath{{0.000122 } } }
\vdef{bdtdefault-11:bgBd2KK:bdRare1E}  {\ensuremath{{0.000093 } } }
\vdef{bdtdefault-11:bgBd2KK:loSideband1:val}   {\ensuremath{{0.000 } } }
\vdef{bdtdefault-11:bgBd2KK:loSideband1:err}   {\ensuremath{{0.000 } } }
\vdef{bdtdefault-11:bgBd2KPi:bsRare0}   {\ensuremath{{0.008300 } } }
\vdef{bdtdefault-11:bgBd2KPi:bsRare0E}  {\ensuremath{{0.001987 } } }
\vdef{bdtdefault-11:bgBd2KPi:bdRare0}   {\ensuremath{{0.083278 } } }
\vdef{bdtdefault-11:bgBd2KPi:bdRare0E}  {\ensuremath{{0.019892 } } }
\vdef{bdtdefault-11:bgBd2KPi:loSideband0:val}   {\ensuremath{{0.033 } } }
\vdef{bdtdefault-11:bgBd2KPi:loSideband0:err}   {\ensuremath{{0.010 } } }
\vdef{bdtdefault-11:bgBd2KPi:bsRare1}   {\ensuremath{{0.007554 } } }
\vdef{bdtdefault-11:bgBd2KPi:bsRare1E}  {\ensuremath{{0.001806 } } }
\vdef{bdtdefault-11:bgBd2KPi:bdRare1}   {\ensuremath{{0.022655 } } }
\vdef{bdtdefault-11:bgBd2KPi:bdRare1E}  {\ensuremath{{0.005403 } } }
\vdef{bdtdefault-11:bgBd2KPi:loSideband1:val}   {\ensuremath{{0.019 } } }
\vdef{bdtdefault-11:bgBd2KPi:loSideband1:err}   {\ensuremath{{0.006 } } }
\vdef{bdtdefault-11:bgBd2PiMuNu:bsRare0}   {\ensuremath{{0.018401 } } }
\vdef{bdtdefault-11:bgBd2PiMuNu:bsRare0E}  {\ensuremath{{0.002720 } } }
\vdef{bdtdefault-11:bgBd2PiMuNu:bdRare0}   {\ensuremath{{0.148929 } } }
\vdef{bdtdefault-11:bgBd2PiMuNu:bdRare0E}  {\ensuremath{{0.017677 } } }
\vdef{bdtdefault-11:bgBd2PiMuNu:loSideband0:val}   {\ensuremath{{2.568 } } }
\vdef{bdtdefault-11:bgBd2PiMuNu:loSideband0:err}   {\ensuremath{{0.770 } } }
\vdef{bdtdefault-11:bgBd2PiMuNu:bsRare1}   {\ensuremath{{0.023103 } } }
\vdef{bdtdefault-11:bgBd2PiMuNu:bsRare1E}  {\ensuremath{{0.004187 } } }
\vdef{bdtdefault-11:bgBd2PiMuNu:bdRare1}   {\ensuremath{{0.057639 } } }
\vdef{bdtdefault-11:bgBd2PiMuNu:bdRare1E}  {\ensuremath{{0.009420 } } }
\vdef{bdtdefault-11:bgBd2PiMuNu:loSideband1:val}   {\ensuremath{{1.032 } } }
\vdef{bdtdefault-11:bgBd2PiMuNu:loSideband1:err}   {\ensuremath{{0.310 } } }
\vdef{bdtdefault-11:bgBd2PiPi:bsRare0}   {\ensuremath{{0.027176 } } }
\vdef{bdtdefault-11:bgBd2PiPi:bsRare0E}  {\ensuremath{{0.002088 } } }
\vdef{bdtdefault-11:bgBd2PiPi:bdRare0}   {\ensuremath{{0.170275 } } }
\vdef{bdtdefault-11:bgBd2PiPi:bdRare0E}  {\ensuremath{{0.005078 } } }
\vdef{bdtdefault-11:bgBd2PiPi:loSideband0:val}   {\ensuremath{{0.003 } } }
\vdef{bdtdefault-11:bgBd2PiPi:loSideband0:err}   {\ensuremath{{0.001 } } }
\vdef{bdtdefault-11:bgBd2PiPi:bsRare1}   {\ensuremath{{0.027110 } } }
\vdef{bdtdefault-11:bgBd2PiPi:bsRare1E}  {\ensuremath{{0.000953 } } }
\vdef{bdtdefault-11:bgBd2PiPi:bdRare1}   {\ensuremath{{0.063379 } } }
\vdef{bdtdefault-11:bgBd2PiPi:bdRare1E}  {\ensuremath{{0.001366 } } }
\vdef{bdtdefault-11:bgBd2PiPi:loSideband1:val}   {\ensuremath{{0.003 } } }
\vdef{bdtdefault-11:bgBd2PiPi:loSideband1:err}   {\ensuremath{{0.001 } } }
\vdef{bdtdefault-11:bgBs2KK:bsRare0}   {\ensuremath{{0.039330 } } }
\vdef{bdtdefault-11:bgBs2KK:bsRare0E}  {\ensuremath{{0.003646 } } }
\vdef{bdtdefault-11:bgBs2KK:bdRare0}   {\ensuremath{{0.200889 } } }
\vdef{bdtdefault-11:bgBs2KK:bdRare0E}  {\ensuremath{{0.009184 } } }
\vdef{bdtdefault-11:bgBs2KK:loSideband0:val}   {\ensuremath{{0.004 } } }
\vdef{bdtdefault-11:bgBs2KK:loSideband0:err}   {\ensuremath{{0.001 } } }
\vdef{bdtdefault-11:bgBs2KK:bsRare1}   {\ensuremath{{0.032820 } } }
\vdef{bdtdefault-11:bgBs2KK:bsRare1E}  {\ensuremath{{0.001713 } } }
\vdef{bdtdefault-11:bgBs2KK:bdRare1}   {\ensuremath{{0.072535 } } }
\vdef{bdtdefault-11:bgBs2KK:bdRare1E}  {\ensuremath{{0.002747 } } }
\vdef{bdtdefault-11:bgBs2KK:loSideband1:val}   {\ensuremath{{0.004 } } }
\vdef{bdtdefault-11:bgBs2KK:loSideband1:err}   {\ensuremath{{0.001 } } }
\vdef{bdtdefault-11:bgBs2KMuNu:bsRare0}   {\ensuremath{{0.042906 } } }
\vdef{bdtdefault-11:bgBs2KMuNu:bsRare0E}  {\ensuremath{{0.001043 } } }
\vdef{bdtdefault-11:bgBs2KMuNu:bdRare0}   {\ensuremath{{0.218773 } } }
\vdef{bdtdefault-11:bgBs2KMuNu:bdRare0E}  {\ensuremath{{0.005214 } } }
\vdef{bdtdefault-11:bgBs2KMuNu:loSideband0:val}   {\ensuremath{{0.880 } } }
\vdef{bdtdefault-11:bgBs2KMuNu:loSideband0:err}   {\ensuremath{{0.264 } } }
\vdef{bdtdefault-11:bgBs2KMuNu:bsRare1}   {\ensuremath{{0.042456 } } }
\vdef{bdtdefault-11:bgBs2KMuNu:bsRare1E}  {\ensuremath{{0.002809 } } }
\vdef{bdtdefault-11:bgBs2KMuNu:bdRare1}   {\ensuremath{{0.097314 } } }
\vdef{bdtdefault-11:bgBs2KMuNu:bdRare1E}  {\ensuremath{{0.007224 } } }
\vdef{bdtdefault-11:bgBs2KMuNu:loSideband1:val}   {\ensuremath{{0.350 } } }
\vdef{bdtdefault-11:bgBs2KMuNu:loSideband1:err}   {\ensuremath{{0.105 } } }
\vdef{bdtdefault-11:bgBs2KPi:bsRare0}   {\ensuremath{{0.048386 } } }
\vdef{bdtdefault-11:bgBs2KPi:bsRare0E}  {\ensuremath{{0.001865 } } }
\vdef{bdtdefault-11:bgBs2KPi:bdRare0}   {\ensuremath{{0.221218 } } }
\vdef{bdtdefault-11:bgBs2KPi:bdRare0E}  {\ensuremath{{0.000832 } } }
\vdef{bdtdefault-11:bgBs2KPi:loSideband0:val}   {\ensuremath{{0.000 } } }
\vdef{bdtdefault-11:bgBs2KPi:loSideband0:err}   {\ensuremath{{0.000 } } }
\vdef{bdtdefault-11:bgBs2KPi:bsRare1}   {\ensuremath{{0.044090 } } }
\vdef{bdtdefault-11:bgBs2KPi:bsRare1E}  {\ensuremath{{0.000556 } } }
\vdef{bdtdefault-11:bgBs2KPi:bdRare1}   {\ensuremath{{0.098426 } } }
\vdef{bdtdefault-11:bgBs2KPi:bdRare1E}  {\ensuremath{{0.000379 } } }
\vdef{bdtdefault-11:bgBs2KPi:loSideband1:val}   {\ensuremath{{0.000 } } }
\vdef{bdtdefault-11:bgBs2KPi:loSideband1:err}   {\ensuremath{{0.000 } } }
\vdef{bdtdefault-11:bgBs2PiPi:bsRare0}   {\ensuremath{{0.050030 } } }
\vdef{bdtdefault-11:bgBs2PiPi:bsRare0E}  {\ensuremath{{0.001699 } } }
\vdef{bdtdefault-11:bgBs2PiPi:bdRare0}   {\ensuremath{{0.221387 } } }
\vdef{bdtdefault-11:bgBs2PiPi:bdRare0E}  {\ensuremath{{0.000175 } } }
\vdef{bdtdefault-11:bgBs2PiPi:loSideband0:val}   {\ensuremath{{0.000 } } }
\vdef{bdtdefault-11:bgBs2PiPi:loSideband0:err}   {\ensuremath{{0.000 } } }
\vdef{bdtdefault-11:bgBs2PiPi:bsRare1}   {\ensuremath{{0.044563 } } }
\vdef{bdtdefault-11:bgBs2PiPi:bsRare1E}  {\ensuremath{{0.000489 } } }
\vdef{bdtdefault-11:bgBs2PiPi:bdRare1}   {\ensuremath{{0.098568 } } }
\vdef{bdtdefault-11:bgBs2PiPi:bdRare1E}  {\ensuremath{{0.000147 } } }
\vdef{bdtdefault-11:bgBs2PiPi:loSideband1:val}   {\ensuremath{{0.000 } } }
\vdef{bdtdefault-11:bgBs2PiPi:loSideband1:err}   {\ensuremath{{0.000 } } }
\vdef{bdtdefault-11:bgLb2KP:bsRare0}   {\ensuremath{{0.052249 } } }
\vdef{bdtdefault-11:bgLb2KP:bsRare0E}  {\ensuremath{{0.000881 } } }
\vdef{bdtdefault-11:bgLb2KP:bdRare0}   {\ensuremath{{0.221605 } } }
\vdef{bdtdefault-11:bgLb2KP:bdRare0E}  {\ensuremath{{0.000087 } } }
\vdef{bdtdefault-11:bgLb2KP:loSideband0:val}   {\ensuremath{{0.000 } } }
\vdef{bdtdefault-11:bgLb2KP:loSideband0:err}   {\ensuremath{{0.000 } } }
\vdef{bdtdefault-11:bgLb2KP:bsRare1}   {\ensuremath{{0.045300 } } }
\vdef{bdtdefault-11:bgLb2KP:bsRare1E}  {\ensuremath{{0.000292 } } }
\vdef{bdtdefault-11:bgLb2KP:bdRare1}   {\ensuremath{{0.098739 } } }
\vdef{bdtdefault-11:bgLb2KP:bdRare1E}  {\ensuremath{{0.000068 } } }
\vdef{bdtdefault-11:bgLb2KP:loSideband1:val}   {\ensuremath{{0.000 } } }
\vdef{bdtdefault-11:bgLb2KP:loSideband1:err}   {\ensuremath{{0.000 } } }
\vdef{bdtdefault-11:bgLb2PMuNu:bsRare0}   {\ensuremath{{0.272315 } } }
\vdef{bdtdefault-11:bgLb2PMuNu:bsRare0E}  {\ensuremath{{0.082664 } } }
\vdef{bdtdefault-11:bgLb2PMuNu:bdRare0}   {\ensuremath{{0.469851 } } }
\vdef{bdtdefault-11:bgLb2PMuNu:bdRare0E}  {\ensuremath{{0.093249 } } }
\vdef{bdtdefault-11:bgLb2PMuNu:loSideband0:val}   {\ensuremath{{1.316 } } }
\vdef{bdtdefault-11:bgLb2PMuNu:loSideband0:err}   {\ensuremath{{0.395 } } }
\vdef{bdtdefault-11:bgLb2PMuNu:bsRare1}   {\ensuremath{{0.126896 } } }
\vdef{bdtdefault-11:bgLb2PMuNu:bsRare1E}  {\ensuremath{{0.030650 } } }
\vdef{bdtdefault-11:bgLb2PMuNu:bdRare1}   {\ensuremath{{0.217519 } } }
\vdef{bdtdefault-11:bgLb2PMuNu:bdRare1E}  {\ensuremath{{0.044618 } } }
\vdef{bdtdefault-11:bgLb2PMuNu:loSideband1:val}   {\ensuremath{{0.498 } } }
\vdef{bdtdefault-11:bgLb2PMuNu:loSideband1:err}   {\ensuremath{{0.149 } } }
\vdef{bdtdefault-11:bgLb2PiP:bsRare0}   {\ensuremath{{0.273142 } } }
\vdef{bdtdefault-11:bgLb2PiP:bsRare0E}  {\ensuremath{{0.000334 } } }
\vdef{bdtdefault-11:bgLb2PiP:bdRare0}   {\ensuremath{{0.469949 } } }
\vdef{bdtdefault-11:bgLb2PiP:bdRare0E}  {\ensuremath{{0.000040 } } }
\vdef{bdtdefault-11:bgLb2PiP:loSideband0:val}   {\ensuremath{{0.000 } } }
\vdef{bdtdefault-11:bgLb2PiP:loSideband0:err}   {\ensuremath{{0.000 } } }
\vdef{bdtdefault-11:bgLb2PiP:bsRare1}   {\ensuremath{{0.127231 } } }
\vdef{bdtdefault-11:bgLb2PiP:bsRare1E}  {\ensuremath{{0.000135 } } }
\vdef{bdtdefault-11:bgLb2PiP:bdRare1}   {\ensuremath{{0.217588 } } }
\vdef{bdtdefault-11:bgLb2PiP:bdRare1E}  {\ensuremath{{0.000028 } } }
\vdef{bdtdefault-11:bgLb2PiP:loSideband1:val}   {\ensuremath{{0.000 } } }
\vdef{bdtdefault-11:bgLb2PiP:loSideband1:err}   {\ensuremath{{0.000 } } }
\vdef{bdtdefault-11:bsRare0}   {\ensuremath{{0.273 } } }
\vdef{bdtdefault-11:bsRare0E}  {\ensuremath{{0.083 } } }
\vdef{bdtdefault-11:bsRare1}   {\ensuremath{{0.127 } } }
\vdef{bdtdefault-11:bsRare1E}  {\ensuremath{{0.031 } } }
\vdef{bdtdefault-11:bdRare0}   {\ensuremath{{0.470 } } }
\vdef{bdtdefault-11:bdRare0E}  {\ensuremath{{0.098 } } }
\vdef{bdtdefault-11:bdRare1}   {\ensuremath{{0.218 } } }
\vdef{bdtdefault-11:bdRare1E}  {\ensuremath{{0.047 } } }
\vdef{default-11Bdt:BsSgEvt0:0:run}   {\ensuremath{{167898 } } }
\vdef{default-11Bdt:BsSgEvt0:0:evt}   {\ensuremath{{1773682763 } } }
\vdef{default-11Bdt:BsSgEvt0:0:chan}   {\ensuremath{{0 } } }
\vdef{default-11Bdt:BsSgEvt0:0:m}   {\ensuremath{{5.402 } } }
\vdef{default-11Bdt:BsSgEvt0:0:pt}   {\ensuremath{{18.133 } } }
\vdef{default-11Bdt:BsSgEvt0:0:phi}   {\ensuremath{{1.751 } } }
\vdef{default-11Bdt:BsSgEvt0:0:eta}   {\ensuremath{{-0.106 } } }
\vdef{default-11Bdt:BsSgEvt0:0:channel}   {barrel }
\vdef{default-11Bdt:BsSgEvt0:0:cowboy}   {\ensuremath{{0 } } }
\vdef{default-11Bdt:BsSgEvt0:0:m1pt}   {\ensuremath{{13.809 } } }
\vdef{default-11Bdt:BsSgEvt0:0:m2pt}   {\ensuremath{{5.008 } } }
\vdef{default-11Bdt:BsSgEvt0:0:m1eta}   {\ensuremath{{-0.039 } } }
\vdef{default-11Bdt:BsSgEvt0:0:m2eta}   {\ensuremath{{-0.273 } } }
\vdef{default-11Bdt:BsSgEvt0:0:m1phi}   {\ensuremath{{1.911 } } }
\vdef{default-11Bdt:BsSgEvt0:0:m2phi}   {\ensuremath{{1.297 } } }
\vdef{default-11Bdt:BsSgEvt0:0:m1q}   {\ensuremath{{-1 } } }
\vdef{default-11Bdt:BsSgEvt0:0:m2q}   {\ensuremath{{1 } } }
\vdef{default-11Bdt:BsSgEvt0:0:iso}   {\ensuremath{{0.952 } } }
\vdef{default-11Bdt:BsSgEvt0:0:alpha}   {\ensuremath{{0.0274 } } }
\vdef{default-11Bdt:BsSgEvt0:0:chi2}   {\ensuremath{{ 0.02 } } }
\vdef{default-11Bdt:BsSgEvt0:0:dof}   {\ensuremath{{1 } } }
\vdef{default-11Bdt:BsSgEvt0:0:fls3d}   {\ensuremath{{18.23 } } }
\vdef{default-11Bdt:BsSgEvt0:0:fl3d}   {\ensuremath{{0.1209 } } }
\vdef{default-11Bdt:BsSgEvt0:0:fl3dE}   {\ensuremath{{0.0066 } } }
\vdef{default-11Bdt:BsSgEvt0:0:docatrk}   {\ensuremath{{0.0142 } } }
\vdef{default-11Bdt:BsSgEvt0:0:closetrk}   {\ensuremath{{1 } } }
\vdef{default-11Bdt:BsSgEvt0:0:lip}   {\ensuremath{{0.0027 } } }
\vdef{default-11Bdt:BsSgEvt0:0:lipE}   {\ensuremath{{0.0058 } } }
\vdef{default-11Bdt:BsSgEvt0:0:tip}   {\ensuremath{{0.0018 } } }
\vdef{default-11Bdt:BsSgEvt0:0:tipE}   {\ensuremath{{0.0040 } } }
\vdef{default-11Bdt:BsSgEvt0:0:pvlip}   {\ensuremath{{0.0027 } } }
\vdef{default-11Bdt:BsSgEvt0:0:pvlips}   {\ensuremath{{0.4736 } } }
\vdef{default-11Bdt:BsSgEvt0:0:pvip}   {\ensuremath{{0.0033 } } }
\vdef{default-11Bdt:BsSgEvt0:0:pvips}   {\ensuremath{{0.6244 } } }
\vdef{default-11Bdt:BsSgEvt0:0:maxdoca}   {\ensuremath{{0.0008 } } }
\vdef{default-11Bdt:BsSgEvt0:0:pvw8}   {\ensuremath{{0.8840 } } }
\vdef{default-11Bdt:BsSgEvt0:0:bdt}   {\ensuremath{{0.2176 } } }
\vdef{default-11Bdt:BsSgEvt0:0:m1pix}   {\ensuremath{{3 } } }
\vdef{default-11Bdt:BsSgEvt0:0:m2pix}   {\ensuremath{{3 } } }
\vdef{default-11Bdt:BsSgEvt0:0:m1bpix}   {\ensuremath{{3 } } }
\vdef{default-11Bdt:BsSgEvt0:0:m2bpix}   {\ensuremath{{3 } } }
\vdef{default-11Bdt:BsSgEvt0:0:m1bpixl1}   {\ensuremath{{1 } } }
\vdef{default-11Bdt:BsSgEvt0:0:m2bpixl1}   {\ensuremath{{1 } } }
\vdef{default-11Bdt:BsSgEvt0:1:run}   {\ensuremath{{167898 } } }
\vdef{default-11Bdt:BsSgEvt0:1:evt}   {\ensuremath{{1487107623 } } }
\vdef{default-11Bdt:BsSgEvt0:1:chan}   {\ensuremath{{0 } } }
\vdef{default-11Bdt:BsSgEvt0:1:m}   {\ensuremath{{5.348 } } }
\vdef{default-11Bdt:BsSgEvt0:1:pt}   {\ensuremath{{10.467 } } }
\vdef{default-11Bdt:BsSgEvt0:1:phi}   {\ensuremath{{-0.036 } } }
\vdef{default-11Bdt:BsSgEvt0:1:eta}   {\ensuremath{{0.090 } } }
\vdef{default-11Bdt:BsSgEvt0:1:channel}   {barrel }
\vdef{default-11Bdt:BsSgEvt0:1:cowboy}   {\ensuremath{{1 } } }
\vdef{default-11Bdt:BsSgEvt0:1:m1pt}   {\ensuremath{{5.526 } } }
\vdef{default-11Bdt:BsSgEvt0:1:m2pt}   {\ensuremath{{5.117 } } }
\vdef{default-11Bdt:BsSgEvt0:1:m1eta}   {\ensuremath{{0.517 } } }
\vdef{default-11Bdt:BsSgEvt0:1:m2eta}   {\ensuremath{{-0.389 } } }
\vdef{default-11Bdt:BsSgEvt0:1:m1phi}   {\ensuremath{{-0.210 } } }
\vdef{default-11Bdt:BsSgEvt0:1:m2phi}   {\ensuremath{{0.154 } } }
\vdef{default-11Bdt:BsSgEvt0:1:m1q}   {\ensuremath{{-1 } } }
\vdef{default-11Bdt:BsSgEvt0:1:m2q}   {\ensuremath{{1 } } }
\vdef{default-11Bdt:BsSgEvt0:1:iso}   {\ensuremath{{1.000 } } }
\vdef{default-11Bdt:BsSgEvt0:1:alpha}   {\ensuremath{{0.0437 } } }
\vdef{default-11Bdt:BsSgEvt0:1:chi2}   {\ensuremath{{ 0.00 } } }
\vdef{default-11Bdt:BsSgEvt0:1:dof}   {\ensuremath{{1 } } }
\vdef{default-11Bdt:BsSgEvt0:1:fls3d}   {\ensuremath{{16.60 } } }
\vdef{default-11Bdt:BsSgEvt0:1:fl3d}   {\ensuremath{{0.1053 } } }
\vdef{default-11Bdt:BsSgEvt0:1:fl3dE}   {\ensuremath{{0.0063 } } }
\vdef{default-11Bdt:BsSgEvt0:1:docatrk}   {\ensuremath{{0.0099 } } }
\vdef{default-11Bdt:BsSgEvt0:1:closetrk}   {\ensuremath{{1 } } }
\vdef{default-11Bdt:BsSgEvt0:1:lip}   {\ensuremath{{0.0041 } } }
\vdef{default-11Bdt:BsSgEvt0:1:lipE}   {\ensuremath{{0.0040 } } }
\vdef{default-11Bdt:BsSgEvt0:1:tip}   {\ensuremath{{0.0020 } } }
\vdef{default-11Bdt:BsSgEvt0:1:tipE}   {\ensuremath{{0.0032 } } }
\vdef{default-11Bdt:BsSgEvt0:1:pvlip}   {\ensuremath{{0.0041 } } }
\vdef{default-11Bdt:BsSgEvt0:1:pvlips}   {\ensuremath{{1.0385 } } }
\vdef{default-11Bdt:BsSgEvt0:1:pvip}   {\ensuremath{{0.0046 } } }
\vdef{default-11Bdt:BsSgEvt0:1:pvips}   {\ensuremath{{1.1885 } } }
\vdef{default-11Bdt:BsSgEvt0:1:maxdoca}   {\ensuremath{{0.0001 } } }
\vdef{default-11Bdt:BsSgEvt0:1:pvw8}   {\ensuremath{{0.8120 } } }
\vdef{default-11Bdt:BsSgEvt0:1:bdt}   {\ensuremath{{0.1393 } } }
\vdef{default-11Bdt:BsSgEvt0:1:m1pix}   {\ensuremath{{2 } } }
\vdef{default-11Bdt:BsSgEvt0:1:m2pix}   {\ensuremath{{3 } } }
\vdef{default-11Bdt:BsSgEvt0:1:m1bpix}   {\ensuremath{{2 } } }
\vdef{default-11Bdt:BsSgEvt0:1:m2bpix}   {\ensuremath{{3 } } }
\vdef{default-11Bdt:BsSgEvt0:1:m1bpixl1}   {\ensuremath{{0 } } }
\vdef{default-11Bdt:BsSgEvt0:1:m2bpixl1}   {\ensuremath{{1 } } }
\vdef{default-11Bdt:SgEvt0:0:run}   {\ensuremath{{167675 } } }
\vdef{default-11Bdt:SgEvt0:0:evt}   {\ensuremath{{405600630 } } }
\vdef{default-11Bdt:SgEvt0:0:chan}   {\ensuremath{{0 } } }
\vdef{default-11Bdt:SgEvt0:0:m}   {\ensuremath{{5.094 } } }
\vdef{default-11Bdt:SgEvt0:0:pt}   {\ensuremath{{19.093 } } }
\vdef{default-11Bdt:SgEvt0:0:phi}   {\ensuremath{{-2.777 } } }
\vdef{default-11Bdt:SgEvt0:0:eta}   {\ensuremath{{-0.667 } } }
\vdef{default-11Bdt:SgEvt0:0:channel}   {barrel }
\vdef{default-11Bdt:SgEvt0:0:cowboy}   {\ensuremath{{0 } } }
\vdef{default-11Bdt:SgEvt0:0:m1pt}   {\ensuremath{{13.396 } } }
\vdef{default-11Bdt:SgEvt0:0:m2pt}   {\ensuremath{{5.621 } } }
\vdef{default-11Bdt:SgEvt0:0:m1eta}   {\ensuremath{{-0.475 } } }
\vdef{default-11Bdt:SgEvt0:0:m2eta}   {\ensuremath{{-1.050 } } }
\vdef{default-11Bdt:SgEvt0:0:m1phi}   {\ensuremath{{-2.785 } } }
\vdef{default-11Bdt:SgEvt0:0:m2phi}   {\ensuremath{{-2.759 } } }
\vdef{default-11Bdt:SgEvt0:0:m1q}   {\ensuremath{{1 } } }
\vdef{default-11Bdt:SgEvt0:0:m2q}   {\ensuremath{{-1 } } }
\vdef{default-11Bdt:SgEvt0:0:iso}   {\ensuremath{{0.942 } } }
\vdef{default-11Bdt:SgEvt0:0:alpha}   {\ensuremath{{0.0138 } } }
\vdef{default-11Bdt:SgEvt0:0:chi2}   {\ensuremath{{ 1.18 } } }
\vdef{default-11Bdt:SgEvt0:0:dof}   {\ensuremath{{1 } } }
\vdef{default-11Bdt:SgEvt0:0:fls3d}   {\ensuremath{{77.12 } } }
\vdef{default-11Bdt:SgEvt0:0:fl3d}   {\ensuremath{{0.8910 } } }
\vdef{default-11Bdt:SgEvt0:0:fl3dE}   {\ensuremath{{0.0116 } } }
\vdef{default-11Bdt:SgEvt0:0:docatrk}   {\ensuremath{{0.0416 } } }
\vdef{default-11Bdt:SgEvt0:0:closetrk}   {\ensuremath{{0 } } }
\vdef{default-11Bdt:SgEvt0:0:lip}   {\ensuremath{{-0.0099 } } }
\vdef{default-11Bdt:SgEvt0:0:lipE}   {\ensuremath{{0.0040 } } }
\vdef{default-11Bdt:SgEvt0:0:tip}   {\ensuremath{{0.0013 } } }
\vdef{default-11Bdt:SgEvt0:0:tipE}   {\ensuremath{{0.0032 } } }
\vdef{default-11Bdt:SgEvt0:0:pvlip}   {\ensuremath{{-0.0099 } } }
\vdef{default-11Bdt:SgEvt0:0:pvlips}   {\ensuremath{{-2.4571 } } }
\vdef{default-11Bdt:SgEvt0:0:pvip}   {\ensuremath{{0.0100 } } }
\vdef{default-11Bdt:SgEvt0:0:pvips}   {\ensuremath{{2.4848 } } }
\vdef{default-11Bdt:SgEvt0:0:maxdoca}   {\ensuremath{{0.0044 } } }
\vdef{default-11Bdt:SgEvt0:0:pvw8}   {\ensuremath{{0.9197 } } }
\vdef{default-11Bdt:SgEvt0:0:bdt}   {\ensuremath{{0.4217 } } }
\vdef{default-11Bdt:SgEvt0:0:m1pix}   {\ensuremath{{3 } } }
\vdef{default-11Bdt:SgEvt0:0:m2pix}   {\ensuremath{{2 } } }
\vdef{default-11Bdt:SgEvt0:0:m1bpix}   {\ensuremath{{3 } } }
\vdef{default-11Bdt:SgEvt0:0:m2bpix}   {\ensuremath{{2 } } }
\vdef{default-11Bdt:SgEvt0:0:m1bpixl1}   {\ensuremath{{1 } } }
\vdef{default-11Bdt:SgEvt0:0:m2bpixl1}   {\ensuremath{{1 } } }
\vdef{default-11Bdt:SgEvt1:0:run}   {\ensuremath{{167281 } } }
\vdef{default-11Bdt:SgEvt1:0:evt}   {\ensuremath{{408306037 } } }
\vdef{default-11Bdt:SgEvt1:0:chan}   {\ensuremath{{1 } } }
\vdef{default-11Bdt:SgEvt1:0:m}   {\ensuremath{{4.988 } } }
\vdef{default-11Bdt:SgEvt1:0:pt}   {\ensuremath{{10.625 } } }
\vdef{default-11Bdt:SgEvt1:0:phi}   {\ensuremath{{-1.176 } } }
\vdef{default-11Bdt:SgEvt1:0:eta}   {\ensuremath{{1.203 } } }
\vdef{default-11Bdt:SgEvt1:0:channel}   {endcap }
\vdef{default-11Bdt:SgEvt1:0:cowboy}   {\ensuremath{{1 } } }
\vdef{default-11Bdt:SgEvt1:0:m1pt}   {\ensuremath{{5.793 } } }
\vdef{default-11Bdt:SgEvt1:0:m2pt}   {\ensuremath{{5.177 } } }
\vdef{default-11Bdt:SgEvt1:0:m1eta}   {\ensuremath{{0.771 } } }
\vdef{default-11Bdt:SgEvt1:0:m2eta}   {\ensuremath{{1.512 } } }
\vdef{default-11Bdt:SgEvt1:0:m1phi}   {\ensuremath{{-1.416 } } }
\vdef{default-11Bdt:SgEvt1:0:m2phi}   {\ensuremath{{-0.906 } } }
\vdef{default-11Bdt:SgEvt1:0:m1q}   {\ensuremath{{-1 } } }
\vdef{default-11Bdt:SgEvt1:0:m2q}   {\ensuremath{{1 } } }
\vdef{default-11Bdt:SgEvt1:0:iso}   {\ensuremath{{1.000 } } }
\vdef{default-11Bdt:SgEvt1:0:alpha}   {\ensuremath{{0.0112 } } }
\vdef{default-11Bdt:SgEvt1:0:chi2}   {\ensuremath{{ 0.08 } } }
\vdef{default-11Bdt:SgEvt1:0:dof}   {\ensuremath{{1 } } }
\vdef{default-11Bdt:SgEvt1:0:fls3d}   {\ensuremath{{25.45 } } }
\vdef{default-11Bdt:SgEvt1:0:fl3d}   {\ensuremath{{0.2983 } } }
\vdef{default-11Bdt:SgEvt1:0:fl3dE}   {\ensuremath{{0.0117 } } }
\vdef{default-11Bdt:SgEvt1:0:docatrk}   {\ensuremath{{0.0842 } } }
\vdef{default-11Bdt:SgEvt1:0:closetrk}   {\ensuremath{{0 } } }
\vdef{default-11Bdt:SgEvt1:0:lip}   {\ensuremath{{0.0001 } } }
\vdef{default-11Bdt:SgEvt1:0:lipE}   {\ensuremath{{0.0056 } } }
\vdef{default-11Bdt:SgEvt1:0:tip}   {\ensuremath{{0.0033 } } }
\vdef{default-11Bdt:SgEvt1:0:tipE}   {\ensuremath{{0.0047 } } }
\vdef{default-11Bdt:SgEvt1:0:pvlip}   {\ensuremath{{0.0001 } } }
\vdef{default-11Bdt:SgEvt1:0:pvlips}   {\ensuremath{{0.0202 } } }
\vdef{default-11Bdt:SgEvt1:0:pvip}   {\ensuremath{{0.0033 } } }
\vdef{default-11Bdt:SgEvt1:0:pvips}   {\ensuremath{{0.7174 } } }
\vdef{default-11Bdt:SgEvt1:0:maxdoca}   {\ensuremath{{0.0014 } } }
\vdef{default-11Bdt:SgEvt1:0:pvw8}   {\ensuremath{{0.7814 } } }
\vdef{default-11Bdt:SgEvt1:0:bdt}   {\ensuremath{{0.1607 } } }
\vdef{default-11Bdt:SgEvt1:0:m1pix}   {\ensuremath{{3 } } }
\vdef{default-11Bdt:SgEvt1:0:m2pix}   {\ensuremath{{3 } } }
\vdef{default-11Bdt:SgEvt1:0:m1bpix}   {\ensuremath{{3 } } }
\vdef{default-11Bdt:SgEvt1:0:m2bpix}   {\ensuremath{{3 } } }
\vdef{default-11Bdt:SgEvt1:0:m1bpixl1}   {\ensuremath{{1 } } }
\vdef{default-11Bdt:SgEvt1:0:m2bpixl1}   {\ensuremath{{1 } } }
\vdef{default-11Bdt:BsSgEvt1:0:run}   {\ensuremath{{167043 } } }
\vdef{default-11Bdt:BsSgEvt1:0:evt}   {\ensuremath{{133697997 } } }
\vdef{default-11Bdt:BsSgEvt1:0:chan}   {\ensuremath{{1 } } }
\vdef{default-11Bdt:BsSgEvt1:0:m}   {\ensuremath{{5.317 } } }
\vdef{default-11Bdt:BsSgEvt1:0:pt}   {\ensuremath{{8.727 } } }
\vdef{default-11Bdt:BsSgEvt1:0:phi}   {\ensuremath{{1.609 } } }
\vdef{default-11Bdt:BsSgEvt1:0:eta}   {\ensuremath{{-1.878 } } }
\vdef{default-11Bdt:BsSgEvt1:0:channel}   {endcap }
\vdef{default-11Bdt:BsSgEvt1:0:cowboy}   {\ensuremath{{1 } } }
\vdef{default-11Bdt:BsSgEvt1:0:m1pt}   {\ensuremath{{4.713 } } }
\vdef{default-11Bdt:BsSgEvt1:0:m2pt}   {\ensuremath{{4.049 } } }
\vdef{default-11Bdt:BsSgEvt1:0:m1eta}   {\ensuremath{{-1.199 } } }
\vdef{default-11Bdt:BsSgEvt1:0:m2eta}   {\ensuremath{{-2.335 } } }
\vdef{default-11Bdt:BsSgEvt1:0:m1phi}   {\ensuremath{{1.515 } } }
\vdef{default-11Bdt:BsSgEvt1:0:m2phi}   {\ensuremath{{1.719 } } }
\vdef{default-11Bdt:BsSgEvt1:0:m1q}   {\ensuremath{{-1 } } }
\vdef{default-11Bdt:BsSgEvt1:0:m2q}   {\ensuremath{{1 } } }
\vdef{default-11Bdt:BsSgEvt1:0:iso}   {\ensuremath{{1.000 } } }
\vdef{default-11Bdt:BsSgEvt1:0:alpha}   {\ensuremath{{0.0188 } } }
\vdef{default-11Bdt:BsSgEvt1:0:chi2}   {\ensuremath{{ 0.01 } } }
\vdef{default-11Bdt:BsSgEvt1:0:dof}   {\ensuremath{{1 } } }
\vdef{default-11Bdt:BsSgEvt1:0:fls3d}   {\ensuremath{{32.52 } } }
\vdef{default-11Bdt:BsSgEvt1:0:fl3d}   {\ensuremath{{1.2500 } } }
\vdef{default-11Bdt:BsSgEvt1:0:fl3dE}   {\ensuremath{{0.0384 } } }
\vdef{default-11Bdt:BsSgEvt1:0:docatrk}   {\ensuremath{{0.1056 } } }
\vdef{default-11Bdt:BsSgEvt1:0:closetrk}   {\ensuremath{{0 } } }
\vdef{default-11Bdt:BsSgEvt1:0:lip}   {\ensuremath{{-0.0004 } } }
\vdef{default-11Bdt:BsSgEvt1:0:lipE}   {\ensuremath{{0.0086 } } }
\vdef{default-11Bdt:BsSgEvt1:0:tip}   {\ensuremath{{0.0235 } } }
\vdef{default-11Bdt:BsSgEvt1:0:tipE}   {\ensuremath{{0.0091 } } }
\vdef{default-11Bdt:BsSgEvt1:0:pvlip}   {\ensuremath{{-0.0004 } } }
\vdef{default-11Bdt:BsSgEvt1:0:pvlips}   {\ensuremath{{-0.0512 } } }
\vdef{default-11Bdt:BsSgEvt1:0:pvip}   {\ensuremath{{0.0235 } } }
\vdef{default-11Bdt:BsSgEvt1:0:pvips}   {\ensuremath{{2.5959 } } }
\vdef{default-11Bdt:BsSgEvt1:0:maxdoca}   {\ensuremath{{0.0013 } } }
\vdef{default-11Bdt:BsSgEvt1:0:pvw8}   {\ensuremath{{0.7732 } } }
\vdef{default-11Bdt:BsSgEvt1:0:bdt}   {\ensuremath{{0.0965 } } }
\vdef{default-11Bdt:BsSgEvt1:0:m1pix}   {\ensuremath{{3 } } }
\vdef{default-11Bdt:BsSgEvt1:0:m2pix}   {\ensuremath{{1 } } }
\vdef{default-11Bdt:BsSgEvt1:0:m1bpix}   {\ensuremath{{3 } } }
\vdef{default-11Bdt:BsSgEvt1:0:m2bpix}   {\ensuremath{{0 } } }
\vdef{default-11Bdt:BsSgEvt1:0:m1bpixl1}   {\ensuremath{{1 } } }
\vdef{default-11Bdt:BsSgEvt1:0:m2bpixl1}   {\ensuremath{{0 } } }
\vdef{default-11Bdt:SgEvt1:1:run}   {\ensuremath{{167039 } } }
\vdef{default-11Bdt:SgEvt1:1:evt}   {\ensuremath{{210818502 } } }
\vdef{default-11Bdt:SgEvt1:1:chan}   {\ensuremath{{1 } } }
\vdef{default-11Bdt:SgEvt1:1:m}   {\ensuremath{{5.004 } } }
\vdef{default-11Bdt:SgEvt1:1:pt}   {\ensuremath{{14.317 } } }
\vdef{default-11Bdt:SgEvt1:1:phi}   {\ensuremath{{1.073 } } }
\vdef{default-11Bdt:SgEvt1:1:eta}   {\ensuremath{{1.366 } } }
\vdef{default-11Bdt:SgEvt1:1:channel}   {endcap }
\vdef{default-11Bdt:SgEvt1:1:cowboy}   {\ensuremath{{0 } } }
\vdef{default-11Bdt:SgEvt1:1:m1pt}   {\ensuremath{{7.864 } } }
\vdef{default-11Bdt:SgEvt1:1:m2pt}   {\ensuremath{{6.873 } } }
\vdef{default-11Bdt:SgEvt1:1:m1eta}   {\ensuremath{{1.089 } } }
\vdef{default-11Bdt:SgEvt1:1:m2eta}   {\ensuremath{{1.576 } } }
\vdef{default-11Bdt:SgEvt1:1:m1phi}   {\ensuremath{{1.293 } } }
\vdef{default-11Bdt:SgEvt1:1:m2phi}   {\ensuremath{{0.820 } } }
\vdef{default-11Bdt:SgEvt1:1:m1q}   {\ensuremath{{-1 } } }
\vdef{default-11Bdt:SgEvt1:1:m2q}   {\ensuremath{{1 } } }
\vdef{default-11Bdt:SgEvt1:1:iso}   {\ensuremath{{0.880 } } }
\vdef{default-11Bdt:SgEvt1:1:alpha}   {\ensuremath{{0.0179 } } }
\vdef{default-11Bdt:SgEvt1:1:chi2}   {\ensuremath{{ 0.11 } } }
\vdef{default-11Bdt:SgEvt1:1:dof}   {\ensuremath{{1 } } }
\vdef{default-11Bdt:SgEvt1:1:fls3d}   {\ensuremath{{30.73 } } }
\vdef{default-11Bdt:SgEvt1:1:fl3d}   {\ensuremath{{0.5418 } } }
\vdef{default-11Bdt:SgEvt1:1:fl3dE}   {\ensuremath{{0.0176 } } }
\vdef{default-11Bdt:SgEvt1:1:docatrk}   {\ensuremath{{0.0157 } } }
\vdef{default-11Bdt:SgEvt1:1:closetrk}   {\ensuremath{{1 } } }
\vdef{default-11Bdt:SgEvt1:1:lip}   {\ensuremath{{-0.0024 } } }
\vdef{default-11Bdt:SgEvt1:1:lipE}   {\ensuremath{{0.0129 } } }
\vdef{default-11Bdt:SgEvt1:1:tip}   {\ensuremath{{0.0083 } } }
\vdef{default-11Bdt:SgEvt1:1:tipE}   {\ensuremath{{0.0051 } } }
\vdef{default-11Bdt:SgEvt1:1:pvlip}   {\ensuremath{{-0.0024 } } }
\vdef{default-11Bdt:SgEvt1:1:pvlips}   {\ensuremath{{-0.1845 } } }
\vdef{default-11Bdt:SgEvt1:1:pvip}   {\ensuremath{{0.0087 } } }
\vdef{default-11Bdt:SgEvt1:1:pvips}   {\ensuremath{{1.4288 } } }
\vdef{default-11Bdt:SgEvt1:1:maxdoca}   {\ensuremath{{0.0012 } } }
\vdef{default-11Bdt:SgEvt1:1:pvw8}   {\ensuremath{{0.9165 } } }
\vdef{default-11Bdt:SgEvt1:1:bdt}   {\ensuremath{{-0.0059 } } }
\vdef{default-11Bdt:SgEvt1:1:m1pix}   {\ensuremath{{2 } } }
\vdef{default-11Bdt:SgEvt1:1:m2pix}   {\ensuremath{{3 } } }
\vdef{default-11Bdt:SgEvt1:1:m1bpix}   {\ensuremath{{2 } } }
\vdef{default-11Bdt:SgEvt1:1:m2bpix}   {\ensuremath{{2 } } }
\vdef{default-11Bdt:SgEvt1:1:m1bpixl1}   {\ensuremath{{1 } } }
\vdef{default-11Bdt:SgEvt1:1:m2bpixl1}   {\ensuremath{{1 } } }
\vdef{default-11Bdt:SgEvt1:2:run}   {\ensuremath{{166408 } } }
\vdef{default-11Bdt:SgEvt1:2:evt}   {\ensuremath{{1091686553 } } }
\vdef{default-11Bdt:SgEvt1:2:chan}   {\ensuremath{{1 } } }
\vdef{default-11Bdt:SgEvt1:2:m}   {\ensuremath{{5.077 } } }
\vdef{default-11Bdt:SgEvt1:2:pt}   {\ensuremath{{39.515 } } }
\vdef{default-11Bdt:SgEvt1:2:phi}   {\ensuremath{{0.153 } } }
\vdef{default-11Bdt:SgEvt1:2:eta}   {\ensuremath{{1.614 } } }
\vdef{default-11Bdt:SgEvt1:2:channel}   {endcap }
\vdef{default-11Bdt:SgEvt1:2:cowboy}   {\ensuremath{{1 } } }
\vdef{default-11Bdt:SgEvt1:2:m1pt}   {\ensuremath{{34.496 } } }
\vdef{default-11Bdt:SgEvt1:2:m2pt}   {\ensuremath{{5.020 } } }
\vdef{default-11Bdt:SgEvt1:2:m1eta}   {\ensuremath{{1.653 } } }
\vdef{default-11Bdt:SgEvt1:2:m2eta}   {\ensuremath{{1.289 } } }
\vdef{default-11Bdt:SgEvt1:2:m1phi}   {\ensuremath{{0.138 } } }
\vdef{default-11Bdt:SgEvt1:2:m2phi}   {\ensuremath{{0.256 } } }
\vdef{default-11Bdt:SgEvt1:2:m1q}   {\ensuremath{{-1 } } }
\vdef{default-11Bdt:SgEvt1:2:m2q}   {\ensuremath{{1 } } }
\vdef{default-11Bdt:SgEvt1:2:iso}   {\ensuremath{{1.000 } } }
\vdef{default-11Bdt:SgEvt1:2:alpha}   {\ensuremath{{0.0180 } } }
\vdef{default-11Bdt:SgEvt1:2:chi2}   {\ensuremath{{ 0.05 } } }
\vdef{default-11Bdt:SgEvt1:2:dof}   {\ensuremath{{1 } } }
\vdef{default-11Bdt:SgEvt1:2:fls3d}   {\ensuremath{{18.15 } } }
\vdef{default-11Bdt:SgEvt1:2:fl3d}   {\ensuremath{{0.4098 } } }
\vdef{default-11Bdt:SgEvt1:2:fl3dE}   {\ensuremath{{0.0226 } } }
\vdef{default-11Bdt:SgEvt1:2:docatrk}   {\ensuremath{{0.0046 } } }
\vdef{default-11Bdt:SgEvt1:2:closetrk}   {\ensuremath{{0 } } }
\vdef{default-11Bdt:SgEvt1:2:lip}   {\ensuremath{{0.0028 } } }
\vdef{default-11Bdt:SgEvt1:2:lipE}   {\ensuremath{{0.0032 } } }
\vdef{default-11Bdt:SgEvt1:2:tip}   {\ensuremath{{0.0003 } } }
\vdef{default-11Bdt:SgEvt1:2:tipE}   {\ensuremath{{0.0017 } } }
\vdef{default-11Bdt:SgEvt1:2:pvlip}   {\ensuremath{{0.0028 } } }
\vdef{default-11Bdt:SgEvt1:2:pvlips}   {\ensuremath{{0.8798 } } }
\vdef{default-11Bdt:SgEvt1:2:pvip}   {\ensuremath{{0.0028 } } }
\vdef{default-11Bdt:SgEvt1:2:pvips}   {\ensuremath{{0.8896 } } }
\vdef{default-11Bdt:SgEvt1:2:maxdoca}   {\ensuremath{{0.0009 } } }
\vdef{default-11Bdt:SgEvt1:2:pvw8}   {\ensuremath{{0.8942 } } }
\vdef{default-11Bdt:SgEvt1:2:bdt}   {\ensuremath{{0.0512 } } }
\vdef{default-11Bdt:SgEvt1:2:m1pix}   {\ensuremath{{3 } } }
\vdef{default-11Bdt:SgEvt1:2:m2pix}   {\ensuremath{{2 } } }
\vdef{default-11Bdt:SgEvt1:2:m1bpix}   {\ensuremath{{3 } } }
\vdef{default-11Bdt:SgEvt1:2:m2bpix}   {\ensuremath{{2 } } }
\vdef{default-11Bdt:SgEvt1:2:m1bpixl1}   {\ensuremath{{1 } } }
\vdef{default-11Bdt:SgEvt1:2:m2bpixl1}   {\ensuremath{{2 } } }
\vdef{default-11Bdt:SgEvt0:1:run}   {\ensuremath{{166408 } } }
\vdef{default-11Bdt:SgEvt0:1:evt}   {\ensuremath{{664224410 } } }
\vdef{default-11Bdt:SgEvt0:1:chan}   {\ensuremath{{0 } } }
\vdef{default-11Bdt:SgEvt0:1:m}   {\ensuremath{{5.571 } } }
\vdef{default-11Bdt:SgEvt0:1:pt}   {\ensuremath{{8.884 } } }
\vdef{default-11Bdt:SgEvt0:1:phi}   {\ensuremath{{-1.675 } } }
\vdef{default-11Bdt:SgEvt0:1:eta}   {\ensuremath{{0.626 } } }
\vdef{default-11Bdt:SgEvt0:1:channel}   {barrel }
\vdef{default-11Bdt:SgEvt0:1:cowboy}   {\ensuremath{{0 } } }
\vdef{default-11Bdt:SgEvt0:1:m1pt}   {\ensuremath{{4.833 } } }
\vdef{default-11Bdt:SgEvt0:1:m2pt}   {\ensuremath{{4.038 } } }
\vdef{default-11Bdt:SgEvt0:1:m1eta}   {\ensuremath{{-0.006 } } }
\vdef{default-11Bdt:SgEvt0:1:m2eta}   {\ensuremath{{1.179 } } }
\vdef{default-11Bdt:SgEvt0:1:m1phi}   {\ensuremath{{-1.650 } } }
\vdef{default-11Bdt:SgEvt0:1:m2phi}   {\ensuremath{{-1.706 } } }
\vdef{default-11Bdt:SgEvt0:1:m1q}   {\ensuremath{{-1 } } }
\vdef{default-11Bdt:SgEvt0:1:m2q}   {\ensuremath{{1 } } }
\vdef{default-11Bdt:SgEvt0:1:iso}   {\ensuremath{{1.000 } } }
\vdef{default-11Bdt:SgEvt0:1:alpha}   {\ensuremath{{0.0873 } } }
\vdef{default-11Bdt:SgEvt0:1:chi2}   {\ensuremath{{ 1.03 } } }
\vdef{default-11Bdt:SgEvt0:1:dof}   {\ensuremath{{1 } } }
\vdef{default-11Bdt:SgEvt0:1:fls3d}   {\ensuremath{{34.31 } } }
\vdef{default-11Bdt:SgEvt0:1:fl3d}   {\ensuremath{{0.3241 } } }
\vdef{default-11Bdt:SgEvt0:1:fl3dE}   {\ensuremath{{0.0094 } } }
\vdef{default-11Bdt:SgEvt0:1:docatrk}   {\ensuremath{{0.0661 } } }
\vdef{default-11Bdt:SgEvt0:1:closetrk}   {\ensuremath{{0 } } }
\vdef{default-11Bdt:SgEvt0:1:lip}   {\ensuremath{{-0.0233 } } }
\vdef{default-11Bdt:SgEvt0:1:lipE}   {\ensuremath{{0.0066 } } }
\vdef{default-11Bdt:SgEvt0:1:tip}   {\ensuremath{{0.0034 } } }
\vdef{default-11Bdt:SgEvt0:1:tipE}   {\ensuremath{{0.0053 } } }
\vdef{default-11Bdt:SgEvt0:1:pvlip}   {\ensuremath{{-0.0233 } } }
\vdef{default-11Bdt:SgEvt0:1:pvlips}   {\ensuremath{{-3.5483 } } }
\vdef{default-11Bdt:SgEvt0:1:pvip}   {\ensuremath{{0.0236 } } }
\vdef{default-11Bdt:SgEvt0:1:pvips}   {\ensuremath{{3.5976 } } }
\vdef{default-11Bdt:SgEvt0:1:maxdoca}   {\ensuremath{{0.0057 } } }
\vdef{default-11Bdt:SgEvt0:1:pvw8}   {\ensuremath{{0.9408 } } }
\vdef{default-11Bdt:SgEvt0:1:bdt}   {\ensuremath{{-0.0232 } } }
\vdef{default-11Bdt:SgEvt0:1:m1pix}   {\ensuremath{{3 } } }
\vdef{default-11Bdt:SgEvt0:1:m2pix}   {\ensuremath{{3 } } }
\vdef{default-11Bdt:SgEvt0:1:m1bpix}   {\ensuremath{{3 } } }
\vdef{default-11Bdt:SgEvt0:1:m2bpix}   {\ensuremath{{3 } } }
\vdef{default-11Bdt:SgEvt0:1:m1bpixl1}   {\ensuremath{{1 } } }
\vdef{default-11Bdt:SgEvt0:1:m2bpixl1}   {\ensuremath{{1 } } }
\vdef{default-11Bdt:SgEvt1:3:run}   {\ensuremath{{166049 } } }
\vdef{default-11Bdt:SgEvt1:3:evt}   {\ensuremath{{72634539 } } }
\vdef{default-11Bdt:SgEvt1:3:chan}   {\ensuremath{{1 } } }
\vdef{default-11Bdt:SgEvt1:3:m}   {\ensuremath{{4.932 } } }
\vdef{default-11Bdt:SgEvt1:3:pt}   {\ensuremath{{11.805 } } }
\vdef{default-11Bdt:SgEvt1:3:phi}   {\ensuremath{{0.886 } } }
\vdef{default-11Bdt:SgEvt1:3:eta}   {\ensuremath{{1.718 } } }
\vdef{default-11Bdt:SgEvt1:3:channel}   {endcap }
\vdef{default-11Bdt:SgEvt1:3:cowboy}   {\ensuremath{{1 } } }
\vdef{default-11Bdt:SgEvt1:3:m1pt}   {\ensuremath{{7.232 } } }
\vdef{default-11Bdt:SgEvt1:3:m2pt}   {\ensuremath{{5.410 } } }
\vdef{default-11Bdt:SgEvt1:3:m1eta}   {\ensuremath{{1.780 } } }
\vdef{default-11Bdt:SgEvt1:3:m2eta}   {\ensuremath{{1.461 } } }
\vdef{default-11Bdt:SgEvt1:3:m1phi}   {\ensuremath{{0.573 } } }
\vdef{default-11Bdt:SgEvt1:3:m2phi}   {\ensuremath{{1.310 } } }
\vdef{default-11Bdt:SgEvt1:3:m1q}   {\ensuremath{{-1 } } }
\vdef{default-11Bdt:SgEvt1:3:m2q}   {\ensuremath{{1 } } }
\vdef{default-11Bdt:SgEvt1:3:iso}   {\ensuremath{{0.904 } } }
\vdef{default-11Bdt:SgEvt1:3:alpha}   {\ensuremath{{0.0137 } } }
\vdef{default-11Bdt:SgEvt1:3:chi2}   {\ensuremath{{ 0.67 } } }
\vdef{default-11Bdt:SgEvt1:3:dof}   {\ensuremath{{1 } } }
\vdef{default-11Bdt:SgEvt1:3:fls3d}   {\ensuremath{{65.08 } } }
\vdef{default-11Bdt:SgEvt1:3:fl3d}   {\ensuremath{{0.8585 } } }
\vdef{default-11Bdt:SgEvt1:3:fl3dE}   {\ensuremath{{0.0132 } } }
\vdef{default-11Bdt:SgEvt1:3:docatrk}   {\ensuremath{{0.1008 } } }
\vdef{default-11Bdt:SgEvt1:3:closetrk}   {\ensuremath{{0 } } }
\vdef{default-11Bdt:SgEvt1:3:lip}   {\ensuremath{{0.0036 } } }
\vdef{default-11Bdt:SgEvt1:3:lipE}   {\ensuremath{{0.0027 } } }
\vdef{default-11Bdt:SgEvt1:3:tip}   {\ensuremath{{0.0057 } } }
\vdef{default-11Bdt:SgEvt1:3:tipE}   {\ensuremath{{0.0026 } } }
\vdef{default-11Bdt:SgEvt1:3:pvlip}   {\ensuremath{{0.0036 } } }
\vdef{default-11Bdt:SgEvt1:3:pvlips}   {\ensuremath{{1.3103 } } }
\vdef{default-11Bdt:SgEvt1:3:pvip}   {\ensuremath{{0.0068 } } }
\vdef{default-11Bdt:SgEvt1:3:pvips}   {\ensuremath{{2.5646 } } }
\vdef{default-11Bdt:SgEvt1:3:maxdoca}   {\ensuremath{{0.0031 } } }
\vdef{default-11Bdt:SgEvt1:3:pvw8}   {\ensuremath{{0.8785 } } }
\vdef{default-11Bdt:SgEvt1:3:bdt}   {\ensuremath{{-0.0775 } } }
\vdef{default-11Bdt:SgEvt1:3:m1pix}   {\ensuremath{{3 } } }
\vdef{default-11Bdt:SgEvt1:3:m2pix}   {\ensuremath{{3 } } }
\vdef{default-11Bdt:SgEvt1:3:m1bpix}   {\ensuremath{{2 } } }
\vdef{default-11Bdt:SgEvt1:3:m2bpix}   {\ensuremath{{2 } } }
\vdef{default-11Bdt:SgEvt1:3:m1bpixl1}   {\ensuremath{{1 } } }
\vdef{default-11Bdt:SgEvt1:3:m2bpixl1}   {\ensuremath{{1 } } }
\vdef{default-11Bdt:BsSgEvt1:1:run}   {\ensuremath{{165993 } } }
\vdef{default-11Bdt:BsSgEvt1:1:evt}   {\ensuremath{{890015250 } } }
\vdef{default-11Bdt:BsSgEvt1:1:chan}   {\ensuremath{{1 } } }
\vdef{default-11Bdt:BsSgEvt1:1:m}   {\ensuremath{{5.338 } } }
\vdef{default-11Bdt:BsSgEvt1:1:pt}   {\ensuremath{{14.497 } } }
\vdef{default-11Bdt:BsSgEvt1:1:phi}   {\ensuremath{{-2.352 } } }
\vdef{default-11Bdt:BsSgEvt1:1:eta}   {\ensuremath{{-1.862 } } }
\vdef{default-11Bdt:BsSgEvt1:1:channel}   {endcap }
\vdef{default-11Bdt:BsSgEvt1:1:cowboy}   {\ensuremath{{0 } } }
\vdef{default-11Bdt:BsSgEvt1:1:m1pt}   {\ensuremath{{9.433 } } }
\vdef{default-11Bdt:BsSgEvt1:1:m2pt}   {\ensuremath{{4.996 } } }
\vdef{default-11Bdt:BsSgEvt1:1:m1eta}   {\ensuremath{{-1.538 } } }
\vdef{default-11Bdt:BsSgEvt1:1:m2eta}   {\ensuremath{{-2.282 } } }
\vdef{default-11Bdt:BsSgEvt1:1:m1phi}   {\ensuremath{{-2.394 } } }
\vdef{default-11Bdt:BsSgEvt1:1:m2phi}   {\ensuremath{{-2.275 } } }
\vdef{default-11Bdt:BsSgEvt1:1:m1q}   {\ensuremath{{1 } } }
\vdef{default-11Bdt:BsSgEvt1:1:m2q}   {\ensuremath{{-1 } } }
\vdef{default-11Bdt:BsSgEvt1:1:iso}   {\ensuremath{{0.903 } } }
\vdef{default-11Bdt:BsSgEvt1:1:alpha}   {\ensuremath{{0.0080 } } }
\vdef{default-11Bdt:BsSgEvt1:1:chi2}   {\ensuremath{{ 2.89 } } }
\vdef{default-11Bdt:BsSgEvt1:1:dof}   {\ensuremath{{1 } } }
\vdef{default-11Bdt:BsSgEvt1:1:fls3d}   {\ensuremath{{17.03 } } }
\vdef{default-11Bdt:BsSgEvt1:1:fl3d}   {\ensuremath{{0.4715 } } }
\vdef{default-11Bdt:BsSgEvt1:1:fl3dE}   {\ensuremath{{0.0277 } } }
\vdef{default-11Bdt:BsSgEvt1:1:docatrk}   {\ensuremath{{0.0288 } } }
\vdef{default-11Bdt:BsSgEvt1:1:closetrk}   {\ensuremath{{1 } } }
\vdef{default-11Bdt:BsSgEvt1:1:lip}   {\ensuremath{{-0.0011 } } }
\vdef{default-11Bdt:BsSgEvt1:1:lipE}   {\ensuremath{{0.0024 } } }
\vdef{default-11Bdt:BsSgEvt1:1:tip}   {\ensuremath{{0.0005 } } }
\vdef{default-11Bdt:BsSgEvt1:1:tipE}   {\ensuremath{{0.0035 } } }
\vdef{default-11Bdt:BsSgEvt1:1:pvlip}   {\ensuremath{{-0.0011 } } }
\vdef{default-11Bdt:BsSgEvt1:1:pvlips}   {\ensuremath{{-0.4685 } } }
\vdef{default-11Bdt:BsSgEvt1:1:pvip}   {\ensuremath{{0.0013 } } }
\vdef{default-11Bdt:BsSgEvt1:1:pvips}   {\ensuremath{{0.4717 } } }
\vdef{default-11Bdt:BsSgEvt1:1:maxdoca}   {\ensuremath{{0.0113 } } }
\vdef{default-11Bdt:BsSgEvt1:1:pvw8}   {\ensuremath{{0.8751 } } }
\vdef{default-11Bdt:BsSgEvt1:1:bdt}   {\ensuremath{{-0.0019 } } }
\vdef{default-11Bdt:BsSgEvt1:1:m1pix}   {\ensuremath{{3 } } }
\vdef{default-11Bdt:BsSgEvt1:1:m2pix}   {\ensuremath{{3 } } }
\vdef{default-11Bdt:BsSgEvt1:1:m1bpix}   {\ensuremath{{3 } } }
\vdef{default-11Bdt:BsSgEvt1:1:m2bpix}   {\ensuremath{{1 } } }
\vdef{default-11Bdt:BsSgEvt1:1:m1bpixl1}   {\ensuremath{{2 } } }
\vdef{default-11Bdt:BsSgEvt1:1:m2bpixl1}   {\ensuremath{{1 } } }
\vdef{default-11Bdt:SgEvt1:4:run}   {\ensuremath{{165548 } } }
\vdef{default-11Bdt:SgEvt1:4:evt}   {\ensuremath{{448725265 } } }
\vdef{default-11Bdt:SgEvt1:4:chan}   {\ensuremath{{1 } } }
\vdef{default-11Bdt:SgEvt1:4:m}   {\ensuremath{{5.827 } } }
\vdef{default-11Bdt:SgEvt1:4:pt}   {\ensuremath{{7.927 } } }
\vdef{default-11Bdt:SgEvt1:4:phi}   {\ensuremath{{-1.105 } } }
\vdef{default-11Bdt:SgEvt1:4:eta}   {\ensuremath{{-1.707 } } }
\vdef{default-11Bdt:SgEvt1:4:channel}   {endcap }
\vdef{default-11Bdt:SgEvt1:4:cowboy}   {\ensuremath{{1 } } }
\vdef{default-11Bdt:SgEvt1:4:m1pt}   {\ensuremath{{4.680 } } }
\vdef{default-11Bdt:SgEvt1:4:m2pt}   {\ensuremath{{4.033 } } }
\vdef{default-11Bdt:SgEvt1:4:m1eta}   {\ensuremath{{-1.978 } } }
\vdef{default-11Bdt:SgEvt1:4:m2eta}   {\ensuremath{{-0.968 } } }
\vdef{default-11Bdt:SgEvt1:4:m1phi}   {\ensuremath{{-0.711 } } }
\vdef{default-11Bdt:SgEvt1:4:m2phi}   {\ensuremath{{-1.568 } } }
\vdef{default-11Bdt:SgEvt1:4:m1q}   {\ensuremath{{1 } } }
\vdef{default-11Bdt:SgEvt1:4:m2q}   {\ensuremath{{-1 } } }
\vdef{default-11Bdt:SgEvt1:4:iso}   {\ensuremath{{0.827 } } }
\vdef{default-11Bdt:SgEvt1:4:alpha}   {\ensuremath{{0.0188 } } }
\vdef{default-11Bdt:SgEvt1:4:chi2}   {\ensuremath{{ 0.10 } } }
\vdef{default-11Bdt:SgEvt1:4:dof}   {\ensuremath{{1 } } }
\vdef{default-11Bdt:SgEvt1:4:fls3d}   {\ensuremath{{18.74 } } }
\vdef{default-11Bdt:SgEvt1:4:fl3d}   {\ensuremath{{0.2147 } } }
\vdef{default-11Bdt:SgEvt1:4:fl3dE}   {\ensuremath{{0.0115 } } }
\vdef{default-11Bdt:SgEvt1:4:docatrk}   {\ensuremath{{0.0437 } } }
\vdef{default-11Bdt:SgEvt1:4:closetrk}   {\ensuremath{{0 } } }
\vdef{default-11Bdt:SgEvt1:4:lip}   {\ensuremath{{0.0003 } } }
\vdef{default-11Bdt:SgEvt1:4:lipE}   {\ensuremath{{0.0029 } } }
\vdef{default-11Bdt:SgEvt1:4:tip}   {\ensuremath{{0.0040 } } }
\vdef{default-11Bdt:SgEvt1:4:tipE}   {\ensuremath{{0.0042 } } }
\vdef{default-11Bdt:SgEvt1:4:pvlip}   {\ensuremath{{0.0003 } } }
\vdef{default-11Bdt:SgEvt1:4:pvlips}   {\ensuremath{{0.1022 } } }
\vdef{default-11Bdt:SgEvt1:4:pvip}   {\ensuremath{{0.0040 } } }
\vdef{default-11Bdt:SgEvt1:4:pvips}   {\ensuremath{{0.9563 } } }
\vdef{default-11Bdt:SgEvt1:4:maxdoca}   {\ensuremath{{0.0019 } } }
\vdef{default-11Bdt:SgEvt1:4:pvw8}   {\ensuremath{{0.9130 } } }
\vdef{default-11Bdt:SgEvt1:4:bdt}   {\ensuremath{{-0.0193 } } }
\vdef{default-11Bdt:SgEvt1:4:m1pix}   {\ensuremath{{4 } } }
\vdef{default-11Bdt:SgEvt1:4:m2pix}   {\ensuremath{{3 } } }
\vdef{default-11Bdt:SgEvt1:4:m1bpix}   {\ensuremath{{2 } } }
\vdef{default-11Bdt:SgEvt1:4:m2bpix}   {\ensuremath{{3 } } }
\vdef{default-11Bdt:SgEvt1:4:m1bpixl1}   {\ensuremath{{1 } } }
\vdef{default-11Bdt:SgEvt1:4:m2bpixl1}   {\ensuremath{{1 } } }
\vdef{default-11Bdt:BsSgEvt0:2:run}   {\ensuremath{{172268 } } }
\vdef{default-11Bdt:BsSgEvt0:2:evt}   {\ensuremath{{69388599 } } }
\vdef{default-11Bdt:BsSgEvt0:2:chan}   {\ensuremath{{0 } } }
\vdef{default-11Bdt:BsSgEvt0:2:m}   {\ensuremath{{5.325 } } }
\vdef{default-11Bdt:BsSgEvt0:2:pt}   {\ensuremath{{36.115 } } }
\vdef{default-11Bdt:BsSgEvt0:2:phi}   {\ensuremath{{-1.714 } } }
\vdef{default-11Bdt:BsSgEvt0:2:eta}   {\ensuremath{{0.049 } } }
\vdef{default-11Bdt:BsSgEvt0:2:channel}   {barrel }
\vdef{default-11Bdt:BsSgEvt0:2:cowboy}   {\ensuremath{{1 } } }
\vdef{default-11Bdt:BsSgEvt0:2:m1pt}   {\ensuremath{{28.910 } } }
\vdef{default-11Bdt:BsSgEvt0:2:m2pt}   {\ensuremath{{7.361 } } }
\vdef{default-11Bdt:BsSgEvt0:2:m1eta}   {\ensuremath{{0.112 } } }
\vdef{default-11Bdt:BsSgEvt0:2:m2eta}   {\ensuremath{{-0.198 } } }
\vdef{default-11Bdt:BsSgEvt0:2:m1phi}   {\ensuremath{{-1.675 } } }
\vdef{default-11Bdt:BsSgEvt0:2:m2phi}   {\ensuremath{{-1.868 } } }
\vdef{default-11Bdt:BsSgEvt0:2:m1q}   {\ensuremath{{1 } } }
\vdef{default-11Bdt:BsSgEvt0:2:m2q}   {\ensuremath{{-1 } } }
\vdef{default-11Bdt:BsSgEvt0:2:iso}   {\ensuremath{{0.870 } } }
\vdef{default-11Bdt:BsSgEvt0:2:alpha}   {\ensuremath{{0.0149 } } }
\vdef{default-11Bdt:BsSgEvt0:2:chi2}   {\ensuremath{{ 0.92 } } }
\vdef{default-11Bdt:BsSgEvt0:2:dof}   {\ensuremath{{1 } } }
\vdef{default-11Bdt:BsSgEvt0:2:fls3d}   {\ensuremath{{30.24 } } }
\vdef{default-11Bdt:BsSgEvt0:2:fl3d}   {\ensuremath{{0.3296 } } }
\vdef{default-11Bdt:BsSgEvt0:2:fl3dE}   {\ensuremath{{0.0109 } } }
\vdef{default-11Bdt:BsSgEvt0:2:docatrk}   {\ensuremath{{0.0645 } } }
\vdef{default-11Bdt:BsSgEvt0:2:closetrk}   {\ensuremath{{0 } } }
\vdef{default-11Bdt:BsSgEvt0:2:lip}   {\ensuremath{{0.0047 } } }
\vdef{default-11Bdt:BsSgEvt0:2:lipE}   {\ensuremath{{0.0032 } } }
\vdef{default-11Bdt:BsSgEvt0:2:tip}   {\ensuremath{{0.0013 } } }
\vdef{default-11Bdt:BsSgEvt0:2:tipE}   {\ensuremath{{0.0035 } } }
\vdef{default-11Bdt:BsSgEvt0:2:pvlip}   {\ensuremath{{0.0047 } } }
\vdef{default-11Bdt:BsSgEvt0:2:pvlips}   {\ensuremath{{1.4832 } } }
\vdef{default-11Bdt:BsSgEvt0:2:pvip}   {\ensuremath{{0.0049 } } }
\vdef{default-11Bdt:BsSgEvt0:2:pvips}   {\ensuremath{{1.5237 } } }
\vdef{default-11Bdt:BsSgEvt0:2:maxdoca}   {\ensuremath{{0.0034 } } }
\vdef{default-11Bdt:BsSgEvt0:2:pvw8}   {\ensuremath{{0.8360 } } }
\vdef{default-11Bdt:BsSgEvt0:2:bdt}   {\ensuremath{{0.4904 } } }
\vdef{default-11Bdt:BsSgEvt0:2:m1pix}   {\ensuremath{{3 } } }
\vdef{default-11Bdt:BsSgEvt0:2:m2pix}   {\ensuremath{{3 } } }
\vdef{default-11Bdt:BsSgEvt0:2:m1bpix}   {\ensuremath{{3 } } }
\vdef{default-11Bdt:BsSgEvt0:2:m2bpix}   {\ensuremath{{3 } } }
\vdef{default-11Bdt:BsSgEvt0:2:m1bpixl1}   {\ensuremath{{1 } } }
\vdef{default-11Bdt:BsSgEvt0:2:m2bpixl1}   {\ensuremath{{1 } } }
\vdef{default-11Bdt:SgEvt1:5:run}   {\ensuremath{{172033 } } }
\vdef{default-11Bdt:SgEvt1:5:evt}   {\ensuremath{{830147524 } } }
\vdef{default-11Bdt:SgEvt1:5:chan}   {\ensuremath{{1 } } }
\vdef{default-11Bdt:SgEvt1:5:m}   {\ensuremath{{5.896 } } }
\vdef{default-11Bdt:SgEvt1:5:pt}   {\ensuremath{{21.839 } } }
\vdef{default-11Bdt:SgEvt1:5:phi}   {\ensuremath{{-0.520 } } }
\vdef{default-11Bdt:SgEvt1:5:eta}   {\ensuremath{{-1.138 } } }
\vdef{default-11Bdt:SgEvt1:5:channel}   {endcap }
\vdef{default-11Bdt:SgEvt1:5:cowboy}   {\ensuremath{{0 } } }
\vdef{default-11Bdt:SgEvt1:5:m1pt}   {\ensuremath{{15.275 } } }
\vdef{default-11Bdt:SgEvt1:5:m2pt}   {\ensuremath{{6.551 } } }
\vdef{default-11Bdt:SgEvt1:5:m1eta}   {\ensuremath{{-0.933 } } }
\vdef{default-11Bdt:SgEvt1:5:m2eta}   {\ensuremath{{-1.511 } } }
\vdef{default-11Bdt:SgEvt1:5:m1phi}   {\ensuremath{{-0.505 } } }
\vdef{default-11Bdt:SgEvt1:5:m2phi}   {\ensuremath{{-0.555 } } }
\vdef{default-11Bdt:SgEvt1:5:m1q}   {\ensuremath{{-1 } } }
\vdef{default-11Bdt:SgEvt1:5:m2q}   {\ensuremath{{1 } } }
\vdef{default-11Bdt:SgEvt1:5:iso}   {\ensuremath{{1.000 } } }
\vdef{default-11Bdt:SgEvt1:5:alpha}   {\ensuremath{{0.0137 } } }
\vdef{default-11Bdt:SgEvt1:5:chi2}   {\ensuremath{{ 0.06 } } }
\vdef{default-11Bdt:SgEvt1:5:dof}   {\ensuremath{{1 } } }
\vdef{default-11Bdt:SgEvt1:5:fls3d}   {\ensuremath{{35.05 } } }
\vdef{default-11Bdt:SgEvt1:5:fl3d}   {\ensuremath{{0.5235 } } }
\vdef{default-11Bdt:SgEvt1:5:fl3dE}   {\ensuremath{{0.0149 } } }
\vdef{default-11Bdt:SgEvt1:5:docatrk}   {\ensuremath{{0.1003 } } }
\vdef{default-11Bdt:SgEvt1:5:closetrk}   {\ensuremath{{0 } } }
\vdef{default-11Bdt:SgEvt1:5:lip}   {\ensuremath{{0.0024 } } }
\vdef{default-11Bdt:SgEvt1:5:lipE}   {\ensuremath{{0.0062 } } }
\vdef{default-11Bdt:SgEvt1:5:tip}   {\ensuremath{{0.0058 } } }
\vdef{default-11Bdt:SgEvt1:5:tipE}   {\ensuremath{{0.0054 } } }
\vdef{default-11Bdt:SgEvt1:5:pvlip}   {\ensuremath{{0.0024 } } }
\vdef{default-11Bdt:SgEvt1:5:pvlips}   {\ensuremath{{0.3899 } } }
\vdef{default-11Bdt:SgEvt1:5:pvip}   {\ensuremath{{0.0063 } } }
\vdef{default-11Bdt:SgEvt1:5:pvips}   {\ensuremath{{1.1343 } } }
\vdef{default-11Bdt:SgEvt1:5:maxdoca}   {\ensuremath{{0.0012 } } }
\vdef{default-11Bdt:SgEvt1:5:pvw8}   {\ensuremath{{0.8369 } } }
\vdef{default-11Bdt:SgEvt1:5:bdt}   {\ensuremath{{0.0080 } } }
\vdef{default-11Bdt:SgEvt1:5:m1pix}   {\ensuremath{{2 } } }
\vdef{default-11Bdt:SgEvt1:5:m2pix}   {\ensuremath{{3 } } }
\vdef{default-11Bdt:SgEvt1:5:m1bpix}   {\ensuremath{{2 } } }
\vdef{default-11Bdt:SgEvt1:5:m2bpix}   {\ensuremath{{3 } } }
\vdef{default-11Bdt:SgEvt1:5:m1bpixl1}   {\ensuremath{{1 } } }
\vdef{default-11Bdt:SgEvt1:5:m2bpixl1}   {\ensuremath{{1 } } }
\vdef{default-11Bdt:BsSgEvt1:2:run}   {\ensuremath{{171446 } } }
\vdef{default-11Bdt:BsSgEvt1:2:evt}   {\ensuremath{{393652751 } } }
\vdef{default-11Bdt:BsSgEvt1:2:chan}   {\ensuremath{{1 } } }
\vdef{default-11Bdt:BsSgEvt1:2:m}   {\ensuremath{{5.333 } } }
\vdef{default-11Bdt:BsSgEvt1:2:pt}   {\ensuremath{{10.104 } } }
\vdef{default-11Bdt:BsSgEvt1:2:phi}   {\ensuremath{{-2.317 } } }
\vdef{default-11Bdt:BsSgEvt1:2:eta}   {\ensuremath{{1.431 } } }
\vdef{default-11Bdt:BsSgEvt1:2:channel}   {endcap }
\vdef{default-11Bdt:BsSgEvt1:2:cowboy}   {\ensuremath{{1 } } }
\vdef{default-11Bdt:BsSgEvt1:2:m1pt}   {\ensuremath{{6.789 } } }
\vdef{default-11Bdt:BsSgEvt1:2:m2pt}   {\ensuremath{{4.538 } } }
\vdef{default-11Bdt:BsSgEvt1:2:m1eta}   {\ensuremath{{1.216 } } }
\vdef{default-11Bdt:BsSgEvt1:2:m2eta}   {\ensuremath{{1.483 } } }
\vdef{default-11Bdt:BsSgEvt1:2:m1phi}   {\ensuremath{{-1.941 } } }
\vdef{default-11Bdt:BsSgEvt1:2:m2phi}   {\ensuremath{{-2.900 } } }
\vdef{default-11Bdt:BsSgEvt1:2:m1q}   {\ensuremath{{1 } } }
\vdef{default-11Bdt:BsSgEvt1:2:m2q}   {\ensuremath{{-1 } } }
\vdef{default-11Bdt:BsSgEvt1:2:iso}   {\ensuremath{{1.000 } } }
\vdef{default-11Bdt:BsSgEvt1:2:alpha}   {\ensuremath{{0.0103 } } }
\vdef{default-11Bdt:BsSgEvt1:2:chi2}   {\ensuremath{{ 0.13 } } }
\vdef{default-11Bdt:BsSgEvt1:2:dof}   {\ensuremath{{1 } } }
\vdef{default-11Bdt:BsSgEvt1:2:fls3d}   {\ensuremath{{37.81 } } }
\vdef{default-11Bdt:BsSgEvt1:2:fl3d}   {\ensuremath{{0.5463 } } }
\vdef{default-11Bdt:BsSgEvt1:2:fl3dE}   {\ensuremath{{0.0144 } } }
\vdef{default-11Bdt:BsSgEvt1:2:docatrk}   {\ensuremath{{0.0795 } } }
\vdef{default-11Bdt:BsSgEvt1:2:closetrk}   {\ensuremath{{0 } } }
\vdef{default-11Bdt:BsSgEvt1:2:lip}   {\ensuremath{{0.0010 } } }
\vdef{default-11Bdt:BsSgEvt1:2:lipE}   {\ensuremath{{0.0040 } } }
\vdef{default-11Bdt:BsSgEvt1:2:tip}   {\ensuremath{{0.0052 } } }
\vdef{default-11Bdt:BsSgEvt1:2:tipE}   {\ensuremath{{0.0039 } } }
\vdef{default-11Bdt:BsSgEvt1:2:pvlip}   {\ensuremath{{0.0010 } } }
\vdef{default-11Bdt:BsSgEvt1:2:pvlips}   {\ensuremath{{0.2391 } } }
\vdef{default-11Bdt:BsSgEvt1:2:pvip}   {\ensuremath{{0.0053 } } }
\vdef{default-11Bdt:BsSgEvt1:2:pvips}   {\ensuremath{{1.3586 } } }
\vdef{default-11Bdt:BsSgEvt1:2:maxdoca}   {\ensuremath{{0.0021 } } }
\vdef{default-11Bdt:BsSgEvt1:2:pvw8}   {\ensuremath{{0.8778 } } }
\vdef{default-11Bdt:BsSgEvt1:2:bdt}   {\ensuremath{{0.1607 } } }
\vdef{default-11Bdt:BsSgEvt1:2:m1pix}   {\ensuremath{{3 } } }
\vdef{default-11Bdt:BsSgEvt1:2:m2pix}   {\ensuremath{{3 } } }
\vdef{default-11Bdt:BsSgEvt1:2:m1bpix}   {\ensuremath{{3 } } }
\vdef{default-11Bdt:BsSgEvt1:2:m2bpix}   {\ensuremath{{3 } } }
\vdef{default-11Bdt:BsSgEvt1:2:m1bpixl1}   {\ensuremath{{1 } } }
\vdef{default-11Bdt:BsSgEvt1:2:m2bpixl1}   {\ensuremath{{1 } } }
\vdef{default-11Bdt:SgEvt0:2:run}   {\ensuremath{{173692 } } }
\vdef{default-11Bdt:SgEvt0:2:evt}   {\ensuremath{{744193364 } } }
\vdef{default-11Bdt:SgEvt0:2:chan}   {\ensuremath{{0 } } }
\vdef{default-11Bdt:SgEvt0:2:m}   {\ensuremath{{4.910 } } }
\vdef{default-11Bdt:SgEvt0:2:pt}   {\ensuremath{{10.674 } } }
\vdef{default-11Bdt:SgEvt0:2:phi}   {\ensuremath{{-0.749 } } }
\vdef{default-11Bdt:SgEvt0:2:eta}   {\ensuremath{{0.509 } } }
\vdef{default-11Bdt:SgEvt0:2:channel}   {barrel }
\vdef{default-11Bdt:SgEvt0:2:cowboy}   {\ensuremath{{1 } } }
\vdef{default-11Bdt:SgEvt0:2:m1pt}   {\ensuremath{{6.680 } } }
\vdef{default-11Bdt:SgEvt0:2:m2pt}   {\ensuremath{{4.500 } } }
\vdef{default-11Bdt:SgEvt0:2:m1eta}   {\ensuremath{{0.726 } } }
\vdef{default-11Bdt:SgEvt0:2:m2eta}   {\ensuremath{{0.084 } } }
\vdef{default-11Bdt:SgEvt0:2:m1phi}   {\ensuremath{{-0.997 } } }
\vdef{default-11Bdt:SgEvt0:2:m2phi}   {\ensuremath{{-0.377 } } }
\vdef{default-11Bdt:SgEvt0:2:m1q}   {\ensuremath{{-1 } } }
\vdef{default-11Bdt:SgEvt0:2:m2q}   {\ensuremath{{1 } } }
\vdef{default-11Bdt:SgEvt0:2:iso}   {\ensuremath{{1.000 } } }
\vdef{default-11Bdt:SgEvt0:2:alpha}   {\ensuremath{{0.0288 } } }
\vdef{default-11Bdt:SgEvt0:2:chi2}   {\ensuremath{{ 0.35 } } }
\vdef{default-11Bdt:SgEvt0:2:dof}   {\ensuremath{{1 } } }
\vdef{default-11Bdt:SgEvt0:2:fls3d}   {\ensuremath{{ 8.20 } } }
\vdef{default-11Bdt:SgEvt0:2:fl3d}   {\ensuremath{{0.0574 } } }
\vdef{default-11Bdt:SgEvt0:2:fl3dE}   {\ensuremath{{0.0070 } } }
\vdef{default-11Bdt:SgEvt0:2:docatrk}   {\ensuremath{{0.0260 } } }
\vdef{default-11Bdt:SgEvt0:2:closetrk}   {\ensuremath{{2 } } }
\vdef{default-11Bdt:SgEvt0:2:lip}   {\ensuremath{{-0.0009 } } }
\vdef{default-11Bdt:SgEvt0:2:lipE}   {\ensuremath{{0.0041 } } }
\vdef{default-11Bdt:SgEvt0:2:tip}   {\ensuremath{{0.0013 } } }
\vdef{default-11Bdt:SgEvt0:2:tipE}   {\ensuremath{{0.0032 } } }
\vdef{default-11Bdt:SgEvt0:2:pvlip}   {\ensuremath{{-0.0009 } } }
\vdef{default-11Bdt:SgEvt0:2:pvlips}   {\ensuremath{{-0.2101 } } }
\vdef{default-11Bdt:SgEvt0:2:pvip}   {\ensuremath{{0.0016 } } }
\vdef{default-11Bdt:SgEvt0:2:pvips}   {\ensuremath{{0.4555 } } }
\vdef{default-11Bdt:SgEvt0:2:maxdoca}   {\ensuremath{{0.0031 } } }
\vdef{default-11Bdt:SgEvt0:2:pvw8}   {\ensuremath{{0.8526 } } }
\vdef{default-11Bdt:SgEvt0:2:bdt}   {\ensuremath{{0.0972 } } }
\vdef{default-11Bdt:SgEvt0:2:m1pix}   {\ensuremath{{3 } } }
\vdef{default-11Bdt:SgEvt0:2:m2pix}   {\ensuremath{{3 } } }
\vdef{default-11Bdt:SgEvt0:2:m1bpix}   {\ensuremath{{3 } } }
\vdef{default-11Bdt:SgEvt0:2:m2bpix}   {\ensuremath{{3 } } }
\vdef{default-11Bdt:SgEvt0:2:m1bpixl1}   {\ensuremath{{1 } } }
\vdef{default-11Bdt:SgEvt0:2:m2bpixl1}   {\ensuremath{{1 } } }
\vdef{default-11Bdt:SgEvt1:6:run}   {\ensuremath{{173663 } } }
\vdef{default-11Bdt:SgEvt1:6:evt}   {\ensuremath{{7993174 } } }
\vdef{default-11Bdt:SgEvt1:6:chan}   {\ensuremath{{1 } } }
\vdef{default-11Bdt:SgEvt1:6:m}   {\ensuremath{{5.034 } } }
\vdef{default-11Bdt:SgEvt1:6:pt}   {\ensuremath{{13.279 } } }
\vdef{default-11Bdt:SgEvt1:6:phi}   {\ensuremath{{-1.641 } } }
\vdef{default-11Bdt:SgEvt1:6:eta}   {\ensuremath{{2.051 } } }
\vdef{default-11Bdt:SgEvt1:6:channel}   {endcap }
\vdef{default-11Bdt:SgEvt1:6:cowboy}   {\ensuremath{{1 } } }
\vdef{default-11Bdt:SgEvt1:6:m1pt}   {\ensuremath{{9.957 } } }
\vdef{default-11Bdt:SgEvt1:6:m2pt}   {\ensuremath{{4.140 } } }
\vdef{default-11Bdt:SgEvt1:6:m1eta}   {\ensuremath{{2.063 } } }
\vdef{default-11Bdt:SgEvt1:6:m2eta}   {\ensuremath{{1.802 } } }
\vdef{default-11Bdt:SgEvt1:6:m1phi}   {\ensuremath{{-1.856 } } }
\vdef{default-11Bdt:SgEvt1:6:m2phi}   {\ensuremath{{-1.100 } } }
\vdef{default-11Bdt:SgEvt1:6:m1q}   {\ensuremath{{-1 } } }
\vdef{default-11Bdt:SgEvt1:6:m2q}   {\ensuremath{{1 } } }
\vdef{default-11Bdt:SgEvt1:6:iso}   {\ensuremath{{1.000 } } }
\vdef{default-11Bdt:SgEvt1:6:alpha}   {\ensuremath{{0.0086 } } }
\vdef{default-11Bdt:SgEvt1:6:chi2}   {\ensuremath{{ 0.09 } } }
\vdef{default-11Bdt:SgEvt1:6:dof}   {\ensuremath{{1 } } }
\vdef{default-11Bdt:SgEvt1:6:fls3d}   {\ensuremath{{ 6.87 } } }
\vdef{default-11Bdt:SgEvt1:6:fl3d}   {\ensuremath{{0.1791 } } }
\vdef{default-11Bdt:SgEvt1:6:fl3dE}   {\ensuremath{{0.0261 } } }
\vdef{default-11Bdt:SgEvt1:6:docatrk}   {\ensuremath{{0.0610 } } }
\vdef{default-11Bdt:SgEvt1:6:closetrk}   {\ensuremath{{0 } } }
\vdef{default-11Bdt:SgEvt1:6:lip}   {\ensuremath{{0.0002 } } }
\vdef{default-11Bdt:SgEvt1:6:lipE}   {\ensuremath{{0.0030 } } }
\vdef{default-11Bdt:SgEvt1:6:tip}   {\ensuremath{{0.0013 } } }
\vdef{default-11Bdt:SgEvt1:6:tipE}   {\ensuremath{{0.0033 } } }
\vdef{default-11Bdt:SgEvt1:6:pvlip}   {\ensuremath{{0.0002 } } }
\vdef{default-11Bdt:SgEvt1:6:pvlips}   {\ensuremath{{0.0616 } } }
\vdef{default-11Bdt:SgEvt1:6:pvip}   {\ensuremath{{0.0013 } } }
\vdef{default-11Bdt:SgEvt1:6:pvips}   {\ensuremath{{0.4046 } } }
\vdef{default-11Bdt:SgEvt1:6:maxdoca}   {\ensuremath{{0.0017 } } }
\vdef{default-11Bdt:SgEvt1:6:pvw8}   {\ensuremath{{0.7919 } } }
\vdef{default-11Bdt:SgEvt1:6:bdt}   {\ensuremath{{0.2564 } } }
\vdef{default-11Bdt:SgEvt1:6:m1pix}   {\ensuremath{{3 } } }
\vdef{default-11Bdt:SgEvt1:6:m2pix}   {\ensuremath{{3 } } }
\vdef{default-11Bdt:SgEvt1:6:m1bpix}   {\ensuremath{{1 } } }
\vdef{default-11Bdt:SgEvt1:6:m2bpix}   {\ensuremath{{2 } } }
\vdef{default-11Bdt:SgEvt1:6:m1bpixl1}   {\ensuremath{{1 } } }
\vdef{default-11Bdt:SgEvt1:6:m2bpixl1}   {\ensuremath{{1 } } }
\vdef{default-11Bdt:BdSgEvt0:0:run}   {\ensuremath{{173389 } } }
\vdef{default-11Bdt:BdSgEvt0:0:evt}   {\ensuremath{{173713433 } } }
\vdef{default-11Bdt:BdSgEvt0:0:chan}   {\ensuremath{{0 } } }
\vdef{default-11Bdt:BdSgEvt0:0:m}   {\ensuremath{{5.263 } } }
\vdef{default-11Bdt:BdSgEvt0:0:pt}   {\ensuremath{{17.605 } } }
\vdef{default-11Bdt:BdSgEvt0:0:phi}   {\ensuremath{{-1.667 } } }
\vdef{default-11Bdt:BdSgEvt0:0:eta}   {\ensuremath{{0.862 } } }
\vdef{default-11Bdt:BdSgEvt0:0:channel}   {barrel }
\vdef{default-11Bdt:BdSgEvt0:0:cowboy}   {\ensuremath{{0 } } }
\vdef{default-11Bdt:BdSgEvt0:0:m1pt}   {\ensuremath{{13.428 } } }
\vdef{default-11Bdt:BdSgEvt0:0:m2pt}   {\ensuremath{{4.941 } } }
\vdef{default-11Bdt:BdSgEvt0:0:m1eta}   {\ensuremath{{0.841 } } }
\vdef{default-11Bdt:BdSgEvt0:0:m2eta}   {\ensuremath{{0.808 } } }
\vdef{default-11Bdt:BdSgEvt0:0:m1phi}   {\ensuremath{{-1.495 } } }
\vdef{default-11Bdt:BdSgEvt0:0:m2phi}   {\ensuremath{{-2.151 } } }
\vdef{default-11Bdt:BdSgEvt0:0:m1q}   {\ensuremath{{-1 } } }
\vdef{default-11Bdt:BdSgEvt0:0:m2q}   {\ensuremath{{1 } } }
\vdef{default-11Bdt:BdSgEvt0:0:iso}   {\ensuremath{{1.000 } } }
\vdef{default-11Bdt:BdSgEvt0:0:alpha}   {\ensuremath{{0.0419 } } }
\vdef{default-11Bdt:BdSgEvt0:0:chi2}   {\ensuremath{{ 1.12 } } }
\vdef{default-11Bdt:BdSgEvt0:0:dof}   {\ensuremath{{1 } } }
\vdef{default-11Bdt:BdSgEvt0:0:fls3d}   {\ensuremath{{14.76 } } }
\vdef{default-11Bdt:BdSgEvt0:0:fl3d}   {\ensuremath{{0.1837 } } }
\vdef{default-11Bdt:BdSgEvt0:0:fl3dE}   {\ensuremath{{0.0124 } } }
\vdef{default-11Bdt:BdSgEvt0:0:docatrk}   {\ensuremath{{0.0791 } } }
\vdef{default-11Bdt:BdSgEvt0:0:closetrk}   {\ensuremath{{0 } } }
\vdef{default-11Bdt:BdSgEvt0:0:lip}   {\ensuremath{{0.0049 } } }
\vdef{default-11Bdt:BdSgEvt0:0:lipE}   {\ensuremath{{0.0066 } } }
\vdef{default-11Bdt:BdSgEvt0:0:tip}   {\ensuremath{{0.0036 } } }
\vdef{default-11Bdt:BdSgEvt0:0:tipE}   {\ensuremath{{0.0052 } } }
\vdef{default-11Bdt:BdSgEvt0:0:pvlip}   {\ensuremath{{0.0049 } } }
\vdef{default-11Bdt:BdSgEvt0:0:pvlips}   {\ensuremath{{0.7350 } } }
\vdef{default-11Bdt:BdSgEvt0:0:pvip}   {\ensuremath{{0.0061 } } }
\vdef{default-11Bdt:BdSgEvt0:0:pvips}   {\ensuremath{{0.9853 } } }
\vdef{default-11Bdt:BdSgEvt0:0:maxdoca}   {\ensuremath{{0.0058 } } }
\vdef{default-11Bdt:BdSgEvt0:0:pvw8}   {\ensuremath{{0.8792 } } }
\vdef{default-11Bdt:BdSgEvt0:0:bdt}   {\ensuremath{{0.0041 } } }
\vdef{default-11Bdt:BdSgEvt0:0:m1pix}   {\ensuremath{{3 } } }
\vdef{default-11Bdt:BdSgEvt0:0:m2pix}   {\ensuremath{{3 } } }
\vdef{default-11Bdt:BdSgEvt0:0:m1bpix}   {\ensuremath{{3 } } }
\vdef{default-11Bdt:BdSgEvt0:0:m2bpix}   {\ensuremath{{3 } } }
\vdef{default-11Bdt:BdSgEvt0:0:m1bpixl1}   {\ensuremath{{1 } } }
\vdef{default-11Bdt:BdSgEvt0:0:m2bpixl1}   {\ensuremath{{1 } } }
\vdef{default-11Bdt:BsSgEvt0:3:run}   {\ensuremath{{173389 } } }
\vdef{default-11Bdt:BsSgEvt0:3:evt}   {\ensuremath{{352631363 } } }
\vdef{default-11Bdt:BsSgEvt0:3:chan}   {\ensuremath{{0 } } }
\vdef{default-11Bdt:BsSgEvt0:3:m}   {\ensuremath{{5.381 } } }
\vdef{default-11Bdt:BsSgEvt0:3:pt}   {\ensuremath{{10.929 } } }
\vdef{default-11Bdt:BsSgEvt0:3:phi}   {\ensuremath{{-0.662 } } }
\vdef{default-11Bdt:BsSgEvt0:3:eta}   {\ensuremath{{-0.039 } } }
\vdef{default-11Bdt:BsSgEvt0:3:channel}   {barrel }
\vdef{default-11Bdt:BsSgEvt0:3:cowboy}   {\ensuremath{{1 } } }
\vdef{default-11Bdt:BsSgEvt0:3:m1pt}   {\ensuremath{{5.757 } } }
\vdef{default-11Bdt:BsSgEvt0:3:m2pt}   {\ensuremath{{5.693 } } }
\vdef{default-11Bdt:BsSgEvt0:3:m1eta}   {\ensuremath{{-0.382 } } }
\vdef{default-11Bdt:BsSgEvt0:3:m2eta}   {\ensuremath{{0.315 } } }
\vdef{default-11Bdt:BsSgEvt0:3:m1phi}   {\ensuremath{{-0.970 } } }
\vdef{default-11Bdt:BsSgEvt0:3:m2phi}   {\ensuremath{{-0.350 } } }
\vdef{default-11Bdt:BsSgEvt0:3:m1q}   {\ensuremath{{-1 } } }
\vdef{default-11Bdt:BsSgEvt0:3:m2q}   {\ensuremath{{1 } } }
\vdef{default-11Bdt:BsSgEvt0:3:iso}   {\ensuremath{{1.000 } } }
\vdef{default-11Bdt:BsSgEvt0:3:alpha}   {\ensuremath{{0.0073 } } }
\vdef{default-11Bdt:BsSgEvt0:3:chi2}   {\ensuremath{{ 2.87 } } }
\vdef{default-11Bdt:BsSgEvt0:3:dof}   {\ensuremath{{1 } } }
\vdef{default-11Bdt:BsSgEvt0:3:fls3d}   {\ensuremath{{29.55 } } }
\vdef{default-11Bdt:BsSgEvt0:3:fl3d}   {\ensuremath{{0.1738 } } }
\vdef{default-11Bdt:BsSgEvt0:3:fl3dE}   {\ensuremath{{0.0059 } } }
\vdef{default-11Bdt:BsSgEvt0:3:docatrk}   {\ensuremath{{0.0652 } } }
\vdef{default-11Bdt:BsSgEvt0:3:closetrk}   {\ensuremath{{0 } } }
\vdef{default-11Bdt:BsSgEvt0:3:lip}   {\ensuremath{{-0.0013 } } }
\vdef{default-11Bdt:BsSgEvt0:3:lipE}   {\ensuremath{{0.0051 } } }
\vdef{default-11Bdt:BsSgEvt0:3:tip}   {\ensuremath{{0.0002 } } }
\vdef{default-11Bdt:BsSgEvt0:3:tipE}   {\ensuremath{{0.0036 } } }
\vdef{default-11Bdt:BsSgEvt0:3:pvlip}   {\ensuremath{{-0.0013 } } }
\vdef{default-11Bdt:BsSgEvt0:3:pvlips}   {\ensuremath{{-0.2445 } } }
\vdef{default-11Bdt:BsSgEvt0:3:pvip}   {\ensuremath{{0.0013 } } }
\vdef{default-11Bdt:BsSgEvt0:3:pvips}   {\ensuremath{{0.2474 } } }
\vdef{default-11Bdt:BsSgEvt0:3:maxdoca}   {\ensuremath{{0.0072 } } }
\vdef{default-11Bdt:BsSgEvt0:3:pvw8}   {\ensuremath{{0.8971 } } }
\vdef{default-11Bdt:BsSgEvt0:3:bdt}   {\ensuremath{{0.0591 } } }
\vdef{default-11Bdt:BsSgEvt0:3:m1pix}   {\ensuremath{{3 } } }
\vdef{default-11Bdt:BsSgEvt0:3:m2pix}   {\ensuremath{{2 } } }
\vdef{default-11Bdt:BsSgEvt0:3:m1bpix}   {\ensuremath{{3 } } }
\vdef{default-11Bdt:BsSgEvt0:3:m2bpix}   {\ensuremath{{2 } } }
\vdef{default-11Bdt:BsSgEvt0:3:m1bpixl1}   {\ensuremath{{1 } } }
\vdef{default-11Bdt:BsSgEvt0:3:m2bpixl1}   {\ensuremath{{1 } } }
\vdef{default-11Bdt:SgEvt1:7:run}   {\ensuremath{{172868 } } }
\vdef{default-11Bdt:SgEvt1:7:evt}   {\ensuremath{{1362756879 } } }
\vdef{default-11Bdt:SgEvt1:7:chan}   {\ensuremath{{1 } } }
\vdef{default-11Bdt:SgEvt1:7:m}   {\ensuremath{{5.853 } } }
\vdef{default-11Bdt:SgEvt1:7:pt}   {\ensuremath{{11.512 } } }
\vdef{default-11Bdt:SgEvt1:7:phi}   {\ensuremath{{0.710 } } }
\vdef{default-11Bdt:SgEvt1:7:eta}   {\ensuremath{{2.078 } } }
\vdef{default-11Bdt:SgEvt1:7:channel}   {endcap }
\vdef{default-11Bdt:SgEvt1:7:cowboy}   {\ensuremath{{1 } } }
\vdef{default-11Bdt:SgEvt1:7:m1pt}   {\ensuremath{{6.958 } } }
\vdef{default-11Bdt:SgEvt1:7:m2pt}   {\ensuremath{{5.431 } } }
\vdef{default-11Bdt:SgEvt1:7:m1eta}   {\ensuremath{{1.719 } } }
\vdef{default-11Bdt:SgEvt1:7:m2eta}   {\ensuremath{{2.282 } } }
\vdef{default-11Bdt:SgEvt1:7:m1phi}   {\ensuremath{{0.372 } } }
\vdef{default-11Bdt:SgEvt1:7:m2phi}   {\ensuremath{{1.151 } } }
\vdef{default-11Bdt:SgEvt1:7:m1q}   {\ensuremath{{-1 } } }
\vdef{default-11Bdt:SgEvt1:7:m2q}   {\ensuremath{{1 } } }
\vdef{default-11Bdt:SgEvt1:7:iso}   {\ensuremath{{1.000 } } }
\vdef{default-11Bdt:SgEvt1:7:alpha}   {\ensuremath{{0.0134 } } }
\vdef{default-11Bdt:SgEvt1:7:chi2}   {\ensuremath{{ 2.61 } } }
\vdef{default-11Bdt:SgEvt1:7:dof}   {\ensuremath{{1 } } }
\vdef{default-11Bdt:SgEvt1:7:fls3d}   {\ensuremath{{19.29 } } }
\vdef{default-11Bdt:SgEvt1:7:fl3d}   {\ensuremath{{0.3726 } } }
\vdef{default-11Bdt:SgEvt1:7:fl3dE}   {\ensuremath{{0.0193 } } }
\vdef{default-11Bdt:SgEvt1:7:docatrk}   {\ensuremath{{0.0672 } } }
\vdef{default-11Bdt:SgEvt1:7:closetrk}   {\ensuremath{{0 } } }
\vdef{default-11Bdt:SgEvt1:7:lip}   {\ensuremath{{-0.0008 } } }
\vdef{default-11Bdt:SgEvt1:7:lipE}   {\ensuremath{{0.0029 } } }
\vdef{default-11Bdt:SgEvt1:7:tip}   {\ensuremath{{0.0037 } } }
\vdef{default-11Bdt:SgEvt1:7:tipE}   {\ensuremath{{0.0033 } } }
\vdef{default-11Bdt:SgEvt1:7:pvlip}   {\ensuremath{{-0.0008 } } }
\vdef{default-11Bdt:SgEvt1:7:pvlips}   {\ensuremath{{-0.2839 } } }
\vdef{default-11Bdt:SgEvt1:7:pvip}   {\ensuremath{{0.0038 } } }
\vdef{default-11Bdt:SgEvt1:7:pvips}   {\ensuremath{{1.1674 } } }
\vdef{default-11Bdt:SgEvt1:7:maxdoca}   {\ensuremath{{0.0072 } } }
\vdef{default-11Bdt:SgEvt1:7:pvw8}   {\ensuremath{{0.8512 } } }
\vdef{default-11Bdt:SgEvt1:7:bdt}   {\ensuremath{{0.0724 } } }
\vdef{default-11Bdt:SgEvt1:7:m1pix}   {\ensuremath{{3 } } }
\vdef{default-11Bdt:SgEvt1:7:m2pix}   {\ensuremath{{3 } } }
\vdef{default-11Bdt:SgEvt1:7:m1bpix}   {\ensuremath{{2 } } }
\vdef{default-11Bdt:SgEvt1:7:m2bpix}   {\ensuremath{{1 } } }
\vdef{default-11Bdt:SgEvt1:7:m1bpixl1}   {\ensuremath{{1 } } }
\vdef{default-11Bdt:SgEvt1:7:m2bpixl1}   {\ensuremath{{1 } } }
\vdef{default-11Bdt:SgEvt1:8:run}   {\ensuremath{{172791 } } }
\vdef{default-11Bdt:SgEvt1:8:evt}   {\ensuremath{{-2105436800 } } }
\vdef{default-11Bdt:SgEvt1:8:chan}   {\ensuremath{{1 } } }
\vdef{default-11Bdt:SgEvt1:8:m}   {\ensuremath{{5.067 } } }
\vdef{default-11Bdt:SgEvt1:8:pt}   {\ensuremath{{15.107 } } }
\vdef{default-11Bdt:SgEvt1:8:phi}   {\ensuremath{{0.890 } } }
\vdef{default-11Bdt:SgEvt1:8:eta}   {\ensuremath{{-1.506 } } }
\vdef{default-11Bdt:SgEvt1:8:channel}   {endcap }
\vdef{default-11Bdt:SgEvt1:8:cowboy}   {\ensuremath{{0 } } }
\vdef{default-11Bdt:SgEvt1:8:m1pt}   {\ensuremath{{9.506 } } }
\vdef{default-11Bdt:SgEvt1:8:m2pt}   {\ensuremath{{6.221 } } }
\vdef{default-11Bdt:SgEvt1:8:m1eta}   {\ensuremath{{-1.340 } } }
\vdef{default-11Bdt:SgEvt1:8:m2eta}   {\ensuremath{{-1.639 } } }
\vdef{default-11Bdt:SgEvt1:8:m1phi}   {\ensuremath{{0.658 } } }
\vdef{default-11Bdt:SgEvt1:8:m2phi}   {\ensuremath{{1.250 } } }
\vdef{default-11Bdt:SgEvt1:8:m1q}   {\ensuremath{{1 } } }
\vdef{default-11Bdt:SgEvt1:8:m2q}   {\ensuremath{{-1 } } }
\vdef{default-11Bdt:SgEvt1:8:iso}   {\ensuremath{{0.918 } } }
\vdef{default-11Bdt:SgEvt1:8:alpha}   {\ensuremath{{0.0153 } } }
\vdef{default-11Bdt:SgEvt1:8:chi2}   {\ensuremath{{ 1.95 } } }
\vdef{default-11Bdt:SgEvt1:8:dof}   {\ensuremath{{1 } } }
\vdef{default-11Bdt:SgEvt1:8:fls3d}   {\ensuremath{{52.93 } } }
\vdef{default-11Bdt:SgEvt1:8:fl3d}   {\ensuremath{{0.7021 } } }
\vdef{default-11Bdt:SgEvt1:8:fl3dE}   {\ensuremath{{0.0133 } } }
\vdef{default-11Bdt:SgEvt1:8:docatrk}   {\ensuremath{{0.0418 } } }
\vdef{default-11Bdt:SgEvt1:8:closetrk}   {\ensuremath{{0 } } }
\vdef{default-11Bdt:SgEvt1:8:lip}   {\ensuremath{{-0.0011 } } }
\vdef{default-11Bdt:SgEvt1:8:lipE}   {\ensuremath{{0.0025 } } }
\vdef{default-11Bdt:SgEvt1:8:tip}   {\ensuremath{{0.0105 } } }
\vdef{default-11Bdt:SgEvt1:8:tipE}   {\ensuremath{{0.0031 } } }
\vdef{default-11Bdt:SgEvt1:8:pvlip}   {\ensuremath{{-0.0011 } } }
\vdef{default-11Bdt:SgEvt1:8:pvlips}   {\ensuremath{{-0.4362 } } }
\vdef{default-11Bdt:SgEvt1:8:pvip}   {\ensuremath{{0.0105 } } }
\vdef{default-11Bdt:SgEvt1:8:pvips}   {\ensuremath{{3.4109 } } }
\vdef{default-11Bdt:SgEvt1:8:maxdoca}   {\ensuremath{{0.0049 } } }
\vdef{default-11Bdt:SgEvt1:8:pvw8}   {\ensuremath{{0.8830 } } }
\vdef{default-11Bdt:SgEvt1:8:bdt}   {\ensuremath{{-0.0512 } } }
\vdef{default-11Bdt:SgEvt1:8:m1pix}   {\ensuremath{{2 } } }
\vdef{default-11Bdt:SgEvt1:8:m2pix}   {\ensuremath{{3 } } }
\vdef{default-11Bdt:SgEvt1:8:m1bpix}   {\ensuremath{{2 } } }
\vdef{default-11Bdt:SgEvt1:8:m2bpix}   {\ensuremath{{2 } } }
\vdef{default-11Bdt:SgEvt1:8:m1bpixl1}   {\ensuremath{{1 } } }
\vdef{default-11Bdt:SgEvt1:8:m2bpixl1}   {\ensuremath{{1 } } }
\vdef{default-11Bdt:SgEvt0:3:run}   {\ensuremath{{180250 } } }
\vdef{default-11Bdt:SgEvt0:3:evt}   {\ensuremath{{588378945 } } }
\vdef{default-11Bdt:SgEvt0:3:chan}   {\ensuremath{{0 } } }
\vdef{default-11Bdt:SgEvt0:3:m}   {\ensuremath{{5.086 } } }
\vdef{default-11Bdt:SgEvt0:3:pt}   {\ensuremath{{11.663 } } }
\vdef{default-11Bdt:SgEvt0:3:phi}   {\ensuremath{{3.113 } } }
\vdef{default-11Bdt:SgEvt0:3:eta}   {\ensuremath{{1.012 } } }
\vdef{default-11Bdt:SgEvt0:3:channel}   {barrel }
\vdef{default-11Bdt:SgEvt0:3:cowboy}   {\ensuremath{{0 } } }
\vdef{default-11Bdt:SgEvt0:3:m1pt}   {\ensuremath{{6.606 } } }
\vdef{default-11Bdt:SgEvt0:3:m2pt}   {\ensuremath{{5.500 } } }
\vdef{default-11Bdt:SgEvt0:3:m1eta}   {\ensuremath{{0.662 } } }
\vdef{default-11Bdt:SgEvt0:3:m2eta}   {\ensuremath{{1.286 } } }
\vdef{default-11Bdt:SgEvt0:3:m1phi}   {\ensuremath{{-2.917 } } }
\vdef{default-11Bdt:SgEvt0:3:m2phi}   {\ensuremath{{2.806 } } }
\vdef{default-11Bdt:SgEvt0:3:m1q}   {\ensuremath{{-1 } } }
\vdef{default-11Bdt:SgEvt0:3:m2q}   {\ensuremath{{1 } } }
\vdef{default-11Bdt:SgEvt0:3:iso}   {\ensuremath{{1.000 } } }
\vdef{default-11Bdt:SgEvt0:3:alpha}   {\ensuremath{{0.0336 } } }
\vdef{default-11Bdt:SgEvt0:3:chi2}   {\ensuremath{{ 2.30 } } }
\vdef{default-11Bdt:SgEvt0:3:dof}   {\ensuremath{{1 } } }
\vdef{default-11Bdt:SgEvt0:3:fls3d}   {\ensuremath{{33.11 } } }
\vdef{default-11Bdt:SgEvt0:3:fl3d}   {\ensuremath{{0.2910 } } }
\vdef{default-11Bdt:SgEvt0:3:fl3dE}   {\ensuremath{{0.0088 } } }
\vdef{default-11Bdt:SgEvt0:3:docatrk}   {\ensuremath{{0.0415 } } }
\vdef{default-11Bdt:SgEvt0:3:closetrk}   {\ensuremath{{0 } } }
\vdef{default-11Bdt:SgEvt0:3:lip}   {\ensuremath{{-0.0057 } } }
\vdef{default-11Bdt:SgEvt0:3:lipE}   {\ensuremath{{0.0025 } } }
\vdef{default-11Bdt:SgEvt0:3:tip}   {\ensuremath{{0.0042 } } }
\vdef{default-11Bdt:SgEvt0:3:tipE}   {\ensuremath{{0.0029 } } }
\vdef{default-11Bdt:SgEvt0:3:pvlip}   {\ensuremath{{-0.0057 } } }
\vdef{default-11Bdt:SgEvt0:3:pvlips}   {\ensuremath{{-2.2714 } } }
\vdef{default-11Bdt:SgEvt0:3:pvip}   {\ensuremath{{0.0071 } } }
\vdef{default-11Bdt:SgEvt0:3:pvips}   {\ensuremath{{2.6599 } } }
\vdef{default-11Bdt:SgEvt0:3:maxdoca}   {\ensuremath{{0.0068 } } }
\vdef{default-11Bdt:SgEvt0:3:pvw8}   {\ensuremath{{0.9195 } } }
\vdef{default-11Bdt:SgEvt0:3:bdt}   {\ensuremath{{0.0141 } } }
\vdef{default-11Bdt:SgEvt0:3:m1pix}   {\ensuremath{{3 } } }
\vdef{default-11Bdt:SgEvt0:3:m2pix}   {\ensuremath{{3 } } }
\vdef{default-11Bdt:SgEvt0:3:m1bpix}   {\ensuremath{{3 } } }
\vdef{default-11Bdt:SgEvt0:3:m2bpix}   {\ensuremath{{3 } } }
\vdef{default-11Bdt:SgEvt0:3:m1bpixl1}   {\ensuremath{{1 } } }
\vdef{default-11Bdt:SgEvt0:3:m2bpixl1}   {\ensuremath{{1 } } }
\vdef{default-11Bdt:BsSgEvt1:3:run}   {\ensuremath{{179889 } } }
\vdef{default-11Bdt:BsSgEvt1:3:evt}   {\ensuremath{{533479508 } } }
\vdef{default-11Bdt:BsSgEvt1:3:chan}   {\ensuremath{{1 } } }
\vdef{default-11Bdt:BsSgEvt1:3:m}   {\ensuremath{{5.311 } } }
\vdef{default-11Bdt:BsSgEvt1:3:pt}   {\ensuremath{{14.914 } } }
\vdef{default-11Bdt:BsSgEvt1:3:phi}   {\ensuremath{{0.571 } } }
\vdef{default-11Bdt:BsSgEvt1:3:eta}   {\ensuremath{{-1.532 } } }
\vdef{default-11Bdt:BsSgEvt1:3:channel}   {endcap }
\vdef{default-11Bdt:BsSgEvt1:3:cowboy}   {\ensuremath{{0 } } }
\vdef{default-11Bdt:BsSgEvt1:3:m1pt}   {\ensuremath{{8.050 } } }
\vdef{default-11Bdt:BsSgEvt1:3:m2pt}   {\ensuremath{{7.620 } } }
\vdef{default-11Bdt:BsSgEvt1:3:m1eta}   {\ensuremath{{-1.616 } } }
\vdef{default-11Bdt:BsSgEvt1:3:m2eta}   {\ensuremath{{-1.333 } } }
\vdef{default-11Bdt:BsSgEvt1:3:m1phi}   {\ensuremath{{0.875 } } }
\vdef{default-11Bdt:BsSgEvt1:3:m2phi}   {\ensuremath{{0.250 } } }
\vdef{default-11Bdt:BsSgEvt1:3:m1q}   {\ensuremath{{-1 } } }
\vdef{default-11Bdt:BsSgEvt1:3:m2q}   {\ensuremath{{1 } } }
\vdef{default-11Bdt:BsSgEvt1:3:iso}   {\ensuremath{{1.000 } } }
\vdef{default-11Bdt:BsSgEvt1:3:alpha}   {\ensuremath{{0.0158 } } }
\vdef{default-11Bdt:BsSgEvt1:3:chi2}   {\ensuremath{{ 0.05 } } }
\vdef{default-11Bdt:BsSgEvt1:3:dof}   {\ensuremath{{1 } } }
\vdef{default-11Bdt:BsSgEvt1:3:fls3d}   {\ensuremath{{18.99 } } }
\vdef{default-11Bdt:BsSgEvt1:3:fl3d}   {\ensuremath{{0.2519 } } }
\vdef{default-11Bdt:BsSgEvt1:3:fl3dE}   {\ensuremath{{0.0133 } } }
\vdef{default-11Bdt:BsSgEvt1:3:docatrk}   {\ensuremath{{0.0486 } } }
\vdef{default-11Bdt:BsSgEvt1:3:closetrk}   {\ensuremath{{0 } } }
\vdef{default-11Bdt:BsSgEvt1:3:lip}   {\ensuremath{{-0.0014 } } }
\vdef{default-11Bdt:BsSgEvt1:3:lipE}   {\ensuremath{{0.0034 } } }
\vdef{default-11Bdt:BsSgEvt1:3:tip}   {\ensuremath{{0.0021 } } }
\vdef{default-11Bdt:BsSgEvt1:3:tipE}   {\ensuremath{{0.0032 } } }
\vdef{default-11Bdt:BsSgEvt1:3:pvlip}   {\ensuremath{{-0.0014 } } }
\vdef{default-11Bdt:BsSgEvt1:3:pvlips}   {\ensuremath{{-0.4098 } } }
\vdef{default-11Bdt:BsSgEvt1:3:pvip}   {\ensuremath{{0.0025 } } }
\vdef{default-11Bdt:BsSgEvt1:3:pvips}   {\ensuremath{{0.7730 } } }
\vdef{default-11Bdt:BsSgEvt1:3:maxdoca}   {\ensuremath{{0.0008 } } }
\vdef{default-11Bdt:BsSgEvt1:3:pvw8}   {\ensuremath{{0.9004 } } }
\vdef{default-11Bdt:BsSgEvt1:3:bdt}   {\ensuremath{{0.1044 } } }
\vdef{default-11Bdt:BsSgEvt1:3:m1pix}   {\ensuremath{{3 } } }
\vdef{default-11Bdt:BsSgEvt1:3:m2pix}   {\ensuremath{{3 } } }
\vdef{default-11Bdt:BsSgEvt1:3:m1bpix}   {\ensuremath{{3 } } }
\vdef{default-11Bdt:BsSgEvt1:3:m2bpix}   {\ensuremath{{3 } } }
\vdef{default-11Bdt:BsSgEvt1:3:m1bpixl1}   {\ensuremath{{1 } } }
\vdef{default-11Bdt:BsSgEvt1:3:m2bpixl1}   {\ensuremath{{2 } } }
\vdef{default-11Bdt:BsSgEvt1:4:run}   {\ensuremath{{179434 } } }
\vdef{default-11Bdt:BsSgEvt1:4:evt}   {\ensuremath{{232993294 } } }
\vdef{default-11Bdt:BsSgEvt1:4:chan}   {\ensuremath{{1 } } }
\vdef{default-11Bdt:BsSgEvt1:4:m}   {\ensuremath{{5.411 } } }
\vdef{default-11Bdt:BsSgEvt1:4:pt}   {\ensuremath{{11.355 } } }
\vdef{default-11Bdt:BsSgEvt1:4:phi}   {\ensuremath{{-0.426 } } }
\vdef{default-11Bdt:BsSgEvt1:4:eta}   {\ensuremath{{1.546 } } }
\vdef{default-11Bdt:BsSgEvt1:4:channel}   {endcap }
\vdef{default-11Bdt:BsSgEvt1:4:cowboy}   {\ensuremath{{0 } } }
\vdef{default-11Bdt:BsSgEvt1:4:m1pt}   {\ensuremath{{6.332 } } }
\vdef{default-11Bdt:BsSgEvt1:4:m2pt}   {\ensuremath{{4.960 } } }
\vdef{default-11Bdt:BsSgEvt1:4:m1eta}   {\ensuremath{{1.863 } } }
\vdef{default-11Bdt:BsSgEvt1:4:m2eta}   {\ensuremath{{0.937 } } }
\vdef{default-11Bdt:BsSgEvt1:4:m1phi}   {\ensuremath{{-0.435 } } }
\vdef{default-11Bdt:BsSgEvt1:4:m2phi}   {\ensuremath{{-0.412 } } }
\vdef{default-11Bdt:BsSgEvt1:4:m1q}   {\ensuremath{{1 } } }
\vdef{default-11Bdt:BsSgEvt1:4:m2q}   {\ensuremath{{-1 } } }
\vdef{default-11Bdt:BsSgEvt1:4:iso}   {\ensuremath{{1.000 } } }
\vdef{default-11Bdt:BsSgEvt1:4:alpha}   {\ensuremath{{0.0129 } } }
\vdef{default-11Bdt:BsSgEvt1:4:chi2}   {\ensuremath{{ 2.27 } } }
\vdef{default-11Bdt:BsSgEvt1:4:dof}   {\ensuremath{{1 } } }
\vdef{default-11Bdt:BsSgEvt1:4:fls3d}   {\ensuremath{{91.16 } } }
\vdef{default-11Bdt:BsSgEvt1:4:fl3d}   {\ensuremath{{1.1407 } } }
\vdef{default-11Bdt:BsSgEvt1:4:fl3dE}   {\ensuremath{{0.0125 } } }
\vdef{default-11Bdt:BsSgEvt1:4:docatrk}   {\ensuremath{{0.0719 } } }
\vdef{default-11Bdt:BsSgEvt1:4:closetrk}   {\ensuremath{{0 } } }
\vdef{default-11Bdt:BsSgEvt1:4:lip}   {\ensuremath{{-0.0039 } } }
\vdef{default-11Bdt:BsSgEvt1:4:lipE}   {\ensuremath{{0.0077 } } }
\vdef{default-11Bdt:BsSgEvt1:4:tip}   {\ensuremath{{0.0112 } } }
\vdef{default-11Bdt:BsSgEvt1:4:tipE}   {\ensuremath{{0.0057 } } }
\vdef{default-11Bdt:BsSgEvt1:4:pvlip}   {\ensuremath{{-0.0039 } } }
\vdef{default-11Bdt:BsSgEvt1:4:pvlips}   {\ensuremath{{-0.4981 } } }
\vdef{default-11Bdt:BsSgEvt1:4:pvip}   {\ensuremath{{0.0119 } } }
\vdef{default-11Bdt:BsSgEvt1:4:pvips}   {\ensuremath{{1.9991 } } }
\vdef{default-11Bdt:BsSgEvt1:4:maxdoca}   {\ensuremath{{0.0083 } } }
\vdef{default-11Bdt:BsSgEvt1:4:pvw8}   {\ensuremath{{0.8368 } } }
\vdef{default-11Bdt:BsSgEvt1:4:bdt}   {\ensuremath{{0.0175 } } }
\vdef{default-11Bdt:BsSgEvt1:4:m1pix}   {\ensuremath{{3 } } }
\vdef{default-11Bdt:BsSgEvt1:4:m2pix}   {\ensuremath{{3 } } }
\vdef{default-11Bdt:BsSgEvt1:4:m1bpix}   {\ensuremath{{2 } } }
\vdef{default-11Bdt:BsSgEvt1:4:m2bpix}   {\ensuremath{{3 } } }
\vdef{default-11Bdt:BsSgEvt1:4:m1bpixl1}   {\ensuremath{{1 } } }
\vdef{default-11Bdt:BsSgEvt1:4:m2bpixl1}   {\ensuremath{{1 } } }
\vdef{default-11Bdt:BsSgEvt1:5:run}   {\ensuremath{{178920 } } }
\vdef{default-11Bdt:BsSgEvt1:5:evt}   {\ensuremath{{240281543 } } }
\vdef{default-11Bdt:BsSgEvt1:5:chan}   {\ensuremath{{1 } } }
\vdef{default-11Bdt:BsSgEvt1:5:m}   {\ensuremath{{5.372 } } }
\vdef{default-11Bdt:BsSgEvt1:5:pt}   {\ensuremath{{34.330 } } }
\vdef{default-11Bdt:BsSgEvt1:5:phi}   {\ensuremath{{-1.919 } } }
\vdef{default-11Bdt:BsSgEvt1:5:eta}   {\ensuremath{{-2.117 } } }
\vdef{default-11Bdt:BsSgEvt1:5:channel}   {endcap }
\vdef{default-11Bdt:BsSgEvt1:5:cowboy}   {\ensuremath{{0 } } }
\vdef{default-11Bdt:BsSgEvt1:5:m1pt}   {\ensuremath{{21.019 } } }
\vdef{default-11Bdt:BsSgEvt1:5:m2pt}   {\ensuremath{{13.886 } } }
\vdef{default-11Bdt:BsSgEvt1:5:m1eta}   {\ensuremath{{-2.133 } } }
\vdef{default-11Bdt:BsSgEvt1:5:m2eta}   {\ensuremath{{-2.063 } } }
\vdef{default-11Bdt:BsSgEvt1:5:m1phi}   {\ensuremath{{-2.041 } } }
\vdef{default-11Bdt:BsSgEvt1:5:m2phi}   {\ensuremath{{-1.732 } } }
\vdef{default-11Bdt:BsSgEvt1:5:m1q}   {\ensuremath{{1 } } }
\vdef{default-11Bdt:BsSgEvt1:5:m2q}   {\ensuremath{{-1 } } }
\vdef{default-11Bdt:BsSgEvt1:5:iso}   {\ensuremath{{0.936 } } }
\vdef{default-11Bdt:BsSgEvt1:5:alpha}   {\ensuremath{{0.0074 } } }
\vdef{default-11Bdt:BsSgEvt1:5:chi2}   {\ensuremath{{ 4.13 } } }
\vdef{default-11Bdt:BsSgEvt1:5:dof}   {\ensuremath{{1 } } }
\vdef{default-11Bdt:BsSgEvt1:5:fls3d}   {\ensuremath{{15.85 } } }
\vdef{default-11Bdt:BsSgEvt1:5:fl3d}   {\ensuremath{{0.6748 } } }
\vdef{default-11Bdt:BsSgEvt1:5:fl3dE}   {\ensuremath{{0.0426 } } }
\vdef{default-11Bdt:BsSgEvt1:5:docatrk}   {\ensuremath{{0.0365 } } }
\vdef{default-11Bdt:BsSgEvt1:5:closetrk}   {\ensuremath{{0 } } }
\vdef{default-11Bdt:BsSgEvt1:5:lip}   {\ensuremath{{0.0010 } } }
\vdef{default-11Bdt:BsSgEvt1:5:lipE}   {\ensuremath{{0.0019 } } }
\vdef{default-11Bdt:BsSgEvt1:5:tip}   {\ensuremath{{0.0025 } } }
\vdef{default-11Bdt:BsSgEvt1:5:tipE}   {\ensuremath{{0.0022 } } }
\vdef{default-11Bdt:BsSgEvt1:5:pvlip}   {\ensuremath{{0.0010 } } }
\vdef{default-11Bdt:BsSgEvt1:5:pvlips}   {\ensuremath{{0.5322 } } }
\vdef{default-11Bdt:BsSgEvt1:5:pvip}   {\ensuremath{{0.0027 } } }
\vdef{default-11Bdt:BsSgEvt1:5:pvips}   {\ensuremath{{1.2522 } } }
\vdef{default-11Bdt:BsSgEvt1:5:maxdoca}   {\ensuremath{{0.0066 } } }
\vdef{default-11Bdt:BsSgEvt1:5:pvw8}   {\ensuremath{{0.8157 } } }
\vdef{default-11Bdt:BsSgEvt1:5:bdt}   {\ensuremath{{-0.0650 } } }
\vdef{default-11Bdt:BsSgEvt1:5:m1pix}   {\ensuremath{{3 } } }
\vdef{default-11Bdt:BsSgEvt1:5:m2pix}   {\ensuremath{{3 } } }
\vdef{default-11Bdt:BsSgEvt1:5:m1bpix}   {\ensuremath{{1 } } }
\vdef{default-11Bdt:BsSgEvt1:5:m2bpix}   {\ensuremath{{1 } } }
\vdef{default-11Bdt:BsSgEvt1:5:m1bpixl1}   {\ensuremath{{1 } } }
\vdef{default-11Bdt:BsSgEvt1:5:m2bpixl1}   {\ensuremath{{1 } } }
\vdef{default-11Bdt:BsSgEvt1:6:run}   {\ensuremath{{163659 } } }
\vdef{default-11Bdt:BsSgEvt1:6:evt}   {\ensuremath{{32869825 } } }
\vdef{default-11Bdt:BsSgEvt1:6:chan}   {\ensuremath{{1 } } }
\vdef{default-11Bdt:BsSgEvt1:6:m}   {\ensuremath{{5.388 } } }
\vdef{default-11Bdt:BsSgEvt1:6:pt}   {\ensuremath{{8.888 } } }
\vdef{default-11Bdt:BsSgEvt1:6:phi}   {\ensuremath{{1.948 } } }
\vdef{default-11Bdt:BsSgEvt1:6:eta}   {\ensuremath{{-1.874 } } }
\vdef{default-11Bdt:BsSgEvt1:6:channel}   {endcap }
\vdef{default-11Bdt:BsSgEvt1:6:cowboy}   {\ensuremath{{1 } } }
\vdef{default-11Bdt:BsSgEvt1:6:m1pt}   {\ensuremath{{5.559 } } }
\vdef{default-11Bdt:BsSgEvt1:6:m2pt}   {\ensuremath{{4.661 } } }
\vdef{default-11Bdt:BsSgEvt1:6:m1eta}   {\ensuremath{{-1.893 } } }
\vdef{default-11Bdt:BsSgEvt1:6:m2eta}   {\ensuremath{{-1.529 } } }
\vdef{default-11Bdt:BsSgEvt1:6:m1phi}   {\ensuremath{{1.479 } } }
\vdef{default-11Bdt:BsSgEvt1:6:m2phi}   {\ensuremath{{2.517 } } }
\vdef{default-11Bdt:BsSgEvt1:6:m1q}   {\ensuremath{{-1 } } }
\vdef{default-11Bdt:BsSgEvt1:6:m2q}   {\ensuremath{{1 } } }
\vdef{default-11Bdt:BsSgEvt1:6:iso}   {\ensuremath{{0.882 } } }
\vdef{default-11Bdt:BsSgEvt1:6:alpha}   {\ensuremath{{0.0080 } } }
\vdef{default-11Bdt:BsSgEvt1:6:chi2}   {\ensuremath{{ 0.05 } } }
\vdef{default-11Bdt:BsSgEvt1:6:dof}   {\ensuremath{{1 } } }
\vdef{default-11Bdt:BsSgEvt1:6:fls3d}   {\ensuremath{{17.02 } } }
\vdef{default-11Bdt:BsSgEvt1:6:fl3d}   {\ensuremath{{0.2694 } } }
\vdef{default-11Bdt:BsSgEvt1:6:fl3dE}   {\ensuremath{{0.0158 } } }
\vdef{default-11Bdt:BsSgEvt1:6:docatrk}   {\ensuremath{{0.0088 } } }
\vdef{default-11Bdt:BsSgEvt1:6:closetrk}   {\ensuremath{{1 } } }
\vdef{default-11Bdt:BsSgEvt1:6:lip}   {\ensuremath{{-0.0005 } } }
\vdef{default-11Bdt:BsSgEvt1:6:lipE}   {\ensuremath{{0.0036 } } }
\vdef{default-11Bdt:BsSgEvt1:6:tip}   {\ensuremath{{0.0011 } } }
\vdef{default-11Bdt:BsSgEvt1:6:tipE}   {\ensuremath{{0.0048 } } }
\vdef{default-11Bdt:BsSgEvt1:6:pvlip}   {\ensuremath{{-0.0005 } } }
\vdef{default-11Bdt:BsSgEvt1:6:pvlips}   {\ensuremath{{-0.1524 } } }
\vdef{default-11Bdt:BsSgEvt1:6:pvip}   {\ensuremath{{0.0013 } } }
\vdef{default-11Bdt:BsSgEvt1:6:pvips}   {\ensuremath{{0.2788 } } }
\vdef{default-11Bdt:BsSgEvt1:6:maxdoca}   {\ensuremath{{0.0011 } } }
\vdef{default-11Bdt:BsSgEvt1:6:pvw8}   {\ensuremath{{0.9045 } } }
\vdef{default-11Bdt:BsSgEvt1:6:bdt}   {\ensuremath{{-0.0288 } } }
\vdef{default-11Bdt:BsSgEvt1:6:m1pix}   {\ensuremath{{3 } } }
\vdef{default-11Bdt:BsSgEvt1:6:m2pix}   {\ensuremath{{3 } } }
\vdef{default-11Bdt:BsSgEvt1:6:m1bpix}   {\ensuremath{{1 } } }
\vdef{default-11Bdt:BsSgEvt1:6:m2bpix}   {\ensuremath{{2 } } }
\vdef{default-11Bdt:BsSgEvt1:6:m1bpixl1}   {\ensuremath{{1 } } }
\vdef{default-11Bdt:BsSgEvt1:6:m2bpixl1}   {\ensuremath{{1 } } }
\vdef{default-11Bdt:SgEvt1:9:run}   {\ensuremath{{178871 } } }
\vdef{default-11Bdt:SgEvt1:9:evt}   {\ensuremath{{24104557 } } }
\vdef{default-11Bdt:SgEvt1:9:chan}   {\ensuremath{{1 } } }
\vdef{default-11Bdt:SgEvt1:9:m}   {\ensuremath{{5.885 } } }
\vdef{default-11Bdt:SgEvt1:9:pt}   {\ensuremath{{16.150 } } }
\vdef{default-11Bdt:SgEvt1:9:phi}   {\ensuremath{{0.975 } } }
\vdef{default-11Bdt:SgEvt1:9:eta}   {\ensuremath{{-1.873 } } }
\vdef{default-11Bdt:SgEvt1:9:channel}   {endcap }
\vdef{default-11Bdt:SgEvt1:9:cowboy}   {\ensuremath{{1 } } }
\vdef{default-11Bdt:SgEvt1:9:m1pt}   {\ensuremath{{9.525 } } }
\vdef{default-11Bdt:SgEvt1:9:m2pt}   {\ensuremath{{6.580 } } }
\vdef{default-11Bdt:SgEvt1:9:m1eta}   {\ensuremath{{-1.516 } } }
\vdef{default-11Bdt:SgEvt1:9:m2eta}   {\ensuremath{{-2.237 } } }
\vdef{default-11Bdt:SgEvt1:9:m1phi}   {\ensuremath{{1.003 } } }
\vdef{default-11Bdt:SgEvt1:9:m2phi}   {\ensuremath{{0.934 } } }
\vdef{default-11Bdt:SgEvt1:9:m1q}   {\ensuremath{{1 } } }
\vdef{default-11Bdt:SgEvt1:9:m2q}   {\ensuremath{{-1 } } }
\vdef{default-11Bdt:SgEvt1:9:iso}   {\ensuremath{{1.000 } } }
\vdef{default-11Bdt:SgEvt1:9:alpha}   {\ensuremath{{0.0106 } } }
\vdef{default-11Bdt:SgEvt1:9:chi2}   {\ensuremath{{ 0.62 } } }
\vdef{default-11Bdt:SgEvt1:9:dof}   {\ensuremath{{1 } } }
\vdef{default-11Bdt:SgEvt1:9:fls3d}   {\ensuremath{{38.30 } } }
\vdef{default-11Bdt:SgEvt1:9:fl3d}   {\ensuremath{{0.6969 } } }
\vdef{default-11Bdt:SgEvt1:9:fl3dE}   {\ensuremath{{0.0182 } } }
\vdef{default-11Bdt:SgEvt1:9:docatrk}   {\ensuremath{{0.0950 } } }
\vdef{default-11Bdt:SgEvt1:9:closetrk}   {\ensuremath{{0 } } }
\vdef{default-11Bdt:SgEvt1:9:lip}   {\ensuremath{{0.0019 } } }
\vdef{default-11Bdt:SgEvt1:9:lipE}   {\ensuremath{{0.0024 } } }
\vdef{default-11Bdt:SgEvt1:9:tip}   {\ensuremath{{0.0038 } } }
\vdef{default-11Bdt:SgEvt1:9:tipE}   {\ensuremath{{0.0029 } } }
\vdef{default-11Bdt:SgEvt1:9:pvlip}   {\ensuremath{{0.0019 } } }
\vdef{default-11Bdt:SgEvt1:9:pvlips}   {\ensuremath{{0.7960 } } }
\vdef{default-11Bdt:SgEvt1:9:pvip}   {\ensuremath{{0.0042 } } }
\vdef{default-11Bdt:SgEvt1:9:pvips}   {\ensuremath{{1.4860 } } }
\vdef{default-11Bdt:SgEvt1:9:maxdoca}   {\ensuremath{{0.0038 } } }
\vdef{default-11Bdt:SgEvt1:9:pvw8}   {\ensuremath{{0.9219 } } }
\vdef{default-11Bdt:SgEvt1:9:bdt}   {\ensuremath{{0.0624 } } }
\vdef{default-11Bdt:SgEvt1:9:m1pix}   {\ensuremath{{3 } } }
\vdef{default-11Bdt:SgEvt1:9:m2pix}   {\ensuremath{{3 } } }
\vdef{default-11Bdt:SgEvt1:9:m1bpix}   {\ensuremath{{3 } } }
\vdef{default-11Bdt:SgEvt1:9:m2bpix}   {\ensuremath{{1 } } }
\vdef{default-11Bdt:SgEvt1:9:m1bpixl1}   {\ensuremath{{1 } } }
\vdef{default-11Bdt:SgEvt1:9:m2bpixl1}   {\ensuremath{{1 } } }
\vdef{default-11Bdt:SgEvt0:4:run}   {\ensuremath{{178871 } } }
\vdef{default-11Bdt:SgEvt0:4:evt}   {\ensuremath{{199599780 } } }
\vdef{default-11Bdt:SgEvt0:4:chan}   {\ensuremath{{0 } } }
\vdef{default-11Bdt:SgEvt0:4:m}   {\ensuremath{{4.929 } } }
\vdef{default-11Bdt:SgEvt0:4:pt}   {\ensuremath{{17.592 } } }
\vdef{default-11Bdt:SgEvt0:4:phi}   {\ensuremath{{-2.149 } } }
\vdef{default-11Bdt:SgEvt0:4:eta}   {\ensuremath{{-1.150 } } }
\vdef{default-11Bdt:SgEvt0:4:channel}   {barrel }
\vdef{default-11Bdt:SgEvt0:4:cowboy}   {\ensuremath{{0 } } }
\vdef{default-11Bdt:SgEvt0:4:m1pt}   {\ensuremath{{11.554 } } }
\vdef{default-11Bdt:SgEvt0:4:m2pt}   {\ensuremath{{6.518 } } }
\vdef{default-11Bdt:SgEvt0:4:m1eta}   {\ensuremath{{-0.978 } } }
\vdef{default-11Bdt:SgEvt0:4:m2eta}   {\ensuremath{{-1.369 } } }
\vdef{default-11Bdt:SgEvt0:4:m1phi}   {\ensuremath{{-2.000 } } }
\vdef{default-11Bdt:SgEvt0:4:m2phi}   {\ensuremath{{-2.416 } } }
\vdef{default-11Bdt:SgEvt0:4:m1q}   {\ensuremath{{-1 } } }
\vdef{default-11Bdt:SgEvt0:4:m2q}   {\ensuremath{{1 } } }
\vdef{default-11Bdt:SgEvt0:4:iso}   {\ensuremath{{1.000 } } }
\vdef{default-11Bdt:SgEvt0:4:alpha}   {\ensuremath{{0.0084 } } }
\vdef{default-11Bdt:SgEvt0:4:chi2}   {\ensuremath{{ 3.97 } } }
\vdef{default-11Bdt:SgEvt0:4:dof}   {\ensuremath{{1 } } }
\vdef{default-11Bdt:SgEvt0:4:fls3d}   {\ensuremath{{18.91 } } }
\vdef{default-11Bdt:SgEvt0:4:fl3d}   {\ensuremath{{0.2735 } } }
\vdef{default-11Bdt:SgEvt0:4:fl3dE}   {\ensuremath{{0.0145 } } }
\vdef{default-11Bdt:SgEvt0:4:docatrk}   {\ensuremath{{0.0408 } } }
\vdef{default-11Bdt:SgEvt0:4:closetrk}   {\ensuremath{{0 } } }
\vdef{default-11Bdt:SgEvt0:4:lip}   {\ensuremath{{-0.0009 } } }
\vdef{default-11Bdt:SgEvt0:4:lipE}   {\ensuremath{{0.0027 } } }
\vdef{default-11Bdt:SgEvt0:4:tip}   {\ensuremath{{0.0016 } } }
\vdef{default-11Bdt:SgEvt0:4:tipE}   {\ensuremath{{0.0027 } } }
\vdef{default-11Bdt:SgEvt0:4:pvlip}   {\ensuremath{{-0.0009 } } }
\vdef{default-11Bdt:SgEvt0:4:pvlips}   {\ensuremath{{-0.3526 } } }
\vdef{default-11Bdt:SgEvt0:4:pvip}   {\ensuremath{{0.0019 } } }
\vdef{default-11Bdt:SgEvt0:4:pvips}   {\ensuremath{{0.6934 } } }
\vdef{default-11Bdt:SgEvt0:4:maxdoca}   {\ensuremath{{0.0081 } } }
\vdef{default-11Bdt:SgEvt0:4:pvw8}   {\ensuremath{{0.9344 } } }
\vdef{default-11Bdt:SgEvt0:4:bdt}   {\ensuremath{{-0.0277 } } }
\vdef{default-11Bdt:SgEvt0:4:m1pix}   {\ensuremath{{3 } } }
\vdef{default-11Bdt:SgEvt0:4:m2pix}   {\ensuremath{{3 } } }
\vdef{default-11Bdt:SgEvt0:4:m1bpix}   {\ensuremath{{3 } } }
\vdef{default-11Bdt:SgEvt0:4:m2bpix}   {\ensuremath{{2 } } }
\vdef{default-11Bdt:SgEvt0:4:m1bpixl1}   {\ensuremath{{1 } } }
\vdef{default-11Bdt:SgEvt0:4:m2bpixl1}   {\ensuremath{{1 } } }
\vdef{default-11Bdt:BdSgEvt0:1:run}   {\ensuremath{{178840 } } }
\vdef{default-11Bdt:BdSgEvt0:1:evt}   {\ensuremath{{859654315 } } }
\vdef{default-11Bdt:BdSgEvt0:1:chan}   {\ensuremath{{0 } } }
\vdef{default-11Bdt:BdSgEvt0:1:m}   {\ensuremath{{5.273 } } }
\vdef{default-11Bdt:BdSgEvt0:1:pt}   {\ensuremath{{12.959 } } }
\vdef{default-11Bdt:BdSgEvt0:1:phi}   {\ensuremath{{-1.993 } } }
\vdef{default-11Bdt:BdSgEvt0:1:eta}   {\ensuremath{{0.993 } } }
\vdef{default-11Bdt:BdSgEvt0:1:channel}   {barrel }
\vdef{default-11Bdt:BdSgEvt0:1:cowboy}   {\ensuremath{{0 } } }
\vdef{default-11Bdt:BdSgEvt0:1:m1pt}   {\ensuremath{{7.587 } } }
\vdef{default-11Bdt:BdSgEvt0:1:m2pt}   {\ensuremath{{6.286 } } }
\vdef{default-11Bdt:BdSgEvt0:1:m1eta}   {\ensuremath{{1.056 } } }
\vdef{default-11Bdt:BdSgEvt0:1:m2eta}   {\ensuremath{{0.792 } } }
\vdef{default-11Bdt:BdSgEvt0:1:m1phi}   {\ensuremath{{-1.663 } } }
\vdef{default-11Bdt:BdSgEvt0:1:m2phi}   {\ensuremath{{-2.394 } } }
\vdef{default-11Bdt:BdSgEvt0:1:m1q}   {\ensuremath{{-1 } } }
\vdef{default-11Bdt:BdSgEvt0:1:m2q}   {\ensuremath{{1 } } }
\vdef{default-11Bdt:BdSgEvt0:1:iso}   {\ensuremath{{0.925 } } }
\vdef{default-11Bdt:BdSgEvt0:1:alpha}   {\ensuremath{{0.0044 } } }
\vdef{default-11Bdt:BdSgEvt0:1:chi2}   {\ensuremath{{ 0.28 } } }
\vdef{default-11Bdt:BdSgEvt0:1:dof}   {\ensuremath{{1 } } }
\vdef{default-11Bdt:BdSgEvt0:1:fls3d}   {\ensuremath{{46.54 } } }
\vdef{default-11Bdt:BdSgEvt0:1:fl3d}   {\ensuremath{{0.4705 } } }
\vdef{default-11Bdt:BdSgEvt0:1:fl3dE}   {\ensuremath{{0.0101 } } }
\vdef{default-11Bdt:BdSgEvt0:1:docatrk}   {\ensuremath{{0.0619 } } }
\vdef{default-11Bdt:BdSgEvt0:1:closetrk}   {\ensuremath{{0 } } }
\vdef{default-11Bdt:BdSgEvt0:1:lip}   {\ensuremath{{0.0013 } } }
\vdef{default-11Bdt:BdSgEvt0:1:lipE}   {\ensuremath{{0.0053 } } }
\vdef{default-11Bdt:BdSgEvt0:1:tip}   {\ensuremath{{0.0006 } } }
\vdef{default-11Bdt:BdSgEvt0:1:tipE}   {\ensuremath{{0.0049 } } }
\vdef{default-11Bdt:BdSgEvt0:1:pvlip}   {\ensuremath{{0.0013 } } }
\vdef{default-11Bdt:BdSgEvt0:1:pvlips}   {\ensuremath{{0.2410 } } }
\vdef{default-11Bdt:BdSgEvt0:1:pvip}   {\ensuremath{{0.0014 } } }
\vdef{default-11Bdt:BdSgEvt0:1:pvips}   {\ensuremath{{0.2660 } } }
\vdef{default-11Bdt:BdSgEvt0:1:maxdoca}   {\ensuremath{{0.0025 } } }
\vdef{default-11Bdt:BdSgEvt0:1:pvw8}   {\ensuremath{{0.9211 } } }
\vdef{default-11Bdt:BdSgEvt0:1:bdt}   {\ensuremath{{0.4245 } } }
\vdef{default-11Bdt:BdSgEvt0:1:m1pix}   {\ensuremath{{3 } } }
\vdef{default-11Bdt:BdSgEvt0:1:m2pix}   {\ensuremath{{3 } } }
\vdef{default-11Bdt:BdSgEvt0:1:m1bpix}   {\ensuremath{{3 } } }
\vdef{default-11Bdt:BdSgEvt0:1:m2bpix}   {\ensuremath{{3 } } }
\vdef{default-11Bdt:BdSgEvt0:1:m1bpixl1}   {\ensuremath{{1 } } }
\vdef{default-11Bdt:BdSgEvt0:1:m2bpixl1}   {\ensuremath{{1 } } }
\vdef{default-11Bdt:SgEvt1:10:run}   {\ensuremath{{178803 } } }
\vdef{default-11Bdt:SgEvt1:10:evt}   {\ensuremath{{264105354 } } }
\vdef{default-11Bdt:SgEvt1:10:chan}   {\ensuremath{{1 } } }
\vdef{default-11Bdt:SgEvt1:10:m}   {\ensuremath{{5.103 } } }
\vdef{default-11Bdt:SgEvt1:10:pt}   {\ensuremath{{10.547 } } }
\vdef{default-11Bdt:SgEvt1:10:phi}   {\ensuremath{{3.041 } } }
\vdef{default-11Bdt:SgEvt1:10:eta}   {\ensuremath{{1.751 } } }
\vdef{default-11Bdt:SgEvt1:10:channel}   {endcap }
\vdef{default-11Bdt:SgEvt1:10:cowboy}   {\ensuremath{{1 } } }
\vdef{default-11Bdt:SgEvt1:10:m1pt}   {\ensuremath{{6.346 } } }
\vdef{default-11Bdt:SgEvt1:10:m2pt}   {\ensuremath{{5.284 } } }
\vdef{default-11Bdt:SgEvt1:10:m1eta}   {\ensuremath{{1.737 } } }
\vdef{default-11Bdt:SgEvt1:10:m2eta}   {\ensuremath{{1.553 } } }
\vdef{default-11Bdt:SgEvt1:10:m1phi}   {\ensuremath{{2.639 } } }
\vdef{default-11Bdt:SgEvt1:10:m2phi}   {\ensuremath{{-2.757 } } }
\vdef{default-11Bdt:SgEvt1:10:m1q}   {\ensuremath{{-1 } } }
\vdef{default-11Bdt:SgEvt1:10:m2q}   {\ensuremath{{1 } } }
\vdef{default-11Bdt:SgEvt1:10:iso}   {\ensuremath{{0.847 } } }
\vdef{default-11Bdt:SgEvt1:10:alpha}   {\ensuremath{{0.0189 } } }
\vdef{default-11Bdt:SgEvt1:10:chi2}   {\ensuremath{{ 3.18 } } }
\vdef{default-11Bdt:SgEvt1:10:dof}   {\ensuremath{{1 } } }
\vdef{default-11Bdt:SgEvt1:10:fls3d}   {\ensuremath{{14.99 } } }
\vdef{default-11Bdt:SgEvt1:10:fl3d}   {\ensuremath{{0.3113 } } }
\vdef{default-11Bdt:SgEvt1:10:fl3dE}   {\ensuremath{{0.0208 } } }
\vdef{default-11Bdt:SgEvt1:10:docatrk}   {\ensuremath{{0.0237 } } }
\vdef{default-11Bdt:SgEvt1:10:closetrk}   {\ensuremath{{1 } } }
\vdef{default-11Bdt:SgEvt1:10:lip}   {\ensuremath{{0.0011 } } }
\vdef{default-11Bdt:SgEvt1:10:lipE}   {\ensuremath{{0.0037 } } }
\vdef{default-11Bdt:SgEvt1:10:tip}   {\ensuremath{{0.0048 } } }
\vdef{default-11Bdt:SgEvt1:10:tipE}   {\ensuremath{{0.0037 } } }
\vdef{default-11Bdt:SgEvt1:10:pvlip}   {\ensuremath{{0.0011 } } }
\vdef{default-11Bdt:SgEvt1:10:pvlips}   {\ensuremath{{0.3032 } } }
\vdef{default-11Bdt:SgEvt1:10:pvip}   {\ensuremath{{0.0050 } } }
\vdef{default-11Bdt:SgEvt1:10:pvips}   {\ensuremath{{1.3558 } } }
\vdef{default-11Bdt:SgEvt1:10:maxdoca}   {\ensuremath{{0.0107 } } }
\vdef{default-11Bdt:SgEvt1:10:pvw8}   {\ensuremath{{0.8959 } } }
\vdef{default-11Bdt:SgEvt1:10:bdt}   {\ensuremath{{0.0452 } } }
\vdef{default-11Bdt:SgEvt1:10:m1pix}   {\ensuremath{{3 } } }
\vdef{default-11Bdt:SgEvt1:10:m2pix}   {\ensuremath{{2 } } }
\vdef{default-11Bdt:SgEvt1:10:m1bpix}   {\ensuremath{{3 } } }
\vdef{default-11Bdt:SgEvt1:10:m2bpix}   {\ensuremath{{2 } } }
\vdef{default-11Bdt:SgEvt1:10:m1bpixl1}   {\ensuremath{{1 } } }
\vdef{default-11Bdt:SgEvt1:10:m2bpixl1}   {\ensuremath{{1 } } }
\vdef{default-11Bdt:SgEvt0:5:run}   {\ensuremath{{178708 } } }
\vdef{default-11Bdt:SgEvt0:5:evt}   {\ensuremath{{214563111 } } }
\vdef{default-11Bdt:SgEvt0:5:chan}   {\ensuremath{{0 } } }
\vdef{default-11Bdt:SgEvt0:5:m}   {\ensuremath{{4.984 } } }
\vdef{default-11Bdt:SgEvt0:5:pt}   {\ensuremath{{23.709 } } }
\vdef{default-11Bdt:SgEvt0:5:phi}   {\ensuremath{{-2.593 } } }
\vdef{default-11Bdt:SgEvt0:5:eta}   {\ensuremath{{-0.494 } } }
\vdef{default-11Bdt:SgEvt0:5:channel}   {barrel }
\vdef{default-11Bdt:SgEvt0:5:cowboy}   {\ensuremath{{1 } } }
\vdef{default-11Bdt:SgEvt0:5:m1pt}   {\ensuremath{{15.039 } } }
\vdef{default-11Bdt:SgEvt0:5:m2pt}   {\ensuremath{{8.808 } } }
\vdef{default-11Bdt:SgEvt0:5:m1eta}   {\ensuremath{{-0.334 } } }
\vdef{default-11Bdt:SgEvt0:5:m2eta}   {\ensuremath{{-0.738 } } }
\vdef{default-11Bdt:SgEvt0:5:m1phi}   {\ensuremath{{-2.537 } } }
\vdef{default-11Bdt:SgEvt0:5:m2phi}   {\ensuremath{{-2.688 } } }
\vdef{default-11Bdt:SgEvt0:5:m1q}   {\ensuremath{{1 } } }
\vdef{default-11Bdt:SgEvt0:5:m2q}   {\ensuremath{{-1 } } }
\vdef{default-11Bdt:SgEvt0:5:iso}   {\ensuremath{{0.921 } } }
\vdef{default-11Bdt:SgEvt0:5:alpha}   {\ensuremath{{0.0033 } } }
\vdef{default-11Bdt:SgEvt0:5:chi2}   {\ensuremath{{ 2.19 } } }
\vdef{default-11Bdt:SgEvt0:5:dof}   {\ensuremath{{1 } } }
\vdef{default-11Bdt:SgEvt0:5:fls3d}   {\ensuremath{{32.27 } } }
\vdef{default-11Bdt:SgEvt0:5:fl3d}   {\ensuremath{{0.3267 } } }
\vdef{default-11Bdt:SgEvt0:5:fl3dE}   {\ensuremath{{0.0101 } } }
\vdef{default-11Bdt:SgEvt0:5:docatrk}   {\ensuremath{{0.0351 } } }
\vdef{default-11Bdt:SgEvt0:5:closetrk}   {\ensuremath{{0 } } }
\vdef{default-11Bdt:SgEvt0:5:lip}   {\ensuremath{{-0.0010 } } }
\vdef{default-11Bdt:SgEvt0:5:lipE}   {\ensuremath{{0.0052 } } }
\vdef{default-11Bdt:SgEvt0:5:tip}   {\ensuremath{{0.0001 } } }
\vdef{default-11Bdt:SgEvt0:5:tipE}   {\ensuremath{{0.0029 } } }
\vdef{default-11Bdt:SgEvt0:5:pvlip}   {\ensuremath{{-0.0010 } } }
\vdef{default-11Bdt:SgEvt0:5:pvlips}   {\ensuremath{{-0.1834 } } }
\vdef{default-11Bdt:SgEvt0:5:pvip}   {\ensuremath{{0.0010 } } }
\vdef{default-11Bdt:SgEvt0:5:pvips}   {\ensuremath{{0.1850 } } }
\vdef{default-11Bdt:SgEvt0:5:maxdoca}   {\ensuremath{{0.0043 } } }
\vdef{default-11Bdt:SgEvt0:5:pvw8}   {\ensuremath{{0.9372 } } }
\vdef{default-11Bdt:SgEvt0:5:bdt}   {\ensuremath{{0.0708 } } }
\vdef{default-11Bdt:SgEvt0:5:m1pix}   {\ensuremath{{3 } } }
\vdef{default-11Bdt:SgEvt0:5:m2pix}   {\ensuremath{{3 } } }
\vdef{default-11Bdt:SgEvt0:5:m1bpix}   {\ensuremath{{3 } } }
\vdef{default-11Bdt:SgEvt0:5:m2bpix}   {\ensuremath{{3 } } }
\vdef{default-11Bdt:SgEvt0:5:m1bpixl1}   {\ensuremath{{1 } } }
\vdef{default-11Bdt:SgEvt0:5:m2bpixl1}   {\ensuremath{{1 } } }
\vdef{default-11Bdt:SgEvt1:11:run}   {\ensuremath{{178421 } } }
\vdef{default-11Bdt:SgEvt1:11:evt}   {\ensuremath{{69800916 } } }
\vdef{default-11Bdt:SgEvt1:11:chan}   {\ensuremath{{1 } } }
\vdef{default-11Bdt:SgEvt1:11:m}   {\ensuremath{{5.565 } } }
\vdef{default-11Bdt:SgEvt1:11:pt}   {\ensuremath{{8.032 } } }
\vdef{default-11Bdt:SgEvt1:11:phi}   {\ensuremath{{1.055 } } }
\vdef{default-11Bdt:SgEvt1:11:eta}   {\ensuremath{{-2.037 } } }
\vdef{default-11Bdt:SgEvt1:11:channel}   {endcap }
\vdef{default-11Bdt:SgEvt1:11:cowboy}   {\ensuremath{{0 } } }
\vdef{default-11Bdt:SgEvt1:11:m1pt}   {\ensuremath{{5.228 } } }
\vdef{default-11Bdt:SgEvt1:11:m2pt}   {\ensuremath{{4.539 } } }
\vdef{default-11Bdt:SgEvt1:11:m1eta}   {\ensuremath{{-1.834 } } }
\vdef{default-11Bdt:SgEvt1:11:m2eta}   {\ensuremath{{-1.866 } } }
\vdef{default-11Bdt:SgEvt1:11:m1phi}   {\ensuremath{{0.497 } } }
\vdef{default-11Bdt:SgEvt1:11:m2phi}   {\ensuremath{{1.711 } } }
\vdef{default-11Bdt:SgEvt1:11:m1q}   {\ensuremath{{1 } } }
\vdef{default-11Bdt:SgEvt1:11:m2q}   {\ensuremath{{-1 } } }
\vdef{default-11Bdt:SgEvt1:11:iso}   {\ensuremath{{1.000 } } }
\vdef{default-11Bdt:SgEvt1:11:alpha}   {\ensuremath{{0.0159 } } }
\vdef{default-11Bdt:SgEvt1:11:chi2}   {\ensuremath{{ 0.04 } } }
\vdef{default-11Bdt:SgEvt1:11:dof}   {\ensuremath{{1 } } }
\vdef{default-11Bdt:SgEvt1:11:fls3d}   {\ensuremath{{20.35 } } }
\vdef{default-11Bdt:SgEvt1:11:fl3d}   {\ensuremath{{0.3340 } } }
\vdef{default-11Bdt:SgEvt1:11:fl3dE}   {\ensuremath{{0.0164 } } }
\vdef{default-11Bdt:SgEvt1:11:docatrk}   {\ensuremath{{0.0216 } } }
\vdef{default-11Bdt:SgEvt1:11:closetrk}   {\ensuremath{{1 } } }
\vdef{default-11Bdt:SgEvt1:11:lip}   {\ensuremath{{-0.0010 } } }
\vdef{default-11Bdt:SgEvt1:11:lipE}   {\ensuremath{{0.0034 } } }
\vdef{default-11Bdt:SgEvt1:11:tip}   {\ensuremath{{0.0035 } } }
\vdef{default-11Bdt:SgEvt1:11:tipE}   {\ensuremath{{0.0048 } } }
\vdef{default-11Bdt:SgEvt1:11:pvlip}   {\ensuremath{{-0.0010 } } }
\vdef{default-11Bdt:SgEvt1:11:pvlips}   {\ensuremath{{-0.2963 } } }
\vdef{default-11Bdt:SgEvt1:11:pvip}   {\ensuremath{{0.0037 } } }
\vdef{default-11Bdt:SgEvt1:11:pvips}   {\ensuremath{{0.7844 } } }
\vdef{default-11Bdt:SgEvt1:11:maxdoca}   {\ensuremath{{0.0010 } } }
\vdef{default-11Bdt:SgEvt1:11:pvw8}   {\ensuremath{{0.9205 } } }
\vdef{default-11Bdt:SgEvt1:11:bdt}   {\ensuremath{{-0.0226 } } }
\vdef{default-11Bdt:SgEvt1:11:m1pix}   {\ensuremath{{4 } } }
\vdef{default-11Bdt:SgEvt1:11:m2pix}   {\ensuremath{{3 } } }
\vdef{default-11Bdt:SgEvt1:11:m1bpix}   {\ensuremath{{3 } } }
\vdef{default-11Bdt:SgEvt1:11:m2bpix}   {\ensuremath{{2 } } }
\vdef{default-11Bdt:SgEvt1:11:m1bpixl1}   {\ensuremath{{1 } } }
\vdef{default-11Bdt:SgEvt1:11:m2bpixl1}   {\ensuremath{{1 } } }
\vdef{default-11Bdt:SgEvt0:6:run}   {\ensuremath{{178420 } } }
\vdef{default-11Bdt:SgEvt0:6:evt}   {\ensuremath{{32586085 } } }
\vdef{default-11Bdt:SgEvt0:6:chan}   {\ensuremath{{0 } } }
\vdef{default-11Bdt:SgEvt0:6:m}   {\ensuremath{{5.128 } } }
\vdef{default-11Bdt:SgEvt0:6:pt}   {\ensuremath{{8.291 } } }
\vdef{default-11Bdt:SgEvt0:6:phi}   {\ensuremath{{1.874 } } }
\vdef{default-11Bdt:SgEvt0:6:eta}   {\ensuremath{{-0.214 } } }
\vdef{default-11Bdt:SgEvt0:6:channel}   {barrel }
\vdef{default-11Bdt:SgEvt0:6:cowboy}   {\ensuremath{{0 } } }
\vdef{default-11Bdt:SgEvt0:6:m1pt}   {\ensuremath{{5.411 } } }
\vdef{default-11Bdt:SgEvt0:6:m2pt}   {\ensuremath{{4.064 } } }
\vdef{default-11Bdt:SgEvt0:6:m1eta}   {\ensuremath{{-0.389 } } }
\vdef{default-11Bdt:SgEvt0:6:m2eta}   {\ensuremath{{0.092 } } }
\vdef{default-11Bdt:SgEvt0:6:m1phi}   {\ensuremath{{1.443 } } }
\vdef{default-11Bdt:SgEvt0:6:m2phi}   {\ensuremath{{2.464 } } }
\vdef{default-11Bdt:SgEvt0:6:m1q}   {\ensuremath{{1 } } }
\vdef{default-11Bdt:SgEvt0:6:m2q}   {\ensuremath{{-1 } } }
\vdef{default-11Bdt:SgEvt0:6:iso}   {\ensuremath{{1.000 } } }
\vdef{default-11Bdt:SgEvt0:6:alpha}   {\ensuremath{{0.0352 } } }
\vdef{default-11Bdt:SgEvt0:6:chi2}   {\ensuremath{{ 0.17 } } }
\vdef{default-11Bdt:SgEvt0:6:dof}   {\ensuremath{{1 } } }
\vdef{default-11Bdt:SgEvt0:6:fls3d}   {\ensuremath{{52.49 } } }
\vdef{default-11Bdt:SgEvt0:6:fl3d}   {\ensuremath{{0.3102 } } }
\vdef{default-11Bdt:SgEvt0:6:fl3dE}   {\ensuremath{{0.0059 } } }
\vdef{default-11Bdt:SgEvt0:6:docatrk}   {\ensuremath{{0.0365 } } }
\vdef{default-11Bdt:SgEvt0:6:closetrk}   {\ensuremath{{0 } } }
\vdef{default-11Bdt:SgEvt0:6:lip}   {\ensuremath{{0.0080 } } }
\vdef{default-11Bdt:SgEvt0:6:lipE}   {\ensuremath{{0.0053 } } }
\vdef{default-11Bdt:SgEvt0:6:tip}   {\ensuremath{{0.0072 } } }
\vdef{default-11Bdt:SgEvt0:6:tipE}   {\ensuremath{{0.0054 } } }
\vdef{default-11Bdt:SgEvt0:6:pvlip}   {\ensuremath{{0.0080 } } }
\vdef{default-11Bdt:SgEvt0:6:pvlips}   {\ensuremath{{1.5078 } } }
\vdef{default-11Bdt:SgEvt0:6:pvip}   {\ensuremath{{0.0108 } } }
\vdef{default-11Bdt:SgEvt0:6:pvips}   {\ensuremath{{2.0167 } } }
\vdef{default-11Bdt:SgEvt0:6:maxdoca}   {\ensuremath{{0.0023 } } }
\vdef{default-11Bdt:SgEvt0:6:pvw8}   {\ensuremath{{0.8727 } } }
\vdef{default-11Bdt:SgEvt0:6:bdt}   {\ensuremath{{-0.0233 } } }
\vdef{default-11Bdt:SgEvt0:6:m1pix}   {\ensuremath{{3 } } }
\vdef{default-11Bdt:SgEvt0:6:m2pix}   {\ensuremath{{3 } } }
\vdef{default-11Bdt:SgEvt0:6:m1bpix}   {\ensuremath{{3 } } }
\vdef{default-11Bdt:SgEvt0:6:m2bpix}   {\ensuremath{{3 } } }
\vdef{default-11Bdt:SgEvt0:6:m1bpixl1}   {\ensuremath{{1 } } }
\vdef{default-11Bdt:SgEvt0:6:m2bpixl1}   {\ensuremath{{1 } } }
\vdef{default-11Bdt:SgEvt1:12:run}   {\ensuremath{{178116 } } }
\vdef{default-11Bdt:SgEvt1:12:evt}   {\ensuremath{{30653590 } } }
\vdef{default-11Bdt:SgEvt1:12:chan}   {\ensuremath{{1 } } }
\vdef{default-11Bdt:SgEvt1:12:m}   {\ensuremath{{5.508 } } }
\vdef{default-11Bdt:SgEvt1:12:pt}   {\ensuremath{{15.520 } } }
\vdef{default-11Bdt:SgEvt1:12:phi}   {\ensuremath{{-1.625 } } }
\vdef{default-11Bdt:SgEvt1:12:eta}   {\ensuremath{{1.508 } } }
\vdef{default-11Bdt:SgEvt1:12:channel}   {endcap }
\vdef{default-11Bdt:SgEvt1:12:cowboy}   {\ensuremath{{0 } } }
\vdef{default-11Bdt:SgEvt1:12:m1pt}   {\ensuremath{{11.838 } } }
\vdef{default-11Bdt:SgEvt1:12:m2pt}   {\ensuremath{{4.559 } } }
\vdef{default-11Bdt:SgEvt1:12:m1eta}   {\ensuremath{{1.517 } } }
\vdef{default-11Bdt:SgEvt1:12:m2eta}   {\ensuremath{{1.292 } } }
\vdef{default-11Bdt:SgEvt1:12:m1phi}   {\ensuremath{{-1.822 } } }
\vdef{default-11Bdt:SgEvt1:12:m2phi}   {\ensuremath{{-1.090 } } }
\vdef{default-11Bdt:SgEvt1:12:m1q}   {\ensuremath{{1 } } }
\vdef{default-11Bdt:SgEvt1:12:m2q}   {\ensuremath{{-1 } } }
\vdef{default-11Bdt:SgEvt1:12:iso}   {\ensuremath{{0.848 } } }
\vdef{default-11Bdt:SgEvt1:12:alpha}   {\ensuremath{{0.0136 } } }
\vdef{default-11Bdt:SgEvt1:12:chi2}   {\ensuremath{{ 4.98 } } }
\vdef{default-11Bdt:SgEvt1:12:dof}   {\ensuremath{{1 } } }
\vdef{default-11Bdt:SgEvt1:12:fls3d}   {\ensuremath{{30.97 } } }
\vdef{default-11Bdt:SgEvt1:12:fl3d}   {\ensuremath{{0.4400 } } }
\vdef{default-11Bdt:SgEvt1:12:fl3dE}   {\ensuremath{{0.0142 } } }
\vdef{default-11Bdt:SgEvt1:12:docatrk}   {\ensuremath{{0.0334 } } }
\vdef{default-11Bdt:SgEvt1:12:closetrk}   {\ensuremath{{0 } } }
\vdef{default-11Bdt:SgEvt1:12:lip}   {\ensuremath{{0.0017 } } }
\vdef{default-11Bdt:SgEvt1:12:lipE}   {\ensuremath{{0.0044 } } }
\vdef{default-11Bdt:SgEvt1:12:tip}   {\ensuremath{{0.0045 } } }
\vdef{default-11Bdt:SgEvt1:12:tipE}   {\ensuremath{{0.0038 } } }
\vdef{default-11Bdt:SgEvt1:12:pvlip}   {\ensuremath{{0.0017 } } }
\vdef{default-11Bdt:SgEvt1:12:pvlips}   {\ensuremath{{0.3829 } } }
\vdef{default-11Bdt:SgEvt1:12:pvip}   {\ensuremath{{0.0048 } } }
\vdef{default-11Bdt:SgEvt1:12:pvips}   {\ensuremath{{1.2293 } } }
\vdef{default-11Bdt:SgEvt1:12:maxdoca}   {\ensuremath{{0.0113 } } }
\vdef{default-11Bdt:SgEvt1:12:pvw8}   {\ensuremath{{0.8691 } } }
\vdef{default-11Bdt:SgEvt1:12:bdt}   {\ensuremath{{-0.0455 } } }
\vdef{default-11Bdt:SgEvt1:12:m1pix}   {\ensuremath{{3 } } }
\vdef{default-11Bdt:SgEvt1:12:m2pix}   {\ensuremath{{3 } } }
\vdef{default-11Bdt:SgEvt1:12:m1bpix}   {\ensuremath{{3 } } }
\vdef{default-11Bdt:SgEvt1:12:m2bpix}   {\ensuremath{{3 } } }
\vdef{default-11Bdt:SgEvt1:12:m1bpixl1}   {\ensuremath{{1 } } }
\vdef{default-11Bdt:SgEvt1:12:m2bpixl1}   {\ensuremath{{1 } } }
\vdef{default-11Bdt:BdSgEvt1:0:run}   {\ensuremath{{178100 } } }
\vdef{default-11Bdt:BdSgEvt1:0:evt}   {\ensuremath{{717632202 } } }
\vdef{default-11Bdt:BdSgEvt1:0:chan}   {\ensuremath{{1 } } }
\vdef{default-11Bdt:BdSgEvt1:0:m}   {\ensuremath{{5.208 } } }
\vdef{default-11Bdt:BdSgEvt1:0:pt}   {\ensuremath{{8.960 } } }
\vdef{default-11Bdt:BdSgEvt1:0:phi}   {\ensuremath{{-0.709 } } }
\vdef{default-11Bdt:BdSgEvt1:0:eta}   {\ensuremath{{-2.038 } } }
\vdef{default-11Bdt:BdSgEvt1:0:channel}   {endcap }
\vdef{default-11Bdt:BdSgEvt1:0:cowboy}   {\ensuremath{{0 } } }
\vdef{default-11Bdt:BdSgEvt1:0:m1pt}   {\ensuremath{{4.877 } } }
\vdef{default-11Bdt:BdSgEvt1:0:m2pt}   {\ensuremath{{4.117 } } }
\vdef{default-11Bdt:BdSgEvt1:0:m1eta}   {\ensuremath{{-2.397 } } }
\vdef{default-11Bdt:BdSgEvt1:0:m2eta}   {\ensuremath{{-1.316 } } }
\vdef{default-11Bdt:BdSgEvt1:0:m1phi}   {\ensuremath{{-0.604 } } }
\vdef{default-11Bdt:BdSgEvt1:0:m2phi}   {\ensuremath{{-0.833 } } }
\vdef{default-11Bdt:BdSgEvt1:0:m1q}   {\ensuremath{{-1 } } }
\vdef{default-11Bdt:BdSgEvt1:0:m2q}   {\ensuremath{{1 } } }
\vdef{default-11Bdt:BdSgEvt1:0:iso}   {\ensuremath{{0.864 } } }
\vdef{default-11Bdt:BdSgEvt1:0:alpha}   {\ensuremath{{0.0146 } } }
\vdef{default-11Bdt:BdSgEvt1:0:chi2}   {\ensuremath{{ 0.44 } } }
\vdef{default-11Bdt:BdSgEvt1:0:dof}   {\ensuremath{{1 } } }
\vdef{default-11Bdt:BdSgEvt1:0:fls3d}   {\ensuremath{{17.80 } } }
\vdef{default-11Bdt:BdSgEvt1:0:fl3d}   {\ensuremath{{0.3326 } } }
\vdef{default-11Bdt:BdSgEvt1:0:fl3dE}   {\ensuremath{{0.0187 } } }
\vdef{default-11Bdt:BdSgEvt1:0:docatrk}   {\ensuremath{{0.0254 } } }
\vdef{default-11Bdt:BdSgEvt1:0:closetrk}   {\ensuremath{{1 } } }
\vdef{default-11Bdt:BdSgEvt1:0:lip}   {\ensuremath{{-0.0009 } } }
\vdef{default-11Bdt:BdSgEvt1:0:lipE}   {\ensuremath{{0.0026 } } }
\vdef{default-11Bdt:BdSgEvt1:0:tip}   {\ensuremath{{0.0033 } } }
\vdef{default-11Bdt:BdSgEvt1:0:tipE}   {\ensuremath{{0.0038 } } }
\vdef{default-11Bdt:BdSgEvt1:0:pvlip}   {\ensuremath{{-0.0009 } } }
\vdef{default-11Bdt:BdSgEvt1:0:pvlips}   {\ensuremath{{-0.3608 } } }
\vdef{default-11Bdt:BdSgEvt1:0:pvip}   {\ensuremath{{0.0034 } } }
\vdef{default-11Bdt:BdSgEvt1:0:pvips}   {\ensuremath{{0.9099 } } }
\vdef{default-11Bdt:BdSgEvt1:0:maxdoca}   {\ensuremath{{0.0046 } } }
\vdef{default-11Bdt:BdSgEvt1:0:pvw8}   {\ensuremath{{0.8820 } } }
\vdef{default-11Bdt:BdSgEvt1:0:bdt}   {\ensuremath{{-0.0594 } } }
\vdef{default-11Bdt:BdSgEvt1:0:m1pix}   {\ensuremath{{3 } } }
\vdef{default-11Bdt:BdSgEvt1:0:m2pix}   {\ensuremath{{3 } } }
\vdef{default-11Bdt:BdSgEvt1:0:m1bpix}   {\ensuremath{{1 } } }
\vdef{default-11Bdt:BdSgEvt1:0:m2bpix}   {\ensuremath{{3 } } }
\vdef{default-11Bdt:BdSgEvt1:0:m1bpixl1}   {\ensuremath{{1 } } }
\vdef{default-11Bdt:BdSgEvt1:0:m2bpixl1}   {\ensuremath{{1 } } }
\vdef{default-11Bdt:SgEvt0:7:run}   {\ensuremath{{177730 } } }
\vdef{default-11Bdt:SgEvt0:7:evt}   {\ensuremath{{803697646 } } }
\vdef{default-11Bdt:SgEvt0:7:chan}   {\ensuremath{{0 } } }
\vdef{default-11Bdt:SgEvt0:7:m}   {\ensuremath{{5.533 } } }
\vdef{default-11Bdt:SgEvt0:7:pt}   {\ensuremath{{14.467 } } }
\vdef{default-11Bdt:SgEvt0:7:phi}   {\ensuremath{{0.315 } } }
\vdef{default-11Bdt:SgEvt0:7:eta}   {\ensuremath{{0.302 } } }
\vdef{default-11Bdt:SgEvt0:7:channel}   {barrel }
\vdef{default-11Bdt:SgEvt0:7:cowboy}   {\ensuremath{{0 } } }
\vdef{default-11Bdt:SgEvt0:7:m1pt}   {\ensuremath{{7.900 } } }
\vdef{default-11Bdt:SgEvt0:7:m2pt}   {\ensuremath{{6.744 } } }
\vdef{default-11Bdt:SgEvt0:7:m1eta}   {\ensuremath{{0.598 } } }
\vdef{default-11Bdt:SgEvt0:7:m2eta}   {\ensuremath{{-0.085 } } }
\vdef{default-11Bdt:SgEvt0:7:m1phi}   {\ensuremath{{0.177 } } }
\vdef{default-11Bdt:SgEvt0:7:m2phi}   {\ensuremath{{0.476 } } }
\vdef{default-11Bdt:SgEvt0:7:m1q}   {\ensuremath{{1 } } }
\vdef{default-11Bdt:SgEvt0:7:m2q}   {\ensuremath{{-1 } } }
\vdef{default-11Bdt:SgEvt0:7:iso}   {\ensuremath{{0.929 } } }
\vdef{default-11Bdt:SgEvt0:7:alpha}   {\ensuremath{{0.0619 } } }
\vdef{default-11Bdt:SgEvt0:7:chi2}   {\ensuremath{{ 0.33 } } }
\vdef{default-11Bdt:SgEvt0:7:dof}   {\ensuremath{{1 } } }
\vdef{default-11Bdt:SgEvt0:7:fls3d}   {\ensuremath{{ 8.28 } } }
\vdef{default-11Bdt:SgEvt0:7:fl3d}   {\ensuremath{{0.0684 } } }
\vdef{default-11Bdt:SgEvt0:7:fl3dE}   {\ensuremath{{0.0083 } } }
\vdef{default-11Bdt:SgEvt0:7:docatrk}   {\ensuremath{{0.0103 } } }
\vdef{default-11Bdt:SgEvt0:7:closetrk}   {\ensuremath{{8 } } }
\vdef{default-11Bdt:SgEvt0:7:lip}   {\ensuremath{{-0.0000 } } }
\vdef{default-11Bdt:SgEvt0:7:lipE}   {\ensuremath{{0.0043 } } }
\vdef{default-11Bdt:SgEvt0:7:tip}   {\ensuremath{{0.0042 } } }
\vdef{default-11Bdt:SgEvt0:7:tipE}   {\ensuremath{{0.0029 } } }
\vdef{default-11Bdt:SgEvt0:7:pvlip}   {\ensuremath{{-0.0000 } } }
\vdef{default-11Bdt:SgEvt0:7:pvlips}   {\ensuremath{{-0.0108 } } }
\vdef{default-11Bdt:SgEvt0:7:pvip}   {\ensuremath{{0.0042 } } }
\vdef{default-11Bdt:SgEvt0:7:pvips}   {\ensuremath{{1.4440 } } }
\vdef{default-11Bdt:SgEvt0:7:maxdoca}   {\ensuremath{{0.0026 } } }
\vdef{default-11Bdt:SgEvt0:7:pvw8}   {\ensuremath{{0.8370 } } }
\vdef{default-11Bdt:SgEvt0:7:bdt}   {\ensuremath{{-0.0299 } } }
\vdef{default-11Bdt:SgEvt0:7:m1pix}   {\ensuremath{{3 } } }
\vdef{default-11Bdt:SgEvt0:7:m2pix}   {\ensuremath{{3 } } }
\vdef{default-11Bdt:SgEvt0:7:m1bpix}   {\ensuremath{{3 } } }
\vdef{default-11Bdt:SgEvt0:7:m2bpix}   {\ensuremath{{3 } } }
\vdef{default-11Bdt:SgEvt0:7:m1bpixl1}   {\ensuremath{{1 } } }
\vdef{default-11Bdt:SgEvt0:7:m2bpixl1}   {\ensuremath{{1 } } }
\vdef{default-11Bdt:BsSgEvt1:7:run}   {\ensuremath{{177139 } } }
\vdef{default-11Bdt:BsSgEvt1:7:evt}   {\ensuremath{{59599661 } } }
\vdef{default-11Bdt:BsSgEvt1:7:chan}   {\ensuremath{{1 } } }
\vdef{default-11Bdt:BsSgEvt1:7:m}   {\ensuremath{{5.339 } } }
\vdef{default-11Bdt:BsSgEvt1:7:pt}   {\ensuremath{{19.643 } } }
\vdef{default-11Bdt:BsSgEvt1:7:phi}   {\ensuremath{{2.475 } } }
\vdef{default-11Bdt:BsSgEvt1:7:eta}   {\ensuremath{{-2.119 } } }
\vdef{default-11Bdt:BsSgEvt1:7:channel}   {endcap }
\vdef{default-11Bdt:BsSgEvt1:7:cowboy}   {\ensuremath{{0 } } }
\vdef{default-11Bdt:BsSgEvt1:7:m1pt}   {\ensuremath{{14.857 } } }
\vdef{default-11Bdt:BsSgEvt1:7:m2pt}   {\ensuremath{{5.391 } } }
\vdef{default-11Bdt:BsSgEvt1:7:m1eta}   {\ensuremath{{-2.015 } } }
\vdef{default-11Bdt:BsSgEvt1:7:m2eta}   {\ensuremath{{-2.275 } } }
\vdef{default-11Bdt:BsSgEvt1:7:m1phi}   {\ensuremath{{2.334 } } }
\vdef{default-11Bdt:BsSgEvt1:7:m2phi}   {\ensuremath{{2.876 } } }
\vdef{default-11Bdt:BsSgEvt1:7:m1q}   {\ensuremath{{1 } } }
\vdef{default-11Bdt:BsSgEvt1:7:m2q}   {\ensuremath{{-1 } } }
\vdef{default-11Bdt:BsSgEvt1:7:iso}   {\ensuremath{{1.000 } } }
\vdef{default-11Bdt:BsSgEvt1:7:alpha}   {\ensuremath{{0.0047 } } }
\vdef{default-11Bdt:BsSgEvt1:7:chi2}   {\ensuremath{{ 0.28 } } }
\vdef{default-11Bdt:BsSgEvt1:7:dof}   {\ensuremath{{1 } } }
\vdef{default-11Bdt:BsSgEvt1:7:fls3d}   {\ensuremath{{33.20 } } }
\vdef{default-11Bdt:BsSgEvt1:7:fl3d}   {\ensuremath{{1.4587 } } }
\vdef{default-11Bdt:BsSgEvt1:7:fl3dE}   {\ensuremath{{0.0439 } } }
\vdef{default-11Bdt:BsSgEvt1:7:docatrk}   {\ensuremath{{0.1915 } } }
\vdef{default-11Bdt:BsSgEvt1:7:closetrk}   {\ensuremath{{0 } } }
\vdef{default-11Bdt:BsSgEvt1:7:lip}   {\ensuremath{{0.0006 } } }
\vdef{default-11Bdt:BsSgEvt1:7:lipE}   {\ensuremath{{0.0024 } } }
\vdef{default-11Bdt:BsSgEvt1:7:tip}   {\ensuremath{{0.0065 } } }
\vdef{default-11Bdt:BsSgEvt1:7:tipE}   {\ensuremath{{0.0035 } } }
\vdef{default-11Bdt:BsSgEvt1:7:pvlip}   {\ensuremath{{0.0006 } } }
\vdef{default-11Bdt:BsSgEvt1:7:pvlips}   {\ensuremath{{0.2435 } } }
\vdef{default-11Bdt:BsSgEvt1:7:pvip}   {\ensuremath{{0.0065 } } }
\vdef{default-11Bdt:BsSgEvt1:7:pvips}   {\ensuremath{{1.8554 } } }
\vdef{default-11Bdt:BsSgEvt1:7:maxdoca}   {\ensuremath{{0.0026 } } }
\vdef{default-11Bdt:BsSgEvt1:7:pvw8}   {\ensuremath{{0.9105 } } }
\vdef{default-11Bdt:BsSgEvt1:7:bdt}   {\ensuremath{{0.4679 } } }
\vdef{default-11Bdt:BsSgEvt1:7:m1pix}   {\ensuremath{{3 } } }
\vdef{default-11Bdt:BsSgEvt1:7:m2pix}   {\ensuremath{{4 } } }
\vdef{default-11Bdt:BsSgEvt1:7:m1bpix}   {\ensuremath{{2 } } }
\vdef{default-11Bdt:BsSgEvt1:7:m2bpix}   {\ensuremath{{2 } } }
\vdef{default-11Bdt:BsSgEvt1:7:m1bpixl1}   {\ensuremath{{1 } } }
\vdef{default-11Bdt:BsSgEvt1:7:m2bpixl1}   {\ensuremath{{1 } } }
\vdef{default-11Bdt:SgEvt0:8:run}   {\ensuremath{{176885 } } }
\vdef{default-11Bdt:SgEvt0:8:evt}   {\ensuremath{{299215348 } } }
\vdef{default-11Bdt:SgEvt0:8:chan}   {\ensuremath{{0 } } }
\vdef{default-11Bdt:SgEvt0:8:m}   {\ensuremath{{5.135 } } }
\vdef{default-11Bdt:SgEvt0:8:pt}   {\ensuremath{{11.483 } } }
\vdef{default-11Bdt:SgEvt0:8:phi}   {\ensuremath{{-1.976 } } }
\vdef{default-11Bdt:SgEvt0:8:eta}   {\ensuremath{{-0.635 } } }
\vdef{default-11Bdt:SgEvt0:8:channel}   {barrel }
\vdef{default-11Bdt:SgEvt0:8:cowboy}   {\ensuremath{{1 } } }
\vdef{default-11Bdt:SgEvt0:8:m1pt}   {\ensuremath{{5.933 } } }
\vdef{default-11Bdt:SgEvt0:8:m2pt}   {\ensuremath{{5.572 } } }
\vdef{default-11Bdt:SgEvt0:8:m1eta}   {\ensuremath{{-0.166 } } }
\vdef{default-11Bdt:SgEvt0:8:m2eta}   {\ensuremath{{-1.032 } } }
\vdef{default-11Bdt:SgEvt0:8:m1phi}   {\ensuremath{{-1.993 } } }
\vdef{default-11Bdt:SgEvt0:8:m2phi}   {\ensuremath{{-1.959 } } }
\vdef{default-11Bdt:SgEvt0:8:m1q}   {\ensuremath{{-1 } } }
\vdef{default-11Bdt:SgEvt0:8:m2q}   {\ensuremath{{1 } } }
\vdef{default-11Bdt:SgEvt0:8:iso}   {\ensuremath{{1.000 } } }
\vdef{default-11Bdt:SgEvt0:8:alpha}   {\ensuremath{{0.0214 } } }
\vdef{default-11Bdt:SgEvt0:8:chi2}   {\ensuremath{{ 0.46 } } }
\vdef{default-11Bdt:SgEvt0:8:dof}   {\ensuremath{{1 } } }
\vdef{default-11Bdt:SgEvt0:8:fls3d}   {\ensuremath{{45.33 } } }
\vdef{default-11Bdt:SgEvt0:8:fl3d}   {\ensuremath{{0.5134 } } }
\vdef{default-11Bdt:SgEvt0:8:fl3dE}   {\ensuremath{{0.0113 } } }
\vdef{default-11Bdt:SgEvt0:8:docatrk}   {\ensuremath{{0.0370 } } }
\vdef{default-11Bdt:SgEvt0:8:closetrk}   {\ensuremath{{0 } } }
\vdef{default-11Bdt:SgEvt0:8:lip}   {\ensuremath{{-0.0077 } } }
\vdef{default-11Bdt:SgEvt0:8:lipE}   {\ensuremath{{0.0072 } } }
\vdef{default-11Bdt:SgEvt0:8:tip}   {\ensuremath{{0.0058 } } }
\vdef{default-11Bdt:SgEvt0:8:tipE}   {\ensuremath{{0.0054 } } }
\vdef{default-11Bdt:SgEvt0:8:pvlip}   {\ensuremath{{-0.0077 } } }
\vdef{default-11Bdt:SgEvt0:8:pvlips}   {\ensuremath{{-1.0727 } } }
\vdef{default-11Bdt:SgEvt0:8:pvip}   {\ensuremath{{0.0097 } } }
\vdef{default-11Bdt:SgEvt0:8:pvips}   {\ensuremath{{1.4553 } } }
\vdef{default-11Bdt:SgEvt0:8:maxdoca}   {\ensuremath{{0.0029 } } }
\vdef{default-11Bdt:SgEvt0:8:pvw8}   {\ensuremath{{0.9423 } } }
\vdef{default-11Bdt:SgEvt0:8:bdt}   {\ensuremath{{0.7747 } } }
\vdef{default-11Bdt:SgEvt0:8:m1pix}   {\ensuremath{{3 } } }
\vdef{default-11Bdt:SgEvt0:8:m2pix}   {\ensuremath{{3 } } }
\vdef{default-11Bdt:SgEvt0:8:m1bpix}   {\ensuremath{{3 } } }
\vdef{default-11Bdt:SgEvt0:8:m2bpix}   {\ensuremath{{3 } } }
\vdef{default-11Bdt:SgEvt0:8:m1bpixl1}   {\ensuremath{{1 } } }
\vdef{default-11Bdt:SgEvt0:8:m2bpixl1}   {\ensuremath{{1 } } }
\vdef{default-11Bdt:SgEvt0:9:run}   {\ensuremath{{176309 } } }
\vdef{default-11Bdt:SgEvt0:9:evt}   {\ensuremath{{-2036084515 } } }
\vdef{default-11Bdt:SgEvt0:9:chan}   {\ensuremath{{0 } } }
\vdef{default-11Bdt:SgEvt0:9:m}   {\ensuremath{{5.712 } } }
\vdef{default-11Bdt:SgEvt0:9:pt}   {\ensuremath{{32.616 } } }
\vdef{default-11Bdt:SgEvt0:9:phi}   {\ensuremath{{2.164 } } }
\vdef{default-11Bdt:SgEvt0:9:eta}   {\ensuremath{{-0.027 } } }
\vdef{default-11Bdt:SgEvt0:9:channel}   {barrel }
\vdef{default-11Bdt:SgEvt0:9:cowboy}   {\ensuremath{{0 } } }
\vdef{default-11Bdt:SgEvt0:9:m1pt}   {\ensuremath{{27.827 } } }
\vdef{default-11Bdt:SgEvt0:9:m2pt}   {\ensuremath{{5.282 } } }
\vdef{default-11Bdt:SgEvt0:9:m1eta}   {\ensuremath{{-0.021 } } }
\vdef{default-11Bdt:SgEvt0:9:m2eta}   {\ensuremath{{-0.061 } } }
\vdef{default-11Bdt:SgEvt0:9:m1phi}   {\ensuremath{{2.237 } } }
\vdef{default-11Bdt:SgEvt0:9:m2phi}   {\ensuremath{{1.765 } } }
\vdef{default-11Bdt:SgEvt0:9:m1q}   {\ensuremath{{-1 } } }
\vdef{default-11Bdt:SgEvt0:9:m2q}   {\ensuremath{{1 } } }
\vdef{default-11Bdt:SgEvt0:9:iso}   {\ensuremath{{1.000 } } }
\vdef{default-11Bdt:SgEvt0:9:alpha}   {\ensuremath{{0.0344 } } }
\vdef{default-11Bdt:SgEvt0:9:chi2}   {\ensuremath{{ 1.63 } } }
\vdef{default-11Bdt:SgEvt0:9:dof}   {\ensuremath{{1 } } }
\vdef{default-11Bdt:SgEvt0:9:fls3d}   {\ensuremath{{13.95 } } }
\vdef{default-11Bdt:SgEvt0:9:fl3d}   {\ensuremath{{0.1045 } } }
\vdef{default-11Bdt:SgEvt0:9:fl3dE}   {\ensuremath{{0.0075 } } }
\vdef{default-11Bdt:SgEvt0:9:docatrk}   {\ensuremath{{0.0061 } } }
\vdef{default-11Bdt:SgEvt0:9:closetrk}   {\ensuremath{{0 } } }
\vdef{default-11Bdt:SgEvt0:9:lip}   {\ensuremath{{-0.0036 } } }
\vdef{default-11Bdt:SgEvt0:9:lipE}   {\ensuremath{{0.0043 } } }
\vdef{default-11Bdt:SgEvt0:9:tip}   {\ensuremath{{0.0002 } } }
\vdef{default-11Bdt:SgEvt0:9:tipE}   {\ensuremath{{0.0021 } } }
\vdef{default-11Bdt:SgEvt0:9:pvlip}   {\ensuremath{{-0.0036 } } }
\vdef{default-11Bdt:SgEvt0:9:pvlips}   {\ensuremath{{-0.8250 } } }
\vdef{default-11Bdt:SgEvt0:9:pvip}   {\ensuremath{{0.0036 } } }
\vdef{default-11Bdt:SgEvt0:9:pvips}   {\ensuremath{{0.8280 } } }
\vdef{default-11Bdt:SgEvt0:9:maxdoca}   {\ensuremath{{0.0077 } } }
\vdef{default-11Bdt:SgEvt0:9:pvw8}   {\ensuremath{{0.8801 } } }
\vdef{default-11Bdt:SgEvt0:9:bdt}   {\ensuremath{{0.1985 } } }
\vdef{default-11Bdt:SgEvt0:9:m1pix}   {\ensuremath{{3 } } }
\vdef{default-11Bdt:SgEvt0:9:m2pix}   {\ensuremath{{3 } } }
\vdef{default-11Bdt:SgEvt0:9:m1bpix}   {\ensuremath{{3 } } }
\vdef{default-11Bdt:SgEvt0:9:m2bpix}   {\ensuremath{{3 } } }
\vdef{default-11Bdt:SgEvt0:9:m1bpixl1}   {\ensuremath{{1 } } }
\vdef{default-11Bdt:SgEvt0:9:m2bpixl1}   {\ensuremath{{1 } } }
\vdef{default-11Bdt:BsSgEvt1:8:run}   {\ensuremath{{176309 } } }
\vdef{default-11Bdt:BsSgEvt1:8:evt}   {\ensuremath{{644985947 } } }
\vdef{default-11Bdt:BsSgEvt1:8:chan}   {\ensuremath{{1 } } }
\vdef{default-11Bdt:BsSgEvt1:8:m}   {\ensuremath{{5.329 } } }
\vdef{default-11Bdt:BsSgEvt1:8:pt}   {\ensuremath{{6.978 } } }
\vdef{default-11Bdt:BsSgEvt1:8:phi}   {\ensuremath{{2.710 } } }
\vdef{default-11Bdt:BsSgEvt1:8:eta}   {\ensuremath{{1.611 } } }
\vdef{default-11Bdt:BsSgEvt1:8:channel}   {endcap }
\vdef{default-11Bdt:BsSgEvt1:8:cowboy}   {\ensuremath{{0 } } }
\vdef{default-11Bdt:BsSgEvt1:8:m1pt}   {\ensuremath{{4.713 } } }
\vdef{default-11Bdt:BsSgEvt1:8:m2pt}   {\ensuremath{{4.061 } } }
\vdef{default-11Bdt:BsSgEvt1:8:m1eta}   {\ensuremath{{1.375 } } }
\vdef{default-11Bdt:BsSgEvt1:8:m2eta}   {\ensuremath{{1.436 } } }
\vdef{default-11Bdt:BsSgEvt1:8:m1phi}   {\ensuremath{{2.114 } } }
\vdef{default-11Bdt:BsSgEvt1:8:m2phi}   {\ensuremath{{-2.864 } } }
\vdef{default-11Bdt:BsSgEvt1:8:m1q}   {\ensuremath{{1 } } }
\vdef{default-11Bdt:BsSgEvt1:8:m2q}   {\ensuremath{{-1 } } }
\vdef{default-11Bdt:BsSgEvt1:8:iso}   {\ensuremath{{0.724 } } }
\vdef{default-11Bdt:BsSgEvt1:8:alpha}   {\ensuremath{{0.0049 } } }
\vdef{default-11Bdt:BsSgEvt1:8:chi2}   {\ensuremath{{ 0.01 } } }
\vdef{default-11Bdt:BsSgEvt1:8:dof}   {\ensuremath{{1 } } }
\vdef{default-11Bdt:BsSgEvt1:8:fls3d}   {\ensuremath{{34.35 } } }
\vdef{default-11Bdt:BsSgEvt1:8:fl3d}   {\ensuremath{{0.3847 } } }
\vdef{default-11Bdt:BsSgEvt1:8:fl3dE}   {\ensuremath{{0.0112 } } }
\vdef{default-11Bdt:BsSgEvt1:8:docatrk}   {\ensuremath{{0.0517 } } }
\vdef{default-11Bdt:BsSgEvt1:8:closetrk}   {\ensuremath{{0 } } }
\vdef{default-11Bdt:BsSgEvt1:8:lip}   {\ensuremath{{0.0001 } } }
\vdef{default-11Bdt:BsSgEvt1:8:lipE}   {\ensuremath{{0.0034 } } }
\vdef{default-11Bdt:BsSgEvt1:8:tip}   {\ensuremath{{0.0019 } } }
\vdef{default-11Bdt:BsSgEvt1:8:tipE}   {\ensuremath{{0.0049 } } }
\vdef{default-11Bdt:BsSgEvt1:8:pvlip}   {\ensuremath{{0.0001 } } }
\vdef{default-11Bdt:BsSgEvt1:8:pvlips}   {\ensuremath{{0.0327 } } }
\vdef{default-11Bdt:BsSgEvt1:8:pvip}   {\ensuremath{{0.0019 } } }
\vdef{default-11Bdt:BsSgEvt1:8:pvips}   {\ensuremath{{0.3836 } } }
\vdef{default-11Bdt:BsSgEvt1:8:maxdoca}   {\ensuremath{{0.0004 } } }
\vdef{default-11Bdt:BsSgEvt1:8:pvw8}   {\ensuremath{{0.8348 } } }
\vdef{default-11Bdt:BsSgEvt1:8:bdt}   {\ensuremath{{0.5175 } } }
\vdef{default-11Bdt:BsSgEvt1:8:m1pix}   {\ensuremath{{3 } } }
\vdef{default-11Bdt:BsSgEvt1:8:m2pix}   {\ensuremath{{2 } } }
\vdef{default-11Bdt:BsSgEvt1:8:m1bpix}   {\ensuremath{{3 } } }
\vdef{default-11Bdt:BsSgEvt1:8:m2bpix}   {\ensuremath{{2 } } }
\vdef{default-11Bdt:BsSgEvt1:8:m1bpixl1}   {\ensuremath{{1 } } }
\vdef{default-11Bdt:BsSgEvt1:8:m2bpixl1}   {\ensuremath{{1 } } }
\vdef{bdtdefault-11:N-EFF-TOT-BPLUS0:val}   {\ensuremath{{0.00142 } } }
\vdef{bdtdefault-11:N-EFF-TOT-BPLUS0:err}   {\ensuremath{{0.000004 } } }
\vdef{bdtdefault-11:N-EFF-TOT-BPLUS0:tot}   {\ensuremath{{0.00012 } } }
\vdef{bdtdefault-11:N-EFF-TOT-BPLUS0:all}   {\ensuremath{{(1.42 \pm 0.12)\times 10^{-3}} } }
\vdef{bdtdefault-11:N-ACC-BPLUS0:val}   {\ensuremath{{0.162 } } }
\vdef{bdtdefault-11:N-ACC-BPLUS0:err}   {\ensuremath{{0.000 } } }
\vdef{bdtdefault-11:N-ACC-BPLUS0:tot}   {\ensuremath{{0.006 } } }
\vdef{bdtdefault-11:N-ACC-BPLUS0:all}   {\ensuremath{{(16.19 \pm 0.57)\times 10^{-2}} } }
\vdef{bdtdefault-11:N-EFF-MU-PID-BPLUS0:val}   {\ensuremath{{0.785 } } }
\vdef{bdtdefault-11:N-EFF-MU-PID-BPLUS0:err}   {\ensuremath{{0.000 } } }
\vdef{bdtdefault-11:N-EFF-MU-PID-BPLUS0:tot}   {\ensuremath{{0.031 } } }
\vdef{bdtdefault-11:N-EFF-MU-PID-BPLUS0:all}   {\ensuremath{{(78.45 \pm 3.14)\times 10^{-2}} } }
\vdef{bdtdefault-11:N-EFF-MU-PIDMC-BPLUS0:val}   {\ensuremath{{0.771 } } }
\vdef{bdtdefault-11:N-EFF-MU-PIDMC-BPLUS0:err}   {\ensuremath{{0.000 } } }
\vdef{bdtdefault-11:N-EFF-MU-PIDMC-BPLUS0:tot}   {\ensuremath{{0.031 } } }
\vdef{bdtdefault-11:N-EFF-MU-PIDMC-BPLUS0:all}   {\ensuremath{{(77.07 \pm 3.08)\times 10^{-2}} } }
\vdef{bdtdefault-11:N-EFF-MU-MC-BPLUS0:val}   {\ensuremath{{0.765 } } }
\vdef{bdtdefault-11:N-EFF-MU-MC-BPLUS0:err}   {\ensuremath{{0.001 } } }
\vdef{bdtdefault-11:N-EFF-MU-MC-BPLUS0:tot}   {\ensuremath{{0.031 } } }
\vdef{bdtdefault-11:N-EFF-MU-MC-BPLUS0:all}   {\ensuremath{{(76.53 \pm 3.06)\times 10^{-2}} } }
\vdef{bdtdefault-11:N-EFF-TRIG-PID-BPLUS0:val}   {\ensuremath{{0.782 } } }
\vdef{bdtdefault-11:N-EFF-TRIG-PID-BPLUS0:err}   {\ensuremath{{0.000 } } }
\vdef{bdtdefault-11:N-EFF-TRIG-PID-BPLUS0:tot}   {\ensuremath{{0.023 } } }
\vdef{bdtdefault-11:N-EFF-TRIG-PID-BPLUS0:all}   {\ensuremath{{(78.22 \pm 2.35)\times 10^{-2}} } }
\vdef{bdtdefault-11:N-EFF-TRIG-PIDMC-BPLUS0:val}   {\ensuremath{{0.827 } } }
\vdef{bdtdefault-11:N-EFF-TRIG-PIDMC-BPLUS0:err}   {\ensuremath{{0.000 } } }
\vdef{bdtdefault-11:N-EFF-TRIG-PIDMC-BPLUS0:tot}   {\ensuremath{{0.025 } } }
\vdef{bdtdefault-11:N-EFF-TRIG-PIDMC-BPLUS0:all}   {\ensuremath{{(82.72 \pm 2.48)\times 10^{-2}} } }
\vdef{bdtdefault-11:N-EFF-TRIG-MC-BPLUS0:val}   {\ensuremath{{0.763 } } }
\vdef{bdtdefault-11:N-EFF-TRIG-MC-BPLUS0:err}   {\ensuremath{{0.001 } } }
\vdef{bdtdefault-11:N-EFF-TRIG-MC-BPLUS0:tot}   {\ensuremath{{0.023 } } }
\vdef{bdtdefault-11:N-EFF-TRIG-MC-BPLUS0:all}   {\ensuremath{{(76.28 \pm 2.29)\times 10^{-2}} } }
\vdef{bdtdefault-11:N-EFF-CAND-BPLUS0:val}   {\ensuremath{{0.980 } } }
\vdef{bdtdefault-11:N-EFF-CAND-BPLUS0:err}   {\ensuremath{{0.003 } } }
\vdef{bdtdefault-11:N-EFF-CAND-BPLUS0:tot}   {\ensuremath{{0.010 } } }
\vdef{bdtdefault-11:N-EFF-CAND-BPLUS0:all}   {\ensuremath{{(98.00 \pm 1.01)\times 10^{-2}} } }
\vdef{bdtdefault-11:N-EFF-ANA-BPLUS0:val}   {\ensuremath{{0.0154 } } }
\vdef{bdtdefault-11:N-EFF-ANA-BPLUS0:err}   {\ensuremath{{0.0005 } } }
\vdef{bdtdefault-11:N-EFF-ANA-BPLUS0:tot}   {\ensuremath{{0.0010 } } }
\vdef{bdtdefault-11:N-EFF-ANA-BPLUS0:all}   {\ensuremath{{(1.54 \pm 0.10)\times 10^{-2}} } }
\vdef{bdtdefault-11:N-OBS-BPLUS0:val}   {\ensuremath{{107046 } } }
\vdef{bdtdefault-11:N-OBS-BPLUS0:err}   {\ensuremath{{341 } } }
\vdef{bdtdefault-11:N-OBS-BPLUS0:tot}   {\ensuremath{{5363 } } }
\vdef{bdtdefault-11:N-OBS-BPLUS0:all}   {\ensuremath{{107046 } } }
\vdef{bdtdefault-11:N-OBS-CBPLUS0:val}   {\ensuremath{{106597 } } }
\vdef{bdtdefault-11:N-OBS-CBPLUS0:err}   {\ensuremath{{448 } } }
\vdef{bdtdefault-11:N-EFF-TOT-BPLUS1:val}   {\ensuremath{{0.00045 } } }
\vdef{bdtdefault-11:N-EFF-TOT-BPLUS1:err}   {\ensuremath{{0.000002 } } }
\vdef{bdtdefault-11:N-EFF-TOT-BPLUS1:tot}   {\ensuremath{{0.00006 } } }
\vdef{bdtdefault-11:N-EFF-TOT-BPLUS1:all}   {\ensuremath{{(0.45 \pm 0.06)\times 10^{-3}} } }
\vdef{bdtdefault-11:N-ACC-BPLUS1:val}   {\ensuremath{{0.111 } } }
\vdef{bdtdefault-11:N-ACC-BPLUS1:err}   {\ensuremath{{0.000 } } }
\vdef{bdtdefault-11:N-ACC-BPLUS1:tot}   {\ensuremath{{0.006 } } }
\vdef{bdtdefault-11:N-ACC-BPLUS1:all}   {\ensuremath{{(11.05 \pm 0.55)\times 10^{-2}} } }
\vdef{bdtdefault-11:N-EFF-MU-PID-BPLUS1:val}   {\ensuremath{{0.782 } } }
\vdef{bdtdefault-11:N-EFF-MU-PID-BPLUS1:err}   {\ensuremath{{0.000 } } }
\vdef{bdtdefault-11:N-EFF-MU-PID-BPLUS1:tot}   {\ensuremath{{0.063 } } }
\vdef{bdtdefault-11:N-EFF-MU-PID-BPLUS1:all}   {\ensuremath{{(78.15 \pm 6.25)\times 10^{-2}} } }
\vdef{bdtdefault-11:N-EFF-MU-PIDMC-BPLUS1:val}   {\ensuremath{{0.827 } } }
\vdef{bdtdefault-11:N-EFF-MU-PIDMC-BPLUS1:err}   {\ensuremath{{0.000 } } }
\vdef{bdtdefault-11:N-EFF-MU-PIDMC-BPLUS1:tot}   {\ensuremath{{0.066 } } }
\vdef{bdtdefault-11:N-EFF-MU-PIDMC-BPLUS1:all}   {\ensuremath{{(82.73 \pm 6.62)\times 10^{-2}} } }
\vdef{bdtdefault-11:N-EFF-MU-MC-BPLUS1:val}   {\ensuremath{{0.775 } } }
\vdef{bdtdefault-11:N-EFF-MU-MC-BPLUS1:err}   {\ensuremath{{0.001 } } }
\vdef{bdtdefault-11:N-EFF-MU-MC-BPLUS1:tot}   {\ensuremath{{0.062 } } }
\vdef{bdtdefault-11:N-EFF-MU-MC-BPLUS1:all}   {\ensuremath{{(77.49 \pm 6.20)\times 10^{-2}} } }
\vdef{bdtdefault-11:N-EFF-TRIG-PID-BPLUS1:val}   {\ensuremath{{0.752 } } }
\vdef{bdtdefault-11:N-EFF-TRIG-PID-BPLUS1:err}   {\ensuremath{{0.001 } } }
\vdef{bdtdefault-11:N-EFF-TRIG-PID-BPLUS1:tot}   {\ensuremath{{0.045 } } }
\vdef{bdtdefault-11:N-EFF-TRIG-PID-BPLUS1:all}   {\ensuremath{{(75.19 \pm 4.51)\times 10^{-2}} } }
\vdef{bdtdefault-11:N-EFF-TRIG-PIDMC-BPLUS1:val}   {\ensuremath{{0.738 } } }
\vdef{bdtdefault-11:N-EFF-TRIG-PIDMC-BPLUS1:err}   {\ensuremath{{0.001 } } }
\vdef{bdtdefault-11:N-EFF-TRIG-PIDMC-BPLUS1:tot}   {\ensuremath{{0.044 } } }
\vdef{bdtdefault-11:N-EFF-TRIG-PIDMC-BPLUS1:all}   {\ensuremath{{(73.75 \pm 4.43)\times 10^{-2}} } }
\vdef{bdtdefault-11:N-EFF-TRIG-MC-BPLUS1:val}   {\ensuremath{{0.575 } } }
\vdef{bdtdefault-11:N-EFF-TRIG-MC-BPLUS1:err}   {\ensuremath{{0.002 } } }
\vdef{bdtdefault-11:N-EFF-TRIG-MC-BPLUS1:tot}   {\ensuremath{{0.035 } } }
\vdef{bdtdefault-11:N-EFF-TRIG-MC-BPLUS1:all}   {\ensuremath{{(57.51 \pm 3.46)\times 10^{-2}} } }
\vdef{bdtdefault-11:N-EFF-CAND-BPLUS1:val}   {\ensuremath{{0.980 } } }
\vdef{bdtdefault-11:N-EFF-CAND-BPLUS1:err}   {\ensuremath{{0.005 } } }
\vdef{bdtdefault-11:N-EFF-CAND-BPLUS1:tot}   {\ensuremath{{0.011 } } }
\vdef{bdtdefault-11:N-EFF-CAND-BPLUS1:all}   {\ensuremath{{(98.00 \pm 1.08)\times 10^{-2}} } }
\vdef{bdtdefault-11:N-EFF-ANA-BPLUS1:val}   {\ensuremath{{0.0093 } } }
\vdef{bdtdefault-11:N-EFF-ANA-BPLUS1:err}   {\ensuremath{{0.0006 } } }
\vdef{bdtdefault-11:N-EFF-ANA-BPLUS1:tot}   {\ensuremath{{0.0008 } } }
\vdef{bdtdefault-11:N-EFF-ANA-BPLUS1:all}   {\ensuremath{{(0.93 \pm 0.08)\times 10^{-2}} } }
\vdef{bdtdefault-11:N-OBS-BPLUS1:val}   {\ensuremath{{32972 } } }
\vdef{bdtdefault-11:N-OBS-BPLUS1:err}   {\ensuremath{{195 } } }
\vdef{bdtdefault-11:N-OBS-BPLUS1:tot}   {\ensuremath{{1660 } } }
\vdef{bdtdefault-11:N-OBS-BPLUS1:all}   {\ensuremath{{32972 } } }
\vdef{bdtdefault-11:N-OBS-CBPLUS1:val}   {\ensuremath{{32634 } } }
\vdef{bdtdefault-11:N-OBS-CBPLUS1:err}   {\ensuremath{{193 } } }
\vdef{bdtdefault-11:N-EXP2-SIG-BSMM0:val}   {\ensuremath{{ 3.70 } } }
\vdef{bdtdefault-11:N-EXP2-SIG-BSMM0:err}   {\ensuremath{{ 0.56 } } }
\vdef{bdtdefault-11:N-EXP2-SIG-BDMM0:val}   {\ensuremath{{0.325 } } }
\vdef{bdtdefault-11:N-EXP2-SIG-BDMM0:err}   {\ensuremath{{0.033 } } }
\vdef{bdtdefault-11:N-OBS-BKG0:val}   {\ensuremath{{10 } } }
\vdef{bdtdefault-11:N-EXP-BSMM0:val}   {\ensuremath{{ 1.03 } } }
\vdef{bdtdefault-11:N-EXP-BSMM0:err}   {\ensuremath{{ 0.66 } } }
\vdef{bdtdefault-11:N-EXP-BDMM0:val}   {\ensuremath{{ 0.69 } } }
\vdef{bdtdefault-11:N-EXP-BDMM0:err}   {\ensuremath{{ 0.44 } } }
\vdef{bdtdefault-11:N-LOW-BD0:val}   {\ensuremath{{5.200 } } }
\vdef{bdtdefault-11:N-HIGH-BD0:val}   {\ensuremath{{5.300 } } }
\vdef{bdtdefault-11:N-LOW-BS0:val}   {\ensuremath{{5.300 } } }
\vdef{bdtdefault-11:N-HIGH-BS0:val}   {\ensuremath{{5.450 } } }
\vdef{bdtdefault-11:N-PSS0:val}   {\ensuremath{{0.869 } } }
\vdef{bdtdefault-11:N-PSS0:err}   {\ensuremath{{0.007 } } }
\vdef{bdtdefault-11:N-PSS0:tot}   {\ensuremath{{0.044 } } }
\vdef{bdtdefault-11:N-PSD0:val}   {\ensuremath{{0.291 } } }
\vdef{bdtdefault-11:N-PSD0:err}   {\ensuremath{{0.015 } } }
\vdef{bdtdefault-11:N-PSD0:tot}   {\ensuremath{{0.021 } } }
\vdef{bdtdefault-11:N-PDS0:val}   {\ensuremath{{0.067 } } }
\vdef{bdtdefault-11:N-PDS0:err}   {\ensuremath{{0.005 } } }
\vdef{bdtdefault-11:N-PDS0:tot}   {\ensuremath{{0.006 } } }
\vdef{bdtdefault-11:N-PDD0:val}   {\ensuremath{{0.646 } } }
\vdef{bdtdefault-11:N-PDD0:err}   {\ensuremath{{0.015 } } }
\vdef{bdtdefault-11:N-PDD0:tot}   {\ensuremath{{0.036 } } }
\vdef{bdtdefault-11:N-EFF-TOT-BSMM0:val}   {\ensuremath{{0.0040 } } }
\vdef{bdtdefault-11:N-EFF-TOT-BSMM0:err}   {\ensuremath{{0.0001 } } }
\vdef{bdtdefault-11:N-EFF-TOT-BSMM0:tot}   {\ensuremath{{0.0003 } } }
\vdef{bdtdefault-11:N-EFF-TOT-BSMM0:all}   {\ensuremath{{(0.40 \pm 0.03)\times 10^{-2}} } }
\vdef{bdtdefault-11:N-ACC-BSMM0:val}   {\ensuremath{{0.248 } } }
\vdef{bdtdefault-11:N-ACC-BSMM0:err}   {\ensuremath{{0.001 } } }
\vdef{bdtdefault-11:N-ACC-BSMM0:tot}   {\ensuremath{{0.009 } } }
\vdef{bdtdefault-11:N-ACC-BSMM0:all}   {\ensuremath{{(24.76 \pm 0.87)\times 10^{-2}} } }
\vdef{bdtdefault-11:N-EFF-MU-PID-BSMM0:val}   {\ensuremath{{0.792 } } }
\vdef{bdtdefault-11:N-EFF-MU-PID-BSMM0:err}   {\ensuremath{{0.001 } } }
\vdef{bdtdefault-11:N-EFF-MU-PID-BSMM0:tot}   {\ensuremath{{0.032 } } }
\vdef{bdtdefault-11:N-EFF-MU-PID-BSMM0:all}   {\ensuremath{{(79.21 \pm 3.17)\times 10^{-2}} } }
\vdef{bdtdefault-11:N-EFF-MU-PIDMC-BSMM0:val}   {\ensuremath{{0.786 } } }
\vdef{bdtdefault-11:N-EFF-MU-PIDMC-BSMM0:err}   {\ensuremath{{0.002 } } }
\vdef{bdtdefault-11:N-EFF-MU-PIDMC-BSMM0:tot}   {\ensuremath{{0.031 } } }
\vdef{bdtdefault-11:N-EFF-MU-PIDMC-BSMM0:all}   {\ensuremath{{(78.56 \pm 3.15)\times 10^{-2}} } }
\vdef{bdtdefault-11:N-EFF-MU-MC-BSMM0:val}   {\ensuremath{{0.701 } } }
\vdef{bdtdefault-11:N-EFF-MU-MC-BSMM0:err}   {\ensuremath{{0.008 } } }
\vdef{bdtdefault-11:N-EFF-MU-MC-BSMM0:tot}   {\ensuremath{{0.029 } } }
\vdef{bdtdefault-11:N-EFF-MU-MC-BSMM0:all}   {\ensuremath{{(70.15 \pm 2.91)\times 10^{-2}} } }
\vdef{bdtdefault-11:N-EFF-TRIG-PID-BSMM0:val}   {\ensuremath{{0.795 } } }
\vdef{bdtdefault-11:N-EFF-TRIG-PID-BSMM0:err}   {\ensuremath{{0.002 } } }
\vdef{bdtdefault-11:N-EFF-TRIG-PID-BSMM0:tot}   {\ensuremath{{0.024 } } }
\vdef{bdtdefault-11:N-EFF-TRIG-PID-BSMM0:all}   {\ensuremath{{(79.51 \pm 2.39)\times 10^{-2}} } }
\vdef{bdtdefault-11:N-EFF-TRIG-PIDMC-BSMM0:val}   {\ensuremath{{0.842 } } }
\vdef{bdtdefault-11:N-EFF-TRIG-PIDMC-BSMM0:err}   {\ensuremath{{0.002 } } }
\vdef{bdtdefault-11:N-EFF-TRIG-PIDMC-BSMM0:tot}   {\ensuremath{{0.025 } } }
\vdef{bdtdefault-11:N-EFF-TRIG-PIDMC-BSMM0:all}   {\ensuremath{{(84.23 \pm 2.53)\times 10^{-2}} } }
\vdef{bdtdefault-11:N-EFF-TRIG-MC-BSMM0:val}   {\ensuremath{{0.832 } } }
\vdef{bdtdefault-11:N-EFF-TRIG-MC-BSMM0:err}   {\ensuremath{{0.007 } } }
\vdef{bdtdefault-11:N-EFF-TRIG-MC-BSMM0:tot}   {\ensuremath{{0.026 } } }
\vdef{bdtdefault-11:N-EFF-TRIG-MC-BSMM0:all}   {\ensuremath{{(83.16 \pm 2.60)\times 10^{-2}} } }
\vdef{bdtdefault-11:N-EFF-CAND-BSMM0:val}   {\ensuremath{{0.980 } } }
\vdef{bdtdefault-11:N-EFF-CAND-BSMM0:err}   {\ensuremath{{0.002 } } }
\vdef{bdtdefault-11:N-EFF-CAND-BSMM0:tot}   {\ensuremath{{0.010 } } }
\vdef{bdtdefault-11:N-EFF-CAND-BSMM0:all}   {\ensuremath{{(98.00 \pm 0.99)\times 10^{-2}} } }
\vdef{bdtdefault-11:N-EFF-ANA-BSMM0:val}   {\ensuremath{{0.028 } } }
\vdef{bdtdefault-11:N-EFF-ANA-BSMM0:err}   {\ensuremath{{0.000 } } }
\vdef{bdtdefault-11:N-EFF-ANA-BSMM0:tot}   {\ensuremath{{0.001 } } }
\vdef{bdtdefault-11:N-EFF-ANA-BSMM0:all}   {\ensuremath{{(2.80 \pm 0.10)\times 10^{-2}} } }
\vdef{bdtdefault-11:N-EFF-TOT-BDMM0:val}   {\ensuremath{{0.0040 } } }
\vdef{bdtdefault-11:N-EFF-TOT-BDMM0:err}   {\ensuremath{{0.0001 } } }
\vdef{bdtdefault-11:N-EFF-TOT-BDMM0:tot}   {\ensuremath{{0.0003 } } }
\vdef{bdtdefault-11:N-EFF-TOT-BDMM0:all}   {\ensuremath{{(0.40 \pm 0.03)\times 10^{-2}} } }
\vdef{bdtdefault-11:N-ACC-BDMM0:val}   {\ensuremath{{0.247 } } }
\vdef{bdtdefault-11:N-ACC-BDMM0:err}   {\ensuremath{{0.001 } } }
\vdef{bdtdefault-11:N-ACC-BDMM0:tot}   {\ensuremath{{0.009 } } }
\vdef{bdtdefault-11:N-ACC-BDMM0:all}   {\ensuremath{{(24.65 \pm 0.87)\times 10^{-2}} } }
\vdef{bdtdefault-11:N-EFF-MU-PID-BDMM0:val}   {\ensuremath{{0.790 } } }
\vdef{bdtdefault-11:N-EFF-MU-PID-BDMM0:err}   {\ensuremath{{0.002 } } }
\vdef{bdtdefault-11:N-EFF-MU-PID-BDMM0:tot}   {\ensuremath{{0.032 } } }
\vdef{bdtdefault-11:N-EFF-MU-PID-BDMM0:all}   {\ensuremath{{(78.95 \pm 3.16)\times 10^{-2}} } }
\vdef{bdtdefault-11:N-EFF-MU-PIDMC-BDMM0:val}   {\ensuremath{{0.782 } } }
\vdef{bdtdefault-11:N-EFF-MU-PIDMC-BDMM0:err}   {\ensuremath{{0.003 } } }
\vdef{bdtdefault-11:N-EFF-MU-PIDMC-BDMM0:tot}   {\ensuremath{{0.031 } } }
\vdef{bdtdefault-11:N-EFF-MU-PIDMC-BDMM0:all}   {\ensuremath{{(78.25 \pm 3.14)\times 10^{-2}} } }
\vdef{bdtdefault-11:N-EFF-MU-MC-BDMM0:val}   {\ensuremath{{0.681 } } }
\vdef{bdtdefault-11:N-EFF-MU-MC-BDMM0:err}   {\ensuremath{{0.011 } } }
\vdef{bdtdefault-11:N-EFF-MU-MC-BDMM0:tot}   {\ensuremath{{0.030 } } }
\vdef{bdtdefault-11:N-EFF-MU-MC-BDMM0:all}   {\ensuremath{{(68.09 \pm 2.95)\times 10^{-2}} } }
\vdef{bdtdefault-11:N-EFF-TRIG-PID-BDMM0:val}   {\ensuremath{{0.792 } } }
\vdef{bdtdefault-11:N-EFF-TRIG-PID-BDMM0:err}   {\ensuremath{{0.003 } } }
\vdef{bdtdefault-11:N-EFF-TRIG-PID-BDMM0:tot}   {\ensuremath{{0.024 } } }
\vdef{bdtdefault-11:N-EFF-TRIG-PID-BDMM0:all}   {\ensuremath{{(79.18 \pm 2.39)\times 10^{-2}} } }
\vdef{bdtdefault-11:N-EFF-TRIG-PIDMC-BDMM0:val}   {\ensuremath{{0.838 } } }
\vdef{bdtdefault-11:N-EFF-TRIG-PIDMC-BDMM0:err}   {\ensuremath{{0.003 } } }
\vdef{bdtdefault-11:N-EFF-TRIG-PIDMC-BDMM0:tot}   {\ensuremath{{0.025 } } }
\vdef{bdtdefault-11:N-EFF-TRIG-PIDMC-BDMM0:all}   {\ensuremath{{(83.80 \pm 2.53)\times 10^{-2}} } }
\vdef{bdtdefault-11:N-EFF-TRIG-MC-BDMM0:val}   {\ensuremath{{0.840 } } }
\vdef{bdtdefault-11:N-EFF-TRIG-MC-BDMM0:err}   {\ensuremath{{0.011 } } }
\vdef{bdtdefault-11:N-EFF-TRIG-MC-BDMM0:tot}   {\ensuremath{{0.027 } } }
\vdef{bdtdefault-11:N-EFF-TRIG-MC-BDMM0:all}   {\ensuremath{{(83.97 \pm 2.74)\times 10^{-2}} } }
\vdef{bdtdefault-11:N-EFF-CAND-BDMM0:val}   {\ensuremath{{0.980 } } }
\vdef{bdtdefault-11:N-EFF-CAND-BDMM0:err}   {\ensuremath{{0.002 } } }
\vdef{bdtdefault-11:N-EFF-CAND-BDMM0:tot}   {\ensuremath{{0.010 } } }
\vdef{bdtdefault-11:N-EFF-CAND-BDMM0:all}   {\ensuremath{{(98.00 \pm 1.00)\times 10^{-2}} } }
\vdef{bdtdefault-11:N-EFF-ANA-BDMM0:val}   {\ensuremath{{0.029 } } }
\vdef{bdtdefault-11:N-EFF-ANA-BDMM0:err}   {\ensuremath{{0.001 } } }
\vdef{bdtdefault-11:N-EFF-ANA-BDMM0:tot}   {\ensuremath{{0.001 } } }
\vdef{bdtdefault-11:N-EFF-ANA-BDMM0:all}   {\ensuremath{{(2.91 \pm 0.11)\times 10^{-2}} } }
\vdef{bdtdefault-11:N-EXP-OBS-BS0:val}   {\ensuremath{{ 5.01 } } }
\vdef{bdtdefault-11:N-EXP-OBS-BS0:err}   {\ensuremath{{ 0.87 } } }
\vdef{bdtdefault-11:N-EXP-OBS-BD0:val}   {\ensuremath{{ 1.48 } } }
\vdef{bdtdefault-11:N-EXP-OBS-BD0:err}   {\ensuremath{{ 0.45 } } }
\vdef{bdtdefault-11:N-OBS-BSMM0:val}   {\ensuremath{{4 } } }
\vdef{bdtdefault-11:N-OBS-BDMM0:val}   {\ensuremath{{2 } } }
\vdef{bdtdefault-11:N-OFFLO-RARE0:val}   {\ensuremath{{ 4.81 } } }
\vdef{bdtdefault-11:N-OFFLO-RARE0:err}   {\ensuremath{{ 1.00 } } }
\vdef{bdtdefault-11:N-OFFHI-RARE0:val}   {\ensuremath{{ 0.02 } } }
\vdef{bdtdefault-11:N-OFFHI-RARE0:err}   {\ensuremath{{ 0.01 } } }
\vdef{bdtdefault-11:N-PEAK-BKG-BS0:val}   {\ensuremath{{ 0.27 } } }
\vdef{bdtdefault-11:N-PEAK-BKG-BS0:err}   {\ensuremath{{ 0.08 } } }
\vdef{bdtdefault-11:N-PEAK-BKG-BD0:val}   {\ensuremath{{ 0.47 } } }
\vdef{bdtdefault-11:N-PEAK-BKG-BD0:err}   {\ensuremath{{ 0.10 } } }
\vdef{bdtdefault-11:N-TAU-BS0:val}   {\ensuremath{{ 0.20 } } }
\vdef{bdtdefault-11:N-TAU-BS0:err}   {\ensuremath{{ 0.01 } } }
\vdef{bdtdefault-11:N-TAU-BD0:val}   {\ensuremath{{ 0.13 } } }
\vdef{bdtdefault-11:N-TAU-BD0:err}   {\ensuremath{{ 0.01 } } }
\vdef{bdtdefault-11:N-EXP2-SIG-BSMM1:val}   {\ensuremath{{ 1.81 } } }
\vdef{bdtdefault-11:N-EXP2-SIG-BSMM1:err}   {\ensuremath{{ 0.27 } } }
\vdef{bdtdefault-11:N-EXP2-SIG-BDMM1:val}   {\ensuremath{{0.155 } } }
\vdef{bdtdefault-11:N-EXP2-SIG-BDMM1:err}   {\ensuremath{{0.015 } } }
\vdef{bdtdefault-11:N-OBS-BKG1:val}   {\ensuremath{{13 } } }
\vdef{bdtdefault-11:N-EXP-BSMM1:val}   {\ensuremath{{ 2.21 } } }
\vdef{bdtdefault-11:N-EXP-BSMM1:err}   {\ensuremath{{ 0.73 } } }
\vdef{bdtdefault-11:N-EXP-BDMM1:val}   {\ensuremath{{ 1.48 } } }
\vdef{bdtdefault-11:N-EXP-BDMM1:err}   {\ensuremath{{ 0.48 } } }
\vdef{bdtdefault-11:N-LOW-BD1:val}   {\ensuremath{{5.200 } } }
\vdef{bdtdefault-11:N-HIGH-BD1:val}   {\ensuremath{{5.300 } } }
\vdef{bdtdefault-11:N-LOW-BS1:val}   {\ensuremath{{5.300 } } }
\vdef{bdtdefault-11:N-HIGH-BS1:val}   {\ensuremath{{5.450 } } }
\vdef{bdtdefault-11:N-PSS1:val}   {\ensuremath{{0.708 } } }
\vdef{bdtdefault-11:N-PSS1:err}   {\ensuremath{{0.012 } } }
\vdef{bdtdefault-11:N-PSS1:tot}   {\ensuremath{{0.038 } } }
\vdef{bdtdefault-11:N-PSD1:val}   {\ensuremath{{0.308 } } }
\vdef{bdtdefault-11:N-PSD1:err}   {\ensuremath{{0.019 } } }
\vdef{bdtdefault-11:N-PSD1:tot}   {\ensuremath{{0.025 } } }
\vdef{bdtdefault-11:N-PDS1:val}   {\ensuremath{{0.166 } } }
\vdef{bdtdefault-11:N-PDS1:err}   {\ensuremath{{0.010 } } }
\vdef{bdtdefault-11:N-PDS1:tot}   {\ensuremath{{0.013 } } }
\vdef{bdtdefault-11:N-PDD1:val}   {\ensuremath{{0.545 } } }
\vdef{bdtdefault-11:N-PDD1:err}   {\ensuremath{{0.021 } } }
\vdef{bdtdefault-11:N-PDD1:tot}   {\ensuremath{{0.034 } } }
\vdef{bdtdefault-11:N-EFF-TOT-BSMM1:val}   {\ensuremath{{0.0025 } } }
\vdef{bdtdefault-11:N-EFF-TOT-BSMM1:err}   {\ensuremath{{0.0001 } } }
\vdef{bdtdefault-11:N-EFF-TOT-BSMM1:tot}   {\ensuremath{{0.0003 } } }
\vdef{bdtdefault-11:N-EFF-TOT-BSMM1:all}   {\ensuremath{{(0.25 \pm 0.03)\times 10^{-2}} } }
\vdef{bdtdefault-11:N-ACC-BSMM1:val}   {\ensuremath{{0.229 } } }
\vdef{bdtdefault-11:N-ACC-BSMM1:err}   {\ensuremath{{0.001 } } }
\vdef{bdtdefault-11:N-ACC-BSMM1:tot}   {\ensuremath{{0.011 } } }
\vdef{bdtdefault-11:N-ACC-BSMM1:all}   {\ensuremath{{(22.94 \pm 1.15)\times 10^{-2}} } }
\vdef{bdtdefault-11:N-EFF-MU-PID-BSMM1:val}   {\ensuremath{{0.780 } } }
\vdef{bdtdefault-11:N-EFF-MU-PID-BSMM1:err}   {\ensuremath{{0.001 } } }
\vdef{bdtdefault-11:N-EFF-MU-PID-BSMM1:tot}   {\ensuremath{{0.062 } } }
\vdef{bdtdefault-11:N-EFF-MU-PID-BSMM1:all}   {\ensuremath{{(78.04 \pm 6.24)\times 10^{-2}} } }
\vdef{bdtdefault-11:N-EFF-MU-PIDMC-BSMM1:val}   {\ensuremath{{0.826 } } }
\vdef{bdtdefault-11:N-EFF-MU-PIDMC-BSMM1:err}   {\ensuremath{{0.001 } } }
\vdef{bdtdefault-11:N-EFF-MU-PIDMC-BSMM1:tot}   {\ensuremath{{0.066 } } }
\vdef{bdtdefault-11:N-EFF-MU-PIDMC-BSMM1:all}   {\ensuremath{{(82.63 \pm 6.61)\times 10^{-2}} } }
\vdef{bdtdefault-11:N-EFF-MU-MC-BSMM1:val}   {\ensuremath{{0.853 } } }
\vdef{bdtdefault-11:N-EFF-MU-MC-BSMM1:err}   {\ensuremath{{0.008 } } }
\vdef{bdtdefault-11:N-EFF-MU-MC-BSMM1:tot}   {\ensuremath{{0.069 } } }
\vdef{bdtdefault-11:N-EFF-MU-MC-BSMM1:all}   {\ensuremath{{(85.30 \pm 6.87)\times 10^{-2}} } }
\vdef{bdtdefault-11:N-EFF-TRIG-PID-BSMM1:val}   {\ensuremath{{0.769 } } }
\vdef{bdtdefault-11:N-EFF-TRIG-PID-BSMM1:err}   {\ensuremath{{0.002 } } }
\vdef{bdtdefault-11:N-EFF-TRIG-PID-BSMM1:tot}   {\ensuremath{{0.046 } } }
\vdef{bdtdefault-11:N-EFF-TRIG-PID-BSMM1:all}   {\ensuremath{{(76.87 \pm 4.62)\times 10^{-2}} } }
\vdef{bdtdefault-11:N-EFF-TRIG-PIDMC-BSMM1:val}   {\ensuremath{{0.756 } } }
\vdef{bdtdefault-11:N-EFF-TRIG-PIDMC-BSMM1:err}   {\ensuremath{{0.003 } } }
\vdef{bdtdefault-11:N-EFF-TRIG-PIDMC-BSMM1:tot}   {\ensuremath{{0.045 } } }
\vdef{bdtdefault-11:N-EFF-TRIG-PIDMC-BSMM1:all}   {\ensuremath{{(75.59 \pm 4.54)\times 10^{-2}} } }
\vdef{bdtdefault-11:N-EFF-TRIG-MC-BSMM1:val}   {\ensuremath{{0.733 } } }
\vdef{bdtdefault-11:N-EFF-TRIG-MC-BSMM1:err}   {\ensuremath{{0.010 } } }
\vdef{bdtdefault-11:N-EFF-TRIG-MC-BSMM1:tot}   {\ensuremath{{0.045 } } }
\vdef{bdtdefault-11:N-EFF-TRIG-MC-BSMM1:all}   {\ensuremath{{(73.33 \pm 4.52)\times 10^{-2}} } }
\vdef{bdtdefault-11:N-EFF-CAND-BSMM1:val}   {\ensuremath{{0.980 } } }
\vdef{bdtdefault-11:N-EFF-CAND-BSMM1:err}   {\ensuremath{{0.002 } } }
\vdef{bdtdefault-11:N-EFF-CAND-BSMM1:tot}   {\ensuremath{{0.010 } } }
\vdef{bdtdefault-11:N-EFF-CAND-BSMM1:all}   {\ensuremath{{(98.00 \pm 1.00)\times 10^{-2}} } }
\vdef{bdtdefault-11:N-EFF-ANA-BSMM1:val}   {\ensuremath{{0.018 } } }
\vdef{bdtdefault-11:N-EFF-ANA-BSMM1:err}   {\ensuremath{{0.000 } } }
\vdef{bdtdefault-11:N-EFF-ANA-BSMM1:tot}   {\ensuremath{{0.001 } } }
\vdef{bdtdefault-11:N-EFF-ANA-BSMM1:all}   {\ensuremath{{(1.77 \pm 0.07)\times 10^{-2}} } }
\vdef{bdtdefault-11:N-EFF-TOT-BDMM1:val}   {\ensuremath{{0.0023 } } }
\vdef{bdtdefault-11:N-EFF-TOT-BDMM1:err}   {\ensuremath{{0.0001 } } }
\vdef{bdtdefault-11:N-EFF-TOT-BDMM1:tot}   {\ensuremath{{0.0003 } } }
\vdef{bdtdefault-11:N-EFF-TOT-BDMM1:all}   {\ensuremath{{(0.25 \pm 0.03)\times 10^{-2}} } }
\vdef{bdtdefault-11:N-ACC-BDMM1:val}   {\ensuremath{{0.226 } } }
\vdef{bdtdefault-11:N-ACC-BDMM1:err}   {\ensuremath{{0.001 } } }
\vdef{bdtdefault-11:N-ACC-BDMM1:tot}   {\ensuremath{{0.011 } } }
\vdef{bdtdefault-11:N-ACC-BDMM1:all}   {\ensuremath{{(22.58 \pm 1.13)\times 10^{-2}} } }
\vdef{bdtdefault-11:N-EFF-MU-PID-BDMM1:val}   {\ensuremath{{0.780 } } }
\vdef{bdtdefault-11:N-EFF-MU-PID-BDMM1:err}   {\ensuremath{{0.002 } } }
\vdef{bdtdefault-11:N-EFF-MU-PID-BDMM1:tot}   {\ensuremath{{0.062 } } }
\vdef{bdtdefault-11:N-EFF-MU-PID-BDMM1:all}   {\ensuremath{{(78.04 \pm 6.25)\times 10^{-2}} } }
\vdef{bdtdefault-11:N-EFF-MU-PIDMC-BDMM1:val}   {\ensuremath{{0.830 } } }
\vdef{bdtdefault-11:N-EFF-MU-PIDMC-BDMM1:err}   {\ensuremath{{0.002 } } }
\vdef{bdtdefault-11:N-EFF-MU-PIDMC-BDMM1:tot}   {\ensuremath{{0.066 } } }
\vdef{bdtdefault-11:N-EFF-MU-PIDMC-BDMM1:all}   {\ensuremath{{(83.03 \pm 6.65)\times 10^{-2}} } }
\vdef{bdtdefault-11:N-EFF-MU-MC-BDMM1:val}   {\ensuremath{{0.836 } } }
\vdef{bdtdefault-11:N-EFF-MU-MC-BDMM1:err}   {\ensuremath{{0.012 } } }
\vdef{bdtdefault-11:N-EFF-MU-MC-BDMM1:tot}   {\ensuremath{{0.068 } } }
\vdef{bdtdefault-11:N-EFF-MU-MC-BDMM1:all}   {\ensuremath{{(83.57 \pm 6.79)\times 10^{-2}} } }
\vdef{bdtdefault-11:N-EFF-TRIG-PID-BDMM1:val}   {\ensuremath{{0.770 } } }
\vdef{bdtdefault-11:N-EFF-TRIG-PID-BDMM1:err}   {\ensuremath{{0.004 } } }
\vdef{bdtdefault-11:N-EFF-TRIG-PID-BDMM1:tot}   {\ensuremath{{0.046 } } }
\vdef{bdtdefault-11:N-EFF-TRIG-PID-BDMM1:all}   {\ensuremath{{(77.02 \pm 4.63)\times 10^{-2}} } }
\vdef{bdtdefault-11:N-EFF-TRIG-PIDMC-BDMM1:val}   {\ensuremath{{0.760 } } }
\vdef{bdtdefault-11:N-EFF-TRIG-PIDMC-BDMM1:err}   {\ensuremath{{0.004 } } }
\vdef{bdtdefault-11:N-EFF-TRIG-PIDMC-BDMM1:tot}   {\ensuremath{{0.046 } } }
\vdef{bdtdefault-11:N-EFF-TRIG-PIDMC-BDMM1:all}   {\ensuremath{{(75.98 \pm 4.58)\times 10^{-2}} } }
\vdef{bdtdefault-11:N-EFF-TRIG-MC-BDMM1:val}   {\ensuremath{{0.677 } } }
\vdef{bdtdefault-11:N-EFF-TRIG-MC-BDMM1:err}   {\ensuremath{{0.016 } } }
\vdef{bdtdefault-11:N-EFF-TRIG-MC-BDMM1:tot}   {\ensuremath{{0.044 } } }
\vdef{bdtdefault-11:N-EFF-TRIG-MC-BDMM1:all}   {\ensuremath{{(67.75 \pm 4.37)\times 10^{-2}} } }
\vdef{bdtdefault-11:N-EFF-CAND-BDMM1:val}   {\ensuremath{{0.980 } } }
\vdef{bdtdefault-11:N-EFF-CAND-BDMM1:err}   {\ensuremath{{0.003 } } }
\vdef{bdtdefault-11:N-EFF-CAND-BDMM1:tot}   {\ensuremath{{0.010 } } }
\vdef{bdtdefault-11:N-EFF-CAND-BDMM1:all}   {\ensuremath{{(98.00 \pm 1.03)\times 10^{-2}} } }
\vdef{bdtdefault-11:N-EFF-ANA-BDMM1:val}   {\ensuremath{{0.019 } } }
\vdef{bdtdefault-11:N-EFF-ANA-BDMM1:err}   {\ensuremath{{0.001 } } }
\vdef{bdtdefault-11:N-EFF-ANA-BDMM1:tot}   {\ensuremath{{0.001 } } }
\vdef{bdtdefault-11:N-EFF-ANA-BDMM1:all}   {\ensuremath{{(1.89 \pm 0.08)\times 10^{-2}} } }
\vdef{bdtdefault-11:N-EXP-OBS-BS1:val}   {\ensuremath{{ 4.15 } } }
\vdef{bdtdefault-11:N-EXP-OBS-BS1:err}   {\ensuremath{{ 0.77 } } }
\vdef{bdtdefault-11:N-EXP-OBS-BD1:val}   {\ensuremath{{ 1.85 } } }
\vdef{bdtdefault-11:N-EXP-OBS-BD1:err}   {\ensuremath{{ 0.49 } } }
\vdef{bdtdefault-11:N-OBS-BSMM1:val}   {\ensuremath{{9 } } }
\vdef{bdtdefault-11:N-OBS-BDMM1:val}   {\ensuremath{{1 } } }
\vdef{bdtdefault-11:N-OFFLO-RARE1:val}   {\ensuremath{{ 1.91 } } }
\vdef{bdtdefault-11:N-OFFLO-RARE1:err}   {\ensuremath{{ 0.41 } } }
\vdef{bdtdefault-11:N-OFFHI-RARE1:val}   {\ensuremath{{ 0.02 } } }
\vdef{bdtdefault-11:N-OFFHI-RARE1:err}   {\ensuremath{{ 0.00 } } }
\vdef{bdtdefault-11:N-PEAK-BKG-BS1:val}   {\ensuremath{{ 0.13 } } }
\vdef{bdtdefault-11:N-PEAK-BKG-BS1:err}   {\ensuremath{{ 0.03 } } }
\vdef{bdtdefault-11:N-PEAK-BKG-BD1:val}   {\ensuremath{{ 0.22 } } }
\vdef{bdtdefault-11:N-PEAK-BKG-BD1:err}   {\ensuremath{{ 0.05 } } }
\vdef{bdtdefault-11:N-TAU-BS1:val}   {\ensuremath{{ 0.20 } } }
\vdef{bdtdefault-11:N-TAU-BS1:err}   {\ensuremath{{ 0.01 } } }
\vdef{bdtdefault-11:N-TAU-BD1:val}   {\ensuremath{{ 0.13 } } }
\vdef{bdtdefault-11:N-TAU-BD1:err}   {\ensuremath{{ 0.01 } } }
\vdef{bdtdefault-11:N-CSBF-TNP-BS0:val}   {\ensuremath{{0.000026 } } }
\vdef{bdtdefault-11:N-CSBF-TNP-BS0:err}   {\ensuremath{{0.000000 } } }
\vdef{bdtdefault-11:N-CSBF-MC-BS0:val}   {\ensuremath{{0.000026 } } }
\vdef{bdtdefault-11:N-CSBF-MC-BS0:err}   {\ensuremath{{0.000000 } } }
\vdef{bdtdefault-11:N-CSBF-BS0:val}   {\ensuremath{{0.000026 } } }
\vdef{bdtdefault-11:N-CSBF-BS0:err}   {\ensuremath{{0.000000 } } }
\vdef{bdtdefault-11:N-CSBF-TNP-BS1:val}   {\ensuremath{{0.000026 } } }
\vdef{bdtdefault-11:N-CSBF-TNP-BS1:err}   {\ensuremath{{0.000001 } } }
\vdef{bdtdefault-11:N-CSBF-MC-BS1:val}   {\ensuremath{{0.000026 } } }
\vdef{bdtdefault-11:N-CSBF-MC-BS1:err}   {\ensuremath{{0.000001 } } }
\vdef{bdtdefault-11:N-CSBF-BS1:val}   {\ensuremath{{0.000027 } } }
\vdef{bdtdefault-11:N-CSBF-BS1:err}   {\ensuremath{{0.000001 } } }
\vdef{bdtdefault-11:N-EFF-TOT-BS0:val}   {\ensuremath{{0.000971 } } }
\vdef{bdtdefault-11:N-EFF-TOT-BS0:err}   {\ensuremath{{0.000007 } } }
\vdef{bdtdefault-11:N-ACC-BS0:val}   {\ensuremath{{0.1145 } } }
\vdef{bdtdefault-11:N-ACC-BS0:err}   {\ensuremath{{0.0002 } } }
\vdef{bdtdefault-11:N-EFF-MU-PID-BS0:val}   {\ensuremath{{0.7877 } } }
\vdef{bdtdefault-11:N-EFF-MU-PID-BS0:err}   {\ensuremath{{0.0005 } } }
\vdef{bdtdefault-11:N-EFF-MU-PIDMC-BS0:val}   {\ensuremath{{0.7767 } } }
\vdef{bdtdefault-11:N-EFF-MU-PIDMC-BS0:err}   {\ensuremath{{0.0006 } } }
\vdef{bdtdefault-11:N-EFF-MU-MC-BS0:val}   {\ensuremath{{0.7737 } } }
\vdef{bdtdefault-11:N-EFF-MU-MC-BS0:err}   {\ensuremath{{0.0023 } } }
\vdef{bdtdefault-11:N-EFF-TRIG-PID-BS0:val}   {\ensuremath{{0.7871 } } }
\vdef{bdtdefault-11:N-EFF-TRIG-PID-BS0:err}   {\ensuremath{{0.0007 } } }
\vdef{bdtdefault-11:N-EFF-TRIG-PIDMC-BS0:val}   {\ensuremath{{0.8335 } } }
\vdef{bdtdefault-11:N-EFF-TRIG-PIDMC-BS0:err}   {\ensuremath{{0.0006 } } }
\vdef{bdtdefault-11:N-EFF-TRIG-MC-BS0:val}   {\ensuremath{{0.7622 } } }
\vdef{bdtdefault-11:N-EFF-TRIG-MC-BS0:err}   {\ensuremath{{0.0026 } } }
\vdef{bdtdefault-11:N-EFF-CAND-BS0:val}   {\ensuremath{{0.9800 } } }
\vdef{bdtdefault-11:N-EFF-CAND-BS0:err}   {\ensuremath{{0.0021 } } }
\vdef{bdtdefault-11:N-EFF-ANA-BS0:val}   {\ensuremath{{0.0151 } } }
\vdef{bdtdefault-11:N-EFF-ANA-BS0:err}   {\ensuremath{{0.0010 } } }
\vdef{bdtdefault-11:N-OBS-BS0:val}   {\ensuremath{{9066 } } }
\vdef{bdtdefault-11:N-OBS-BS0:err}   {\ensuremath{{107 } } }
\vdef{bdtdefault-11:N-EFF-TOT-BS1:val}   {\ensuremath{{0.000279 } } }
\vdef{bdtdefault-11:N-EFF-TOT-BS1:err}   {\ensuremath{{0.000004 } } }
\vdef{bdtdefault-11:N-ACC-BS1:val}   {\ensuremath{{0.0760 } } }
\vdef{bdtdefault-11:N-ACC-BS1:err}   {\ensuremath{{0.0002 } } }
\vdef{bdtdefault-11:N-EFF-MU-PID-BS1:val}   {\ensuremath{{0.7797 } } }
\vdef{bdtdefault-11:N-EFF-MU-PID-BS1:err}   {\ensuremath{{0.0005 } } }
\vdef{bdtdefault-11:N-EFF-MU-PIDMC-BS1:val}   {\ensuremath{{0.8281 } } }
\vdef{bdtdefault-11:N-EFF-MU-PIDMC-BS1:err}   {\ensuremath{{0.0005 } } }
\vdef{bdtdefault-11:N-EFF-MU-MC-BS1:val}   {\ensuremath{{0.7664 } } }
\vdef{bdtdefault-11:N-EFF-MU-MC-BS1:err}   {\ensuremath{{0.0037 } } }
\vdef{bdtdefault-11:N-EFF-TRIG-PID-BS1:val}   {\ensuremath{{0.7525 } } }
\vdef{bdtdefault-11:N-EFF-TRIG-PID-BS1:err}   {\ensuremath{{0.0019 } } }
\vdef{bdtdefault-11:N-EFF-TRIG-PIDMC-BS1:val}   {\ensuremath{{0.7404 } } }
\vdef{bdtdefault-11:N-EFF-TRIG-PIDMC-BS1:err}   {\ensuremath{{0.0019 } } }
\vdef{bdtdefault-11:N-EFF-TRIG-MC-BS1:val}   {\ensuremath{{0.5707 } } }
\vdef{bdtdefault-11:N-EFF-TRIG-MC-BS1:err}   {\ensuremath{{0.0050 } } }
\vdef{bdtdefault-11:N-EFF-CAND-BS1:val}   {\ensuremath{{0.9800 } } }
\vdef{bdtdefault-11:N-EFF-CAND-BS1:err}   {\ensuremath{{0.0040 } } }
\vdef{bdtdefault-11:N-EFF-ANA-BS1:val}   {\ensuremath{{0.0089 } } }
\vdef{bdtdefault-11:N-EFF-ANA-BS1:err}   {\ensuremath{{0.0013 } } }
\vdef{bdtdefault-11:N-OBS-BS1:val}   {\ensuremath{{2611 } } }
\vdef{bdtdefault-11:N-OBS-BS1:err}   {\ensuremath{{70 } } }

\newcommand{\base}{default-11}
\newcommand{\noA}{NoData}
\newcommand{\noB}{NoData2011}
\newcommand{\noMc}{NoMc}
\newcommand{\sample}{NoData}
\newcommand{\samplet}{NoData}
\newcommand{\channel}{A}
\newcommand{\StrutSmall}{\rule[-1.0mm]{0pt}{4.7mm}}
\newcommand{\StrutLarge}{\rule[-2.3mm]{0pt}{6.3mm}}
\newcommand{\mcA}{SgMcPU-APV0}
\newcommand{\mcB}{SgMcPU-APV1}
\newcommand{\daA}{SgData-APV0}
\newcommand{\daB}{SgData-APV1}

\newboolean{george}
\setboolean{george}{true}
\ifthenelse{\boolean{george}}{
\renewcommand{\W}{\PW}
\renewcommand{\Z}{\mathrm{Z}}
\renewcommand{\p}{\Pp}
\renewcommand{\t}{\ensuremath{\mathrm{t}}}
\renewcommand{\b}{\ensuremath{\mathrm{b}}}
\renewcommand{\c}{\ensuremath{\mathrm{c}}}
\renewcommand{\s}{\ensuremath{\mathrm{s}}}
\renewcommand{\d}{\ensuremath{\mathrm{d}}}
\renewcommand{\u}{\ensuremath{\mathrm{u}}}
\renewcommand{\B}{\PB}
\renewcommand{\Bz}{\PBz}
\renewcommand{\Bs}{\PBzs}
\renewcommand{\Bp}{\PBp}
\renewcommand{\mup}{\Pgmp}
\renewcommand{\mun}{\Pgmm}
\renewcommand{\h}{\mathrm{h}}
\renewcommand{\K}{\PK}
\renewcommand{\Kp}{\PKp}
\renewcommand{\Km}{\PKm}
\renewcommand{\Kstar}{\PK^*}
\renewcommand{\Dstarp}{\mathrm{D}^{*+}} %% George, change this yourself!
\renewcommand{\Dz}{\PDz}
\renewcommand{\pim}{\Pgpm}
\renewcommand{\pip}{\Pgpp}
\renewcommand{\bbbar}{\ensuremath{\mathrm{b}\mathrm{\overline{b}}}}
\renewcommand{\bsmm}{\ensuremath{\PBzs\to\Pgmp\Pgmm}}
\renewcommand{\bdmm}{\ensuremath{\PBz\to\Pgmp\Pgmm}}
\renewcommand{\bupsik}{\ensuremath{\PBp\to\cPJgy\PKp}}
\renewcommand{\bspsiphi}{\ensuremath{\PBzs\to\cPJgy\phi}}
\renewcommand{\bpsikst}{\ensuremath{\PB\to\cPJgy\Kstar}}
\renewcommand{\jpsi}{\ensuremath{\cPJgy}}

\title{Search for $\mathrm{B}_\mathrm{s}^0\to\mu^+\mu^-$ and $\mathrm{B^0}\to\mu^+\mu^-$ decays}
\abstract{A search for the rare decays $\mathrm{B}_\mathrm{s}^0\to\mu^+\mu^-$ and
$\mathrm{B^0}\to\mu^+\mu^-$ is performed in pp collisions at
$\sqrt{s}=7\,\mathrm{TeV}$, with a data sample corresponding to an
integrated luminosity of $5\,\mathrm{fb}^{-1}$ collected by the CMS
experiment at the LHC.  In both decays, the number of events observed
after all selection requirements is consistent with the expectation
from background plus standard model signal predictions.  The resulting
upper limits on the branching fractions are ${\cal
B}(\mathrm{B}_\mathrm{s}^0\to\mu^+\mu^-) < 7.7\times10^{-9}$ and ${\cal
B}(\mathrm{B^0}\to\mu^+\mu^-) < 1.8\times10^{-9}$ at 95\% confidence level.}

}{

\title{Search for $B_s^0\to\mu^+\mu^-$ and $B^0\to\mu^+\mu^-$ decays}
\abstract{A search for the rare decays $B_s^0\to\mu^+\mu^-$ and
$B^0\to\mu^+\mu^-$ is performed in $pp$ collisions at
$\sqrt{s}=7\,\mathrm{TeV}$, with a data sample corresponding to an
integrated luminosity of $5\,\mathrm{fb}^{-1}$ collected by the CMS
experiment at the LHC.  In both decays, the number of events observed
after all selection requirements is consistent with the expectation
from background plus standard model signal predictions.  The resulting
upper limits on the branching fractions are ${\cal
B}(B_s^0\to\mu^+\mu^-) < 7.7\times10^{-9}$ and ${\cal
B}(B^0\to\mu^+\mu^-) < 1.8\times10^{-9}$ at 95\% confidence level.}

}

\date{\today}

\hypersetup{%
pdfauthor={CMS Collaboration},%
pdftitle={Search for B_s to mu^+ mu^- and B^0 to mu^+ mu^- decays},%
pdfsubject={CMS},%
pdfkeywords={CMS, physics, software, computing}}

\maketitle %maketitle comes after all the front information has been supplied

\newlength\figwid
\ifthenelse{\boolean{cms@external}}{\setlength\figwid{0.23\textwidth}}{\setlength\figwid{0.45\textwidth}}

\section{Introduction}
\label{s:intro}

The decays  $\Bs(\Bz)\to \mup\mun$  
are highly suppressed in the standard model (SM) of particle physics, 
which predicts the branching fractions to be $\cbf(\bsmm)=(3.2\pm0.2)\times10^{-9}$ and 
$\cbf(\bdmm)=(1.0\pm0.1)\times10^{-10}$~\cite{Buras:2010wr}. 
This suppression is due to the flavor-changing neutral current transitions $\b \to \s(\d)$,
which are forbidden at tree level and can only proceed via high-order diagrams that are described by 
electroweak penguin and box diagrams at the one-loop level. 
Additionally, the decays are helicity suppressed by a factor 
of $m_{\mu}^2/m_{\B}^2$, where $m_{\mu}$  and $m_\B$ are the masses of the muon and $\B$ meson,
 respectively (the symbol $\B$ is used to denote $\Bz$ or $\Bs$ mesons).
Furthermore, these 
decays also require an internal quark annihilation within the $\B$ meson that reduces the decay rate by an additional factor 
 of $f_{\B}^2/m_{\B}^2$, where $f_\B$ is the decay constant of the $\B$ meson. The leading theoretical uncertainty is due to 
 incomplete knowledge of $f_\B$, which is constrained by measurements of the mixing mass difference 
$\Delta m_{\s}$ ($\Delta m_{\d}$) for \Bs (\Bz) mesons. 

Several extensions of the SM predict enhancements to the branching fractions for these rare decays. 
In supersymmetric models with non-universal Higgs masses~\cite{Ellis:2006jy} and in specific models containing
leptoquarks~\cite{Davidson:2010uu}, for example, the \bsmm\ and \bdmm\ branching fractions can be 
enhanced. 
In the minimal supersymmetric extension of the SM,  the rates are strongly enhanced at large values of $\tan\beta$, 
which is the ratio of the two vacuum expectation values of the two Higgs boson doublets~\cite{Choudhury:2005rz,Parry:2005fp} . 
However, in most models of new physics, the decay rates can also be suppressed for specific choices of model 
parameters~\cite{Ellis:2007kb}.

At the Tevatron, the D0 experiment has published an upper limit of $\cbf(\bsmm) < 5.1\times10^{-8}$~\cite{Abazov:2010fs}
at 95\% confidence level (CL). 
The CDF experiment has set a limit of $\cbf(\bsmm) < 4.0\times10^{-8}$ and
$\cbf(\bdmm) < 6.0\times10^{-9}$, and also reported an excess of \bsmm\
events, corresponding to $\cbf(\bsmm) = (1.8^{+1.1}_{-0.9})\times10^{-8}$~\cite{Aaltonen:2011fi}. 
At the Large Hadron Collider (LHC), two experiments have published results:
$\cbf(\bsmm) < 1.9\times10^{-8}$ and $\cbf(\bdmm) < 4.6\times10^{-9}$
by the Compact Muon Solenoid (CMS) Collaboration~\cite{Chatrchyan:2011kr}, and 
$\cbf(\bsmm) <1.4\times10^{-8}$ and $\cbf(\bdmm) <3.2\times10^{-9}$ 
by the LHCb  Collaboration~\cite{LHCb:2011ac}.

This paper reports on a new simultaneous search for \bsmm\ and \bdmm\ decays using data  
collected in 2011 by the CMS experiment in $\p\p$ collisions at $\sqrt{s}=7$\,TeV at the LHC.
The dataset corresponds to an integrated luminosity of $\vuse{lumiTot} \invfb$.
An event-counting experiment is performed in dimuon mass regions
around the \Bs\ and \Bz\ masses. 
To avoid potential bias, a ``blind'' analysis approach is applied where the signal region is not 
observed until all selection criteria are established. 
Monte Carlo (MC) simulations are used to estimate backgrounds due to $\B$ decays. 
Combinatorial backgrounds are evaluated from the data in dimuon invariant mass ($m_{\mu\mu}$) sidebands. 
In the CMS detector, the mass resolution, which influences the separation between \bsmm\ and \bdmm\ decays, 
depends on the pseudorapidity $\eta$ of the reconstructed particles. 
The pseudorapidity is defined as $\eta = -\ln[\tan(\theta/2)]$, 
where $\theta$ is the polar angle with respect to the counterclockwise proton beam direction.
The background level also depends significantly on the $\eta$ of the $\B$ candidate. 
Therefore, the analysis is performed separately in two channels, ``barrel'' and ``endcap'', and then combined for the final result. 
The barrel channel contains the candidates where both muons have $|\eta| < 1.4$ and 
the endcap channel contains those where at least one muon has $|\eta| > 1.4$.

A ``normalization'' sample of events with \bupsik decays (where $\jpsi\to\mup\mun$) is used 
to remove uncertainties related to the $\bbbar$ production cross section and the integrated luminosity. 
The signal and normalization efficiencies are determined through MC simulation studies. 
To validate the simulation distributions, such as the \Bs\ transverse momentum (\pt) spectrum, 
and to evaluate potential effects resulting from differences in the fragmentation of \Bp\ and \Bs, a ``control'' sample of
reconstructed \bspsiphi\ decays (with $\jpsi\to\mup\mun$ and $\phi\to\Kp\Km$) is used.

The dataset includes periods of high instantaneous luminosity conditions, 
with an average of 8 interactions per bunch crossing (later referred to as ``pileup'').
The analysis algorithms and the selection criteria have been optimized to mitigate the effects of pileup by  
reducing the influence of tracks coming from additional interactions in the event, as explained in Section~\ref{s:selection}. 
In parallel with the LHC luminosity increase, the CMS event triggering requirements also changed during the data-taking period.
The analysis and simulations take these changes into account so that all MC samples  
incorporate the appropriate mixture of the trigger conditions, 
and the selection requirements applied in the data reconstruction are more restrictive than the most stringent trigger criteria. 

The limits on the branching fractions depend on both systematic and statistical uncertainties.
Several sources of systematic uncertainties can influence the estimated efficiency: detector acceptance,
and analysis, muon identification and triggering efficiencies. 
The evaluation of the individual values are presented in the sections below when discussing the relevant efficiencies and then 
are combined in Section~\ref{s:results}.

The data analyzed here include the event sample corresponding to an integrated luminosity of $1.14\,\mathrm{fb}^{-1}$,
which was used to obtain the earlier CMS result~\cite{Chatrchyan:2011kr}. 
The present analysis differs in several ways:
the total dataset is almost five times larger;
new selection variables are added to the analysis;
the selection criteria are optimized for higher pileup and varying trigger requirements;
and the description of rare backgrounds is improved. 
All these changes result in a better signal sensitivity.  %  and a higher expected signal-to-background ratio.

\section{Monte Carlo simulation}
\label{s:mcsimulation}
    
Simulated events are used to determine the efficiencies for the signal and normalization samples.
We split the efficiency into four parts: detector acceptance, analysis efficiency, and muon identification and trigger efficiencies.
The detector acceptance combines the geometrical detector acceptance and the tracking efficiency, 
and is defined as tracks within $|\eta| < 2.4$ and satisfying $\pt > 1\gev$ ($\pt > 0.5\gev$) for muons (kaons).
The acceptance is about $25\%$ ($23\%$) for signal events in the barrel (endcap) channels.  
In the \pt\ range relevant for this analysis the tracking efficiency for isolated muons and kaons is 
above 99.5\%~\cite{CMS-PAS-TRK-10-002}.
The analysis efficiency refers to the selection requirements described in Section~\ref{s:selection}, 
and is for signal events about $2.0\%$ ($1.2\%$) in the barrel (endcap) channels.
The muon identification and trigger efficiencies are presented in Sections~\ref{s:cmsdetector} and~\ref{s:muons}, respectively.
The analysis, muon identification, and trigger efficiencies are all obtained from simulation and checked in data.  
Good agreement is found, and the residual differences are used to estimate systematic uncertainties on the efficiency estimates.

The simulated samples are also used to estimate the background from rare B decays where one or two hadrons are misidentified as muons. 
These decays include a variety of channels of the type $\B\to \h^-\mup\nu$ and $\B\to \h^+\h^-$, where $\h$ is a  $\pi$, $\K$ or $\p$ 
and $\B$ stands for $\Bz, \Bs$ mesons or  $\Lambda_\b$ baryons. 
The most important backgrounds are from
$\Bs\to\Km\Kp$, $\Bz\to\Kp\pim$  and from the semileptonic decays  
$\Bz\to \pim\mup\nu$,
$\Bs\to \K^-\mup\nu$, and
$\Lambda^{0}_{\b} \rightarrow \p\mu^{-}\bar{\nu}$.

The samples of simulated events are generated with {\sc Pythia~6.424 (Tune Z2)}~\cite{Sjostrand:2006za}, the unstable particles 
are decayed via {\sc EvtGen}~\cite{Lange:2001uf}, and the detector response is simulated with {\sc Geant4}~\cite{Ivanchenko:2003xp}. 
The signal and background events are selected from generic quantum chromodynamic (QCD) $2 \rightarrow 2$ sub-processes 
and provide a mixture of gluon-fusion, flavor-excitation, and gluon-splitting production. 
The evolution of the triggers used to collect the data is incorporated in the reconstruction of the simulated events. 
The number of simulated events in all the channels approximately match the expected number given the integrated luminosity.

\section{The CMS detector}
\label{s:cmsdetector}
The CMS detector is a general-purpose detector designed and built to study physics at the TeV scale. A detailed description
can be found in Ref.~\cite{Adolphi:2008zzk}. For this analysis, the main subdetectors used are a silicon tracker, composed of 
pixel and strip detectors within a 3.8\,T axial magnetic field, and a muon detector, which is divided into a barrel section and two
endcaps, consisting of gas-ionization detectors embedded in the steel return yoke of the solenoid. 
The silicon tracker detects charged particles within the pseudorapidity range $|\eta| < 2.5$. The pixel detector is composed of 
three layers in the barrel region and two disks located on each side in the forward regions of the detector. 
In total, the pixel detector contains about 66 million 100$\mum \times 150\mum$ pixels. 
Further from the interaction region is a microstrip detector, 
which is composed of ten barrel layers, and three inner and nine outer disks on either end of the detector, 
with a strip pitches between 80 and 180$\mum$. 
In total, the microstrip detector contains around 10 million strips and, together with the pixel
detector, provides an impact parameter resolution of $\sim 15 \mum$. Due to the high granularity of the silicon tracker and to 
the strong magnetic field, a \pt\ resolution of about 1.5\%~\cite{Khachatryan:2010pw} is obtained for the charged particles 
in the \pt\ range relevant for this analysis. 
The systematic uncertainty on the hadronic track reconstruction efficiency is estimated to be 4\%~\cite{Khachatryan:2010pw}. 
Muons are detected in the pseudorapidity range $|\eta| < 2.4$ by detectors made of three technologies: drift tubes, cathode strip chambers, 
and resistive plate chambers.
The analysis is nearly independent of pileup because of the high granularity of the CMS silicon tracker and the excellent 
three-dimensional (3D) hit resolution of the pixel detector.

The dimuon candidate events are selected with a two-level trigger system,
the first level only uses the muon detector information, 
while the high-level trigger (HLT) uses additional information from the pixel and strip detectors. 
The first-level trigger requires two muon candidates without any explicit \pt\ requirement, 
but there is an implicit  selection since muons must reach the muon detectors
(about $3.5\gev$ in the barrel and $2\gev$ in the endcap).
The HLT imposes a \pt\ requirement and uses additional information from the silicon tracker. 
As the LHC instantaneous luminosity increased, the trigger requirements were gradually tightened. 
This change in trigger requirements is also included in the trigger simulations.  
The most stringent HLT selection requires two muons each with $\pt>4\gev$, 
the dimuon $\pt>3.9\gev$ ($5.9\gev$ in the endcap),
dimuon invariant mass within $4.8 < m_{\mu\mu} < 6.0\gev$, 
and a 3D distance of closest approach to each other of $d'_\mathrm{ca} < 0.5\cm$.  
For the entire dataset, the offline analysis selection is more restrictive than the most stringent trigger selections. 

For the normalization (\bupsik) and control (\bspsiphi) samples, the data are collected by requiring the following:
two muons each with $\pt > 4\gev$, 
dimuon $\pt > 6.9\gev$, $|\eta|<2.2$,  
invariant mass within $2.9 < m_{\mu\mu} <3.3\gev$, $d'_\mathrm{ca} < 0.5\cm$, 
and the probability of the $\chi^2$ per degree of freedom ($\chidof$) of the dimuon vertex fit greater than 15\%. 
To reduce the rate of prompt  \jpsi\ candidates, two additional requirements are imposed in the transverse plane: 
(i) the pointing angle $\alpha_{xy}$ between  the dimuon momentum and the vector from the beamspot 
(defined as the average interaction point) to the dimuon vertex must fulfill $\cos\alpha_{xy} > 0.9$; and (ii) the flight distance significance $\ell_{xy}/\sigma(\ell_{xy})$ must be 
larger than 3, where $\ell_{xy}$ is the two-dimensional distance between the primary and dimuon vertices and $\sigma(\ell_{xy})$ is its uncertainty.

The trigger efficiencies for the various samples are determined from the MC simulation. 
They are calculated after all muon identification selection criteria, as discussed in Section~\ref{s:muons}, have been applied. 
For the signal events the average trigger efficiency is 84\% (74\%) in the barrel (endcap) channel. 
The trigger efficiency for the normalization and control samples varies from 77\% in the barrel channel  to 60\% in the endcap channel.
This analysis depends on the ratio of the signal efficiency to the normalization sample efficiency. 
The systematic uncertainty on the trigger efficiency ratio is estimated as the sum in quadrature of two components.
The first component is defined as the variation of the 
efficiency ratio 
when varying the muon \pt\ threshold from $4$ to $8\gev$ in the MC simulation.
The second one is the difference between the ratios determined in data and MC simulations using 
the tag-and-probe approach (described in Section~\ref{s:muons}). 
The systematic uncertainty on the ratio is estimated to be 3\% in the barrel channel and 6\% in the endcap channel.

\section{Muon identification}
\label{s:muons}
Muon candidates are reconstructed  by combining tracks found in the silicon tracker and the 
muon detector ~\cite{CMS-PAS-MUO-10-002,Khachatryan:2010xn}. 
In order to ensure high-purity muons, the following additional requirements are applied: 
(i) muon candidates must have at least two track segments in the muon stations;  
(ii) they must have more than 10 hits in the silicon tracker, of which at least one must be in the pixel detector; 
(iii) the combined track must have $\chidof < 10$; 
and (iv) the impact parameter in the transverse plane $d_{xy}$, calculated with respect to the beamspot, 
must be smaller than $0.2\cm$. 
The systematic uncertainty on the muon track reconstruction efficiency is 2\%~\cite{CMS-PAS-TRK-10-002} and is included in the 
uncertainty of the total efficiency.

The ratio of the muon identification efficiencies between the signal and normalization samples is used in this analysis. This ratio 
is determined in two ways. First, the MC event samples contain a full simulation of the muon detector, which allows an efficiency 
determination by counting the events that pass or fail the muon identification algorithm. 
Second, the muon identification efficiency is determined with a tag-and-probe method~\cite{CMS-PAS-MUO-10-002}, 
which is applied to both data and MC event samples. 
To study the single-muon identification efficiency, the decays $\jpsi\to\mup\mun$ are used. 
In the tag-and-probe method, a ``tag'' muon, satisfying strict muon criteria, is paired with a``probe'' track,
where together they combine to give the $\jpsi$ invariant mass, thus indicating the probe is in fact a muon. 
The single-muon efficiency is determined by the number of probe tracks passing or failing the muon identification algorithm. 
Dedicated trigger paths constructed using the tag muon and either a silicon track or a signal in the muon chambers are employed for this study,
which ensures large event samples while avoiding potential bias of the efficiency measurement from using events triggered by the probe.

The muon identification efficiency is calculated after all selection criteria, including the detector acceptance, have been applied. 
For the signal events, the average efficiency is 71\% (85\%) in the barrel (endcap) channel based on the MC simulation. 
For the normalization and control samples, the muon identification efficiency is about 77\% (78\%) in the barrel (endcap). 
Pair-correlation effects influence these numbers~\cite{CMS-PAS-MUO-10-002}. 
The dimuon efficiency can be altered with respect to the product of single-muon efficiencies depending on the 
mutual proximity of the two muons in the muon system. 
This effect is included in the efficiency calculations in the detailed MC simulation of the muon detectors.
The systematic uncertainty on the identification efficiency ratio is estimated in the same way as for the 
muon trigger efficiency ratio (Section \ref{s:cmsdetector}), and is 4\% in the barrel and 8\% in the endcap.

\section{Analysis}
\label{s:selection}

The reconstruction of $\B\to\mup\mun$ candidates requires two oppositely-charged muons that originate
from a common vertex and have an invariant mass in the range $4.9 <m_{\mu\mu} < 5.9\gev$.
A fit of the $\B$-candidate vertex is performed and its $\chidof$ is evaluated.
The two daughter muon tracks are combined to form the $\B$-candidate track.

The primary vertex associated with a $\B$ candidate is chosen from all
reconstructed primary vertices as the one which has minimal separation
along the z axis from the z intercept of the extrapolated $\B$ candidate
track.
Reconstruction effects due to pileup are largely eliminated by the primary vertex matching procedure.
The position of this primary vertex is then refit without the tracks of the $\B$ candidate with an adaptive vertex
fit~\cite{Khachatryan:2010pw}, where tracks are assigned a weight $0 <w < 1$ based on their proximity to the primary vertex.
After the refit, $\B$ candidates with badly reconstructed primary vertices are eliminated by requiring the average
track weight $\langle w\rangle > 0.6$.
The 3D impact parameter of the $\B$ candidate $\ip$, its uncertainty $\sigma(\delta_{3D})$,  and its
significance $\ips$ are measured with respect to the primary vertex.

The isolation of the $\B$ candidate is an important criterion in separating the
signal from background. Three variables are used for this purpose:
\begin{itemize}
\item The isolation variable $I = \pt(\B)/(\pt(\B) + \sum_{\mathrm{trk}}\pt)$
  is calculated from the transverse momentum
  of the $\B$ candidate $\pt(\B)$ and the transverse momenta of all
  other charged tracks satisfying $\Delta R = \sqrt{(\Delta\eta)^2 +
    (\Delta\phi)^2} < 0.7$, where $\Delta\eta$ and $\Delta\phi$ are the
  differences in pseudorapidity and azimuthal angle between a charged
  track and the $\B$-candidate momentum.  The sum includes all tracks
  with $\pt>0.9\GeV$ that are (i)~consistent with originating from the
  same primary vertex as the $\B$ candidate or (ii)~have a distance of
  closest approach $\dca < 0.05\cm$ with respect to the $\B$
  vertex and are not associated with any other primary vertex .
\item The number of tracks \closetrk\, with $\pt>0.5\gev$ and $\dca<0.03\cm$ with
  respect to the $\B$-candidate's vertex.
\item The minimum distance of closest approach between tracks and the $\B$-candidate's vertex,
 \docatrk, for all tracks in the event that are either associated with the same
  primary vertex as the $\B$-candidate or not associated with any other primary vertex.
\end{itemize}
The first variable describes the isolation primarily with respect to tracks coming from
the primary vertex itself.
The latter two variables quantify the isolation of the $\B$ vertex.
They help to reject partly reconstructed $\B$ decays where there are other tracks in addition to the two muons
associated with the $\B$-candidate vertex.

The distributions of the variables described above are shown in Fig.~\ref{fig:SgData-SgMc} for
signal events from the MC simulation and for data background events.
These include the momenta of the higher-momentum (leading) and lower-momentum (sub-leading) muons
$p_{T,\mu1}$ and  $p_{T,\mu2}$, $\pt(\B)$,
the 3D pointing angle $\alpha_{\mathrm{3D}}$,
the 3D flight length significance
$\ell_{3\mathrm{D}}/\sigma(\ell_{3\mathrm{D}})$, the \chidof, and the
isolation variables ($I$, \closetrk, and \docatrk).
The data background events are defined as $\B$ candidates with a dimuon mass in the sidebands
covering the range $4.9 < m_{\mu\mu} < 5.9\gev$, excluding the (blinded) signal
windows from $5.20 < m_{\mu\mu} < 5.45\gev$.
Events shown in Fig.~\ref{fig:SgData-SgMc} must pass a tight selection that is close to the final one:
muon $\pt > 4\gev$,
$\pt(\B)>7.5\gev$,
$\alpha<0.05$, $\chidof<2$, $\fls>15$,
$\ip<0.008\cm$, $\ips<2$, $I>0.8$, $\docatrk > 0.015\cm$, and
$\closetrk <2$ tracks.
For each distribution, the selection requirements for all variables, apart from the one plotted, are applied.
This figure illustrates the differences in the distributions of signal and background events,
and shows which variables are effective in reducing the background events, e.g., $\ell_{3\mathrm{D}}/\sigma(\ell_{3\mathrm{D}})$.
The analysis efficiency for each selection requirement is determined from the simulated events.

\renewcommand{\sample}{SgData-A}
\renewcommand{\channel}{SgMc-A}

\setlength\fwidth{0.32\textwidth}
\begin{figure}[htbp]
  \begin{centering}
    \includegraphics[width=\fwidth]{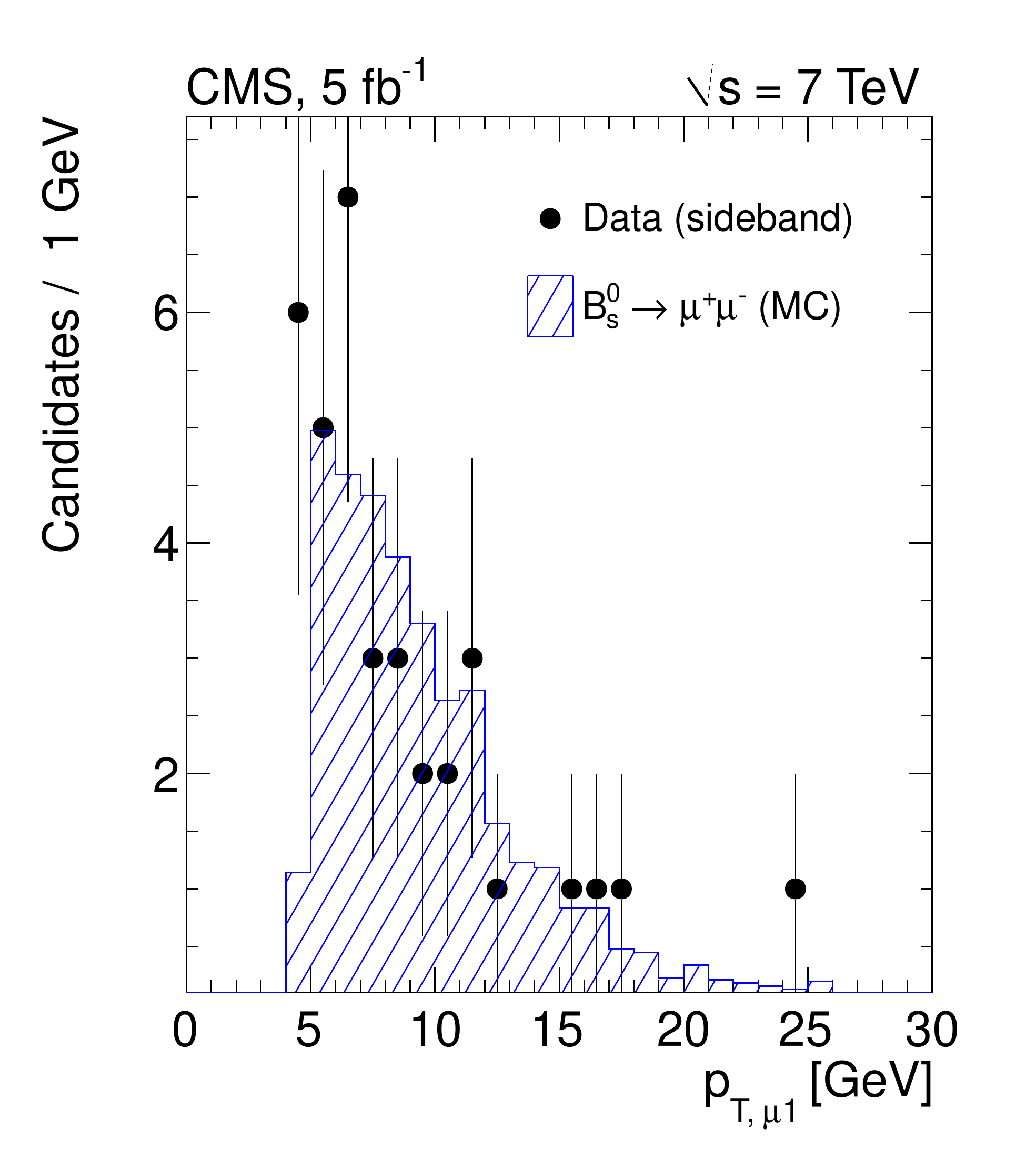}
    \includegraphics[width=\fwidth]{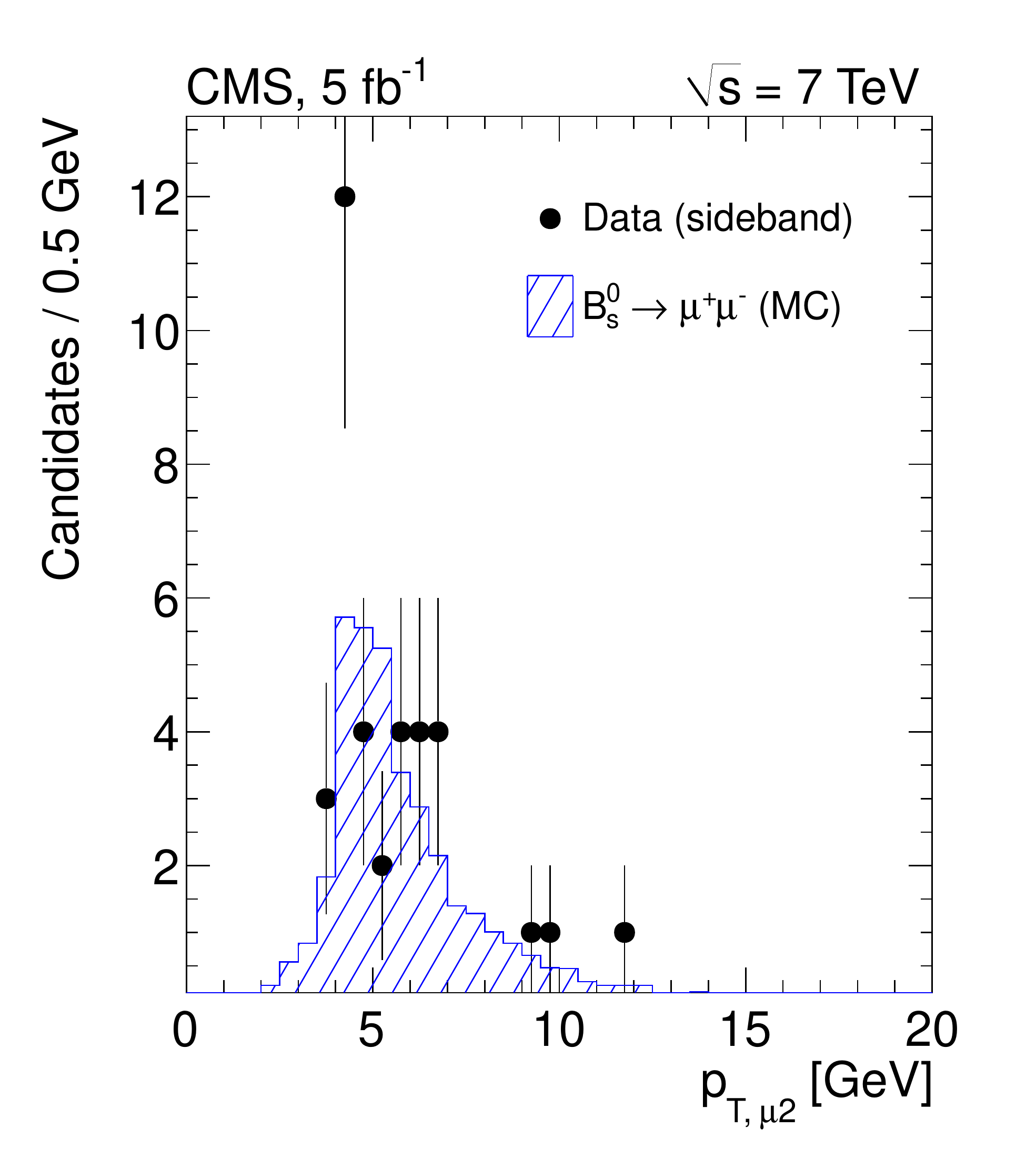}
    \includegraphics[width=\fwidth]{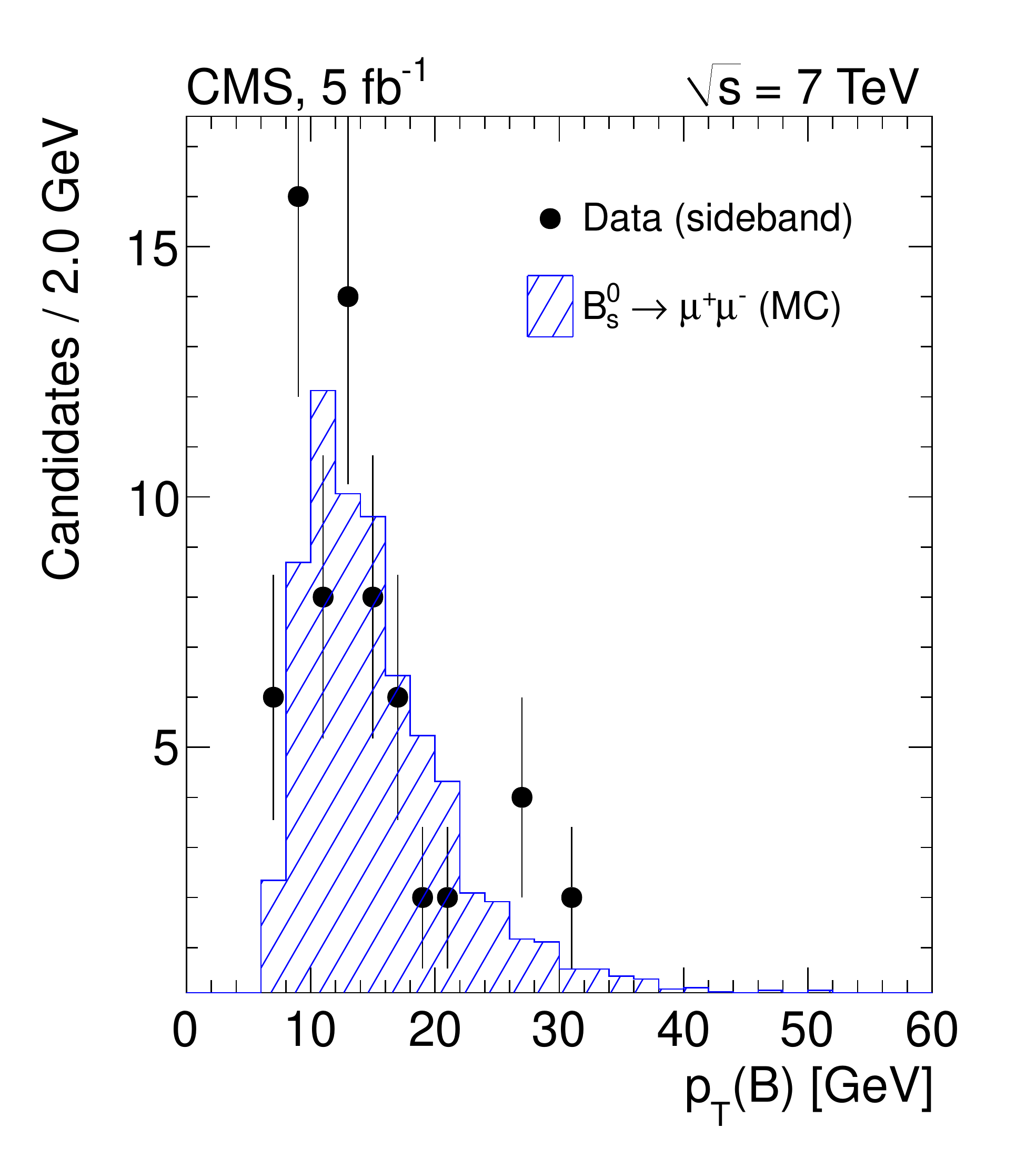}
    \includegraphics[width=\fwidth]{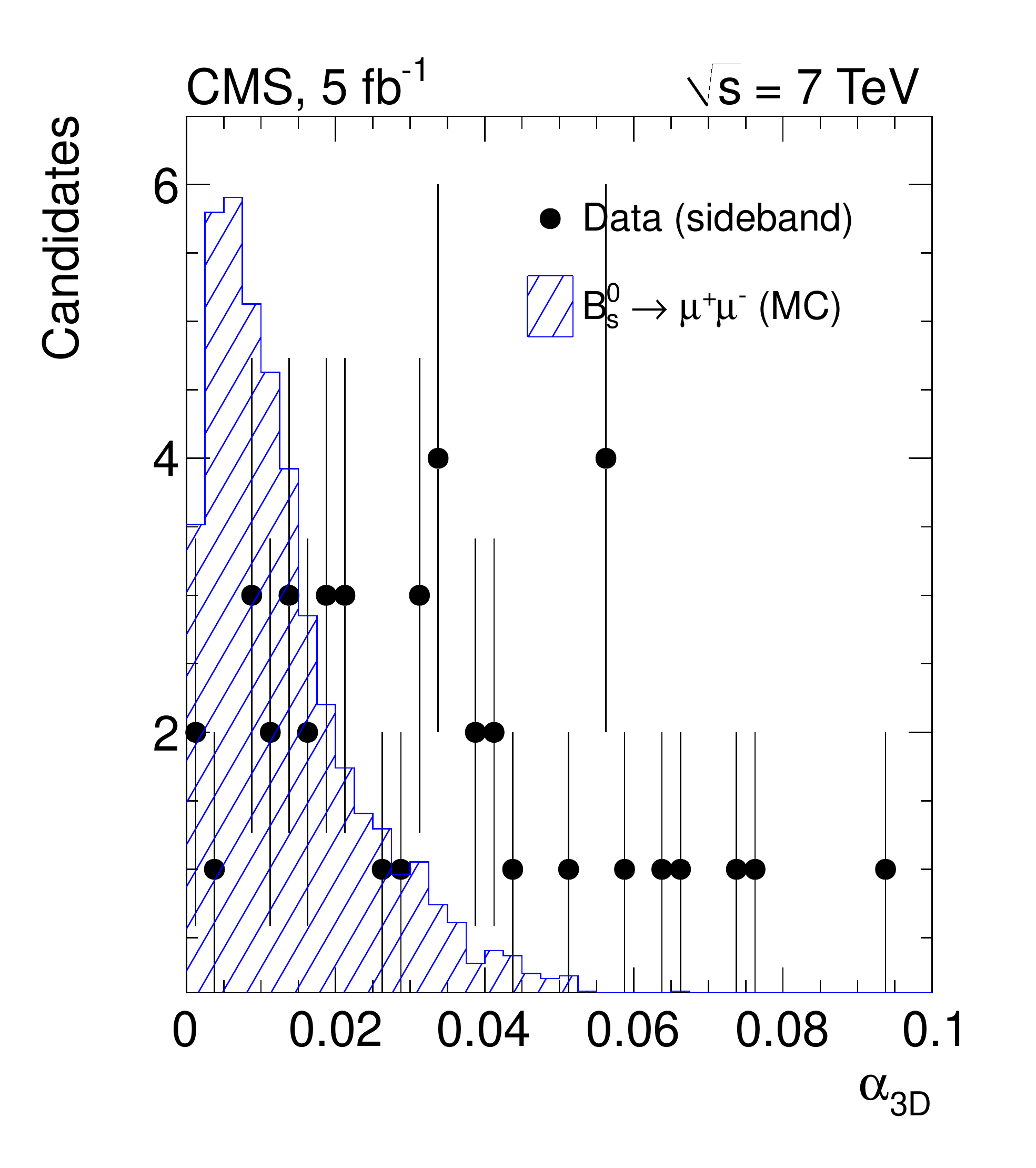}
    \includegraphics[width=\fwidth]{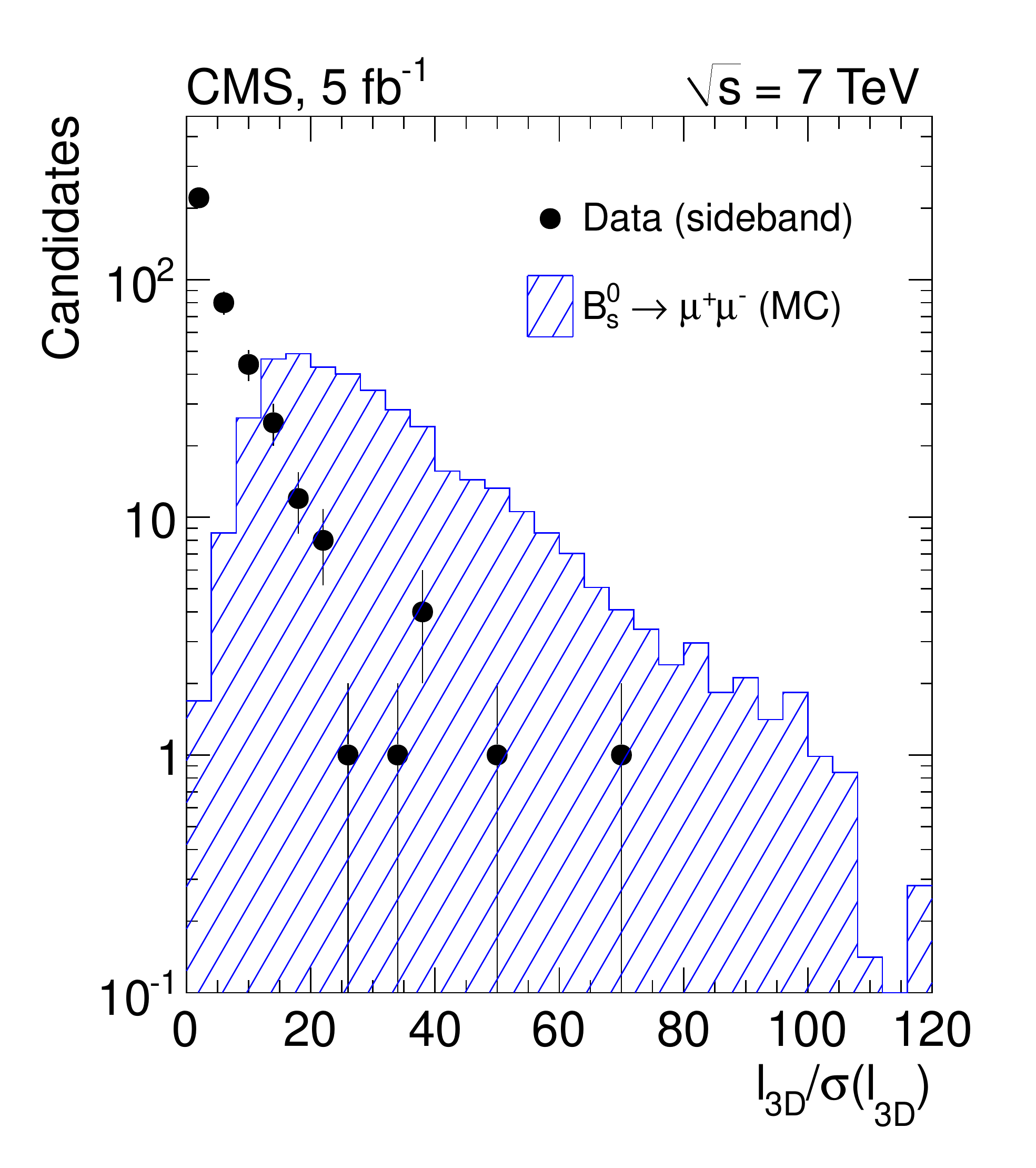}
    \includegraphics[width=\fwidth]{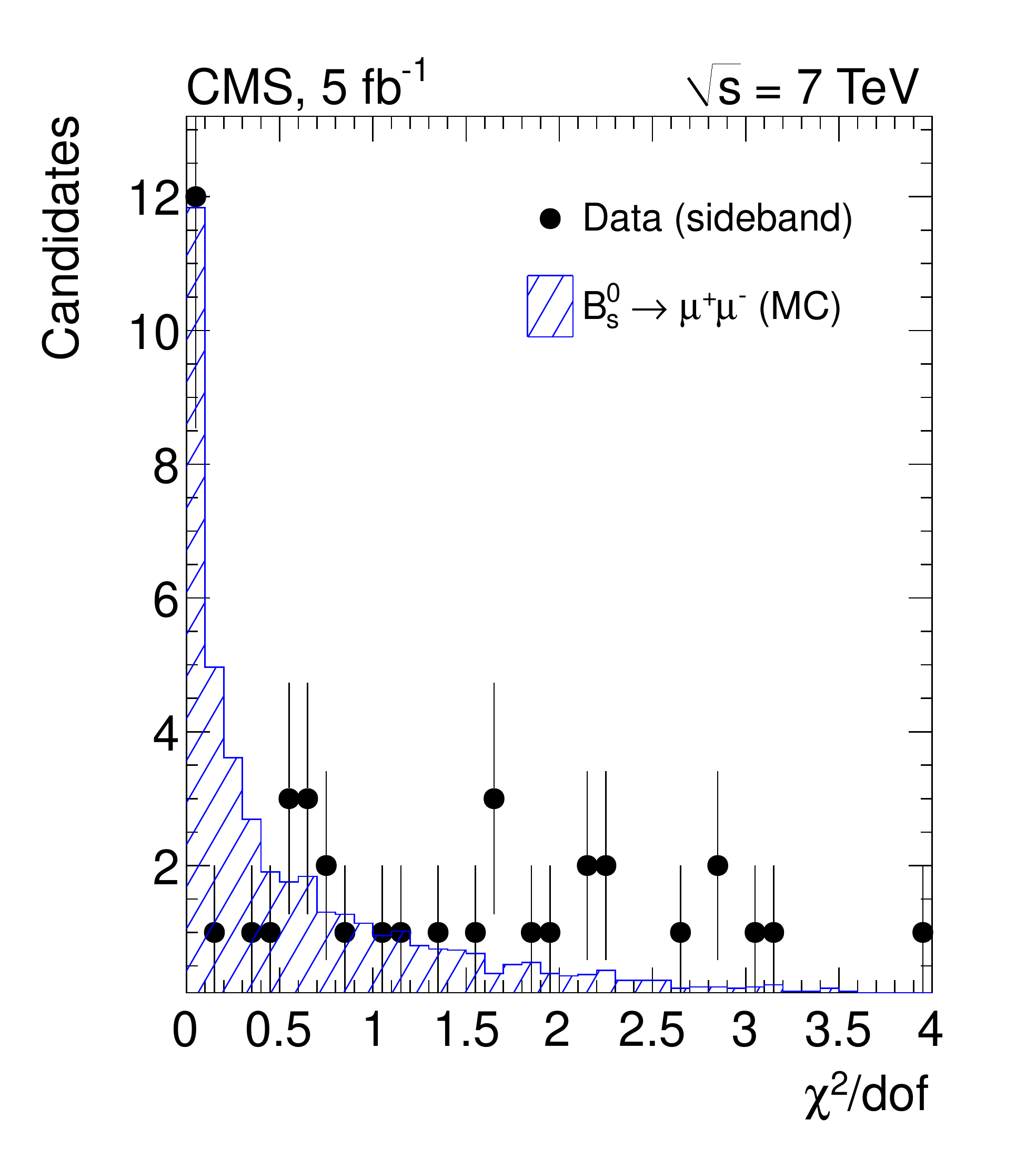}
    \includegraphics[width=\fwidth]{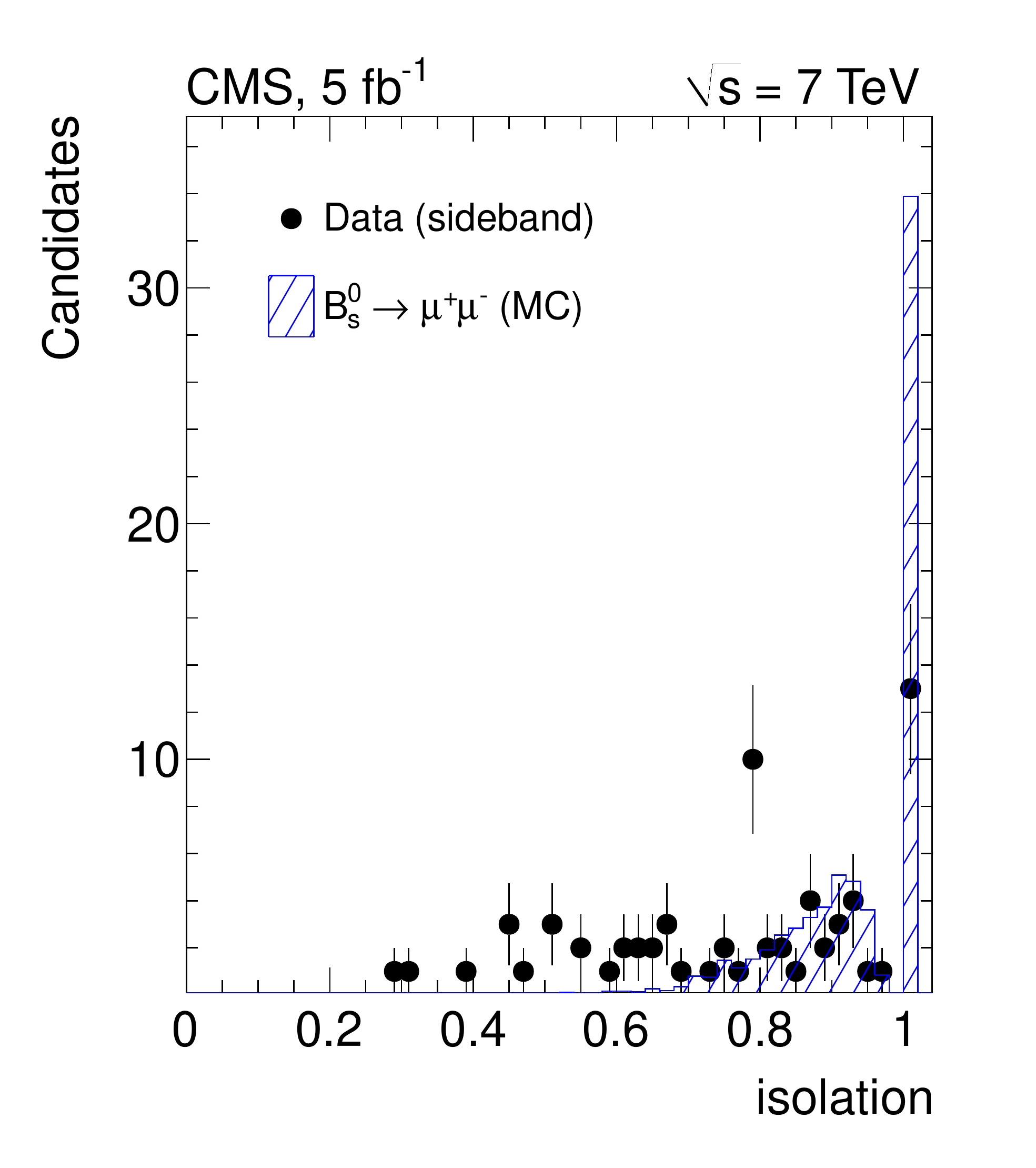}
    \includegraphics[width=\fwidth]{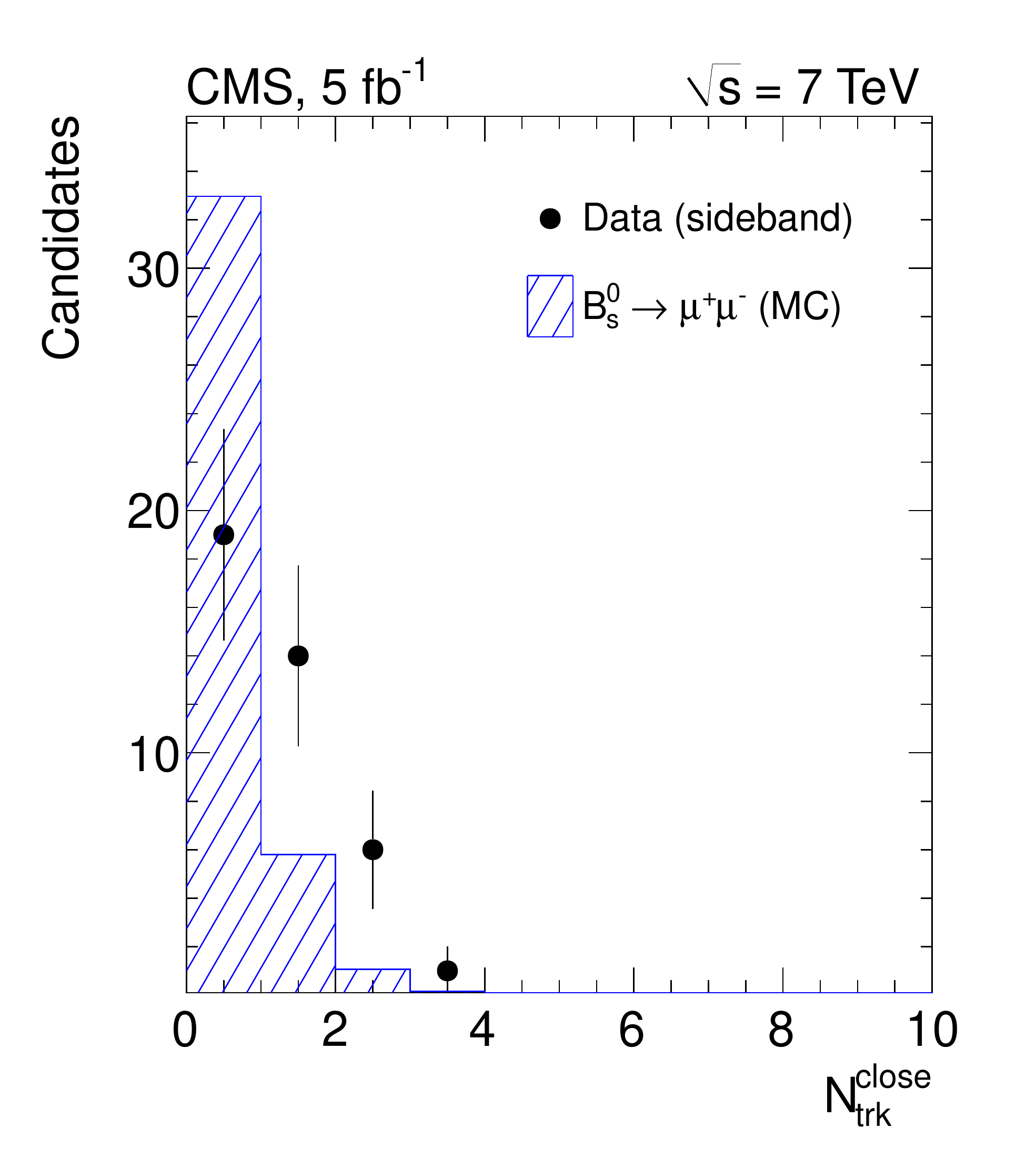}
    \includegraphics[width=\fwidth]{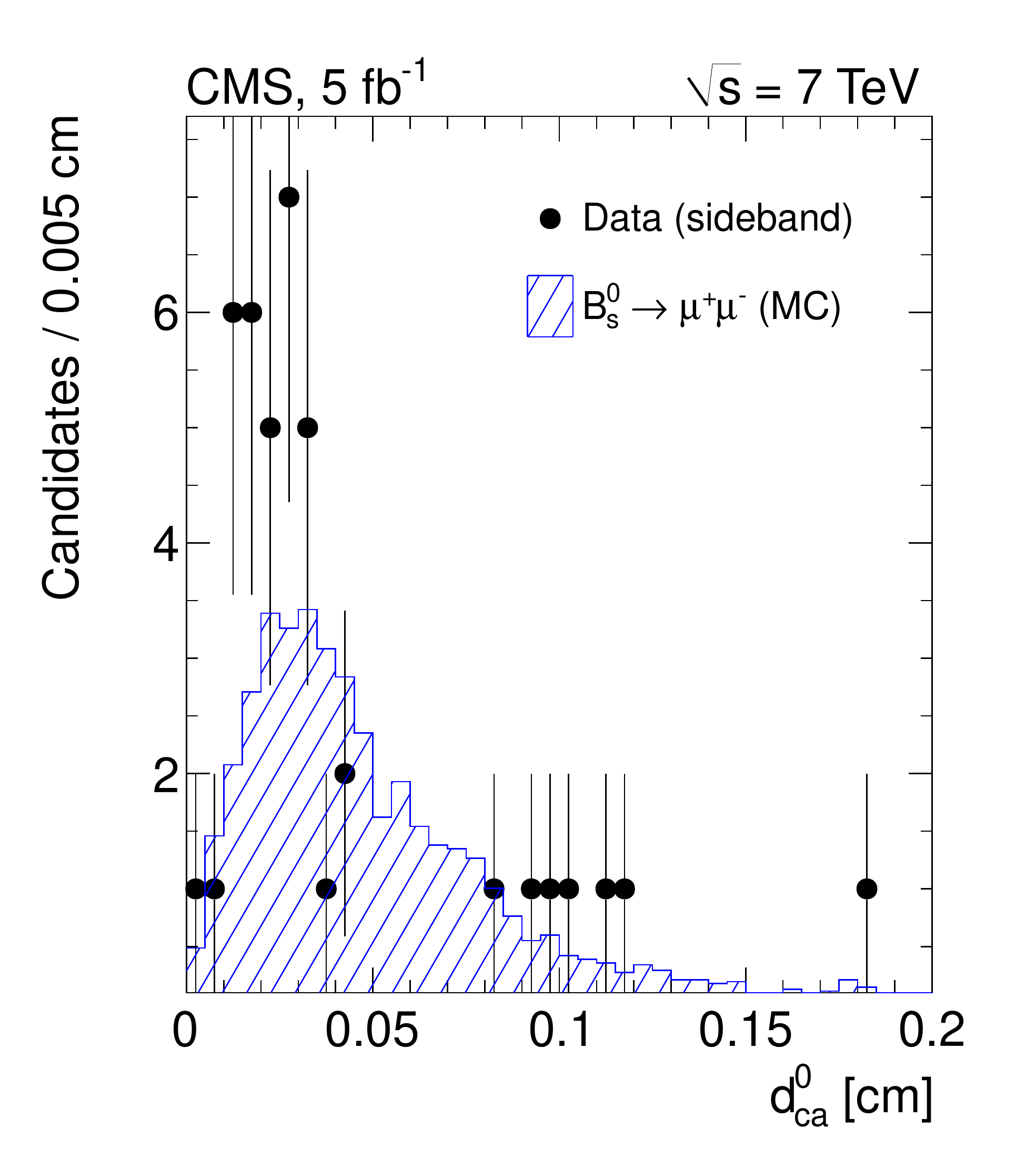}
    \caption{Comparison of simulated \bsmm\ decays and background
      dimuon distributions as measured in the mass sidebands.
      Top row: the transverse momentum for the leading muon, sub-leading muon, and $\B$-candidate;
      middle row: the 3D pointing angle, flight length significance, and $\B$-candidate's vertex \chidof;
      bottom row: the isolation variables $I$, \closetrk, and \docatrk.
      The MC histograms are normalized to the number of events in the data.}
    \label{fig:SgData-SgMc}
  \end{centering}
\end{figure}

The reconstruction of $\bupsik\to\mup\mun \Kp$  and $\bspsiphi\to\mup\mun\Kp\Km$ events is
very similar to the reconstruction of $\B\to\mup\mun$ events.
Candidates with two oppositely-charged muons sharing a common vertex and
with invariant mass in the range 3.0--3.2\gev are reconstructed.
The selected candidates must have a dimuon $\pt > 7\gev$.
Then they are combined with one or two tracks each assumed to be a kaon, with
$\pt > \vuse{default-11:TRACKPTLO:cutValue} \gev$ and
$|\eta| < \vuse{default-11:TRACKETAHI:cutValue} $.
The 3D distance of closest approach between all pairs among the three (four) tracks is required to be less than $0.1\cm$.
For \bspsiphi\ candidates, the two assumed kaon tracks must form an
invariant mass in the range 0.995--1.045\gev and have $\Delta R(\Kp,\Km) <0.25$.
The three (four)  tracks from the decay are used in the vertex fit.
All requirements listed above for $\B\to\mup\mun$ events are also applied here, including the
vertex-fit selection $\chidof<2$, which eliminates poorly reconstructed candidates.
Only $\B$ candidates with an invariant mass in the range 4.8--6.0\gev are considered.

Figures~\ref{fig:NoData-NoMc} and \ref{fig:CsData-CsMc} show the MC simulation and sideband-subtracted
data distributions for a number of variables for the \bupsik\ and \bspsiphi\ candidates, respectively.
For each distribution, the selection requirements for all variables, apart from the one plotted, are applied.
The relative efficiency for each selection requirement is determined separately in data and MC simulation
and compared.
The largest relative differences are 2.5\% for the isolation selection
in the normalization sample and 1.6\% for the \chidof\ selection in the control sample.
We combine in quadrature the differences for all distributions to estimate the systematic uncertainty related to the
selection efficiency and obtain 4\% (3\%) for the normalization (control) sample.
The control sample uncertainty is used for the signal sample.

\renewcommand{\sample}{NoData-A}
\renewcommand{\channel}{NoMc-A}
\setlength\fwidth{0.32\textwidth}
\begin{figure}[htbp]
  \begin{centering}
    \includegraphics[width=\fwidth]{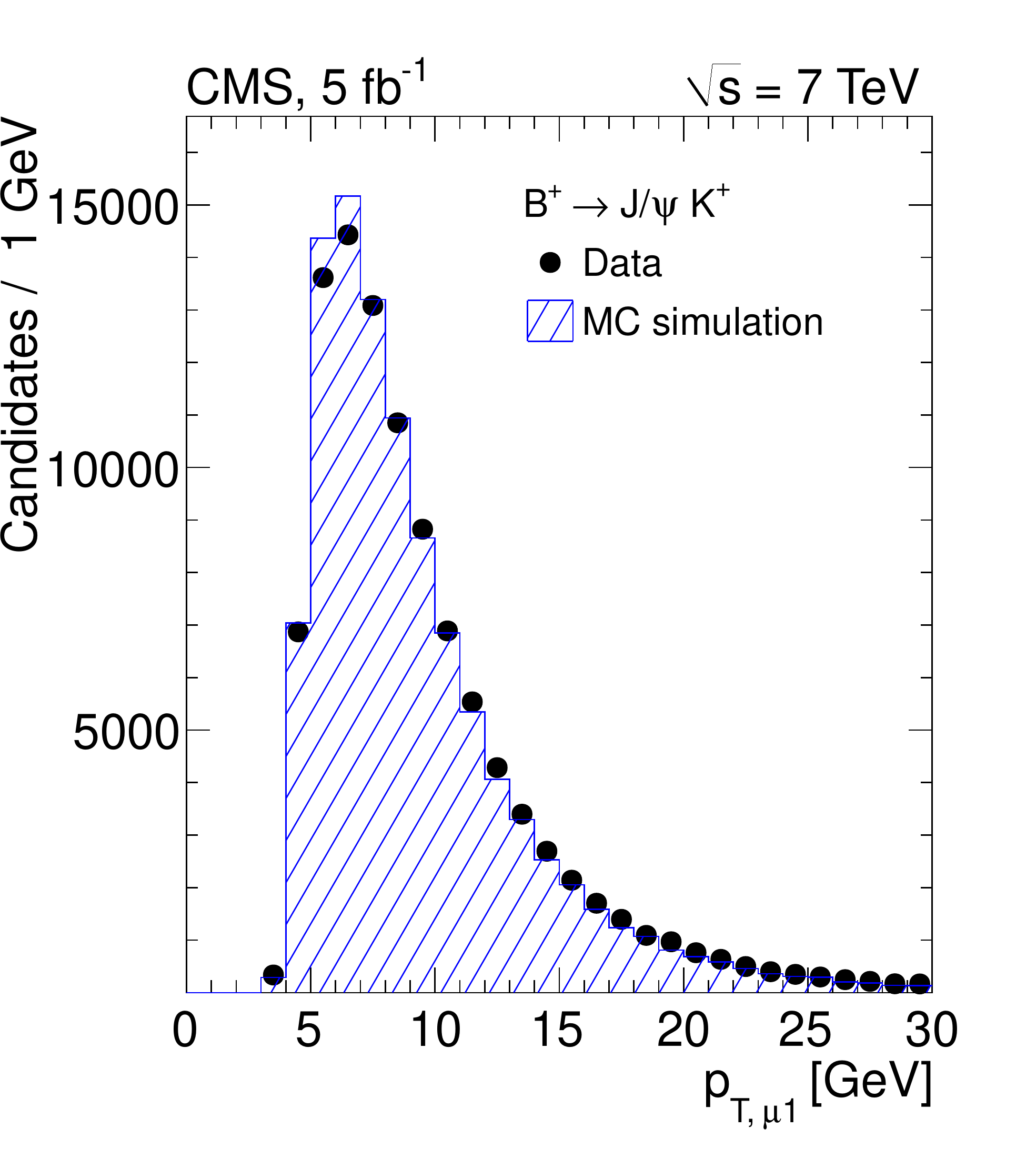}
    \includegraphics[width=\fwidth]{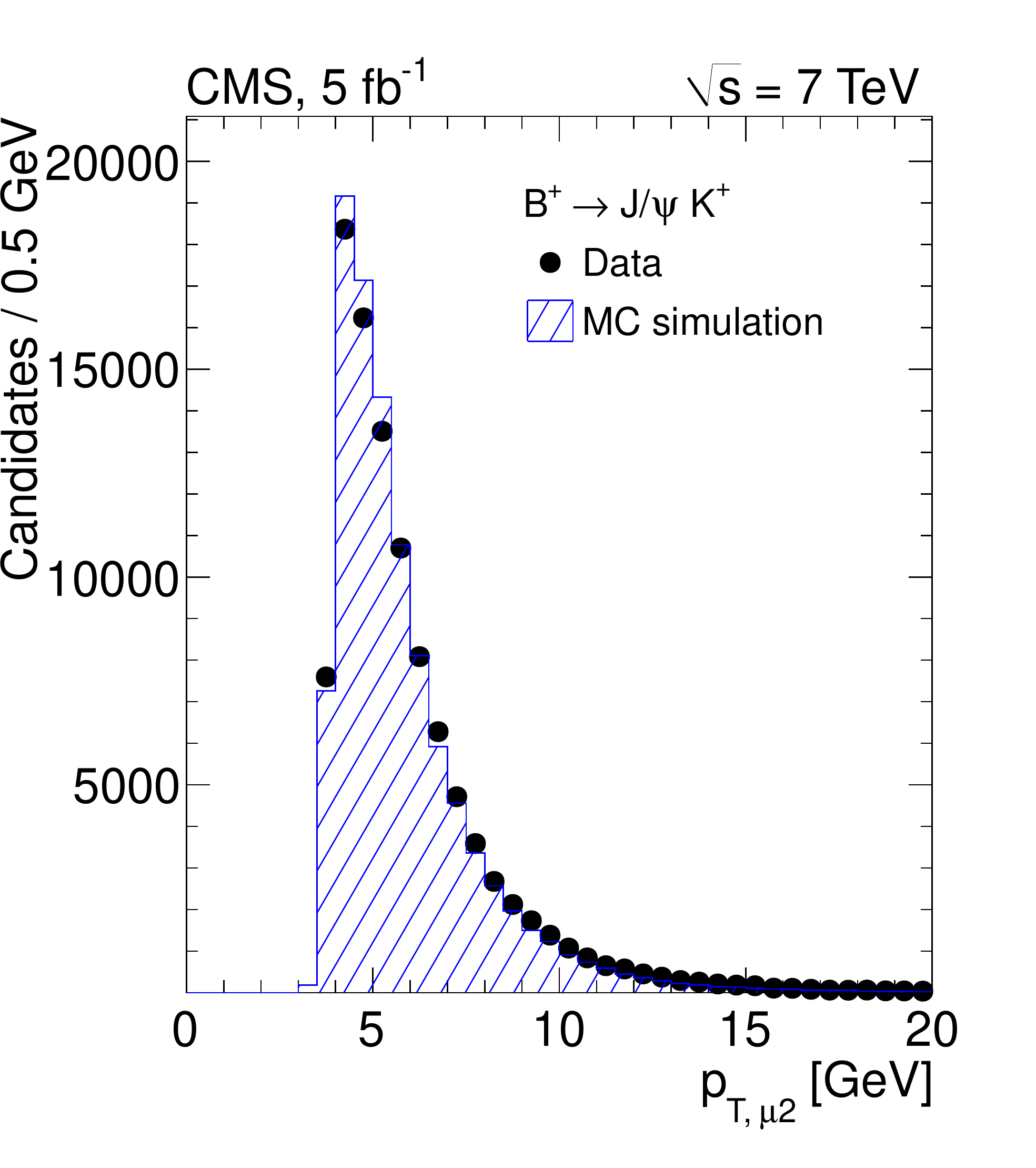}
    \includegraphics[width=\fwidth]{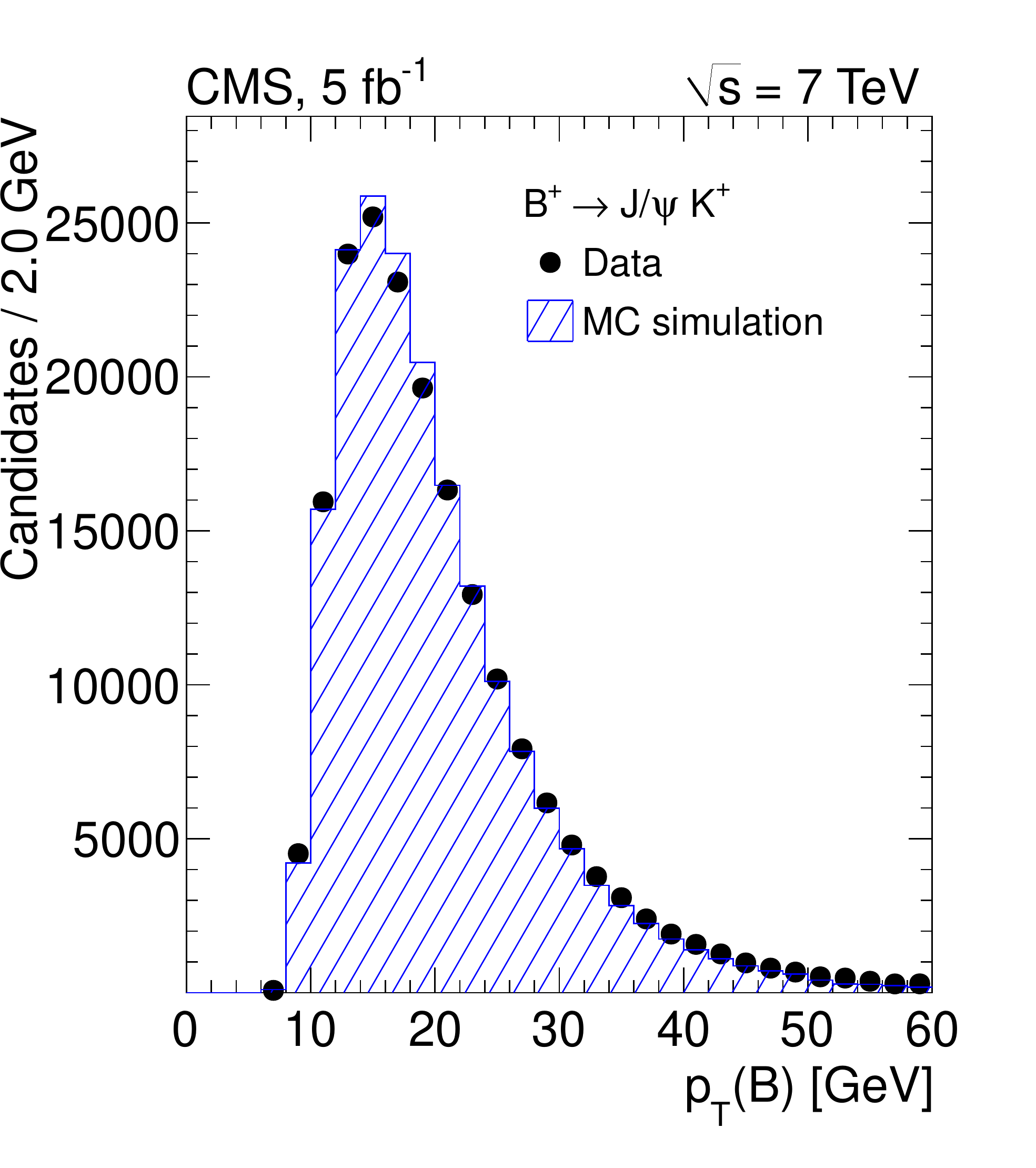}
    \includegraphics[width=\fwidth]{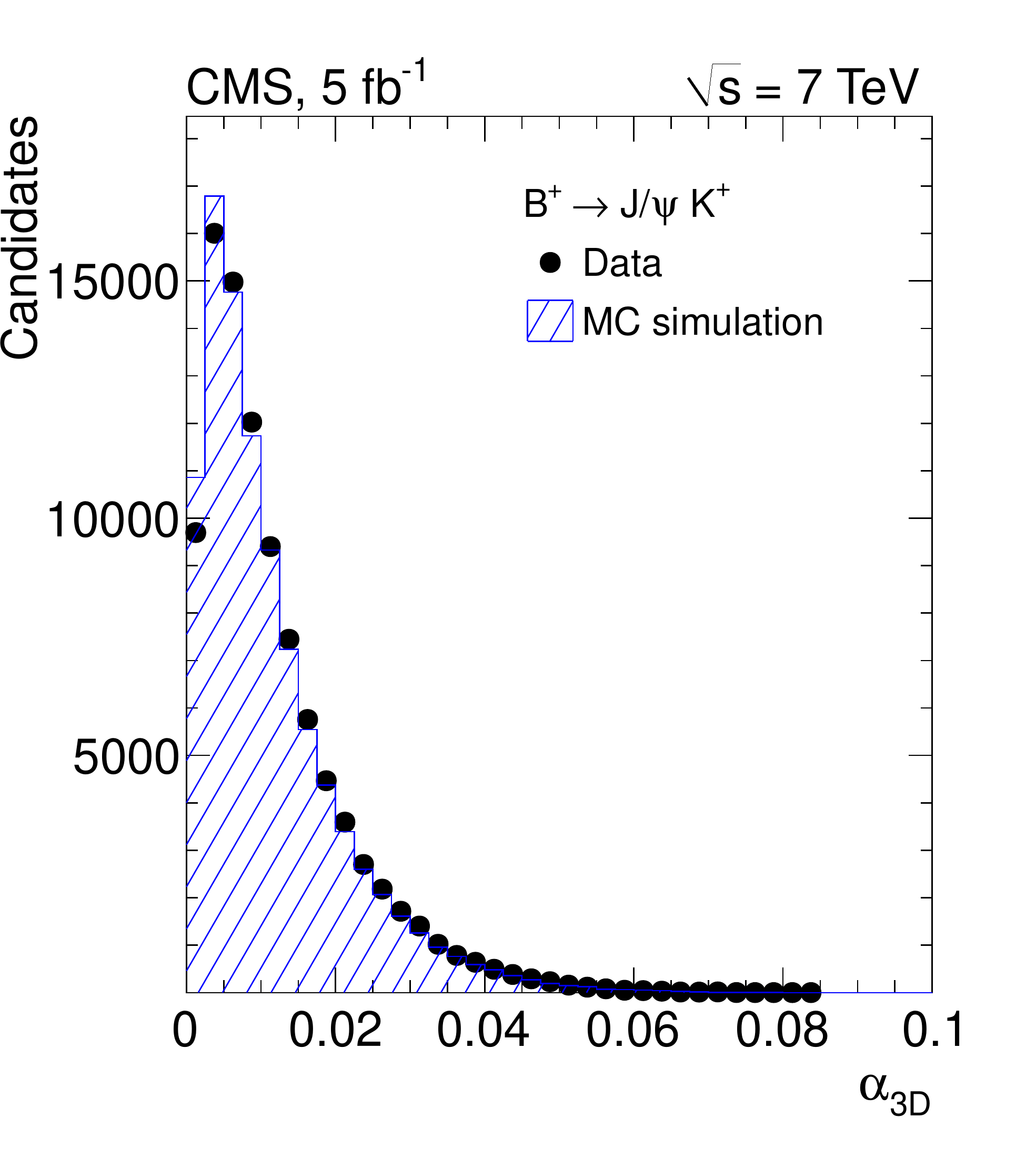}
    \includegraphics[width=\fwidth]{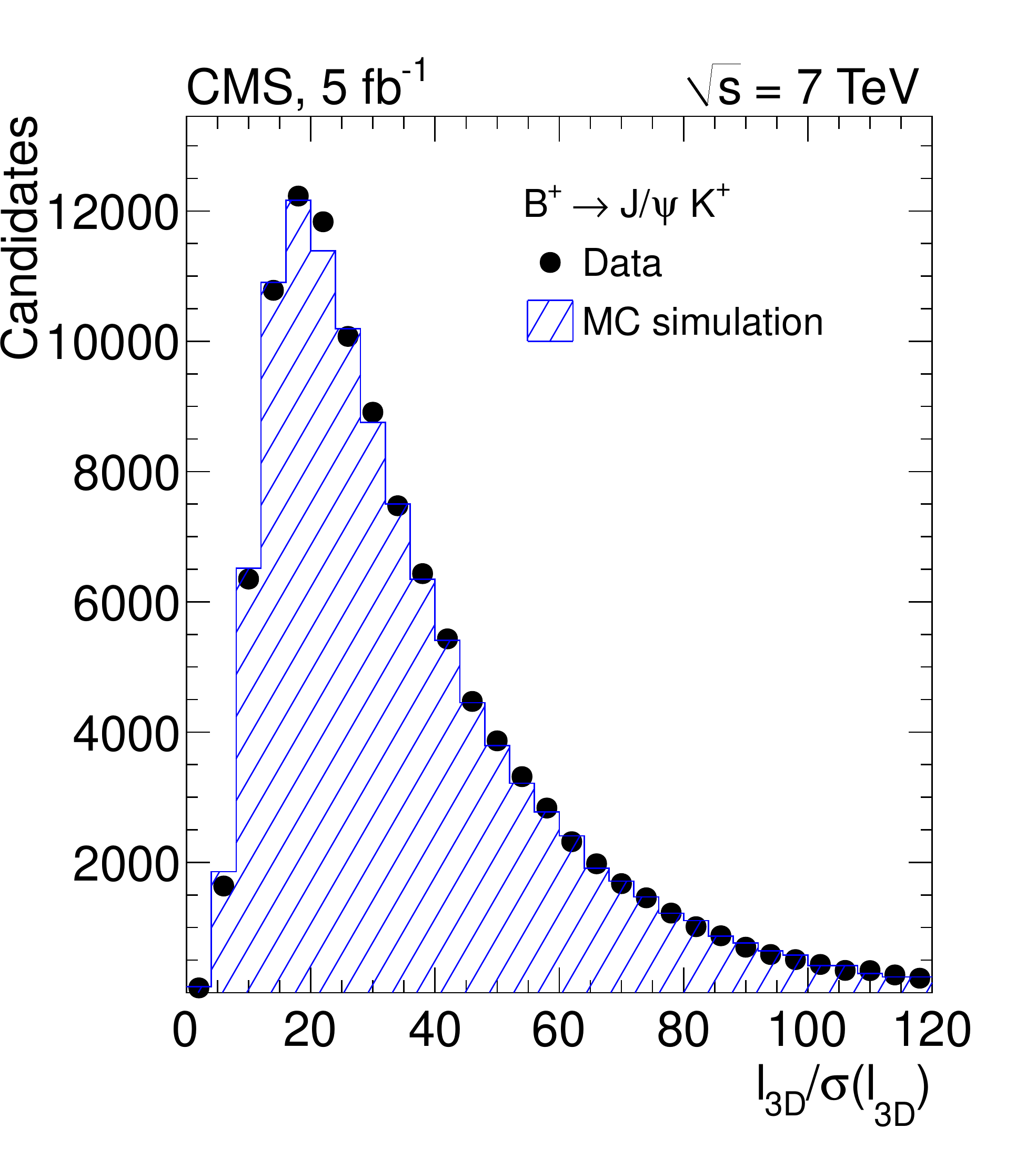}
    \includegraphics[width=\fwidth]{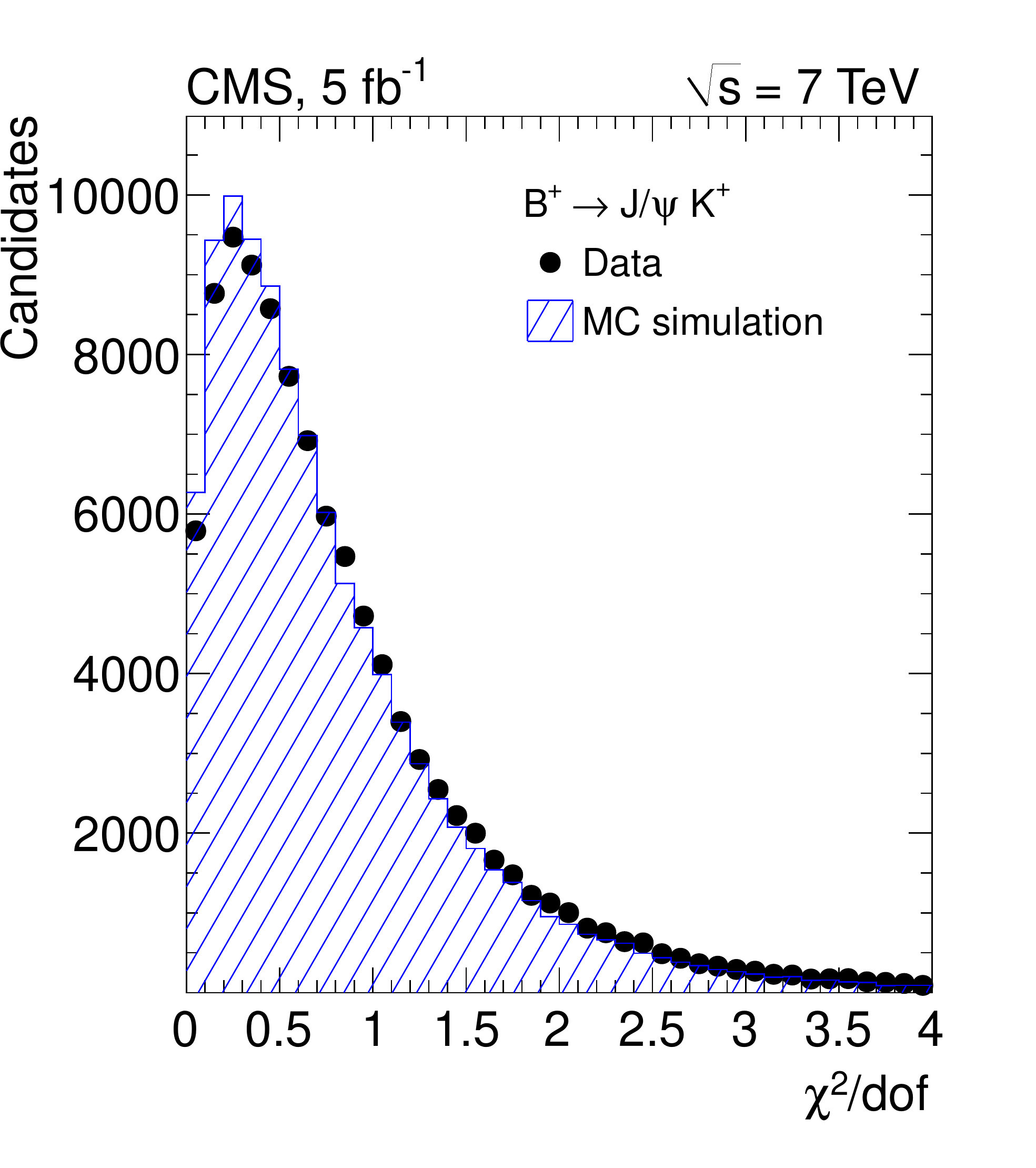}
    \includegraphics[width=\fwidth]{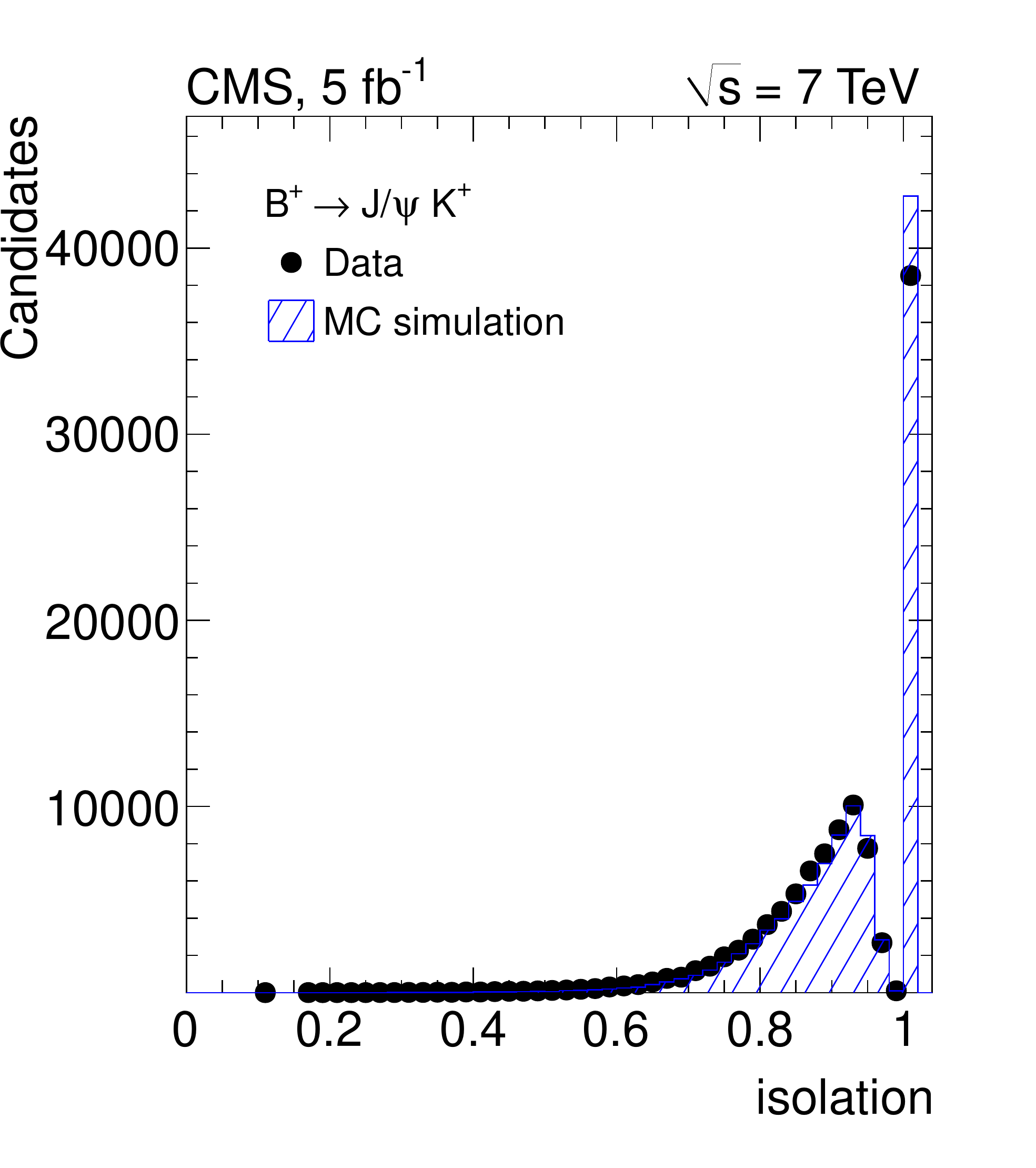}
    \includegraphics[width=\fwidth]{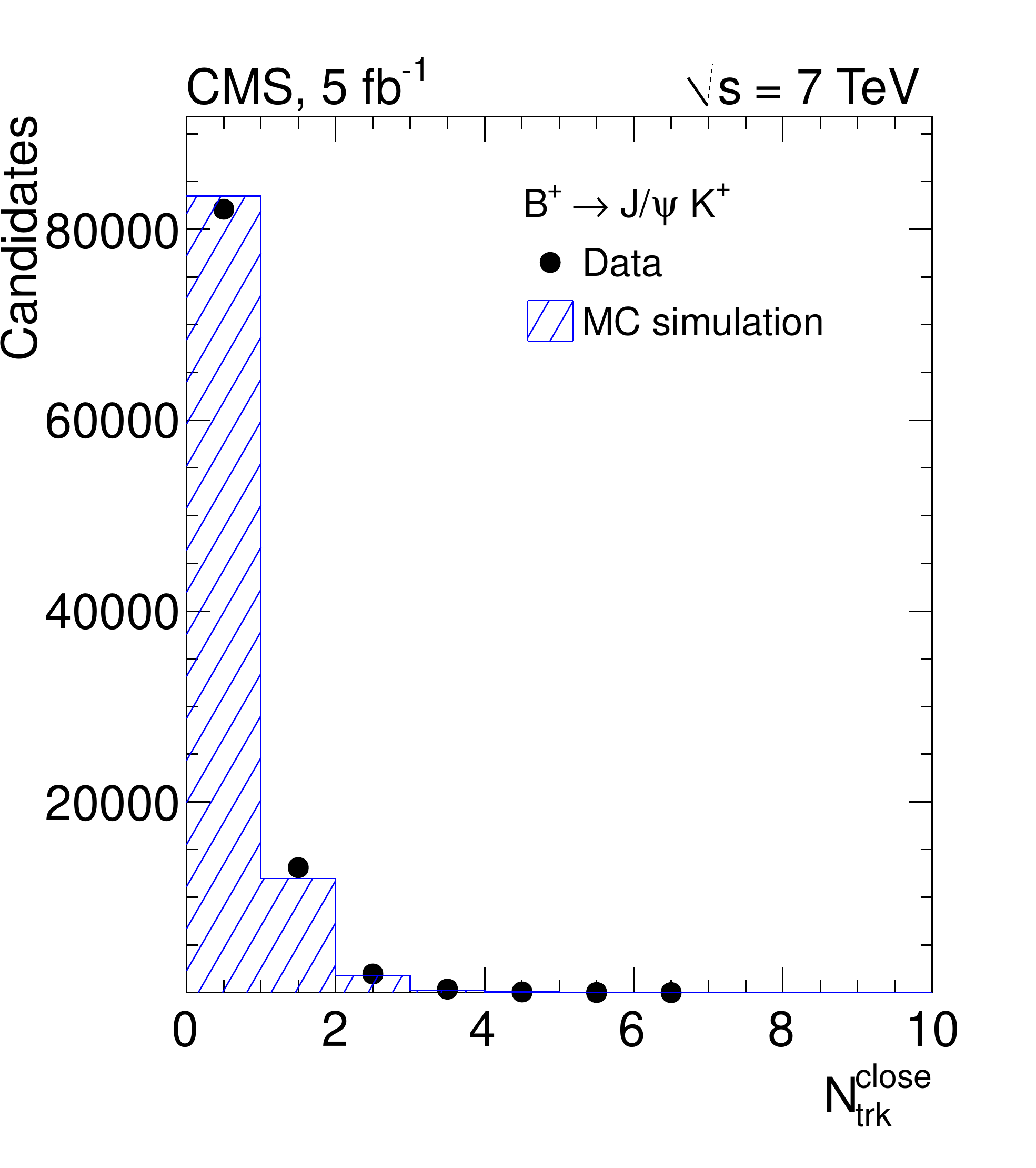}
    \includegraphics[width=\fwidth]{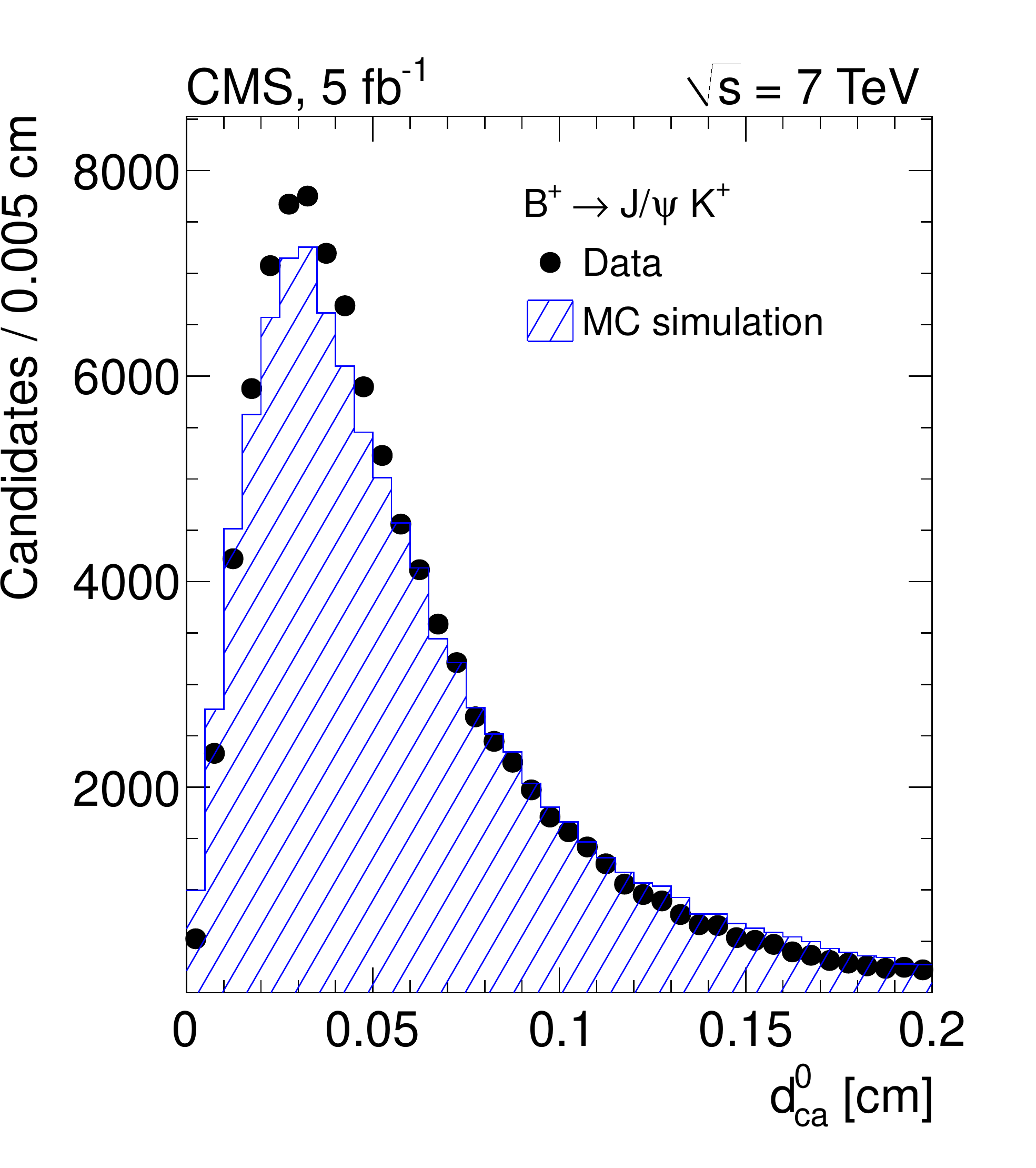}
    \caption{Comparison of measured and simulated \bupsik distributions.
      Top row: the transverse momentum for the leading muon, sub-leading muon, and $\B$-candidate;
      middle row: the 3D pointing angle, flight length significance, and $\B$-candidate's vertex \chidof;
      bottom row: the isolation variables $I$, \closetrk, and \docatrk.
      The MC histograms are normalized to the number of events in the data.}
   \label{fig:NoData-NoMc}
  \end{centering}
\end{figure}

\renewcommand{\sample}{CsData-A}
\renewcommand{\channel}{CsMc-A}
\setlength\fwidth{0.32\textwidth}
\begin{figure}[htbp]
  \begin{centering}
    \includegraphics[width=\fwidth]{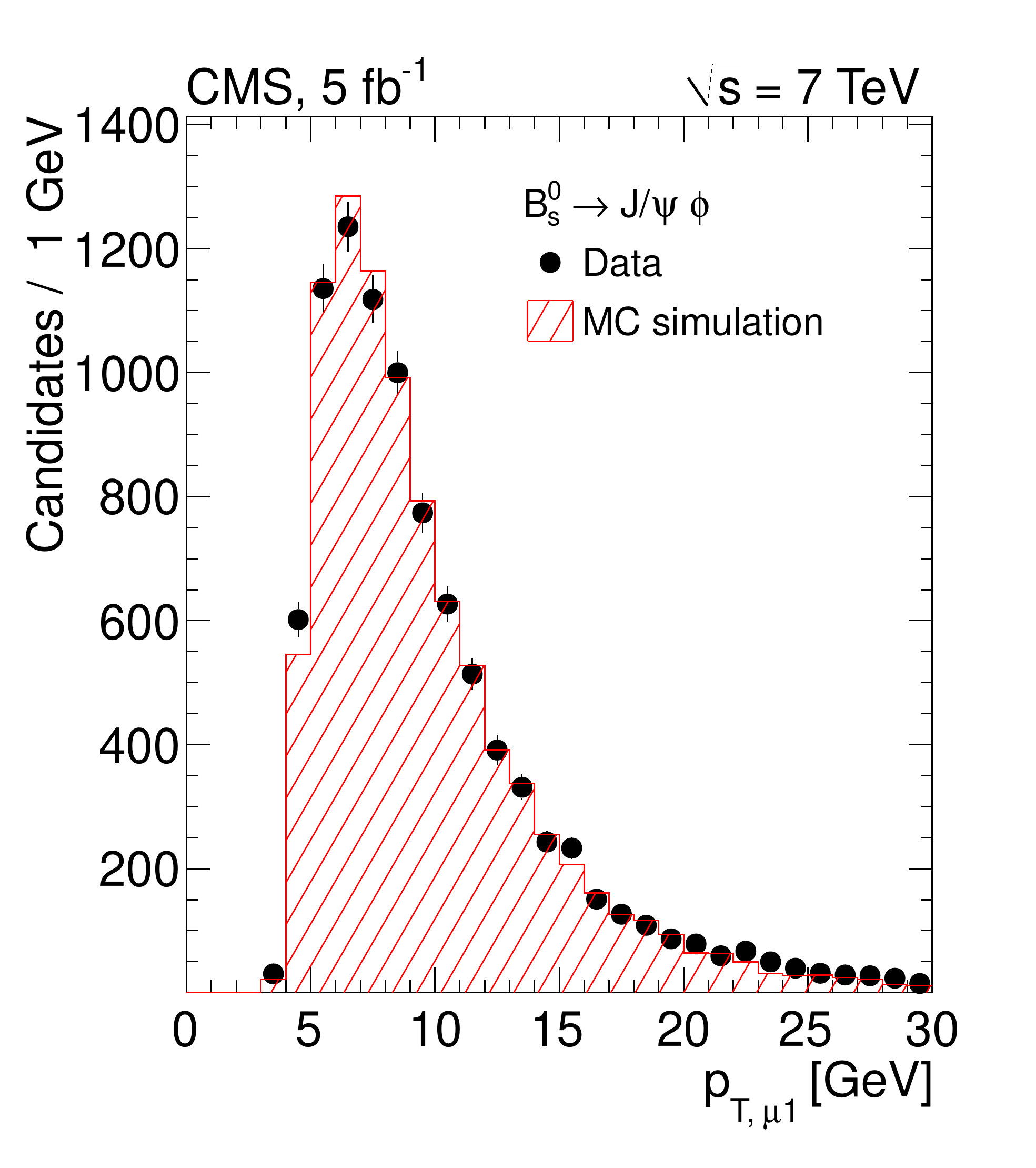}
    \includegraphics[width=\fwidth]{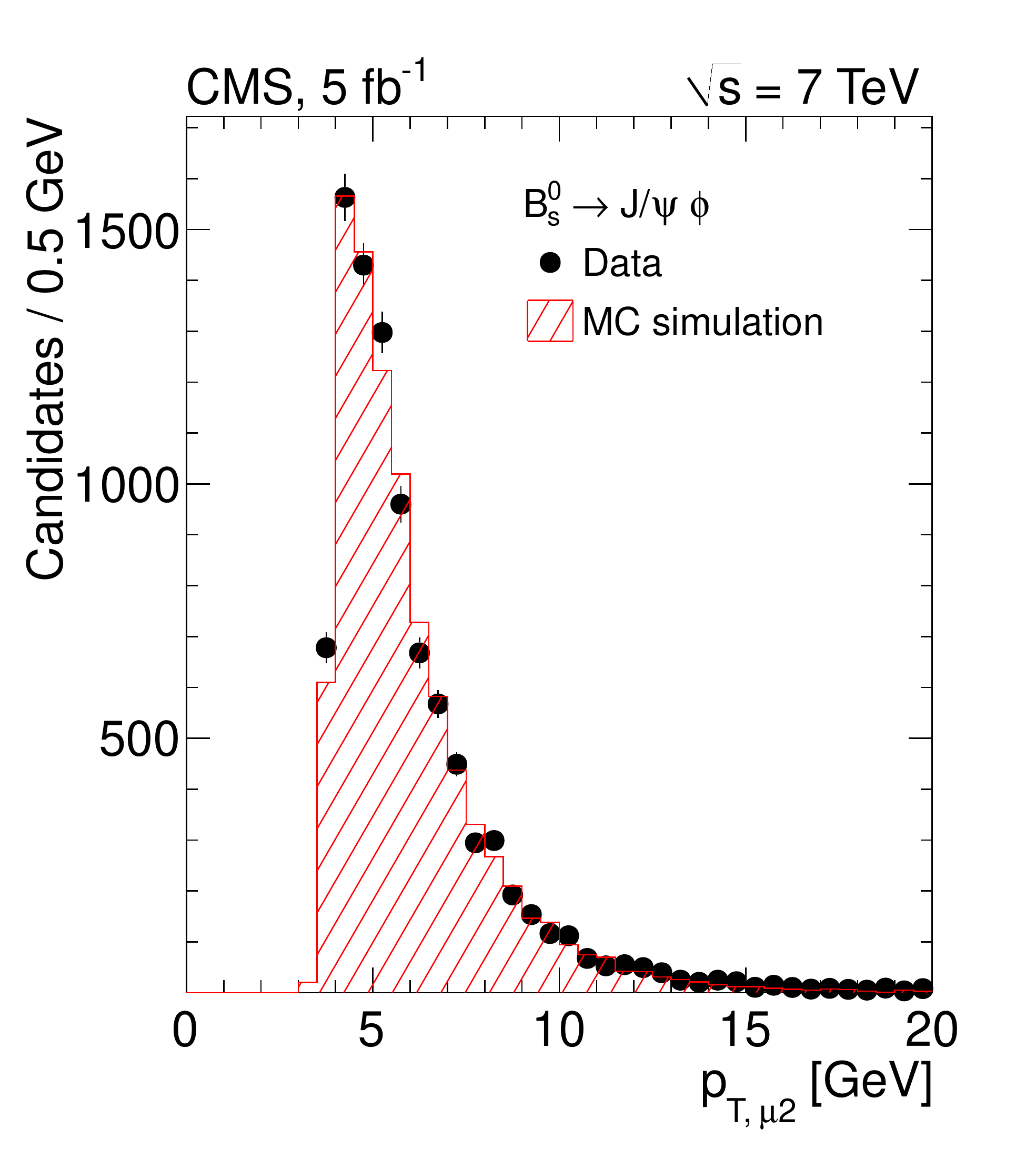}
    \includegraphics[width=\fwidth]{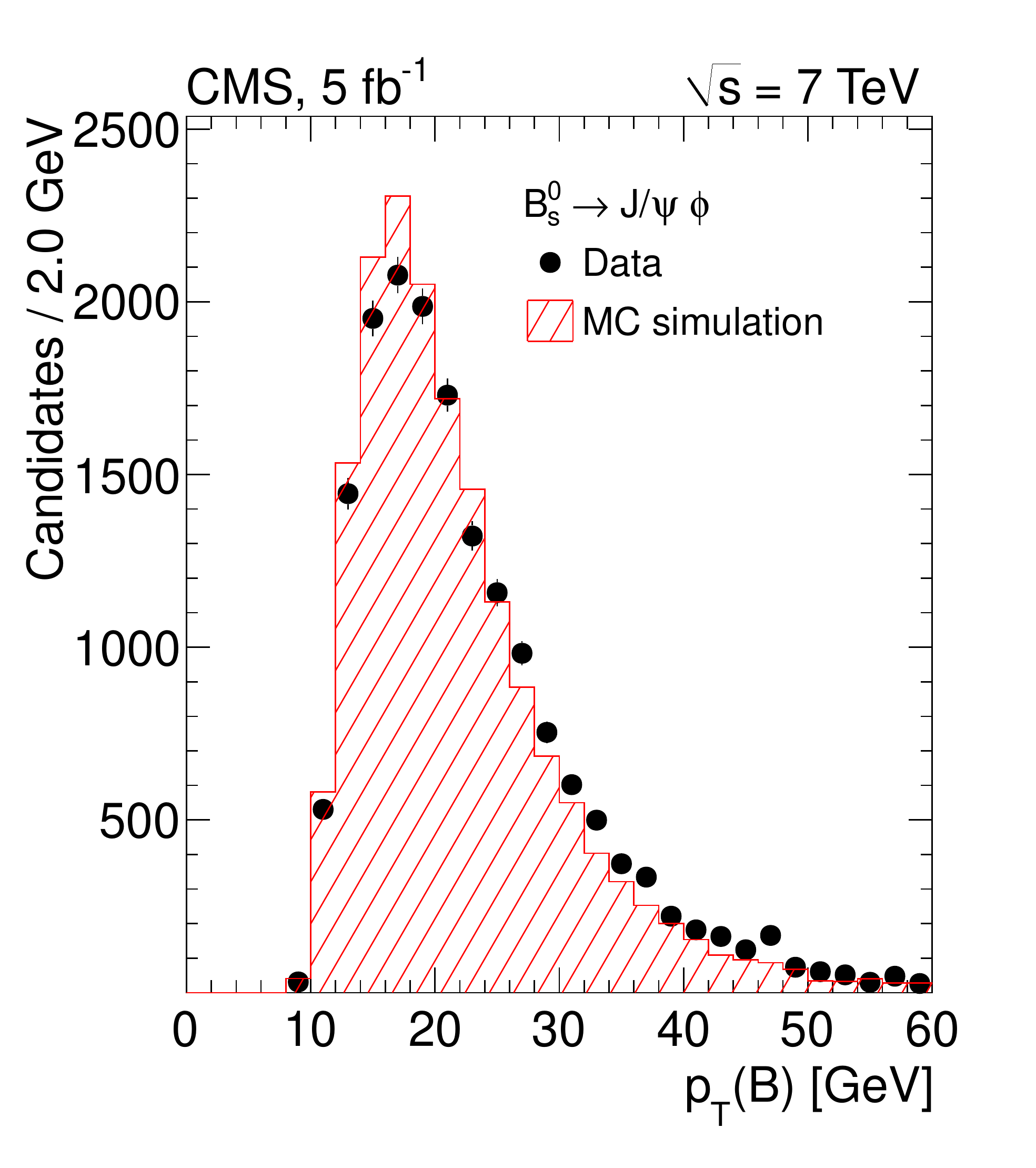}
    \includegraphics[width=\fwidth]{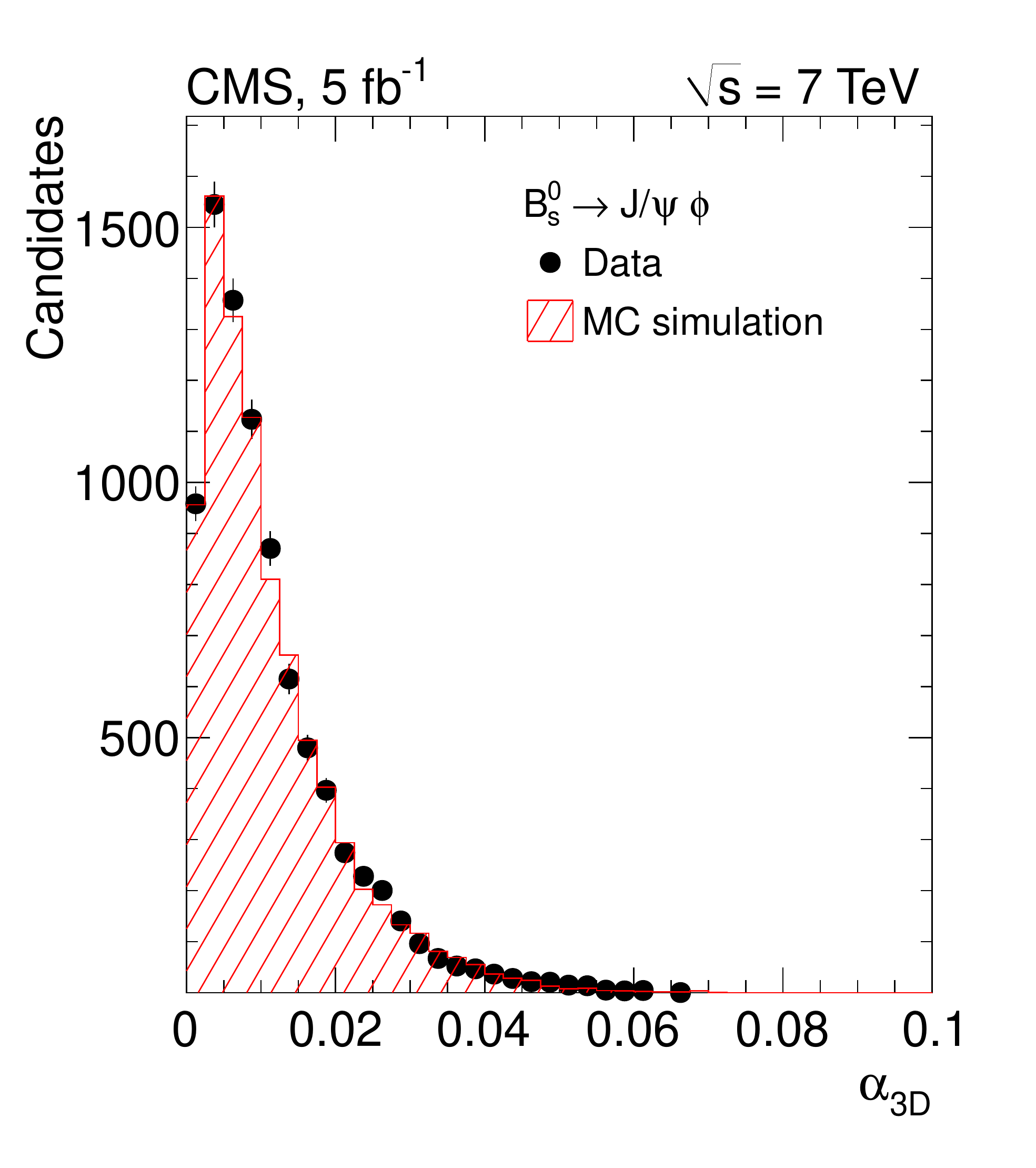}
    \includegraphics[width=\fwidth]{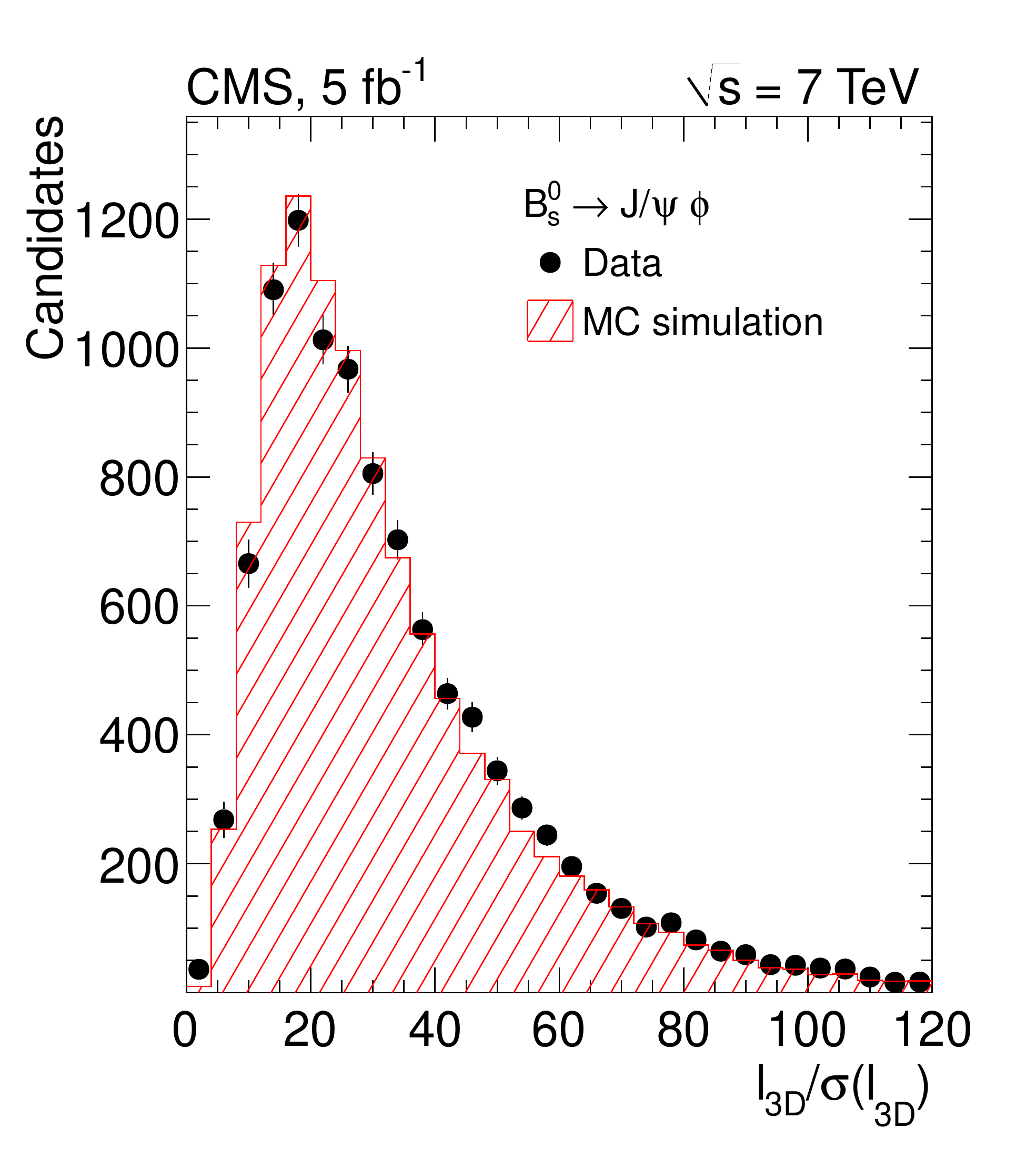}
    \includegraphics[width=\fwidth]{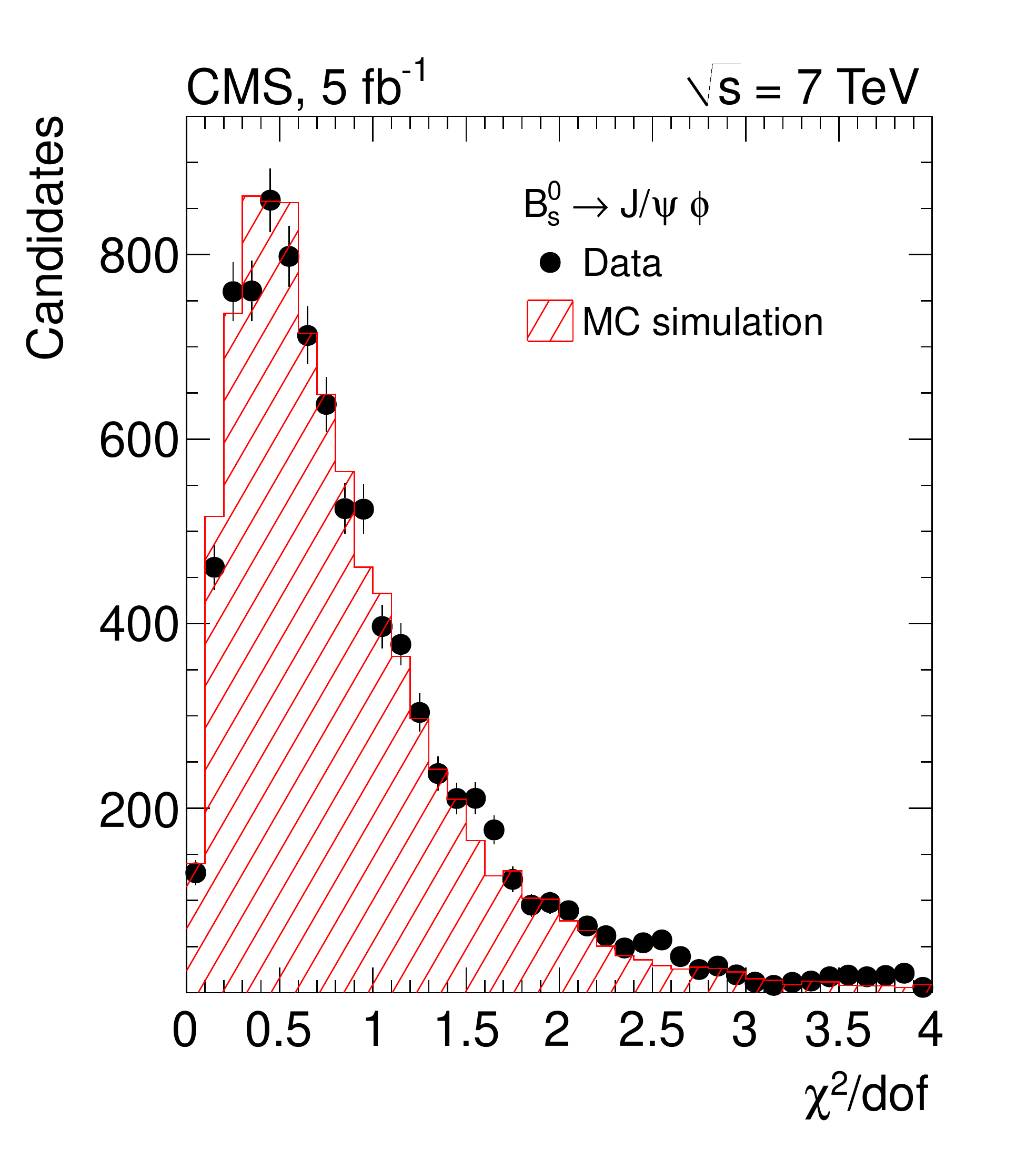}
    \includegraphics[width=\fwidth]{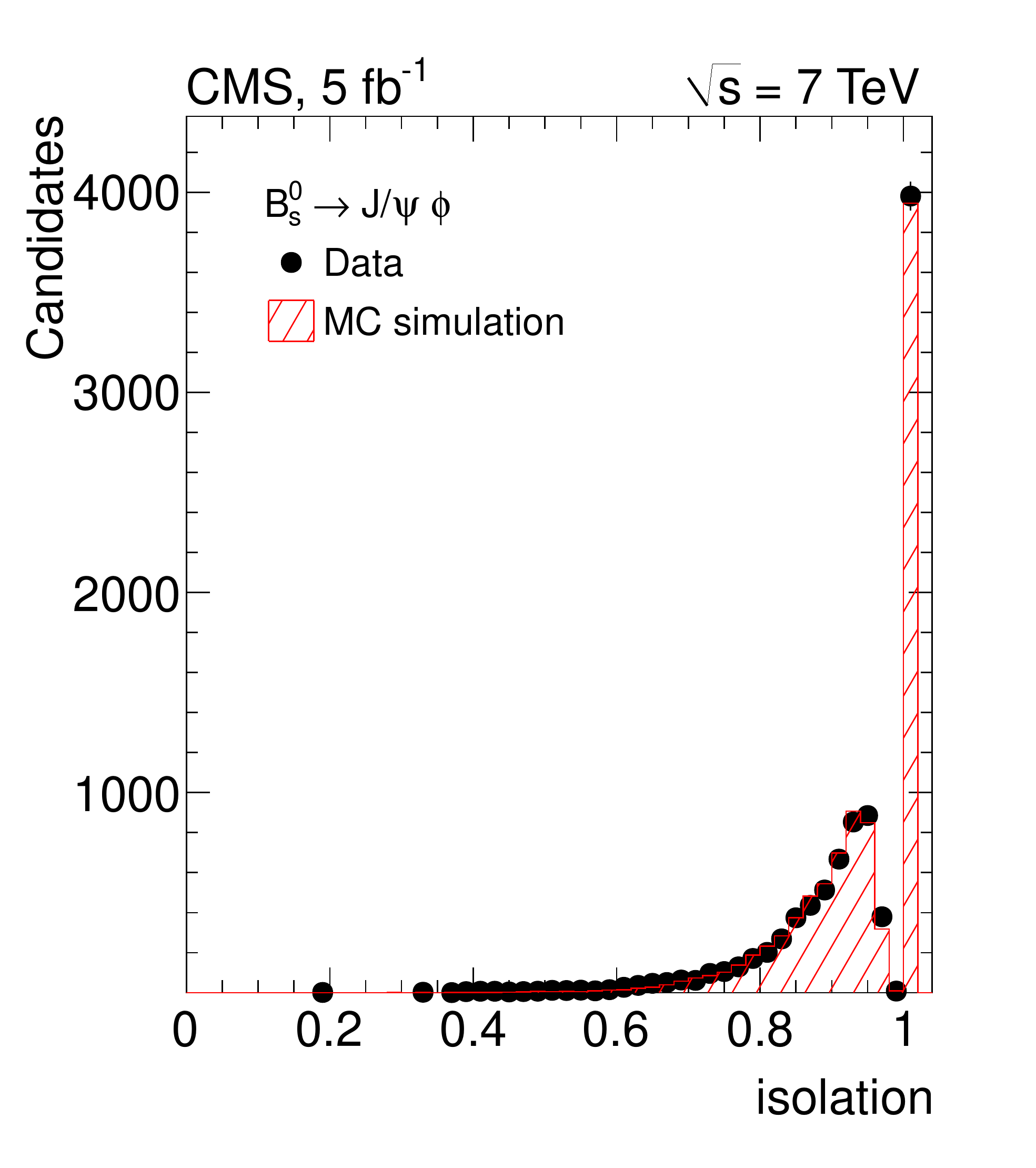}
    \includegraphics[width=\fwidth]{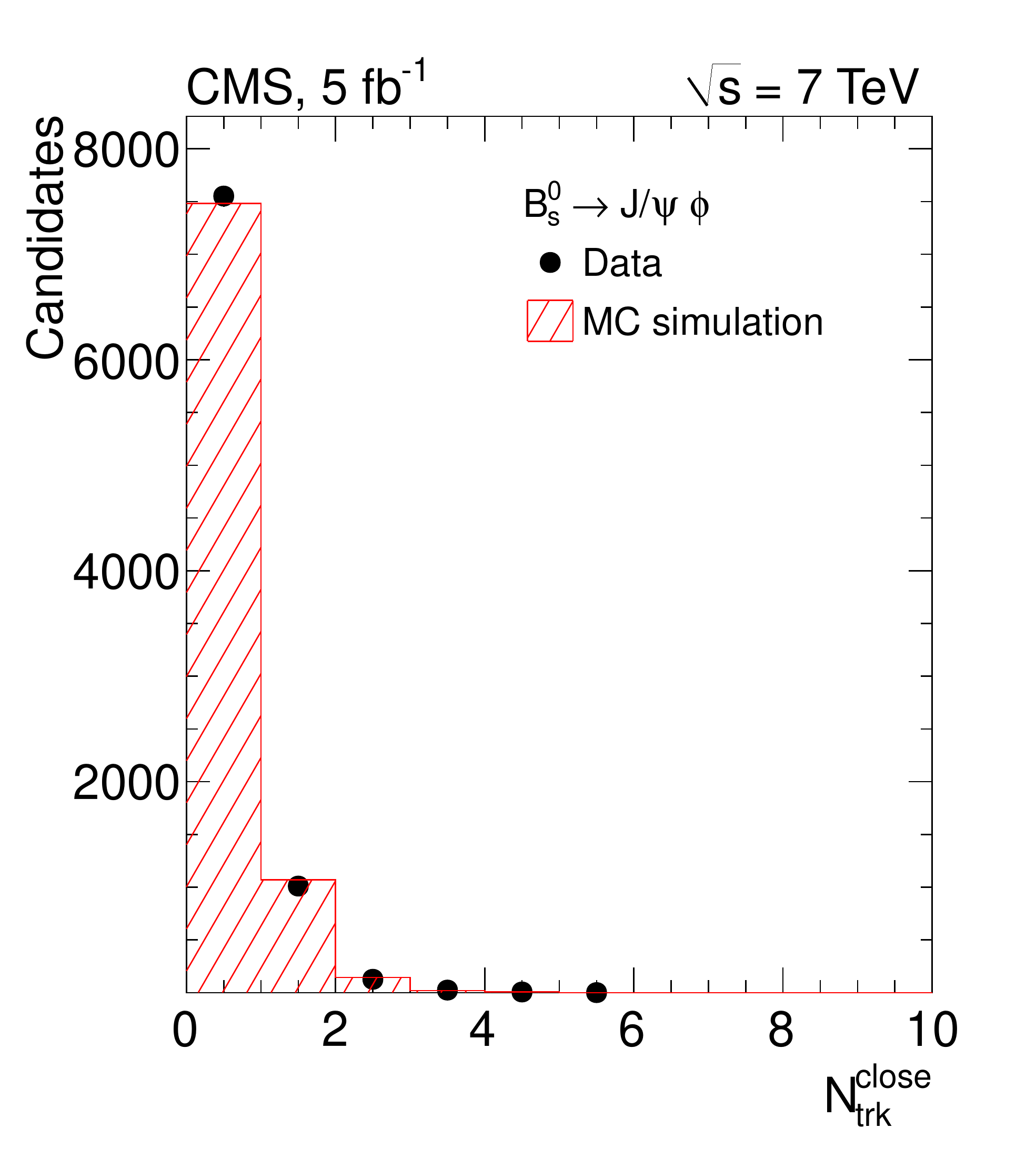}
    \includegraphics[width=\fwidth]{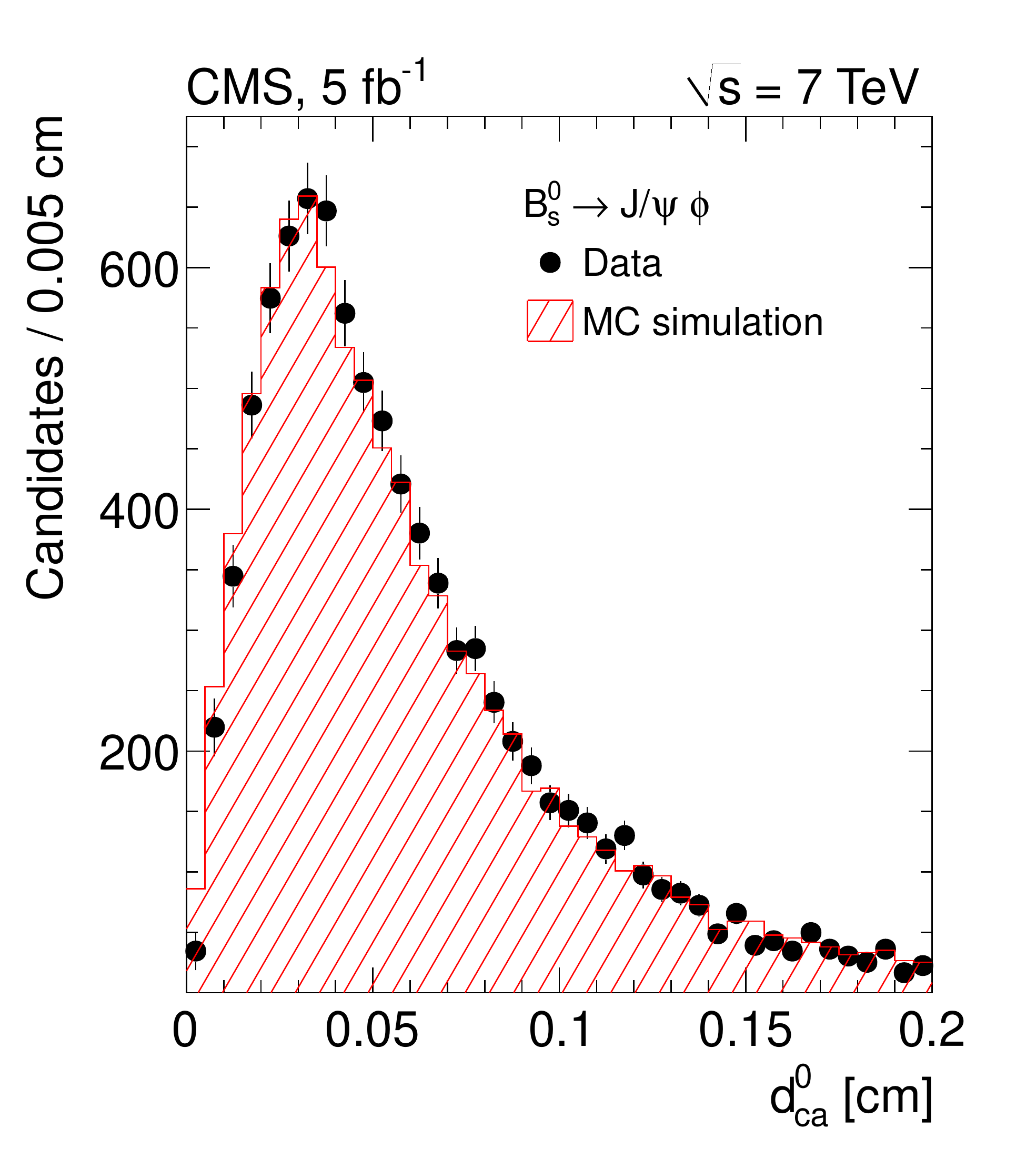}
    \caption{Comparison of measured and simulated \bspsiphi\ distributions.
      Top row: the transverse momentum for the leading muon, sub-leading muon, and $\B$-candidate;
      middle row: the 3D pointing angle, flight length significance, and $\B$-candidate's vertex \chidof;
      bottom row: the isolation variables $I$, \closetrk, and \docatrk.
      The MC histograms are normalized to the number of events in the data.}
   \label{fig:CsData-CsMc}
  \end{centering}
\end{figure}

\begin{figure}[htbp]
  \begin{centering}
    \includegraphics[width=0.3\textwidth]{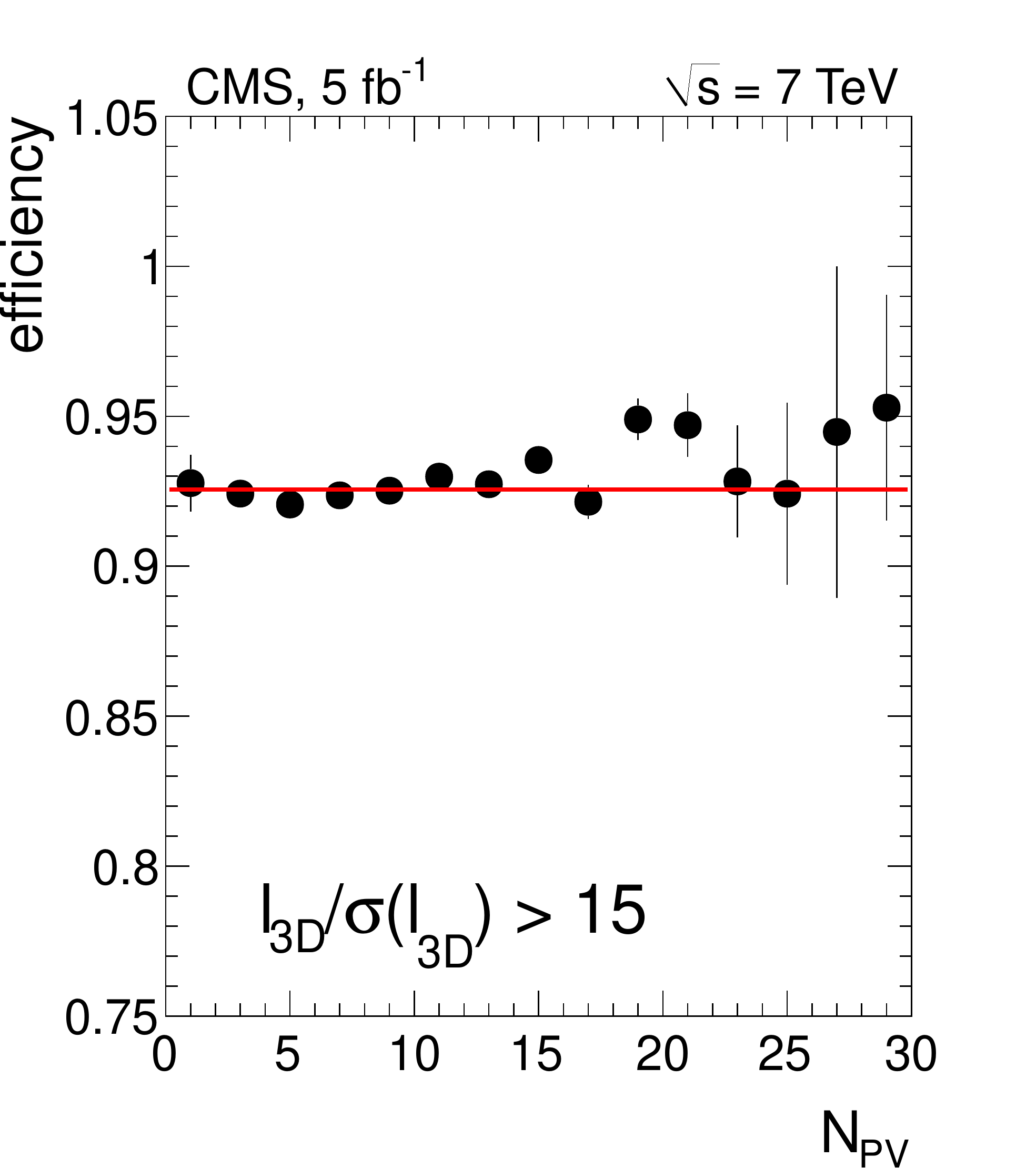}
    \includegraphics[width=0.3\textwidth]{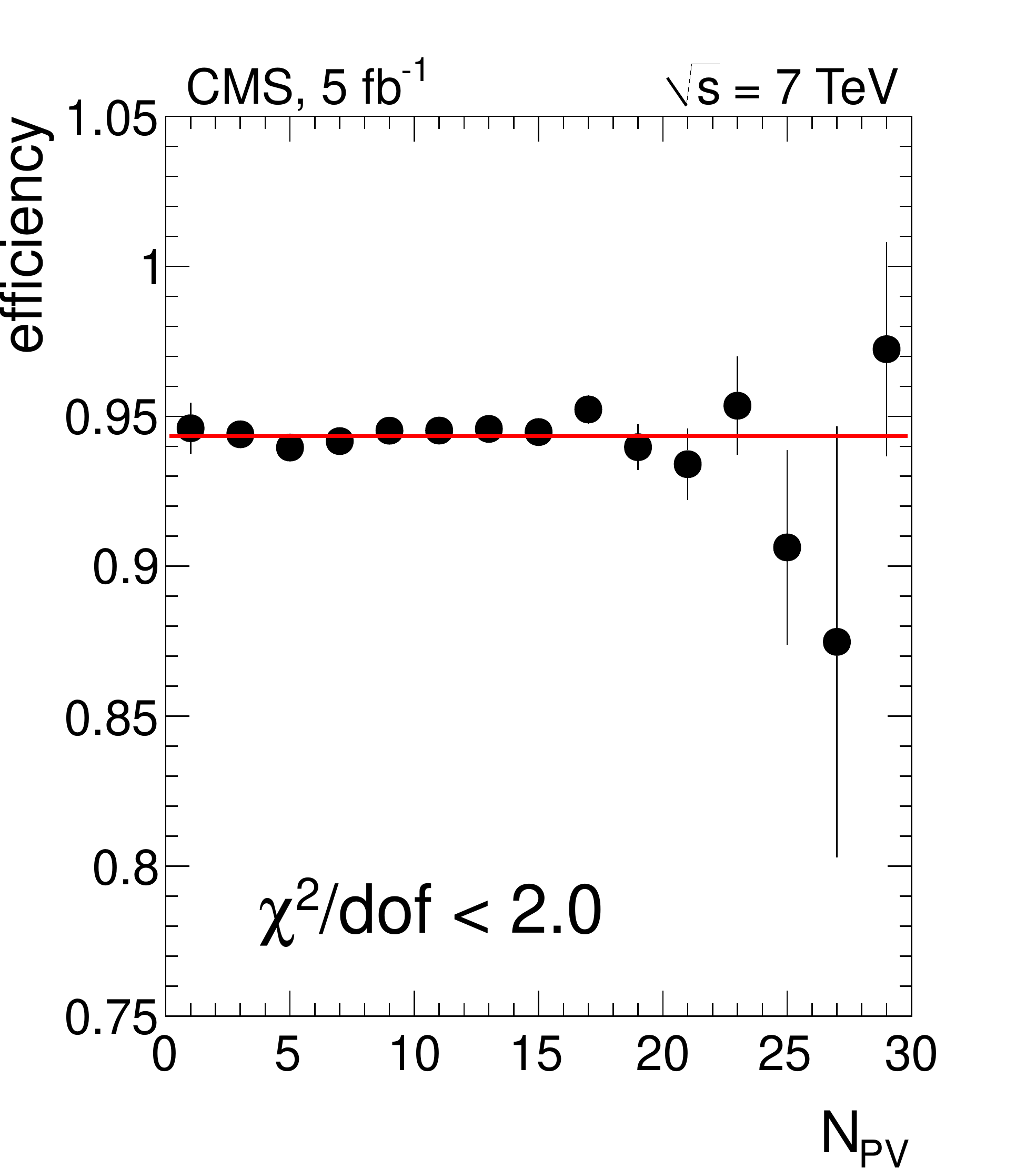}
    \includegraphics[width=0.3\textwidth]{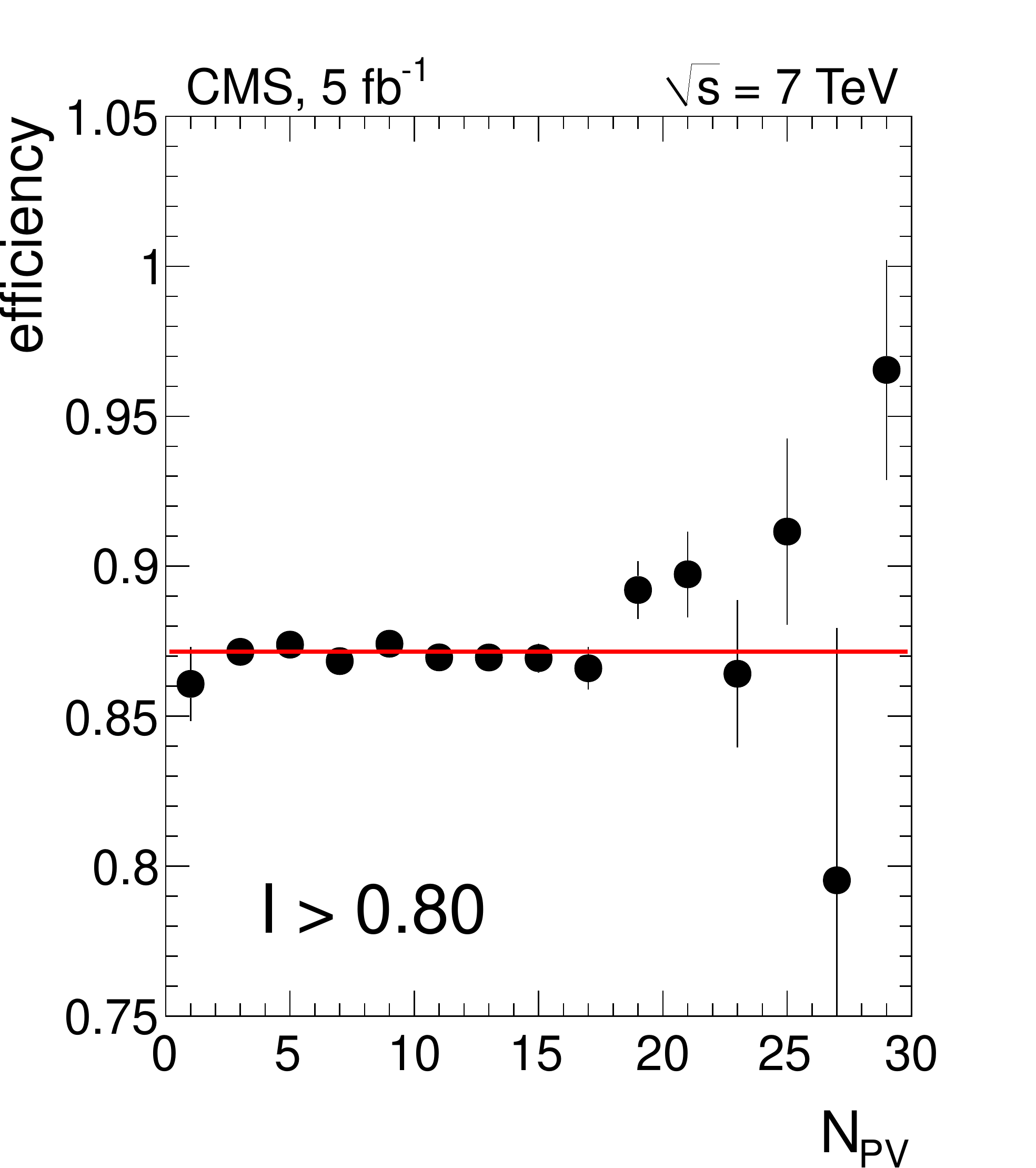}
    \includegraphics[width=0.3\textwidth]{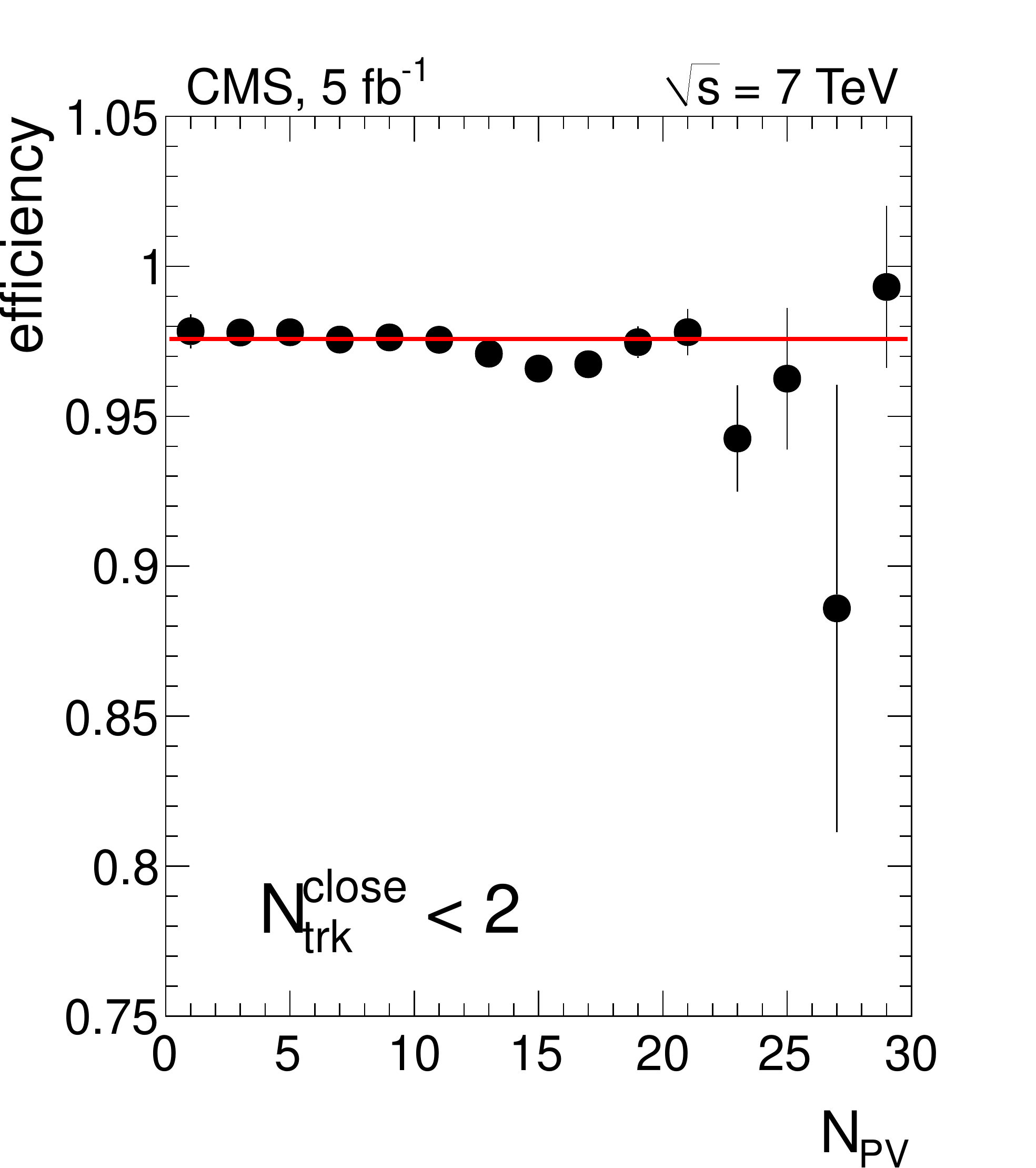}
    \includegraphics[width=0.3\textwidth]{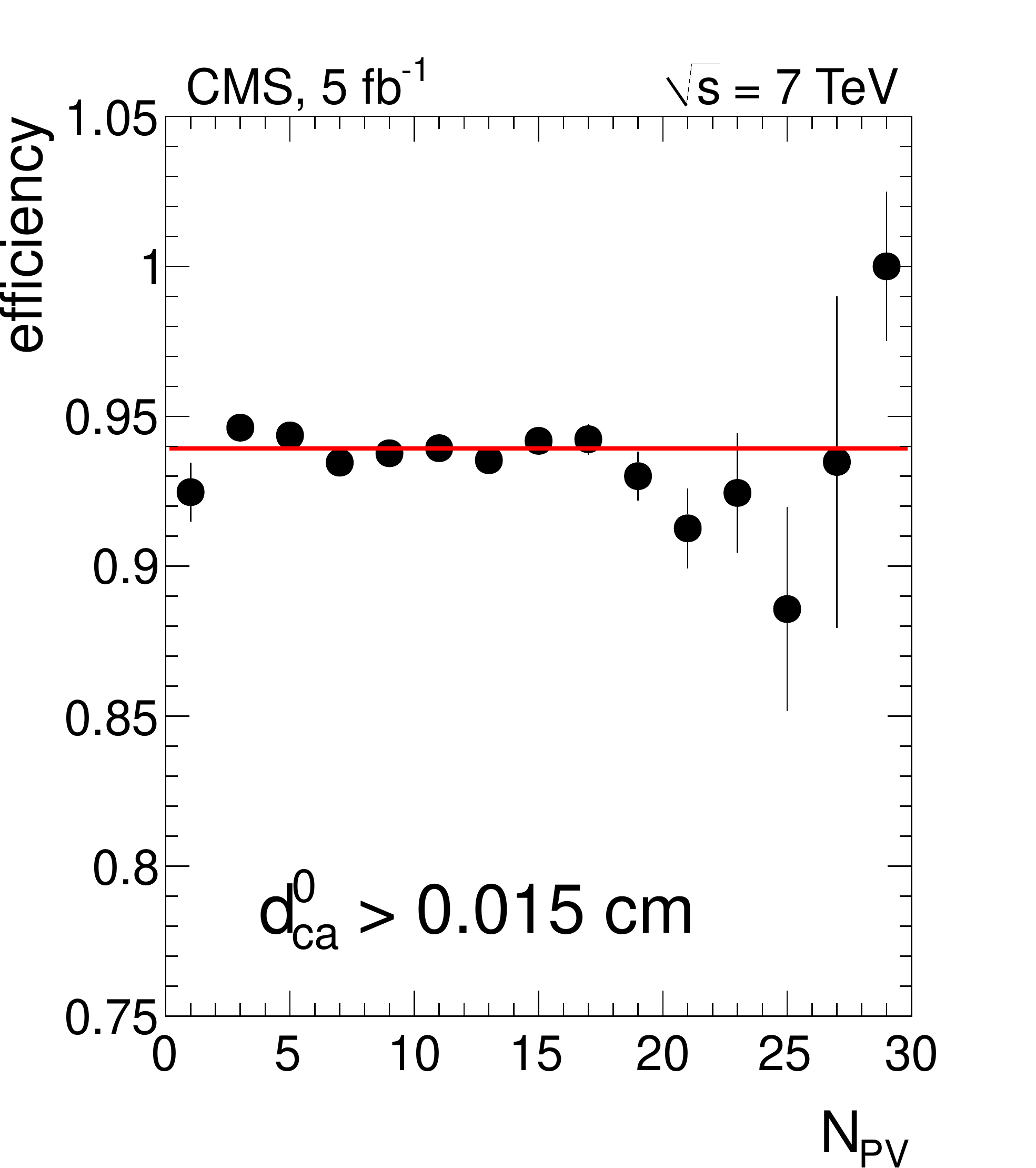}
    \includegraphics[width=0.3\textwidth]{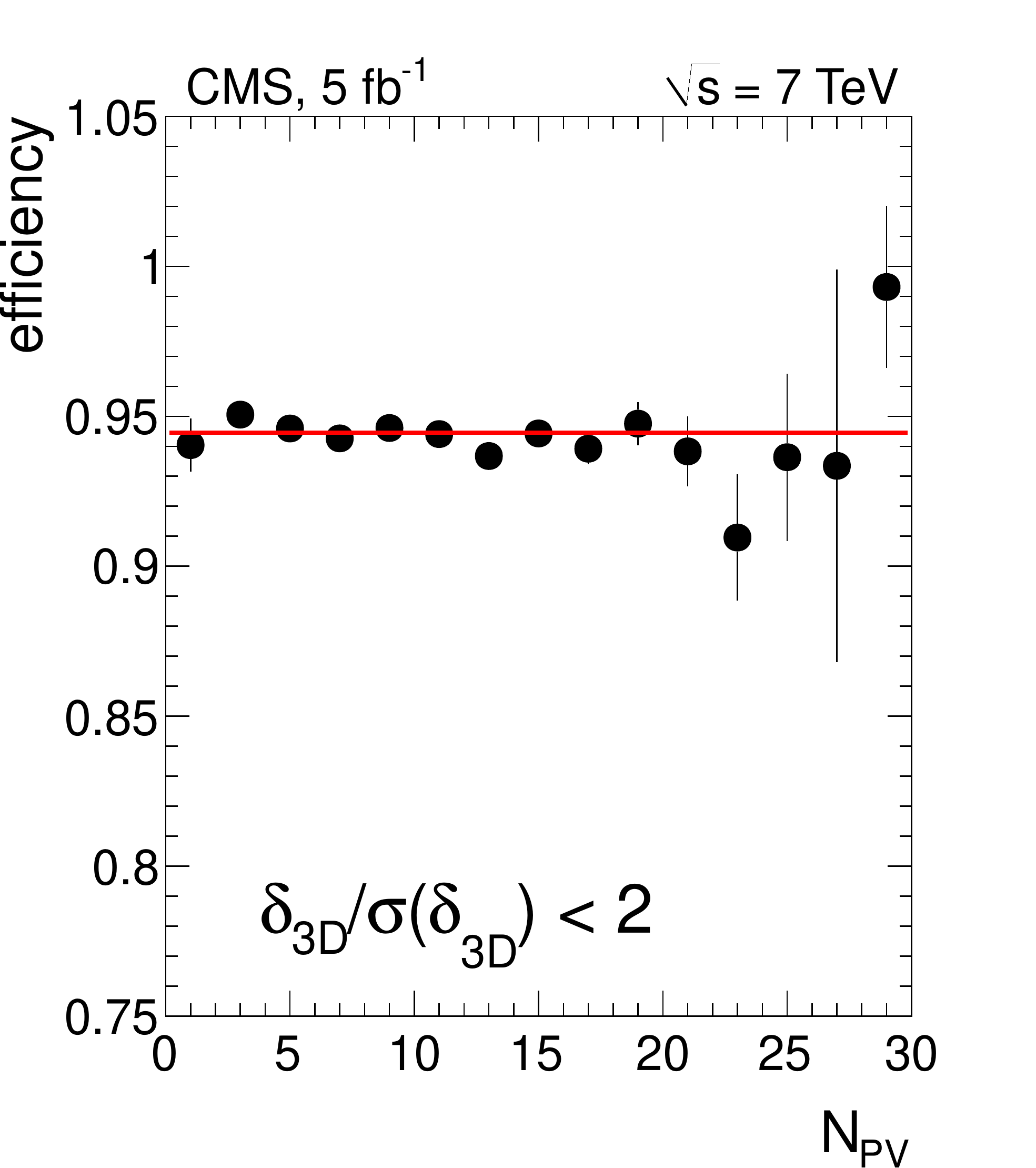}
    \caption{Efficiency versus number of primary vertices, measured with
      \bupsik\ candidates in data for the requirements $\fls > 15$,
      $\chidof < 2$, $I>0.8$, $\closetrk < 2$, and $d_{ca}^{0}>
      0.015\cm$, and $\ips < 2$ (top left to bottom right). The line
      indicates a fit to a constant. The error bars represent the statistical uncertainty only.}
    \label{fig:puEffNo}
  \end{centering}
\end{figure}

The dataset used in this analysis is affected by pileup, which includes an average of 8 reconstructed
primary vertices per event.
The distribution of the primary vertex $z$ position has a Gaussian shape with an RMS of
approximately 5.6\cm.
To study a possible dependence on the amount of pileup, the
efficiencies of all selection criteria are calculated as a function of
the number of reconstructed primary vertices.
In Fig.~\ref{fig:puEffNo} this dependence is shown for the normalization sample.
A horizontal line is superimposed to guide the eye indicating that no significant dependence is observed.
The same conclusion is also obtained in the MC simulation by comparing the selection efficiency for events
with less than six primary vertices to those with more than ten primary vertices.
Similar studies of the control sample, albeit with less precision,
lead to the same conclusion: the analysis is not affected by pileup.

\renewcommand{\sample}{NoData}
\renewcommand{\samplet}{candAnaBu2JpsiK}

Variables sensitive to the underlying production processes (gluon
fusion, flavor excitation, or gluon splitting) are also studied to
validate the production process mixture in the MC simulation. The
clearest distinction among the three processes is obtained by
studying (i) the $\Delta R$ distribution between the $\B$ candidate and
another muon and (ii) the \pt\ spectrum of this muon.
The MC simulation ({\sc Pythia}) describes these distributions adequately.

\section{Results}
\label{s:results}
The present analysis differs significantly from the previous one~\cite{Chatrchyan:2011kr}:
\begin{itemize}
\item The muon identification algorithm has changed. A tighter selection is used,
 which significantly decreases the rate at which kaons and pions are misidentified as muons.
\item The definition of  isolation is different and two additional isolation variables are used,
 which reduces the influence of event pileup and also lowers the combinatorial background.
\item The requirement, for the normalization and control samples, that the two muons bend away from each other
is removed, making the selection of these samples more similar to that for the signal.
\item The rare backgrounds, discussed below, are taken into account when calculating the combinatorial background,
 thus improving the background estimate in the signal window.
\end{itemize}

The variables discussed in Section \ref{s:selection} are optimized to obtain the best expected
upper limit using MC signal events and data sideband events for the background.
The optimization procedure is based on a random-grid search of about
$1.4\times10^6$ analysis selections.
During this search, eleven variables are randomly sampled within predefined ranges. %  with fixed grid sizes.
The resulting optimized requirements, which are used to obtain the final results, are summarized in Table~\ref{tab:ulcuts}. 
These requirements were established before observing the number of data events in the signal region. 
Hence, the analysis was blind to the signal events in the
$5.20 < m_{\mu\mu} < 5.45\gev$ mass range.
In the endcap regions the selection is tighter than in the barrel because of the substantially larger background.
The signal efficiencies $\varepsilon_{\mathrm{tot}}$ for these selections are shown in Table~\ref{t:allTheNumbers}.
They include all selection requirements: the detector acceptance, and the analysis, muon identification, and trigger efficiencies.
The quoted errors include all the systematic uncertainties.
In general, the present analysis uses more strict selection requirements than in the earlier analysis~\cite{Chatrchyan:2011kr},
resulting in a higher sensitivity and a better signal-to-background ratio, but also a lower signal efficiency.
As an additional test, the optimization was repeated to maximize the ratio $S/\sqrt{S+B}$, where $S$ is the number signal events
and $B$ is the number of background events.
This resulted in a similar set of parameters to the ones listed in Table~\ref{tab:ulcuts}, but without an improvement in the expected upper limit.

To evaluate a possible bias due to the optimization of the selection criteria in the data
sidebands and to validate our background expectation, the following crosscheck is performed.
All candidates with $I < 0.7$, including those within the blinded region,
are selected (``inverted isolation'' selection), which generates a background-enriched sample with a very small expected signal contribution.
From this sample, the candidate yields in the sidebands and in the blinded region are determined.
The sideband yields are used to predict, through interpolation, the number of background candidates in the blinded region.
Then the number of predicted background events can be compared to the number of observed candidates in the blinded region.
This comparison is performed separately for the barrel and endcap channels and for the \Bs\ and \Bz\ signal windows.
Within statistical uncertainties, good agreement is found for all four cases, which means that no significant biases are present
in the background interpolation.

\begin{table}[!htb]
  \begin{center}
    \caption{Selection criteria for \bsmm\ and \bdmm\ search in the barrel and endcap. }
    \vspace{0.1in}
    \label{tab:ulcuts}
\begin{tabular}{|l|c|c|l|}
  \hline
  Variable      &Barrel &Endcap &units  \\
  \hline
  $\pt_{\mu,1} >$ &$\vuse{default-11:m1pt:0} $    &$\vuse{default-11:m1pt:1} $    & GeV\\
  $\pt_{\mu,2} >$ &$\vuse{default-11:m2pt:0} $    &$\vuse{default-11:m2pt:1} $    & GeV\\
  $\pt_{\B} >$    &$\vuse{default-11:pt:0} $      &$\vuse{default-11:pt:1} $      & GeV\\
  $\delta_{3D}<$  &$\vuse{default-11:pvip:0} $ &$\vuse{default-11:pvip:1} $ & cm\\
  $\delta_{3D}/\sigma(\delta_{3D})<$  &$\vuse{default-11:pvips:0} $ &$\vuse{default-11:pvips:1} $ &\\
  $\alpha <$    &$\vuse{default-11:alpha:0} $   &$\vuse{default-11:alpha:1} $   & rad\\
  $\chidof <$  &$\vuse{default-11:chi2dof:0} $ &$\vuse{default-11:chi2dof:1} $ &\\
  $\ell_{3d}/\sigma(\ell_{3d}) >$                 &$\vuse{default-11:fls3d:0} $   &$\vuse{default-11:fls3d:1} $ & \\
  $I >$         &$\vuse{default-11:iso:0} $     &$\vuse{default-11:iso:1} $    & \\
  $\dca^0 >$   &$\vuse{default-11:docatrk:0} $ &$\vuse{default-11:docatrk:1} $ & cm\\
  $\closetrk <$    &$\vuse{default-11:closetrk:0} $ &$\vuse{default-11:closetrk:1} $ & tracks\\
  \hline
\end{tabular}
\end{center}
\end{table}

The simulated dimuon mass resolution for signal events
depends on the pseudorapidity of the $\B$ candidate and ranges from $37\MeV$
for $\eta\sim 0$ to $77\MeV$ for $|\eta| > 1.8$.
The dimuon mass scale and resolution in the MC simulation are compared with the measured
detector performance by studying $\jpsi\to\mup\mun$ and $\OneS\to\mup\mun$ decays.
The residual differences between simulation and data are small and the uncertainty on the efficiency
coming from these effects is estimated to be 3\%.

Branching fractions are measured separately in the barrel and endcap channels using
the following equation
\begin{eqnarray}
  \cbf(\bsmm)
  &=&  \frac{N_\mathrm{S}}{N_{\mathrm{obs}}^{\Bp}} \,
  \frac{f_{\u}}{f_{\s}} \,
  \frac{\varepsilon_{\mathrm{tot}}^{\Bp}}{\varepsilon_{\mathrm{tot}}} \,
  \cbf(\Bp)\label{eq:schema},
\end{eqnarray}
where $\varepsilon_\mathrm{tot}$ is the total signal efficiency,
$N_{\mathrm{obs}}^{\Bp}$  is the number of reconstructed \bupsik decays,
$\varepsilon_{\mathrm{tot}}^{\Bp}$ is the total efficiency of \Bp\ reconstruction,
$\cbf(\Bp)$ is the branching fraction for $\bupsik\to\mup\mun \Kp$,
$f_\u/f_\s$ is the ratio of the \Bp\ and \Bs\ production cross sections,
and $N_\mathrm{S}$ is the background-subtracted number of observed \bsmm\ candidates
in the signal window $5.30 < m_{\mu\mu} < 5.45\gev$.
The width of the signal windows is adjusted to maximize the efficiency for the \bsmm decay,
and it is approximately equal to twice the expected mass resolution in the endcap region.
We use the value $f_\s/f_\u= 0.267\pm0.021$, measured by LHCb for $2<\eta<5$~\cite{Aaij:2011jp}
and $\cbf(\Bp) \equiv \cbf(\Bp\to\jpsi\Kp\to\mup\mun\Kp) =(6.0\pm0.2)\times10^{-5}$~\cite{Nakamura:2010zzi}.
An analogous equation is used to measure the $\bdmm$ branching fraction, with the signal window
$5.2 <m_{\mu\mu} < 5.3\GeV$ and the ratio $f_\d/f_\u$ taken to be 1.

\begin{figure}[htbp]
  \begin{centering}
\includegraphics[width=\figwid]{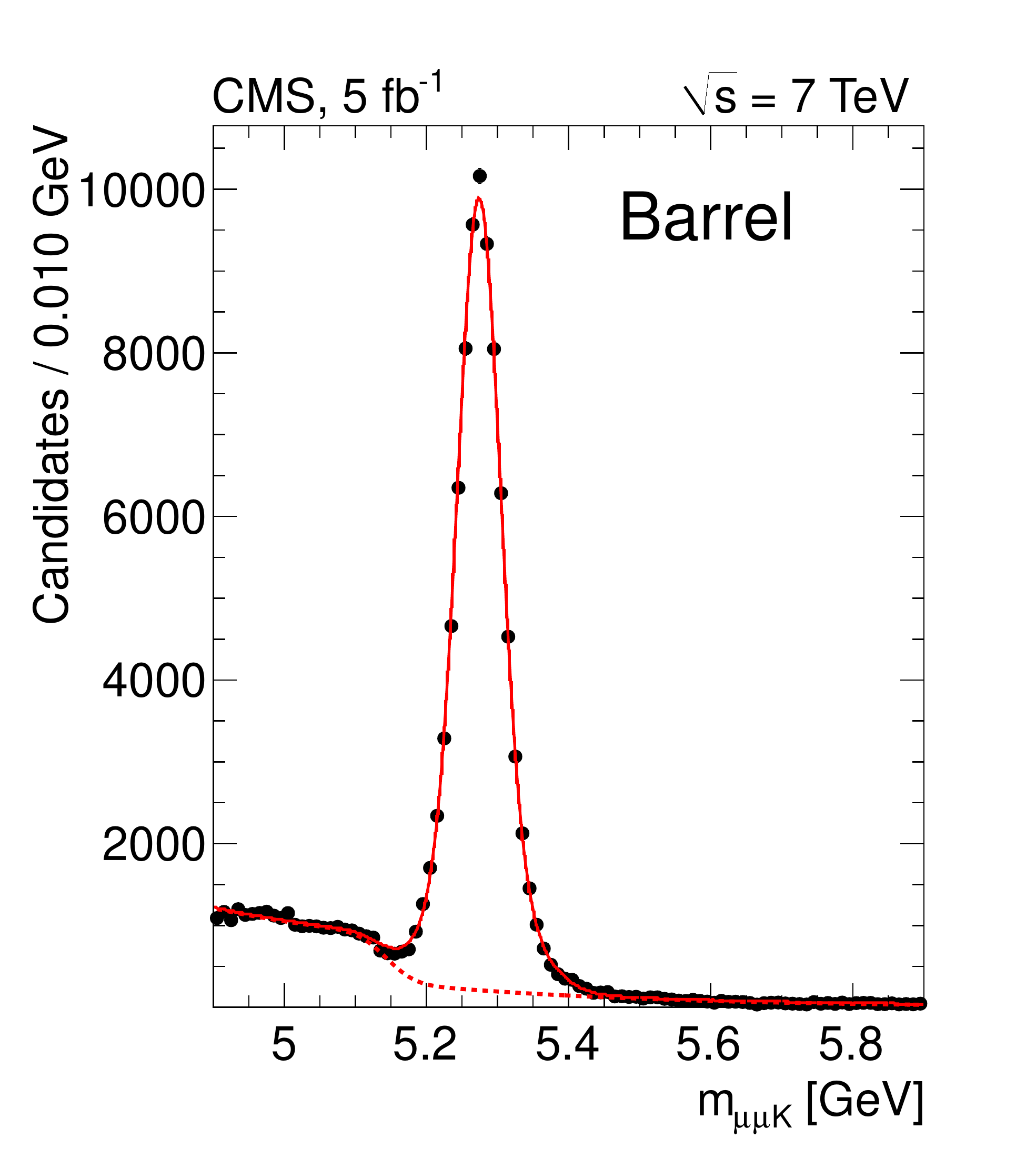}
\includegraphics[width=\figwid]{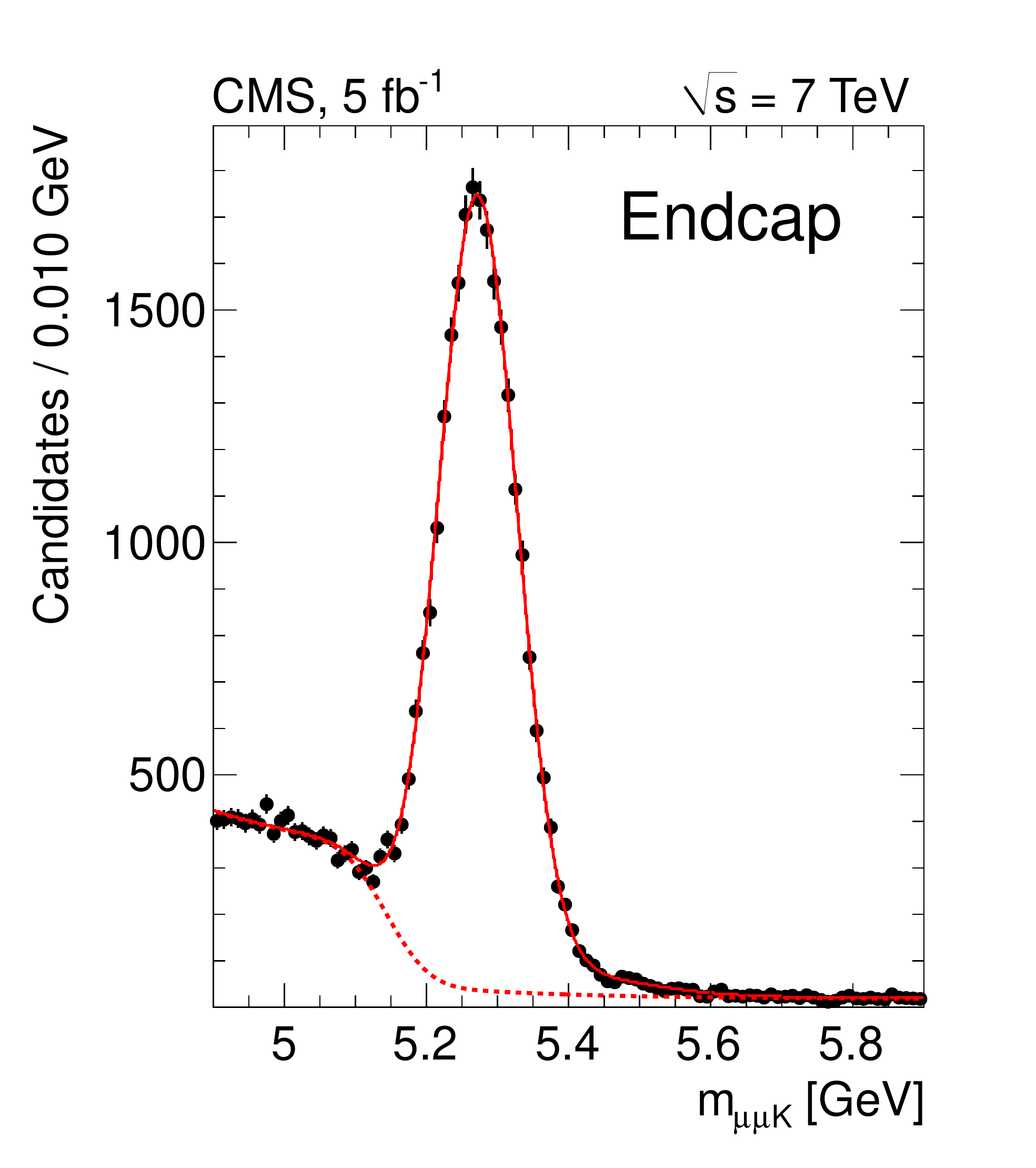}
   \caption{\bupsik\ invariant-mass distributions in the barrel (left)
     and endcap (right) channels. The solid (dashed) lines show the fits
     to the data (background).}
   \label{fig:resNorm}
  \end{centering}
\end{figure}

The number of reconstructed \Bp\ mesons $N_{\mathrm{obs}}^{\Bp}$ is
$(82.7\pm4.2)\times10^{3}$ in the barrel and $(23.8\pm1.2)\times10^{3}$ in the endcap.
The invariant mass distributions are fit with a double-Gaussian function for the signal and
an exponential plus an error function for the background, as shown in Fig.~\ref{fig:resNorm}.
Partially reconstructed $\Bz$ decays (e.g., $\bpsikst$ with one of the $\Kstar$ decay products not reconstructed)
lead to a step function-like behavior at a mass of $m \approx 5.15\gev$.
This background shape was studied in detail in MC simulation
and is parametrized with an error function of different width in the barrel and endcap.
The systematic uncertainty on the fit yield, 5\% in the barrel and in the endcap,
is estimated by considering alternative fitting functions and by performing a fit with the dimuon invariant mass
constrained to the \jpsi\ mass.
The total efficiency $\varepsilon_{\mathrm{tot}}^{\Bp}$, including the detector acceptance,
is determined from MC simulation to be
$(11.0\pm0.9)\times10^{-4}$ for the barrel and $(3.2\pm0.4)\times10^{-4}$ for the endcap,
where the errors include statistical and systematical uncertainties.
The detector acceptance part (which includes the track finding efficiency)
of the total efficiency has a systematic uncertainty of 3.5\% (5.0\%) in the barrel (endcap).
It is estimated by comparing the values obtained separately with three different \bbbar\
production mechanisms: gluon splitting, flavor excitation, and flavor creation.

The branching fraction for the control decay \bspsiphi\, which was analyzed in parallel with the normalization and signal decays,
has also been evaluated using Equation~\ref{eq:schema}.
The resulting branching ratio is in agreement with the world average ~\cite{Nakamura:2010zzi}.
Moreover, the results for the barrel and endcap channels agree within the statistical uncertainties,
showing the validity of extending the $f_\s/f_\u$ measurement from~\cite{Aaij:2011jp} to the barrel region.

Events in the signal window have several sources:
(i) genuine signal decays,
(ii) decays of the type $\B\to \h\h'$, where $\h,\h'$ are charged hadrons misidentified
as muons (referred to as ``peaking'' background),
(iii) rare semileptonic decays $\B \to \h\mu\nu$, where $\h$ is
misidentified as a muon, and
(iv) combinatorial background.
Note that events from the third category predominantly populate the lower sideband.

The expected numbers of signal events $N_{\mathrm{signal}}^{\mathrm{exp}}$ for the barrel and
endcap channels are shown in Table~\ref{t:allTheNumbers}.
They are calculated assuming the SM branching fractions ~\cite{Buras:2010wr} and are normalized to the measured \Bp\ yield.

The expected numbers of rare semileptonic decays and peaking background
events, $N_{\mathrm{peak}}^{\mathrm{exp}}$, are also shown in Table~\ref{t:allTheNumbers}.
They are evaluated from a MC simulation, which is normalized to the measured \Bp\ yields,
and from muon misidentification rates measured in $\Dstarp \to \Dz\pip$,
$\Dz\to\Km\pip$, and $\Lambda\to \p\pim$ samples~\cite{CMS-PAS-MUO-10-002}.
The average misidentification probabilities in the kinematic range of
this analysis are $(0.10\pm0.02)\%$ for pions and kaons, and
$(0.05\pm0.01)\%$ for protons, where the uncertainties are statistical.
The systematic uncertainty on the background includes the uncertainties on
the production ratio (for \Bs\ and $\Lambda_\b$ decays), the branching fraction, and the
misidentification probability.

Also shown in Table~\ref{t:allTheNumbers} are the expected numbers of combinatorial background events
$N_{\mathrm{comb}}^{\mathrm{exp}}$.
They are evaluated by interpolating into the signal window the number of events observed in the sideband regions,
after subtracting the expected rare semileptonic background.
The interpolation procedure assumes a flat background
shape and has a systematic uncertainty of 4\%, which is evaluated by varying
the flight-length significance selections and by using a linear background shape with a variable slope.

\begin{figure}[htbp]
  \begin{centering}
\includegraphics[width=\figwid]{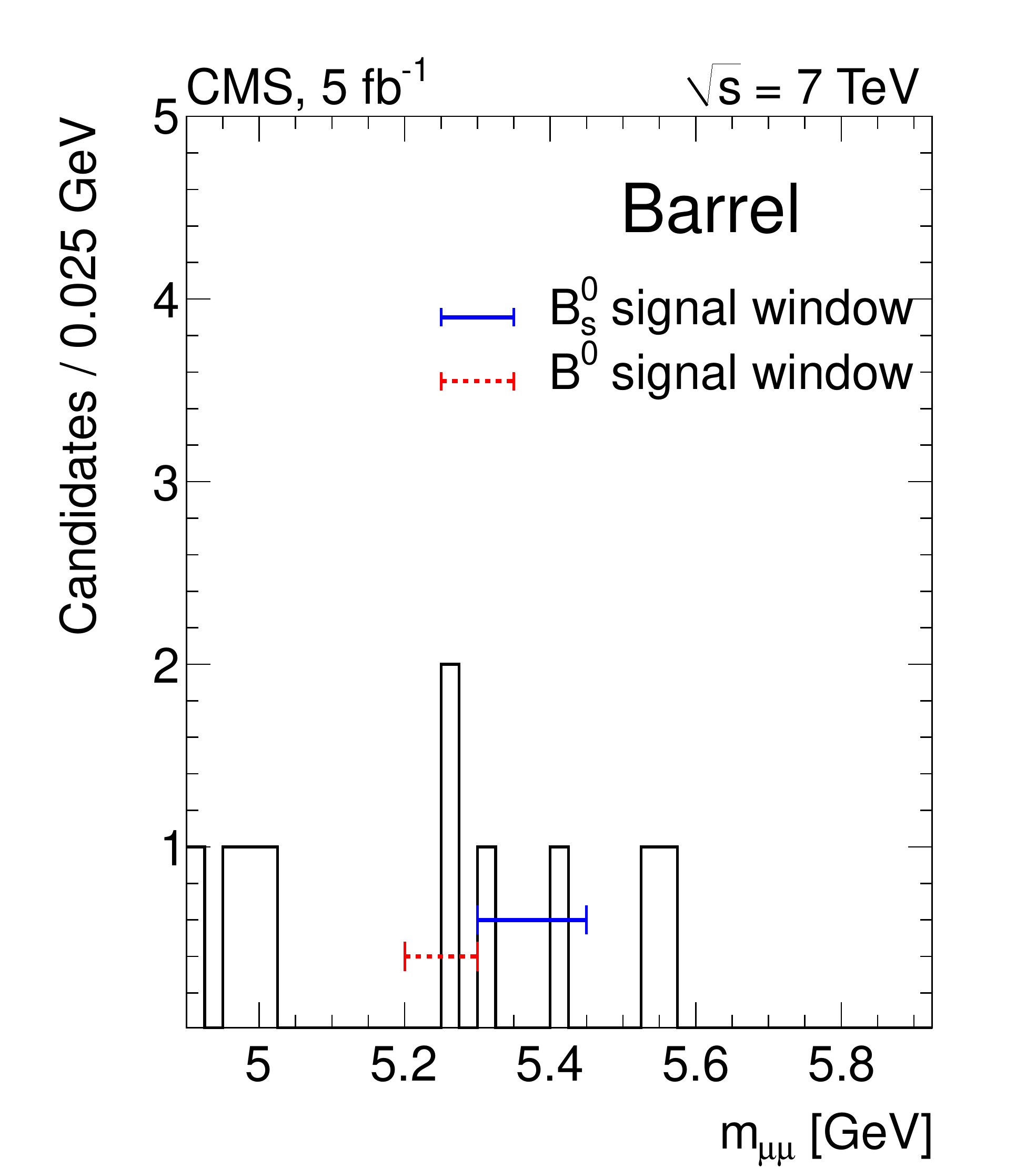}
\includegraphics[width=\figwid]{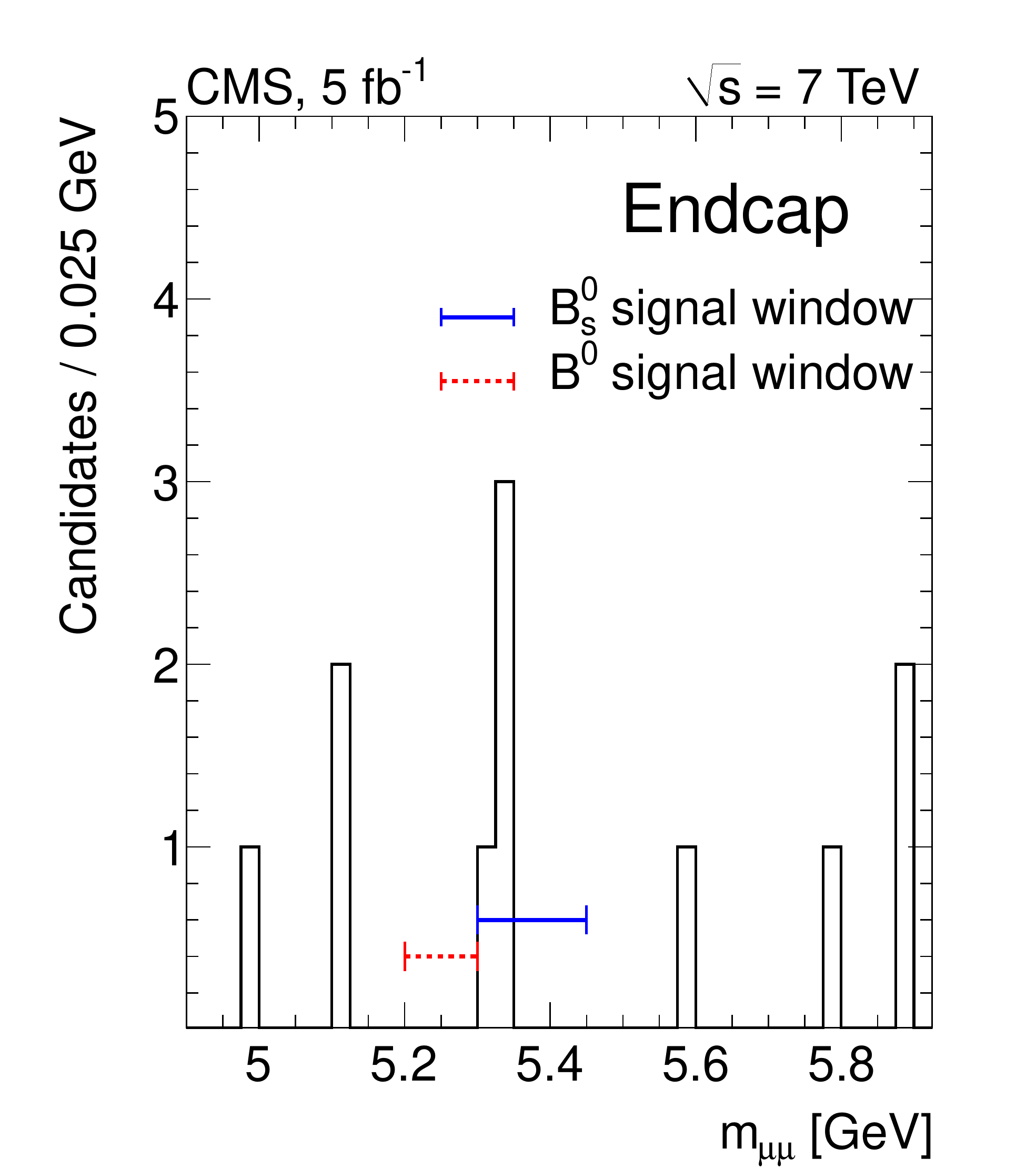}
\caption{Dimuon invariant-mass distributions in the barrel (left) and
  endcap (right) channels. The signal windows for \Bs\ and \Bz\ are
  indicated by horizontal lines. }
   \label{fig:result}
  \end{centering}
\end{figure}

Figure~\ref{fig:result} shows the measured dimuon invariant-mass distributions.
In the sidebands the observed number of events is equal to six (seven)
for the barrel (endcap) channel.
Six events are observed in the \bsmm\ signal windows (two in the barrel and four in the
endcap), while two events are observed in the \bdmm\ barrel channel
and none in the endcap channel.
As indicated by the numbers shown in Table~\ref{t:allTheNumbers},
this observation is consistent with the SM expectation for signal plus background.

Upper limits on the $\bsmm$ and $\bdmm$ branching fractions are determined using the $\mathrm{CL_s}$
method~\cite{Junk:1999kv,Read:2002hq}.  Table~\ref{t:allTheNumbers}
lists all the values needed for the extraction of the results
for both the barrel and endcap channels.
The combined upper limits for the barrel and endcap channels
are ${\cal B}(\bsmm) <
7.7\times10^{-9}$ $(6.4\times10^{-9})$ and ${\cal B}(\bdmm) < 1.8\times10^{-9}$ $(1.4\times10^{-9})$ at 95\% (90\%) CL.
The median expected upper limits at 95\% CL are $8.4\times10^{-9}$ ($1.6\times10^{-9}$)
for \bsmm (\bdmm), where the number of expected signal events is based on the SM value.
Including cross-feed between the  \Bz and \Bs decays, the background-only $p$ value is 0.11 (0.24) for \bsmm
(\bdmm), corresponding to $1.2$ ($0.7$) standard deviations.
The $p$ value for the background plus SM signal hypotheses is 0.71 (0.86) for \bsmm  (\bdmm).

\renewcommand{\base}{default-11}

\begin{table}[!htb]
  \begin{center}
    \caption{The event selection efficiency for signal events $\varepsilon_{\mathrm{tot}}$,
 the SM-predicted number of signal events $N_{\mathrm{signal}}^{\mathrm{exp}}$,
 the expected number of peaking background events $N_{\mathrm{peak}}^{\mathrm{exp}}$ and
 combinatorial background events $N_{\mathrm{comb}}^{\mathrm{exp}}$,
 and the number of observed events $N_{\mathrm{obs}}$ in the barrel and
 endcap channels for \bsmm\ and \bdmm.
The quoted errors include both, the statistical and the systematic uncertainties.
}
    \label{t:allTheNumbers}
    \vspace{0.01in}
\begin{tabular}{|l|c|c|c|c|}
  \hline
  Variable    &\bdmm\ Barrel &\bsmm\ Barrel &\bdmm\ Endcap &\bsmm Endcap
  \\
  \hline
  $\varepsilon_{\mathrm{tot}}$  \StrutSmall
  &$\vuse{\base:N-EFF-TOT-BDMM0:val} \pm \vuse{\base:N-EFF-TOT-BDMM0:tot} $
  &$\vuse{\base:N-EFF-TOT-BSMM0:val} \pm \vuse{\base:N-EFF-TOT-BSMM0:tot} $
  &$\vuse{\base:N-EFF-TOT-BDMM1:val} \pm \vuse{\base:N-EFF-TOT-BDMM1:tot} $
  &$\vuse{\base:N-EFF-TOT-BSMM1:val} \pm \vuse{\base:N-EFF-TOT-BSMM1:tot} $
  \\
  \hline
  $N_{\mathrm{signal}}^{\mathrm{exp}}$  \StrutSmall
  &$0.24\pm0.02$
  &$2.70\pm0.41$
  &$0.10\pm0.01$
  &$1.23\pm0.18$
  \\
  $N_{\mathrm{peak}}^{\mathrm{exp}}$
  &$0.33\pm0.07$
  &$0.18\pm0.06$
  &$0.15\pm0.03$
  &$0.08\pm0.02$
  \\
  $N_{\mathrm{comb}}^{\mathrm{exp}}$
  &$0.40\pm0.34$
  &$0.59\pm0.50$
  &$0.76\pm0.35$
  &$1.14\pm0.53$
  \\
  \hline
  $N_{\mathrm{total}}^{\mathrm{exp}}$   \StrutLarge
  &$0.97\pm0.35$
  &$3.47\pm0.65$
  &$1.01\pm0.35$
  &$2.45\pm0.56$
  \\
  \hline
  $N_{\mathrm{obs}}$  \StrutSmall
  &2&2&0&4
  \\
  \hline
\end{tabular}
\end{center}
\end{table}

\section{Summary}
\label{s:conclusions}

An analysis searching for the rare decays $\bsmm$ and
$\bdmm$ has been performed in $\p\p$ collisions at $\sqrt{s}=7\tev$.
A data sample corresponding to an integrated luminosity of $\vuse{lumiTot} \fbinv$ has been used.  
This result supersedes our previous measurement~\cite{Chatrchyan:2011kr}. 
Stricter selection requirements were applied,
resulting in a better sensitivity and a higher expected signal-to-background ratio.
The observed number of events is consistent with background plus SM signals.
The resulting upper limits on the branching fractions are  
${\cal B}(\bsmm) < 7.7\times10^{-9}$ and ${\cal B}(\bdmm) < 1.8\times10^{-9}$
at 95\% CL.
These upper limits can be used to improve bounds on the parameter
space for a number of potential extensions to the standard model.

\section*{Acknowledgments}
\hyphenation{Bundes-ministerium Forschungs-gemeinschaft Forschungs-zentren} We congratulate our colleagues in the CERN accelerator departments for the 
excellent performance of the LHC machine. We thank the technical and administrative staff at CERN and other CMS institutes. This work was supported by the Austrian Federal Ministry of Science and Research; the Belgium Fonds de la Recherche Scientifique, and Fonds voor Wetenschappelijk Onderzoek; the Brazilian Funding Agencies (CNPq, CAPES, FAPERJ, and FAPESP); the Bulgarian Ministry of Education and Science; CERN; the Chinese Academy of Sciences, Ministry of Science and Technology, and National Natural Science Foundation of China; the Colombian Funding Agency (COLCIENCIAS); the Croatian Ministry of Science, Education and Sport; the Research Promotion Foundation, Cyprus; the Ministry of Education and Research, Recurrent financing contract SF0690030s09 and European Regional Development Fund, Estonia; the Academy of Finland, Finnish Ministry of Education and Culture, and Helsinki Institute of Physics; the Institut National de Physique Nucl\'eaire et de Physique des Particules~/~CNRS, and Commissariat \`a l'\'Energie Atomique et aux \'Energies Alternatives~/~CEA, France; the Bundesministerium f\"ur Bildung und Forschung, Deutsche Forschungsgemeinschaft, and Helmholtz-Gemeinschaft Deutscher Forschungszentren, Germany; the General Secretariat for Research and Technology, Greece; the National Scientific Research Foundation, and National Office for Research and Technology, Hungary; the Department of Atomic Energy and the Department of Science and Technology, India; the Institute for Studies in Theoretical Physics and Mathematics, Iran; the Science Foundation, Ireland; the Istituto Nazionale di Fisica Nucleare, Italy; the Korean Ministry of Education, Science and Technology and the World Class University program of NRF, Korea; the Lithuanian Academy of Sciences; the Mexican Funding Agencies (CINVESTAV, CONACYT, SEP, and UASLP-FAI); the Ministry of Science and Innovation, New Zealand; the Pakistan Atomic Energy Commission; the Ministry of Science and Higher Education and the National Science Centre, Poland; the Funda\c{c}\~ao para a Ci\^encia e a Tecnologia, Portugal; JINR (Armenia, Belarus, Georgia, Ukraine, Uzbekistan); the Ministry of Education and Science of the Russian Federation, the Federal Agency of Atomic Energy of the Russian Federation, Russian Academy of Sciences, and the Russian Foundation for Basic Research; the Ministry of Science and Technological Development of Serbia; the Ministerio de Ciencia e Innovaci\'on, and Programa Consolider-Ingenio 2010, Spain; the Swiss Funding Agencies (ETH Board, ETH Zurich, PSI, SNF, UniZH, Canton Zurich, and SER); the National Science Council, Taipei; the Scientific and Technical Research Council of Turkey, and Turkish Atomic Energy Authority; the Science and Technology Facilities Council, UK; the US Department of Energy, and the US National Science Foundation.

Individuals have received support from the Marie-Curie programme and the European Research Council (European Union); the Leventis Foundation; the A. P. Sloan Foundation; the Alexander von Humboldt Foundation; the Belgian Federal Science Policy Office; the Fonds pour la Formation \`a la Recherche dans l'Industrie et dans l'Agriculture (FRIA-Belgium); the Agentschap voor Innovatie door Wetenschap en Technologie (IWT-Belgium); the Council of Science and Industrial Research, India; and the HOMING PLUS programme of Foundation for Polish Science, cofinanced from European Union, Regional Development Fund. 

\bibliography{auto_generated}
\cleardoublepage \appendix\section{The CMS Collaboration \label{app:collab}}\begin{sloppypar}\hyphenpenalty=5000\widowpenalty=500\clubpenalty=5000\textbf{Yerevan Physics Institute,  Yerevan,  Armenia}\\*[0pt]
S.~Chatrchyan, V.~Khachatryan, A.M.~Sirunyan, A.~Tumasyan
\vskip\cmsinstskip
\textbf{Institut f\"{u}r Hochenergiephysik der OeAW,  Wien,  Austria}\\*[0pt]
W.~Adam, T.~Bergauer, M.~Dragicevic, J.~Er\"{o}, C.~Fabjan, M.~Friedl, R.~Fr\"{u}hwirth, V.M.~Ghete, J.~Hammer\cmsAuthorMark{1}, N.~H\"{o}rmann, J.~Hrubec, M.~Jeitler, W.~Kiesenhofer, M.~Krammer, D.~Liko, I.~Mikulec, M.~Pernicka$^{\textrm{\dag}}$, B.~Rahbaran, C.~Rohringer, H.~Rohringer, R.~Sch\"{o}fbeck, J.~Strauss, A.~Taurok, P.~Wagner, W.~Waltenberger, G.~Walzel, E.~Widl, C.-E.~Wulz
% ...
\vskip\cmsinstskip
\dag:~Deceased\\
1:~~Also at CERN, European Organization for Nuclear Research, Geneva, Switzerland\\
% ...

\end{sloppypar}
\end{document}